\documentclass[11pt,makeidx]{report}

\usepackage{float}
\usepackage{url}
\usepackage{amsmath}
\usepackage{amssymb}
\usepackage{makeidx}
\usepackage[dvips]{graphicx,epsfig,color}

\usepackage{mycaptions}

\usepackage{psfrag}
\usepackage{bm}
\usepackage{dcolumn}
\usepackage{subfigure}
\usepackage{indentfirst}       
\usepackage{afterpage}  
\usepackage{lscape}

\usepackage{slashbox}

\usepackage{wrapfig}

\usepackage[linktocpage]{hyperref}

\makeindex

\renewcommand{\bibname}{{\Huge References}}

\renewcommand{\arraystretch}{1.2}

\allowdisplaybreaks[2]

\renewcommand{\arraystretch}{1.2}

\newcommand{\gev}{\operatorname{GeV}}
\newcommand{\mev}{\operatorname{MeV}}

\newcommand{\fm}{\operatorname{fm}}

\newcommand{\mb}{\operatorname{mb}}
\newcommand{\fb}{\operatorname{fb}}

\newcommand{\jpsi}{J\mskip -2mu/\mskip -0.5mu\Psi}

\newcommand{\lsim}{\raisebox{-4pt}{%
    $\,\stackrel{\textstyle <}{\sim}\,$}}
\newcommand{\gsim}{\raisebox{-4pt}{%
    $\,\stackrel{\textstyle >}{\sim}\,$}}
    
\newcommand{\zh}{z}
\newcommand{\bpar}{\mathcal{B}_T}

\newcommand{\ms}{\mskip 1.5mu}
\newcommand{\open}{{<\kern -0.3em{\scriptscriptstyle )}}}

\newcommand{\pom}{\mathbb{P}}

\newcounter{comment}

{\refstepcounter{comment}%
\begin{quote}
\ttfamily\small$\blacksquare$ \textbf{\underline{Comment} $\sharp$\thecomment:}}%
{\end{quote}}

{
\begin{quote}
\ttfamily\small$\blacktriangleright$ \textbf{\underline{Reply} $\sharp$\thecomment:}}%
{\end{quote}}

\newcommand{\la}{\langle}
\newcommand{\ra}{\rangle}

\newcommand{\ben}{\begin{displaymath}}
\newcommand{\een}{\end{displaymath}}
\newcommand{\be}{\begin{equation}}
\newcommand{\ee}{\end{equation}}
\newcommand{\bea}{\begin{eqnarray}}
\newcommand{\eea}{\end{eqnarray}}

\newcommand{\eq}[1]{Eq.~(\ref{#1})}

\newcommand{\fig}[1]{Fig.~\ref{#1}}

\newcommand{\BMprp}{\perp}
\newcommand{\BMkei}{k}
\newcommand{\BMkeiT}{\mathbf{\BMkei}_\BMprp}
\newcommand{\BMelll}{l}
\newcommand{\BMelllT}{\mathbf{\BMelll}_\BMprp}
\newcommand{\BMSpin}{S}
\newcommand{\BMSpinT}{\mathbf{\BMSpin}_\BMprp}

\newcommand{\BMqfield}{q}
\newcommand{\BMFTx}{\mathcal{F}_x}
\newcommand{\BMFTT}{\mathcal{F}_\perp}

\long\def\comment#1{ }

\newcommand{\del}{\partial}

\def\empile#1\over#2{\mathrel{\mathop{\kern 0pt#1}\limits_{#2}}}
\newcommand{\slvarepsilon}{\raise.15ex\hbox{$/$}\kern-.53em\hbox{$\varepsilon$}}
\newcommand{\slL}{\raise.15ex\hbox{$/$}\kern-.53em\hbox{$L$}}
\newcommand{\slP}{\raise.15ex\hbox{$/$}\kern-.53em\hbox{$P$}}
\newcommand{\slp}{\raise.1ex\hbox{$/$}\kern-.63em\hbox{$p$}}
\newcommand{\slq}{\raise.1ex\hbox{$/$}\kern-.53em\hbox{$q$}}
\newcommand{\slv}{\raise.1ex\hbox{$/$}\kern-.63em\hbox{$v$}}
\newcommand{\slR}{\raise.15ex\hbox{$/$}\kern-.53em\hbox{$R$}}
\newcommand{\slQ}{\raise.15ex\hbox{$/$}\kern-.53em\hbox{$Q$}}
\newcommand{\slK}{\raise.15ex\hbox{$/$}\kern-.53em\hbox{$K$}}
\newcommand{\slk}{\raise.15ex\hbox{$/$}\kern-.53em\hbox{$k$}}
\newcommand{\slSigma}{\raise.15ex\hbox{$/$}\kern-.53em\hbox{$\Sigma$}}
\newcommand{\slcalP}{\raise.15ex\hbox{$/$}\kern-.63em\hbox{$\cal P$}}
\newcommand{\slA}{\raise.15ex\hbox{$/$}\kern-.73em\hbox{$A$}}
\newcommand{\slbfA}{\raise.15ex\hbox{$/$}\kern-.73em\hbox{${\imb A}$}}
\newcommand{\slpartial}{\raise.15ex\hbox{$/$}\kern-.53em\hbox{$\partial$}}
\newcommand{\sla}{\raise.15ex\hbox{$/$}\kern-.53em\hbox{$a$}}
\newcommand{\slb}{\raise.15ex\hbox{$/$}\kern-.53em\hbox{$b$}}
\newcommand{\slc}{\raise.15ex\hbox{$/$}\kern-.53em\hbox{$c$}}
\newcommand{\slC}{\raise.15ex\hbox{$/$}\kern-.63em\hbox{$C$}}

\def\l{{\boldsymbol l}}

\def\m{{\boldsymbol m}}
\def\x{{\boldsymbol x}}
\def\u{{\boldsymbol u}}
\def\y{{\boldsymbol y}}
\def\X{{\boldsymbol X}}
\def\r{{\boldsymbol r}}

\def\b{{\boldsymbol b}}

\def\bs{\boldsymbol}

\def\d{\mathrm{d}}
\newcommand{\Pomeron}{I\!\!P}

\newcommand{\un}[1]{\underline{#1}}
\newcommand{\as}{\alpha_s}

\def\peq#1{{(\ref{#1})}}

\newcommand{\amu}{\alpha_\mu}
\def\ud{\underline}

\newcommand{\oone}{
\begin{picture}(10,8)
\put(5,5){\circle{8}}
\put(2.9,2.5){{\scriptsize 1}}
\end{picture}
}

\newcommand{\am}{\alpha_\mu}

\def\Pom{{ I\!\!P}}
 \newcommand\beq{\begin{equation}}
 \newcommand\eeq{\end{equation}}
 \newcommand\beqn{\begin{eqnarray}}
 \newcommand\eeqn{\end{eqnarray}}
 
\def\be{\begin{equation}}
\def\ee{\end{equation}}
\def\bea{\begin{eqnarray}}
\def\eea{\end{eqnarray}}

\def\f{\frac}

\def\g{\gamma}
\newcommand{\xp}{x_{\mathbb P}}
\newcommand{\xpom}{x_{\mathbb P}}

\newcommand{\beqs}{\begin{eqnarray*}}
\newcommand{\eeqs}{\end{eqnarray*}}

\begin{document}


\pagenumbering{roman}

\pagestyle{empty}

\vspace{4.0\baselineskip}

{\sffamily
\begin{center}
\Large The EIC Science case: a report on the joint BNL/INT/JLab program
\\[1em]
\LARGE Gluons and the quark sea at high energies:
\\
distributions, polarization, tomography
\\[2em]
\large Institute for Nuclear Theory, University of Washington, USA
\\[0.2em]
September 13 to November 19, 2010
\\
\vspace{8.0\baselineskip}
\large Editors:
\\[0.8em]
\parbox{0.68\textwidth}{
D.~Boer, Universiteit Groningen, The Netherlands \\
M.~Diehl, Deutsches Elektronen-Synchroton DESY, Germany \\
R.~Milner, Massachusetts Institute of Technology, USA \\
R.~Venugopalan, Brookhaven National Laboratory, USA \\
W.~Vogelsang, Universit\"at T\"ubingen, Germany
}
\vfill
Brookhaven National Laboratory, Upton, NY \\
Institute for Nuclear Theory, Seattle, WA \\
Thomas Jefferson National Accelerator Facility, Newport News, VA 
\end{center}
}

\newpage


\noindent{}The EIC Science case: a report on the joint BNL/INT/JLab program \\
Gluons and the quark sea at high energies: \\
distributions, polarization, tomography \\
Seattle, Washington, September 13 to November 19, 2010

\vspace{2\baselineskip}

\noindent{}Program homepage: \\
\url{http://www.int.washington.edu/PROGRAMS/10-3}

\vspace{\baselineskip}

\noindent{}Online proceedings at: \\
\url{http://arxiv.org/abs/1108.1713}

\vspace{2\baselineskip}

\noindent{}DISCLAIMER: \\
This report was prepared as an account of work sponsored by an agency of
the United States Government.  Neither the United States Government nor
any agency thereof, nor any of their employees, nor any of their
contractors, subcontractors, or their employees, makes any warranty,
express or implied, or assumes any legal liability or responsibility for
the accuracy, completeness, or any third party's use or the results of
such use of any information, apparatus, product, or process disclosed, or
represents that its use would not infringe privately owned rights.
Reference herein to any specific commercial product, process, or service
by trade name, trademark, manufacturer, or otherwise, does not necessarily
constitute or imply its endorsement, recommendation, or favoring by the
United States Government or any agency thereof or its contractors or
subcontractors.  The views and opinions of authors expressed herein do not
necessarily state or reflect those of the United States Government or any
agency thereof.

\vspace{2\baselineskip}


\vspace{2\baselineskip}

\vfill

\noindent{}Published by: \\
Brookhaven National Laboratory, USA \\
Institute of Nuclear Theory, University of Washington, USA \\
Thomas Jefferson National Accelerator Facility, USA \\
August 2011

\vspace{1\baselineskip}

\noindent{}BNL-96164-2011\\
INT-PUB-11-034 \\
JLAB-THY-11-1373

\vspace{2\baselineskip}
\noindent{}Printed at Brookhaven National Laboratory

\newpage


\pagestyle{plain}

\section*{Foreword}

The study of the fundamental structure of nuclear matter is a central thrust of physics research in the United States.  As indicated in Frontiers of Nuclear Science, the 2007 Nuclear Science Advisory Committee long range plan, consideration of a future Electron-Ion Collider (EIC) is a priority and will likely be a significant focus of discussion at the next long range plan.  We are therefore pleased to have supported the ten week program in fall 2010 at the Institute of Nuclear Theory which examined at length the science case for the EIC.  This program was a major effort; it attracted the maximum allowable attendance over ten weeks.\\

\noindent This report  summarizes the current understanding of the physics and articulates important open questions that can be addressed by an EIC.  It converges towards a set of ``golden'' experiments that illustrate both the science reach and the technical demands on such a facility, and thereby establishes a firm ground from which to launch the next phase in preparation for the upcoming long range plan discussions.   We thank all the participants in this productive program.  In particular, we would like to acknowledge the leadership and dedication of the five co-organizers of the program who are also the co-editors of this report.
\\

\noindent David Kaplan, Director, National Institute for Nuclear Theory\\
Hugh Montgomery, Director, Thomas Jefferson National Accelerator Facility\\
Steven Vigdor, Associate Lab Director, Brookhaven National Laboratory

\section*{Preface}

This volume is based on a ten-week program on ``Gluons and the quark sea
at high energies'', which took place at the Institute for Nuclear Theory
(INT) in Seattle from September 13 to November 19, 2010.  The principal
aim of the program was to develop and sharpen the science case for an
Electron-Ion Collider (EIC), a facility that will be able to collide
electrons and positrons with polarized protons and with light to heavy
nuclei at high energies, offering unprecedented possibilities for in-depth
studies of quantum chromodynamics.  Guiding questions were
\begin{itemize}
\item What are the crucial science issues?
\item How do they fit within the overall goals for nuclear physics?
\item Why can't they be addressed adequately at existing facilities?
\item Will they still be interesting in the 2020's, when a suitable
  facility might be realized?
\end{itemize}
The program started with a five-day workshop on ``Perturbative and
Non-Perturbative Aspects of QCD at Collider Energies'', which was followed
by eight weeks of regular program and a concluding four-day workshop on
``The Science Case for an EIC''.

More than 120 theorists and experimentalists took part in the program over
ten weeks.  It was only possible to smoothly accommodate such a large
number of participants because of the extraordinary efforts of the INT
staff, to whom we extend our warm thanks and appreciation.  We thank the
INT Director, David Kaplan, for his strong support of the program and for
covering a significant portion of the costs for printing this volume.  We
gratefully acknowledge additional financial support provided by BNL and
JLab.

The program was structured along several subtopics, which roughly
correspond to the chapters in this report.  For each topic, convenors were
appointed, who played an important role in the scientific organization of
the program weeks and in editing the corresponding chapters.  We
gratefully thank them for their work.  Special thanks are due to Matt
Lamont and Marco Stratmann, who took on the lion's share in the
painstaking task of merging the different chapters and making final edits.

Last but not least, we thank all participants of the INT program and all
authors of this report for the work and enthusiasm they put into their
contributions.  Thanks to their efforts, much progress has been achieved,
and we hope that the community will keep this momentum going in the
continuing effort to build a compelling case for an Electron-Ion Collider.

\vspace{\baselineskip}

\begin{flushleft}
  August 2011 \\
  The program organizers \\[\baselineskip]
  Dani\"el Boer \\
  Markus Diehl \\
  Richard Milner \\
  Raju Venugopalan \\
  Werner Vogelsang
\end{flushleft}


\section*{Convenors and chapter editors}

\setlength{\parindent}{0pt}

A. Accardi \\
\textsl{Hampton University, Hampton, VA 23668, USA} \\
\textsl{Thomas Jefferson National Accelerator Facility, Newport News, VA 23606, USA} \\

E. C. Aschenauer \\
\textsl{Brookhaven National Laboratory, Upton, NY 11973, USA} \\

M. Burkardt \\
\textsl{New Mexico State University, Las Cruces, NM 88003, USA} \\

R. Ent \\
\textsl{Thomas Jefferson National Accelerator Facility, Newport News, VA 23606, USA} \\

V. Guzey \\
\textsl{Thomas Jefferson National Accelerator Facility, Newport News, VA 23606, USA} \\

D. Hasch \\
\textsl{INFN, Laboratori Nazionali di Frascati, 00044 Frascati, Italy} \\

K. Kumar \\
\textsl{University of Massachusetts, Amherst, MA 01003, USA} \\

M. A. C. Lamont \\
\textsl{Brookhaven National Laboratory, Upton, NY 11973, USA} \\

Y. Li \\
\textsl{Brookhaven National Laboratory, Upton, NY 11973, USA} \\

W. Marciano \\
\textsl{Brookhaven National Laboratory, Upton, NY 11973, USA} \\

C. Marquet \\
\textsl{Physics Department, Theory Unit, CERN, CH-1211 Gen\`eve 23, Switzerland} \\ 

F. Sabati\'e \\
\textsl{CEA Saclay, Service de Physique Nucl\'eaire, 91191 Gif-sur-Yvette, France} \\

M. Stratmann \\
\textsl{Brookhaven National Laboratory, Upton, NY 11973, USA} \\

F. Yuan \\
\textsl{Lawrence Berkeley National Laboratory, Berkeley, CA 94720, USA}

\setlength{\parindent}{17pt}

\clearpage



\section*{List of Authors}


\setlength{\parindent}{0pt}


R. Sassot, P. Zurita \\
\textsl{Instituto de F\'{\i}sica de Buenos Aires and 
Departamento de F\'{\i}sica, Facultad de Ciencias Exactas y Naturales,
Universidad de Buenos Aires, Buenos Aires, Argentina} \\

I. O. Cherednikov \\
\textsl{Universiteit Antwerpen, 2020 Antwerpen, Belgium} \\

V. P. Gon\c{c}alves \\
\textsl{Instituto de F\'{\i}sica e Matem\'atica, Universidade Federal de Pelotas, Pelotas, RS, Brazil} \\

R. Sandapen \\
\textsl{D\'epartement de Physique et d'Astronomie, Universit\'e de Moncton, Canada} \\

B. Z. Kopeliovich \\
\textsl{Departamento de F\'{\i}sica, Universidad T\'ecnica Federico Santa Mar\'{\i}a and
Instituto de Estudios Avanzados en Ciencias e Ingenier\'{\i}a and
Centro Cient\'ifico-Tecnol\'ogico de Valpara\'iso, Valpara\'iso, Chile} \\

J.-H. Gao \\
\textsl{Department of Modern Physics, University of Science and Technology of China, Hefei, Anhui 230026, China} \\

Z.-T. Liang \\
\textsl{Department of Physics, Shandong University, Jinan, Shandong 250100, China} \\

K.~Passek-Kumeri\v{c}ki \\
\textsl{Theoretical Physics Division, Rudjer Bo{\v s}kovi{\'c} Institute,  Zagreb, Croatia} \\

K.~Kumeri\v{c}ki \\
\textsl{Department of Physics, University of Zagreb, Zagreb, Croatia} \\

T. Lappi \\
\textsl{Department of Physics, 40014 University of Jyv\"askyl\"a, Jyv\"askyl\"a and
Helsinki Institute of Physics, 00014 University of Helsinki, Helsinki, Finland} \\

S. Wallon \\
\textsl{LPT, Universit\'e d'Orsay, CNRS, 91404 Orsay and
 UPMC, Univ. Paris 06, Facult\'e de Physique, 75252 Paris, France} \\

B. Pire \\
\textsl{CPHT, \'Ecole Polytechnique, CNRS, 91128 Palaiseau, France} \\

R.~G\'eraud, H.~Moutarde, F.~Sabati\'e \\
\textsl{CEA, Centre de Saclay, Irfu/Service de Physique Nucl\'eaire, 91191 Gif-sur-Yvette, France} \\

\newpage
F. Gelis, G. Soyez \\
\textsl{Institut de Physique Th\'eorique, CEA/Saclay, 91191 Gif-sur-Yvette Cedex, France} \\

M.~Me\v{s}kauskas, D.~M\"uller, N. G. Stefanis \\
\textsl{Institut f\"{u}r Theoretische Physik II, Ruhr-Universit\"{a}t Bochum, 44780 Bochum, Germany} \\

K. Gallmeister, U. Mosel \\
\textsl{Institute for Theoretical Physics, Giessen University, 35392 Giessen, Germany} \\

M. Diehl \\
\textsl{Deutsches Elektronen-Synchroton DESY, 22603 Hamburg, Germany} \\

J. Bartels \\
\textsl{Institute of Theoretical Physics, Universit\"at Hamburg, 22761 Hamburg, Germany} \\

H. J. Pirner \\
\textsl{Institute of Theoretical Physics, Universit\"at Heidelberg, 69120 Heidelberg, Germany} \\

P. H\"agler\\
\textsl{Institut f\"{u}r Kernphysik, Johannes-Gutenberg-Universit\"{a}t, 55099 Mainz, Germany}\\

B. J\"{a}ger, H. Spiesberger\\
\textsl{Institut f\"{u}r Physik (THEP), Johannes-Gutenberg-Universit\"{a}t, 55099 Mainz, Germany}\\

T.~Lautenschlager, A. Sch{\"a}fer \\
\textsl{Institut f\"ur Theoretische Physik, Universit\"at Regensburg, 93040 Regensburg, Germany} \\

F. Ringer, W. Vogelsang \\
\textsl{Institut f\"{u}r Theoretische Physik, Universit\"{a}t T\"{u}bingen, Auf der Morgenstelle 14, 72076 T\"{u}bingen, Germany} \\

P. Kroll \\
\textsl{Fachbereich Physik, Universit\"at Wuppertal, 42097 Wuppertal, Germany} \\

S. Alekhin, J Bl\"{u}mlein, S.-O. Moch \\
\textsl{Deutsches Elektronen-Synchrotron DESY, Platanenallee 6, 15738 Zeuthen, Germany}\\

C. Pisano \\
\textsl{Dipartimento di Fisica, Universit\`a di Cagliari, and INFN, Sezione di Cagliari, Italy} \\

D. Hasch \\
\textsl{INFN Laboratori Nazionali di Frascati, 00044 Frascati, Italy}\\

J. Rojo \\
\textsl{Dipartimento di Fisica, Universit\`a di Milano and 
Instituto Nazionale di Fisica Nucleare, Sezione di Milano, Milano, Italy} \\

A. Bacchetta, B. Pasquini, M. Radici \\
\textsl{Instituto Nazionale di Fisica Nucleare, Sezione di Pavia and
Dipartimento di Fisica, Universit\`a di Pavia, Pavia, Italy} \\

C. Ciofi~degli~Atti, C. B. Mezzetti, L. P. Kaptari \\
\textsl{Instituto Nazionale di Fisica Nucleare, Sezione di Perugia and
Department of Physics, University of Perugia, Perugia, 06123, Italy} \\

M. Anselmino \\
\textsl{Universit\`a di Torino and INFN, Sezione di Torino, 10125 Torino, Italy} \\

K. Tanaka \\
\textsl{Department of Physics, Juntendo University, Inzai, Chiba, Japan} \\

Y. Koike \\
\textsl{Department of Physics, Niigata University, Niigata, Japan} \\

S. Kumano \\
\textsl{KEK Theory Center, Institute of Particle and Nuclear Studies and 
Department of Particle and Nuclear Studies, Graduate University for Advanced Studies, Tsukuba, Japan} \\

L. Motyka \\
\textsl{Institute of Physics, Jagiellonian University, Cracow, Poland} \\

K. Golec-Biernat, A. M. Sta\'sto \\
\textsl{Institute of Nuclear Physics, Polish Academy of Sciences, Cracow, Poland} \\

K. Golec-Biernat \\
\textsl{Institute of Physics, University of Rzesz\'ow, Rzesz\'ow, Poland} \\

L. Szymanowski \\
\textsl{Soltan Institute for Nuclear Studies, Warsaw, Poland} \\

I. O. Cherednikov, L. P. Kaptari, A. Radyushkin \\
\textsl{Bogoliubov Laboratory of Theoretical Physics, JINR, 141980 Dubna, Russia} \\

S. Alekhin \\
\textsl{Institute for High Energy Physics, 142281 Protvino, Moscow Region, Russia} \\

A. Kondratenko \\
\textsl{NTL Zaryad, Novosibirsk, Russia} \\

W. A. Horowitz \\
\textsl{Department of Physics, University of Cape Town, Rondebosch 7701, South Africa}\\

\newpage
G. Schnell \\
\textsl{Department of Theoretical Physics, University of the Basque
  Country UPV/EHU, 48080 Bilbao and 
IKERBASQUE, Basque Foundation for Science, 48011 Bilbao, Spain} \\

C. Marquet \\
\textsl{Physics Department, Theory Unit, CERN, 1211 Gen\`eve 23, Switzerland} \\

P. Chevtsov \\
\textsl{Paul Scherrer Institut, 5232 Villigen PSI, Switzerland} \\

P. J. Mulders, T. C. Rogers \\
\textsl{Department of Physics and Astronomy, Vrije Universiteit, Amsterdam, The Netherlands} \\

D. Boer \\
\textsl{Theory Group, KVI, University of Groningen, The Netherlands} \\

J. R. Forshaw \\
\textsl{Particle Physics Group, University of Manchester, Manchester, UK} \\

A. Cooper-Sarkar \\
\textsl{University of Oxford, Denys Wilkinson Bldg., Keble Road, Oxford OX1 3RH, UK} \\


G. A. Chirilli, D.~M\"uller, X.-N. Wang, F. Yuan \\
\textsl{Nuclear Science Division, Lawrence Berkeley National Laboratory, Berkeley, CA 94720, USA} \\

X. Qian \\
\textsl{California Institute of Technology, Pasadena, CA 91125, USA} \\

S. J. Brodsky \\
\textsl{SLAC National Accelerator Laboratory, Stanford University, Stanford, CA 94309, USA} \\

P. Schweitzer \\
\textsl{Department of Physics, University of Connecticut, Storrs, CT 06269, USA} \\

T. Horn\\
\textsl{Catholic University of America, Washington, DC 20064, USA}\\

K. Tuchin \\
\textsl{Department of Physics and Astronomy, Iowa State University, Ames,
  IA 50011, USA} \\

R. Dupr\'e, B. Erdelyi, S. Manikonda, P. N. Ostrumov\\
\textsl{Physics Division, Argonne National Laboratory, Argonne, IL 60439, USA} \\

S. Abeyratne, B. Erdelyi\\
\textsl{Northern Illinois University, De Kalb, IL 60115, USA} \\

A. Vossen \\
\textsl{Department of Physics, Indiana University, Bloomington, IN 47408, USA} \\

K. Kumar, S. Riordan \\
\textsl{Department of Physics, University of Massachusetts, Amherst, MA 01002, USA} \\

E. Tsentalovich\\
\textsl{MIT Bates Linear Accelerator Center, Middleton, MA 01949, USA}  \\

G. R. Goldstein \\
\textsl{Department of Physics and Astronomy, Tufts University, Medford, MA 02155, USA} \\

E. Pozdeyev \\
\textsl{FRIB, Michigan State University, East Lansing, MI 48824, USA}  \\

M. Huang \\
\textsl{Department of Physics, Duke University, Durham, NC 27708, USA}\\

M. Burkardt \\
\textsl{New Mexico State University, Las Cruces, NM 88003, USA}\\

C. Aidala \\
\textsl{Physics Division, Los Alamos National Laboratory, Los Alamos, NM 87545, USA} \\

A. Dumitru \\
\textsl{Department of Natural Sciences, Baruch College, New York, NY 10010, USA} \\

F. Dominguez \\
\textsl{Department of Physics, Columbia University, New York, NY 10027, USA} \\

I. Ben-Zvi, A. Deshpande, C. Faroughy, L. Hammons, Y. Hao, E. C. Johnson,
V. N. Litvinenko, S. Taneja, N. Tsoupas, S. Webb\\
\textsl{Department of Physics $\&$ Astronomy, Stony Brook University, Stony Brook, NY 11794-3400, USA} \\

J. Beebe-Wang, S. Belomestnykh, I. Ben-Zvi, M. M. Blaskiewicz,
R. Calaga, X. Chang, A. Fedotov, D. Gassner, H. Hahn, L. Hammons, 
Y. Hao, P. He, W. Jackson, A. Jain, E. C. Johnson, D. Kayran, 
J. Kewisch, V. N. Litvinenko, Y. Luo, G. Mahler, G. McIntyre, W. Meng, M. Minty, 
B. Parker, A. Pikin, V. Ptitsyn, T. Rao, T. Roser, B. Sheehy, J. Skaritka, 
S. Tepikian, Y. Than, D. Trbojevic, N. Tsoupas, J. Tuozzolo,
G. Wang, S. Webb, Q. Wu, W. Xu, A. Zelenski\\
\textsl{Collider-Accelerator Department, Brookhaven National Laboratory, Upton, NY 11973, USA}\\

E. C. Aschenauer, G. Beuf, T. Burton, R. Debbe, S. Fazio, M. A. C. Lamont, Y. Li, W. J. Marciano, J.-W. Qiu, M. Stratmann, T. Toll, T. Ullrich\\ 
\textsl{Physics Department, Brookhaven National Laboratory, Upton, NY 11973, USA}\\

\newpage
A. Deshpande, A. Dumitru, Z.-B. Kang, A. M. Sta\'sto, F. Yuan \\
\textsl{RIKEN BNL Research Center, Brookhaven National Laboratory, Upton, NY 11973, USA} \\

Y. V. Kovchegov, A Majumder \\
\textsl{Department of Physics, The Ohio State University, Columbus, OH 43210, USA} \\

A. Metz, J. Zhou \\
\textsl{Department of Physics, Temple University, Philadelphia, PA 19122, USA} \\

L. Gamberg \\
\textsl{Penn State University-Berks, Reading, PA 19610, USA} \\

A. M. Sta\'sto, M. Strikman, B.-W. Xiao \\
\textsl{Physics Department, Pennsylvania State University, State College, PA 16802, USA} \\

M. Guzzi, P. Nadolsky, F. Olness\\
\textsl{Southern Methodist University, Dallas, TX 75275, USA} \\

H. BC \\
\textsl{Department of Physics, University of Texas El Paso, El Paso, TX 79968, USA} \\

S. Liuti \\
\textsl{Department of Physics, University of Virginia, Charlottesville, VA 22904, USA} \\

A. Accardi \\
\textsl{Hampton University, Hampton, VA 23668, USA} \\

S. Ahmed, A. Bogacz, Ya. Derbenev, A. Hutton, G. Krafft, R. Li, F. Marhauser, 
V. Morozov, F. Pilat, R. Rimmer, T. Satogata, M. Sullivan, M. Spata, B. Terzi{\'c}, H. Wang, B. Yunn, Y. Zhang\\
\textsl{Accelerator Division, Thomas Jefferson National Accelerator Facility, Newport News, VA 23606, USA}\\

A. Accardi, H. Avakian, R. Ent, V. Guzey, B. Musch, P. Nadel-Turonski, A. Prokudin, A. Radyushkin,  C. Weiss\\
\textsl{Thomas Jefferson National Accelerator Facility, Newport News, VA 23606, USA}\\

G. Krafft, A. Radyushkin, H. Sayed \\
\textsl{Department of Physics, Old Dominion University, Norfolk, VA 23529, USA} \\

G. P. Gilfoyle \\
\textsl{Physics Department, University of Richmond, Richmond, VA 23173, USA} \\

I. C. Clo\"et, G. Miller \\
\textsl{Department of Physics, University of Washington, Seattle, WA 98195-1560, USA} \\

M. Gonderinger \\
\textsl{Department of Physics, University of Wisconsin-Madison, Madison, WI 53706, USA} \\
\vspace*{-1cm}
\mbox{}
\setlength{\parindent}{17pt}


\setcounter{tocdepth}{1}
\vspace{-10cm}
\tableofcontents

\clearpage



\setcounter{page}{1}
\pagenumbering{arabic}

\chapter*{Executive summary}
\addcontentsline{toc}{chapter}{\numberline{}Executive summary}

\hspace{\parindent}\parbox{0.92\textwidth}{\slshape
  Dani\"el Boer, Markus Diehl, Richard Milner,
  Raju Venugopalan, Werner Vogelsang}

\index{Diehl, Markus}
\index{Boer, Dani\"el}
\index{Milner, Richard}
\index{Venugopalan, Raju}
\index{Vogelsang, Werner}

\section*{Introduction}

Understanding the fundamental structure of matter in the physical universe
is one of the central goals of scientific research.  Strongly bound atomic
nuclei predominantly constitute the matter from which humans and the observable
physical world around us are formed.  In the closing decades of the twentieth
century, physicists developed a beautiful theory, Quantum Chromodynamics (QCD),
which explains all strongly interacting matter in terms of point-like quarks
interacting by the exchange of gauge bosons, known as gluons.  Experiments have
verified QCD quantitatively in  processes involving a very large momentum
exchange between the sub-atomic participants. Further confidence is obtained from significant progress in numerical computations of the 
static properties of the theory, in particular the excellent agreement of theory with the mass spectrum of low lying hadron resonances. 

However, more than thirty years after QCD was first proposed as the fundamental theory of the strong force, and despite
impressive theoretical and experimental progress made in the intervening decades, the understanding of how QCD works
in detail remains an outstanding problem in physics. Very little is known about the dynamical basis of hadron structure in terms of 
the fundamental quark and gluon fields of the theory. How do these fundamental degrees of freedom dynamically generate the mass, spin, motion, and spatial distribution of color charges inside hadrons with varying momentum resolution and energy scales? Deep Inelastic Scattering (DIS) experiments at the HERA collider revealed clearly that at high momentum resolution and energy scales, the proton is a complex, many-body system of gluons and sea quarks, a picture very different from a more familiar view of the proton as a few point-like partons (a term that collectively refers to both quarks and gluons), each carrying a large fraction of its momentum. This picture, which is confirmed at hadron colliders, raises more questions than it answers about the dynamical structure of matter. For instance, how is the spin-1/2 of the proton distributed in this many-body system of sea quarks and gluons? In the early universe, how did the many-body plasma of quarks and gluons cool into hadrons with several simple structural properties? Recreating key features of this quark-hadron transition in heavy ion collisions has been a major activity in nuclear physics, with several surprising findings including the realization that this matter flows with very little resistance as a nearly perfect fluid. A deep understanding of the two cited examples, among many others, ultimately requires detailed knowledge of the quark-gluon structure of hadrons and nuclei.


This report on the science case for an Electron-Ion Collider (EIC) is the result of a ten-week program at the 
Institute for Nuclear Theory (INT) in Seattle (from September 13-November 19, 2010), motivated by the need to develop a strong case for the continued study of 
the QCD description of hadron structure in the coming decades. Hadron structure in the valence quark region will be studied extensively with 
the Jefferson Lab 12 GeV science program, the subject of an INT program the previous year. The focus of the INT program was on understanding the role of gluons and sea quarks, the important dynamical degrees of freedom describing  hadron structure at high energies. 
Experimentally, the  most direct and precise way to access the dynamical structure of hadrons and nuclei at high energies is with a high luminosity lepton
probe in collider mode. An EIC with 
optimized detectors offers enormous potential as the next
generation accelerator to address many of the most important, open questions about the fundamental structure of matter. 
The goal of the INT program, as captured in the writeups in this report, was to 
articulate these questions and to identify golden experiments that have the greatest potential to provide definitive answers to these questions.

At resolution scales where quarks and gluons become manifest as degrees of freedom, the structure of the nucleon and of nuclei is intimately connected with unique features of 
QCD dynamics, such as confinement and the self-coupling of gluons. Information on hadron sub-structure in DIS is obtained in the form of ``snapshots" by the ``lepton microscope" of the dynamical many-body hadron system, over different momentum resolutions and energy scales.   These femtoscopic snapshots, at the simplest level, provide distribution functions which are extracted over the largest accessible kinematic range to
assemble fundamental dynamical insight into hadron and nuclear sub-structure.  For the proton, the EIC would be the brightest femtoscope scale lepton-collider ever, 
exceeding the intensity of the HERA collider a thousand fold. HERA, with its center-of-mass (CM) energy of 320 GeV, was built to search for quark substructure.  An EIC, with its scientific focus on studying QCD in the regime where the sea quarks and gluons dominate, would have a lower CM energy. In a staged EIC design, the CM energy will range from $50$-$70$ GeV in stage I to approximately twice that for the full design. In addition to being the first lepton collider exploring the structure of polarized protons, an EIC will also be the first electron-nucleus collider, probing the gluon and sea quark structure of nuclei for the first time.


Following the same structure as the scientific discussions at the INT, this report is organized around the following 
four major themes:
\begin{itemize}
\item 	The spin and flavor structure of the proton
\item 	Three dimensional structure of nucleons and nuclei in momentum and configuration space
\item 	QCD matter in nuclei 
\item 	Electroweak physics and the search for physics beyond the Standard Model\end{itemize}
In this executive summary, we will briefly outline the outstanding physics
questions in these areas and the suite of measurements that are available
with an EIC to address these. The status of accelerator and detector
designs is addressed  at the end of the summary. Tables of golden
measurements for each of the key science areas outlined are presented on 
page~\pageref{sec:golden-tables}.
In addition, each chapter in the report contains a comprehensive overview of the science topic addressed. Interested readers are encouraged to read these and the individual contributions for more details on the present status of EIC science.

\section*{The spin and flavor structure of the proton}
\label{sec:example}

To understand how the constituents of the proton carry the proton's 
spin has been a defining question in hadron structure for several decades 
now. The proton spin problem presents the formidable 
challenge of understanding an essential feature of how a complex strongly-interacting many-body system organizes itself to 
produce a simple result. It goes directly to the heart of exploring and understanding the QCD dynamics of matter. From the surprising finding by the 
European Muon Collaboration that very little of the proton spin is 
provided by the spins of quarks and anti-quarks combined, the 
exploration of nucleon spin structure has by now developed into 
a world-wide quest central to nuclear and particle physics. 
To provide definitive answers in this area will be among the key 
tasks of an EIC.

Significant progress can be expected from the unique capability of an EIC to 
reach small momentum fractions $x$ and large momentum resolution scales $Q$, 
with high precision. A suite of measurements will be available. A golden  measurement of nucleon spin structure 
at an EIC will be the precision study of the proton's spin structure 
function $g_1^p(x,Q^2)$ and its scaling violations, over wide ranges 
in $x$ and $Q^2$. As studies in this report will demonstrate,  global analyses of spin-dependent parton
distributions will determine the gluon helicity distribution $\Delta g$ 
and the quark singlet $\Delta \Sigma$ down to values of
$x$ of about $10^{-4}$. This vastly extended reach should allow for the determination of the gluon and 
quark/anti-quark spin contributions to the proton spin to about 
$10\%$ accuracy or better. 
The accuracy to which processes such as deeply-virtual Compton scattering can independently provide information on the remaining orbital 
angular momentum contributions will be addressed further in the section on spatial imaging. 

An EIC will provide unprecedented insight into the flavor structure
of the nucleon, a key element in mapping the ``landscape'' of hadron structure. There
are two powerful golden measurements available at an EIC to achieve this. 
One of these methods, {\it Semi-Inclusive Deep-Inelastic Scattering} (SIDIS) has been 
used in previous fixed-target lepton scattering experiments HERMES 
and COMPASS. (Polarized proton-proton collisions at RHIC employ $W$-boson production for flavor identification.) 
 At an EIC, semi-inclusive measurements would extend
to much higher $Q^2$ than in fixed-target scattering, where the 
reaction becomes significantly cleaner, less contaminated with higher-twist 
effects (a technical term for contributions power suppressed in $1/Q^2$), and therefore more tractable theoretically.
The kinematic coverage for SIDIS in $x$ and $Q$ will be similar overall to what
can be achieved in inclusive DIS. With the high luminosity of an EIC, extractions of 
the light-flavor helicity distributions $\Delta u$, $\Delta d$ and 
their anti-quark distributions from SIDIS will be possible with 
exquisite precision. With dedicated studies of kaon production, 
the strange and anti-strange distributions will also be accessible.
All this will likely give insights into the 
question why it is that the combined quark and anti-quark spin 
contribution to the proton spin turns out to be so small. 

The other independent method for accessing the quark and antiquark helicity distributions at an EIC is electroweak DIS. At high $Q^2$, the 
DIS process also proceeds significantly via the exchange of $Z$ and $W^\pm$ 
bosons. This gives rise to novel structure functions that are sensitive
to various different combinations of the proton's helicity 
distributions. Studies show that both neutral current and charged
current interactions would be observable at an EIC.
To fully exploit the 
potential of an EIC for such measurements, positron beams are 
required, albeit not necessarily polarized. Besides the new 
insights into nucleon structure this would provide, studies of 
spin-dependent electroweak scattering at short distances with an EIC 
would be interesting physics in and of itself, much in the line of past and 
ongoing electroweak measurements at HERA, Jefferson Lab, and RHIC. 

Polarized electron-proton physics can be expected to take center stage at an EIC because these would be the first such collider measurements. However, as studies in this 
report show, there is a large potential for unpolarized physics at an EIC. Thanks to its high luminosity and the feasibility for an energy scan, an EIC would vastly improve upon HERA data on measurements of the longitudinal structure function $F_L$.  This quantity is a key observable for studies of gluon structure and the possible 
transition to a high parton density or saturation regime in the proton. At an EIC, several SIDIS measurements of flavor distributions and multi-particle correlations will be possible for the first time. In particular, pinning down the strange quark and antiquark content of the proton would close one of the last notable gaps in our knowledge of unpolarized parton densities. Extended rapidity coverage will also allow for detailed studies of the rapidity gap structure of hard diffractive final states. In addition, the very high luminosities will bring a vast improvement in the precision of measurements of the charm and beauty contributions to nucleon structure. 

\section*{Three dimensional structure of hadrons and nuclei:  Transverse momentum distributions}
Partons can have a momentum 
component transverse to the direction of their parent nucleon and there exists experimental evidence to support an average transverse momentum of a few hundred MeV/c. However, much of our understanding of nucleon structure is in terms of 
integrated parton distributions that 
are only sensitive to the momentum resolution of the probe. 
A rigorous theoretical framework for parton transverse momentum distributions (TMDs) has been developed recently which allows for a 
description of specific scattering cross sections in terms of these distributions.
TMDs are an essential step toward a more comprehensive understanding of the parton structure of the nucleon in QCD. An EIC will enable precise and detailed  measurements of TMDs over a broad kinematic range. 

For the scattering processes of interest, the large scale $Q^2$ justifies, in a leading twist approximation, 
 the factorized description of the cross section in terms of several calculable or measurable factors, yielding a predictive framework. TMDs are examples of such measurable factors. In such descriptions 
not only does the magnitude of the  parton transverse momentum enter, but also the transverse momentum direction, yielding strikingly asymmetric distributions. 
Several recently observed angular asymmetries 
 are most naturally described by asymmetric, spin direction dependent TMDs. 

A golden measurement at an EIC will be 
the Sivers 
asymmetry, a particular angular correlation between the target polarization and the direction of a produced final state hadron in polarized SIDIS. At the parton level, the Sivers effect is a spin-orbit coupling effect in QCD  and is described by a TMD that 
quantifies how strongly the transverse momentum from orbital motion is coupled to spin. 
The Sivers effect is especially interesting because it is a consequence of phase interference peculiar to the gauge structure of QCD. 
The gauge invariant Sivers TMD is non-zero only if gluonic initial or final state interactions are taken into account. There is a calculable process dependence, most strikingly evident  in 
SIDIS and Drell-Yan lepton pair production where the polarized Sivers function in the former is equal in magnitude but opposite in sign to the latter.  
Factorization breaking is also expected in  more complicated processes, such as hadron-hadron collisions with hadronic final states. This process dependence has not yet been demonstrated 
but several such experiments, in particular at RHIC, will study the Sivers and other TMD effects. 
The comparison of these results with complementary information from an EIC will allow a detailed understanding of the nature and extent of factorization breaking for TMDs. 

A goal at an EIC is to obtain a flavor-separated extraction of the 
Sivers TMD in an energy regime where its theoretical interpretation is unambiguous.
 Percent level azimuthal asymmetries measured by HERMES, COMPASS and at Jefferson Lab at rather modest $Q^2$ have enabled rough first estimates of the magnitude of the Sivers effect. 
With the 12 GeV upgrade program at Jefferson Lab, the valence (large $x$) region will be explored in detail, whereas sea quark  and gluon contributions at small $x$ (down to $10^{-4}$) will be 
mapped out with an EIC. The large $Q^2$ reach of an EIC will allow for extensive study of evolution effects in TMDs, and at large $x$  ($x\sim 0.2$) will have overlap with preceding experiments. 
High energies and high precision will enable a good understanding of the $x$ dependence of the Sivers functions for each quark flavor, including antiquarks and gluons. 
In addition, the larger transverse momentum range of final state particles at an EIC allows for studies of weighted asymmetries that are cleaner to interpret theoretically but are beyond the reach of fixed target experiments. 
The extensive transverse momentum range will for the first time in polarized SIDIS, allow studies of the transition region between the TMD description at low transverse momentum and the description in terms of collinear quark-gluon-quark correlation functions (known as the Qiu-Sterman mechanism) at high transverse momentum. Finally, with respect to previous SIDIS experiments and future Jefferson Lab experiments, a larger variety of final states can be considered at an EIC, such as (multiple) jets or D-mesons, all of great interest in isolating quark and gluon contributions to the various TMD effects. 

Now that angular asymmetries consistent with the TMD framework have been observed, 
the road towards full-fledged experimental studies of TMDs can be mapped out and the essential role of an 
EIC identified. Besides the Sivers effect, essential information on the unpolarized TMD $f_1$ is obtained from unpolarized scattering cross sections.
For reasons we shall outline, this extraction of $f_1$ can be classified as another golden measurement. 
This TMD determines the $Q^2$ dependence of the unpolarized cross section, which has been predicted but not yet verified. 
Predictions of the $x$, transverse momentum, scale and flavor dependence of $f_1$ allow for non-trivial checks of the fundamental TMD formalism 
corroborating and complementing what one learns from the Sivers and other spin TMD effects.
The unpolarized SIDIS measurements at an EIC will give detailed information on the difference between sea 
and valence quark contributions, and on the role of gluons. 
Extracting unpolarized gluon TMDs at small $x$ is especially interesting 
because of the recently discovered agreement between predictions in the TMD framework and previous computations of the same in the Color Glass Condensate formalism as we shall discuss later.  

The proposed silver experiments are 1) the distribution of transversely
polarized quarks inside transversely polarized hadrons, 2) spin-orbit
correlations inside unpolarized hadrons (the Boer-Mulders TMD), and 3) the
Collins TMD fragmentation function, which describes a similar spin effect
in the fragmentation of quarks into unpolarized hadrons. All three
quantities involve transverse quark spin, which distinguishes them from
the Sivers effect which deals with unpolarized partons inside a
transversely polarized proton.  An EIC will be able to provide
multi-dimensional representations of all these quantities and the
observables they give rise to. The TMD chapter illustrates by means of
concrete examples and calculations how much further TMD studies can be
pushed with an EIC compared to the present status. A prime example is
shown in figure~\ref{fig:prokudin_sivers_eic_low_energy} on
page~\pageref{fig:prokudin_sivers_eic_low_energy}. 

\section*{Three dimensional structure of nucleons and nuclei:  Spatial imaging}

The high luminosity and large kinematic reach of an EIC offers unique
possibilities for exploring the spatial distribution of sea quarks and
gluons in the nucleon and in nuclei.  The ``imaging'' of partons is
possible in suitable exclusive reactions.  The transverse position of the
quark or gluon on which the scattering took place is obtained by a Fourier
transform from the transverse momentum of the scattered nucleon or
nucleus.  At the same time, the longitudinal momentum loss of the target
is correlated with the longitudinal momentum fraction $x$ of the parton.
By choosing particular final states, measurements at an EIC will be able to
selectively probe the spatial distribution of sea quarks and gluons in a
wide range of $x$.  Such `tomographic images' will provide essential insight into QCD
dynamics inside hadrons, such as the interplay between sea quarks and 
gluons, the role of pion degrees of freedom at
large transverse distances and, from a more general perspective, the mechanism for confinement in QCD.

The quantities that encode this tomographic information are generalized
parton distributions (GPDs).
The formalism of
GPDs is applicable in the full range of $x$. An alternative description at small $x$ is the 
dipole formalism, which is expressed in terms of the amplitude for small color dipoles to scatter off gluons in the hadron target. GPDs allow direct
comparison of tomographic images for sea quarks and gluons with their
counterparts in the valence quark region, where the 12 GeV program at
Jefferson Lab will obtain information of unprecedented accuracy.

Potential golden measurements for parton imaging at an EIC are deeply virtual
Compton scattering and photo- or electro-production of $J/\psi$ mesons.
For Compton scattering, there are a large number of observables that can be
calculated with high precision, whereas a unique advantage of $J/\psi$
production is its sensitivity to gluons.  A suite of further reaction
channels play the role of ``silver measurements'', which will provide
complementary information and in particular help separate different quark
flavors.  Among those exclusive channels whose cross sections grow with
energy, deeply virtual Compton scattering demands the highest luminosity.
Simulations performed during the INT program indicate that precise and
multi-differential measurements of this process can be envisaged with the
projected EIC luminosity (see figures~\ref{dvcs}, \ref{dvcs_ht} and
\ref{bsa_dvcs} on pages ~\pageref{dvcs}, \pageref{dvcs_ht} and
\pageref{bsa_dvcs}).
Detailed studies including detector effects will be required to establish
the achievable experimental accuracy.

The envisaged configuration of an EIC interaction region and detector
will provide data in a wide enough range of transverse momentum transfer
to permit a Fourier analysis of observables.  With this, exclusive cross sections
and angular or polarization asymmetries will give direct quantitative
information about the spatial distribution of partons in a specified range
of $x$.  Estimates indicate that transverse distances ranging from about
$0.1$ fm  to $2$ fm or higher will be accessible, provided that a good
enough momentum resolution can be achieved experimentally.  Such data will
provide the basis for reconstructing generalized parton distributions and,
ultimately, the joint distribution of partons in transverse position $b$
and longitudinal momentum fraction $x$.  For this second step, an EIC's large
lever arm in photon virtuality $Q^2$ at a given photon energy will be
essential, since it is the scale evolution in $Q^2$ that carries the most
detailed information about the longitudinal parton momentum.

Our current knowledge about the helicity distributions of quarks and gluons indeed suggests that the orbital angular momentum of partons plays a prominent role in the
 nucleon.  Exclusive scattering on a transversely polarized target gives access to this degree of freedom in parton tomography and allows one to study spin-orbit correlations at the parton level.  An especially interesting aspect is the relation between a polarization induced asymmetry in transverse parton position and the Sivers asymmetry in transverse parton momentum.  Such a relation is profoundly dynamical, and its quantitative exploration in the sea quark and gluon domain will be a highlight of exploring hadron structure and dynamics at an EIC.  Deeply virtual Compton scattering will again play an essential role in this context, along with vector meson production channels.  Quantitative estimates of the achievable statistical and systematic accuracy were not made during the INT program, but the necessary tools are now in place and results should be available soon.

Ji's angular momentum sum rule condenses the connection between generalized parton distributions and parton angular momentum into a single 
number for each quark flavor and for the gluon.  To evaluate this sum rule from exclusive measurements is truly challenging for several reasons.  The most serious among them is that one needs to reconstruct the full $x$ dependence of GPDs from observed scaling violations in $Q^2$.  As already mentioned, the large kinematic coverage of an EIC provides a good starting point for such a program, but it remains to be seen which accuracy can be attained for the angular momentum.  We regard this as a long-term endeavor, which will profit from the progress one can expect in the coming years from the 12 GeV program at Jefferson Lab.

\section*{Physics opportunities in electron-nucleus collisions}

An EIC would be the world's first e+A collider. It will significantly extend parton studies of nuclear structure into the regime dominated by sea quarks and gluons. Prior fixed target DIS measurements on nuclei revealed that the ratio of nuclear to nucleon cross sections is 
significantly less than unity (normalized by the atomic mass number) both at large $x$ (the EMC effect) and at small $x$ (shadowing).  
These interesting nuclear phenomena were however only observed for valence and (to a lesser  extent) sea quarks. The nuclear gluon distribution is very poorly constrained at all $x$ values, especially at $x<0.01$ where it is completely unknown.
An EIC could reveal surprises in our fundamental understanding of the parton structure of nuclei in this {\it terra incognita}. 


A fundamental feature of QCD is gluon saturation, which arises as a consequence of the fact that gluon distributions at a fixed $Q^2$ cannot grow rapidly indefinitely with decreasing $x$. 
The properties of matter in this novel saturation  regime of strong color fields in QCD is described by a saturation scale which grows both with decreasing $x$ and with increasing nuclear size.  Model estimates of this nuclear ``oomph'' give a saturation scale in a large nucleus at EIC energies to be of the same magnitude as the saturation scale in a proton at a TeV scale electron-proton collider; electron-nucleus collisions therefore provide an efficient method to explore saturation in QCD.


As a consequence of asymptotic freedom, the large saturation scale (relative to the intrinsic QCD scale $\Lambda_{QCD}$) accessible at an electron-nucleus collider implies that the properties of saturated gluon matter at small $x$ can be computed systematically using weak coupling techniques and compared to experimental data. One such weak coupling approach is the Color Glass Condensate (CGC). Renormalization group (RG) methods in the CGC are used to compute observables in electron-nucleus collisions that are sensitive to the energy evolution of particular many-body gluon correlators. These correlators, classified as ``dipole", ``quadrupole" and ``multipole'' effective degrees of freedom from their color structure, are universal. Final states in proton-nucleus and nucleus-nucleus collisions can also be expressed in terms of these objects.
Properties of multipole degrees of freedom can be inferred from measurements of cross-sections for specific final states in one of these reactions and used as input in computations of cross-sections for other final states, thereby providing an important test of the validity  and limits of the CGC effective theory. 
 A further interesting possibility is that multipole correlators at very high energies become independent of the initial conditions specific to a particular nucleus that are inputs at a given $x$ scale to the RG evolution equations. While it appears unlikely that an EIC would have sufficient energy to access this asymptotic regime,  DIS off different nuclei can provide important constraints on pre-asymptotic trends in that direction.

At large $x$ in nuclei,  DIS corresponds to the virtual photon scattering off quarks, with the nucleus acting as an extended colored medium that interacts with the hard colored probe. Because the energy and momentum resolution of the probe can be accurately controlled in DIS, 
one can quantitatively address, with a precision unmatched at hadron colliders, 
interesting questions about the nature of multiple scattering and $p_\perp$ broadening, energy loss and fragmentation, and the propagation of heavy quarks and jets in colored media. Perturbatively calculable short distance physics can be isolated from the hadronization mechanism by tuning the energy and momentum resolution of the virtual photon probe to shed new light on the latter both in medium and in the vacuum. While some such studies have been performed previously at fixed target DIS facilities and in proton-nucleus collisions, the extended kinematic reach, collider geometry and precision probes will vastly add to their scope, allowing for definitive answers to enduring questions about in-medium properties of QCD. For instance, the propagation of heavy charm and beauty quarks in medium will be quantitatively studied in DIS for the first time. In addition to being interesting in their own right, DIS studies of parton propagation in ``cold" QCD media are an important benchmark for a quantitative understanding of their role in the hot QCD medium produced at RHIC and the LHC. 

An important opportunity to understand the role of gluons in the structure of short range nuclear forces is made possible by exclusive measurements with an EIC of open heavy flavor and quarkonium in DIS off light nuclei. Other interesting studies at large $x$ where the kinematic reach of an EIC will complement the Jefferson Lab 12 GeV program, including the EMC effect and generalized parton distributions for nuclei. 

A number of experimental observables have been identified that can shed light on the compelling physics issues outlined. One set of golden measurements include the inclusive structure functions $F_2$ and $F_L$ for light and heavy nuclei. 
They will provide the first ever unambiguous measurements of nuclear gluon distributions. Studies of the evolution of quark singlet and gluon distributions with $x$ and $Q^2$ for light and heavy nuclei can systematically uncover the breakdown of leading twist evolution, the onset and development of non-linear saturation dynamics and enable the extraction of the corresponding saturation scale. Another set of golden measurements are provided by semi-inclusive DIS  (SIDIS) off nuclei. Di-hadron correlations in particular, are very sensitive to non-linear QCD evolution, and allow for a clean extraction of the saturation scale. They will  corroborate (or invalidate) claims of saturation seen in di-hadron correlations in deuteron-gold collisions; more generally, they enable the previously discussed tests of universality of multipole correlators at small $x$.
Golden measurements at large $x$ are semi-inclusive production of light
and heavy flavors and jets. These provide unique insight into energy loss
and parton shower development in an extended colored medium, as well as
into the dynamics of hadronization in this many-body environment. The
heavy flavor and jet measurements will be the first 
of their kind   
in nuclear DIS; we note that feasibility studies for
them  
are still in a preliminary stage. 

In addition to these golden measurements, there are several important measurements classified as ``silver'' instead of gold only in a relative sense. The most important among these  are the diffractive structure functions $F_{2,D}$ and $F_{L,D}$ which will be extracted for nuclei for the very first time. At HERA, these structure functions for protons constituted more than $15$\% of the cross-section; the predictions of saturation models is that this fraction will be significantly larger in nuclei. Exclusive production of vector mesons and deeply virtual Compton scattering probe the spatial distribution of partons in nuclei; at small $x$, they can help clarify the interplay between saturation and the effects of chiral symmetry breaking and confinement. 


Finally, a frequently posed question is whether proton/deuteron-nucleus scattering can provide the same information content as electron-nucleus collisions. In the former, the  computation of final states, in leading twist kinematics, contains convolutions over parton distributions in the nucleon projectile as well as that in the target. In addition, for a number of final states, a large number of parton scattering reactions are likely to contribute. This significantly compromises the accuracy to which one determines the parton structure of the target. For fundamental questions regarding the spatial distribution of partons and color singlet structures exchanged in hard diffractive scattering, there are essential qualitative differences in hadron-hadron and lepton-hadron processes arising from the lack of universality in key aspects of  the dynamical structure of nucleons and nuclei. Thus while proton/deuteron-nucleus scattering at high energies has the strong potential to be a discovery machine for new QCD physics, uncovering the origins of such physics and its implications for our fundamental understanding of the parton structure of nuclei, will require an EIC.

\section*{Electroweak interactions and physics beyond the Standard Model}

While the physics of an EIC is primarily motivated by the study of strong
interactions, its physics case is strengthened by its potential to contribute to electroweak studies as well.
Experience has shown that a new accelerator that pushes the frontiers 
either in energy, and/or luminosity and intensity, is of  
interest for studies of electroweak physics. We have already mentioned
that precision studies of (parity-violating) electroweak spin
structure functions would be possible at an EIC, giving new
insights into nucleon spin structure. However, the electroweak 
physics case for an EIC is broader as it would also allow 
measurements of parameters of electroweak theory. Studies
presented in detail in the INT report suggest that for high energy and luminosity, 
there would be excellent prospects for extractions of the Weinberg angle, 
which should even be possible over a fairly wide range in $Q^2$ 
so that its running can be further studied in detail. In this way,
an EIC would complement the precise LEP and SLD measurements on the $Z$-pole, 
atomic parity-violation measurements,
the SLAC E158 M{\o}ller scattering data,  and
the NuTeV data whose final value is in fact around three standard 
deviations above the SM prediction. A comparison of EIC results 
for $\sin^2\theta_W$ with those on the $Z$-pole in particular can be used 
to search for new physics effects. Some of the experimental systematics involved at an EIC 
are broadly understood, but may still need further work to clarify.  A full ``global survey'' 
of electroweak parameters from EIC data -- much in the spirit of the 
approach also taken at HERA -- is still outstanding but planned. 
In addition, an EIC might possibly be able to open a direct window on 
beyond-Standard Model physics, assuming that conditions are favorable. 
Studies indicate that the EIC might be able to perform a sensitive 
search for a third generation leptoquark in electron-tau conversion 
$ep\to \tau X$, with potential reach well beyond that in previous 
studies at HERA.

\section*{EIC Accelerator Design}

Two substantial, focused efforts at developing a design for an electron-ion collider in the U.S. based on existing 
accelerators are underway at Brookhaven National Laboratory   and Thomas 
Jefferson National Accelerator Facility.  At BNL, the eRHIC design
utilizes a new linear electron accelerator to collide with the existing polarized proton and 
ion beams of the operating Relativistic Heavy Ion Collider (RHIC).  At JLab, the ELIC design employs new electron 
and ion storage rings together with the 12 GeV upgraded existing CEBAF.  Although based on two different, 
existing accelerators, because they are driven by the same science objectives, the two U.S. EIC design efforts have 
similar characteristics.  The most important include:   

\begin {itemize}

\item{} highly polarized ($>$ 70\%) electron and nucleon beams
\item{} ion beams from deuterium to the heaviest nuclei - uranium or lead
\item{} center of mass energies: from about 20 GeV to about 150 GeV
\item{} maximum collision luminosity $\sim 10^{34}$ cm$^{-2}$ s$^{-1}$
\item{} non-zero crossing angle of colliding beams without loss of luminosity (so-called {\it crab crossing})
\item{} cooling of the proton and ion beams to obtain high luminosity 
\item{} staged designs where the first stage would reach CM energies of about 70 GeV
\item{} the possibility to have multiple interaction regions 

\end{itemize}

It is clear from the EIC physics studies that with a luminosity of $\sim 10^{33}$ cm$^{-2}$ s$^{-1}$, and operating for about a decade, ground breaking new experiments to probe our understanding of QCD will become feasible. This would require delivery of order 50 fb$^{-1}$ with polarized nucleon and heavy ion beams to experiments in about a decade.  This would be 100 times more integrated luminosity
than recorded over a decade at the only previous electron-proton collider, HERA at DESY.
With a luminosity of $\sim 10^{34}$ cm$^{-2}$ s$^{-1}$, precision imaging and electroweak experiments become feasible at an EIC.

The EIC accelerator designs being considered will require significant R\&D for realization. The cooling of the hadron
beam is essential to attain the luminosities demanded by the science. The development of a new technique, {\it coherent electron cooling}, 
is underway at BNL while conventional electron cooling is being pushed to high RF power at JLab. Energy recovery linear accelerators
at high energy and intensity are a key technology for an EIC. Further, the eRHIC design demands an increase in the intensity produced by 
polarized electron sources of over an order of magnitude beyond what is available at present.  The ELIC design utilizes novel figure-8
storage rings for both electrons and ions. 

In Europe, two electron-ion collider accelerators are under consideration.  At the Large Hadron Collider at CERN, physicists are considering colliding an electron beam (either a linac or ring) with an energy of about 70 GeV with the existing unpolarized proton and heavy-ion beams.  The present LHeC design can reach a CM energy of about 1.4 TeV with a luminosity of $10^{33}$ cm$^{-2}$ s$^{-1}$.
At GSI in  Germany, an Electron-Nucleon Collider (ENC) would be realized by
colliding electrons in a 3 GeV storage ring with 15 GeV protons in the High Energy Storage Ring of the planned Facility for Antiproton and Ion Research (FAIR). The CM energy at an ENC is about 14 GeV and the expected luminosity is about $10^{32}$ cm$^{-2}$ s$^{-1}$.  Thus, the two European colliders differ in CM energy by about two orders of magnitude, in colliding luminosity by about one order of magnitude, and have very different scientific objectives.

\section*{EIC Detectors}

Optimized detectors are essential to carry out the ground breaking experiments planned at an EIC. 
The design of EIC detectors is intimately connected to the design of the accelerator interaction regions (IR) through the location of magnets, configuration of crossing angles, and available space. A particular challenge is to detect forward-going scattered protons from exclusive reactions, as well as decay neutrons from the break-up of ions in incoherent diffraction. Past experience at colliders with lepton beams has shown that synchrotron radiation generated by bending the electron beam close to the IR can produce challenging backgrounds for detectors.

Detector concepts for an EIC are being developed and are guided both by the demands of the scientific program and by the experience with ZEUS and H1 at HERA.  The EIC detector will certainly include a large central detector likely containing a solenoidal magnetic field (of order 4 T); trackers for momentum and angular resolution; electromagnetic and hadronic calorimetry; particle identification involving Cerenkov detectors, and vertex detectors.  Further, detectors in the forward and backward directions will be required to augment the large central detector. These are necessary to detect hadrons from low $x$ processes and will require particle identification, calorimetry (both electron and hadron) and possibly magnetic field. With multiple interaction regions, it may be more advantageous to consider different detectors ({\it e.g.} forward/backward {\it vs.} central, high luminosity {\it vs.} low luminosity) for different IRs.

Minimizing the effects of systematic uncertainties is an important aspect of detector design.  Absolute and relative luminosity determination is a key to extracting important observables, for instance the longitudinal structure function or small polarization asymmetries.  Measurement of the polarization of electron and hadron beams has a high priority. As with the accelerator, R\&D for EIC detectors will be essential.


\clearpage

\phantomsection
\addcontentsline{toc}{section}{\numberline{}Tables of golden measurements}

\begin{center}
\textbf{\Large Tables of golden measurements}
\label{sec:golden-tables}
\end{center}


\renewcommand{\textfraction}{0.05}

\begin{table}[h]
\centering
\noindent\makebox[\textwidth]{%
\footnotesize
\begin{tabular}{|c|c|c|c|c|}
\hline
\multicolumn{4}{|c|}{Spin and flavor structure of the nucleon} \\
\hline
Deliverables & Observables & What we learn & Requirements \\
\hline
\hline
polarized gluon          &  scaling violations   & gluon contribution  & coverage down to $x\simeq 10^{-4}$; \\
distribution $\Delta g$  &  in inclusive DIS     & to proton spin      & ${\cal{L}}$ of about $10\;\mathrm{fb}^{-1}$\\
\hline
polarized quark and    &  semi-incl.\ DIS for       & quark contr.\ to proton spin;      & similar to DIS;    \\
antiquark densities    &  pions and kaons  & asym.\ like $\Delta \bar{u}-\Delta\bar{d}$; $\Delta s$  & good particle ID  \\
\hline
novel electroweak        &  inclusive DIS  &  flavor separation             & $\sqrt{s}\geq 100\,\mathrm{GeV}$; ${\cal{L}}\geq 10\;\mathrm{fb}^{-1}$ \\
spin structure functions &  at high $Q^2$  &  at medium $x$ and large $Q^2$ & positrons; polarized $^{3}$He beam\\
\hline
\end{tabular}}

\vspace{1.8\baselineskip}


\centering
\noindent\makebox[\textwidth]{%
\footnotesize
\begin{tabular}{|c|c|c|c|c|} 
\hline
\multicolumn{5}{|c|}{Three-dimensional structure of the nucleon and
  nuclei: transverse momentum dependence} \\ 
\hline
Deliverables & Observables & What we learn & Phase I & Phase II\\
\hline\hline
Sivers and &  SIDIS with transv. & quantum interference
&valence+sea  & 3D Imaging of\\
unpolarized &polarization/ions; & multi-parton and  & quarks, overlap &
quarks and gluon;  \\
TMDs for & di-hadron (di-jet) &spin-orbit & with fixed target  &  $Q^2$
($P_\perp$) range\\
quarks and gluon & heavy flavors &correlations & experiments & QCD dynamics\\
\hline
\end{tabular}}

\vspace{1.8\baselineskip}


\centering
\footnotesize
\begin{tabular}{|c|c|c|c|}
\hline
\multicolumn{4}{|c|}{Three-dimensional structure of the nucleon and
  nuclei: spatial imaging} \\ 
\hline
Deliverables  & Observables & What we learn & Requirements \\
\hline\hline
   sea quark and    & DVCS and $J/\psi, \rho, \phi$  & transverse images of & $\mathcal L\geq 10^{34}$~cm$^{-2}$s$^{-1}$, \\
    gluon GPDs      & production cross sect.      &  sea quarks and gluons  & Roman Pots  \\
                    & and asymmetries             &  in nucleon and nuclei;  & wide range of $x_B$ and $Q^2$ \\ 
                    &                             & total angular momentum;        & polarized $e^-$ and $p$ beams \\
                    &                             & onset of saturation & $e^+$ beam for DVCS \\
 \hline
\end{tabular}

\vspace{1.8\baselineskip}


\centering
\noindent\makebox[\textwidth]{%
\footnotesize
\begin{tabular}{|c|c|c|c|c|}
\hline
\multicolumn{5}{|c|}{QCD matter in nuclei} \\
\hline
Deliverables & Observables & What we learn & Phase I & Phase II \\
\hline
\hline
integrated gluon  &  $F_{2,L}$  & nuclear wave function;    & gluons at   &  explore sat. \\
distributions     &            & saturation, $Q_s$ & $10^{-3}\leq x \leq 1$  &  regime \\
\hline
$k_T$-dep. gluons; &   di-hadron        &  non-linear QCD     &  onset of     &  RG evolution\\
gluon correlations    &  correlations     &  evolution/universality  &    saturation; $Q_s$ &   \\
\hline
transp. coefficients &  large-$x$ SIDIS;  &  parton energy loss,  & light flavors, charm&  precision rare \\
in cold matter       &  jets              &  shower evolution;    & bottom; jets        &   probes; \\
                     &                    &  energy loss mech.  &                    &  large-$x$ gluons \\
\hline
\end{tabular}}

\vspace{1.8\baselineskip}


\centering
\footnotesize
\begin{tabular}{|c|c|c|c|c|}
\hline
\multicolumn{5}{|c|}{Electroweak interactions and physics beyond the
  Standard Model} \\
\hline
Deliverables & Observables & What we learn & Phase I & Phase II\\
\hline\hline
Weak mixing &  Parity violating & physics behind electroweak & good precision & high precision \\
angle & asymmetries in & symmetry breaking & over limited & over wide range \\
  &  $ep$- and $ed$-DIS & and BSM physics & range of scales & of scales \\
\hline
$e$-$\tau$ conversion  & $ep \rightarrow \tau,X$  & flavour violation
& challenging & very promising \\
& & induced by BSM physics  & &  \\
\hline
\end{tabular}
\end{table}

\renewcommand{\textfraction}{0.2}


\chapter{The spin and flavor structure of the proton}

\noindent
{\Large Convenor and chapter editor: \\[1em]
M. Stratmann}

\newpage


\section{Introduction and chapter overview}

\hspace{\parindent}\parbox{0.92\textwidth}{\slshape 
  Marco Stratmann
%
}

\index{Stratmann, Marco}

\vspace{\baselineskip}

Two weeks of the INT program on ``Gluons and the Quark Sea at High Energies''
were devoted to the physics of unpolarized and polarized parton distribution functions.
A compelling set of physics opportunities at an EIC 
has emerged from lively discussions among the participants
and subsequent interactions with the hadron structure community.
This Chapter outlines the identified open fundamental questions in hadronic physics
and the ``golden measurements'' and experimental requirements
to thoroughly address them at a future EIC. The anticipated results will
have a profound influence on our understanding of the spin and flavor structure of nucleons.

Sixteen years of operations at DESY-HERA had a transformational impact on the way we view
the internal partonic content of nucleons and have led to various new developments
in the field of Quantum Chromodynamics.
The experiments have left a rich legacy of results, the most prominent ones being
the strong rise of the gluon density at small momentum fractions $x$, the large
portion of diffractive events, and the transition from high to low momentum transfer $Q$
for various processes.  
Likewise, vigorous experimental programs with polarized beams and targets in the 
past twenty-five years at all major laboratories have brought us closer to
pinpoint the various contributions to the proton's spin.
They also revealed novel, often puzzling phenomena which initiated new directions of
research in spin physics such as transverse-momentum dependent parton densities;
see Chapter 2.

In each case, the experimental progress was matched by considerable theoretical efforts
in Quantum Chromodynamics. Most notable in this context are the level of precision
reached in higher-order calculations in perturbative QCD and the much refined 
global analysis tools to reliably extract information on parton densities from data and 
to determine their uncertainties.
Yet, there is still a significant lack of understanding on quite a few outstanding issues.
An EIC will prove crucial in addressing them by making use of the anticipated
high luminosities and the variability of beam energies.

Of course, due to the lower center-of-mass system energies of an EIC as compared to HERA
one cannot extend the kinematic reach towards smaller values of $x$ for unpolarized 
electron-proton collisions.
Also, over the next couple of years the CERN-LHC will provide a great deal of 
information on helicity-averaged parton densities in a broad range of $x$ from various different
hard scattering processes up to very large resolution scales $Q$. The 12 GeV upgrade of
the CEBAF facility at Jefferson Laboratory is designed to map parton distributions up to
very large values of $x$ at scales $Q$ of a few GeV 
to test how well, for instance, counting rules apply.  Therefore,
we expect that most aspects of unpolarized parton densities will be sufficiently
well known by the time an EIC is expected to turn on, with some important
exceptions to be discussed below.

The situation is rather different for spin physics 
where the bulk of experimental information stems from fixed-target
lepton-nucleon scattering experiments at rather low energies.
Ideas to turn HERA into a polarized electron-proton collider never materialized.
Existing experiments studying the helicity structure of the nucleon, 
like PHENIX and STAR at RHIC, will continue to add data in the
next couple of years. In particular, measurements of double-spin asymmetries for
di-jets in $pp$ collisions at $500\,\mathrm{GeV}$ should improve the current
constraints on the polarized gluon density $\Delta g(x)$ 
and extend the covered $x$ range towards somewhat smaller values.
Parity-violating, single-spin asymmetries for $W$ boson production
should reach a level where they help to constrain the spin-dependent
$u$ and $d$ quark and antiquark densities at medium-to-large $x$.
At JLab-12 the focus is again on the large $x$ frontier 
at moderate values of $Q$ to address to what extent quarks obey
helicity retention which predicts that in the limit $x\to 1$ quark
and nucleon spins become fully aligned.
Ultimately, all these efforts are limited by their kinematic coverage
both in $x$ and in $Q$.
Since the most fundamental open questions in spin physics 
concern the polarization of wee partons, see below,
there are many opportunities for a high-energy polarized EIC 
to contribute significantly due to its unique capabilities
to access values of $x$ down to about $10^{-4}$.
This is central to finally determine and understand 
the role of quarks and gluons in the spin decomposition of the nucleon.

Factorization of experimental observables into non-perturbative parton densities and calculable hard scattering 
cross sections is the cornerstone for the theoretical application of QCD at high energies
within perturbative methods.
Available QCD calculations for inclusive and 
semi-inclusive deep-inelastic scattering processes will allow us to confront future 
high-statistics EIC data with theory at the necessary very high level of precision. 
A brief account of the status of perturbative QCD calculations for most of 
the key measurements at an EIC is given in Sec.~\ref{sec:pqcd}.

Since the EIC is a natural extension of the physics program carried out at HERA
both in terms of the anticipated significant increase in luminosity and the possibility to have
polarized beams, we summarize the latest status of HERA data based on the recent combination of
results from the H1 and ZEUS experiments in Sec.~\ref{sec:cooper-sarkar}.
This discussion also helps to expose the open questions about the structure of unpolarized nucleons 
an EIC can elucidate and which cannot be answered solely by measurements at the LHC. 
The most compelling ones comprise 
\begin{itemize}
\item the longitudinal structure function $F_L$, 
\item the elusive strangeness and anti-strangeness densities,
\item and heavy flavor contributions to deep-inelastic scattering.
\end{itemize}
A detailed account, including other second tier opportunities is given in Sec.~\ref{sec:olness-pdfs}.

An EIC could make the first precise measurement of $F_L$ in a kinematic range that overlaps both 
previous fixed-target and HERA data, none of which are very precise. 
$F_L$ is particularly sensitive to the gluon distribution and QCD dynamics at small $x$
which makes it a promising candidate to study the transition to the high parton density regime, 
i.e., the phenomenon of saturation, with an inclusive observable. While one does not expect
non-linear effects to be of significant relevance in electron-proton collisions at an EIC, 
a measurement of $F_L$ provides the baseline for similar studies in electron-heavy ion collisions.
Here, the onset of saturation effects is expected already 
at $x\simeq 10^{-3}$ which elevates $F_L$
to one of the golden measurements to be performed at the EIC; see Chapter 5 on 
QCD matter under extreme conditions for details.
The determination of $F_L$ relies on an accurate measurement of the variation of the
so-called reduced cross section for fixed values of $x$ and $Q$ at different c.m.s.\ energies
$\sqrt{s}$. The large variability of beam energies at sustained large luminosities is a
particular strength of an EIC and proves critical for this measurement. 
A first feasibility study for electron-proton collisions can be found in Sec.~\ref{sec:marco-fl}.

Semi-inclusive deep-inelastic production of identified pions and kaons is expected to be the most viable 
and promising way to determine differences among parton distribution functions for different
quark flavors or between quarks and anti-quarks.
Such measurements make use of the different probabilities for producing a certain hadron
species from a given quark flavor or gluon and
have been successfully performed at fixed-target experiments such as HERMES. 
The EIC offers unprecedented opportunities to extend the kinematic reach toward small $x$ or
large $Q$. In particular,
the elusive strangeness density and a possible asymmetry between strangeness and anti-strangeness 
distributions can be deduced from charged kaon production yields. 
Prerequisites are excellent
particle identification in most of the phase space and a thorough theoretical understanding of
the hadronization of quarks and gluons into the observed hadrons.
In collinear factorization, the latter information is encoded in non-perturbative
fragmentation functions which are constrained by a wealth of available experimental data
on single-inclusive hadron yields. Further significant progress on the quality of such
fits is expected once upcoming data from $B$ factories and the LHC are included.
In Sec.~\ref{sec:marco-sidis} we present a first feasibility study for charged kaon production
at the EIC. 

Heavy flavors, in particular charm quarks, can give a sizable contribution
to deep-inelastic scattering structure functions. Within the foreseen EIC kinematics, charm yields
up to $10\div 15\%$ of the inclusive cross section. The theoretical framework for heavy quark
production is much more complex than for light (massless) quarks 
due to the presence of multiple scales. The mass of the heavy quarks prevents them
from having a partonic interpretation, and they can be only produced externally, for instance, 
by photon-gluon fusion.
This framework yields a very good description of all available HERA data within
the present uncertainties and is expected to be
relevant also in the entire kinematic regime of an EIC. 
Nonetheless, one may introduce heavy quark densities for 
asymptotically large scales, i.e., $Q\gg m$, and smooth interpolation schemes have been
devised which incorporate the correct threshold and asymptotic behavior.
The relevant theoretical framework and recent progress on higher order calculations is 
briefly reviewed in Sec.~\ref{sec:bluemlein-hq}.

The charm contribution to the longitudinal structure function $F_L$ is expected to be 
particularly sensitive to mass effects and has never been measured before. 
A first feasibility study within the kinematics of an EIC can be found in Sec.~\ref{sec:flcharm}.
An EIC is also well suited to address the long-standing question of a possible relevance of
a non-perturbative ``intrinsic'' charm contribution in the nucleon wave-function,
mainly concentrated at large momentum fractions. Quantitative estimates based on
models for an intrinsic charm contribution are promising
and can be found in Sec.~\ref{sec:intrcharm}.

The physics opportunities with polarized lepton and proton beams are even more multifaceted
and will address some of the most fundamental open questions in hadronic physics
for which one has been seeking answers for more the two decades now.
Thus, the anticipated results will have far-reaching impact on our understanding of the nucleon's
spin structure. The unique capability of the EIC to reach small momentum fractions $x$ 
or large scales $Q$ in longitudinally polarized electron-proton collisions with high luminosity will enable us to explore in detail
\begin{itemize}
\item the polarized gluon distribution and its contribution to the proton's spin,
\item the individual light quark helicity distributions in a broad kinematic range,
\item novel electroweak structure functions,
\item and the strangeness and anti-strangeness polarizations.
\end{itemize}
The latest status of global QCD fits to helicity dependent parton densities, which
is not expected to improve much by the time the EIC would turn on,
and the set of questions we want to address at the EIC are laid out in some detail
in Sec.~\ref{sec:polstatus}.

Precise measurements of the polarized structure function $g_1$ in a wide kinematic range will be
a flagship measurement for the EIC. The gluon helicity distribution $\Delta g$ is strongly correlated
with QCD scaling violations, i.e., the $Q$ dependence of $g_1$ at a given $x$. This will allow
for a determination of $\Delta g$ down to unprecedented small values of $x$ of about $10^{-4}$.
This in turn will eventually pinpoint the elusive gluon contribution to the spin of the proton,
given by the integral of $\Delta g$ over all momentum fractions $x$, to about $10\%$ accuracy or better.
The striking quantitative impact on extractions of $\Delta g$ based on projected EIC data is 
demonstrated in Sec.~\ref{sec:rodolfo-spin}.
The same set of inclusive measurements will also provide a significantly better determination of the total
quark contribution $\Delta \Sigma$ both as function of $x$ and the integral relevant for the
nucleon spin sum.

Like in the unpolarized case, see Sec.~\ref{sec:marco-sidis}, the best strategy to achieve a 
full flavor and quark-antiquark separation of polarized helicity densities is based again on 
semi-inclusive deep-inelastic hadron production. The kinematic coverage in $x$ and $Q$ is similar to what
can be achieved in inclusive DIS, with the extra theoretical complication of the need for
fragmentation functions to model hadronization. 
At medium-to-large values of $x$ one can address with precision certain interesting
asymmetries in the polarized quark sea like
$\Delta \bar{u} - \Delta \bar{d}$  (from charged pion yields) and perhaps even
$\Delta s - \Delta \bar{s}$ (from charged kaon yields).
The first quantity is predicted to be sizable in several model
calculations of the nucleon 
but the precision of current experiments only gives a first hint of a 
possible non-zero asymmetry;
the latter quantity may help to understand why the sum 
$\Delta s + \Delta \bar{s}$ appears to be much smaller 
in current experiments than expected.
If $\Delta s$ and $\Delta \bar{s}$ have their spins anti-aligned, 
their sum could be small but the asymmetry would be sizable.
Constraints from hyperon decay matrix elements and arguments based on 
SU(3) symmetry predict a significantly negative total ($x$ integrated) 
strange quark polarization. To address the validity of this constraint
and to access to what extent SU(3) symmetry is broken, one needs to determine
$\Delta s$ down to small values of $x$ to obtain a reliable estimate
of its $x$ integral. This is another unique measurement to be performed at the EIC.

First simulations of electroweak neutral and charged current deep-inelastic scattering
at the EIC in Sec.~\ref{sec:electroweak}
show that such measurements become feasible already with relatively modest
integrated luminosities. The corresponding structure functions for polarized protons
have never been measured before and probe combinations of quark flavors other than
in one-photon-exchange dominating at low $Q$.
To fully exploit the potential of the EIC for such measurements, 
positron beams are required, albeit not
necessarily polarized. An effective source of polarized neutrons such as a Helium-3 beam
would be highly desirable. When combined, these measurements will greatly aid the 
flavor decomposition of polarized parton densities
at medium-to-large $x$, free of any hadronization ambiguities.
At the highest c.m.s.\ energies and luminosities also photon-$Z$ boson interference
contributions to structure functions should be accessible at the EIC.
The production of charmed mesons in charged current DIS events is an alternative probe
for the strange and anti-strange densities both unpolarized and polarized.
This is discussed in Sec.~\ref{sec:marco-cc-charm}.

\begin{table}[htdp]
\centering
\noindent\makebox[\textwidth]{%
\footnotesize
\begin{tabular}{|c|c|c|c|c|}
\hline
Deliverables & Observables & What we learn & Requirements \\
\hline
\hline
polarized gluon          &  scaling violations   & gluon contribution  & coverage down to $x\simeq 10^{-4}$; \\
distribution $\Delta g$  &  in inclusive DIS     & to proton spin      & ${\cal{L}}$ of about $10\mathrm{fb}^{-1}$\\
\hline
polarized quark and    &  semi-incl.\ DIS for       & quark contr.\ to proton spin;      & similar to DIS;    \\
antiquark densities    &  pions and kaons  & asym.\ like $\Delta \bar{u}-\Delta\bar{d}$; $\Delta s$  & good particle ID  \\
\hline
novel electroweak        &  inclusive DIS  &  flavor separation             & $\sqrt{s}\gtrsim 100\,\mathrm{GeV}$; ${\cal{L}}\gtrsim 10\mathrm{fb}^{-1}$ \\
spin structure functions &  at high $Q^2$  &  at medium $x$ and large $Q^2$ & positrons; polarized $^{3}$He beam\\
\hline
\end{tabular}}
\caption{Golden measurements in polarized $ep$ collisions at an EIC.}
\label{tab:ep_golden}
\end{table}
Table~\ref{tab:ep_golden} summarizes the identified golden measurements, science deliverables, and
experimental requirements in spin-dependent lepton-proton collisions at an EIC.
Other, second tier measurements with polarization involve the currently unknown
charm contribution to the deep-inelastic structure function $g_1$ which offers sensitivity to $\Delta g$ through
photon-gluon fusion. Some expectations can be found in Sec.~\ref{sec:rodolfo-spin}.
If an effective neutron beam is available one can also attempt to determine the fundamental
Bjorken sum rule at a few percent level. The Bjorken sum is probably one of the most precisely
calculated quantities in perturbative QCD and provides an interesting link to the
Adler $D$ function in electron-positron annihilation through the Crewther relation.

Finally, the production of hadronic final states in electron-proton collisions is dominated
by the exchange of photons of almost zero virtuality. Photoproduction measurements 
and, in particular, the exploration of kinematic regimes where ``resolved
photon'' contributions dominate was one of the great successes of the HERA physics program.
Resolved processes, where the photon interacts with the proton through its
non-perturbative source of partons, offer a fresh look at these densities
which are so far mainly determined from imprecise LEP data. 
Given the anticipated high luminosity, an EIC can elevate these studies to a level
of unprecedented precision, and, thanks to the polarized beams, allows one to investigate for the
first time also the non-perturbative structure of circularly polarized photons.
A good knowledge of the partonic structure
of photons is essential for part of the physics program of a possible future linear collider.
The general framework for photoproduction and two examples of 
physics studies are presented in Secs.~\ref{sec:photoproduction}-\ref{sec:polphoto}.

To summarize, the physics goals of the EIC should be ambitious and must offer
detailed answers to all the open fundamental questions concerning the spin
and flavor structure of nucleons laid out above.
The following sections will outline the path to achieve these goals.
The program bears significant experimental challenges which all need to be carefully addressed
to reach the desired unprecedented level of precision. With the exception of some of the
electroweak structure function measurements, most observables will be quickly limited
by systematic uncertainties, intrinsic ambiguities of the extraction method
like, for instance, the Rosenbluth separation for $F_L$, and the way how well we can control QED radiative corrections to unfold the information one is actually interested in. Experimental aspects are discussed in Chapter 7.

\section{Status of perturbative QCD calculations}
\label{sec:pqcd}


\hspace{\parindent}\parbox{0.92\textwidth}{\slshape 
  Sven-Olaf Moch
%
}

\index{Moch, Sven-Olaf}





\subsection{Introduction}

Deep-inelastic scattering (DIS) and the observed scaling violations are at the very
center of the formulation of QCD as the gauge theory of the strong interactions~\cite{Gross:1973ju,Gross:1974cs}.

Over the decades the experiments using lepton and neutrino
scattering off fixed targets at CERN, FNAL, SLAC, and JLAB 
as well as electron-proton collisions at the HERA collider at DESY 
have provided unique insight into the nucleon structure with the 
available high precision experimental data spanning a large kinematical range.
Dramatic further improvements 
can be expected from the planned electron-ion collider EIC.

The key observables are either inclusive structure functions or differential cross sections 
in the semi-inclusive case, which parametrize the hard hadronic interaction
in the QCD improved parton model.
The particle data group (PDG)~\cite{Nakamura:2010zzi} provides a very readable
account of DIS, including the definitions of kinematic variables, etc.

\begin{figure}[h!]
\hspace*{0.3cm}
\begin{center}
\includegraphics[width=0.90\textwidth]{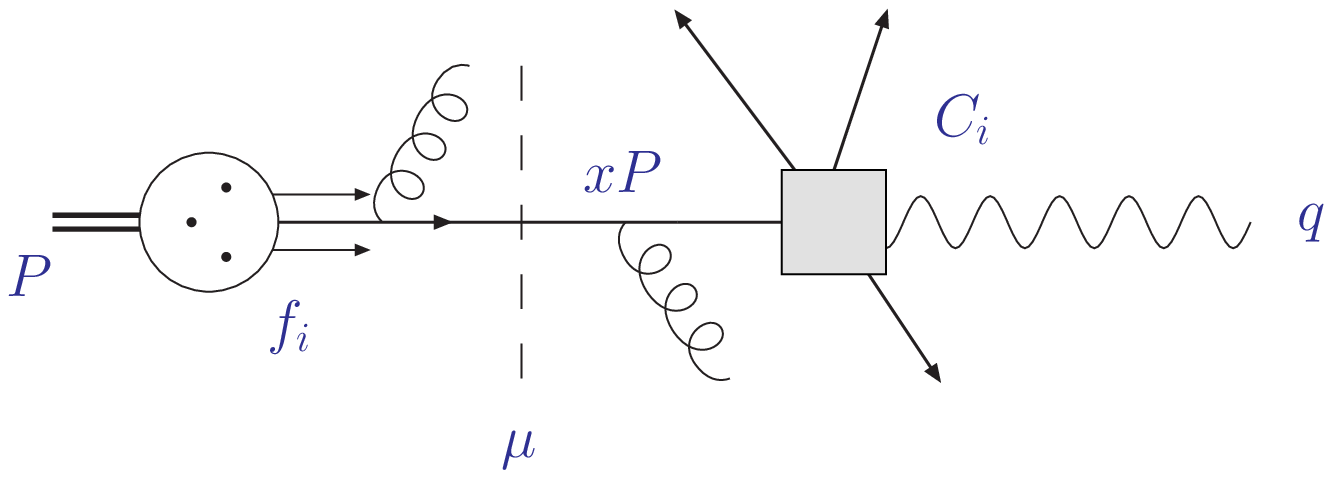}
\end{center}
\vspace*{-3mm}
\caption{\label{fig:ep-rest} 
  {QCD factorization of the cross section for the scattering 
    of a deeply virtual boson with (space-like) momentum $q$ 
    ($-q^2 = Q^2 > 0$) off a proton with momentum $P$ 
    in their center-of-mass frame, see Eq.~(\ref{eq:F2factorisation}). 
}}
\end{figure}
Precision predictions in perturbative QCD rest on the fact that we can separate 
the sensitivity to dynamics from different scales, i.e., 
the physics at scale of the proton mass from hard, high-energy scattering at a large scale $Q^2$. 
For lepton-proton DIS in the one-boson exchange approximation 
this is depicted in Fig.~\ref{fig:ep-rest}.
For unpolarized DIS, this factorization at a scale $\mu$ allows to express the structure functions 
$F_k$ ($k=2,3,L$) as convolutions of parton distributions (PDFs) 
$f_{i}$ ($i=q,{\bar q},g$) and short-distance Wilson coefficient functions $C_{k,i}$,
\begin{eqnarray}
  \label{eq:F2factorisation}
  F_k(x,Q^2) &=& \sum\limits_{i=q,{\bar q},g} \, 
  \int_x^1 dz \, f_{i}\left({x \over z},\mu^2\right)\, 
  C_{k,i}\left(z,Q^2,\alpha_s(\mu),\mu^2 \right) 
  \, , 
\end{eqnarray}
up to corrections of higher twist ${\cal O}(1/Q^2)$.
The coefficient functions $C_{k,i}$ are calculable perturbatively 
in QCD in powers of the strong coupling constant $\alpha_s$, 
\begin{eqnarray}
\label{eq:C-exp-alphas}
  C &=& 
  C^{(0)} + \alpha_s\, C^{(1)} + \alpha_s^2\, C^{(2)} + \alpha_s^3\, C^{(3)} + \, \ldots
  \, ,
\end{eqnarray}
with the expansion coefficients $C^{(0)}$ denoted as the leading order (LO), 
$C^{(1)}$ the next-to-leading order (NLO) and so on. 
The PDFs $f_i$ describe the fraction $x=Q^2/(2P \cdot q)$ of the nucleon momentum carried by the quark or gluon.
PDFs are non-perturbative objects and have to be obtained from global fits to experimental data 
or determined, e.g., by lattice computations.
Perturbation theory, however, provides information about their scale
dependence, i.e., the well-known evolution equations,
\begin{equation}
\label{eq:evolution}
\frac{d}{d \ln \mu^2}\,
  \left( \begin{array}{c} f_{q_i}(x,\mu^2)  \\ f_{g}(x,\mu^2)  \end{array} \right) 
\: =\:
\sum_{j}\, \int\limits_x^1\, {dz \over z}\, 
\left( \begin{array}{cc} P_{q_iq_j}(z) & P_{q_ig}(z) \\ P_{gq_j}(z) & P_{gg}(z) \end{array} \right)\,
  \left( \begin{array}{c} f_{q_j}(x/z,\mu^2)  \\ f_{g}(x/z,\mu^2) \end{array} \right)
\, .
\end{equation}
The splitting functions $P_{ij}$ are universal quantities in QCD and describe 
the different possible parton splittings in the collinear limit.
Like the $C_{k,i}$ also the $P_{ij}$ can be computed in a power series in $\alpha_s$, 
\begin{eqnarray}
\label{eq:P-exp-alphas}
  P & = &
  \alpha_s\, P^{(0)} + \alpha_s^2\, P^{(1)} + \alpha_s^3\, P^{(2)} + \, \ldots
  \, .
\end{eqnarray}

Analogous formulae hold for the polarized DIS structure functions. 
In particular, for $g_1$ one may apply the obvious replacements $f_{i} \to \Delta f_{i}$, 
$C_{k,i} \to \Delta C_{{g_1},i}$, and $P_{ij} \to \Delta P_{ij}$
in Eqs.~(\ref{eq:F2factorisation})--(\ref{eq:P-exp-alphas}).
QCD factorization has also been established for (semi-)inclusive deep-inelastic scattering (SIDIS), 
where the cross section $d^2 \sigma/ dx dQ^2$ is subject to a decomposition similar to Eq.~(\ref{eq:F2factorisation}).
Although, in that case, the process dependent hard parton scattering cross sections need
to be augmented by an additional prescription for the final state parton, 
e.g., a jet algorithm or fragmentation functions.

\subsection{Current status}

QCD predictions for DIS observables have reached over the years an unprecedented level of precision. 
All quantities in Eqs.~(\ref{eq:F2factorisation})--(\ref{eq:P-exp-alphas}) have been 
computed to higher orders in perturbation theory so that the effect of
radiative corrections on those observables is well understood and largely under control.
In the case of unpolarized DIS, the splitting functions $P_{ij}$ 
are known to NNLO~\cite{Moch:2004pa,Vogt:2004mw} 
and, likewise, the coefficient functions $C_{k,i}$~\cite{vanNeerven:1991nn,Zijlstra:1992qd,Zijlstra:1992kj,Moch:1999eb}.
For photon and charged current $W^\pm$-boson exchange, even the hard corrections at order 
${\cal O}(\alpha_s^3)$ are available~\cite{Vermaseren:2005qc,Moch:2008fj}.
In the case of polarized DIS, the spin dependent splitting functions $\Delta P_{ij}$ 
at two loop order have been obtained some time ago~\cite{Mertig:1995ny,Vogelsang:1996im}.
At NNLO, the polarized splitting functions $\Delta P_{qq}$ and $\Delta P_{qg}$
have been reported~\cite{Vogt:2008yw}, and the coefficient functions $\Delta C_{{g_1},i}$ are 
available from~\cite{Zijlstra:1993sh}.
For semi-inclusive observables, the QCD corrections are typically known to NLO.
This corresponds to ${\cal O}(\alpha_s^2)$ since the underlying Born 
cross section behaves as $d^2 \sigma^{(0)}/ dx dQ^2 \sim {\cal O}(\alpha_s)$ due to the additional final state parton. 
Processes considered include, for instance, the electro-production of hadrons with high transverse
momentum~\cite{Daleo:2004pn,Kniehl:2004hf} or single inclusive DIS jet cross sections~\cite{Daleo:2005gk}.


The currently available QCD predictions for inclusive DIS and SIDIS put 
us in comfortable position to confront experimental data with theory at a very high level of precision.
In these comparisons, we no longer test QCD. 
Rather we use perturbative QCD as an essential and established part of our theory toolkit 
to deduce important information about PDFs or the value of the strong coupling constant $\alpha_s(M_Z)$.
Of course, this is a situation that, generally, needs to be addressed also beyond DIS, since  
experimental data from the unpolarized (anti-)proton-proton colliders Tevatron at FNAL and the LHC at CERN 
as well as from the polarized proton-proton collider RHIC at BNL help to further constrain 
the non-perturbative input to QCD precision predictions.
See, e.g., the analyses of unpolarized PDFs to NNLO in Refs.~\cite{Alekhin:2009ni,Aaron:2009wt,JimenezDelgado:2008hf,Martin:2009iq} 
or recent studies of polarized 
PDFs in~\cite{Blumlein:2010rn,Gluck:2000dy,deFlorian:2008mr,deFlorian:2009vb,Hirai:2008aj,Leader:2010rb}.

Given the current status of perturbative QCD, experimental data from a future program of electron-ion collisions, EIC, 
can help to address and clarify a number of still open and  yet very relevant questions;
see also Secs.~\ref{sec:olness-pdfs} and \ref{sec:rodolfo-spin}.
For the case of unpolarized PDFs improvements can be made with respect 
to the flavor asymmetry of sea quarks at low $x$  and the valence quarks at large $x$, 
by studying, e.g., electron-deuteron collisions. 
Much of the physics case here had already been investigated in an assessment of 
the experimental prospects of electron-deuteron scattering at HERA some time ago~\cite{Alexopoulos:2003aa}.
More generally, the high luminosity of an EIC would further constrain PDFs, especially the gluon at low $x$ and $Q^2$. 
In this context, a precision measurements of the longitudinal structure function $F_L$, 
which is an observable predominantly driven by the gluon PDF is of high interest 
as it would complement and, eventually even supersede, existing experimental data, see, e.g.,~\cite{Collaboration:2010ry}.
New high statistics DIS experiments can also improve the current precision of
strong coupling constant $\alpha_s$ measurements in space-like kinematics.

For polarized DIS, a very fundamental question still remains the understanding
of the proton spin, in particular, whether the polarized gluon PDF $\Delta f_g$ provides a significant contribution. 
To that end, an extension of the kinematical coverage in $x$ and $Q^2$, as it
could be achieved by an electron-ion collider, is of paramount importance.
This would help to access higher scales in $Q^2$ in order to test the 
perturbative evolution Eq.~(\ref{eq:evolution}).
Likewise, access to an extended $x$-range allows for a better determination of moments of the $\Delta f_i$. 
They also enter, e.g., in the Bjorken sum rule for polarized electro-production, 
which is again an observable very well-known in perturbative QCD~\cite{Larin:1991tj,Baikov:2010je}.
Other issues of interest for polarized DIS in electron-ion collisions 
concern a reliable extraction of flavor structure as well as a
study of strangeness PDFs, $\Delta f_s$.

\subsection{Summary}

We have briefly summarized the current status of perturbative QCD predictions for DIS experiments. 
To date, we can build on a very mature understanding of the theory, which
could be confronted with experimental data from a future electron-ion collider in
order to improve our knowledge about the fundamental structure of matter and
the important dynamics of quarks and gluons in nucleons.

%
\section{Unpolarized proton structure - HERA's legacy}
\label{sec:cooper-sarkar}

\hspace{\parindent}\parbox{0.92\textwidth}{\slshape 
  Amanda Cooper-Sarkar (for the H1 and ZEUS Collaborations)
%
}

\index{Cooper-Sarkar, Amanda}




\subsection{Introduction}
HERA data provide the most 
insight into the behaviour of unpolarized parton distribution functions (PDFs) 
at present and as such represent an integral part of all global QCD analyses.
The H1 and ZEUS experiments are combining their various sub-sets of
data so as to provide a legacy of HERA results.
The combination of inclusive cross section data from HERA-I and the 
PDF fit based on these data are already 
published~\cite{Aaron:2009wt}. In 2010 further data have been combined and 
PDF fits to the augmented data sets have been made available in 
preliminary form. In Sec.~\ref{sec:cs-inc1} results from the published 
combination are reviewed. In Sec.~\ref{sec:cs-charm} results from a combination of
$F_2^{c\bar{c}}$ data are presented and their sensitivity to the mass of the
charm quark and the choice of the heavy flavor scheme adopted in the
global PDF fit is discussed.
In Sec.~\ref{sec:cs-lowenergy} results from 
the combination of inclusive cross section data taken at lower proton beam 
energies are discussed. Finally, in Sec.~\ref{sec:cs-inc2} an updated combination
of all inclusive data from HERA-I and HERA-II running is shown and 
a PDF fit to these data is presented.

\subsection{Inclusive data from HERA-I running (1992-2001)}
\label{sec:cs-inc1}
\begin{figure}[tbp]
\vspace{-2.0cm}
\begin{center}
\begin{tabular}{ll}
\psfig{figure=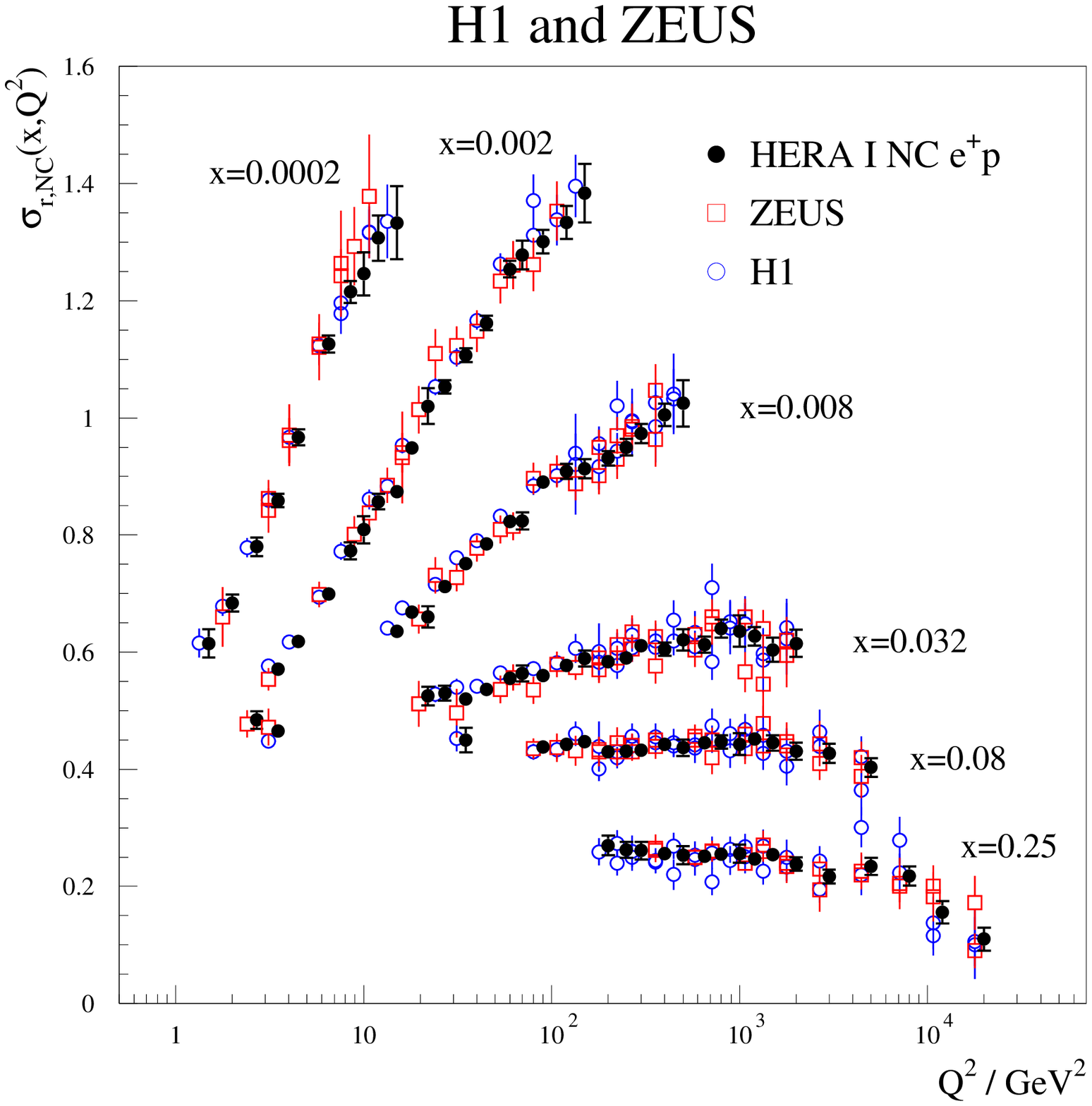,width=0.53\textwidth}
\end{tabular} 
\caption{
HERA combined data points for the NC $e^+p$ cross section as a function of
$Q^2$ in selected bins of $x$, compared to the separate ZEUS and H1 data sets input to the combination.}
\label{fig:herapdf10}
\end{center}
\vspace*{-0.3cm}
\end{figure}

The inclusive cross section data, 
from the HERA-I running period, for Neutral Current (NC) and 
Charged Current (NC), $e^+p$ and $e^-p$ scattering have been 
combined~\cite{Aaron:2009wt}. The combination procedure pays particular 
attention to the correlated systematic uncertainties of the data sets 
such that resulting combined data benefits from the best features of each 
detector. The combined data set has systematic 
uncertainties which are smaller than its statistical errors and the total 
uncertainties are small ($1-2\%$)
over a large part of the kinematic plane. The combined 
data is compared to the separate input data sets of ZEUS and H1 in 
Fig.~\ref{fig:herapdf10}.

These data are used as the sole input to a PDF fit called the 
HERAPDF1.0~\cite{Aaron:2009wt}. The motivations for performing a HERA-only 
fit are firstly, that the combination of the HERA data yields a very accurate 
and consistent data set such that the experimental uncertainties on the PDFs 
may be estimated from the conventional $\chi^2$ criterion $\Delta\chi^2=1$.
Global fits which include dats sets from many different experiments often use 
inflated $\chi^2$ tolerances in order to account for marginal consistency of 
the input data sets. Secondly, the HERA data are proton target data so that 
there is no uncertainty from heavy target corrections or deuterium corrections 
and there is no need to assume that $d$ in the proton is the same as $u$ in 
the neutron since the $d$-quark PDF may be extracted from $e^+p$ CC data.
Thirdly, the HERA inclusive data give information  
on the gluon, the Sea and the $u$- and $d$-valence PDFs over a wide kinematic region: 
the low-$Q^2$ NC $e^+p$ cross-section data are closely related 
to the low-$x$ Sea PDF and the low-$x$ gluon PDF is derived from its scaling 
violations; the high-$x$ $u-$ and $d$-valence PDFs are closely 
related to the high-$Q^2$ NC $e^{\pm}p$, CC $e^-p$, and CC $e^+p$ cross sections, 
respectively; the difference between the high-$Q^2$ $e^-p$ and $e^+p$ 
cross-sections gives the valence shapes down to low $x$, $x\sim 10^{-2}$. 

HERAPDF provides model and parametrisation uncertainties on the PDFs as well as
experimental uncertainties; for details, see Ref.~\cite{Aaron:2009wt}.
A major contribution to the total uncertainties in the HERAPDF1.0 set 
comes from the model uncertainty on the charm mass value.
This can be improved using information from data on $F_2^{c\bar{c}}$.

\subsection{Charm data from HERA-I and II running}
\label{sec:cs-charm}
\begin{figure}[htbp!]
\vspace*{-0.20cm}
\begin{center}
\includegraphics[width=0.495\textwidth]{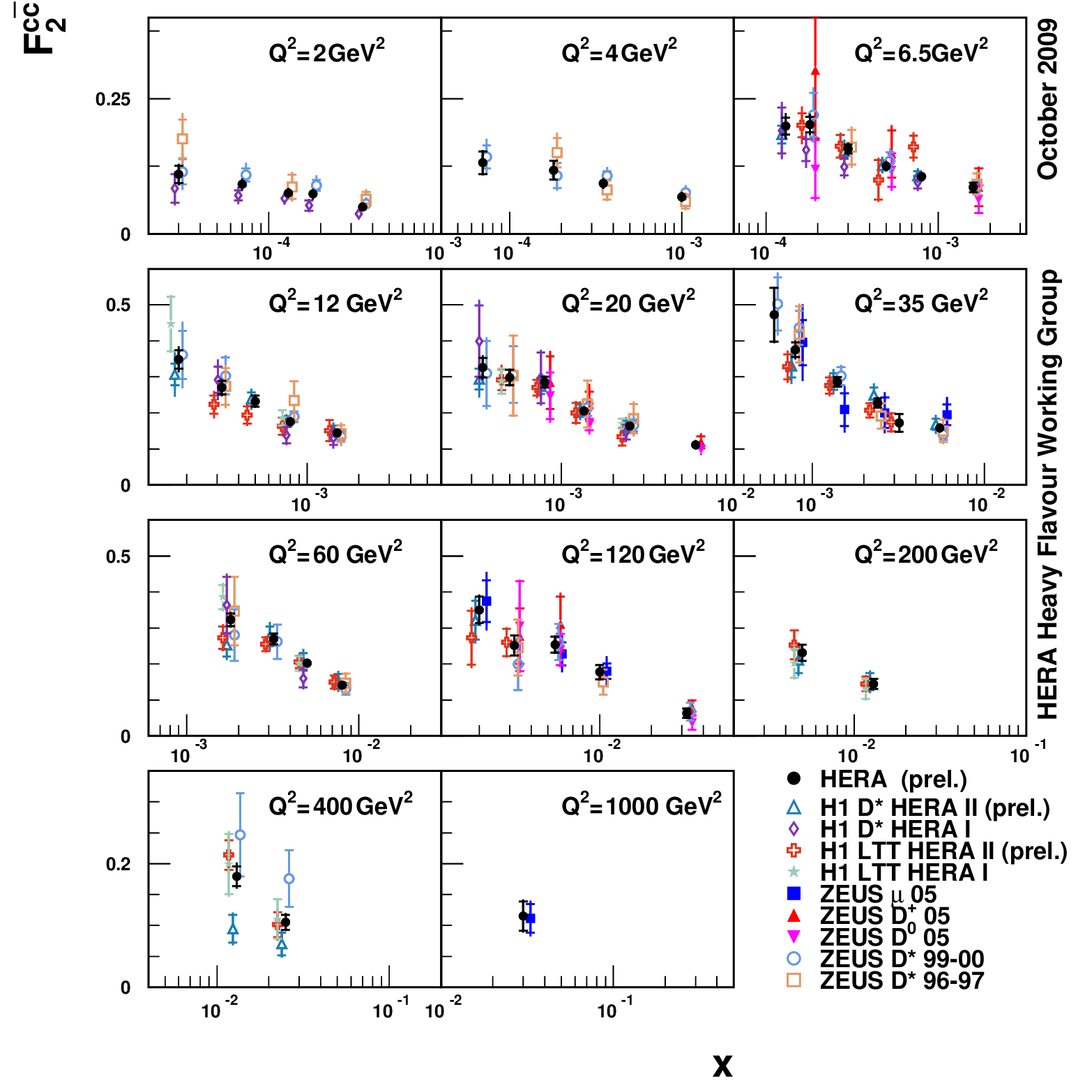}
\includegraphics[width=0.44\textwidth]{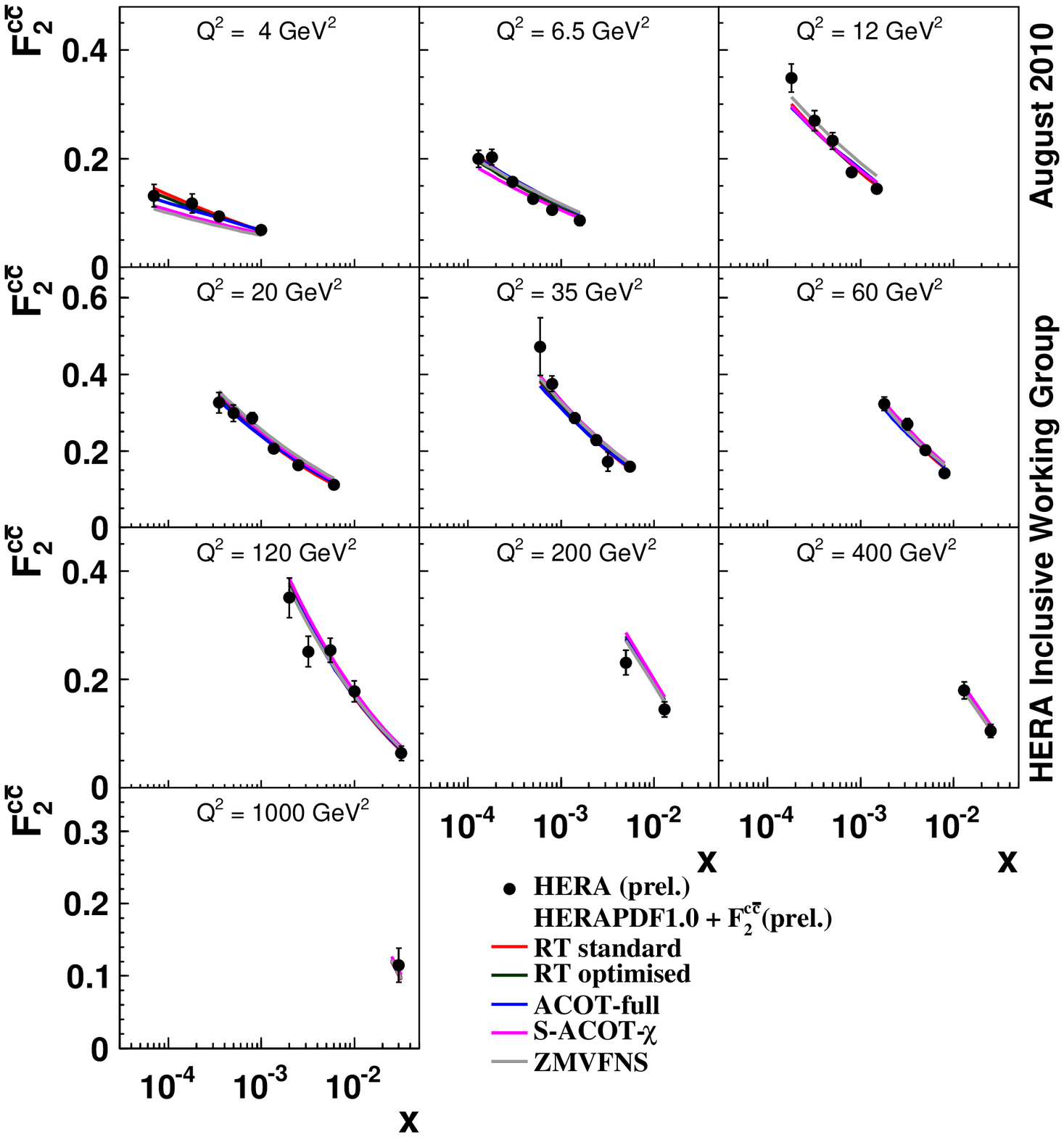}
\caption{Left: HERA combined data points for $F_2^{c\bar{c}}$
compared to the separate ZEUS and H1 data sets. 
Right: HERA combined data points for $F_2^{c\bar{c}}$ compared to
HERAPDF fits to these plus the inclusive DIS data, 
for various different heavy-quark-mass schemes.    }
\label{fig:charmdata}
\end{center}
%
\vspace*{0.90cm} 
\centerline{
\psfig{figure=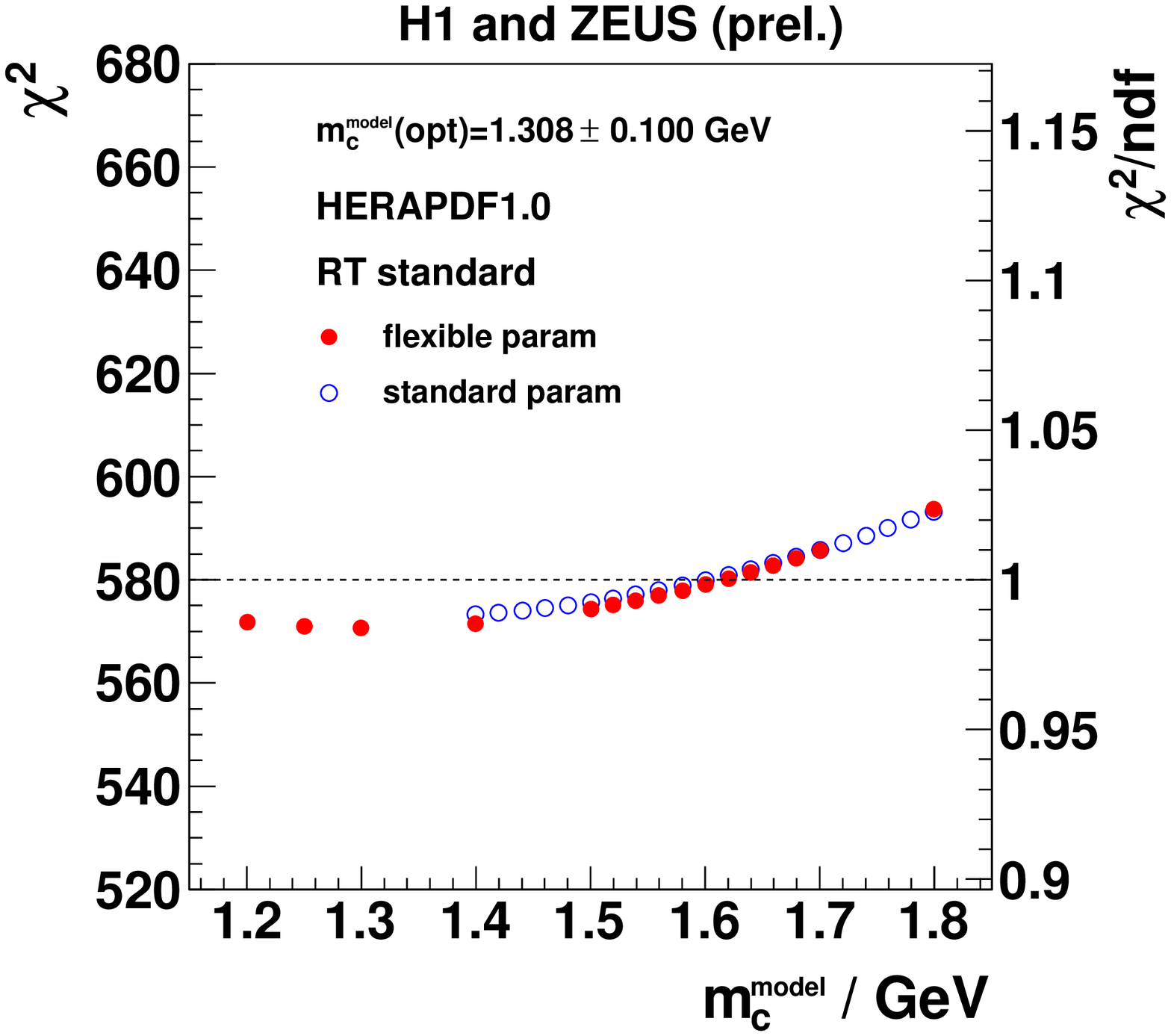,width=0.32\textwidth}~~ 
\psfig{figure=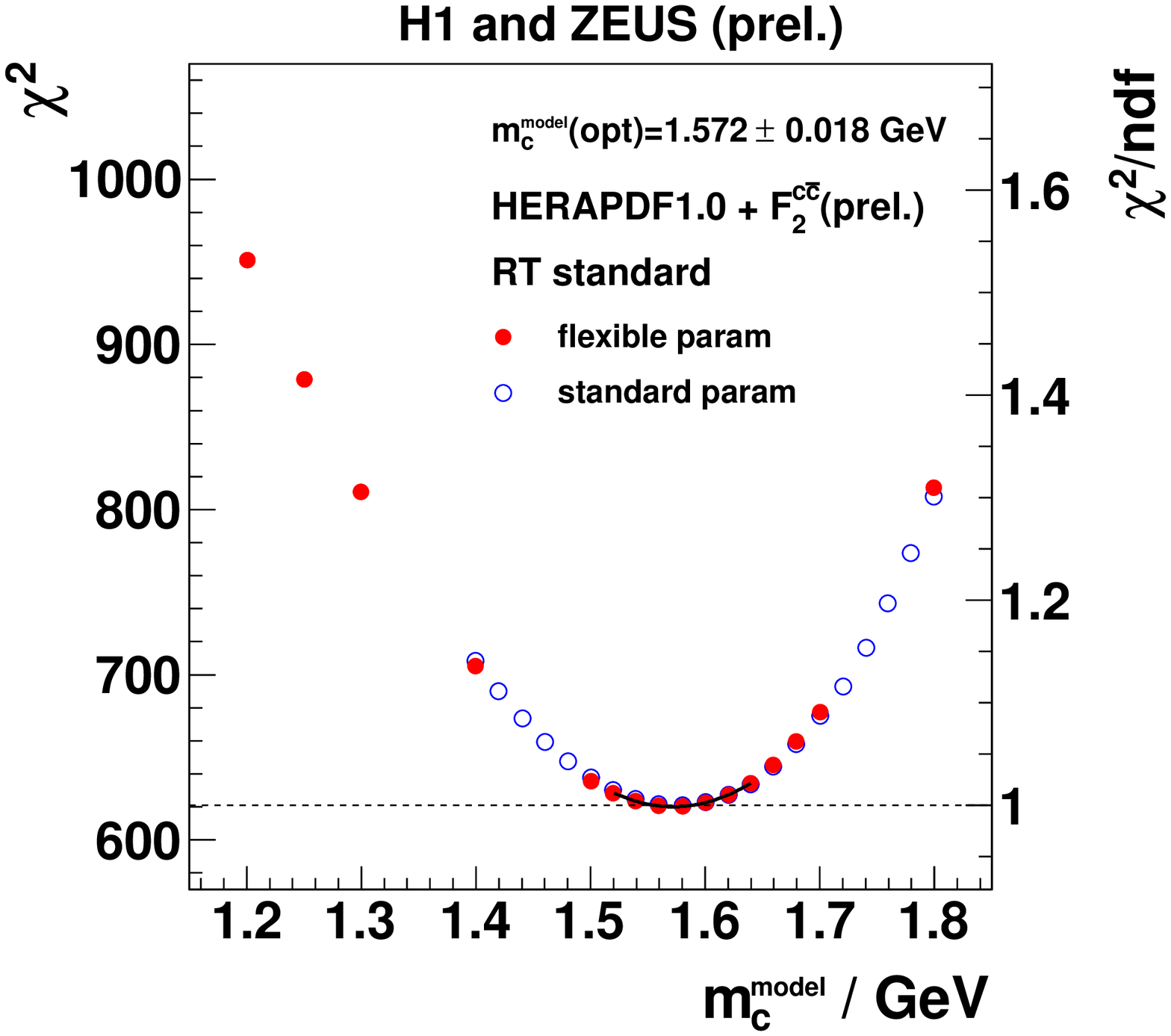,width=0.32\textwidth}~~
\psfig{figure=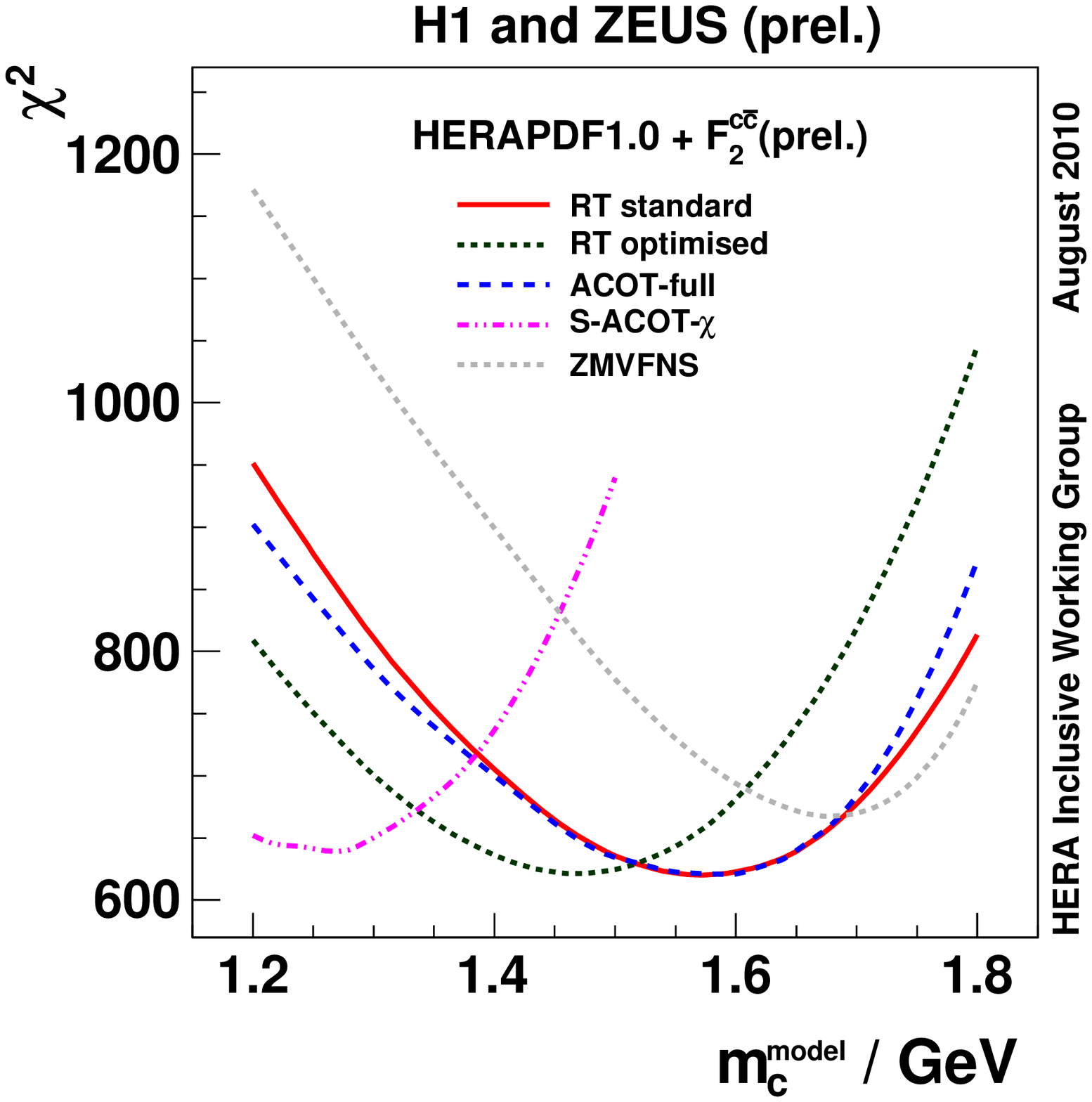,width=0.32\textwidth}} 
\caption {The $\chi^2$ of the HERAPDF fit as a 
function of the charm mass $m_c^{model}$. Left and Middle: using the 
RT-standard scheme, when 
$F_2^{c\bar{c}}$ data are not included and included in the fit, respectively.
Right: results for using various mass schemes in the fit to $F_2^{c\bar{c}}$
data.}
\label{fig:charmpred}
\vspace*{-0.4cm}
\end{figure}

H1 and ZEUS have also combined their data on 
$F_2^{c\bar{c}}$~\cite{Aaron:2009wt}.
In Fig.~\ref{fig:charmdata} the combined data are compared to the separate 
data sets which go into the combination.
These data are input to the HERAPDF fit together with the inclusive data 
which were used for HERAPDF1.0. The $\chi^2$ of this fit is sensitive to the 
value of the charm quark mass. Fig.~\ref{fig:charmpred} compares the $\chi^2$, 
as a function of this mass, for a fit which includes these data (left)
to that for the HERAPDF1.0 fit (middle). 
However, it would be premature to conclude that the data 
can be used to determine the charm pole-mass. The HERAPDF formalism uses 
the Thorne-Roberts (RT) variable-flavour-number (VFN) scheme for heavy 
quarks. This scheme is not unique,
specific choices are made for threshold behaviour. In Fig.~\ref{fig:charmpred} 
(right) the $\chi^2$ profiles for the standard and the 
optimized versions of this scheme are compared to two alternative ACOT VFN 
schemes and the Zero-Mass VFN scheme. Each of these schemes 
favours a different value for the charm quark mass, and the fit to the data 
is equally good for all the heavy quark mass schemes; see Fig.~\ref{fig:charmdata} (right). 
However, the Zero-Mass scheme is $\chi^2$ disfavoured; see 
Ref.~\cite{Aaron:2009wt} for further details.

\subsection{Low energy proton beam data from 2007}
\label{sec:cs-lowenergy}
\begin{figure}[tbp] 
\begin{center}
\includegraphics[width=0.495\textwidth]{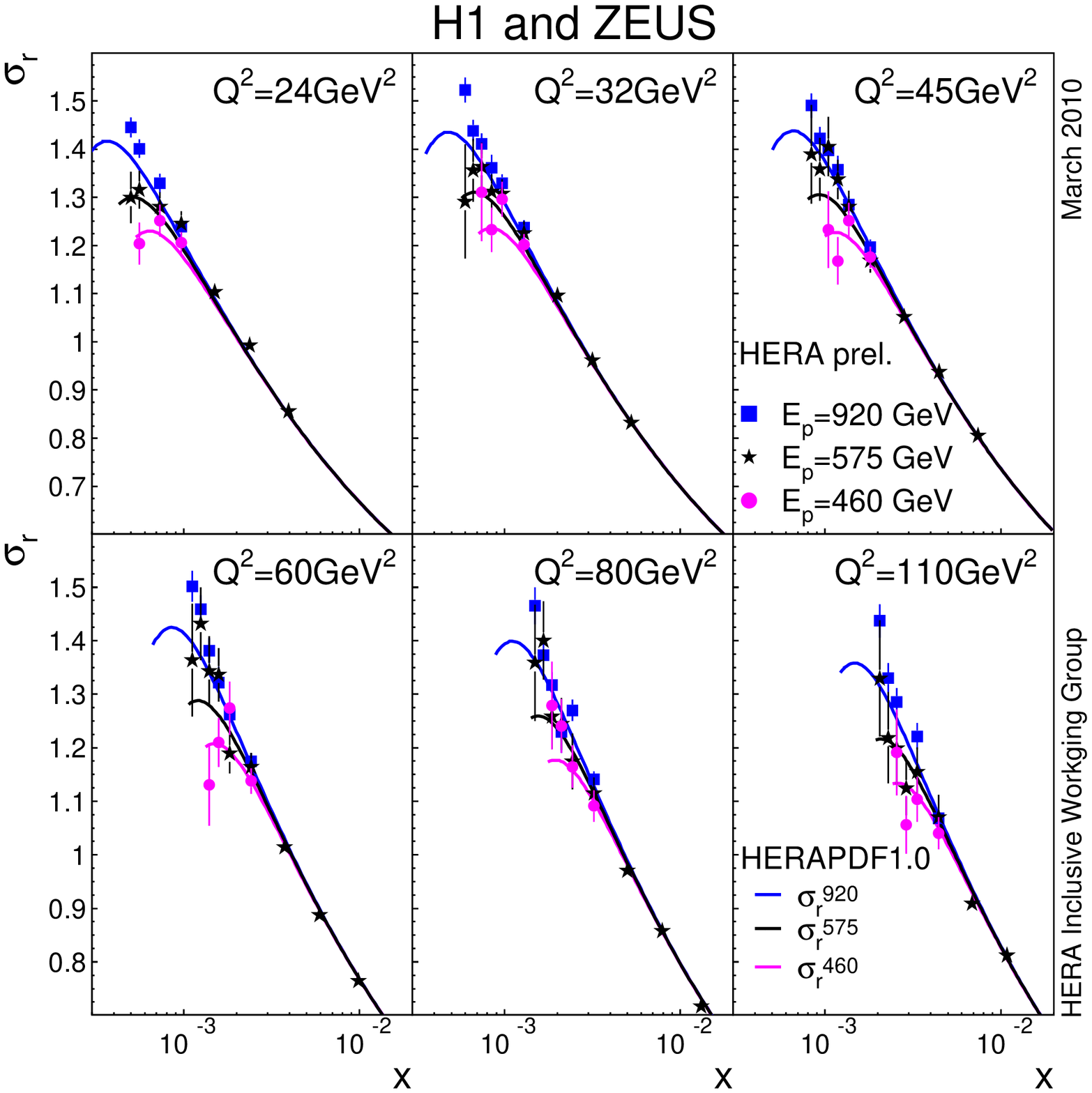}
\includegraphics[width=0.495\textwidth]{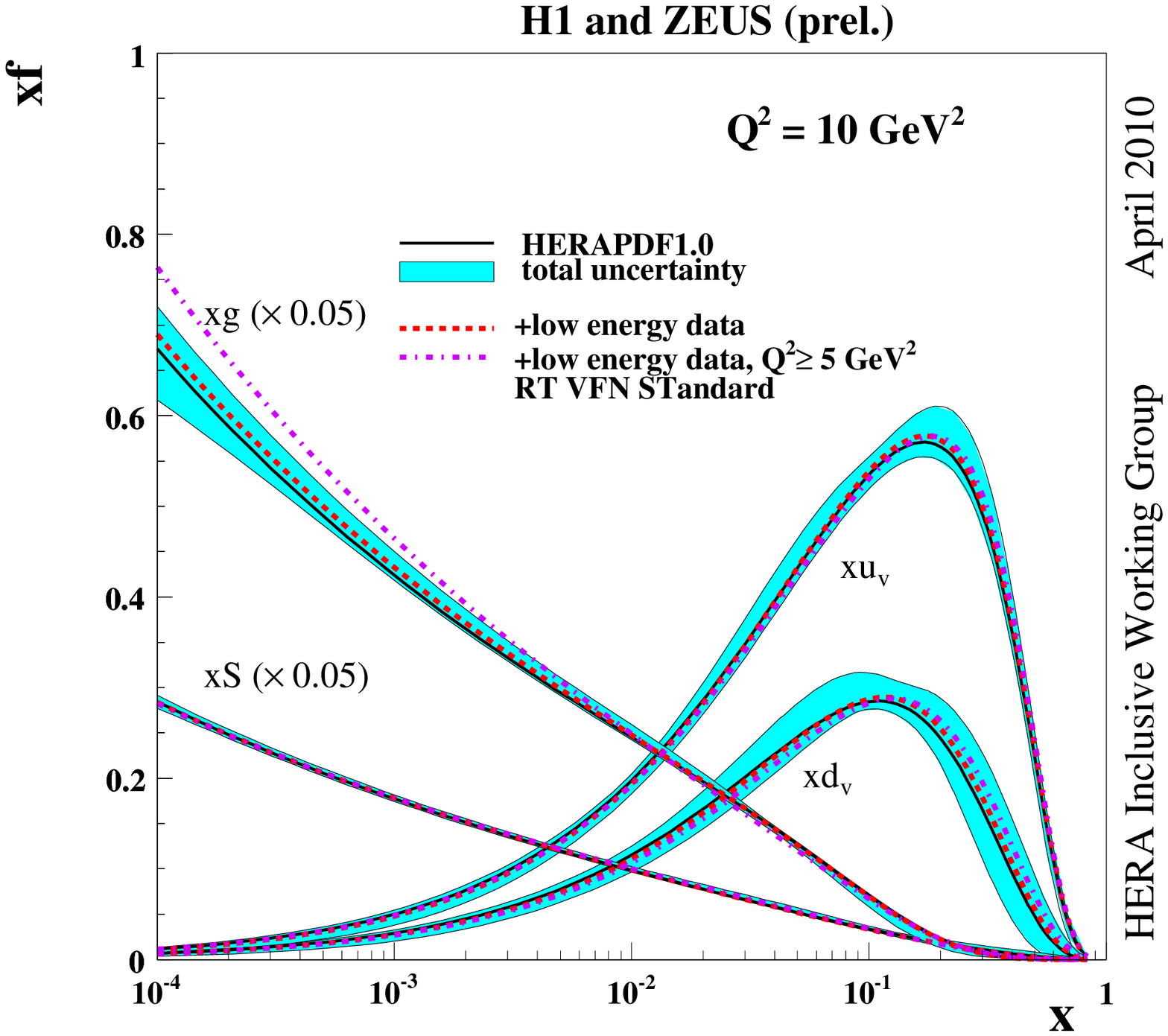}
\caption{Left: HERA combined data points for the NC $e^+p$ cross-section 
for three different proton beam energies. 
Right: PDFs, $xu_v$, $xd_v$, $xS=2x(\bar{U}+\bar{D})$, 
and $xg$ at $Q^2=10~$GeV$^2$, for HERAPDF1.0 and for a HERAPDF fit which also 
includes the low-energy proton beam data, with the standard $Q^2$ cut, 
$Q^2 > 3.5~$GeV$^2$, and for $Q^2 > 5.0~$GeV$^2$.  }
\label{fig:lowenergy}
\end{center}
\end{figure}

In 2007 NC $e^+p$ data were taken at two lower values of the proton beam energy in order
to determine the longitudinal strucure function $F_L$. Some of the H1 and ZEUS 
data sets from these runs have now been combined~\cite{Aaron:2009wt} 
and the results for the NC  $e^+p$ cross section are shown in Fig.~\ref{fig:lowenergy}.
These data have been input to the HERAPDF fit together with the inclusive data 
from HERA-I. The resulting PDFs are compared with those of 
HERAPDF1.0 in Fig.~\ref{fig:lowenergy}. The low energy data are sensitive 
to the choice of minimum $Q^2$ (standard cut $Q^2 > 3.5~$GeV$^2$) 
for data entering the fit. If a 
somewhat harder cut, $Q^2 > 5~$GeV$^2$, is made, a steeper gluon distribution 
results, see Fig.~\ref{fig:lowenergy}, whereas for the HERAPDF1.0 this 
variation 
of cuts results in PDFs which lie within the PDF uncertainty bands. This 
sensitivity is also present 
if an $x$ cut, $x > 5\times 10^{-4}$, or a ``saturation inspired'' cut,
$Q^2 > 0.5~x^{-0.3}$, is made. This sensitivity may indicate the breakdown of 
the DGLAP formalism at low $x$~\cite{lowestudy}.
%

\subsection{High-$Q^2$ data from HERA-II running}
\label{sec:cs-inc2}
\begin{figure}[hbp]
\vspace{-0.20cm} 
\begin{center}
\includegraphics[width=0.495\textwidth]{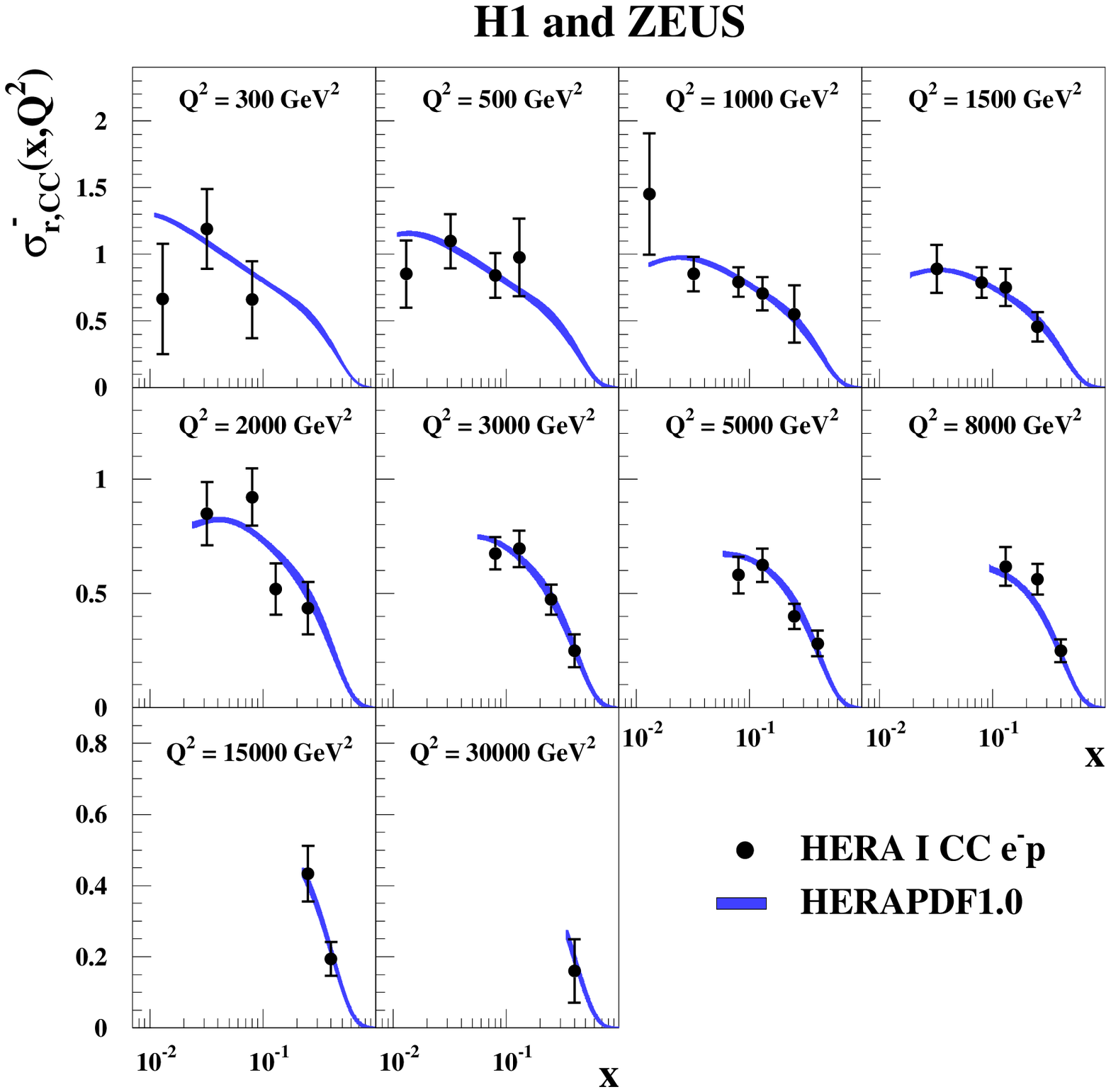}
\includegraphics[width=0.495\textwidth]{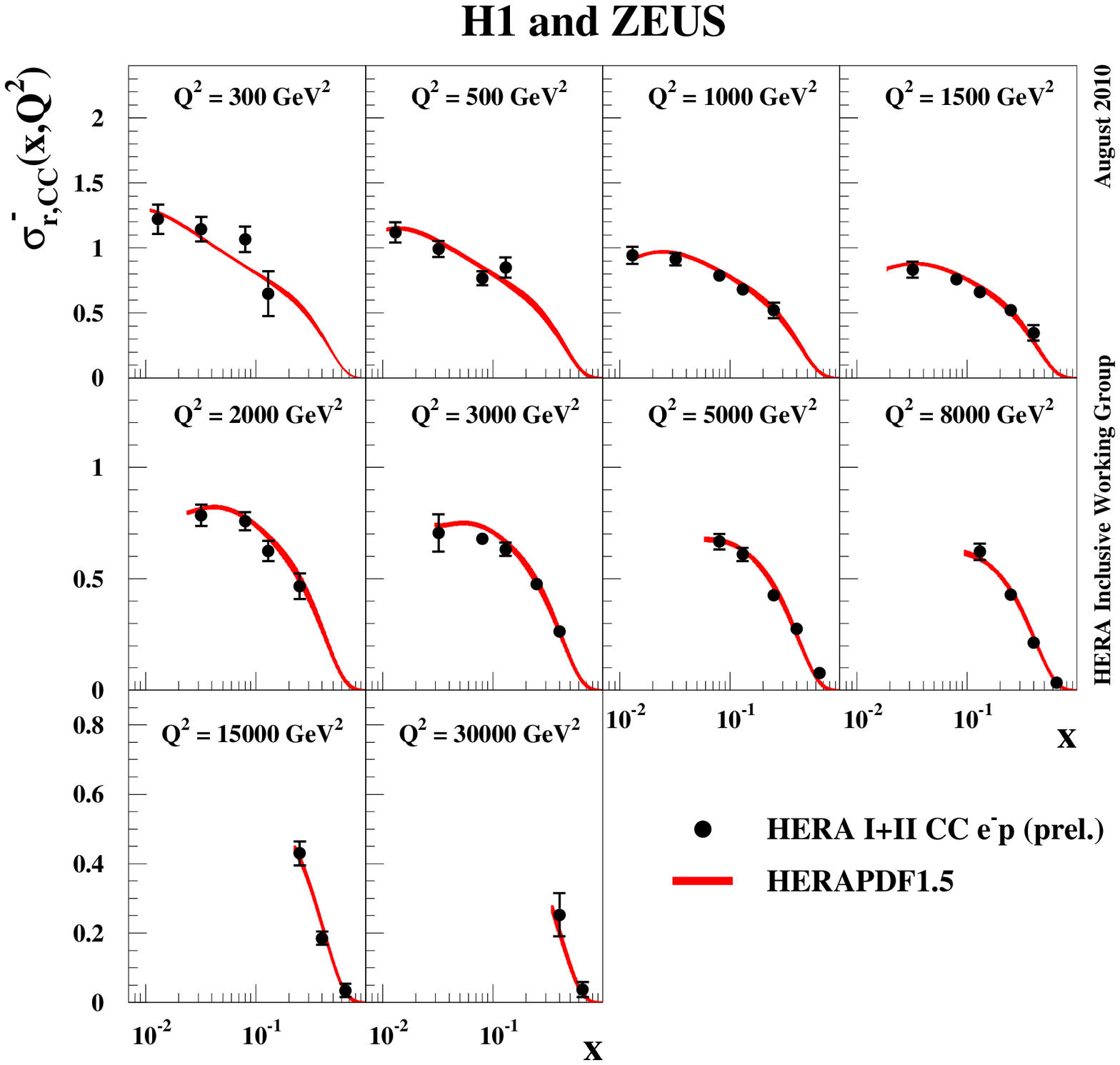}
\caption{HERA combined data points for the CC $e^-p$ cross-section. 
Left: from the HERA-I run period. Right: 
from the HERA-I and II run periods. On each plot the HERAPDF fit which includes
the corresponding data is illustrated: the HERAPDF1.0 fit on the left hand plot and the
HERAPDF1.5 on the right hand plot.}
\label{fig:ccem1015}
\end{center}
%
\vspace{-0.30cm} 
\begin{center}
\includegraphics[width=0.495\textwidth]{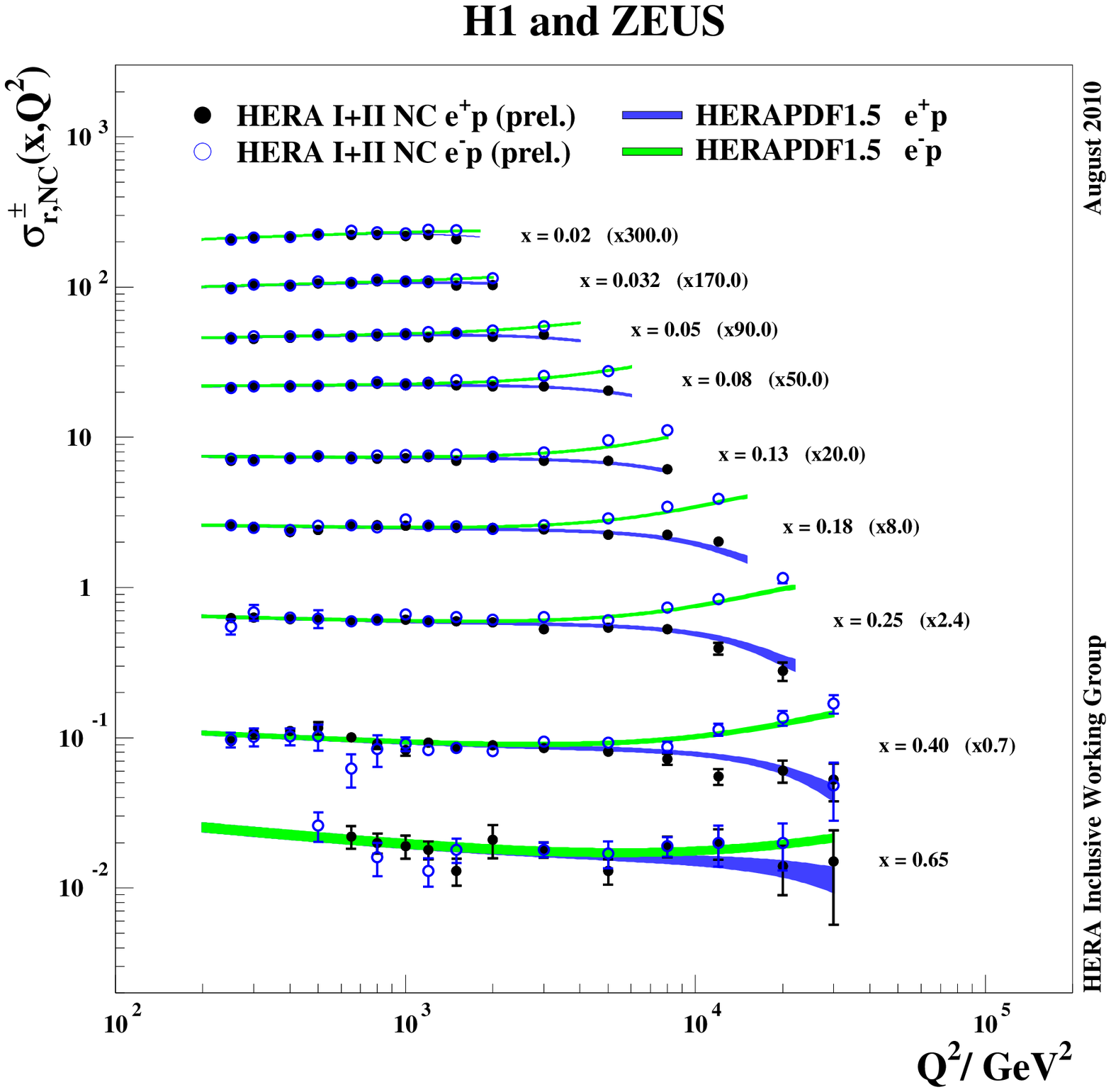}
\includegraphics[width=0.495\textwidth]{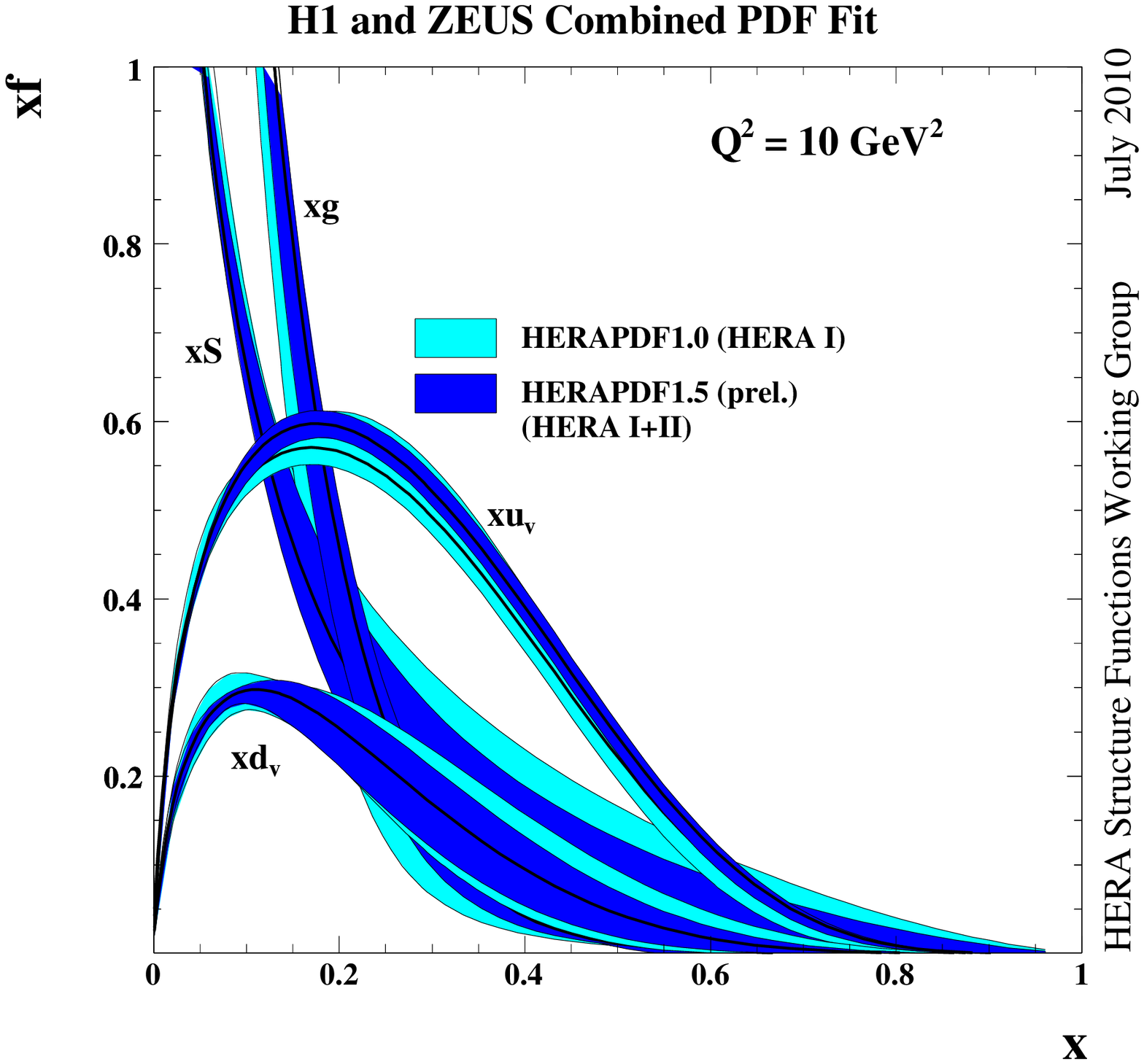}
\caption{Left: HERA combined data points for the NC $e^{\pm}p$ cross-sections 
for data from the HERA-I and II run periods. 
The HERAPDF1.5 fit to these data is also shown on the plot.
Right: Parton distribution functions from HERAPDF1.0 and HERAPDF1.5; $xu_v$, 
$xd_v$,$xS=2x(\bar{U}+\bar{D})$ and $xg$ at $Q^2=10~$GeV$^2$.}
\label{fig:herapdf15}
\end{center}
\vspace*{-0.8cm}
\end{figure}

Preliminary H1 data on NC and CC $e^+p$ and $e^-p$ inclusive cross-sections
and published ZEUS data on NC and CC $e^-p$ and CC $e^+p$ data, from 
HERA-II running, have been combined with the HERA-I data to yield an inclusive 
data set with improved accuracy at high $Q^2$ and high $x$~\cite{highq2}. 
The HERA-I data set and the new HERA I+II data sets are compared for 
CC $e^-p$ data in Fig.~\ref{fig:ccem1015}. This new data set is used as the 
sole input to a new PDF fit called the HERAPDF1.5 which uses the same formalism 
and assumptions as the HERAPDF1.0 fit~\cite{hiq2study}. 
These fits are superimposed on the corresponding data sets in the figure.
Fig.~\ref{fig:herapdf15} (left) shows the combined data for NC $e^{\pm}p$ 
cross-sections with the HERAPDF1.5 fit superimposed. The PDFs
from HERAPDF1.0 and HERAPDF1.5 are compared in Fig.~\ref{fig:herapdf15} (right). 
The improvement in precision at high $x$ is clearly visible.
%
\subsection{Summary}
The status of the combinations of H1 and ZEUS data has been discussed.
HERA leaves rich legacy of results which are the basis for all 
present QCD analyses of unpolarized PDFs and define the goals for
any future DIS experiment.

\section{Unpolarized parton distribution functions:\\ questions to be
  addressed at an EIC}
\label{sec:olness-pdfs}

\hspace{\parindent}\parbox{0.92\textwidth}{\slshape 
  Marco Guzzi, Pavel Nadolsky, Fredrick Olness
%
}

\index{Guzzi, Marco} 
\index{Nadolsky, Pavel} 
\index{Olness, Fredrick} 



\subsection{Introduction}
The Electron-Ion Collider (EIC) will operate at a time when the
Large Hadron Collider (LHC) has established a new ``gold standard'' for perturbative
QCD by measuring a variety of hard-scattering processes. High-luminosity
EIC measurements will be very complementary to those at the LHC, as
they will accurately probe various aspects of hadronic structure using
independent experimental techniques. In the next few years, 
when next-to-next-to-leading
order (NNLO) accuracy of QCD calculations becomes the norm, a variety
of perturbative and nonperturbative effects need to be taken into
account to match the precision of multi-loop radiative contributions.
Some of these effects can be constrained solely by the LHC data; others
need independent measurements, not affected by systematical uncertainties
present at the LHC. With an integrated luminosity of $10\mbox{ fb}^{-1}$ or more,
the EIC will disentangle many such effects, including modifications
of the nucleon structure within heavy-nuclei targets, flavor dependence
of parton distribution functions (PDFs), and QCD dynamics
at very large or small $x$.

As compared with previous lepton--nucleus experiments, 
the EIC will probe to smaller $x$ values with high precision.
In contrast to the HERA $ep$ collider, which explores the same $\{x,Q^2\}$ region, 
heavy-ion scattering will
achieve much higher partonic densities that are a prerequisite for
the onset of saturation. It will help delineate the kinematical
boundary between the DGLAP factorization 
and saturated dynamics in the nuclear medium. 

The $Q^{2}$ range of the EIC will cover the transition region from
the perturbative to the non-perturbative regime. Here, we wish to
learn how the perturbative parton-scattering picture valid at large
momentum transfers matches on nonperturbative models describing the
strongly-coupled resonance region. Understanding of this region is
important for hadronic experiments at the intensity frontier.

\subsection{Open Questions}
\begin{figure}
\centering{}\includegraphics[width=0.55\textwidth]{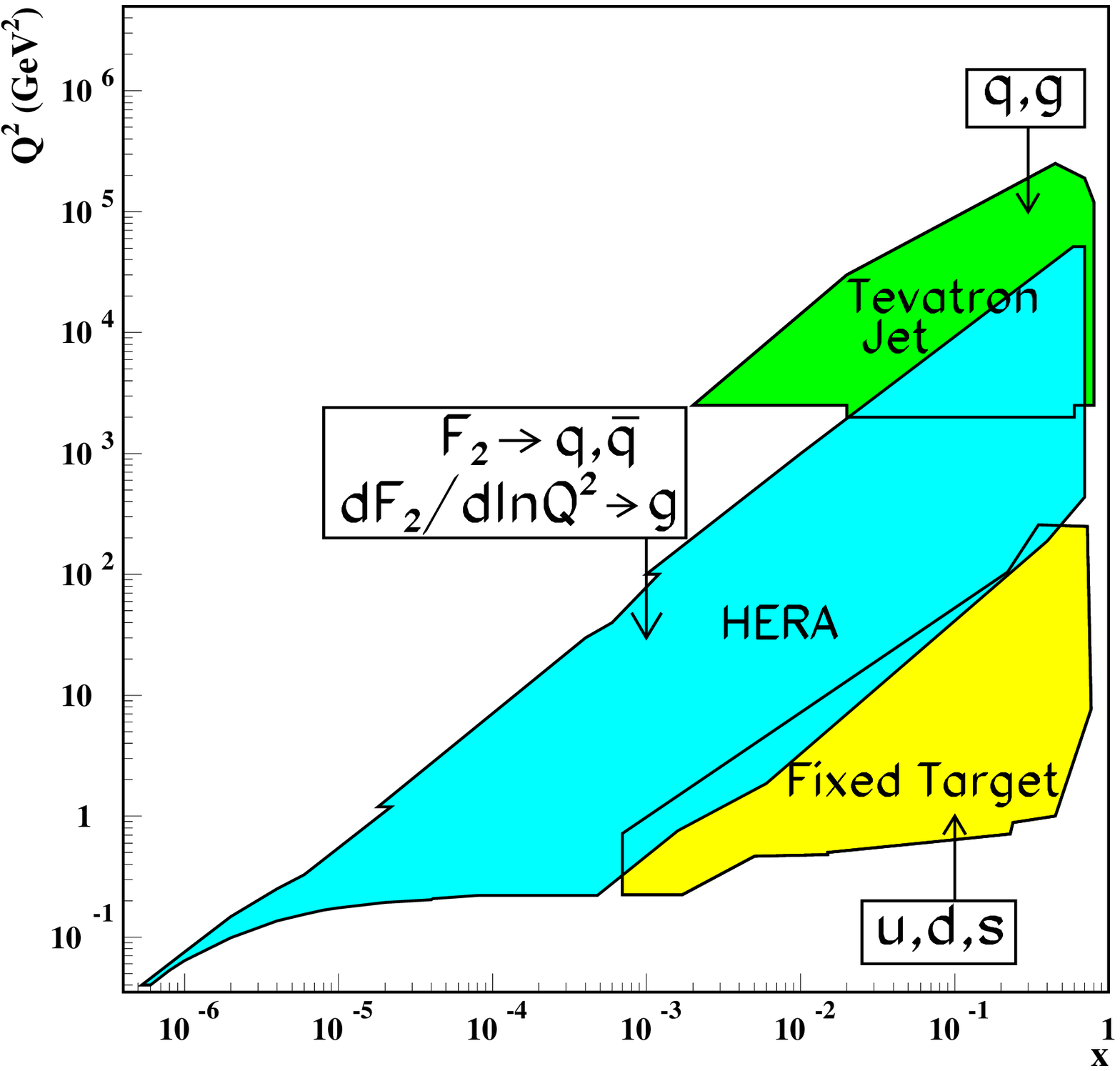}
\caption{Kinematic domains in $x$ and $Q^{2}$ probed by fixed-target and
collider experiments, shown together with the PDFs that are most strongly
constrained by the indicated regions \cite{Nakamura:2010zzi}.
DIS data off nuclear targets exist only in the fixed-target region.
\label{fig:olness:kin}}
\end{figure}

Several questions about PDFs will likely remain open at the time of
the EIC operation. Figure~\ref{fig:olness:kin} shows the kinematic
domains in $x$ and $Q^2$ probed by current experiment and the
PDFs that are most strongly constrained in these reqions.

\textbf{Nuclear PDFs}. 
Several groups extract nuclear PDFs and
their uncertainties by analyzing the global data on nuclear 
targets \cite{Schienbein:2009kk,Kovarik:2010uv,Hirai:2007sx,deFlorian:2003qf,Eskola:2009uj}.
In their studies,
they find that the nuclear corrections depend on the type of the nucleus
(its atomic number $A$), flavor of the probed parton, and even the
type of the probing boson. 
For example, it was found recently \cite{Schienbein:2007fs,Kovarik:2010uv} that
the nuclear correction factors preferred by the $\nu\mbox{Fe}$ DIS data
by NuTeV \cite{Tzanov:2005kr} are surprisingly different from predictions
based on the $\ell^{\pm}\mbox{Fe}$ charged-lepton results. 

By performing deep inelastic scattering (DIS) both on proton and heavy-nuclei
targets, the EIC can distinguish between intrinsic properties of the proton
and those of the extended nuclear medium.
A high-intensity EIC could use a variety of nuclear beams to precisely map the
$A$-dependent nuclear correction factors in the $\{x,Q^{2}\}$ kinematic
plane and clarify the behavior of nuclear corrections to NC
DIS. Such information is of importance for determining the
proton PDFs, in particular, the strange quark PDF that is constrained
largely from the NuTeV data . The nuclear correction affects
the uncertainty in $s(x,Q),$ which is large at present and may limit
the precision of electroweak studies in $W$ and $Z$ boson production
at the LHC \cite{Nadolsky:2008zw}.

The topics of nuclear PDFs and saturation will be extensively discussed
in the Chapter 5 devoted to $eA$ physics at an EIC.

\textbf{Better constraints on the strangeness PDF.} Despite extensive
investigation, there remain large uncertainties in flavor differentiation
of sea-quark PDFs both in the proton and nuclei. In particular, the
strange quark+antiquark distribution in the proton, $s_{+}(x)=s(x)+\bar{s}(x)$,
and its asymmetry, $s_{-}(x)=s(x)-\bar{s}(x)$, are still poorly known
\cite{Olness:2003wz,Lai:2007dq,Lai:2010vv,Martin:2009iq,Ball:2010de},
despite their significance for understanding of the nucleon structure.
Existing constraints on the strangeness come predominantly 
from neutrino (semi-)inclusive DIS \cite{Yang:2000ju,Tzanov:2005kr}. 
At the EIC, both $s_{+}(x)$ and $s_{-}(x)$ can be probed in semi-inclusive
DIS production of kaons; see Sec.~\ref{sec:marco-sidis} for some
quantitative studies.
This measurement will rely on a good understanding
of fragmentation functions, which will be known much better by the
time an EIC turns on. 

\textbf{The $\mathbf{d/u}$ ratio at large $\mathbf{x}$.} Because of its intermediate
energy and high beam intensity, the EIC is ideal for studying parton
distributions at large Bjorken $x$ ($x>0.1$), where separation of
parton flavors is not fully understood despite many years of 
experiments. For example, even the ratio $d(x,Q)/u(x,Q)$
of the dominant up and down quark proton PDFs at $x>0.3$ has been
recently put in doubt by contradicting constraints from
DIS on deuteron targets \cite{Benvenuti:1989rh,Amaudruz:1992bf}
and charged lepton asymmetry 
at the Tevatron \cite{Abazov:2007pm,Abazov:2008qv}.
While the PDF analysis groups labor to understand these
differences \cite{Martin:2009iq,Lai:2010vv,Ball:2010gb} (and new clean
LHC measurements of the $d/u$ ratio in \emph{proton} scattering are
in the queue), the EIC will help to resolve this controversy
by extracting the ratio $F_{2}^{n}(x,Q)/F_{2}^{p}(x,Q)$
from DIS data on various nuclear targets. 
Such measurement will help to separate several types 
of kinematical and nuclear corrections (\cite{Accardi:2009br}, and
references therein) that influence the $F_{2}^{n}/F_{2}^{p}$ ratio 
derived from nuclear-target DIS. 

\textbf{Gluon PDF in the proton and charm production at large $\mathbf{x}$.} 
Even more uncertainty exists in the gluon PDF $g(x,Q)$ at large
$x,$ where it can be larger than the down-quark $d(x,Q)$ at $x>0.5$
in some recent parametrizations for proton PDFs \cite{Pumplin:2009nk}.
This ambiguity will be reduced by upcoming high-$p_T$ jet production at
the LHC, but significant systematic limitations of both experimental
and theoretical nature may persistent at the largest $x$, where the EIC
could independently contribute. Production of heavy-quark ($c,$ $b$)
pairs or heavy mesons ($J/\psi,\Upsilon$) in deep-inelastic scattering could
accurately probe the large-$x$ gluon PDF.
The EIC detectors will have excellent charm tagging efficiency,
in a relatively clean scattering environment as compared to the LHC.

\textbf{Inclusive charm production} is interesting in its own right,
given that large radiative contributions are known to exist near the
heavy-quark production threshold, i.e., at $Q$ comparable
to the charm quark mass; see Sec.~\ref{sec:bluemlein-hq} for a detailed account
of heavy quark contributions to DIS structure functions.
The rate for charm production at large $x$, $x\gtrsim 0.1$,
can be increased by up to an order of magnitude by nonperturbative \textbf{intrinsic
charm} production suggested by light-cone models \cite{Brodsky:1980pb,Brodsky:1981se}. 
An EIC will be a unique opportunity to cleanly
test for the presence of intrinsic charm contributions;
see Sec.~\ref{sec:intrcharm} for some quantitative studies.

\textbf{Transition to the high-density regime. }There is a long-standing
question of partonic saturation and recombination in the small-$x$
region. As a related phenomenon, 
BFKL \cite{Kuraev:1976ge,Kuraev:1977fs,Balitsky:1978ic} effects
from large $\ln\left[1/x\right]$ contributions may supersede the
usual DGLAP evolution in the small-$x$ regime. The EIC should be
capable of probing the transition from DGLAP factorization to BFKL/saturation
dynamics, particularly using heavy nuclei beams in order to produce
large partonic densities; see Chapter 5 for details on $eA$ physics.

\textbf{Perturbative-nonperturbative QCD boundary.} The general kinematic
parameters of an EIC would span across both the perturbative (large
$Q^{2}$) region and the non-perturbative (small $Q^{2}$) region.
The theoretical description of the physics in these two regions is
very different, and precise EIC data might enable us to better connect
these two disparate theoretical descriptions. 

\textbf{The longitudinal structure function.}
The longitudinal structure function $F_{L}=F_{2}-2xF_{1}$ is of special
interest, in view that its leading ${\cal{O}}(1)$ term vanishes according 
to the Callan-Gross relation. 
The first non-vanishing, leading order contribution is of ${\cal{O}}(\alpha_s)$
and dominated by photon-gluon fusion. Hence, $F_L$ is
particularly sensitive to the gluon distribution $g(x,Q^2)$.
Corrections up to ${\cal{O}}(\alpha_s^3)$ are known \cite{Vermaseren:2005qc}, 
allowing for a consistent analysis of $F_L$ at NNLO accuracy.
An EIC could make the first precise measurements of $F_{L}$ in a kinematic
range that overlaps both the fixed-target and HERA collider data
\cite{Collaboration:2010ry} which have large statistical uncertainties;
see Sec.~\ref{sec:marco-fl} for more details on such a measurement
at an EIC.

\textbf{Electroweak contributions to proton PDFs.} Some, if not all,
NLO electroweak effects will be included in future PDF analyses,
as their magnitude is comparable to the size of NNLO QCD radiative
contributions that will be routinely included. The QCD+EW PDFs require
additional experimental input to constrain nonperturbative parametrizations
for photon PDFs, as well as charge asymmetry effects (\textbf{isospin
violation}) between PDFs for up-type quarks and down-type quarks at
the initial scale $Q\approx1$ GeV. An EIC has the potential to contribute
toward improving limits on electroweak PDF terms either directly or
in combination with neutrino DIS measurements. 

When extracting information about the proton PDFs from scattering 
on nuclear targets, we generally make use of isospin symmetry
to relate the proton and neutron PDFs via a $u\leftrightarrow d$ interchange. 
While the isospin symmetry is elegant, 
it is nonetheless
approximate and can be violated at the level of a few percent
\cite{Boros:1999fy,Ball:2000qd,Kretzer:2001mb,Martin:2001es,Boros:1998es,Boros:1998qh,Baldit:1994jk,Schienbein:2007fs,Adams:2009kp}. 
Violation of the exact $p\leftrightarrow n$ isospin symmetry,
or charge symmetry violation (CSV), invalidates 
the parton model relations that reduce the number of 
independent nonperturbative distributions; 
e.g., $u^{n}(x)\not\equiv d^{p}(x)$
and $u^{p}(x)\not\equiv d^{n}(x)$.
It is important to be aware of the potential magnitude of isospin symmetry 
violation and its consequences for flavor separation of proton PDFs. 

It is noteworthy that isospin symmetry is automatically violated both
perturbatively and nonperturbatively. This is because the photon couples
to the up quark distribution $u^{p}(x)$ differently than to the down
quark distribution $d^{n}(x)$. These terms can be comparable to the
NNLO DGLAP evolution effects \cite{Martin:2004dh,Roth:2004ti,Gluck:2005xh}.

Some combinations of structure functions, such as 
$\Delta F_{2}\equiv\frac{5}{18}\, F_{2}^{CC}(x,Q^{2})-F_{2}^{NC}(x,Q^{2})$
and $\Delta xF_{3}=xF_{3}^{W^{+}}-xF_{3}^{W^{-}}$, can be particularly
sensitive to isospin violations, and an EIC can contribute to their
measurement. For example, the EIC is capable of measuring precisely 
the structure function 
$F_{2}^{NC}$ mediated by the neutral-current $\gamma/Z$ exchange processes. 
Measurement of $F_{2}^{CC}$, mediated by the charged-current
$W^{\pm}$ exchange, would rely on compensating
the $M_{W}^{2}/Q^{2}$ suppression of the $W$ boson propagator with high 
intensity of the beams; see Sec.~\ref{sec:electroweak} for more
details on electroweak structure function measurements at an EIC.

In separate experiments, $\Delta xF_{3}$ can be measured precisely
via the neutrino-nucleon DIS process; as these measurements are performed
with heavy nuclear targets, the nuclear correction factors can be
the limiting factor as to the derived CSV constraint. Since an EIC
will use a variety of nuclear targets, it can obtain very precise
nuclear correction factors; this information could, in principle,
be used together with the neutrino-nucleon DIS data to extract improved
CSV limits. 

The structure functions $\Delta F_{2}$ and $\Delta xF_{3}$ receive
contributions from both heavy flavors as well as CSV contributions;
improved understanding of the heavy-quark components (discussed previously)
can indirectly contribute to better CSV limits \cite{Adams:2009kp}. 

The combination of high-statistics EIC measurements and constraints 
could thus yield important information on the fundamental charge symmetry.

\subsection{Acknowledgments}


FIO thanks I.~Schienbein, J.Y.~Yu, K.~Kovarik, C.~Keppel, J.G.~Morfin,
J.F.~Owens, K.~Park, and T.~Stavreva for valuable discussions.

\section{Flavor separation from semi-inclusive DIS}
\label{sec:marco-sidis}

\hspace{\parindent}\parbox{0.92\textwidth}{\slshape 
Elke-Caroline Aschenauer, Marco Stratmann
%
}

\index{Aschenauer, Elke}
\index{Stratmann, Marco}



\subsection{Motivation and Method}
The strangeness distribution and a possible asymmetry between strangeness and
anti-strange\-ness densities have been identified as two of the most compelling open
questions in hadronic physics which are difficult to address without an EIC;
see Sec.~\ref{sec:olness-pdfs}.

\begin{figure}[th!]
\begin{center}
\psfig{figure=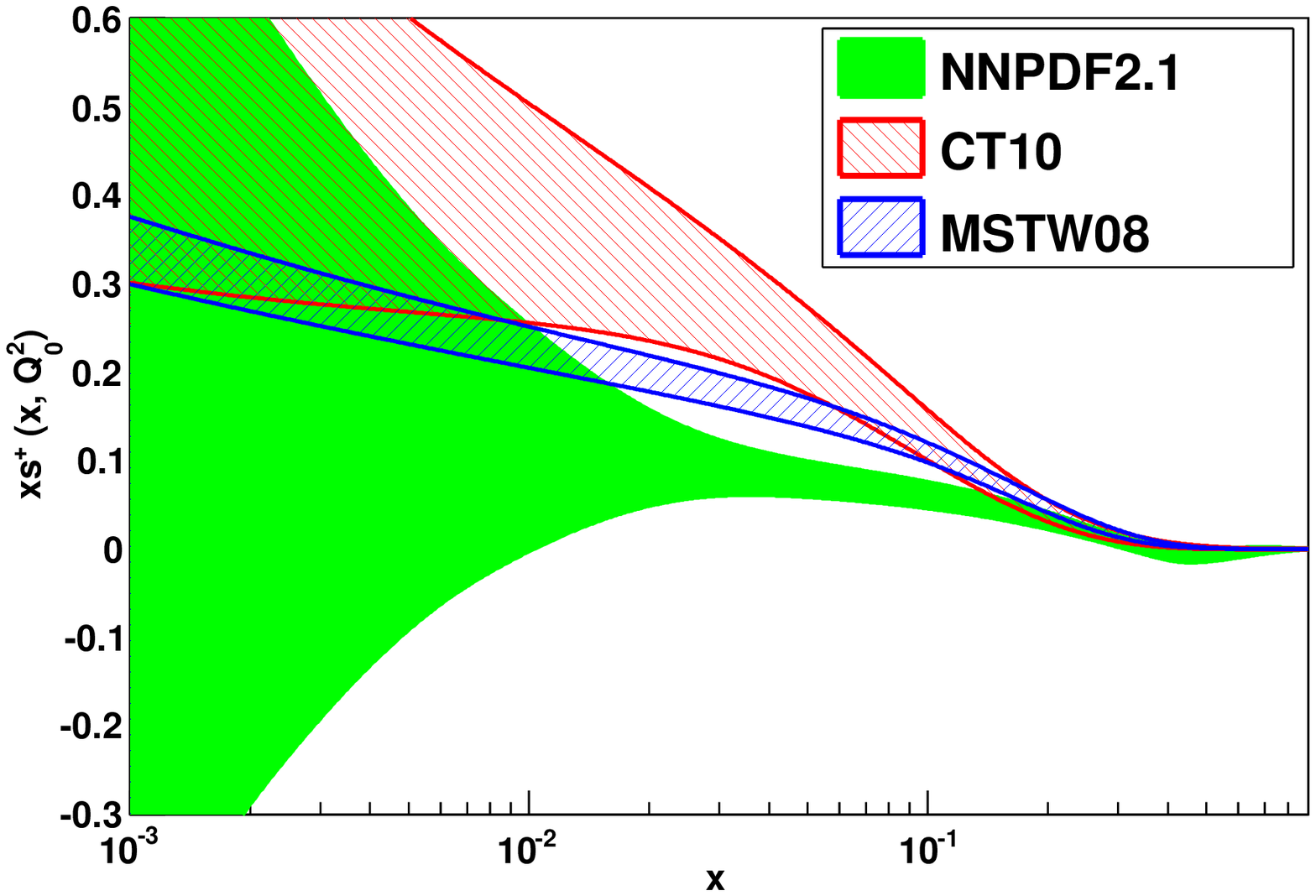,width=0.32\textwidth}
\psfig{figure=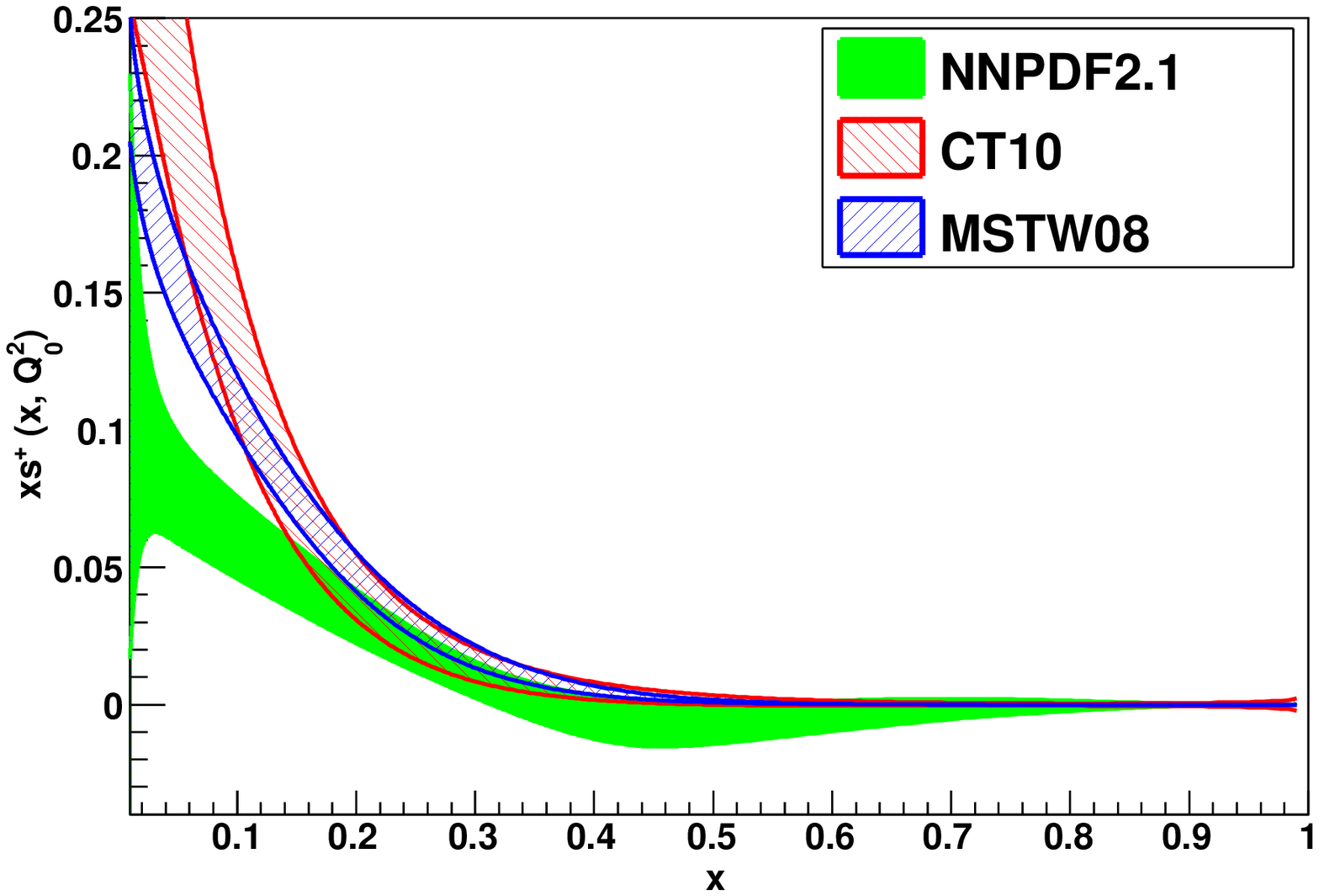,width=0.32\textwidth}
\psfig{figure=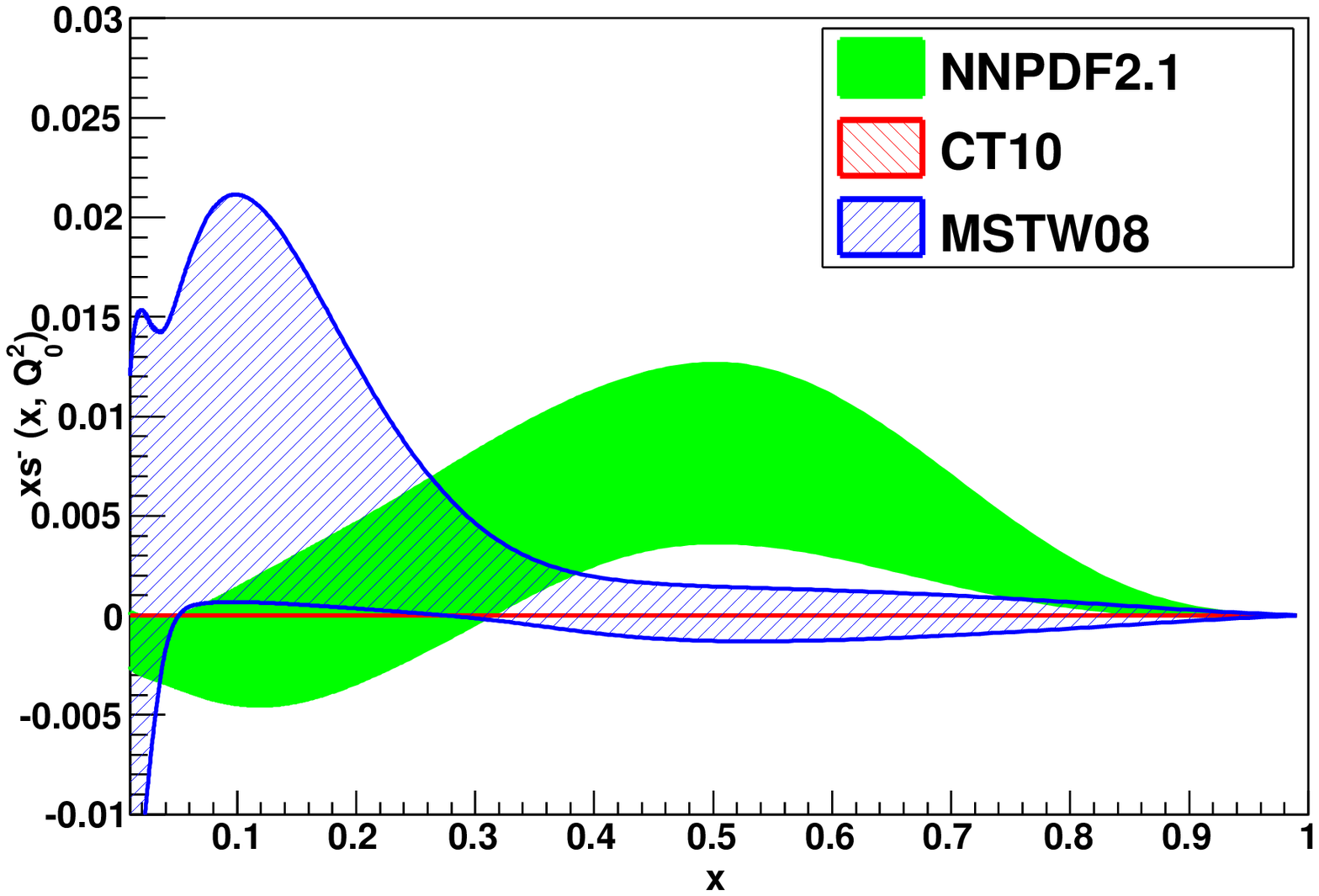,width=0.32\textwidth} 
\end{center}
\vspace*{-0.3cm}
\caption{Uncertainty bands for $s_\pm$ at $Q^2=2\,\mathrm{GeV^2}$ for recent fits.
Figure taken from \cite{Ball:2011mu}.}
\label{fig:strangeness-status}
\end{figure}

Existing constraints in global fits come predominantly 
from neutrino (semi-)inclusive DIS \cite{Yang:2000ju,Tzanov:2005kr}
but both $s_{+}(x)\equiv s(x)+\bar{s}(x)$ and $s_{-}(x)\equiv s(x)-\bar{s}(x)$
are still only poorly known \cite{Lai:2010vv,Martin:2009iq,Ball:2011mu}.
Figure~\ref{fig:strangeness-status} summarizes recent uncertainty estimates for $s_{\pm}$
from three global QCD fits.

Semi-inclusive DIS with identified charged kaons is expected to be a viable method
to determine the elusive strange quark density and perhaps a possible asymmetry $s_-$
experimentally. One can access basically the same a broad kinematic range in $x$ and $Q^2$
as in inclusive DIS. The HERMES collaboration has successfully performed such a measurement in
the range $0.02<x<0.6$ at an average $Q^2$ of about $2.5\,\mathrm{GeV}$ \cite{Airapetian:2008qf}.
Compared to $s(x)$ from most global PDF fits, they find a softer strangeness distribution
in their LO analysis. Clearly, more data in a larger
range of $x$ and $Q^2$ are necessary to clarify this issue.

The SIDIS measurement relies, however, on a good understanding of the 
hadronization mechanism which is encoded in 
non-perturbative, collinear parton-to-hadron fragmentation functions (FFs) 
$D^H_i$ if factorization is
assumed in a pQCD calculation. 
Like PDFs, FFs are extracted from global QCD analyses. One can resort to a wealth of 
single-inclusive hadron production data obtained at different c.m.s.\ energies 
in $e^+e^-$ annihilation and in $ep$ and $pp$ ($p\bar{p}$) scattering. 
Pion FFs are currently known best with uncertainties of
about $5\div 10\%$ depending on the flavor of the fragmenting parton \cite{deFlorian:2007aj}. 
Ambiguities for kaon FFs are about twice as large \cite{deFlorian:2007aj}.
Significant progress on the quality of fits to FFs
is expected once data from $B$ factories and the LHC become available. Also,
NNLO evolution kernels are expected to become available in the near future \cite{Moch:2007tx},
which will help to reduce theoretical scale ambiguities further.

All relevant SIDIS cross sections are known at least to NLO accuracy 
\cite{Altarelli:1979kv,Furmanski:1981cw,Nason:1993xx,deFlorian:1997zj}, and the
analytical expressions are relatively simple and easy to implement into global fits
of PDFs, see, e.g., \cite{Stratmann:2001pb}. Schematically the unpolarized SIDIS cross section for the production of a
hadron $H$ in the current fragmentation region reads
\begin{equation}
\frac{d\sigma^H}{dx dy dz} = \frac{ 2\pi\alpha^2}{Q^2} \left[
\frac{1+(1-y)^2}{y} 2 F_1^H(x,z,Q^2) + \frac{2(1-y)}{y} F_L^H(x,z,Q^2) \right]
\label{eq:sidis}
\end{equation}
with $x$ and $y$ denoting the usual DIS variables, $-q^2=Q^2=S x y$, and $z=p_H\cdot p/p\cdot q$
the momentum fraction taken by the hadron $H$. 
Assuming factorization, the structure functions $F_{1,L}^H$ at a factorization scale $\mu\sim Q$ 
can be expressed as convolutions of non-perturbative PDFs $f_j(x,\mu)$ and FFs $D_i^H(z,\mu)$ with short-distance 
Wilson coefficients $C_{ij}^{1,L}(x,z,\mu)$.

\subsection{Expectations for Charged Kaon Production at an EIC} 
\begin{figure}[th!]
\vspace*{-0.3cm}\begin{center}
\psfig{figure=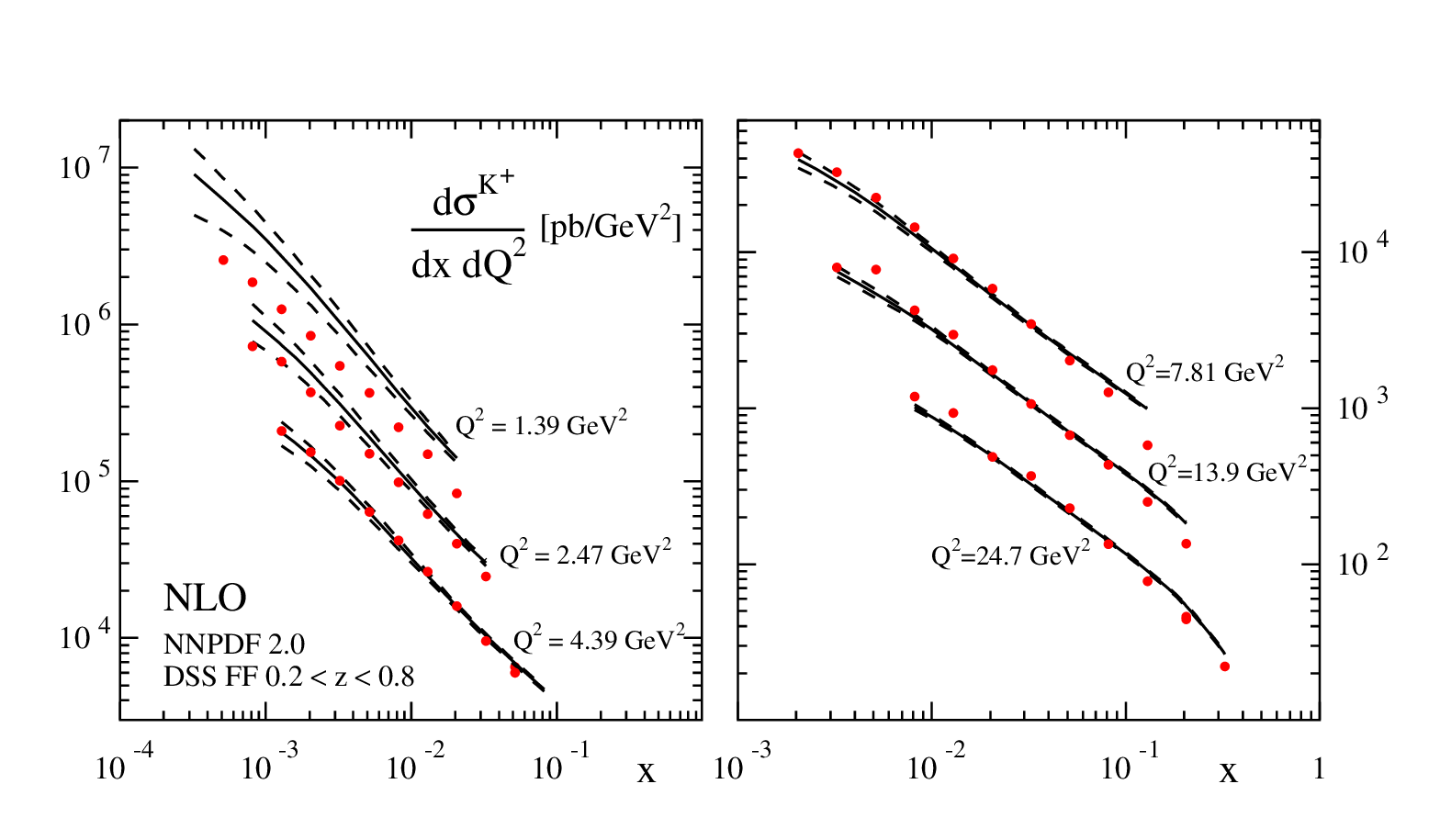,width=0.76\textwidth}
\end{center}
\vspace*{-0.3cm}
\caption{SIDIS cross section for $K^+$ production at NLO accuracy using NNPDF2.0 PDFs \cite{Ball:2010de}.
The dashed lines denote the PDF uncertainties.
Also shown (points) are the results from a PYTHIA simulation (see text).}
\label{fig:kplus}
\vspace*{-0.3cm}
\begin{center}
\psfig{figure=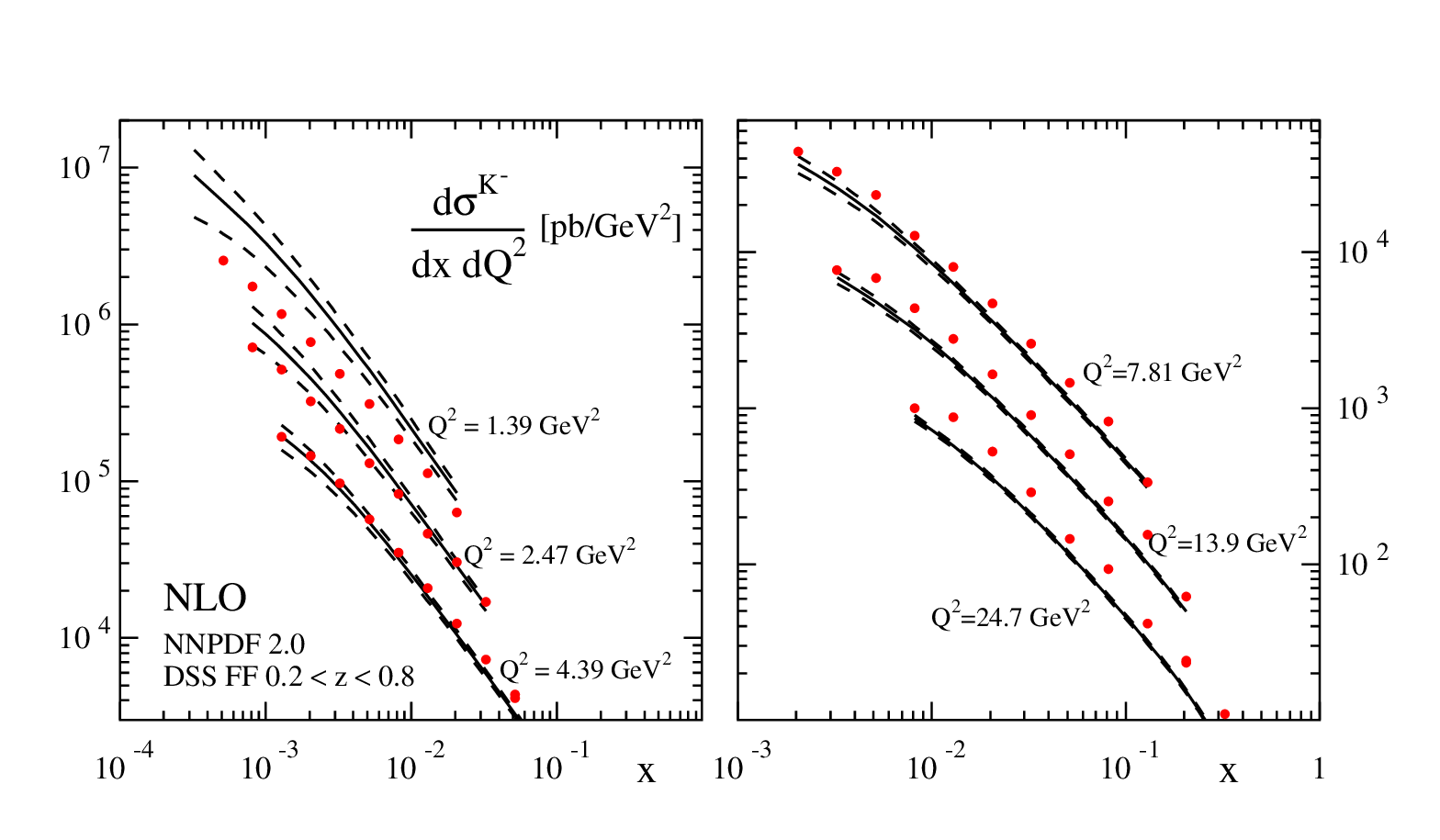,width=0.76\textwidth}
\end{center}
\vspace*{-0.3cm}\caption{
Same as in Fig.~\ref{fig:kplus} but now for $K^-$ production.}
\label{fig:kminus}
\vspace*{-0.3cm}
\end{figure}

%
Figures~\ref{fig:kplus} and \ref{fig:kminus} show expectations for the $K^+$ and $K^-$ production
cross section (\ref{eq:sidis}) at NLO accuracy, respectively, as a function of $x$ in bins of $Q^2$,
using $0.01\le y\le 0.95$ and $\sqrt{S}=70.7\,\mathrm{GeV}$ (i.e., $5\times 250$ GeV collisions at an EIC).
To reduce uncertainties from kaon FFs, $z$ is integrated in the range $0.2\le z\le 0.8$.  The DSS
set \cite{deFlorian:2007aj} is used.
The solid lines are the statistical average over 100 replicas in the NNPDF2.0 
neural network analysis \cite{Ball:2010de} and the dashed lines reflect the corresponding PDF uncertainties.

Also shown in Figs.~\ref{fig:kplus} and \ref{fig:kminus} are simulations based on the PYTHIA~\cite{Sjostrand:2006za}
event generator in the same kinematic range. 
Here, the CTEQ6L set of PDFs \cite{Pumplin:2002vw} has been used. The hadronic
final state was simulated using JETSET based on LEP fragmentation settings and a suppression of 
$s\bar{s}$ pair production from the vacuum of 0.3 [PARJ (2)]
compared to $u\bar{u}$ or $d\bar{d}$ creation. 
The results turn out to be remarkably similar to the NLO calculations based on collinear factorization
despite the very different way hadronization is implemented in PYTHIA and the fact that only
LO matrix elements are used, albeit matched with a parton shower.  
This gives us quite some confidence that the PYTHIA generator can be used to provide very
reasonable estimates of yields for DIS-type processes at an EIC. 
In addition, it also tells us that the current DSS kaon FFs are doing a good job and
include a realistic amount of ``strangeness suppression''.
Already after one month of operation, corresponding to an integrated luminosity of about $20\mathrm{fb}^{-1}$
the measurement will be limited by systematic uncertainties which need to be carefully studied.
The statistical accuracy is significantly better than indicated by size of the points shown in the figures. 

If one compares the results for $K^+$ and $K^-$ in Figs.~\ref{fig:kplus} and \ref{fig:kminus}
one finds hardly any difference at the smallest $x$ values in each $Q^2$ bin. 
At larger $x$ values, where $s_-$ is largest, see Fig.~\ref{fig:strangeness-status},
the yields for $K^-$ are significantly lower than the ones for $K^+$. An EIC should be able to provide
accurate measurements of both $s_+$ and $s_-$ in a broad kinematic range up to $Q^2$ values of a
few hundred GeV.

Within the neural network approach it is in principle fairly straightforward to quantify by how much
a new data set will reduce present PDF uncertainties.
The original ensemble of replicas is constructed in such a way that all have the same weight.
Information contained in new data sets can be incorporated without the need for refitting
by reweighting each PDF in the ensemble by the probability that it agrees with the 
new data \cite{Ball:2010de,Ball:2010gb}. 
Sets with small weights will become largely irrelevant in statistical averages. 
If too many sets receive small weights the accuracy of results from the new PDF ensemble will
deteriorate, and the reweighting procedure becomes unreliable, necessitating a full refit.
One reason for this to happen is, that the new data set contains significant new information
which leads to much smaller uncertainties in certain kinematic regions. This is exactly what
happens when one applies the reweighting method to the SIDIS data shown in Figs.~\ref{fig:kplus} and \ref{fig:kminus}
even if one assigns a fictitious ${\cal{O}}(5\%)$ systematic uncertainty to each data point.

There are many other things which can be studied in SIDIS at an EIC. For instance, one can also bin
in $z$ which makes the measurement more sensitive to the shape of the kaon FFs.
This will provide a more stringent check whether FFs are universal functions in
$e^+e^-$, $ep$, and $pp$ scattering. Pion yields will allow one to study other interesting and
relevant PDF combinations such as $\bar{u}(x)-\bar{d}(x)$.
Similar measurements can be also done with longitudinally polarized beams which will give access
to the helicity-dependent quark and antiquark densities, see Sec.~\ref{sec:rodolfo-spin}.
Detailed quantitative studies including more time-consuming global QCD analyses with simulated SIDIS
data for various c.m.s.\ energies are planned to quantify the impact of such measurements
on our understanding of the spin and flavor structure of the nucleon.
These studies should include also some estimates of the various sources of systematic uncertainties, like
detector resolution, uncertainties in the particle identification, luminosity, and polarization measurements, details
on these can be found in Sec.~\ref{sec:detector}. 

\section{The longitudinal structure function $F_L$ at an EIC}
\label{sec:marco-fl}

\hspace{\parindent}\parbox{0.92\textwidth}{\slshape 
  Elke C.\ Aschenauer, Ramiro Debbe, Marco Stratmann
%
}

\index{Aschenauer, Elke}
\index{Debbe, Ramiro}
\index{Stratmann, Marco}



\subsection{Motivation and Current Status of $F_L$ Results}
%
The DIS reduced cross section $\sigma_r$ for one-photon-exchange can be represented as the sum of 
two independent structure functions $F_2$ and $F_L$ as follows
\begin{equation}
\sigma_r\equiv \frac{Q^4 x}{2\pi \alpha_{em}^2 Y_+} 
\frac{d^2\sigma}{dx dQ^2} = F_2(x,Q^2) - \frac{y^2}{Y_+} F_L(x,Q^2)
\label{eq:redxsec}
\end{equation}
where $Y_+\equiv 1+(1-y)^2$ depends on the inelasticity $y=Q^2/(s x)$ of the process.

$F_L$ is proportional to the cross section for probing the 
proton with a longitudinally polarized virtual photon and vanishes 
in the naive Quark Parton Model due to helicity conservation.
Starting from ${\cal{O}}(\alpha_s)$, the longitudinal structure function differs from zero, 
receiving contributions from both quarks and gluons.

At low $x$, the gluon contribution due to photon-gluon fusion
greatly exceeds the quark contribution. 
Therefore, measuring $F_L$ provides a rather direct way of studying the gluon density and QCD
dynamics at small $x$, i.e., the transition to the high parton density regime.
Measurements can be used to test several phenomenological and QCD models describing the low $x$ behavior 
of the DIS cross section, including color dipole models 
\cite{GolecBiernat:1998js,Iancu:2003ge,Albacete:2009fh} and expectations from DGLAP fits
performed at NLO and NNLO accuracy of QCD. Possible deviations from the DGLAP behavior in the small $x$, low $Q^2$
region can be studied by varying kinematic cuts to the data used in the fits.

\begin{figure}[tbp]
\vspace{-2.0cm}
\begin{center}
\begin{tabular}{ll}
\psfig{figure=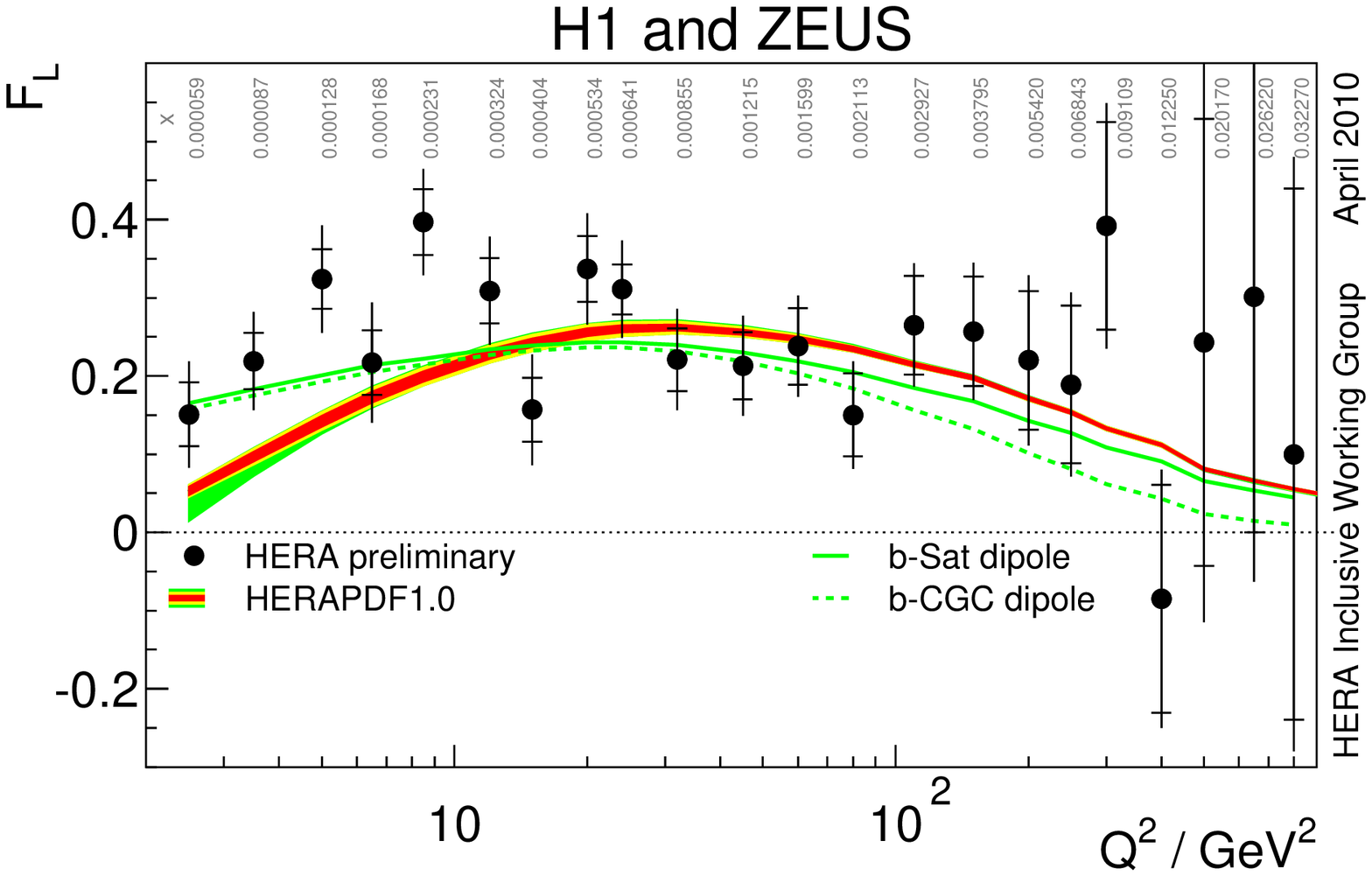,width=0.70\textwidth}
\end{tabular} 
\caption{
Combined H1 and ZEUS extraction of $F_L$ \cite{Aaron:2009wt}
as a function of $Q^2$ averaged over $x$ compared to
the HERAPDF1.0 fit and predictions from dipole models.}
\label{fig:fl-hera-status}
\end{center}
\vspace*{-0.3cm}
\end{figure}

The longitudinal structure function, or the equivalent cross section ratio 
$R = \sigma_L/\sigma_T= F_L/(F_2 - F_L)$, was first measured in fixed target experiments and found to be small 
at large $x$, $x \ge 0.01$, see, e.g., Ref.~\cite{Arneodo:1996qe}.
H1 \cite{Collaboration:2010ry} and ZEUS \cite{Chekanov:2009na} have recently combined their measurements of $\sigma_r$ for three different proton beam energies \cite{Aaron:2009wt}, $E_p=920, 575,$ and 460 GeV, 
see Fig~\ref{fig:lowenergy} in Sec.~\ref{sec:cooper-sarkar}.
The extracted $F_L$, shown in Fig.~\ref{fig:fl-hera-status}, covers a wide kinematic range, spanning
$2.5<Q^2<800\,\mathrm{GeV}^2$ and $0.0006<x<0.0036$.
As can be seen, $F_L$ is clearly non-zero, and there is some mild tension with the HERAPDF1.0 
fit based on DGLAP evolution \cite{Aaron:2009wt} at the lowest values of $x$ and $Q^2$ where one expects non-linear effects to be relevant;
see Chapter 5 on $eA$ physics.
In this regime, predictions from the dipole model provide a better description of the data.
However, the achieved statistical precision of the combined H1 and ZEUS
measurement is too limited to be conclusive.
%

\subsection{Measurement Strategy and Experimental Challenges}
%
The measurement of $F_L$ relies on an accurate determination of the variation of the reduced 
cross section (\ref{eq:redxsec}) for common values of the $(x,Q^2)$ bin centers 
at different beam energies, i.e., c.m.s.\ energies $\sqrt{s}$.
Relative normalizations and systematic uncertainties of the different data sets for $\sigma_r$
have to be well under control.

$F_L$ and $F_2$ can be extracted simultaneously from $\sigma_r$ by plotting 
$\sigma_r$ for fixed values of $(x,Q^2)$ as a function of $y^2/Y_+$. 
$F_L$ is then determined as the slope of the
line fitted to the measurements of $\sigma_r$ for different values of $\sqrt{s}$:
$F_L(x,Q^2)=-\partial\sigma_r(x,Q^2,y)/\partial(y^2/Y_+)$. Likewise, $F_2$ is the intercept
of the fitted line with the $y$ axis: $F_2(x,Q^2)=\sigma_r(x,Q^2,y=0)$.
All measurements at HERA are observed to be consistent with the expected linear 
dependence \cite{Collaboration:2010ry,Chekanov:2009na,Aaron:2009wt}.
At any given value of $Q^2$, the lowest possible $x$ values are only accessed by the highest
$\sqrt{s}$, and the slope related to $F_L$ cannot be determined. Hence, the Rosenbluth
separation limits the kinematic coverage of $F_L$ at small $x$.
At larger values of $x$, measurements of $\sigma_r$ for various different $\sqrt{s}$ are available
and the slopes can be straightforwardly extracted.

The contribution of $F_L$ to the reduced cross section (\ref{eq:redxsec}) can be sizable only at large values 
of $y$. For low values of $y$, $\sigma_r$ is very well approximated by the structure function $F_2$
\cite{Collaboration:2010ry,Chekanov:2009na,Aaron:2009wt}. Low $y$ data can be used to normalize
data sets taken at different c.m.s.\ energies relative to each other.
For measurements at high $y$ the reconstruction of the DIS kinematics using the scattered lepton, 
the so called ``electron method'', has the best resolution and was used at HERA.

In the large $y$ region, $y\gtrsim 0.5$, and low $x$ the electron method is prone to large 
QED radiative corrections which can reach a level of more than $50\%$ of the Born cross section.
Studies based on the DJANGO \cite{Schuler:1991yg} and HECTOR \cite{Arbuzov:1995id} programs 
for HERA kinematics show that the largest radiative contributions
arise because of hard initial-state radiation (ISR) from the incoming lepton \cite{Collaboration:2010ry}.
The radiated photon usually escapes in the beam pipe and the $E-P_z$ of the event is reduced.
Therefore, hard ISR can be efficiently suppressed to a level of about $10\%$ at HERA
with only a slight residual dependence on $y$
by requiring $E-P_z$ close to the nominal value of twice the electron beam energy implied by
energy-momentum conservation \cite{Collaboration:2010ry}.
$E-P_z$ can be reconstructed from the measured final-state
particles. At the highest $y$, $y\gtrsim 0.7$, corrections increase due to QED Compton events which can be
rejected by certain topological cuts.
All cross section measurements at HERA are corrected for QED radiation up to ${\cal{O}}(\alpha_{em})$ 
using HERACLES \cite{Kwiatkowski:1990es} which is included in the DJANGOH package; further details
can be found in Sec.~\ref{sec:detector}. 

Kinematically, for low $Q^2$, large values of $y$ correspond to low energies of the scattered lepton. 
Selecting high $y$ events is thus further complicated due to a possibly large background from 
energy deposits of hadronic final state particles leading to fake electron signals. However, the cut
on $E-P_z$ also suppresses such type of backgrounds. In addition, electron tracking, which is foreseen for an EIC
detector, will largely eliminate fake electron signals as an additional cut on $E/p\simeq 1$ can be placed
to identify the lepton.

Extractions of $F_L$ are certainly the most demanding inclusive structure function measurements but
an EIC will have many advantages compared to HERA, in particular, the possibility to vary $\sqrt{s}$ in a wide
range for high luminosity collisions. Also, much better detector capabilities, for instance, concerning the electron, 
are foreseen. One can also take advantage of all the analysis techniques and Monte Carlo codes
developed for HERA to deal with QED radiative corrections.

\begin{figure}[tbp]
\vspace{-2.0cm}
\begin{center}
\begin{tabular}{ll}
\psfig{figure=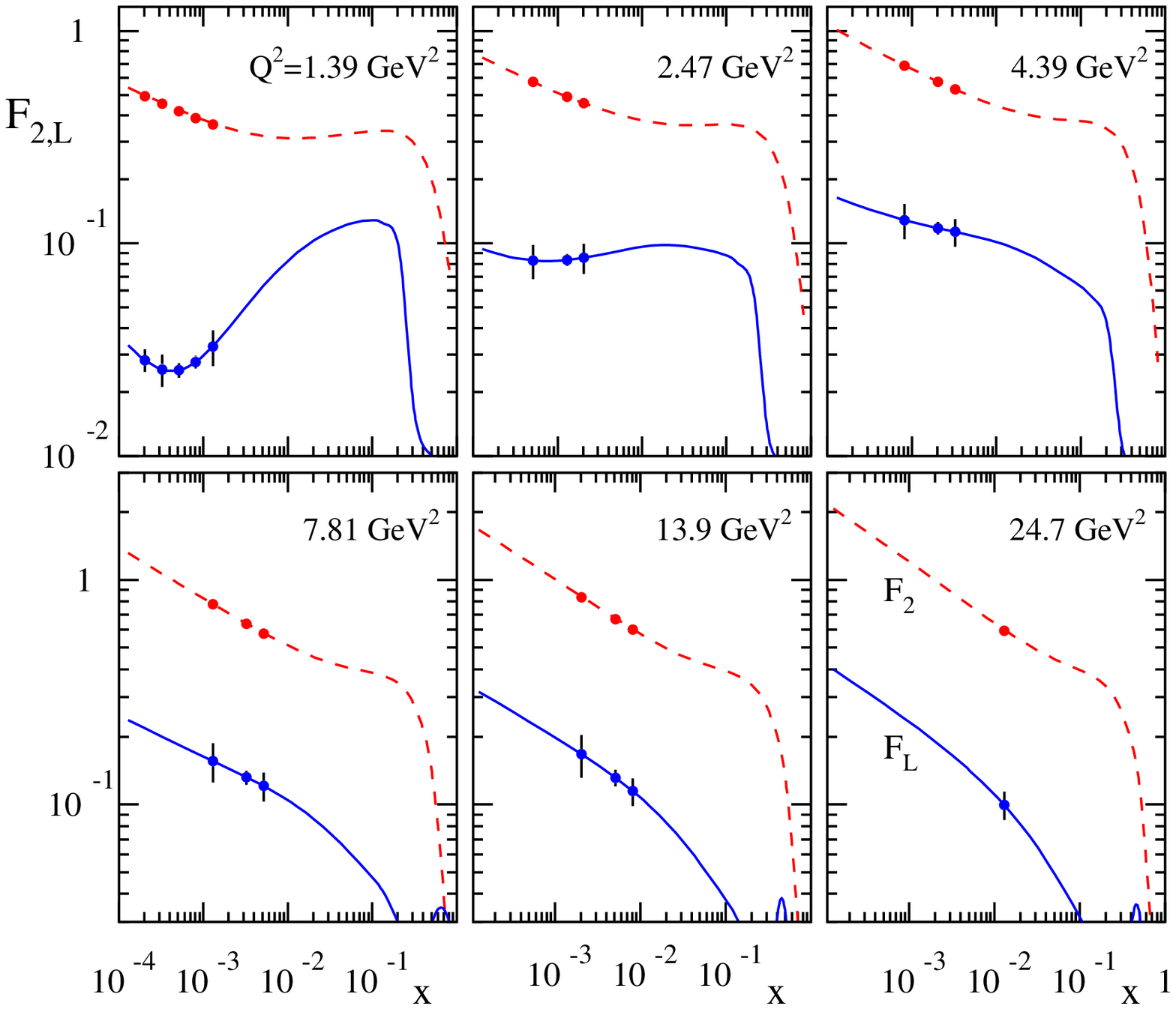,width=0.875\textwidth}
\end{tabular} 
\caption{Projected uncertainties for an extraction of $F_2$ and $F_L$ from a Rosenbluth separation
for data taken at three different c.m.s.\ energies. Also shown are theoretical expectations at NNLO
based on the ABKM09 set of PDFs \cite{Alekhin:2009ni} (see text).}
\label{fig:fl-eic}
\end{center}
\vspace*{-0.3cm}
\end{figure}
\subsection{Expectations for the EIC}
Pseudo-data for the reduced cross section (\ref{eq:redxsec})
have been generated using the Monte Carlo generator LEPTO \cite{Ingelman:1996mq}
for the first stage of an EIC (5 GeV electrons on 100, 250, and 325 GeV protons).
The CTEQ6L set of PDFs \cite{Pumplin:2002vw} has been used in the simulations. 
The hadronic final state was simulated using JETSET \cite{Sjostrand:2006za}. 
We note that the current pseudo-data do not include any simulations of QED radiative effects
and reflect statistical uncertainties which could be achieved by running one month at each of the beam energy
settings with the projected luminosities for eRHIC. In addition, a $1\%$ systematic uncertainty is added.
 
Figure \ref{fig:fl-eic} shows the structure functions $F_L$ and $F_2$ extracted from the 
pseudo-data of the reduced cross section by means of a Rosenbluth separation, requiring 
a minimum scattered lepton momentum of 0.5 GeV, $Q^2>1\,\mathrm{GeV}^2$, $0.01<y<0.90$, and 
$0.5^\circ < \theta < 179.5^\circ$. To guide the eye, the expected uncertainties are placed on 
theoretical expectations for $F_{2,L}$ at NNLO accuracy using the ABKM09 set of PDFs \cite{Alekhin:2009ni}. One should note that
these PDFs use only data with $Q^2>2.5\,\mathrm{GeV}^2$ in their fit and, hence, the behavior of 
$F_{2,L}$ in the lowest $Q^2$ bin must be taken with a grain of salt and are only for illustration.
The extracted uncertainties take detector smearing of the scattered electron momentum into account. 
The momentum resolution was taken from ZEUS, i.e., $\delta p/p = 0.85\% + 0.25\%\times p$. 

\subsection{Summary and To-Do Items}
Like for most inclusive and semi-inclusive measurements at the EIC, an
extraction of $F_L$ will be dominated by systematic uncertainties which need to be 
thoroughly addressed. This is work in progress.
It is planned to study the unfolding of $F_L$ in great detail both in $ep$ and $eA$ scattering,
including QED radiative corrections and a full simulation of the detector. This will elucidate 
to what extent the methods developed and used at HERA \cite{Collaboration:2010ry,Chekanov:2009na,Aaron:2009wt} 
are suited for high precision measurements of $F_L$ aimed at the EIC.
In any case, it will be crucial to design the relevant detector components very carefully to optimize
\begin{itemize}
\item the luminosity measurement and its relative calibration for running at different c.m.s.\ energies,
\item the lowest lepton momentum we can detect (0.5 GeV would be desirable), 
\item the identification of the scattered lepton to suppress potential background from misidentified hadrons,
\item the resolution in momentum and scattering angle of the scattered lepton, and
\item the acceptance for the hadronic final state to suppress events which 
have a photon radiated from the incoming or outgoing lepton as well as quasi real photo-production events.
\end{itemize}
Details on the design of the detector are given in Sec.~\ref{sec:detector}.
Also, it will be possible to extract F$_L$ from the EIC data alone, but the combination of the EIC 
reduced cross section measurements with the ones from HERA may provide an even better lever arm 
in a larger $x, Q^2$ range. This needs to be investigated.

Finally, we note that even for statistically very precise measurements of $\sigma_r$, the Rosenbluth
separation of $F_L$, i.e., the determination of the slope with respect to $y^2/Y_+$, can lead to
significantly larger uncertainties if the measured values of $\sigma_r$ have very similar
$y^2/Y_+$. This source of uncertainties needs be minimized by optimizing the binning in $y$ and the 
set of different c.m.s.\ energies $\sqrt{s}$. Studies is this direction are ongoing as well.
\section{Theoretical status of inclusive heavy quark production in DIS}
\label{sec:bluemlein-hq}

\hspace{\parindent}\parbox{0.92\textwidth}{\slshape 
  Sergey Alekhin, Johannes Bl\"umlein, Sven-Olaf Moch
}

\index{Alekhin, Sergey}
\index{Bl\"umlein, Johannes}
\index{Moch, Sven-Olaf}





\subsection{Introduction}

Heavy quark production gives a sizable contribution to the unpolarized 
DIS structure functions at small $x$, see, e.g., \cite{:2009ut,Aaron:2010ib,Chekanov:2009kj,Abramowicz:2010zq}. 
For the foreseen EIC 
kinematics of DIS it yields up to 10\% of the inclusive cross section.
Therefore in order to employ the full 
potential of the small-$x$ EIC data for phenomenology 
one has to provide an accurate theoretical description of 
heavy-quark electro-production within perturbative QCD. 

For the light-parton 
contributions to DIS structure functions a theoretical accuracy of $O$(few \%) is achieved, with 
the complete QCD corrections up to 3-loops being available, see also Sec.~\ref{sec:pqcd}.
In the case of the fixed-flavor-number scheme (FFNS) the heavy flavor corrections are available
only to $O(\alpha_s^2)$. This can be a bottleneck for the analysis of 
high-precision data. Therefore, progress in the higher-order 
calculations of heavy-quark-production coefficient functions is quite important for  
the EIC phenomenology. For the variable-flavor-number scheme (VFNS) the 
massive quarks are considered on the same footing as the massless ones. 
Furthermore, the heavy-quark PDFs appearing in the VFNS are derived 
from the light-parton PDFs and the appropriate massive operator-matrix elements (OMEs). 
The VFNS coefficient functions are known up to 3-loop 
accuracy due but the massive OMEs are only available to the NLO corrections. 
This limits the theoretical accuracy of the VFNS as well. 

In the following we summarize the state-of-art in calculations of 
the NNLO corrections to the unpolarized heavy-quark coefficient functions 
and to the massive OMEs.
The FFNS and VFNS are compared to the available 
HERA data and to each other.
We also discuss the implementation of the running-mass scheme for 
the NLO and NNLO heavy-quark coefficient functions
and the resulting improvement in the perturbative stability related to this definition.

\vspace*{-3mm}
\subsection{General framework}

The heavy flavor corrections to deep-inelastic structure functions emerge in the
Wilson coefficients for the respective processes, i.e., they contribute
in terms of virtual and final state effects. Heavy quarks have no strict partonic 
interpretation since partons 
are massless, and by virtue of 
this, infinitely long lived, with the possibility to move collinear to each other.
Adopting this picture, heavy quarks can be singly or pair produced from 
massless partons and the gauge bosons of the Standard Model as final states. This 
description is called FFNS, which is the genuine scheme in any
quantum-field theoretic calculation. The DIS structure functions $F_i(x,Q^2)$ obey the 
representation
\begin{eqnarray}
F_i(x,Q^2) = \left[\sum_{k = q_l,g} \left[ 
C_{i,\sf light}^k(x,Q^2/\mu^2) +
C_{i,\sf heavy}^k(x,Q^2/\mu^2, m_h^2/\mu^2)\right] \otimes f^k(\mu^2)\right](x)~,
\end{eqnarray}
where $q_l$ and $g$ label the massless quarks and gluons, $f^k(\mu^2)$ are the 
PDFs, $C_{i,\sf light (heavy)}^k$ the massless (massive) Wilson coefficients,
$h=c,b$ the charm and bottom quarks, and
$\otimes$ denotes the Mellin convolution. 
Other approaches derive from this description.

In case of unpolarized DIS the LO contributions
were given in \cite{Witten:1975bh,Babcock:1977fi,Novikov:1977yc,Leveille:1978px}
and the NLO corrections were 
calculated in semi-analytic form in \cite{Laenen:1992zk,Riemersma:1994hv}. For asymptotic values $Q^2 \gg m_h^2$
one may obtain the massive Wilson coefficients in analytic form. This is due to a 
factorization theorem \cite{Buza:1995ie} relating the massive Wilson coefficients 
$C_{i,\sf heavy}^k$ to 
universal massive OMEs and the massless Wilson coefficients 
\cite{Furmanski:1981cw,vanNeerven:1991nn,Zijlstra:1991qc,Vermaseren:2005qc}.
As comparisons up to NLO showed \cite{Buza:1995ie}, these representations are valid 
for the structure function $F_2(x,Q^2)$ if $Q^2/m_h^2 \gsim 10$.
To $O(\alpha_s^2)$ the Wilson coefficients were obtained in 
\cite{Buza:1995ie,Bierenbaum:2007qe,Bierenbaum:2007dm} at general values of the Mellin variable $N$. A first 
contribution to the 3-loop corrections was given in \cite{Bierenbaum:2008yu} by the
$O(\alpha_s^2 \varepsilon)$ terms which contribute to the logarithmic terms
$O(\ln^k(Q^2/m_h^2)),~k = 1,2,3,$ in $O(\alpha_s^3)$. A large number of even 
Mellin-moments for all unpolarized 3-loop massive OMEs have 
been calculated in  \cite{Bierenbaum:2009mv} up to $N = 10\ldots 14$ depending on the
respective channel. 
For the structure function $F_L(x,Q^2)$ the 
asymptotic 3-loop corrections were given in \cite{Blumlein:2006mh} for general 
values of $N$. However, they are valid at 1\% accuracy at much higher scales 
of $Q^2/m_h^2 \gsim 800$ only. All logarithmic terms at $O(\alpha_s^3)$ 
for the heavy flavor Wilson coefficients contributing to the structure function
$F_2(x,Q^2)$ are known \cite{Bierenbaum:2010jp,Ablinger:2011pb}. More than this, all the contributions  to the 
constant terms emerging from lower order contributions by renormalization have been 
calculated, cf.~\cite{Bierenbaum:2009mv} for details. 
Due to the size of the constant 
contributions phenomenological applications for the kinematic range available at 
HERA and the EIC cannot be based on only the logarithmic contributions. QCD corrections
to charged current heavy flavor production have been considered in \cite{Gluck:1996ve,Buza:1997mg,Blumlein:2011zu}.

\subsection{FFNS and VFNS}

The logarithmic contributions in the heavy flavor Wilson coefficients $\propto
\ln^k(Q^2/m_h^2)$ never become large enough in the kinematic region of HERA 
or the EIC that their resummation 
would be required \cite{Gluck:1993dpa}. Nonetheless one may introduce 
a description changing the number of light flavors effectively, which
refers to the {\it universal} contributions to the 
heavy flavor Wilson coefficients,
consisting of the twist-2 parton densities and the massive OMEs
\cite{Bierenbaum:2009zt,Bierenbaum:2009mv,Buza:1996wv,Collins:1998rz}. This requires the 
knowledge of also the gluonic OMEs 
to 3-loop order \cite{Bierenbaum:2009mv}. 

By matching at typical scales $\mu_f$ one performs 
the transition from $n_f$ to $n_f+1$ massless flavors using the asymptotic relations.
In this way one may introduce a heavy quark density. 
The corresponding representation, which is obtained
in terms of a reformulation of the FFNS, is called zero mass variable flavor number
scheme (ZMVFNS). It is unique up to the choice of the matching point(s). 
An important issue is the choice of the scale $\mu_f$, for which very often $\mu_f 
\simeq m_{h}$ is used. In Ref.~\cite{Blumlein:1998sh} it was shown, however, 
comparing exact and flavor number matched calculations that this scale is process 
dependent and often very different scales have to be chosen. In this 
context various problems arise. Because of the value of
the charm to bottom mass ratio, 
$m_c^2/m_b^2 \sim 1/9$, power corrections due to 
$m_c^2$ usually cannot be neglected at scales $\mu^2 \simeq m_b^2$. Therefore, sequential decoupling 
of both charm and bottom quarks is problematic. Furthermore, starting at  
$O(\alpha_s^3)$, Feynman diagrams with both bottom and charm quarks contribute, 
which cannot be attributed to either the charm or the bottom quark PDF 
\cite{Blumlein:2010yc}. The description of the FFNS, on the other hand, is still 
possible. Therefore, representations based on the ZMVFNS remain approximations
to which one may refer for specific applications. 
Furthermore, it applies only for the asymptotic case $Q^2 \gg m_h^2$.

\vspace*{-5mm}
\begin{center}
\begin{figure}[th] 
\epsfig{figure=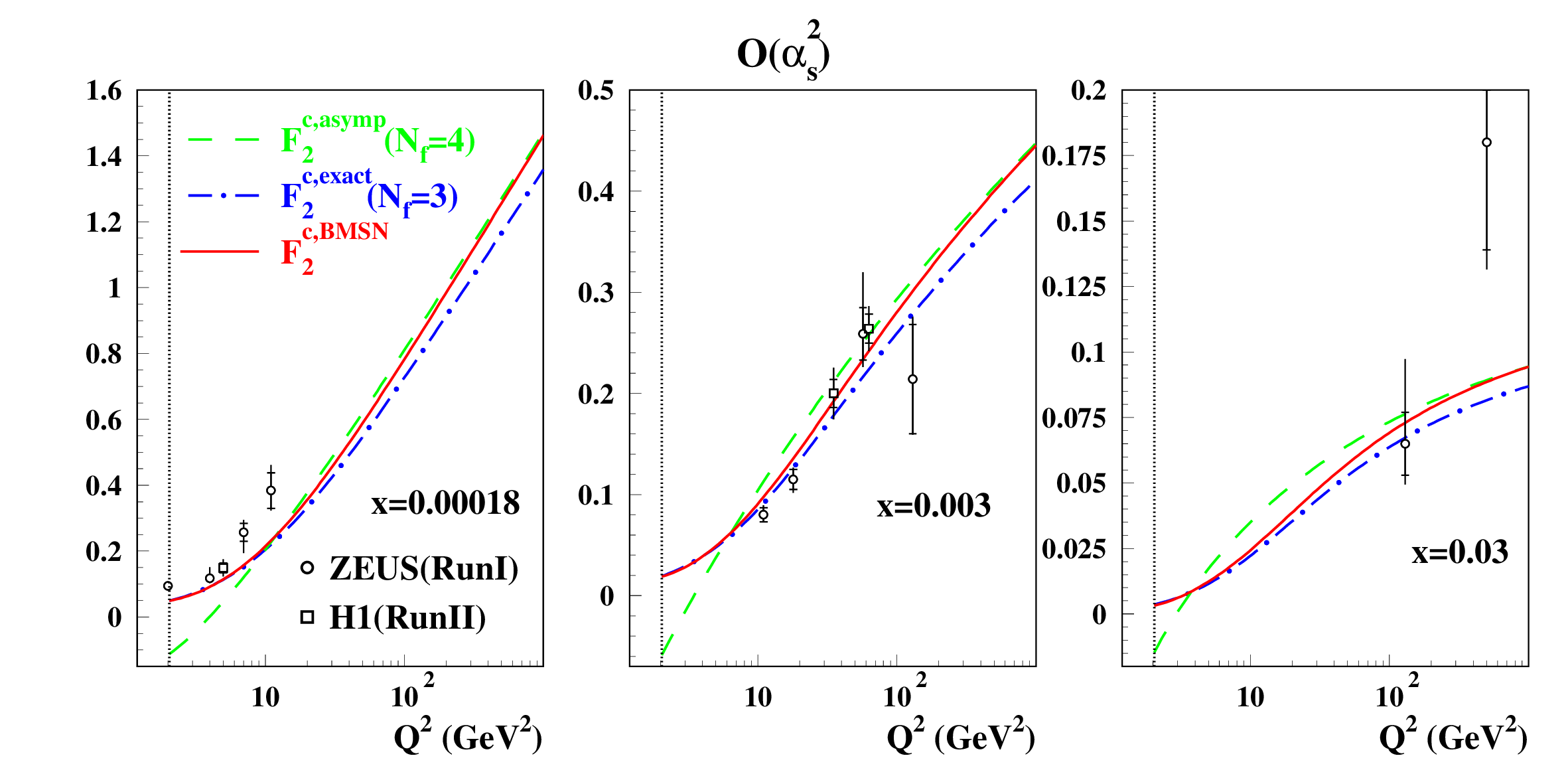,width=\linewidth} 
\vspace*{-1mm}
\caption{
\label{fig:hqfig1}
\small 
Comparison of $F^c_2$ computed in different
schemes to H1 and ZEUS data: GMVFNS
in the BMSN prescription (solid lines), 3-flavor scheme (dot-dashed lines), 
and 4-flavor scheme (dashed lines). The vertical
dotted line denotes the position of $m_c$ = 1.43~GeV. Taken from 
Ref.~\cite{Alekhin:2009ni}.}
\end{figure}
\end{center}
%
For the description of data
one would 
like to have a smooth description of the structure functions at both large 
and low values of $Q^2$, which is called the general mass variable flavor number scheme 
(GMVFNS). 
Here, a smooth interpolation is 
provided by the BMSN scheme~\cite{Buza:1996wv,Alekhin:2009ni} given by
\begin{eqnarray}
\label{eq3}
F_2^{h,\sf BMSN}(n_f+1) = 
F_2^{h,\sf exact}(n_f)
+ F_2^{h,\sf ZMVFNS}(n_f+1)
- F_2^{h,\sf asymp}(n_f)~,
\end{eqnarray}
where {\sf exact} corresponds to \cite{Laenen:1992zk,Riemersma:1994hv}, {\sf asymp} to its asymptotic form for
$Q^2 \gg m_h^2$, and {\sf ZMVFNS} to the value in the zero mass variable flavor scheme.
In Fig.~\ref{fig:hqfig1} the transition is shown for values of $x$ between 0.00018 and 
0.03 for the kinematics at HERA according to (\ref{eq3}) 
(see Ref.~\cite{Ajaltouni:2009zz} for phenomenological variants of the GMVFNS). 

\subsection{The massive NNLO corrections and the running mass}

The radiative corrections to the massive Wilson coefficients are known to be sizable. 
In particular, near the production threshold $s \simeq 4 m_h^2$, 
where large Sudakov double logarithms $\alpha_s^k \ln^{2k}(1 - 4m_h^2/s)$ dominate at each order, 
one may wish to apply resummations; see Refs.~\cite{Laenen:1998kp,Alekhin:2008hc,Presti:2010pd}
for details.
%

\begin{center}
\begin{figure}[H] 
\epsfig{figure=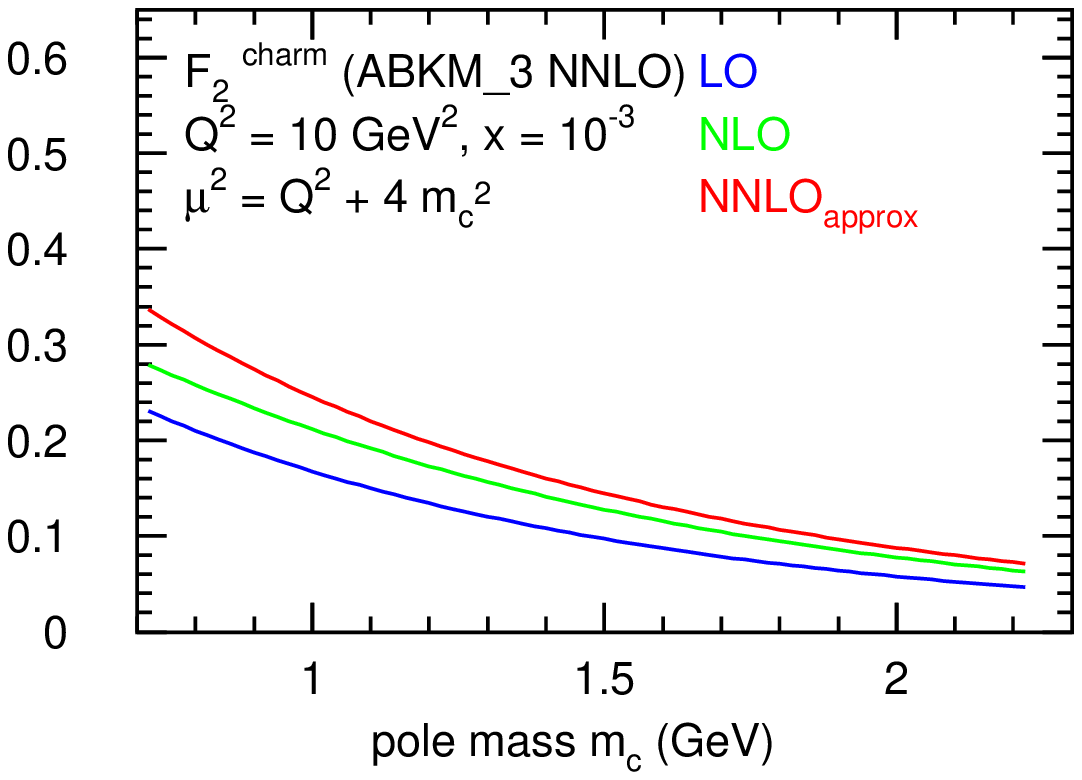,width=0.48\linewidth} \hspace*{2mm}
\epsfig{figure=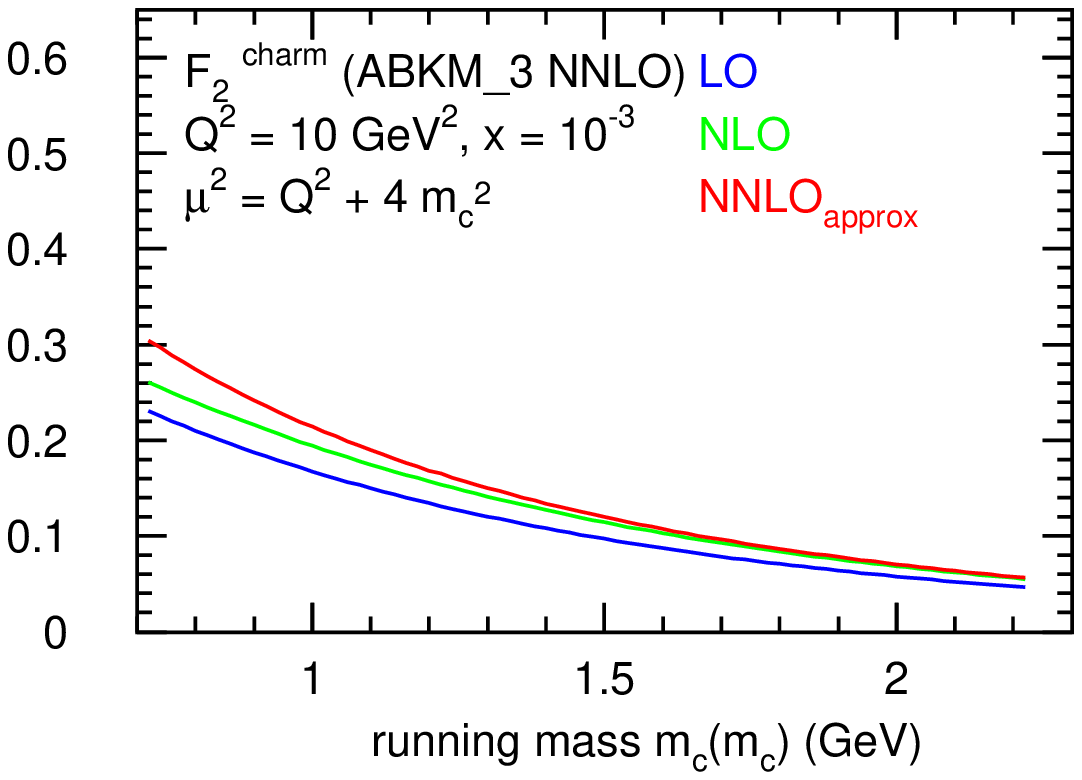,width=0.48\linewidth}
\caption[]{
\label{FIG3}
\small 
The mass dependence of $F^c_2$
for $Q^2$ = 10~GeV$^2$, $x = 10^{-3}$, and $\mu_r = \mu_f = \sqrt{Q^2 +4 m^2_c}$
using the PDFs of \cite{Alekhin:2009ni}. $m_c$ is taken in the on-shell
scheme (left) and in the $\overline{\mathrm{MS}}$ scheme (right) 
at LO (blue), NLO (green), and NNLO$_{\mathrm{approx}}$ (red); from Ref.~\cite{Alekhin:2010sv}.}
\end{figure}
\end{center}

Another aspect at higher orders concerns the definition of the heavy quark mass, 
since it is a scheme dependent quantity.
It is of particular interest to investigate which
choice of scheme leads to the best convergence of the perturbative series. 
Upon conversion of the conventionally used on-shell (pole) mass for heavy quark DIS to the
running mass $m_h(\mu)$ in the $\overline{\rm MS}$-scheme, one observes a considerable
improvement of scale stability and convergence of the perturbative expansion.
The latter aspect is demonstrated in Fig.~\ref{FIG3}. 
Here one uses the Wilson coefficients to NLO and refers to the approximate result
valid in the threshold region~\cite{Laenen:1998kp,Alekhin:2008hc,Presti:2010pd} to give an estimate for the NNLO value,
see Ref.~\cite{Alekhin:2010sv}. %
\vspace*{-5mm}
\begin{center}
\begin{figure}[H] 
\epsfig{figure=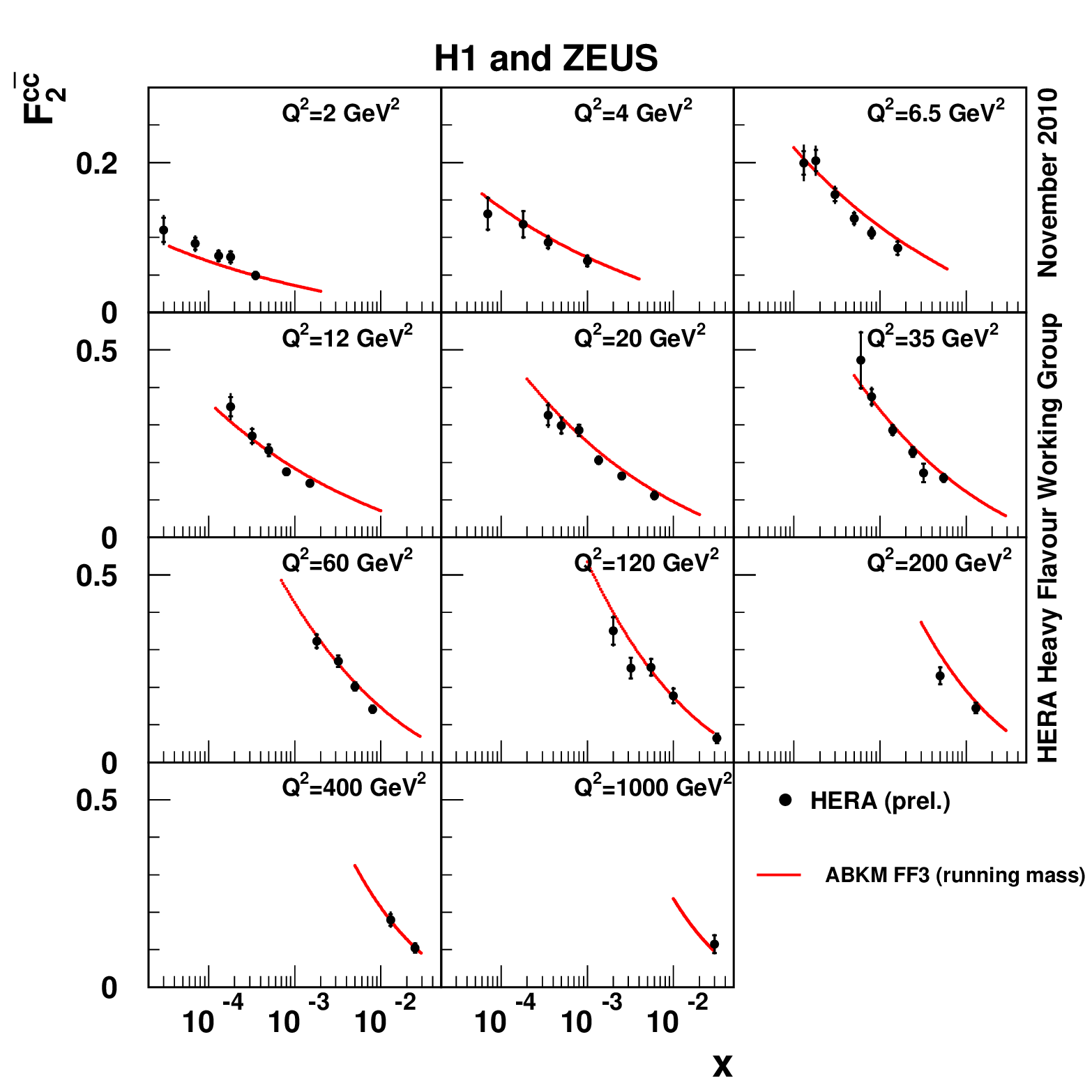,width=\linewidth,height=11cm} 
\vspace*{-1mm}
\caption[]{
\label{FIG5}
\small 
The combined HERA data on $F_2^{c}$ 
in comparison with the prediction of a fit~\cite{Alekhin:2010sv} performed in 
the running-mass scheme~\cite{Lipka}.}
\end{figure}
\end{center}
%

\begin{center}
\begin{figure}[thb] 
\epsfig{figure=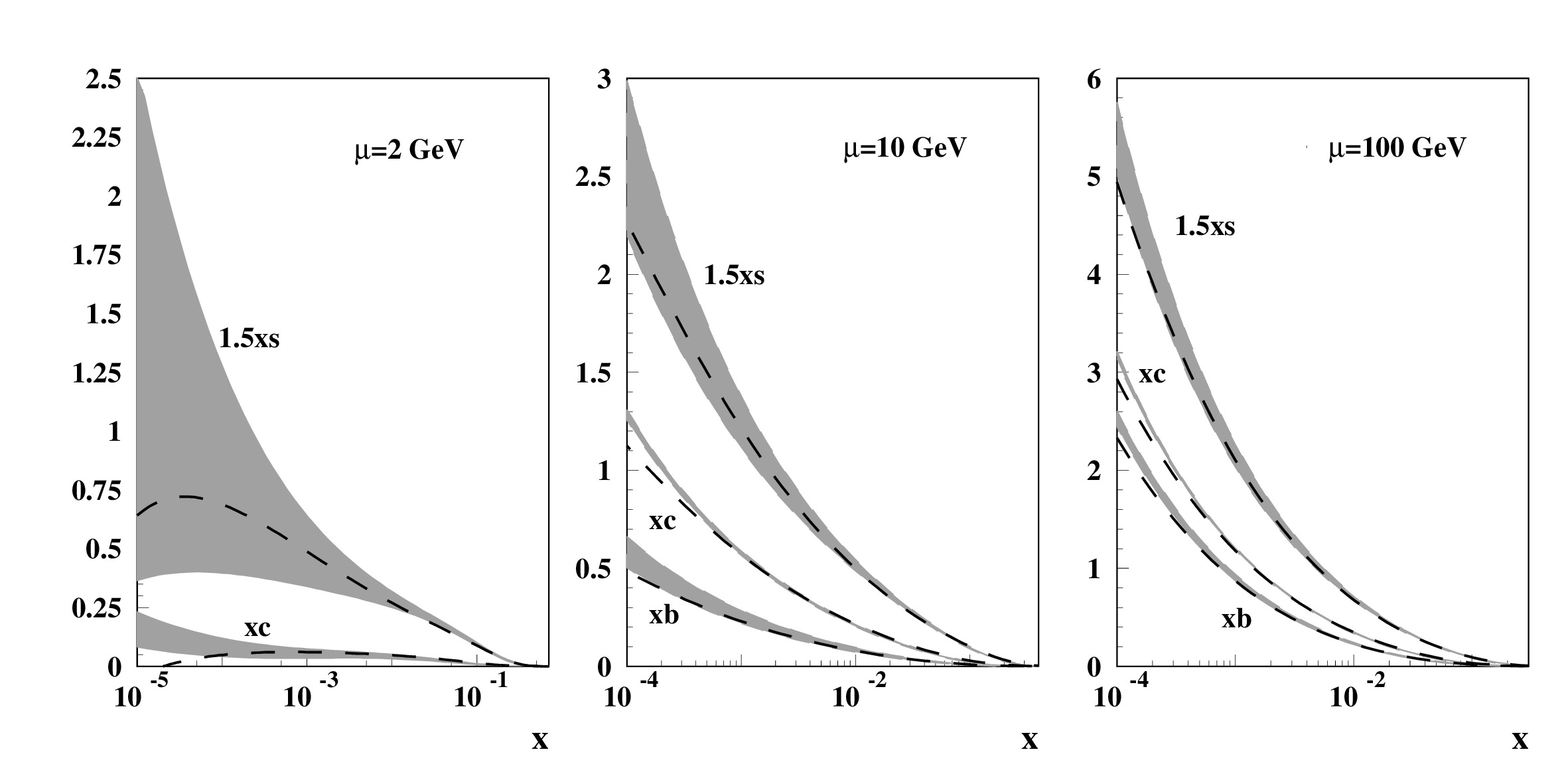,width=\linewidth}
\vspace*{-5mm}
\caption[]{
\label{FIG2}
\small 
The 1-$\sigma$ error bands (shaded area) for our NNLO 4-flavor (left panel) and 
5-flavor (central and right panels) $s$, $c$, and $b$ quark distributions in comparison to the 
corresponding MSTW2008 NNLO PDFs \cite{Martin:2009iq} (dashed lines); from 
Ref.~\cite{Alekhin:2009ni}.}
\end{figure}
\end{center}

The phenomenological impact of the mass scheme re-definition was checked for 
the ABKM fit of Ref.~\cite{Alekhin:2009ni}. In a variant of this fit~\cite{Alekhin:2010sv}
the heavy-quark electro-production was considered in the running mass scheme and 
with the approximate NNLO corrections taken into account.    
$m_c$ was fitted to the DIS data simultaneously 
with the PDG world average \cite{Nakamura:2010zzi} added to the fit as an additional constraint. 
In this way the value of 
$m_c(m_c)=1.18\pm0.06~{\rm GeV}$ was obtained. The corresponding predictions for the 
semi-inclusive structure function $F_2^{c\overline{c}}$  are in good agreement with the 
preliminary HERA data in a wide kinematical region, cf.~Fig.~\ref{FIG5}. This result gives an additional justification of the validity of the
FFNS up to $Q^2\sim 1000~{\rm GeV}^2$, i.e., in the entire kinematic range relevant for an EIC.

\subsection{Heavy-flavor PDFs}

For applications at high-energy hadron colliders, schemes with 4- and 5-light
flavors need to be considered. The necessary charm- and bottom PDFs are
generated perturbatively.
In Fig.~\ref{FIG2} the results for the $s$, $c$, and $b$ quark
flavors are shown at NNLO accuracy as determined in two global fits to the world data 
\cite{Alekhin:2009ni,Martin:2009iq}. The 1-$\sigma$ error bands correspond to the 
analysis of \cite{Alekhin:2009ni}. The central values of the MSTW08 distributions 
turn out to lie below  
those found in the ABKM09 analysis for the $c$ and $b$ quark distributions in the 
whole kinematic range of HERA due to the smaller gluon density \cite{Martin:2009iq}. 
The strange quark distribution still exhibits large errors; see also 
Sec.~\ref{sec:marco-sidis}.
Measurements at the EIC are expected to 
considerably improve both the strange and charm quark densities 
thanks to the much higher luminosities than at HERA.


\section{$F_{2,L}$(charm) at an EIC}
\label{sec:flcharm}

\hspace{\parindent}\parbox{0.92\textwidth}{\slshape 
  Elke C.\ Aschenauer, Marco Stratmann
%
}

\index{Aschenauer, Elke}
\index{Stratmann, Marco}

\vspace{\baselineskip}

Section~\ref{sec:bluemlein-hq} gave an outline of the theoretical
status of heavy flavor contributions to DIS structure function and a comparison
to HERA data.
The mass $m_h$ of the heavy quark introduces extra theoretical complications including
the need for a smooth prescription to cover both the threshold ($Q\simeq m_h$) and
the asymptotic ($Q\gg m_h$) region, the scheme used for $m_h$ (on-shell or $\overline{\mathrm{MS}}$),
and the actual value of $m_h$ used in the calculations.

Detailed experimental results from the EIC, in particular, for the
so far unmeasured charm contribution to $F_L$, will help to refine the 
current theoretical understanding. In the entire kinematic domain of the EIC one expects
the FFNS to be applicable for $F_2^c$; see Sec.~\ref{sec:bluemlein-hq}.
Differences between the exact, massive FFNS results, and the ZMVFNS
are expected to be much more pronounced for $F_L^c$, see, e.g., Fig.~7 in \cite{Gluck:1993dpa}, 
than for $F_2^c$ shown in Fig.~\ref{fig:hqfig1}. 

\begin{figure}[htbp]
\vspace{-0.5cm}
\begin{center}
\begin{tabular}{ll}
\psfig{figure=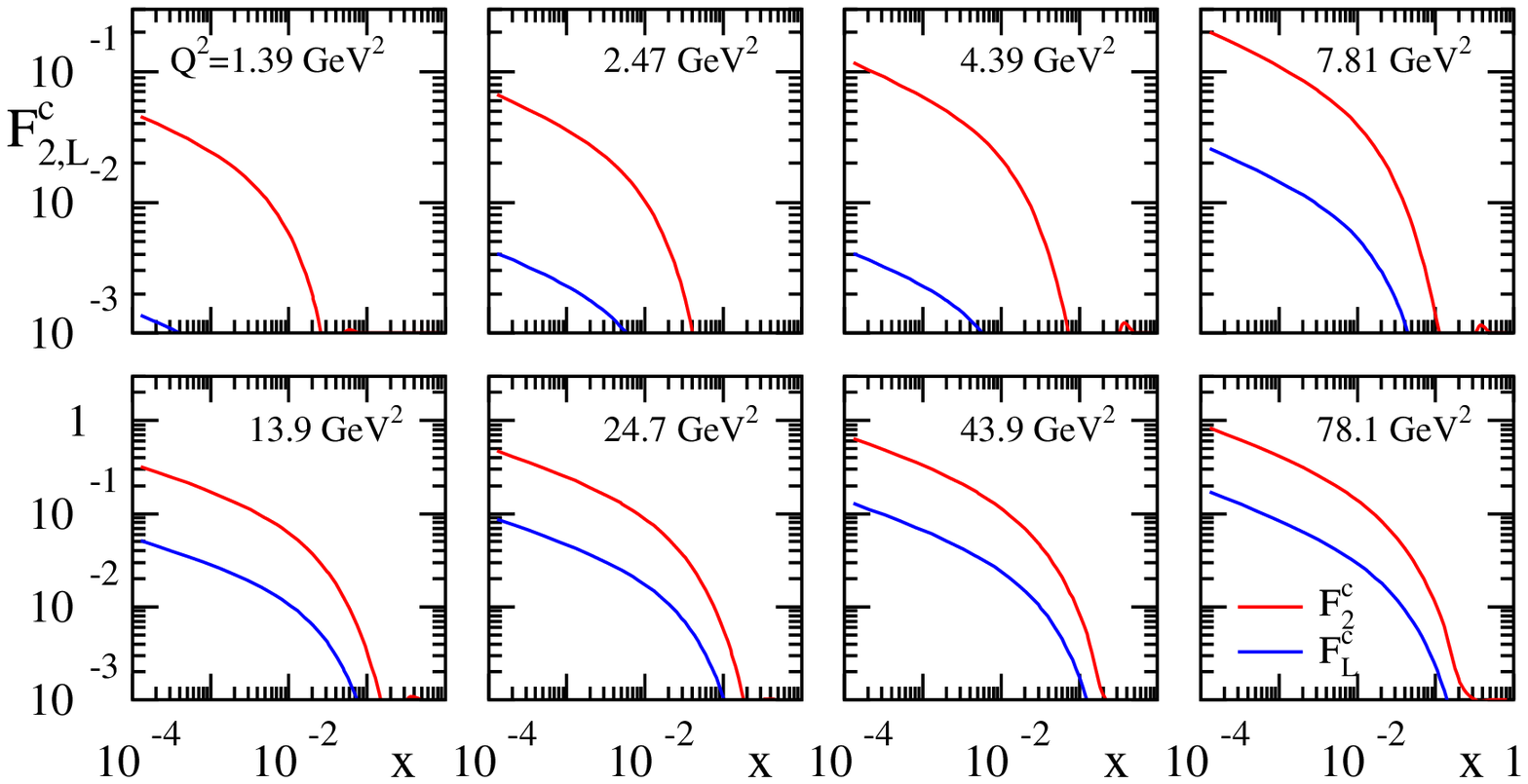,width=0.78\textwidth}
\end{tabular} 
\vspace*{-0.5cm}
\caption{Expectations for $F_{2,L}^c(x,Q^2)$ in bins of $Q^2$ using the ABKM set of PDFs \cite{Alekhin:2009ni}.}
\label{fig:flc-eic}
\end{center}
\vspace*{-0.3cm}
\end{figure}

The extraction of $F_L^c$ requires a Rosenbluth separation and should proceed along
very similar lines as discussed already in Sec.~\ref{sec:marco-fl}. 
The extra experimental complication is the requirement to detect a charm quark in the
final state. A quantitative feasibility study is still ongoing.
We note that the detection of charmed mesons is important also for other physics 
topics. Therefore the design of the detector foresees to have particle 
identification for pions and kaons to fully reconstruct charmed mesons via 
their $K\pi$ decay channel. In addition, a micro-vertex detector is expected to
provide a vertex resolution of $5\,\mu{\mathrm{m}}$ to separate charmed mesons 
from B- and other mesons by measuring a displaced decay vertex.
Using such techniques for a measurement of $F_L$ requires to detect 
a second decay lepton with a displaced vertex in addition to the 
scattered lepton. This, together with good lepton identification, should 
provide a high charmed meson detection efficiency.
The required luminosities for a precise measurement of $F_{2,L}^c$
will scale with the achieved charm detection efficiency of the EIC detectors and the
smaller reduced cross section for charm as compared to the fully inclusive
$\sigma_r$ studied in Sec.~\ref{sec:marco-fl}.
To illustrate the relative size of $F_L^c$ and $F_2^c$ we present in 
Fig.~\ref{fig:flc-eic} some theoretical expectations at NLO accuracy based on the ABKM
set of PDFs \cite{Alekhin:2009ni}; see Sec.~\ref{sec:bluemlein-hq} for details.

\section{Probing intrinsic charm at the EIC}
\label{sec:intrcharm}

\hspace{\parindent}\parbox{0.92\textwidth}{\slshape 
Marco Guzzi, Pavel Nadolsky, Fredrick Olness
%
}

\index{Guzzi, Marco} 
\index{Nadolsky, Pavel} 
\index{Olness, Fredrick}

\vspace{\baselineskip}


In the variable flavor number (VFN) factorization scheme \cite{Collins:1986mp,Aivazis:1993pi,Adams:2009kp},
heavy quark flavors are actively included in the PDF evolution via
gluon splitting to a heavy quark pair $g\to Q\bar{Q}$. While the
heavy quark PDF $f_{Q}(x,\mu)$ is often taken to vanish below the
mass threshold $(\mu<m_{Q})$, there is the possibility that the proton
contains non-vanishing heavy quark constituents even for scales below
$m_{Q}$; this component of the heavy quark PDF is identified as the
\emph{intrinsic} parton distribution \cite{Brodsky:1980pb,Brodsky:1981se,Harris:1995jx,Pumplin:2007wg},
in contrast to the extrinsic distribution generated by gluon splitting
$g\to Q\bar{Q}$.

While we can introduce intrinsic parton distributions for both charm
and bottom quarks, we will focus here on the intrinsic charm (IC).
Operationally, the total charm PDF is then composed as $f_{c}(x,\mu)=f_{c}^{ext}(x,\mu)+f_{c}^{int}(x,\mu)$.
For the extrinsic component, we generally take the boundary condition
$f_{c}^{ext}(x,\mu)=0$ for $\mu<m_{c}$, i.e., we do not need to assume an 
initial functional form for $f_{c}^{ext}$, as it is determined purely by the gluon evolution. 

\begin{figure}[bht!]
\begin{centering}
\includegraphics[clip,width=0.325\textwidth]{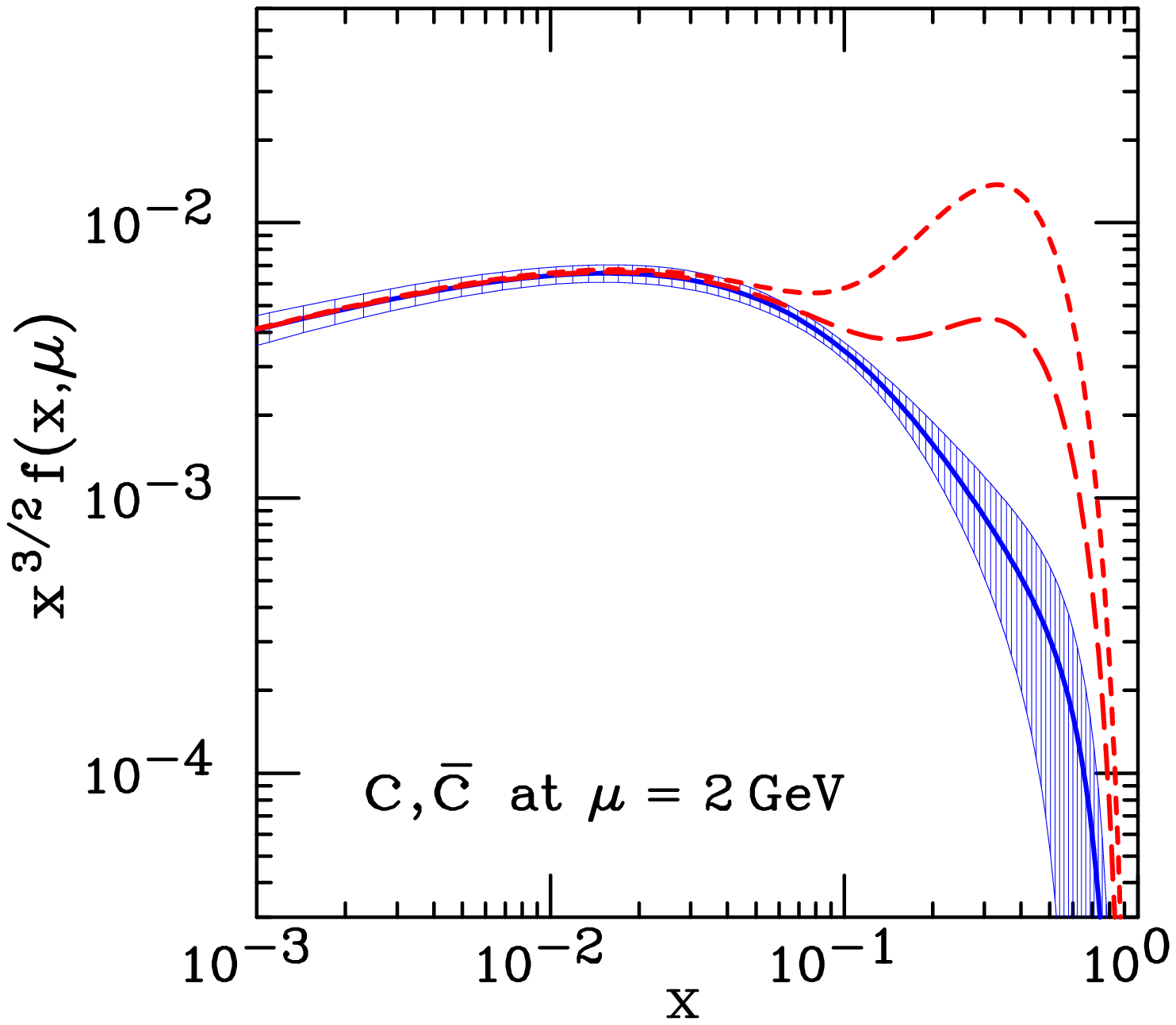}
\includegraphics[clip,width=0.325\textwidth]{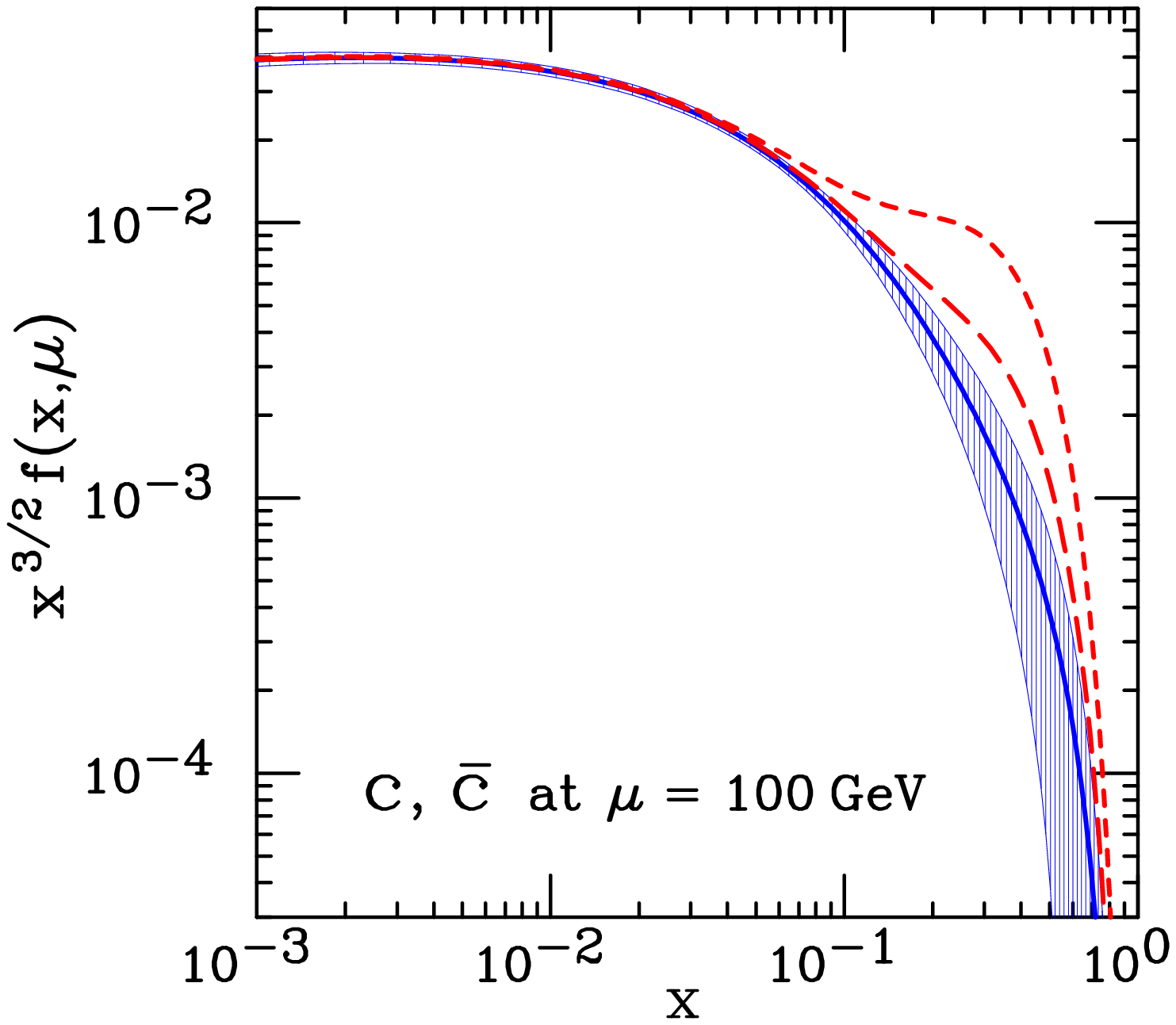}
\includegraphics[clip,width=0.325\textwidth]{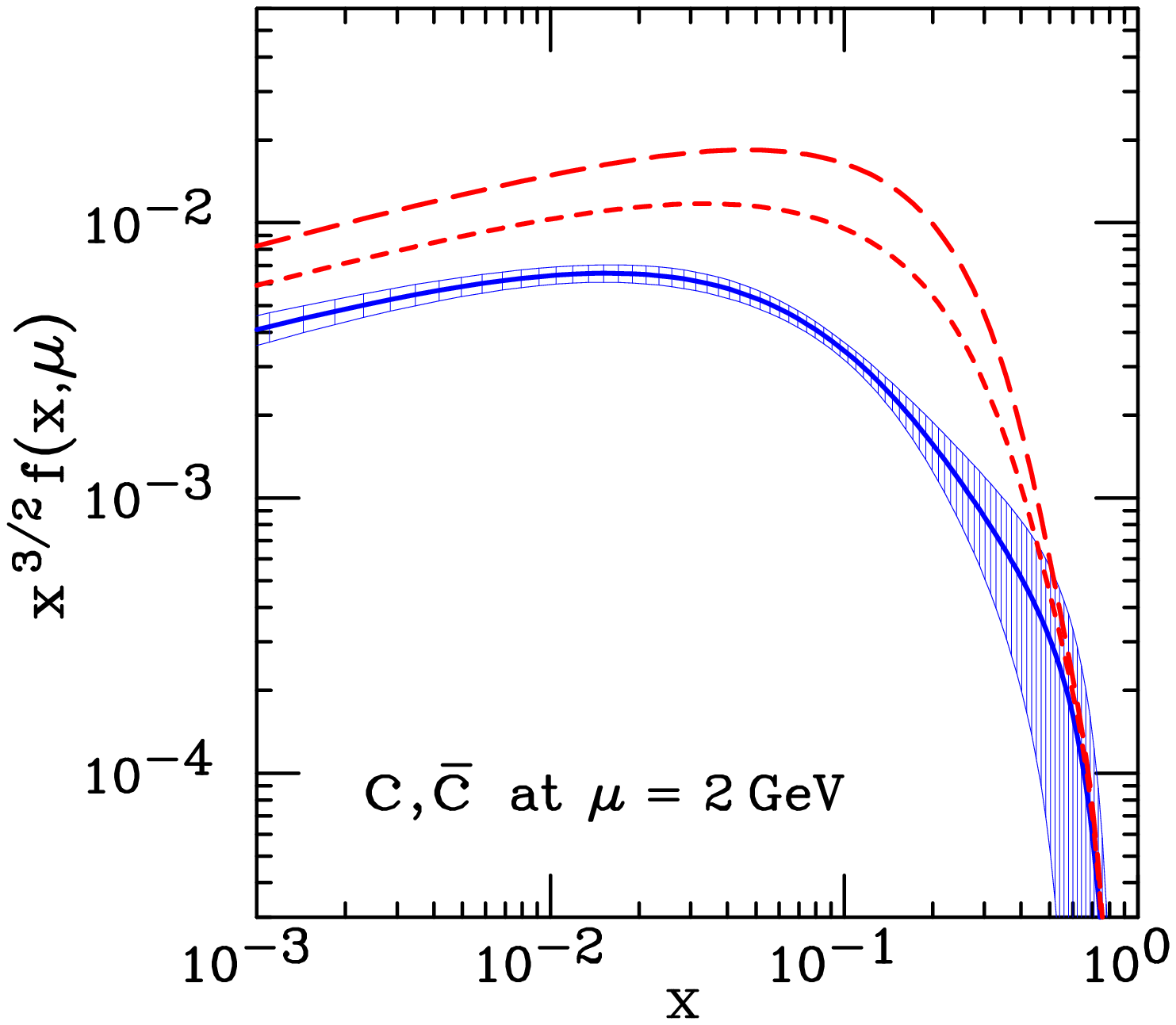} 
\par\end{centering}
\caption{Left, middle: charm PDFs for the BHPS model, at $\mu=2$ and 100 GeV.
The upper dashed curve is for a momentum fraction of 2\%, and the
lower for 0.57\%. The shaded band is the CTEQ6.5 PDF uncertainty.
Right: charm PDFs for the sea-like model. The upper curve is for a momentum
fraction of 2.4\%, and the lower for 1.1\%. Figs.\ are taken from \cite{Pumplin:2007wg}.
\label{fig:olness:icPDF}}
\end{figure}

Conversely, for the IC component $f_{c}^{int}$ we do need to assume
a functional form. Here, we consider two typical shapes
of $f_{c}^{int}$ at the initial scale $\mu=m_{c}$, assuming $m_{c}=1.3$ GeV.
\begin{itemize}
\item In the BHPS model \cite{Brodsky:1980pb,Brodsky:1981se,Pumplin:2005yf},
the intrinsic charm is concentrated at large $x$.
\item In sea-like models \cite{Pumplin:2007wg}, the intrinsic charm is
spread over all $x$ values.
\end{itemize}
Sample distributions of IC PDFs were obtained in a global QCD fit
of hadronic data \cite{Pumplin:2007wg}. We display them in Fig.~\ref{fig:olness:icPDF}. 
In these models, the momentum fraction carried by the charm can be varied in some range. 
Roughly, an intrinsic momentum fraction of $2\%$ or $3\%$ is at the outer limit of what
is allowed in the context of a global fit.

\begin{figure}[t]
\begin{centering}
\includegraphics[clip,width=0.49\columnwidth]{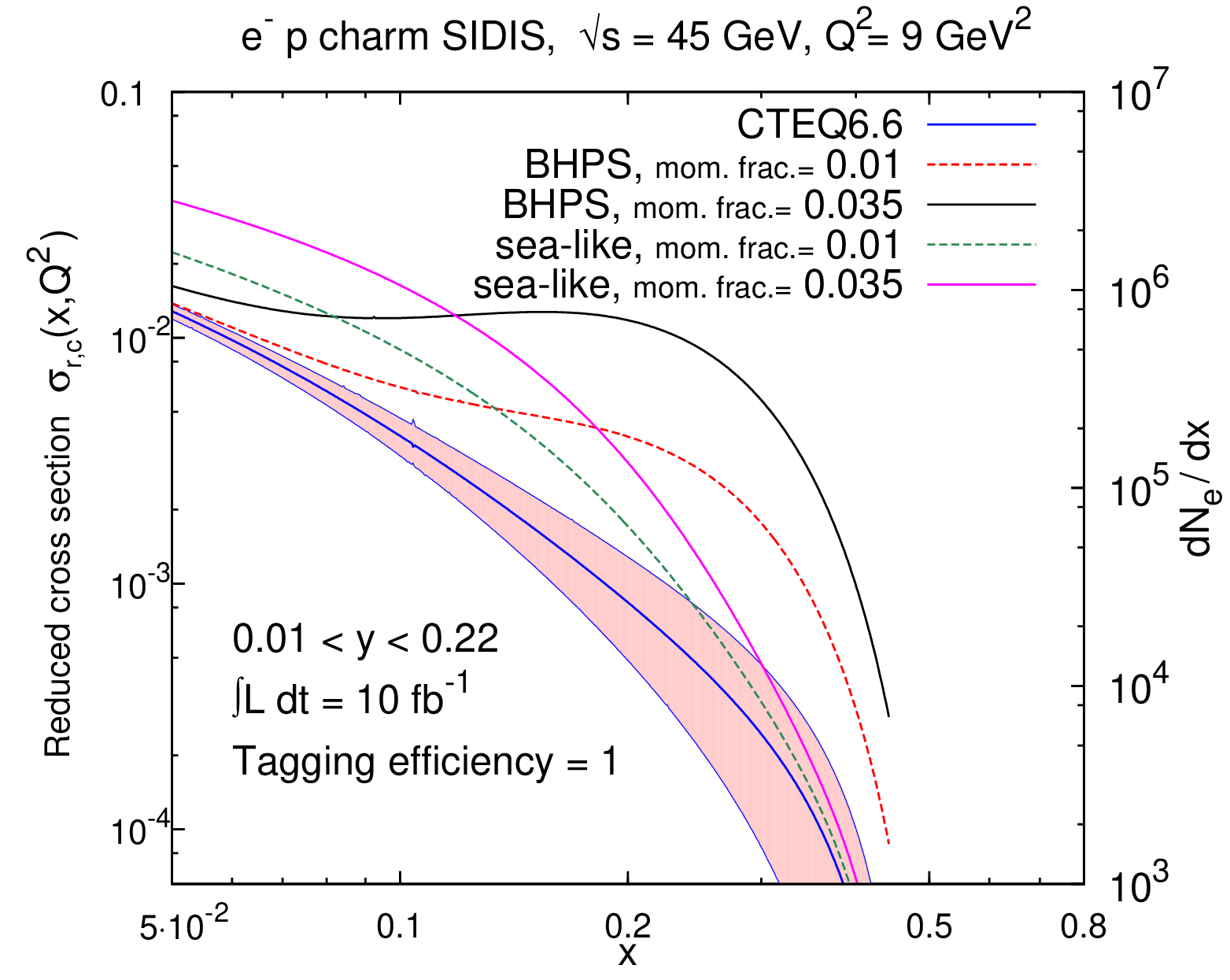} ~\includegraphics[clip,width=0.49\columnwidth]{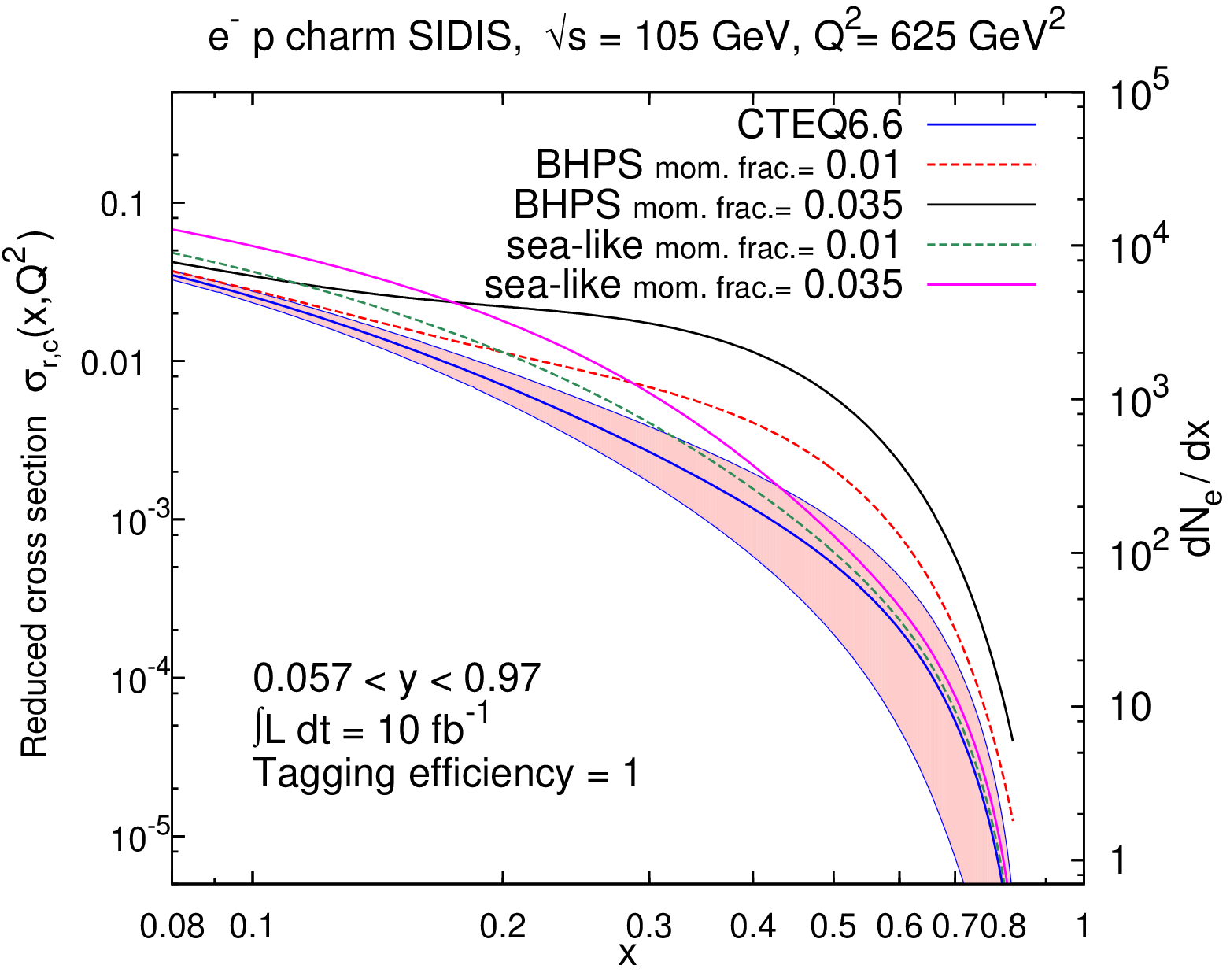} 
\par\end{centering}
\caption{Charm contribution to
the reduced NC $e^{-}p$ DIS cross section at $\sqrt{s}=45$
and $105$~GeV. For each IC model, curves for charm momentum
fractions of 1\% and 3.5\% are shown. For comparison we display the
number of events $dN_{e}/dx$ for $10\,\mbox{fb}^{-1}$, assuming
perfect charm tagging efficiency. \label{fig:olness:charm2} }
\end{figure}

For heavy quark production in the threshold region ($\mu\sim m_{Q}$), the magnitude
of the intrinsic component will be large on the relative scale compared
to the extrinsic contribution. 
At higher $\mu$ scales, the DGLAP evolution will increase the extrinsic
component via $g\to Q\bar{Q}$ splitting. However, the distinctive shape of the BHPS
distribution, with its characteristic large-$x$ enhancement, remains clearly
evident even at much higher scales $\mu\gg m_c$.

We now consider two different c.m.s.~energies for an EIC \cite{EICparams} 
and investigate the degree to which one can distinguish the IC component based on measurements 
of the charm contribution to the DIS cross section. 
Alternatively, the IC can be searched for by measuring the longitudinal structure function $F_{L}$
or angular distributions~\cite{Ivanov:2010qp}. 
In Fig.~\ref{fig:olness:charm2} we display the reduced cross section $\sigma_{r,c}$ for
semi-inclusive DIS charm production at an EIC.
The reduced charm cross section is defined as in Eq.~(\ref{eq:redxsec}).
The probed ranges of $y$ are displayed in the figures. 

The number of events for a conservative integrated luminosity
${\cal L}=10\mbox{ fb}^{-1}$ has been computed as 
$dN_{e}/dx ={\mathcal{L}}\langle d\sigma_{c}/dx \rangle$
where $\langle d\sigma_c/dx\rangle$ is the average cross section
in a $Q$ bin of size 0.15 GeV, evaluated at NLO accuracy.
The shaded band represents the error on the cross section induced by
the CTEQ6.6 PDF uncertainty \cite{Nadolsky:2008zw}. 

For both BHPS and sea-like IC, we observe that the cross
sections significantly exceed the nominal CTEQ6.6 values. 
While a momentum fraction of $3.5\%$ is easily distinguished,
even the intrinsic charm models with $1\%$ can be resolved with
moderate integrated luminosities.


\subsection*{Acknowledgments}

We appreciate a discussion with J. Pumplin.

     
\section{Status of helicity-dependent PDFs and\\ 
open questions to be addressed at an EIC}
\label{sec:polstatus}


\hspace{\parindent}\parbox{0.92\textwidth}{\slshape 
  Rodolfo Sassot, Marco Stratmann
%
}

\index{Sassot, Rodolfo}
\index{Stratmann, Marco}




\subsection{Introduction}
%
Helicity-dependent or polarized PDFs (pPDFs) tell us precisely 
how much quarks and gluons with a given momentum fraction $x$ tend to have their 
spins aligned with the spin direction of a nucleon in a helicity eigenstate. 
Their knowledge is essential in the quest to answer one of the most basic and
fundamental questions in hadronic physics, namely how the spin of a nucleon
is composed of the spins and orbital angular momenta of its constituents.

The nucleon spin structure can be best understood in high-energy scattering
experiments where quarks and gluons behave as almost free particles at scales $\mu\gg\Lambda_{QCD}$.
The relevance of pPDFs or research in spin physics in general is reflected in more than a dozen 
vigorous experimental programs in the wake of the unexpected finding that only very little of the proton 
spin is actually carried by its three valence quarks almost twenty-five years ago.
The experiments have measured with increasing precision various observables sensitive 
to different combinations of quark and gluon polarizations in the nucleon.
This progress was matched by advancements in corresponding theoretical 
higher order calculations in the framework of pQCD and 
phenomenological analyses of available data.
Potentially large sea quark and/or gluon polarizations were initially thought to be
ways to account for the ``missing'' proton spin, but at the same time, both turned out
to be challenging to access experimentally. 

The most comprehensive global fits include all available data taken in spin-dependent
DIS, semi-inclusive DIS (SIDIS) with identified pions and kaons, and 
proton-proton collisions. They allow for extracting sets of pPDFs consistently at NLO accuracy 
along with estimates of their uncertainties \cite{deFlorian:2008mr,deFlorian:2009vb}.
Contributions from the orbital angular momenta of quarks and gluons completely decouple from
such type of experimental probes and need to be quantified by other means. Here, transverse 
momentum-dependent PDFs or generalized PDFs appear to be the most promising approaches which
will be discussed elsewhere in Chapters 2 and 3, respectively.
 
Despite the impressive progress made in the past couple of years
both experimentally and theoretically many fundamental questions
related to the proton's helicity structure
still remain unanswered and shall be summarized below; 
addressing them and providing answers is a prime target for an EIC.

Present fixed-target experiments suffer from their very limited kinematic coverage 
in $x$ and $Q^2$,
which is insufficient to precisely study, for instance, QCD scaling violations for the
polarized DIS structure function $g_1(x,Q^2)$ which 
in turn can be linked to the $x$ dependence of the polarized gluon density $\Delta g(x)$.
There are numerous other opportunities for an EIC to further
our understanding of the nucleon spin structure which will be listed below and 
discussed in some details in Secs.~\ref{sec:rodolfo-spin}, \ref{sec:electroweak}, and \ref{sec:marco-cc-charm}.

\begin{figure}[th!]
\vspace*{-0.5cm}
\begin{center}
\includegraphics[width=0.475\textwidth]{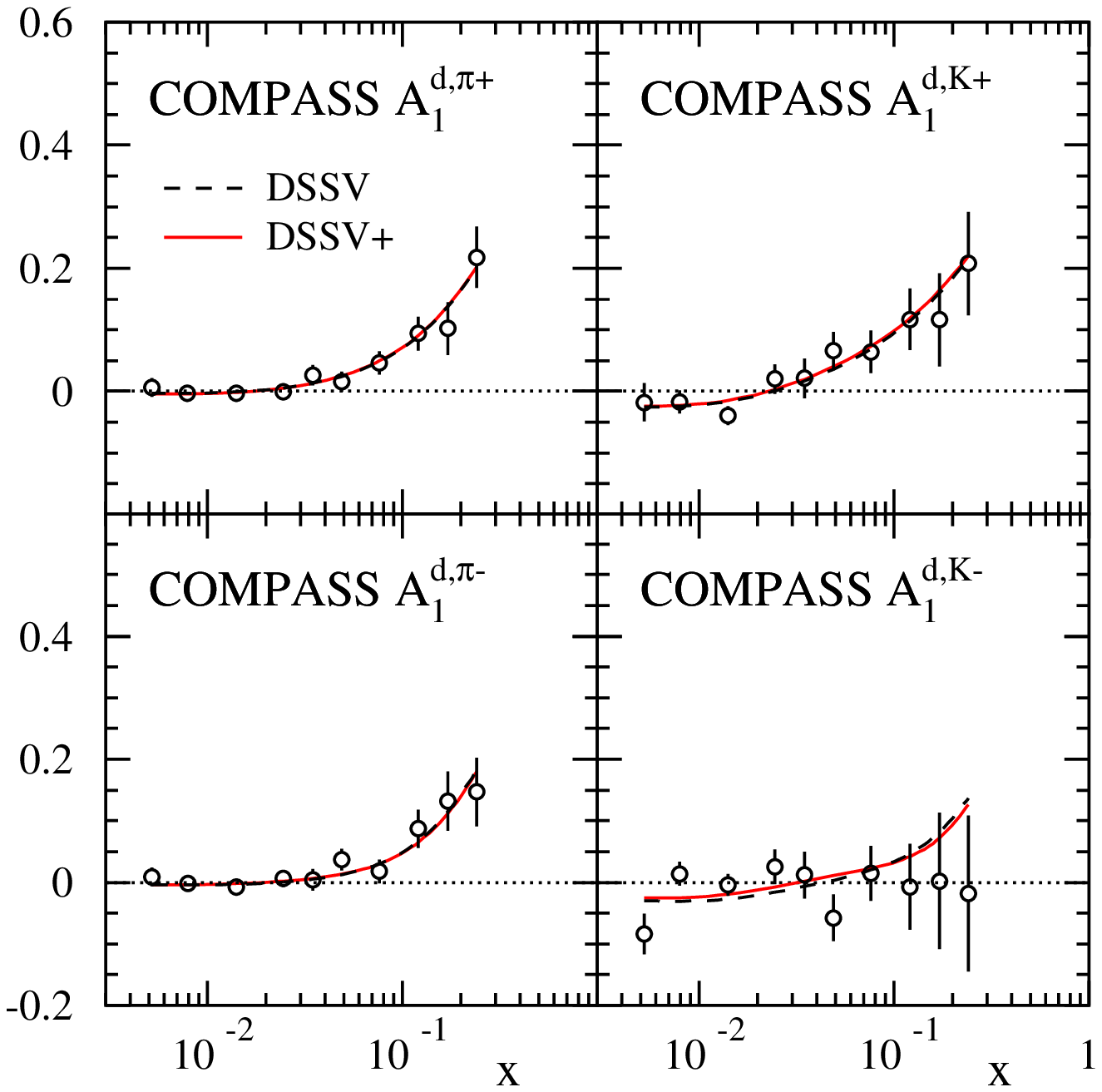}
\includegraphics[width=0.475\textwidth]{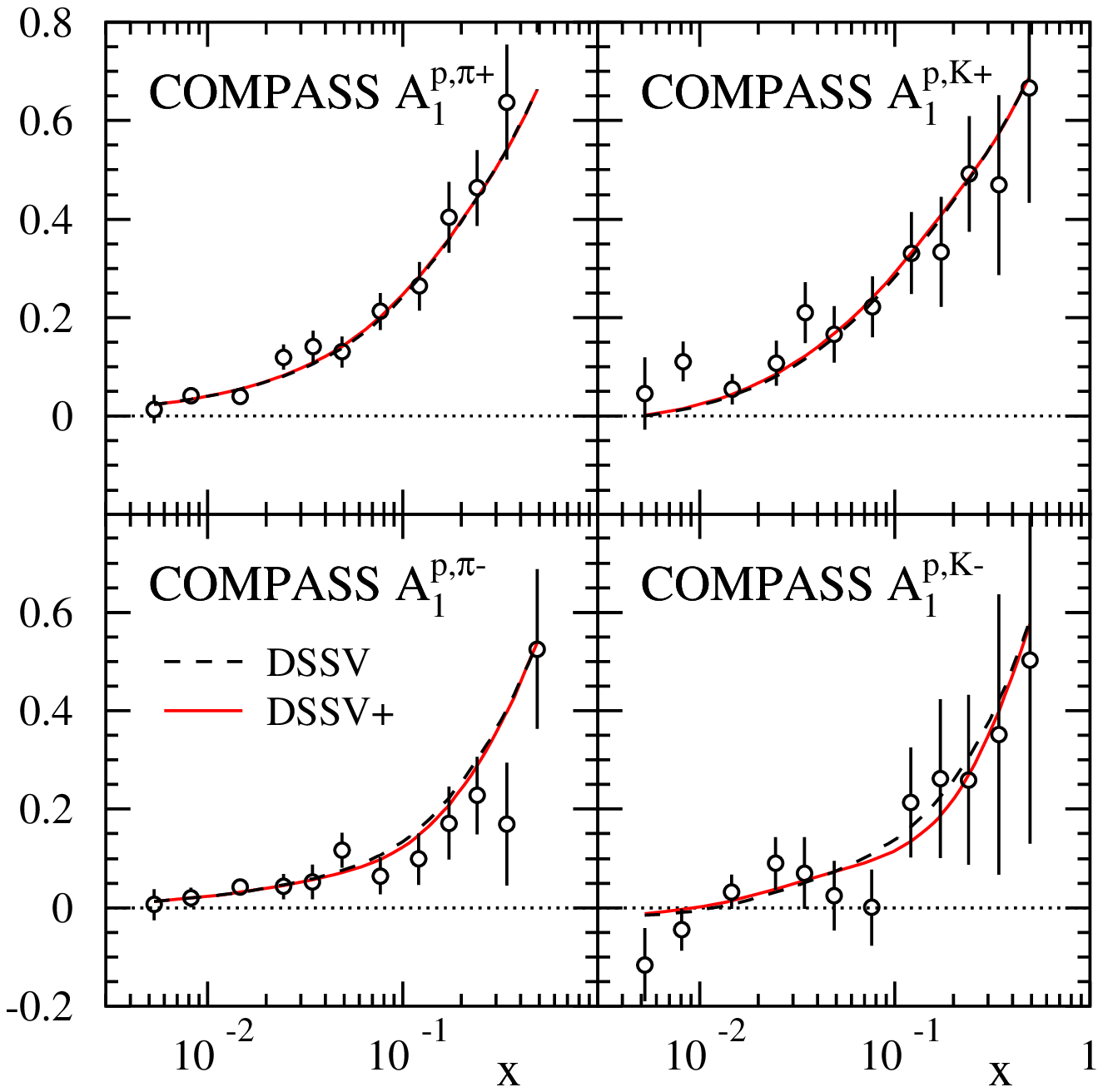}
\end{center}
\vspace*{-0.4cm}
\caption{\label{fig:newsidis} COMPASS results~\cite{Alekseev:2009ci,Alekseev:2010ub} for SIDIS spin asymmetries on
a deuteron (left) and proton target (right) compared to DSSV and DSSV+ fits (see text).}
\end{figure}
%
\subsection{Current status of global pPDF fits - baseline for EIC projections}
%
Unlike unpolarized PDF fits, where a separation of different quark flavors is obtained 
from inclusive DIS data taken with neutrino beams, differences in polarized quark and antiquark
densities are at present determined exclusively from SIDIS data and hence require knowledge 
of fragmentation functions. 
Recently published SIDIS data from the COMPASS collaboration~\cite{Alekseev:2009ci,Alekseev:2010ub}
extend the coverage in $x$ down to about $x\simeq 5\times10^{-3}$, almost
an order of magnitude lower than the kinematic reach of the HERMES data 
used in the DSSV global analysis of 2008~\cite{deFlorian:2008mr,deFlorian:2009vb}.
For the first time, the new results comprise measurements of identified pions and kaons in the final state
taken with a longitudinally polarized proton target.
Clearly, these data can have a significant impact on fits of pPDFs and estimates of their
uncertainties.

In particular, the COMPASS kaon data will serve as an important check of the validity of the strangeness density
obtained in the DSSV analysis, which instead of favoring a negative polarization as in most fits 
based exclusively on DIS data, prefers a vanishing or perhaps even slightly positive $\Delta s$
in the measured range of $x$. One reason for concern is the dependence on fragmentation functions.
Even though pion fragmentation functions are rather well 
constrained~\cite{deFlorian:2007aj} by data, kaon fragmentation 
functions suffer from much larger uncertainties, and this could explain the unexpected
result for $\Delta s$ obtained in the DSSV analysis.

Figure~\ref{fig:newsidis} shows a comparison between the new SIDIS spin
asymmetries from COMPASS~\cite{Alekseev:2009ci,Alekseev:2010ub} and
the DSSV fit of 2008 \cite{deFlorian:2008mr,deFlorian:2009vb}.
Also shown is the result of re-analysis at NLO accuracy
based on the updated data set. This fit, henceforth called ``DSSV+'', 
will serve as baseline pPDFs when quantifying the potential
impact of projected EIC data on our knowledge of the 
nucleon spin structure in Sec.~\ref{sec:rodolfo-spin}.
The differences between the original and the updated fit are hard to notice for both identified pions and kaons.
\begin{figure}[th!]
\vspace*{-0.6cm}
\begin{center}
\includegraphics[width=0.60\textwidth]{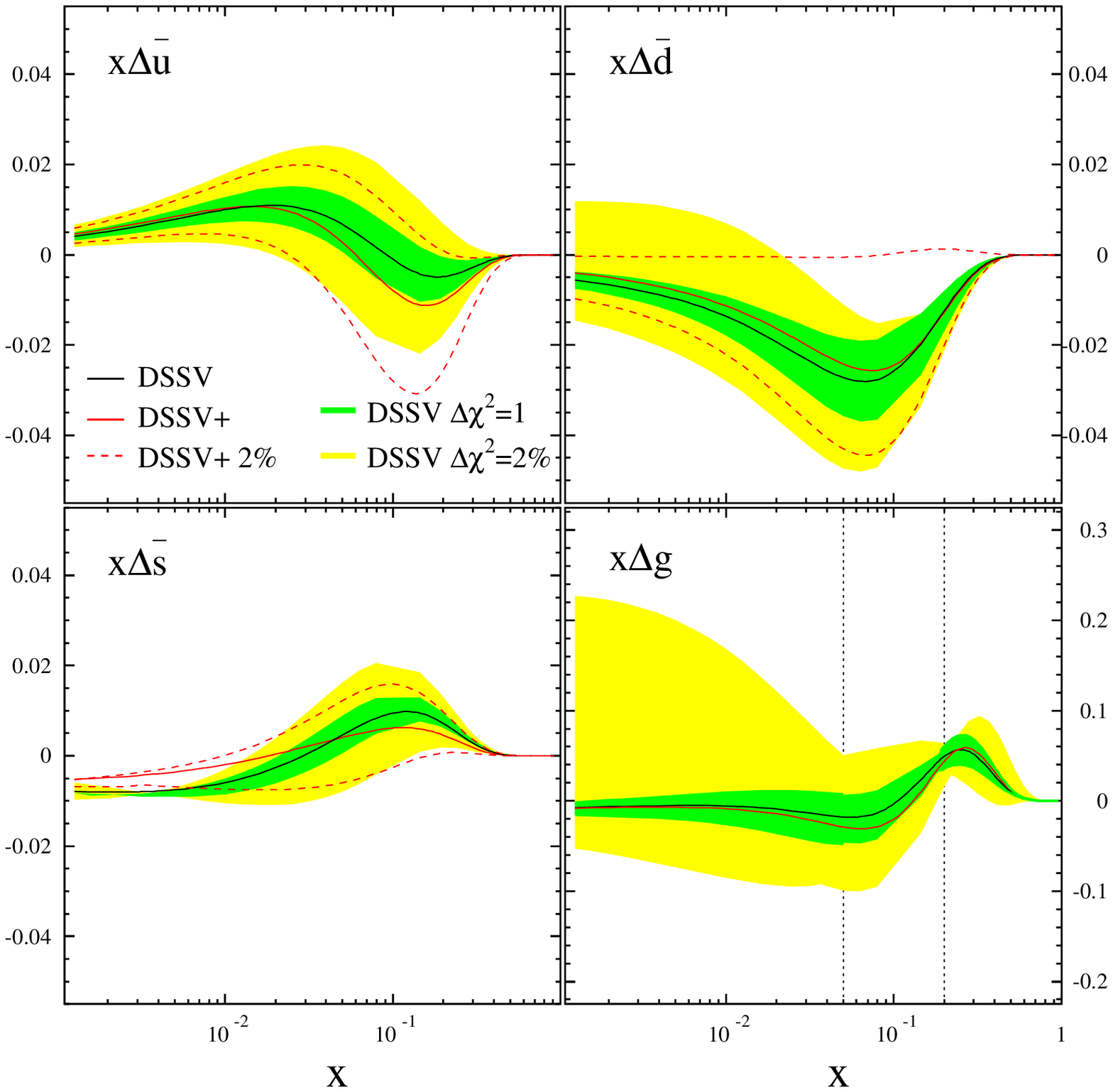}
\end{center}
\vspace*{-0.4cm}
\caption{\label{fig:newpdfs} 
DSSV and DSSV+ sea quark pPDFs and uncertainty bands at $Q^2=10\,\mathrm{GeV}^2$.
Also shown is $\Delta g$. The vertical lines indicate the 
$x$ region constrained by RHIC $pp$ data.}
\end{figure}
In terms of $\chi^2$ values, the original DSSV analysis amounts to 392 units for 
the original set of 467 data points used in the fit~\cite{deFlorian:2007aj}.
Adding both deuteron and proton data from COMPASS (88 points) it goes up to 456 and drops by about 4 units 
upon refitting (DSSV+), which is not really a significant improvement for a PDF analysis
in view of non-Gaussian theoretical uncertainties.
Recall that in the DSSV analysis a $\Delta \chi^2\simeq 9$ (corresponding to $\Delta \chi^2/\chi^2=2\%$) 
was tolerated as a faithful, albeit conservative estimate of PDF uncertainties.

In Fig.~\ref{fig:newpdfs} we compare the individual sea quark densities obtained in the original and
updated DSSV analyses. As can be seen, except for $\Delta s$,
the new central fits fall well within the $\Delta \chi^2 =1$ uncertainty bands of DSSV. 
The gluon distribution is hardly affected by the new SIDIS data.
For DSSV+ we only give the new uncertainty bands (dashed lines) referring to
the $\Delta \chi^2/\chi^2=2\%$ tolerance criterion.

\begin{figure}[th!]
\vspace*{-0.6cm}
\begin{center}
\includegraphics[width=0.55\textwidth]{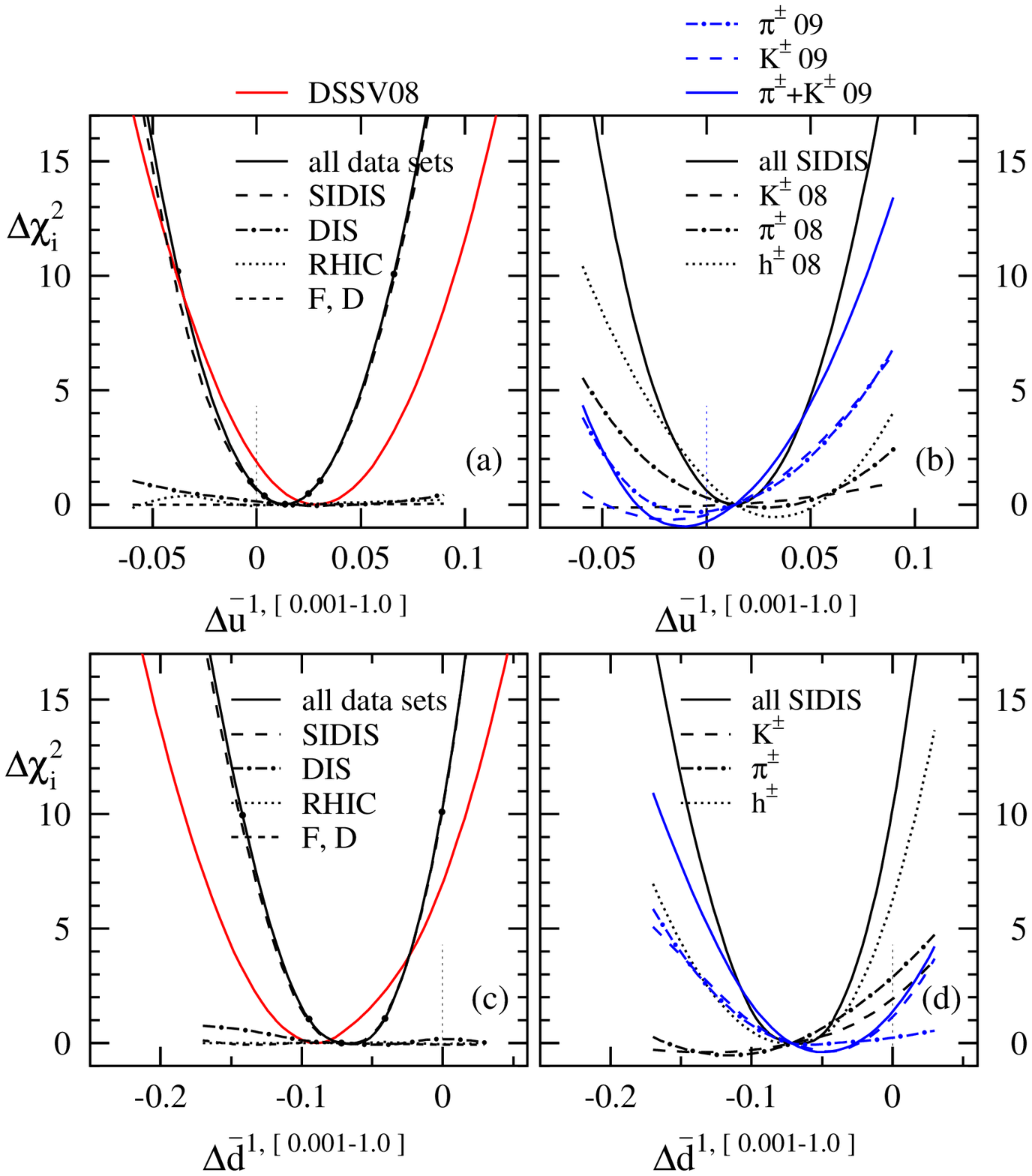}
\end{center}
\vspace*{-0.4cm}
\caption{\label{fig:profiles-ud} 
$\chi^2$ profiles for the first moments of $\Delta \bar{u}$ and $\Delta \bar{d}$
truncated to $0.001\le x \le 1$.}
\end{figure}

Although it may seem that the new SIDIS data have little impact on the fit, this is not 
the case if one studies individual $\chi^2$ profiles in more detail.
Figure~\ref{fig:profiles-ud} shows the contributions to $\Delta \chi^2$ from various data
sets against variations of the truncated first moments for $\Delta \bar{u}$ and $\Delta \bar{d}$ in the range $0.001\le x \le 1$.
Compared to the original DSSV fit one notices a trend towards smaller net polarization as the
best fit values shift towards zero. This is induced by the new COMPASS SIDIS data. 
Both pions and kaons pull in the same direction and to a common smaller best fit value.
There is, however, some mild tension with older SIDIS sets, but this is well within the tolerance 
of the fit and most likely caused by the different $x$ ranges covered by the different data sets. 
In addition, one finds a significant reduction in the uncertainties, as determined by the width of the $\chi^2$ profiles at a given $\Delta \chi^2$.

\begin{figure}
\vspace*{-0.2cm}
\begin{center}
\includegraphics[width=0.485\textwidth]{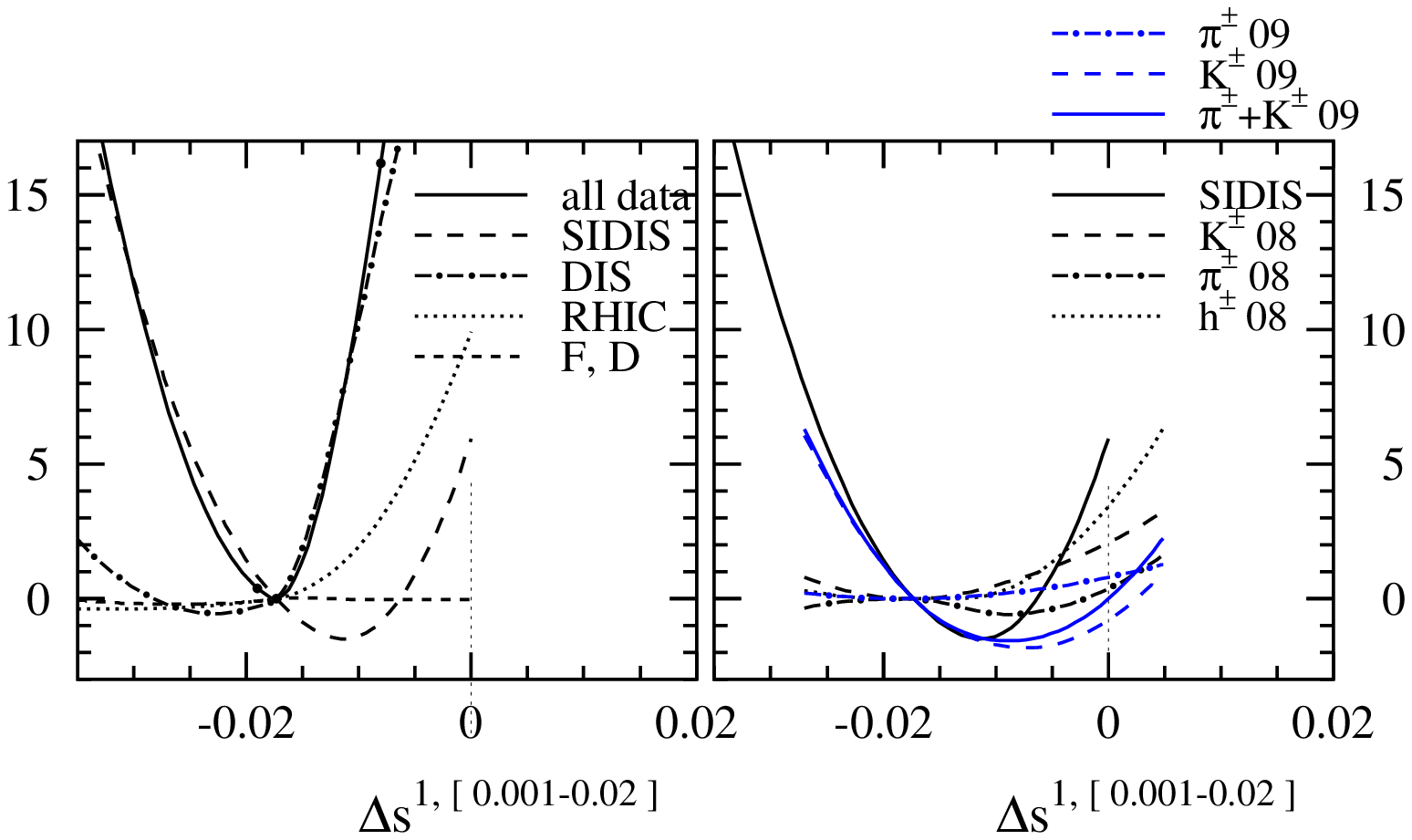}
\includegraphics[width=0.485\textwidth]{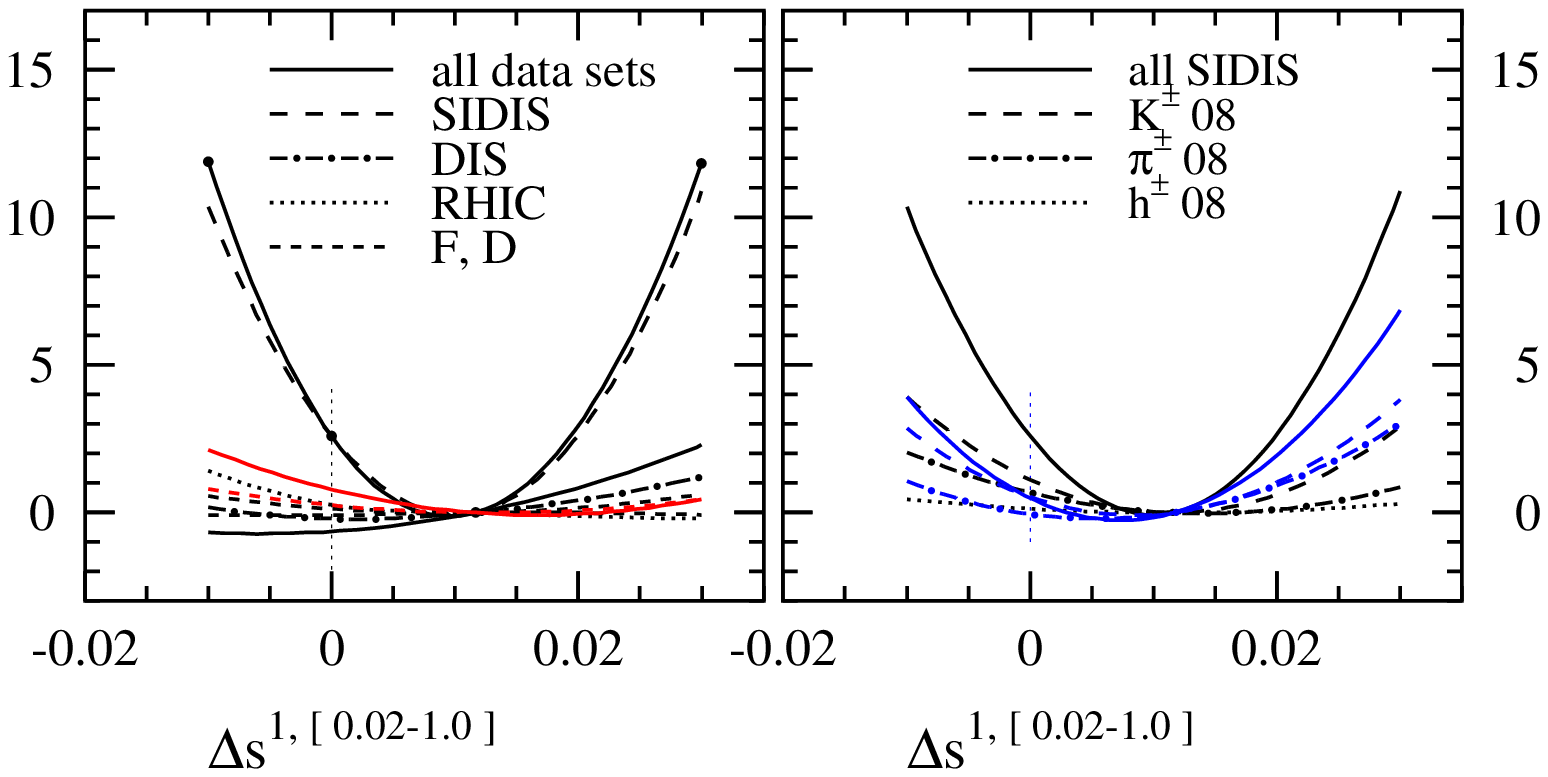}
\end{center}
\caption{\label{fig:profiles-s} $\chi^2$ profiles for the truncated first moment of $\Delta s$ in
two different $x$ intervals.}
\end{figure}
A much debated feature of the strangeness pPDF obtained in the DSSV fit is 
its unexpected small value at medium-to-large $x$ which, 
when combined with a node at intermediate $x$, 
still allows for acquiring a significant negative first moment at small $x$,
in accordance with expectations from SU(3) symmetry (hyperon decay constants $F$ and $D$)
and fits to DIS data only (see, e.g., Ref.~\cite{Blumlein:2010rn}).
To investigate the possibility of a node in $\Delta s(x)$ further we present
in Fig.~\ref{fig:profiles-s}
the $\chi^2$ profiles for two different intervals in $x$: $0.001\le x \le 0.02$ and $0.02\le x \le 1$.
Again, the new COMPASS SIDIS data have quite some impact on the profiles but the central
value for the combined range, $0.001\le x \le 1$, does not shift 
from its original DSSV value.

The profiles in Fig.~\ref{fig:profiles-s} clearly show that for $0.001\le x \le 0.02$
the result for $\Delta s$ is a compromise between DIS and SIDIS data, the latter favoring much 
less negative values. For $0.02\le x \le 1$ everything is determined by SIDIS data and all 
sets consistently ask for a small, slightly positive strange quark polarization. 
There is no hint of a tension with DIS data as they do not provide a useful constraint at 
medium-to-large $x$.
We note that at low $x$, most SIDIS sets give indifferent results except the new
COMPASS data which extend towards the smallest $x$ values so far and actually do show some preference for
a slightly negative value for $\Delta s$. This exemplifies the need for measurements at small $x$.
Clearly, all current extractions of $\Delta s$ from SIDIS data show a significant dependence on kaon FFs,
see, e.g., Ref.~\cite{Alekseev:2009ci,Alekseev:2010ub}. Better determinations 
of $D^K(z)$ are highly desirable, but should be possible with forthcoming data from $B$-factories,
DIS multiplicities, and LHC data. 
We also notice that in the range $x\gtrsim 0.001$ the hyperon decay constants, the so-called
$F$ and $D$ values, do not play a significant role in constraining $\Delta s$ as can be deduced
from their relative contribution to $\Delta \chi^2$ in Fig.~\ref{fig:profiles-s}.
Computations of SU(3) breaking effects in axial current matrix elements
\cite{Savage:1996zd,Zhu:2002tn}, and, more recently, also first lattice results
for the first moment of $\Delta s+\Delta \bar{s}$ \cite{Collins:2010gr}
point towards a sizable breaking of SU(3) symmetry. 
To study its validity of one needs to probe $\Delta s(x)$ at smaller values 
of $x$ at an EIC.

An interesting recent development is that the LSS group produced an 
update of their pPDF fit using for the first time 
DIS and SIDIS data simultaneously~\cite{Leader:2010rb}. 
As in the DSSV analysis they also utilize DSS fragmentation functions~\cite{deFlorian:2007aj}.
Their functional form is also very similar to the one used in DSSV and DSSV+.
As in their previous analyses they carefully include target mass corrections 
and phenomenological higher twist corrections for inclusive DIS data.
Nevertheless, their obtained pPDFs are very similar to the best fit of DSSV shown in
Fig.~\ref{fig:newpdfs}. Their strange quark polarization also changes sign as in DSSV
but is overall slightly smaller in magnitude.
LSS finds non negligible higher twist corrections to
inclusive DIS data, however, these conclusions are not fully shared by another recent
analysis of polarized DIS data \cite{Blumlein:2010rn}.
Ref.~\cite{Blumlein:2010rn} also provides an extraction of $\alpha_s$ from polarized
DIS data. 
There are also interesting first attempts to perform a pPDF analysis based on neural networks
\cite{DelDebbio:2009sq,Rojo:2010gt} similar to successful global fits of unpolarized data \cite{Ball:2010de}. 
This would provide independent estimates
of pPDF uncertainties not biased by the choice of a particular functional form.

\subsection{Open Questions}
%
The status of pPDFs outlined above will likely not change much until the time of EIC operations.
Most of the remaining, compelling open questions in spin physics related to pPDFs
will be still with us and can be only addressed by extending the kinematic coverage 
to smaller values of $x$; see the items listed below.

Existing experiments, like PHENIX and STAR at RHIC, will continue to add data in the
next couple of years. Parity-violating, single-spin asymmetries for $W$ boson production
should reach a level where they help to constrain $\Delta u$, $\Delta \bar{u}$,
$\Delta d$, and $\Delta \bar{d}$ at large $x$, $0.07\le x \le 0.4$ at scales $Q\simeq M_W$
much larger than typically probed in SIDIS \cite{deFlorian:2010aa}. Measurements of double spin asymmetries for
di-jets in $pp$ collisions at $500\,\mathrm{GeV}$ should improve the current
constraints on $\Delta g(x)$ and extend them towards somewhat smaller values of $x$.
The strangeness polarization is, however, very hard to access in polarized $pp$ collisions.
In the future, JLab12 will add very precise DIS data at large $x$. They will allow us
to challenge ideas like helicity retention \cite{Brodsky:1994kg,Avakian:2007xa} which predict that 
$\Delta f(x)/f(x)\rightarrow 1$ as $x\rightarrow 1$.
Currently, only $\Delta u/u$ exhibits this trend, while $\Delta d/d$ remains negative up to
$x\simeq 0.6$.

We expect an EIC to make significant contributions on the following topics:

\begin{figure}[th!]
\begin{center}
\includegraphics[width=0.45\textwidth]{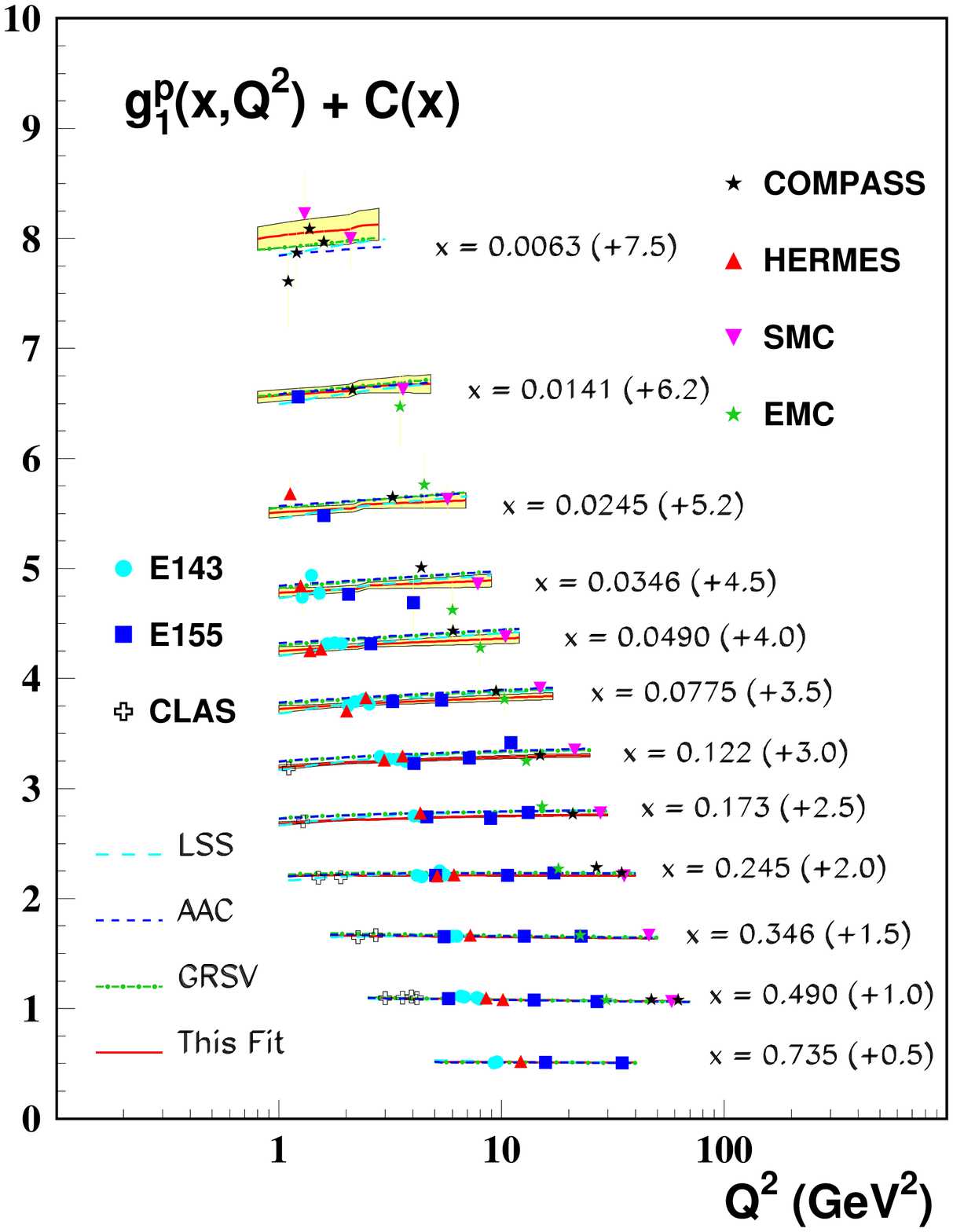}
\end{center}
\vspace*{-0.4cm}
\caption{\label{fig:g1scaling} 
Scaling violations for the structure function $xg_1^p$ in bins of $x$.
Experimental data are compared to various fits at NLO accuracy.
Figure taken from \cite{Blumlein:2010rn}.}
\end{figure}
{\bf Polarized gluon density $\mathbf{\Delta g(x)}$:} precise data for 
the DIS structure function $F_2$ in a broad kinematic range in $x$ and $Q^2$
from HERA provide the world's best and theoretically cleanest constraint on the 
unpolarized gluon density; see Sec.~\ref{sec:cooper-sarkar}.
One of the most important results of HERA was to establish the strong rise of the gluon density at small $x$
which could not be anticipated from previous fixed-target results.

Figure~\ref{fig:g1scaling} summarizes the current
situation for polarized DIS. The kinematic coverage is limited to the fixed-target
regime. There are no data below $x\simeq 0.005$, and the lever-arm in $Q^2$ is very
limited, in particular, for the smallest $x$ values. As a consequence, $\Delta g(x)$ is basically
unconstrained at small $x$ as is reflected in the large uncertainty band shown in 
Fig.~\ref{fig:newpdfs}. 
There are theoretical arguments that $\Delta g(x)\simeq x g(x)$ at small $x$ \cite{Brodsky:1994kg} but
they cannot be verified experimentally due to the lack of data

The fact that current RHIC data favor a very small gluon density in
$0.05\lesssim x\lesssim 0.2$ \cite{deFlorian:2008mr}, perhaps with a node, 
also greatly complicates the determination of the first moment,
$\int_0^1 \Delta g(x,Q^2) dx$, which enters in the fundamental proton spin sum rule
in its light-cone gauge formulation \cite{Jaffe:1989jz,Bashinsky:1998if}. 
Since contributions to the moment largely cancel in the measured $x$ range,
the unmeasured small $x$ region may contribute significantly even up to one unit of $\hbar$.

Precise measurements of the structure function $g_1(x,Q^2)$ in a wide kinematic
range will be a flagship measurement for an EIC. The polarized gluon density is
strongly correlated with QCD scaling violations, $dg_1(x,Q^2)/d\ln Q^2 \simeq - \Delta g(x,Q^2)$,
i.e., a large positive $\Delta g$ at small $x$ is expected to drive $g_1$ towards
large negative values for $x\simeq 10^{-(3\div 4)}$. A precise DIS measurement will also constrain the 
quark singlet density $\Delta \Sigma(x,Q^2)$ and its first moment, i.e.,
the total quark spin contribution to the proton spin, much better.

{\bf Complete flavor separation:}
given the significant impact present SIDIS data already have in global analyses of pPDFs,
it is easy to imagine that an EIC with its extended kinematic coverage
can turn SIDIS measurements into a precision tool for detailed studies of
$\Delta u$, $\Delta \bar{u}$, $\Delta d$, $\Delta \bar{d}$, $\Delta s$, and $\Delta \bar{s}$.
For instance, a precise determination of a possible asymmetry in the light quark sea,
$\Delta \bar{u}(x)-\Delta \bar{d}(x)$ will challenge expectations from model calculations.
Again, current QCD fits have revealed rather complicated functional forms with possible
nodes for the quark densities which need to be studied more precisely. 

Prerequisites are a detector with excellent particle ID in an as large as possible portion
of phase space and an improved theoretical knowledge of FFs, in particular,
for kaons. For the latter, significant progress will be made by the
time the EIC turns on. In any case, there will be also plenty of opportunities to
further constrain them at an EIC if necessary.

{\bf Novel electroweak probes in DIS:}
At large enough $Q^2$ and with the envisioned luminosities 
of up to $10^{34}\,\mathrm{cm}^{-2}\mathrm{s}^{-1}$ an EIC has the unique opportunity to access
polarized electroweak structure functions via charged and neutral current DIS measurements.
These novel probes depend on various combinations of polarized quark PDFs and provide
an alternative way of separating different quark flavors for $x\gtrsim 10^{-2}$.
Prerequisites are both electron and positron beams to fully exploit charged current (CC) DIS,
i.e., the pPDF combinations probed in the exchange of $W^-$ and $W^+$ bosons.
Also, one needs to be able to reconstruct $x$ and $Q^2$ from the final state hadrons in
the absence of a scattered lepton in CC DIS.

{\bf Strangeness polarization, $\mathbf{\Delta s - \Delta \bar{s}}$, and SU(3) symmetry:}
As mentioned already, the surprisingly small strangeness density
determined from SIDIS data has triggered a lot of discussions recently.
It is certainly of outmost importance to precisely map $\Delta s(x)$ and $\Delta \bar{s}(x)$
down to sufficiently small values of $x$ to reliably determine their first moments.
If SU(3) symmetry is approximately valid, one expects a significantly
negative first moment for strangeness; if, on the other hand, SU(3) symmetry 
is badly broken at a $20\div30\%$ level, $\Delta s(x)$ can remain small and perhaps even slightly positive 
down to small $x$.
Ideas have been put forward that $\Delta s(x)$ and $\Delta \bar{s}(x)$ may have
opposite polarizations which could explain the smallness of 
$\Delta s + \Delta \bar{s}$ in DIS but would result in a potentially 
sizable $\Delta s - \Delta \bar{s}$.

At an EIC there are different strategies to determine $\Delta s$ and $\Delta \bar{s}$.
The most promising one is through SIDIS production of charged kaons. Once $K^+$
and $K^-$ yields are known with high precision and uncertainties for kaon FFs are well
understood one can attempt an extraction of $\Delta s(x)$ and $\Delta \bar{s}(x)$
in a large range of $x$. Alternatively, one can study charm production in CC DIS
with a polarized proton target. If one has electron and positron beams available, 
the yields of $D$ and $\bar{D}$ mesons should be related to $\Delta s(x)$ and $\Delta \bar{s}(x)$,
respectively.

{\bf Heavy flavor contributions to $\mathbf{g_1}$:}
for presently available data, any contribution from heavy quarks, i.e., charm and bottom, can be safely ignored. 
From HERA we know, however, that
at sufficiently small values of $x$ and large enough $Q^2$, charm quarks can contribute as much as $20\div 25\%$
to a measurement of $F_2$. It is important to determine the charm contribution to $g_1$ at
small $x$ experimentally and to properly include it in future global analyses.
Since $g_1^c$ is mainly driven by photon-gluon-fusion it can be also a viable probe of $\Delta g$ in
the small $x$ region. 

{\bf Bjorken sum rule:}
the Bjorken sum rule is certainly one of the best known quantities in perturbative QCD.
Corrections up to ${\cal{O}}(\alpha_s^4)$ have been calculated \cite{Baikov:2010je}.
There is also a nontrivial connection to Adler's $D(Q^2)$ function defined in $e^+e^-$ annihilation
through the generalized Crewther relation \cite{Crewther:1997ux,Baikov:2010je}
involving the QCD $\beta$ function which incorporates the deviation from
the limit of exact conformal invariance.
It is certainly important and legitimate to ask to what level of precision an EIC
can verify this fundamental sum rule.

Since the Bjorken sum rule relates the moments of the $g_1$ structure functions for protons and neutrons,
it first of all requires an ``effective neutron target'' such as Helium-3. Perhaps the biggest challenge is
then to develop a polarimeter to control its polarization with high accuracy. Most likely this
will be the limited factor for a measurement of the Bjorken sum.

In addition, the sum rule involves the first moments of $g_1$, i.e., one has to worry about possible
extrapolation uncertainties for $x\rightarrow 0$. However, since the Bjorken sum is a non-singlet 
quantity, contributions from the small $x$ region should be under control up to a $1\div 2\%$
once a measurement down to $x\simeq 10^{-4}$ can be performed.
At this level of accuracy one may also expect contributions to matter which break isospin symmetry.

\section{Opportunities in spin physics at an EIC}
\label{sec:rodolfo-spin}


\hspace{\parindent}\parbox{0.92\textwidth}{\slshape 
  Elke C.\ Aschenauer, Rodolfo Sassot, Marco Stratmann
%
}

\index{Aschenauer, Elke}
\index{Sassot, Rodolfo}
\index{Stratmann, Marco}

\vspace{\baselineskip}

Here, we demonstrate how an EIC can address the fundamental open questions concerning the
proton's helicity structure raised in the previous Section. 
A detailed, quantitative discussion of novel electroweak effects in polarized DIS 
can be found in Secs.~\ref{sec:electroweak} and \ref{sec:marco-cc-charm}

\subsection{Scaling violations in inclusive DIS and their impact on $\mathbf{\Delta g(x)}$}
A precise determination of the polarized gluon distribution $\Delta g(x,Q^2)$ in a
broad kinematic regime is a primary goal for the EIC. 
Current determinations of $\Delta g$ suffer from both  
a limited $x$ coverage and fairly large theoretical scale ambiguities
in polarized $pp$ collisions for inclusive (di)jet \cite{deFlorian:1998qp,Jager:2004jh} and 
pion production \cite{Jager:2002xm,deFlorian:2002az}.
Several channels are sensitive to
$\Delta g$ in $ep$ scattering at collider energies
such as DIS jet \cite{Mirkes:1994nt,Feltesse:1996ic} or charm 
\cite{Watson:1981ce,Bojak:1998bd,Bojak:1998zm} production but QCD scaling violations in 
inclusive polarized DIS have been identified as the golden measurement.

The inclusive structure function $g_1(x,Q^2)$ is the most straightforward
probe in spin physics and has been determined in various 
fixed-target experiments at medium-to-large values of $x$ in the last two decades.
It is also the best understood quantity from a theoretical point of view.
Unlike for most other processes, full NNLO corrections of the relevant hard scattering coefficient functions
are available \cite{Zijlstra:1993sh}, and partial results
for the polarized splitting functions at NNLO have been reported in \cite{Vogt:2008yw} recently.
A consistent framework up to NNLO accuracy will be in place by the time of first EIC operations and
is required in order to limit the size of residual theoretical scale uncertainties to the anticipated 
unprecedented level of precision for a polarized DIS experiment. 
To achieve the latter, systematic uncertainties need to be controlled extremely well which imposes
stringent requirements on the detector performance, acceptance, and the design of the interaction region.
Necessary, on-going studies comprise the detection of scattered electrons down to small momenta of
${\cal{O}}(0.5\,\mathrm{GeV})$ to access small $x$, 
the required resolution in momentum and angle of the scattered lepton,
and the unfolding of QED radiative corrections, see Sec.~\ref{sec:detector}.

For studying scaling violations $dg_1(x,Q^2)/d\log Q^2$ 
efficiently, it is not only essential to have good precision but also
to cover the largest possible range in $Q^2$ for any given fixed value of $x$. 
The accessible range in $Q^2$ is again linked (via the inelasticity $y$)
to the capabilities of detecting electrons in an as 
wide as possible range of momenta and scattering angles.
For a detailed discussion of the kinematic coverage at the EIC see Sec.~\ref{sec:detector}.

\begin{figure}[th!]
\begin{center}
\includegraphics[width=0.4\textwidth]{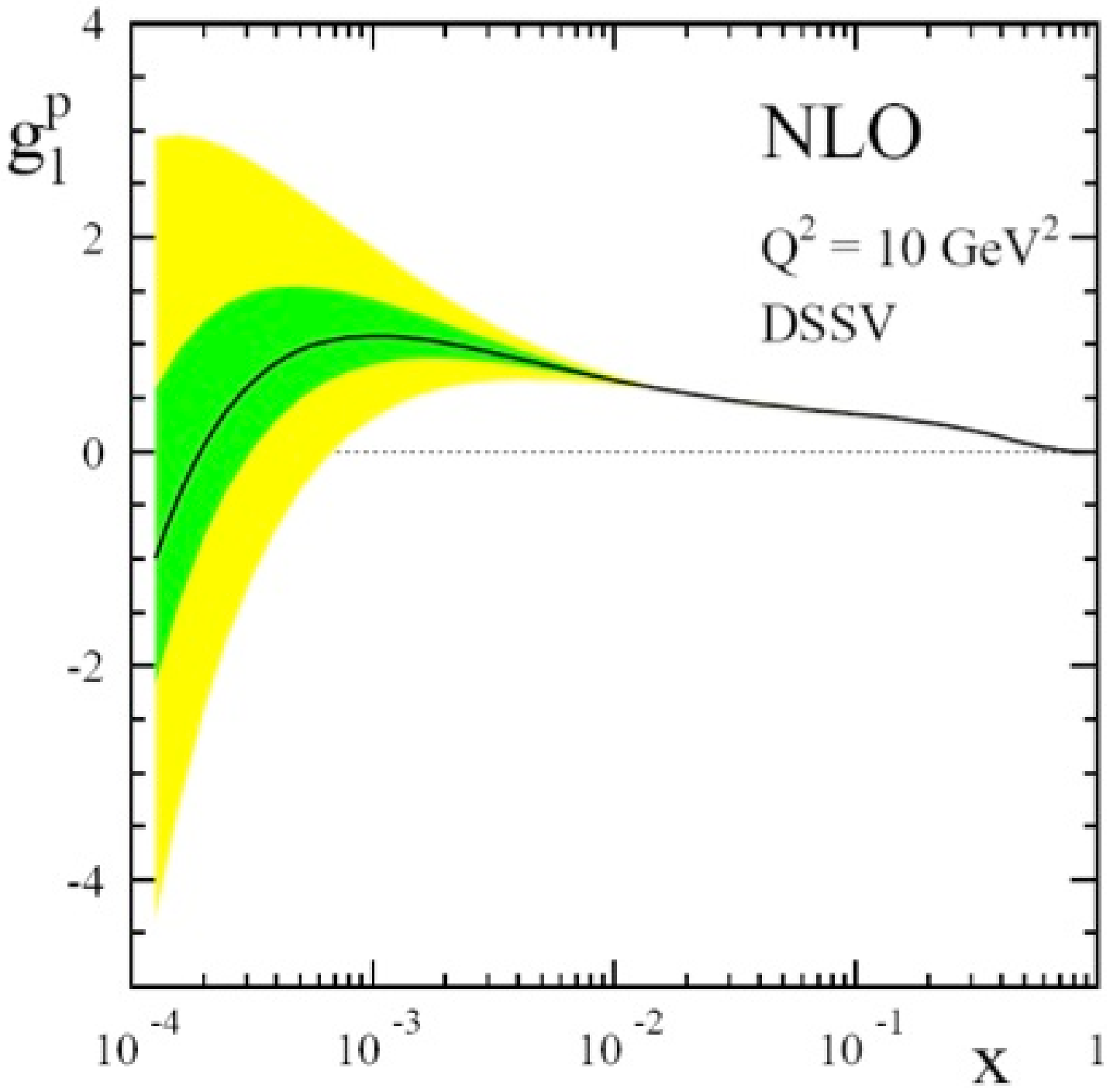}
\end{center}
\vspace*{-0.4cm}
\caption{\label{fig:g1-scal-viol}Spread of predictions for $g_1(x)$ 
induced by the current uncertainty in $\Delta g(x)$.}
\end{figure}
Figure~\ref{fig:g1-scal-viol} highlights the main motivation for a measurement
of $g_1$ at the EIC. The significant uncertainty in $\Delta g(x,Q^2)$ at $x\lesssim  0.01$
shown in Fig.~\ref{fig:newpdfs} translates into a large spread of predictions for 
the behavior of $g_1$ at small $x$. The spin-dependent scale evolution is such that
$dg_1(x,Q^2)/d\log Q^2$ at low $x$ is strongly correlated with the negative of $\Delta g(x,Q^2)$,
i.e., a positive gluon distribution drives $g_1$ at small $x$ to more and more negative
values as $Q^2$ increases, and vice versa.
Hence, a precision measurement of $g_1$ and its logarithmic scale dependence will
determine $\Delta g(x,Q^2)$ at small $x$, hereby dramatically reducing the extrapolation uncertainties of
the integral $\int_0^1 \Delta g(x,Q^2) dx$ entering the proton spin sum rule.
Depending on the shape of $\Delta g(x,Q^2)$ in the unmeasured region, it is currently still possible
to accommodate up to one unit of $\pm\hbar$ at small $x$ \cite{deFlorian:2008mr,deFlorian:2009vb}, i.e., twice the proton spin!
Having determined the functional form of $\Delta g(x,Q^2)$ down to about $10^{-4}$, even extreme 
extrapolations to $x\to 0$ are not expected to contribute anymore significantly to the integral $\int_0^1 \Delta g(x,Q^2) dx$. 

To quantify the impact of polarized DIS measurements on our knowledge of the
gluon density we have performed a series of global QCD analyses based on realistic pseudo-data
for various c.m.s.\ energies at a first stage of eRHIC: 5 GeV electrons on 50, 100, 250, and
325 GeV protons. The simulations are based on the PEPSI Monte Carlo code~\cite{Mankiewicz:1991dp}
using the GRSV ``std'' set of polarised PDFs \cite{Gluck:2000dy}. 
The statistical precision of the data sets for $100-325$ GeV protons
corresponds to about two months of running at the anticipated luminosities for eRHIC with 
an assumed operations efficiency of $50\%$.
For $5\times 50\,\mathrm{GeV}$ an integrated luminosity of $5\,\mathrm{fb}^{-1}$ was assumed.
Demanding a minimum $Q^2$ of $1\,\mathrm{GeV}^2$, $W^2>10\,\mathrm{GeV}^2$,
the depolarization factor of the virtual photon to be $D(y)>0.1$, 
and $0.1\le y\le 0.95$, the highest
$\sqrt{s} \simeq 70\div 80 \,\mathrm{GeV}$ allows one to access $x$ values down to about
$2\times 10^{-4}$. 
As can be seen from the kinematic plots in Sec.~\ref{sec:detector}, the lever-arm in $Q^2$ more and more
diminishes if smaller $x$ values are probed. 
For instance, choosing $Q^2_{\min}=2\,\mathrm{GeV}^2$ would limit the $x$ range to 
$x\gtrsim 4\times 10^{-4}$ at the first stage of eRHIC.
Clearly, one wants to utilize $Q^2$ values as low as possible in a QCD analysis but
once actual EIC data become available one needs to systematically
study how far down $Q^2_{\min}$ can be pushed before the pQCD framework breaks down.
We plan to investigate the impact of the $Q^2_{\min}$ cut on constraining $\Delta g$
based on analyses with the pseudo-data. At small enough $x$ one may observe also
deviations from standard DGLAP evolution as we will discuss briefly below.
A full eRHIC with energies of up to 30 GeV electrons on 325 GeV protons is 
certainly desirable as it would cover the most interesting kinematic 
region around $x=10^{-4}$ at larger values of $Q^2$.

\begin{figure}[th!]
\begin{minipage}[b]{0.45\linewidth}
\centering
\includegraphics[scale=0.4]{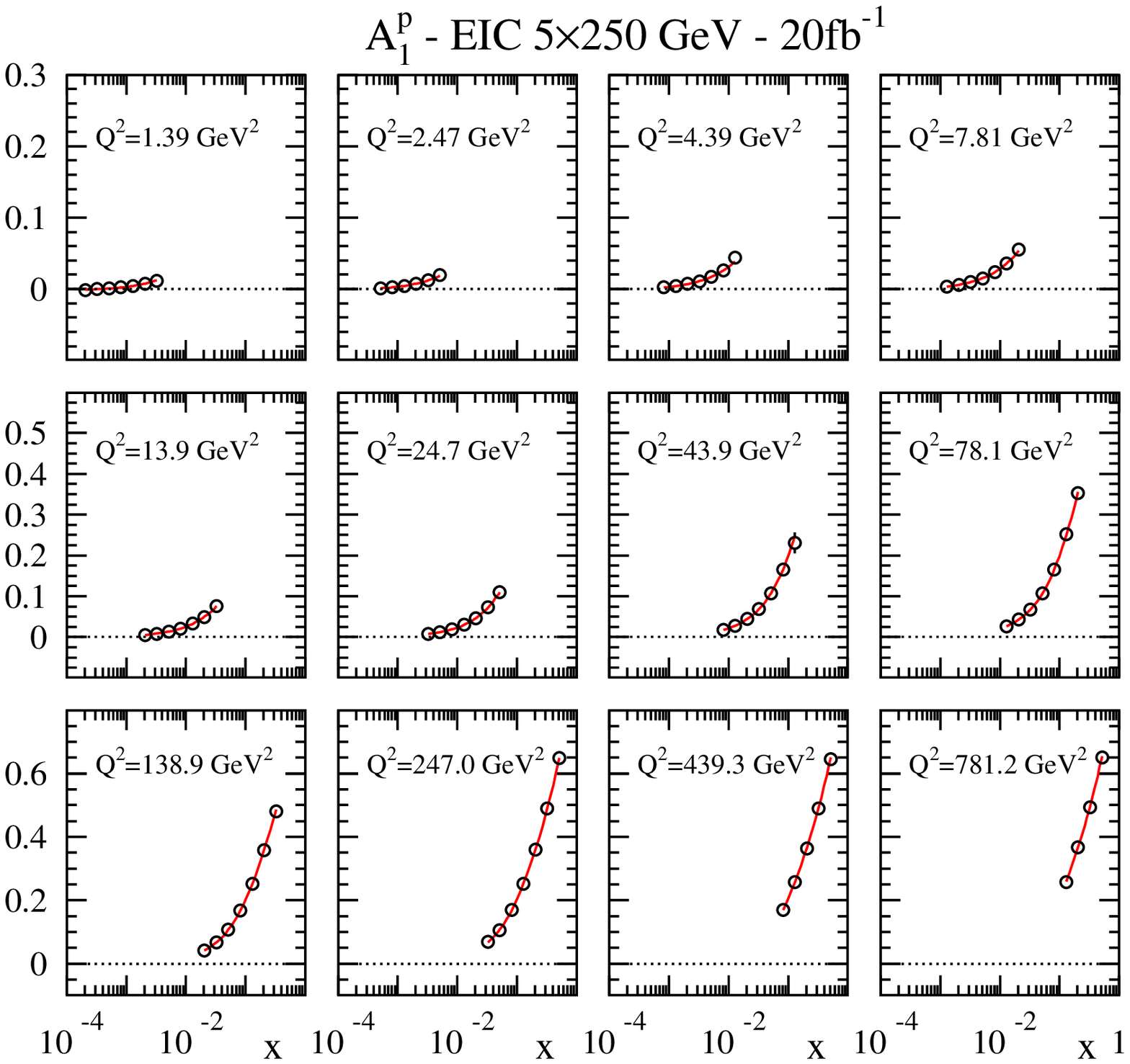}
\end{minipage}
\hspace{0.4cm}
\begin{minipage}[b]{0.52\linewidth}
\centering
\includegraphics[scale=0.47]{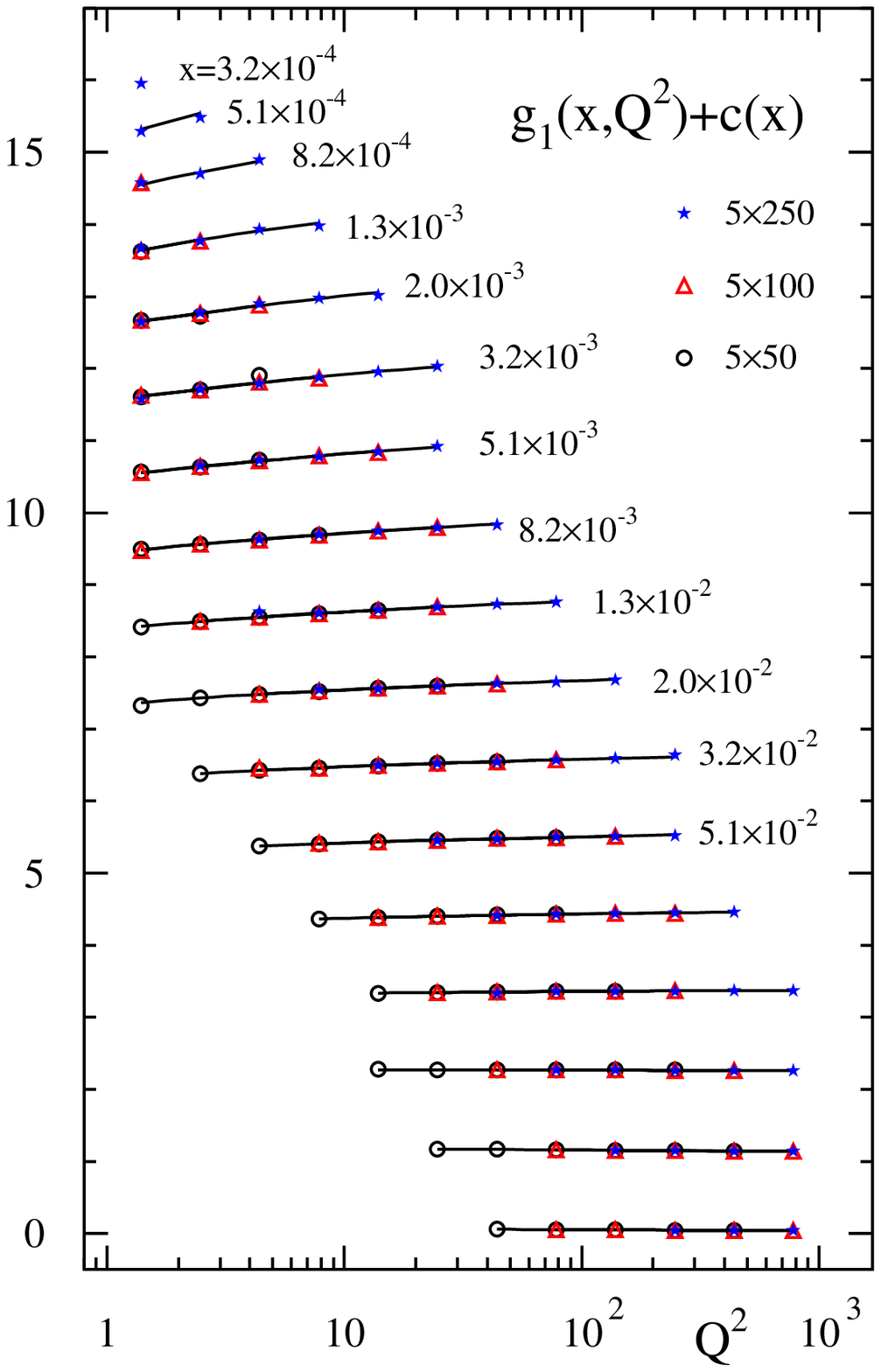}
\end{minipage}
\vspace*{-0.4cm}
\caption{\label{fig:g1-pseudo}{\bf left:} generated pseudo-data for $A_1$ in bins of $Q^2$ for
$5\times 250\,\mathrm{GeV}$ collisions; {\bf right:} $g_1$ as a function of $Q^2$ for fixed $x$
for 5 GeV electrons on three different proton energies.} 
\end{figure}
The l.h.s.\ of Fig.~\ref{fig:g1-pseudo} shows the $x$ and $Q^2$ coverage for one 
of the simulated data sets for the spin asymmetry $A_1$. The statistical uncertainties
are in general way too small to be visible.
For the smallest $x$ and $Q^2$ values, the expected size of the asymmetries
is of the order of a few times $10^{-3}$, which sets the scale for the required 
experimental precision.
On the r.h.s.\ of Fig.~\ref{fig:g1-pseudo} we show the $Q^2$ dependence of the
structure function $g_1$ for various bins in $x$. As can be seen, combining the data sets for
the different c.m.s.\ energies extends the coverage in $x$ and $Q^2$.
We note that present fixed-target data, cf.\ Fig.~\ref{fig:g1scaling}, 
all fall in the lower right corner of the plot but have some
overlap with the projected EIC data.

\begin{figure}[th!]
\begin{center}
\includegraphics[width=0.42\textwidth]{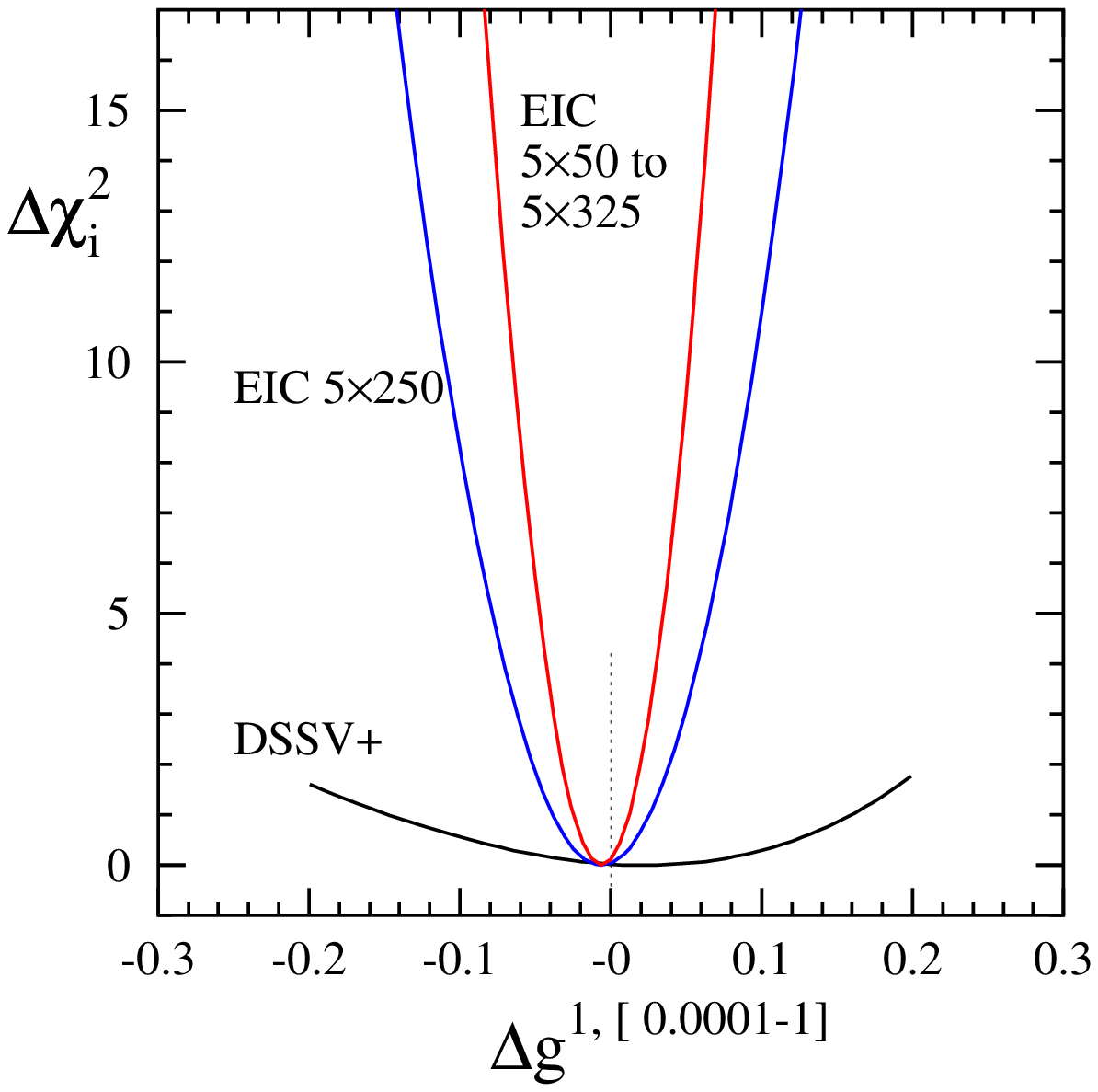}
\includegraphics[width=0.42\textwidth]{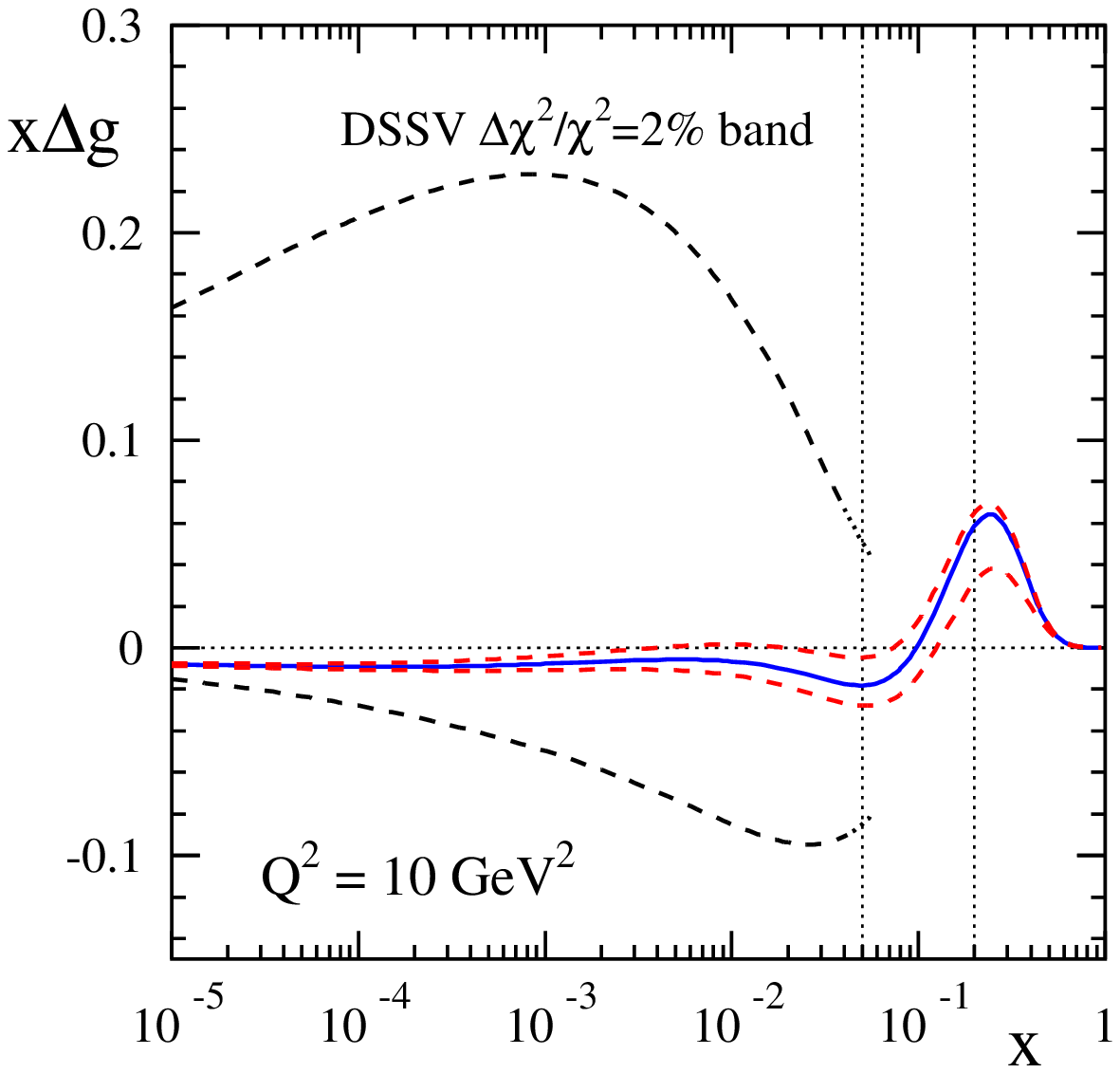}
\end{center}
\vspace*{-0.4cm}
\caption{\label{fig:deltag-impact} $\chi^2$ profiles for the
truncated $x$ integral of $\Delta g$ (l.h.s.) and uncertainty bands for $x\Delta g$ 
referring to $\Delta \chi^2/\chi^2=2\%$ (r.h.s.) with and without including the generated 
EIC pseudo-data in the fit.}
\end{figure}
The pseudo-data for the spin asymmetry $A_1$ have been added to a global QCD fit of 
helicity-dependent PDFs based on the DSSV framework \cite{deFlorian:2008mr,deFlorian:2009vb}. 
We have used the projected uncertainties to randomize the pseudo-data by one sigma 
around their central values determined by the DSSV set of PDFs.
To demonstrate the impact of the generated EIC data on $\Delta g$, we show on the l.h.s.\ of 
Fig.~\ref{fig:deltag-impact} the $\chi^2$ profile for the first moment of $\Delta g$ truncated
to the range $10^{-4}\le x <1$ where EIC DIS data with 
$Q^2>1\,\mathrm{GeV}^2$ can potentially constrain
its value. As can be inferred from the plot, the fit based on all presently available DIS, SIDIS, 
and RHIC $pp$ data set (labeled as ``DSSV+'' and described in the previous Section) only very marginally constrains
the integral. Adding in the projected data for $5\times 250$ GeV, shown in 
Fig.~\ref{fig:g1-pseudo}, already greatly improves the $\chi^2$ profile. Including all four
EIC data sets determines the integral very well; recall that the width of the profile
determines the uncertainty for a given, tolerated increase $\Delta \chi^2$.
To achieve such a level of accuracy, the data sets with the highest
$\sqrt{s} \simeq 70\div 80 \,\mathrm{GeV}$ are most critical in the fit as they probe the smallest $x$ values.

Even more impressive is the reduction of the ambiguities on the $x$ shape of $\Delta g(x,Q^2)$ shown on the r.h.s.\
of Fig.~\ref{fig:deltag-impact}. The currently completely undetermined shape for
$x\lesssim 0.01$ can be mapped precisely to an accuracy of about $\pm 10\%$ (or better)
for $\gtrsim 10^{-4}$. Below $\approx 2\times 10^{-4}$ the shown $\Delta g(x,Q^2)$ 
and its uncertainties are not constrained by the
projected EIC data and merely result from an extrapolation of the used functional form.
We note that since one needs to control all sources of uncertainties extremely well it might be advantageous
to measure and analyze polarized cross sections instead of spin asymmetries
traditionally used so far. This should
greatly simplify the theoretical analysis as one does not need any information on
unpolarized PDFs or the ratio of $\sigma_L/\sigma_T$ anymore.  
There are also first, very interesting attempts to analyze polarized DIS data 
within the methodology of neural networks \cite{DelDebbio:2009sq,Rojo:2010gt}, which provides a less 
biased way to estimate PDF uncertainties than standard approaches based on pre-defined functional forms.  

As was mentioned above, one expects to find deviations from DGLAP evolution at sufficiently small
values of $x$. In contrast to the unpolarized case, the dominant contribution of gluons 
mixes with quarks also at $x\ll 1$. From DGLAP evolution one expects for the small $x$ behavior
of gluons and quarks 
\begin{equation}
\Delta q(x,Q^2), \Delta g(x,Q^2) \simeq \exp \left[ \mathrm{const}\times
\alpha_s \ln(Q^2/\mu^2) \ln (1/x)\right]^{1/2}
\label{eq:g1smallx}
\end{equation}
assuming for simplicity a fixed coupling $\alpha_s$.
In \cite{Bartels:1995iu,Bartels:1996wc} it was demonstrated that this simple behavior
can strongly underestimate the rise at small $x$ due to other 
potentially large double logarithmic contributions
of the type $\alpha_s \ln^2(1/x)^n$ in the $n$-th order of $\alpha_s$ 
which are beyond the standard DGLAP framework.
This gives rise to a power-like behavior of $g_1$ at small $x$ of the form
$g_1(x,Q^2)\sim(1/x)^{{\cal{O}}(\alpha_s)}$. There are qualitative arguments that
in the polarized case the relevance of these logarithms in $1/x$ is larger than the
difference between DGLAP and BFKL evolution in the unpolarized case \cite{Bartels:1995iu,Bartels:1996wc}.
However, more detailed quantitative studies are still lacking, and it remains to be seen
if the kinematic reach of an EIC is large enough to actually observe deviations from
DGLAP in polarized DIS. Clearly, any such estimate will strongly depend upon the initial
input distributions, and eventually one needs data to clarify the relevance of small $x$ enhancements.
Finally, we note that in Ref.~\cite{Ermolaev:2005ny} the leading small $x$ logarithms were combined with
DGLAP evolution, and some effects of running coupling were addressed in \cite{Ermolaev:2003zx}.

\subsection{Charm Contribution to $\mathbf{g_1}$}
\begin{figure}[th!]
\begin{center}
\includegraphics[width=0.43\textwidth]{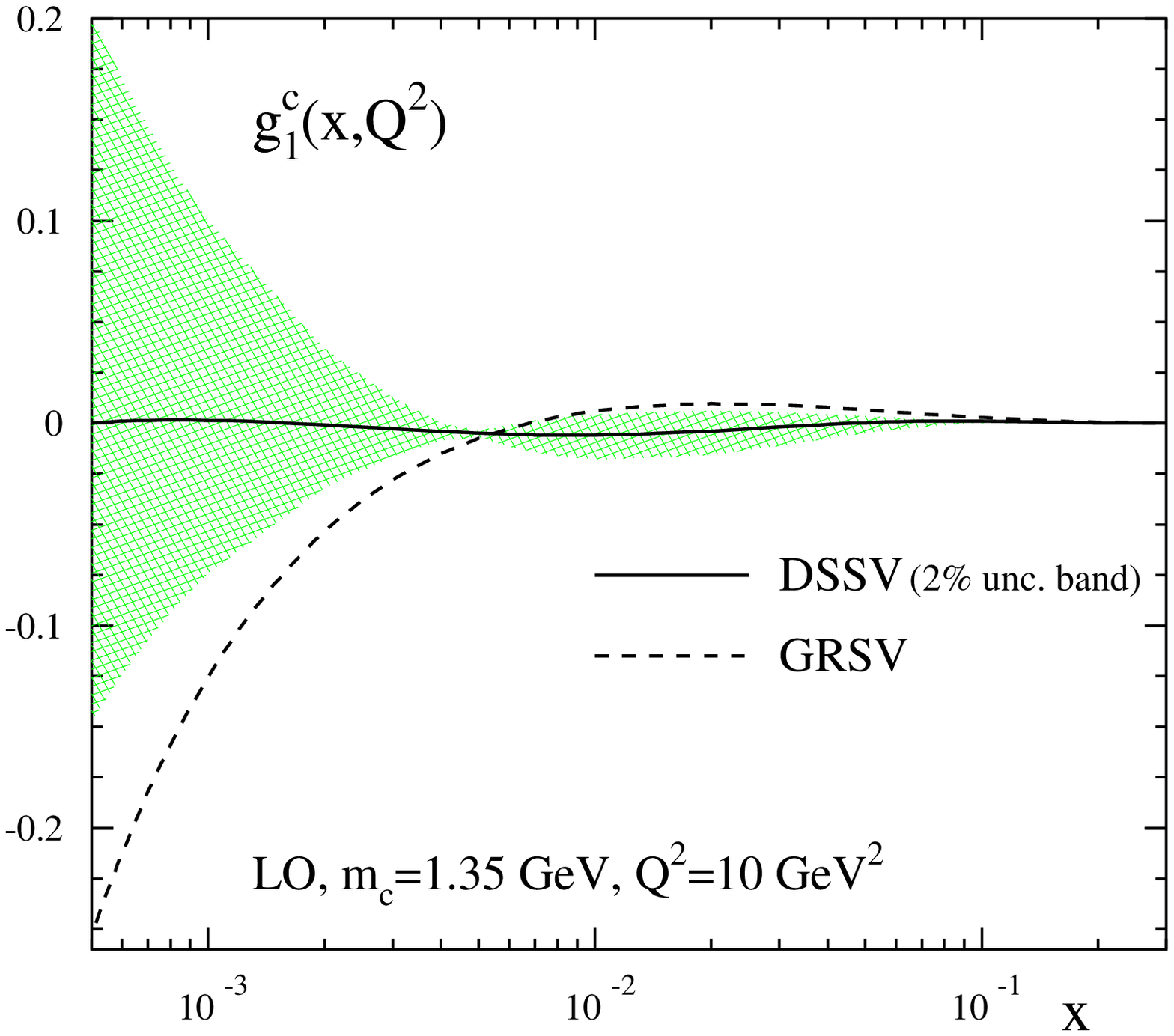}
\includegraphics[width=0.43\textwidth]{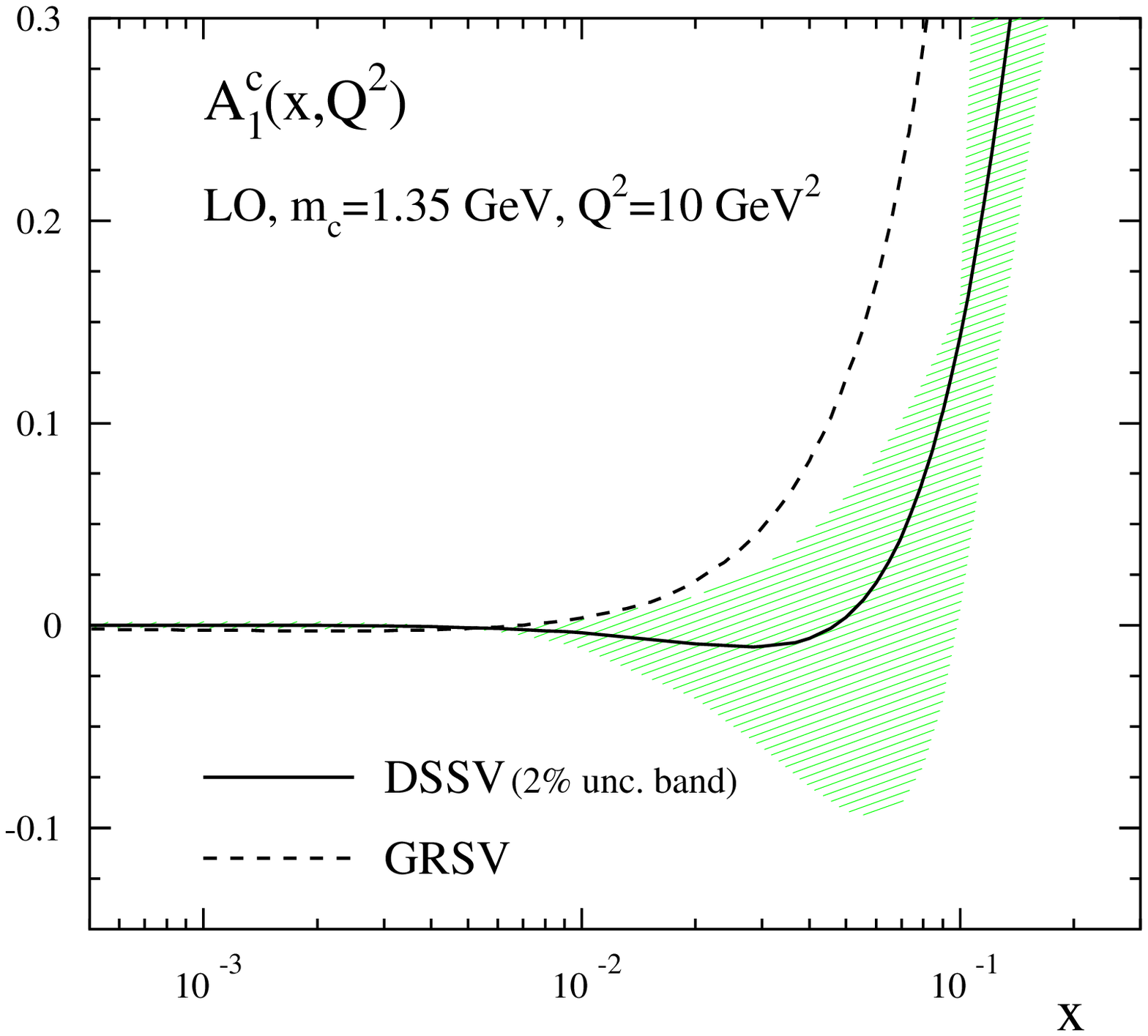}
\end{center}
\vspace*{-0.4cm}
\caption{\label{fig:g1charm} LO expectations for $g_1^c$ (l.h.s.) and $A_1^c$ (r.h.s) for the
$Q^2=10\,\mathrm{GeV}^2$, $m_c=1.35\,\mathrm{GeV}$, and using the
DSSV and GRSV ``std'' sets of PDFs. The shaded band corresponds to the $\Delta \chi^2/\chi^2=2\%$
uncertainty estimate of DSSV.}
\end{figure}
As discussed in Sec.~\ref{sec:bluemlein-hq} in the context of unpolarized DIS structure functions, the contributions
from heavy flavors require a special theoretical framework. For the kinematic regime covered
at the EIC it is expected that effects of the finite heavy quark mass play an important role and should not be neglected.
This is, of course, particularly relevant not too far from threshold, i.e., for $Q^2$ less than
a few times $m_h^2$.

For all presently available DIS data, the charm contribution to $g_1$ can be safely neglected and, hence, is
usually not included in any of the QCD analyses except for the fit presented in Ref.~\cite{Blumlein:2010rn}.
The relevant coefficient functions for $\gamma^* g \to c\bar{c}X$ have been calculated only to LO 
accuracy \cite{Watson:1981ce} so far which is not sufficient for
the anticipated experimental precision. The computation of the NLO corrections is, however,
work in progress and results should become available for more detailed quantitative studies soon.

For spin dependent DIS the heavy quark contributions 
are expected to be smaller than in the helicity-averaged case 
but, of course, will very much depend on the
currently unknown size of $\Delta g(x,Q^2)$ at small $x$.
There is also an interesting constraint on the  
gluonic Wilson coefficient for heavy quark production, demanding a
vanishing first moment when regulated dimensionally or with a quark mass
\cite{Bodwin:1989nz,Vogelsang:1990ug}.
This leads to a non-trivial oscillating pattern for $g_1^c$
depending on the sign of $\Delta g$ which
will look rather different in the case that $\Delta g$ itself changes
sign within the $x$ range probed.

Figure~\ref{fig:g1charm} shows some expectations for the spin asymmetry $A_1^c$ for DIS charm
production (r.h.s.) and the charm contribution to the structure function $g_1$ (l.h.s.)
both computed at LO accuracy with two different polarized gluon distributions. For a small
$\Delta g$ with a node, as in the best fit of DSSV, the charm contribution turns out to be at most at the
percent level even at collider kinematics, and the corresponding spin asymmetry is most likely too
small, ${\cal{O}}(\mathrm{few}\times 10^{-5})$, to be measured directly.
For a larger gluon distribution at small $x$, as in the GRSV fit, or for a gluon within the 
current uncertainty band of DSSV, asymmetries can be significantly larger, reaching 
${\cal{O}}(\mathrm{few}\times 10^{-3})$, and at $x=10^{-3}$ and $Q^2\simeq 10\,\mathrm{GeV}^2$
charm quarks can contribute about $10\div 15\%$ to the inclusive $g_1$.
The experimental aspects for detecting charmed mesons have beed discussed
already in Sec.~\ref{sec:flcharm} and apply also here.

\subsection{Remark on the Bjorken sum rule}
The Bjorken sum rule
\begin{equation}
\label{eq:bjsum}
\int_0^1 dx \left[ g_1^p(x,Q^2)-g_1^n(x,Q^2) \right] =
\frac{1}{6} C_{Bj}\left[\alpha_s(Q^2)\right] g_A
\end{equation}
is not only one of the most fundamental relations in QCD 
but presumably also
one of the best known quantities in pQCD. Corrections up to
${\cal{O}}(\alpha_s^4)$ have been calculated 
\cite{Larin:1990zw,Larin:1991tj,Baikov:2010je}.
Given the anticipated precision of DIS measurements at the EIC, it is
natural to ask what can be achieved concerning the Bjorken sum.
The major obstacle is, of course, the need for an effective, longitudinally polarized neutron beam.
One conceivable option would be to run with $^3\mathrm{He}$ but developing
a method to measure its polarization to the required percent level is certainly an
extremely challenging R\&D task requiring novel ideas.
From the theoretical side it might be advantageous to analyze the data not in
terms of PDFs but directly on the structure function level with the help
of so called ``physical anomalous dimensions'' \cite{Bluemlein:2002be}. This reduces not only the number
of parameters but also theoretical scale uncertainties.

\begin{figure}[th!]
\begin{center}
\includegraphics[width=0.42\textwidth]{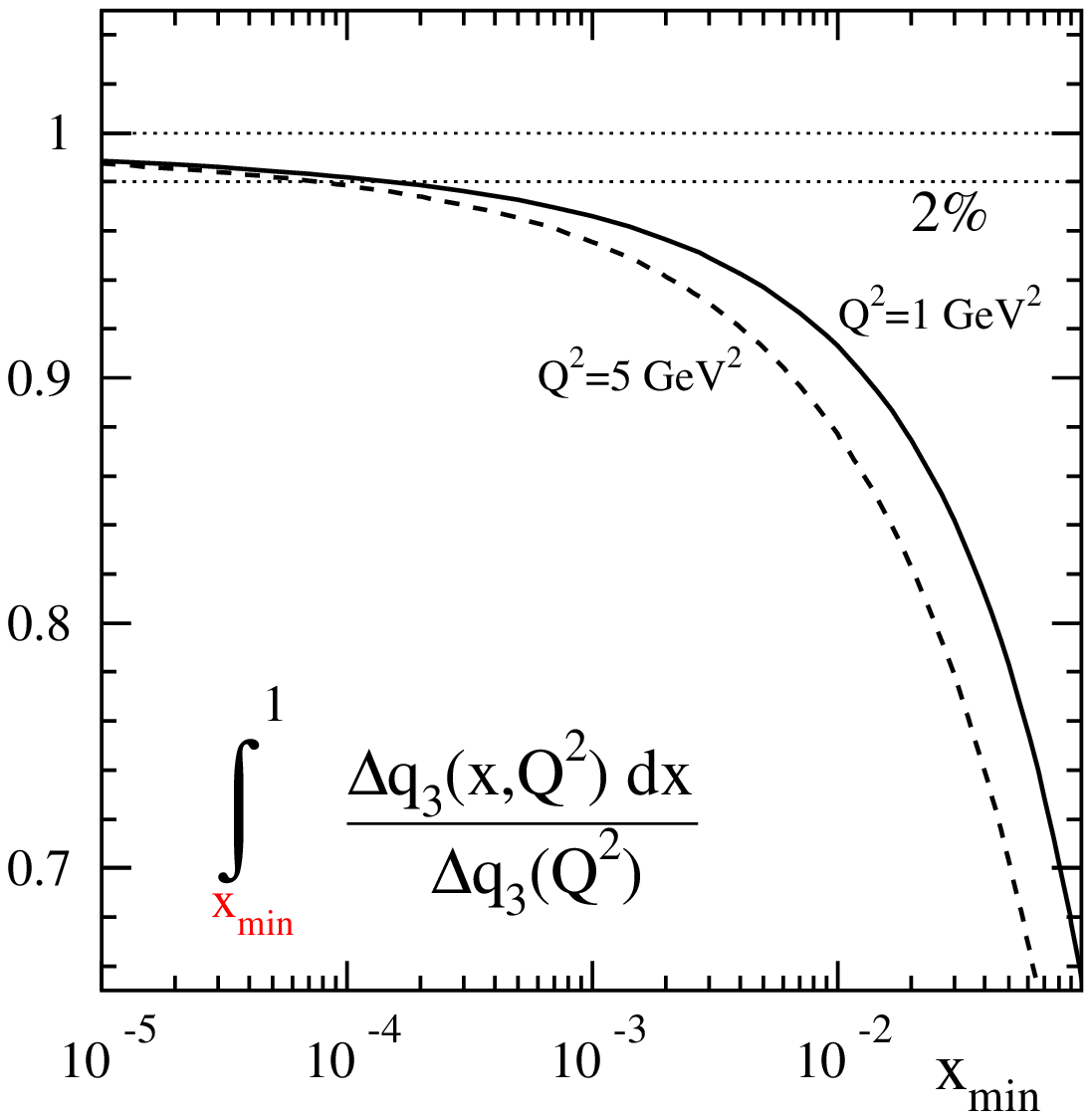}
\end{center}
\vspace*{-0.4cm}
\caption{\label{fig:bjsum} 
The truncated (``running'') $x$ integral for the non-singlet combination $\Delta q_3$ related to
the Bjorken sum normalized to the full first moment for two values of $Q^2$.}
\end{figure}
From present fixed target experiments the sum rule is currently verified
to about $10\%$, which sets the target for any future measurement
to the $1\div 2$ percent level.
One of the current limitations is the extrapolation uncertainty
from the unmeasured small $x$ region. Since the Bjorken sum probes a non-singlet (NS)
quark combination, the small $x$ uncertainties are considerably less severe than
for $\Delta g(x,Q^2)$, but to reduce them to a level of about $2\%$,
measurements of $g_1^{p,n}$ down to $x\simeq 10^{-4}$ are required. 
This is illustrated in Fig.~\ref{fig:bjsum} where we show the ``running'' $x$ integral for the relevant
NS quark combination $\Delta q_3$ normalized to its full first
moment, assuming the functional form from the DSSV analysis.
At the required $1\div 2\%$ level of accuracy one might start to see deviations from (\ref{eq:bjsum})
due to isospin and charge symmetry violations. Very little is known about these effects,
and, if experimentally feasible, measurements could reveal genuine new insights into the hadronic
structure.

\begin{figure}[th!]
\begin{center}
\includegraphics[width=0.56\textwidth]{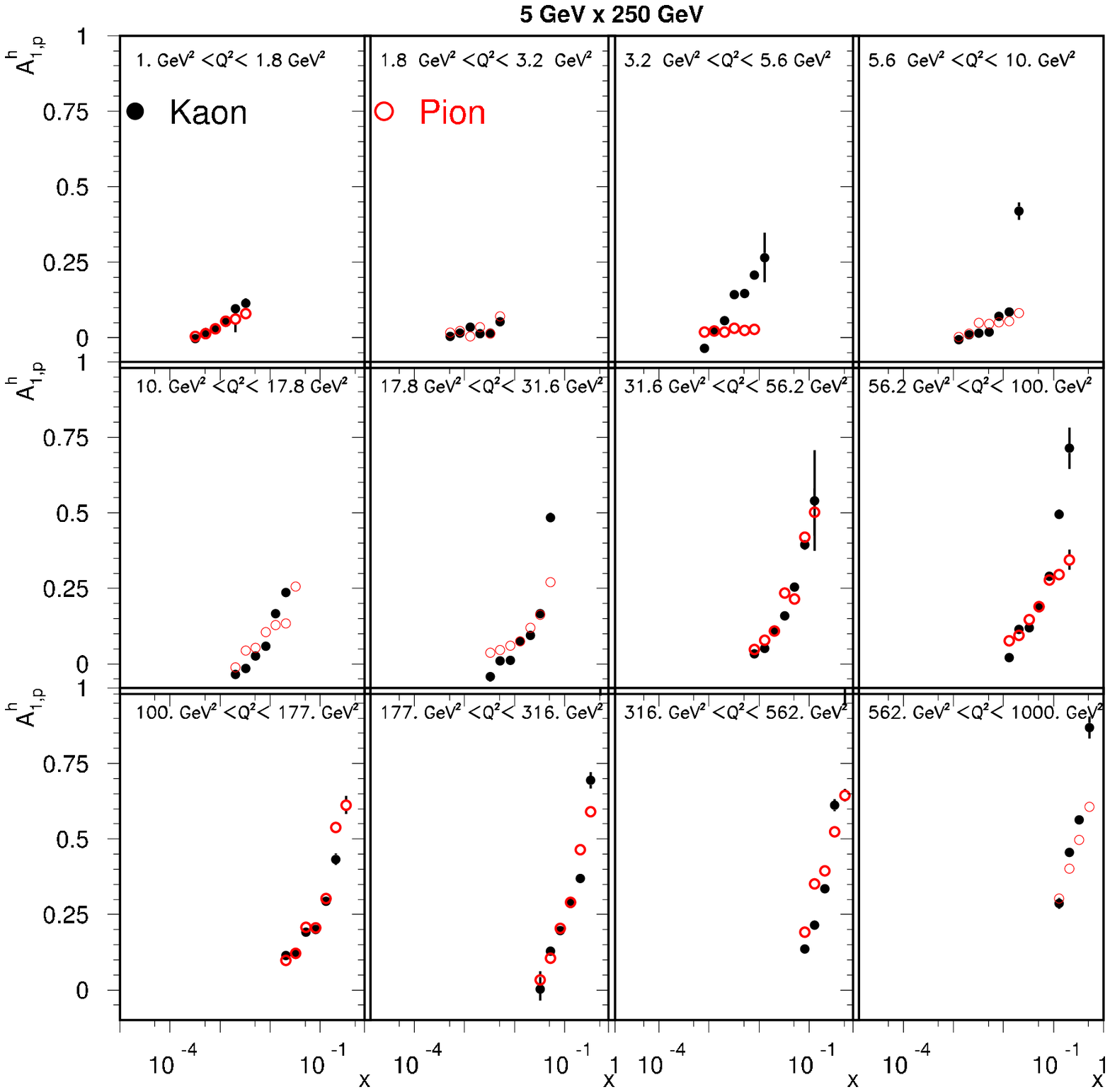}
\end{center}
\vspace*{-0.3cm}
\caption{\label{fig:polsidis} 
Projected spin asymmetries for pion and kaon production in SIDIS
for beam energies of $5\times 250$ GeV and various bins in $Q^2$.}
\end{figure}
The fundamental relation (\ref{eq:bjsum}) between a high-energy measurement
of DIS structure functions and a low-energy quantity like the axial charge $g_A$
by itself warrants an experimental exploration at the EIC. 
From a more theoretical perspective one might argue that since ${\cal{O}}(\alpha_s^4)$
corrections are available, a precision measurement of the Bjorken sum can be turned into
one of the most accurate determinations of $\alpha_s$.
One can easily convince oneself, however, that this does not work out. Changing
$\alpha_s$ by about one percent, translates only in a $0.1\%$ change of 
the Bjorken sum, which is impossible to resolve experimentally.
Perhaps more interesting is the non-trivial connection of the Bjorken sum rule to the
Adler $D(Q^2)$ function which naturally appears, for instance, 
in the $e^+e^-$ annihilation into hadrons \cite{Adler:1974gd}.
These two, seemingly unrelated quantities are connected through the generalized
Crewther relation \cite{Crewther:1997ux,Baikov:2010je}.
For large enough $Q^2$, the Adler function can be expanded as a power series in
$\alpha_s$ like $C_{Bj}\left[\alpha_s(Q^2)\right]$ in (\ref{eq:bjsum}), and results
are available up to ${\cal{O}}(\alpha_s^4)$ as well \cite{Baikov:2008jh}. The Crewther
relation then states for the NS part of the $D$ function that
\begin{equation}
\label{eq:crewther}
D[\alpha_s(Q^2)] \,C_{Bj}[\alpha_s(Q^2)] = 3 \left[ 1+ \frac{\pi\beta(\alpha_s)}{\alpha_s}
K[\alpha_s(Q^2)]\right]
\end{equation}
where $\beta$ denotes the QCD beta function, and the first four terms in the
expansion of $K[\alpha_s(Q^2)]$ are known. 
The term proportional to $\beta$
in (\ref{eq:crewther}) describes the deviation from the limit of
exact conformal invariance of QCD \cite{Crewther:1997ux,Braun:2003rp}.
We also note that since the Bjorken sum rule can be measured down to small values of $Q^2$
it provides a way to define an effective strong coupling constant \cite{Grunberg:1980ja,Deur:2005cf} 
which is by construction gauge and scheme invariant and approaches the standard running of $\alpha_s$ in the
perturbative domain.

\subsection{Opportunities in semi-inclusive DIS}
As has been mentioned in Sec.~\ref{sec:polstatus}, the flavor separation of polarized PDFs
in current fits is largely based on pion and kaon yields in SIDIS. An EIC can easily extend the
existing kinematic coverage in the same way as for inclusive DIS. 
Prerequisites for exploiting SIDIS as a precision tool at the EIC, 
such as good particle identification and well constrained fragmentation functions,
have been already discussed in Sec.~\ref{sec:marco-sidis} for the unpolarized case.

Figure~\ref{fig:polsidis} shows projected data for the longitudinal spin asymmetry
in SIDIS with identified pions and kaons in the same $Q^2$ bins as used for
inclusive DIS studies in Fig.~\ref{fig:g1-pseudo}. 
The simulation is based on the PEPSI Monte Carlo \cite{Mankiewicz:1991dp}
using the GRSV ``std'' set of polarised PDFs \cite{Gluck:2000dy}. 
The following cuts have been applied to model some detector and acceptance effects:
$Q^2>1\,\mathrm{GeV}^2$, $0.1<y<0.95$, photon depolarization factor $D(y)>0.1$, $W^2>10\,\mathrm{GeV}^2$, 
$0.2<z<0.8$, $p_H>1.5\mathrm{GeV}$, and $1^{\circ}<\theta_H<179^{\circ}$. 
The momentum cut on the detected hadron $H$ is placed to ensure 
to be above the PID Cherenkov threshold.
The statistical precision reflects
one month of running at the luminosities anticipated for the first stage of eRHIC.
Again, these measurements will be limited by systematic uncertainties, which have
to be addressed in detail. In addition to the sources of systematic uncertainties
present for inclusive DIS, the detector performance for the identification of different
produced hadron species is most critical for SIDIS.  
Additional sets of data have been generated for other combinations of electron and
proton beam energies. They are currently being implemented into the same global QCD analysis
framework used to analyze the projected inclusive DIS data above. 
Plots similar to those for the $\chi^2$ profile of the truncated $x$ integral 
and the $x$ dependent uncertainty bands for $\Delta g(x,Q^2)$ in Fig.~\ref{fig:deltag-impact} will be 
prepared to quantify the impact of SIDIS data on our knowledge of helicity-dependent quark densities.
We expect that all light quark and anti-quark flavors, i.e.,
$\Delta u$, $\Delta \bar{u}$, $\Delta d$, $\Delta \bar{d}$, $\Delta s$, and $\Delta \bar{s}$,
can be determined with a precision close to the one obtained for $\Delta g(x,Q^2)$
in Fig.~\ref{fig:deltag-impact}.

Although knowledge of individual quark and anti-quark flavors is in principle not required for 
an understanding of the proton spin sum rule, where only the total quark singlet $\Delta \Sigma$ enters,
it would provide deeper insight into the question why the observed total quark polarization is considerably
smaller than in naive quark models. Here, it is essential to understand in detail how sea quarks are
polarized, i.e., whether they have a preference for spinning ``against'' the direction of the proton
spin thereby diluting the total quark polarization. Current QCD fits \cite{deFlorian:2008mr,deFlorian:2009vb} 
start to reveal rather complicated patters of polarization at medium-to-large $x$ with possible sign changes but
the statistical precision and kinematic reach of the fixed-target data is not sufficient for any definitive conclusions.

\begin{figure}[!ht]
\begin{center}
\vspace*{-0.8cm}
\includegraphics[width=0.5\textwidth]{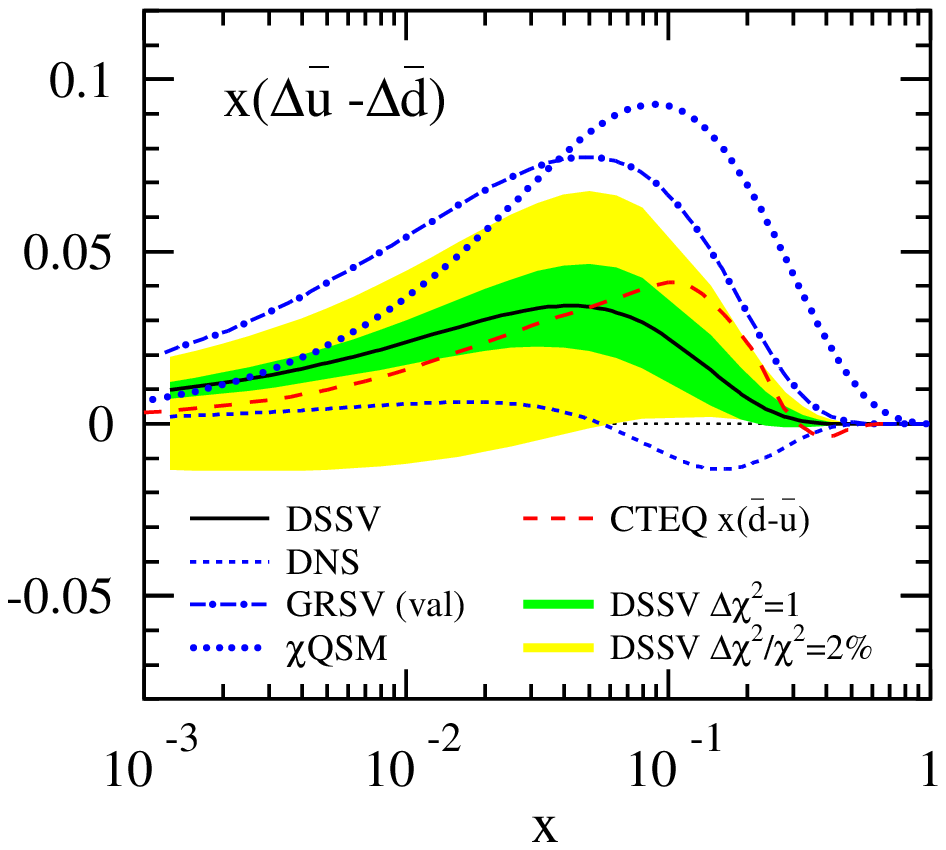}
\end{center}
\vspace*{-0.85cm}
\caption{$x(\Delta \bar{u}-\Delta \bar{d})$ at $Q^2=10$~GeV$^2$ along with the uncertainty
bands from DSSV, results from earlier global fits, and predictions from 
the chiral quark soliton model \cite{Diakonov:1996sr,Diakonov:1997vc}.
\label{fig:ubardbar}}
\vspace*{0.5cm}
\end{figure}
To give an example, Fig.~\ref{fig:ubardbar} shows the current significance of 
a possible asymmetry in the light quark sea, $\Delta \bar{u}(x)-\Delta \bar{d}(x)$.
Given the well-established 
pronounced difference between $\bar{u}$ and $\bar{d}$ in the spin-averaged  
case, a precise determination of $\Delta \bar{u}(x)-\Delta \bar{d}(x)$
is of of particular interest.
Different patterns of symmetry breaking in the light anti-quark sea polarizations have been predicted 
qualitatively by a number of models of nucleon structure. 
For instance, within the large-$N_c$ limit of QCD as incorporated in the
chiral quark soliton model \cite{Diakonov:1996sr,Diakonov:1997vc,Wakamatsu:1998rx,Wakamatsu:2003wg}
one expects $|\Delta\bar{u}-\Delta\bar{d}| > |\bar{u}-\bar{d}|$. 
In addition, charged kaon data should help to clarify issues related to SU(3) symmetry and the 
polarized strangeness density $\Delta s(x,Q^2)$ by providing sufficient input to determine
its first moment reliably. 

\section{Electroweak structure functions at the EIC}
\label{sec:electroweak}


\hspace{\parindent}\parbox{0.92\textwidth}{\slshape 
Abhay Deshpande, Krishna Kumar, Felix Ringer,
Seamus Riordan, Swadhin Taneja, 
Werner Vogelsang
%
}

\index{Deshpande, Abhay}
\index{Kumar, Krishna}
\index{Ringer, Felix}
\index{Riordan, Seamus}
\index{Taneja, Swadhin}
\index{Vogelsang, Werner}





\subsection{Motivation and Introduction}

The use of charged leptons to probe the structure of nucleons through 
electroweak interactions has proven to be an invaluable tool in our 
exploration of the strong force. Experiments on deep inelastic scattering 
(DIS) $ep\to eX$, which dominantly proceeds via the exchange of a virtual 
photon between the electron and the nucleon, have established the 
existence of quarks and provided detailed studies of the short range 
aspects of the strong coupling.

It is well known that {\it neutral 
current} (NC) interactions can also be mediated by the $Z$-bosons of 
the weak interactions, and their interference with the photon.  
This gives rise to parity violating effects, which offer 
complementary access to nucleon structure. This has been a theme
at parity violating electron scattering experiments, both at fixed target 
facilities~\cite{Kumar:2000eq,Nakamura:2010zzi} and at HERA~\cite{Zhang:2007pp,Diaconu:2010zz}. 
For an unpolarized target, the NC parity violating asymmetry
is given by
\begin{equation}
A_{\mathrm{beam}}\equiv \frac{\sigma_R-\sigma_L}{\sigma_R+\sigma_L} \;,
 \label{eq:anc}
\end{equation}
where $\sigma_R$ ($\sigma_L$) denotes the cross section for right- (left-)
handed electrons. For fixed-target experiments, where the virtuality $Q$
of the exchanged boson is typically
much smaller than the $Z$-boson mass $M_Z$, 
only $\gamma Z$-interference is relevant, and one obtains
\begin{equation}
A_{\mathrm{beam}} \sim
\frac{G_F M_Z^2}{2 \sqrt{2}\pi\alpha}\frac{Q^2}{Q^2+M_Z^2} 
\underset{Q^2\ll M_Z^2}{\simeq} 10^{-4}Q^2 [\mathrm{GeV}^2]\;,
 \label{eq:anc1}
\end{equation}
with the Fermi constant $G_F$ and the fine structure constant $\alpha$.  
At modern fixed target facilities, measured asymmetries were typically 
of the order of $10^{-4}$ or less~\cite{Kumar:2000eq}. At HERA, on the other hand,
with its enormous kinematic reach in $Q^2$, also contributions by
pure $Z$-exchange play a role~\cite{Zhang:2007pp}. 

{\it Charged current} (CC) interactions in DIS lepton scattering measurements have 
been performed at HERA in $e^\pm p$ collisions~\cite{Zhang:2007pp} and at 
various neutrino scattering experiments~\cite{Tzanov:2009zz}. They are 
inaccessible at fixed target charged lepton beam facilities
where $Q^2 \ll M_W^2$. 

An EIC provides a number of advantages in the study of 
structure functions through electroweak interactions over previous and 
existing facilities.  As the asymmetries and relative likelihood of 
$Z^0$ and $W^\pm$ exchange monotonically increase with $Q^2$, larger 
c.m.s.\ energies are more favorable for such measurements.  
Additionally, advances in accelerator and source technologies should 
provide luminosities on the order of $\sim 10^{34}$~cm$^{-2}$~s$^{-1}$,
two orders of magnitude higher than what was available
at HERA. A new feature will be the ability for bunch-by-bunch variation of the 
sign of the longitudinal polarization of both the electron and hadron beams.  
A broader $Q^2$ and $y$ acceptance than at fixed target 
facilities, and variable beam energy, also allow for separation 
of the various structure functions.  High precision is possible over
a broad range in Bjorken-$x$, $0.01\lesssim x \lesssim 0.4$, 
whereas fixed target facilities typically are sensitive only to $x>0.1$.

\subsubsection{Polarized Hadrons}

Arguably the most important feature at the EIC is the availability of 
polarized $^1$H, and potentially $^2$H and $^3$He, 
beams with rapid polarization flips, 
which offers access to electroweak spin structure 
functions that may provide additional constraints on polarized PDFs. 
The counterpart of $A_{\mathrm{beam}} $ in 
(\ref{eq:anc}) with polarized protons has never been measured before,
and neither have spin asymmetries in CC interactions.
Both would in principle be accessible at the EIC. 

The theoretical study of electroweak spin-dependent structure functions 
dates back to the seventies~\cite{Nash:1971aw,Dicus:1972pq,Wray:1972yb,Derman:1973sp,Ahmed:1976ee,Joshipura:1976sd,Kaur:1977ce,Bartelski:1979wc,Hochberg:1984en}. 
Renewed interest arose in the nineties
in the context of a possible polarized $ep$ program at HERA~\cite{Lampe:1989um,Vogelsang:1990ka,Ravishankar:1990tz,Mathews:1991sn,Mathews:1992pv,Ji:1992kv,deFlorian:1994wp,Stratmann:1995fn,Blumlein:1996tp,Blumlein:1996vs,Anselmino:1993tc,Anselmino:1996cd,Contreras:1997fc}, 
and later in terms of studies for a neutrino factory~\cite{Forte:2001ph}. 
Parity-violating spin structure functions were shown to contain 
rich information on polarized PDFs. For example,
as we shall discuss in more detail in the next section, 
for CC interactions via $W^-$ exchange in the 
parton model, two structure functions $g_1^{W^-}$ and $g_5^{W^-}$ 
contribute to the spin asymmetry~\cite{Anselmino:1993tc,Anselmino:1996cd}:
\begin{equation}\label{a5}
A^{W^-} = \frac{2b g_1^{W^-}+a g_5^{W^-}}
{a F_1^{W^-}+b F_3^{W^-}} \; ,
\end{equation}
where $a=2(y^2-2y+2)$, $b=y (2-y)$, and
\begin{equation}\label{g5lo}
g_1^{W^-} (x) = \Delta u (x) + \Delta\bar{d}(x)+\Delta c
+\Delta\bar{s} (x) \; , \; \; \; 
g_5^{W^-} (x) = -\Delta u (x) + \Delta\bar{d}(x)-\Delta c
+\Delta\bar{s} (x) \; .
\end{equation}
In Eq.~(\ref{a5}), $F_1^{W^-}$ and $F_3^{W^-}$ are the corresponding 
unpolarized CC structure functions. Extraction of $g_1^{W^-}$ 
and $g_5^{W^-}$ hence offers new and independent constraints on the 
quark and anti-quark helicity distributions, with $g_1^{W^-}$ 
measuring singlet contributions, while $g_5^{W^-}$ is a flavor
non-singlet. If additionally positrons and polarized neutrons are 
available, which is possible at the EIC, one could obtain a 
full flavor decomposition of the nucleon polarized quark and anti-quark
sector. For instance, for proton scattering $g_1^{W^-}+g_1^{W^+}$ 
provides the full quark singlet distribution $\Delta \Sigma$, whose 
first moment gives the quark and anti-quark spin contribution to the 
proton's spin. Likewise, $g_5^{W^-}+g_5^{W^+}$ determines the
``valence'' distributions $\Delta q-\Delta\bar{q}$. Adding neutrons,
one has, for example, $g_{5,p}^{W^+,p}-g_{5,n}^{W^+,n}=
\Delta u+\Delta \bar{u}-(\Delta d+\Delta \bar{d})$, which 
satisfies a sum rule equally fundamental as the Bjorken sum rule:
\begin{equation}
\int_0^1 dx \left[ g_5^{W^+,p}-g_5^{W^+,n} \right] 
= \left( 1-\frac{2 \alpha_s}{3 \pi} \right) g_A \; ,
\label{Wsumrule}
\end{equation}
where we have included the first-order QCD correction~\cite{Stratmann:1995fn}. 
NC structure functions offer independent insights
into nucleon structure. For example, for the $\gamma$-$Z$ interference
contribution, the structure function $g_1$ becomes
to good approximation $g_1^{\gamma Z}\propto \Delta u+\Delta \bar{u} + 
\Delta d+\Delta \bar{d} +\Delta s+\Delta \bar{s}$ and thus again probes
the full quark and anti-quark singlet. 
The structure function $g_5$, on the other hand, probes the
valence densities: $g_5^{\gamma Z}\propto 2 \Delta u_v + 
\Delta d_v$. 

We present a few first studies of the prospects
for measurements of electroweak spin structure functions in 
CC and NC scattering at an EIC. 
These are not meant to present an exhaustive assessment of
all the opportunities the EIC would provide in this area. 

\subsection{Electroweak Deep Inelastic Scattering \label{sec2}}

\subsubsection{Structure Functions and Parton Model Expressions}

\noindent
In the determination of cross sections and asymmetries, we follow closely 
the PDG review~\cite{Nakamura:2010zzi}. The spin-averaged DIS
cross section for $Q^2\gg M^2$, where $M$ is the mass of the 
nucleon, is given by
\begin{equation}
\frac{d^2\sigma^i}{dx dy} = \frac{2\pi\alpha^2}{xyQ^2} \eta^i \Bigg[
Y_+F_2^i \mp Y_- xF_3^i - y^2 F_L^i \Bigg]\;,
\end{equation}
where $i$ is for NC or CC  and $Y_\pm = 1 \pm (1-y)^2$. We have introduced the longitudinal 
structure function $F^i_L = F_2^i -2xF_1^i$, which vanishes to lowest
order according to the Callan-Gross relation. The NC structure functions
for $e^\pm N$ scattering can be represented as the sums of the photon, 
$Z^0$, and interference contributions:
\begin{equation}
F_2^\mathrm{NC} = F_2^\gamma - (g_V^e \pm \lambda g_A^e) \eta_{\gamma Z} 
F_2^{\gamma Z} + (g_V^e{}^2 + g^e_A{}^2 \pm 2\lambda g_V^e g_A^e) \eta_Z F_2^Z
\end{equation}
and 
\begin{equation}
xF_3^\mathrm{NC} = -(g_A^e\pm\lambda g_V^e)\eta_{\gamma Z}x F_3^{\gamma Z} + 
[2g_V^e g_A^e \pm \lambda(g_V^e{}^2 + g_A^e{}^2)] \eta_Z x F_3^Z.
\end{equation}
Here and above, the sign $\pm$ is commensurate to the lepton charge. We have
\begin{equation}
\eta_\gamma = 1;\quad \eta_{\gamma Z} = \left( \frac{G_F M_Z^2}{2\sqrt{2} 
\pi\alpha} \right) \left( \frac{Q^2}{Q^2 + M_Z^2} \right);\quad \eta_Z = 
\eta^2_{\gamma Z},
\end{equation}
and $g_V^e = -\frac{1}{2} + 2\sin^2\theta_W$, $g_A^e = -\frac{1}{2}$. 
$\lambda=\pm 1$ is the electron/positron helicity.

The spin-averaged structure functions can be written as
\begin{eqnarray}
\left[ F_2^\gamma,\, F_2^{\gamma Z},\, F_2^Z \right] & = & 
x \sum_{q} \left[ e_q^2,\, 2e_q g_V^q,\, g_V^q{}^2 + g_A^q{}^2 \right] 
(q+\bar{q}), \nonumber\\
\left[ F_3^\gamma,\, F_3^{\gamma Z},\, F_3^Z \right] & = & 
\sum_{q} \left[ 0,\, 2e_q g_A^q,\, 2g_V^q g_A^q \right] (q-\bar{q})\;,
\label{eq:unpolsf}
\end{eqnarray}
where $e_q$ is the fractional electric charge of the quark, 
$g_V^q = \pm\frac{1}{2}-2e_q\sin^2\theta_W$, and 
$g_A^q = \pm\frac{1}{2}$, with the $+$ sign for up-type quarks and the 
$-$ sign for down-type quarks. 

For $Q^2 \ll M_Z^2$, the pure $Z$ contribution can be neglected,
and one finds in this limit
\begin{equation}
A_\mathrm{beam} = \frac{G_F Q^2}{2 \sqrt{2} \pi \alpha} 
\left[ g^e_A \frac{F_1^{\gamma Z}}{F_1^{\gamma}} + g^e_V \frac{Y_-}{2Y_+} 
\frac{F_3^{\gamma Z}}{F_1^\gamma} \right].
\label{Abeam}
\end{equation}

For the case of a polarized target, there are similar spin dependent 
structure functions. The difference $\Delta \sigma$ of cross sections for 
the two nucleon helicity states is
\begin{equation}
\frac{d^2\Delta\sigma^i}{dx dy} = \frac{8\pi\alpha^2}{xyQ^2} \eta^i \Bigg[
Y_+xg_5^i \pm Y_- xg_1^i - y^2 g_L^i \Bigg],
\label{dsigm}
\end{equation}
where again $i$ is for NC or CC and where $g_L^i = g_4^i - 2xg_5^i$. 
We note that, like $F_L$, the latter quantity 
vanishes to ${\cal{O}}(\alpha_s^0)$~\cite{Dicus:1972pq}. 
The NC spin dependent structure functions are
\begin{eqnarray}
g_5^\mathrm{NC} & = & -(g_V^e \pm \lambda g_A^e) \eta_{\gamma Z} 
g_5^{\gamma Z} + (g_V^e{}^2 + g^e_A{}^2 \pm 2\lambda g_V^e g_A^e) \eta_Z 
g_5^Z, \nonumber\\
g_1^\mathrm{NC} & = & \lambda g_1^\gamma - (g_A^e \pm \lambda g_V^e) 
\eta_{\gamma Z} g_1^{\gamma Z} + ( 2 g_V^e g_A^e \pm \lambda(g_V^e{}^2 + 
g^e_A{}^2) ) \eta_Z g_1^Z.
\end{eqnarray}
Their components can be written as
\begin{eqnarray}
\left[ g_1^\gamma,\, g_1^{\gamma Z},\, g_1^Z \right] & = & 
\frac{1}{2} \sum_{q} \left[ e_q^2,\, 2e_q g_V^q,\, 
g_V^q{}^2 + g_A^q{}^2 \right] (\Delta q+\Delta\bar{q}), \nonumber\\[2mm]
\left[ g_5^\gamma,\, g_5^{\gamma Z},\, g_5^Z \right] & = & 
\sum_{q} \left[ 0,\, e_q g_A^q,\, g_V^q g_A^q \right] 
(\Delta q-\Delta \bar{q}).
\label{pol:sfs}
\end{eqnarray}

The spin asymmetry for scattering an unpolarized lepton 
off a polarized nucleon is then given by
\begin{equation}
A_\mathrm{L} = \eta^{\gamma Z} \left[ g^e_V \frac{g_5^{\gamma Z}}{F_1^\gamma} 
\mp \frac{Y_-}{Y_+} g^e_A \frac{g_1^{\gamma Z}}{F_1^\gamma} \right].
\end{equation}
In the CC case, we have
\begin{equation}
\eta_\mathrm{CC} = (1\pm\lambda)^2\eta_W  = \frac{(1\pm\lambda)^2}{2} 
\left( \frac{G_F M_W}{4\pi\alpha} \frac{Q^2}{Q^2 + M_W^2}\right)^2 \;.
\label{eq:acc}
\end{equation}
For $W^-$ exchange (electron scattering), the structure functions 
(assuming four active flavors) are in the parton model:
\begin{eqnarray}
F_2^{W-}  =  2x(u + \bar{d} + \bar{s} + c),&&\;\;
F_3^{W-}  =  2(u + \bar{d} + \bar{s} + c),\nonumber\\[2mm]
g_1^{W-}  =   \Delta u + \Delta \bar{d} + \Delta \bar{s} + \Delta c,&&\;\;
g_5^{W-}  =  -\Delta u + \Delta \bar{d} + \Delta \bar{s} - \Delta c.
\end{eqnarray}
For $W^+$ exchange, one replaces $u\leftrightarrow d$ and 
$s\leftrightarrow c$. The spin asymmetries for electron and positron 
scattering then take the simple parton model forms
\begin{equation}
A_{W^-}  =  \frac{ \Delta u + \Delta c - (1-y)^2 (\Delta \bar{d} + 
\Delta \bar{s} )}{ u+c + (1-y)^2 (\bar{d} + \bar{s} )}, \;\;
A_{W^+}  =  \frac{ (1-y)^2 (\Delta d + \Delta s) - 
\Delta \bar{u} - \Delta \bar{c}}{  (1-y)^2(d+s) + \bar{u} + \bar{c} }.
\label{eq:Acc}
\end{equation}
By measuring over a range in $y$, one can perform a separation of the 
$\Delta u+\Delta c$, $\Delta d+\Delta s$ quark or anti-quark combinations.

\subsubsection{Next-to-leading Order QCD Corrections \label{nlosec} }

\noindent
The NLO QCD corrections to the 
spin-dependent structure functions have been computed 
in Refs.~\cite{deFlorian:1994wp,Stratmann:1995fn}. To NLO, 
the expression for a given structure function can be cast into the generic 
form~\cite{Forte:2001ph} 
\begin{eqnarray}
g^{{\rm NLO}}_1(x,Q^2)&=&\Delta C_{q,1}\otimes
g^{{\rm LO}}_1+f_\Sigma\,\Delta C_g\otimes\Delta g\;,\nonumber \\[2mm]
\frac{g^{{\rm NLO}}_4(x,Q^2)}{2x}&=&\Delta C_{q,4}\otimes \left[ 
\frac{g^{{\rm LO}}_4}{2x}\right]\; ,\nonumber \\[2mm]
g^{{\rm NLO}}_5(x,Q^2)&=&\Delta C_{q,5}\otimes g^{{\rm LO}}_5\;,\label{sfsnlo}
\end{eqnarray}
where the symbol $\otimes$ denotes a convolution, and 
$g_i^{{\rm LO}}$ is the LO (parton model) 
expression for the respective structure function.
The coefficient functions to NLO in the 
$\overline{{\mathrm{MS}}}$ scheme can be found in \cite{deFlorian:1994wp,Stratmann:1995fn}.
The factor $f_\Sigma$ in Eq.~(\ref{sfsnlo}) is the sum over
the coefficient of each quark or anti-quark distribution in the
LO expression for $g_1$. For example, for the electromagnetic
$g_1^\gamma$ with four flavors, $f_\Sigma=10/9$, while
for $g_1^{W^-}$ one has $f_\Sigma=4$. Needless to say that 
when including the NLO corrections in the calculation of the 
structure functions, one also has to perform the evolution
of the polarized PDFs 
to NLO~\cite{Mertig:1995ny,Vogelsang:1995vh,Vogelsang:1996im}.
For the most part of our study, we will only use the LO
expressions for the structure functions, which are expected to be
entirely sufficient for estimating the sensitivities at an EIC.
We will, however, briefly investigate the typical size of the 
NLO corrections in Figs.~\ref{fig:ccsf} and~\ref{fig:ncsf} below.

\subsection{Measurements of Parton Distribution Functions}

In the following, we will present estimates for rates and spin
asymmetries for electroweak DIS at an EIC. For the spin-averaged
case, we use the CTEQ6.5~\cite{Tung:2006tb} unpolarized PDFs. 
For the helicity PDFs we 
use the ones of \cite{deFlorian:2008mr}. We note that the latter
do not contain a charm quark distribution. 

\begin{figure}[t]
\begin{center}
\includegraphics[scale=0.65,angle=90]{WRITEUP-PDF-SECTION/FIGURES-PDF/sqsdep.epsi}
\end{center}
\caption{Total NC and CC cross sections for $Q^2>1$~GeV$^2$ as functions
of the $ep$ $\sqrt{s}$.
\label{fig:crsec}}
\end{figure}

\subsubsection{Basic kinematics and machine considerations}

Proposed EIC parameters allow for electron energies of $5-30~\mathrm{GeV}$ 
and ion energies of $50-325~\mathrm{GeV}$.  Figure~\ref{fig:crsec} shows
the spin-averaged NC and CC total cross sections for electron and positron 
scattering, as functions of the $ep$ c.m.s.\ energy $\sqrt{s}$. 
We have integrated over all $Q^2>1$~GeV$^2$, based on a simple
theoretical LO calculation. One can see that the cross 
section of course rises with energy, but relatively mildly so. Therefore,
measurements of electroweak structure functions may well be feasible 
in collisions at energies significantly lower than those at HERA.

\begin{figure}[p]
\begin{center}
\includegraphics[scale=0.4,angle=90]{WRITEUP-PDF-SECTION/FIGURES-PDF/q2dist.epsi}
\hspace*{1cm}
\includegraphics[scale=0.4,angle=90]{WRITEUP-PDF-SECTION/FIGURES-PDF/xdist.epsi}

\vspace*{1cm}
\hspace*{-0.6cm}
\includegraphics[scale=0.4,angle=90]{WRITEUP-PDF-SECTION/FIGURES-PDF/q2dist_nc.epsi}
\hspace*{0.8cm}
\includegraphics[scale=0.4,angle=90]{WRITEUP-PDF-SECTION/FIGURES-PDF/xdist_nc.epsi}

\vspace*{1cm}
\includegraphics[scale=0.4,angle=90]{WRITEUP-PDF-SECTION/FIGURES-PDF/q2dist_nc_PVe-.epsi}
\hspace*{1cm}
\includegraphics[scale=0.4,angle=90]{WRITEUP-PDF-SECTION/FIGURES-PDF/xdist_nc_PVe-.epsi}
\end{center}
\caption{Top row: Distributions of the CC spin-averaged cross section
in $Q^2$ (left) and $x$ (right). We have applied the cuts
$Q^2\geq 1$~GeV$^2$ and $0.1\leq y \leq 0.9$. Center row: same 
for the NC case. Bottom row: Same for the NC parity-violating electron 
beam-helicity difference of cross sections.
\label{fig:distrib}}
\end{figure}

The upper two plots in Figure~\ref{fig:distrib} show distributions
of the CC cross section in $\log(Q^2)$ and $\log(x)$, respectively,
at three different c.m.s.\ energies. One can see that the largest 
statistical weight
 would be at $x\sim 0.1$ and
$Q^2\sim 1000$~GeV$^2$, which is a consequence of the $W$-propagator
factor in~Eq.~(\ref{eq:acc}). Binning in $x$ and $Q^2$ of course
allows to investigate more detailed distributions, see below.
For NC interactions, the $\gamma$-exchange contribution dominates
the spin-averaged cross section and strongly pushes the $Q^2$ distribution
towards $Q^2\to 0$ (see center row of the figure). 
Taking the parity-violating electron beam-helicity  
difference of cross sections, however, essentially singles out the 
$\gamma Z$-interference contribution. For this piece, which of 
course is much smaller than the full spin-averaged cross section,
the $Q^2$ distribution levels off towards $Q^2\to 0$, as follows
from the expressions in Sec.~\ref{sec2} and as shown in the
bottom row of Fig.~\ref{fig:distrib}. 

In CC electron scattering, $e^-p \rightarrow 
\nu_eX$, the neutrino remains undetected.  To identify a CC
event and to reconstruct $x$ and $Q^2$, the final-state hadrons must then be reconstructed instead. 
The detectors must hence be optimized to detect resulting hadronic jet 
formation. There will likely be some additional detection and 
reconstruction efficiency associated with this type of analysis.
The discussion of the specific requirements is 
beyond the scope of this study, and we will assume that this 
reconstruction is possible. 
In practice, CC
measurements could be performed simultaneously with the NC
ones, though at a reduced duty factor if the electron 
helicity is flipped, as the interaction is purely $V-A$. We 
also assume that polarized positron beams would be available at 
an EIC. 

For the following analysis, we will consider configurations of 
$E_e [\mathrm{GeV}] \times E_\mathrm{ion} [\mathrm{GeV}]$ with 
$20\times325$ and $20\times250$.  For each of these, a luminosity 
of about $\sim 1\times10^{34}~/\mathrm{s}/\mathrm{cm}^2$ was considered, 
with estimates for machine availabilities, detector acceptance and 
efficiency, and beam polarization.  Based on an expected five year 
run time, we consider a realistic effective integrated luminosity of 
$100~\mathrm{fb}^{-1}$ for NC processes and 
$10~\mathrm{fb}^{-1}$ for CC. For the studies below, 
a Monte Carlo simulation framework
was developed to evaluate rates and asymmetries of both the NC 
and CC processes. No detector responses have yet been included,
and a full azimuthal acceptance was assumed.  In all analyses we 
consider a minimum scattered electron energy of $2~\mathrm{GeV}$ 
within $3^\circ < \theta < 177^\circ$ scattering angle. 
The smaller integrated luminosity for CC studies is because of a factor of 2 loss 
due to helicity flips 
and also because efficiency 
of hadron jet and kinematic reconstruction has not yet been studied.

\begin{figure}[t]
\begin{center}
\includegraphics[scale=0.37]{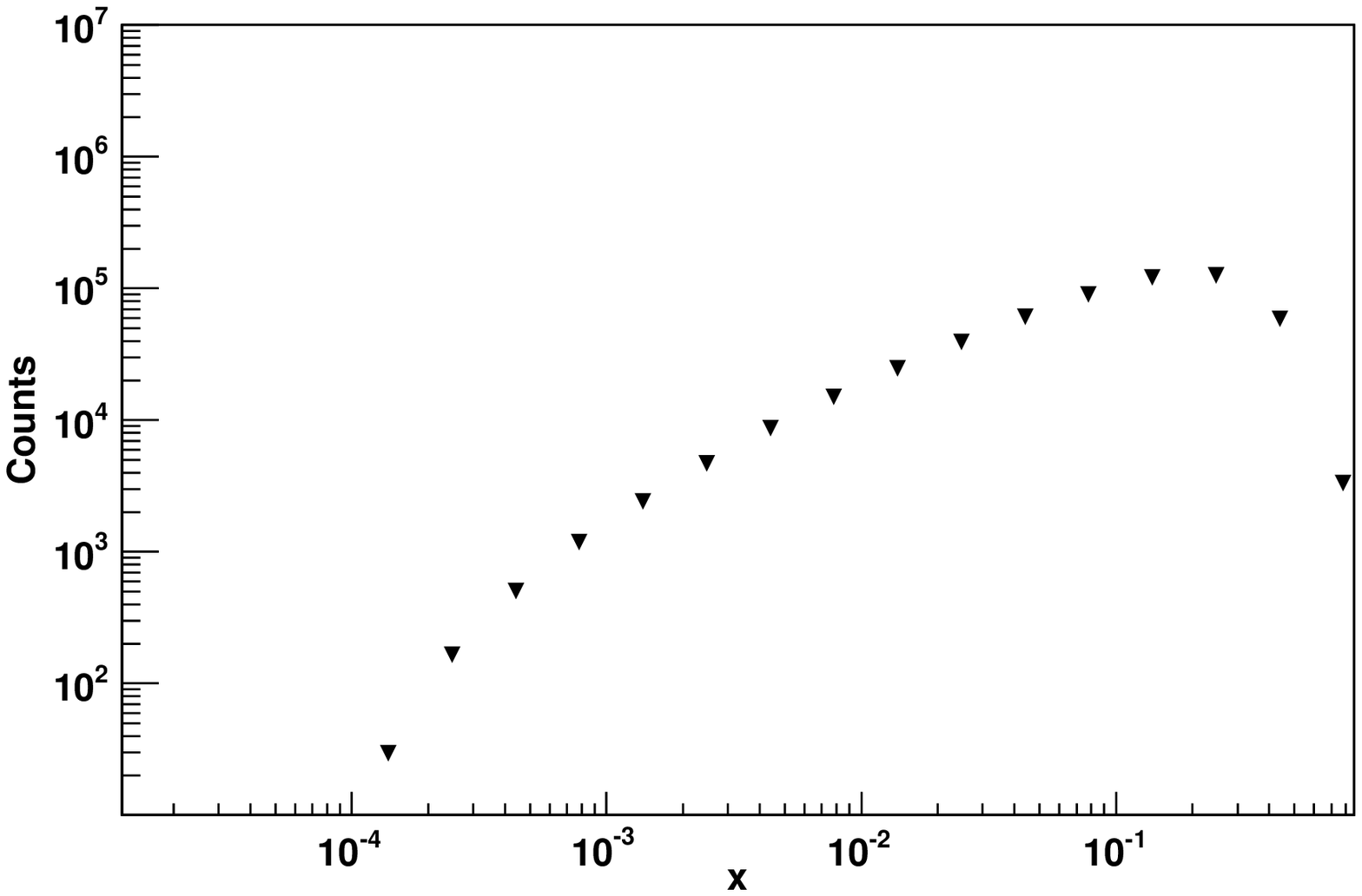}
\includegraphics[scale=0.35]{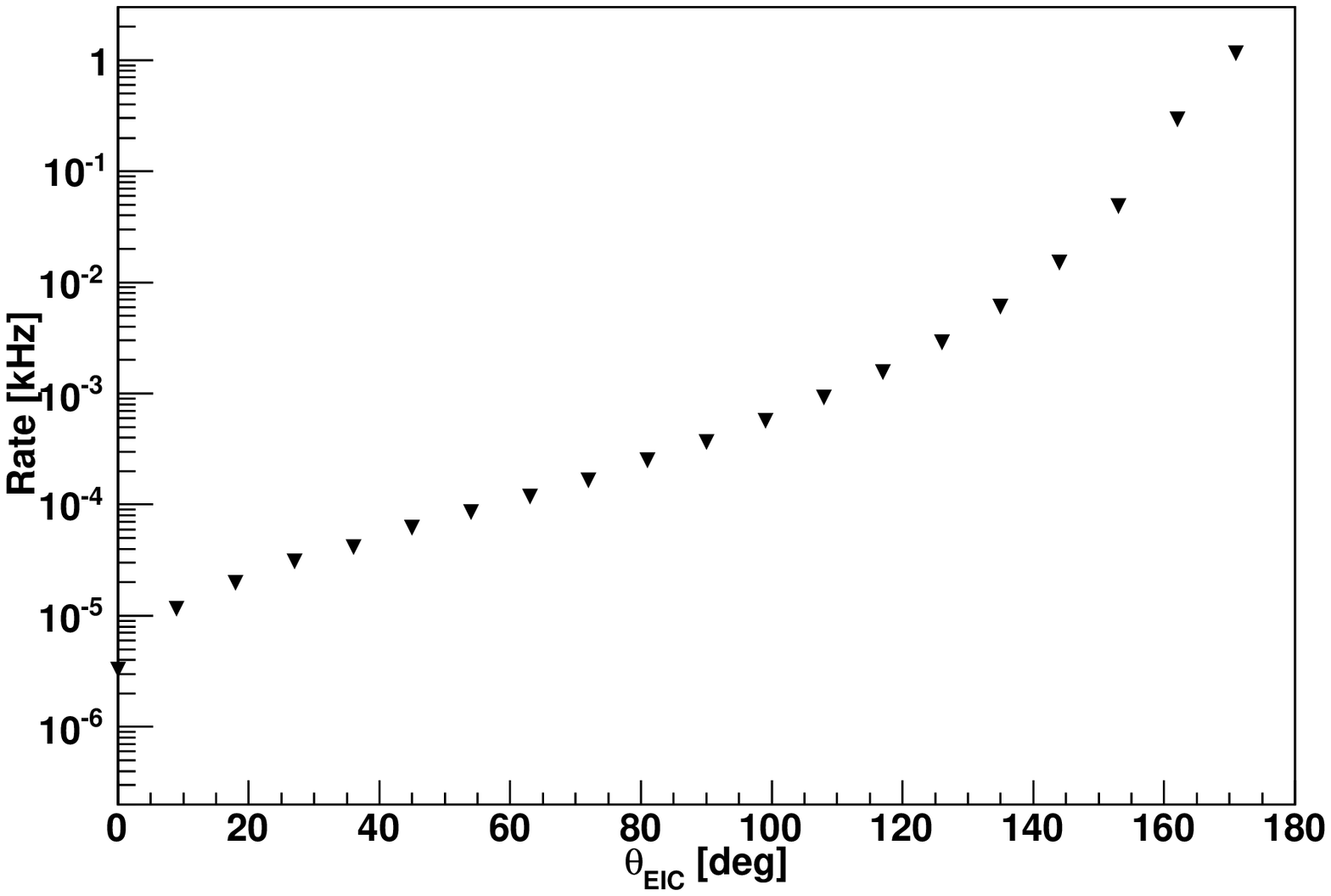}
\end{center}
\caption{Left: Total number of CC events for 
$20\times250$ $e^-p$ scattering for an integrated luminosity of 
$10~\mathrm{fb}^{-1}$. Right: Binned NC event rate as
function of the electron scattering angle, for $20\times 325$ $e^-p$
collisions at ${\cal L}=1\times10^{33}~/\mathrm{s}/\mathrm{cm}^2$.}
\label{fig:cccnt}
\end{figure}

Of practical importance is to evaluate how well a separation of 
the structure functions can be done at individual points in $x$, though it 
remains for a future 
Monte Carlo study to evaluate the $x$ resolution after reconstruction.  
We bin all data in 20 $x$ bins logarithmically spaced from $10^{-5}$ 
to $1$.  When binned in $Q^2$, we use 20 bins from $2$ to 
$5\times10^{4}~\mathrm{GeV}^2$.  These $Q^2$ bins were also used 
in determining any $y$ dependence. Figure~\ref{fig:cccnt} (left) 
shows the total number of events expected for CC interactions in 
$e^-p$ scattering at $\sqrt{s}=141$~GeV and ${\cal L}=10~\mathrm{fb}^{-1}$, 
binned in $x$. 

Typical rates in NC scattering are up to $1~\mathrm{kHz}$, as
shown in the right part of Fig.~\ref{fig:cccnt} for 
$\sqrt{s}=161$~GeV. The highest rate occurs in the forward direction 
of the electron beam. Here, pipeline electronics will likely be 
necessary in order to avoid significant deadtime effects.  

\subsubsection{Polarized Parton Distributions 
from CC Interactions}
As follows from Eq.~(\ref{eq:Acc}), CC processes in 
electron scattering off polarized targets offer a unique 
method to extract combinations of $\Delta u + \Delta c$ and 
$\Delta \bar{d} + \Delta \bar{s}$.  With positron beams, 
one could also extract $\Delta d + \Delta s$ and 
$\Delta \bar{u} + \Delta \bar{c}$. 
For the present analysis, we have assumed a $100\%$ polarized 
electron/positron source. 
As mentioned before, we have assumed only $10~\mathrm{fb}^{-1}$ integrated 
luminosity, making our estimates  somewhat conservative.   

\begin{figure}
\vspace*{4mm}
\begin{center}
\includegraphics[scale=0.35,angle=90]{WRITEUP-PDF-SECTION/FIGURES-PDF/sfs.epsi}
\end{center}
\vspace*{-3mm}
\caption{CC spin dependent structure functions 
$g_1^{W^-}$, $g_5^{W^-}$, and $g_4^{W^-}/2x$, at $Q^2=100~\mathrm{GeV}^2$.
The dashed lines show the LO results (the one for $g_4^{W^-}/2x$ is not
shown in this case, since it coincides with that for $g_5^{W^-}$), while
the solid curves are NLO. For comparison, we also show
the electromagnetic $g_1^\gamma$.}
\label{fig:ccsf}
\end{figure}

\begin{figure}
\vspace*{4mm}
\begin{center}
\includegraphics[scale=0.35]{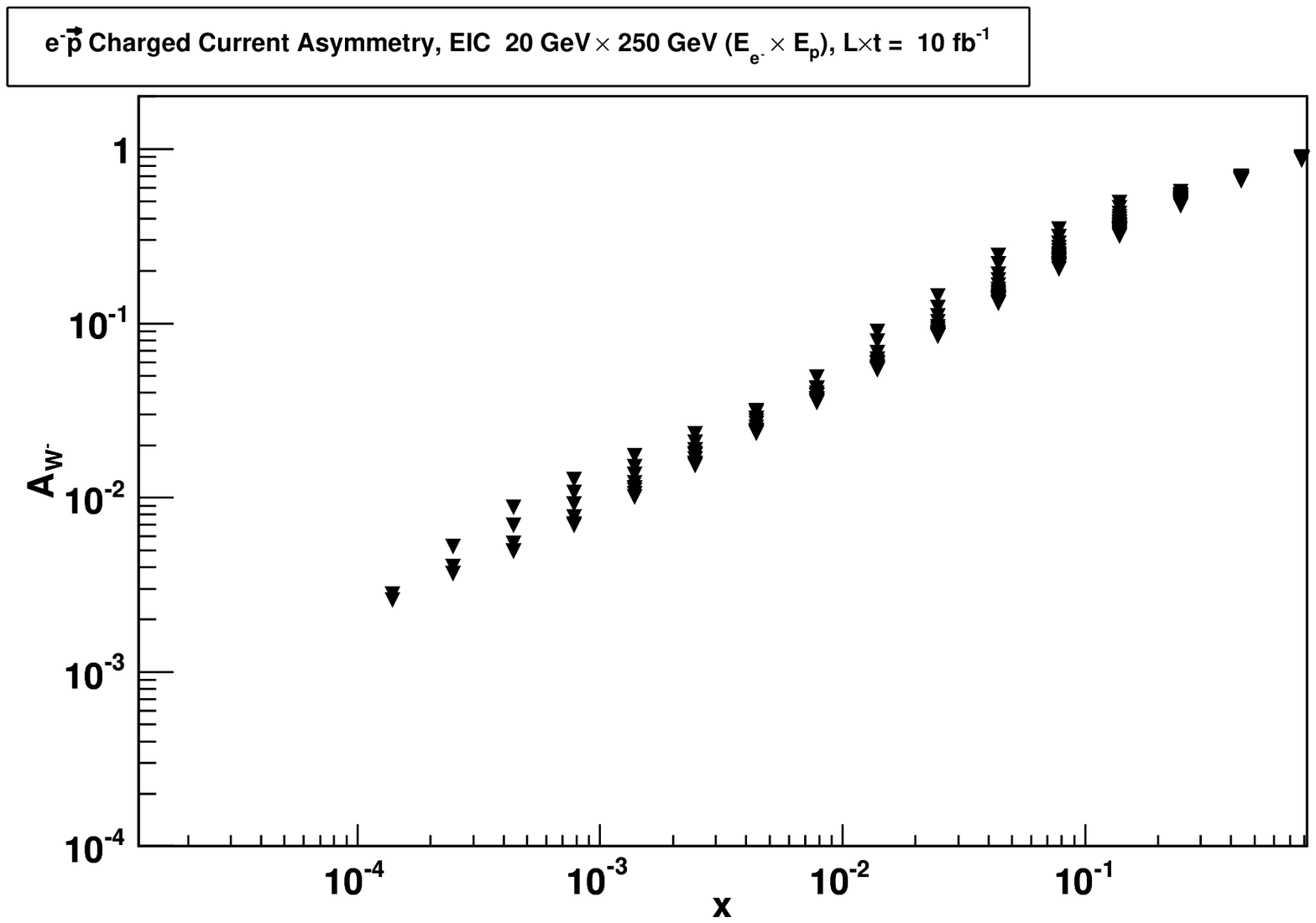}
\includegraphics[scale=0.35]{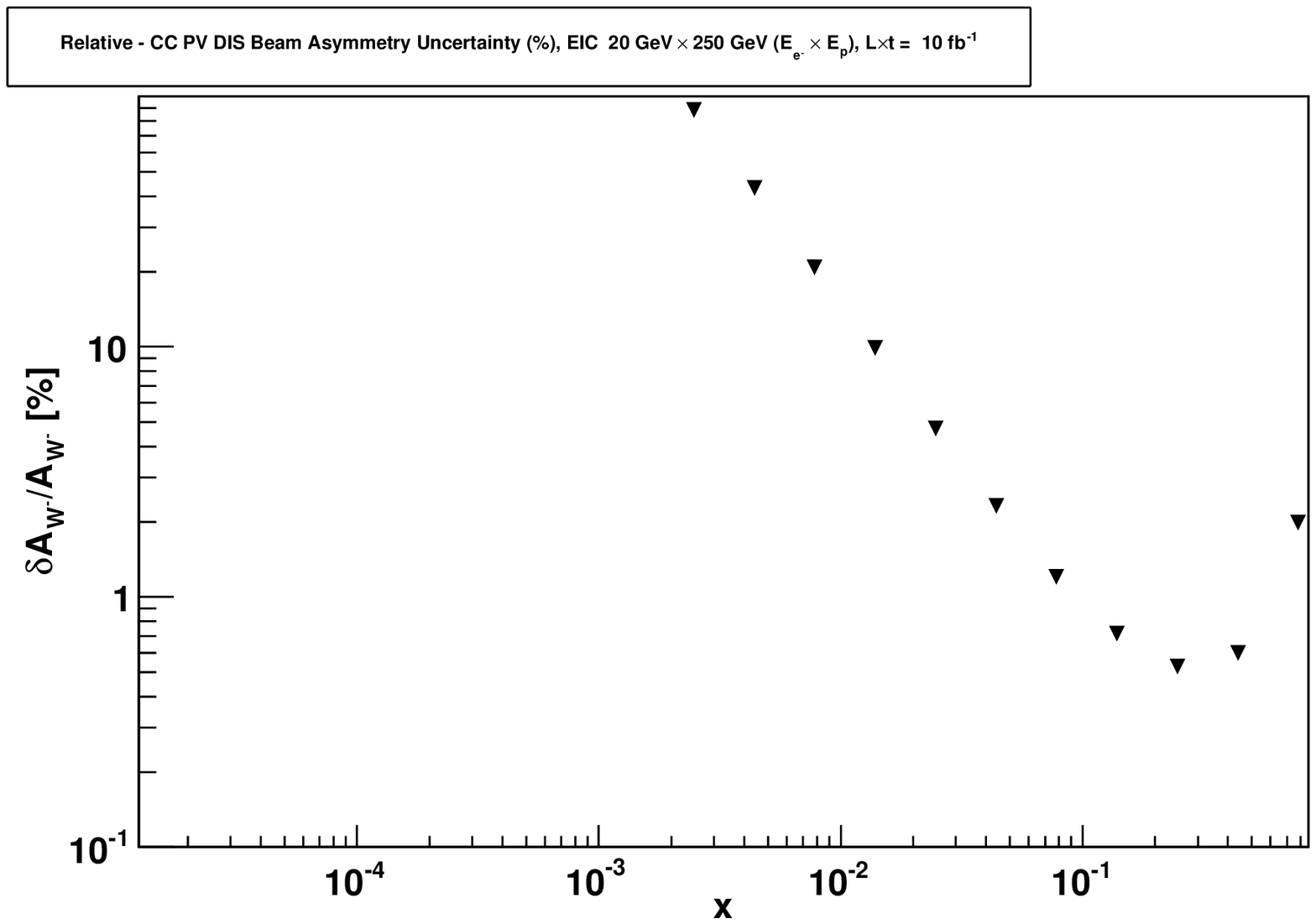}
\end{center}
\vspace*{-4mm}
\caption{Left: spin asymmetry for CC $e^{-}\vec{p}$ 
scattering, as function of $x$ for various bins in $Q^2$. Right: 
resulting relative uncertainties of the asymmetry.}
\label{fig:ccA}
\end{figure}

In Fig.~\ref{fig:ccsf} we show the spin structure functions 
$g_1^{W^-}$, $g_5^{W^-}$, and $g_4^{W^-}/2x$, at $Q^2=100$~GeV$^2$, 
using the PDFs of~\cite{deFlorian:2008mr}. 
Results are shown both at LO (dashed) and at NLO (solid). 
One observes that 
the NLO corrections are well under control. 
To guide the eye, also the ordinary electromagnetic 
structure function $g_1^\gamma$ is shown. 
Figure~\ref{fig:ccA} (left) displays the asymmetry $A_{W^-}$ for CC
$e^-\vec{p}$ scattering, as function of $x$. Different data 
points at same $x$ correspond to different bins in $Q^2$. As mentioned 
above, we have chosen here 20 bins in $Q^2$, spaced logarithmically from 
$2~\mathrm{GeV}^2$ to $5000~\mathrm{GeV}^2$. The lower 
asymmetries correspond to the lower bins in $Q^2$. 
Thanks to the simple 
structure of the LO expressions for the cross sections,
the asymmetries in CC interactions become very large in the
valence region, much larger than those in the NC case
to be discussed below. On the other hand, as we saw in 
Figs.~\ref{fig:distrib} and~\ref{fig:cccnt},
event rates are much more suppressed at lower $Q^2$ and therefore 
$x$. The right 
part of Fig.~\ref{fig:ccA} gives the resulting values for the 
relative uncertainty $\delta A_{W^-}/ A_{W^-}$ of the asymmetry. 
Here we have summed over all $Q^2$ bins. The results shown look 
very promising, with better than $10\%$ measurements appearing
feasible all the way down to $x\sim 10^{-2}$. It is worth keeping 
in mind that relative polarimetry uncertainties at an EIC are also
expected to be at the $0.5-1\%$ level for electrons and $2-3\%$ level for hadrons, 
so that these might become
the dominant sources of uncertainty in the regions where the
statistical $\delta A_{W^-}/ A_{W^-}$ is very small, especially at high $x$. 

\begin{figure}
\begin{center}
\includegraphics[scale=0.35]{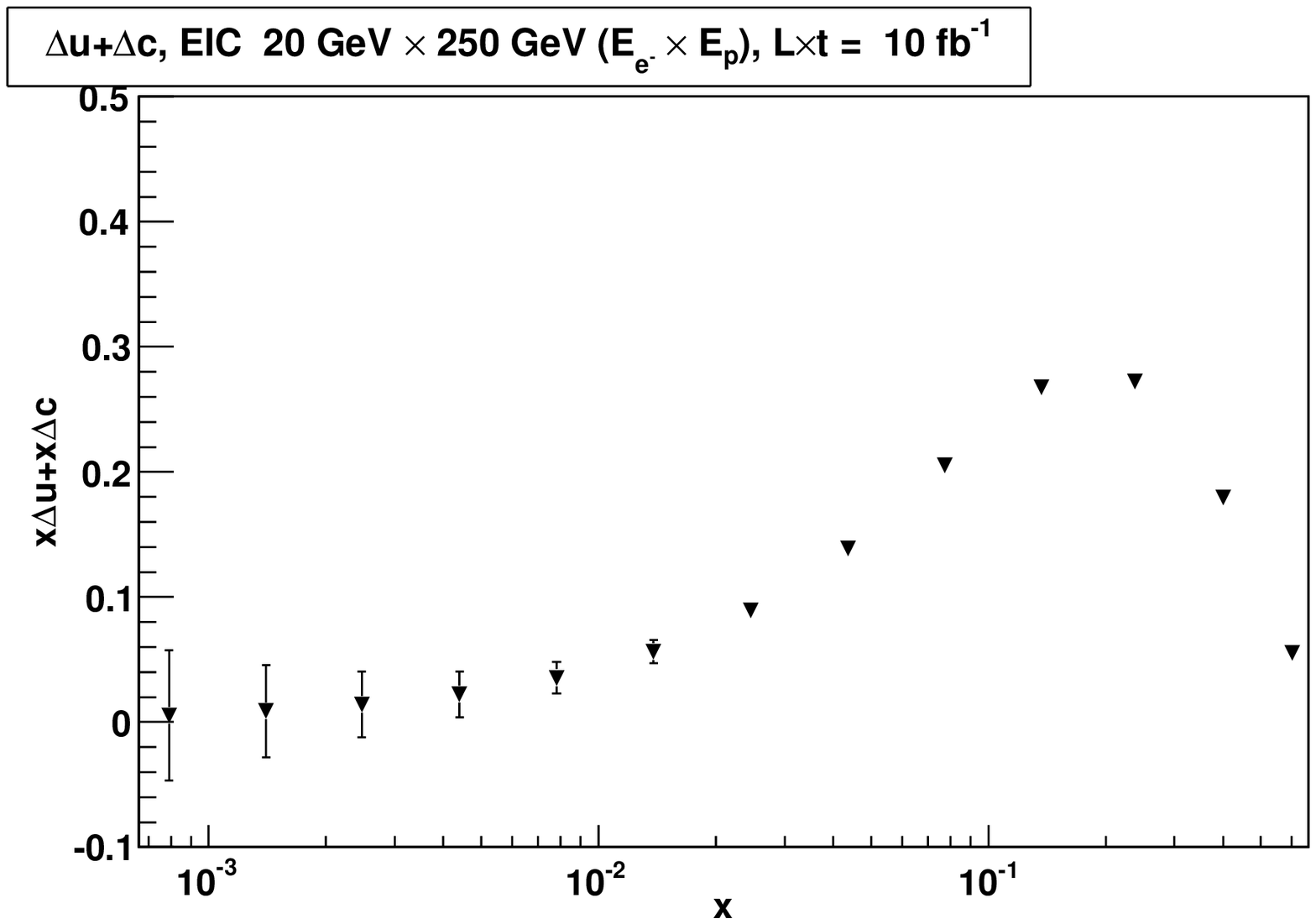}
\includegraphics[scale=0.35]{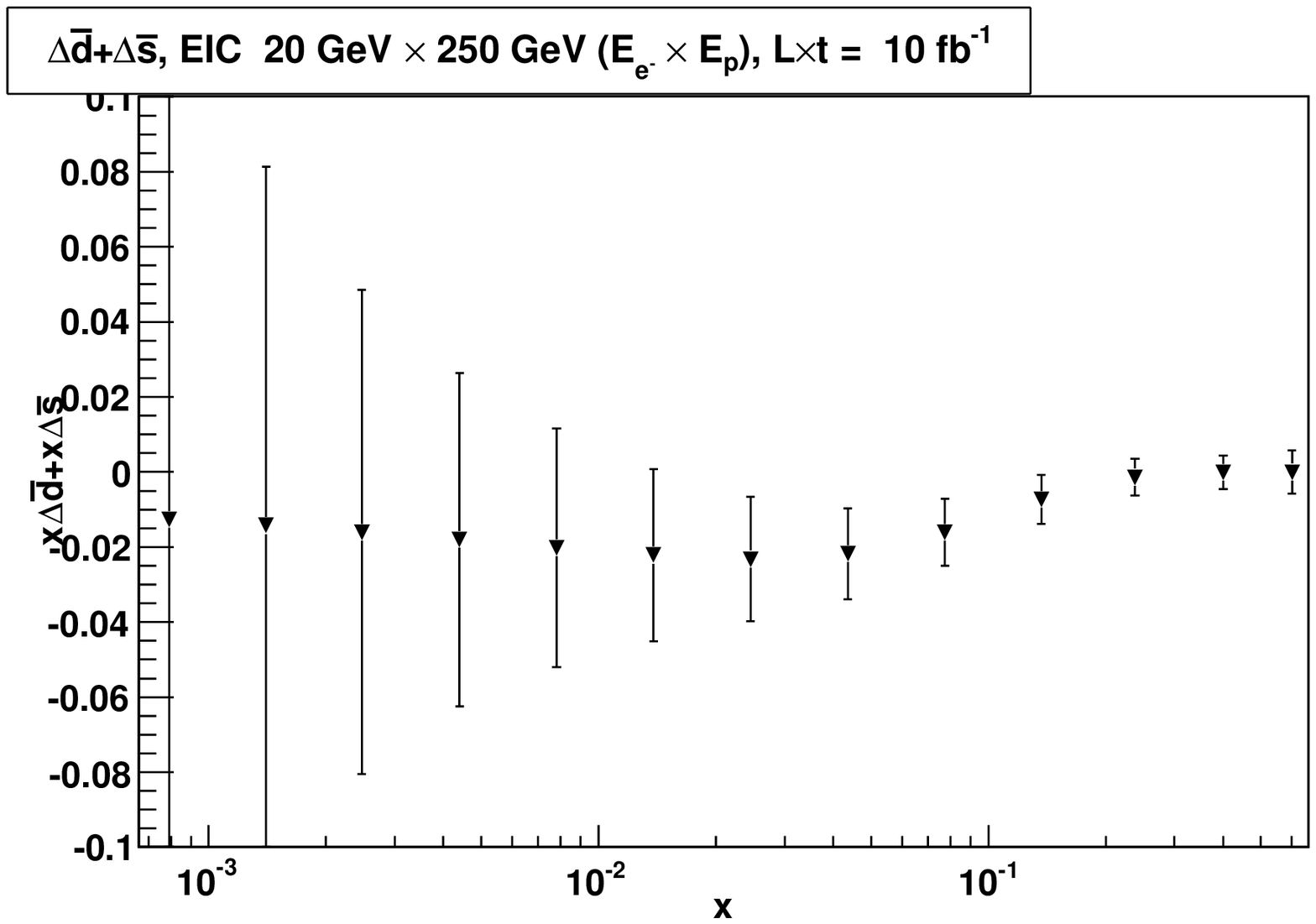}

\vspace*{4mm}
\includegraphics[scale=0.35]{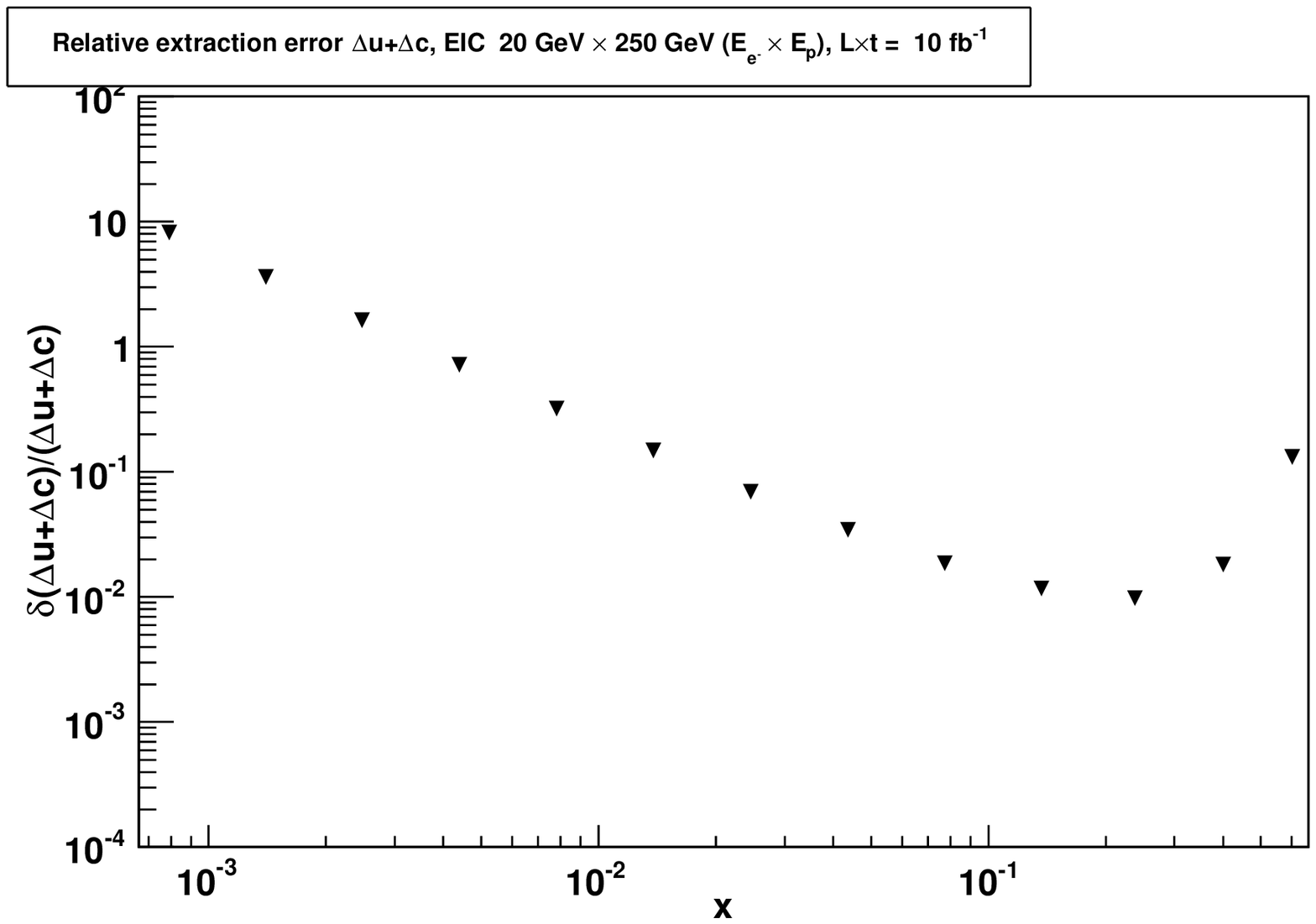}
\includegraphics[scale=0.35]{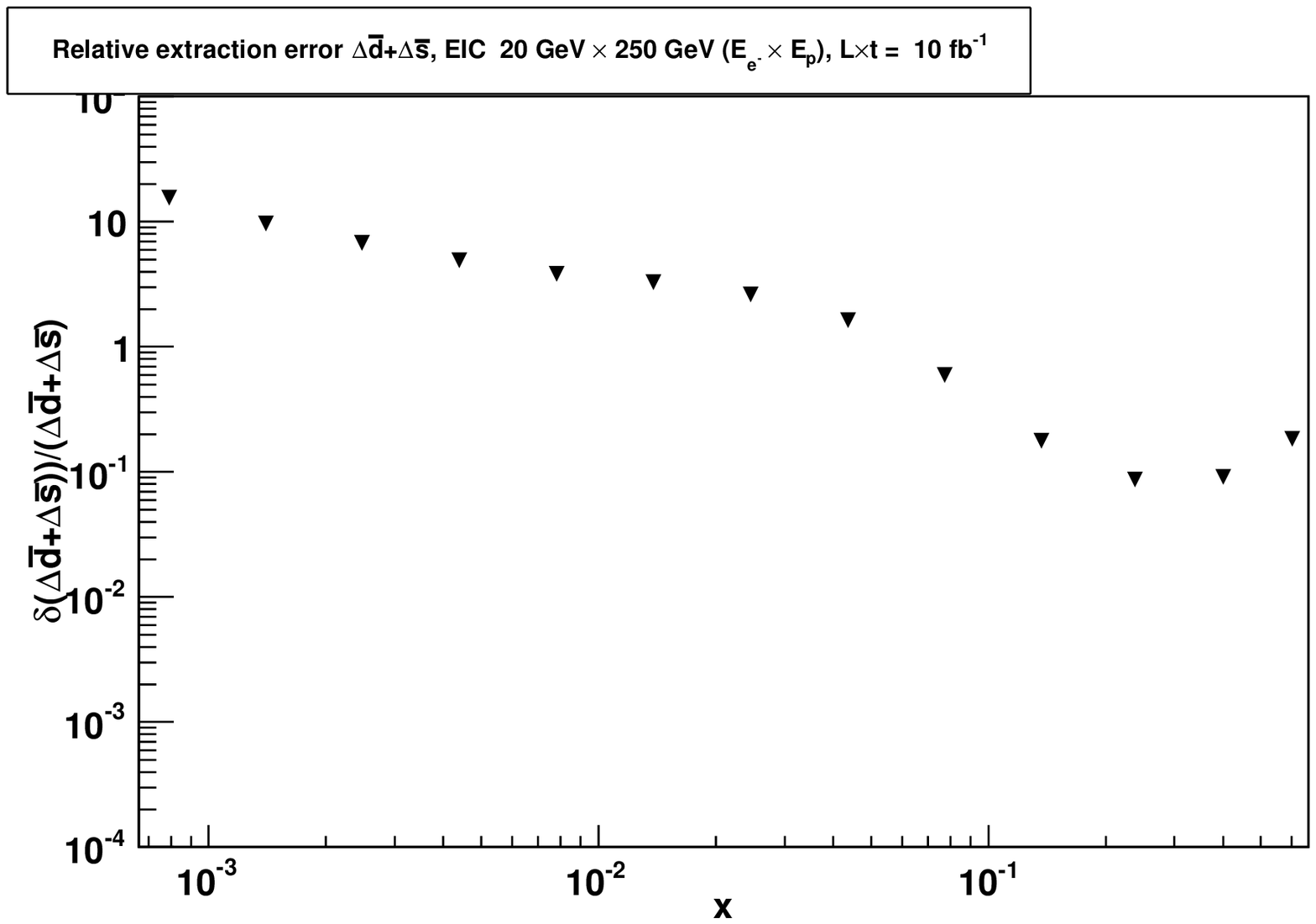}
\end{center}
\vspace*{-3mm}
\caption{Top: LO extraction of polarized quark and anti-quark 
distributions from the spin asymmetry for CC $e^{-}\vec{p}$ 
scattering. Bottom: Corresponding relative uncertainties of the 
extracted distributions.}
\label{fig:ccmextract}
\end{figure}

\begin{figure}
\vspace*{3mm}
\begin{center}
\includegraphics[scale=0.35]{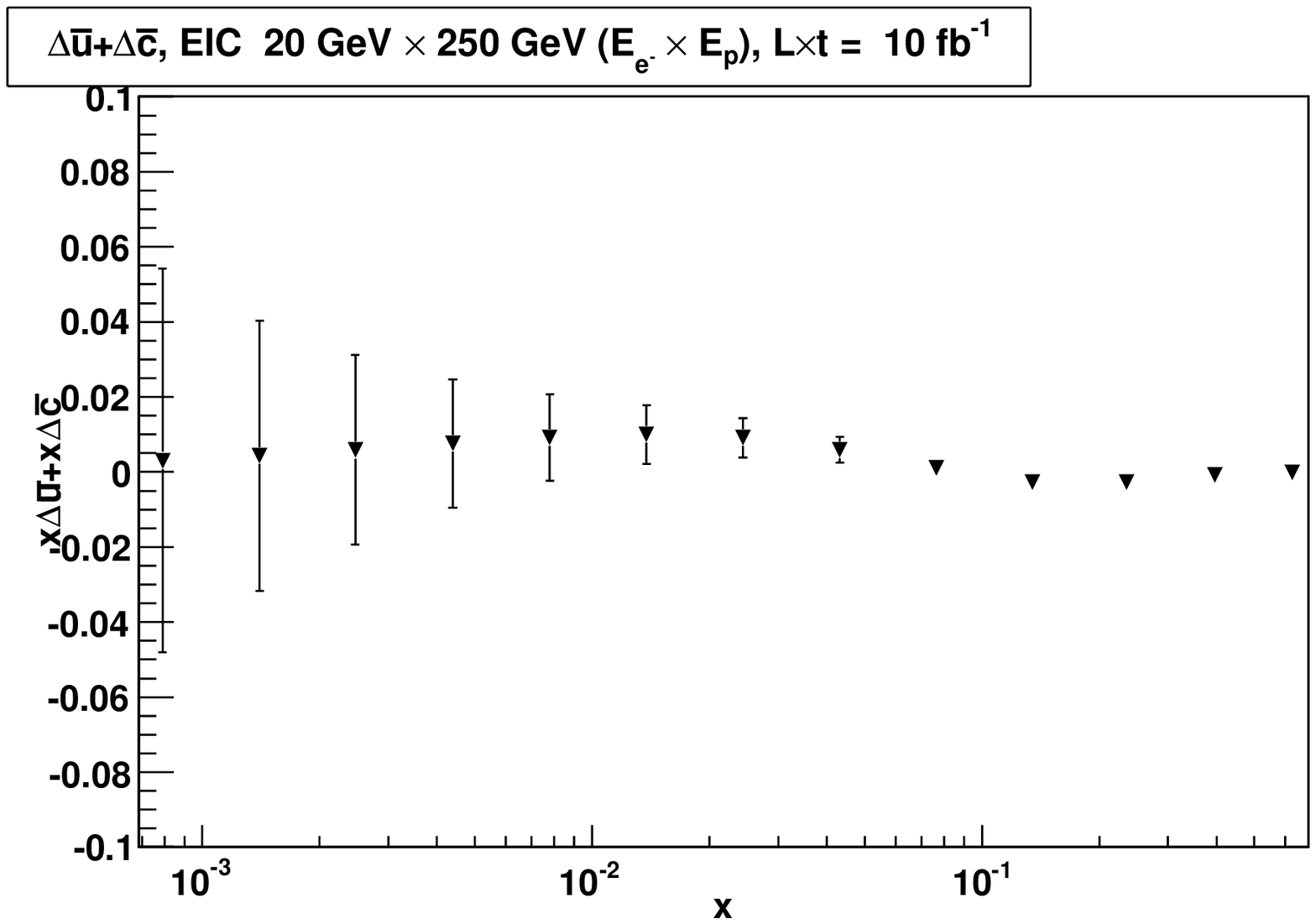}
\includegraphics[scale=0.35]{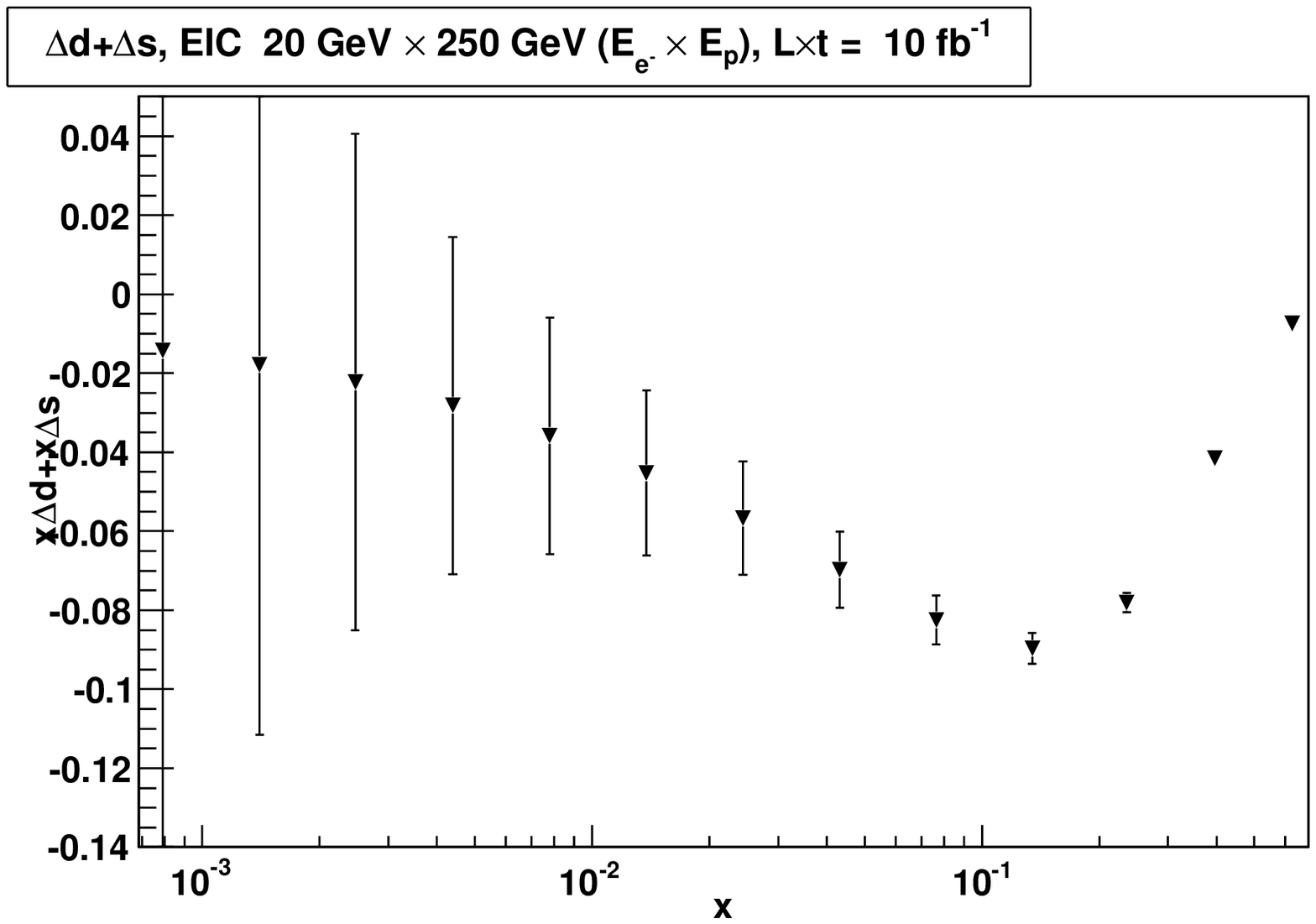}

\vspace*{4mm}
\includegraphics[scale=0.35]{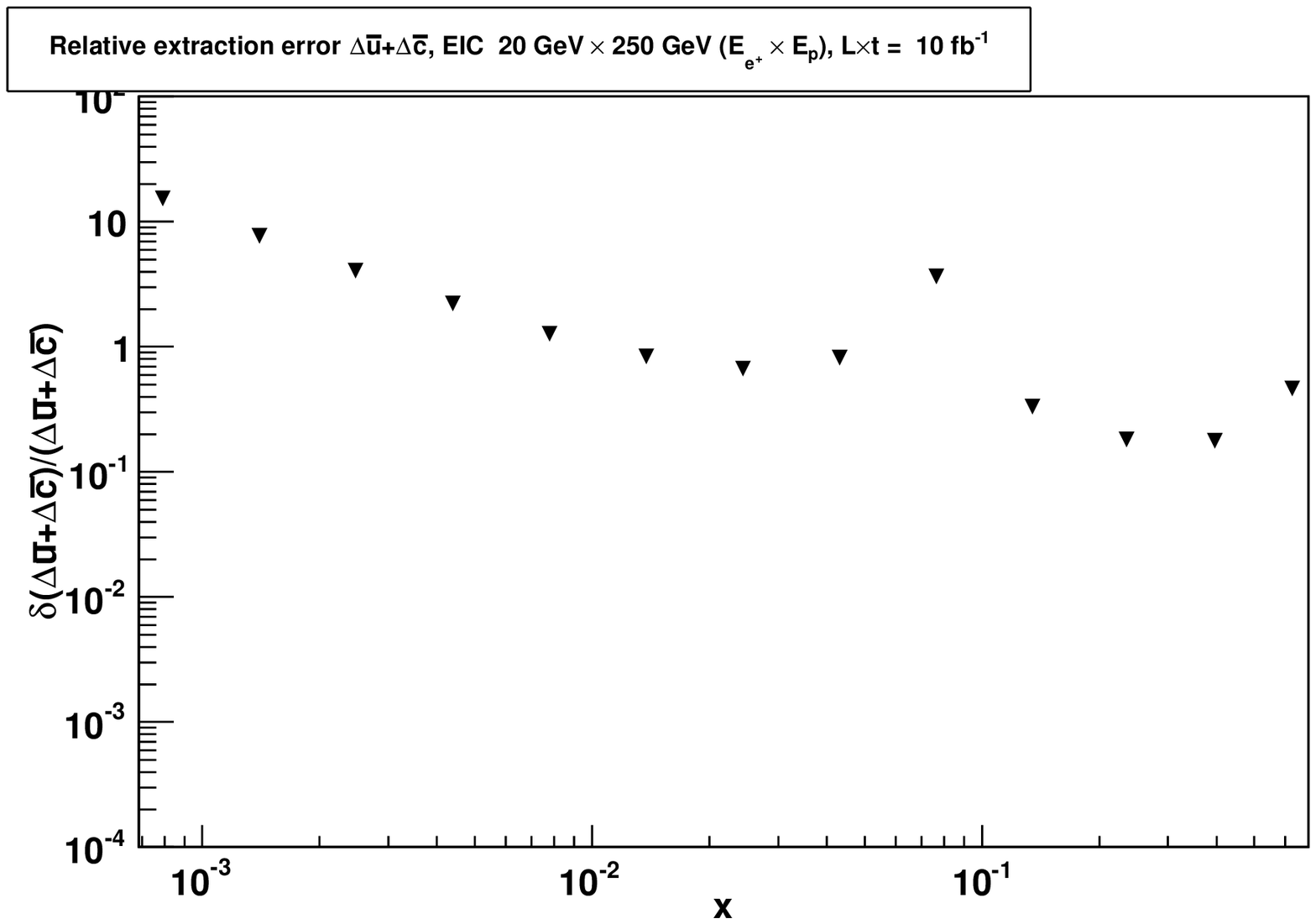}
\includegraphics[scale=0.35]{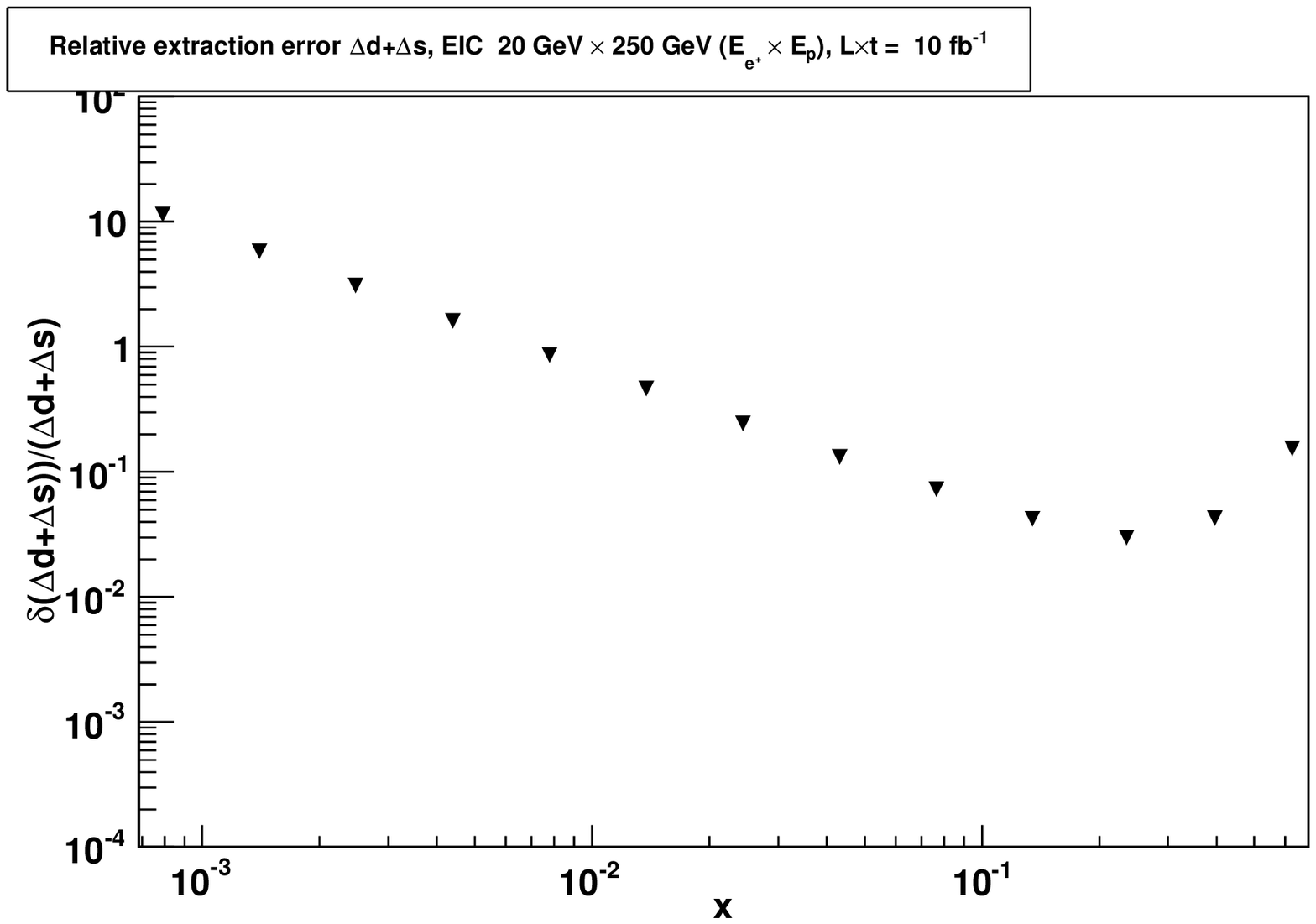}
\end{center}
\vspace*{-3mm}
\caption{Same as~ Fig.~\ref{fig:ccmextract}, but for $e^{+}\vec{p}$ 
scattering.}
\label{fig:ccpextract}
\end{figure}

Using Eq.~(\ref{eq:Acc}), the asymmetries give direct access to the 
polarized quark and anti-quark distributions. As we discussed,
higher-order QCD corrections (and also Cabibbo-suppressed contributions)
will somewhat modify the expressions in Eq.~(\ref{eq:Acc}). However, 
for a first estimate use of Eq.~(\ref{eq:Acc}) as a means to gauge the
sensitivity to the distributions is justified. The additional 
contributions will not make a qualitative difference and can be 
systematically included in future studies. If furthermore full 
knowledge of the unpolarized parton distributions is assumed, 
then extraction of the sums of the two up-type quarks and 
down-type anti-quarks can be performed by a linear fit in $(1-y)^2$.  
The results of such fits are shown for electron and positron running
in Figs.~\ref{fig:ccmextract} and~\ref{fig:ccpextract}, respectively.
We note that if a polarized deuterium or ${}^3$He beam were available,
additional opportunities would arise; $e^{-}\vec{n}$ scattering
would probe the combinations $\Delta u + \Delta d + 2 \Delta c$ and 
$\Delta \bar{d} +\Delta \bar{u}+ 2\Delta \bar{s}$. At larger $x$ where 
the sea quarks are suppressed relative to the valence quarks, 
$e^-\vec{p}$ and $e^-\vec{n}$ scattering could be used to separate the 
valence polarizations.

\subsubsection{Structure Functions and Polarized PDFs from
NC Interactions}

Again we first show the spin-dependent structure functions; 
see Figure~\ref{fig:ncsf}. As the contributions 
from pure $Z$-exchange are small, we only consider the electromagnetic 
$g_1^\gamma$, and the $\gamma$-$Z$ interference contributions
$g_1^{\gamma Z}$ and $g_{4,5}^{\gamma Z}$, whose expressions were
given in Eq.~(\ref{pol:sfs}). 

\begin{figure}
\begin{center}
\includegraphics[scale=0.45,angle=90]{WRITEUP-PDF-SECTION/FIGURES-PDF/sfs_nc.epsi}
\end{center}
\caption{NC spin-dependent structure functions for
$\gamma$-$Z$ interference, at $Q^2=100~\mathrm{GeV}^2$, calculated at 
LO (dashed) and NLO (solid), using the polarized 
PDFs of~\cite{deFlorian:2008mr}.}
\label{fig:ncsf}
\end{figure}

\begin{figure}
\vspace*{-2mm}
\begin{center}
\includegraphics[scale=0.335]{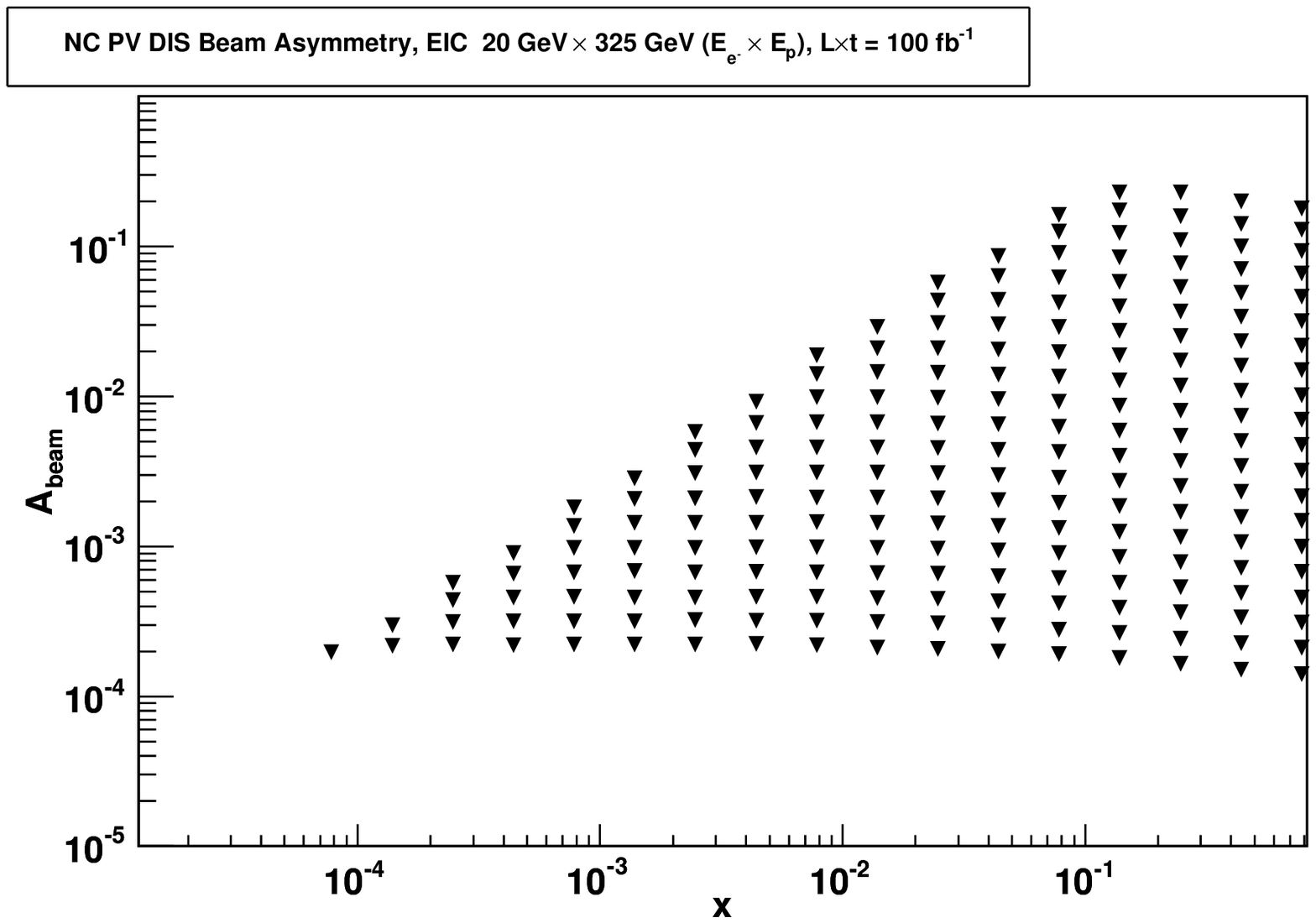}
\includegraphics[scale=0.335]{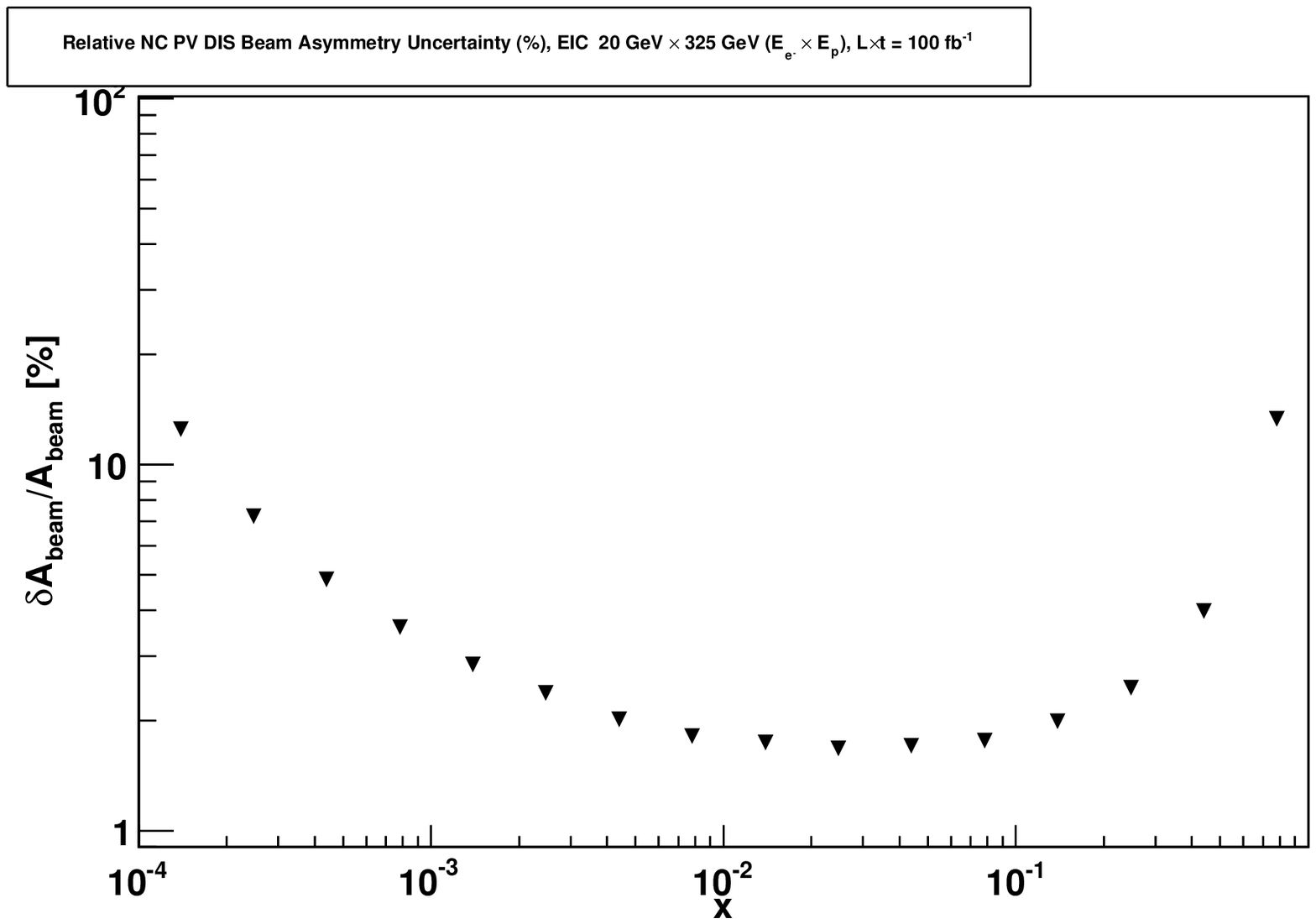}
\end{center}
\vspace*{-4mm}
\caption{Left: 
Parity violating NC spin asymmetries for polarized electrons on 
unpolarized protons, binned logarithmically in $x$ and $Q^2$. Right: 
Resulting relative uncertainties of the asymmetry.}
\label{fig:20x325A}
\end{figure}

The left part of Fig.~\ref{fig:20x325A} shows the parity-violating
spin asymmetry in Eq.~(\ref{Abeam}), obtained for a polarized lepton 
beam scattering off an unpolarized proton beam, as function of $x$ 
in various different $Q^2$ bins. The lower (upper) asymmetries 
correspond to $Q^2\sim 2~\mathrm{GeV}^2$ ($Q^2\sim 4000~\mathrm{GeV}^2$). 
As one can see, typical asymmetries range from $10^{-4}$ to $0.1$. The 
right part of the figure gives the resulting values for the 
relative uncertainty $\delta A_\mathrm{beam}/A_\mathrm{beam}$ of the 
asymmetry. Here we have summed over all $Q^2$ bins and assumed
an integrated luminosity of ${\cal L}=100$~fb$^{-1}$. The relative 
uncertainty is found to be near  $2\%$ over a relatively wide range 
in $x$; the relative electron polarization uncertainty achievable with modern 
polarimetry techniques should be better than this.

\begin{figure}
\vspace*{-4mm}
\begin{center}
\includegraphics[scale=0.335]{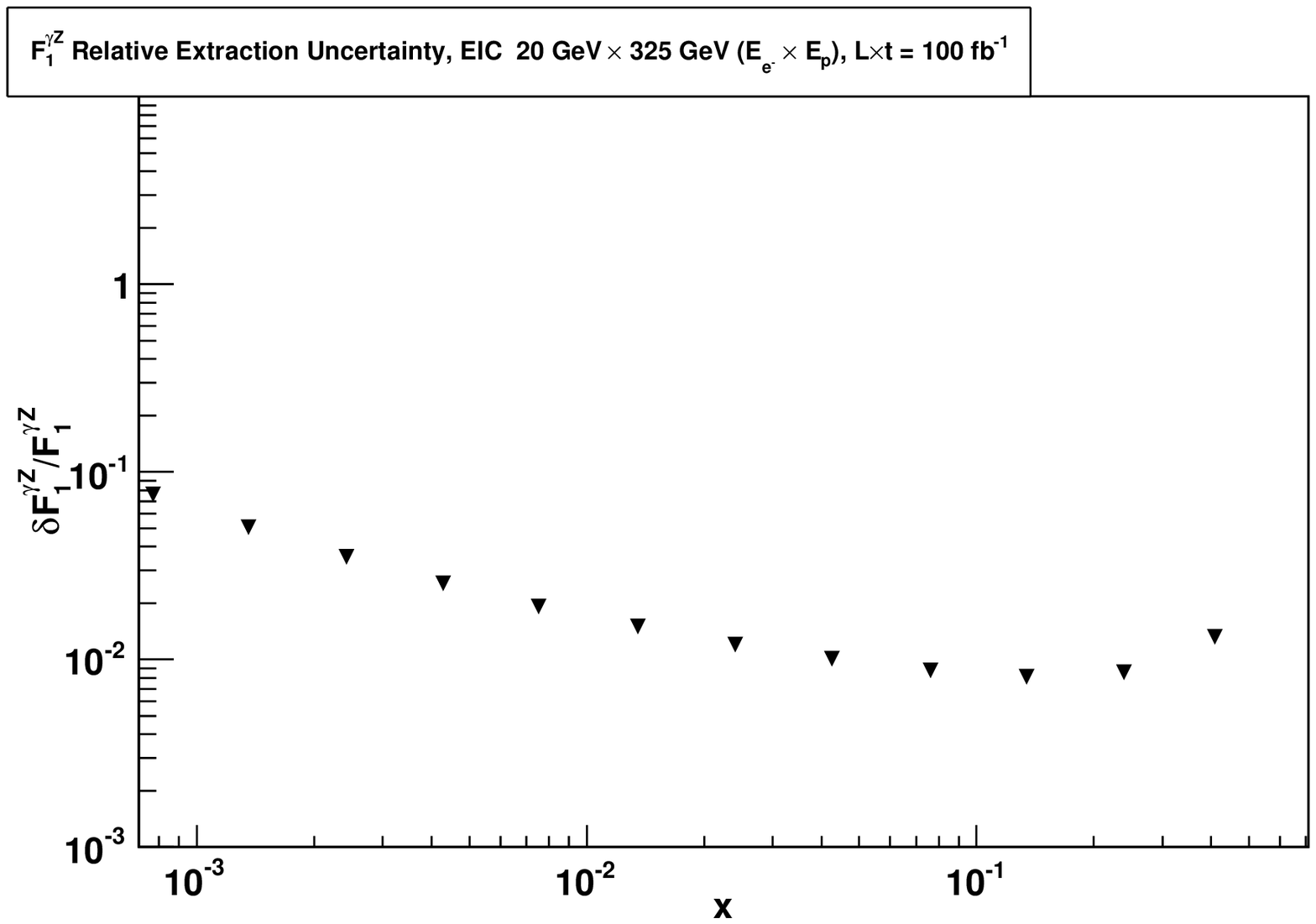}
\includegraphics[scale=0.335]{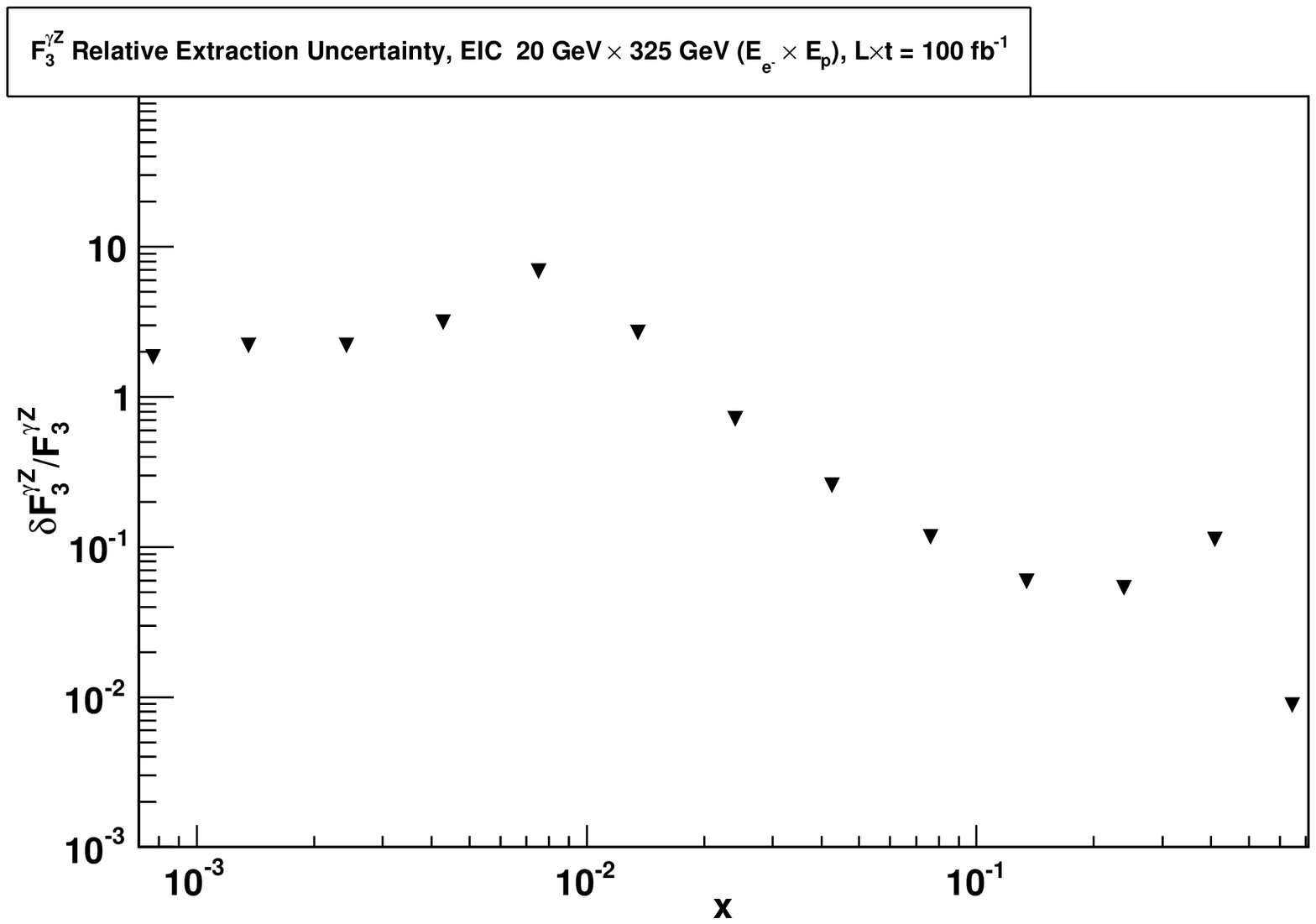}
\end{center}
\vspace*{-4mm}
\caption{Relative uncertainties of $F_1^{\gamma Z}$ (left) and
$F_3^{\gamma Z}$ (right) extracted from NC $\vec{e}^-p$ scattering.}
\label{fig:f1f3relunc}
\end{figure}

\begin{figure}
\begin{center}
\includegraphics[scale=0.4]{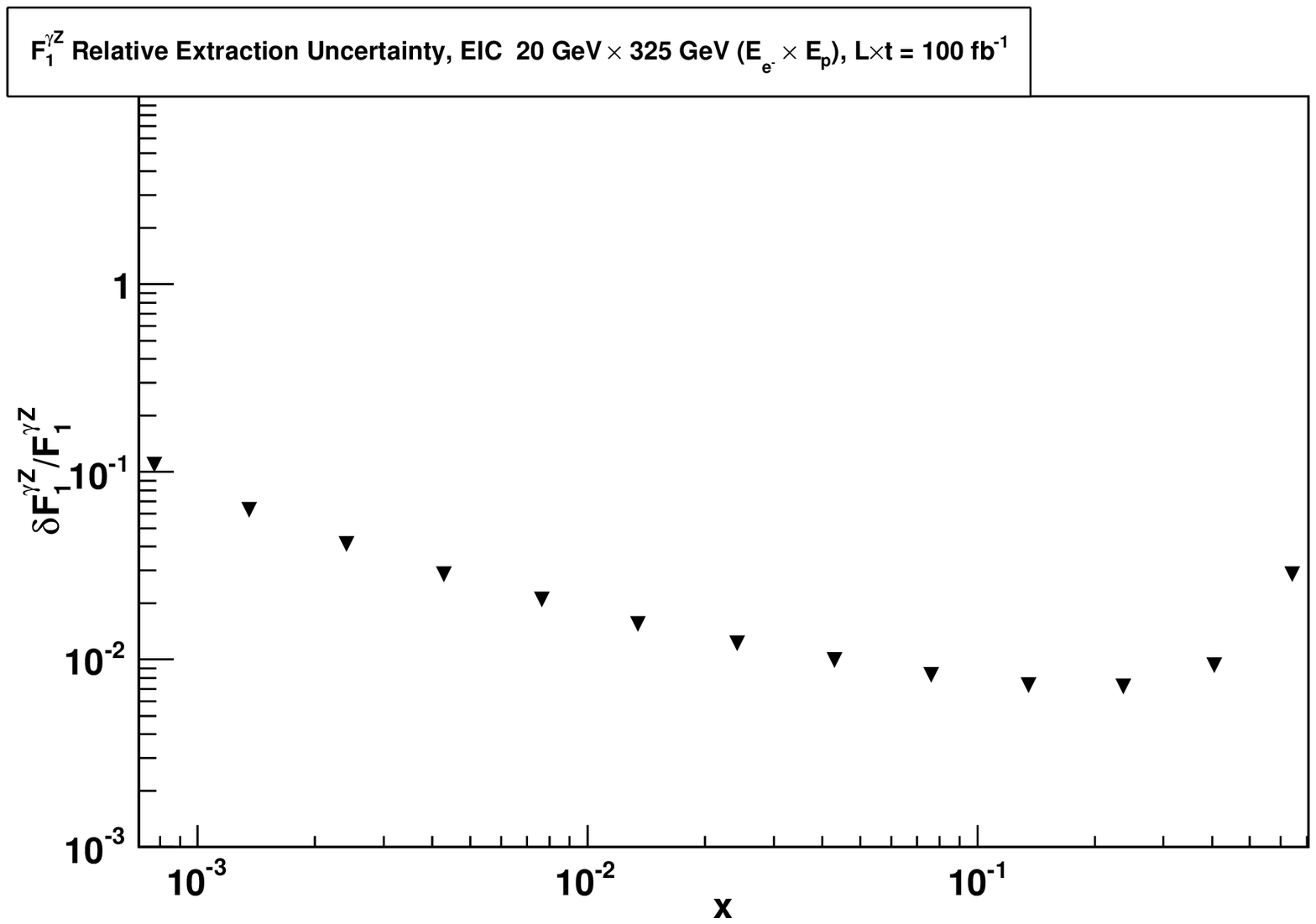}
\end{center}
\vspace*{-4mm}
\caption{Same as left part of Fig.~\ref{fig:f1f3relunc}, but for 
$\vec{e}^{\,-}\mathrm{D}$ scattering.}
\label{fig:f1Drelunc}
\end{figure}

According to Eq.~(\ref{Abeam}), measurement of the asymmetry $A_\mathrm{beam}$
gives access to $F_1^{\gamma Z}$ and $F_3^{\gamma Z}$. Figure~\ref{fig:f1f3relunc}
presents the expected relative uncertainties for these structure functions,
corresponding to the results shown in Fig.~\ref{fig:20x325A}. 
Figure~\ref{fig:f1Drelunc} shows the corresponding result for 
the case of $\vec{e}^{\,-}D$ scattering, for the structure function 
$F_1^{\gamma Z}$. Due to the suppression by the 
electron vector coupling, the uncertainty of $F_3^{\gamma Z}$ 
is about an order of magnitude worse than that of $F_1^{\gamma Z}$. The
sensitivity is maximized in the region of $x\sim0.01-0.4$. 
The approved PVDIS experiment using the SoLID spectrometer 
in Hall A at Jefferson Lab~\cite{solid} anticipates achieving an 
extraction of $A_\mathrm{beam}$ with relative accuracy $\approx 0.5-1\%$ 
over several bins in $x$ in the range of $0.2\leq x \leq 0.7$, both from
proton and deuterium targets. The products of the quarks' electric 
charges and their vector charges are approximately equal for up-type 
and down-type quarks, $e_u g_V^u \approx e_d g_V^d \approx 0.1$.
Therefore, one has from Eq.~(\ref{eq:unpolsf}) that 
$F^{\gamma Z}_1 \propto u+\bar{u}+d+\bar{d}+s+\bar{s}$,
{\it both} for proton and deuterium. On the other hand, for the
corresponding products of the charges and axial charges one finds
$e_u g_A^u \approx 2 e_d g_A^d$, and hence in the valence region 
$F^{\gamma Z}_3 \propto 2 u_v + d_v$ for protons and $\propto
u_v + d_v$ for deuterium.  While $F_3^{\gamma Z}$ could thus give a 
clean separation of the $u$ and $d$ valence distributions, 
its contribution to the beam asymmetry is unfortunately suppressed.

\begin{figure}
\vspace*{2mm}
\begin{center}
\includegraphics[scale=0.35]{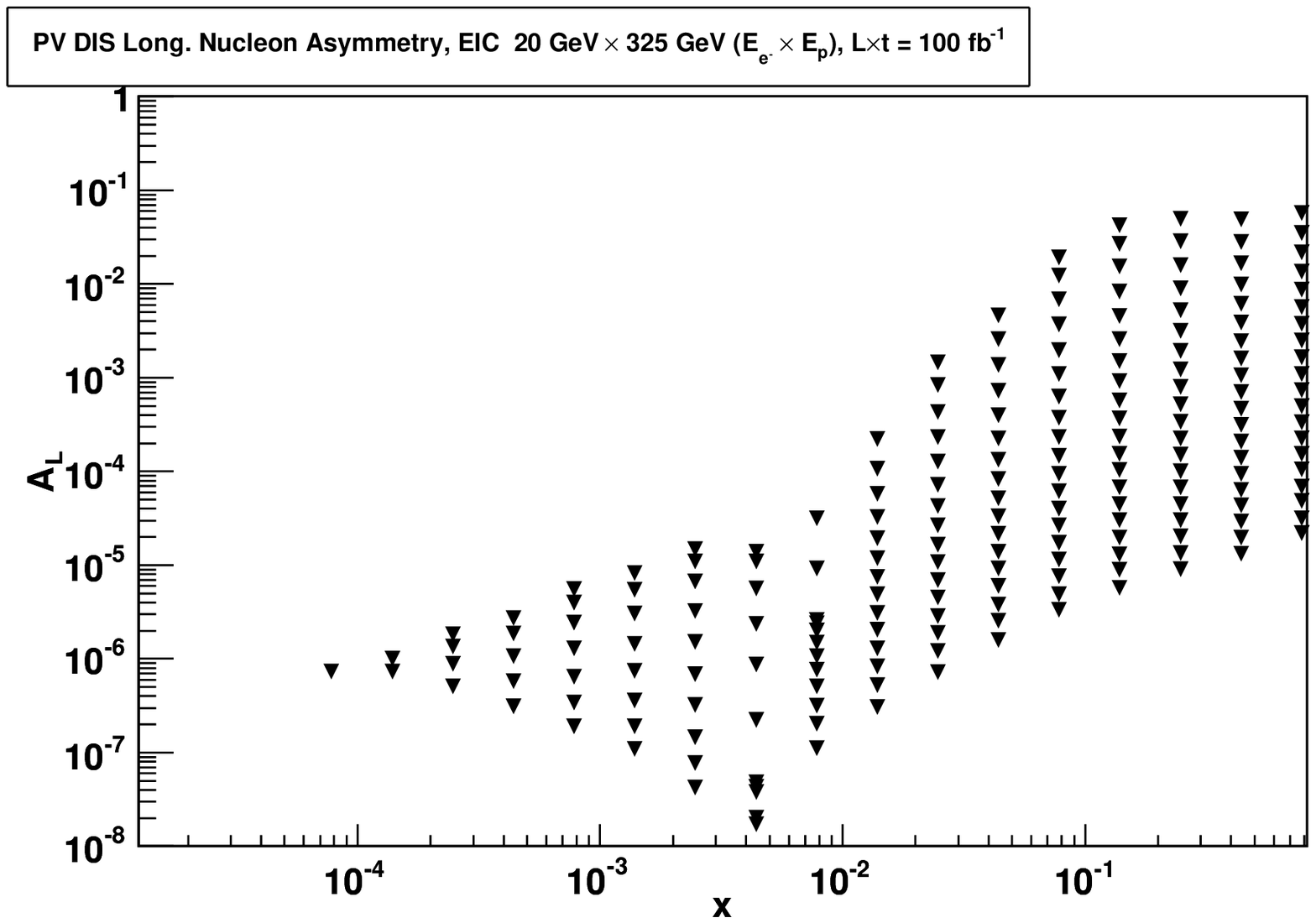}
\includegraphics[scale=0.35]{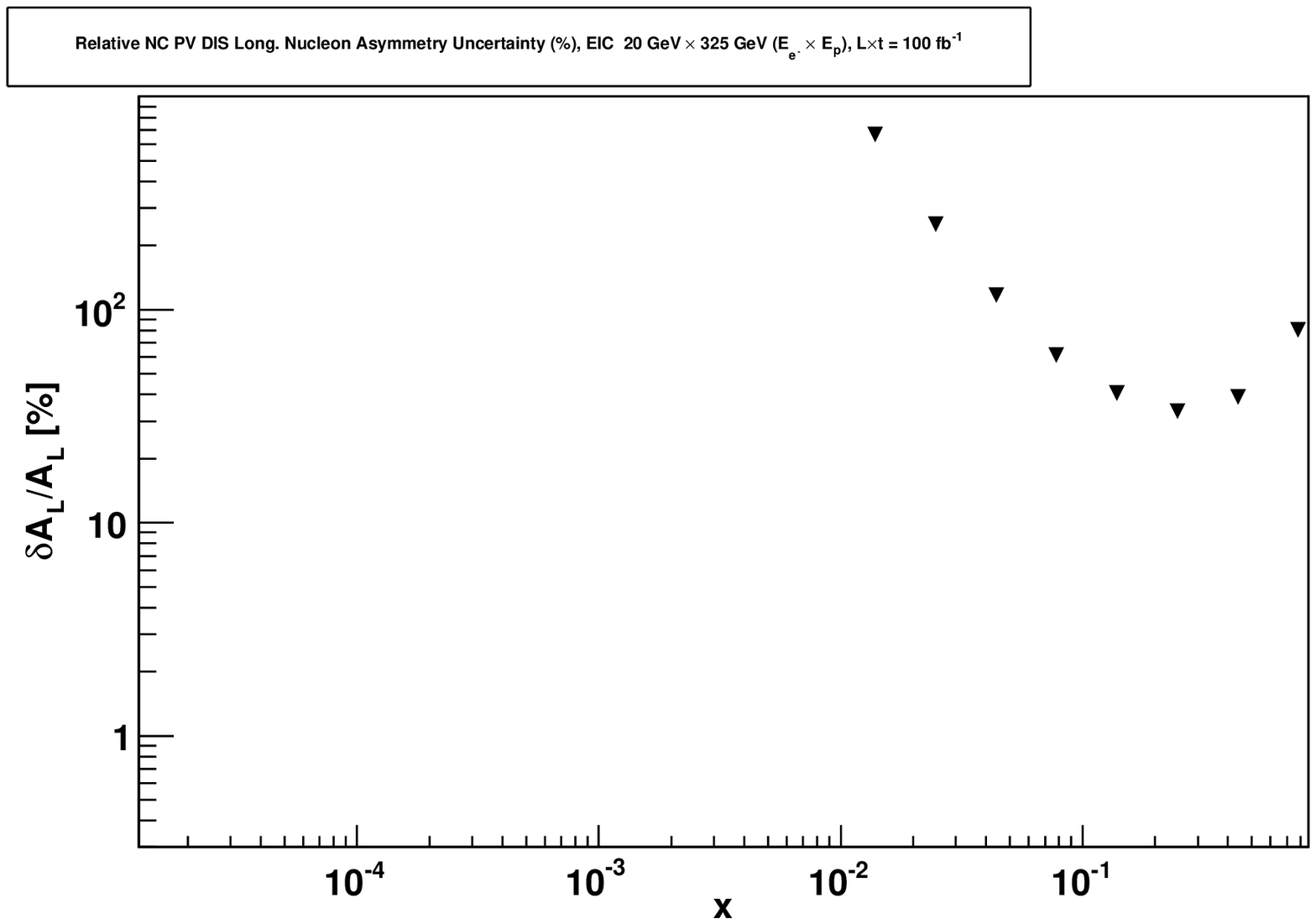}
\end{center}
\vspace*{-4mm}
\caption{Same as Fig.~\ref{fig:20x325A}, but for unpolarized 
electrons on polarized protons.}
\label{fig:AL}
\end{figure}
\begin{figure}
\vspace*{3mm}
\begin{center}
\includegraphics[scale=0.35]{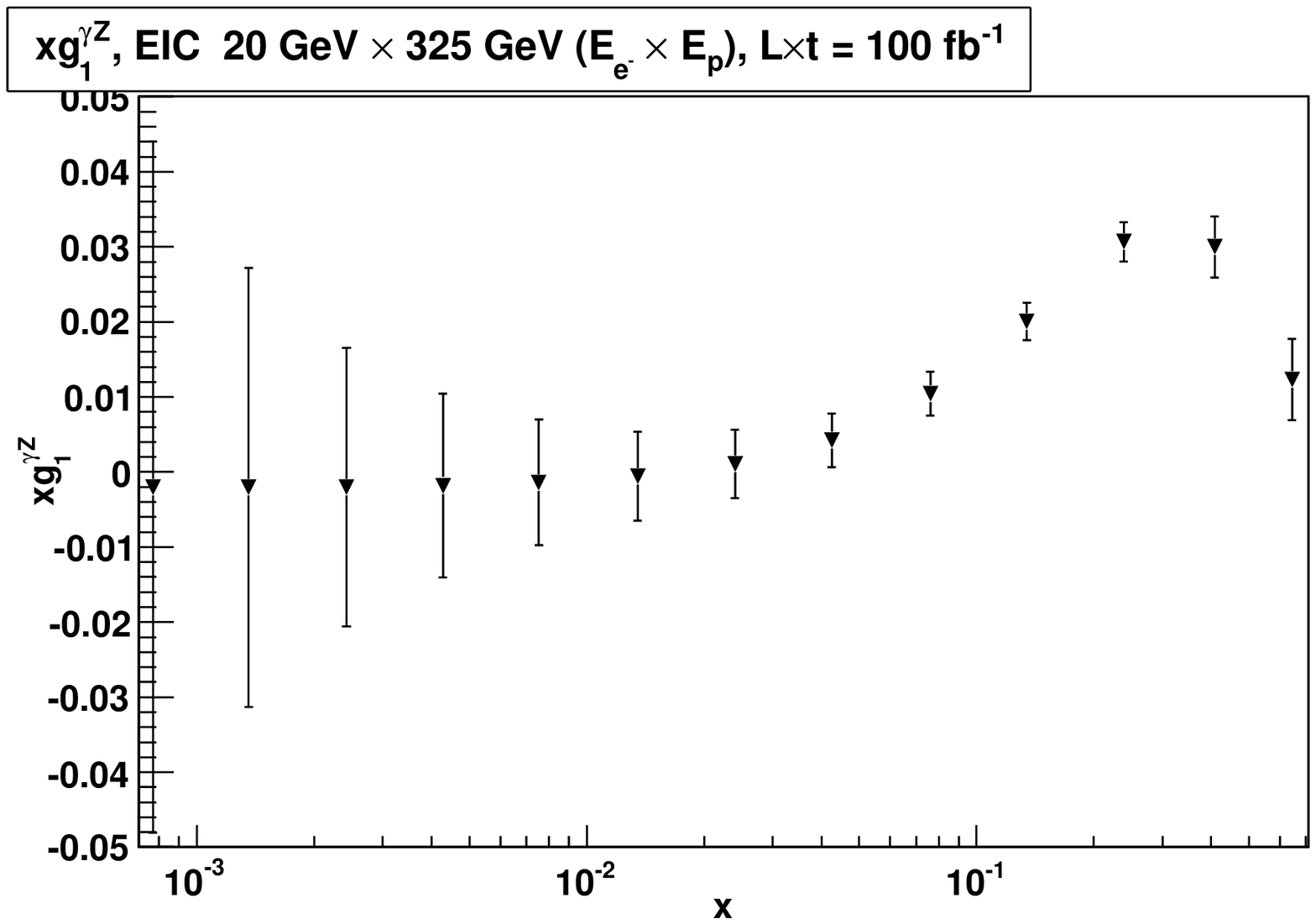}
\includegraphics[scale=0.35]{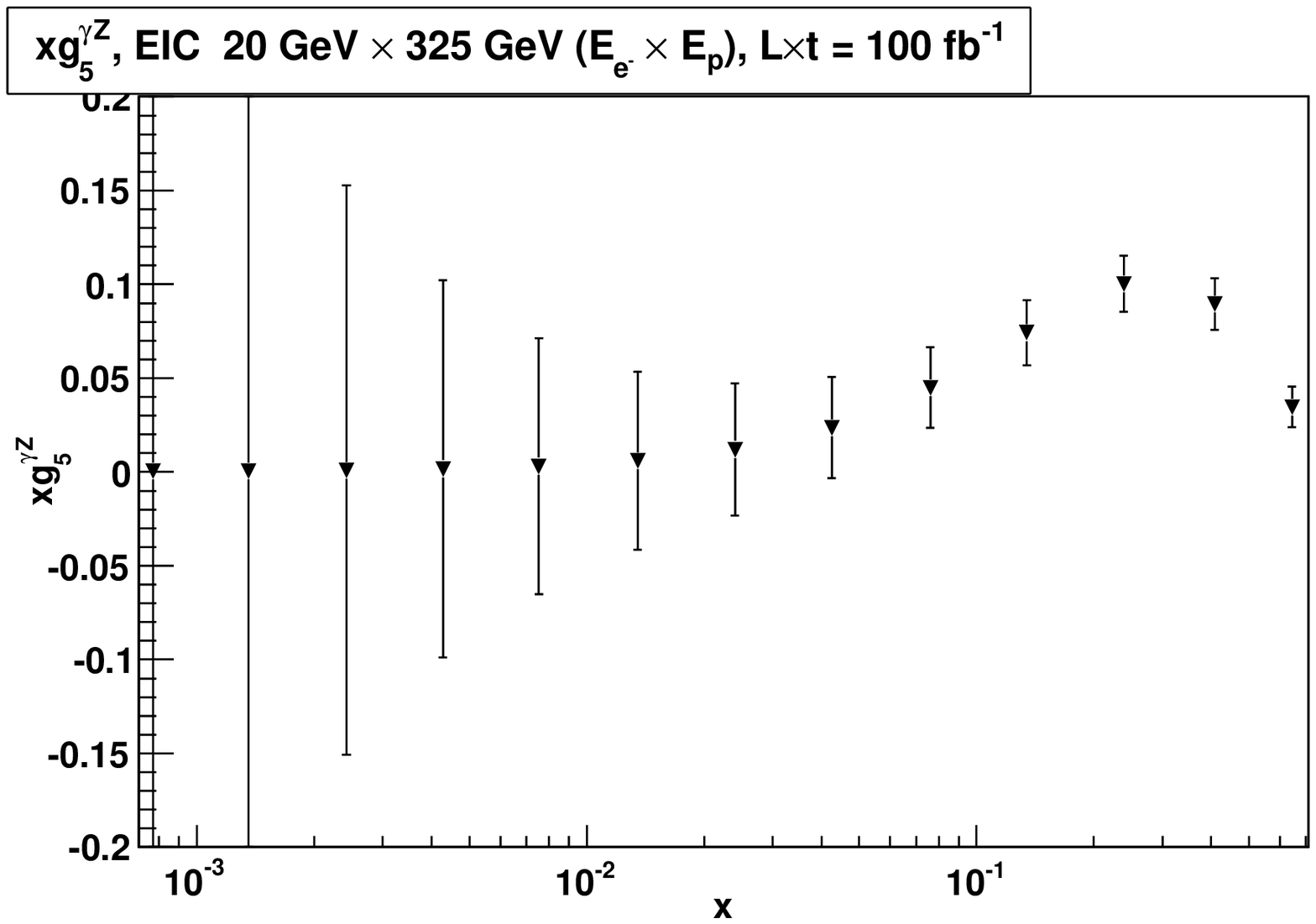}

\vspace*{4mm}
\includegraphics[scale=0.35]{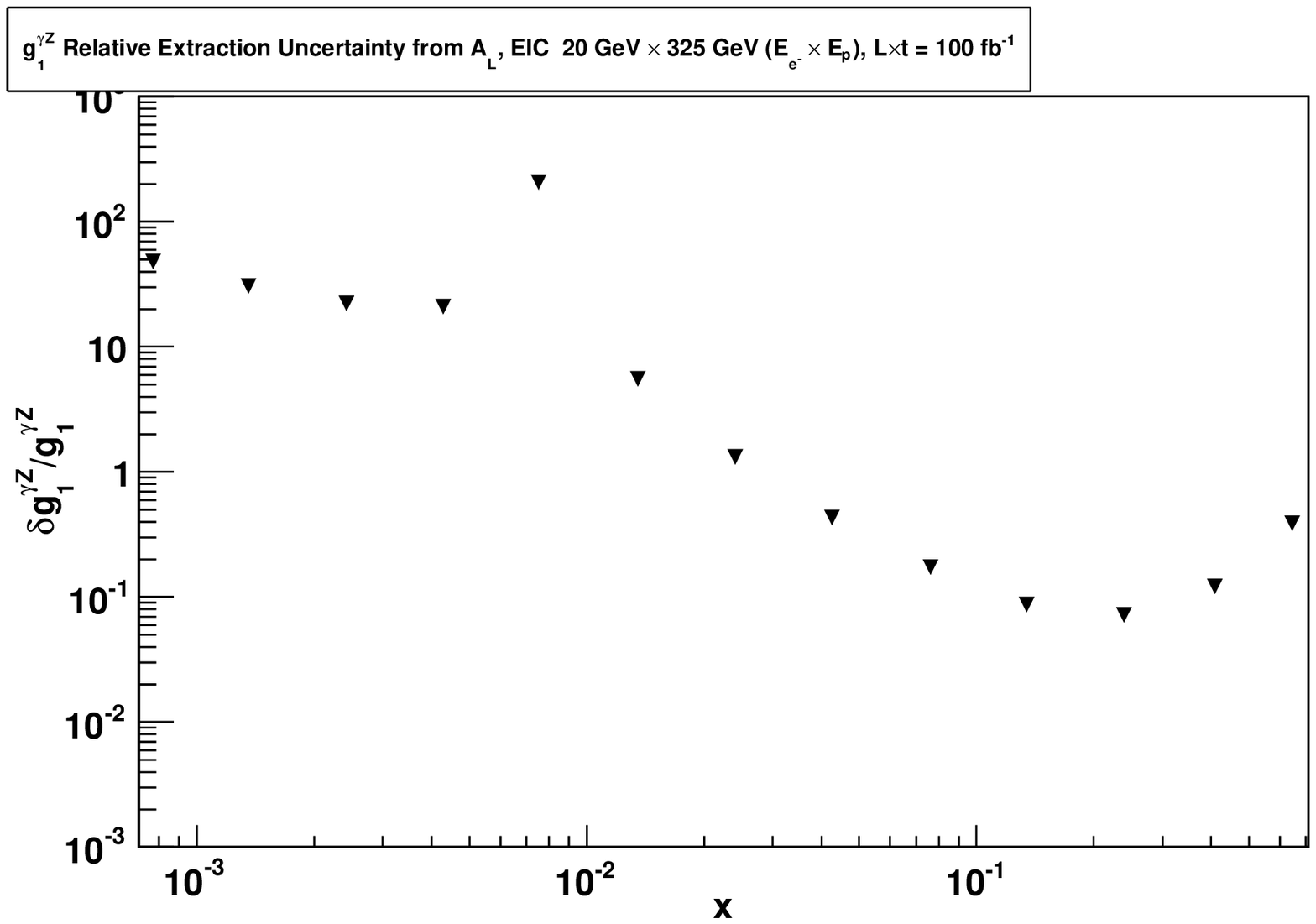}
\includegraphics[scale=0.35]{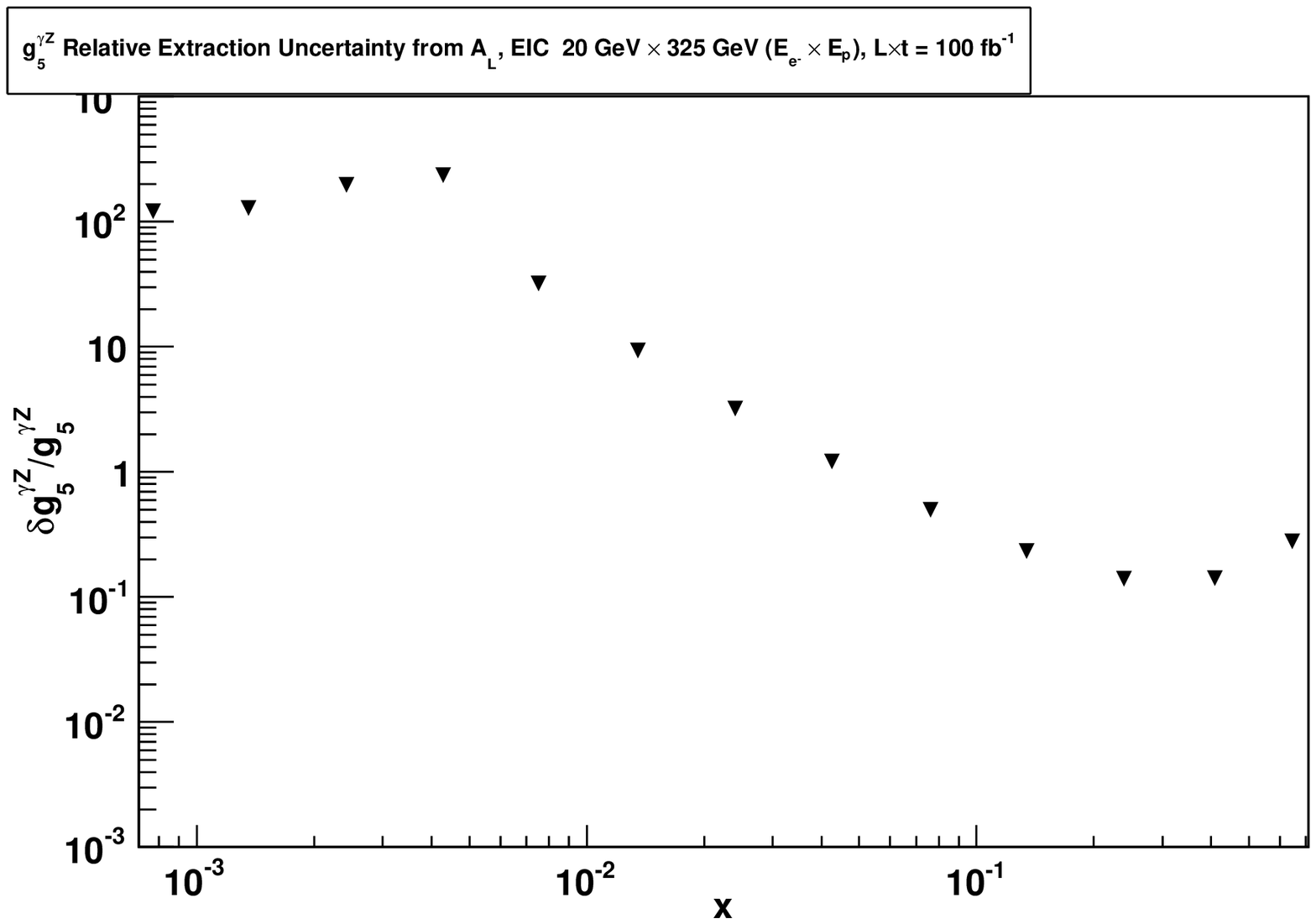}
\end{center}
\vspace*{-3mm}
\caption{Structure functions $g_1^{\gamma Z}$ and $g_5^{\gamma Z}$ (top) 
and their relative uncertainties resulting from Fig.~\ref{fig:AL} (bottom).}
\label{fig:g1g5extract}
\end{figure}

Of significant interest are measurements of $g_1^{\gamma Z}$ and $g_5^{\gamma Z}$,
which contain complementary information on the polarized PDFs.
Similarly to what we discussed for the case of $F_1^{\gamma Z}$, one finds 
that to a good approximation $g_1^{\gamma Z}\propto \Delta u+\Delta \bar{u} + 
\Delta d+\Delta \bar{d} +\Delta s+\Delta \bar{s}$, which would in principle
make this structure function an complementary probe of the quark and
anti-quark singlet and spin contribution to the proton spin. Furthermore,
$g_5^{\gamma Z}$ offers probes of the valence regime. 
According to Eq.~(\ref{dsigm}), $g_1^{\gamma Z}$ and $g_5^{\gamma Z}$ 
may be accessed by flipping the proton helicity while leaving the 
electron polarization unchanged. The corresponding spin asymmetries, 
obtained after summing over the electron helicities, are unfortunately 
overall much smaller than their counterparts with polarized electron and 
unpolarized proton. They are shown in Fig.~\ref{fig:AL}, along with the
their expected relative uncertainties, computed again for 
${\cal L}=100$~fb$^{-1}$.
The best sensitivity is in the valence quark region, $x>0.1$. Even 
here, it remains at the $10\%$ level. This directly translates into
similar uncertainties for the structure functions $g_1^{\gamma Z}$ and 
$g_5^{\gamma Z}$, which are shown in Fig.~\ref{fig:g1g5extract}. 
In the valence region,  
where sea quarks are irrelevant, we have $g_1^{\gamma Z}\propto 
\Delta u_v + \Delta d_v$ and $g_5^{\gamma Z}\propto 2 \Delta u_v + 
\Delta d_v$, which may provide a separation of $\Delta u$ and $\Delta d$.

\begin{figure}
\begin{center}
\includegraphics[scale=0.375]{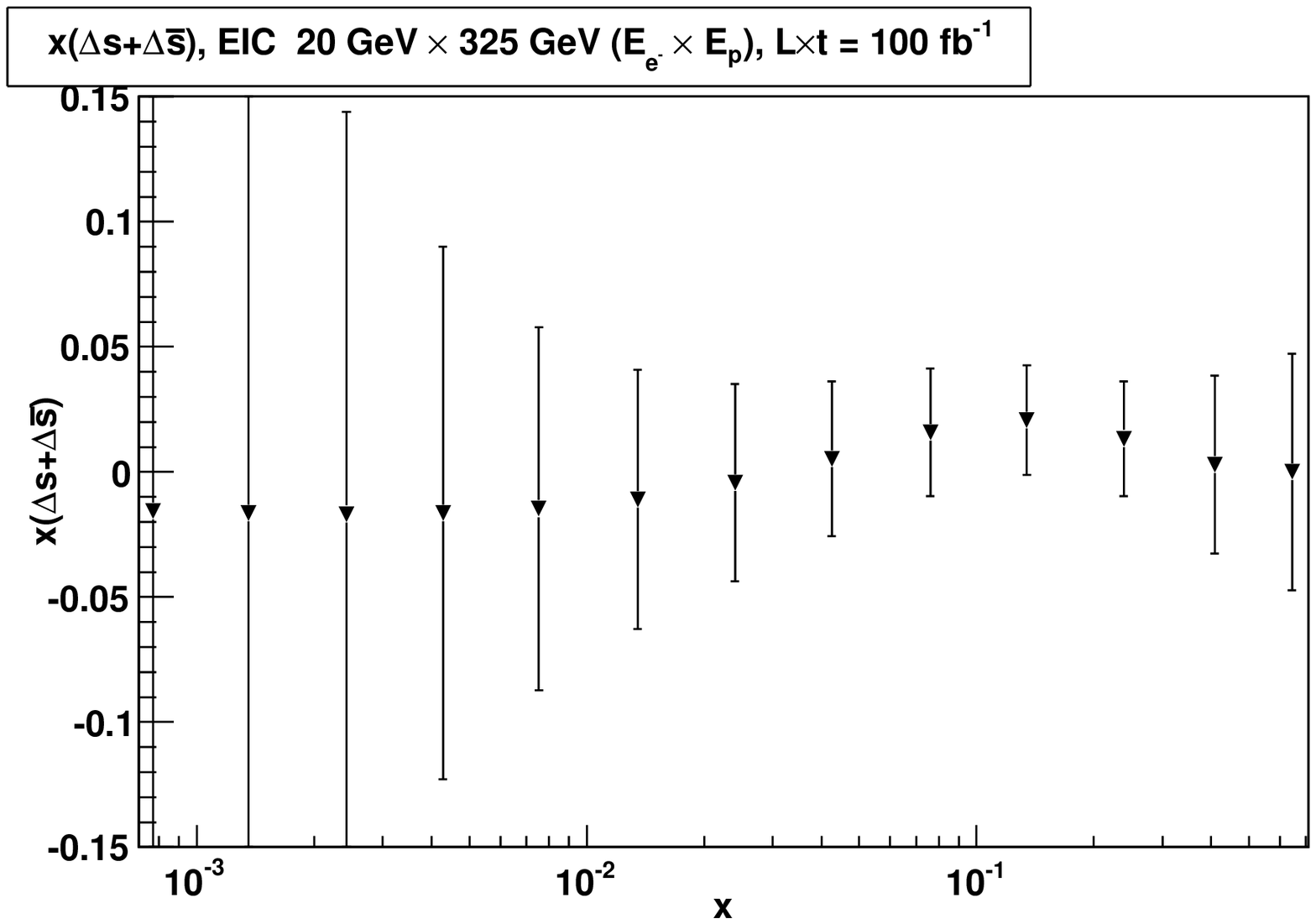}
\end{center}
\caption{Results for the $x(\Delta s+\Delta \bar{s})$ distribution extracted
from the $A_{\mathrm{L}}$ spin asymmetry under the assumption that all other
helicity distributions are known.}
\label{fig:deltas}
\end{figure}

Finally, assuming perfect knowledge of $\Delta u$ and $\Delta d$ and their
anti-quark distributions from other sources, one might ask if an extraction
of $\Delta s+\Delta \bar{s}$ from $g_1^{\gamma Z}$ and $g_5^{\gamma Z}$ 
could be possible. This quantity, and in particular its integral,
is a key ingredient to nucleon spin structure and for understanding
why quarks and anti-quarks combined appear to carry little of
the proton spin. Constraints on $\Delta s+\Delta \bar{s}$ are presently
available from an SU(3) symmetry analysis of hyperon $\beta$-decays,
and from kaon production in semi-inclusive DIS, which are both 
inflicted with sizable uncertainties and in fact show some tension
(for discussion, see~\cite{deFlorian:2009vb}). The result for the extraction 
of $\Delta s+\Delta \bar{s}$ from electroweak DIS at the EIC is shown 
in Fig.~\ref{fig:deltas}. As can be seen, a non-zero measurement would be 
challenging for the assumed 100 fb$^{-1}$ integrated luminosity. Nevertheless, 
this measurement might become interesting if independent methods of extracting 
$\Delta s+\Delta \bar{s}$ were to provide surprising results.  If
this measurement is deemed sufficiently interesting and important, 
larger integrated luminosities will indeed help, since the measurement will continue to remain
statistics limited, provided relative hadron polarization errors can be kept at the 3\%\ level or better.

\subsection{Summary}

We have performed a basic analysis of the potential of an  EIC
in terms of measurements of structure functions in electroweak NC
and CC scattering. Precise measurements of the 
CC functions $F_1^W$, $F_3^W$, $g_1^W$, and $g_5^W$ become
feasible with a relatively modest integrated luminosity. These measurements will
greatly aid the flavor decomposition of polarized and unpolarized PDFs 
in the region $x\gtrsim 0.01$. NC structure functions become
accessible with good precision at high integrated luminosities. 
Measurements of $F_1^{\gamma Z}$ and  $F_3^{\gamma Z}$  
seem to be of limited use in improving present or approved measurements. At the
highest luminosities and center of mass energies, $g_1^{\gamma Z}$ and $g_5^{\gamma Z}$
become accessible; these structure functions have never before been measured. 
The combined analysis of the new CC and NC structure functions with 
electrons and positrons as well as with polarized protons and neutrons at 
these highest luminosities could potentially open a new window into precision QCD tests of the spin structure of the nucleon; this will be the focus of future
experimental and theoretical investigations.



\section{Charged current charm production and the strange sea}
\label{sec:marco-cc-charm}

\hspace{\parindent}\parbox{0.92\textwidth}{\slshape 
  Marco Stratmann
%
}

\index{Stratmann, Marco}



\subsection{Basic idea}
The leading order contribution to CC charm production in $e^+p$ DIS is given by the
${\cal{O}}(\alpha_s^0)$ parton model process $W^+ s' \rightarrow c$,
where $s'$ denotes the Cabibbo-Kobayashi-Maskawa (CKM) ``rotated'' combination
$s' \equiv \left|V_{cs}\right|^2 s + \left|V_{cd}\right|^2 d$. 
Due to the smallness of $\left|V_{cd}\right|^2$ \cite{Nakamura:2010zzi}
the process is expected to be essentially sensitive to the strange sea content. 
Only at large $x$, where quark sea contributions are less relevant, 
the $\left|V_{cd}\right|^2$ suppression is balanced by the 
valence enhancement of the well-known $d(x)$ density. 
Likewise, in $e^-p$ DIS, the process $W^- \bar{s}' \rightarrow \bar{c}$ predominantly probes
the anti-strange density $\bar{s}(x)$.
With a polarized proton beam one can access also $\Delta s(x)$ and $\Delta \bar{s}(x)$.

Current determinations of $s(x)$ rely mainly on fixed-target neutrino scattering off
nuclear targets with potentially large uncertainties, see Fig.~\ref{fig:strangeness-status}
in Sec.~\ref{sec:marco-sidis}.
Much less is known about
the longitudinally polarized $\Delta s(x)$ so far, see Sec.~\ref{sec:polstatus}.
Due to the limited luminosity and charm detection efficiency, charm production in CC DIS 
could not be studied at HERA.
CC DIS would provide an independent way to extract the unpolarized and polarized 
strange sea distributions at much larger scales, typically $Q\sim M_W$, than probed in 
semi-inclusive kaon production, cf.\ Sec.~\ref{sec:marco-sidis}.
On the downside, such a measurement requires also a positron beam, though not polarized.

Next-to-leading order QCD corrections also complicate the simple picture for CC charm production
and may deteriorate the sensitivity to strangeness.
Apart from the ${\cal{O}}(\alpha_s)$ corrections to the LO process $W^+ s' \rightarrow c$,
the genuine NLO, gluon induced subprocess $W^+ g \rightarrow c \bar{s}'$
has to be taken into account as well. It contributes significantly to the
charm production cross section in certain regions of phase space and hence
dilutes the sensitivity to the strange sea. 
In addition, a proper theoretical calculation also needs to take into account
the mass of the produced heavy (charm) quark, as was also discussed in the context of
$F_{2,L}^c$ in Sec.~\ref{sec:bluemlein-hq}.
In order to make contact with experiment, a fully inclusive calculation \cite{Gottschalk:1980rv,Gluck:1996ve}
is not entirely sufficient, and
one should compute also the momentum $z$ spectrum of the detected charmed $D$ mesons.
In the unpolarized case this was achieved in \cite{Gluck:1997sj}.
The corresponding polarized results can be found in Ref.~\cite{Kretzer:1999nn}.
Imposing a lower cut $z_{\min}$ on the $D$ meson momentum fraction was shown to considerably
reduce gluon-initiated NLO contributions and enhance the sensitivity to the strange sea.

Concerning the mass $m_c$ of the charm quark, it turns out that the naive 
``rescaling prescription'' \cite{Barnett:1976ak}, i.e., $s(x)\rightarrow s(\xi)$ where $\xi\equiv x(1+m_c^2/Q^2)$,  
applies also at NLO accuracy as it allows for a consistent factorization of all
initial-state collinear singularities.

\subsection{Sensitivity to the Strange Sea}
So far, detailed phenomenological studies have been provided only for HERA kinematics \cite{Kretzer:1999nn},
and they still need to be updated for EIC kinematics. However, these projections are sufficient to demonstrate
the idea of the measurement and give a rough estimate of the size of cross sections and spin asymmetries.
\begin{figure}[thb!]
\begin{centering}
\includegraphics[clip,width=0.8\textwidth]{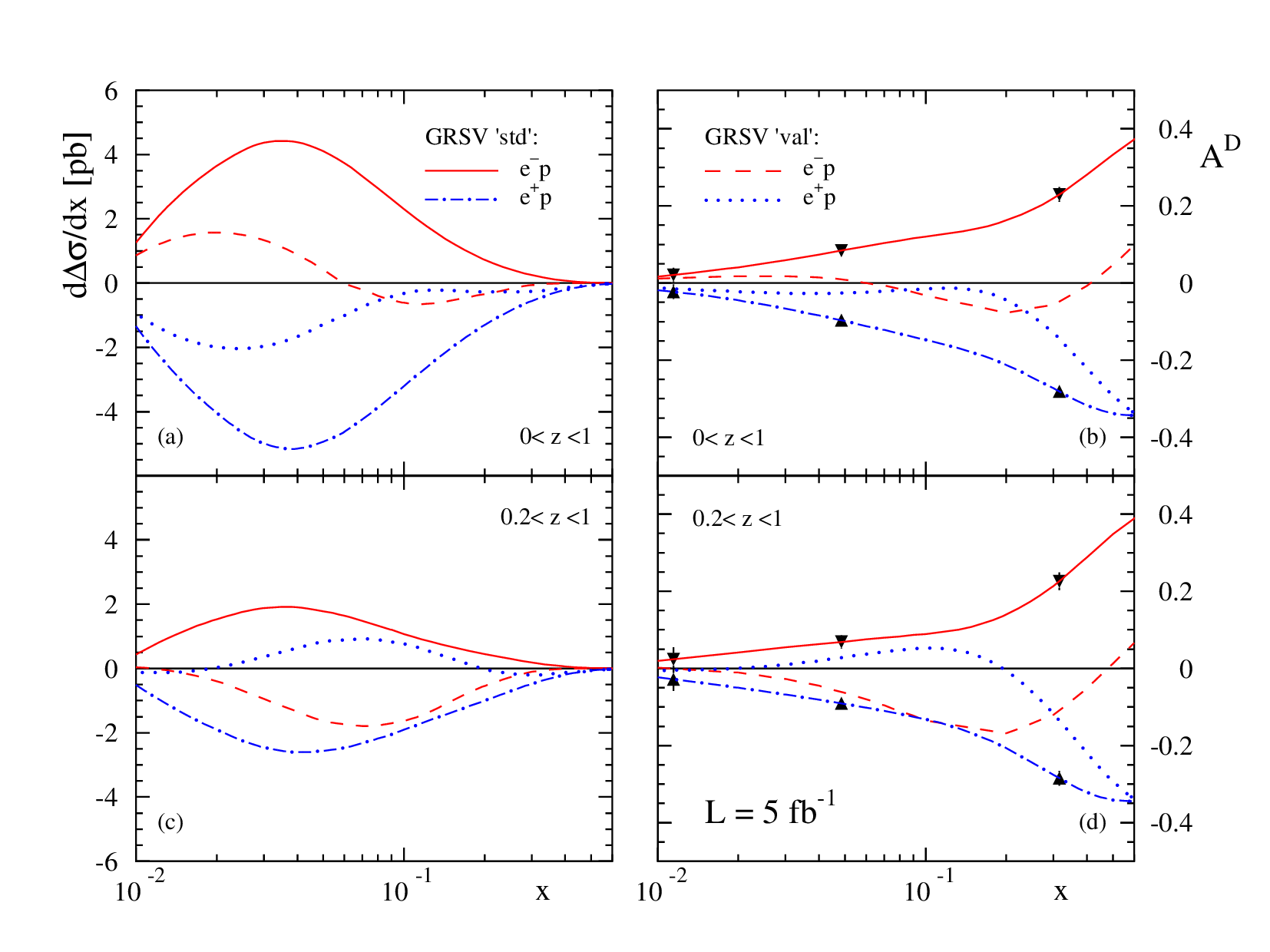}
\par\end{centering}
\caption{The $z$ integrated polarized cross section
for CC charm production in $e^-p$ and $e^+p$ collisions
and the corresponding spin asymmetry $A^D$ for 
{\bf (a,b)}: $0<z<1$, {\bf (c,d)}: $0.2<z<1$,
using the GRSV ``std'' and ``val'' sets of PDFs.
Projected uncertainties are for 70\% polarization, 100\% charm
detection efficiency, and an integrated luminosity of $5\,\mathrm{fb}^{-1}$.
\label{fig:cc-charm}}
\end{figure}
From the studies of inclusive CC electroweak DIS structure functions in Sec.~\ref{sec:electroweak}
we already know that such measurements appear to be feasible at an EIC despite its lower c.m.s.\ energy than HERA
even with moderate integrated luminosities of about $10\mathrm{fb}^{-1}$.

As an example, Fig.~\ref{fig:cc-charm} shows the sensitivity 
of CC charm ($D$ meson) production in
$e^-p$ and $e^+p$ collisions at $\sqrt{S}=300\,\mathrm{GeV}$,
$Q^2>500\,\mathrm{GeV}^2$, and $0.01\le y\le0.9$,
to the choice of $\Delta s$. The momentum fraction of the detected
$D$ meson has been integrated using $z_{\min}=0$ (upper row) and $0.2$
(lower row). The GRSV valence set \cite{Gluck:2000dy} has a
very small positive $\Delta s(x)$ in the relevant region $x\gtrsim 0.01$,
roughly comparable to what is nowadays obtained from fixed target SIDIS
data, e.g., in the DSSV analysis \cite{deFlorian:2008mr,deFlorian:2009vb}; see Sec.~\ref{sec:polstatus}.
On the contrary, the GRSV standard set has a sizable negative
strangeness polarization as favored by fits including only
inclusive DIS data \cite{Blumlein:2010rn}. Other PDFs, in particular the gluon density, 
are very similar in both GRSV sets. Note that $\Delta s(x)=\Delta\bar{s}(x)$
is assumed in all current polarized PDF analyses due to the lack of data constraining them
separately.

The solid and dashed lines in Fig.~\ref{fig:cc-charm} show
the results for $e^-p$ scattering for GRSV standard and valence PDFs, respectively.
Within the projected statistical uncertainties, obtained for
70\% proton polarization, 100\% charm
detection efficiency, and an integrated luminosity of $5\,\mathrm{fb}^{-1}$,
differences in $\Delta \bar{s}(x)$ can be easily resolved.
The dot-dashed and dotted lines show the results for a corresponding
measurement with 
positron beams. Having results for both $W^-$ and $W^+$ exchange, one
should be able to study a possible asymmetry in $\Delta s(x)-\Delta\bar{s}(x)$.
The results presented here need to be backed up with more detailed 
simulations of CC charm production for EIC kinematics.

\section{Photoproduction processes at an EIC}
\label{sec:photoproduction}


\hspace{\parindent}\parbox{0.92\textwidth}{\slshape 
  Hubert Spiesberger, Marco Stratmann
%
}

\index{Spiesberger, Hubert}
\index{Stratmann, Marco}

\vspace{\baselineskip}





\begin{figure}[b!]
\begin{center}
\includegraphics[width=0.45\textwidth]{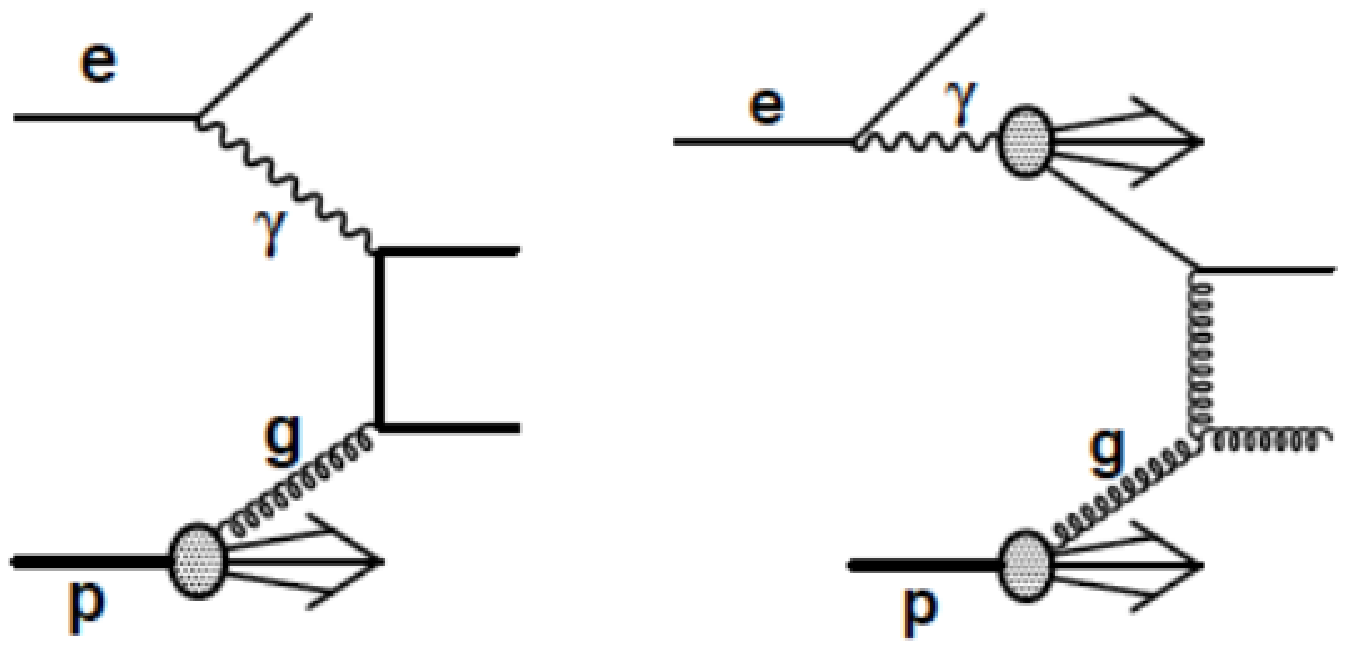}
\end{center}
\caption{\label{fig:PP:FD} Example of a gluon-initiated direct and resolved 
contributions to photoproduction at LO (taken from Ref.~\cite{zeus:2007qt}).
}
\end{figure}

The production of hadronic final states in $ep$ 
collisions is dominated by photoproduction where the electron 
is scattered by a small angle producing photons of almost zero 
virtuality ($Q^2 \simeq 0$).  At LO of pQCD, the dominant process for the production of high-$p_T$
hadrons, jets, or heavy quarks is often photon-gluon fusion,
$\gamma g \rightarrow q \bar{q}$. 
Here, the photon interacts directly with a gluon from 
the nucleon. Besides this so-called ``direct'' photoproduction channel, 
the scattering can proceed also via ``resolved'' processes. 
In this case, the photon acts as a source of partons which 
interact with the partons in the nucleon through any of the
standard $2\rightarrow 2$ LO QCD hard scattering processes such
as $gg\rightarrow gg$ or $q\bar{q}\rightarrow q\bar{q}$.
The large number of possible subprocesses can 
make the resolved contribution sizable in certain regions of phase space. 
Examples for a direct and a resolved process are shown in Fig.~\ref{fig:PP:FD}.
%

At LO, the two interaction mechanisms in Fig.~\ref{fig:PP:FD}
both contribute at ${\cal{O}}(\alpha_{em}\alpha_s)$ but
otherwise appear to be independent. Starting from NLO, however, 
the separation into direct and resolved contributions becomes 
factorization scheme dependent. This is due 
to soft and collinear singularities appearing in a perturbative 
approach. These singularities have to be identified and 
consistently factorized into non-perturbative PDFs of the nucleon and the photon. 
This procedure is not unique, and it is therefore important that the direct and resolved 
parts are treated together consistently. Only their sum is an experimentally meaningful and
measurable cross section. For a theoretical review on photoproduction, see, e.g., Ref.~\cite{Klasen:2002xb}.

The differential cross section for electron-nucleon scattering, 
$d \sigma_{eN}$, at a c.m.s.\ energy $\sqrt{s}$ is related to the 
photoproduction cross section $d \sigma_{\gamma N}$
through
\begin{equation}
d\sigma_{eN}(\sqrt{s}) 
= 
\int_{y_{\rm min}}^{y_{\rm max}} dy\,
f_{e\gamma}(y)\, d\sigma_{\gamma N}(y\sqrt{s})\, .
\label{gpcrosss}
\end{equation}
Here, $f_{e\gamma}$ is the energy spectrum of the exchanged photon 
which in the Weizs\"acker-Williams approximation is given by
\begin{equation}
f_{e\gamma}(y) 
=
\frac{\alpha_{em}}{2\pi} 
\left[\frac{1+(1-y)^2}{y}
\ln \frac{(1-y) Q_{\rm max}^2}{y^2 m_e^2} 
+ 2 (1-y) \left(\frac{y m_e^2}{(1-y) Q_{\rm max}^2}-\frac{1}{y}\right)
\right] \, .
\label{fegwwa}
\end{equation}
The photon flux $f_{e\gamma}$ depends $y = E_{\gamma}/E_e$.
$Q_{\rm max}$ and the range $y_{\rm min} \leq y \leq y_{\rm max}$ 
are determined by cuts in the experimental 
analysis. Typically, a lower cut $y_{\rm min} = {\cal O}(0.1)$ is 
applied in order to exclude low-mass hadronic final states, and an 
upper limit on $y$, e.g., $y_{\rm max} = 0.7 \div 0.9$, is used 
to reduce the kinematic range where radiative corrections are 
expected to be large. 

The photoproduction cross section 
is then obtained as the sum of its direct and resolved parts,
$d\sigma_{\gamma N}=d\sigma^{\rm dir}_{\gamma N}+d\sigma^{\rm res}_{\gamma N}$,
as convolutions $\otimes$ of the appropriate partonic hard scattering cross sections $d\sigma_{ab}$
with the PDFs $f_{a/\gamma} (x_{\gamma})$ and $f_{b/N} (x_N)$ of the photon and 
nucleon, respectively, at a factorization scale $\mu_f$, i.e.,
\begin{equation}
d \sigma^{\rm res}_{\gamma N}
= 
\sum_{a, b} 
f_{a/\gamma} (x_{\gamma},\mu_f) \otimes f_{b/N} (x_N,\mu_f) \otimes
d\sigma_{ab} (x_{\gamma},x_N,\mu_f) 
\, .
\label{eq:PP:sigres}
\end{equation}
$d\sigma^{\rm dir}_{\gamma N}$ can be obtained from \eqref{eq:PP:sigres} 
by replacing the photon PDFs by a $\delta$-function and considering 
only photon-parton scattering processes $d\sigma_{\gamma b}$ in the sum.

The resolved process is accompanied by a hadronic remnant of the photon 
which carries the fraction $1 - x_{\gamma}$ of the photon energy.
At LO, the presence of a hadronic  
remnant could be used to distinguish different event topologies 
for the two mechanisms. In addition, for two-jet final states
$x_{\gamma}$ can be reconstructed 
experimentally from the measured transverse momenta and rapidities of the jets. 
It is customary to define 
\begin{equation}
x_{\gamma}^{\rm obs} \equiv
\left(
E_{\rm T}^{\rm jet_1} e^{-\eta^{\rm jet_1}}
+
E_{\rm T}^{\rm jet_2} e^{-\eta^{\rm jet_2}}
\right)
/ 
\left(2 y E_e \right)
\, .
\end{equation}
However, at higher orders of pQCD, initial- and final-state 
radiation of additional partons will also give rise to hadrons 
emitted in the direction of the incoming photon. Moreover, 
non-perturbative hadronization may contribute to the appearance 
of hadrons in the same kinematic region. Both effects lead to 
a reduction of the experimentally determined value of $x_{\gamma}$. 
Therefore a unique separation of the direct and resolved parts 
is not possible anymore. Nevertheless, the variable $x_{\gamma}$ 
can still be used to define kinematic regimes where direct (large 
$x_{\gamma}$) or resolved (small $x_{\gamma}$) contributions 
dominate.

At HERA, photoproduction has been used to test pQCD and the
presence of both direct and resolved photon processes for
final-states comprising hadrons, jets, prompt photons, and
heavy quarks.
Generally, the data are well described by NLO calculations in 
regimes expected to be dominated by the direct process. Kinematic 
regions where resolved processes are sizable are somewhat
less well described; for a review see, e.g., \cite{Klein:2008di}.
This is mainly due to the fact that the photon PDFs 
needed for the calculation of the resolved contribution 
are significantly less well constrained by data than the
partonic structure of protons. Only data for inclusive DIS off a 
quasi-real photon target, i.e., $\gamma^*(Q^2)\gamma$ scattering in $e^+e^-$
\cite{Nisius:1999cv}, 
have been used in fits of photon PDFs so far, see, e.g., \cite{Gluck:1991jc}. 
No attempts
have been made to perform global analyses or to quantify
uncertainties at a level similar to current fits of proton PDFs.
Any additional, more precise data are therefore of vital importance 
for an improved understanding of the theoretical description of 
photoproduction processes and a reliable determination of photon PDFs.
The latter are of great phenomenological relevance at a possible future
linear $e^+e^-$ collider to describe processes involving quasi-real photons.

The next two sections show some examples how an EIC can contribute to
further our knowledge of photoproduction processes both in unpolarized
and in polarized electron-proton scattering.

\section{Expectations for charm quark photoproduction}
\label{sec:heavy_quarks}


\hspace{\parindent}\parbox{0.92\textwidth}{\slshape 
  Hubert Spiesberger 
%
}

\index{Spiesberger, Hubert}

\vspace{\baselineskip}





The description of heavy quark production in the framework of 
perturbative QCD is complicated due to the presence of several 
large scales, like the transverse momentum $p_T$ of the produced 
charmed meson, the momentum transfer $Q$ in DIS, or the 
mass of the produced heavy hadron. Depending 
on the kinematic range considered, the mass $m_c$ of the charm quark 
may have to be taken into account. Different calculational schemes 
(see, e.g.~\cite{Kramer:2003jw,Kniehl:2004fy}, and references 
therein) have been developed to obtain predictions from pQCD, 
depending on the specific kinematical region and the relative 
importance of the different scales.  

In the case of relatively small transverse momentum, $p_T \lesssim m_c$, 
the fixed-flavor number scheme (FFNS) is usually applied. Here 
one assumes that the light quarks and the gluon are the only 
active flavors and the charm quark appears only 
in the final state. The charm quark mass can explicitly be 
taken into account together with the $p_T$ of the 
produced heavy meson; this approach is therefore expected to be 
reliable when $p_T$ and $m$ are of the same order of magnitude. 


In the complementary kinematical region where $p_T \gg m_c$, 
calculations are usually based on the zero-mass 
variable-flavor-number scheme (ZM-VFNS) where $m_c=0$
and the charm quark acts as an active parton with its own
PDF; see also Sec.~\ref{sec:bluemlein-hq}.
The charmed meson is produced not 
only by fragmentation from the charm quark but 
also from the light quarks and the gluon. 
The fragmentation 
process is described with the help of scale-dependent fragmentation 
functions (FFs), $D(z,\mu)$, which determine the probability that 
the produced heavy meson carries the fraction $z$ of the momentum 
of the parton it is produced from. The predictions obtained in this 
scheme are expected to be reliable only in the region of large 
$p_T$ since all terms of the order $m_c^2/p_T^2$ are neglected in 
the hard scattering cross section. 

A unified scheme that combines the virtues of the FFNS and the 
ZM-VFNS is the so-called general-mass variable-flavour-number 
scheme (GM-VFNS)~\cite{Kramer:2003jw,Kniehl:2004fy}. In this 
approach the large logarithms $\ln(p_T^2/m_c^2)$ 
are factorized into the PDFs and FFs and summed to all orders by the 
well-known DGLAP evolution equations. At the 
same time, mass-dependent power corrections are retained in the 
hard-scattering cross sections, as in the FFNS. 
In order to conform with standard $\overline{\rm MS}$ factorization, finite 
subtraction terms must be supplemented to the results of the 
FFNS. 
As in the ZM-VFNS, one has to take into account processes 
with incoming charm quarks, as well as light quarks and gluons 
in the final state which fragment into the heavy meson. It is 
expected that this scheme is valid not only in the region 
$p_T^2 \gg m_c^2$, but also in the kinematic region where $p_T$ 
is only a few times larger than $m_c$. 
The basic features of the GM-VFNS are described in Ref.~\cite{Kniehl:2009mh}. 
Analytic results for the required hard scattering cross sections 
can be found in Refs.~\cite{Kniehl:2004fy,Kramer:2001gd,Kramer:2003cw,Kniehl:2005mk}.


Next, we present theoretical predictions \cite{Kniehl:2009mh}
for the photoproduction of $D^{*}$-mesons in $ep$ 
scattering at the EIC. We assume an experimental analysis 
with $Q_{\max}=1\,\mathrm{GeV}$ in Eq.~(\ref{fegwwa}).
Since the cross section is dominated by 
low $Q^2$, our results should not depend too strongly on the 
precise value of $Q_{\max}$.
The relevant direct and resolved hard scattering cross sections
are calculated at NLO accuracy. For the 
photon PDFs we use the parametrization of Ref.~\cite{Aurenche:2005da} 
with the standard set of parameter values, and for the proton PDF 
we have chosen the CTEQ6.5 set~\cite{Tung:2006tb}.
For the FFs we use the Global-GM set of Ref.~\cite{Kneesch:2007ey} 
based on a fit to the combined Belle~\cite{Seuster:2005tr}, 
CLEO~\cite{Artuso:2004pj}, ALEPH~\cite{Barate:1999bg}, and 
OPAL~\cite{Alexander:1996wy,Ackerstaff:1997ki} data. 
We choose the renormalization and factorization scales
to be equal and use $\mu_r=\mu_f= m_T$, where $m_T = \sqrt{m_c^2+p_T^2}$ 
is the transverse mass and $m_c = 1.5$ GeV. 
In Ref.~\cite{Kniehl:2009mh} we studied scale 
uncertainties for photoproduction at HERA, as well as 
ambiguities due to various possible choices for input variables, 
such as the proton and photon PDFs, the $D^{*}$ FFs, and
the dependence on $m_c$. 

\begin{figure}[t!]
\begin{center}
\includegraphics[width=0.45\textwidth]{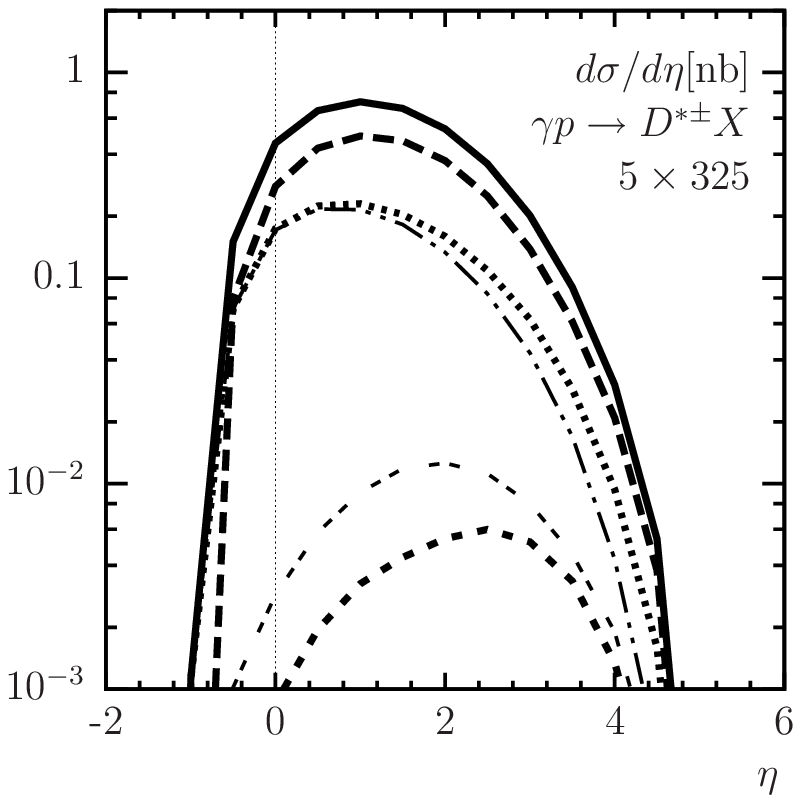}
\includegraphics[width=0.45\textwidth]{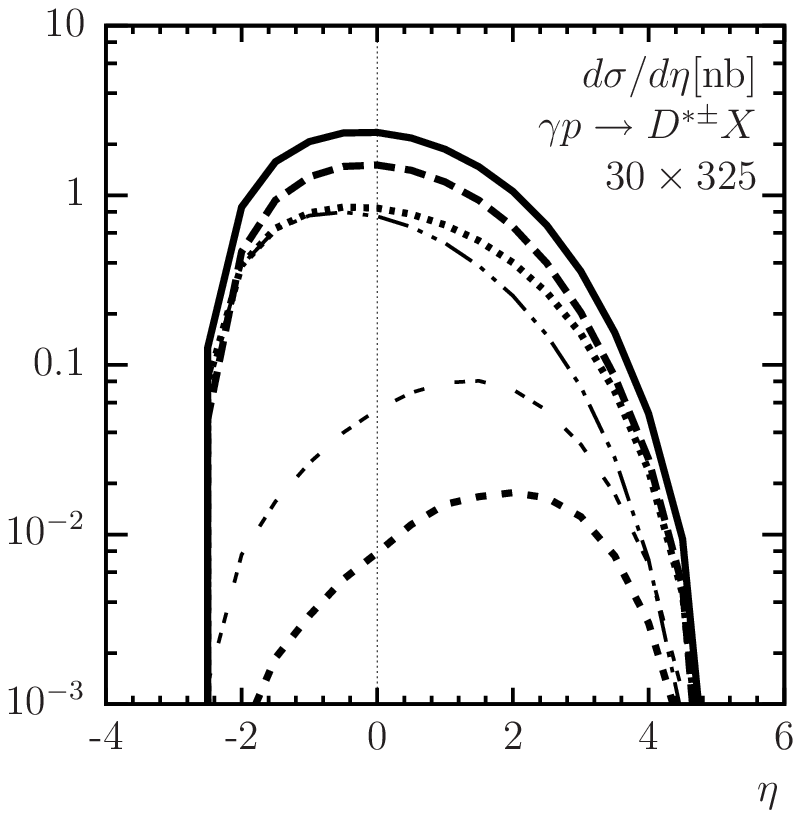}
\end{center}
\vspace*{-0.3cm}
\caption{\label{fig:HQ:dst_eta} $d\sigma/d\eta$ 
for the production of $D^{\ast}$ mesons at the EIC for two settings of
beam energies integrated over transverse momenta $3~\mbox{GeV} \leq 
p_T \leq 5~\mbox{GeV}$. The different curves are explained in the text.
}
\end{figure}
%
In our calculation of the differential cross section $d\sigma/d\eta$ 
(where $\eta$ is the rapidity of the observed heavy meson, $D^{\ast \pm}$) 
we use $E_p = 325$ GeV and consider two choices for the energy of the 
electron beam: $E_e = 5$ GeV (left panel of Fig.~\ref{fig:HQ:dst_eta}) and 
$E_e = 30$ GeV (right panel). 
The transverse momentum $p_T$ 
is integrated over the range $3 < p_T < 5$ GeV. 
The results show that 
the higher electron beam energy would lead to an increase of the 
cross section by roughly a factor of three and the rapidity 
distribution is shifted towards the backwards region, as expected. 

The figure shows a split-up of the total cross section into 
contributions from different subprocesses. From top to bottom, the 
curves correspond to the total cross section (full line), the direct 
contribution (long dashed), the total resolved part (dotted), 
the contribution due to charm in the photon (dash-dot-dotted) 
and charm in the proton (long double-dashed), and, finally, the 
part due to resolved subprocesses with light partons in the initial 
state. The direct contribution, which is sensitive mainly to the gluon 
distribution in the proton, is dominating throughout the shown
range of $p_T$ and $\eta$. The resolved part is mainly due to the 
charm content of the photon, in particular, at negative rapidities. 
Here one may hope that measurements at an EIC, in particular, for the 
option with the highest $\sqrt{s}$, will contribute to a better 
determination of the photon PDFs. 

The total cross sections for charm production at an EIC are not 
very different from those measured at HERA; however, 
an increase in the precision of corresponding measurements can be 
expected due to the higher luminosity. Apart from providing a better
testing-ground for pQCD, one may expect that 
the experimental information will contribute to an improved 
determination of the charm content of the 
proton and, perhaps, the charm FFs. 

\section{Polarized photoproduction at an EIC}
\label{sec:polphoto}

\hspace{\parindent}\parbox{0.92\textwidth}{\slshape 
Barbara J\"{a}ger, Marco Stratmann
%
}

\index{J\"{a}ger, Barbara}
\index{Stratmann, Marco}

\vspace{\baselineskip}


The framework for photoproduction outlined in Sec.~\ref{sec:photoproduction} can be readily
extended to longitudinally polarized $ep$ collisions by replacing all unpolarized
hard scattering cross sections and PDFs with their helicity-dependent counterparts.
The energy spectrum of circularly polarized photons is given by \cite{deFlorian:1999ge}
\begin{equation}
\label{eq:polww}
\Delta f_{e \gamma}(y) =\frac{\alpha_{em}}{2\pi} 
                \Bigg[
                \frac{1-(1-y)^2}{y} 
                \ln\frac{
                Q_{\max}^2(1-y)}{m_{e}^2 y^2}
+ 2m_{e}^2 y^2\left(\frac{1}{Q_{\max}^2}
                        -\frac{1-y}{m_{e}^2 y^2}\right)\Bigg]\;.
\end{equation}

The polarized beams available at an EIC 
offer unique opportunities for studying the spin structure of circularly
polarized photons in photoproduction processes.
Such measurements could yield also valuable, complementary information on the 
gluon helicity density of the proton as we shall demonstrate below.

To study the sensitivity of an EIC to the parton content of polarized photons,
which is completely unmeasured so far, we consider two extreme
models \cite{Stratmann:1996an} based on the current knowledge of the unpolarized $f^{\gamma}(x,\mu_0)$ \cite{Gluck:1991jc}
and the positivity constraint 
$|\Delta f^\gamma(x,\mu_0)|\leq f^\gamma(x,\mu_0)$.
In the ``minimal'' scenario we assume  $\Delta f^\gamma(x,\mu_0)=0$ at a scale
$\mu_0\simeq 1\,\mathrm{GeV}$ and
we saturate the bound in
the ``maximal'' scenario, i.e., $\Delta f^\gamma(x,\mu_0)= f^\gamma(x,\mu_0)$.
 
We present results of NLO calculations for single-inclusive jet photoproduction
at a c.m.s.\ energy of $\sqrt{s}=100\,\mathrm{GeV}$.
In order to compute the cross section for jet production,
an algorithm has to be specified describing the formation of jets by the 
final-state partons produced in the hard scattering.
A frequently adopted choice is to
define a jet as the deposition of the total transverse energy  
of all final-state partons that fulfill
$(\eta-\eta^i)^2+(\phi-\phi^i)^2\leq R^2$,
where $\eta^i$ and $\phi^i$ denote the pseudo-rapidities and azimuthal angles of
the particles and $R$ the jet cone aperture.
We work in the so-called ``small-cone approximation''
\cite{Sterman:1977wj,Furman:1981kf,Aversa:1988vb,Aversa:1989xw,Aversa:1990ww}
which can be considered as an expansion of the jet cross
section in terms of $R$ of the form $A\log{R} + B + {\cal{O}}(R^2)$.
Neglecting ${\cal{O}}(R^2)$ pieces, the evaluation and phase-space integration of
the partonic cross sections can be performed analytically.
This approximation has been shown~\cite{Aversa:1990ww,Aversa:1990uv,Guillet:1990ez,Jager:2004jh}
to account extremely well for jet observables up to cone sizes of 
about $R\approx 0.7$ in related $pp$-scattering reactions by explicit comparison to
calculations that take $R$ fully into account. 

\begin{figure}[th]
\begin{center}
\epsfig{figure=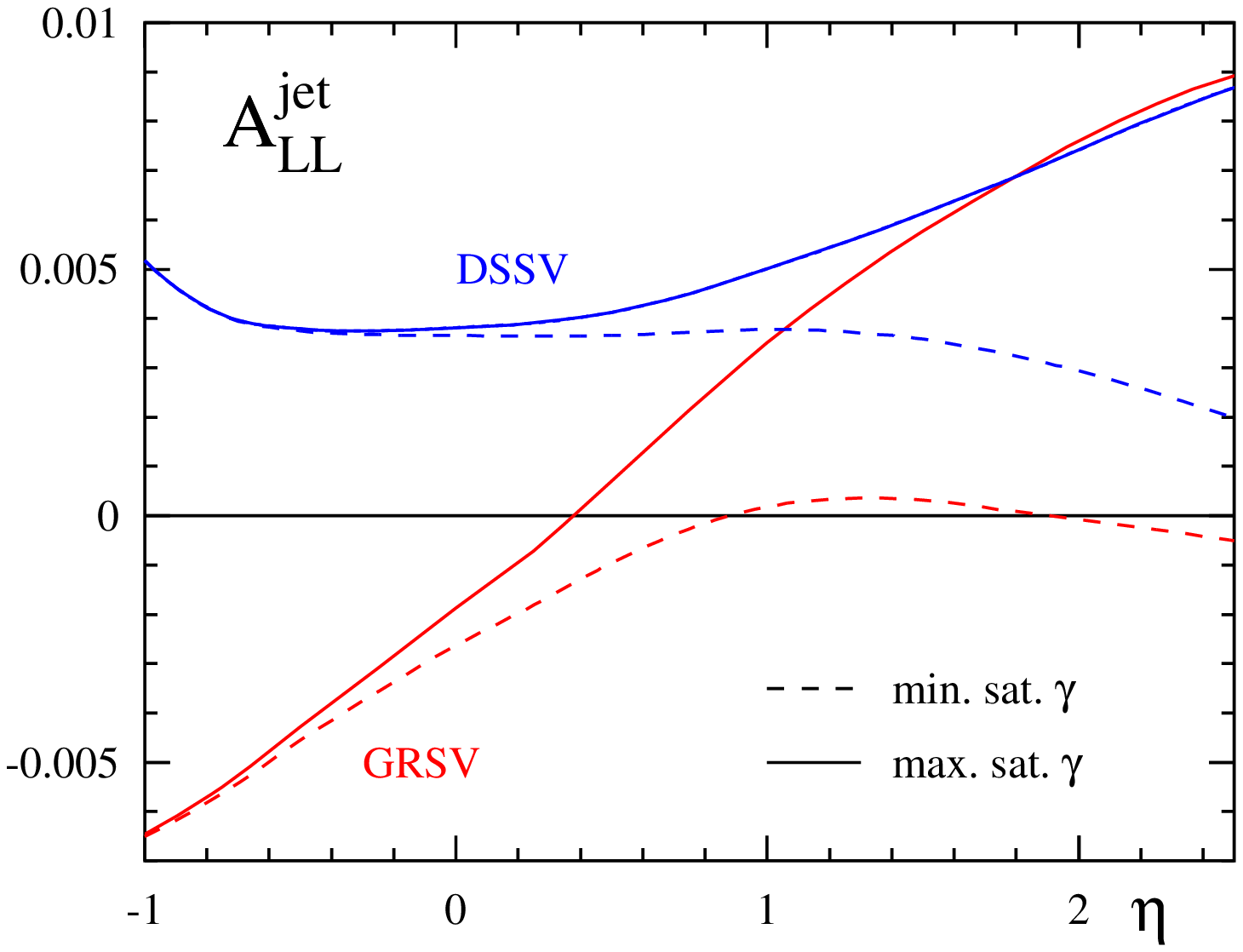,width=0.75\textwidth,clip=}
\end{center}
\vspace*{-0.7cm}
\caption{
\label{fig:erhic-asy}
Pseudo-rapidity dependence of the NLO QCD spin asymmetry for single-inclusive jet photoproduction 
at $\sqrt{S}=100$~GeV integrated over $p_T>4$~GeV for two different choices 
of proton helicity PDFs and two extreme sets of polarized photon densities. 
Taken from Ref.~\cite{Jager:2008qm}.}
\end{figure}
Figure~\ref{fig:erhic-asy} presents our results \cite{Jager:2008qm} for the expected NLO double-spin
asymmetry $A_{LL}^{jet}$ for single-inclusive jet photoproduction at $\sqrt{S}=100$~GeV
for two different choices of proton helicity densities \cite{Gluck:2000dy,deFlorian:2008mr}
and the two extreme sets of polarized photon densities introduced above.
In (\ref{eq:polww}) we chose
$Q_{\max}^2=1$~GeV$^2$ and the range of photon energies is limited to $0.2\leq y\leq 0.85$;
see also Sec.~\ref{sec:photoproduction}. The jet transverse momentum is integrated over for
$p_T>4$~GeV, and the factorization and renormalization scales are chosen to be $p_T$.

For single-inclusive observables, the rapidity-differential cross sections and the spin asymmetry are 
particularly interesting, since the relevant ranges of momentum fractions of the 
partons in the photon and the proton are related to the rapidity of the observed jet. 
As explained, e.g., in Ref.~\cite{Stratmann:1996xy}, if counting positive rapidity in the forward direction 
of the proton, large momentum fractions $x_\gamma\simeq 1$ are probed at large negative values 
of $\eta$. In this region, the direct contribution is expected to be largest 
and the photon structure is dominated by the purely
perturbative ``pointlike'' QED part \cite{Stratmann:1996an} which does not depend on the 
unknown non-perturbative input. As can be seen in Fig.~\ref{fig:erhic-asy},
measurements of $A_{LL}^{jet}$ for negative $\eta$ can provide valuable information
on the proton's spin structure, in particular, the gluon helicity density due to the 
dominance of gluon-induced processes.
On the other hand, at large positive rapidities, $A_{LL}^{jet}$ is particularly sensitive to the
parton content of the resolved photon, $x_{\gamma}\ll 1$, 
as is also exemplified in the figure. The size of $A_{LL}^{jet}$ increases if the lower cut
for the jet transverse momentum is raised to larger values. The range in $p_T$ where jets can
be reliably reconstructed at an EIC still needs to be investigated in detail.

If one has determined the proton helicity PDFs from
elsewhere, see Sec.~\ref{sec:rodolfo-spin}, the prospects for learning about the parton content of 
polarized photons are excellent. We note that the latter may become relevant in estimates
of photon induced cross sections at a future linear collider if the lepton beams will be
longitudinally polarized. Resolved photon contributions also complicate current extractions
of $\Delta g(x,\mu)$ in polarized-lepton nucleon scattering experiments at fixed-target energies
\cite{Jager:2005uf,Ageev:2005pq}.
We have estimated the expected size of statistical uncertainties
in case of the related single-inclusive pion photoproduction at an EIC
in Ref.~\cite{Jager:2003vy}. Measurements appear to be feasible already with very
moderate integrated luminosities of a few $\mathrm{fb}^{-1}$ thanks to the sizable 
cross sections for small $Q^2$. 

We note that other promising observables, like di-jet production where one has a better
control of the range of $x_{\gamma}$ probed, see Sec.~\ref{sec:photoproduction}, or
heavy quark production still need to be studied. Some theoretical results and simulations,
mainly for HERA energies, can be found in Refs.~\cite{Stratmann:1996xy,Butterworth:1997as,deFlorian:1999ge}.






\chapter[Three-dimensional structure of the proton and nuclei: transverse
\\ momentum]{\huge Three-dimensional structure of the proton and nuclei:
  transverse momentum \label{chap:tmd}}

\noindent
{\Large Convenors and chapter editors: \\[1em]
D. Hasch, F. Yuan}

\newpage

\section{Introduction and chapter summary}
\label{Sec-I:general-introduction}

\hspace{\parindent}\parbox{0.92\textwidth}{\slshape
 Mauro Anselmino, Andreas Metz, Peter Schweitzer}

\index{Anselmino, Mauro} 
\index{Metz, Andreas} 
\index{Schweitzer, Peter}

\vspace{\baselineskip}

%

The exploration of the internal structure of the nucleon in terms
of quarks and gluons, the fundamental degrees of freedom of
Quantum Chromodynamics (QCD), has been and still is at the
frontier of hadronic high energy physics research. After four
decades of Deep Inelastic Scattering (DIS) experiments of high
energy leptons off nucleons, our knowledge of the nucleon
structure has made impressive progress. 
To leading order in the electromagnetic coupling constant $\alpha_{\mbox{\tiny
QED}}\sim \frac{1}{137}$ the lepton with initial momentum $l$
interacts via one photon exchange with the quarks inside the
nucleon. By observing the momentum $l^\prime$ of the lepton in the
final state one obtains information about the quark and gluon content 
of the nucleon.

This information is encoded in the Parton Distribution Function
(PDF) $f_1^a(x,Q^2)$ where $x=Q^2/(2P\cdot q)$ is the fraction of
the nucleon momentum $P$ which is carried by the parton with
$Q^2=-q^2$ and $q=l-l^\prime$. This PDF can be interpreted as the
number density of partons of type $q$ inside the nucleon, carrying
a momentum fraction $x$. Similar information  has been obtained
about the number density of longitudinally polarized partons
inside longitudinally polarized nucleons, the helicity
distribution $g_1^a(x,Q^2)$. The successful prediction of the
scale ($Q^2$) dependence of the PDFs is one of the great triumphs
of QCD.

However consolidated our understanding of the nucleon structure
from DIS experiments is, it is basically one-dimensional. From DIS
we `only' learn about the longitudinal motion of partons in a fast
moving nucleon or, which is equivalent, about their momentum
distributions along the light-cone direction singled out by the
hard momentum flow in the process (i.e., in DIS, of the virtual
photon). In DIS the nucleon is seen as a bunch of fast-moving quarks, 
antiquarks and gluons, whose transverse momenta are not resolved.
A fast moving nucleon is Lorentz-contracted but its transverse size is still about
$1\,\fm$, which is a large distance on the strong interaction
scale.

It makes therefore sense to ask questions like: how are quarks
spatially distributed inside the nucleon? How do they move in the
transverse plane? Do they orbit, and carry orbital angular
momentum? Is there a correlation between orbital motion of quarks,
their spin and the spin of the nucleon? How can we access
information on such spin-orbit correlations, and what will this
tell us about the nucleon? Recent theoretical progress has put
many of these questions on a firm field-theoretical basis. We do
not know all answers, yet, but we have now a much better idea on
how to get them. The past decade has also witnessed tremendous
experimental achievements which lead to fascinating new
phenomenological insights into the structure of the nucleon.

The above questions address two complementary aspects of the
nucleon structure: the description of quarks in the transverse
plane in momentum space and in coordinate space. The
field-theoretical tools adequate to describe the former are the
Transverse Momentum Dependent Parton Distribution Functions (TMD
PDFs, or, shortly, TMDs). The field-theoretical objects tailored
to describe the spatial distributions of quarks in the transverse
plane are the Generalized Parton Distributions (GPDs), which are 
discussed in chapter~\ref{sec:imaging_executive_summary}. 
The focus of this chapter is on the TMDs, their theoretical properties and
phenomenological implications.

Several fascinating topics are related to the study of
TMDs:
\begin{itemize}

\item {\it 3D-imaging.} The TMDs depend on the intrinsic motion of
partons inside the nucleon and allow the reconstruction of the
nucleon structure in momentum space. Such an information, when
combined with the analogous information on the parton spatial
distribution from GPDs, leads to a complete 3-dimensional imaging
of the nucleon.

\item {\it Orbital motion.} Most TMDs would vanish in the absence
of parton orbital angular momentum. The possibility of learning
about the orbital motion of quarks inside a nucleon emerges from
the study of TMDs.

\item {\it Spin-orbit correlations.} Most TMDs and related,
observable, azimuthal asymmetries, are due to couplings of the
transverse momentum of quarks with the nucleon (or the quark)
spin. Spin-orbit correlations, similar to those in hydrogen atoms,
can therefore be studied.

\item {\it QCD gauge invariance and universality.} The origin of
some TMDs and the related spin asymmetries, when considered at
partonic level, reveal fundamental properties of QCD, mainly its
color gauge invariance. This interpretation leads to expect some
clear differences, between TMDs, in different processes
(universality breaking). A test of such ideas is crucial for our
understanding of QCD at work.

\end{itemize}

\subsection{~\label{sec:what-are-TMDs} What are TMDs?}

The `simplest' TMD is the unpolarized function $f_1^q(x,k_\perp)$
which describes, in a fast moving nucleon, the probability to find
a quark carrying the longitudinal momentum fraction $x$ of the
nucleon momentum, and a transverse momentum
$k_\perp=|\mbox{\boldmath $k$}_\perp|$. It is formally related to
the collinear (`integrated') PDF by $\int{\rm d}^2 \mbox{\boldmath
$k$}_\perp \, f_1^q(x,k_\perp)=f_1^q(x)$ (notice that, for
brevity, the dependence of TMDs and PDFs on auxiliary scales is
often not indicated).

This and other quark TMDs are defined in terms of the unintegrated
quark-quark correlator~\cite{Collins:1981uk,Collins:1981uw}
\begin{equation} \label{eq:corr}
  \Phi_{ij}^q(x,\mbox{\boldmath $k$}_\perp,\mbox{\boldmath $S$})_\eta
  = \int \frac{dz^-d^2z_\perp}{(2 \pi)^3}\,
  \mathrm{e}^{ik\cdot z}\langle \mbox{\boldmath $P$},\mbox{\boldmath $S$}\,|
  \,\bar{\psi}_j^q(0) \, \mathcal{W}_\eta(0,z) \, \psi_i^q(z)\,
  |\mbox{\boldmath $P$},\mbox{\boldmath $S$}\rangle \Big|_{z^+=0}\>,
\end{equation}
in which the gauge link operator $\mathcal{W}_\eta(0,z)$ ensures
the color gauge invariance of the matrix element.
$\mathcal{W}_\eta(0,z)$ depends on a path. Factorization theorems
give the prescription along which path the positions $0$ and $z$
of the quark fields have to be connected, and the index $\eta$
indicates that strictly speaking $\mathcal{W}_\eta(0,z)$ depends
on the process, as it will be further discussed. The light-cone
coordinates are defined as $a^\mu=(a^-,a^+,\mbox{\boldmath
$a$}_\perp)$ with $a^\pm=\frac{1}{\sqrt{2}}(a^0\pm a^3)$ and
$\mbox{\boldmath $a$}_\perp=(a^1,a^2)$.

The power and rich possibilities of the TMD approach arise from
the simple fact that $\mbox{\boldmath $k$}_{\perp}$ is a vector,
which allows various correlations with the other vectors involved: 
the nucleon momentum \mbox{\boldmath $P$} and the nucleon
spin \mbox{\boldmath $S$}. A systematic description of the
information content of the correlator was initiated in
\cite{Kotzinian:1994dv,Mulders:1995dh,Boer:1997nt}. Of particular
importance are `leading-twist' TMDs, i.e.\ TMDs which enter in
observables without power suppression. In this context, a TMD or
observable is said to be twist-t if its contribution to a cross
section is suppressed by the factor $(M/Q)^{t-2}$
\cite{Jaffe:1996zw} in addition to kinematic overall factors
($M$ represents a generic hadronic scale including the transverse
momentum.).

The leading-twist TMDs are associated with the large $+$ component
of the nucleon momentum (in a frame where the nucleon moves fast).
For a spin $\frac12$ particle like the nucleon there are 8
leading-twist TMDs, namely (we suppress the
$\eta$ process dependence label)
\begin{eqnarray}
  \frac{1}{2}\,\mathrm{tr}\left[
  \gamma^{+}\,\Phi^q(x,\mbox{\boldmath $k$}_\perp,
  \mbox{\boldmath $S$}) \,\right]
  &=& f_{1}^q(x,k_\perp)-\frac{\varepsilon ^{jk}\,k_\perp^{j}\,S_T^{k}}{M}
      \,f_{1T}^{\perp q}(x,k_\perp),  \label{Eq:f1-f1Tperp} \\
  \frac{1}{2}\,\mathrm{tr}\left[
  \gamma^{+}\gamma_{5}\,\Phi^q(x,\mbox{\boldmath $k$}_\perp,
  \mbox{\boldmath $S$})\,\right]
  &=& S_{L}\,g_{1L}^q(x,k_\perp)+\frac{\mbox{\boldmath $k$}_\perp\cdot\mbox{\boldmath $S$}_{T}}{M}
      g_{1T}^q(x,k_\perp), \label{Eq:g1-g1Tperp} \\
  \hspace{-8mm}
  \frac{1}{2}\,\mathrm{tr}\left[ i\sigma ^{j+}\gamma _{5} \,
  \Phi^q (x,\mbox{\boldmath $k$}_\perp,\mbox{\boldmath $S$})\right]
  &=& S_{T}^{j}\,h_1^q(x,k_\perp)
   +S_{L}\,\frac{k_\perp^{j}}{M}\,h_{1L}^{\perp q}(x,k_\perp) \nonumber\\
   &+&\label{Eq:h1-h1Lperp-h1Tperp}
   \frac{(k_\perp^{j}\,k_\perp^{k}-\frac{1}{2}\,\mbox{\boldmath $k$}_\perp^{\:2}\,
     \delta^{jk})S_{T}^{k}}{M^{2}}\,h_{1T}^{\perp q}(x,k_\perp)
   +\frac{\varepsilon^{jk}\,k_\perp^{k}}{M}\,h_1^{\perp q}(x,k_\perp).
\end{eqnarray}
Dirac structures other than those above yield higher twist TMDs
\cite{Goeke:2005hb,Bacchetta:2006tn}. TMDs of antiquarks and
gluons are defined similarly in terms of correlators analogous to
(\ref{eq:corr}). The notation used in
Eqs.~(\ref{Eq:f1-f1Tperp})--(\ref{Eq:h1-h1Lperp-h1Tperp}) follows
\cite{Mulders:1995dh,Boer:1997nt,Jaffe:1996zw}, where the common
subscript $1$ is used to indicate twist-2 TMDs. (Notice that in
the TMD literature also a different notation is often used, in
which, for instance, $\Delta^N
f_{q/p^\uparrow}(x,k_\perp)=-(2k_\perp/M)\,f_{1T}^{\perp
q}(x,k_\perp)$. We refer to \cite{Anselmino:2005sh} for an
overview.)

The leading twist TMDs
(\ref{Eq:f1-f1Tperp}--\ref{Eq:h1-h1Lperp-h1Tperp}) have partonic
interpretations. The gamma-structures signal the quark
polarizations. $\gamma^+$ describes unpolarized quarks, thus
Eq.~(\ref{Eq:f1-f1Tperp}) gives the number density of unpolarized
quarks inside an unpolarized (first term) or transversely
polarized (second term) proton. $\gamma^+\gamma_5$, which appears
in Eq.~(\ref{Eq:g1-g1Tperp}), singles out longitudinally polarized
quarks, either in a longitudinally (first term) or transversely
polarized (second term) proton. Finally, in
Eq.~(\ref{Eq:h1-h1Lperp-h1Tperp}), the gamma-factor
$i\,\sigma^{+j}\gamma_5$ selects transversely polarized quarks
inside transversely polarized (first and third terms),
longitudinally polarized (second term) or unpolarized (fourth
term) protons.


\vskip 6pt
\subsection {Partonic interpretation and properties of the TMDs}
\vskip 6pt

As they are the central focus of interest in this Chapter, let us
further elaborate on the leading order TMDs and their partonic
interpretation. We also introduce the Transverse Momentum
Dependent Fragmentation Functions (TMD FFs). The TMDs contain
information on the longitudinal and transverse (or intrinsic)
motion of quarks and gluons inside a fast moving nucleon. When
adding the spin degree of freedom they link the parton spin (say a
quark, $\mbox{\boldmath $s$}_q$) to the parent proton spin
($\mbox{\boldmath $S$}$) and to the intrinsic motion
($\mbox{\boldmath $k$}_\perp$). The correlator (\ref{eq:corr})
restricted to leading twist defines the most general spin
dependent TMD, which we denote by $f_1^q(x, \mbox{\boldmath
$k$}_\perp; \mbox{\boldmath $s$}_q, \mbox{\boldmath $S$})$, and
may depend on all possible combinations of the pseudo-vectors
$\mbox{\boldmath $s$}_q, \mbox{\boldmath $S$}$  and the vectors
$\mbox{\boldmath $k$}_\perp, \mbox{\boldmath $P$}$ which are
allowed by parity invariance. At leading order in $1/Q$, there are
eight such combinations, leading to the eight independent TMDs in
Eqs.~(\ref{Eq:f1-f1Tperp}--\ref{Eq:h1-h1Lperp-h1Tperp}).

A similar correlation between spin and transverse motion can occur
in the fragmentation process of a transversely polarized quark,
with spin vector $\mbox{\boldmath $s$}_q$ and three-momentum
$\mbox{\boldmath $k$}_q$, into a hadron with longitudinal momentum
fraction $z$ and transverse momentum $\mbox{\boldmath $P$}_\perp$
(with respect to the quark direction); such a mechanism is called
the Collins effect \cite{Collins:1992kk} and appears in the
fragmentation function via a $\mbox{\boldmath $s$}_q \cdot
(\mbox{\boldmath $k$}_q \times \mbox{\boldmath $P$}_\perp)$ term.
For a quark fragmentation into a spinless hadron there are two
independent leading-twist transverse momentum dependent
fragmentation functions.

We briefly list here the eight leading-twist Transverse Momentum
Dependent Partonic Distributions of a proton and the two
Fragmentation Functions (for a final spinless hadron), which are
the main objects in our investigation of the nucleon momentum
structure.
%
\begin{itemize}
\item
$f_1^a(x, k_\perp)$ 
is the unpolarized, $k_\perp$ dependent distribution of parton $a$
inside a proton. Its integrated version is the usual PDF measured
in DIS. Common notations are $q(x) = \int \!\! d^2\mbox{\boldmath
$k$}_\perp \> f_1^q(x, k_\perp)$, and $g(x) = \int \!\!
d^2\mbox{\boldmath $k$}_\perp \> f_1^g(x, k_\perp) $ for quarks of
flavor $q$ and gluons respectively.

Most experimental and theoretical efforts have so far been
dedicated to $q(x,Q^2)$ and $g(x,Q^2)$; these are by now the best
known partonic distributions, and the comparison of the predicted
$Q^2$ dependence with data has been a great success for
perturbative QCD.

\item $g_{1L}^a(x, k_\perp)$ (or simply $g_1^a$) is the
unintegrated helicity distribution: the difference between the
number density of partons $a$ with the same and opposite helicity
of the parent proton. Common notations for the integrated helicity
distributions are
$\Delta q(x) = \int \!\! d^2\mbox{\boldmath $k$}_\perp \>
g_{1L}^q(x, k_\perp)$ for quarks and similarly $\Delta g(x)$
for gluons. 
See the relevant discussions in section~\ref{sec:polstatus}.

The $\Delta q(x)$'s are not so well known as the corresponding
$q(x)$, as they require polarized DIS, but have been measured by
several experiments. The least known of the helicity distributions
is the gluon one, $\Delta g(x)$, despite some attempts to measure
it.

\item $h_1^q(x, k_\perp)$ is the analogue of the helicity
distribution, for transverse nucleon spin, {\it i.e.} the
transversity distribution. The integrated version has several
notations in the literature
$\Delta _\perp q(x) =  h_1^q(x) = \int \!d^2\mbox{\boldmath
$k$}_\perp \, h_{1}^q(x, k_\perp)$
for quarks of flavor $q$. There is no transversity distribution
for gluons in a spin $\frac12$ hadron.

The unpolarized, the helicity and the transversity distributions
are the only three independent PDFs which survive in the collinear
limit, $\mbox{\boldmath $k$}_\perp = 0$. The transversity
distribution is chiral-odd and needs to be coupled to another
chiral-odd quantity to be observed. So far only one extraction of
the $u$ and $d$ quark transversities is available in the
literature \cite{Anselmino:2007fs}, obtained by a combined fit of
SIDIS and $e^+e^-$ data.

A good knowledge of the transversity distributions for quarks and
antiquarks would allow computation of the tensor charge, given by
$ \int_0^1 \! dx \> [h_1^q(x) - h_1^{\bar q}(x)] $,
a non perturbative quantity for which lattice and model
computations exist.

\item $f_{1T}^{\perp a}(x, k_\perp)$ is the Sivers function
\cite{Sivers:1989cc}, appearing in the distribution of unpolarized
partons $a$ inside a polarized proton. It links the parton
intrinsic motion to the proton spin:
\begin{equation}
f_{1}^a(x, \mbox{\boldmath $k$}_\perp; \mbox{\boldmath $S$}) =
f_1^{a}(x, k_\perp) - \frac{k_\perp}{M} \, f_{1T}^{\perp a}(x,
k_\perp) \> \mbox{\boldmath $S$} \cdot  (\hat{\mbox{\boldmath
$P$}} \times \hat{\mbox{\boldmath $k$}}_\perp) \> .
\end{equation}
The Sivers function offers new information and plays a crucial
role in our understanding of the nucleon structure. Its
observation, already confirmed, is a clear indication of parton
orbital motion; the opposite values for $u$ and $d$ quarks is
argued to be linked to the nucleons' anomalous magnetic moments; its
very origin and expected process dependence are related to
fundamental QCD effects. Due to its importance the Sivers TMD for
quarks will be discussed at length in Sec.~\ref{sec:Sivers-example} and for gluons in
Sec.~\ref{sec:TMD-gluons}. Theoretical issues concerning $f_{1T}^{\perp a}$, its
origin and relation with basic QCD properties like the color gauge
links and color gauge invariance will be treated in Sec.~\ref{sec:TMD-theory}.

\item $h_{1}^{\perp q}(x, k_\perp)$ is the Boer-Mulders function
\cite{Boer:1997nt}, appearing in the distribution of polarized
quarks $q$ inside an unpolarized proton:
\begin{eqnarray}
f_1^{q}(x, \mbox{\boldmath $k$}_\perp; \mbox{\boldmath $s$}_q)
\!\! &=& \!\! \frac 12 \, f_1^{q}(x, k_\perp) - \frac{k_\perp}{2M}
\, h_{1}^{\perp q}(x, k_\perp) \> \mbox{\boldmath $s$}_q \cdot
(\hat{\mbox{\boldmath $P$}} \times \hat{\mbox{\boldmath
$k$}}_\perp) \;.\end{eqnarray}
This function has the striking peculiarity that it might give
unexpected spin effects even in unpolarized processes, as it
singles out polarized quarks from unpolarized protons and
neutrons. It will be discussed in Sec.~\ref{secV:chiralodd}.

\item The remaining three TMDs, $g_{1T}^{a}(x, k_\perp),
h_{1L}^{\perp q}(x, k_\perp)$ and $h_{1T}^{\perp q}(x, k_\perp)$
are related to double spin correlations in the PDFs; respectively,
the amount of longitudinally polarized partons in a transversely
polarized proton, of transversely polarized quarks in a
longitudinally polarized proton, and of transversely polarized
quarks in a transversely (but in a different direction) polarized
proton. Neglecting higher-twist terms, some approximate
relationships with the other TMDs can be obtained
\cite{Avakian:2007mv}. They will briefly be discussed in Sec.~\ref{Sec-IV:overview-on-TMDs}.

\item
$D_1^a(z, P_\perp)$ (also denoted as $D_{h/a})$ 
is the unpolarized, $P_\perp$ dependent, parton $a$ fragmentation
function (into a hadron $h$). Its integrated version
$
D_{1h}^a(z) = \int \! d^2\mbox{\boldmath $P$}_\perp \, D_1^a(z,
P_\perp) $ is the usual FF.

\item $H_1^{\perp q}(z, P_\perp)$ is the Collins function
\cite{Collins:1992kk}, describing the fragmentation of a polarized
quark into a spinless (or unpolarized) hadron:
\begin{eqnarray}
D_1^q(z, \mbox{\boldmath $P$}_\perp; \mbox{\boldmath $s$}_q) \!\!
&=&\!\! D_1^q(z, P_\perp) + \frac{P_\perp}{zM_h} \, H_{1}^{\perp
q}(z, P_\perp) \> \mbox{\boldmath $s$}_q \cdot
(\hat{\mbox{\boldmath $p$}}_q \times \hat{\mbox{\boldmath
$P$}}_\perp) \;.
\end{eqnarray}

The Collins effect has been observed by several experiments and is
well established. It is considered as a universal property of the
quark hadronization process and it plays a crucial role in many
spin effects. Its chiral-odd nature makes it the ideal partner to
access chiral-odd TMDs like the transversity distribution and the
Boer-Mulders function. All these will be discussed in Sec.~\ref{secV:chiralodd}.

\end{itemize}

\subsection{\label{secI:info-about-TMDs}
How do we obtain information on TMDs?}

Our guiding experiments involve again lepton-nucleon scattering at
high energy, with the difference, with respect to the usual DIS,
that one observes in the final state a hadron in addition to the
scattered lepton, $\ell(l) + N(P) \to \ell(l') + h(P_h) + X$, the
so-called Semi-Inclusive Deep-Inelastic Scattering (SIDIS). In
this case the hadron, which results from the fragmentation of a
scattered quark, `remembers' the original motion of the quark,
including the transverse one, and offers new information.
\begin{figure}[t]
\begin{center}
\includegraphics[width=8.0cm]{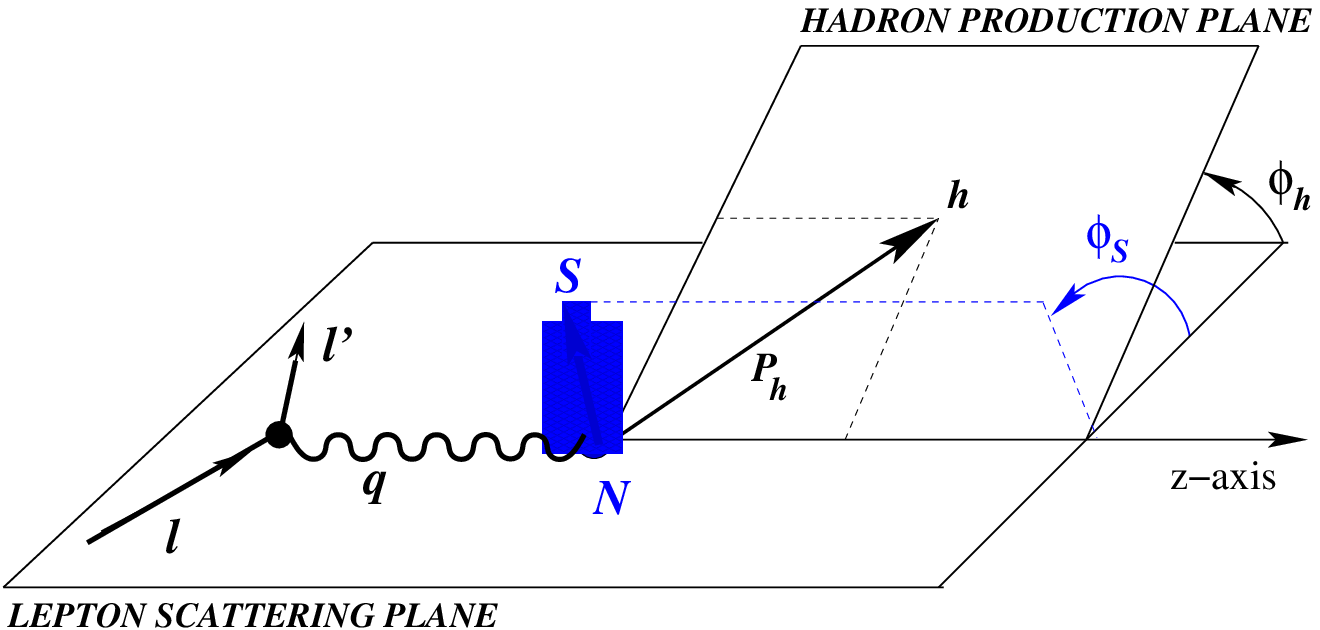}
\caption{ \label{f:anglestrento} Illustration of the kinematics,
especially the azimuthal angles, for SIDIS in the target rest
frame~\cite{Bacchetta:2004jz}. $\mbox{\boldmath $P$}_{hT}$ and
$\mbox{\boldmath $S$}_T$ are the transverse parts of
$\mbox{\boldmath $P$}_h$ and $\mbox{\boldmath $S$}$ with respect
to the virtual photon momentum $\mbox{\boldmath $q$}=\mbox{\boldmath
$l$}-\mbox{\boldmath $l^\prime$}$.}
\end{center}
\end{figure}

In general, SIDIS depends on six kinematic variables. In
addition to the variables for inclusive DIS, $x$, $y = (P \cdot q)
/ (P \cdot l)$, and the azimuthal angle $\phi_S$ describing the
orientation of the target spin vector for transverse polarization,
one has three variables for the final state hadron, which we
denote by $z = (P \cdot P_h) / (P \cdot q)$ (longitudinal hadron
momentum), $P_{hT}$ (magnitude of transverse hadron momentum), and
the angle $\phi_h$ for the orientation of $\mbox{\boldmath
$P$}_{hT}$ (see also Fig.~\ref{f:anglestrento}). In the one-photon
exchange approximation, the SIDIS cross section can be decomposed
in terms of structure
functions~\cite{Kotzinian:1994dv,Bacchetta:2006tn,Diehl:2005pc,Anselmino:2011ch} where, largely following the notation
of~\cite{Bacchetta:2006tn}, one has
\begin{align}
\frac{d\sigma}{dx_B \, dy\, d\phi_S \,dz_h\, d\phi_h\, d P_{hT}^2}
\propto & \Big\{ F_{UU ,T} + \varepsilon \, \cos(2\phi_h) \,
F_{UU}^{\cos 2\phi_h} \nonumber \\ & + S_\parallel \, \varepsilon
\, \sin(2\phi_h) \, F_{UL}^{\sin 2\phi_h} + S_\parallel \,
\lambda_\ell \, \sqrt{1-\varepsilon^2}\, F_{LL} \phantom{\Big[}
\nonumber \\  & + |\mbox{\boldmath $S$}_\perp| \, \Big[
  \sin(\phi_h-\phi_S)\,
F_{UT,T}^{\sin\left(\phi_h -\phi_S\right)} + \varepsilon \,
\sin(\phi_h+\phi_S) \, F_{UT}^{\sin\left(\phi_h +\phi_S\right)}
\nonumber \\ & \hspace{1.3cm} + \varepsilon \,
\sin(3\phi_h-\phi_S) \, F_{UT}^{\sin\left(3\phi_h -\phi_S\right)}
\Big] \nonumber \\  & + |\mbox{\boldmath $S$}_\perp| \, \lambda_e
\,
  \sqrt{1-\varepsilon^2} \, \cos(\phi_h-\phi_S) \,
F_{LT}^{\cos(\phi_h -\phi_S)} + \ldots \Big\} \,.
\label{e:crossmaster}
\end{align}
In Eq.~(\ref{e:crossmaster}), $\varepsilon$ is the degree of
longitudinal polarization of the virtual photon which can be
expressed through $y$~\cite{Bacchetta:2006tn},
$S_\parallel$ denotes longitudinal target polarization, and
$\lambda_e$ is the lepton helicity. The structure functions
$F_{XY}$ ($X$ and $Y$ refer to the lepton and the nucleon,
respectively: $U$ = unpolarized; $L, T$ = longitudinally,
transversely polarized) merely depend on $x$, $z$, and $P_{hT}$. 
The third subscript $F_{XY,T}$ specifies the polarization of the
virtual photon.
By choosing specific polarization states and weighting with the
appropriate azimuthal dependence, one can extract each structure
function in~(\ref{e:crossmaster}) as pioneering experiments have already
unambiguously shown.

For TMD studies one is interested in the kinematic region
defined by
\begin{equation}
P_{hT} \simeq \Lambda_{\rm QCD} \ll Q \,,
\end{equation}
for which the structure functions can be written as certain
convolutions of TMDs. In this region, the components in
Eq.~(\ref{e:crossmaster}) appear at leading order when expanding
the cross section in powers of $1/Q$, while additional ones show
up at sub-leading order~\cite{Kotzinian:1994dv,Bacchetta:2006tn,Diehl:2005pc,Anselmino:2011ch}. 
Measuring the structure
functions in Eq.~(\ref{e:crossmaster}) allows one to obtain
information on all eight leading quark TMDs. To be specific, one
has (for a spinless final state
hadron)~\cite{Bacchetta:2006tn,Anselmino:2011ch},
\begin{eqnarray} \label{eq:sf_tmd}
&&F_{UU} \sim \sum_q e_q^2 \> f_1^q \otimes D_1^q \hskip 54pt
F_{LT}^{\cos(\phi - \phi_S)} \sim \sum_q e_q^2 \> g_{1T}^{q}
\otimes D_1^q
\label{eq:sf_tmd1} \\
&&F_{LL} \sim \sum_q e_q^2 \> g_{1L}^q \otimes D_1^q \hskip 50pt
F_{UT}^{\sin(\phi - \phi_S)} \sim \sum_q e_q^2 \> f_{1T}^{\perp q}
\otimes D_1^q
\label{eq:sf_tmd12} \\
&&F_{UU}^{\cos(2\phi)} \sim \sum_q e_q^2 \> h_{1}^{\perp q}
\otimes H_1^{\perp q} \hskip 24pt F_{UT}^{\sin(\phi + \phi_S)}
\sim \sum_q e_q^2 \> h_{1T}^{q} \otimes H_1^{\perp q}
\label{eq:sf_tmd3} \\
&&F_{UL}^{\sin(2\phi)} \sim \sum_q e_q^2 \> h_{1L}^{\perp q}
\otimes H_1^{\perp q} \hskip 24pt F_{UT}^{\sin(3\phi - \phi_S)}
\sim \sum_q e_q^2 \> h_{1T}^{\perp q} \otimes H_1^{\perp q} \,,
\label{eq:sf_tmd4}
\end{eqnarray}
where $e_q$ is the charge of the struck quark in units of the
elementary charge. Notice that the four chiral-even TMDs couple to
the well known unpolarized fragmentation function $D_1$, while the
chiral-odd TMDs couple to the (chiral-odd) Collins function
$H_1^\perp$. In the subsequent sections the major focus will be on
$F_{UT}^{\sin(\phi - \phi_S)}$ containing the Sivers function.

The factorized expressions for the structure functions in
Eqs.~(\ref{eq:sf_tmd1})-(\ref{eq:sf_tmd4}) hold in this form in
the parton model approximation. If loop corrections are included,
one not only obtains a nontrivial higher order term describing the
hard scattering part of the process but also a leading-twist
contribution arising from soft gluon emission (soft
factor)~\cite{Collins:1981uk,Ji:2004wu,Ji:2004xq,Collins:2004nx,Collins:2008ht,Aybat:2011zv}. 
In the case of inclusive DIS such soft gluon
effects cancel between real and virtual radiative corrections, but
they survive in the SIDIS cross section for $P_{hT} \simeq
\Lambda_{\rm{QCD}}$. While the hard coefficient enters the
structure functions in a simple multiplicative way, the soft
factor gets convoluted with the parton distributions and the
fragmentation functions.
The presence of uncanceled soft gluon emission also requires to
somewhat generalize the field-theoretical definition of TMDs given
above.
More details about this point will
be presented in Sec.~\ref{sec:TMD-theory}.

Almost all existing analyses of TMD-observables are based on the
parton model approximation. This is sufficient for getting a good
first idea about the general features of the TMDs and also at the
present stage of the data, which often are plagued by considerable
uncertainties. However, precision studies will be necessary to reveal 
features of QCD dynamics. The parton model approach will then be 
no longer
appropriate, and one will
have to deal with soft gluon effects, especially when high quality
data from the EIC become available that will cover a large
kinematic range.

\subsection{\label{secI:theory}
Gauge invariance, universality, and beyond}

Local gauge invariance is the underlying principle of the Standard
Model of Particle Physics. In the case of QCD it is the SU(3)
gauge invariance associated with the color degree of freedom of
the quarks which matters. This color gauge invariance plays a
particularly crucial role for TMDs. Here a brief introduction to
this topic is given, while especially in Sec.~\ref{sec:TMD-theory} more details
about this very active and fascinating field can be found.

As discussed in Sec.~\ref{sec:what-are-TMDs}, in order to have a gauge invariant
definition of TMDs a gauge link (Wilson line) has to be inserted
between the two quark fields showing up in the correlator in
Eq.~(\ref{eq:corr}). This is not specific for TMDs but applies
also to, e.g., ordinary PDFs. However, two features {\it are
unique} in the case of TMDs: first, certain TMDs are non-zero only
if the Wilson line is taken into
account~\cite{Brodsky:2002cx,Collins:2002kn,Belitsky:2002sm,Boer:2003cm}.
Second, the Wilson line depends on the process, which leads to a
nontrivial universality behavior of TMDs~\cite{Collins:2002kn}.

The mere existence of two TMDs depends on the presence of the
Wilson line --- the Sivers function $f_{1T}^\perp$ and the
Boer-Mulders function $h_1^\perp$. They are also denoted as naive
time-reversal odd (T-odd) functions. (This term is not related to
real violation of T-invariance but, roughly speaking, is
associated with a nontrivial phase at the amplitude level of a
process.)

The Wilson line is automatically generated when carrying out
factorization. In the case of SIDIS, it arises due to the exchange
of (infinitely many) gluons between the active struck quark and
the remnants of the target. Since in DIS these exchanges happen
{\it after} the virtual photon strikes the quark one also talks
about final state interactions (FSI). On the other hand, for the
Drell-Yan process, there exist corresponding gluon exchanges {\it
before} the photon-quark interaction, which we call initial state
interactions (ISI). As a consequence, the Wilson-lines for the two
processes are running along different paths. This in turn
endangers the universality (process-independence) of TMDs, which
is a crucial prerequisite for factorization being of any
practical use.

Although the paths of the Wilson lines are different, the TMDs for
both processes can be related by using the parity and
time-reversal transformation~\cite{Collins:2002kn}. One finds that
the six T-even TMDs are actually universal, while the T-odd TMDs
are non-universal. However, this non-universality is well under
control and `merely' consists of a sign
change~\cite{Collins:2002kn},
\begin{equation} \label{eq:univ}
f_{1T}^\perp \big|_{\rm{DY}} = - f_{1T}^\perp \big|_{\rm{DIS}} \,,
\qquad \qquad h_{1}^\perp \big|_{\rm{DY}} = - h_{1}^\perp
\big|_{\rm{DIS}} \,.
\end{equation}
In other words, the predictive power of factorization is
maintained. The experimental check of this sign change is
currently one of the outstanding topics in hadronic physics.

We are now in a position to further motivate why the study of the
Sivers effect should play a central role in the EIC science case.
First, the Sivers function not only tells us something about the
three-dimensional structure of the nucleon, a particular spin-orbit correlation,
etc. Its physics is also intimately related to the gauge
invariance of QCD. Second, existing data for non-zero transverse
single-spin asymmetries in SIDIS and in proton-proton collisions
can be explained on the basis of the Sivers effect. In other words, 
the physics of FSI/ISI is the key to describing these asymmetries 
(which can be as large as 40\%) in QCD.
Third, according to our present knowledge, in SIDIS the Sivers function is easier to
measure than the Boer-Mulders function. Fourth, the check of the
predicted sign reversal in~(\ref{eq:univ}), strictly speaking, is
more direct for $f_{1T}^\perp$ than for the chiral-odd
$h_1^{\perp}$. In the latter case input from models is required.

Quite some progress was made in recent years to further elucidate
this physics associated with the underlying gauge structure of
QCD. In particular, for hadron-hadron collisions with hadronic
final states the presence of both ISI and FSI may unable any kind
of (standard)
TMD-factorization~\cite{Bomhof:2004aw,Bacchetta:2005rm,Bomhof:2006dp,Bomhof:2007xt,Qiu:2007ar,Collins:2007nk,Collins:2007jp,Rogers:2010dm}.
The consequences of a breakdown of TMD-factorization are
far-reaching. For instance, in such a case also the
so-called QCD resummation technique~\cite{Collins:1984kg}, which
is widely used whenever there is more than one physical momentum
scale in a process, becomes questionable. Moreover, if the sign
reversal of the Sivers function in Eq.~(\ref{eq:univ}) is not
confirmed by experiment, the general procedure of applying QCD to
hard scattering processes may have to be revisited. 
Further striking
developments in this rather new field can be expected, and only
the close interplay between lepton-nucleon scattering and hadronic
collisions will allow us to fully explore this physics, as is also
obvious from the relations~(\ref{eq:univ}).

\subsection{TMDs and orbital angular momentum}
\label{Sec:Intr-TMDS-OAM}

The helicity PDFs $g_1^a(x)$ are still not well known, especially
in the sea quark and gluon sector, but by now one fact seems
clear: the spin of quarks and gluons accounts only for a part of
the nucleon spin. A substantial fraction of the nucleon spin must
be due to orbital angular momentum (OAM). It is important to keep
in mind that in gauge theories there is no unique decomposition of
the nucleon spin into contributions due to the spin and OAM of
quarks and gluons \cite{Jaffe:1989jz,Ji:1996ek}. Nevertheless it
is possible \cite{Ji:1996ek,Hoodbhoy:1998yb} to learn about OAM
from GPDs which describe the dynamics of partons in the transverse
plane in {\sl position space}.

TMDs provide complementary information on the dynamics of partons
in the transverse plane in {\sl momentum space}, and one naturally
expects TMDs to teach us about parton OAM. That the OAM of partons
plays an important role is well known: in the light-cone wave
function of the nucleon components with OAM $L_z\neq 0$ must be
present in order to have a non-zero anomalous magnetic moment
\cite{Brodsky:1980zm,Burkardt:2005km}, and the situation is similar for several
other quantities \cite{Ji:2002xn}. Model calculations have also
shown that the leading twist TMDs $f_{1T}^{\perp q}$,  $g_{1T}^q$,
$h_1^{\perp q}$, $h_{1L}^{\perp q}$, $h_{1T}^{\perp q}$ and many
sub-leading twist TMDs would vanish without different
components in the nucleon wave function with $\Delta L_z\neq 0$.
But although OAM seems to play a crucial role also for many TMDs,
so far no rigorous connection between the OAM contribution of
partons and the nucleon spin could be established.


\subsection{Further important topics}
In this subsection some further important aspects about TMDs are
briefly discussed; more details will be presented in the other
Sections of this TMD Chapter.

\subsubsection{Models and lattice QCD}
Model calculations have had a particularly strong impact on the
TMD field. It suffices to recall the calculations in the
quark-diquark model \cite{Brodsky:2002cx} which helped to
establish the existence of the Sivers effect within QCD and the
TMD factorization framework \cite{Collins:2002kn}. Models may
allow to see more clearly the relevant aspects of TMDs which are
obscured in the much more complicated QCD dynamics. We encountered
one promising instance of that above, in
Sec.~\ref{Sec:Intr-TMDS-OAM}. Model results have, however, also
very practical applications. Nearly nothing is known about most of
the TMDs. Models provide information on the sign and magnitude of
TMDs, or possible (model) relations among different TMDs. This
information can be applied to make predictions for the planned
experiments, and in this way help to better explore the
opportunities of the available and planned facilities. The
importance of model studies is discussed in Sec.~\ref{sec:TMD-theory}.


Lattice QCD is in principle a powerful approach. 
What can be handled presently in lattice studies are calculations of 
the matrix element in the integrand of the correlator in Eq.~(\ref{eq:corr}),
i.e., TMDs in Fourier-space. 
Most readily accessible is information on $x$-integrated TMDs such as
$\int dx\ f_1^q(x,k_\perp)$~\cite{Hagler:2009mb,Musch:2010ka}. 
The caveat is that lattice results presently available have been obtained 
with a simplified gauge-link in the correlator~(\ref{eq:corr}). 
This simplified gauge-link differs from the link-geometry dictated by factorization 
in a particular scattering process.
Investigations with more realistic gauge-links are ongoing.

\subsubsection{Gluon TMDs}
In addition to the eight TMDs for quarks, there also exist eight
TMDs for
gluons~\cite{Anselmino:2005sh,Mulders:2000sh,Meissner:2007rx}. The
most prominent one is the unpolarized gluon TMD, which is a widely
used ingredient of many calculations in high-energy processes.
Because of the initial and final state interactions,
the universality of this object is nontrivial and
has attracted renewed interest lately~\cite{Dominguez:2010xd}.
Moreover, linearly polarized gluons for an unpolarized nucleon
can, in principle, be explored through, e.g., heavy quark pair production
in $\ell \, p$-collisions~\cite{Boer:2010zf}. A particularly
important role is played by the Sivers function for gluons, which
will be discussed in quite some detail in Sec.~\ref{sec:TMD-gluons}.
Experimentally, the sector of gluon TMDs is largely unexplored so
far, and the EIC could provide extremely valuable information in
this respect.

\subsubsection{Moments of TMDs}
Momentum moments of some of the TMDs are of particular interest
because of their relation to certain collinear 3-parton
correlators, which appear in the QCD-description of, e.g., SIDIS
structure functions at large $P_{hT} \simeq Q$ or weighted
asymmetries (see Sec.~\ref{sec:Sivers-example}). 
For instance, in the case of the
Sivers function one can consider the
moment~\cite{Boer:2003cm,Ma:2003ut}
\begin{equation} \label{eq:siv_mom}
f_{1T}^{\perp (1)}(x) \equiv \int d^2 \mbox{\boldmath $k$}_\perp
\, \frac{\mbox{\boldmath $k$}_\perp^{\,2}}{2 M^2} \,
f_{1T}^{\perp}(x,\mbox{\boldmath $k$}_\perp^{\,2}) = \pi \,
T_F(x,x) \,,
\end{equation}
where $T_F$ represents a quark-gluon-quark correlator.
These correlation functions were
also introduced in the literature to describe the single-spin
asymmetries in hard scattering processes in the collinear
factorization
framework~\cite{Efremov:1981sh,Efremov:1984ip,Qiu:1991pp,Qiu:1998ia}.
Equation~(\ref{eq:siv_mom}) is a model-independent result which
allows one to relate different observables. A corresponding
relation holds for the Boer-Mulders
function~\cite{Boer:2003cm,Ma:2003ut}. Also the moments
$g_{1T}^{(1)}$ and $h_{1L}^{\perp (1)}$ can be expressed through
collinear 3-parton correlators~\cite{Zhou:2008mz}.

\subsubsection{Integrated/weighted observables}
In Sec.~\ref{secI:info-about-TMDs} leading-twist soft gluon effects were mentioned.
Such effects can cancel if the components in
Eqs.~(\ref{e:crossmaster}) are integrated upon the transverse
momentum $P_{hT}$ of the hadron. For instance, a cancellation
occurs for the unpolarized structure function $F_{UU}$, and also
for the term associated with $F_{UT}^{\sin(\phi_h-\phi_S)}$ which
is related to the Sivers effect~\cite{Vogelsang:2009pj}. In the
latter case the integration needs to be done with a proper weight
factor (a more elaborate account on this topic will be given in
Sec.~\ref{sec:Sivers-example}). Such weighted observables are therefore rather
attractive from a theoretical point of view. They depend on
moments of the TMDs just discussed above and as such provide
additional complementary information. The EIC would be ideal for
seriously studying these interesting observables.

\subsubsection{Structure functions from low to high transverse momenta}

While at low $P_{hT}$ the SIDIS structure functions can be
described by means of TMD-factorization, for $P_{hT} \simeq Q$
collinear factorization is the appropriate framework. Recently, a
lot of progress has been made to understand the quantitative
relation between TMD-factorization on the one hand and collinear
factorization on the other in the region $\Lambda_{\rm{QCD}} \ll
P_{hT} \ll Q$ where both approaches
apply~\cite{Ji:2006ub,Ji:2006vf,Ji:2006br,Koike:2007dg,Bacchetta:2008xw,Yuan:2009dw}. An extended discussion of these
aspects, with a focus on the EIC, will also be given in Sec.~\ref{sec:Sivers-example}.

\subsubsection{Higher twist TMDs}
The focus of present research is on the leading-twist TMDs.
However, there is also a lot of important information encoded in
twist-3 TMDs, which contain detailed information on the
quark-gluon correlators. Experimentally, such twist-3 effects can
be explored by measuring sub-leading structure functions appearing
in the general decomposition of the SIDIS cross
section~(\ref{e:crossmaster})~\cite{Kotzinian:1994dv,Bacchetta:2006tn,Diehl:2005pc,Anselmino:2011ch}. 
In fact, the first clear
single-spin phenomena in SIDIS, which crucially vitalized the
field, were sub-leading twist observables. Although studied in
numerous works, these first data on single-spin asymmetries in
SIDIS remain basically unexplained. Some aspects of the
interesting topic of higher twist TMDs will be discussed in more
detail in Sec.~\ref{Sec-IV:overview-on-TMDs}.

\subsection{TMDs and the EIC}
Despite the tremendous progress in understanding TMDs and the
related physics, without a new lepton-hadron collider many aspects
of this fascinating field will remain untouched or at least on a
qualitative level. Existing facilities either suffer from a much
too restricted kinematic coverage or from low luminosity or from both.
Based on the present status of research we see the following
potential in an EIC:
\begin{itemize}

\item clean quantitative measurements of TMDs in the valence
region due to high luminosity, and ability to go to sufficiently
large $Q^2$ in order to suppress potential higher twist
contaminations. Primordial orbital motion is expected for valence
quarks.

\item related to the wide kinematic coverage and the high
luminosity, ability to provide multi-dimensional representations of 
the observables, which is basically impossible on the basis of current experiments.

\item production and possible observation of jets with
significantly larger particle multiplicities, allowing for the
study a larger variety of hadronic final states.

\item first access to TMDs for antiquarks.

\item (first) access to TMDs for gluons, for instance through
dihadron correlations, dijet correlations, or semi-inclusive
production of quarkonium.

\item systematic study of perturbative QCD techniques (for
polarization observables). Tests and studies of QCD evolution
properties of TMDs.

\end{itemize}

We strongly believe that the EIC will bring our knowledge of the
partonic structure of the nucleon to an entirely new level.
Keeping in mind deeply QCD rooted effects, like the (potential)
sign-change of the Sivers function, the EIC can be expected to
stimulate further developments in the application of perturbative
QCD to other hard scattering processes. A series of ``golden'' and ``silver''  measurements are outlined in table~\ref{tab:TMD-sciencematrix}.  The significance of these
points is 
further enhanced by newly planned (polarized)
Drell-Yan experiments, which will study complementary
physics aspects.


\begin{table}[t]
\small
\begin{center}
\begin{tabular}{|c|c|c|c|c|} \hline
Deliverables & Observables & What we learn & Phase I & Phase II\\
\hline\hline
Sivers + unp. &  SIDIS with Tran. & Quant. Interf.
&valence+sea  & 3D Imaging of\\
TMD quarks &polarization/ion; & Multi-parton $\&$  & quarks, overlap &
quarks $\&$ gluon;  \\
and gluon & di-hadron (di-jet) &Spin-Orbit & with the fixed  &  $Q^2$
($P_\perp$) range\\
& heavy flavor &correlations & target exp.& QCD dynamics\\
\hline
\hline
Chiral-odd &  SIDIS with Tran. & 3${}^{\rm rd}$ basic quark &valence+sea  &
$Q^2$ ($P_\perp$) range\\
functions: &polarization/ion; & PDF; novel & quarks, overlap &   for
detailed \\
Transversity; & di-hadron  &hadronization & with the fixed  & QCD dynamics
\\
Boer-Mulders & production & effects & target exp.& \\
\hline
\end{tabular}
\caption{\label{tab:TMD-sciencematrix}
\small Science Matrix for TMD physics: 3D structure in transverse momentum
space:  golden
measurements  (upper part) and  silver measurements  (lower part).
}
\end{center}
\end{table}


%



\section{Sivers function}
\label{sec:Sivers-example}


%

\hspace{\parindent}\parbox{0.92\textwidth}{\slshape
Christine Aidala, Elke Aschenauer, Alessandro Bacchetta, Thomas Burton, Leonard Gamberg, Delia Hasch, Min Huang, Zhong-Bo Kang,  Yuji Koike, Bernhard Musch, Alexei Prokudin, Xin Qian,  Gunar Schnell, Kazuhiro Tanaka, Anselm Vossen, Feng Yuan}

\index{Aidala, Christine} 
\index{Aschenauer, Elke} 
\index{Bacchetta, Alessandro} 
\index{Burton, Thomas} 
\index{Boer, Dani\"el}  
\index{Gamberg, Leonard} 
\index{Hasch, Delia} 
\index{Kang, Zhong-Bo} 
\index{Koike, Yuji}
\index{Huang, Min} 
\index{Musch, Bernhard}
\index{Prokudin, Alexei} 
\index{Qian, Xin} 
\index{Schnell, Gunar} 
\index{Tanaka, Kazuhiro} 
\index{Vossen, Anselm} 

\vspace{\baselineskip}


We choose the example of the Sivers function to illustrate  the physics case 
for TMD distributions at the EIC.
This function incorporates all new facets and intriguing physical aspects of TMD distributions 
outlined in the introduction and discussed in more detail in the following sections.
We start this discussion with a brief review of the peculiarities of the Sivers function
thereby illustrating the crucial role TMDs play in our understanding of the 
nucleon structure.

\begin{figure}[h]
\begin{center}
\mbox{
\hspace{-0.5cm}
\includegraphics[width=.48\textwidth]{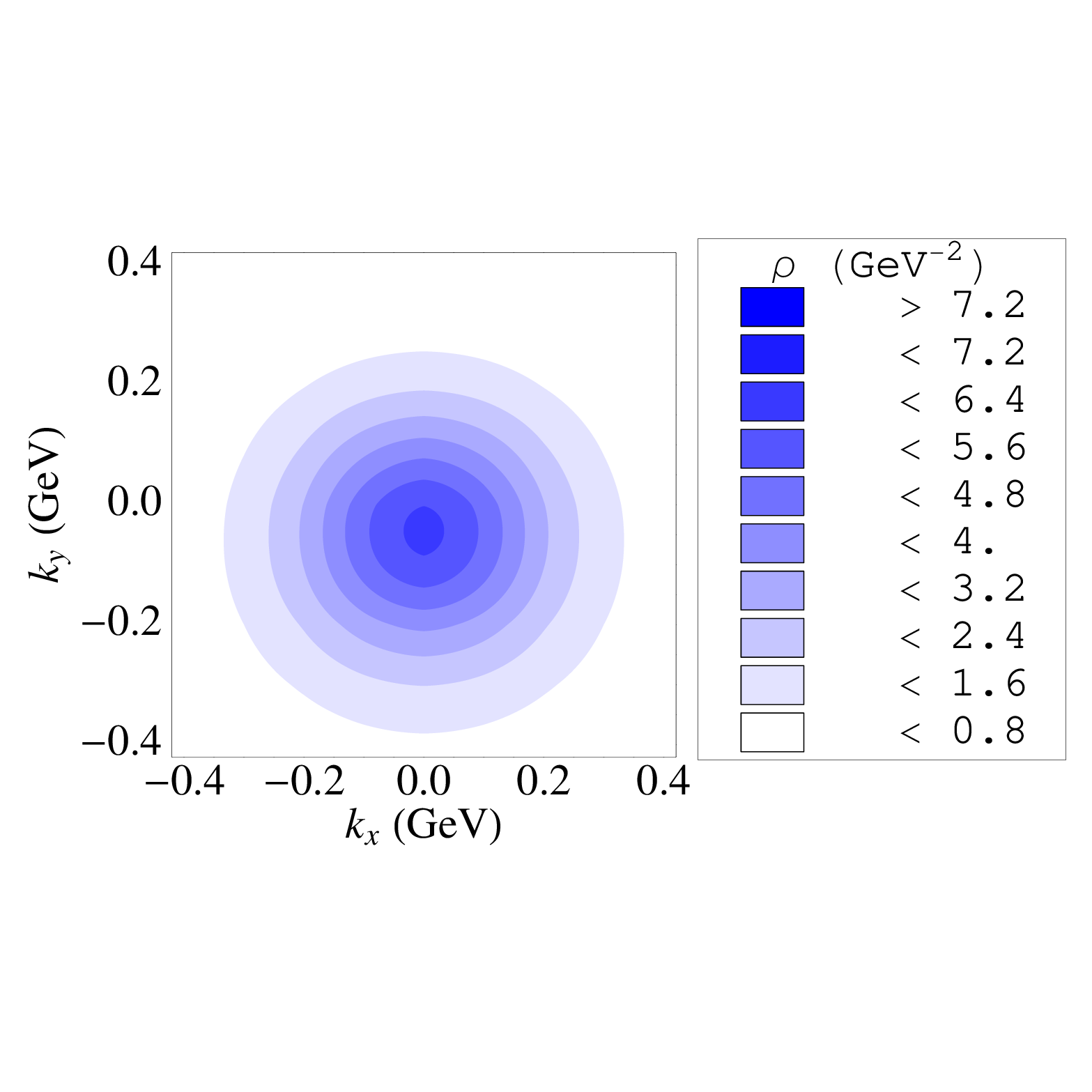}
\includegraphics[width=.48\textwidth]{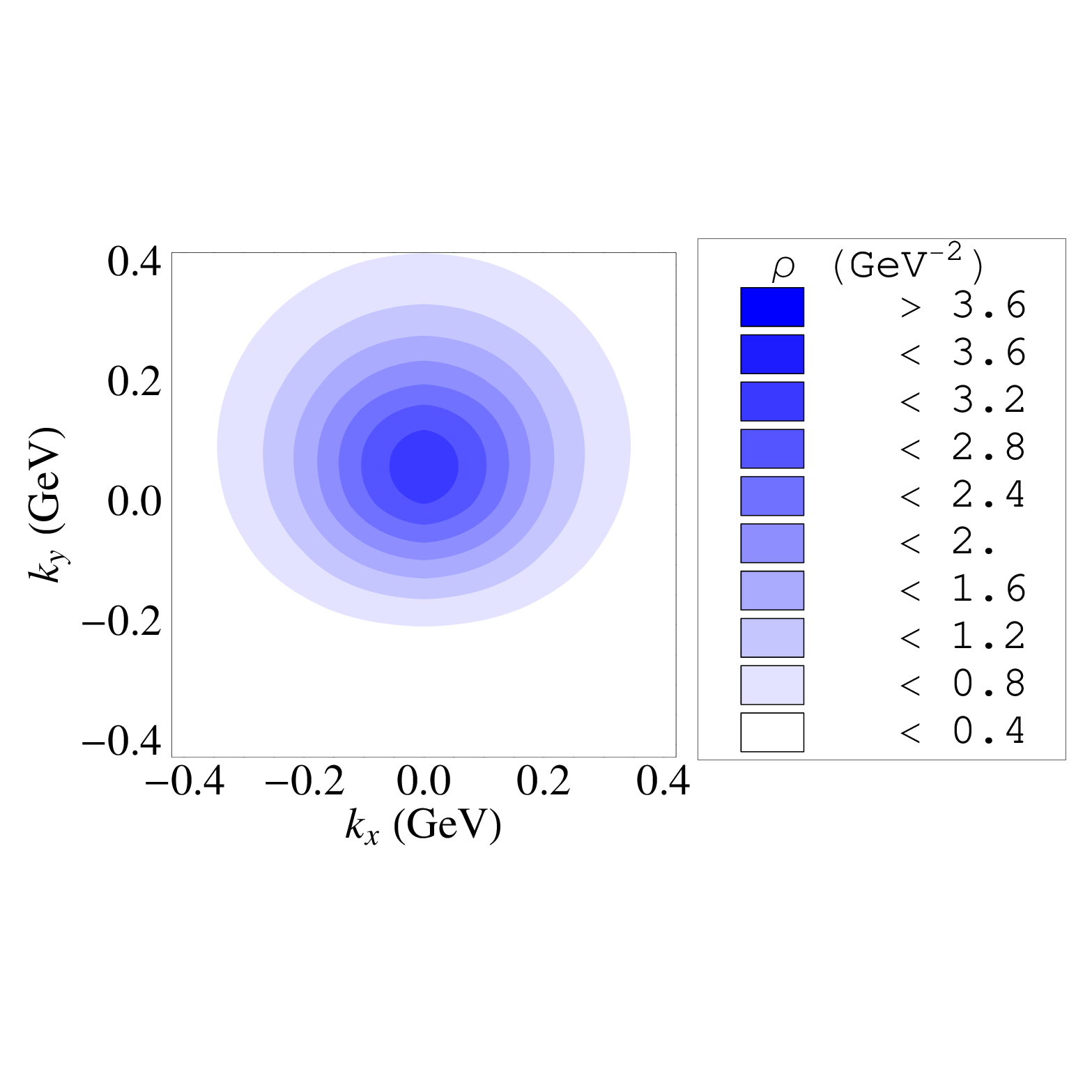}
}
\vspace{-1.5cm}
\caption{
Spin density in the transverse-momentum plane for unpolarized quarks in a 
transversely polarized nucleon, as described by the Sivers function. 
The left panel is for up quarks and the right one for down quarks.
The model calculation of Ref.~\cite{Pasquini:2010af} was used.
}
\label{fig:alexei-dipole}
\end{center}
\end{figure}

The Sivers function $f_{1T}^{\perp a}(x, k_\perp)$, appearing in the distribution of unpolarized
partons $a$ inside a polarized nucleon: 
\begin{equation}
f_{1}^a(x, \mbox{\boldmath $k$}_\perp; \mbox{\boldmath $S$}) =
f_1^{a}(x, k_\perp) - \frac{k_\perp}{M} \, f_{1T}^{\perp a}(x,
k_\perp) \> \mbox{\boldmath $S$} \cdot  (\hat{\mbox{\boldmath
$P$}} \times \hat{\mbox{\boldmath $k$}}_\perp) \> \,
\label{eq:sivers-expression}
\end{equation}
describes the correlation between the momentum direction of the struck parton and 
the spin of its parent nucleon and is hence related to the orbital motion of
partons inside the nucleon. 
This correlation generates a dipole pattern in the transverse $k_\perp$-plane.
We illustrate this fascinating aspect of certain TMDs in providing a three-dimensional 
imaging of the nucleon in momentum space by choosing a specific configuration
for the vectors involved in Eq.~(\ref{eq:sivers-expression}).
Taking for example
$\hat{\mbox{\boldmath$P$}} \equiv \frac{\bf P}{|\bf P|}= (0,0,-1)$
and the spin of the proton along the $y$ direction, so that ${\boldmath S} = (0,1,0)$
and the transverse momentum of the parton ${\boldmath k_\perp } = (k_{\perp x}, k_{\perp y}, 0)$, 
yields a typical ``dipole'' modulation of the distribution:
\begin{equation}
f_{1}^a(x, \mbox{\boldmath $k$}_\perp; \mbox{\boldmath $S$}) =
f_1^{a}(x, k_\perp) + \frac{k_{\perp x}}{M} \, f_{1T}^{\perp a}(x,
k_\perp) \,.
\end{equation}
The $f_1$ term provides an axially symmetric contribution, 
while the second term containing $f_{1T}^\perp$ gives rise to the dipole pattern. 
A superposition of both effects results in a distribution that is shifted
away from the center (distorted) in the $k_\perp$-plane as shown in 
fig.~\ref{fig:alexei-dipole}. 
This distortion turns out to be of opposite sign for up and down quarks.

The Sivers function manifests the importance of initial and final
state interaction effects in hard scattering processes as the presence 
of these effects is required for the existence of a non-zero Sivers function. 
Their inclusion in the TMD factorization approach yields a
peculiar breaking of the universality of the Sivers function.
As introduced in sec.~\ref{secI:theory} and detailed in sec.~\ref{secIV:mulders-rogers},
this non-universality is well under control and `merely' consists of a sign
change of the Sivers function when appearing in the Drell-Yan process as compared
to DIS.  
The experimental verification of this sign change is currently one of the outstanding 
topics in hadronic physics and presents a crucial test for our understanding of hadron
production in high-energy reactions.
We will therefore briefly review the prospects for measurements 
of the Sivers effect in Drell-Yan in sec.~\ref{sec:DY}. 

A further intriguing aspects of the Sivers function is its connection to the 
orbital angular momentum in the nucleon.
A non-zero quark Sivers function involves a transition
between initial and final nucleon states that differ by one unit of
orbital angular momentum.
This property together with the 
potential for a three-dimensional imaging, puts 
the Sivers function in close relation to the GPD $E$ discussed in 
chapter~\ref{chap:gpd}.
In particular, it was proposed that there is a dynamical relation 
called ``chromodynamic lensing", where the spatial distortion of 
the transverse quark distribution (in a transversely polarized proton) 
leads to a distortion in transverse momentum distribution
described by the Sivers function~\cite{Burkardt:2002ks,Burkardt:2003je,Burkardt:2007rv}.

\subsection{What do we know so far from experiments?}
\label{sec:sivers-status}
 
Though the Sivers function was first suggested to explain the surprisingly large
single-spin asymmetries measured in $pp$ collisions,
our guiding experiments for obtaining unambiguous information about this function, 
and most of the other TMDs, involve high-energy lepton-nucleon scattering 
with the observation of one or more hadrons in coincidence with the scattered lepton 
(semi-inclusive DIS).
In addition, model calculations of TMDs, discussed in sec.~\ref{sec:TMD-theory}, 
guide Ans\"atze for global fits of TMD parameterizations and provide an
interpretation of the various aspects of TMDs.

In this section, after a brief review of the results from $pp$ collisions, 
we will summarize available semi-inclusive DIS measurements of observables 
related to the Sivers effect and present phenomenological extractions of 
the Sivers function from data.
The following section will then highlight the potential of an EIC for a detailed and 
systematic exploration of the various aspects of the quark Sivers functions 
illustrative of TMDs in general.

\subsubsection{Transverse-spin effects in proton-proton collisions}

Historically, the surprisingly large left-right asymmetries observed in
hadronic reactions with transversely polarized protons initiated the idea about 
a transverse momentum dependence of quark distributions in polarized protons.
The pioneering measurements~\cite{Klem:1976ui,Antille:1980th}   
of these large (up to 0.3-0.4 in magnitude) transverse-spin asymmetries in inclusive
forward production of pions in $pp$ collisions $ p^\uparrow p \rightarrow \pi + X$, 
have been extensively confirmed by experiments at FermiLab
~\cite{Adams:1991rx,Adams:1991ru,Adams:1991cs,Adams:1994yu,Bravar:1996ki} 
and at RHIC (BNL) at much higher center-of-mass energies of up to $\sqrt{s} = 200$ GeV
~\cite{Adams:2003fx}.
The observation of such asymmetries was frequently quoted as a puzzle or
challenge for theory. 
In fact, for a long time, transverse single-spin asymmetries were assumed to be negligible in hard
scattering processes~\cite{Kane:1978nd}.
The work of~\cite{Sivers:1989cc} introduced a transverse momentum dependent
quark distribution, now termed the Sivers function,
which provides a mechanism for the observed asymmetries that does not vanish at high energies.

A rich variety of single-spin asymmetries for identified hadrons 
($\pi^{\pm},\pi^{0},K^{\pm},p,\bar{p}$)
measured over a wide kinematic range is now available from the BRAHMS, PHENIX
and STAR experiments at RHIC 
(BNL)~\cite{Arsene:2008mi,Lee:2009ck,Adare:2010bd,Adler:2005in,Dharmawardane:2010zz,Abelev:2007ii,Abelev:2008qb,Eun:2010zz}.
The results exhibit a general pattern: 
sizable asymmetries are measured at forward-rapidity and for positive Feynman $x_F > 0.3$ 
which increase in magnitude with increasing $x_F$ and $P_{hT}$.
In contrast,
for negative  $x_F$ and at mid-rapidity all asymmetries are found to be 
consistent with zero.

Several mechanisms have been suggested to explain these asymmetries.
At large values of \(P_{hT}\) collinear factorization involving twist-3 
distributions can be applied. 
However, the intrinsic prediction of a \(1/P_{hT}\) fall-off has yet to be confirmed. 
An alternative approach using a generalized parton model that takes intrinsic 
transverse momentum dependences into account has been used to describe existing data,
achieving a fairly successful description of the observed asymmetries for pion production in
$pp$ collisions~\cite{D'Alesio:2007jt}.
If less inclusive measurements are performed, with an observed soft momentum scale in
addition to a hard scale, one can attempt to describe the data using a TMD approach in pQCD.
However, as discussed in sec.~\ref{secIV:mulders-rogers},
the presence of both initial and final state interactions in hadron-hadron collisions
may prevent any kind of (standard) TMD-factorization.  
More insight might be gained regarding the intricate color structure of $pp$ reactions for example
by measuring di-jet production. 
In di-jet production both large scales (e.g., jet \(p_T\)) and small scales
(e.g., \(\Delta p_T\) of nearly back-to-back jets)  can be observed.
To assess
factorization breaking due to color interactions in $pp$ collisions, the experimental
measurements can be compared to calculations using TMDs extracted from DIS and Drell-Yan,
for which TMD-factorization has been demonstrated.  
Little experimental information
currently exists on these processes, but they are part of the physics program at RHIC.

Many questions still need to be answered, but it is clear that for a strict assessment of whether
the TMD Ansatz is indeed possible and appropriate to describe results from hadronic
collisions, more precise parameterizations of the Sivers function and, hence, more precise
data on the Sivers effect in a well-understood process like DIS is needed.

\subsubsection{Semi-inclusive Deep-Inelastic Scattering}

\begin{figure}[t]
\begin{center}
\includegraphics[width=0.75\textwidth]{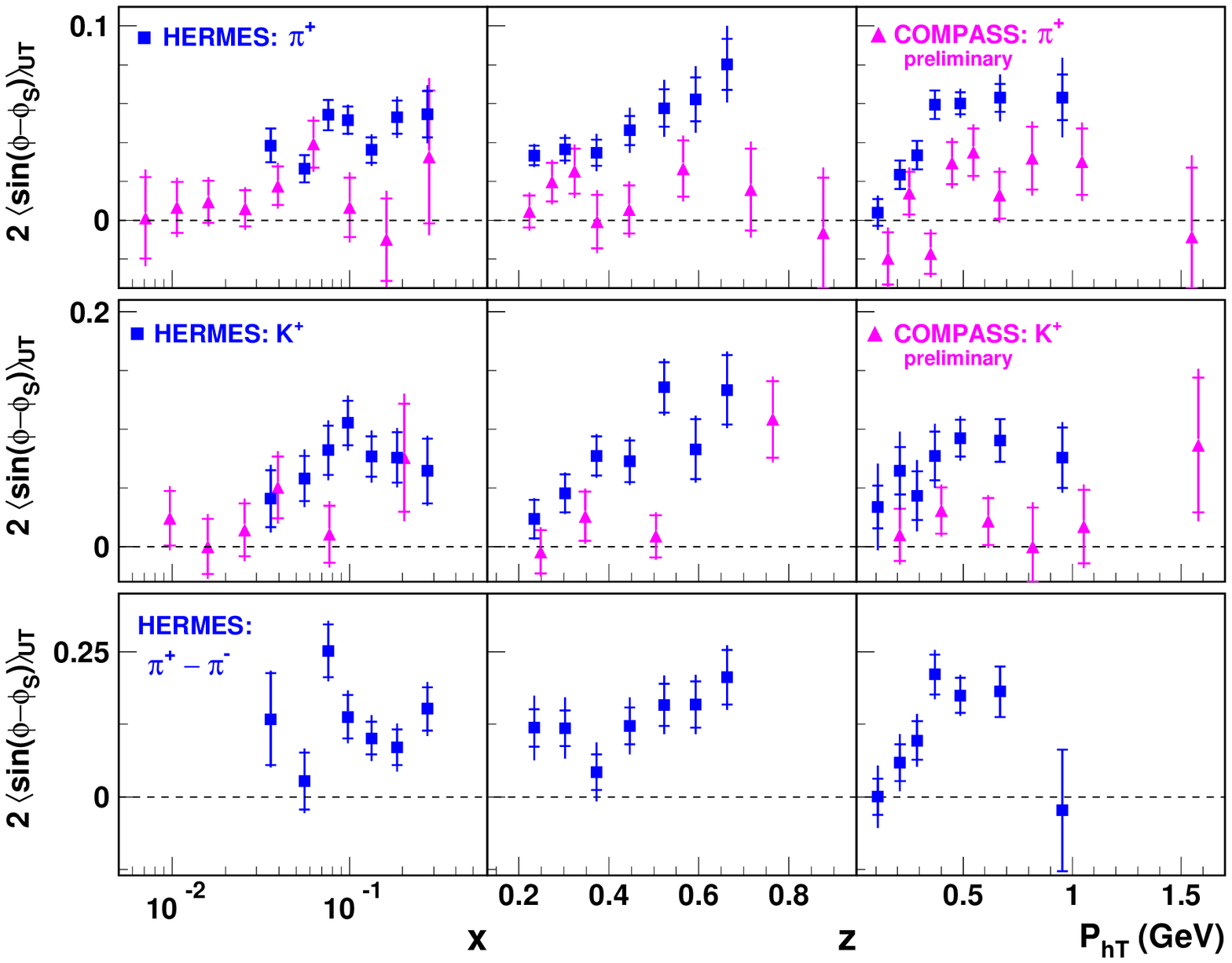}
\caption{\label{f:sivers-amplitude-hermes-compass}
\small
Sivers amplitudes for $\pi^+$, $K^+$ and the pion-difference (as denoted in the
panels) from HERMES~\cite{Airapetian:2009ti} 
and for $\pi^+$ and $K^+$ from COMPASS~\cite{Pesaro:2011zz} measured with a proton target.
Inner error bars present statistical uncertainties and full error bars the quadratic
sum of statistical and systematic uncertainties.
Note that the average kinematics in each bin differs for HERMES and COMPASS.  
}
\end{center}
\end{figure}

In semi-inclusive DIS, the Sivers function leads to single-spin asymmetries in the
distribution of hadrons in the azimuthal angles illustrated in fig.~\ref{f:anglestrento}.
The azimuthal modulations of the SIDIS cross section are given in Eq.~(\ref{e:crossmaster}).
The Sivers effect manifests itself as a $\sin(\phi_h-\phi_S)$ modulation and requires 
transverse polarization of the target nucleon.
The additional information provided by the azimuthal angle $\phi_S$ of the transverse
component of the target-proton spin about the virtual photon direction
allows for an unambiguous extraction of the Sivers effect.
Experimentally, the so-called Sivers 
amplitude $2 \langle \sin(\phi_h -\phi_S) \rangle_{UT}^h$~\cite{Bacchetta:2004jz},
which projects 
out the structure function $F_{UT,T}^{\sin(\phi_h -\phi_S)}$ in Eq.~(\ref{e:crossmaster}) for
a specific hadron $h$, is extracted from the asymmetry
\begin{equation}
A_{UT}^h(\phi_h,\phi_S) \equiv \frac{1}{|{\bf S}_T|} \frac{d\sigma^h(\phi_h,\phi_S) - d\sigma^h(\phi_h,\phi_S+\pi)}{d\sigma^h(\phi_h,\phi_S) + d\sigma^h(\phi_h,\phi_S+\pi)}\;,
\end{equation}
where the subscript $U$ indicates an unpolarized lepton beam and $T$ a transversely polarized
target nucleon.
This amplitude has so far been extracted by three polarized fixed-target experiments as summarized 
in Tab.~\ref{t:exp-sivers-measurements}.
\begin{table}[t]
\begin{center} \small
\begin{tabular}{lcccc}\hline
experiment (laboratory) & $\sqrt{s}$ in GeV& target type & hadron types & references \\\hline
COMPASS (CERN) & 18 & deuteron & $h^{\pm},\pi^{\pm},K^{\pm},K^0$ & ~\cite{Ageev:2006da,Alekseev:2008dn}\\
 &  & proton & $h^{\pm}$ & ~\cite{Alekseev:2010rw}\\
 &  & proton & $\pi^{\pm},K^{\pm}$ & prelim.~\cite{Pesaro:2011zz}\\\hline 
HERMES (DESY) & 7.4  & proton & $\pi^{\pm}$ & \cite{Airapetian:2004tw} \\ 
 &  & proton & $\pi^{\pm},(\pi^+ - \pi^-),\pi^{0},K^{\pm}$ & \cite{Airapetian:2009ti} \\\hline
HallA (JLab) & 3.5  & neutron & $\pi^\pm$ & prelim.~\cite{HallA-sivers-neutron-2010}  \\\hline
\end{tabular}
\caption{\small
Summary of currently available measurements of Sivers asymmetry amplitudes from
lepton-nucleon DIS experiments, their center-of-mass energy, transversely polarized target type, 
and analyzed hadron types.
}
\label{t:exp-sivers-measurements}
\end{center}
\end{table}
From these measurements, fig.~\ref{f:sivers-amplitude-hermes-compass} shows a selection of results 
that are significantly non-zero and help in determining the shape of the Sivers function.
All other asymmetry amplitudes listed in Tab.~\ref{t:exp-sivers-measurements} are small or
consistent with zero.
%

The results have so far been interpreted in the parton model as a convolution of distribution
and fragmentation functions, where the Sivers amplitude can be approximated by
\begin{equation}
2\langle\sin{(\phi_h-\phi_S)}\rangle^h_{UT} (x_B,y,z_h,P_{hT}) 
= -\frac{\sum_q e_q^2 \,\, f^{\perp q}_{1T}(x,k_\perp^2) \otimes_{\mathcal W} D_1^q(z,P_\perp^2)}{\sum_q e_q^2 \,\, f_1^q(x,k_\perp^2) \otimes D_1^q(z,P_\perp^2)}.
\label{e:QPM-sivers}
\end{equation}
Here the sums run over the quark flavors, the \(e_q\) are the quark
charges, and $f_1(x,k_\perp^2)$ and $D_1(z,P_\perp^2)$ are the spin-independent quark
distribution and fragmentation functions, respectively. 
The symbol $\otimes$ ($\otimes_{\mathcal W}$) represents a (weighted) convolution 
integral over intrinsic and fragmentation transverse momenta,
$\boldsymbol{k}_\perp$ and $\boldsymbol{P}_\perp$ respectively, as explicitely given 
in~(\ref{eq:convolution_our_main}). 

A qualitative picture of the Sivers function can already be derived from the measured
asymmetry amplitudes.
The non-zero results shown in fig.~\ref{f:sivers-amplitude-hermes-compass} are obtained 
with a proton target.
As scattering off $u$ quarks dominates these data due to the charge factor, 
the positive Sivers amplitudes for $\pi^+$ and $K^+$
suggest a large and negative Sivers function for up quarks. 
This is supported by the positive amplitudes of the pion difference asymmetry, 
which originates mainly from the difference ($f^{\perp d_v}_{1T} - 4f^{\perp u_v}_{1T}$)
in the Sivers functions for valence down and up quarks
and is dominated by the contribution from valence $u$ quarks.
The vanishing amplitudes for $\pi^-$ require cancellation effects, e.g. from a 
$d$ quark Sivers function opposite in sign to the  $u$ quark Sivers function.
Such cancellation effects between Sivers functions for up and down quarks are supported 
by the vanishing asymmetry amplitudes extracted from deuteron
data by the COMPASS collaboration.
An interesting facet of the data shown in fig.~\ref{f:sivers-amplitude-hermes-compass}
is the magnitude of the $K^+$ amplitudes, which are nearly twice as large as those of
the $\pi^+$.
Again, on the basis of $u$ quark dominance,
one might naively expect that the $\pi^+$ and $K^+$  amplitudes should be similar.
Their difference in size may thus point to a significant role of other quark flavors, 
e.g. sea  quarks.

Phenomenological analyses of HERMES and COMPASS data
~\cite{Anselmino:2010bs,Efremov:2004tp,Anselmino:2005ea,Vogelsang:2005cs,Collins:2005ie,Anselmino:2008sga},
confirm the picture drawn above as discussed in the following.
So far, only the analysis of Ref.~\cite{Anselmino:2010bs} makes use of a subset of the most recent data 
listed in Tab.~\ref{t:exp-sivers-measurements}
and all fits have yet to be updated
for the results from proton data from COMPASS and the first neutron data from HallA.

\subsubsection{Phenomenological extractions and models of  the Sivers function}
\label{sec:sivers-phenomenology-models}

The strong impact and success of model calculations and lattice QCD on the TMD field
is discussed in detail in sec.~\ref{sec:TMD-theory} and 
sec.~\ref{ch:lattice-sec:intro}, respectively. 
Models provide information on the magnitudes and signs of TMDs and guide
Ans\"atze for global fits of TMD parameterizations.
For example, from chiral models~\cite{Drago:2005gz} and the QCD limit of a large number of
colours (large $N_c$ limit)~\cite{Pobylitsa:2003ty}
a Sivers function for up and down quarks of equal size but with opposite sign 
($f_{1T}^{\perp u} = -f_{1T}^{\perp d}$) is predicted.

Phenomenological analyses provide extractions of TMDs from data.
As discussed in sec.~\ref{secI:info-about-TMDs},
existing analyses of TMD observables are so far based on the parton model approximation,
where the measured amplitudes of the SIDIS cross section in Eq.~(\ref{e:crossmaster}),
are expressed as convolutions of distribution $f^q$ and fragmentation functions $D^q$.
For the Sivers amplitude it reads
\begin{equation}
F_{UT,T}^{\sin (\phi_h - \phi_S)}
 \propto \sum_q e_q^2 \,\, f^{\perp q}_{1T}(x,k_\perp^2) \otimes_{\mathcal W} D_1^q(z,P_\perp^2)
\end{equation}
where $\otimes_{\mathcal W}$ is defined as
%
\begin{equation}
 \otimes_{\mathcal W}
\equiv
\, \int d^2 {\boldsymbol k}_{\perp}\,d^2 {\boldsymbol P}_{\perp}\,
\delta^{(2)} \left(z {\boldsymbol k}_{\perp}  + {\boldsymbol P}_{\perp}- {{\boldsymbol P}_{h T}}  \right)\,
\,{\mathcal W} \,\, , 
\label{eq:convolution_our_main}
\end{equation}
with the kinematic factor $\mathcal{W}$ depending on the involved transverse momenta.
This convolution can be resolved by either employing a particular model for
the transverse momentum dependence or by integrating over the transverse 
momentum $P_{hT}$ using a proper weight factor in the extraction of the asymmetry
amplitudes which involves $P_{hT}$, building for example 
$2\langle \frac{P_{hT}}{M_p}\sin{(\phi_h-\phi_S)}\rangle^h_{UT}$.
The latter approach is very attractive but experimentally challenging for measurements at current  
fixed target facilities as it requires full $P_{hT}$ coverage, which cannot be obtained
at any of the existing experiments.
An EIC would be the ideal facility to study such weighted asymmetries and to seriously
explore the advantages of these observables, as further discussed in 
sec.~\ref{sec:TMD-weighted-asymmetries}.

An intuitive and common Ansatz for the transverse momentum dependence of distribution and
fragmentation functions, which provides an  analytic solution of (\ref{eq:convolution_our_main}), 
is a Gaussian distribution like
\begin{equation}
f^{\perp q}_{1T}(x,k_{\perp}^2) = f^{\perp q}_{1T}(x)\frac{1}{\pi \langle k_{\perp}^2 \rangle}\exp\left( -\frac{{\boldsymbol k}_{\perp}^2}{\langle k_{\perp}^2 \rangle} \right) , \,\,\,
D^q_1(z,P_{\perp}^2) = D^q_1(z) \frac{1}{\pi \langle P_{\perp}^2 \rangle}\exp\left( -\frac{{\boldsymbol P}_{\perp}^2}{\langle P_{\perp}^2 \rangle} \right)
\label{eq:TMD-gaussian-ansatz}
\end{equation}
with typical values for $\langle k_{\perp}^2 \rangle$ and $\langle P_{\perp}^2 \rangle$
of 0.2 to 0.3 GeV$^2$.

The Sivers function was among the first to be extracted from data, as it
couples to the usual unpolarized fragmentation function $D_{1}^q$.
This fragmentation function is reasonably well parameterized~\cite{Kretzer:2000yf,Hirai:2007cx} 
using precise data from electron-positron annihilation into charged hadrons 
and, most recently, also from single-hadron production in $pp$ collisions and
semi-inclusive DIS~\cite{deFlorian:2007aj}, which provide complementary information on the 
flavour dependence of the fragmentation process. 
\begin{figure}[t]
\begin{center}
\includegraphics[width=0.4\textwidth]{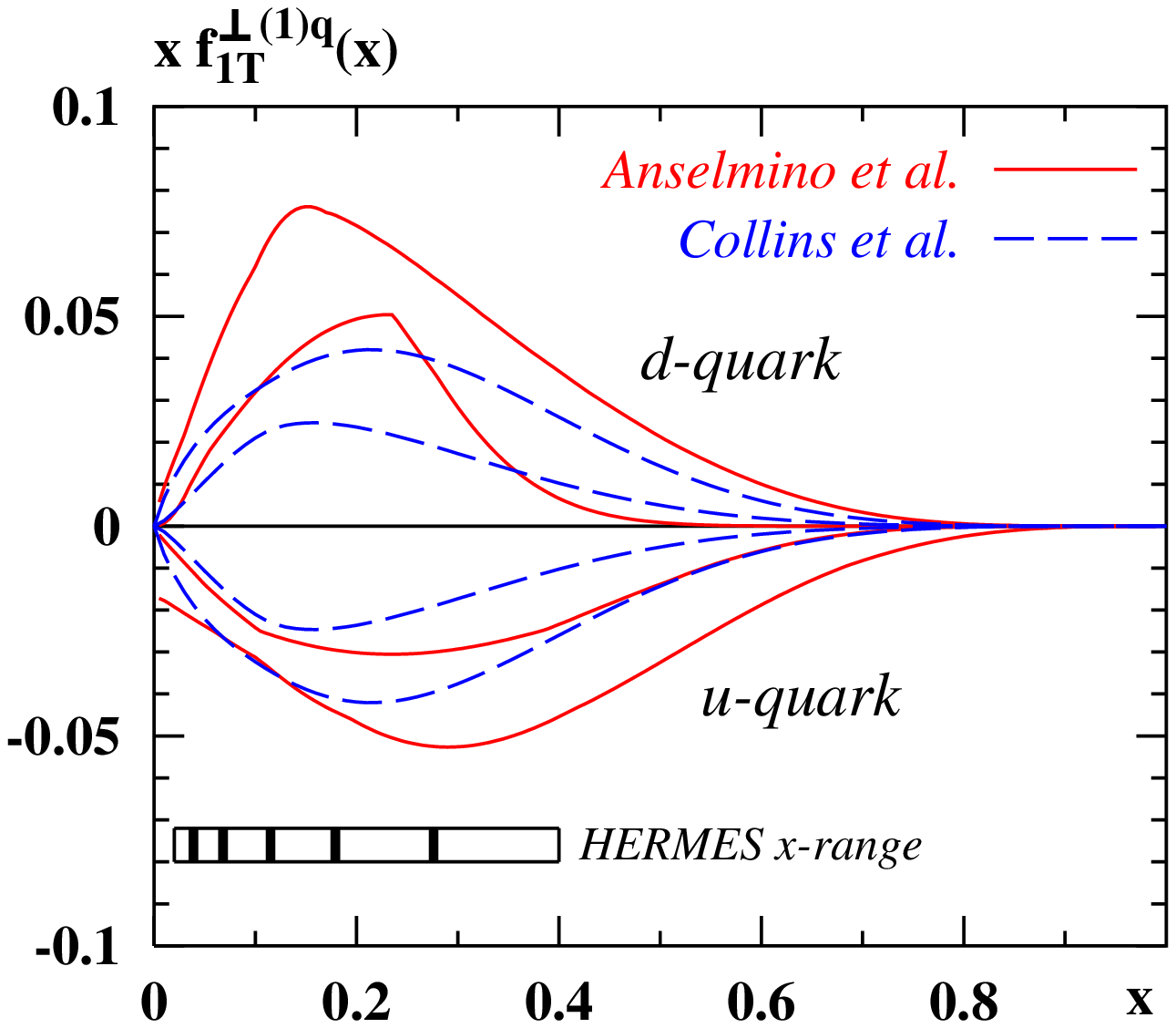} \qquad
\includegraphics[width=0.4\textwidth]{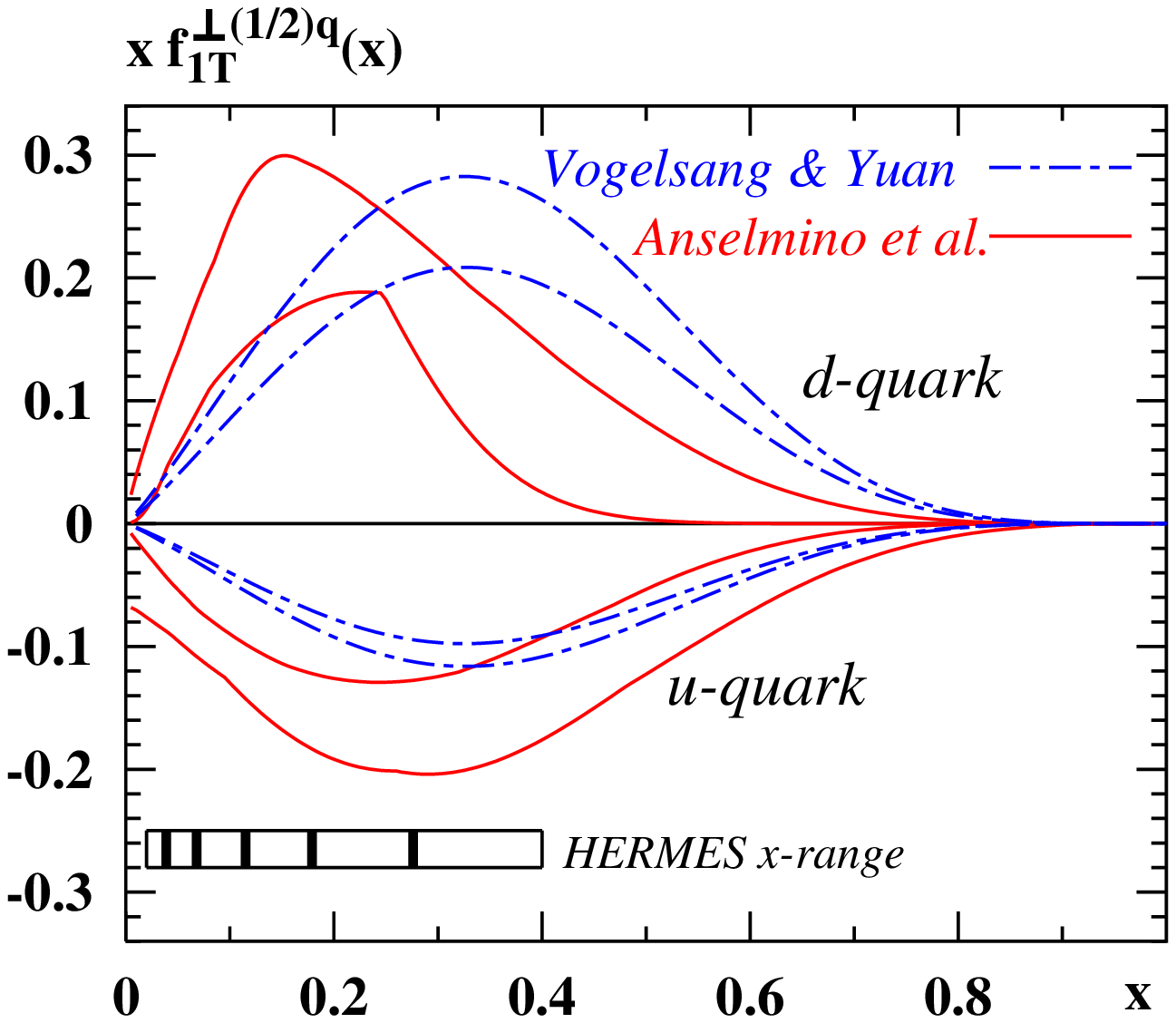}
\caption{\label{f:sivers-first-extractions}
\small
Up and down quark Sivers distributions extracted from HERMES (and for
the full line also from COMPASS) data using three different parameterizations
~\cite{Anselmino:2005ea,Vogelsang:2005cs,Collins:2005ie} (see text).
The left and right panels show, respectively, the first and the 1/2 moment.
The curves indicate the 1-sigma regions of the various parameterizations.
None of three parameterizations makes use of the latest experimental results
listed in Tab.~\ref{t:exp-sivers-measurements}.
}
\end{center}
\end{figure}

Figure~\ref{f:sivers-first-extractions}
 shows the extraction of the up and down quark Sivers
distributions using three different parameterizations for the Sivers function
~\cite{Anselmino:2005ea,Vogelsang:2005cs,Collins:2005ie},
presenting 
$k_\perp$-moments defined as
\begin{equation}
f^{(1)}(x) \equiv \int d^2 {\boldsymbol k}_{\perp} \frac{{\boldsymbol k}_{\perp}^2}{2M^2} f(x, {k}_{\perp}^2) \quad \mbox{and} \quad
f^{(1/2)}(x) \equiv \int d^2 {\boldsymbol k}_{\perp} \frac{|{\boldsymbol k}_{\perp}|}{2M} f(x, {k}_{\perp}^2)\;. \\
\label{eq:TMD-moments}
\end{equation}
The parameterization from Ref.~\cite{Anselmino:2005ea} (full line) is based on a 
combined fit to previous
HERMES and COMPASS data, while the other two fit HERMES data only but describe
the COMPASS data well when using the obtained parameters to calculate the asymmetries
for COMPASS kinematics.
All three extractions use the parameterization from Ref.~\cite{Kretzer:2000yf}
for the unpolarized fragmentation function.
The two curves of each set indicate the 1-sigma regions of the various parameterizations,
taking into account solely statistical uncertainties of the data sets employed in the fit.
The three approaches describe the HERMES Sivers asymmetries equally well.
The differences in size and shape of the extracted Sivers up and down quark distributions
hence reflect the model dependence of the fit results.
The parameterization of~\cite{Collins:2005ie} imposes the constraint from the large $N_c$
limit, which results in
the symmetric parametrization of up and down Sivers distributions, shown in the left
panel of fig.~\ref{f:sivers-first-extractions}
 with dashed lines.
None of the extractions involve parameterizations for sea quarks as they could not be
constrained by the data used in the fits.

%
\begin{figure}[t]
\centering \includegraphics[width=0.35\textwidth,angle=270]{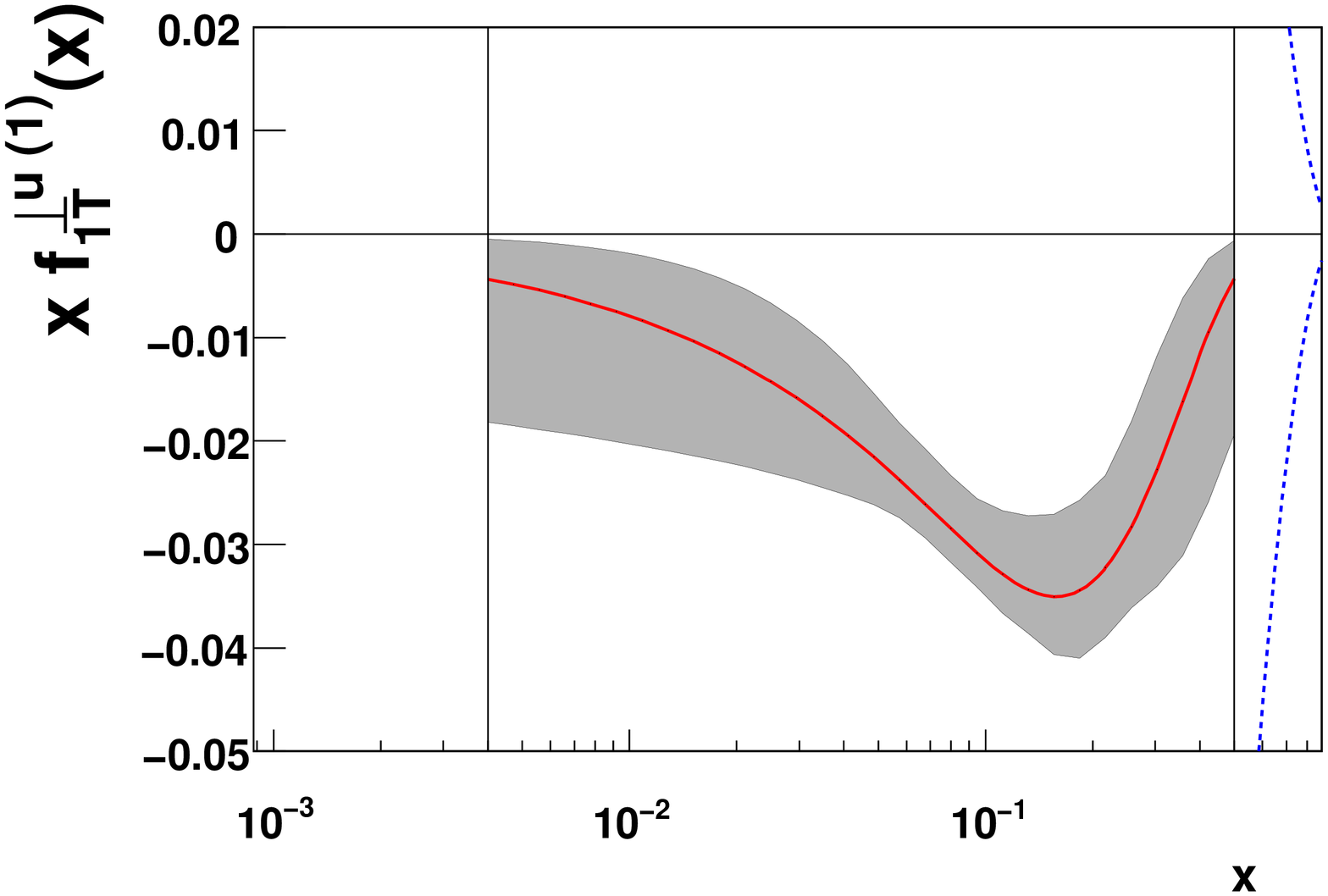}
\caption{\label{fig:prokudin_sivers_eic}
\small
The Sivers function for $u$ quarks extracted from recent experimental data~\cite{Anselmino:2010bs}. 
Vertical lines indicate the region where experimental data are available.
The band represents the 2-sigma range for the chosen parameterization.
The dashed blue lines indicate the positivity bound. 
}
\end{figure}
%

However, the recent, surprisingly large, Sivers asymmetry amplitudes for $K^+$ measured by HERMES,
which were found to be nearly twice as large as those of the  $\pi^+$, might hint at a possibly 
important role of sea quarks.
In Ref.~\cite{Anselmino:2010bs}, the sensitivity of these data to sea quark contributions was tested.
A fit including Sivers functions for only up and down quarks was compared with a second fit that 
allowed also for sea quark contributions ($\bar{u},\bar{d},s,\bar{s}$) to the Sivers amplitude.
Both fits describe the data with equally good $\chi^2$, demonstrating that their precision is not yet
good enough to independently constrain the Sivers function for six quark flavours.
In this analysis, the usage of new parameterizations of the fragmentation functions from 
Ref.~\cite{deFlorian:2007aj} was essential for obtaining a good description of the kaon data.

The available parameterizations of the Sivers function for up and down 
quarks~\cite{Anselmino:2010bs,Efremov:2004tp,Collins:2005ie,Anselmino:2008sga}
agree, within their large uncertainties, 
with calculations based on a light-cone model~\cite{Pasquini:2010af} and on a diquark spectator
model~\cite{Bacchetta:2008af,Gamberg:2007wm},
while predictions based on the bag model~\cite{Courtoy:2008dn}
appear to be too small in magnitude for both the up and down quark Sivers function
(see also sec.~\ref{sec:TMD-theory}).

\subsubsection{Open issues in extractions of the Sivers function}

Figure~\ref{fig:prokudin_sivers_eic} illustrates our current knowledge of the Sivers function.
So far, only the up and  down quark Sivers functions can be constrained with relatively large 
uncertainties within the range $0.004 < x < 0.5$ using basic parameterizations for 
their shapes.

The precision of current data permits neither constraints of the Sivers functions for sea quarks nor 
an employment of more flexible functional forms, which would also allow for a sign change as suggested by 
a spectator model (see fig.~\ref{fig:ps-tmd-odd-models} in sec.~\ref{sec:TMD-theory}).
The band in fig.~\ref{fig:prokudin_sivers_eic} represents the 2-sigma range for 
the chosen parameterization and reflects the precision of the data,
%
%
but does not account for model uncertainties or for variations of the functional form of the 
parameterizations.
Also not estimated so far, is any uncertainty stemming from the Gaussian Ansatz used to 
resolve the convolution in (\ref{eq:convolution_our_main}).
For example, the average value, $\langle k_{\perp}^2 \rangle$, of the quark intrinsic transverse momentum 
used in this Ansatz might be flavour dependent, and both  $\langle k_{\perp}^2 \rangle$ and 
$\langle P_{\perp}^2 \rangle$ dependent on the energy scale.
The latter is particularly relevant for the fragmentation functions, which are 
extracted from data collected at much higher energy than the available SIDIS asymmetry data 
used in the fits.
The EIC would provide both TMD observables at substantially higher scale than any fixed
target DIS experiment and unique data sets of hadron production for a flavour tagging 
in the fragmentation process and a study of its transverse momentum dependence. 

At this stage of analysis, also specific known issues of experimental data are ignored.
For example,
the limited precision of currently available SIDIS data usually allows
only for presenting the results as a function of one kinematic variable while integrating 
over the others within the experimental acceptance.
Hence, the asymmetry amplitudes from a specific experiment, presented for different kinematic 
variables are correlated.
Moreover, the experimental acceptance usually does not provide a full coverage in $P_{hT}$. 
Thus, the 'unweighthed' asymmetry amplitudes extracted as function of $x$ or $z$ present only 
partial $P_{hT}$ moments in contrast to theoretical considerations.
A fully differential analysis of SIDIS data, which requires high statistic datasets, 
would resolve these issues.

Turning our essentially qualitative picture of the Sivers function and the related physics
into a quantitative description, 
which goes beyond the tree-level approximation,
requires new facilities providing high precision polarized data
over a wide kinematic range as discussed in the following section.

\subsection{The Sivers function at the EIC}
\label{sec:sivers-at-EIC}

%
\begin{figure}[t]
\centering \includegraphics[width=0.75\textwidth]{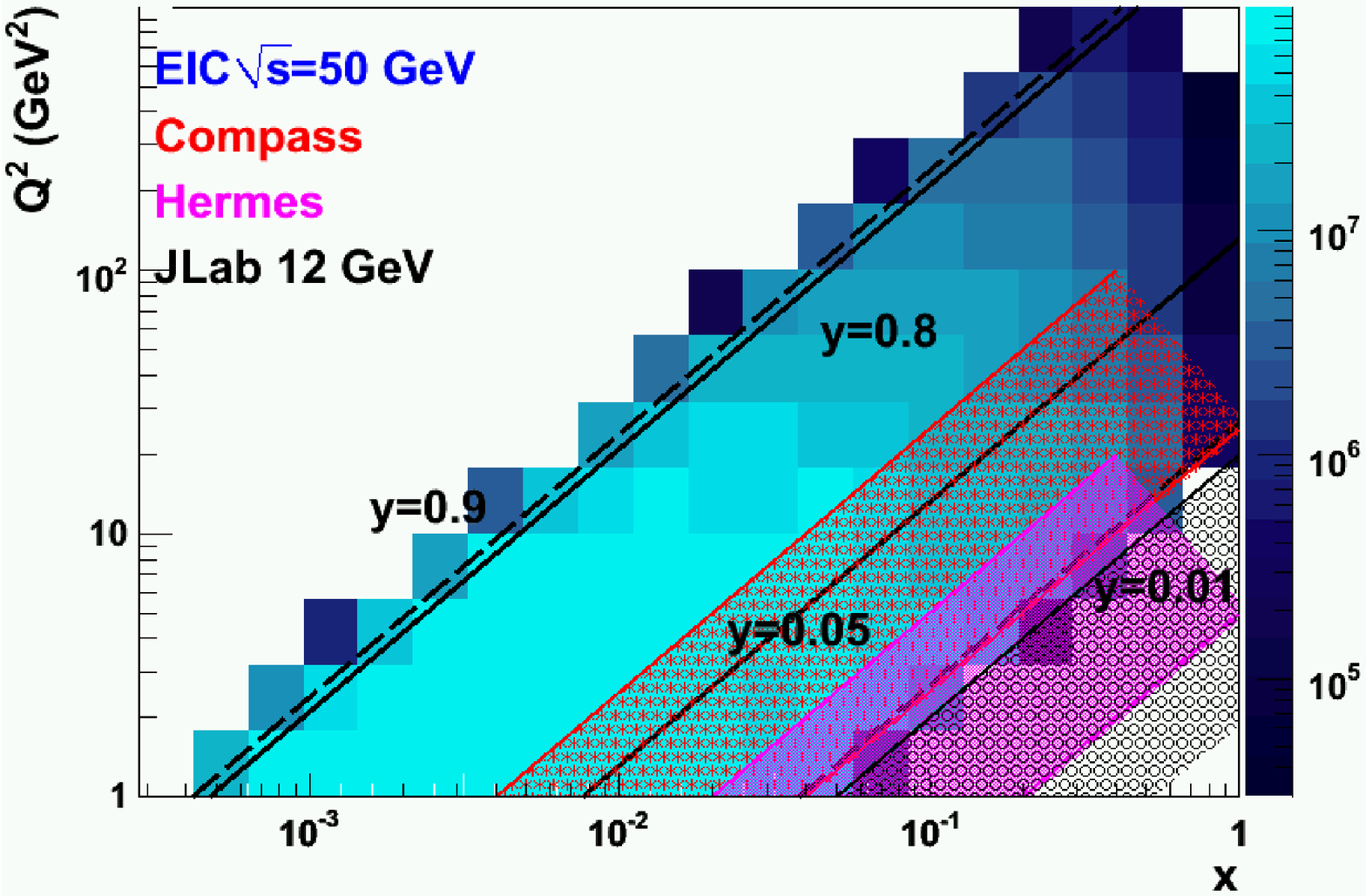} 
\caption{\label{fig:TMD-EIC-kine-ranges}
[color online]
Kinematic coverage in $x$ and $Q^2$ for the EIC for an energy setting of $\sqrt{s} = 50$ GeV
compared to the coverage of COMPASS, HERMES and future JLab12 experiments represented by the
red, purple and black hatched areas, respectively.
}
\end{figure}
%
A systematic and detailed study of the Sivers function, and TMDs in general, can only be 
performed on the basis of precise spin- and azimuthal-asymmetry amplitude measurements 
 in semi-inclusive DIS over a wide kinematic range.
The availability of experimental results that are fully differential in the kinematic variables 
$x$, $Q^2$, $z$ and $P_{hT}$ would be a great asset for phenomenological analyses,
as they permit testing the underlying perturbative QCD techniques and assumptions.
Particle identification over the full momentum range and measurements with both proton and (effective) neutron 
targets would allow for a full flavour separation of the distribution functions under study.

Planned experiments at the upcoming JLab12 facility aim at providing high precision 
semi-inclusive DIS data in the {\it valence} quark region at relatively low $Q^2$, taken 
with transversely polarized 
neutrons (HallA)~\cite{Gao:2010av}, protons and deuterons (CLAS12)~\cite{JLab12-Clas}.
The expected high luminosities should allow for fully differential extractions of the 
relevant azimuthal and transverse-spin asymmetries. 
The kinematic range of JLab12 experiments will be complementary to COMPASS 
measurements~\cite{compass-II}, 
partially overlap with those of HERMES, and provide data in the so-far unexplored high-$x$ region.     

The kinematic coverage of these experiments is compared in fig.~\ref{fig:TMD-EIC-kine-ranges} with
the coverage of an EIC for an energy setting of $\sqrt{s} = 50$ GeV. 
As discussed in sec.~\ref{sec:eRHIC-design} and sec.~\ref{sec:MEIC_design}, 
the ability to vary the energy of both the
electron and proton (ion) beams at the EIC provides variable energy in the range 
$\sqrt{s} = 15-65$ GeV or $\sqrt{s} = 45-200$ GeV depending on the realization options under 
discussion. 
This ability puts the EIC in the unique position of accessing the valence region at much larger 
$Q^2$ than current and near-future experiments (thereby suppressing potential higher twist
contaminations) while also accessing low $x$ down to values of about $10^{-5}$, where sea quarks 
and gluons could be studied in detail.
The expected high luminosity will allow for a fully differential analysis over almost the 
whole wide kinematic range.
In this section we will illustrate this potential for fully differential analyses of 
TMD observables
and test the sensitivity to sea quark distributions.
The unique features of the EIC for access to TMDs for gluons, a study of the evolution
properties of TMDs, and of the transition from low to high transverse momenta will be discussed, 
using the Sivers function as an example, in secs.~\ref{sec:TMD-gluons},~\ref{sec:TMD-evolution}
and~\ref{sec:TMD-matching}, respectively.    

\subsubsection{Generation of pseudo-data}

%
\begin{figure}[t]
\begin{center}
\includegraphics[width=0.65\textwidth]{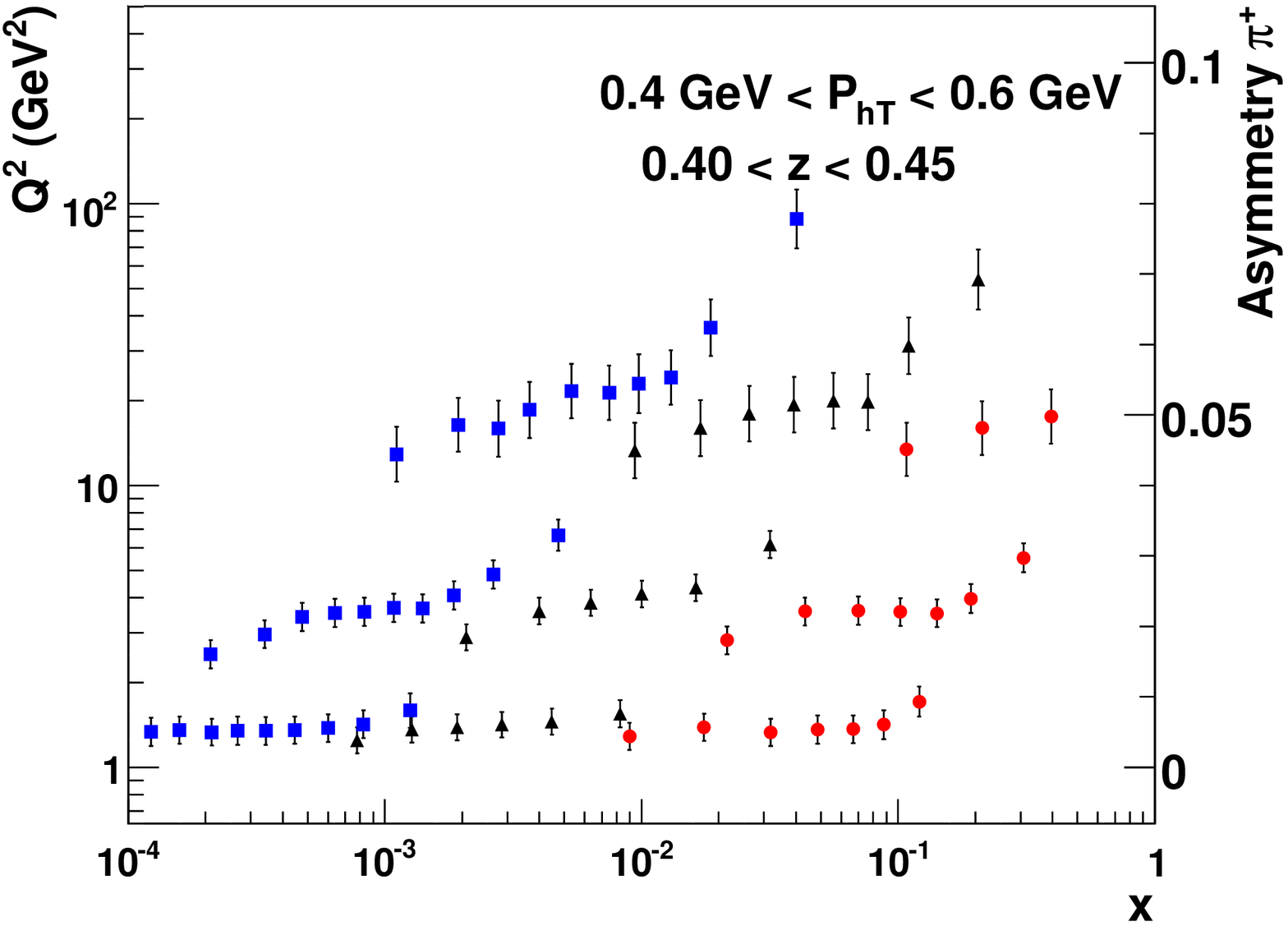}
\caption{\label{fig:TMD-asymm-simulation-pip-zoom}
\small
Projected accuracy for $\pi^+$ production in semi-inclusive DIS off the proton
for a particular $P_{hT}$ and $z$ range as indicated in the figure.
The position of each point is according to its $Q^2$ and $x$ value,
within the range $0.05 < y < 0.9$.   
The projected event rate, represented by the error bar, is scaled to the (arbitrarily chosen)
asymmetry value at the right axis.
The blue squares, black triangles and red dots represent the 
 $\sqrt{s} = 140 $ GeV,  $\sqrt{s} = 50 $ GeV and $\sqrt{s} = 15 $ GeV
EIC configurations, respectively. 
Event counts correspond to an integrated luminosity of 30 fb$^{-1}$ for 
each of the three configurations.
}
\end{center}
\end{figure}
%

The projections presented in the following for the Sivers asymmetry where estimated using
either modified existing Monte Carlo generators or 
standard parameterizations of the unpolarized parton distribution and fragmentation functions. 
%
%
Events were generated for $Q^2 > $ 1 GeV$^2$, 0.01 $< y <$ 0.9 and 0.1 $< z <$ 0.9, 
over the full kinematically allowed range in $x$.
At this stage no cuts were applied on the scattered electron or produced hadron.
Events were divided into four-dimensional ($x$, $Q^2$, $z$, $P_{hT}$) bins and the mean asymmetry 
in each bin was evaluated.
Full acceptance in azimuth was assumed and statistical uncertainties of 
$\sqrt{2/N}$ 
were assigned in each bin.
More details about the simulations can be found in~\cite{Anselmino:2011ay}.
For all projections shown in the following, no losses due to detector acceptance were applied, 
but an overall operational efficiency of 50\% was assumed.
The transverse proton beam polarization is set to 70\%.
No estimate of systematic uncertainties is applied.
%
%

Most of the projections will be given for an integrated luminosity of 4 fb$^{-1}$ or 30 fb$^{-1}$. 
These statistics would be achieved in approximately one week to one month 
(4 fb$^{-1}$) or one month to six month (30 fb$^{-1}$)
for luminosities ranging from 
1 $\times$ 10$^{34}$ cm$^{-2}$s$^{-1}$ to 3 $\times$ 10$^{33}$ cm$^{-2}$s$^{-1}$.
Therefore the statistical precision in the figures presented here should be understood 
as that achievable 
in a relatively brief period of operation for an EIC.
%
\\

\subsubsection{Four-dimensional mapping of the phase space}

%
\begin{figure}[t]
\begin{center}
\includegraphics[width=0.99\textwidth]{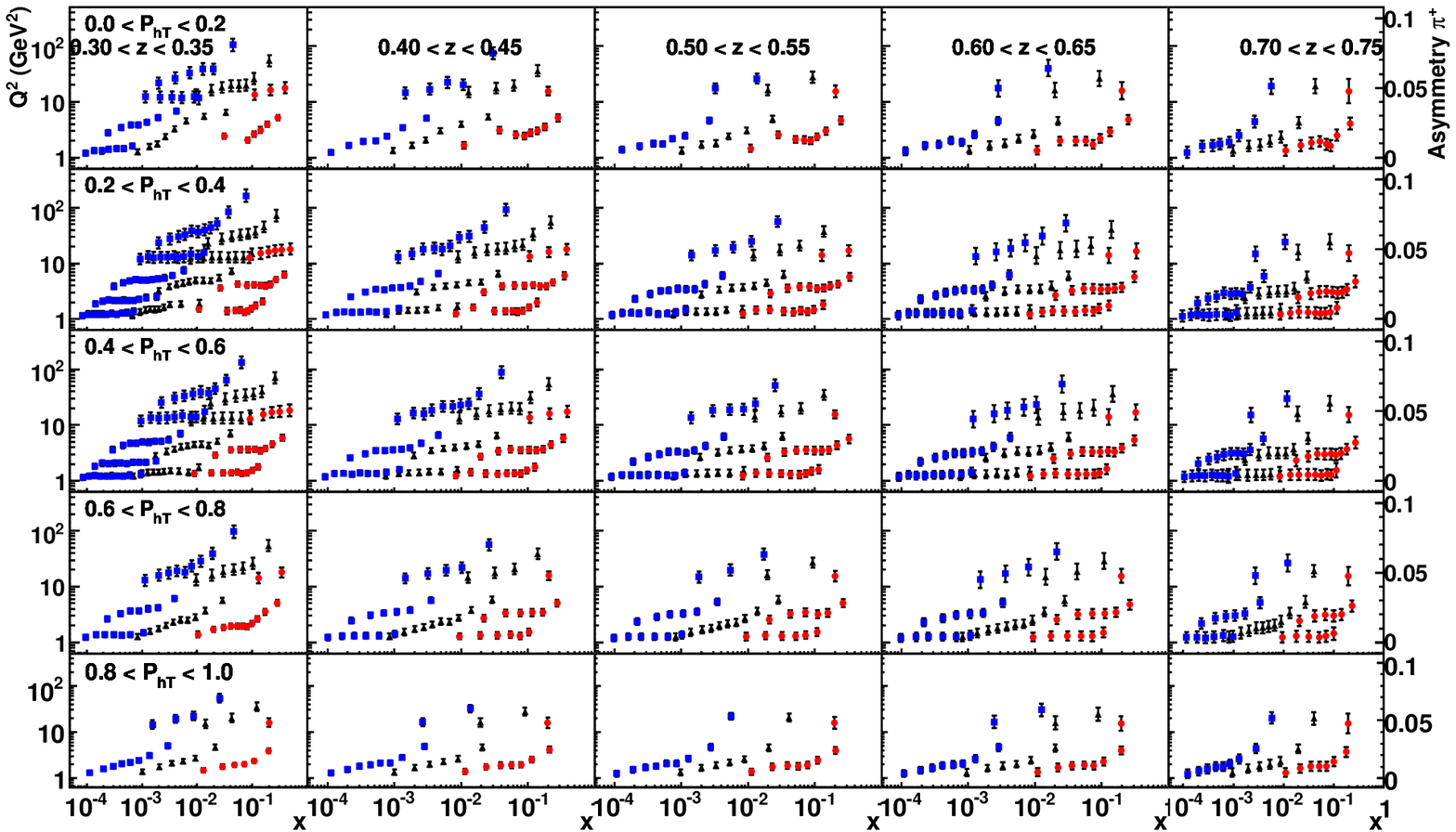}
\caption{\label{fig:TMD-asymm-simulation-pip}
\small
Four-dimensional representation of the projected accuracy for $\pi^+$ production in 
semi-inclusive DIS off the proton.
Each panel corresponds to a specific $z$ bin with increasing value from left to right and
a specific $P_{hT}$ bin  with increasing value from top to bottom, with values given in the figure.
The position of each point is according to its $Q^2$ and $x$ value,
within the range $0.05 < y < 0.9$.   
The projected event rate, represented by the error bar, is scaled to the (arbitrarily chosen)
asymmetry value at the right axis.
Blue squares, black triangles and red dots represent the 
 $\sqrt{s} = 140 $ GeV,  $\sqrt{s} = 50 $ GeV and $\sqrt{s} = 15 $ GeV
EIC configurations, respectively. 
Event counts correspond to an integrated luminosity of 30 fb$^{-1}$ for 
each of the three configurations.
}
\end{center}
\end{figure}

The great potential of the EIC for obtaining a fully differential mapping of almost the
entire phase space relevant for TMD studies is illustrated in
figs.~\ref{fig:TMD-asymm-simulation-pip-zoom} and~\ref{fig:TMD-asymm-simulation-pip}.
A wide $x$ and $Q^2$ range can be mapped using different beam energies. 
The projected accuracy for single $\pi^+$ production is given for a four-dimensional 
binning in the kinematic variables $x$, $Q^2$, $z$ and $P_{hT}$,
using three different energy configurations for the EIC ($\sqrt{s} = $ 15, 50 and 140 GeV) 
and an integrated luminosity of 30 fb$^{-1}$ for each configuration. 
Events are selected for $0.05 < y < 0.9$ and $W^2 > 5$ GeV$^2$.
For a clearer view and explanation of the presented projections, we show in 
fig.~\ref{fig:TMD-asymm-simulation-pip-zoom} one of the panels from
fig.~\ref{fig:TMD-asymm-simulation-pip} corresponding to a specific  $z$ and $P_{hT}$
range.
In both figures, the position of each point is according to its $x$ and $Q^2$ value 
(abscissa and left ordinate, respectively) and
each panel is for a specific $z$ and $P_{hT}$ bin as indicated in the figure.  
The projected event rate is
represented by the error bar scaled with respect to the (arbitrarily chosen) asymmetry value given 
at the right ordinate.

The simulations demonstrate that
a four-dimensional mapping of TMD observables for pions over the whole phase space of 
main interest, meaning $P_{hT}$ values of up to about \mbox{1 GeV},  
could be achieved in about 3-5 month of running for each energy configuration. 
Kaon rates are typically a factor 4-5 lower than those for pions and
a similar quality of data can be achieved within a correspondingly longer running time.


The strategy for a full flavour separation of the Sivers distribution, and TMD distributions 
in general, involves both pion and kaon identification over almost the whole momentum range
and measurements with proton and effective neutron targets.
For the latter, the usage of polarized $^3$He ions is foreseen for both EIC concepts.
Compared to the projections shown in fig.~\ref{fig:TMD-asymm-simulation-pip}, the
dilution factor of 1/3 has to be compensated with higher luminosities 
(respectively longer running times).
The resulting different phase space for the neutron measurements compared with the proton case 
due to the $Z/A$ factor entering the momentum distribution and the expected lower 
center-of-mass energy (by about 2/3) 
because of the different rigidities of the beams can be compensated to a large
extent by using the different beam energy settings.

\begin{figure}[t]
\begin{center}
\includegraphics[width=0.95\textwidth]{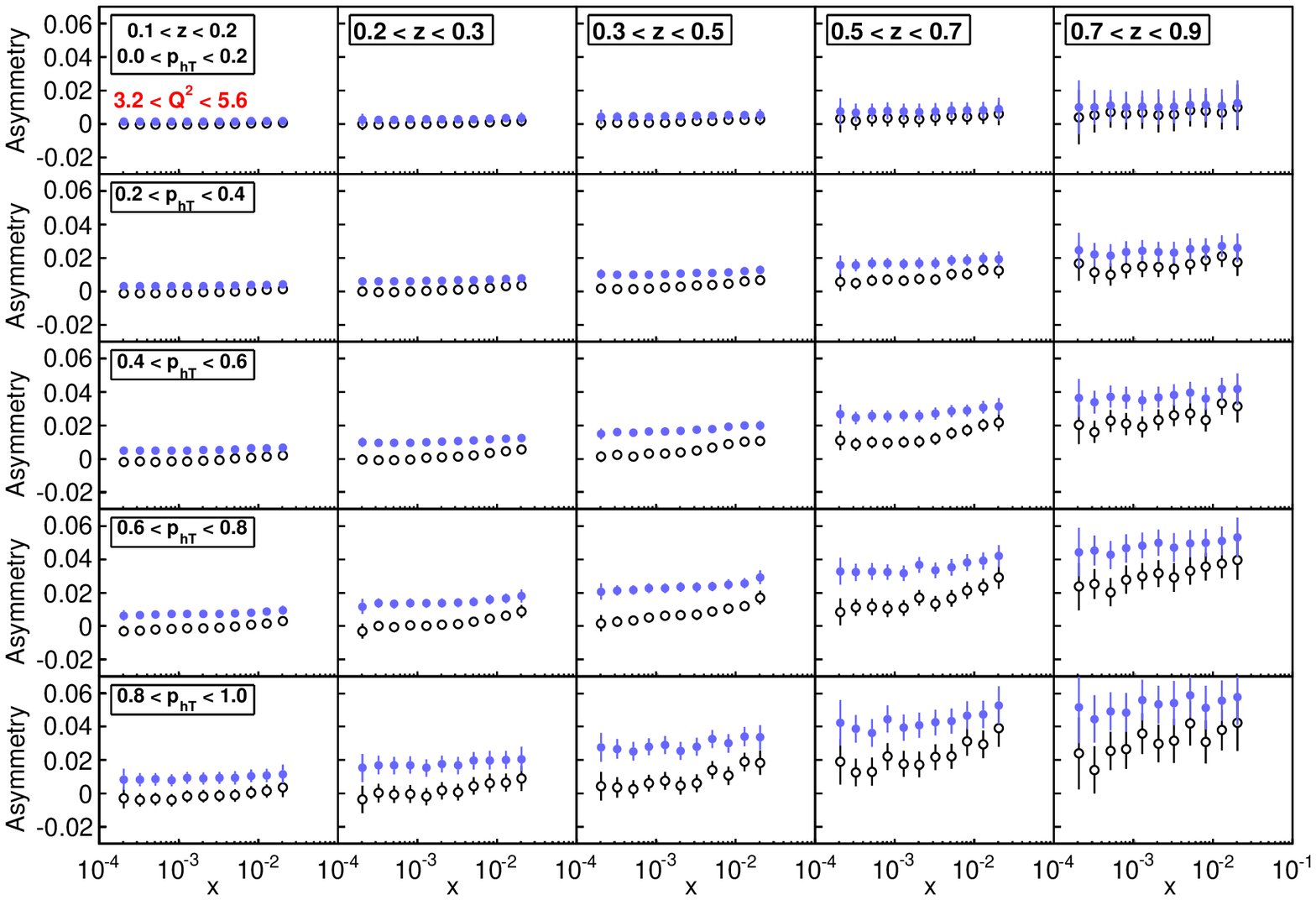}
\caption{\label{fig:TMD-Sivers-sea-pip}  
\small
Simulated Sivers asymmetry amplitudes for $\pi^+$, obtained with an 
energy of $\sqrt{s} = 140$ GeV, as a function of $x$ in bins in $z$, $P_{hT}$ and 
for a single bin in $Q^2$ as given in the panels. 
Closed blue (open black) dots correspond to (non-)zero Sivers functions for sea quarks.
Error bars represent the projected accuracy corresponding to an integrated luminosity 
of 4 fb$^{-1}$.
}
\end{center}
\end{figure}

In addition, valuable and necessary information about the transverse momentum dependence of 
the fragmentation process will be obtained from the same data using a 
fully differential extraction of the individual hadron multiplicities.

\subsubsection{Sensitivity to sea quarks}

\begin{figure}[t]
\begin{center}
\includegraphics[width=0.95\textwidth]{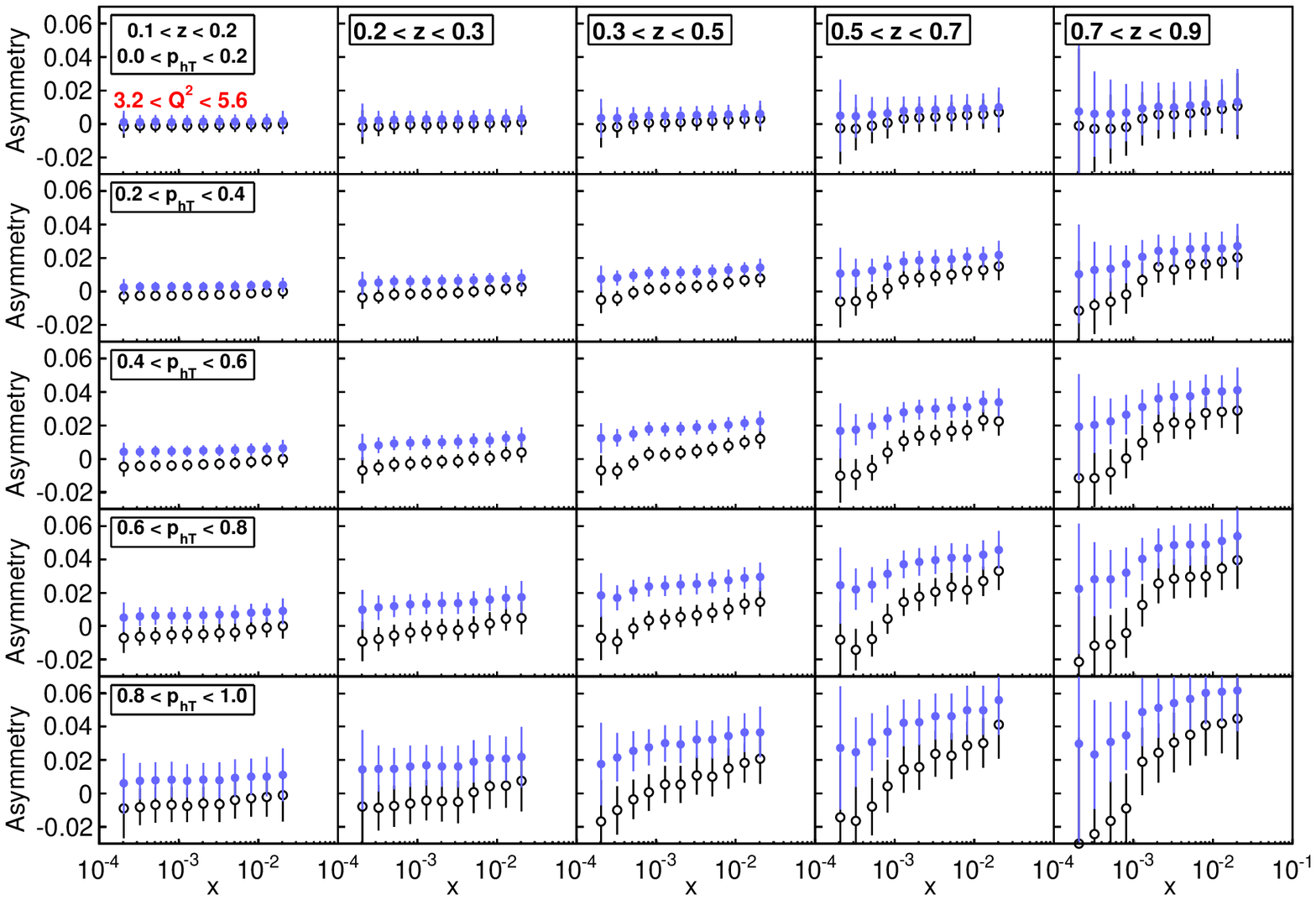}
\caption{\label{fig:TMD-Sivers-sea-Kpl}  
\small
Simulated Sivers asymmetry amplitudes for $K^+$, obtained with an 
energy of $\sqrt{s} = 140$ GeV, as a function of $x$ in bins in $z$, $P_{hT}$ and 
for a single bin in $Q^2$ as given in the panels. 
Closed blue (open black) dots correspond to (non-)zero Sivers functions for sea quarks.
Error bars represent the projected accuracy corresponding to an integrated luminosity 
of 4 fb$^{-1}$.
}
\end{center}
\end{figure}

Among the unique features of the EIC is its sensitivity for an exploration of the 
Sivers function for sea quarks, which are expected to play an important role 
in the lower $x$ region.
We investigate this sensitivity 
by generating two sets of events, one with and one without contributions from sea quarks.
As the Sivers distribution is essentially unknown, it was parameterized via a constant multiplied 
by the unpolarized PDF with independent constants for $u$, $d$ and sea quarks.
The Sivers asymmetry is returned by the generator on an event-by-event basis.
The unpolarized PDFs of \cite{Pumplin:2002vw} and the fragmentation functions of \cite{deFlorian:2007aj} 
were used.

In both cases, the same parameterization for up and down quark Sivers functions was used, which
were set equal to 25\% of the unpolarized distribution, but with opposite sign, i.e. 
$f_{1T}^{\perp u}(x) = -0.25f_1^u(x)$ and $f_{1T}^{\perp d}(x) = 0.25f_1^d(x)$.
In the first data set, the Sivers functions for sea quarks were also set to 25\% of the 
corresponding unpolarized distribution.
In the second data set the sea quark Sivers distributions were fixed to zero.
This allowed for a comparison of the case in which the sea quark Sivers function was significant 
compared to that of the valence quarks with the case of a vanishing sea quark contribution.

Figure~\ref{fig:TMD-Sivers-sea-pip} shows the asymmetry amplitudes for $\pi^+$, 
obtained with an energy of $\sqrt{s} = 140$ GeV, for a single bin 
in $Q^2$ as a function of $x$, binned in $z$ and $P_{hT}$ as indicated in the panels.
Open black dots represent the case of non-zero Sivers functions for sea quarks and 
closed blue dots the case of vanishing contributions.
Error bars correspond to an integrated luminosity of 4 fb$^{-1}$, already yielding sufficient
precision to resolve small resulting differences in the asymmetry.
Because of their different quark content,  kaon production is expected to have a higher
sensitivity to sea quark contributions.
Figure~\ref{fig:TMD-Sivers-sea-Kpl} shows the asymmetry amplitudes for $K^+$ where indeed
both scenarios are more distinct.
As for $\pi^+$, the estimate is based on an integrated luminosity of 4 fb$^{-1}$ and obtained 
with an energy of $\sqrt{s} = 140$ GeV.

The study demonstrates that even a relatively brief running of the EIC provides the potential to 
distinguish zero and non-zero Sivers functions for sea quarks.
Note that these parameterizations are intended \emph{not} as a prediction of what asymmetries 
will actually be seen at an EIC, but as an indicator of sensitivity given the expected 
statistical precision.

\subsubsection{Impact of the EIC}

The EIC will be the unique facility for exploring the Sivers function (and TMDs in general) for
sea quarks and the gluon, to study the evolution properties of TMD distributions and to investigate
experimentally the transition from low to high transverse momenta.
As discussed in sec.~\ref{sec:sivers-status},
our current knowledge is restricted to an essentially qualitative picture of the Sivers 
function. 
Available data permit to constrain parameterizations for up and down quarks only,
employing relatively simple functional forms.

\begin{figure}[t]
\begin{center}
\hspace*{-0.5cm}
\mbox{
\includegraphics[width=0.33\textwidth,angle=-90]{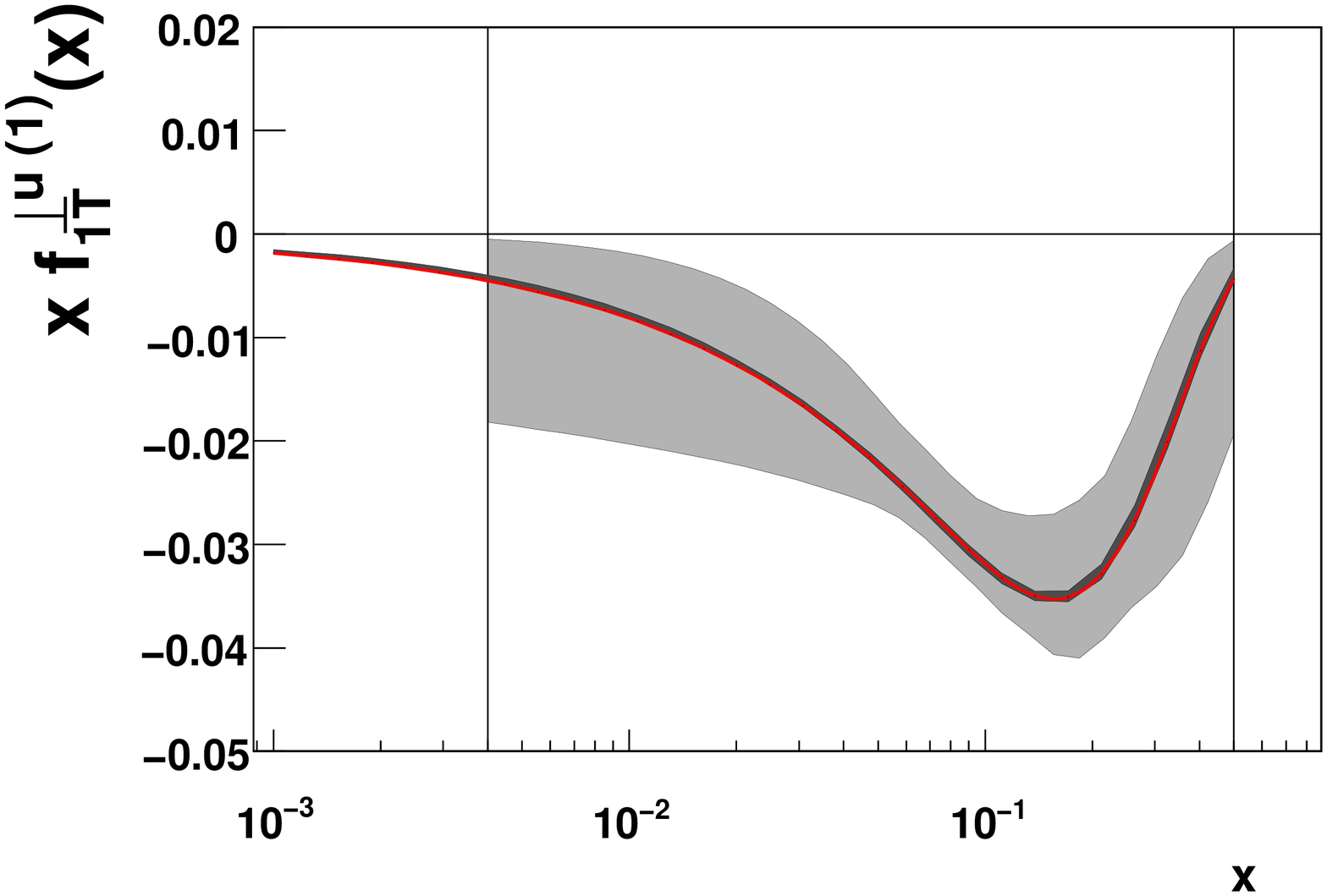} 
\includegraphics[width=0.33\textwidth,angle=-90]{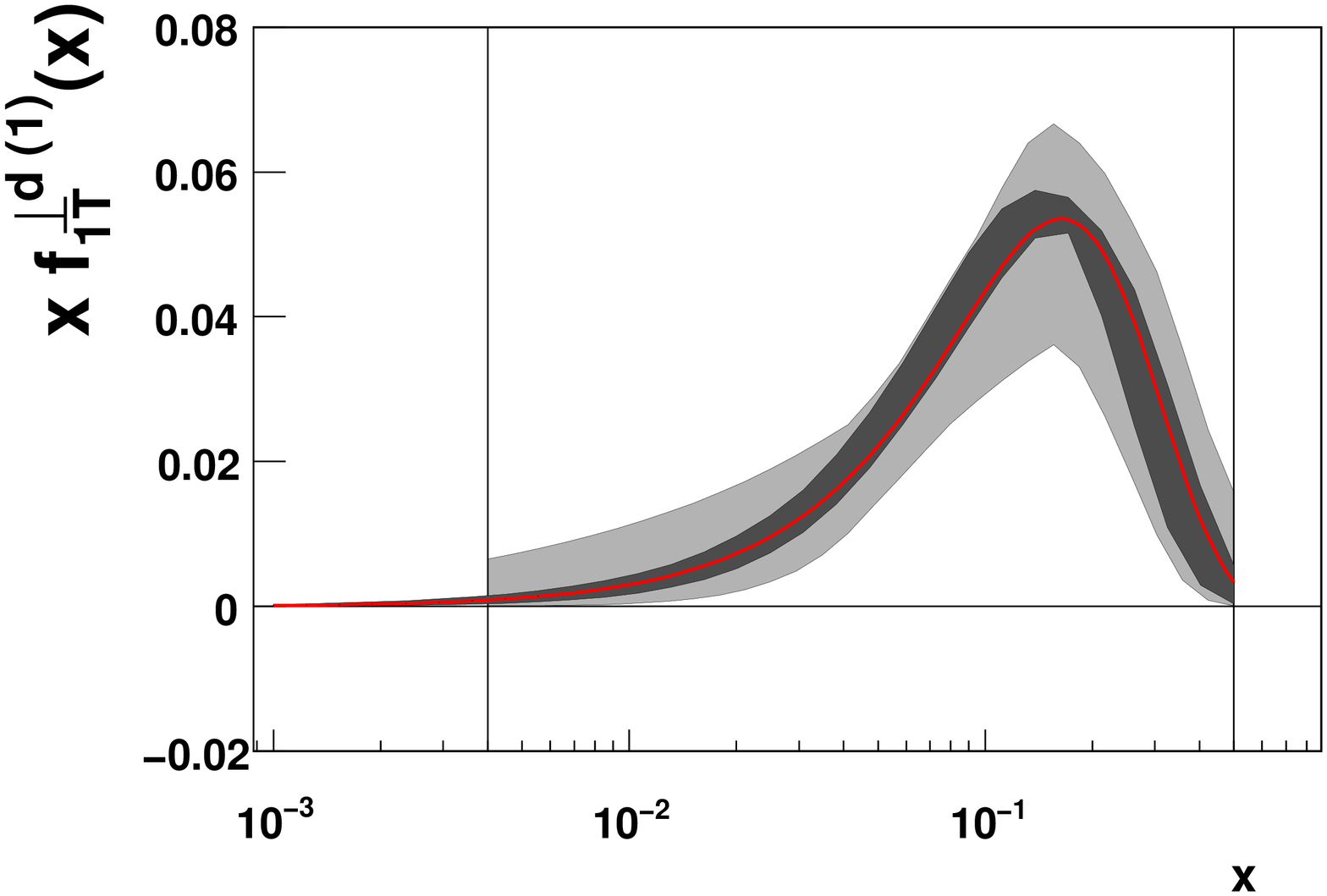} 
}
\caption{\label{fig:prokudin_sivers_eic_low_energy}
[color online]
Comparison of the precision (2-$\sigma$ uncertainty) of extractions of the Sivers function 
for $u$ quarks (left) and $d$ quarks (right) from currently available 
data~\cite{Anselmino:2010bs} (grey band)
and from pseudo-data generated for the EIC with energy setting of $\sqrt{s} = 45$ GeV
and an integrated luminosity of 4 fb$^{-1}$ (dark grey band around the red line).
The uncertainty estimates are for the specifically chosen underlying functional form
(see text for details).
}
\end{center}
\end{figure}
We illustrate the expected impact of data from the EIC using the parameterization from 
Ref.~\cite{Anselmino:2010bs}
as an arbitrarily chosen model of the Sivers function.
This parameterization, denoted $theor_i = F(x_i, z_i, P_{hT}^i, Q^2_i; {\bf a_0})$ 
with the $M$ parameters ${\bf a_0} = \{a_1^0, ..., a_M^0\}$ fitted to existing data,
serves to generate a set of pseudo-data in each kinematic bin $i$
\begin{equation}
f(value_i; theor_i, \sigma_i^2) = \frac{1}{\sqrt{2\pi \sigma_i^2}} e^{-(value_i-theor_i)^2/2\sigma_i^2}. 
\label{ea:normal_pdf}
\end{equation}
In each $x_i$, $Q^2_i$, $z_i$ and $P^i_{hT}$ bin, the obtained values, $value_i$, 
for the Sivers function are distributed using a
Gaussian smearing with a width  $\sigma_i$ corresponding to the simulated event rate
at an energy of $\sqrt{s} = 45$ GeV obtained with an integrated luminosity of 4 fb$^{-1}$.
For illustration of the obtainable statistical precision the event rate for  
the production of $\pi^+$ in semi-inclusive DIS was used.

This new set of pseudo-data was then analysed like the real data in Ref.~\cite{Anselmino:2010bs}.
Figure~\ref{fig:prokudin_sivers_eic_low_energy} shows the result
for the extraction of the Sivers function for  $u$ and  $d$ quarks. 
The central value of $f_{1T}^{\perp u}$, represented by the red line, follows by construction 
the underlying model.
The 2-sigma uncertainty of this extraction, valid for the specifically chosen
functional form, is indicated by the
dark grey area, which is hardly seen around the red line. 
This precision, obtainable with an integrated luminosity of 4 fb$^{-1}$, 
 is compared with 
the uncertainty of the extraction from existing data, represented by the light grey band
and shown before in  fig.~\ref{fig:prokudin_sivers_eic}.

Remembering that the event rate of the generated pseudo-data
is achievable in a brief period of 
operation for an EIC, the impressive impact of the EIC on studies of TMDs is greatly 
illustrated. 

\subsection{\label{sec:DY} TMDs in Drell-Yan processes}

One of the intriguing facets of the Sivers effect is its peculiar breaking of 
universality, as discussed in secs.~\ref{secI:theory} and~\ref{sec:TMD-theory}.
The symmetry properties of QCD require
a reversal of sign of the Sivers function when appearing 
in the Drell-Yan process, the production of di-lepton pairs in the collision of two hadrons,
as compared to DIS. The important test of this fundamental QCD prediction 
remains outstanding, its invalidation would have profound consequences for our understanding of 
high-energy reactions involving hadrons. 
It is thus not surprising to see the Drell-Yan process appear as a milestone measurement in 
the update for the future spin program at RHIC~\cite{rhic-spin-DY}.  

The Drell-Yan process with \emph{un}polarized hadrons has been studied at numerous 
fixed-target experiments~\cite{Falciano:1986wk,Guanziroli:1987rp,Anassontzis:1987hk,Zhu:2006gx,Zhu:2008sj}. 
There are several proposals for future polarized Drell-Yan measurements, 
either at fixed-target experiments 
(CERN, FermiLab, GSI, and J-Parc), but also at colliders (BNL, GSI). 
So far no measurement exists for Drell-Yan with transverse hadron polarization to 
isolate the Sivers effect, 
unlike the case for the related mechanism of the Boer-Mulders function. 
Being a naive-T-odd distribution
the latter also involves a reversal of sign when going from DIS to Drell-Yan.
For the Boer-Mulders function data from the Drell-Yan process exist. 
In particular the violation
of the Lam-Tung relation~\cite{Lam:1978pu} 
is a substantial hint of the Boer-Mulders effect,
as discussed in sec.~\ref{SubSec-V_2:Boer-Mulders-function}. 
However, 
being also a chiral-odd  distribution, presents an additional challenge for experimental 
measurements and their interpretation,
given that a second, presently poorly constrained, chiral-odd function is needed.  
In the case of Drell-Yan the other chiral-odd function is a second Boer-Mulders function, 
making it especially tricky to look for the sign change between Drell-Yan and DIS.

Among the proposed measurements of the Sivers effect in Drell-Yan two have  timescales of a 
few years from now. 
One is an experiment set at IP2 of RHIC (BNL) where transversely polarized ``beam" protons 
will interact with effectively unpolarized ``target" 
protons\footnote{The proton beams at IP2 are transversely polarized but due to rapid spin 
flips they can be spin balanced to get unpolarized protons.}~\cite{rhic-spin-DY}. 
At the COMPASS experiment at CERN it is not the beam---in this case consisting of pions---that will 
be polarized but the target~\cite{compass-II}.
This configuration is the theoretically more challenging one of the two as the partonic 
structure of the pion enters besides the structure of the proton. 

\begin{figure}[t]
  \centering
  \includegraphics[width=1.0\textwidth]{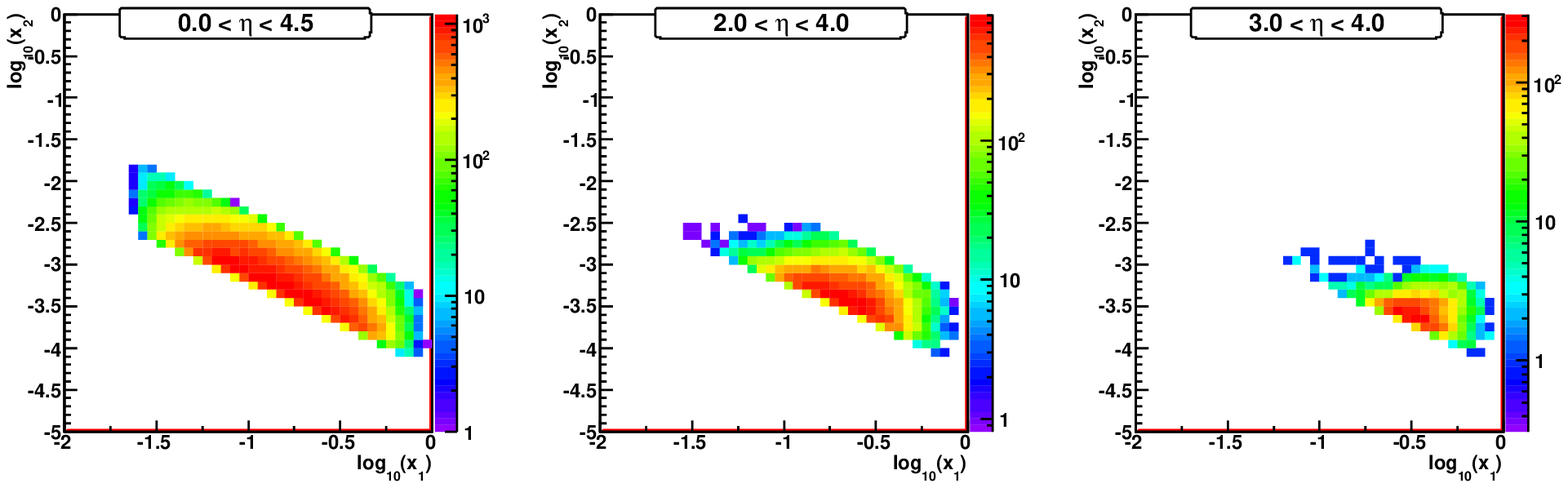}
  \caption{
\small
The correlation of the quark and antiquark
momentum fractions, $x_1$ and $x_2$, in Drell-Yan for different rapidity bins in 
proton-proton collisions at $\sqrt{s}$ = 500 GeV.
   \label{fig:DYkinematics}
}
\end{figure}

\begin{figure}[t]
  \centering
  \includegraphics[width=0.5\textwidth]{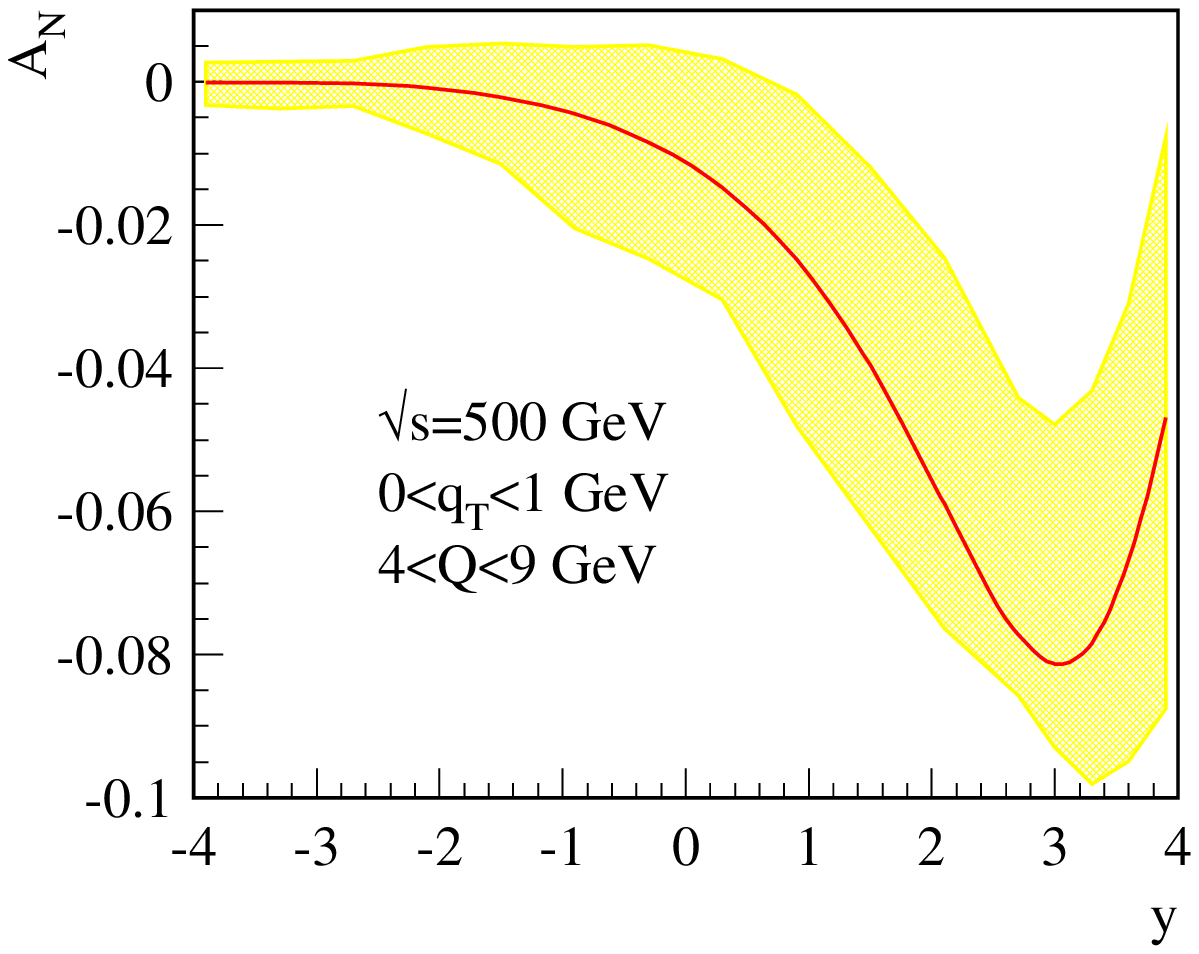}
  \caption{
\small
Sivers asymmetries for the Drell-Yan
process at RHIC, as a function rapidity for  \(\sqrt{s}=500\)~GeV~\cite{Kang:2009sm}. 
   \label{fig:DYprediction}
}
\end{figure}

The choice of measuring Drell-Yan single-spin asymmetries at a collider like RHIC has 
various advantages.  
Among others, the asymmetries depend only weakly on the partonic momentum \(x_2\) 
of the (anti)quark in the unpolarized nucleon. 
When integrated over \(x_2\)  the cross section increases with the center-of-mass energy 
\(\sqrt{s}\) as one can reach lower values of \(x_2\) where anti-quarks are more 
abundant. 
Furthermore, it is easier to differentiate between ``forward" and ``backward'' production 
at a collider allowing easy access to the valence region of the (transversely polarized) 
beam nucleon. 
In fig.~\ref{fig:DYkinematics}, we show the correlation of the quark and antiquark momentum fractions, 
$x_1$ and $x_2$, in the Drell-Yan (DY) process for different rapidity bins in proton-proton collisions at 
$\sqrt{s}$ = 500 GeV. The plot assumes an invariant mass range of the DY lepton 
pair between the $J/\psi$ and the Upsilon. To select DY at masses below the 
$J/\psi$ and/or at rapidities below 2.0 will be experimentally extremely challenging 
due to the dominance of the QCD $2\rightarrow 2$ processes ($>10^8$).
The expected single-spin asymmetry, $A_N$, 
is presented in fig.~\ref{fig:DYprediction} as a function of rapidity, $y$, for 
\(\sqrt{s}=500\)~GeV and integrated over the range \(4\div9\)~GeV in the invariant mass 
of the di-lepton pair~\cite{Kang:2009sm}.
The estimate makes use of a recent ``DIS'' Sivers function 
parameterization from fits to COMPASS and HERMES data~\cite{Anselmino:2008sga}.
Asymmetries of this size should be readily measurable with a limited data set.
Nevertheless, one should keep in mind that the change of sign applies to the 
flavor-dependent Sivers function. 
For a stringent test of this sign change it is therefore 
of utter importance not only to measure the Sivers effect in DIS and Drell-Yan,
but to perform a flavor-decomposition of the Sivers effect as well.
In \(pp\) collisions one will be mainly sensitive to the \(u\)-quark Sivers function 
due to the charge factor. 
Using pion beams one can vary the sensitivity to the various quark flavors via the 
choice of the pion charge as the valence anti-quark flavor in the pion will either 
be an anti-\(u\) or an anti-\(d\). 
This will help in a subsequent flavor decomposition of
the Sivers effect in Drell-Yan.


\subsection{Single-spin asymmetry in the collinear factorization:
Twist-three mechanism} \label{sec:twist3}





The quark Sivers function discussed in the last subsection is also
closely related to the twist-3 quark-gluon-quark correlation
functions in the collinear factorization approach which can
generate large single spin asymmetries in hard scattering process, 
in particular, in inclusive hadron production in $pp$ 
collisions. The single-transverse spin asymmetry in the process like $pp\to
\pi X$ is among the simplest spin observables in hadronic scattering.
One scatters a beam of transversely polarized protons off unpolarized
protons and measures the numbers of pions produced to either the
left or the right of the plane spanned by the momentum and spin directions of the initial
polarized protons.  Measurements of single-spin asymmetries in hadronic scattering
experiments over the past three decades have shown spectacular
results. Large asymmetries of up to several tens of percents were
observed at forward (with respect to the polarized initial beam)
angles of the produced pion. 
Despite the conceptual simplicity of $A_N$, the theoretical
analysis of single-spin asymmetries in hadronic scattering is
remarkably complex. The reason for this is that the asymmetry for
a single-inclusive reaction like $p p\to  \pi X$ is
power-suppressed as $1/\ell_\perp$ in the hard scale set by the
observed large pion transverse momentum. 
Power-suppressed contributions to hard-scattering processes are
generally much harder to describe in QCD than leading-twist ones.
In the case of the single-spin asymmetry,  a
complete and consistent framework could be developed
\cite{Efremov:1981sh,Qiu:1991pp,Qiu:1998ia,Eguchi:2006mc}. 
It is based on a collinear factorization theorem at non-leading 
twist that relates the single-spin cross
section to convolutions of twist-three quark-gluon correlation
functions for the polarized proton with the usual parton
distributions for the unpolarized proton and the pion fragmentation
functions, and with hard-scattering functions calculated
from an interference of two partonic scattering amplitudes:
one with a two-parton initial state and the other with a three-parton
initial state.

In the following, we briefly describe the collinear factorization formalism
for the twist-3 single-spin-dependent cross section in the
semi-inclusive deep inelastic scattering, $ep\to ehX$. This factorization
applies when the transverse momentum of the final state hadron
is large compared to the non-perturbative scale $\Lambda_{\rm QCD}$.
The usual leading twist spin-average cross section for this process can be schematically
written as
\begin{eqnarray}
d\sigma\sim \sum_{a,b}f_a(x,\mu)\otimes D_{h/b}(z,\mu) \otimes \hat{\sigma}^{ab}(x,z,Q,\mu),
\label{tw2}
\end{eqnarray}
where $f_a(x,\mu)$ and $D_{h/b}(z,\mu)$ ($a,b= q, \bar{q}, g$)
are, respectively, the parton density in the nucleon and
the fragmentation function for $b \to h$, convoluted with the hart part $\hat{\sigma}^{ab}$.
The twist-3 cross section relevant for SSA in $ep^\uparrow \to ehX$ takes the factorized form,
\begin{eqnarray}
&&d\sigma^{\rm tw3}\sim \sum_{a,b} G_a^{(3)}(x_1,x_2,\mu)\otimes D_{h/b}(z,\mu) \otimes
\hat{\sigma}_1^{ab}(x_1,x_2,z,Q,\mu)
\nonumber\\
&&\qquad\quad+\sum_{a,b} \delta f_a(x,\mu)\otimes D_{h/b}^{(3)}(z_1,z_2,\mu) \otimes
\hat{\sigma}_2^{ab}(x,z_1,z_2,Q,\mu) ,
\label{tw3}
\end{eqnarray}
where $\otimes$ represents the appropriate convolution, similarly
as the twist-2 factorization formula (\ref{tw2}), with the
relevant momentum fractions $x_{1,2}$, $z$, $x$, $z_{1,2}$
integrated over. $G_a^{(3)}(x_1,x_2,\mu)$ is the twist-3
distribution function in the transversely-polarized nucleon
$p^\uparrow$, and $D_{h/b}^{(3)}(z_1,z_2,\mu)$ is the twist-3
fragmentation function for the hadron $h$; the latter function is
chiral-odd, combined with the chiral-odd transversity distribution
$\delta f_a(x,\mu)$ for $p^\uparrow$. (In the TMD approach, the
first term in (\ref{tw3}) is described in terms of the Sivers
function, and the second term is described using the Collins
function.) These twist-3 distribution and fragmentation functions
describe the multi-parton correlations in the nucleon and in the
fragmentation process, respectively, and thus provides us with an
opportunity to reveal the more detailed internal structure of
hadrons beyond the parton-model picture.
Each twist-3 function has its own logarithmic scale dependence,
which differs from that of the twist-2 functions; for the
corresponding $\mu$-dependence, see section (\ref{sec:evolution}).

For $G_a^{(3)}(x_1,x_2,\mu)$ with $a=q$ in (\ref{tw3}), two
independent quark-gluon correlation functions, $G_F(x_1, x_2)$ and
$\widetilde{G}_F(x_1, x_2)$, participate. They are defined as
dimensionless, real, Lorentz-scalar functions in terms of nucleon
matrix element associated with the gluon field strength tensor
$F^{\alpha\beta}$ as well as the quark field $\psi$ on the
light-cone
\cite{Eguchi:2006mc,Eguchi:2006qz}.
Similarly, the twist-3 purely gluonic correlation functions
$O(x_1,x_2)$ and $N(x_1,x_2)$
as $G_{g}^{(3)}(x_1,x_2,\mu)$ in (\ref{tw3}), are defined through
the gauge-invariant lightcone correlation of three field-strength
tensors~\cite{Beppu:2010qn}.
Thus, a complete set of the twist-3 correlation functions in the
transversely-polarized nucleon is now provided by $G_F(x_1, x_2)$,
$\widetilde{G}_F(x_1, x_2)$, $O(x_1, x_2)$ and $N(x_1, x_2)$,
taking into account all symmetry constraints in QCD. We note that
the twist-3 correlation functions, $T_F(x_1, x_2)$,
$T_G^{(f,d)}(x,x)$, etc., used in the
literature~\cite{Kouvaris:2006zy,Kang:2008qh,Kang:2008ih} can be
expressed by the above correlation functions.

Another origin of SSA is in the fragmentation process for the
final hadron, as represented in terms of the twist-3 fragmentation
function $D_{h/b}^{(3)}(z_1,z_2,\mu)$ of (\ref{tw3}), which is
also defined as a multi-parton light-cone correlation function
(see~\cite{Yuan:2009dw}).


For SIDIS, $ep \to e h X$, the large transverse-momentum
$P_{hT}$ of the  hadron $h$ should come from a perturbative
mechanism, i.e.\ from the recoil from the hard (unobserved) final-state
partons. Then, the factorization formula (\ref{tw3}) is derived in
the LO perturbative QCD, manifesting their gauge invariance at the
twist-3 level, and a practical procedure to calculate the relevant
partonic hard part $\hat{\sigma}_{i}^{ab}$ is provided in
\cite{Eguchi:2006mc,Beppu:2010qn,Yuan:2009dw}: an extra gluon,
which emanates from nonperturbative multi-parton correlation and
carries the momentum fraction $x_2 -x_1$,
participates in the partonic hard scattering. The coupling of this
gluon allows an internal propagator in the partonic subprocess to
be on-shell, and this produces the required imaginary phase.
The results for those partonic subprocesses
imply~\cite{Eguchi:2006mc,Koike:2006qv,Koike:2009yb},
\begin{eqnarray}
\!\!\!\!\!
\frac{d^5\sigma^{\rm tw3}}{dx_{B}dQ^2 dz_h dP_{hT}^2 d\phi_h}
&& \!\!\!\!\!\!
\!\!\!\!
=\sin(\phi_h-\phi_S)\,F^{\sin(\phi_h-\phi_S)}
+\sin(2\phi_h-\phi_S)\,F^{\sin(2\phi_h-\phi_S)}
\nonumber\\
&&
\!\!\!\!\!\!\!\!\!\!\!\!\!\!\!\!\!\!\!\!\!\!\!\!\!\!\!\!
+\sin\phi_S\,F^{\sin\phi_S}
+\sin(3\phi_h-\phi_S)\,F^{\sin(3\phi_h-\phi_S)}
+\sin(\phi_h+\phi_S)\,F^{\sin(\phi_h+\phi_S)},
\label{tw3azim}
\end{eqnarray}
with the azimuthal angles $\phi_h$ and $\phi_S$ of $P_{hT}$
and $S_\perp$, respectively, measured from the lepton plane; the
five azimuthal dependences in (\ref{tw3}) are similar as those in
the TMD approach. Here, each structure function $F^{\sin(\cdots)}$
is expressed in a factorized form, convoluted with $G_F(x,x)$ and
$dG_F(x,x)/dx$.
The similar twist-3 effects from $G_F$ and $\widetilde{G}_F$ have
been investigated for SSA in Drell-Yan and direct photon
productions, and hadron production in $pp$ collisions.

Charm production in SIDIS and $pp$ collisions is useful to
study the twist-2 gluon distributions in the nucleon, since the
$c\bar{c}$-pair creation through the photon-gluon or gluon-gluon
fusion is their driving subprocess. Likewise, the three-gluon
correlation functions
can be probed by SSA in these processes. From this point of view,
the three-gluon contribution to SSA in $D$-meson production processes,
$ep^\uparrow\to eDX$ and $p^\uparrow p\to DX$, have been studied
in \cite{Kang:2008qh,Kang:2008ih,Beppu:2010qn,Koike:2010jz}. For
both processes, the twist-3 cross sections for SSA can be derived
entirely as the gluonic pole contribution leading to $x_1=x_2$,
and thus receive the contributions $O(x,x)$, $O(x,0)$, $N(x,x)$
and $N(x,0)$ (and their
derivatives)~\cite{Beppu:2010qn,Koike:2010jz}. The result for
$ep^\uparrow\to eDX$ has five azimuthal dependences like 
in (\ref{tw3azim})~\cite{Beppu:2010qn}.

So far, RHIC at BNL reported a significant amount of data of $A_N$
for $p^\uparrow p\to hX$ ($h=\pi,K,\eta,D,\jpsi$). Given that the
NLO QCD in collinear factorization can provide a reasonable
description of the corresponding unpolarized cross section, we
expect that one can apply the above twist-3 formalism to analyze
the $A_N$ data~\cite{Qiu:1998ia,Kouvaris:2006zy,Kanazawa:2010au}.
The complete LO QCD formula for $A_N$ from the twist-3 quark-gluon
correlation functions 
to $p^\uparrow p\to hX$ has been derived: It consists of the
contribution associated with $G_F(x,x)$ and $dG_F(x,x)/dx$
($\widetilde{G}_F(x,x)=0$)~\cite{Kouvaris:2006zy}, and the
contribution~\cite{Koike:2009ge} associated with $G_F(x,0)$ and
$\widetilde{G}_F(x,0)$. Phenomenological analysis of RHIC data
shows that both contributions are important, although the main
contribution comes from the $G_F(x,x)$
contribution~\cite{Qiu:1998ia,Kouvaris:2006zy}, the $G_F(x,0)$
($\widetilde{G}_F(x,0)$) contribution also plays an important
role, and the combination of both contributions provides a
reasonable description of the RHIC data, shedding light on
the behavior of $G_F$ and
$\widetilde{G}_F$~\cite{Kanazawa:2010au}. There are also some
initial efforts to calculate the twist-3 fragmentation
contribution to $A_N$~\cite{Kang:2010zzb}. Global analysis of
RHIC and future EIC data is expected to reveal more details on the
role of the multi-parton correlations, including the three-gluon
correlation functions.
%
%
%

\begin{figure}
\begin{center}
\mbox{
\hspace*{-0.5cm}\includegraphics[width=0.55\textwidth]{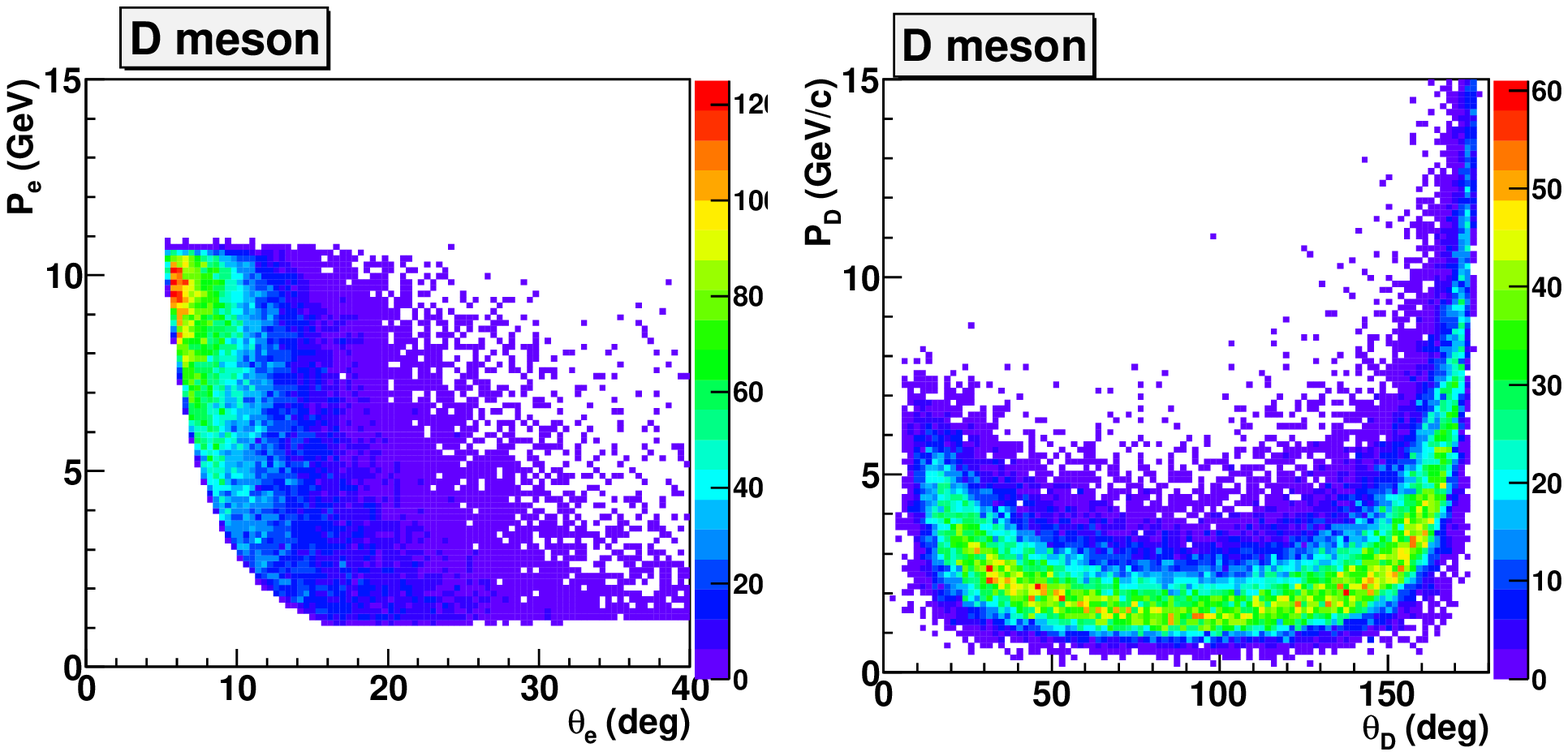}
\hspace*{0.3cm}\includegraphics[width=0.45\textwidth]{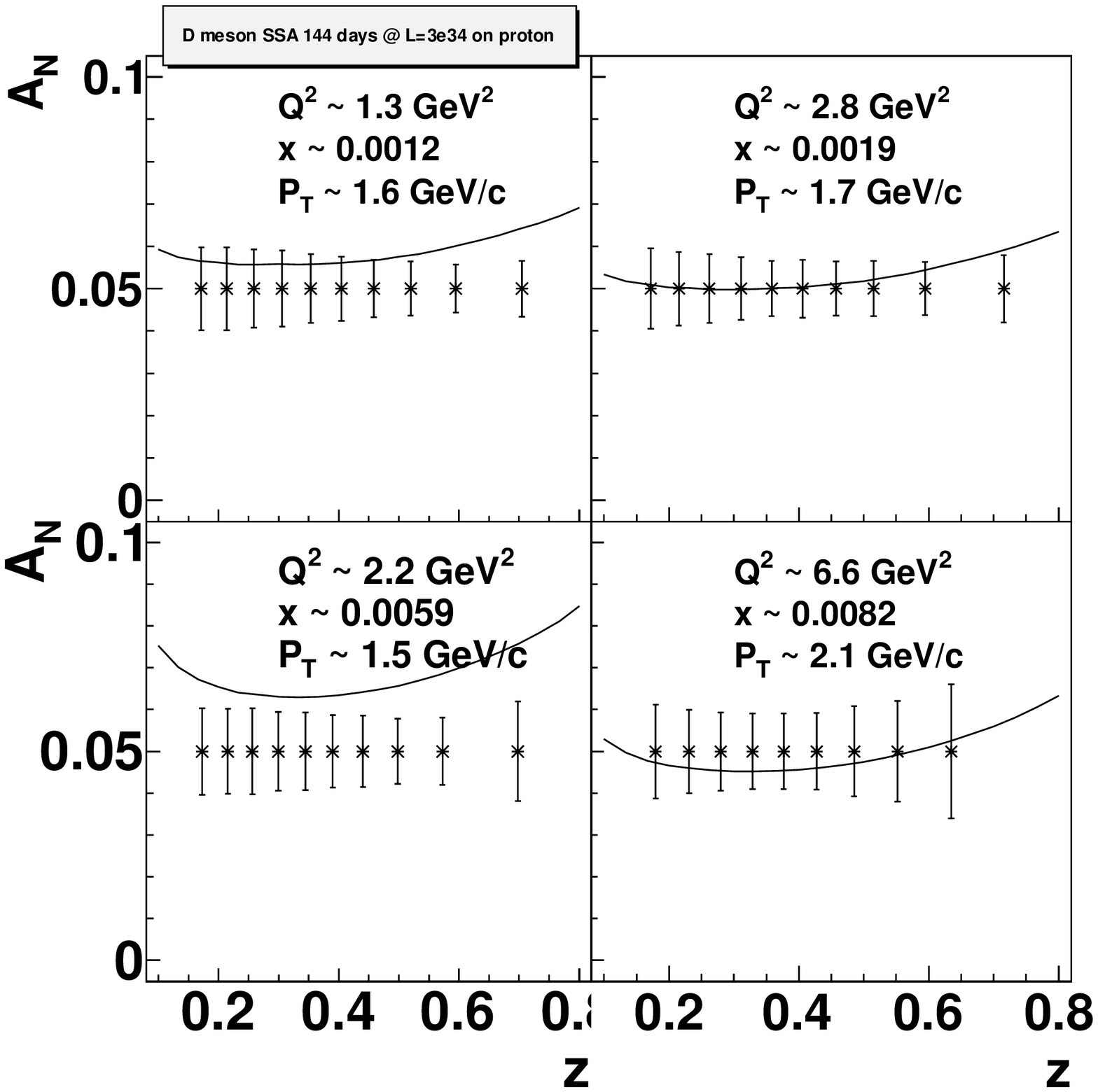}
}
\end{center}
\caption{
\label{fig:simulation-Dmeson-tw3} 
\small
Left: kinematics for $D$-meson events showing the momentum vs. polar angle
distribution for the electron and $D$ (or equivalently $\bar{D}$) meson in the laboratory frame.
Right:
Projected accuracy for transverse single-spin asymmetries from single $D$ meson
production using an energy of $\sqrt{s} = 50$ GeV 
and an integrated luminosity of 370 fb$^{-1}$.
}
\end{figure}

The potential of the EIC for measuring transverse single-spin asymmetries in 
charm production is illustrated in fig.~\ref{fig:simulation-Dmeson-tw3}.
In the simulation, based on the PYTHIA event generator, the main decay channel for $D$ mesons,
$D \rightarrow \pi^+K^-$, with a branching ratio of $3.8 \pm 0.1 \%$ is investigated.
Events are selected for $P_{hT} > 1$ GeV and $Q^2 > 1$ GeV within $0.05 < y < 0.9$ and 
$1.86 < M_D < 1.87$. 
The signal-to-background ratio for the reconstructed $D$ mesons strongly depends on the detector 
resolution.
In this study, we assume a momentum resolution of 
$0.8\% \cdot \frac{p}{10\ \mbox{GeV}}$ and a resolution of the polar and azimuthal angles 
of 0.3 mrad and 1 mrad, respectively. 
The resulting resolution of the reconstructed invariant mass of the $D$ meson 
is 1.8 MeV yielding an overall signal-to-background ratio of about 1.6 to 1. 
The overall detection efficiency for this triple coincidence process is assumed 
to be 60\%. The polarization of the proton beam is set to 80\%.

The projected accuracy for measuring transverse single-spin asymmetries in single $D$ meson
production is shown in fig.~\ref{fig:simulation-Dmeson-tw3} (right) as a function
of $z$ for different regions in $Q^2$, $x$ and $P_{hT}$, as indicated in the figure, together
with model calculations of the asymmetry from Ref.~\cite{Kang:2008qh}.
An  energy of $\sqrt{s} = 50$ GeV 
and an integrated luminosity of 370 fb$^{-1}$ were used.  
The study demonstrates a very promising feasibility of extracting observables
involving charm production.
It will significantly benefit from higher energies up to  $\sqrt{s} = 200$ GeV.

In summary, the twist-3 collinear factorization framework provides
us with a systematic way for describing SSA in the region of large
transverse-momentum $P_{hT}$ of the final hadron, and is thus
complementary to the TMD description of SSA which is valid in the
low $P_{hT}$ region. For the twist-3 distribution functions in
the transversely-polarized nucleon, relevant to SSA, there are two
independent quark-gluon correlation functions and the two
independent three-gluon correlation functions, all of which are
process-independent. 
Twist-3 cross section formulae
for SSA are available for many important processes, which can be
used for confronting with the RHIC and EIC data and may
serve to reveal multi-parton correlation effects in QCD hard
processes.

\subsection{Unifying the Mechanisms for the Sivers effect}
\label{sec:TMD-matching}


Recent developments have shown that the TMD approach and the
collinear factorization approach can be unified to describe the
Sivers effect for the single transverse-spin asymmetries in semi-inclusive DIS.
The TMD approach covers the
kinematic region $P_{hT} \ll Q$ where $Q \gg \Lambda_{\rm
QCD}$, while the twist-3 approach covers the large $P_{hT}$
region, $P_{hT} \gg \Lambda_{\rm QCD}$. A natural question
here is whether the two mechanisms give rise to equivalent (or
consistent) SSA in the overlapping region, $\Lambda_{\rm QCD} \ll
P_{hT} \ll Q$. To address this issue, we first recall the
relation between the Sivers function $f_{1T}^\perp(x,k_\perp)$ and
the quark-gluon correlation function $G_F(x,x)$ 
\cite{Boer:1997bw}:
$\int\,d\boldsymbol{k}_\perp^2\,\boldsymbol{k}_\perp^2f_{1T}^\perp(x,k_\perp)=\pi
M_N^2 G_F(x,x)$, which indicates that the two mechanisms are
closely related.

A more explicit relation for the SSA in the two approaches has also
been derived for the Sivers cross section,
$F^{\sin(\phi_h-\phi_S)}$ in
(\ref{tw3azim})~\cite{Ji:2006ub,Ji:2006br,Koike:2007dg}: In the
TMD approach, $F^{\sin(\phi_h-\phi_S)}$ is expressed in terms of
the Sivers function $f_{1T}^\perp(x,k_\perp)$. In the large
$k_\perp$-region, relevant to $\Lambda_{\rm QCD} \ll P_{hT}
\ll Q$, the $k_\perp$-dependence of $f_{1T}^\perp(x,k_\perp)$ can
be generated perturbatively, such that $f_{1T}^\perp(x,k_\perp)$
is expressed as the convolution of the corresponding perturbative
coefficient functions with the nonperturbative correlation
functions $G_F$ and $\widetilde{G}_F$.
By inserting this form of $f_{1T}^\perp(x,k_\perp)$ into the TMD
factorization formula for $F^{\sin(\phi_h-\phi_S)}$, one obtains
the cross section written in terms of $G_F$ and $\widetilde{G}_F$,
and this expression turns out to be identical to the leading
$P_{hT}$ behavior of the twist-3 mechanism for
$F^{\sin(\phi_h-\phi_S)}$ in the overlap region $\Lambda_{\rm QCD}
\ll P_{hT} \ll Q$. From these studies, the two mechanisms for single-spin asymmetries
represent a unique QCD effect over the entire $P_{hT}$ region.
The same equivalence was also shown for the SSA in the Drell-Yan
process. It should be noted that the sign of the Sivers function
changes from SIDIS to the Drell-Yan case, while the twist-3
quark-gluon correlation functions
are process-independent. The connection between the two mechanisms
is also consistent with such
process-(in)dependence~\cite{Ji:2006vf}.

The contribution from the twist-3 fragmentation function in
(\ref{tw3}) gives rise to the structure function
$F^{\sin(\phi_h+\phi_S)}$ in (\ref{tw3azim}), and dominates the
leading $P_{hT}$ behavior of $F^{\sin(\phi_h+\phi_S)}$
compared to that from the quark-gluon correlation functions. This
leading $P_{hT}$ behavior in $F^{\sin(\phi_h+\phi_S)}$ turns
out to be identical to the corresponding contribution from the
Collins function in the TMD approach, similarly as the above
equivalence for $F^{\sin(\phi_h-\phi_S)}$~\cite{Yuan:2009dw}.

These are nontrivial and important results, which demonstrate that
we indeed have a unique picture for single transverse-spin
asymmetries in DIS and hadronic collisions. The discussion can be
further generalized to other structure functions in SIDIS as well.


To analyze the general power behavior of the structure functions,
it is important to realize that the power expansions are done in
two different ways in the above two descriptions. At low $q_T$,
first we expand in $(q_T/Q)^{n-2}$ and neglect terms with $n$
bigger than a certain value (so far, analyses have been carried
out only up to $n=3$, i.e., twist-3). To study the behavior at
intermediate $q_T$ we further expand in $(M/q_T)^k$. Vice versa,
at high $q_T$ we first expand in $(M/q_T)^n$ (also in this case,
analyses are available up to $n=3$, i.e., twist-3). To study the
intermediate-$q_T$ region, we further expand in $(q_T/Q)^{k-2}$.
We can encounter two different situations:
\begin{itemize}
\item{ Type-I observables, where the leading terms at high and low
transverse momentum  have the same behavior. For instance,
\begin{equation}
  \label{low-ex1}
F(q_T,Q) =
  A\, \biggl[\frac{q_T}{Q}\biggr]^{0}\,
            \biggl[\frac{M}{q_T}\biggr]^{2}
+ B\, \biggl[\frac{q_T}{Q}\biggr]^{2}\,
            \biggl[\frac{M}{q_T}\biggr]^{2}
+ \ldots \,,
\end{equation}
where the term $A$ is leading
  in both the low- and high-$q_T$ calculations.
In this case, the calculations at high and low transverse momentum
must yield exactly the same result at intermediate transverse
momentum~\cite{Collins:1984kg,Ji:2006ub}. If a mismatch occurs, it
means that one of the calculations is incorrect or incomplete. }
\item{ Type-II observables, where the leading terms at high and
low transverse momentum have different behavior. For instance,
\begin{equation}
F(q_T,Q) =
  A'\, \biggl[\frac{q_T}{Q}\biggr]^{0}\,
            \biggl[\frac{M}{q_T}\biggr]^{4}
+ B'\, \biggl[\frac{q_T}{Q}\biggr]^{2}\,
            \biggl[\frac{M}{q_T}\biggr]^{2}
+ \ldots \,. \label{e:typeII}
\end{equation}
where the first term is leading and the second term sub-leading in
the low-$q_T$ calculation, whereas the reverse holds in the
high-$q_T$ calculation. In this case, if the calculations at high
and low transverse momentum are performed at their respective
leading order, they describe two different mechanisms and will not
lead to the same result at intermediate transverse momentum. In
order
 to
``match'', the calculations should be carried out in both regimes
up to the sub-subleading order. We could call this situation an
``expected mismatch'', since it is simply due to the difference
between the two expansions. }
\end{itemize}

In Tab.~\ref{tab:overview} we list the power behavior of the
structure functions at intermediate transverse momentum, as
obtained from the limits of the low-$q_T$ and high-$q_T$
calculation. For details of the calculation, we refer
to~\cite{Bacchetta:2008xw}.
The structure functions with a ``yes'' or ``no'' in the last
column of Tab.~\ref{tab:overview}   are type-I observables, where
on the basis of power counting we know that two calculations
describe the same physics and should therefore exactly match. In
these cases, the high-$q_T$ calculation describes the perturbative
tail of the low-$q_T$ effect. The two mechanisms need not be
distinguished. Using resummation it should be possible to
construct expressions for these observables that are valid at any
$q_T$.
Six of these structure functions have been calculated explicitly.

For the functions  identified as type-II in the last column of Tab.~\ref{tab:overview},
the low-$q_T$ and high-$q_T$ calculations at
leading order pick up two different components of the
full structure function. They therefore describe two different
mechanisms and do not match.
For such type-II observables, if one aims at studying the leading-twist
contribution from transverse momentum distributions, some
considerations have to be kept in mind:
\begin{itemize}
\item{the leading contribution from the high-$q_T$ calculation
(often referred
    to as a pQCD or radiative correction) is a competing
    effect that has to be taken into
    account~\cite{Oganesian:1997jq,Barone:2008tn,Barone:2009hw};}
\item{$q_T$-weighted asymmetries enhance the high-$q_T$ mechanism
and thus are
    not appropriate to extract type-II TMDs;}
\item{it is at present impossible to construct an expression that
extends the
    high-$q_T$ calculation to $q_T \approx M$, since this requires a smooth
    merging into unknown twist-4 contributions, which most probably cannot be
    factorized (see also Ref.~\cite{Berger:2007jw});}
\item{it is desirable from the experimental point of view to build
observables
  that are least sensitive to the effect of radiative corrections.}
\end{itemize}
We stress that the above considerations apply not only to
semi-inclusive DIS, but also to Drell-Yan and $e^+ e^-$
annihilation~\cite{Boer:2008fr}, which have been already used to
extract the Boer--Mulders and Collins
functions~\cite{Barone:2009hw,Lu:2009ip}.

\begin{table}[t]
\renewcommand{\arraystretch}{1.3}
\begin{tabular}{|l|c|c|c|} \hline
structure & {low-$q_T$}
 & {high-$q_T$} & exact \\
function & power & power & match\\ \hline $F^{}_{UU,T}$ &
$1/q_T^{2}$
              &  $1/q_T^{2}$ & yes \\
$F^{\cos 2\phi_h}_{UU}$ &  $1/q_T^{4}$
                        &  $1/Q^{2}$ & type\! II \\
$F^{\sin 2\phi_h}_{UL}$ &  $1/q_T^{4}$ &  & (type\! II)   \\
$F^{}_{LL}$ &  $1/q_T^{2}$
            &  $1/q_T^{2}$ &  yes \\
\hline
\end{tabular}
\hfill
\begin{tabular}{|l|c|c|c|} \hline
structure & {low-$q_T$}
 & {high-$q_T$} & exact \\
function & power & power & match\\ \hline
$F^{\sin(\phi_h-\phi_S)}_{UT,T}$ &  $1/q_T^{3}$
                                 &  $1/q_T^{3}$ &  yes \\
$F^{\sin(\phi_h+\phi_S)}_{UT}$ &  $1/q_T^{3}$
                               &  $1/q_T^{3}$ &  yes \\
$F^{\sin(3\phi_h-\phi_S)}_{UT}$
                      & $1/q_T^{3}$
                      &  $1/(Q^{2}\ms q_T)$ & type\! II \\
$F^{\cos(\phi_h-\phi_S)}_{LT}$ &  $1/q_T^{3}$ &  & (yes) \\
\hline
\end{tabular}
\caption{\label{tab:overview} 
\small
Behavior of SIDIS structure
functions in
  the region $M\ll q_T \ll Q$, as deduced from the low-$q_T$ calculation based
  on TMD factorization and the high-$q_T$ calculation based on collinear
  factorization. Empty fields indicate that no calculation is available.
  The last column indicates whether the expressions match exactly, do not match
  exactly, or should not
  be expected to match. In parentheses: expected answers based on
  analogy, rather than actual calculation.
}
\end{table}

%
%

In summary, at the moment there is the hope to build descriptions
of the structure functions that go from low to high transverse
momentum for the five structure functions with a ``yes'' in the
last column of Tab.~\ref{tab:overview}.


\subsection{From low to high transverse momentum}
\label{sec:TMD-resum}


Based on the above results, we can write down a unique formula for
the transverse momentum dependence. Following the procedure
of~\cite{Collins:1984kg}, the differential cross section for the
spin dependent SIDIS process can be written as,
\begin{eqnarray}
\frac{d\Delta \sigma(S_\perp)}{dydx_Bdz_hd^2P_{hT}}=
\frac{d\Delta \sigma^{\rm
TMD}}{dydx_Bdz_hd^2P_{hT}}+\left(\frac{d\Delta \sigma^{\rm
CO}}{dydx_Bdz_hd^2P_{hT}}-\frac{d\Delta \sigma^{\rm
CO}}{dydx_Bdz_hd^2P_{hT}}|_{P_{hT}\ll Q}\right),
\end{eqnarray}
which is valid in the whole transverse momentum region at leading power
of $1/Q^2$. In the above equation, the first
term comes from the TMD factorization formalism, and the second term
from the collinear factorization, CO, with the twist-three quark-gluon
correlations contributions. The second term will dominate the SSA
at large transverse momentum, and its $q_T$-dependence can be
calculated from perturbative QCD. On the other hand, at low
transverse momentum $P_{hT}\ll Q$, the second term vanishes,
because the two contributions are exactly the same in this limit,
and cancel each other out. 
Experimentally, if we can study the
transverse momentum dependence of the SSA for a wide range, we
shall explore the transition from the perturbative region to the
nonperturbative region.

%
%
\begin{figure}[t]
\begin{center}
\includegraphics[width=0.85\textwidth]{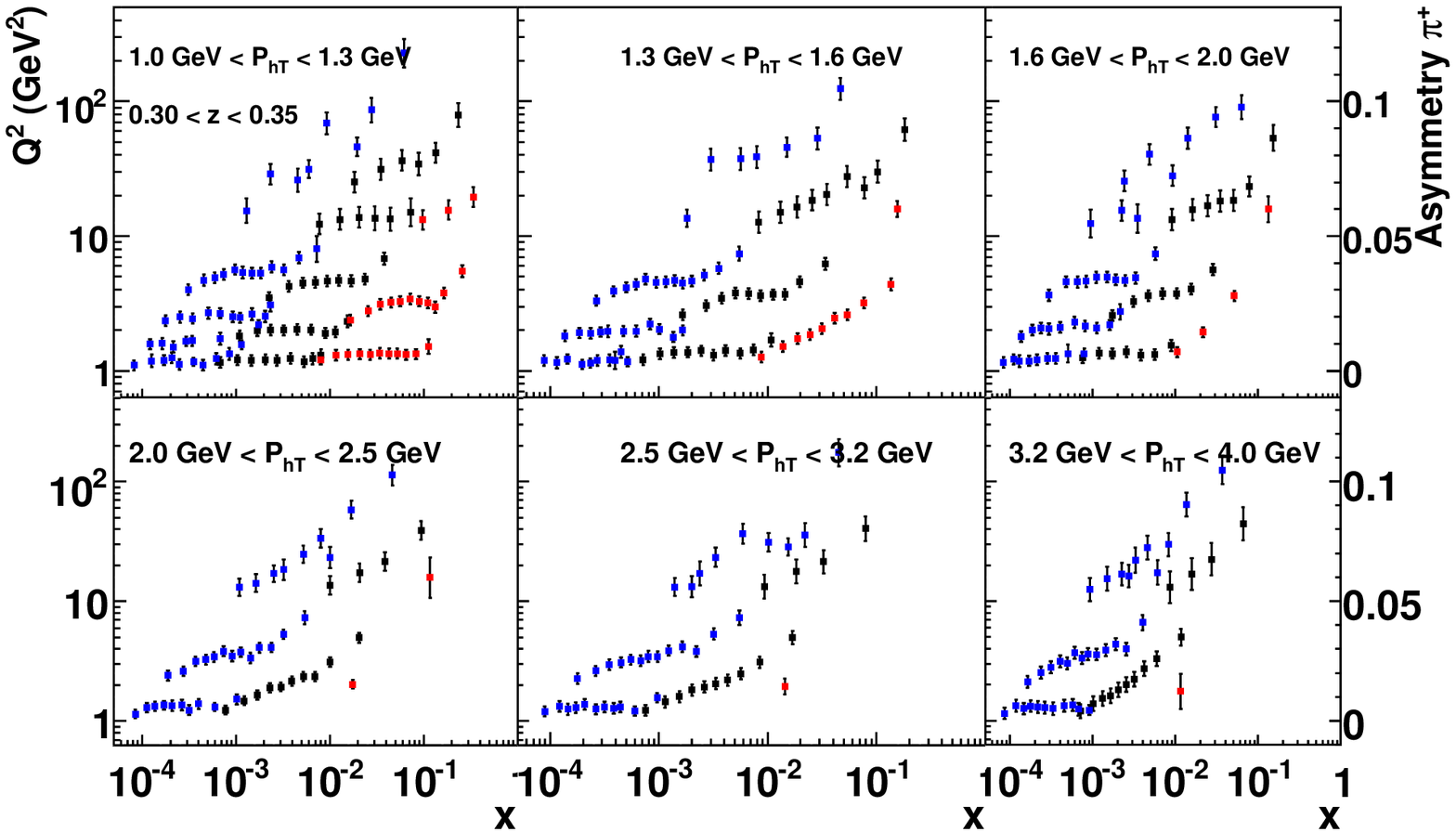}
\end{center}
\caption{\label{fig:simulation-hpt} 
\small
Four-dimensional representation of the projected accuracy for single $\pi^+$ production in
semi-inclusive DIS off the proton focussing on the transition region from low to   
high $ P_{hT}$ ($q_T \approx P_{hT}/z_h$) as indicated in the panels.
The position of each point is according to its $Q^2$ and $x$ value for a specific
bin in $z$ of $0.30 < z < 0.35$ and within the range $0.05 < y < 0.9$.
The projected event rate, represented by the error bar, is scaled to the (arbitrarily chosen)
asymmetry value at the right axis.
Blue squares, black triangles and red dots represent the
 $\sqrt{s} = 140 $ GeV,  $\sqrt{s} = 50 $ GeV and $\sqrt{s} = 15 $ GeV
EIC configurations, respectively.
Event counts correspond to an integrated luminosity of 120 fb$^{-1}$ for
each of the three configurations.
}
\end{figure}
The potential of the EIC for a study of this transition is illustrated in 
fig.~\ref{fig:simulation-hpt}, which shows the projected accuracy for 
single $\pi^+$ production for a four-dimensional
binning in the kinematic variables $x$, $Q^2$, $z$ and $P_{hT}$,
using three different energy configurations for the EIC ($\sqrt{s} = $ 15, 50 and 140 GeV)
and an integrated luminosity of 120 fb$^{-1}$ for each configuration.
Events are selected for $0.05 < y < 0.9$ and $W^2 > 5$ GeV$^2$ and
for the $z$ range of $0.30 < z < 0.35$, as example.
An overall detection efficiency of 50\% and a beam polarization of 70 \% are assumed.
The position of each point is according to its $x$ and $Q^2$ value
(abscissa and left ordinate, respectively) and
each panel is for a $P_{hT}$ bin as indicated in the figure.
The projected event rate is
represented by the error bar scaled with respect to the (arbitrarily chosen) asymmetry value given
at the right ordinate.
The parameterization of Ref.~\cite{Anselmino:2006rv} was used to simulate the cross section in the
transition region.
The  simulation demonstrates that the transition region $q_T \approx P_{hT}/z_h \sim 4 \div 8$ GeV
can be explored in great detail.
Energies up to $\sqrt{s} = 200 $ GeV and longer running times will allow for exploring  
even higher values of $ P_{hT}$. 













The most important example to study the transition between low and
high transverse momentum and the role of resummation is the
structure function $F_{UU,T}$. The double-longitudinal structure
function $F_{LL}$ is the only other example where the theoretical
framework has been developed at the same
level~\cite{Koike:2006fn}.

\begin{figure}
\begin{center}
\includegraphics[width=0.5\textwidth]{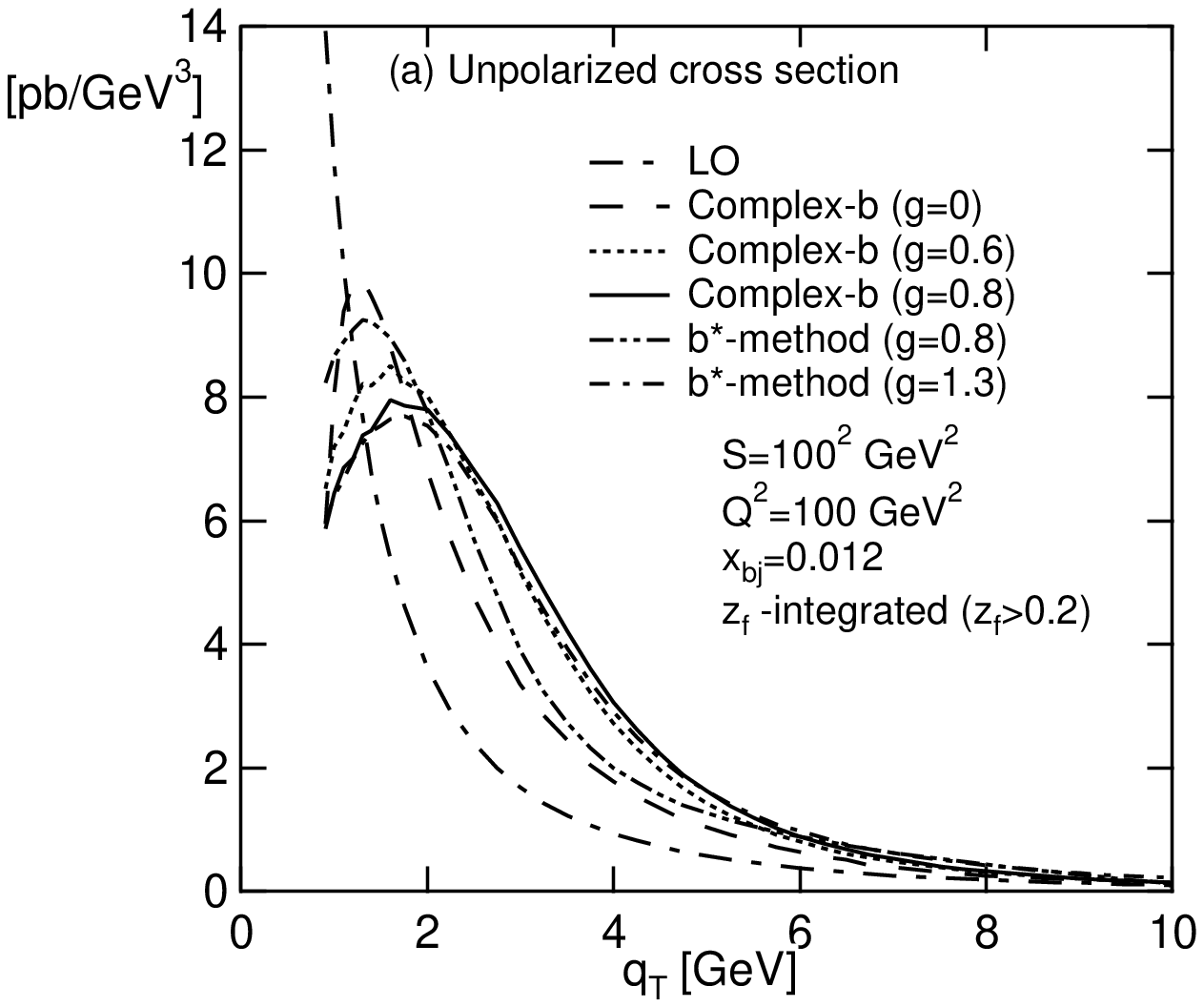}
\end{center}
\caption{\label{fig:EICkoike} 
\small
Unpolarized SIDIS cross section for
EIC kinematics from Ref.~\cite{Koike:2006fn}. Shown are: the
fixed-order result, 
the resummation results 
with different high-$b$ regularizations and different values of
the nonperturbative Sudakov factor.
}
\end{figure}
%




Fig.~\ref{fig:EICkoike} shows an example of resummation results for
DIS at a high-energy EIC option. These results give us an idea of
the extension of the region of intermediate transverse momentum
(and therefore also of the regions of high and low transverse
momentum). This extension obviously depends on experimental
kinematics, in particular on $Q^2$. As a lower boundary of this
region we can consider the values of $q_T$ where the
nonperturbative component of the Sudakov factor becomes relevant.
As an upper boundary we can consider the values of $q_T$ for which
the fixed-order cross section 
becomes comparable to the resummed cross section. 
From fig.~\ref{fig:EICkoike} we can estimate that the
intermediate-transverse-momentum region corresponds to $4 \text{
GeV} \lesssim q_T \lesssim 8 \text{ GeV}$.

A lot remains to be done to better pin down the nonperturbative
Sudakov factors, their functional form, their flavor dependence,
and their errors. This should be a high-priority task for the EIC.
The same is true for the doubly-longitudinally polarized case,
where the nonperturbative components are unknown.

To conclude this section, we mention that the same program of
resumming radiative contributions should be pursued also for the
Sivers, Collins, and $F^{\cos(\phi_h-\phi_S)}_{LT}$ structure
functions. At the moment the only discussion of similar topics is
done in sec.~9 of Ref.~\cite{Boer:2008fr}. However,  we can expect
developments in this direction in the near future and hope to
obtain an expression of the above-mentioned structure functions
that includes transverse-momentum resummation and describes the
physics in the whole transverse-momentum spectrum.


\subsection{Weighted Asymmetries} \label{sec:TMD-weighted-asymmetries}

Currently, experimental studies in semi-inclusive DIS have limited access to
single-spin asymmetries at
large transverse momentum, and most of the data are in the
low transverse momentum region, where the TMD formalism dominates.
In phenomenological studies, in order to compare with the
experimental data, one has to make model assumptions for the
transverse momentum dependence of the distribution and fragmentation
functions.
However, there is a class of observables that does not require detailed model assumptions about
transverse momentum dependence. These are transverse momentum
weighted single-spin asymmetries, which transform the convolutions in the factorized cross section
into simple products~\cite{Boer:1997nt,Kotzinian:1997wt}.

Staying for the moment in the framework of collinear factorization, an example for a weighted differential
cross section at leading order in $\alpha_s$ is
\begin{eqnarray}
\int d^2P_{hT}
\frac{P_{hT}}{z_hM_P}\sin(\phi_h-\phi_S)\frac{d\Delta\sigma^{\rm
TMD}(S_\perp)}{dx_Bdydz_hd^2\vec
{P}_{hT}}=\sigma_0\sum_q e_q^2 \frac{g_s}{2M_P} T_F^q(x)D(z)\ ,
\end{eqnarray}
where $e_q$ is the electric charge for a quark of flavor $q$,
and where $T_F(x)$ is the Qiu-Sterman matrix element of the
quark-gluon correlation function, and has been defined above.
With the standard choice of $P_{hT}$-weights $w_1 =  P_{h T}/ z_h M_P$
for the numerator and $w_0=1$ for the denominator, the
$P_{hT}$-weighted Sivers-asymmetry thus becomes
\begin{equation}
\frac{\langle \frac{P_{hT}}{z_hM_P}\sin(\phi_h-\phi_S)\rangle
_{\rm UT}}{\langle 1\rangle_{\rm
UU}}=\frac{\frac{1}{Q^4}\left(1-y+\frac{y^2}{2}\right)
\frac{x_B}{2M_P}\sum_q e_q^2g_sT_F^q(x)D(z)}{
\frac{1}{Q^4}\left(1-y+\frac{y^2}{2}\right)x_B\sum_q
e_q^2f_{1}(x)D(z)} \ .
\label{co_WA}
\end{equation}
We can go beyond the above leading order results and establish a
collinear factorization formalism for the weighted single
transverse spin dependent cross section. A similar study has been
performed for the Drell-Yan lepton pair production process, where
a next-to-leading order perturbative corrections have been
obtained~\cite{Vogelsang:2009pj}. We expect similar calculations for SIDIS shall
appear soon.

Recently, a generalization to employ Bessel functions as weights
$w_{n} \propto J_n(|\mathbf{P}_{hT}| \mathcal{B}_T)$
has been suggested \cite{BGMPtbp}. The Sivers asymmetry with generalized weights reads
\begin{align}
\frac{
\langle  \frac{{2 J}_1(|\mathbf{P}_{hT}|\bpar )}{\zh M \bpar} \sin(\phi_h - \phi_s)\rangle_{UT}}{\langle J_0(
P_{hT} \mathcal{B}_T )\rangle}
&= \nonumber \\
 & \hspace{-3cm} -2 \frac{\frac{1}{Q^4}\left(1 -y+\frac{y^2}{2}\right)\sum_q e_q^2\,  \tilde{f}_{1T}^{
\perp(1)q} (x,z^2 \bpar^2)\,  \tilde D(z,\bpar^2) }{ \frac{1}{Q^4}\left(1 -y+\frac{y^2}{2}\right)\sum_
q e_q^2\, \tilde f_{1}(x,z^2 \bpar^2)\, \tilde D_{1}(z,\bpar^2) },
\label{eq:ssa_sivers_final}
\end{align}
where now $\tilde{f}_{1T}^{\perp(1)q}$, $\tilde f_{1}^{q}$ and $\tilde D$ are TMDs and TMD FFs
Fourier transformed with respect to transverse momentum.
In the asymptotic limit $\mathcal{B}_T \rightarrow 0$, we recover the conventional
 weighted asymmetry
Eq.\ \eqref{co_WA}, and the Fourier transformed TMDs and FFs can be identified with the moments in 
that equation.

An important advantage of the generalized weights is that a non-zero choice of the parameter $\mathcal
{B}_T$ can reduce the sensitivity to large transverse momenta. This property also applies to the 
Fourier transformed TMDs and TMD FFs entering the asymmetries. The new approach thus 
avoids the problem of divergent $k_\perp$-integrals that affects moments of TMDs and 
TMD FFs.  Additionally, the analysis in Ref.~\cite{BGMPtbp} shows that soft factors 
appearing beyond tree level cancel out of the weighted asymmetry.

We conclude that an EIC presents a
unique opportunity to obtain the necessary coverage and resolution
in $P_{h T}$ to  explore the nucleon spin structure in
the language  of weighted asymmetries.


\section{Transverse polarization effects with gluons}
\label{sec:TMD-gluons}

\hspace{\parindent}\parbox{0.92\textwidth}{\slshape
 Dani\"el Boer,  Stanley J. Brodsky, Piet J. Mulders, Cristian Pisano, 
Markus Diehl, Bo-Wen Xiao, Feng Yuan}

\index{Boer, Dani\"el}
\index{Brodsky, Stanley J.}
\index{Mulders, Piet J.}
\index{Pisano, Cristian}
\index{Diehl, Markus}
\index{Xiao, Bo-Wen}
\index{Yuan, Feng}

\vspace{\baselineskip}



The gluon Sivers function shares the same characteristic features
as its counterpart in the quark sector, the quark Sivers function,
as discussed in the last section. Among the important information
we can obtain from this distribution is the spin-orbit
correlation of gluons inside the nucleon, which will help us to
understand the gluon spin contribution to the proton spin. The
EIC is the unique machine to
%
map out in much detail the gluon distribution,
including the spin-dependent and spin-averaged transverse momentum
dependent distributions. In this section, we will focus on
the gluon Sivers function.
The study of this distribution is strongly related to other
measurements such as the gluon GPDs and the unintegrated gluon
distributions of nucleon/nucleus at small-$x$.

Various processes in DIS can be used to probe the transverse
momentum dependent gluon distributions, such as heavy quark
and quarkonium production. Also the dijet/di-hadron correlation has
been proposed as a promising probe for the gluon Sivers function
and other TMD gluon distributions.


In Ref.~\cite{Boer:2003tx}, it was suggested to use the
dijet-correlation to study the gluon Sivers function in $pp$
collisions. However, because of both initial and final state
interaction effects involved in $pp$ scattering, the factorization
of this process is shown to be broken (see detailed discussions
in next section). On the other hand, for the DIS processes,
because only one hadron is involved in the initial state, the
dijet-correlation process could be factorized in the same spirit as
the semi-inclusive hadron production discussed in the previous
sections.

\begin{figure}[tbp]
\begin{center}
\includegraphics[width=2cm]{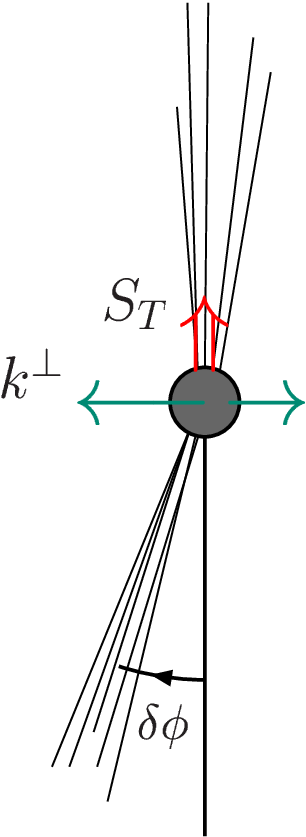} \\ 
\end{center}
\caption[*]{Back-to-back dijet correlation can be used to probe
the TMD gluon distributions.} \label{gluon-dijet}
\end{figure}

We consider here the dijet/quark-antiquark production in DIS
\begin{equation}
\gamma^{\ast }N^{\uparrow} \rightarrow H_1(k_{1}) + H_2(k_{2})+X\
,\label{dis-dijet}
\end{equation}
where $N$ represents the transversely polarized nucleon, $H_1$ and
$H_2$ are the two final state particles with momenta $k_1$ and
$k_2$, respectively. We are interested in the kinematic region
where the transverse momentum {\it imbalance} between them is much
smaller than the individual transverse momenta: $k_{\perp}=|\boldsymbol{k}_{1T
}+\boldsymbol{k}_{2T}|\ll P_{JT}$ where $\boldsymbol{P}_{JT}$ is
defined as $(\boldsymbol{k}_{1T}-\boldsymbol{k}_{2T})/2$. This is
referred to as the (back-to-back) correlation limit. An important advantage 
of taking this correlation limit is
that we can apply the power counting method to obtain the leading
order contribution of $k_{\perp}/P_{JT}$ where the
differential cross section directly depends on the TMD gluon
distribution. As illustrated in Fig.~\ref{gluon-dijet}, with
transverse spin in the dijet plane, the correlation between the
two jets will lead to a preferred direction in the transverse
plane. This will signal the gluon Sivers effect if the process is
dominated by the gluonic subprocesses.

As demonstrated in Ref.~\cite{Dominguez:2010xd}, the TMD gluon
distribution in the quark-antiquark jet correlation in the DIS
process of (\ref{dis-dijet}) follows the original gluon
distribution definition of Ref.~\cite{Collins:1981uw},
\begin{eqnarray}
xf_1^g(x,k_{\perp })&=&\int \frac{d\xi ^{-}d^2\xi _{\perp }}{(2\pi )^{3}P^{+}}%
e^{ixP^{+}\xi ^{-}-ik_{\perp }\cdot \xi _{\perp }} \langle
P|F^{+i}(\xi
^{-},\xi _{\perp })\mathcal{L}_{\xi }^{\dagger }\mathcal{L}%
_{0}F^{+i}(0)|P\rangle \ ,\label{tmd-g1}
\end{eqnarray}%
where $F^{\mu \nu }$ is the gauge field strength tensor $F_a^{\mu
\nu }=\partial ^{\mu }A_a^{\nu }-\partial ^{\nu }A_a^{\mu
}-gf_{abc}A_b^\mu A_c^\nu$ with $f_{abc}$ the antisymmetric
structure constants for $SU(3)$, and the gauge-link follows the
similar definition as that for the quark distribution but in the
adjoint representation. The physics behind this factorization is
the following. The virtual photon scatters on the nucleon target
and produces a quark-antiquark pair through the partonic process
$\gamma^*g\to q \bar q$. In the correlation limit, the
quark-antiquark pair stays close in the coordinate space, and act
as a color-octet object,  which effectively behaves like a single gluon.
In particular, the net effect of the final state
interactions between the nucleon target and the quark-antiquark 
pair is exactly the same antisymmetric structure
$f_{abc}$ as in the TMD gluon definition of Eq.~(\ref{tmd-g1}).
This is totally different from the analogous QED process where the
final state interactions cancel out completely with the
fermion-antifermion pair.

In the following, we will present some recent phenomenological
studies on the gluon TMDs from the quark-antiquark correlation in
DIS processes. We expect more interesting results shall be
obtained in the near future.

\subsection{The gluonic Sivers effect in dihadron production}
\label{sec:dijetgluonsivers}






The production of a pair of hadrons with high transverse momenta in DIS is
sensitive to the transverse-momentum dependent gluon distribution.  In
particular, it has a transverse target spin asymmetry due to the gluon
Sivers function.  The relevant parton-level subprocess is $\gamma^* g \to
q\bar{q}$, and to eliminate contributions from $\gamma^* q\to q g$ and
$\gamma^* \bar{q} \to \bar{q} g$ we focus on charm production.

As a straight forward generalization of the unpolarized case recently
studied in \cite{Dominguez:2010xd}, the cross section for the dijet/$c\bar{c}$
production from a nucleon with transverse polarization $\boldsymbol{S}_\perp$ can
be written as
\begin{equation}
  \label{xiao:x-sect}
\frac{d\sigma^{\gamma^{\ast}_{T,L} p\rightarrow c\bar{c}+X}}{%
dz\, d^{2}k_{1T}\, d^{2}k_{2T}}
= \frac{H^{\gamma^{\ast}_{T,L} g\rightarrow c\bar{c}}}{z \bar{z}}\,
  \biggl[ f_1^g(x,k_\perp) +
           \frac{(\boldsymbol{S_\perp} \times \boldsymbol{k}_\perp)^3}{M}\,
           f_{1T}^{g \perp}(x,k_\perp)\, \biggr] \,.
\end{equation}
Here, $f_1^g$ is the usual gluon TMD,
$f_{1T}^{g \perp}$ the gluon Sivers distribution and $\bar{z} = 1 - z$.
Again, we are interested in the back-to-back correlation limit.
The gluon momentum fraction is
then given by $x /x_B \approx 1 + (\smash{P_{JT}^2} + m_c^2)
/(z\bar{z} Q^2)$, where $m_c$ is the charm quark mass.  The
hard-scattering cross sections $H^{\gamma^{\ast}_{T,L}
g\rightarrow c\bar{c}}$ for transverse and longitudinal photons
depend on $P_{hT}^2$, $Q^2$, $z$ and $m_c$ and can be found in
\cite{Dominguez:2010xd}.

It may be possible to study the cross section \eqref{xiao:x-sect}
experimentally through the production of two heavy-quark jets, but the
interpretation of this process requires a quantitative understanding of
the relative transverse momentum between a reconstructed jet and the heavy
quark it originates from.  As an alternative, we consider here the
production of two heavy hadrons, e.g.\ $D$ mesons.  Its cross section
reads
\begin{align}
& \frac{d\sigma^{\gamma^{\ast}_{T,L} p\rightarrow h_1 h_2 +X}}{%
  dz_{1}\, dz_{2}\, d^{2}P_{h_1T}\, d^{2}P_{h_2T}}
 = \int_{z_{1}}^{1-z_{2}} dz\,
   \frac{H^{\gamma^{\ast}_{T,L} g\rightarrow c\bar{c}}}{z^2 \bar{z}^2}
   \int d^2\lambda_{1T}\, d^2\lambda_{2T}\,
   \biggl[ f_1^g(x,k_\perp)
\nonumber \\[0.3em]
&\qquad + \frac{(\boldsymbol{S}_\perp \times \boldsymbol{k}_\perp)^3}{M}\,
     f_{1T}^{g \perp}(x,k_\perp)\, \biggr] \,
 D^{h_1/c}\Bigl( \frac{z_{1}}{z}, \frac{z_{1}}{z} \lambda_{1T} \Bigr) \,
 D^{h_2/\bar{c}}\Bigl( \frac{z_{2}}{\bar{z}},
         \frac{z_{2}}{\bar{z}} \lambda_{2T} \Bigr) \,,
  \label{xiao:siversgf}
\end{align}
where
\begin{align}
\boldsymbol{k}_{1T} &= \boldsymbol{\lambda}_{1T}
                  + \frac{z}{z_{1}}\, \boldsymbol{P}_{h_1T} \,,
&
\boldsymbol{k}_{2T} &= \boldsymbol{\lambda}_{2T}
                  + \frac{\bar{z}}{z_{2}}\, \boldsymbol{P}_{h_2T} \,.
\end{align}
Here $h_1$ is the hadron containing a $c$ quark and $h_2$ the one
containing a $\bar{c}$, with $\boldsymbol{P}_{h_1T}$, $\boldsymbol{P}_{h_2T}$
denoting their transverse momenta and $z_{1}$, $z_{2}$ their momentum
fractions w.r.t.\ the virtual photon.  The fragmentation functions
$D(z,P_\perp)$ depend on the momentum fraction $z$ and the relative
transverse momentum $P_\perp$ of the hadron with respect to the quark or
antiquark.

The parton-level variables $\boldsymbol{k}_{1T}$, $\boldsymbol{k}_{2T}$
  and $z$ are not directly measurable,
 but a detailed analysis of the kinematics \cite{Diehl:2011}
reveals that they can be partly determined from the hadronic final state.
  In particular, one can define variables
  $\boldsymbol{k}_{\perp}^{\prime}$, $\boldsymbol{P}_{T}^{\prime}$ and
  $z'$ that are measurable and closely related to
  $\boldsymbol{k}_{\perp} = \boldsymbol{k}_{1T} + \boldsymbol{k}_{2T}$,
  $\boldsymbol{P}_{T} = (\boldsymbol{k}_{1T} - \boldsymbol{k}_{2T}) /2$
  and $z$, respectively.
The cross product $(\boldsymbol{S}_\perp \times \boldsymbol{k}_\perp)$ in
\eqref{xiao:siversgf} gives rise to an angular modulation
\begin{equation}
  \label{xiao:angular-dep}
  \frac{d\sigma^{\gamma^{\ast} p\rightarrow h_{1}h_{2}+X}}{%
    dk'\, d\phi _{S,k^{\prime }}}
  \approx A(k'_{\perp}) + B(k'_{\perp}) \sin(\phi_{S k'} + \gamma ) \,,
\end{equation}
where $\phi_{S k'}$ is the azimuthal angle between $\boldsymbol{S}_\perp$ and
$\boldsymbol{k}^{\,\prime}_{\perp}$.  The coefficient $B(k')$ depends on the
gluon Sivers function, as well as the phase $\gamma$.



\begin{figure}[tbp]
\begin{center}
\includegraphics[width=0.6\textwidth,bb=10 10 680 540,%
  clip=true]{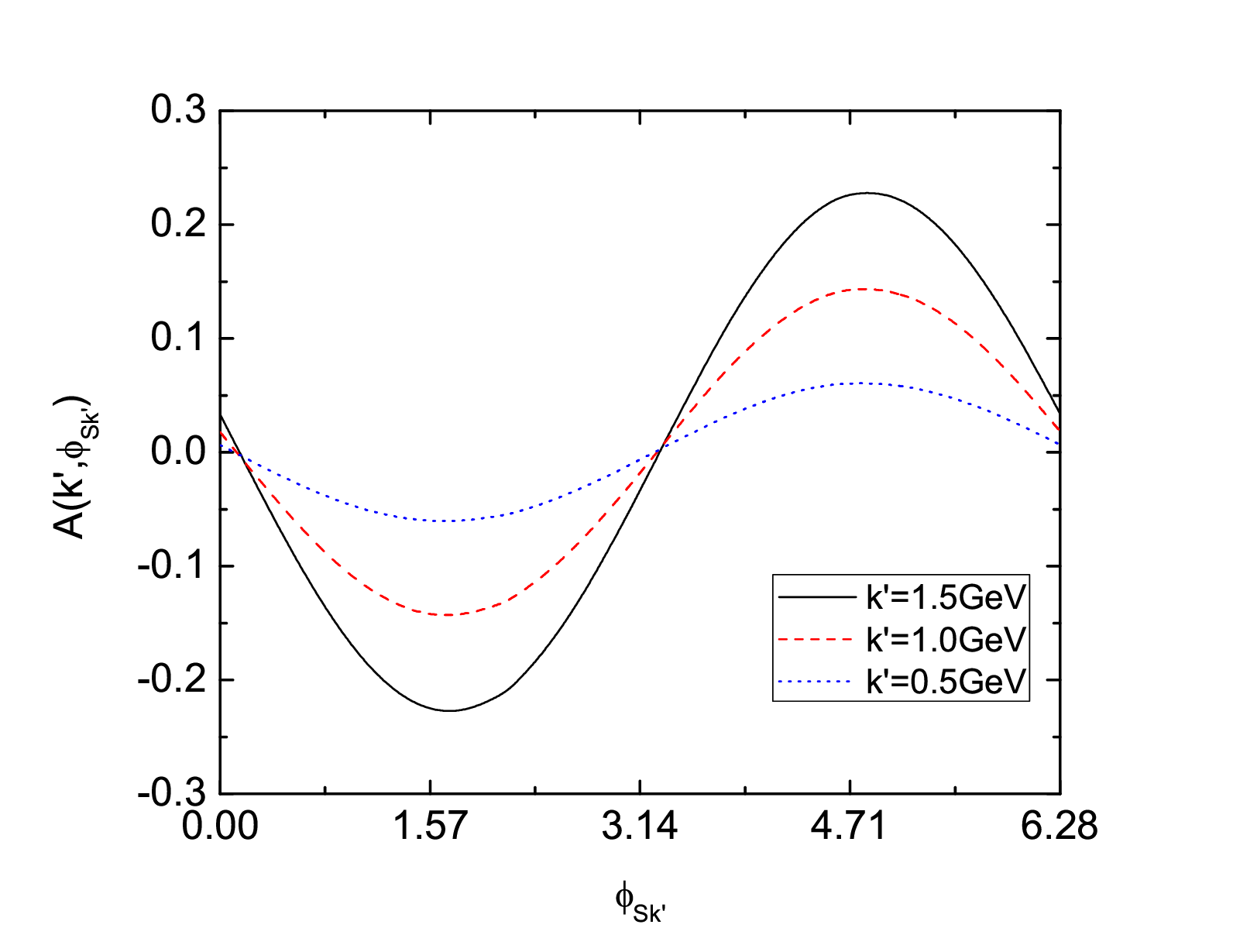}
\end{center}
\caption{\label{xiao:gsivers} The transverse target asymmetry
  \protect\eqref{xiao:asy} for $\gamma^{\ast} p\to D^0 \bar{D}^0 + X$.
  The kinematics is specified by $W = 100 \gev$, $Q^2 = 16 \gev^2$, $z_{1}
  = z_{2} = 0.3$, $0.25 < z' < 0.75$ and $5\gev < P^{\,\prime}_{T} <
  40 \gev$.}
\end{figure}

To estimate the possible size of the Sivers asymmetry, we follow
\cite{Anselmino:2004nk} and assume
\begin{align}
  \label{xiao:gluon-ansatz}
f_{1T}^{g\perp}(x, k_{\perp}) &=
  \frac{2 \sigma M}{k_{\perp}^2 + \sigma^2}\, f_1^g(x, {k}_{\perp}) \,,
&
f_1^g(x, {k}_{\perp}) &=
  \frac{e^{- k_{\perp}^2 /\sigma^2}}{\pi \sigma^2}\, f_1^g(x)
\end{align}
with $\sigma = 800 \mev$ and the integrated gluon distribution $f_1^g(x)$
from MSTW 2008 \cite{Martin:2009iq}.  This Ansatz saturates the positivity
bound $\frac{k_{\perp}}{M} \big| f_{1T}^{g\perp}(x, k_{\perp}) \big| \le
f_1^g(x, {k}_{\perp})$ at $k_{\perp} = \sigma$ and undershoots it for all
other values of $k_{\perp}$.  We consider the production of $D$ meson
pairs and take a fragmentation function $D(z, P_{\perp}) = D(z)\,
e^{-P_{\perp}^2 /\sigma^2} /(\pi \sigma^2)$ with the same Gaussian
width as in \eqref{xiao:gluon-ansatz}.  We take $D(z) \propto z^\alpha
(1-z)^\beta\, e^{\gamma z (1-z)}$ with $\alpha = 2.86, \beta = 1.57,
\gamma = 5.66$, which gives a fair description of the $D^0$ spectrum
observed in $e^+e^-$ annihilation \cite{Artuso:2004pj,Seuster:2005tr}.
In Fig.~\ref{xiao:gsivers} we show the transverse target spin asymmetry
\begin{equation}
  \label{xiao:asy}
A(k'_{\perp}, \phi_{Sk'}) =
 \frac{d\sigma(k'_{\perp}, \phi_{Sk'})
       - d\sigma(k'_{\perp}, \phi_{Sk'} + \pi)}{%
       d\sigma(k'_{\perp}, \phi_{Sk'})
       + d\sigma(k'_{\perp}, \phi_{Sk'} + \pi)}
\end{equation}
for the process $\gamma^{\ast} p\to D^0 \bar{D}^0 + X$ summed over
transverse and longitudinal photon polarization.  We find that the
phase shift $\gamma$ in \eqref{xiao:angular-dep} is tiny.
The asymmetry is found to be sizable with our Ansatz, which
suggests that DIS production of heavy meson pairs at EIC has good
sensitivity to the gluon Sivers function.

\subsection{Probing the linear polarization of gluons in
unpolarized hadrons} \label{BBMPsec:lin_pol_glu}





Gluons inside unpolarized hadrons can be linearly polarized
provided they have a non-zero transverse momentum. The simplest and
theoretically safest way to probe this TMD distribution of
linearly polarized gluons is through $\cos 2\phi$ asymmetries in
heavy quark pair or dijet production in electron-hadron
collisions. Future EIC or LHeC experiments are ideally suited for
this purpose. Here we estimate the maximum asymmetries for EIC
kinematics.


Linearly polarized gluons in an unpolarized hadron, carrying a
light-cone momentum fraction $x$ and transverse momentum
$\boldsymbol{k}_\perp$ w.r.t.\ to the parent's momentum, are
described by the TMD
$h_1^{\perp\,g}(x,k_\perp)$
\cite{Mulders:2000sh,Boer:2010zf,Boer:2009nc}. Unlike the quark
TMD
 $h_1^{\perp \, q}$ of transversely polarized quarks inside an
unpolarized hadron (also frequently referred to as Boer-Mulders
function) \cite{Boer:1997nt}, $h_1^{\perp\,g}$ is chiral-even and
$T$-even. This means it does not require initial or final state
interactions (ISI/FSI) to be non-zero. Nevertheless, as any TMD,
$h_1^{\perp\, g}$ can receive contributions from ISI or FSI and
therefore can be process dependent, in other words, non-universal,
and its extraction can be hampered in non-factorizing cases.

Thus far no experimental studies of $h_1^{\perp\,  g}$ have been
performed. As recently pointed out, it is possible to obtain an
extraction of $h_1^{\perp\, g}$ in a simple and theoretically safe
manner, since unlike $h_1^{\perp\, q}$ it does not need to appear
in pairs \cite{Boer:2010zf}. Here we will discuss observables that
involve only a single $h_1^{\perp\, g}$ in semi-inclusive DIS to
two heavy quarks or to two jets, which allow for TMD factorization
and hence a safe extraction. The corresponding hadroproduction
processes  run into the problem of factorization breaking
\cite{Rogers:2010dm,Boer:2010zf}.

\begin{figure}[t]
\begin{center}
 \includegraphics[angle=0,width=0.43\textwidth]{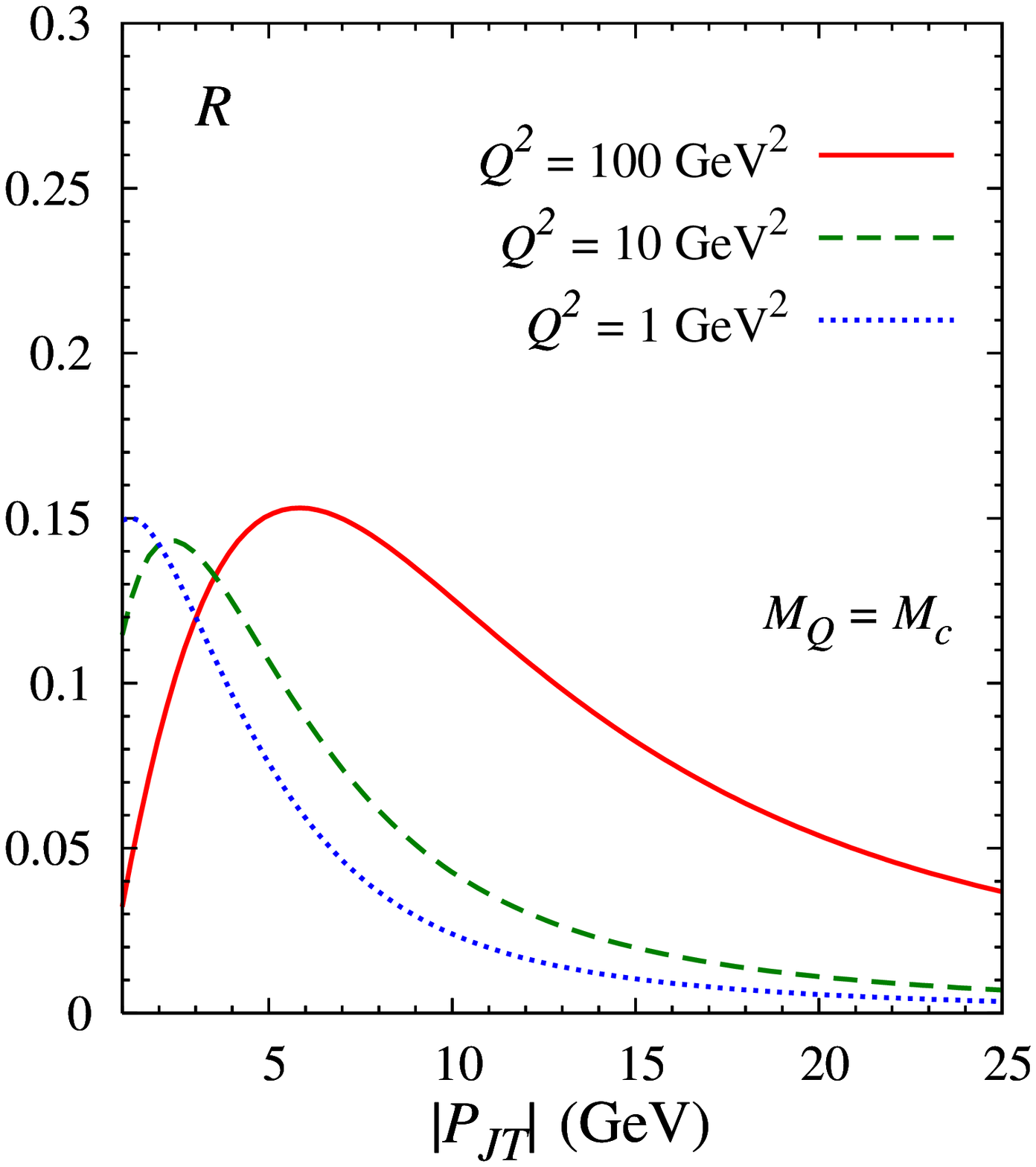}
 \includegraphics[angle=0,width=0.43\textwidth]{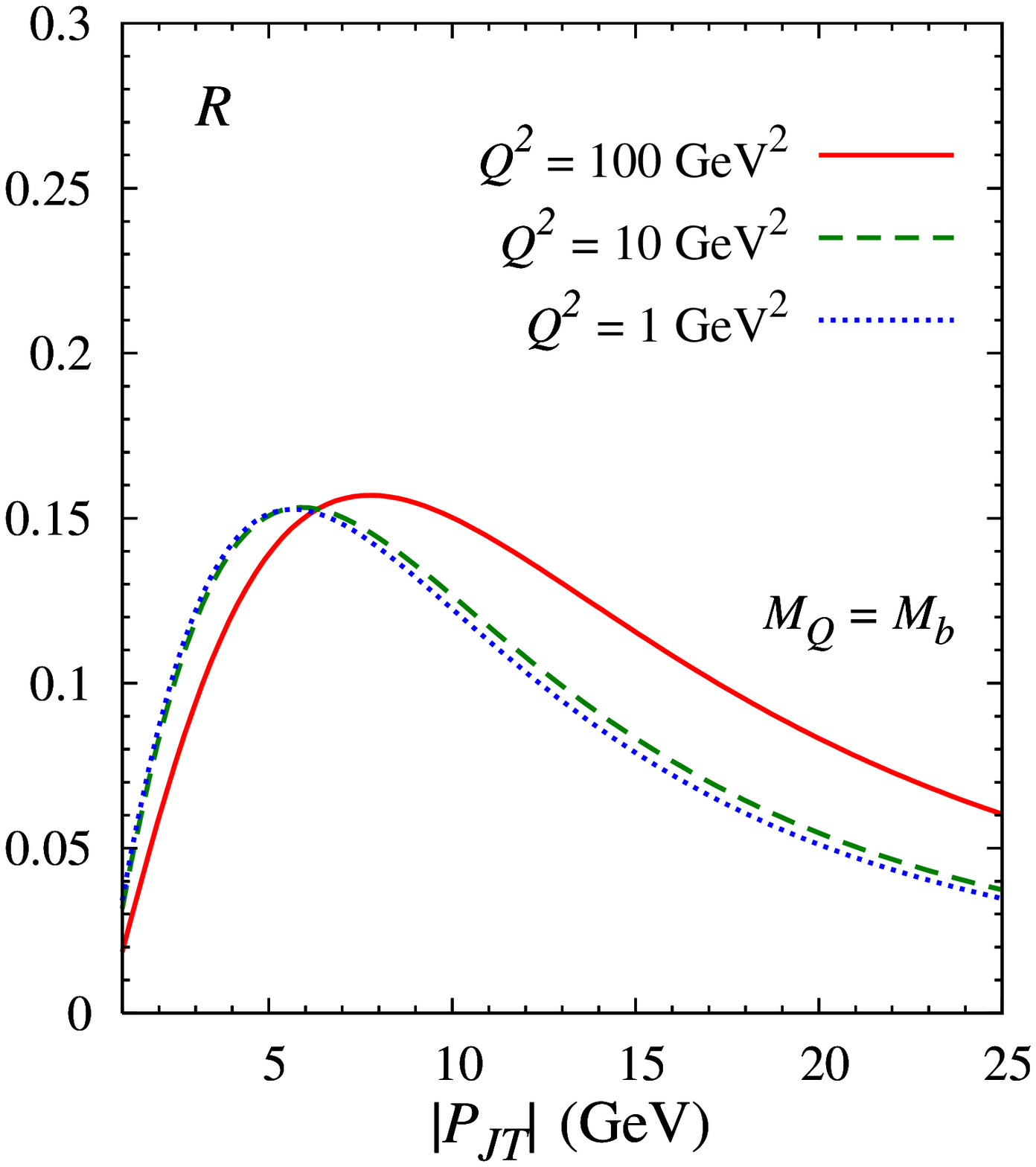}
 \caption{Upper bounds of the asymmetry ratio $R$ in equation \eqref{BBMPeq:bound} as a function of 
$\vert \boldsymbol P_{JT} \vert$ at different values of $Q^2$, with $y=0.01$ and $z=0.5$.
\label{BBMPfig:asy} }
\end{center}
\end{figure}

Again, we consider heavy quark production, $e
(\ell)${+}$h(P)$$\to$$e(\ell^\prime)${+}$Q(k_1)${+}$\bar{Q}(k_2)${+}$X$,
where the four-momenta of the particles are given within brackets,
and the heavy quark-antiquark pair in the final state is almost
back-to-back in the plane perpendicular to the direction of the
exchanged photon and hadron. The calculation proceeds along the
lines explained in Refs.\ \cite{Boer:2009nc,Boer:2007nd}. We
obtain for the cross section integrated over the angular
distribution of the back-scattered electron $e(\ell^\prime)$:
\begin{eqnarray}
\frac{d\sigma} {dy_1\,dy_2\,dy\,dx_{\scriptscriptstyle
B}\,d^2\boldsymbol{k}_{\perp}\, d^2\boldsymbol{P}_{JT}}\,\,
=\,\, \frac{\alpha^2\alpha_s}{\pi s
  M_T^2}\, \frac{(1+y x_{\scriptscriptstyle B})}{ y^5 x_{\scriptscriptstyle B}}\,  \bigg( A +
\frac{{\boldsymbol k}_{\perp}^2}{M^2}\, B \, \cos 2 \phi \bigg)
\,\delta(1 - z_1 - z_2)\, . \label{BBMPeq:cso}
\end{eqnarray}
The kinematics are the same as in the last subsection with the heavy quark mass $M_Q$,
$M_{iT}^2 \approx M_T^2 = M_Q^2 + \boldsymbol P_{JT}^2$ and the rapidities $y_i$ for the 
quark momenta along photon-target direction. 
The azimuthal angles of
$\boldsymbol{k}_{\perp}$ and $\boldsymbol{P}_{JT}$ are denoted
by $\phi_\perp$ and $\phi_T$, respectively, 
and $\phi \equiv \phi_\perp - \phi_T$. 
The functions $A$ and $B$ depend on $y, z (\equiv z_2),
Q^2/M_{T}^2, M_Q^2/M_{T}^2$, and $\boldsymbol{k}_{\perp}^2$. 
The angular independent part $A$
involves only the unpolarized TMD gluon distribution $f_1^g$,
while the magnitude $B$ of the $\cos 2 \phi$ asymmetry is
determined by $h_1^{\perp\, g}(x,k_\perp)$. Since $h_1^{\perp\, g}$ is
completely unknown, we estimate the maximum asymmetry that is
allowed by the bound \cite{Boer:2010zf}
\begin{equation}
|h_1^{\perp\, g (2)}(x)| \leq \frac{\langle k_\perp^2\rangle}{2M^2} f_1^g(x)\, ,
\label{BBMPbound}
\end{equation}
where the superscript $(2)$ denotes the $n=2$ transverse moment
(defined as $f^{(n)}(x) \equiv \int  d^2 \boldsymbol{k}_{\perp}\;
\left(\boldsymbol{k}_{\perp}^2/2 M^2\right)^n \;f(x,
\boldsymbol{k}_{\perp}^2)$). The maximal (absolute) value of the
asymmetry ratio
\begin{equation}
R = \left| \frac{\int d^2 \boldsymbol{k}_{\perp} \,
\boldsymbol{k}_{\perp}^2 \, \cos 2 (\phi_{\perp}-\phi_T) \, 
d\sigma}{\int d^2 \boldsymbol{k}_{\perp} \,
\boldsymbol{k}_{\perp}^2 \, d\sigma}\right| = \frac{\int
d\boldsymbol{k}_{\perp}^2 \ \boldsymbol{k}_{\perp}^4\, | B |}{ 2
M^2 \int d\boldsymbol{k}_{\perp}^2 \ \boldsymbol{k}_{\perp}^2 \,
A} \label{BBMPeq:bound}
\end{equation}
is depicted in Fig.~\ref{BBMPfig:asy} as a function of 
$\vert \boldsymbol P_{JT}\vert$ at different values of
$Q^2$ for charm (left panel) and bottom (right panel) production,
where we have selected $y= 0.01$, $z=0.5$, and taken $M_c^2=$ 2
$\gev^2$, $M_b^2=$ 25 $\gev^2$. Such large asymmetries, together
with the relative simplicity of the suggested measurement
(polarized beams are not required), would probably allow an
extraction of $h_1^{\perp\, g}(x,k_\perp)$ at the EIC (or LHeC).

\section{Theory highlights}
\label{sec:TMD-theory}

\hspace{\parindent}\parbox{0.92\textwidth}{\slshape
Igor O. Cherednikov, Zhong-Bo Kang, Piet J. Mulders, Barbara Pasquini, Ted Rogers, Peter
Schweitzer,  Nicolaos G. Stefanis, Jian-Wei Qiu }

\index{Cherednikov, Igor O.}
\index{Kang, Zhong-Bo}
\index{Mulders, Piet J.}
\index{Pasquini, Barbara}
\index{Rogers, Ted}
\index{Schweitzer, Peter}
\index{Stefanis, Nicolaos G.}
\index{Qiu, Jian-Wei}

\vspace{\baselineskip}

The candidates for the {\it golden} measurement at the EIC are the
spin-dependent Sivers function $f_{1T}^{\perp}$, as well as the
unpolarized quark distribution $f_{1}$. The proposed {\it silver}
candidates are the transversity, the Boer-Mulders, and the Collins
functions. All these objects are transverse-momentum dependent
parton densities that describe the inner structure of hadrons by
taking into account the longitudinal {\it and} the transversal
partonic degrees of freedom.

In the last few years, there has been tremendous progress on the
theory developments for the transverse momentum dependent parton
distributions. In particular, there have been intensive
investigations on the QCD factorization and the associated
universality of the TMD parton distributions in various hard
processes; the energy scale dependence for the TMD distributions
and related quark-gluon correlation functions.
In this section, we will highlight these developments.

\subsection{\label{secIV:mulders-rogers} 
Gauge-links, TMD-factorization, and TMD-factorization breaking} 



\vspace{0.7\baselineskip}
In this section, we discuss some basic
features of transverse momentum dependent
parton distribution
functions.
%
%
%
In hard processes, parton distribution functions and fragmentation
functions are expressed as matrix elements of nonlocal
combinations of quark or gluon fields. In the collinear situation
that all transverse momenta of partons are integrated over in the
definitions, the nonlocality is in essence light-like. These
correlation functions are convoluted with the squared amplitude
for the partonic subprocess (in essence the partonic cross
section) of a hard process.  When the transverse momenta of
partons are involved, the non-locality in the matrix elements
includes a transverse separation, and a \emph{transverse momentum
dependent} (TMD) factorization theorem is needed. In all cases the
definitions of the non-perturbative functions include gluon
contributions resummed into gauge-links (or Wilson lines) that bridge the
nonlocality.

It is important to realize that the appearance of the gauge-links
is a consequence of the systematic resummation of extra gluon contributions
in the derivations of factorization, so their structure is
dictated by the requirements of factorization.

In processes like $\ell + H \longrightarrow \ell^\prime + h + X$
(semi-inclusive DIS), $\ell + \bar\ell \longrightarrow h_1 + h_2 + X$
(annihilation process) or $H_1 + H_2 \longrightarrow \ell + \bar \ell + X$
(Drell-Yan process) one has,
at leading power in the hard scale, a simple underlying hard
process, which is a virtual photon (or weak boson) coupling to a
parton line. The color flow from the hard part to
collinear or soft parts is simple.
Additional gluons with polarizations collinear to the parton momenta
are resummed into gauge-links, which exhibit the interesting behavior
that for transverse momentum dependent functions they bridge the
transverse separation between the non-local field combinations at
lightcone past or future infinity. Which gauge-link is relevant in
a particular non-perturbative function depends on the color flow
in the full process. For a quark distribution
function one has a link via (future) lightcone $+\infty$ if the color
flows into the final state, and a link via (past) lightcone $-\infty$
if the color is annihilated by another incoming parton.

QCD factorization theorems are central to understanding
high energy hadronic scattering cross sections
in terms of the fundamentals of perturbative QCD.
In addition to providing a practical prescription for order-by-order calculations,
derivations of factorization provide a solid theoretical underpinning for
concepts like PDFs and FFs which
are crucial in the quest to expand the basic understanding of hadronic structure.
The most natural first attempt at a TMD-factorization formula is simply
to extend the classic parton model intuition familiar from collinear factorization.
For the semi-inclusive deep inelastic scattering (SIDIS) cross section, for example, the cross section
might be written schematically as
\begin{equation}
\label{eq:basic} d \sigma \sim | \mathcal{H} |^2 \otimes
\Phi(x,{\boldsymbol k}_\perp) \otimes D(z,{\boldsymbol P}_\perp) \; \delta^{(2)}
({\boldsymbol q}_T  + {\boldsymbol k}_\perp - {\boldsymbol P}_\perp ).
\end{equation}
Here $\Phi(x,{\boldsymbol k}_\perp)$ is the TMD PDF while $D(z,{\boldsymbol
P}_\perp)$ is the TMD FF, with the usual probability
interpretations, and $| \mathcal{H} |^2$ represents the hard part.
The momentum ${\boldsymbol q}_T$ is the small momentum sensitive to
intrinsic transverse momenta, ${\boldsymbol k}_\perp$ and ${\boldsymbol P}_\perp$,
carried by the colliding proton and the produced hadron. The
$\otimes$ symbol denotes all relevant convolution integrals, and
the $x$ and $z$ arguments are the usual longitudinal momentum
fractions.

In a perturbative derivation of factorization, a small-coupling
perturbative expansion of the cross section is analyzed in terms
of ``leading regions'', and the sum is shown order-by-order to
separate into the factors of Eq.~\eqref{eq:basic}. The
precise field theoretic definitions of the correlation functions,
$ \Phi(x,{\boldsymbol k}_\perp)$ and $D(z,{\boldsymbol P}_\perp)$, should emerge
naturally from the requirements of factorization. In the hard part
$| \mathcal{H} |^2$, all propagators must be off-shell by order
the hard scale $Q$ so that asymptotic freedom applies, and
small-coupling perturbation theory is valid, with non-factorizing
higher-twist contributions suppressed by powers of $Q$. Such
factorization theorems are well-established for inclusive
processes that utilize the standard \emph{integrated} correlation
functions (see~\cite{Collins:1989gx} and references therein), but
TMD-factorization theorems involve other subtleties, particularly
with regard to the definitions of the TMD PDFs and FFs and
their associated gauge-links.

In cases where there is a more complex color flow such as is
often
the case
when the underlying hard process involves multiple color flows and/or
if the incoming partons are gluons, this
can potentially lead to a more complex
gauge-link structure including traced closed loops or looping gauge-links.
For situations in which only one
TMD correlation function
is studied, these structures have been examined
in~\cite{Bomhof:2004aw,Bomhof:2006dp,Bomhof:2006ra,Bacchetta:2007wc}
for two-to-two partonic subprocesses.
In situations that involve several TMD functions,
factorization using separate TMD functions fails completely.

\begin{figure*}
\centering
  \begin{tabular}{c@{\hspace*{5mm}}c}
    \includegraphics[scale=0.6]{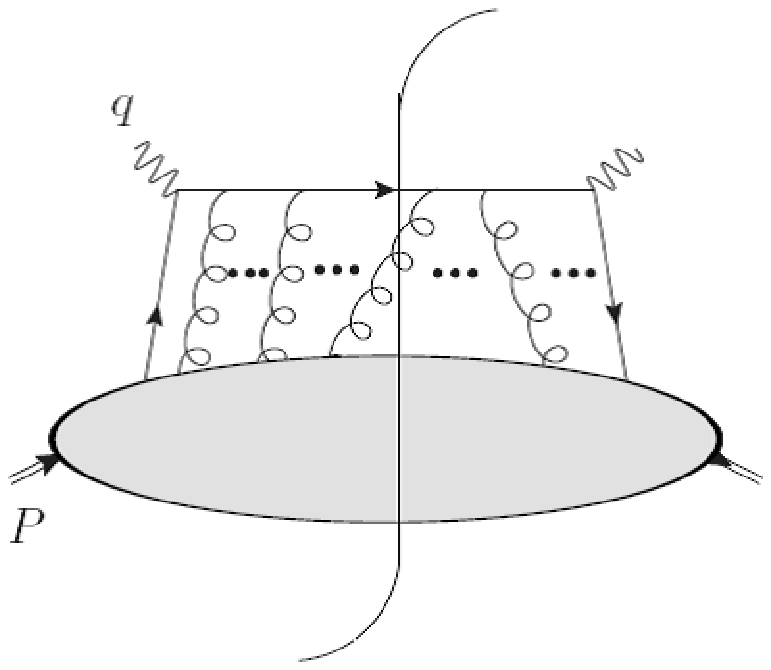}
    &
    \includegraphics[scale=0.6]{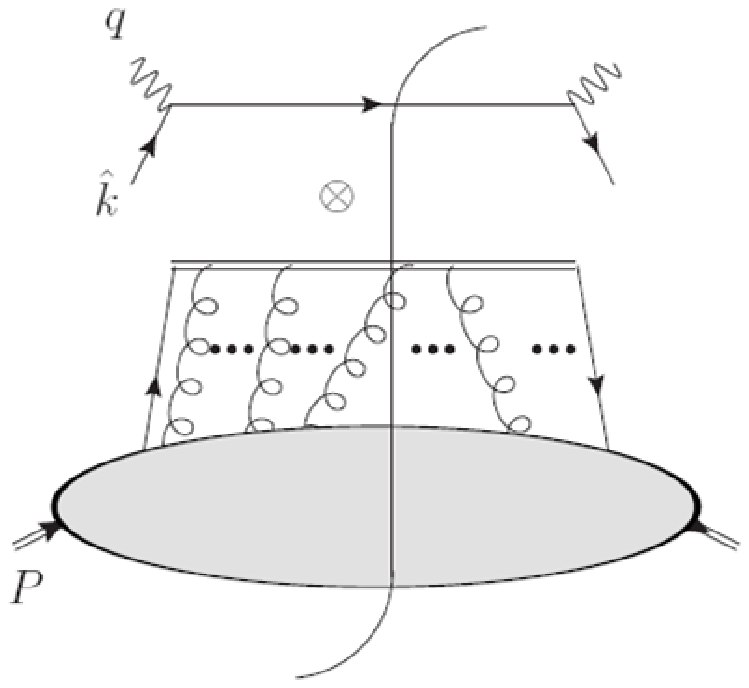}
  \\
  (a) & (b)
  \end{tabular}
\caption{(a) Target-collinear gluons in a graph for SIDIS.  (b) Factorization of extra gluons into
gauge-link contributions.}
\label{fig:basicfact}
\end{figure*}

To understand the issues that arise in defining TMDs, it is
instructive to start with a review of the definition of the
standard \emph{integrated} quark PDF.
It is
\begin{equation}
\label{eq:intDEF}
f(x;\mu) = {\rm F.T.} \, \langle p | \, \bar{\psi}(0,w^-,{\boldsymbol
0}_t) \gamma^+ V_{[0,w]} (u_{\rm J}) \,  \psi(0) | p \rangle \ ,
\end{equation}
where ``${\rm F.T.}$'' stands for the Fourier transform from
coordinate space to momentum space.
The above definition contains UV divergences which must be
renormalized. This gives dependence on an extra scale $\mu$, and
ultimately results in the well-known DGLAP evolution equations for
the integrated PDF.  
For a gauge invariant definition, the PDF must contain a path
ordered exponential of the gauge field that connects the points
$0$ and $(0,w^-,{\boldsymbol 0}_t)$. This is the gauge-link and its formal
definition is
\begin{equation}
\label{eq:wilsonline}
V_{[0,w]} (u_{\rm J}) = P \exp \left(-igt^a \int_{0}^{w^-} \, d \lambda \,  u_J \cdot A^a(\lambda u_J) \right).
\end{equation}
The path of the gauge-link is determined by the
light-like vector $u_{\rm J} = (0,1,{\boldsymbol 0}_t)$.
That is, the gauge-link follows a straight path connecting $0$
and $(0,w^-,{\boldsymbol 0}_t)$ along the exactly light-like minus direction.
In Feynman graph calculations, the contribution
from the gauge-link corresponds to the so-called ``eikonal factors,''
which have definite Feynman rules that
follow naturally from factorization proofs.
After a sum over graphs, and the application of appropriate approximations and Ward identity arguments,
extra collinear gluons like those shown in Fig.~\ref{fig:basicfact}(a) for SIDIS factor into gauge-link contributions.
In Fig.~\ref{fig:basicfact}(b),
the eikonal factors are shown as gluon attachments from the target-collinear bubble to a double line.



The most natural first try at extending the PDF definition
in Eq.~(\ref{eq:intDEF}) to the TMD case is to simply
leave the integration over transverse momentum in the TMD PDF
definition undone.  That is, instead of Eq.~(\ref{eq:intDEF}) one may try
\begin{equation}
\label{TMDdef1}
\Phi(x,{\boldsymbol k}_t) =  {\rm F.T.} \, \langle p | \, \bar{\psi}(0,w^-,{\boldsymbol w}_t) \gamma^+ U_{[0,w]}(u_J) \psi(0) \, | p \rangle.
\end{equation}
The separation is now $0$ and $(0,w^-,{\boldsymbol w}_t)$ --- it has
acquired a transverse component and the Fourier transform is now
in both $w^-$ and ${\boldsymbol w}_t$.  As a result, the structure of the
gauge-link $U_{[0,w]}(u_J)$ must also be modified from the simple
straight light-like $V_{[0,w]} (u_{\rm J})$ gauge-link of
Eq.~(\ref{eq:intDEF}). The eikonal attachments on either side
of the cut in Fig. \ref{fig:basicfact} still give
minus-direction Wilson lines, but now in order to have a closed
link there must also be a small transverse detour at light-cone
infinity. This detour arises naturally from boundary terms that
are needed as subtractions to make higher twist contributions
gauge invariant~\cite{Belitsky:2002sm,Boer:2003cm}.

The gauge-link structure in Eq.~(\ref{TMDdef1}), with its two
exactly light-like legs and a transverse link at infinity is
commonly cited as the gauge-link that is necessary for the
definition of the TMD PDFs.  However, there are a number of
further subtleties, and we will find that the definition needs to
be modified. One complication is that rapidity divergences, which
in collinear factorization would cancel in the sum of graphs,
remain uncanceled in the definition of the TMD correlation
functions. Rapidity divergences correspond to gluons moving with
infinite rapidity in the direction opposite the containing hadron,
and remain even when infrared gluon mass regulators are included.
(For a more complete review of these and related issues, see for
example~\cite{Collins:2008ht,Collins:2003fm}.) The most common way
to regularize the light-cone divergences is to make the gauge
links slightly non-light-like. In the coordinate space picture,
the gauge-link therefore becomes more like the \emph{tilted} hook
shape.
This introduces a new arbitrary rapidity parameter -- the ``tilt''
of the gauge-link.  A generalization of renormalization group
techniques is needed to recover predictability in the
factorization formula.  A system of evolution equations for the
TMD case was developed by Collins, Soper and Sterman (CSS) and has
been successfully applied to specific
processes~\cite{Collins:1981uk,Collins:1981uw,Collins:1984kg}.

A complete treatment of TMD-factorization involves soft gluons, which
give rise to an extra ``soft factor'' $S({\boldsymbol q})$
in the factorization formula of Eq.~(\ref{eq:basic}).  The TMD-factorization formula then becomes
\begin{equation}
\label{eq:basic2}
d \sigma \sim | \mathcal{H} |^2 \otimes \Phi(x,{\boldsymbol k}_\perp) \otimes D(z,{\boldsymbol P}_\perp) \otimes S({\boldsymbol h}_T) \;
\delta^{(2)} ({\boldsymbol q}_T + {\boldsymbol k}_\perp - {\boldsymbol P}_\perp - {\boldsymbol h}_T).
\end{equation}
The soft factor describes the role of gluons with
nearly zero center-of-mass rapidity.  One difficulty
with the usual presentation of the CSS formulation is
that the explicit appearance of a soft factor seems
somewhat counter to the basic parton model intuition
wherein all non-perturbative effects are associated
with functions for each external hadron with simple and specific probabilistic interpretations.
A natural hope is that, with an appropriate sequence of redefinitions,
the role of the soft gluons can be absorbed into
the definitions of the PDFs and FFs.  The recent work of
Collins~\cite{Collins_book2011} has shown how this is possible.  Indeed, this treatment of the soft factor
is necessary for a completely correct factorization derivation
with fully consistent definitions for the correlation functions.

While the CSS formalism has been implemented for specific spin
independent processes (see, for example, \cite{Landry:2002ix}),
much work remains to be done in tabulating and classifying the
TMDs.  This is especially true for cases that involve spin. Work
in this direction has been started in~\cite{Aybat:2011zv}.


\vskip 0.5cm

\noindent {\bf TMD-factorization breaking}

\vskip 0.5cm

The discussion 
has focussed on situations where factorization is known to hold.
There are also, however, situations where TMD-factorization is now
known to break
down~\cite{Bomhof:2004aw,Bomhof:2006dp,Bomhof:2007xt,Collins:2007nk,Collins:2007jp,Rogers:2010dm,Bomhof:2006ra,Bacchetta:2007wc}. The
key issue is the failure of the usual Ward identity arguments that
ordinarily allow eikonalized gluons to be factorized and
identified with a particular gauge-link structure in the
definitions of the TMDs.  A hint of what leads to
TMD-factorization breaking is already suggested by the well-known
overall relative sign flip in the Sivers function for SIDIS as
compared to the Drell-Yan (DY)
process~\cite{Brodsky:2002cx,Collins:2002kn}. The difference
comes because in the SIDIS TMD-factorization formula, the gauge
link in
the Sivers function 
is future pointing, whereas it is past pointing in the DY
case.  At the level of Feynman graphs, the difference can be seen
in the fact that the ``extra'' gluons which contribute to the
gauge-link attach before the hard scattering in one case, and
after the hard scattering in the other.
This illustrates that the direction of the flow of color through
the eikonal lines is
a critical factor in the definition of the correlation functions.

In the more complicated hadro-production processes, $H_1 + H_2 \to H_3 + H_4 + X$, where $H_3$ and $H_4$ may
be either jets or hadrons, a reasonable first approach would be to trace
the flow of color through the eikonal factors and use analogous arguments
to what we used for SIDIS and DY in the previous section.
One finds that the resulting structures are not simply the future
or past pointing gauge-links familiar from SIDIS or DY, but rather
are complicated and highly process dependent
objects~\cite{Bomhof:2004aw,Bomhof:2006dp,Bomhof:2006ra,Bacchetta:2007wc}.
That this corresponds (at least) to a breakdown of universality is
most directly seen in an explicit spectator model calculation. For
example, one may consider an Abelian scalar-quark / Dirac
spectator model with multiple flavors as in~\cite{Collins:2007nk}.
Then, in addition to the standard gauge-link attachments, there
are extra gluon attachments that do not cancel in a simple Ward
identity argument, and which give contributions that are not
consistent with having a simple gauge-link like what is found
SIDIS
or DY
(opposite pointing).

Therefore, it is clear that there is \emph{at least} a violation
of universality in the hadro-production of hadrons.  The natural
next approach to try is to maintain a basic factorization
structure, but to loosen the requirement that the TMDs be
universal, resulting in  a kind of ``generalized''
TMD-factorization formalism.  That is, the cross section might
still be expected to factorize order-by-order into a hard part and
well-defined, albeit non-universal, matrix elements for each
separate external hadron~\cite{Bomhof:2007xt}. However, a careful
order-by-order consideration of multiple gluons in the derivation
of TMD-factorization shows that even this is not
possible~\cite{Rogers:2010dm}. If, for example, one extends the
model of~\cite{Collins:2007nk} to allow the gluons to carry color
(while still considering a hard part that involves only the
exchange of a colorless boson) then it is straightforward to see
that the flow of color spoils the possibility of factorizing the
graph into TMD PDFs with separate gauge-links for each TMD,
regardless of what kind of gauge-link geometries are allowed.
Therefore, the problem with factorization in the hadro-production of hadrons
is more than just a problem with universality -- separate correlation
functions cannot even be defined in a way that is consistent with factorization.

The root of the problem is a failure of Ward identity arguments,
which normally allow ``extra'' gluons to be factorized after a sum
over graphs.  The Ward identity arguments are only valid after an
appropriate sequence of contour deformations on the momentum
integrals.  In the case of hadro-production of hadrons the
necessary deformations are prohibited.  In other cases where the
direction of color flow may at first appear to pose a problem for
factorization (such as in $e + p \to h_1 + X$ and $e + p \to h_1 +
h_2 + X$), the necessary contour deformations are possible and
factorization holds.  (See the explanation in chapter 12
of~\cite{Collins_book2011}.)


To summarize, we list the status of TMD-factorization for various
well-known processes with a check mark for processes where
factorization appears to be valid and $!!$ where it has been shown
to fail:

\begin{center}
\renewcommand{\labelitemi}{$\checkmark$}
\begin{itemize}
\item Semi-inclusive DIS ($e + p \to e^\prime + h_1 + X$).
\end{itemize}
\renewcommand{\labelitemi}{$\checkmark$}
\begin{itemize}
\item Drell-Yan (up to overall minus signs for some spin-dependent TMDs).
\end{itemize}
\renewcommand{\labelitemi}{$\checkmark$}
\begin{itemize}
\item Back-to-Back hadron or jet production in $e^+ e^-$ annihilation.
\end{itemize}
\renewcommand{\labelitemi}{$\checkmark$}
\begin{itemize}
\item Back-to-back hadron or jet production in DIS ($e + p \to e^\prime + h_1 + h_2 + X$).
\end{itemize}
\renewcommand{\labelitemi}{$!!$}
\begin{itemize}
\item Hadro-production of back-to-back jets or hadrons ($H_1 + H_2 \to H_3 + H_4 + X$).
\end{itemize}
\end{center}

In cases where TMD-factorization is valid, there is still much
work left to be done (and much potential insight to be gained) in
terms of implementing the evolution of precisely defined
TMDs~\cite{Aybat:2011zv}.  Much already exists for the case of
unpolarized scattering, but even here the most complete and formal
identification of evolution effects with separate TMDs has only
recently been clarified in~\cite{Collins_book2011}.  For polarization
dependent functions, it is also important to include evolution,
but to date there has been very little work that accounts for
evolution in actual fits to data.

Finally, the experimental search for TMD-factorization breaking
effects opens the possibility of new and exciting insights into
the transverse dynamics of hadronic collisions.  The breakdown of
TMD-factorization in the hadro-production of hadrons implies that
unexpected and exotic correlations between partons in
\emph{different} hadrons can exist. Calculations that allow for
experiments to distinguish between factorization and
factorization-breaking scenarios are therefore very important, and
a quantitative understanding of factorization (via the methods
of~\cite{Dominguez:2010xd}, for example) are part of the next step
toward understanding hadronic structure in high energy collisions.






\subsection{Evolution of transverse-momentum-dependent densities}
\label{sec:TMD-evolution}


Much of the success of QCD collinear factorization relies on our
ability to calculate the short-distance partonic dynamics in QCD
perturbation theory order-by-order in powers of $\alpha_s$ and the
universality as well as the scale evolution of the
non-perturbative collinear parton distribution and correlation
functions.� With its dependence on the parton's transverse momentum,
TMDs carry much richer information on the partonic structure of a
hadron than what collinear PDFs could provide.� Like the case of
collinear factorization, the predictive power of the TMD factorization
formalism also requires our ability to calculate the
short-distance dynamics and the evolution of TMDs.�� However, the
theoretical framework for calculating the evolution of TMDs and
radiative corrections to short-distance dynamics has not been
fully established.
All existing parameterizations of TMDs are extracted from SIDIS
data at relatively low $Q^2$.� The available hard scale $Q^2$ at a
future EIC is expected to be much larger.� The TMDs, like PDFs,
depend on the momentum scale $Q^2$ where they are probed.
Understanding the $Q^2$ dependence of the TMDs is crucial for
testing the TMD factorization formalism and for extracting correct
information on the partonic structure of hadrons at the EIC.
However, the $Q^2$-dependence of TMDs in the existing TMD
factorization formalism is very different from the factorization
scale $\mu_F^2$ dependence of the PDFs. The factorization scale
is not a physical scale. Any factorized physical cross section
should not be sensitive to the choice of the factorization scale.
The perturbatively calculated factorization scale dependence of
PDFs is necessarily compensated by the same scale dependence in
the high order short-distance partonic dynamics. On the other
hand, the TMDs in the existing proved TMD factorization formalism
are effectively physical quantities.� They are connected to a
physical observable by a partonic scattering cross section without
strong interaction and a soft factor which can be absorbed into
the redefinition of TMDs \cite{Collins_book2011}. Unlike the DGLAP
evolution equation of PDFs, the $Q^2$-dependence of TMDs cannot be
derived by a simple renormalization group equation.
The $Q^2$-dependence of TMDs was systematically studied in the
context of the transverse momentum ($q_T$) distribution of the Drell-Yan
process and the two-jet momentum imbalance in $e^+e^-$ collisions
\cite {Collins:1984kg}. The $Q^2$-dependence was derived by
resumming $\ln^2(Q^2/q_T^2)$-type large logarithms perturbatively
in the impact parameter $b_T$-space (a Fourier transform of the
parton's transverse momentum space). The CSS formalism was
extended to SIDIS \cite{Meng:1995yn,Nadolsky:1999kb} as well as
spin observables \cite{Ji:2004xq,Boer:2001he}.� With the proof
that the soft factor of the TMD factorization formalism could be
absorbed into the redefinition of TMDs \cite{Collins_book2011}, the CSS
resummation formalism was recently applied to the TMDs directly
\cite{Aybat:2011zv}. Within the CSS formalism, it is not the $Q^2$-dependence of
TMDs that is derived but rather the $Q^2$-dependence of the Fourier transformed TMDs
at small $b_\perp$.� In order to obtain the
$Q^2$-dependence of TMDs, one has to perform the Fourier transform
from the impact parameter $b_\perp$-space to the parton's transverse
momentum $k_T$-space.� The procedure of Fourier transform requires
necessarily input from the nonperturbative large $b_\perp$ region,
which could significantly reduce the predictive power of the TMDs
\cite{Qiu:2000ga}.� Various treatments/models for the
extrapolation into the large $b_\perp$ region have been proposed to
fit the existing data \cite{Landry:2002ix}.
For the precision study of TMDs at the EIC, it is very important
to examine the universality of the nonperturbative extrapolation
to the large $b_\perp$ region and its dependence on the observed
kinematic variables; and most important, the predictive power
of the formalism \cite{Qiu:2000ga}.� In order to understand the
$Q^2$-dependence of spin-dependent TMDs, a careful generalization
of the CSS resummation formalism to $\boldsymbol{k}_\perp$-dependent TMDs is
needed \cite{Boer:2001he}, which is necessary for the study of
asymmetries generated by the TMDs at the EIC.

\subsection{QCD Evolution for the Correlation Functions}
\label{sec:evolution}



As introduced in Sec.~\ref{sec:Sivers-example}, a collinear factorization formalism at
twist-3 is relevant for describing the SSAs of high
$P_{hT}$ particle production. Even though the phenomenological
applications of this approach have been successful, the
theoretical calculations so far have been mainly limited to the
``bare'' parton model, that is, to the zeroth order of
perturbation theory without any QCD corrections. These leading
order (LO) calculations have some disadvantages: they strongly
depend on the choice of the renormalization as well as the
factorization scale, while the physically observed SSAs should not
depend on the choice of these scales. The strong dependence on the
choice of these scales is an artifact of the LO perturbative
calculation, and a significant cancellation of the scale
dependence between the leading and the next-to-leading (NLO)
contribution is expected from the QCD factorization theorem. As
demonstrated by many examples, NLO contributions are typically
very important in hadronic processes, and often offer a more
comprehensive test of the relevant QCD factorization formalism.

To move forward to the NLO QCD dynamics, it is necessary to study
the evolution (or the scale dependence) of the universal long
distance distributions and to evaluate the perturbative
short-distance contribution beyond the LO. The evolution equation
of the twist-3 distribution functions have been derived by
different
groups~\cite{Zhou:2008mz,Vogelsang:2009pj,Kang:2008ey,Braun:2009mi}.
Recently the evolution equations for the twist-3 fragmentation
functions have also become available~\cite{Kang:2010xv}. A first
NLO calculation for the short-distance hard part function has been
presented in~\cite{Vogelsang:2009pj}.

As emphasized in Sec.\ 2.3, there are close connections between the
twist-3 collinear factorization formalism and the TMD
factorization formalism. The twist-3 correlation functions are
closely related to the relevant TMD functions. Even though the
Collins-Soper evolution equations have been derived for all the
leading-twist TMD functions~\cite{Idilbi:2004vb}, these evolution
equations are available in $b$-space ($b$ is conjugate to the
transverse momentum $k_\perp$). How these evolution equations are
transformed into the scale (or energy) dependence of the SSAs
(thus leading to a similar Collins-Soper-Sterman transverse momentum
resummation) is not yet fully understood.

The evolution equations of twist-3 distribution functions,
particularly for the so-called soft-gluonic-pole correlation
functions have been derived
in~\cite{Zhou:2008mz,Vogelsang:2009pj,Kang:2008ey,Braun:2009mi}.
Among them,
$T_F(x_1, x_2)$ and $T_F^{(\sigma)}(x_1, x_2)$ are the most
discussed ones and they are related to the Sivers and Boer-Mulders
functions~\cite{Boer:2003cm}:

\vspace{-1.3em}

\begin{align}
T_F(x, x) &=-\int d^2k_\perp \frac{|\boldsymbol{k}_\perp|^2}{M_p}
f_{1T}^{\perp}(x, k_\perp^2)|_{\rm DIS},
\nonumber \\[-0.1em]
T_F^{(\sigma)}(x,x) &=-\int d^2k_\perp
\frac{|\boldsymbol{k}_\perp|^2}{M_p} h_{1}^{\perp}(x, 
k_\perp^2)|_{\rm DIS},
\end{align}

\vspace{-0.8em}
\noindent
where $M_p$ is the nucleon mass.
The evolution equations for both $T_F(x, x)$ and
$T_F^{(\sigma)}(x, x)$ have the following generic form:
\begin{eqnarray}
\frac{\partial T(x, x,\mu^2)}{\partial\ln
\mu^2}=\frac{\alpha_s}{2\pi}\int \frac{dx'}{x'}\left[A(\hat
\xi)T(x', x',\mu^2)+B(x, x')T(x, x', \mu^2)\right],
\label{kang:tf}
\end{eqnarray}
where $T$ represents either $T_F$ or $T_F^{(\sigma)}$, and $\hat
\xi=x/x'$. As can be seen in \eqref{kang:tf}, the evolution
equation for the diagonal correlation function ($x_1=x_2=x$) is
not a closed equation since it also depends on the off-diagonal
piece (the $B(x, x')$ term). The diagonal $A(\hat \xi)$ terms are
typically similar to the relevant twist-2 splitting kernel: for
$T_F$, it is the same as the $q\to q$ splitting kernel for the
unpolarized distribution functions; for $T_F^{(\sigma)}$, it is
the same as the splitting kernel for the transversity
distribution. It might be worth pointing out that there are some
discrepancies for the evolution equation of $T_F$ in the literature:
Ref.~\cite{Braun:2009mi} contains additional contributions
compared to~\cite{Zhou:2008mz,Vogelsang:2009pj,Kang:2008ey}. One
additional piece corresponds to a contribution from the mixing
between a gluon state and quark-antiquark state, which are missing
in~\cite{Zhou:2008mz,Vogelsang:2009pj,Kang:2008ey} and could be
easily reproduced. Another term $[-N_c T_F(x, x)]$ seems difficult
to reconcile at the moment, and further study is needed to resolve
this discrepancy.

Similarly, one could study the evolution of the three-gluon
correlation functions. For an initial effort,
see~\cite{Kang:2008ey}. They receive contributions from
themselves, as well as from the quark-gluon correlation functions
$T_F$. Even though our information on three-gluon correlation
functions is very scarce, one can not rule out the possibility
that they might be large since they could be generated through the
QCD radiation from the quark-gluon correlation. It is also worth
pointing out that we now have data from PHENIX on the SSA of
$\jpsi$~\cite{Adare:2010bd}, which turns out to be non-zero and
gives some indication that three-gluon correlation functions might
be sizable. It has been suggested that open charm production in a
future Electron Ion Collider (EIC) with broader kinematics could
be used to unravel the three-gluon correlation functions.

Within the same method, one could study the evolution equations
for the twist-3 fragmentation functions. The two most important
ones are related to the first transverse-momentum-moment of the
Collins function $H_1^\perp(z, z^2k_\perp^2)$ and the polarizing
fragmentation function $D_{1T}^\perp(z,
z^2k_\perp^2)$~\cite{Yuan:2009dw,Boer:2010ya}:
\begin{eqnarray}
\hat H(z)=-z^3\int d^2k_\perp \frac{|\boldsymbol{k}_\perp|^2}{M_h}
H_1^\perp(z, z^2k_\perp^2), \quad \hat T(z)=-z^3\int d^2k_\perp
\frac{|\boldsymbol{k}_\perp|^2}{M_h} D_{1T}^\perp(z, z^2k_\perp^2),
\end{eqnarray}
with both $H_1^\perp$ and $D_{1T}^\perp$ from the convention
in~\cite{Mulders:1995dh}. These twist-3 fragmentation functions
belong to the more general two-argument fragmentation functions
denoted as $\hat H_F(z, z_1)$ and $\hat T_F(z, z_1)$, for details
on the operator definitions, see~\cite{Kang:2010xv}. The evolution
equation for $\hat H(z)$ takes the following generic form (same
form for $\hat T(z)$):
\begin{align}
\frac{\partial \hat{H}(z_h, \mu^2)}{\partial \ln\mu^2}
=\frac{\alpha_s}{2\pi}\int \frac{dz}{z} \left[A(\hat{z})
\hat{H}(z, \mu^2) +\int \frac{dz_1}{z_1^2}\,{\rm
PV}\left(\frac{1}{\frac{1}{z}-\frac{1}{z_1}}\right)B(z_h, z, z_1)
\hat H_F(z, z_1, \mu^2) \right],
\end{align}
where $\hat z=z/z_h$, and in the case of $\hat H(z_h, \mu^2)$,
$A(\hat z)$ is the same as the evolution kernel for the
transversity distribution; while for $\hat T(z_h, \mu^2)$, $A(\hat z)$ is the
same as the $q\to q$ splitting kernel for the unpolarized
fragmentation function.

We have reviewed the evolution equations
for the twist-3 distribution and fragmentation functions.
Particularly for those related to the first
transverse-momentum-moment of the Sivers and Boer-Mulders
function, and Collins and polarizing fragmentation function. These
evolution equations are generally not a closed set of equations.
However, the diagonal pieces are very similar to those appearing in
the evolution of leading-twist distribution and fragmentation
functions. For the Sivers function and polarizing fragmentation
function, this piece is the same as for the unpolarized
distribution functions. For the Boer-Mulders function and Collins
function, this piece is the same as for transversity.
The evolution equations of these functions will transform into the
scale dependence of the spin observables, which could be studied
at EIC. With a wide coverage in $x$ and $Q^2$,
EIC offers a great opportunity to study these scale dependences - a
direct test of QCD dynamics.














\label{sec:tmds}




\subsection{Non-perturbative studies of TMDs in effective approaches}
\label{sec:TMD-models}


TMDs are matrix elements of certain non-local QCD light-front
operators in hadron states and can only be calculated using
non-perturbative frameworks. Several low-energy QCD-inspired
models have been employed. Although they all have in common that
they strongly oversimplify the complexity of the QCD dynamics in
hadrons, studies in different models based on often complementary
assumptions, help to unravel non-perturbative aspects of TMDs.
Insights into non-perturbative properties are of particular
interest when confirmed in various models. The practical value of
model results is that they can be used to predict new observables,
or to guide educated Ans\"atze for fits of TMD parameterizations.
Especially in the context of TMDs one should not underestimate the
conceptual importance of model calculations. Model calculations
demonstrated the existence of effects \cite{Brodsky:2002cx}, paved
the way towards an understanding of universality in the
fragmentation process \cite{Metz:2002iz}, established new TMDs
\cite{Afanasev:2003ze,Metz:2004je}, see \cite{Metz:2004ya} for a
review. The distinction of T-even and T-odd TMDs is important also
from the point of view of modeling. In order to model the former
it is sufficient to use a model with explicit quark degrees of
freedom. In contrast, the modeling of T-odd TMDs requires the
explicit presence of gauge-field degrees of freedom.

In the following we will briefly review TMD models, though a
detailed classification of all models in which TMDs have been
studied would go far beyond the scope of this section.

\subsubsection{Models of TMDs}

An interesting model is QCD in the {\sl multicolor limit}, i.e.\
one works with $N_c\to\infty$ instead of $N_c=3$ colors. In the
large-$N_c$ limit the nucleon can be described as a classical
soliton of the chiral field \cite{Witten:1979kh}. Also for
$N_c\to\infty$ QCD cannot be solved (in $3+1$ dimensions). But
certain symmetry properties of the soliton field are known
\cite{Witten:1979kh} and can be used to derive relations which
compare the relative magnitudes of different flavor combinations
\cite{Pobylitsa:2003ty},
\begin{eqnarray}
&(f_1^u+f_1^d)                      \gg|f_1^u-f_1^d| \, ,&
 |f_{1T}^{\perp u}-f_{1T}^{\perp d}|\gg|f_{1T}^{\perp u}+f_{1T}^{\perp d}|\,,
\nonumber\\
&|g_1^u-g_1^d|                      \gg|g_1^u+g_1^d| \, ,&
 |g_{1T}^{\perp u}-g_{1T}^{\perp d}|\gg|g_{1T}^{\perp u}+g_{1T}^{\perp d}|\,,
\nonumber\\
&|h_1^u-h_1^d|  \;                  \gg|h_1^u+h_1^d| \, ,&
 |h_{1L}^{\perp u}-h_{1L}^{\perp d}|\gg|h_{1L}^{\perp u}+h_{1L}^{\perp d}|\,,
\nonumber\\
&|h_1^{\perp u}+h_1^{\perp d}|      \gg|h_1^{\perp u}-h_1^{\perp
d}|\,,&
 |h_{1T}^{\perp u}-h_{1T}^{\perp d}|\gg|h_{1T}^{\perp u}+h_{1T}^{\perp d}|\,,
\label{Eq:large-Nc}
\end{eqnarray}
where the not indicated arguments of the TMDs scale with $N_c$ as
$x\sim 1/N_c$ and $k_\perp\sim N_c^0$. Analogous relations hold for
antiquarks \cite{Pobylitsa:2003ty}. In (\ref{Eq:large-Nc}) the
respectively `large' flavor combinations are one order in $N_c$
enhanced compared to the `small' ones. For known distribution
functions the hierarchies in (\ref{Eq:large-Nc}) are roughly
supported in nature \cite{Efremov:2000ar}. The large-$N_c$
prediction \cite{Pobylitsa:2003ty} also proved useful as a
guideline for a first extraction of the Sivers function from SIDIS
\cite{Efremov:2004tp}. Conclusions about gluon TMDs can also be
drawn. For instance, $f_{1T}^{\perp g}$ is predicted to be one
order in $N_c$ suppressed with respect to the quark Sivers
distributions \cite{Efremov:2004tp}, which seems supported by
phenomenology~\cite{Brodsky:2006ha,Anselmino:2006yq}.

The first quark model to give practical results on T-even TMDs was
the {\sl quark-diquark spectator model}~\cite{Jakob:1997wg}. The
basic idea of this model is to make a spectral decomposition of
the correlation function which defines the TMDs, and to evaluate
it in the spectator approximation, i.e.\  by truncating the sum
over intermediate states to a single on-shell spectator with
definite mass. The spectator can have the quantum numbers of a
scalar (spin 0) isoscalar or axial-vector (spin 1) iso-vector
diquark, and it plays the role of an effective particle which
effectively takes into account non-perturbative effects related to the sea
and gluon content of the nucleon. The nucleon-quark-diquark
coupling is described by an effective vertex which may contain a
model-dependent form factor. This class of models with various
vertex functions and different choices for the axial-vector
diquark polarization states have been used extensively in
literature
\cite{Bacchetta:2008af,Gamberg:2003ey,She:2009jq,Lu:2010dt}. These
results for TMDs can also be interpreted in terms of overlap of
light-cone wave functions (LCWFs) for the
diquark~\cite{Brodsky:2000ii}. The advantage of the spectator
model is that the complicated many-particle system can be
effectively treated by a simple two-particle technique. However,
the price to pay is that basic properties like the momentum and
quark-number sum rules cannot be satisfied simultaneously, since
the number of quarks ``seen'' in the spectator model is only one.
This fundamental limitation  can be resolved only by considering
the diquark not as an elementary particle, but as formed by two
quarks which play the role of active particles (see, e.g.,
ref.~\cite{Cloet:2007em}).

A different approach consists in exploiting LCWFs to model the
three-quark structure of the nucleon. The three-quark LCWFs encode
the bound state quark properties of hadrons, including their
momentum, spin and flavor correlations, in the form of universal
process- and frame-independent amplitudes. Such amplitudes have
also the important property to be eigenstates of the total quark
orbital-angular momentum $L_z^q$ \cite{Ji:2002xn,Brodsky:2000ii}
and therefore, allow for mapping in a transparent way the multipole
pattern in $\boldsymbol k_\perp$ associated with each
TMD~\cite{Pasquini:2008ax,Pasquini:2009bv}. In
particular, $f_1^q,$ $g_{1L}^q$ and  $h_1^q$ describe monopole
distributions with $\Delta L_z^q=0$ between the initial and final
nucleon states, with  $f_1^q,$ $g_{1L}^q$ containing $S, P$ and
$D$ wave contributions, and $h_1^q$ only $S$ and $P$ waves. The
other twist-2 T-even TMDs are non-diagonal in the orbital angular
momentum, with $g_{1T}^q$ and $h_{1L}^q$ describing dipole
distributions due to the interference of $S-P$ and $P-D$ waves,
and  $h_{1T}^{\perp q}$ being related to a quadrupole shape  due
to a transfer of two units of orbital angular
momentum~\cite{Miller:2007ae,Burkardt:2007rv}. Two phenomenologically
successful models were used to compute the quark LCWFs: the {\sl
light-cone constituent quark model} (LCCQM) \cite{Pasquini:2008ax}
and the {\sl chiral quark-soliton model}
($\chi$QSM)~\cite{Petrov:2002jr,Lorce:2007fa,Pasquini:2010pa,Lorce:2010tv}.
In the LCCQM one describes the baryon state in terms of three free
on-shell valence quarks. The three-quark state is however not
on-shell, i.e.\ $M\neq\sum_i\omega_i$, where $\omega_i$ is the
energy of free quark $i$ and $M$ is the physical mass of the bound
state. The motion of the quarks inside the nucleon is described by
a momentum-dependent function which is assumed to have a simple
analytical expression, with free parameters fitted, e.g.,  to the
anomalous magnetic moments and the axial charge of the nucleon. In
the $\chi$QSM quarks are not free but bound by a relativistic
chiral mean field (semi-classical approximation). This field
creates a discrete level in the one-quark spectrum and distorts at
the same time the Dirac sea. Despite the different model
assumptions in LCCQM and $\chi$QSM, it turns out that the
corresponding LCWFs are very similar in structure. It should be
noticed that the $\chi$QSM naturally incorporates higher Fock
states and it has been applied to describe the unpolarized TMD for
both quark and antiquarks \cite{Wakamatsu:2009fn}.

\begin{figure}[h]
\begin{center}
\includegraphics[width=0.55\textwidth]{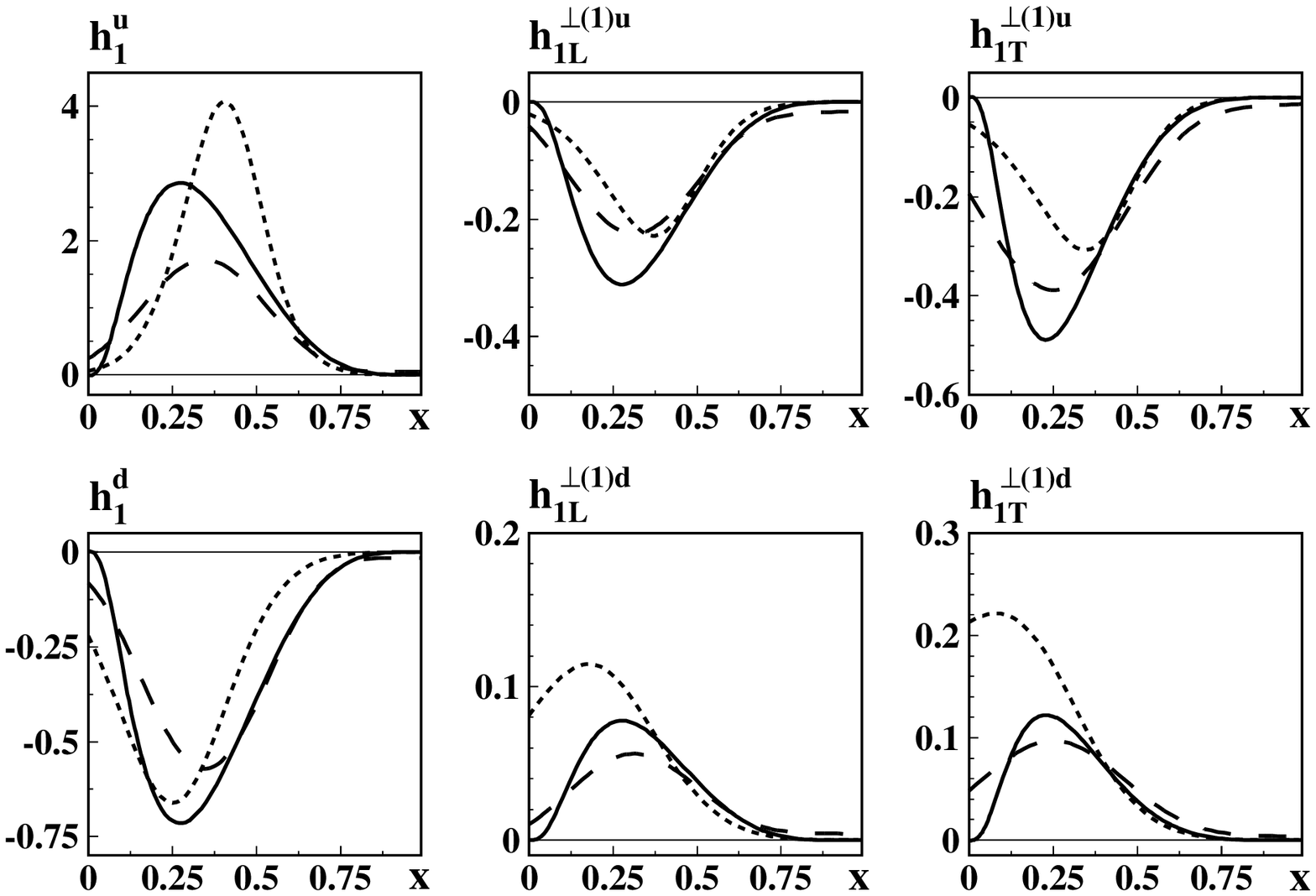}
\end{center}
\vspace{-1.3em}
\caption{\label{fig:ps-tmd-even-models} Results for $h_1^q(x)$
(left panels), $h_{1L}^{\perp(1)q}$ (middle panels) and
$h_{1T}^{\perp(1)q}$ (right panels) as functions of $x$ within
different models at low scales for up (upper panels) and down
quarks (lower panels). {\sl Dashed curves:} spectator model of
ref.~\cite{Jakob:1997wg}. {\sl Dotted curves:} bag model of
ref.~\cite{Avakian:2010br}. {\sl Solid curves:} light-cone
constituent quark model of ref.~\cite{Pasquini:2010af}.}
\end{figure}
A different model used to compute TMDs is the {\sl bag model}. In
its simplest version it describes the nucleon as three
non-interacting massless quarks confined inside a sphere.
This is therefore the only quark model discussed so far which
incorporates confinement, which is modeled by the bag boundary
condition, i.e.\ in some sense the boundary condition mimics
gluons \cite{Jaffe:1991ra}. All twist-2 and twist-3 T-even TMDs
were studied in this model in
~\cite{Avakian:2008dz}, and a complete set of linear
and non-linear relations among them was derived.
Another remarkable insight was that the bag model strongly
supports the Gaussian $k_\perp$-dependence of TMDs observed in
phenomenology \cite{Schweitzer:2010tt}.

A physical picture nearly ``opposite'' to the bag model is
provided by the covariant parton model
~\cite{Efremov:2009ze,Efremov:2010cy,Efremov:2010mt}. In this
approach the partons are free, and assumed to be described in
terms of 3D spherically symmetric momentum distributions in the
nucleon rest frame. Compliance of the model with relations derived
from QCD equations of motion allows the existence of only two such
covariant momentum distributions: one describes unpolarized and
the other polarized quarks. All twist-2 TMDs are described in
terms of these two covariant distributions. This also implies
relations among TMDs discussed in~\cite{Efremov:2009ze}. The most
interesting aspect of the model is that the symmetry of the
covariant momentum distributions tightly connects longitudinal and
transverse parton momenta. As a consequence, it is possible to
predict the $x$- and $k_\perp$-dependence of TMDs from the
$x$-dependence of known PDFs \cite{Efremov:2010mt}. Interestingly,
also this model supports the Gaussian $k_\perp$-dependence. An
important feature is that the covariant parton model yields
results which refer to a large scale. Other parton model
approaches in the context of TMDs were discussed in
\cite{Bourrely:2005tp,Bourrely:2010ng,D'Alesio:2009kv}.

TMDs in the {\sl non-relativistic limit} were studied for an
arbitrary number of colors $N_c$ in \cite{Efremov:2009ze}. In this
context we recall the popular non-relativistic model prediction
$h_1^q(x)=g_1^q(x)$. The non-relativistic model makes similar
predictions for other TMDs. In particular, it naturally explains
why in many models the integrated pretzelosity function,
$h_{1T}^{\perp q}(x)$, is so large compared to other TMDs.

Results for selected T-even TMDs computed within different models
are shown in Fig.~\ref{fig:ps-tmd-even-models}. In order to model
T-odd TMDs one needs to invoke also gauge-boson degrees of
freedom. We shall devote a separate section to that. But before
that we discuss relations among~TMDs.


In QCD all TMDs are independent functions. However, in a large
class of quark
models~\cite{Pasquini:2008ax,Jakob:1997wg,She:2009jq,Lorce:2007fa,Pasquini:2010pa,Lorce:2010tv,Avakian:2008dz,Efremov:2009ze,Efremov:2010cy,Efremov:2010mt}
there appear relations among different TMDs. In fact, certain
relations, the so-called `LIRs' (`Lorentz-invariance relations')
must hold in any consistent quark model framework without
gauge-field degrees of freedom. The 14 T-even leading- and
subleading-twist TMDs can be expressed in terms of 9 independent
`quark-nucleon scattering amplitudes' which implies the relations
\cite{Mulders:1995dh,Tangerman:1994bb} (see
\cite{Teckentrup:2009tk} for a review).

\subsubsection{T-odd TMDs}

T-odd TMDs emerge from the gauge-link structure of the parton
correlation functions which describe initial/final-state
interactions (ISI/FSI)  via soft-gluon exchanges between the
struck parton and the target remnant. Here we will summarize the
status of model calculations for the two leading-twist T-odd TMDs,
namely the Sivers function $f_{1T}^\perp$ and the Boer-Mulders
function $h_{1}^{\perp}$. Both these functions require orbital
angular momentum in the nucleon, since they involve a transition
between initial and final nucleon states whose orbital angular
momentum differ by $\Delta L_z^q=\pm 1$. Following the first
calculation which explicitly predicted a non-zero Sivers function
within a scalar-diquark model~\cite{Brodsky:2002cx}, more refined
calculation of the T-odd TMDs were performed in the spectator
models with both scalar and axial-vector
diquark~\cite{Bacchetta:2008af,Gamberg:2003ey,Goldstein:2002vv,Bacchetta:2003rz,Lu:2004au,Goeke:2006ef,Lu:2006kt,Ellis:2008in}.
Other model calculations include the bag
model~\cite{Courtoy:2008dn,Yuan:2003wk,Courtoy:2009pc}, the
non-relativistic constituent quark model~\cite{Courtoy:2008vi} and
a light-cone constituent quark model~\cite{Pasquini:2010af}.
Within all these models, the FSI/ISI are approximated by taking
into account only the leading contribution due to the one-gluon
exchange mechanism. As a result, the final expressions for the
T-odd functions are proportional to the strong coupling constant,
which plays the role of a global normalization factor with
different values depending on the intrinsic hadronic scale of the
model. Meanwhile, we  also notice  that it may be not  appropriate
to use a perturbative coupling for these non-perturbative
calculations. A non-perturbative approach was studied in
refs.~\cite{Cherednikov:2006zn,Hoyer:2005ev}, where T-odd
distributions were obtained
 from the non-perturbative chromomagnetic quark-gluon interaction induced by
instantons. A complementary approach is also to take into account
the physics of the FSI/ISI by constructing augmented LCWFs which
incorporate the rescattering effects by acquiring an imaginary
(process-dependent) phase~\cite{Brodsky:2010vs}. Finally we remark
that an interesting way to circumvent the no-go theorem concerning
the modeling of T-odd TMDs in chiral quark models
\cite{Pobylitsa:2002fr} was discussed in \cite{Drago:2005gz} where
the role of gluons is played by a `hidden vector-meson gauge
symmetry'.

Recently, interesting studies were presented, which go beyond the
one-gluon exchange approximation by resumming all order
contributions~\cite{Gamberg:2007wm,Gamberg:2009ma,Gamberg:2009uk}.
This is achieved using approximate relations between TMDs and
GPDs. In particular, the T-odd TMDs are described via
factorization of the effects of FSIs, incorporated in a so-called
``chromodynamics lensing function'', and a spatial distortion of
impact parameter space parton
distributions~\cite{Burkardt:2003uw,Burkardt:2002ks,Burkardt:2003je}.
While such relations are fulfilled from lowest order contributions
in spectator models~\cite{Meissner:2007rx,Burkardt:2003je}, they
are not expected to hold in
general~\cite{Meissner:2009ww,Meissner:2008ay}. However, the
interesting novelty  in the approach of
refs.~\cite{Gamberg:2007wm,Gamberg:2009ma,Gamberg:2009uk} is the
calculation of the lensing function using non-perturbative eikonal
methods which permit to take into account higher order gluonic
contributions from the gauge-link.

A non trivial constraint in modeling or fitting  the Sivers
function is given by the Burkardt sum rule~\cite{Burkardt:2004ur}.
This sum rule is related to momentum conservation, which requires
that the first transverse-momentum moment of the Sivers function,
i.e. the net transverse momentum due to final state interactions,
should vanish. In the bag model this sum rule is violated by a few
percent~\cite{Courtoy:2008dn,Yuan:2003wk}, since the bag states
are not good momentum eigenstates. Analogously, the
non-relativistic calculation in constituent quark models leads to
a small violation of the sum rule. In spectator models, the sum
rule is expected to be fulfilled  only when taking into account
both the quark and the diquark as explicit degrees of
freedom~\cite{Goeke:2006ef}. On the other side, it was proven to
hold in light-cone constituent quark
models~\cite{Pasquini:2010af}.

\begin{figure}[t]
\begin{center}
\includegraphics[width=0.54\textwidth]{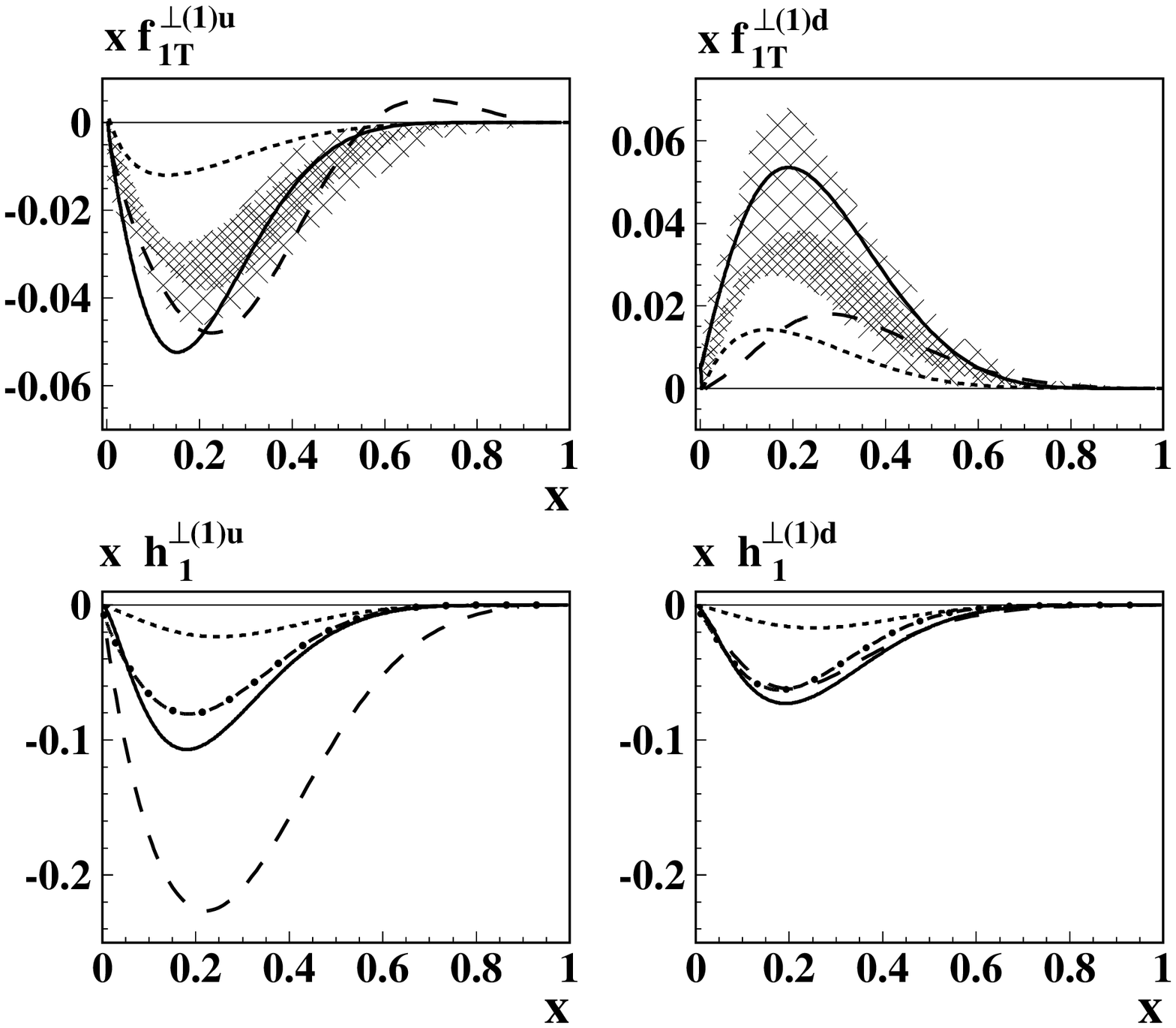}
\end{center}
\vspace{-2em}
\caption{\label{fig:ps-tmd-odd-models} Results for the (1)-moments
of the quark Sivers (upper panels) and Boer-Mulders (lower panels)
functions as function of $x$. The different curves correspond to
the results after (approximate) evolution from the model scale to
$Q^2=2.5$ GeV$^2$. {\sl Solid curves:} light-cone constituent
quark model of ref.~\cite{Pasquini:2010af}. {\sl Dashed curves:}
spectator model of ref.~\cite{Bacchetta:2008af}. {\sl Dotted
curves:} bag model of ref.~\cite{Courtoy:2008dn,Courtoy:2009pc}.
In the case of the Sivers function, the lighter and darker shaded
areas indicate statistical uncertainties of the parameterizations
of ref.~\cite{Anselmino:2008sga} and
\cite{Efremov:2004tp,Collins:2005ie}. For the Boer-Mulders
function the dashed-dotted curves are the results of the
phenomenological parametrization of
refs.~\cite{Barone:2008tn,Barone:2009hw}.}
\end{figure}

In fig.~\ref{fig:ps-tmd-odd-models} the results from different
models for the first transverse-momentum moment of the Sivers  and
Boer-Mulders functions are compared with phenomenological
parametrizations~\cite{Efremov:2004tp,Anselmino:2008sga,Collins:2005ie},
valid at an average scale of $Q^2=2.5$ GeV$^2$, extracted by a fit
to available experimental data for pion and kaon production in
semi-inclusive deep inelastic scattering. The model results are
evolved from the corresponding hadronic scale to $Q^2=2.5$
GeV$^2$, by employing those evolution equations which seem most
promising to be able to simulate the correct evolution, which is
presently not available. In particular, we evolved the (1)-moment
of the Sivers function by means of the evolution pattern of the
unpolarized parton distribution, while for the (1)-moment
Boer-Mulders function we used the evolution pattern of the
chiral-odd transversity. Within the large error bar, the results
of both the LCCQM and spectator model for the Sivers function are
compatible with the parameterizations for both up and down quark,
although the shapes of the distributions and the magnitude of the
up- and down-quark contributions  are quite different. On the
other hand, the bag model predicts much smaller results, for both
the Sivers and Boer-Mulders functions. In all the models the
Boer-Mulders function  has the same sign for both the up and down
contributions, confirming theoretical
expectations~\cite{Pobylitsa:2003ty,Burkardt:2007xm}. Furthermore, the up and down
contributions to the Boer-Mulders function are expected to have
the same  order of magnitude within the available
parametrizations~\cite{Barone:2008tn,Barone:2009hw,Lu:2009ip,Zhang:2008ez}.
This is confirmed from the predictions of the LCCQM and bag model,
while it is at variance with the spectator model where the up
distribution is more than twice bigger than the down distribution.
However, we note that the available data do not allow yet a full
fit of $h_1^\perp$ with its $x$ and $k_\perp^2$ dependence and the
available phenomenological parameterizations are only first
attempts to extract information on this distribution. New
experimental data will play a crucial role to better constrain
these analyses.

\section{Chiral-odd partonic densities}
\label{secV:chiralodd}

\hspace{\parindent}\parbox{0.92\textwidth}{\slshape
 Harut Avakian, Alessandro Bacchetta, Andreas Metz, Marco Radici}

\index{Bacchetta, Alessandro} \
\index{Radici, Marco}
\index{Avakian, Harut} 
\index{Metz, Andreas}

\vspace{\baselineskip}



Half of the leading-twist TMDs are denoted by the letter $h$, which means that
they describe the distribution of transversely polarized partons. In the helicity
basis for a spin $\textstyle{\frac{1}{2}}$ nucleon, where the unpolarized
distribution $f_1$ and the helicity distribution $g_1$ have their well known
probabilistic interpretation, transverse polarization states are given by
linear combinations of positive and negative helicity states.
Since helicity and chirality are the same at leading
twist~\cite{Jaffe:1996zw}, they are called \emph{chiral-odd}
distributions.

One of the four leading-twist chiral-odd TMDs, the transversity
distribution $h_1$, survives the integration upon transverse
momentum.
From the experimental point of view, transversity is quite an
elusive object.
In any observable the chiral-odd transversity needs to be coupled
to a chiral-odd nonperturbative partner. In SIDIS, as discussed in
Sec.~\ref{Sec-I:general-introduction}, $h_1$ can appear in the leading-twist part of the cross
section together with the chiral-odd Collins fragmentation
function $H_1^{\perp}$,
which can be determined separately, e.g., by measuring azimuthal
asymmetries of the distribution of back-to-back pions in two-jet
events in electron-positron annihilations, i.e.\ $e^+e^- \to \pi^+
\pi^- X$~\cite{Anselmino:2007fs,Boer:2008fr}.
Another promising approach to access transversity is
semi-inclusive production of pion pairs, $ep^{\uparrow} \to e'
(\pi^+ \pi^-)X$~\cite{Collins:1993kq}, where the chiral-odd
partner of $h_1$ is represented by the chiral-odd Dihadron
Fragmentation Function (DiFF)
$H_1^{\open}$~\cite{Bianconi:1999cd}.

Among the remaining chiral-odd quark distributions, the
so-called Boer-Mulders function attracted great interest from
both experiment and theory. It shares some common features
as the quark Sivers function discussed in Sec.~\ref{sec:Sivers-example}. 
In this section,
we will dedicate one subsection to briefly describe this function,
including the unique opportunity of exploring it using unpolarized
hadrons.

\subsection{The quark transversity distribution}

At leading twist, three collinear distribution functions are
needed to describe the quark distribution in the nucleon.
Transversity is a leading-twist collinear PDF and enjoys the same
status as $f_1$ and $g_1$~\cite{Ralston:1979ys,Jaffe:1991ra}. An
important difference between $h_1$ and $g_1$ is that in
spin-$\textstyle{\frac{1}{2}}$ hadrons there is no gluonic
function analogous to transversity. The most important consequence
is that $h_1^q$ for a quark with flavor $q$ does not mix with
gluons in its evolution and it behaves as a non-singlet quantity;
this has been verified up to NLO, where chiral-odd evolution
kernels have been studied so
far~\cite{Hayashigaki:1997dn,Kumano:1997qp,Vogelsang:1997ak}.

The tensor charge of the nucleon is defined as the sum of the Mellin moments
$\delta q (Q^2) = \int dx \left[ h_1^q(x,Q^2) - h_1^{\overline q}
(x,Q^2)\right] $.
Contrary to the axial charge --- which is related to $g_1^q(x,Q^2)$ ---
it has a nonvanishing anomalous dimension: it evolves with
the hard scale $Q^2$~\cite{Jaffe:1991ra}. It has been calculated on
the lattice~\cite{Gockeler:2006zu} and in various
models~\cite{Cloet:2007em,Wakamatsu:2007nc,He:2003qs,Pasquini:2006iv,Gamberg:2001qc},
and was found to be sizable.
%
%
For a more comprehensive review, we refer to Ref.~\cite{Barone:2003fy}.

The extraction of transversity is of fundamental interest for obtaining a complete 
description of the nucleon structure even for the case when internal transverse momenta
are integrated over.
To achieve this goal, it is crucial to cover the widest possible range in $(x,Q^2)$,
to measure the related asymmetries differential in the relevant kinematic variables
and to be able to perform a flavor separation.

\subsubsection{The Collins effect}
\label{sec:radici-SIDIS}

\begin{figure}[t]
\begin{center}
\includegraphics[width=0.75\textwidth]{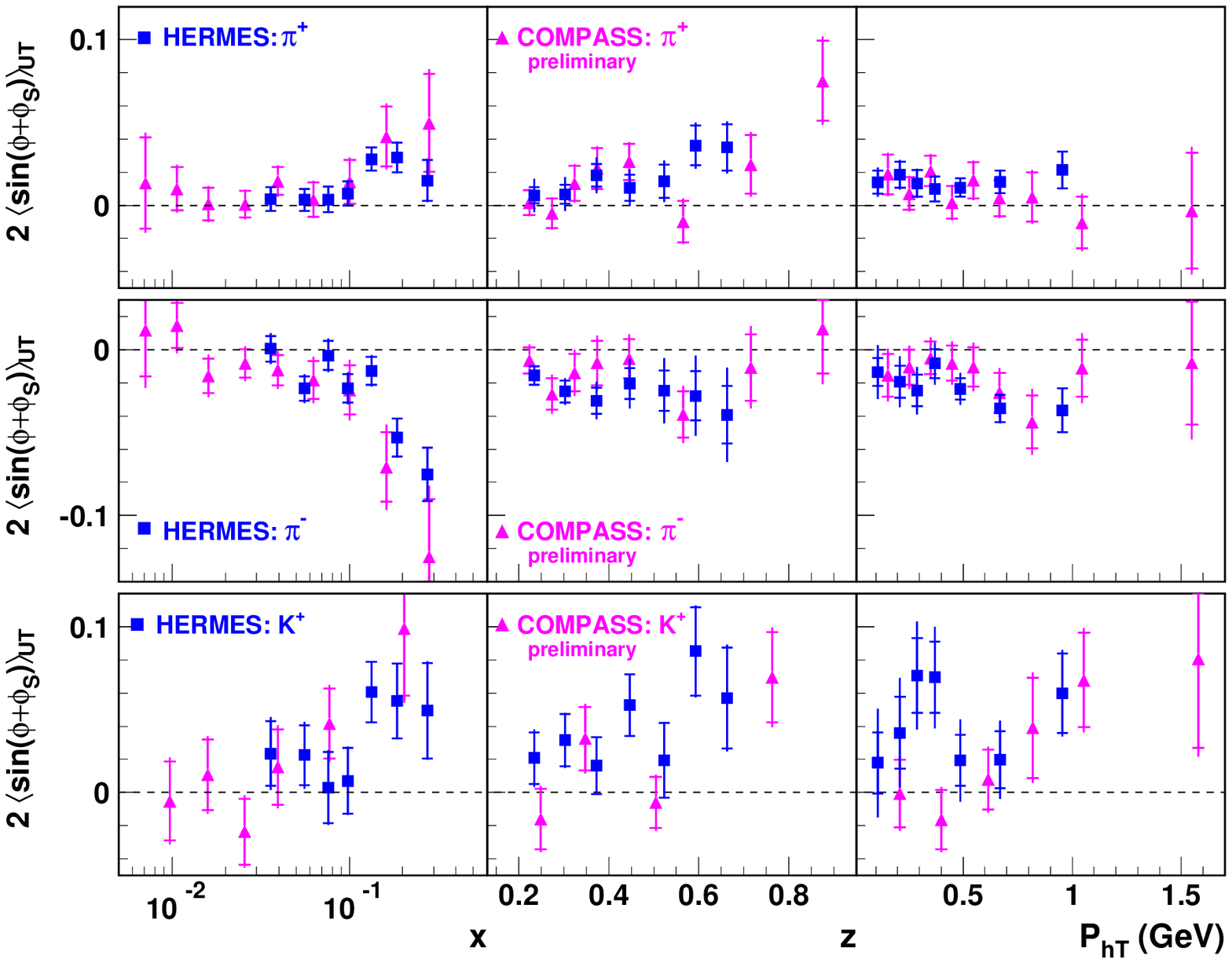}
\caption{\label{f:collins-amplitude-hermes-compass}
Collins amplitudes for $\pi^+$, $\pi^-$ and $K^+$ (as denoted in the
panels) from HERMES~\cite{Airapetian:2010ds} 
and COMPASS~\cite{Pesaro:2011zz} measured with a proton target.
Inner error bars present statistical uncertainties and full error bars the quadratic
sum of statistical and systematic uncertainties.
Note that the average kinematics in each bin differs for HERMES and COMPASS and the sign of the 
COMPASS asymmetries have been reversed.
}
\end{center}
\end{figure}

As discussed in Sec.~1, at tree-level and leading-twist, the SIDIS
$F_{UT}^{\sin (\phi_h+\phi_S)}$ structure function of
Eq.~\eqref{e:crossmaster} can be described as a  convolution
between the transversity $h_{1T}^q$ and the Collins fragmentation
function $H_1^{\perp\,q}$, i.e.,
%
%
%
\begin{equation}
F_{UT}^{\sin \left( \phi_h +\phi_S \right)}
\sim \sum_q e_q^2 \,\, h_{1T}^{q} \otimes H_1^{\perp q} \,.
\label{eq:radici-CollinsSF}
\end{equation}
%
In order to project out the structure function $F_{UT,T}^{\sin(\phi_h +\phi_S)}$ in Eq.~(\ref{e:crossmaster}),
the so-called Collins amplitude $2 \langle \sin(\phi_h +\phi_S) \rangle_{UT}^h$
for a specific hadron $h$ is extracted from the asymmetry
\begin{equation}
A_{UT}^h(\phi_h,\phi_S) \equiv \frac{1}{|{\bf S}_T|} \frac{d\sigma^h(\phi_h,\phi_S) + d\sigma^h(\phi_h,\phi_S+\pi)}{d\sigma^h(\phi_h,\phi_S) + d\sigma^h(\phi_h,\phi_S+\pi)}\;,
\end{equation}
where the subscript $U$ indicates an unpolarized lepton beam and $T$ a transversely polarized
target nucleon.
The azimuthal angles are illustrated in Fig.~\ref{f:anglestrento}.
This amplitude has so far been extracted by three polarized fixed-target experiments as summarized 
in Table~\ref{t:exp-collins-measurements}.
\begin{table}[t]
\begin{center}
\begin{tabular}{lcccc}\hline
experiment (laboratory) & $\sqrt{s}$ in GeV& target type & hadron types & references \\\hline
COMPASS (CERN) & 18 & deuteron & $h^{\pm},\pi^{\pm},K^{\pm},K^0$ & ~\cite{Ageev:2006da,Alekseev:2008dn}\\
 &  & proton & $h^{\pm}$ & ~\cite{Alekseev:2010rw}\\
 &  & proton & $\pi^{\pm},K^{\pm}$ & prelim.~\cite{Pesaro:2011zz}\\\hline 
HERMES (DESY) & 7.4  & proton & $\pi^{\pm}$ & \cite{Airapetian:2004tw} \\ 
 &  & proton & $\pi^{\pm},\pi^{0},K^{\pm}$ & \cite{Airapetian:2010ds} \\\hline
HallA (JLab) & 3.5  & neutron & $\pi^\pm$ & prelim.~\cite{HallA-sivers-neutron-2010}  \\\hline
\end{tabular}
\caption{
Summary of currently available measurements of Collins asymmetry amplitudes from
lepton-nucleon DIS experiments, their center-of-mass energy, transversely polarized target type, 
and analyzed hadron types.
}
\label{t:exp-collins-measurements}
\end{center}
\end{table}
From these measurements, Fig.~\ref{f:collins-amplitude-hermes-compass} shows a selection of results 
that are significantly non-zero and help in determining both the shape of transversity 
and the relative size and sign of the Collins fragmentation function.
All other asymmetry amplitudes listed in Table~\ref{t:exp-collins-measurements} are small or
consistent with zero.
%

For the second unknown in Eq.~(\ref{eq:radici-CollinsSF}), the Collins fragmentation function,  
model calculations are available
~\cite{Bacchetta:2007wc,Artru:1995bh,Bacchetta:2001di,Bacchetta:2002tk,Gamberg:2003eg,Bacchetta:2003xn,Amrath:2005gv,Artru:2010st}.
However, for a model-independent extraction of transversity from
the SIDIS asymmetry amplitudes we need to determine the Collins function
from an independent source. This is represented by the measurement
of azimuthal asymmetries in the distribution of back-to-back pions
in two-jet events in electron-positron annihilations, i.e. $e^+e^-
\to \pi^+ \pi^- X$~\cite{Boer:1997mf}.

\begin{figure}
\begin{center}
\mbox{
\hspace*{-0.3cm}
\includegraphics[width=0.50\textwidth]{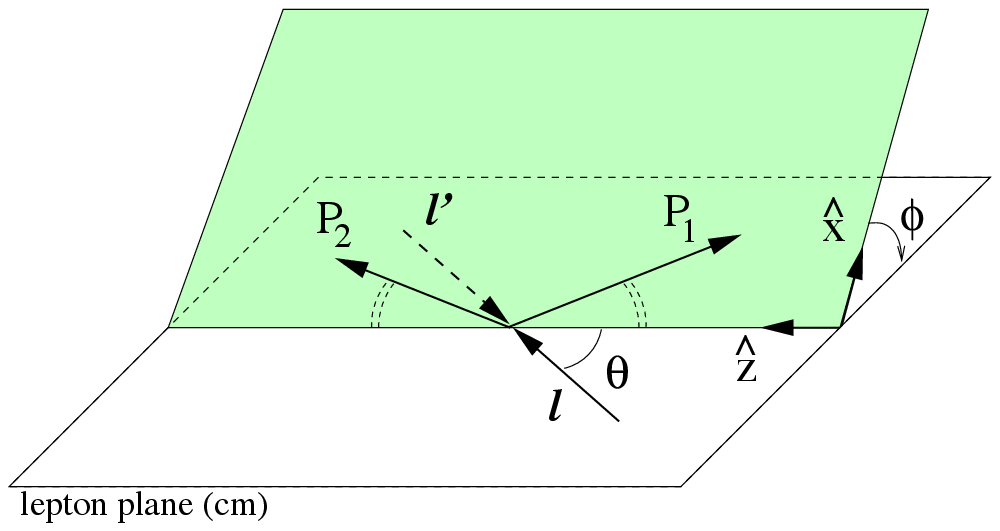}\hspace{.3cm}
\includegraphics[width=0.47\textwidth]{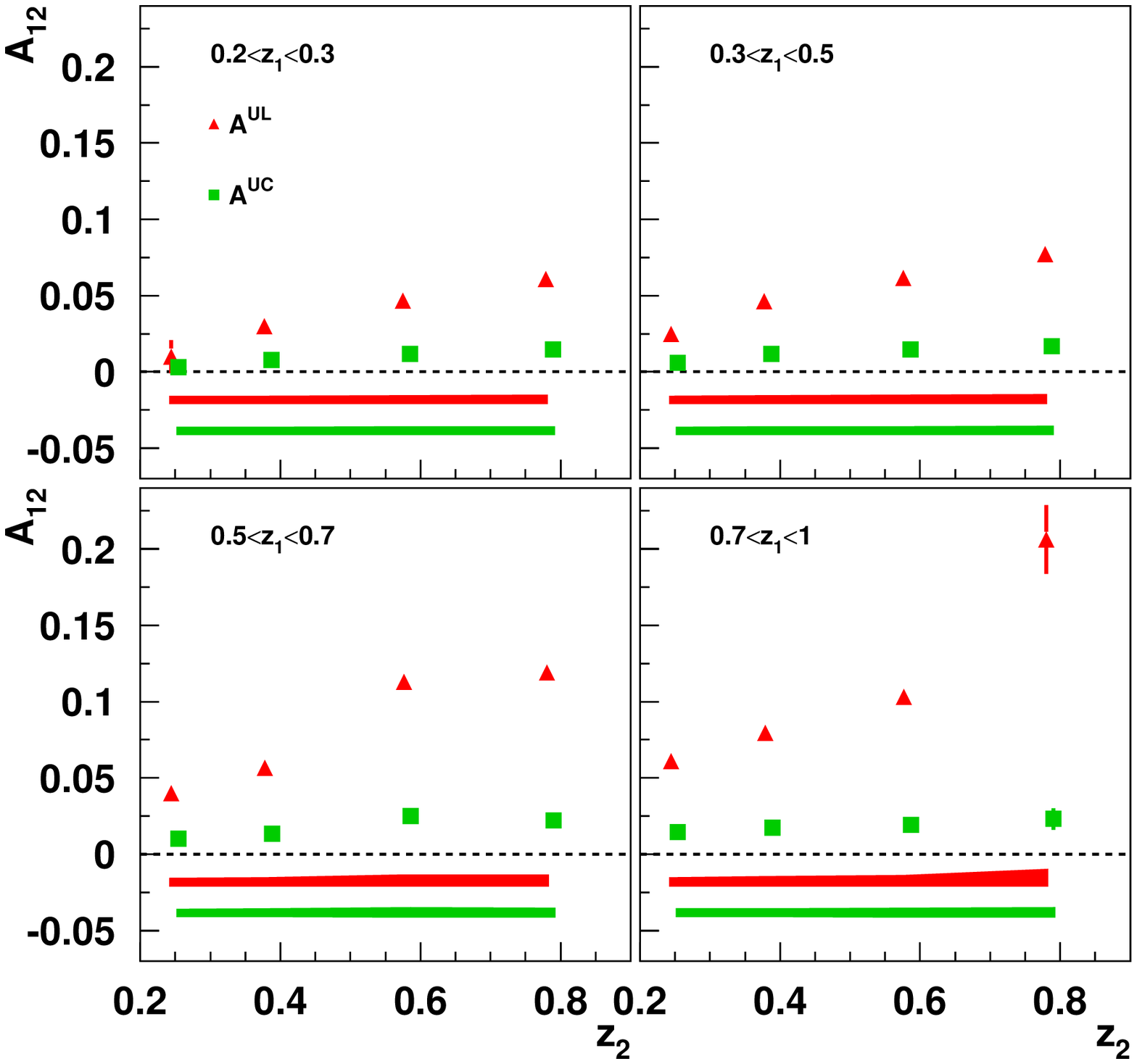}
}
\end{center}
\caption{\label{fig:belle-Collins} 
Left: the kinematics of $e^+e^-$ annihilation
leading to back-to-back jets along the $\hat{\bm{z}}$ axis (jet frame), 
$P_1$ is the momentum of a hadron in one jet, $P_2$ is the momentum of a
hadron in the other jet. 
Right:
Collins asymmetry $A_{12}$ for the double ratios for like-sign (L), unlike-sign (U) and 
any charged (C) pion pairs as function of $z_2$ in bins of $z_1$ from 
BELLE~\cite{Seidl:2008xc}.
$A^{UL}$ and $A^{UC}$ are sensitive to different combinations of the favored and
unfavored Collins fragmentation functions.
}
\end{figure}

The relevant vectors and angles involved in $e^+e^-$ annihilations leading to back-to-back jets
are depicted in Fig.~\ref{fig:belle-Collins} (left panel).
The following asymmetry can be measured~\cite{Anselmino:2007fs,Boer:2008fr}
\begin{equation} 
A_{12} (z_1,z_2,\theta_2,\phi_{1}+\phi_{2}) =
1+\frac{\sin^2\theta}{1+\cos^2\theta}\,\cos (\phi_{1}+\phi_{2})\,
\frac{\sum_q\,e_q^2\,H_1^{\perp\,(1)\,q} (z_1) \,
H_1^{\perp\,(1)\,{\overline q}} (z_2) }{\sum_q\,e_q^2\,D_1^q(z_1)\,
D_1^{\overline q}(z_2)} \; .
\label{eq:radici-e+e-LEPA}
\end{equation}
%

Pioneering measurements of this spin-dependent fragmentation function have been
performed by the BELLE Collaboration (KEK)~\cite{Abe:2005zx,Seidl:2008xc}.
Experimentally, double ratios of asymmetries for like-sign (L), unlike-sign (U) and 
any charged (C) pion pairs are built in order to cancel (to a large extent) contributions from 
the experimental acceptance and radiative effects.
The resulting asymmetries, $A^{UL}$ and  $A^{UC}$,
are then sensitive to different combinations of the favored and unfavored Colllins
fragmentation functions as given in~\cite{Seidl:2008xc}.
These asymmetries are presented in Fig.~\ref{fig:belle-Collins} as function of $z_2$ for
four bins of $z_1$ for the light quarks ($u,d,s$), where $z_1$ and  $z_2$ are for a
hadron in each of the back-to-back jets.

   
The experimental results shown in Figs.~\ref{f:collins-amplitude-hermes-compass}
and~\ref{fig:belle-Collins} are striking.
First, they clearly demonstrate that the Collins effect as a
manifestation of chiral-odd and na\"ive T-odd mechanisms is
different from zero and not suppressed, both in SIDIS and in
$e^+e^-$ annihilations. 
Second, the results for oppositely charged pions (hadrons) in 
Fig.~\ref{f:collins-amplitude-hermes-compass} suggest a very peculiar feature
for the Collins fragmentation function.
As scattering off $u$ quarks dominates these data due to the charge factor,
the large magnitude of $\pi^-$ amplitudes being of similar size
than the  $\pi^+$ ones but having opposite sign, 
can only be understood if the  \emph{dis}favored Collins function $H_1^{\perp\,\mathrm{unfav}}$
is large and of opposite sign to the favored one.
Opposite signs for the  favored and unfavored Collins functions are also 
supported by the different size of $A^{UL}$ and $A^{UC}$ asymmetries from BELLE 
in Fig.~\ref{fig:belle-Collins}.
They can be understood in light of the string model of
fragmentation~\cite{Artru:1995bh} (and also of the
Sch\"afer--Teryaev sum rule~\cite{PhysRevD.61.077903}).
If a favored pion is created at the string end by the first break, 
an unfavored pion from the next break is likely to inherit transverse momentum
in the opposite direction.

\begin{figure}
\begin{center}
\includegraphics[width=0.31\textwidth]{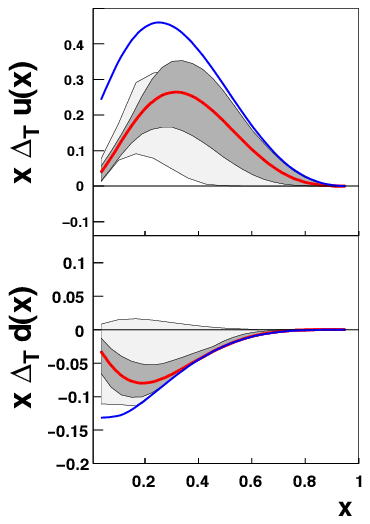}\hspace{1.5cm}
\includegraphics[width=0.31\textwidth]{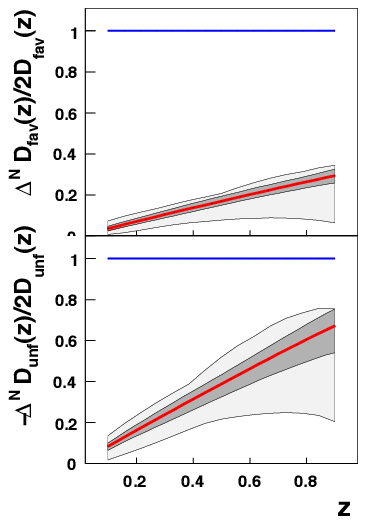}
\end{center}
\vspace{-1cm}
\caption{\label{fig:radici-h1H1params}
Left: transversity $x h_1^q(x)$ for $u$ (upper panel) and $d$ (lower panel) quarks.
Right:  the normalized Collins functions
$\sqrt{2}\, H_1^{\perp\,(1/2)}(z)/D_1(z)$ for favored (upper panel) and unfavored
(lower panel) fragmentation. 
The light grey band represents the uncertainty for the extraction in 
Ref.~\protect{\cite{Anselmino:2007fs}} and the dark grey band from the updated
analysis~\protect{\cite{Anselmino:2008jk}}.
Blue lines indicate the  Soffer and positivity bound for transversity and
Collins function, respectively.
}
\end{figure}

The extraction of transversity and Collins functions from available data faces the same
issues as discussed for the Sivers function in Sec.~\ref{sec:sivers-phenomenology-models}
for resolving the convolution in Eq.~(\ref{eq:radici-CollinsSF}) and the same strategies
are applied here.
Employing the Gaussian Ansatz in Eq.~(\ref{eq:TMD-gaussian-ansatz}) both 
transversity and Collins function have been extracted~\cite{Anselmino:2007fs,Anselmino:2008jk}
from (part of) the experimental data discussed before.
The new COMPASS proton or Hall-A neutron data are not yet included in this fit.
The results of this global analysis are presented in Fig.~\ref{fig:radici-h1H1params}
for $u$ and $d$ transversity distributions (left panel) and favored and unfavored
Collins fragmenation functions (right panel).
The decrease in the presented uncertainties for the specifically chosen parametrization, 
which is the same as in~\cite{Anselmino:2007fs,Anselmino:2008jk}, 
is due to the new BELLE and HERMES data. 
The extracted favored and unfavored Collins functions confirm the features discussed
before.

\subsubsection{Dihadron Fragmentation Functions}
\label{sec:radici-DiFF}

A complementary approach to transversity is provided by semi-inclusive
two-hadron production, $ep^{\uparrow} \to e' (h_1\,h_2 ) X$, where the two
unpolarized hadrons with momenta $P_1$ and $P_2$ 
emerge from the fragmentation of the struck quark.
The underlying mechanism differs from the Collins mechanism in that the 
transverse spin of the fragmenting quark is transferred to the \emph{relative}
orbital angular momentum of the hadron pair. 
Consequently, this mechanism does not require transverse momentum of the hadron
pair and collinear factorization applies.

Dihadron fragmentation functions were introduced in Ref.~\cite{Konishi:1978yx}
and studied for the polarized case in Refs.~\cite{Collins:1993kq,Jaffe:1997hf,Artru:1995zu}.
The decomposition of the SIDIS cross section in terms of quark
distributions and dihadron fragmentation functions was carried out to leading twist in
Ref.~\cite{Bianconi:1999cd} and to sub-leading twist in
Ref.~\cite{Bacchetta:2003vn}. 

\begin{figure}
\begin{center}
\includegraphics[width=0.45\textwidth]{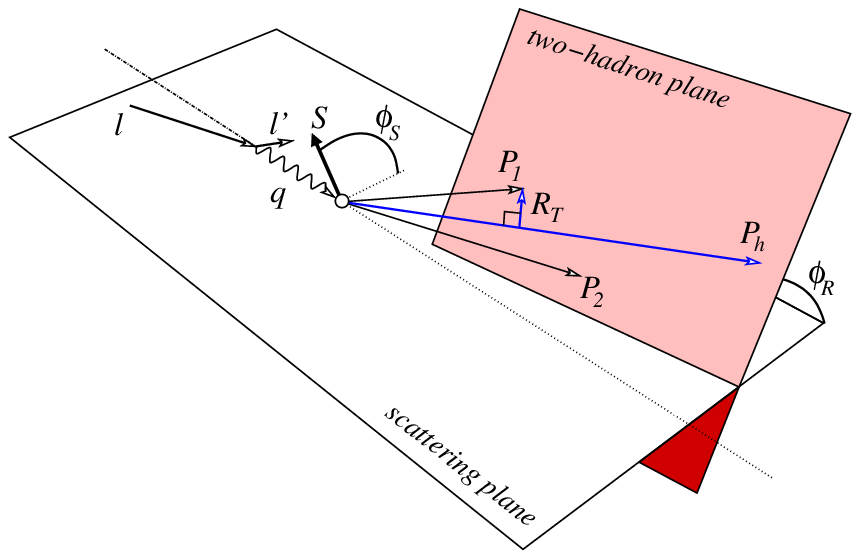}
\includegraphics[width=0.4\textwidth]{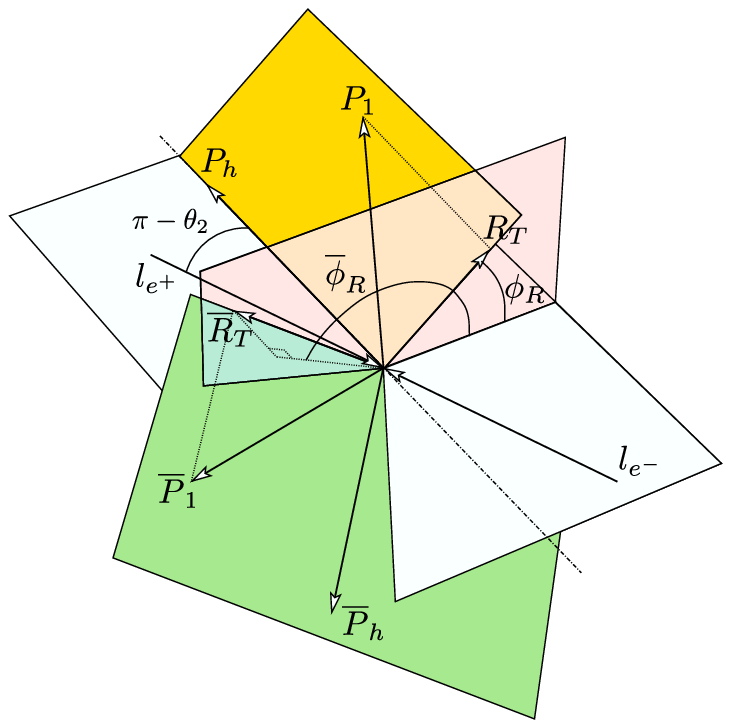} 
\end{center}
\caption{\label{fig:radici-diffkin} 
Kinematics for the production of two hadrons (left) and for the 
$e^+e^- \to (\pi^+ \pi^-)_{\mathrm{jet1}}\,(\pi^+ \pi^-)_{\mathrm{jet2}} X$ process (right). 
}
\end{figure}

The kinematics is similar to the one in single-hadron SIDIS
except for the final hadronic state, where now $z=z_1+z_2$ is the fractional energy carried by the hadron pair and we introduce the vectors $P_h=P_1+P_2$ and $R=(P_1-P_2)/2$ (see
Fig.~\ref{fig:radici-diffkin}), together with the pair
invariant mass $M_h$, which must be considered much smaller than the hard
scale (e.g., $P_h^2 = M_h^2 \ll Q^2$).
We shall often use the quantity~\cite{Bacchetta:2002ux},
\begin{equation}
|\bm{R}| =
\frac{1}{2}\, \sqrt{M_h^2-2\,(M_1^2+M_2^2)+(M_1^2-M_2^2)^2} \quad ,
\label{eq:radici-DiFFRT}
\end{equation}
where $P_1^2 =M_1^2,\, P_2^2=M_2^2$ and $R_T^2$ is related to $M_h^2$~\cite{Bacchetta:2002ux}.

In analogy with the Collins function,
the expression for unpolarized hadrons $(h_1,\, h_2)$ produced by
a transversely polarized quark reads
\begin{equation}
D_{h_1 h_2/q^{\uparrow}} (z,M_h^2,\bm{R}_{T}) = D_1^q(z,M_h^2) -
H_{1 sp}^{\open\, q} (z,M_h^2)\, \frac{\bm{S}_{\perp\,q} \cdot ( \hat{\bm{p}}
\times \bm{R}_{T} )}{M_h} \quad .
\label{eq:radici-DiFFdensity}
\end{equation}
Choosing $\hat{\bm{p}} \parallel \hat{z}$ and $\bm{S}_{\perp\,q}
\parallel \hat{y}$, a positive $H_{1 sp}^{\open\, q}$ means that
hadron $h_1$ is preferentially emitted along $-\hat{x}$ and hadron
$h_2$ along $\hat{x}$.

Since $\bm{R}_T = \bm{R}\,\sin\theta$, where in the c.m.\ frame of the hadron pair $\theta$ is the angle between $P_1$ and the direction of $P_h$ in the laboratory frame (for more details, see refs.~\cite{Bacchetta:2002ux,Bacchetta:2006un,Airapetian:2008sk,Bacchetta:2008wb}),
the relevant asymmetry that should be measured in SIDIS is
\begin{eqnarray}
A_{UT}^{\sin(\phi_R^{} + \phi_S^{})\sin \theta}
&\equiv
&2\, \frac{\int d \cos\theta d\phi_R d\phi_S \, \sin (\phi_R + \phi_S )\, [
d\sigma (\phi_R,\,\phi_S) - d\sigma (\phi_R,\,\phi_S+\pi) ]\, / \sin\theta}{\int d \cos\theta
d\phi_R d\phi_S \, [ d\sigma (\phi_R,\,\phi_S) + d\sigma (\phi_R,\,\phi_S+\pi) ]} \nonumber \\
&\sim & \frac{|\bm{R}|}{M_h}\, \frac{\sum_q e_q^2\,\,h_1^q(x)\
H_{1\,sp}^{\open \,q}(z, M_h^2)}{\sum_q e_q^2\,\,f_1^q(x)\, D_1^q (z, M_h^2)}
\quad .
\label{eq:radici-DiFFas}
\end{eqnarray}

As in the single-hadron production case,
transversity can be extracted from the
asymmetry~\eqref{eq:radici-DiFFas} only if the unknown $H_{1
sp}^{\open}$ is independently determined from the $e^+e^-$
annihilation producing, in this case, two hadron pairs: $e^+e^-
\to (\pi^+ \pi^-)_{\mathrm{jet1}}\,(\pi^+ \pi^-)_{\mathrm{jet2}}
X$ with kinematics depicted in Fig.~\ref{fig:radici-diffkin} (right).
The relevant signal is similar to that of the Collins
function, except that each transverse polarization of the
quark-antiquark pair is now correlated to the azimuthal
orientation of the plane formed by the momenta of the
corresponding hadron pairs, suggesting that $H_1^{\open}$ is
related to the concept of handedness of the jet containing a
specific pair~\cite{Efremov:1992pe,Stratmann:1992gu,Boer:2003ya}.


The leading-twist cross section of this process
contains many terms~\cite{Boer:2003ya}, among which there is one
involving the product of $H_{1 sp}^{\open\, q}$ for the quark $q$
and of $\overline{H}_{1\, sp}^{\open\, q}$ for the ${\overline q}$
partner, weighted by $\cos (\phi_R + {\overline \phi}_R)$. Thus,
we can properly weight the cross section and extract this
contribution by defining the so-called Artru--Collins azimuthal
asymmetry~\cite{Boer:2003ya,Bacchetta:2008wb}
\begin{align}
A^{\cos (\phi_R + {\overline \phi}_R)}(\cos\theta_2, z,
M_h^2, \overline{z}, \overline{M}_h^2) &=
\frac{\sin^2 \theta_2}{1+\cos^2 \theta_2}
\frac{\pi^2}{32} \frac{|\bm{R}|\,|\overline{\bm{R}}|}{M_h\,\overline{M}_h}
\frac{\sum_q e_q^2  H_{1\,sp}^{\open q}(z, M_h^2) \overline{H}_{1\,
sp}^{\open q}(\overline{z}, \overline{M}_h^2)}{\sum_q e_q^2 D_1^q (z,
M_h^2)  \overline{D}_1^q (\overline{z}, \overline{M}_h^2) } \quad ,
\label{eq:radici-DiFFe+e-as}
\end{align}
where the dihadron fragmentation functions $D_1^q$ and $H_{1\, sp}^{\open\, q}$ are the
same universal functions appearing in the SIDIS asymmetry of equation~\eqref{eq:radici-DiFFas}.



\begin{figure}
\begin{center}
\includegraphics[width=0.6\textwidth]{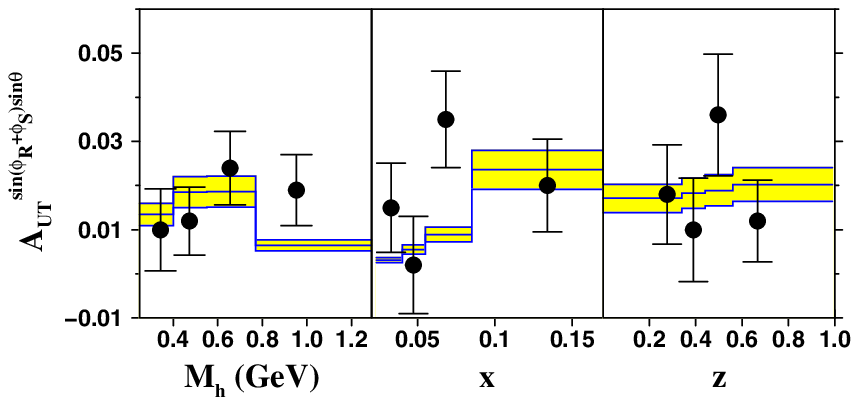}
\end{center}
\vspace{-.7cm}
\caption{\label{fig:radici-diffHermes} The spin asymmetry for the
semi-inclusive production of a pion pair in deep-inelastic scattering on a
transversely polarized proton~\protect{\cite{Airapetian:2008sk}}.
The grey band presents a fit to the data involving the dihadron FF calculated in
the spectator model of
Ref.~\protect{\cite{Bacchetta:2006un}} and on the  parametrization for $h_1$ from
Ref.~\protect{\cite{Anselmino:2008jk}}. }
\end{figure}
%
%

Pioneering measurements of $A_{UT}^{\sin(\phi_R^{} + \phi_S^{})\sin\theta}$ 
from HERMES~\cite{Airapetian:2008sk} gave evidence for a non-zero
dihadron fragmentation function $H_{1\,sp}^{\open\, q}$ as shown in Fig.~\ref{fig:radici-diffHermes}.
The $M_h$ dependence does not exhibit any sign change and rules out the model of 
Ref.~\cite{Jaffe:1997hf}: interference patterns in semi-inclusive $\pi^+\pi^-$ production 
are different from those in $\pi^+\pi^-$ elastic scattering. 
Calculations based on the spectator model~\cite{Bacchetta:2006un,She:2007ht} are 
compatible with data.
They, however, overestimate the asymmetries 
if $h_1^q$ is taken from the parametrization~\cite{Anselmino:2008jk} 
discussed in Fig.~\ref{fig:radici-h1H1params}.
This estimate is presented in Fig.~\ref{fig:radici-diffHermes} by the grey band
where the model $H_{1\,sp}^{\open\, q}$ is reduced by a factor $\alpha = 0.32
\pm 0.06$ in order to reproduce the magnitude of the asymmetry.

Preliminary SIDIS data are also available from the COMPASS Collaboration
using transversely polarized deuteron and
hydrogen~\cite{Wollny:2009eq} targets. 
While $A_{UT}^{\sin(\phi_R^{} +\phi_S^{})\sin\theta}$ is basically vanishing on the deuteron, 
the proton data show a signal larger than the HERMES results in
Fig.~\ref{fig:radici-diffHermes}, which might be due to different kinematics.

Last but not least, results from pioneering measurements of the 
 $A^{\cos (\phi_R + {\overline \phi}_R)}$ asymmetry related to the
dihadron fragmentation function became recently available 
from the BELLE Collaboration in Ref.~\cite{Vossen:2011fk}.

For a real breakthrough of this promising approach to transversity, much
more data over wide kinematic range are needed.
We only mention that the SIDIS cross section does now depend on nine kinematic
variables compared to six for the single-hadron case, which calls even more for 
a multi-dimensional analysis for a bias-free extraction of the asymmetries.

\subsubsection{Collins effect at EIC} \label{sec:radici-eic}

The exploration of chiral-odd structures using the Collins effect
is far from being complete. 
Several aspects need to be significantly improved.
The $x$ dependence is largely unconstrained due to the lack of 
SIDIS asymmetries outside the range $0.005 \lsim x \lsim 0.3$.
%
The antiquark and sea-quark content of transversity in the proton
is completely unknown. 
Together with the loose constraints on the $x$
dependence, this missing piece of information makes the
calculation of the tensor charge still unsatisfactory. 
Also the transverse momentum dependence of both the transversity and the
Collins function has a significant degree of arbitrariness.
Lastly, the $Q^2$ range of HERMES and COMPASS measurements is
approximately the same: it would be desirable to study the
$A_{UT}^{\sin (\phi_h+\phi_S)}(Q^2)$ dependence in a wide range
of $Q^2$.

All these remarks call for more data in order to enlarge the phase
space and perform a multi-dimensional analysis in all relevant
kinematic variables simultaneously.
An ambitious program is planned at JLab12, that would aim for exploring
$A_{UT}^{\sin (\phi_h+\phi_S)}$ in the valence region with high 
luminosity~\cite{Gao:2010av,JLab12-Clas}.
The EIC would be the ideal facility to carry out this program over a uniquely wide
range in $x$ and $Q^2$.
This potential for a mapping of the multi-dimensional phase-space 
in an unprecedented kinematic range is illustrated by the
studies presented in Sec.~\ref{sec:sivers-at-EIC} and are equally valid for transversity.

\begin{figure}
\begin{center}
\includegraphics[width=0.95\textwidth]{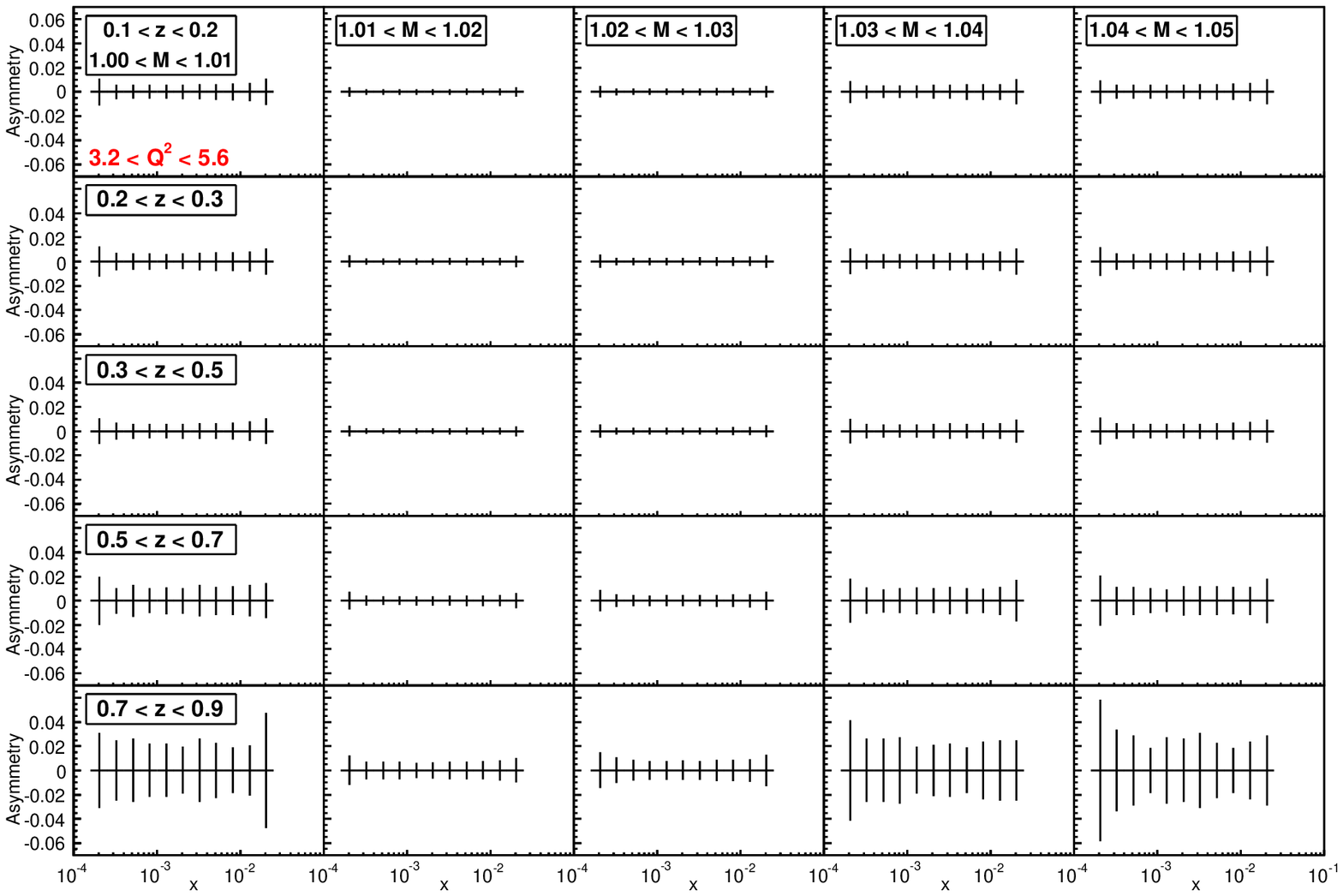}
\end{center}
\caption{\label{fig:dihadron-at-EIC}
Projected accuracy, represented by the error bars, 
for semi-inclusive kaon pair production obtained with an 
energy of $\sqrt{s} = 140$ GeV for an integrated luminosity of 30 fb$^{-1}$, 
as a function of $x$ in bins in $z$, $M_{KK}$ and 
for a single bin in $Q^2$ as indicated in the panels.
}
\end{figure}

The promising and complementary approach of extracting transversity with help
of the dihadron fragmentation function will even more profit from the 
high energy option of an EIC.
Fig.~\ref{fig:dihadron-at-EIC} shows the projected accuracy for semi-inclusive
kaon pair production at an energy $\sqrt{s} = 140$ GeV and for an integrated 
luminosity of 30 fb$^{-1}$. 
The PYTHIA event generator has been used to obtain the SIDIS event rate, and an 
overall detection efficiency of 50$\%$ and beam polarization of 70$\%$ were assumed. 
Data are shown as function of $x$ for the various different $z$ and $M_{KK}$ bins indicated
in the panels. The invariant mass range of the kaon pair, $M_{KK}$, is chosen for 
the vicinity of the $\phi$ meson, which 
provides unique access to strange quark distributions. 

Furthermore, the general picture obtained so far would significantly
profit from data available over a wide $Q^2$ range which can only be
provided by the EIC.
This picture obtained so far, is based on
a tree-level analysis of transverse-momentum dependent azimuthal
(spin) asymmetries occurring at very different energies: while the
average scale of SIDIS experiments is approximately 2.5 GeV$^2$,
the BELLE measurement was performed at the typical bottonium mass,
i.e. $Q^2 \sim 100$ GeV$^2$. Beyond tree level, the evolution
effects with running scale were included (at LO) only in the
modification of the $x$ and $z$ dependence of the various
functions. At low $\bm{P}_{hT}^2/Q^2$ ($Q_T^2/Q^2$ for $e^+e^-$
annihilation, where $Q_T=|\bm{q}_T|$ is the transverse momentum of
the virtual photon), the correct $Q^2$ dependence beyond tree
level of transverse-momentum dependent structure functions should
be studied extending the Collins--Soper--Sterman formalism
mentioned in Sec.~\ref{sec:TMD-resum}~\cite{Collins:1984kg}. A
quantitative attempt to go in this direction was presented in
Ref.~\cite{Boer:2008fr}, where it was estimated that
transverse-momentum resummation produces a suppression of the tree
level result by almost a factor 5 at BELLE energies. Therefore,
the extraction of the Collins function using the tree level
formula 
could significantly underestimate its actual magnitude. In order
to fit the available SIDIS asymmetries, a
larger $H_1^{\perp}$ would automatically imply a transversity
smaller than that one illustrated in
Fig.~\ref{fig:radici-h1H1params}.

\subsection{Boer-Mulders function}
\label{SubSec-V_2:Boer-Mulders-function}





%

The Boer-Mulders function $h_1^\perp$~\cite{Boer:1997nt} can be
considered as the counterpart of the Sivers function
$f_{1T}^\perp$: while $f_{1T}^\perp$ describes the distribution of
unpolarized quarks in a transversely polarized target, $h_1^\perp$
describes the distribution of transversely polarized quarks in an
unpolarized target. Both functions are T-odd, and therefore vanish
if the gauge-link is not taken into account in their operator
definition, which makes them somewhat unique among the TMDs. Put
it differently, their existence depends on the presence of initial
and/or final state interactions between the active partons of a
process and the target remnants (see the corresponding discussion
in Sec.~\ref{secIV:mulders-rogers}). It is expected that both
TMDs change their sign when going from SIDIS to the Drell-Yan
process~\cite{Collins:2002kn}. There is, however, one important
difference between them. The Sivers function is chiral-even,
whereas the Boer-Mulders function is chiral-odd. Since the
elementary interactions of the Standard Model do not change the
chirality (helicity) of fermions, one has to couple the
Boer-Mulders function --- like any other chiral-odd object too ---
to another nonperturbative chiral-odd correlator in order to
generate a non-zero observable. 
This implies
that $h_1^\perp$, in general, is harder to measure than
$f_{1T}^\perp$.

On the other hand, in the case of the Boer-Mulders function no
polarized target is required, which makes this distribution rather
attractive. In fact, it is believed that the Boer-Mulders effect
is essential for understanding data on the angular distribution of
the unpolarized Drell-Yan process~\cite{Boer:1999mm}. To be more
specific, the general structure of the Drell-Yan cross section
reads (see~\cite{Boer:2006eq} and references therein)
\begin{equation} \label{e:dy_cs}
\frac{1}{\sigma_{DY}} \frac{d \sigma_{DY}}{d \Omega} = \frac{3}{4
\pi} \frac{1}{\lambda + 3} \Big( 1 + \lambda \cos^2 \theta + \mu
\sin 2 \theta \cos \phi + \frac{\nu}{2} \sin^2 \theta \cos 2\phi
\Big) \,,
\end{equation}
where the angles $\theta$ and $\phi$ characterize the orientation
of the lepton pair in a dilepton rest frame like the Collins-Soper
frame~\cite{Collins:1977iv}. What attracted particular attention
is the so-called Lam-Tung relation between the coefficients
$\lambda$ and $\nu$~\cite{Lam:1978pu,Collins:1978yt},
\begin{equation} \label{e:lamtung}
\lambda + 2 \nu = 1 \,.
\end{equation}
This relation is exact if one computes the Drell-Yan process to
${\cal O}(\alpha_s)$ in the standard collinear perturbative QCD
framework. Even at ${\cal O}(\alpha_s^2)$ the numerical violation
of~(\ref{e:lamtung}) is small~\cite{Mirkes:1994dp}. However, data
for $\pi^- \, N \to \mu^- \, \mu^+ \, X$ taken at
CERN~\cite{Falciano:1986wk,Guanziroli:1987rp} and at
Fermilab~\cite{Conway:1989fs} were found to clearly violate the
Lam-Tung relation. In particular, an unexpectedly large $\cos
2\phi$ modulation of the cross section was observed. Various
explanations of this experimental result have been put forward,
with the most favorable one being based on intrinsic transverse
motion of partons leading to the Boer-Mulders
effect~\cite{Boer:1999mm}. The product of two Boer-Mulders
functions --- one for each initial state hadron --- contributes to
the $\cos 2\phi$ term in the cross section
in~(\ref{e:dy_cs})~\cite{Boer:1999mm}. An ultimate understanding
of the angular distribution in~(\ref{e:dy_cs}), and thus also of
the role played by the Boer-Mulders function, is of crucial
importance if one keeps in mind that, from a theoretical point of
view, the Drell-Yan process is the cleanest hard hadron-hadron
reaction.


Several model calculations have been carried out for the
Boer-Mulders function of both the
nucleon~\cite{Meissner:2007rx,Pasquini:2010af,Bacchetta:2008af,Gamberg:2007wm,Gamberg:2003ey,Goldstein:2002vv,Yuan:2003wk,Courtoy:2009pc}
and the
pion~\cite{Gamberg:2009uk,Meissner:2008ay,Lu:2004hu}, where the
treatments for the nucleon comprise spectator models, the MIT bag
model, and constituent quark models. In the case of the nucleon
two general features emerge: first, the Boer-Mulders function
comes out to be as large as the Sivers function or even larger.
Second, it has the same sign for up-quarks and down-quarks. This
finding nicely agrees with a model-independent analysis according
to which $h_1^{\perp u} = h_1^{\perp d}$ to leading order of an
expansion in powers of $1/N_c$, with $N_c$ being the number of
colors~\cite{Pobylitsa:2003ty}.

A lot of attention has been paid to an intuitive relation between
the Boer-Mulders function and (a specific linear combination of)
chiral-odd Generalized Parton Distributions in impact parameter
space~\cite{Diehl:2005jf,Burkardt:2005hp}. (This connection
between two types of parton distributions is the analogue of a
corresponding relation involving the Sivers function which was
proposed earlier~\cite{Burkardt:2003uw,Burkardt:2002ks}.) The
intuitive picture is compatible with the two general results from
model calculations discussed above. In particular, it also
suggests a significant size for the Boer-Mulders function in the
valence region. In Quantum Field Theory one can make such a
relation quantitative in the framework of simple spectator
models~\cite{Meissner:2007rx,Lu:2006kt,Burkardt:2003je}. However,
according to current knowledge, a general model-independent
relation cannot exist~\cite{Meissner:2009ww,Meissner:2008ay}.

The Boer-Mulders function describes the strength of a correlation
between the transverse momentum and the transverse spin of the
active quark. This correlation generates a dipole pattern in the
transverse $k_\perp$-plane --- like the correlations associated
with $f_{1T}$, $g_{1T}$, and $h_{1L}^\perp$ do. One way of
visualizing the Boer-Mulders effect is by looking at the density
\begin{equation} \label{e:bm_density}
\rho^q_{h_{1}^\perp}(\boldsymbol{k}_\perp,\boldsymbol{s}_\perp)=
\int dx \, \frac{1}{2} \bigg[ f_1^q(x,\boldsymbol{k}_\perp^{\,2})
+ \frac{\epsilon_\perp^{ij} s_\perp^i k_\perp^j}{M} \, h_1^{\perp
q}(x,\boldsymbol{k}_\perp^{\,2}) \bigg]
\end{equation}
describing the distribution of transversely polarized quarks in an
unpolarized nucleon~\cite{Pasquini:2010af}. The quark polarization
is specified by the spin vector $\boldsymbol{s}_\perp$. Note that
the longitudinal momentum fraction has been integrated over. In
Eq.~(\ref{e:bm_density}), the $f_1$ term provides an axially
symmetric contribution, while the second term containing
$h_1^\perp$ gives rise to the mentioned dipole pattern. If both
effects are superimposed, the resulting distribution is shifted
away from the center (distorted) in the $k_\perp$-plane.

The Boer-Mulders function can also be studied in SIDIS and
therefore at the EIC. In this process it couples to the chiral-odd
Collins fragmentation function $H_1^\perp$~\cite{Collins:1992kk}
and gives rise to a $\cos 2 \phi_h$-modulation of the cross
section. The pertinent structure function takes the generic form
\begin{equation} \label{e:sf_cos2phi}
F_{UU}^{\cos 2\phi_h} \sim \sum_q e_q^2
 \bigg( h_1^{\perp q} \otimes H_1^{\perp q} +
       \frac{C}{Q^2} \, f_1^q \otimes D_1^q
 + \ldots \bigg) \,,
\end{equation}
where $C$ is a kinematic factor. The second term on the right
hand side of~(\ref{e:sf_cos2phi}) is the so-called Cahn
effect~\cite{Cahn:1978se,Cahn:1989yf}, which is also caused by
intrinsic transverse parton motion. It is a kinematic twist-4
contribution, i.e., it is suppressed by a factor $1/Q^2$ relative
to the first term. Theoretical estimates of this effect are still
plagued by large uncertainties, mainly related to the insufficient
knowledge of the transverse momentum dependence of $f_1^q$ and
$D_1^q$. The explicit form of all potential additional (dynamical)
twist-4 effects in this structure function is presently not known.
These considerations show that a reliable extraction of the
Boer-Mulders function from SIDIS requires data in a kinematic
region for which the (largely unknown) higher-twist contributions
can be neglected.
Since  $f_1^q \gg h_1^{\perp q}$ and $D_1^q \gg H_1^{\perp q}$,
the suppression of the Cahn effect requires very large $Q^2$. 

The SIDIS structure function $F_{UU}^{cos 2\phi_h}$ has already
been measured by the CLAS Collaboration at JLab~\cite{Osipenko:2008rv}, 
the HERMES Collaboration at
DESY~\cite{Giordano:2009hi}, and the COMPASS Collaboration at
CERN~\cite{Bressan:2009eu}.
More precisely, typically data are
shown for the relevant azimuthal asymmetry given by $F_{UU}^{\cos
2\phi_h} / F_{UU}$.
However, due
to the limited range in $Q^2$ the present SIDIS data allow at most
a qualitative extraction of $h_1^\perp$, as is also obvious from a
first exploratory study~\cite{Barone:2009hw}. Moreover, the
Boer-Mulders function for antiquarks is not at all constrained by
the available data from SIDIS. Some information about antiquarks
is available from recent Fermilab data on
proton-deuteron~\cite{Zhu:2006gx} and
proton-proton~\cite{Zhu:2008sj} Drell-Yan, though the
uncertainties are again significant and not the least due to the
presently large uncertainties for the Boer-Mulders function of
quarks~\cite{Lu:2009ip,Zhang:2008nu,Barone:2010gk}.

Even without
further detailed reasoning it is clear that a {\it quantitative} knowledge about the
Boer-Mulders function can only be obtained with data from new 
facilities.
Measurements of the structure function  $F_{UU}^{\cos2\phi_h}$
in the valence region in electroproduction of pions and
kaons compose an important part of the upgraded JLab program on
TMD studies.
However, the $Q^2$ range obtainable with JLab12 will not be sufficient to 
suppress the contribution from the Cahn effect.

Only the unprecedented wide kinematic range of the EIC would provide clean 
measurements of the Boer-Mulders function for valence and sea quarks, and will
allow for studying both, its $Q^2$ evolution and transition behavior from
low to high $P_{hT}$. 


Finally, there also exists a Boer-Mulders function for gluons,
$h_1^{\perp g}$, describing the distribution of linearly polarized
gluons in an unpolarized
hadron~\cite{Anselmino:2005sh,Mulders:2000sh,Meissner:2007rx}. In
contrast to the Boer-Mulders function for quarks, $h_1^{\perp g}$
is T-even. See the relevant discussions in Sec.~3.



\section{Overview on other TMDs}
\label{Sec-IV:overview-on-TMDs}

\hspace{\parindent}\parbox{0.92\textwidth}{\slshape
 Harut Avakian, Alessandro Bacchetta, Andreas Metz, Peter Schweitzer}

\index{Avakian, Harut}
\index{Bacchetta, Alessandro} 
\index{Metz, Andreas}
\index{Schweitzer, Peter}

\vspace{\baselineskip}


In previous Sections, we discussed the unpolarized TMD $f_1$, the Sivers
distribution $f_{1T}^{\perp}$, the transversity distribution $h_1$, and the
Boer--Mulders distribution $h_{1}^{\perp}$.
They have been given more emphasis because at the present state of our
knowledge they seem to be the most attractive and promising for EIC studies.

Nevertheless, interesting physics is embodied also in all other TMDs.
Only the combination of information from all TMDs will fully explore the
information contained in the unintegrated quark correlator, and provide a
complete picture of the parton structure of the nucleon in transverse momentum space.
This wealth of information may become one of the biggest legacies of the EIC.

In this Section, we briefly discuss the leading-twist TMDs that have not been
analyzed in previous Sections and some of the sub-leading twist TMDs.

\begin{table}[h]
\begin{center}
\begin{tabular}{ccc}
\includegraphics[height=3.5cm]{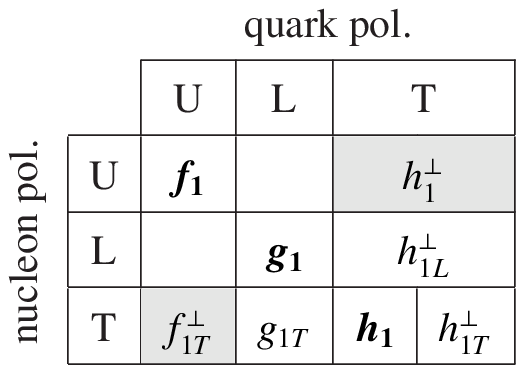}
&
\qquad
&
\includegraphics[height=3.5cm]{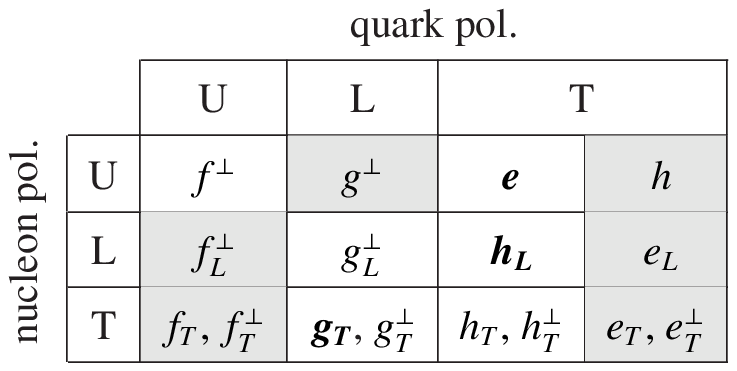}
\\
(a) && (b)
\end{tabular}
  \caption{
Transverse momentum dependent 
(a) twist-2, (b) twist-3 distribution functions.
The U,L,T correspond to unpolarized, longitudinally polarized and transversely
polarized nucleons (rows) and quarks (columns). Functions in boldface survive
transverse momentum integration. Functions in gray cells are T-odd.}
\label{tmdTable}
\end{center}
\end{table}

\subsection{Other leading-twist TMDs}

\begin{figure}
\begin{center}
\includegraphics[width=0.5\textwidth]{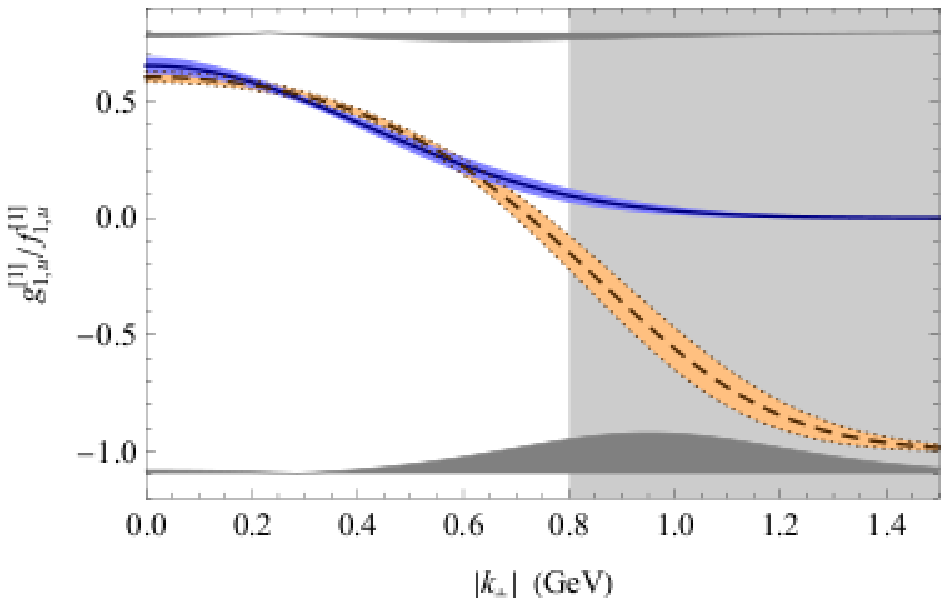}
\end{center}
\caption{
Ratio between the helicity distribution and the unpolarized distribution for
  up quarks based on lattice QCD computations~\cite{Musch:2010ka}:
  the significant $k_\perp$ dependence of the two curves 
(corresponding to two different parameterizations)
suggests that quarks with different
  spin orientation have different transverse momentum distributions.
}
\label{f:g1_pt}
\end{figure}

Table~\ref{tmdTable}a summarizes the full list of leading-twist
TMDs. The helicity distribution $g_1$, together with $f_1$ and
$h_1$, survives integration over transverse momentum and has been
already discussed extensively. Here we mention
the importance of also studying its transverse momentum
dependence. It may be possible that the transverse momentum
distribution of quarks with spin antiparallel to the nucleon is
different from that of quarks with spin parallel to the nucleon
as suggested by lattice calculations~\cite{Musch:2010ka}
shown in Fig.~\ref{f:g1_pt}.
The
structure function $F_{LL}$, involving the transverse-momentum
dependence of $g_1$, is the only one where transverse-momentum
resummation studies have been carried out to a level similar to
$F_{UU,T}$~\cite{Koike:2006fn}, but no extraction of the
nonperturbative component has ever been attempted. The EIC will be
an ideal machine to address this question.

The chiral-odd T-even TMD $h_{1T}^{\perp}$ appears in the SIDIS
structure function $F_{UT}^{\sin(3\phi-\phi_S)}$. This function
may be interpreted as the distribution of quarks with a
polarization transverse but orthogonal to that of a transversely
polarized nucleon. The popular name ``pretzelosity'' is due to the
fact that this distribution has a quadrupole shape, vaguely
reminiscent of a pretzel~\cite{Miller:2007ae,Burkardt:2007rv}.
This TMD
has attracted a lot of interest in the literature recently because
of its possible connection with orbital angular momentum (see
detailed discussion in Sec.~\ref{sec:TMD-models}). It is also
interesting that, in a number of nonperturbative models,
$h_{1T}^{\perp}$ is just the difference between the quark helicity
and the transversity distribution~\cite{Avakian:2008dz}. Moreover,
in simple spectator models of the nucleon it can be related to a
particular linear combination of chiral-odd generalized parton
distributions~\cite{Meissner:2007rx}. In general, $h_{1T}^{\perp}$
involves an interference between light-cone wave function
components that differ by two units of orbital angular momentum.
Preliminary data from COMPASS~\cite{Kotzinian:2007uv}
and from HERMES~\cite{Pappalardo:2010zz}
taken with transversely polarized deuterons or protons, respectively,
showed an effect compatible with zero, however, within large experimental uncertainties.

The TMDs $g_{1T}$ and $h_{1L}^{\perp}$  appear in the structure
functions $F_{LT}^{\cos(\phi-\phi_S)}$ and $F_{UL}^{\sin2\phi}$,
respectively. The chiral-even (chiral-odd) $g_{1T}$
($h_{1L}^{\perp}$) describes longitudinally (transversely)
polarized quarks in a transversely (longitudinally) polarized
nucleon. Since both functions link two perpendicular spin
directions, they are sometimes named ``worm-gear'' functions. 
Both functions are related
to quark orbital motion inside nucleons. They represent the real
part of an interference between nucleon wave functions that differ
by one unit of orbital angular momentum, while the imaginary parts
are related to the Sivers and Boer--Mulders
functions~\cite{Ji:2002xn,Bacchetta:1999kz}. Because of this, they
appear in positivity bounds together with the Sivers and
Boer--Mulders function~\cite{Bacchetta:1999kz}. They do not depend
on final-state interactions and may offer cleaner insights into
orbital angular momentum compared to the Sivers and Boer--Mulders
functions. Interestingly, these functions are the first TMDs that
have been computed on the
lattice~\cite{Hagler:2009mb,Musch:2010ka}. The results (with the
due caveats) indicate that they are sizable, $g_{1T}^{u_v}> 0$,
$g_{1T}^{d_v}< 0$, and $g_{1T} \approx - h_{1L}^{\perp}$. These
general findings also agree with some model calculations, see
Sec.~\ref{sec:TMD-models}.

Notice that due to their chirality properties, $g_{1T}$ couples through evolution
to its analogous function for gluons (named $\Delta G_T$ in Ref.~\cite{Mulders:2000sh}
and $g_{1T}^g$ in Ref.~\cite{Meissner:2007rx}), while this is not true for
$h_{1L}^{\perp}$.
This difference will be particularly relevant at the EIC, where gluons will play an
important role.

By exploring QCD equations of motion, and neglecting ``pure twist-3''
quark-gluon correlators and current quark mass terms, one can express
$g_{1T}$ ($h_{1L}^{\perp}$) in terms of $g_1$ ($h_1$)
(see, e.g.,~\cite{Mulders:1995dh,Avakian:2007mv,Tangerman:1994bb,Metz:2008ib}
and references therein).
This is similar in spirit to the classic Wandzura--Wilczek
approximation~\cite{Wandzura:1977qf} for the twist-3 distribution
function $g_T^q(x)\approx\int_x^1{\rm d} y \,g_1^q(y)/y$,
(which is supported by
the instanton QCD vacuum model
~\cite{Balla:1997hf,Dressler:1999hc} and
lattice QCD
~\cite{Gockeler:2000ja,Gockeler:2005vw}).
At an initial stage, it may be convenient to exploit such
Wandzura--Wilczek-type
approximations, in order to make estimates for planned
experiments~\cite{Avakian:2007mv,Metz:2008ib,Kotzinian:2006dw}.
In fact, existing data suggest that they are reasonable~\cite{Metz:2008ib},
even though at present there are no compelling grounds for
supporting
their validity~\cite{Jaffe:1990qh,Harindranath:1997qn,Accardi:2009au}.
In the end, these approximations should be tested and twist-3 effects should be
extracted from the data, as we will also argue in the next subsection.

\subsection{Subleading-twist TMDs}
\label{Sec:twist-3-WW-type}

Eight out of the 18 structure functions providing the complete
description of the SIDIS cross section are leading twist and were discussed
in detail in previous sections. However, 10 structure functions are
higher twist, where the underlying twist-classification follows
~\cite{Jaffe:1996zw}: ``an observable is twist-$t$ if its effect
is effectively suppressed by  $(M/Q)^{t-2}$.''

Higher twist functions, see Table~\ref{tmdTable}b for a full list of twist-3 TMDs,
are of interest for several reasons. Their understanding is required
not only to complete the description of the SIDIS process. Besides
being indispensable to correctly extract twist-2 parts from data, the
knowledge of higher twists will also offer important tools to access
the physics of the largely unexplored quark-gluon correlations
which provide direct and unique insights into the dynamics inside
hadrons, see, e.g.,~\cite{Jaffe:1989xx}.
The EIC, which will span a large $Q$-range, will be an ideal tool
to identify higher-twist effects, which fall off as powers of $1/Q$.

Although suppressed with respect to twist-2 observables by $1/Q$,
twist-3 observables are not small in the kinematics of
fixed target experiments. Indeed, the first unambiguously measured
single spin phenomena in SIDIS which triggered important theoretical
developments, were the sizable longitudinal target
($A^{\sin \phi}_{UL}$) and beam ($A^{\sin \phi}_{LU}$) spin asymmetries
observed at HERMES and JLab
~\cite{Airapetian:1999tv,Airapetian:2001eg,Airapetian:2002mf,Avakian:2003pk,Airapetian:2005jc,Airapetian:2006rx}.
Further data on twist-3 spin asymmetries are underway
~\cite{Kotzinian:2007uv,Gohn:2009zz,Savin:2010zz}.
In unpolarized SIDIS, the sizable twist-3 effects ($A^{\cos \phi}_{UU}$)
are known since EMC ~\cite{Aubert:1983cz,Arneodo:1986cf}, see also
recent results from JLab, HERMES and COMPASS
~\cite{Giordano:2009hi,Mkrtchyan:2007sr,Osipenko:2008rv,Kafer:2008ud}.
At high energies $A^{\cos \phi}_{UU}$ can be described in perturbative
QCD, and the unique possibilities of EIC could
bridge~\cite{Anselmino:2006rv} the gap to high energy data
~\cite{Adams:1993hs,Breitweg:2000qh,Chekanov:2002sz,Chekanov:2006gt}.
The understanding of the ``matching'' of the TMD formalism and the
large-$p_T$ collinear description is of fundamental importance, see
Sec.~\ref{sec:TMD-matching} and references therein.

The theoretical description of twist-3 observables is challenging.
A good illustration of this point is that in spite of the enormous
dedicated theoretical and phenomenological effort
~\cite{Afanasev:2003ze,Metz:2004je,Bacchetta:2002tk,Oganessian:1998ma,Kotzinian:1999dy,Boglione:2000jk,Ma:2000ip,Wakamatsu:2000fd,Efremov:2002sd,Ma:2002ns,Efremov:2004hz,Anselmino:2000mb,DeSanctis:2000fh,Efremov:2000za,Oganessian:2000um,Efremov:2001cz,Ma:2001ie,Efremov:2001ia,Efremov:2002ut,Efremov:2003tf,Efremov:2003eq,Yuan:2003gu,Schweitzer:2003yr,Gamberg:2006ru}
to explain the first single spin phenomena in SIDIS,
$A^{\sin \phi}_{UL}$ and $A^{\sin \phi}_{LU}$, these observables
are still not understood.
The theoretical challenge is that presently it is not understood
how to control light-cone divergences in SIDIS at $1/Q$ order
~\cite{Gamberg:2006ru}.
This does not necessarily mean there is no factorization,
but it indicates that possibly new techniques are needed
to pave the way towards a factorization proof in SIDIS at $\mbox{twist-3}$.
If one {\sl assumes} $\mbox{twist-3}$ TMD factorization,
the phenomenological challenge is that each twist-3 observable
receives contributions from several unknown twist-3 TMDs or
fragmentation functions~\cite{Bacchetta:2006tn}.
The situation simplifies in semi-inclusive jet production,
a promising process to study at EIC energies,
which could provide valuable complementary
information on twist-3 TMDs~\cite{Bacchetta:2004zf}.

An important process which can provide independent
information on twist-3 (and, of course, also twist-2)
TMDs are interference functions
~\cite{Bianconi:1999cd,Bacchetta:2003vn,Bacchetta:2002ux,Bacchetta:2006un,Jaffe:1997hf,Radici:2001na,Ceccopieri:2007ip}.
The advantage of this approach is that here collinear factorization
applies, i.e.\ one cannot access TMDs. However, those
functions which ``survive'' the $k_\perp$-integration
of the quark correlator can be studied, and this
includes at the twist-3 $e^a(x)$, $g_T^a(x)$, $h_L^a(x)$.
These functions contribute to observables in convolution with specific
interference fragmentation functions, which can be inferred from azimuthal
asymmetries in $e^+e^-$ annihilations~\cite{Boer:2003ya}.

There is no doubt that experimental, phenomenological and
theoretical efforts to go beyond twist-2 are worth. Twist-3
functions describe multiparton distributions corresponding to the
interference of higher Fock components in the hadron wave functions,
and as such have no probabilistic partonic interpretations.
Yet they offer fascinating insights into the nucleon
structure ~\cite{Burkardt:2008ps}.
The Mellin moment $\int{\rm d}x\,x^2\tilde{g}_T^a(x)$
of the pure twist-3 piece in $g_T^a$ describes the transverse
impulse the active quark acquires after being struck by the
virtual photon due to the color Lorentz force.
The Mellin moment $\int{\rm d}x\,x^2\tilde{e}^a(x)$
of the pure twist-3 piece in $e^a(x)$
describes the average transverse force acting on a transversely
polarized quark in an unpolarized target after interaction with
the virtual photon.

Twist-3 TMDs are closely related to projections of different
combinations of the collinear twist-3 correlation functions
$G_F(x,x\prime)$ and $\tilde{G}_F (x, x\prime)$ discussed in
Sec.~\ref{sec:twist3}, which are involved in the evolution equations of twist-3 collinear
PDFs~\cite{Koike:1994st,Koike:1996bs,Belitsky:1997by,Belitsky:1997zw,Ali:1991em,Balitsky:1996uh,Kodaira:1998jn}, and play important roles
also in derivations of the evolution equations for transverse
moments of TMDs
~\cite{Zhou:2008mz,Vogelsang:2009pj,Kang:2008ey,Braun:2009mi,Kang:2010xv},
calculations of processes at high transverse momentum ~\cite{Eguchi:2006mc},
or calculations of the high transverse momentum tails of TMDs
~\cite{Ji:2006ub,Koike:2007dg}.
Ultimately, through a global study of all of these observables,
one could simultaneously obtain better knowledge of twist-3
collinear functions and twist-2 TMDs, and at the same time test
the validity of the formalism. Gathering as much information as
one can on the quark-gluon-quark correlator is essential to reach
this goal.





\chapter{Three-dimensional structure of the proton and nuclei: spatial
  imaging \label{chap:gpd}}

\noindent
{\Large Convenors and chapter editors: \\[1em]
M. Burkardt, V. Guzey, F. Sabati\'e}

\newpage

\section{Spatial imaging of sea quarks and gluons: summary}
\label{sec:imaging_executive_summary}


\hspace{\parindent}\parbox{0.92\textwidth}{\slshape 
  V.~Guzey, F.~Sabati\'e, M. Burkardt}
%
\index{Guzey, Vadim}
\index{Sabati\'e, Franck}
\index{Burkardt, Matthias}

The internal landscape of the nucleon and nuclei in terms of the fundamental quarks and gluons
can be studied in different hard processes and can be characterized by different quantities (distributions). Hard exclusive reactions such as deeply virtual Compton scattering (DVCS) and exclusive
production of mesons give an access to the aspects of the hadron structure that are encoded in
generalized parton distributions (GPDs) and dipole amplitudes.

GPDs generalize the well-known form factors, distribution amplitudes and parton distributions 
and quantify various correlations/distributions of quarks and gluons
in terms of their momentum fractions and positions in the transverse plane.
Thus, GPDs provide a rigorous framework for studies of the three-dimensional parton 
structure of hadrons as well as many additional important aspects of the hadron structure such as the parton angular momentum and the related ``spin puzzle'', spin and flavor content, the role of chiral symmetry, and many more.

At the moment, our knowledge about GPDs is mostly limited to valence quark GPDs
(Hermes, Compass, Jefferson Lab 6 GeV and also Jefferson Lab 12 GeV in the near future)
and rather low precision data from HERA. A high-energy high-luminosity 
Electron-Ion Collider (EIC) will be an ideal machine for the studies of hard exclusive
 reactions and sea quark and gluon GPDs as summarised in table~\ref{Tab:Imaging}.\\
 
 \begin{table} [hbt] 
\small
\label{matrix}
\begin{center}
\begin{tabular}{|c|c|c|c|}
\hline
Deliverables  & Observables & What we learn & Requirements \\
\hline\hline
   sea quark and    & DVCS and $J/\psi, \rho, \phi$  & transverse images of & $\mathcal L\geq 10^{34}$~cm$^{-2}$s$^{-1}$, \\
    gluon GPDs      & production cross sect.      &  sea quarks and gluons  & Roman Pots  \\
                    & and asymmetries             &  in nucleon and nuclei;  & wide range of $x_B$ and $Q^2$ \\ 
                    &                             & total angular momentum;        & polarized $e^-$ and $p$ beams \\
                    &                             & onset of saturation & $e^+$ beam for DVCS \\ \hline
   sea and valence & cross sections for          & flavor decomposition and & $\mathcal L\geq 10^{34}$~cm$^{-2}$s$^{-1}$ \\
    quark GPDs      & $\pi^+, K, K^*, \rho^+$     &  polarization of quarks    & Roman Pots \\
                    & electroproduction           & in the transverse plane        & high $Q^2$ \\
                    &                             &                    & range of beam energies \\
                    &                             &                    &  for $\sigma_L/\sigma_T$ separation \\ \hline
\hline
\end{tabular}
\end{center}
\begin{caption}
{\small Science Matrix for Exclusive Processes at EIC.}
\end{caption}
\label{Tab:Imaging}
\end{table}

(i) One essential aspect of the GPD program is obtaining the transverse image 
of quarks and gluons in the nucleon/nucleus through the measurement of the $t$ dependence of cross sections of 
various exclusive processes (DVCS, production of $J/ \psi$, $\phi$, $\pi$, $K$, etc.~mesons) in 
a wide range of $t$.
In the nucleon case, covering the interval $0 \approx |t| \leq 2$ GeV$^2$ 
will enable one to map out the parton distributions in the transverse plane of the impact
parameter $b$ down to as low as $b \approx 0.1$ fm. \\
(ii) One area where an EIC shines is the large range in $Q^2$ available in the
full $x_B$ interval. QCD evolution equations of GPDs, similarly to the PDF case,
allow one to globally fit the data using flexible parameterizations of GPDs and
to extract accurate and model-independent information on GPDs.
One also will use the large lever arm in $Q^2$ to establish the reaction mechanisms
(scaling properties, higher twist effects).\\
(iii) Another clear advantage of an EIC
is the availability of different polarizations for the lepton and proton beams
that allows
one to fully disentangle the various GPDs from the experimental observables. 
While DVCS is sensitive to singlet quark and gluon GPDs, other exclusive diffractive
processes (electroproduction of $\rho$, $J/\psi$, $\phi$, etc.) and non-diffractive 
processes (electroproduction of $\pi^+$, $K^+$, etc.) will allow one to
access the spin and flavor dependences of GPDs.
Note that the non-diffractive processes push the requirements for high luminosity 
much further than DVCS or other diffractive processes. \\
(iv) Exclusive processes with nuclei in a collider 
and, subsequently, the spatial image of sea quarks and gluons in nuclei
will be studied for the first time.
All the processes mentioned above will benefit from the high luminosity of
an EIC (of the order of 10$^{34}$~cm$^{-2}$s$^{-1}$) as well as excellent detection capabilities
and particle identification guaranteeing exclusivity.\\
The contributions below describe in detail various aspects of the rich program of spatial
imaging of sea quarks and gluons at an EIC.
In conclusion, a high-energy high-luminosity EIC, studying various
deep exclusive processes through cross sections and polarization observables, would
uniquely extend and complement our knowledge of the 3D partonic structure of the
nucleon/nucleus to the sea of quarks and gluons.










\section{Basics of generalized parton distributions}
\label{sec:Radyushkin}


\hspace{\parindent}\parbox{0.92\textwidth}{\slshape 
  Anatoly Radyushkin}
%

\index{Radyushkin,  Anatoly}





\subsection{Introduction}

 The fundamental physics to be accessed via the    generalized parton 
 distributions  (GPDs)  \cite{Ji:1996ek,Mueller:1998fv,Ji:1996nm,Radyushkin:1996nd,Radyushkin:1996ru,Radyushkin:1997ki,Collins:1996fb} 
  is the  structure of hadrons.  This is a rather 
 general  statement, and we  may want  to have 
 a more specific   one. A classic 
   example  of  such a specific case 
 is the 
  search for the Higgs boson (HB)  performed currently at the Large Hadron Collider 
(LHC).
   The motivation for the search is  that HB is supposed to 
   be responsible for generation of masses, in particular, quark  masses.  
   However, by far, the largest  part of visible mass
 is due to the nucleons, and  out of 940 MeV of the nucleon mass, less than 
    $ 30$ MeV (current quark masses) 
    may be related to   HB. The remaining 
    97\% of the nucleon mass is due to gluons -- which are massless!
    This is a characteristic illustration of the  situation 
in hadron physics:\\
$i)$ All  the relevant particles are already established, i.e., no  ``higgses'' 
to find. \\
 $ii)$ The QCD Lagrangian is known. \\
$iii)$ However,  we still need to understand how QCD works, 
i.e., to understand hadronic structure  in terms of 
quark and gluon fields.

Projecting  quark and gluon fields 
$ q (z_1) \ , \ q(z_2) \ , \ \ldots \ $  onto hadronic states 
$|p,s \rangle$  gives  matrix  elements:
\begin{equation}
 \langle  \, 0 \, | \, \bar q_\alpha (z_1) \, q_\beta (z_2) \, 
| \, M(p),s \, \rangle \  \  \  \ ,  \  \  \  \ 
\langle  \, 0 \, | \,  q_\alpha (z_1) \ 
q_\beta (z_2) \, q_\gamma (z_3)| \, B(p),s \, \rangle 
\label{eq:radyush_1}
\end{equation}
that  can be interpreted as hadronic wave functions.
In particular,
      in   the light-cone (LC) formalism \cite{Brodsky:1989pv},
        a  hadron is described 
      by its Fock components in the  infinite-momentum frame. 
    For the nucleon, one can schematically  write:
\begin{equation}  
    | P \, \rangle = 
   \Psi_{qqq}  |q(x_1 P, k_{1 \perp}) q(x_2 P, k_{2 \perp}) q(x_3 P, k_{3 \perp}) \rangle 
     + \Psi_{qqqG}  |qqqG \rangle + \Psi_{qqq \bar q q}  |qqq \bar qq \rangle + \ldots \,,
\label{eq:radyush_2}
\end{equation}  
     where  $x_i$ are  momentum fractions  satisfying  $\sum_i x_i=1$; 
         $k_{i \perp} $  are transverse momenta, $\sum_i k_{i \perp} =0$;
$\Psi$ are light-cone wave functions.
 In principle, solving  the 
  bound-state  equation $H  | P \, \rangle =  E | P \,\rangle $
   one should get the  wave function   $| P \, \rangle$ that  contains  complete
  information about the hadron structure.
  In practice,  however, the equation 
  (involving an infinite number of Fock components) has not been solved yet
  in the  realistic 4-dimensional case.
  Moreover, the LC wave functions are not directly accessible 
  experimentally.
  
   The way out 
of
this situation    is the  description of hadron
  structure in terms of phenomenological functions. 
  Among the ``old'' functions used   for a long time we can list  
 form factors, 
   usual parton densities, and 
   distribution amplitudes. 
 The ``new''  functions, 
   generalized parton distributions (for  reviews, see 
   \cite{Goeke:2001tz,Diehl:2003ny,Belitsky:2005qn,Boffi:2007yc}),   
  are  hybrids of form factors, parton densities and distribution amplitudes.
Furthermore, the  ``old'' functions are limiting cases of  the ``new''  ones.
  
\subsection{Form factors}

The form factors are defined   
  through matrix elements 
  of electromagnetic (EM) and weak  currents between  hadronic states. 
    In particular, the nucleon electromagnetic  form factors are given by 
 \begin{equation}
{  \langle \, p', \, s' \, 
 |  \, J^\mu (0)  \, |  \, p, \, s \, \rangle 
 =\bar u (p',s') \left [\gamma^\mu F_1 (t)  + \frac{r^\nu \sigma^{\mu \nu}}
 {2 m_N} F_2 (t) \right ] u(p,s)  \ ,
 }\end{equation}
 where $r= p-p'$ is the momentum transfer and $t=r^2$.
  The electromagnetic current is given by the sum of its 
 flavor components:
\begin{equation}
J^\mu (z) = \sum_{f} 
  \, e_f \bar \psi_f (z) \gamma^\mu \psi_f (z) \,.
\end{equation}
 The nucleon helicity non-flip form factor $F_1 (t)$ 
   can also be written as a sum  $ \sum_f  \, e_f
  F_{1f} (t)$. 
  A similar decomposition holds for the helicity flip form factor $F_2 (t) = \sum_f  \, e_f
  F_{2f} (t)$.
  At  $t=0$, these functions have well known limiting values. In particular,  
  $F_1 (t=0)  = e_N = \sum_f N_f e_f$
  gives total electric charge of the nucleon 
 ($N_f$ is the  number of valence quarks of flavor {$f$}) 
 and  
 $F_2 (t=0)  = \kappa_N  $  
gives its {anomalous magnetic moment}.
The form factors are  measurable 
 through elastic $eN$ scattering.
 

\begin{figure}[h]
\begin{center}
\includegraphics[width=0.25\textwidth]{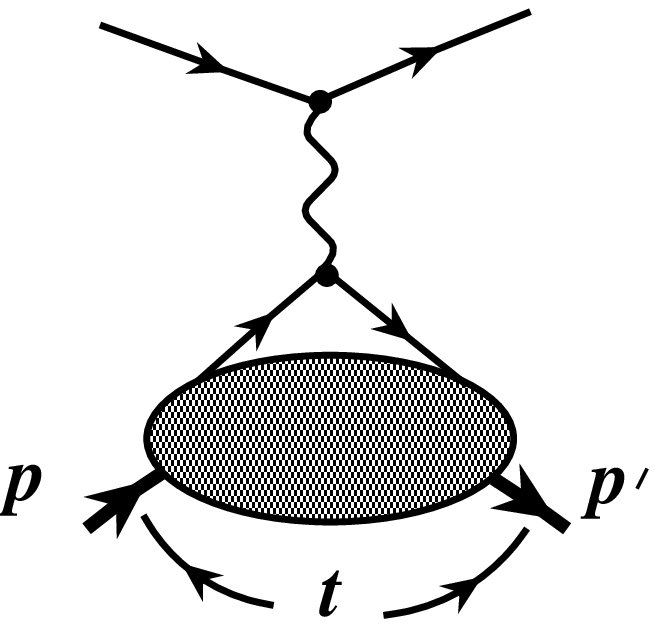}
  \vspace{-0.5cm} 
  \caption{\small Elastic $eN$ scattering  in the one-photon exchange approximation.}
\end{center}
   \end{figure}
         

\subsection{Usual parton densities}

 The parton densities are defined   
     through forward matrix elements 
  of quark/gluon fields separated by light-like 
  distances. 
  In particular, in the unpolarized case  we have 
    \begin{equation}  \left. \langle \, p \, 
 |  \,   \bar \psi_a (-z/2)  \gamma^\mu \psi_a (z/2) \, |  \, p \, \rangle
 \right |_{z^2=0} 
  =2 p^\mu  \int_0^1 \left [ e^{-ix (pz)} f_a (x) - e^{ix (pz)} f_{\bar a} (x)\right ]
 dx \  .
 \end{equation} 
  In the local limit  $z=0$, the operators in this definition
  coincide with  the operators 
  contributing   into the non-flip form factor $F_1$.
  Since $t=0$ for the forward matrix element,
  we obtain the 
  sum rule for the numbers of valence quarks: 
 \begin{equation} 
\int_0^1 \left [f_a (x) -  f_{\bar a} (x) \right ]
 dx  = N_a \,  . 
\end{equation}
 
 The definition of parton densities has the form
 of the plane wave decomposition.
 This observation allows one to give the momentum   space interpretation: 
 $f_{a (\bar a)} (x)$ is the probability  to find  $a \, (\bar a)$-quark 
  with momentum $xp$ inside a nucleon with momentum $p$. 
The classic process to access the usual parton densities is 
deep inelastic scattering (DIS) 
 $\gamma^{\ast} N \to X$. 
 
\begin{figure}[htb]
\begin{center}
\includegraphics[width=0.45\textwidth]{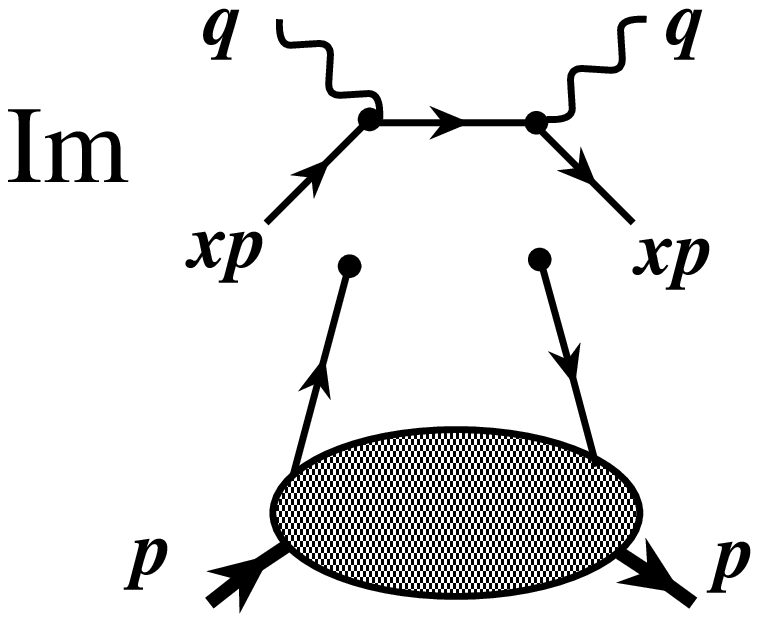}
     \caption{\small Lowest order pQCD factorization for DIS.}
\label{fig:radyush_ftdis_1}
\end{center}
     \end{figure}

   Using the optical theorem, the $\gamma^{\ast} N \to X$ cross section is given by 
   the imaginary part of the forward virtual Compton scattering amplitude. 
   The momentum transfer $q$ is spacelike  $q^2 \equiv -Q^2$,
   and when it is sufficiently large,
   perturbative QCD factorization works.
   At the leading order, one deals with the so-called handbag diagram, see 
figure~\ref{fig:radyush_ftdis_1}.
   
    
  Through simple algebra,   $\frac1{\pi} {\rm Im} \  1/{(q+xp)^2} 
   \approx   \delta (x-x_{B}) /{2(pq)}$, 
  one finds that DIS measures parton densities 
  at   the point  $x=x_ {B}$, where the parton momentum fraction equals the 
   Bjorken variable $x_{B} = {Q^2}/{2(pq)}$. 
Comparing parton densities  to form factors, we note that the latter have a 
point vertex instead of a light-like separation  
and $p \neq p'$.

  \subsection{Nonforward parton densities}

 ``Hybridization'' of different parton distributions 
 is the key idea of the GPD approach. As the first step,  we can 
 combine  form factors with parton densities~\cite{Radyushkin:1998rt}
 and write the flavor components $F_{1a} (t)$ of  form factors   
 as  integrals 
 over the momentum fraction variable $x$: 
 \begin{equation}  
F_{1a} (t) = \int_0^1  \, 
 \left [ {\cal F}_{a}(x,t) -  {\cal F}_{\bar a} (x,t) \right ] dx  \  .
 \label{local}
\end{equation}

    In the forward limit  $t=0$, the new objects---nonforward 
 parton densities ${\cal F}_{a (\bar a)}(x,t) $ (NPDs)---coincide
 with the usual (``forward'')  densities:   
\begin{equation}
{\cal F}_{a(\bar a)} (x,t=0) = 
 f_{a(\bar a)} (x) \,.
\end{equation}
NPDs  can be also treated as  Fourier 
   transforms of  the impact parameter
  $b_\perp$ distributions  
  $f(x,b_\perp)$  describing the variation of parton densities  
in the transverse plane~\cite{Burkardt:2000za,Burkardt:2002hr}.

  A nontrivial   question is the interplay 
  between $x$ and $t$ dependencies of ${\cal F}_{a(\bar a)}(x,t)$.
  The simplest factorized ansatz   
  $ {\cal F}_{a} (x,t) = f_{a}(x) F_1 (t)$ 
  satisfies both the forward constraint, 
  $ {\cal F}_{a}(x,t=0)=f_{a}(x)$,
  and also the  local constraint~(\ref{local}).  
  The reality may be more complicated:  light-cone  wave functions 
with  Gaussian $k_{\perp}$ dependence 
\begin{equation}
\Psi (x_i, k_{i \perp}) \sim 
\exp \left [-\frac1{\lambda^2} \sum_i k^2_{i \perp}/x_i  \right ]
\end{equation}
suggest  that
\begin{equation}
{\cal F}^a(x,t) = f_a(x) e^{\bar x t /2 x \lambda^2 }  \,,
\end{equation}
where ${\bar x} \equiv 1-x$.
 Taking $f_a(x)$ from existing parametrizations 
 and adjusting  $ \lambda^2$  to provide the standard value
 of the quark intrinsic transverse momentum
 $\langle k^2_{\perp}\rangle \approx (300 \, {\rm MeV})^2$
  gives a rather reasonable  description of the proton 
  form factor $F_1(t)$ in a wide range of momentum transfers 
  $-t \sim 1-10$ GeV$^2$~\cite{Radyushkin:1998rt}.  
To comply with the Regge behavior, one may  wish to change 
$ e^{\bar x t /2 x \lambda^2 }  \to x^{-\alpha' t}$, where $\alpha'$ is the 
Regge trajectory slope. The modified 
Regge ansatz, 
\begin{equation}
{\cal F}^a(x,t) = f_a(x) x^{-\alpha' (1-x)t} \,,
\end{equation}
allows one to easily fit  electromagnetic form factors for the proton 
and  neutron~\cite{Guidal:2004nd}.
A similar model  was proposed in Ref.~\cite{Diehl:2004cx}.

\begin{figure}[htb]
\begin{center}
\includegraphics[width=0.45\textwidth]{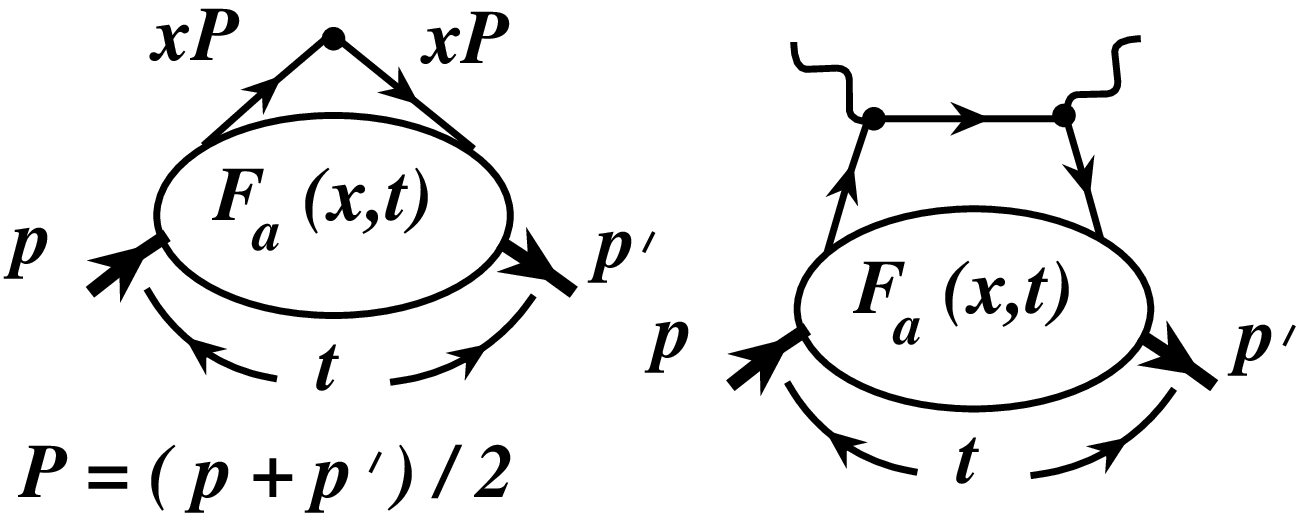}
    \caption{\small Form factor and  wide-angle 
   Compton scattering amplitude in terms of nonforward parton densities.}
\label{fig:radyush_ftkwacs_1}
\end{center}
   \end{figure}

  The same nonforward parton densities 
    appear in the handbag diagrams for the  wide-angle 
    real Compton scattering, see figure~\ref{fig:radyush_ftkwacs_1}.  
    
   The handbag contribution is approximately  given by
   the product of a new form factor, $R_V^a(t)$,  and the 
    cross section of the Compton scattering  
   off an elementary fermion (given by Klein--Nishina expression):
    \begin{equation}    \frac{d \sigma}{dt} = 
    \left [\sum_a e_a^2   R_V^a (t)  \right ]^2 
   \left. \frac{d \sigma}{dt} \right |_{KN}     \quad  {\rm with } 
  \quad  R_V^a (t) = \int_0^1 \frac{ {\cal F}^a(x,t)}{x} \, dx \ .\end{equation}

The predictions based on handbag dominance
and NPDs~\cite{Radyushkin:1998rt,Diehl:1998kh} 
are  in much better  agreement with the existing data~\cite{Hamilton:2004fq}
than the predictions based on two-gluon hard exchange 
mechanism of asymptotic perturbative QCD:  the predicted cross section is  too small
in the latter case. 
 The absolute normalization for predictions is 
 settled by the form  of the nonperturbative 
 functions (NPDs in the handbag approach and nucleon 
 distribution amplitudes in the pQCD approach) 
 which were fixed   by  fitting the $F_1$ form factor data.
Still, when there is an uncertain overall  factor,
it is risky to make strong statements.  
Remarkably, the 
 perturbative QCD hard scattering mechanism 
and soft handbag mechanism give drastically different predictions
 for the polarization asymmetry $A_{LL}$~\cite{Diehl:1998kh}.
  Experiment E-99-114  performed at Jefferson Lab~\cite{Hamilton:2004fq}
strongly favors {  handbag }  mechanism  that  predicts 
the value close to the asymmetry 
for the scattering on a single quark. 
 
\subsection{Distribution amplitudes} 
   
\begin{figure}[h]
\begin{center}
\includegraphics[width=0.6\textwidth]{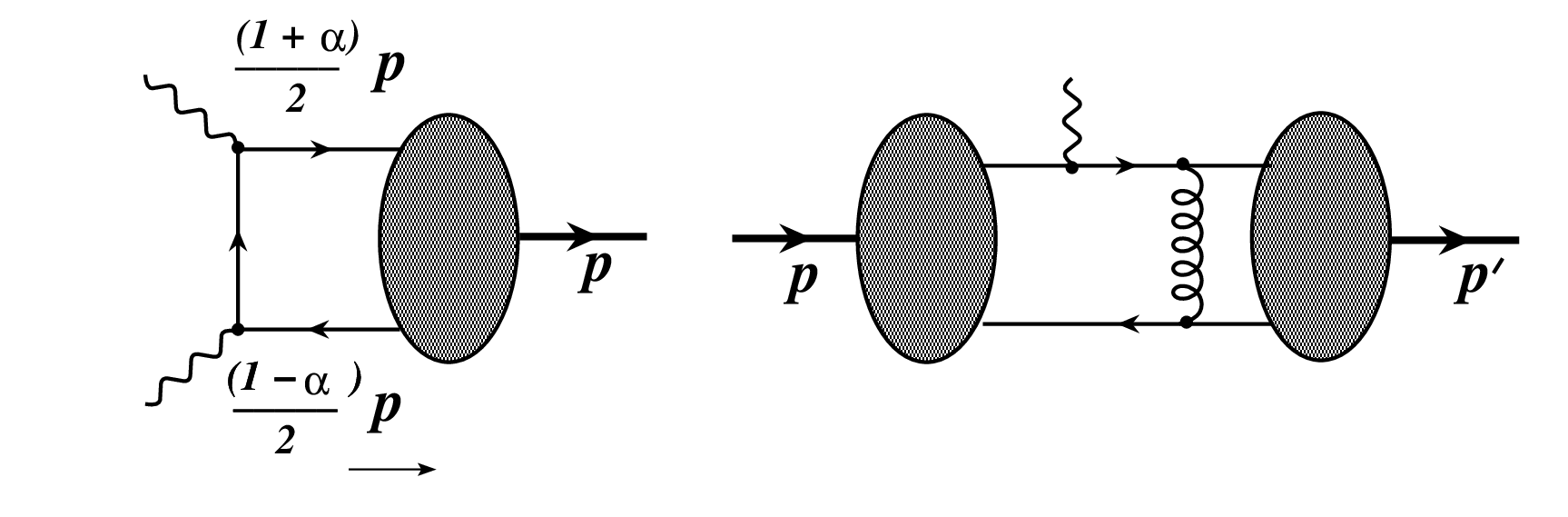}
 \caption{\small Lowest-order pQCD factorization for $\gamma^* \gamma \to \pi^0$ transition
 amplitude and for the pion electromagnetic form factor.} 
\label{fig:radyush_ftffs_1}
\end{center}
  \end{figure}

   Another  example of nonperturbative functions describing the hadron structure 
   are the distribution amplitudes (DAs).
   They can be interpreted   as light cone  wave functions integrated over transverse
   momentum, 
  or as $\langle 0 | \ldots  |p \rangle $ matrix elements  of light cone  operators.
  In the case of the  pion, we have
   \begin{equation} 
  { \left. \langle \, 0 \, 
 |  \,   \bar \psi_d (-z/2)\gamma_5  \gamma^\mu 
 \psi_u (z/2) \, |  \, \pi^+(p) \, \rangle
 \right |_{z^2=0}} 
  = i p^\mu f_\pi \int_{-1}^1  e^{-i\alpha (pz)/2}  \varphi_\pi (\alpha) \,  
 d \alpha  
 \,, 
\end{equation} 
 with  $x_1 =(1+\alpha)/2, \  x_2 =(1-\alpha)/2$
 being the fractions of the pion momentum carried 
 by the quarks. 
 The distribution amplitudes  describe the hadrons 
 in situations when the pQCD hard scattering 
 approach is applicable  to exclusive processes.
 The classic  example is the 
  $\gamma^{\ast} \gamma \to \pi^0$ transition; 
  its amplitude is proportional to
 the $1/(1-\alpha^2)$ moment of  $\varphi_\pi (\alpha)$, 
see figure~\ref{fig:radyush_ftffs_1}, left.
 The predictions for the $\gamma^{\ast} \gamma \to \pi^0$
 form factor based on  two competing models for the pion DA, 
 the asymptotic $\varphi_\pi^{\rm as}(\alpha) = 
 \frac34 (1-\alpha^2) $ and Chernyak-Zhitnitsky DA
  $\varphi_\pi^{\rm CZ} (\alpha) = 
   \frac{15}{4} \alpha^2 (1-\alpha^2)$ differ by factor of $5/3$, 
   and the hope was that  this  difference would allow 
    for an experimental discrimination between them.   
 Indeed, the  comparison with CLEO and CELLO data   for
  $Q^2 F_{\gamma^{\ast} \gamma \pi^0}(Q^2)$ 
  that extend to $Q^2 \lesssim 10$\, GeV$^2$
  favors DAs that are  closer  to $\varphi^{\rm as} (\alpha)$.
  However, recent {\sc BaBar} data covering the range up to 
  $Q^2 \sim 40$\, GeV$^2$ show the  increase of 
  $Q^2 F_{\gamma^{\ast} \gamma \pi^0}(Q^2)$ 
for $Q^2 \gtrsim 10$\, GeV$^2$.
  To explain this increase, the scenarios were
  proposed in which the pion DA does not vanish at the end-points, e.g., 
$\varphi_\pi^{\rm flat}(\alpha) =  1$.

 Another classic application of 
  pQCD to exclusive processes is the  pion electromagnetic 
  form factor, see figure~\ref{fig:radyush_ftffs_1}, right.
    With the asymptotic pion DA $\varphi^{\rm as}_{ \pi} (\alpha)$, the hard pQCD
    contribution to $ F_{ \pi}(Q^2)$ 
    is  $(2\alpha_s /\pi) (0.7 \,{\rm GeV}^2)/Q^2$, which is 
    less than 1/3 of the experimental value.
    Taking wider DAs formally increases the size of the
    one-gluon-exchange contribution,
    but it is dominated then by the regions where
    the gluon virtuality  is too small 
to
    be treated perturbatively.
    So, in this case we deal with the dominance 
of the competing soft mechanism which
is described by nonforward parton densities,
exactly in the same way as  the proton form factor $F_1^p(t)$
discussed in the previous section. 
 
\subsection{Hard electroproduction processes}

\begin{figure}[htb]
\begin{center}
\includegraphics[width=0.5\textwidth]{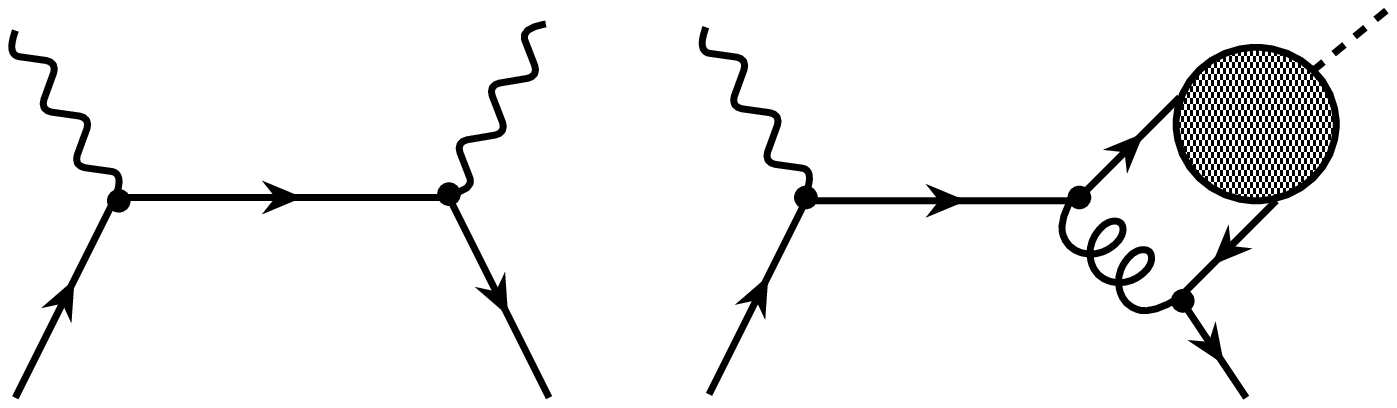}
 \caption{\small Lowest-order factorization for 
 deeply virtual photon and meson production.} 
\label{fig:radyush_ftdvpc_1}
\end{center}
  \end{figure}

An attempt to use  perturbative QCD
to extract  new information 
about hadronic structure is the study of 
deep exclusive photon \cite{Ji:1996ek} or meson 
\cite{Radyushkin:1996nd,Collins:1996fb} electroproduction reactions.
In the hard kinematics when both $Q^2$ and  $s \equiv (p+q)^2$ are large while 
the momentum transfer  
$t \equiv (p-p')^2$ is small,  
 one can use pQCD factorization which represents the amplitudes 
 as a convolution of a perturbatively calculable short-distance 
 amplitude and nonperturbative parton functions describing the hadron structure.
The  hard pQCD subprocesses in these two cases have different structure,
see figure~\ref{fig:radyush_ftdvpc_1}.
Since the photon is a pointlike particle, the  deeply virtual Compton scattering 
(DVCS)
amplitude has the structure similar to that of the $\gamma^{\ast} \gamma \pi^0$
 form factor: the pQCD hard term is of zero order in $\alpha_s$ (the handbag mechanism),
 and there is no competing soft contribution.
Thus, we can expect  that pQCD works  from 
 $Q^2 \sim 2\, {\rm GeV}^2$. On the other hand, the deeply virtual 
 meson production process is similar to the pion EM form factor: 
the  hard term has a $O(\alpha_s /\pi) 
 \sim 0.1$ suppression factor. As a result, the 
dominance of the hard pQCD term  
  may be postponed to $Q^2 \sim 5 - 10  \,{\rm GeV}^2$. 
  
  
  One should also have in mind that the competing soft mechanism 
  can mimic the same power-law $Q^2$-behavior 
  (just like in case of pion and nucleon EM form factors).
 Hence, a mere observation of a ``right''  power-law behavior of the cross section 
  may be insufficient to claim that pQCD is already working.
  One should look at other characteristics of the reaction,
  especially its spin properties, to make 
  strong statements about the reaction mechanism. 
 
\subsection{Deeply virtual Compton scattering and generalized parton distributions} 

 It is convenient to visualize DVCS in the   $\gamma^{\ast} N $ center-of-mass frame,
 with the initial hadron and the virtual photon moving 
 in opposite directions along the $z$-axis. Since the momentum transfer
 $t$ is small, the hadron and the real photon in the final state
 also move close to the $z$-axis.
This means that the virtual photon momentum $q= q' - x_{B}p$ has 
the component $- x_{B}p$  canceled by the momentum transfer $r$. 
In other words, the momentum  transfer  $r$ has the 
 longitudinal component 
 $r^+= x_{B} p^+$, where  $x_{B}= {Q^2}/{2(pq)}$
 is the DIS Bjorken variable. One can say that DVCS has a skewed kinematics 
 in which  the final hadron has the  ``plus'' 
 momentum $(1-\zeta)p^+$ that is smaller than that of the initial hadron.
In the particular case of DVCS, we have $\zeta = x_{B}$.

\begin{figure}[htb] 
\begin{center}
\includegraphics[width=0.6\textwidth]{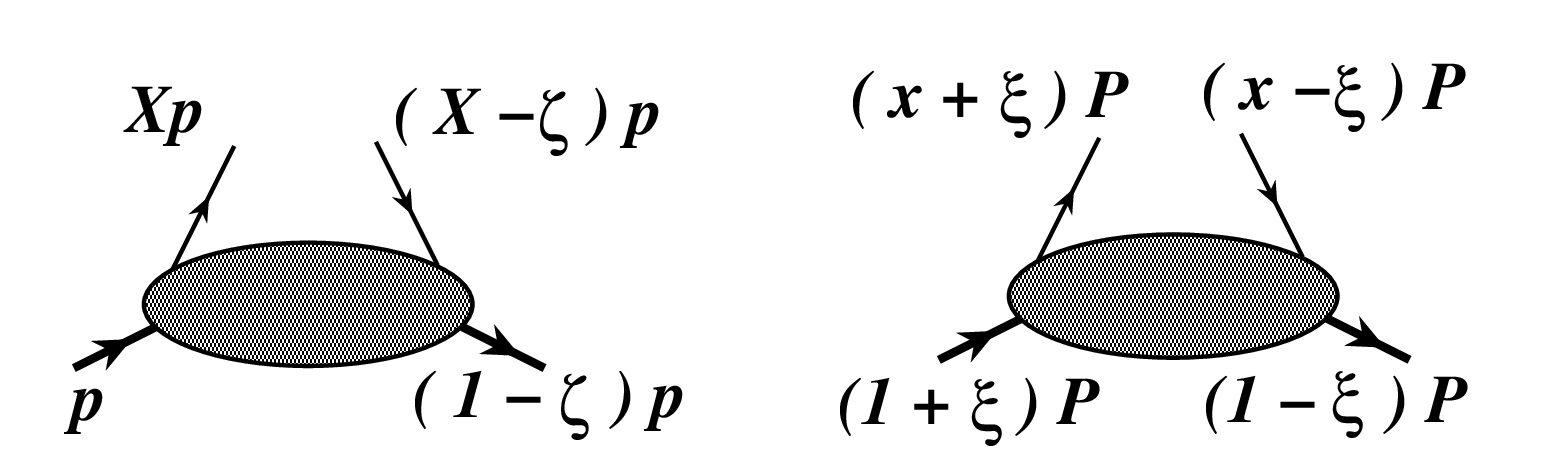}
 \caption{\small Comparison of NFPDs and OFPDs. } 
\label{fig:radyush_ftnfof}
\end{center}
  \end{figure}

The parton picture for DVCS  
has some similarity to that of DIS, with the main difference 
that the plus-momenta of  the incoming and outgoing quarks
in DVCS are not equal; they are $Xp^+$ and $(X-\zeta)p^+$, see
figure~\ref{fig:radyush_ftnfof}.
Another difference is that the invariant momentum transfer 
$t$ in DVCS is nonzero: the matrix element of partonic fields is
essentially nonforward.

Thus, the nonforward parton distributions (NFPDs) 
${\cal F}_{\zeta}(X,t)$  describing the hadronic structure 
in DVCS depend on $X$ (the  fraction of $p^+$  carried by the outgoing quark),
$\zeta$ (the skewness parameter characterizing the
difference between   initial and final hadron  momenta),
and $t$ (the invariant 
momentum transfer). 
In the forward $r=0$ limit, we have a reduction formula
\begin{equation}
{\cal F}^a_{\zeta=0}(X,t=0) = f_a(X) 
\end{equation}
relating NFPDs with the usual parton densities. 
The nontriviality of this relation is that ${\cal F}_{\zeta}(X,t)$
appear in the amplitude of the 
exclusive DVCS process, 
 while the usual parton densities are measured
 from the cross section of the   inclusive  DIS 
reaction.

 Another limit for NFPDs is zero skewness $\zeta =0 $, where they 
 correspond to  nonforward parton densities: 
 ${\cal F}^a_{\zeta=0}(X,t) ={\cal F}^a(X,t)$.
 The local limit relates NFPDs  to form factors:
 \begin{equation} 
\int_0^1 {\cal F}^a_{\zeta}(X,t) \, \frac{dX}{1-\zeta/2}
  =F_1^a(t) \,. 
\end{equation} 

The description in terms of NFPDs has the advantage of using the variables most close
to those of the usual parton densities.
However, the initial and final hadron momenta are
not treated symmetrically in this scheme.
Ji \cite{Ji:1996ek} proposed to use  symmetric variables in which 
the plus-momenta of the hadrons are 
$(1+\xi)P^+$ and  $(1-\xi)P^+$, and those 
of the active partons are $(x+\xi)P^+$ and $(x-\xi)P^+$,
$P$ being the  average  momentum $P= (p+p')/2$, see figure~\ref{fig:radyush_ftnfof}.
In the simplified case of scalar  fields,
the GPD parametrization of the nonforward matrix element is 
\begin{align}
&  \langle P+r/2 |   \psi(-z/2) \psi (z/2)|P-r/2 \rangle  
 =  
\int_{-1}^1 \,  e^{-ix(Pz) } H(x,\xi) \, dx +  {\cal O} (z^2) \,. 
\label{GPDdef}
\end{align}


To take into account the spin properties of hadrons 
and quarks, one needs four  off-forward parton distributions 
$H, E, \tilde H, \widetilde E$, each of which is a function  
 of $x$, $\xi$, and $t$.
The skewness parameter $\xi \equiv r^+/2 P^+$ can be expressed in terms of the 
Bjorken variable,
 $\xi = x_{B}/(2-x_{B})$, but it does not coincide with it.
 
Depending on the value of $x$, each GPD has  3 
distinct regions.
When  $\xi < x < 1$, GPDs  are analogous to  usual  quark distributions; when 
$-1 <x< -\xi$,  they are similar to  antiquark distributions.
In the region 
 $-\xi < x < \xi,$   the ``returning''  quark 
 has a negative momentum and should be treated as 
 an outgoing antiquark with momentum $(\xi-x)P$. 
 The total $q\bar q$ pair momentum $r=2\xi P$ is shared by the quarks in 
 fractions $r(1+x/\xi)/2 $ and $r(1-x/\xi)/2 $.
 Hence, a GPD in the region $-\xi < x < \xi $ 
 is similar to a distribution amplitude $\Phi (\alpha)$
 with $\alpha = x/\xi$.
  
 In the local limit, GPDs  reduce to elastic form factors:
\begin{equation}  
\sum_a e_a \int \limits_{-1}^1   
 {H}^a (x, \xi;t)
 \,  dx  =F_1(t) \quad ,  \quad  \sum_a e_a \int \limits_{-1}^1   
 {E}^a (x, \xi;t)
 \,  dx  =F_2(t) \,. 
\end{equation}  
The  $E$ function, like $F_2(t)$,   comes with  the $r_{\mu} $ factor. Hence,  it is 
 invisible
in DIS  
described 
by  the  forward $r=0$ Compton
amplitude. However, the 
$t=0, \xi =0 $ limit of  $E$  exists: 
\begin{equation}
E^{a,\bar a} (x, \xi =0 ;t=0)\equiv \kappa^{a,\bar a} (x) \,.
\end{equation}
In particular, its integral gives the  proton anomalous magnetic moment
$\kappa_p$,
\begin{equation} 
 \sum_a e_a \int \limits_{-0}^1   
 (\kappa^a(x) -   \kappa^{\bar a}(x))
 \,  dx  = \kappa_p \, ,
  \end{equation}
 while its first moment enters Ji's sum rule for 
the total quark  contribution $J_q$ to 
  the proton spin:
 \begin{equation} 
 J_q = \frac12 \sum_a  \int \limits_{-0}^1   
x \,  [f^a(x) +f^{\bar a}(x)+ \kappa^a(x) + \kappa^{\bar a}(x)]  
 \,  dx  \ .  \end{equation}
 Note that only valence quarks contribute to
$\kappa_p$, while $J_q$ involves also sea quarks. 
Furthermore, the  values of  $\kappa_{p,n}$
(unlike $e_{p,n} \equiv F_1^{p,n} (0)$)   
strongly  depend  on dynamics, e.g., $\kappa_N \sim 1/m_q$ 
in constituent quark models.

\subsection{Double distributions} 

 To model GPDs, two approaches are used:  a 
direct calculation 
in  specific  dynamical models: 
bag model, chiral soliton model, 
light-cone formalism, etc.,
and a phenomenological construction
based on the 
relation of GPDs to usual parton densities 
$f_a(x), \Delta f_a(x)$   and form factors
$F_1(t), F_2(t), G_A(t), G_P(t)$.
The key question in the second approach 
is the interplay between $x,\xi$ and $t$ 
dependencies of GPDs.  There are not so many cases
in which the pattern of the  interplay is 
evident. 
One example is 
 the function  
$\widetilde E(x,\xi,t)$ which is related to the $G_P(t)$  form factor 
and is dominated 
for small $t$ by the pion pole term $1/(t-m_{\pi}^2)$.  It 
is also proportional to the pion distribution amplitude 
$\varphi_{\pi} (\alpha)$
taken at $\alpha = x/\xi$. 
  The   construction  of   self-consistent  models for other GPDs
 can be  performed using   
an
 ansatz based on the 
formalism of  double distributions (DD)~\cite{Radyushkin:1998es}. 

\begin{figure}[htb]
\begin{center}
\includegraphics[width=0.67\textwidth]{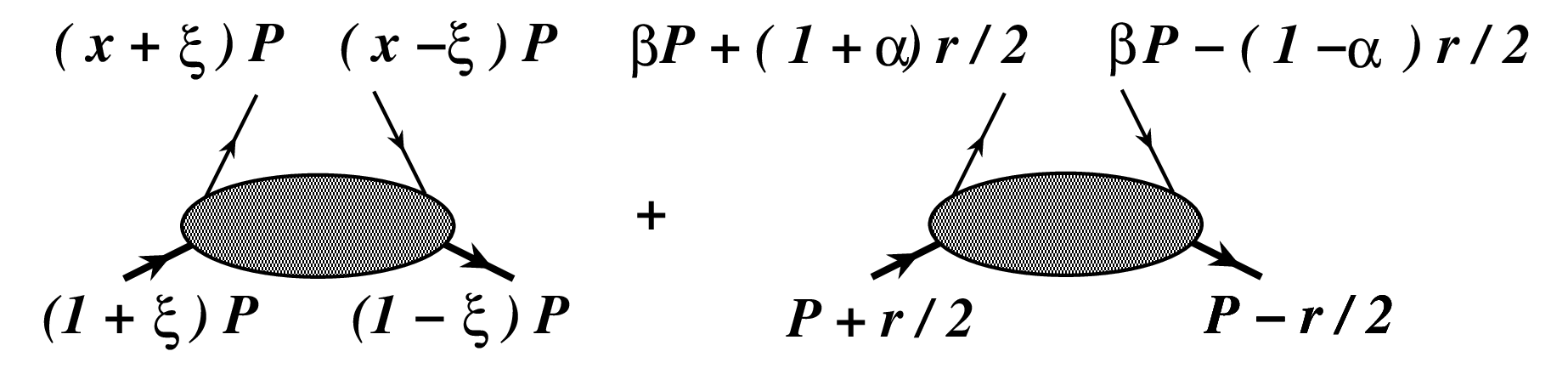}
  \caption{\small Comparison of GPD and DD descriptions.}
\label{fig:radyush_ftspdd}
\end{center}
\end{figure}

 The main idea behind the double distributions
 is a ``superposition'' 
of $P^+$ and $r^+$ momentum flows,
i.e., the representation of the parton 
momentum $k^+ = \beta P^+ + (1+\alpha) r^+/2$  
as the sum of a component $\beta P^+$ due to 
the average hadron momentum $P$ (flowing in the $s$-channel)
and a component $(1+\alpha) r^+/2$ due to the $t$-channel
momentum $r$, see figure~\ref{fig:radyush_ftspdd}.
In the simplified case of scalar fields, the DD parametrization reads 
\begin{align}
& \langle P-r/2 |   \psi(-z/2) \psi (z/2)|P+r/2 \rangle 
=  
\int_{\Omega}  F(\beta, \alpha) \,  e^{-i \beta (Pz) -i\alpha (rz)/2} \, d\beta \, d\alpha
 +  {\cal O} (z^2) \   . 
\label{DDF}
\end{align}
Thus, the double distribution $f(\beta,\alpha)$ 
(we  consider here for simplicity the $t=0$ limit) 
looks like a usual parton density with respect to $\beta$
and like a distribution amplitude with respect to $\alpha$.
The connection between the DD variables $\beta, \alpha$
and  the GPD variables $x,\xi$  is obtained from 
$r^+ = 2 \xi P^+$, which results in the basic relation
$ x = \beta + \xi \alpha $.
The formal  connection between  DDs and GPDs is 
\begin{align}
& H(x,\xi) =  
\int_{\Omega}  F(\beta, \alpha) \,  \delta (x - \beta -\xi \alpha)  \, d\beta \, d\alpha  \  \,.
\label{GPD}
\end{align}

The forward limit $\xi = 0, t=0$ corresponds to $ x = \beta $,
and  gives  the relation between DDs and the 
usual parton densities:
 \begin{equation}  \int_{-1 + |\beta |}^{1-|\beta |}  
 F_a(\beta ,\alpha;t=0) \, d \alpha =  f_a(\beta )
\ .
 \end{equation} 
 The DDs live on  the { rhombus} $ |\alpha| + |\beta| \leq 1$
[denoted by $\Omega$ in~(\ref{DDF}) and (\ref{GPD})]
and are symmetric functions of the ``DA'' variable $\alpha$:  
$f_a(\beta ,\alpha;t) = f_a(\beta ,-\alpha;t)$
(``Munich'' symmetry~\cite{Mankiewicz:1997uy}). 
These restrictions suggest a factorized representation  for a DD in the form of a 
product of a
usual  parton density
in the $\beta$-direction and a distribution 
 amplitude 
in the $\alpha$-direction:
\begin{align}
F(\beta,\alpha) = f(\beta)\, h(\beta,\alpha)   \   ,  \  h_{N}(\beta,\alpha) \sim \frac{[(1-|\beta|)^2-\alpha^2]^N}{(1-|\beta|)^{2N+1} }  \  , \
 \int_{-1+|\beta |} ^{1  -|\beta|}     h(\beta,\alpha) \,   d\alpha =1 \,.
\label{FDDA}
\end{align}
  
To obtain usual parton densities from DDs, one should integrate (scan) them
over the vertical lines $\beta = x = {\rm const}$. 
To obtain the GPD $H(x,\xi)$ with nonzero $\xi$ from DDs $f(\beta,\alpha)$, 
one should integrate (scan) 
DDs along the parallel lines $\alpha = (x-\beta)/\xi$ 
with a $\xi$-dependent slope. One can call this process the 
DD-tomography.   
 The basic feature of GPDs  $H(x,\xi)$ resulting from  DDs
is that 
for $\xi =0 $  they reduce to usual parton densities,
and for   $\xi =1 $ they have a shape like a meson  
distribution amplitude.
  A more complete truth is that such a  DD modeling 
misses terms invisible in the forward limit: 
meson-exchange contributions and so-called D-term,
which can be interpreted as $\sigma$-exchange.
The inclusion of the D-term  induces nontrivial
behavior in the central $|x| < \xi$ region
 (for details, see~\cite{Polyakov:1999gs}).  
  
 \subsection{GPDs and the structure of hadrons}
 
Hadronic structure  is a complicated subject, and it 
 requires a study from many sides and in many different 
 types of experiments. 
 The description   of specific aspects of 
 hadronic structure 
 is  provided by several different functions:
 form factors, usual parton densities,  
  distribution amplitudes.
Generalized parton distributions  provide a unified description: 
all these functions can  be treated 
as particular or limiting cases  of GPDs $H(x,\xi,t)$. 

{\it Usual parton densities}   $f(x)$ correspond to the case
 $\xi = 0, t=0$. 
They describe  a hadron  in terms of  probabilities 
 $\sim |\Psi|^2$. However, 
 QCD is a quantum theory: GPDs with $\xi \neq 0$ describe
correlations   $\sim \Psi_1^{\ast} \Psi_2$. 
Taking only the point $t=0$ corresponds to integration over impact parameters 
$b_{\perp}$ ---  information about the transverse structure is lost. 

{ \it Form factors} $F(t)$ 
contain  information about the 
 distribution of partons in the transverse plane, but $F(t)$ 
involve  integration over   momentum
fraction $x$ 
 --- information about   longitudinal structure is lost.

A simple ``hybridization'' of usual densities
and form factors in terms of NPDs ${\cal F}(x,t)$ 
(GPDs with $\xi=0$) 
shows that the behavior of $F(t)$ 
is governed   both by transverse 
and longitudinal distributions.
 GPDs provide adequate description of 
nonperturbative soft  mechanism. They  also allow to study
 transition from soft to hard mechanism.
 
 {\it  Distribution amplitudes}  $\varphi (x)$
provide quantum-level information about the longitudinal 
structure of hadrons.   
   In principle, they are  accessible in exclusive 
 processes at  large momentum transfer,
 when hard scattering mechanism dominates.  
 GPDs have DA-type structure  in the central region 
$|x| < \xi$.

{\it Generalized parton distributions} $H(x,\xi,t)$
 provide a 3-dimensional picture of hadrons.
GPDs also provide some novel  possibilities,
such as   ``magnetic distributions'' related to the
spin-flip GPD $E(x,\xi,t)$.  In particular, the structure of nonforward 
density $E(x,\xi=0,t)$
determines the $t$-dependence of $F_2(t)$. 
 Recent JLab data give $F_2(t) / F_1 (t) \sim 1/\sqrt{- t}$
rather than $1/t$ expected in  hard pQCD and many models 
--- a puzzle waiting to be resolved. The forward reductions 
$\kappa^a (x)$ of $E(x,\xi,t)$ look as fundamental
as $f^a(x)$  and $\Delta f^a(x)$:
 Ji's sum rule involves $\kappa^a (x)$
on equal footing with $f(x)$. 
Magnetic properties of hadrons are strongly sensitive 
to dynamics providing a   testing ground for models.
Another novel possibility is  the study of flavor-nondiagonal  distributions, e.g.,  
 proton-to-neutron GPDs accessible  through   processes like 
exclusive charged pion electroproduction, 
proton-to-$\Lambda$ GPDs (they appear in 
kaon electroproduction), and 
proton-to-$\Delta$ GPDs --- these    
can be related to form factors of  proton-to-$\Delta$
transition (another puzzle for hard pQCD). The GPDs for 
$N \to N + {\rm soft} \ \pi$  processes  can be used for testing  the soft
pion theorems and physics of chiral symmetry breaking.

An interesting problem is the separation and flavor decomposition 
of GPDs.  The 
 DVCS amplitude involves all four types of GPDs, $H,E, \widetilde H, \widetilde E$, 
so we need  
to study other processes involving different  combinations
of GPDs.  An important observation is that, in hard electroproduction of mesons, the 
spin nature of produced meson dictates the type
of GPDs involved, e.g.,  for pion electroproduction, only   
$\widetilde H, \widetilde E$ appear, with 
$\widetilde E$  dominated  by the pion pole at small $t$. 
This gives an access to 
(generalization of) polarized parton densities 
without polarizing the target. 


In summary,
the  structure of hadrons is the fundamental physics
to be accessed via GPDs. 
 GPDs  describe hadronic structure on the 
{ quark-gluon} level and 
 provide a  three-dimensional picture 
{(``tomography'')}  
of the hadronic structure. 
GPDs adequately reflect  the  quantum-field  
nature of QCD {(correlations, interference)}.  
They also  provide new insights into spin structure 
of hadrons (spin-flip  distributions, 
{ orbital} angular momentum). 
GPDs are sensitive to { chiral 
symmetry breaking} effects,
a fundamental property of QCD. 
Furthermore,  
GPDs { unify}  existing ways of describing 
hadronic structure. 
   The GPD formalism provides nontrivial relations between 
 different exclusive reactions and also between 
 { exclusive and inclusive}  processes.


\section{GPDs and transverse nucleon structure at collider energies}
\label{sec:Weiss}


\hspace{\parindent}\parbox{0.92\textwidth}{\slshape 
C.~Weiss}

\index{Weiss, Christian}

\vspace{1\baselineskip}




Generalized parton distributions (GPDs) have emerged as a key concept 
in nucleon structure and the theory of high momentum--transfer processes 
in QCD. They unify the traditional notions of parton densities 
and elastic form factors and describe the transverse
spatial distribution of quarks and gluons in a fast--moving hadron.
A general introduction to GPDs and hard exclusive processes is given in 
section~\ref{sec:Radyushkin}. Here we summarize the properties of GPDs at 
collider energies, where the parton picture can be combined with 
methods specific to high--energy scattering (``small--$x$ physics'').
This includes the transverse spatial structure of the nucleon 
at small $x$; gluon and quark imaging with hard exclusive processes at 
$ep$ colliders (HERA, EIC); the correspondence with the QCD dipole model
and the role of transverse nucleon structure in saturation at small $x$; 
and the application of GPDs to high--energy $pp$ collisions with hard 
processes (Tevatron, LHC).

%
%
\begin{figure}[b]
\includegraphics[width=0.84\textwidth]{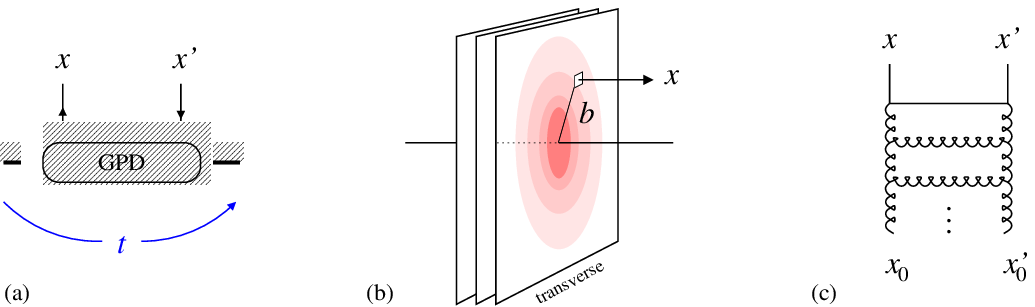}
\caption{\small (a) GPD and partonic variables. 
(b) Transverse spatial distribution of partons.
(c) QCD evolution generates small $x, x^{\prime}$ 
from the quasi--diagonal GPD at lower scale.
\label{weiss:fig:gpd}}
\end{figure}
GPDs are defined as the transition matrix elements of the QCD twist--2 
operators between nucleon states of different momenta. They are functions
of the longitudinal momentum fractions of the partons, $x$ and $x'$,
and the invariant momentum transfer $t$, as well as the resolution
scale $Q^2$ (see figure~\ref{weiss:fig:gpd}a). 
Of particular interest is the ``diagonal'' limit $x = x'$, 
where the momentum transfer is in the transverse direction only, 
$t = -|\bm{\Delta}|^2$, and the GPD can be regarded as the form factor 
of partons carrying longitudinal momentum fraction $x$.
Its two--dimensional Fourier transform 
\begin{equation}
f(x, b , Q^2) \; \equiv \; \int\!\frac{d^2 \Delta}{(2 \pi)^2}
\; e^{-i (\bm{\Delta} \bm{b})}
\; \textrm{GPD}(x, t = -\bm{\Delta}^2 , Q^2) 
\label{weiss:f_b}
\end{equation}
describes the transverse spatial distribution of partons with momentum
fraction $x$ and thus provides a ``tomographic'' image of the structure
of the fast--moving nucleon 
(see figure~\ref{weiss:fig:gpd}b) \cite{Burkardt:2002hr}. 
The coordinate $b$ measures the distance from the transverse 
center--of--mass (CM), defined as the average of the transverse positions 
of all constituents weighted with their longitudinal momentum fractions.
In general, the removal of a parton with momentum fraction $x$ 
changes the position of the CM, and this effect must be taken into
account in interpreting the coordinate distributions at $x \sim 1$.
At $x \ll 1$ however, the contribution of the removed parton to the 
CM is negligible and one can think of the $b$--distributions of~(\ref{weiss:f_b}) as referring to a fixed transverse center of the
nucleon. This considerably simplifies the spatial interpretation
of GPDs at small $x$.

%
%


\begin{figure}
\begin{minipage}[b]{0.44\columnwidth}
\centerline{\includegraphics[width=0.89\columnwidth]{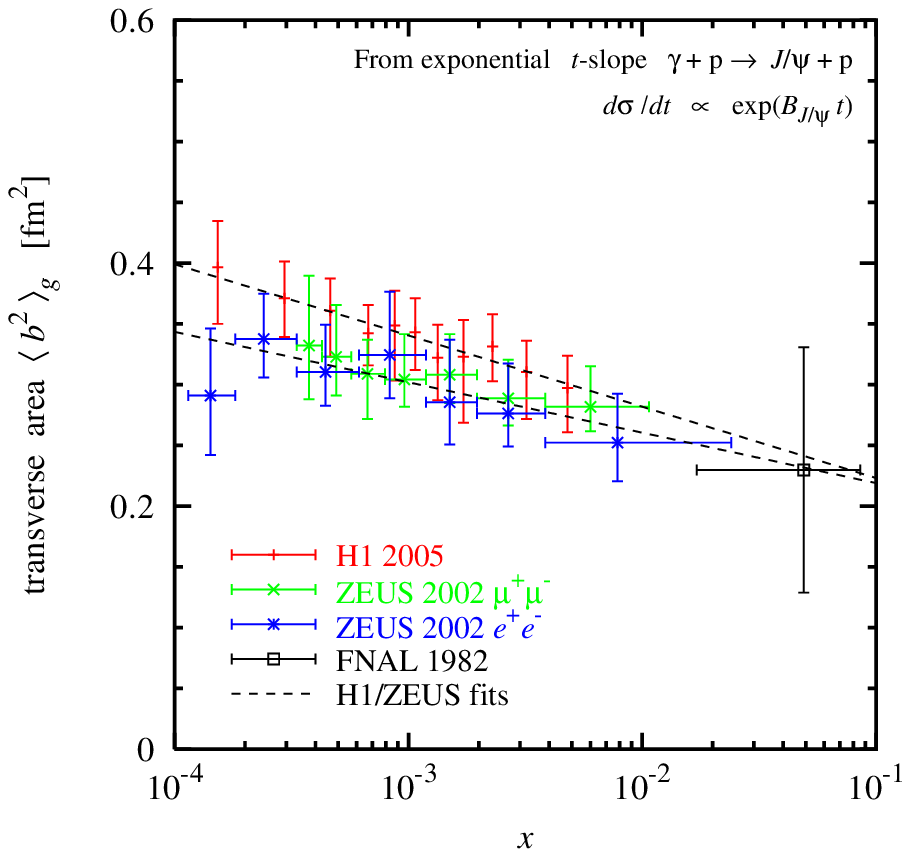}}
\end{minipage}
\begin{minipage}[b]{0.44\columnwidth}
\hspace{0.75cm}
\vspace{1.5cm}
\centerline{\includegraphics[width=0.89\columnwidth]{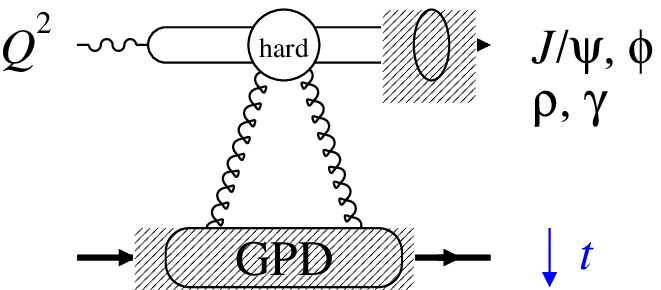}}
\end{minipage}
\caption{{\small (left) Exclusive $J/\psi$ production as a probe of the gluon GPD. 
(right) Average transverse gluonic size of the nucleon 
$\langle b^2 \rangle_g$ extracted from $J/\psi$ photoproduction
at HERA~\cite{Chekanov:2002xi,Aktas:2005xu} and FNAL~\cite{Binkley:1981kv}
(adapted from \cite{Frankfurt:2010ea}). The effective scale at which
the GPD is probed is $Q^2_{\rm eff} \approx 3 \, \textrm{GeV}^2$.
\label{weiss:fig:jpsi}}}
\end{figure}

\begin{wrapfigure}{r}{0.5\textwidth}
\begin{center}
\includegraphics[width=0.48\textwidth]{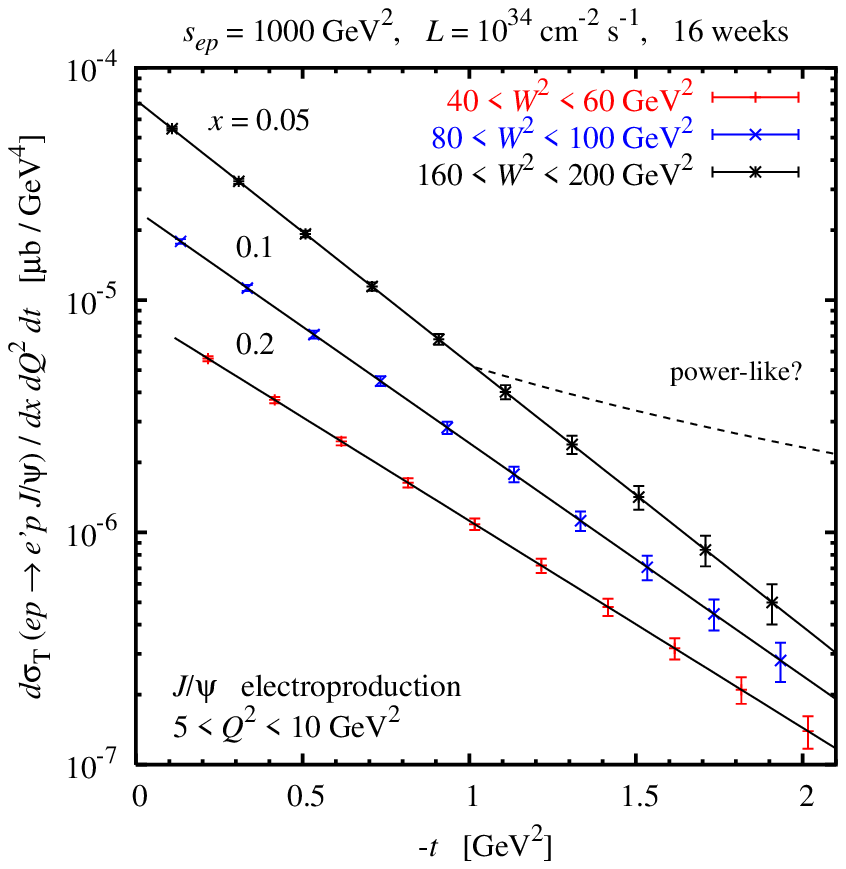}
\end{center}
\caption{\small A simulated measurement of exclusive $J/\psi$ electro-production
with a medium--energy EIC for an integrated luminosity of 100 fb$^{-1}$. The expected statistical errors in 
the $t$--dependence of the $J/\psi$ dilepton cross section 
in a fully differential measurement in $W, Q^2$ and $t$ are shown.
The values of $x \equiv M_{J/\psi}^2/W^2$ 
in the bins are indicated above the curves, corresponding approximately
to the $x$--values where the gluon GPD is probed.
Such measurements can 
image the transverse
distribution of 
gluons at $x > 0.1$ and 
explore the unknown $t$--dependence at $|t| > 1 \, \textrm{GeV}^2$.
\label{weiss:fig:eic}}
\end{wrapfigure} 

Hard exclusive processes require a non--zero longitudinal momentum 
transfer to the nucleon and probe the GPDs at $x - x' \equiv 2\xi \neq 0$, 
where the ``skewness'' is related to the Bjorken variable by 
$\xi = x_B/(2 - x_B)$. Models or additional assumptions are generally
needed to extract the diagonal GPD from the data. However, at $x_B \ll 1$
and sufficiently large $Q^2$ the ``skewed'' GPD can 
approximately be reconstructed from the diagonal 
limit~\cite{Frankfurt:1997ha,Martin:2009zzb}. 
In this case QCD evolution generates the GPD with $x$ and $x'$ from 
configurations at a lower scale with momentum fractions 
$x_0, x_0' \gg x, x'$; because the difference of the parton momentum 
fractions is preserved under evolution, the lower--scale GPD is 
effectively evaluated in the diagonal limit $x_0 - x_0' \ll x_0, x_0'$
(see figure~\ref{weiss:fig:gpd}c). This approximation allows one to relate 
the measured $t$--dependence of the differential cross sections
directly to the transverse structure of the nucleon at fixed $x$.

The transverse spatial distribution of partons changes with the
momentum fraction $x$ and the scale $Q^2$. The valence quarks and 
gluons at $x > 0.1$ are concentrated at small transverse distances 
$b \ll 1 \, \textrm{fm}$, as can be inferred from the nucleon 
axial form factor and exclusive processes at large $x$. Below 
$x < M_\pi / M_N$ chiral dynamics gives rise to a distinct 
large--distance contribution to the parton density at 
$b \sim 2/M_\pi$ \cite{Strikman:2003gz}. At even smaller values of 
$x$ the nucleon's transverse size is expected to grow as a result
of Gribov diffusion in the successive 
parton branchings
 building up the 
small--$x$ parton density. The transverse distribution also
shrinks with increasing $Q^2$ as a result of DGLAP evolution
\cite{Frankfurt:2003td}. Overall, much interesting information 
on nucleon structure and non-perturbative dynamics can be obtained 
from the study of the transverse spatial distributions of 
quarks and gluons.

The transverse spatial distribution of gluons can be measured 
cleanly through exclusive $J/\psi$ photo-- or electroproduction
$\gamma^{(\ast)} N \rightarrow J/\psi + N$, or electroproduction
of $\phi$ mesons at $Q^2 \gtrsim 10\, \textrm{GeV}^2$
(see figure~\ref{weiss:fig:jpsi}a). Measurements at HERA have confirmed 
the applicability of QCD factorization, with corrections for 
the finite size of the produced meson, and tested the universality 
of the gluon GPD; see~\cite{Frankfurt:2005mc} for a review. 
The data show that the nucleon's transverse gluonic radius at $x < 0.01$ 
is substantially smaller than the transverse charge radius (see figure~\ref{weiss:fig:jpsi}b).
It increases only moderately with decreasing $x$, 
with a logarithmic slope much smaller than that of the Pomeron
trajectory, $\alpha'_P = 0.25 \, \textrm{GeV}^{-2}$,
showing that Gribov diffusion is suppressed for partons with
virtualities $\sim \textrm{few GeV}^2$. Both observations are of 
central importance for nucleon structure and small--$x$ physics.

While the HERA experiments have provided basic information on the 
nucleon's transverse gluonic size at small $x$, many important
questions remain unanswered:
\begin{itemize}
\item How are the gluons at $x > 10^{-2}$ distributed in transverse space?
Global PDF fits indicate a substantial 
momentum density of gluons in that $x$-range at low scales $Q^2 \sim $ few GeV$^2$.
Knowledge of their spatial distribution would help to
explain their dynamical origin, one of the key issues of nucleon structure
in QCD.
\item Do singlet quarks and gluons have the same transverse distribution?
This can be studied by comparing the $t$--dependence of $J/\psi$ and
$\phi$ with $\rho^0$ and $\gamma$ electroproduction. 
A larger radius for quarks than gluons is expected from
non-perturbative effects \cite{Strikman:2009bd}.
\item How are non-singlet sea quarks distributed in transverse space?
The non-singlet sea at $x < 0.1$ reveals non-perturbative QCD interactions
(vacuum fluctuations, mesonic degrees of freedom) in the nucleon.
This component is probed in exclusive $\pi, K, \rho^+$ or 
$K^\ast$ production --- non--diffractive processes involving quantum
number exchange.
\item How does the nucleon's gluon GPD behave at $|t| \sim \textrm{few GeV}^2$?
The large--$|t|$ behavior of GPDs is important not only to obtain accurate 
images at small $b$, but also to understand how soft Regge--like dynamics
is connected to QCD at short distances.
\item What is the probability for a nucleon to break up into a low--mass
hadronic state ($M_H \sim \textrm{few GeV}$) in an exclusive process
at small $x_B$? Such ``diffractive dissociation'' reveals the quantum 
fluctuations of the nucleon's gluon density -- new information going 
beyond the average densities described by the GPDs \cite{Frankfurt:2008vi}.
\end{itemize}

An EIC would enable a comprehensive program of transverse imaging of gluons 
and sea quarks in the nucleon. Measurements of $J/\psi$ photo-- and 
electroproduction, as well as $\phi$ meson electroproduction at
$Q^2 > 10\, \textrm{GeV}^2$, would cleanly map the transverse 
distribution of gluons, including the 
gluons at 
$x > 0.1$ (see the example in figure~\ref{weiss:fig:eic}). 
They could also explore the unknown $t$--dependence of the GPD 
at $|t| > 1 \, \textrm{GeV}^2$. Measurements of $\rho^0$ and $\gamma$ 
production (DVCS) would provide additional information on the 
singlet quarks. With a high--luminosity EIC, even the non--diffractive
channels ($\pi, K, \rho^+, K^\ast$) could be measured for the first
time down to $x \sim 0.01$, providing detailed information on the spatial 
distribution of the non-singlet sea, including its spin and flavor
composition (see section~\ref{sec:Horn}).

\begin{figure}
\begin{center}
\includegraphics[width=0.55\textwidth]{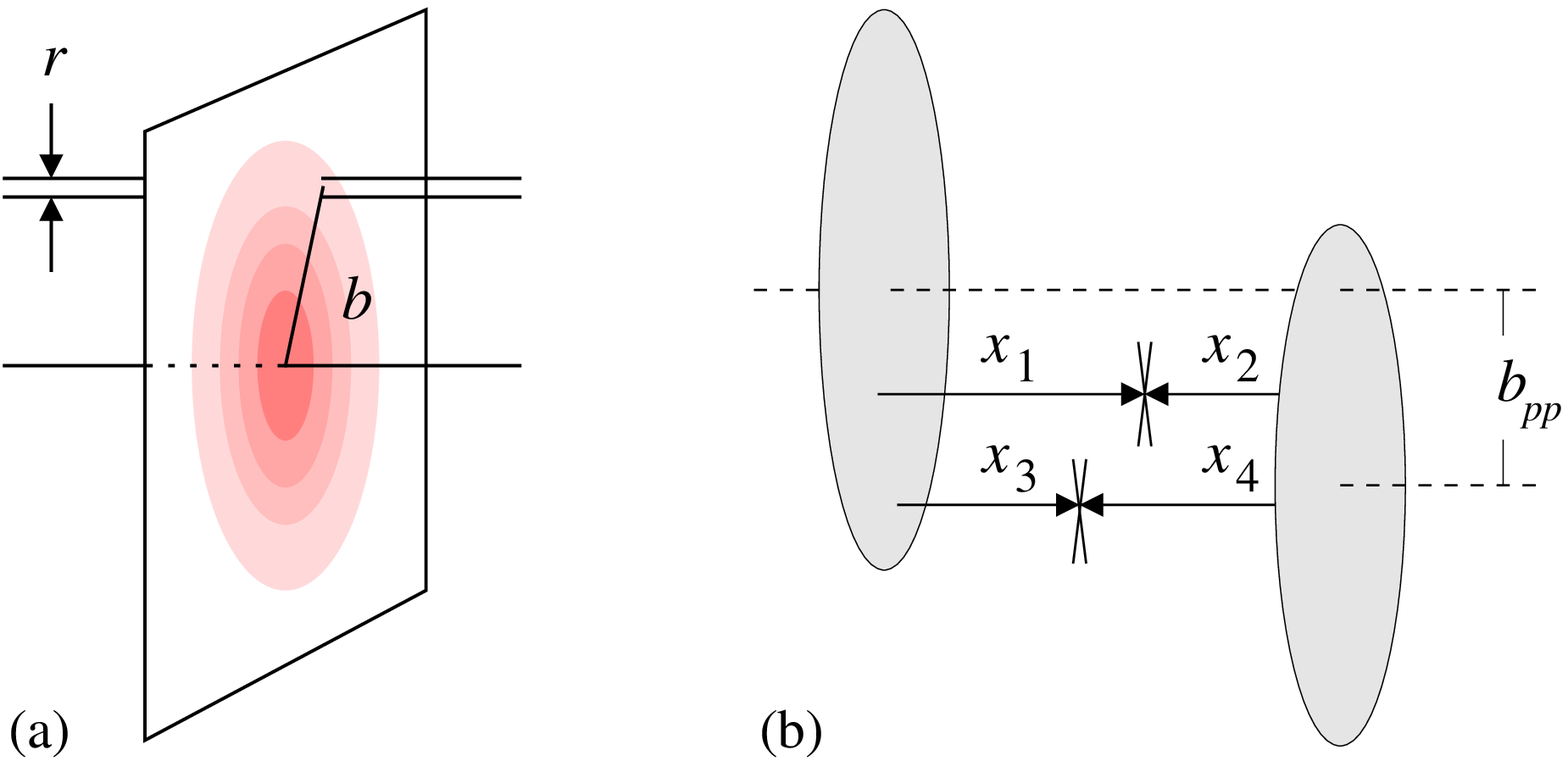}
{\caption{\small (a) Dipole picture of
high--energy scattering in the target rest frame.
(b) Multiparton processes in high--energy $pp$ collisions.}}
\end{center}
\label{weiss:fig:dipole}
\end{figure}

The QCD factorization theorem for hard exclusive processes 
at small $x$ (figure~\ref{weiss:fig:jpsi}a) is equivalent to the dipole 
picture of exclusive processes in the nucleon rest frame in 
the leading $\alpha_s \log Q^2$ approximation \cite{Frankfurt:1996ri}. 
The scattering amplitude for a dipole of size $r$ with impact parameter 
$b$ is proportional to the $b$--dependent gluon density of~(\ref{weiss:f_b}) at a scale $Q^2 \approx \pi^2/r^2$ 
(see figure~\ref{weiss:fig:dipole}a). This correspondence relates GPDs to the 
dipole model phenomenology of small--$x$ physics \cite{Frankfurt:2005mc}. 
In particular, the transverse 
spatial distribution of gluons is an essential input 
to studies of the unitarity limit in hard processes at small $x$
(``black--disk regime''). It defines the spatial profile of the
initial conditions of non-linear QCD evolution equations leading
to gluon saturation at small $x$. Detailed studies of saturation
in the dipole model have used the transverse gluonic size extracted
from the HERA data (see figure~\ref{weiss:fig:jpsi}b)~\cite{Kowalski:2003hm,Rogers:2003vi}; better knowledge 
of the transverse profile would help to accurately predict the
$x$ and $b$--dependence of the saturation scale.
%
%

The transverse distribution of partons also plays an important 
role in high--energy $pp$ collisions with hard processes.
It determines the probability of hard parton--parton processes
as a function of the $pp$ impact parameter. Using knowledge of
the transverse distribution of partons from $ep$ scattering
one can explain many features of the underlying event in 
$pp$ collisions with hard processes \cite{Frankfurt:2010ea}. 
In particular, one can
predict the rate of multiparton processes (see figure~\ref{weiss:fig:dipole}b),
which form a potentially large background to new physics events at the LHC.
The enhancement of such processes beyond their geometric probability
signals dynamical correlations between partons, the study of which 
represents a new frontier of nucleon structure.

\section{How large can the distributions $E^q$ and $E^g$ be?}
\label{sec:gpd-e}


\hspace{\parindent}\parbox{0.92\textwidth}{\slshape 
  Markus Diehl}
%

\index{Diehl, Markus}





\subsection{Positivity bounds}

The generalized parton distributions $E$ for quarks and gluons play a key
role in the study of nucleon structure through exclusive processes.  In
the following I focus on the case of zero skewness, $\xi=0$, where the
physics interpretation is most intuitive and where constraints on these
distributions are most easily obtained.  The density of unpolarized quarks
in a proton polarized along the $x$-axis is given by
\begin{equation}
  \label{diehl:qXdef}
q^X(x,\vec{b}) = q(x,b^2)
  - \frac{b^y}{m} \, \frac{\partial}{\partial b^2}\, e_q(x,b^2) \,,
\end{equation}
where $m$ is the proton mass.  The quarks have impact parameter $\vec{b}$
and move in the $z$-direction with momentum fraction $x$.  The term with
\begin{equation}
e_q(x,b^2) = \int \frac{d^2 \Delta}{(2\pi)^2}
  e^{-i \vec{b} \vec{\Delta}}\,
  E^q(x,\xi=0,t=-\vec{\Delta}^2)
\end{equation}
quantifies the transverse shift of the density due to the proton
polarization.  The density interpretation of \eqref{diehl:qXdef} (together
with its analog for longitudinal quark and proton polarization) entails a
positivity bound \cite{Burkardt:2003ck}:
\begin{equation}
  \label{diehl:b-bound}
\frac{b^2}{m^2} 
\biggl[ \frac{\partial}{\partial b^2}
  e_q(x,b^2) \biggr]^2
\le \bigl[ q(x,b^2) + \Delta q(x,b^2) \bigr]
     \bigl[ q(x,b^2) - \Delta q(x,b^2) \bigr] \,.
\end{equation}
The theoretical status of this bound is the same as for the positivity of
unpolarized parton densities and for the Soffer inequality: they hold in
the parton model and are preserved by leading-order DGLAP evolution to
higher scales, but they can be violated by higher-order evolution effects
or at very low scales.  Since so little is known about $E$, I suggest to
use \eqref{diehl:b-bound} as a guide, with proper caution.  A consequence
of \eqref{diehl:b-bound} is that $(\partial /\partial b^2)\, e_q$ must
decrease faster with $b$ than $\sqrt{q^2 - \Delta q^2}$.  This has
immediate consequences for parameterizations: using Gaussian forms $E^q
\propto e^{B_e t}$ and $\sqrt{q^2 - \Delta q^2} \propto e^{B_q t}$ for the
momentum-space distributions at $\xi=0$, one must have $B_e < B_q$, and
with power laws $E^q \propto (1 - t/M_e^2)^{-3}$ and $\sqrt{q^2 - \Delta
  q^2} \propto (1 - t/M_q^2)^{-2}$, one must have $1/M_e < 1/M_q$, with
equality of the parameters not being allowed in either case.
Starting from \eqref{diehl:b-bound} one can also derive a bound
\cite{Burkardt:2003ck} for the integrated distribution:
\begin{equation}
e_q(x) = \int d^2b\, e_q(x,b^2) = E^q(x,\xi=0,t=0) \,.
\end{equation}
That bound constrains the large $x$ behavior of $e_q(x)$, but numerically
turns out to be rather weak for $x$ below $0.5$, see e.g.\
\cite{Diehl:talk1108}.

Analogous definitions and bounds apply to antiquark and gluon
distributions $e_{\bar q}$ and $e_g$.


\subsection{Sum rules}

\begin{figure}
\begin{center}
\includegraphics[width=0.32\textwidth,bb=120 50 345 300]{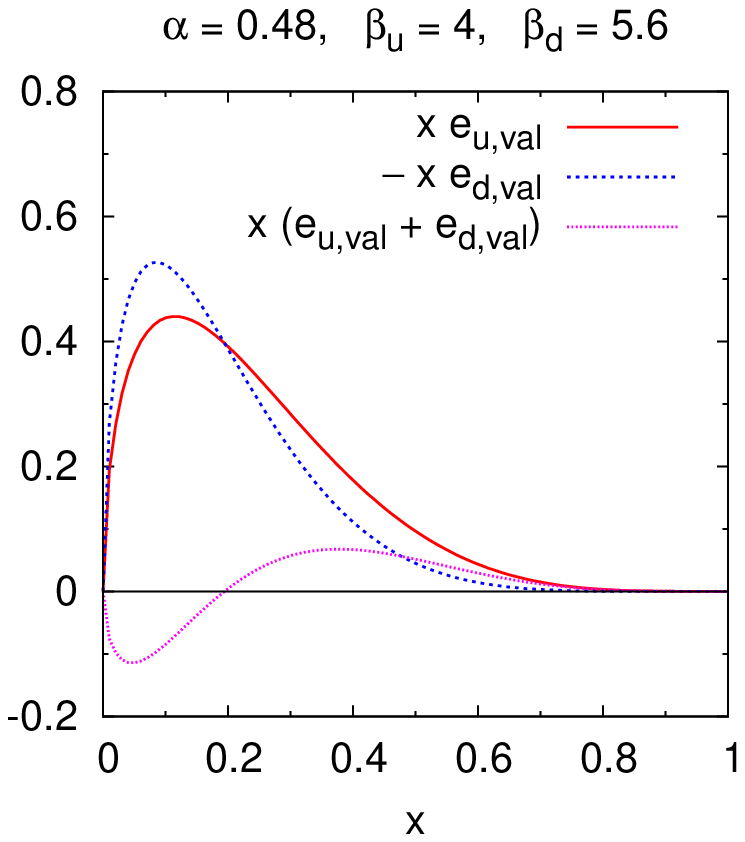}
\includegraphics[width=0.32\textwidth,bb=120 50 345 300]{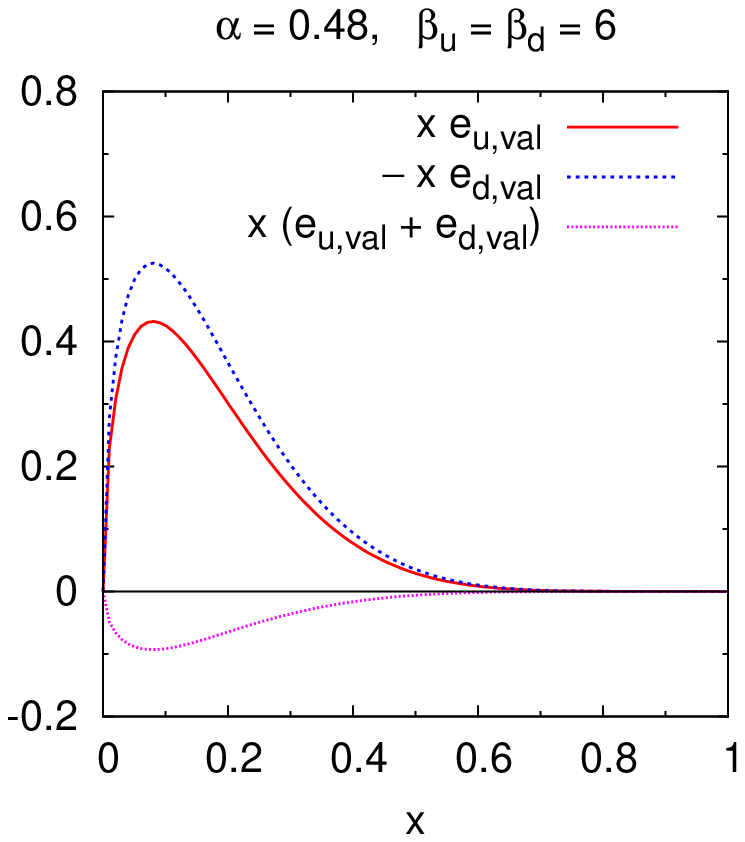}
\includegraphics[width=0.32\textwidth,bb=120 50 345 300]{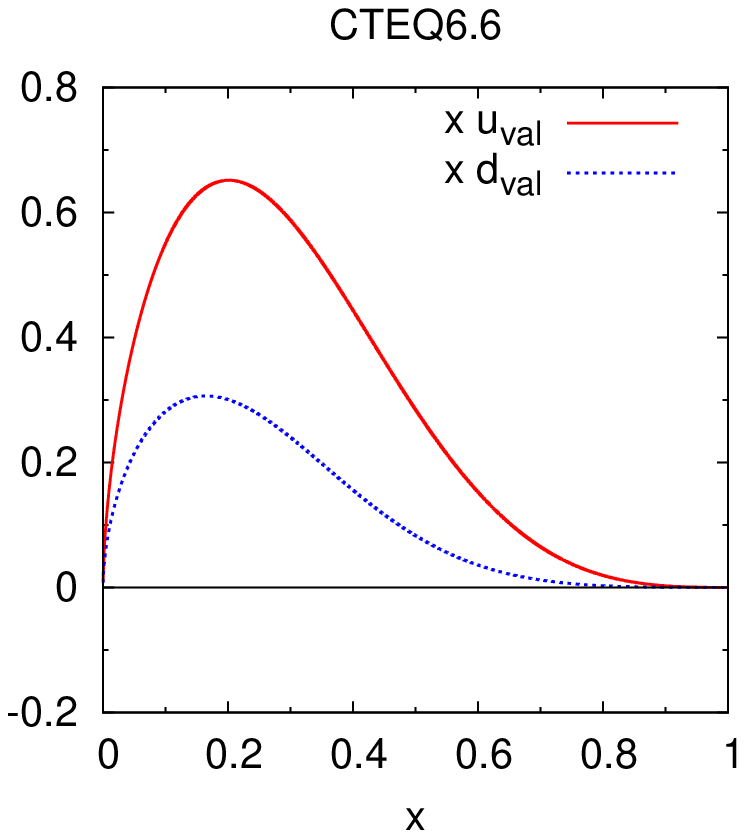}
\end{center}
\caption{\label{diehl:fig-1} \small GPDs in the forward limit obtained in a
  phenomenological fit \protect\cite{Diehl:2004cx} to the nucleon form
  factors.  The first two panels correspond to two parameter sets giving a
  good fit.  The valence quark distributions $u_{\text{val}} = u -
  \bar{u}$ and $d_{\text{val}} = d - \bar{d}$ in the third panel are shown
  for comparison.  All distributions are shown at the scale $\mu= 2\gev$.}
\end{figure}

An important constraint follows from the sum rule
\begin{align}
  \label{diehl:1st-sum-rule}
\kappa_q &= {\textstyle\int\limits_0^1} dx\,
   [e_q(x) - e_{\bar{q}}(x) \big] \,,
\end{align}
where $\kappa_q$ is the contribution of quark flavor $q$ to the anomalous
magnetic moment of the proton.  From the magnetic moments of proton and
neutron one obtains $\kappa_u - \kappa_d = 3.71$ and $\kappa_u + \kappa_d
+ \kappa_s = -0.36$.  Under the reasonable assumption that $\kappa_s$ is
small compared with $\kappa_u$ and $\kappa_d$, these numbers imply that
$\kappa_u$ and $\kappa_d$ are both large but have opposite signs and
largely cancel in the flavor sum.  As a consequence, the functions
$e{}_{u, \text{val}}(x) = e_u(x) - e_{\bar{u}}(x)$ and $e_{d,
  \text{val}}(x) = e_d(x) - e_{\smash{\bar{d}}}(x)$ must be large at least
in some region of $x$.  This is illustrated in figure~\ref{diehl:fig-1},
which shows distributions obtained by fitting a model ansatz for $u$ and
$d$ quark GPDs to the electromagnetic nucleon form factors
\cite{Diehl:2004cx} (neglecting strange-quark contributions).  The fit
suggests that $e_{u, \text{val}}$ and $e_{d, \text{val}}$ are of similar
size as the unpolarized valence distributions, whereas $e_{u, \text{val}}
+ e_{d, \text{val}}$ is small and poorly known, to the point that we do
not know whether it has zero crossings.

The second moments of $e(x)$ appear in Ji's angular momentum sum rules,
\begin{align}
2 J^q &= {\textstyle\int\limits_0^1}
             dx\, x \bigl[ q(x) + \bar{q}(x) \bigr]
       + {\textstyle\int\limits_0^1}
             dx\, x \bigl[ e_q(x) + e_{\bar{q}}(x) \bigr] \,,
&
2 J^{g} &= {\textstyle\int\limits_0^1} dx\, x g(x)
         + {\textstyle\int\limits_0^1} dx\, x e_g(x) \,,
\end{align}
where they give ``nontrivial'' contributions in addition to the
``trivial'' ones from the momentum integrals of quarks and gluons (whose
values are well known).  Summed over all partons, the momentum integrals
add up to 1 and the angular momenta to $\frac{1}{2}$, so that
\begin{align}
  \label{diehl:2nd-sum-rule}
{\textstyle\int\limits_0^1} dx\, x e_{\text{sing}}(x) + 
 {\textstyle\int\limits_0^1} dx\, x e_g(x) = 0 \,,
\end{align}
where $e_{\text{sing}}(x) = \sum_q\, [ e_q(x) + e_{\bar{q}}(x) ]$.  Note
that both \eqref{diehl:1st-sum-rule} and \eqref{diehl:2nd-sum-rule} are
exact relations in QCD, in contrast to the positivity bound
\eqref{diehl:b-bound}.  The scale dependence of $e_g(x)$ and
$e_{\text{sing}}(x)$ is governed by coupled DGLAP equations, with the same
kernels as for the unpolarized gluon and quark singlet distributions.
With \eqref{diehl:2nd-sum-rule} one finds that to leading order in
$\alpha_s$
\begin{equation}
  \label{diehl:evolution}
{\textstyle\int\limits_0^1} dx\, x e_g(x, \mu) 
= \biggl( \frac{\alpha_s(\mu)}{\alpha_s(\mu_0)} \biggr)^{\!\gamma}\;
  {\textstyle\int\limits_0^1} dx\, x e_g(x, \mu_0) \,,
\end{equation}
where $\gamma = 50/81$ for $n_f=3$ and $56/75$ for $n_f=4$ active flavors.
All numbers in the following refer to $\mu=2\gev$; the evolution of
\eqref{diehl:evolution} to higher scales is rather slow.

With the distributions in \cite{Diehl:2004cx} one finds that $\int dx\, x
e_{\text{sing}}$ has a very small valence part $\int dx\, x [e_u -
e_{\bar{u}} + e_d - e_{\smash{\bar{d}}}]$ between $-0.042$ and $0.068$.  A
similar situation is found in lattice calculations, which obtain a small
contribution to $\int dx\, x [e_u + e_{\bar{u}} + e_d +
e_{\smash{\bar{d}}}]$ from \emph{connected} graphs, with values between
$-0.077(16)$ and $0.015(11)$ for different extrapolations to the physical
quark masses \cite{Bratt:2010jn}.

\begin{figure}[h]
\begin{center}
\includegraphics[width=0.44\textwidth,bb=60 50 400 300]{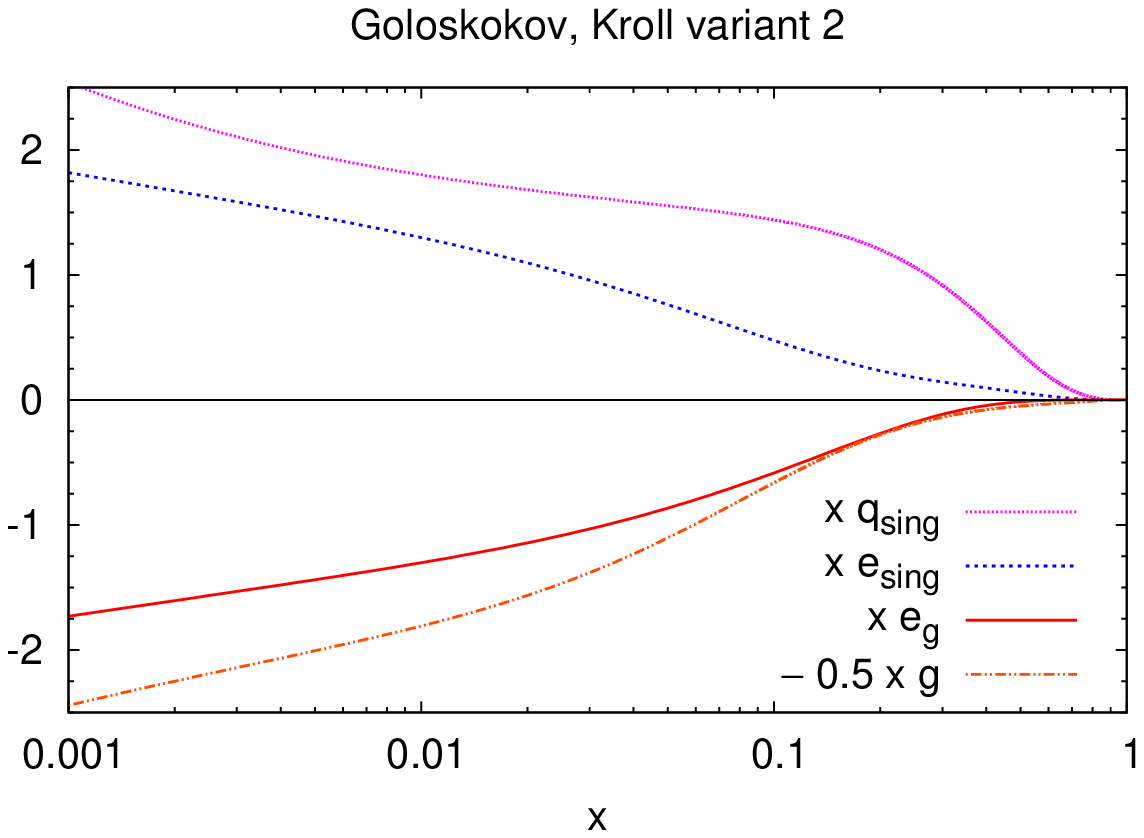}
\includegraphics[width=0.44\textwidth,bb=60 50 400 300]{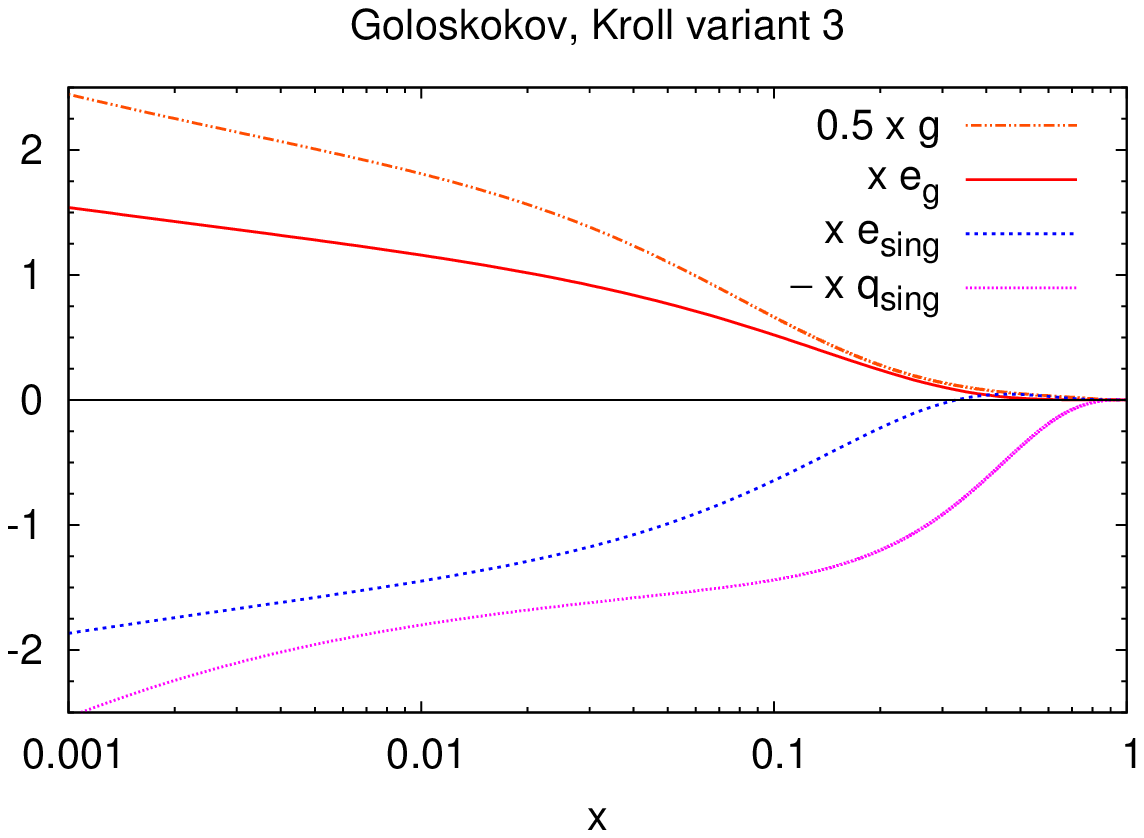}
\end{center}
\caption{\label{diehl:fig-2} \small Two variants of model distributions $e_g$ and
  $e_{\text{sing}}$ at $\mu=2\gev$ from \protect\cite{Goloskokov:2008ib}.
  The distributions of the quark singlet $q_{\,\text{sing}} = \sum_q (q +
  \bar{q})$ and the gluon are shown for comparison.}
\end{figure}

Assuming that $\int dx\, x [e_u - e_{\bar{u}} + e_d -
e_{\smash{\bar{d}}}]$ is indeed small (and barring the possibility of an
implausibly large $e_{s} - e_{\bar{s}}$) we find that the sum $\int dx\, x
e_g + \int dx\, x e_{\text{sea}}$ of second moments must be small, where
$e_{\text{sea}} = 2 \sum_q e_{\bar{q}}$.  This still leaves us with a
number of possible scenarios:
\begin{enumerate}
\item both $e_g(x)$ and $e_{\text{sea}}(x)$ are small (note that this does
  not exclude large $e_{\bar{q}}(x)$ for individual quark flavors: only
  the flavor sum must be small),
\item $e_g(x)$ and $e_{\text{sea}}(x)$ are both large but have opposite
  signs, 
\item both distributions are large but have nodes such that their second
  moments are small.
\end{enumerate}
Scenario 2 is illustrated in figure \ref{diehl:fig-2}, which shows two
variants of model distributions proposed in \cite{Goloskokov:2008ib}.  The
absolute size of the distributions is limited by the bound
\eqref{diehl:b-bound} and its analogs for $e_{\bar{q}}$ and $e_g$, and the
opposite signs of $e_g$ and $e_{\text{sea}}$ ensure that
\eqref{diehl:2nd-sum-rule} can be fulfilled.  We see that scenarios where
both $e_g$ and $e_{\text{sea}}$ are large cannot be ruled out with our
present knowledge.  If the above model distributions are evolved to higher
scales, $e_g$ becomes even larger and steeper at small $x$
\cite{Diehl:talk1108}.


\subsection{Exclusive processes}

Up to now I discussed $E^q$, $E^{\bar{q}}$ and $E^g$ at zero skewness
$\xi=0$, but in exclusive processes like DVCS and meson production $\xi$
is always nonzero.  Nevertheless, experience from phenomenology and models
suggests that GPDs at $\xi=0$ are closely enough related to those at
$\xi\neq 0$ to serve as a guide for their overall size, see e.g.\
\cite{Boffi:2007yc,Muller:talk1108}.

Note that even the large model distributions $e_g$ and $e_{\text{sea}}$ in
figure~\ref{diehl:fig-2} result in small values for the transverse target
spin asymmetry $A_{UT}$ in exclusive $\rho$ electroproduction
\cite{Goloskokov:2008ib}.  This is in part due to cancellations in the sum
over $u$ and $d$ quarks in this process (the same distributions give a
larger asymmetry for $\omega$ production).  Moreover, $A_{UT}$ in
exclusive meson production is proportional to $\operatorname{Im}
(\mathcal{H}\, \mathcal{E}^*)$, where $\mathcal{H}$ and $\mathcal{E}$ are
the scattering amplitudes associated with $H$ and $E$ distributions,
respectively.  Hence $A_{UT}$ is also small when both amplitudes are large
but have a small relative phase.  The transverse target asymmetry in DVCS
is therefore of special importance, because the interference between
Compton scattering and the Bethe-Heitler process is linear in
$\operatorname{Im} \mathcal{E}$.



\section{Imaging transverse distributions}
\label{sec:Miller}


\hspace{\parindent}\parbox{0.92\textwidth}{\slshape 
  Gerald A. Miller}
%

\index{Miller, Gerald A.}





\subsection{Introduction}

Much effort has gone into measuring electromagnetic form factors, which are related to the charge and magnetization densities within the nucleons. The  influence of relativistic motion of the quarks within the nucleon causes  the standard textbook interpretation of form factors as three-dimensional Fourier transforms to be wrong~\cite{Miller:2009sg}.
The use of transverse densities  \cite{Miller:2007uy,Miller:2010nz} avoids various  difficulties by  working in the infinite
momentum frame and taking the spacelike momentum transfer to be in the
direction transverse to that of the infinite momentum. In this case,  the different momenta of the initial and final
nucleon states are  accommodated by using 
two-dimensional Fourier transforms and  transverse charge and magnetization densities  are constructed from
density operators that are the absolute square of quark-field operators.

The transverse charge  density is given by~\cite{Miller:2007uy,Soper:1976jc}
\begin{equation}
\rho(b) = \int dx^-\rho(x^-,b)
 = {1\over 2\pi}\int QdQ J_0(Q b)F_1(Q^2) \,,
\label{rhodef}
\end{equation}
where $\rho(x^-,b)$ is the three dimensional spatial density.

The transverse charge densities are shown in \cite{Miller:2007uy,Miller:2010nz}. 
The interesting feature is that the central neutron charge density is negative.
An interpretation of this finding based on the impact parameter distribution~\cite{Burkardt:2002hr,Diehl:2002he}
   was presented in \cite{Miller:2008jc}.
   All models of these quantities  are based on the Drell--Yan--West relation, which connects large values of $x$ with large values of $Q^2$.  These models tell us that 
the $d$ quarks that dominate deep inelastic scattering from the neutron  at large values of $x$
dominate the neutron center. It is also possible that the negatively charge pionic cloud may penetrate the center~\cite{Rinehimer:2009sz}.
 
 The transverse anomalous magnetization density is obtained from the matrix element of the magnetization density operator
 ${1\over 2}\vec{b} \times \vec{j}$, where $\vec{j}$ is taken in the $z$-direction:
\begin{equation}
\rho_M(b)={\sin^2\phi\over 2M}b \int {Q^2dQ\over 2\pi}F_2(Q^2)J_1(Qb) \,.
\end{equation}
 The integral $\int d^2b \rho_M(b)$ gives the anomalous magnetic moment.

\subsection{Realistic transverse images of the proton charge and magnetic densities}

The word ``realistic" refers to the ability to know the uncertainty  in the transverse densities derived from experiment.
The previously obtained transverse densities are derived from various parameterizations of the form factors.
A more detailed treatment is needed to be able to extract uncertainties. The following discussion is
based on the analysis~\cite{Venkat:2010by}.
 
\begin{figure}
\begin{center}
\includegraphics[width=0.5\textwidth]{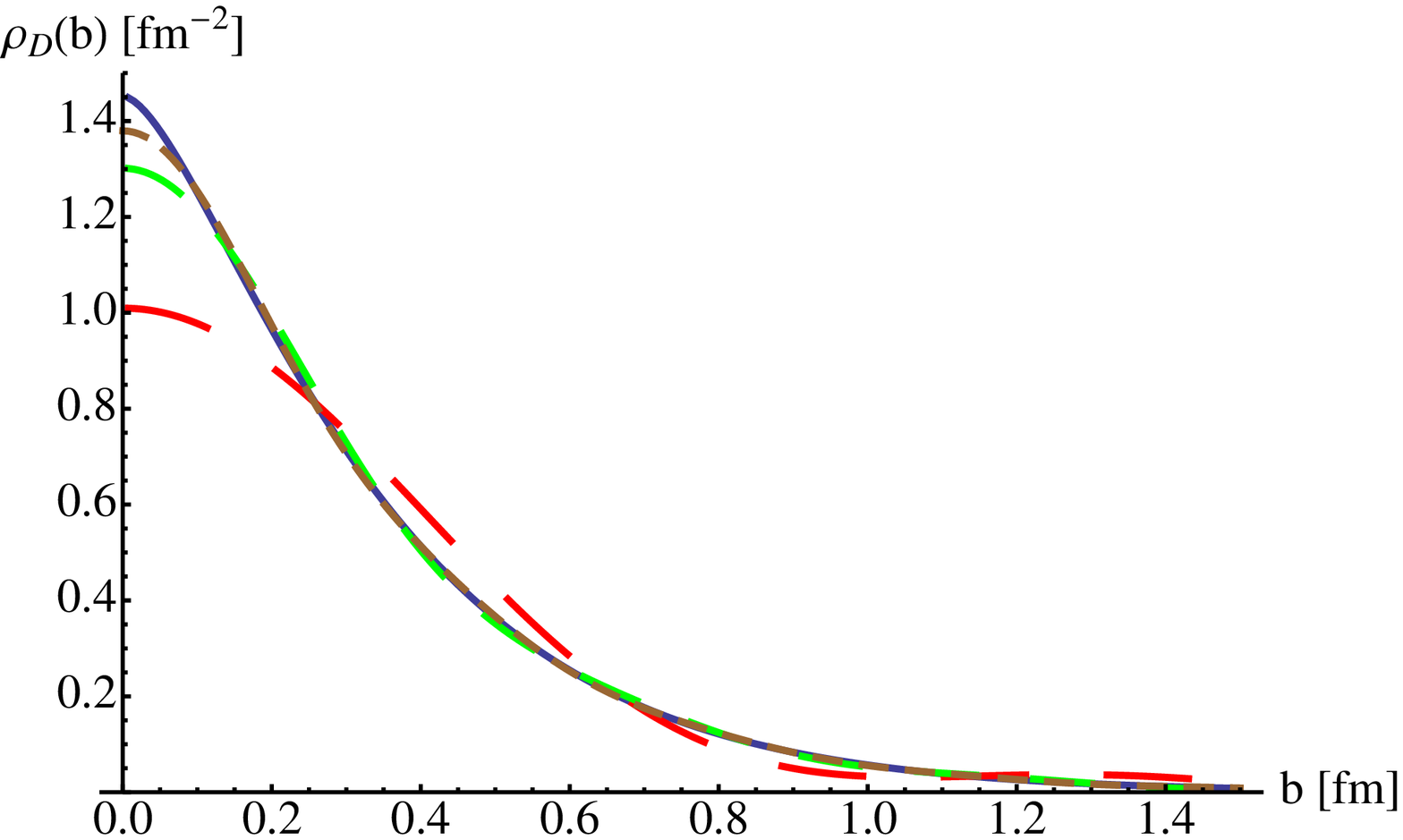}
\end{center}
\caption{\label{fig:dipole} \small Plot of $\rho_D$ (solid), 5 term approximation (red, long dash), 10
term approximation (green, medium dash) and 15 term approximation (brown, short
dash). From Ref.~\cite{Venkat:2010by}.}
 
\end{figure}

The basic idea behind our approach is to use the observation that
$\rho(b)\approx0$ for $b \geq R$, where $R$ is a finite distance.
Since the functions $\rho$ and $F$ are  Fourier transforms,  $F$ is band-limited.
We proceed in the spirit of the Nyquist-Shannon sampling theorem and expand the function $\rho$
as 
\begin{equation} 
\rho(b)=\sum_{n=1}^{\infty}\frac{1}{2\pi}\frac{2}{R^2 J_1(X_n)^2} F(Q_n^2)J_0\left(X_n\frac{b}{R}\right) \,, 
\label{fra}
\end{equation}
where $X_n$ is the $n$-th zero of the regular cylindrical Bessel function of order 0, $J_0$;
$Q_n  \equiv X_n/R$ 
and $X_n\approx (n+3/4)\pi$. Equation~(\ref{fra}) defines the so-called finite radius approximation 
(FRA). Using, for example, $R=3$ fm and $n=10$, $Q_n^2\approx 4$ GeV$^2$. Thus, 
the measurement up to $Q^2=4$ GeV$^2$ determines the first ten terms of the expansion. 
As an example, let us consider the expansion~(\ref{fra}) for  the dipole form factor: $F_D(Q^2)=1/(1+Q^2/\Lambda^2)^2$ with $\Lambda^2=0.71 $ GeV$^2$.
The results shown in figure~\ref{fig:dipole} indicate that relatively few terms suffice to give an accurate representation.

The relationship between  the FRA and the usual expansion into a complete set of functions is examined in~\cite{Venkat:2010by} where it is shown that the FRA is very accurate.
The available data set consists of $ep$ scattering up to 31 GeV$^2$ and $G_{E,M}$ are separately extracted for up to 10   GeV$^2$.
The form factors $G_E$ and $G_M$ have been extracted from a global analysis of
the world's cross section and polarization data, including corrections for
two-photon exchange corrections~\cite{Blunden:2005ew}. The analysis
is largely identical to that of~\cite{Arrington:2007ux},
although additional high $Q^2$ form factor results~\cite{Puckett:2010ac} have
been included.  In addition, the slopes of $G_E$ and $G_M$ at $Q^2=0$ were
constrained in the global fit based on a dedicated analysis of the low $Q^2$
data.  In writing $G_E(Q^2)=1-Q^2R_E^2/6$, the value of
$R_E$ was constrained to be 0.878~fm and $R_M$ was constrained to be 0.860~fm.
This is important in the extraction of the large scale structure of the
density.  The fit is given in~\cite{Venkat:2010by}.

\begin{figure}
\begin{center}
\includegraphics[width=0.5\textwidth]{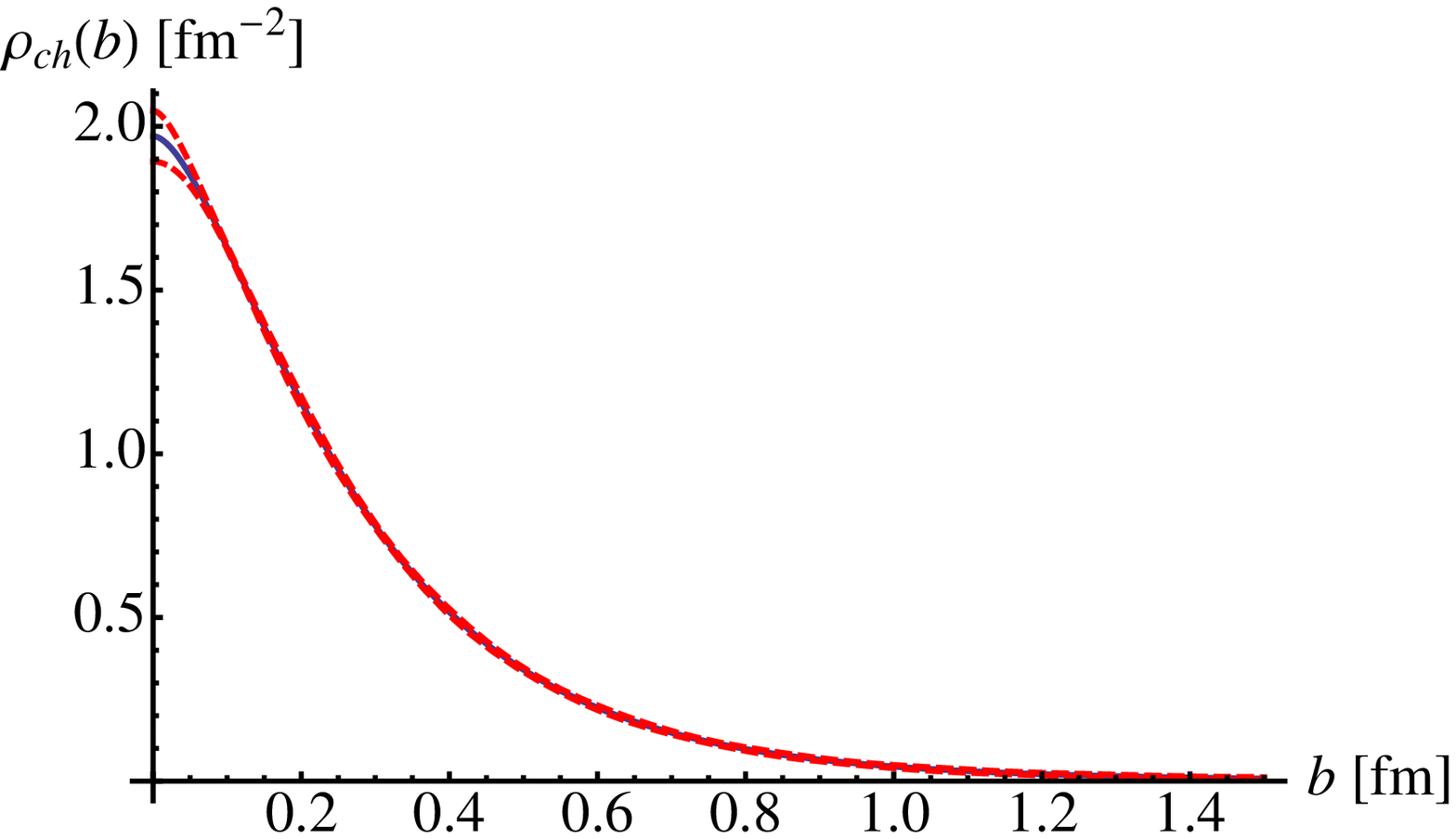}
\end{center}
\caption{\label{fig:rho}  \small (Color online) $\rho_{ch}$ (solid, blue) with error bands (short dashed, red). From Ref.~\cite{Venkat:2010by}. }
\end{figure}

We then use the fit and uncertainties for $G_E$ and $G_M$ to extract
$F_1$ and $F_2$, treating the uncertainties in $G_E$ and $G_M$ as uncorrelated,
yielding:
\begin{eqnarray}
&&(dF_1)^2=(\frac{1}{1+\tau})^2 (dG_E)^2+(\frac{\tau}{1+\tau})^2 (dG_M)^2 \label{dF1} \,,
\nonumber\\
&&(dF_2)^2=(\frac{1}{1+\tau})^2 (dG_E)^2+(\frac{1}{1+\tau})^2 (dG_M)^2 \label{dF2} \,.
\end{eqnarray}
For $Q^2<30$ GeV$^2$, we use $dF_1$ above in the FRA to get $d\rho(b)$. 
For $Q^2>30$ GeV$^2$, we use the FRA and take $dF_1=\pm|F_1({\rm fit})|$. This corresponds to a maximum value of $n=30$.    
The resulting transverse charge density is shown in figure~\ref{fig:rho}. The proton transverse charge density is now very well known.

Our FRA  technique can be exploited to image
other quantities that depend on the transverse position. Suppose there is a
transverse quantity $\rho^{(\lambda)}(b)$ that is a two-dimensional Fourier
transform of an experimental observable $F^{(\lambda)}(Q^2)$ such that 
\begin{equation}
\rho^{(\lambda)}(b)={1\over2\pi}\int QdQJ_\lambda(Qb)F^{(\lambda)}(Q^2) \,.
\end{equation}
An example, discussed in detail in~\cite{Venkat:2010by},  is the magnetization
density $\rho_M$.  The index
$(\lambda)$ is associated with a given number of units of the orbital angular
momentum. The extraction of $\rho^{(\lambda)}(b)$ is facilitated by using the
expansion 
\begin{equation} 
\rho^{(\lambda)}(b)=\sum_{n=1}^\infty
{2\over R^2
J_{\lambda+1}(X_{\lambda,n})^2}F^{(\lambda)}(Q_{\lambda,n}^2),J_\lambda\left(X_{\lambda,n}{b\over R}\right) \,,
\label{gen}
\end{equation} 
where
$X_{\lambda,n}$ is the $n$-th zero of the Bessel function of order $\lambda$;
 $Q_{\lambda,n}=X_{\lambda,n}/R$.
The result~(\ref{gen}) can be used
to relate accessible kinematic ranges with transverse regions.

\subsection{Summary}

Much data for form factors exist and JLab12 will further improve the data set. 
The  charge density is not a three-dimensional Fourier 
transform of $G_E$. One can  
interpret form factors as determining transverse 
charge and magnetization densities. The 
nucleon transverse densities are known now to high 
precision. 
The  new FRA technique 
can be used for other quantities that depend on transverse position, 
in particular, for the exclusive scattering amplitudes and generalized 
parton distributions discussed in this chapter.

\noindent
{\it Acknowledgments}.
I thank S.~Venkat, J.~Arrington, and X.~Zhan for their extensive efforts in producing the
paper~\cite{Venkat:2010by} on which this presentation is based.
I also wish to thank Jefferson Laboratory for its hospitality during a visit while this work was being completed. 

\section{From transverse-momentum spectra to transverse images}
\label{sec:bspace}


\hspace{\parindent}\parbox{0.92\textwidth}{\slshape 
   Elke-Caroline Aschenauer,
   Markus Diehl,
  Salvatore Fazio}
%

\index{Aschenauer, Elke}
\index{Diehl, Markus}
\index{Fazio, Salvatore}





\subsection{Imaging partons in the transverse plane}

The principle of ``parton imaging'' using exclusive processes such as DVCS
or hard exclusive meson production is rather simple.  The key variable to
measure is the transverse momentum transfer $\vec{\Delta}_T$ to the target
proton or nucleus in the $\gamma^{\ast}$-target c.m.  The invariant momentum
transfer is then given by
\begin{align}
  \label{bspace:t-Delta}
t &= - \frac{x^2 m^2 + \vec{\Delta}_T^{\, 2}}{1-x}  &
\text{with}~~~ x &= \frac{Q^2 + M_V^2}{Q^2 + W^2} \,,
\end{align}
where $m$ is the target mass and $M_V$ the mass of the produced meson.
For DVCS one should omit $M_V$, so that $x$ coincides with the Bjorken
variable.  In the limit of large $Q^2 + M_V^2$, the $\gamma^{\ast} p$ scattering
amplitude is a linear combination of generalized parton distributions
convoluted with hard-scattering kernels.  The distribution of partons in
the transverse plane is obtained by a Fourier transform w.r.t.\
$\vec{\Delta}_T$ \cite{Burkardt:2002hr,Diehl:2002he}.  In the simple case
where the unpolarized quark or gluon GPDs $H^i$ dominate the $\gamma^{\ast} p$
cross section $d\sigma/dt$, the impact parameter profile is
\begin{align}
  \label{bspace:master-formula}
F(b,x,Q^2) & \propto \frac{1}{(2\pi)^2} \int d^2\vec{\Delta}_T\,
         e^{-i \vec{b} \vec{\Delta}_T}\, \sqrt{\frac{d\sigma}{dt}}
       = \frac{1}{2\pi} \int_0^\infty d\Delta_T\,
             \Delta_T\, J_0(b \Delta_T)\,
         \sqrt{\frac{d\sigma}{dt}} \,,
\end{align}
where $\Delta_T = |\vec{\Delta}_T|$ and $b = |\vec{b}|$.  For simplicity
we drop the information from the absolute size of the cross section in
this contribution and focus our attention on the normalized $b$-space
profile, which satisfies $\int d^2 b\, F(b,x,Q^2) = 1$.  For polarization
asymmetries and for the interference term between DVCS and the
Bethe-Heitler process, the extraction of the relevant $\gamma^{\ast} p$
amplitudes is more involved, but the principle of Fourier transforming
these amplitudes w.r.t.\ $\vec{\Delta}_T$ remains the same.

In the present contribution, we estimate how accurately one can hope to
determine $F(b,x,Q^2)$ from cross section measurements for DVCS on the
proton.  Firstly, $d\sigma/dt$ will have statistical and systematic
errors.  Secondly, the range of $\Delta_T$ in a measurement will be
restricted both from above and from below, so that an extrapolation is
required in order to perform the Fourier integral in
\eqref{bspace:master-formula}.


\subsection{Acceptance in transverse momentum}

\begin{figure}
\centerline{\includegraphics[width=0.85\textwidth]{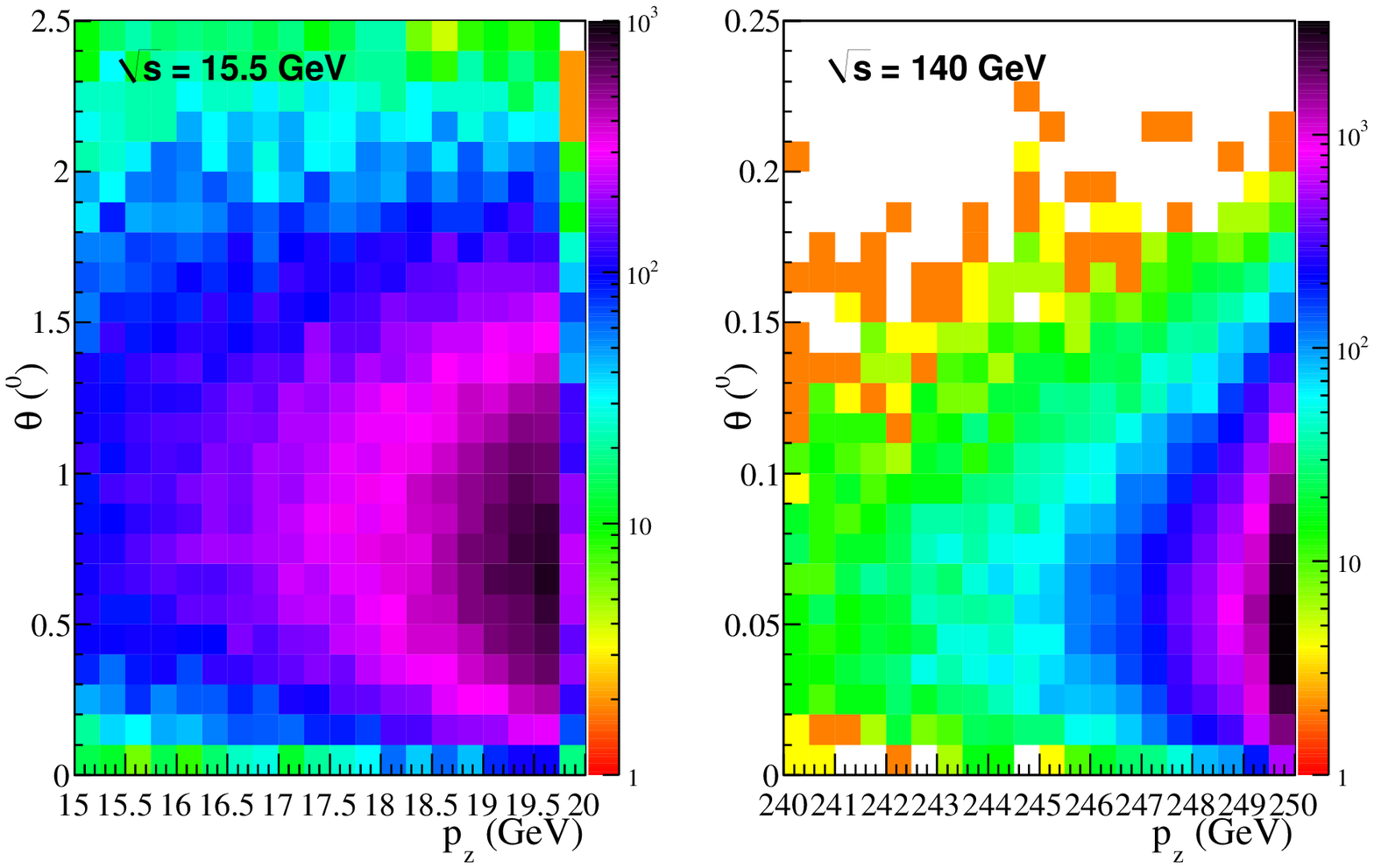}}
\caption{\label{plot:exclproton} \small (Color online) The longitudinal momentum
  $p_z$ of the scattered proton in exclusive reactions vs.\ its scattering
  angle $\theta$ for an $ep$ center-of-mass energy of $15.5 \gev$ (left)
  and $145 \gev$ (right).}
\end{figure}

To achieve the precision discussed below for imaging partons in the impact
parameter space, it is critical to integrate from the beginning the
detection of the scattered proton into the detector and interaction region
design.  The scattered proton in exclusive reactions is characterized by
carrying almost the full beam momentum and a transverse momentum
$\Delta_T$ between several $\mev$ and a few $\gev$, corresponding to very
small scattering angles.  Figure~\ref{plot:exclproton} shows the relation
between the longitudinal momentum of the protons and their scattering
angle for two different $ep$ center-of-mass energies.

The commonly used method to detect these protons is to integrate ``Roman
pots'' in the machine lattice. The standard technologies for such
detectors are silicon strip detectors or scintillating fiber detectors.
The acceptance for protons with the transverse momentum in the $\mev$ region
is limited by the requirement that Roman pots must have a beam clearance
distance of $10$ times the beam emittance.  The upper transverse
acceptance is given by the apertures of the magnets that the protons have
to transverse.  For transverse momenta above $1 \gev$, the proton can be
detected in the main solenoidal detector.  Details on the solutions for
the eRHIC and ELIC interaction region designs are given in section
\ref{sec:detector}.


\subsection{Precision of the measurement}

\begin{figure}
\centerline{\includegraphics[width=0.55\textwidth]{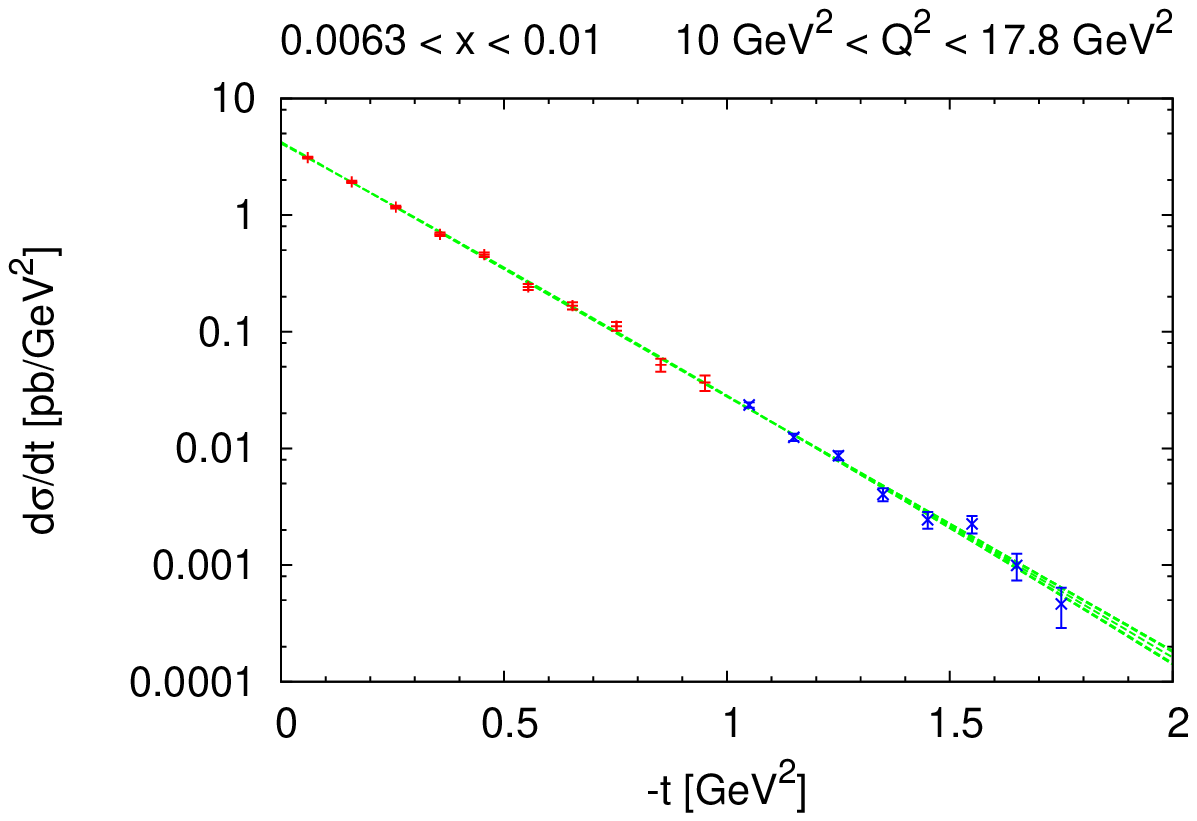}}
\caption{\label{bspace:Xsect} \small (Color online) Generated $t$ spectrum for
  the DVCS cross section in a selected bin of $x$ and $Q^2$.  The errors
are statistical only and correspond to the integrated
  luminosity of $11.9 \fb^{-1}$ for $|t| < 1 \gev^2$ and to $151 \fb^{-1}$
  for $|t| > 1 \gev^2$.  The curve represents a fit explained in the
  text.}
\end{figure}

A detailed simulation of DVCS events is described in
section~\ref{sec:fazio}.  To illustrate the expected statistical accuracy
of a measurement, we show $d\sigma/dt$ for a selected bin of $x$ and $Q^2$
in figure~\ref{bspace:Xsect}.  The value of $y$ in this bin ranges from
$0.05$ to $0.14$.  For bins with lower $x$ or lower $Q^2$, the statistical
errors are smaller, except for kinematics where the $y > 0.01$ cut applied
in the simulation becomes relevant.

The $t$ spectrum shown in the figure~\ref{bspace:Xsect} was generated with
an exponential dependence $d\sigma/dt \propto \exp( B t )$
with $B = 5 \gev^{-2}$.  An exponential fit to the generated spectrum 
gives $B = 5.02 \gev^{-2}$ with an error below $1\%$.  Data of this
quality also allows one to explore possible deviations from an exponential
spectrum.  To this end, we have also fitted to $d\sigma/dt \propto \exp( B
t - C t^2 )$.  This fit and its $1\sigma$ error band is shown in the
figure and gives $B = (4.92 \pm 0.10) \gev^{-2}$ and $C = (0.079 \pm
0.076) \gev^{-4}$.  Although the relative uncertainty on the extra
parameter $C$ is large, the term $C t^2$ in the exponential is small
compared with $B t$ in the fitted $t$ range (as it should be for a
spectrum generated with a pure exponential law).  The logarithmic $t$
slope at $|t| = 1.75 \gev^2$ in this fit is $(5.20 \pm 0.18) \gev^{-2}$.

We conclude at this point that with the projected luminosity available at
an EIC, the $t$ spectrum for the DVCS cross section will be dominated by
systematic uncertainties and not by statistics, even if one measures
differentially in $x$ and $Q^2$.  Systematic uncertainties, for instance
due to momentum resolution, strongly depend on details of the experimental
setup and have not been studied yet.  We note that the normalized $b$
space profile $F(b,x,Q^2)$ is not affected by errors on the overall
luminosity and acceptance.


\subsection{Uncertainty from the extrapolation in $t$}

We now estimate the uncertainty in the impact parameter profile $F(b)$ due
to the lack of knowledge of the scattering amplitude for all $t$.  Since
the projected statistical errors are so small, 
we do not include them in this exercise.

For the extrapolation to large $|t|$, we assume a measured $t$-spectrum
$d\sigma/dt \propto \exp( B t )$ with $B = 4 \gev^{-2}$ up to
$|t|_{\text{max}} = 1$ or $2 \gev^2$.  Larger values of $B$ give a smaller
cross section at high $|t|$ and thus a smaller extrapolation uncertainty
in the Fourier integral \eqref{bspace:master-formula}.  In turn, the
statistical errors on the cross section at high $|t|$ are then larger, so
that in $F(b)$ there is a tradeoff between the uncertainties from the
measured $t$ spectrum and those from its extrapolation.

\begin{figure}[p]
\begin{center}
\includegraphics[bb=50 63 410 300, height=11.3em]{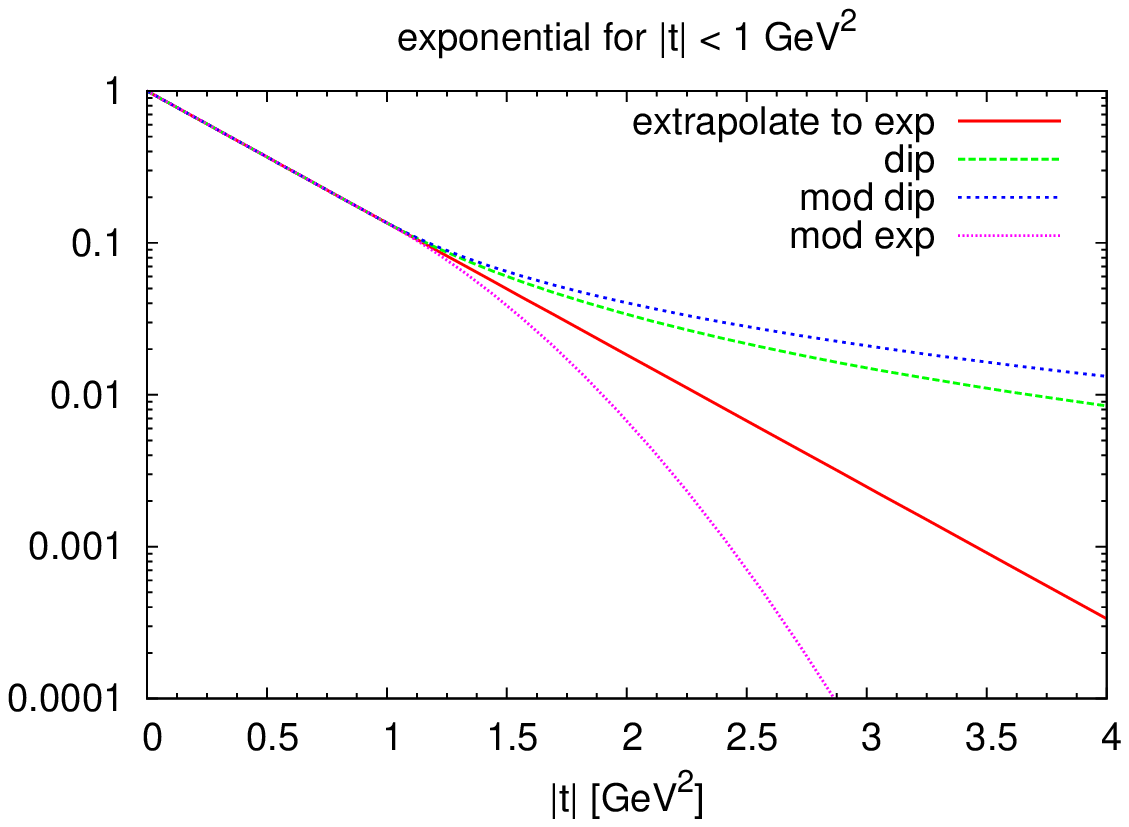}
\includegraphics[height=11.3em]{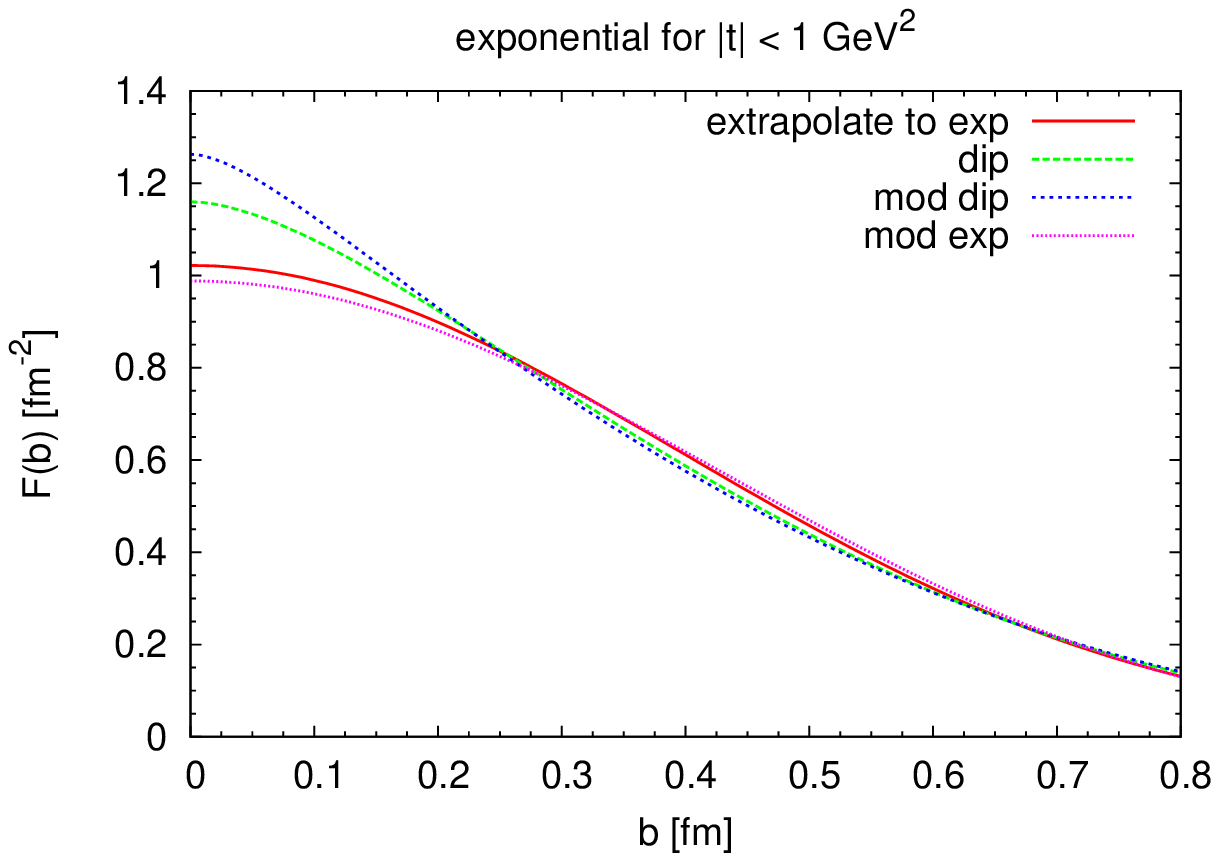} \\[0.5em]
\includegraphics[bb=50 63 410 300, height=11.3em]{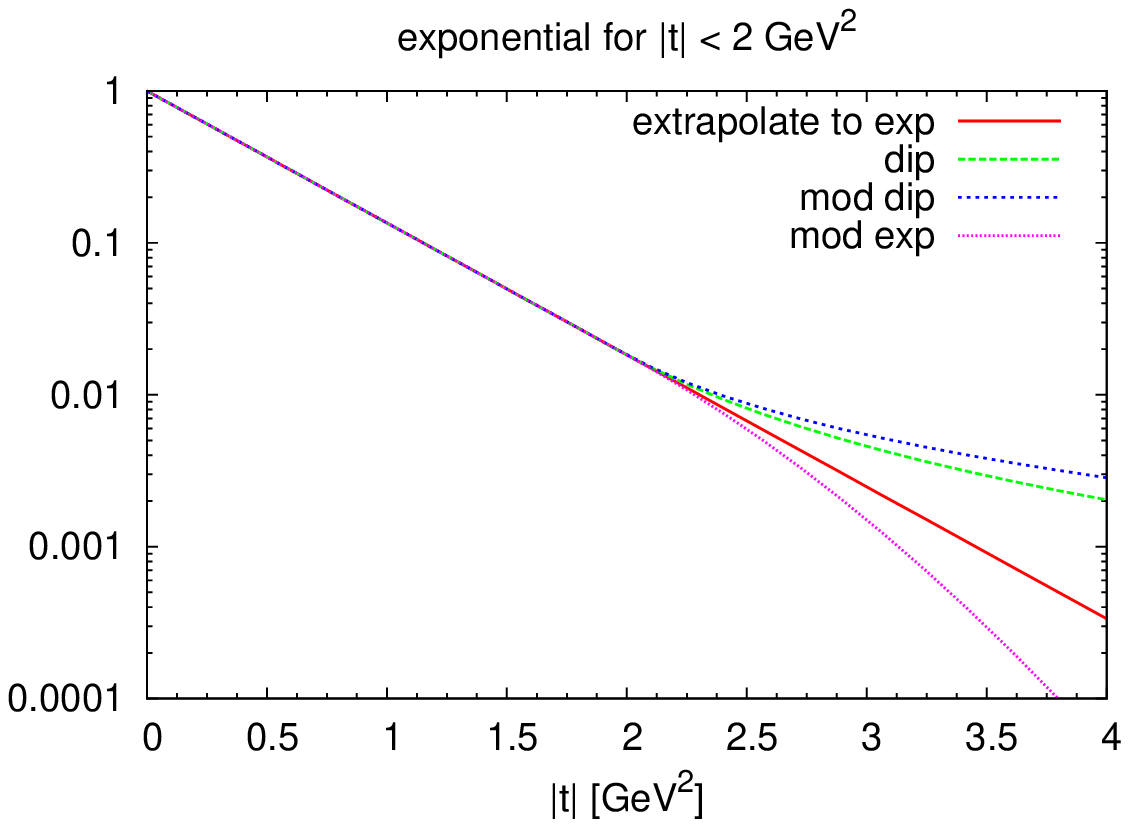}
\includegraphics[height=11.3em]{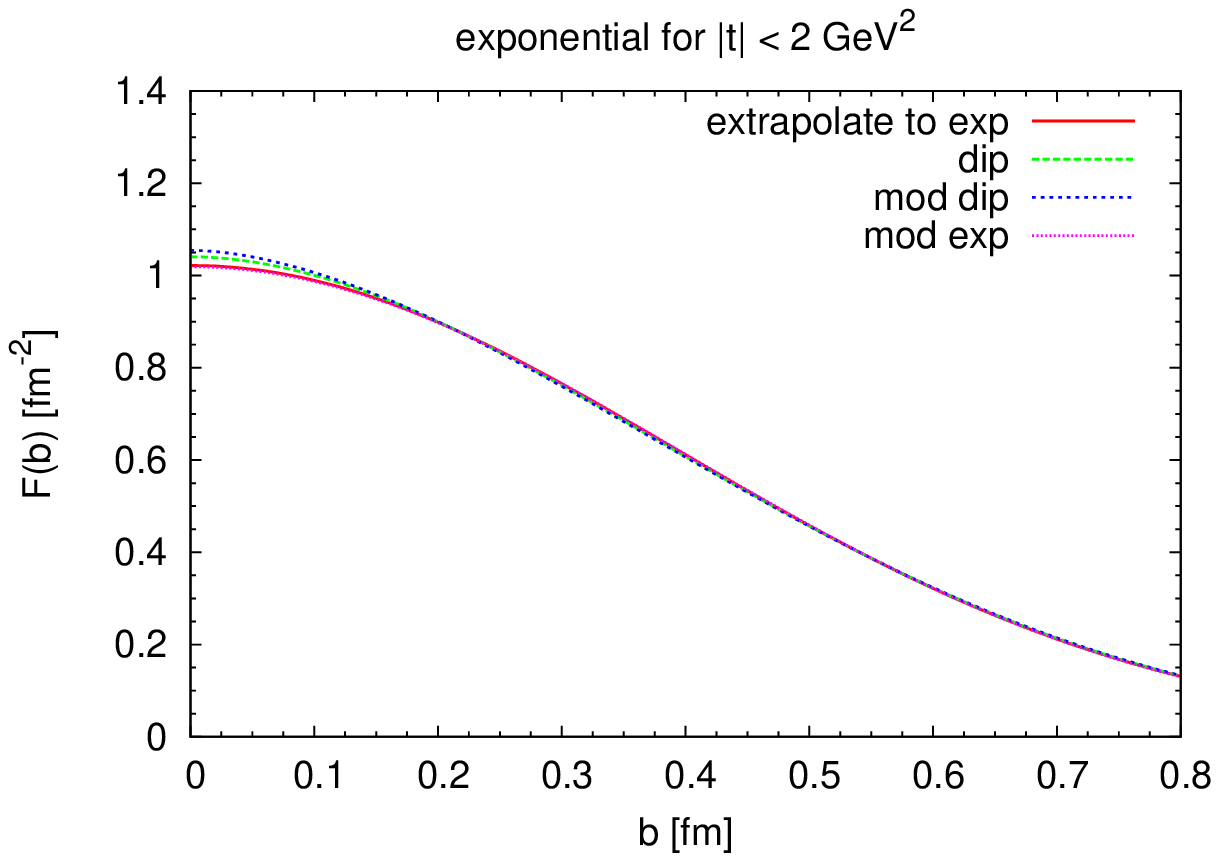}
\end{center}
\caption{\label{bspace:high-t-extrap} (Color online) Examples for
  normalized amplitudes (left) with different extrapolations to
  large~$|t|$, together with their Fourier transforms to impact parameter
  space (right).}
\end{figure}

To estimate the extrapolation uncertainty, we adopt a strategy similar to
that in \cite{Munier:2001nr} and assume different forms for the scattering
amplitude (i.e., for $\sqrt{d\sigma/dt}$) at $|t| > |t|_{\text{max} \,}$:
\begin{enumerate}
\item an exponential  $\propto \exp( B t /2)$, labeled ``exp'' in figure
  \ref{bspace:high-t-extrap},
\item a dipole form $\smash{\propto \bigl( 1 + |t| /M^2 \bigr)^{-2}}$,
  labeled ``dip'',
\item a modified dipole form $\smash{\propto \bigl( 1 + 0.05\, |t| /M^2
    \bigr)^{-1} \bigl( 1 + 0.45\, |t| /M^2 \bigr)^{-1}}$, labeled ``mod
  dip'',
\item a modified exponential $\propto \exp(- D t^2)$, labeled ``mod exp''.
\end{enumerate}
In each case we require the amplitude and its first derivative to be
continuous at $|t| = |t|_{\text{max}}$.  Note that in the measured $t$
region, forms 2 to 4 would give unacceptable fits to the simulated
spectrum in figure~\ref{bspace:Xsect}.  Forms 3 and 4 should be regarded
as examples for functions falling off especially slowly or especially fast
and do not claim to be particularly realistic.  When performing the
Fourier transform \eqref{bspace:master-formula}, we neglect the term $x^2
m^2$ in \eqref{bspace:t-Delta}, which is justified in a large region of
phase space.

In figure~\ref{bspace:high-t-extrap} we show the resulting scattering
amplitude (normalized to unity at $\vec{\Delta}_T = \vec{0}$) and its
Fourier transform $F(b)$.  We observe that the curves in $b$ space are
close together in a wide region and rather quickly start to differ below a
certain critical value $b_{\text{cr}}$.  For $|t|_{\text{max}} = 1 \gev^2$,
we find $b_{\text{cr}} \sim 0.25\fm$ and an appreciable spread of $F(b)$
at lower $b$.  This would be a serious limitation for studying the central
region of the proton.  Interesting physical effects like the variation of
$F(b)$ with $x$ or $Q^2$ are typically expected to be only logarithmic
(see, e.g., the estimates in \cite{Diehl:2007zu}) and hence require
sufficiently precise measurements.  Clearly, there is a very significant
gain of accuracy in impact parameter space if $|t|_{\text{max}}$ can be
raised from $1$ to $2 \gev^2$, i.e., if a scattered proton in the
corresponding kinematics can be seen in the main detector.  We then find
$b_{\text{cr}} \sim 0.1\fm$ and a small uncertainty even at $b=0$.

\begin{figure}[p]
\begin{center}
\includegraphics[bb=50 63 410 300, height=11.3em]{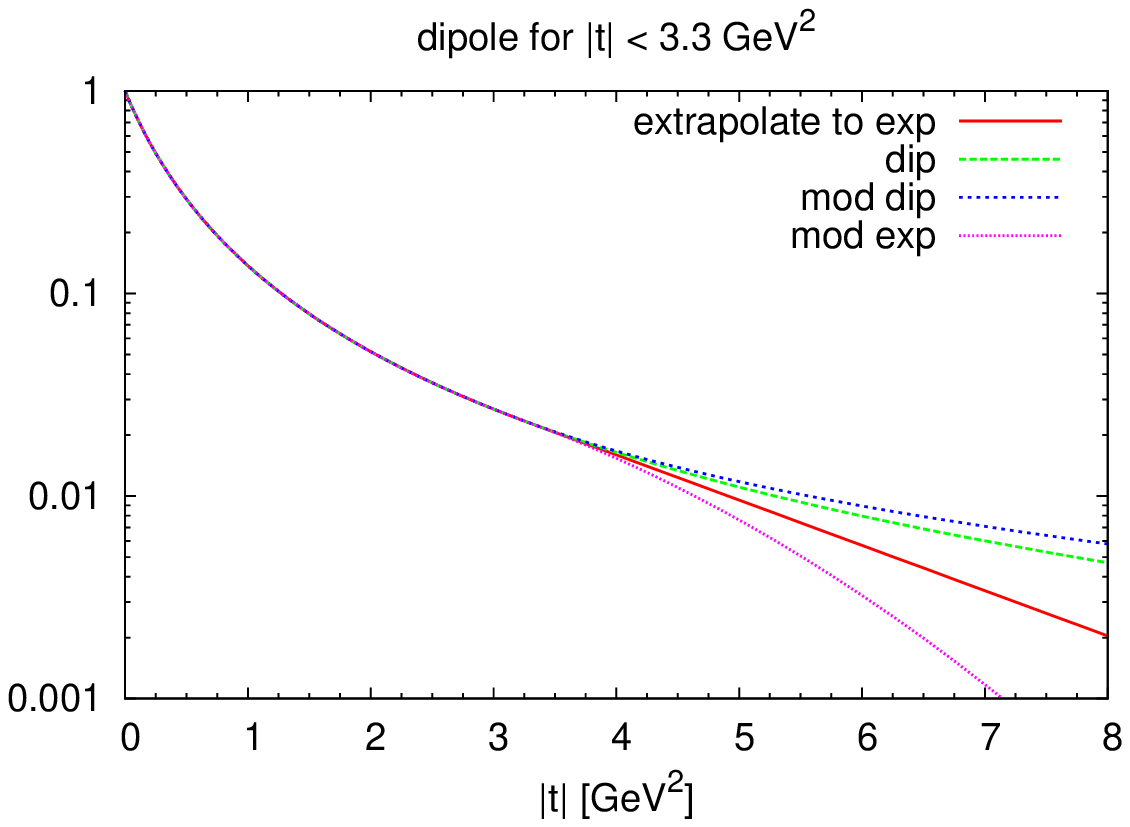}
\includegraphics[height=11.3em]{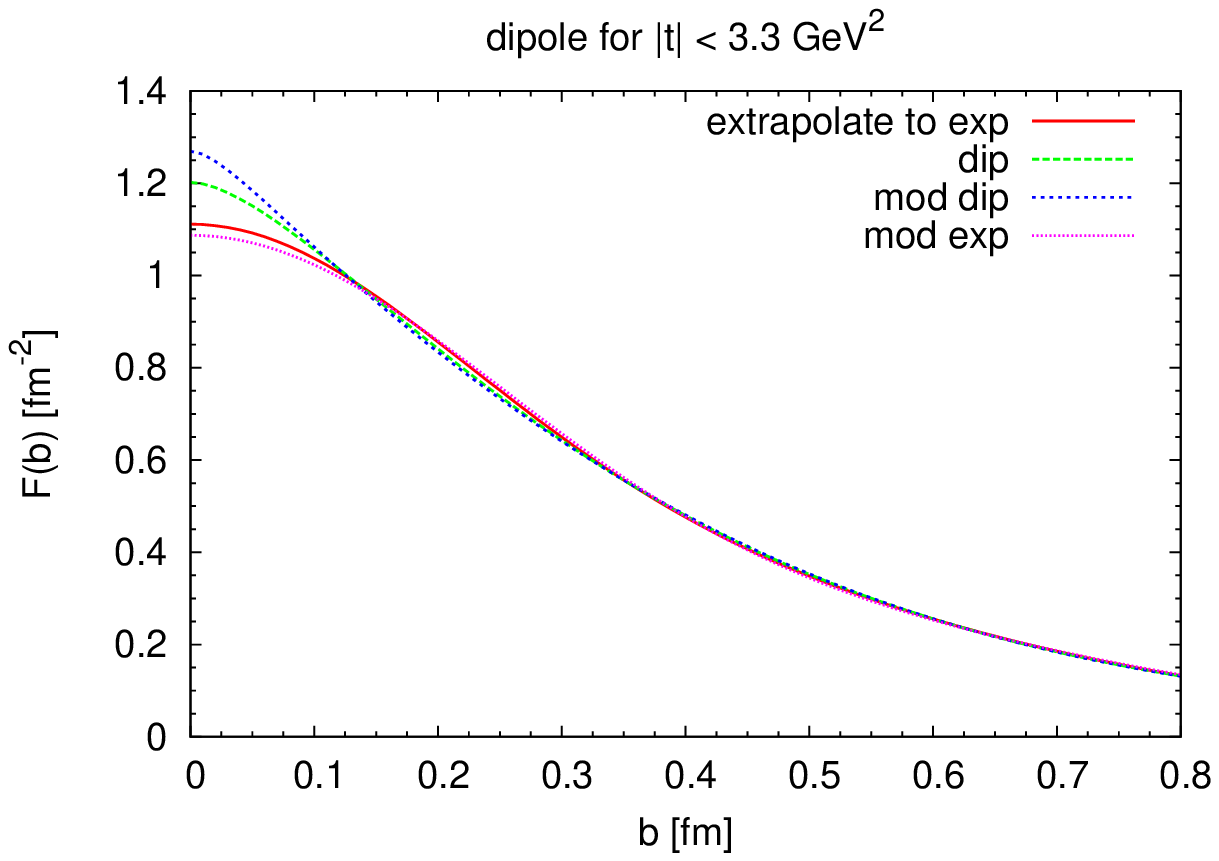}
\end{center}
\caption{\label{bspace:high-t-dip} (Color online) As in
  figure~\protect\ref{bspace:high-t-extrap} but with a dipole form of the
  amplitude up to $|t| = 3.3 \gev^2$.}
\end{figure}

As an alternative scenario we assume a dipole form instead of an
exponential $t$ dependence in the measured region,\footnote{We note that
  the measurement of $\jpsi$ photoproduction at HERA
  \protect\cite{Aktas:2005xu,Chekanov:2004mw} strongly favors an
  exponential $t$ dependence at $|t|$ below $1 \gev^2$, but the
  behavior of exclusive hard scattering cross sections at larger $t$ is
  poorly known.  For a conservative error estimate, we do not want to rule
  out a dipole behavior at $|t| > 1 \gev^2$.}
with a dipole mass $M = 770\mev$ that gives the same scattering amplitude
at $|t| = 1 \gev^2$ as the exponential with $B= 4 \gev^{-2}$.  The
extrapolation error is larger in the dipole scenario, but since the cross
section decreases much more slowly, it can be measured out to higher
values of $|t|$ before statistics becomes an issue.  We recall however
that a description in terms of generalized parton distributions requires
$|t| \ll Q^2 + M_V^2$.  As seen in figure~\ref{bspace:high-t-dip}, a
measurement up to $|t|_{\text{max}} = 3.3 \gev^2$ in the dipole scenario
gives a very precise $F(b)$ down to $b_{\text{cr}} \sim 0.1\fm$.  The
extrapolation uncertainty at lower $b$ is larger than for
$|t|_{\text{max}} = 2 \gev^2$ in the exponential scenario.

\begin{figure}
\begin{center}
\includegraphics[bb=50 63 410 300, height=11.3em]{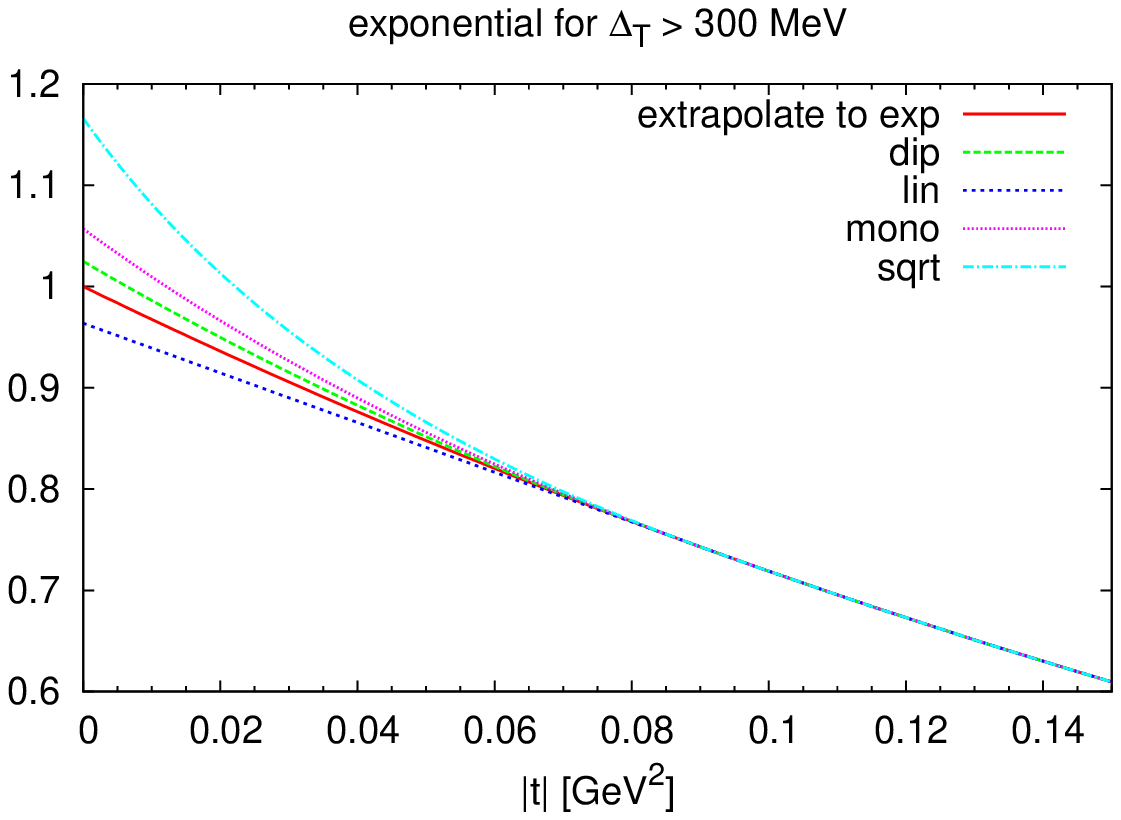}
\includegraphics[height=11.3em]{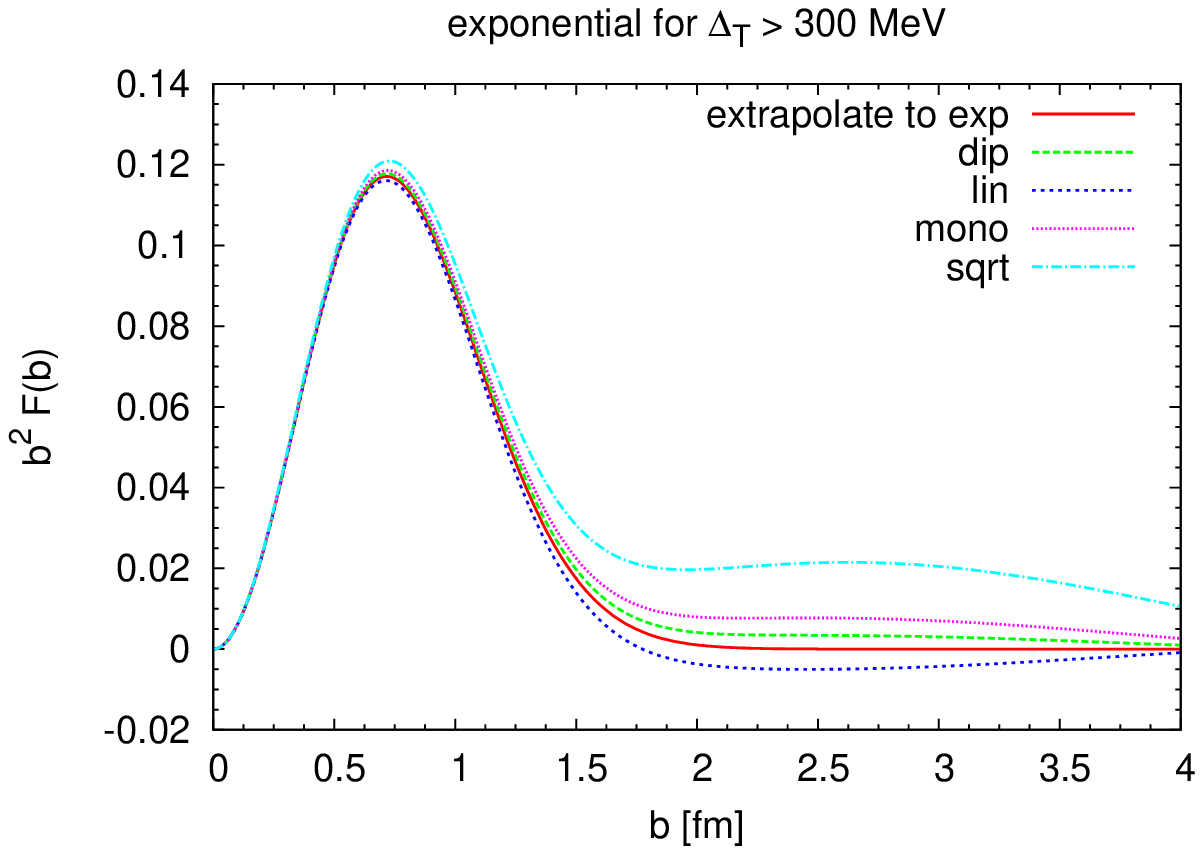} \\[0.5em]
\includegraphics[bb=50 63 410 300, height=11.3em]{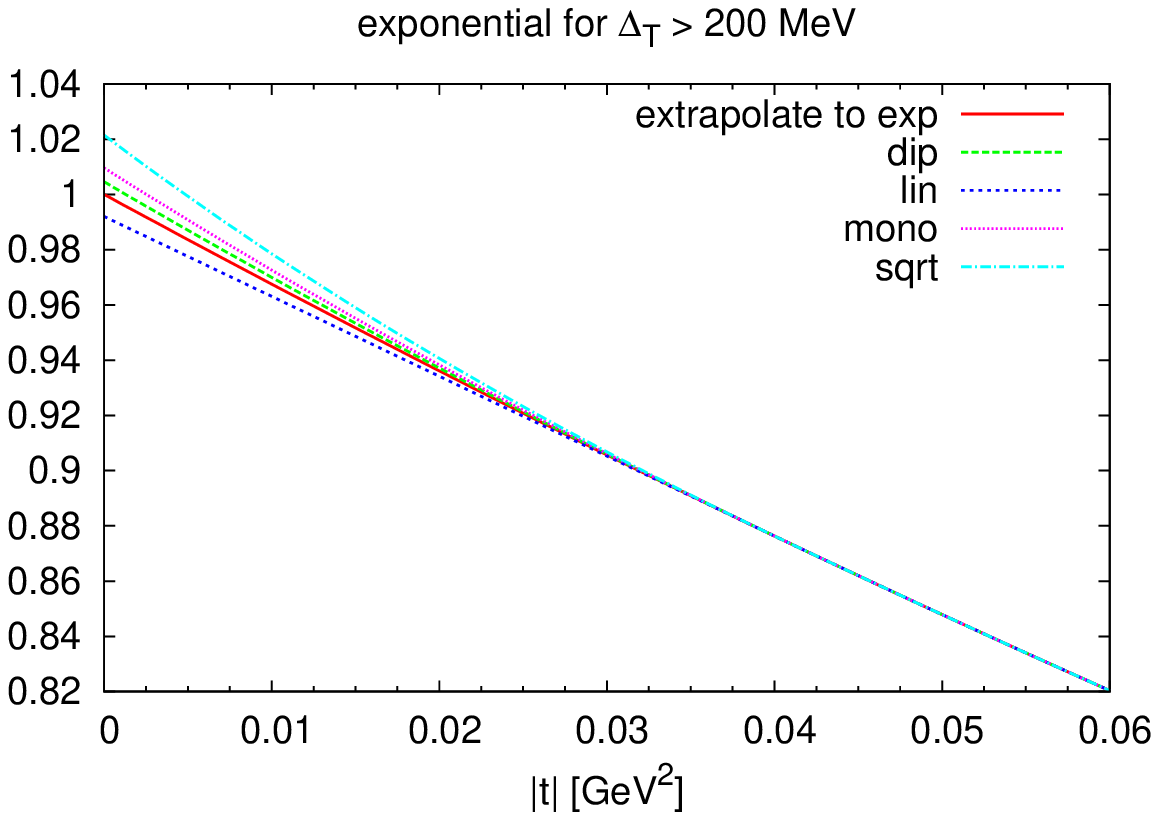}
\includegraphics[height=11.3em]{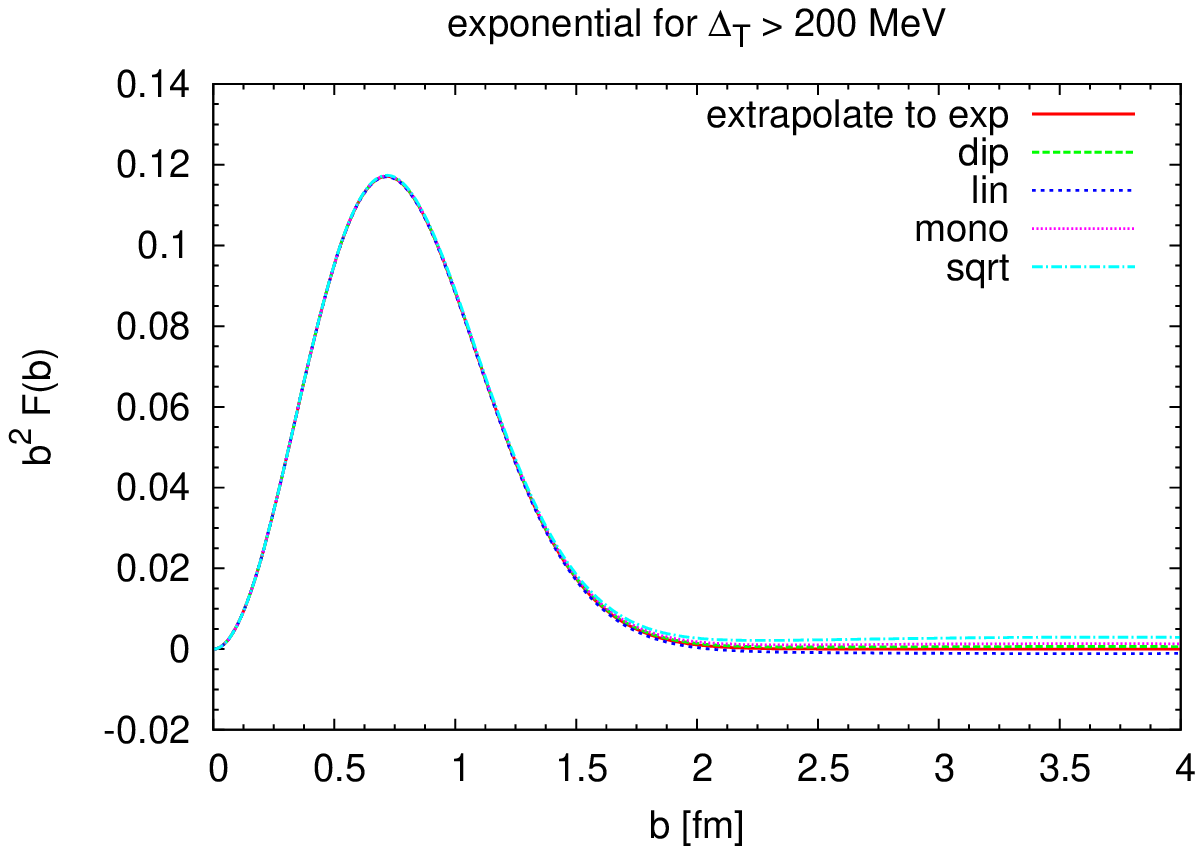}
\end{center}
\caption{\label{bspace:low-t-extrap} \small  (Color online) As in
  figure~\protect\ref{bspace:high-t-extrap}, but with extrapolation to
  small $|t|$, i.e.\ small $\Delta_T$.  The impact parameter profile
  $F(b)$ is multiplied with $b^2$ in order to make the large $b$ behavior
  visible.}
\end{figure}

Let us now investigate the extrapolation to small $|t|$.  We assume again
an exponential cross section $d\sigma/dt \propto \exp( B t )$, but now
with a larger slope $B = 6.6 \gev^{-2}$ in order to maximize the
importance of low $|t|$ in the Fourier integral.  We consider either $300
\mev$ or $200 \mev$ as minimum measured values of $\Delta_T$, and take the
following extrapolations for $\Delta_T$ down to zero:
\begin{enumerate}
\item an exponential in $t$, labeled ``exp'' in figure
  \ref{bspace:low-t-extrap},
\item a dipole form $\smash{\propto \bigl( 1 + |t| /M^2 \bigr)^{-2}}$,
  labeled ``dip'',
\item a linear function in $t$, labeled ``lin'',
\item a monopole form $\smash{\propto \bigl( 1 + |t| /M^2 \bigr)^{-1}}$,
  labeled ``mono'',
\item an inverse square root $\smash{\propto \bigl( 1 + |t| /M^2
    \bigr)^{-1/2}}$, labeled ``sqrt''.
\end{enumerate}
We see in figure \ref{bspace:low-t-extrap} that with a measurement down to
$\Delta_T = 300 \mev$, one has a rapidly growing extrapolation uncertainty
for $b$ above about $1.25 \fm$.  The situation dramatically improves if
one has to extrapolate only below $\Delta_T = 200 \mev$.  Repeating this
study with a dipole form in the measured region yields the same conclusion
\cite{Diehl:talk-11-1}.  Whether a measurement down to even lower
$\Delta_T$ can still improve the accuracy of $b$ space images can only be
decided after an estimate of experimental uncertainties.

Let us recall the specific physics interest of the impact parameter
profile of the proton at very large $b$.  This is the region where the
dynamics of chiral symmetry breaking should manifest itself.  A
description in terms of virtual pion fluctuations yields definite
predictions, such as a behavior $F(b) \propto b^{-1} e^{-\kappa b}$ with
$\kappa \approx 2 m_\pi \approx (0.7 \fm)^{-1}$ at large $b$
\cite{Strikman:2003gz}.  This translates into a small $|t|$ behavior given
by the inverse square root law in point 5 (with $M^2 = \kappa^2$).  These
predictions should be tested quantitatively.

In summary, we find that with the parameters we have assumed, neither
statistics nor acceptance in $t$ will seriously limit $b$ space imaging at
an EIC, with an accessible $b$ range from $0.1\fm$ up to $1.5\fm$ or
larger.  Detailed estimates of experimental uncertainties will be
necessary to assess the limiting factors of accuracy in this endeavor.


\section{GPDs from DVCS}
\label{sec:Burkardt}


\hspace{\parindent}\parbox{0.92\textwidth}{\slshape 
  Matthias Burkardt, Hikmat BC}
%

\index{Burkardt, Matthias}
\index{BC, Hikmat}





\subsection{Introduction}

GPDs are linked to many processes and observables
involving hadrons~\cite{Burkardt:2008jw}, but their most intuitive application is in the
context of Ji's angular momentum decomposition (see section~\ref{sec:gpd-e})
 and in 
three-dimensional imaging (see section~\ref{sec:bspace}).
Both involve GPDs in the $\xi=0$ limit (the $\xi$-dependence drops out in the Ji sum rule).
At the same time, the DVCS amplitude ${\cal A}_{\rm DVCS}$ 
provides direct access only to GPDs along the ''diagonal'' $x=\xi$
(through the imaginary part of the DVCS amplitude)
as well as to a convolution integral involving GPDs 
(through the real part of ${\cal A}_{\rm DVCS}$). In the leading order (LO)
factorization, one finds
\begin{eqnarray}
\Im m \,{\cal A}_{\rm DVCS} &\longrightarrow GPD^{(+)}(\xi,\xi,t) \,, \nonumber\\
\Re e\, {\cal A}_{\rm DVCS} &\longrightarrow \int_{-1}^1 dx
\frac{GPD^{(+)}(x,\xi,t)}{x-\xi} \,.
\label{eq:DVCS}
\end{eqnarray}
The '$(+)$' superscript in \eqref{eq:DVCS} emphasizes
that DVCS is sensitive only charge-even (i.e., quark+antiquark) combinations of GPDs. Moreover, the accessible 
range in $\xi$ is limited,
$\xi_{\rm min}<\xi<\xi_{\rm max}$.
The lower limit $\xi_{\rm min}$ is defined by the DIS kinematics. The upper limit $\xi_{\rm max}$
follows from the relation
\begin{equation}
-t = \frac{4\xi^2M^2+{\bf \Delta}_\perp^2}{1-\xi^2} \,,
\end{equation}
and the positivity of ${\bf \Delta}_\perp^2$.
Thus, even in an idealized DVCS experiment
(fixed $Q^2$), where angular dependencies as well as spin asymmetries have been used to disentangle
 different GPDs and the proton and 'neutron' targets have been used to accomplish
the flavor decomposition, one can at best expect a determination
of the observables in~\eqref{eq:DVCS} for
$\xi_{min}<\xi<\xi_{max}$.
One of the key question in the context of DVCS is whether this 
information will allow an unambiguous and model-independent extraction of GPDs.

\subsection{Constraints on GPDs: polynomiality, dispersion relations and QCD evolution}

GPDs are not only constrained by DVCS, but
also by DIS and form factor data. However, the form factor
data constrains only charge-odd distributions                                                                   
and helps only in kinematical regimes where antiquark contributions
are negligible. While DIS data is sensitive to charge-even
distributions, there is no DIS data that would constrain
the forward ($\xi=0$, $t=0$) limit of $E^q(x,\xi,t)$.

Fortunately, multiple theoretical constraints exist that
will be helpful in determining GPDs from DVCS data.
For example, Lorentz invariance implies the polynomiality
conditions on GPDs~\cite{Ji:1996ek,Goeke:2001tz}:
\begin{equation}
\int_{-1}^1 dx\, x^n GPD(x,\xi,t) = A_{n,0}(t) + A_{n,2}(t)\xi^2+...
+ A_{n,n+1}\xi^{n+1} \,,
\end{equation}
where the highest power $\xi^{n+1}$ is only present when $n$ is odd. 
These polynomiality conditions imply that the dependence of
GPDs on the variables $x$ and $\xi$ cannot be independent. This imposes
significant and rigorous constraints on any GPD extraction from DVCS data.

Rigorous dispersion relations exist for the DVCS amplitude ${\cal A}(\nu,t,Q^2)$:
\begin{equation}
\Re e\, {\cal A}(\nu,t,Q^2) = \frac{\nu^2}{\pi} \int_0^\infty
\frac{d\nu^{\prime 2}}{\nu^{\prime 2}} 
\frac{\Im m \,{\cal A}(\nu^\prime,t,Q^2)}{\nu^{\prime 2} - \nu^2}
+ \Delta(t,Q^2)\,,
\end{equation}
where $\Delta(t,Q^2)$ is a possible subtraction that can be identified with the
$D$-form factor~\cite{Polyakov:1999gs}.
In combination with the leading order (LO) factorization
\eqref{eq:DVCS}, this implies for GPDs
\begin{equation}
\Re e \,{\cal A}(\xi,t,Q^2) \sim \int_{-1}^1 dx \frac{GPD^{(+)}(x,\xi,t,Q^2)}{x-\xi} = 
\int_{-1}^1 dx \frac{GPD^{(+)}(x,x,t,Q^2)}{x-\xi}+
\Delta(t,Q^2) \,.
\label{eq:DR}
\end{equation}
Although its derivation from dispersion relations is more physical, \eqref{eq:DR} was 
first derived from polynomiality \cite{Goeke:2001tz}.

One of the consequences of \eqref{eq:DR} is that it allows to '``condense''
the information from the DVCS amplitude (including the real part) into GPDs along the 
diagonal
$x=\xi$ plus the D-form factor.
However, it should be emphasized that this does {\em not} render measurements of the
real part of the DVCS amplitude redundant. Indeed, measurements of $\Im m \,{\cal A}_{\rm DVCS}$
for a given beam energy do not cover the whole region $0<\xi<1$ that enters~(\ref{eq:DR}). 
This implies that one can, for example, use 
$\Re e\,{\cal A}(\xi,t,Q^2)$ at fixed $Q^2$ to constrain $GPD(\xi,\xi,t,Q^2)$ for the 
values of $\xi$ that
are not accessible directly through the measurement of $\Im m\,{\cal A}(\xi,t,Q^2)$.
In summary, a DVCS experiment at fixed $Q^2$ (large enough for GPD factorization to hold)
should in principle allow for the determination of GPDs
along the diagonal $x=\xi$ as well as the D-form factor, which  (through the polynomiality condition) 
impose some constraints on GPDs for $x\neq \xi$.


Additional important constraints on GPDs come from their QCD evolution.
The $Q^2$ evolution equations can be ''diagonalized'' by expanding
GPDs in terms of Gegenbauer polynomials $C_n^{3/2}(x)$:
\begin{equation}
GPD(x,\xi,t,Q^2) = (1-x^2)\sum_{n=0}^\infty C_n^{3/2}(x)
\sum_{m=0 (even)}^n a_{nm}(\xi) {\cal C}_{n-m}(\xi,t,Q^2) \,,
\label{eq:master1}
\end{equation}
where $a_{nm}(\xi)$ are known polynomials. The
coefficients ${\cal C}_{k}(\xi,t,Q^2)$  are {\sl a priori}
unknown, but their $Q^2$ evolution is known.
This allows one (in principle) to determine ${\cal C}_{k}(\xi,t,Q^2)$
model independently. For this purpose, let us consider
$x=\xi$, where GPDs can be measured directly.
Upon relabeling $k=n-m$, \eqref{eq:master1} reads
\begin{equation}
GPD(\xi,\xi,t,Q^2) = (1-\xi^2) \sum_{k=0}^\infty
{\cal C}_k(\xi,t,Q^2) f_k(\xi) \,,
\label{eq:master2}
\end{equation}
where $f_k(\xi)=\sum_{m=0(even)}^\infty a_{m+k,m}(\xi)
C^{3/2}_{m+k}(\xi)$ are known functions.
For any fixed $\xi$, each term in \eqref{eq:master2} 
evolves differently
and, thus,  a measurement over a wide range of $Q^2$
should allow for the  determination of ${\cal C}_k(\xi,t,Q^2)$
as well as 
the GPDs for $x\neq \xi$ [via  \eqref{eq:master1}].
At an EIC with its wide $Q^2$ range and high luminosity, 
it may be possible for the first time to carry out
a model-independent extraction of GPDs. More detailed numerical 
studies will be required to quantify this expectation.



\section{Accessing GPDs from experiment: potential of a high-luminosity EIC}
\label{sec:Mueller}


\hspace{\parindent}\parbox{0.92\textwidth}{\slshape
 K.~Kumeri\v{c}ki, T.~Lautenschlager, D.~M\"uller,
 K.~Passek-Kumeri\v{c}ki, A.~Sch\"{a}fer, M.~Me\v{s}kauskas}
%

\index{Kumeri\v{c}ki, Kre\v{s}imir}
\index{Lautenschlager, Tobias}
\index{M\"uller, Dieter}
\index{Passek-Kumeri\v{c}ki, Kornelija}
\index{Sch\"afer, Andreas}
\index{Me\v{s}kauskas, Mantas}





\subsection{Introduction}

Generalized parton distributions (GPDs)~\cite{Mueller:1998fv,Ji:1996nm,Radyushkin:1996nd} have received much attention from both the theoretical and experimental sides. 
This was triggered by the hope to solve the ``spin puzzle'' that refers to 
the mismatch between the quark contribution to the proton spin extracted from polarized 
DIS and the one given by the constituent quark model.
We view the ``spin puzzle'' first and foremost
as a quest to quantify the partonic structure of the nucleon
in terms of quark and gluon angular momenta~\cite{Ji:1996ek}. 
Furthermore, it has been realized that GPDs allow for a three-dimensional
imaging of nucleons and nuclei~\cite{Ralston:2001xs}, providing, in the
zero-skewness case ($\xi=0$), a probabilistic interpretation in terms of partonic degrees
of freedom~\cite{Burkardt:2000za}. 
In fact, GPDs build up a
whole framework for description of hadron
structure~\cite{Diehl:2003ny,Belitsky:2005qn}, with the  ``spin puzzle'' being
just one interesting aspect.

In phenomenology, GPDs are used for modeling elastic form factors and 
the description of hard exclusive leptoproduction and even photoproduction. 
For hard exclusive processes, factorization theorems have been proven 
in the collinear framework at twist-two level~\cite{Collins:1996fb,Collins:1998be}. 
In the last decade, various hard exclusive processes have been measured
by the H1 and ZEUS collaborations (DESY) in the small $x_{B}$
region and by HERMES (DESY), CLAS (JLAB), and Hall
A (JLAB) in the moderate $x_{B}$ region in the fixed-target experiments.

Deeply virtual Compton scattering (DVCS) off nucleon is considered
as the theoretically cleanest process offering access to GPDs.
Its amplitude can  be parameterized by twelve Compton
form factors (CFFs)~\cite{Belitsky:2001ns}, which are given in terms of twist-two
(including gluon transversity) and twist-three GPDs.
For instance, at leading order (LO), parity-even twist-two CFFs, ${\cal H}$ and
${\cal E}$,
can be expressed through quark GPDs $H$ and $E$:
\begin{eqnarray}
\label{KLMSPM-DVCS-HS}
\left\{ {\cal H}  \atop {\cal E} \right\}(x_{B},t,{\cal Q}^2)  \stackrel{\rm
LO}{=}  \int_{-1}^1\!dx\, \frac{2x}{\xi^2-x^2- i \epsilon}
 \left\{ H \atop E\right\}(x,\eta=\xi,t,{\cal Q}^2)\,,
\end{eqnarray}
where both quark and anti-quark GPDs  are  defined in the
region  $ x \in [ -\xi,1 ]$; $x_{B}=2\xi/(1+\xi)$. 
Similar expressions can be written for twist-two parity-odd CFFs
$\widetilde{\cal H}$ and $\widetilde{\cal E}$, while
for other CFFs they are a bit more intricate \cite{Belitsky:2001ns}.
Analogous formulae hold for the LO description of
$\gamma^{\ast} N \to M N$ transition form factors (TFFs),
measurable in deeply virtual electroproduction of mesons (DVEM).
Here, in addition to GPDs, the non-perturbative meson distribution amplitude
enters, which describes the transition of a quark-antiquark state into the
final meson. This induces an additional uncertainty in the GPD phenomenology.

Let us briefly clarify which GPD information can be extracted from experimental measurements.
Neglecting radiative and higher twist-contributions, one might view the GPD on the $\eta=x$ cross-over line  as a ``spectral function", which provides also the real part of the CFF via the ``dispersion relation"~\cite{Teryaev:2005uj,Kumericki:2007sa,Diehl:2007jb,Kumericki:2008di}:
\begin{eqnarray}
\label{KLMSPM-DR-Im}
\Im{\rm m}  {\cal F}(x_{B},t,{\cal Q}^2)  & \stackrel{\rm LO}{=} & \pi  F (\xi,\xi,t,{\cal Q}^2)\,, \quad F= \{H, E, \widetilde H, \widetilde E\}\,,
\\
\label{KLMSPM-DR-Re}
\Re{\rm e}  \!
\left\{\! {\cal H}  \atop {\cal E} \!\right\}\!(x_{B},t,{\cal Q}^2) & \stackrel{\rm LO}{=} &
{\rm PV}\! \int_{0}^1\!dx\, \frac{2x}{\xi^2-x^2}\!  \left\{\! H \atop E \! \right\}\! (x,x,t,{\cal Q}^2)
\pm {\cal D}(t,{\cal Q}^2).
\end{eqnarray}
The GPD support properties ensure that (\ref{KLMSPM-DR-Im}) and (\ref{KLMSPM-DR-Re}) are in {\em one-to-one} correspondence to the perturbative formula (\ref{KLMSPM-DVCS-HS}), where  the subtraction constant $\cal D$, which is related in a specific GPD representation to the so-called
$D$-term \cite{Polyakov:1999gs}, can be calculated from either $H$ or $E$.   
However, we note that the ``dispersion relation" (\ref{KLMSPM-DR-Re}) is given in terms of 
partonic variables and compared to the dispersion relation formulated in physical variables
it differs by power suppressed contributions.
To pin down the GPD in the outer region $y \ge  \eta=x$, one might employ evolution. 
For instance, in the non-singlet case, the change of the GPD on the cross-over line is governed by (the equation in the whole outer region is needed)
\begin{eqnarray}
\label{KLMSPM-evolution}
\mu^2 \frac{d}{d\mu^2} F(x,x,t,\mu^2) =
 \int_x^1 \frac{dy}{x} V(1,y/x, \alpha_s(\mu)) F(y,x,t,\mu^2)\,,
\end{eqnarray}
where $V$ is the evolution kernel~\cite{Mueller:1998fv}.
Unfortunately, a large enough ${\cal Q}^2$ range is not available in fixed target experiments. Hence, we must conclude that in such measurements, essentially only the GPD on the cross-over line [thanks to (\ref{KLMSPM-DR-Re}), also outside of the experimentally accessible part of this line~\cite{Kumericki:2008di}] and the subtraction constant $\cal D$ can be accessed. Moments, such as those entering the spin sum rule, can only be obtained from a GPD model, fitted to data, or more generally with help of some ``holographic'' mapping~\cite{Kumericki:2008di}:
\begin{eqnarray}
\label{KLMSPM-hol-pro}
\left\{F(x, \eta=0, t,Q^2),\, F(x, \eta=x, t,Q^2)\right\}  \quad \Longrightarrow \quad   F(x, \eta, t,Q^2)  \,.
\end{eqnarray}
Here, $F(x,\eta=0,t,{\cal Q}^2)$ are constrained from form factor measurements and, additionally, GPDs ${\widetilde H}$  ($H$) by
(un)polarized phenomenological PDFs. Of course, a given `holographic' mapping  holds only for a specific class of GPD models.

\subsection{GPD modeling}

The implementation of radiative corrections, even including LO evolution (\ref{KLMSPM-evolution}), requires
to model CFFs or TFFs in terms of GPDs.
This can be done in different representations,
which should be finally considered as equivalent.
However, for a specific purpose a particular representation
may be more suitable than the others.

Neglecting positivity constraints, we model GPDs by means
of a conformal SL(2,$\mathbb{R}$) partial wave expansion, which can be written as a
Mellin-Barnes integral~\cite{Mueller:2005ed}:
\begin{eqnarray}
F(x,\eta,t,\mu^2) = \frac{i}{2}  \int_{c -i \infty}^{c+i\infty} dj\, \frac{ p_j(x,\eta) }{\sin(\pi j)} F_j(\eta,t,\mu^2)\,.
\end{eqnarray}
Here, $p_j(x,\eta) $ are the partial waves given in terms of associated Legendre functions of the first and second kind, and the integral conformal GPD moments  $F_j(\eta,t,\mu^2)$
are even polynomials in $\eta$ of order $j$ or $j+1$.
Other representations of GPDs based on the SL(2,$\mathbb{R}$) partial wave expansion
include the so-called ``dual'' parameterization \cite{Polyakov:2002wz,Polyakov:2007rv,SemenovTianShansky:2008mp,Polyakov:2008aa}.

In the Mellin-Barnes representation, the CFFs possess a rather convenient form,
e.g., (\ref{KLMSPM-DVCS-HS}) can be rewritten in the 
following form \cite{Kumericki:2007sa,Mueller:2005nz}:
\begin{eqnarray}
\label{KLMSPM-DVCS-MB}
\left\{ {\cal H}  \atop {\cal E} \right\}(x_{B},t,Q^2)
\!\!\!&\stackrel{\rm LO}{=}&\!\!\!
\frac{1}{2 i} \int_{c-i\infty}^{c+i\infty} dj
\, \xi^{-j-1} \left[i +\tan\left(\frac{\pi j}{2}\right) \right]
\nonumber \\[0.2cm] & &
\times \,
\frac{2^{j+1} \Gamma(j+5/2)}{\Gamma(3/2)\Gamma(j+3)}
\,
 \left\{ H_j \atop E_j\right\}(\eta=\xi,t,Q^2)\Big|_{\xi =\frac{x_{B}}{2-x_{B}}}\,.
\end{eqnarray}
This  integral is numerically implemented in an efficient routine in two different factorization schemes,
including the standard minimal subtraction ($\overline{\rm MS}$) one at next-to-leading order (NLO) accuracy.
Further advantages of this representation are:
\begin{itemize}
\item[(i)] The conformal moments evolve autonomously at LO.
\item[(ii)] One can employ conformal symmetry to obtain next-to-next-to-leading order (NNLO) corrections to the DVCS amplitude~\cite{Mueller:2005nz,Kumericki:2006xx}.
\item [(iii)] PDF and form factor constraints can be straightforwardly implemented. 
Namely,   $F_j(\eta=0,t=0,\mu^2)$
are the Mellin moments of PDFs, $F_{j=0}$ are partonic contributions to elastic form factors,
$H_{j=1}$ and $E_{j=1}$ are the energy-momentum tensor form factors, and for general $j$ one immediately makes contact to lattice measurements.
\end{itemize}

To parameterize the degrees of freedom that
can be accessed in hard exclusive reactions, one can expand the conformal 
moments in terms of  $t$-channel SO(3) partial waves expressed in terms of the
Wigner rotation matrices $\hat d_j(\eta)$ ($\hat d_j(\eta=0) =1$)~\cite{Polyakov:1998ze}.
An effective GPD model at given input scale $Q^2_0$ is
provided by taking into account three partial waves,
\begin{eqnarray}
\label{KLMSPM-mod-nnlo}
F_j(\eta,t) =\hat d_j(\eta) f_j^{j+1}(t) +\eta^2 \hat d_{j-2}(\eta) f_j^{j-1}(t) + \eta^4 \hat d_{j-4}(\eta) f_j^{j-3}(t) \,,
\end{eqnarray}
which is valid for integral $j\ge 4$. In the simplest version of such a model, one might introduce just
two additional parameters by setting the non-leading partial wave amplitudes to:
\begin{eqnarray}
\label{KLMSPM-mod-nnlo1}
f_j^{j-k}(\eta,t) = s_k f_j^{j+1}(\eta,t)\,, \quad k={2,4,\dots}.
\end{eqnarray}
Such a model allows us to control the size of the GPD on the cross-over line 
and its $Q^2$-evolution, see fig.~\ref{KLMSPM-fig-CFF}.
A flexible parameterization of the skewness effect in the large $x$ region requires to decorate the skewness parameters
$s_k$ with some $j$ dependence and for more convenience one might replace Wigner`s rotation matrices by some effective SO(3) partial waves.

\subsection{GPDs from hard exclusive measurements}

Based on the experimental data set from the collider experiments H1 and ZEUS at DESY, the fixed target experiment HERMES at DESY,
and the Hall A, CLAS, and Hall C experiments at JLAB,   GPDs have been accessed from hard exclusive meson and photon electroproduction  in the last few years.
Favorably, DVCS enters as a subprocess into the hard photon electroproduction where its interference with the Bethe-Heitler (BH) bremsstrahlung process provides variety of handles on the real and imaginary part of twist-two and twist-three 
CFFs~\cite{Belitsky:2001ns,Diehl:1997bu}. However, switching from a proton to a neutron target allows only for a partial flavor separation, which is much more intricate than in DIS.
On the other hand, DVEM can be used as a flavor filter, however, here one expects that both radiative~\cite{Belitsky:2001nq,Ivanov:2004zv,Diehl:2007hd} and (non-factorizable) higher-twist contributions might be rather important. The onset of the collinear description remains here an issue which should be 
explored.

\begin{figure}[t]
\begin{center}
\includegraphics[width=13cm]{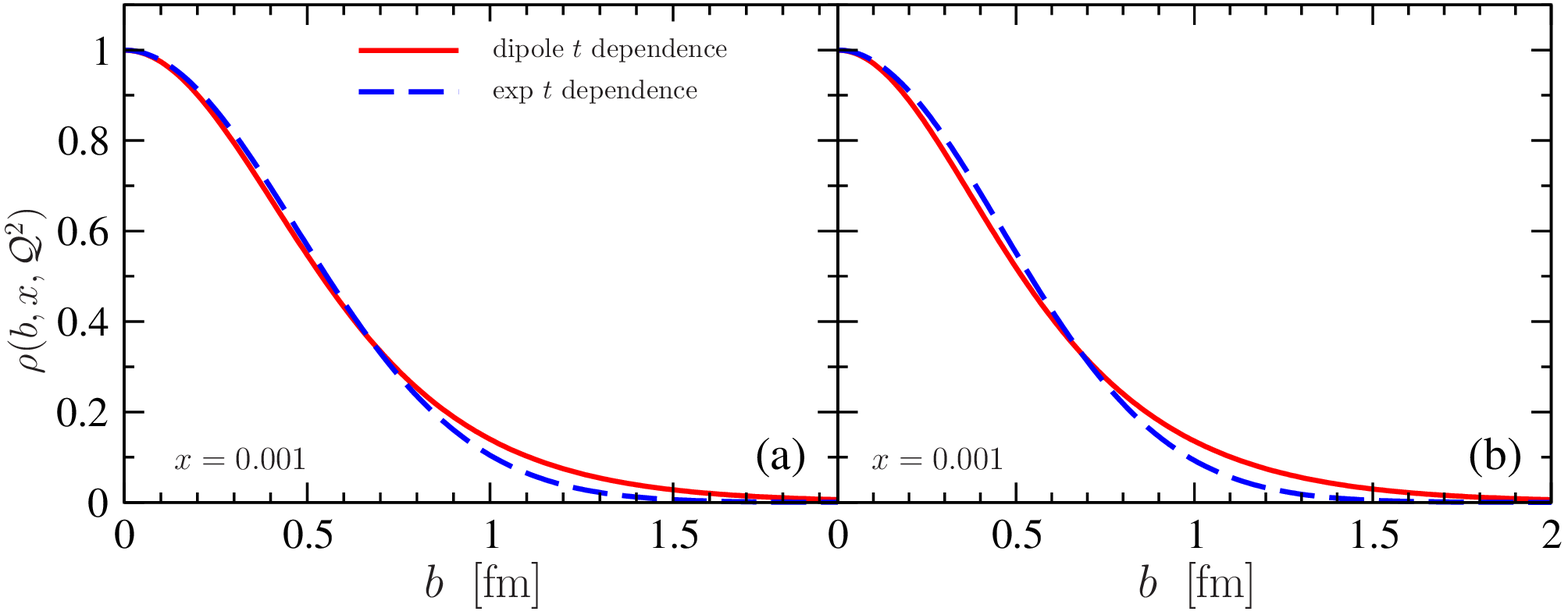}
\end{center}
\vspace{-0.7cm}
\caption{\small
Quark (a) and gluon (b) transverse  profile function
(\ref{KLMSPM-eq:TraProFun}) for $Q^2=4\,{\rm GeV}^2$ and $x=10^{-3}$
from a six parameter DVCS fit~\cite{Kumericki:2009uq}.
}
\label{KLMSPM-fig:TraDis}
\end{figure}

For the DVCS process, the collinear factorization approach has been employed
in a specific scheme up to NNLO in the small $x_{B}$ region~\cite{Kumericki:2007sa,Mueller:2005nz,Kumericki:2006xx}. It turns out
that NLO corrections are moderate, while NNLO ones are becoming much
smaller \cite{Kumericki:2007sa}.
Experimentally, the unpolarized DVCS cross section has been provided
by the H1 and ZEUS collaborations \cite{Chekanov:2003ya,Aktas:2005ty,:2007cz,Chekanov:2008vy}. 
In the collider kinematics, the DVCS cross section is primarily given in terms of two 
CFFs, ${\cal H}$ and ${\cal E}$:
\begin{eqnarray}
\label{KLMSPM-Def-CroSec}
\frac{d\sigma^{\rm DVCS}}{dt}(W,t,Q^2) \approx
\frac{\pi \alpha^2 }{{\cal Q}^4} \frac{W^2 x_{B}^2}{W^2+Q^2}
\left[\left| {\cal H} \right|^2  - \frac{t}{4 M^2_{p}}
\left| {\cal E} \right|^2
\right]
\left(x_{B},t,{\cal Q}^2\right)\Big|_{x_{B}\approx\frac{Q^2}{W^2+Q^2}}\,.
\end{eqnarray}
Although at a fixed scale and to LO accuracy the CFFs are given
by (dominant sea) quark GPDs, evolution will induce
a  gluonic contribution, too. Indeed, the
experimental lever arm $3\,{\rm GeV}^2 \lesssim Q^2 \lesssim 80\,{\rm GeV}^2$  is sufficiently large to access the gluonic GPD. In our fitting procedure,
 the Mellin-Barnes integral was utilized within a SO(3) partial wave ansatz for the conformal moments and good fits ($\chi^2/{\rm d.o.f.} \approx 1$) could be obtained at LO to NNLO accuracy, exemplifying that flexible GPD models were at hand. From such fits, one can then obtain the image of quark and gluon distributions.
 It is illustrated in figure~\ref{KLMSPM-fig:TraDis} that in impact space,
the (normalized) transverse profiles,
\begin{eqnarray}
\label{KLMSPM-eq:TraProFun}
\rho(b,x,Q^2) =
\frac{
\int_{-\infty}^{\infty}\! d^2\vec{\Delta}\;
e^{i \vec{\Delta} \vec{b}} H(x,\eta=0,t=-\vec{\Delta}^2,Q^2)
}{
\int_{-\infty}^{\infty}\! d^2\vec{\Delta}\;
H(x,\eta=0,t=-\vec{\Delta}^2,Q^2)
} \;,
\end{eqnarray}
determined for dipole and exponential $t$-dependence of $H$, mainly differ for
distances larger than the disc radius of the proton,
 i.e., for $b> 0.6$ fm.
Hence, the larger values of the transverse widths,
$\sqrt{\langle\vec{b}^2 \rangle}_{\rm sea} \approx 0.9 \, {\rm fm}$
and  $\sqrt{\langle\vec{b}^2 \rangle}_{\rm G} \approx 0.8\, {\rm fm}$ for the
dipole ansatz, arise from the long-range tail of the profile
function, see the solid curves in figure~\ref{KLMSPM-fig:TraDis}.
For an exponential ansatz, we find slightly smaller values
$\sqrt{\langle\vec{b}^2 \rangle}_{\rm sea} \approx 0.7 \, {\rm fm}$
and  $\sqrt{\langle\vec{b}^2 \rangle}_{\rm G} \approx 0.6\, {\rm fm}$,
where the gluonic one is compatible with the analysis of $J/\psi$ 
production~\cite{Strikman:2003gz}.
Note that the model uncertainty in the
extrapolation of the GPD to $t=0$ corresponds to the uncertainty
in the long-range tail. Moreover, the model uncertainty of the extrapolation into the region $-t > 1\, {\rm GeV}^2$ is
essentially canceled in the profile (\ref{KLMSPM-eq:TraProFun}) normalized at $b=0$.

We also note that at LO the gluonic GPD (as the gluonic
PDF) is rather steep and radiative corrections might provide a large GPD/PDF reparameterization effect, which will be studied in more detail in the future.
Our first successful LO description of DVCS within a flexible GPD model~\cite{Kumericki:2009uq} 
is in agreement with
aligned-jet model considerations \cite{Frankfurt:1997at}.
We also mention that an attempt has been undertaken to access the ${\cal E}$  CFF from the
beam charge asymmetry measurement \cite{:2009vda}, proportional to the combination $\Re{\rm e}\left[F_1(t) {\cal H}-\frac{t}{4 M^2} F_2(t) {\cal E}\right]$.
Unfortunately, the size of the experimental uncertainties does not allow one
to separate the $\cal H$ and $\cal E$ contributions.

An approach analogous to the one employed for DVCS~\cite{Kumericki:2007sa}
is also suitable for LO and NLO analysis of DVEM. Hence, one can simultaneously make use
of DVCS and DVEM measurements in a global fitting procedure, which is in progress.

GPD studies were also performed for the DVCS process in the fixed target kinematics to LO accuracy. In this region, relying on the scaling hypothesis, one might directly ask for the value of the GPDs on their cross-over line. For instance, for valence quarks we use the following generically motivated ansatz:
\begin{eqnarray}
\label{KLMSPM-ansHval}
H^{\rm val}(x,x,t)  =
\frac{1.35\,  r}{1+x} \left(\frac{2 x}{1+x}\right)^{-\alpha(t)}
\left(\frac{1-x}{1+x}\right)^{b}
\left(1-  \frac{1-x}{1+x} \frac{t}{M^{\rm val}}\right)^{-1}\,,
\end{eqnarray}
where 
$r=\lim_{x\to 0}H(x,x)/H(x,0)$ is the skewness ratio;  
$\alpha(t) =  0.43 + 0.85\, t/{\rm GeV}^2$; $b$  controls the  $x\to1$ limit and
$M^{\rm val}$ controls the residual $t$-dependence, which we set to $M^{\rm val}=0.8\,{\rm GeV}$.
For the forward limit $q(x)=H(x,0)$, we used the LO parameterization of Alekhin~\cite{Alekhin:2002fv}.  The generic $(-t)^{-2}$ fall-off at large $-t$  for generalized form factors is indirectly encoded in the Regge-trajectory and the residual 
$t$ dependence is modeled by a monopole form with an $x$-dependent cut-off mass. 
The subtraction constant~(\ref{KLMSPM-DR-Re}) is taken an a dipole form:
\begin{eqnarray}
\label{KLMSPM-ansD}
{\cal D}(t) =  d\left(1- \frac{t}{M_d^2}\right)^{-2}\,.
\end{eqnarray}

In a first global fit~\cite{Kumericki:2009uq} to hard exclusive photon electroproduction off unpolarized proton, we took sea quark and gluon GPD models with two SO(3) partial waves at small $x$, reparameterized the outcome from H1 and ZEUS DVCS fits at $Q^2 = 2\, {\rm GeV}^2$, and employed it in fits of fixed target data within the scaling hypothesis. To relate the CFFs with the observables, we employed the BKM formulas~\cite{Belitsky:2001ns} within the `hot-fix' convention~\cite{Belitsky:2008bz} and used the Sachs parameterization for the electromagnetic form factors. Thereby, we utilized the ``dispersion relation" (\ref{KLMSPM-DR-Im},\ref{KLMSPM-DR-Re}), where the ansatz (\ref{KLMSPM-ansHval})
specifies a valence-like GPD on the cross-over line. Besides the subtraction constant (\ref{KLMSPM-ansD}),
we also included the parameter-free pion-pole model for the $\tilde E$ GPD~\cite{Penttinen:1999th} and parameterized  the $\widetilde H$ GPD rather analogously to~(\ref{KLMSPM-ansHval}) with $b=3/2$. For the fixed target fits, we chose two data sets resulting in two fits (\emph{KM09a} and \emph{KM09b}). 
Out fit gives:
\begin{eqnarray}
\hskip -1cm
\mbox{\em KM09a:}&& b^{\rm sea} =3.09\,,\quad
r^{\rm val} =0.95\,,\;\; b^{\rm val}=0.45\,,
\quad
 d=-0.24\,,\;\; M_d=0.5\,{\rm GeV}\,,
\nonumber\\
 \mbox{\em KM09b:}&&
b^{\rm sea} =4.60\,,\quad
r^{\rm val} =1.11\,,\;\; b^{\rm val}=2.40\,,
\quad d=-6.00\,,\;\; M_d=1.5 \,{\rm GeV}\,.
\end{eqnarray}
These values of the fit parameters
are compatible with our generic expectations: the skewness effect at small $x$ should be small, i.e., $r\sim 1$, the subtraction constant should be 
negative~\cite{Goeke:2001tz,Goeke:2007fp}, and, according to counting rules~\cite{Yuan:2003fs}, $b$ should be smaller than the corresponding $\beta$ value of the relevant PDF~\cite{Kumericki:2009uq,Kumericki:2010fr}.

To improve the models that we just described, we now use a hybrid technique where the 
sea quark and gluon GPDs  are represented in terms of conformal moments, while, for convenience, the valence quarks are still modeled in momentum fraction space and within the ``dispersion integral" approach. Also, the residue of the pion-pole contribution is now considered as a parameter, and the Hall A data forces a roughly three times larger value than expected from the model~\cite{Penttinen:1999th}. Optionally, we might also use the improved formulae from~\cite{Belitsky:2010jw} applicable for a longitudinally polarized target. The new parameters
read:
\begin{eqnarray}
\hskip -2 cm
\mbox{\em KM10a:} &&
r^{\rm val} =0.88\,,\;\;M^{\rm val}=1.5\,{\rm GeV}\,,\;\;
b^{\rm val}=0.40\,, \ d=-1.72\,,\;\; M_d=2.0\,{\rm GeV}\,,
\nonumber\\
\mbox{\it KM10b:} &&
r^{\rm val} =0.81\,,\;\;M^{\rm val}=0.8\,{\rm GeV}\,,\;\; b^{\rm val}=0.77\,,
\ d=-5.43\,,\;\; M_d= 1.33 \,{\rm GeV}\,.
\end{eqnarray}
Note that 
for the valence part of the $H$ GPD, these results are qualitatively compatible with those from
the pure {\em KM09}  ``dispersion relation" fits.

We also performed an additional fit where we directly used the harmonics of beam spin
sums and differences measured by Hall A (fit {\em KM10}).
The results of our two ``dispersion-relation" fits and three hybrid model fits are
available as a computer program providing the four-fold cross section of
polarized lepton scattering on unpolarized proton for a given kinematics, see
\url{http://calculon.phy.hr/gpd/}.  Unlike ``dispersion-relation" fits, the
hybrid model fits, where  LO
evolution of sea quark and gluon GPDs has been taken into account, are suitable
for estimates in the small $x_{B}$ region.

\begin{figure}[t]
\begin{center}
\includegraphics[width=15cm]{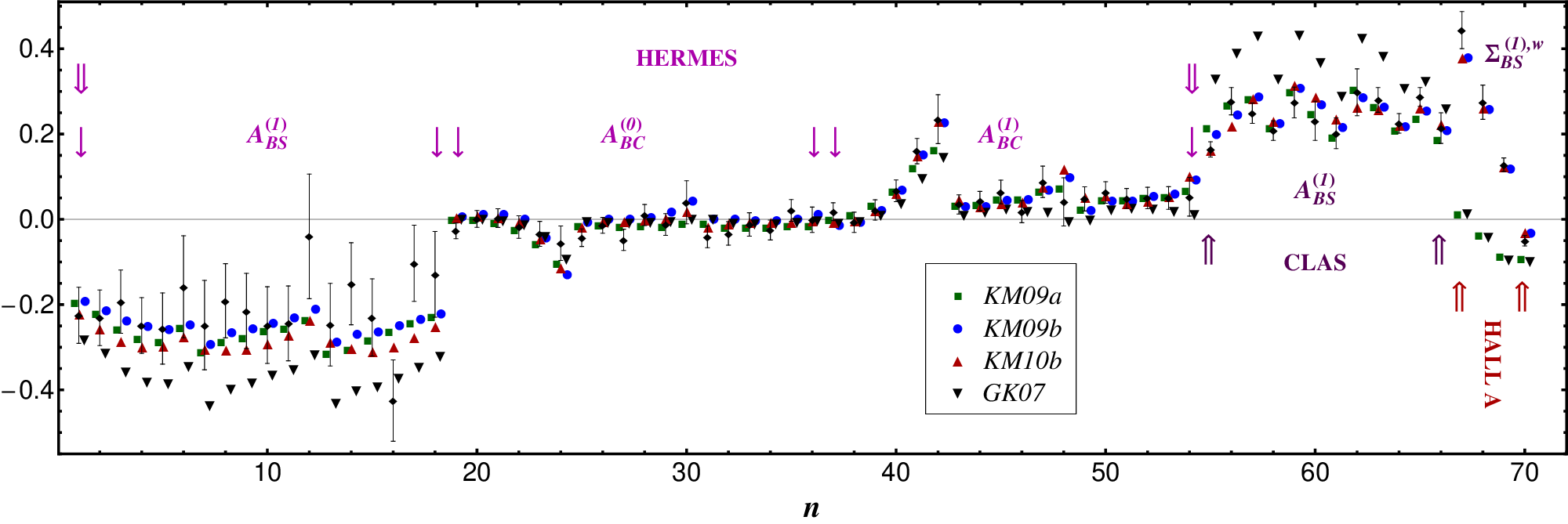}%
\end{center}
\vspace*{-0.5cm}
\caption{\small
Experimental  measurements for fixed target kinematics (circles) labeled by data point number $n$: $A^{(1)}_{\rm BS}$ (1-18),  $A^{(0)}_{\rm BC}$  (19-36),
$A^{(1)}_{\rm BC}$ (37-54)
from~\cite{:2009rj}; $A^{(1)}_{\rm BS}$  (55-66)
and $\Sigma_{\rm BS}^{(1),w}$  (67-70) are derived from~\cite{:2007jq}
and \cite{Camacho:2006hx}. Model results  are from the ``dispersion-relation" fits {\em KMO9a} without Hall A data \cite{Kumericki:2009uq} (squares, slightly shifted to the left) and {\em KMO9b} with the Hall A data
(circles, slightly shifted to the right), hybrid model fit {\em KM10b} (triangles-up), and a hand-bag prediction {\em GK07} from hard vector meson production (triangles-down, slightly shifted to the r.h.s.)~\cite{Goloskokov:2007nt}.
}
\label{KLMSPM-Fig_modelvsdata}
\vspace{-4pt}
\end{figure}

In figure~\ref{KLMSPM-Fig_modelvsdata} we confront our fit results ($\chi^2/{\rm d.o.f.}\approx 1$ w.r.t.~the employed data sets) to experimental data:
 {\em KM09a} (squares), {\em KM09b} (circles), and the hybrid model fit {\em KM10b}
(triangles-up) in which we now utilized the improved formulae set
\cite{Belitsky:2010jw} and the Kelly form factor parameterization~\cite{Kelly:2004hm}.  
We also include the predictions from the {\em GK07} model~\cite{Goloskokov:2007nt}
(triangles-down), where we adopt the hypothesis of $H$ dominance.  Qualitatively, these predictions are consistent with a VGG\footnote{
VGG refers to a computer code originally written by M.~Vanderhaeghen, P.~Guichon, and M.~Guidal. To our best knowledge, the code for DVCS presently used by experimentalists employs a model that adopts Radyushkin's DD ansatz~\cite{Goeke:2001tz}.} code estimate, which tends
to over-estimate the BSAs \cite{:2007jq,:2009rj} and describes the BCAs from HERMES rather well without the $D$-term \cite{:2008jga}. This is perhaps not astonishing, since the employed $H$ GPD model relies on Radyushkin's DD ansatz, too. We would like to emphasize that at LO, 
the {\em GK07} model is in reasonable agreement with the H1 and ZEUS DVCS data ($\chi^2/{\rm d.o.f.}\approx 2$), essentially thanks to the rather small and stable skewness ratio $r^{\rm sea}$ of sea quarks.

Longitudinally polarized target data from CLAS~\cite{Chen:2006na}  and HERMES~\cite{:2010mb} provide a handle on $\widetilde H$~\cite{Belitsky:2001ns},  where the mean values of CFF fits~\cite{Guidal:2010ig} in the JLAB kinematics give two to three times bigger $\widetilde H$ contribution compared to our expectations ($r_{\widetilde H}\simeq 1, b_{\widetilde H}\simeq2$).
These findings are one to two standard deviations away from our big $\widetilde H$ ad hoc scenario of the {\em KM09b} fit,  which is indeed disfavored by the
longitudinally polarized proton data. We like to add that with our present hybrid model a reasonable global fit, such as {\em KM10} above, is possible. In such a fit, the Hall A data require a rather large pion pole contribution, inducing a large DVCS cross section contribution. Still, we have not included the transversal target data from the HERMES collaboration~\cite{:2008jga} or the neutron data from Hall A~\cite{:2007vj}.

\begin{figure}[t]
\includegraphics[width=16cm]{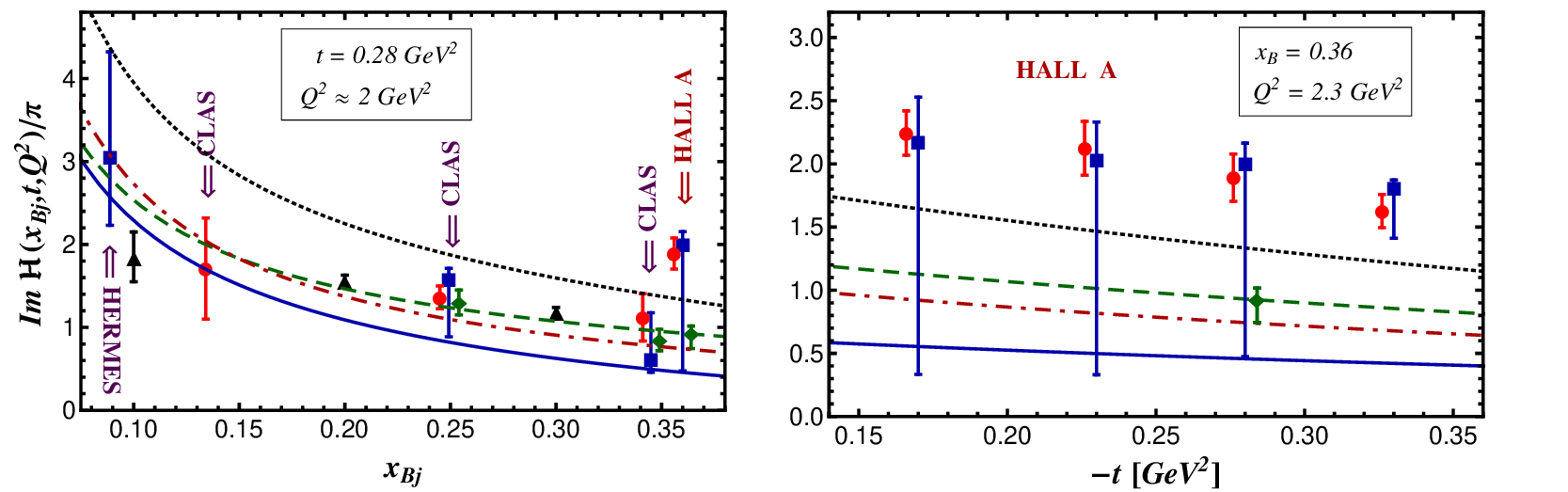}
\vspace{-4pt}
\caption{\small
$\Im{\rm m} {\cal H}/\pi$ from different strategies: our DVCS fits [dashed (solid) curve excludes (includes) Hall A data
from "dispersion relation" {\em KM09a} ({\em KM09b})~\cite{Kumericki:2009uq} and hybrid {\em KM10b} (dash-dotted) models],  {\em GK07} model from DVEM (dotted),
seven-fold CFF fit~\cite{Guidal:2008ie,Guidal:2009aa} with boundary conditions  (squares),  $\cal H$, $\widetilde {\cal H}$ CFF fit~\cite{Guidal:2010ig} (diamonds), smeared conformal partial wave model fit~\cite{Moutarde:2009fg} within $H$ GPD (circles).
The triangles result from our neural network fit, cf.~figure~\ref{KLMSPM-fig3}.
}
\vspace{-4pt}
\label{KLMSPM-fig2}
\end{figure}
So far we did not study model uncertainties or experimental error propagation, since both tasks might be rather intricate.  To illuminate this, in figure~\ref{KLMSPM-fig2} we compare  
our results for $\Im{\rm m}{\cal H}(x_{B},t)/\pi$ with the results that do provide error estimates.
The squares arise from constrained least squares fits~\cite{Guidal:2008ie,Guidal:2009aa} at given kinematic means of HERMES and JLAB  measurements on unpolarized proton, where the imaginary and real parts of twist-two CFFs are taken as parameters.  
The huge size of the error bars shows the limited accuracy with which $H$ can be extracted from unpolarized proton data alone~\cite{Belitsky:2001ns}. A pure $H$ GPD model fit~\cite{Moutarde:2009fg} (circles) to JLAB data provides much smaller errors, arising from error propagation and some estimated model uncertainties. All three of our curves are compatible with the findings~\cite{Guidal:2008ie,Guidal:2009aa} and the $H$ GPD model analysis~\cite{Moutarde:2009fg} of CLAS data. However, for Hall A kinematics, 
the deviation of the two predictions that are based
on the $H$ dominance hypothesis (the dashed curve and circles in the right panel) 
are obvious and are explained by our underestimation of the cross section normalization by about 50\%.  Moreover, the quality of fit~\cite{Moutarde:2009fg}, $\chi^2/{\rm d.o.f.} \sim 1.7$, might provide another indication that CLAS and Hall A data are not compatible, when this hypothesis is assumed, see, e.g., the two rightmost circles in the left panel for CLAS
($x_{B}=0.34$, $t=-0.3 {\rm GeV}^2 $, $Q^2=2.3$ GeV$^2$) and Hall A ($x_{\rm Bj}=0.36$, $t=-0.28 {\rm GeV}^2 $, $Q^2=2.3$ GeV$^2$).
While the pure ${\cal H}$ and $\widetilde{\cal H}$ CFF fit~\cite{Guidal:2010ig} (diamonds), including longitudinally polarized target data, is within error bars inconsistent with the $H$ dominated scenario~\cite{Moutarde:2009fg} (circles), it (accidentally) reproduces our dashed curve.

Another source of uncertainties are twist-three contributions and perhaps also gluon transversity related contributions, which might be strongly affected by
twist-four effects~\cite{Kivel:2001rw}.

\begin{figure}[t]
\centerline{\includegraphics[scale=0.6]{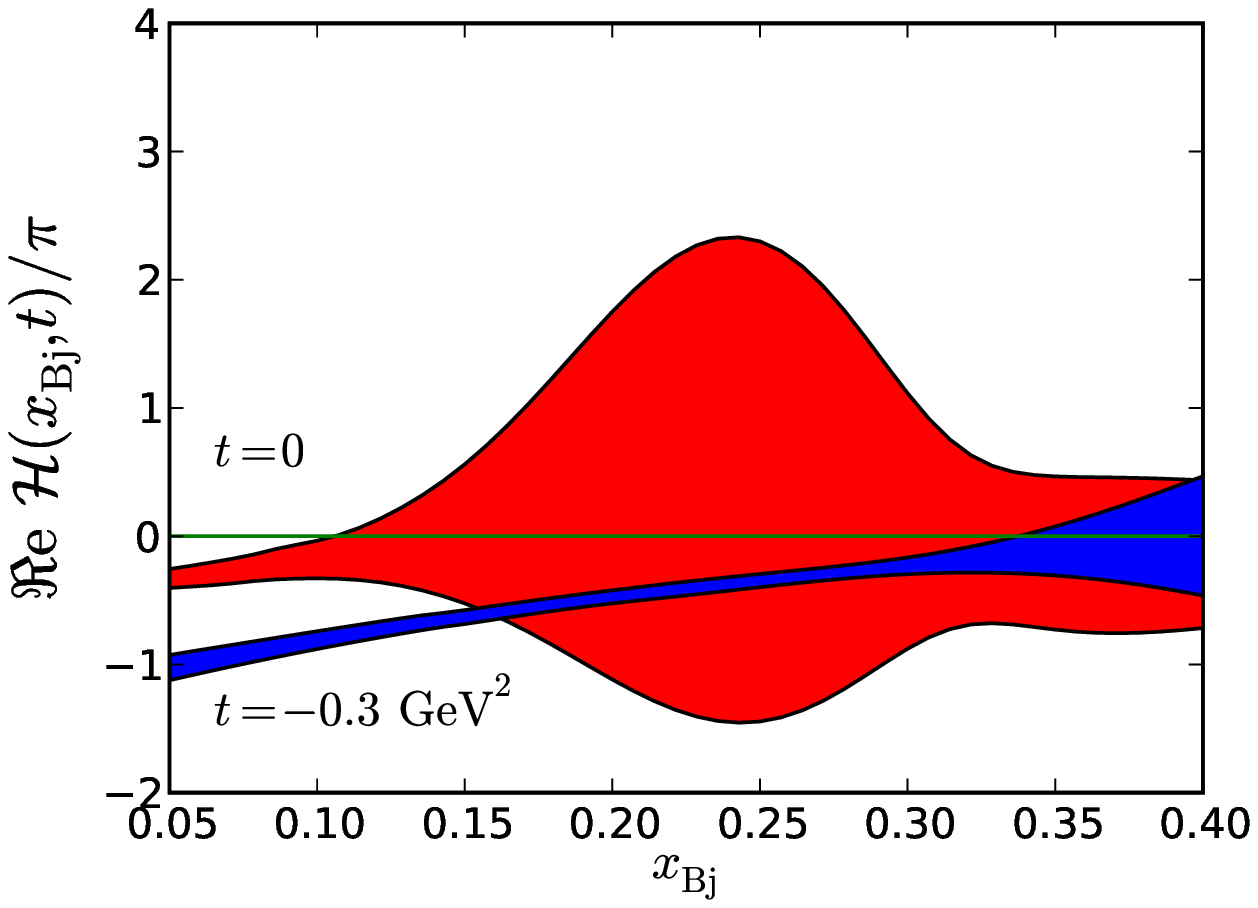}}%
\vspace{-4pt}
\caption{\small
Neural network extraction of $\Re{\rm e}\,{{\cal H}(x_{B},t)}/\pi$ from BCA~\cite{:2009rj} and BSA~\cite{:2007jq} data.}
\label{KLMSPM-fig3}
\vspace{-4pt}
\end{figure}
All this exemplifies that within (strong) assumptions and the present set of
measurements, the propagated experimental errors cannot be taken as an estimate
of GPD uncertainties. An error estimation in model fits might be based on
twist-two sector projection technique~\cite{Belitsky:2001ns},  boundaries for the
unconstrained  model degrees of freedom, and error propagation in the twist-two sector.
Alternatively, neural networks, already successfully used for PDF fits~\cite{Ball:2010de}, may
be an ideal tool to extract CFFs or GPDs. In figure~\ref{KLMSPM-fig3},
we present a first example in
which, within the $H$-dominance hypothesis, ${\cal H}$ is extracted using
a procedure similar to the one of~\cite{Forte:2002fg}.  Here, 50
feed-forward neural nets with two hidden layers were trained
using HERMES BCA~\cite{:2009rj} and CLAS
BSA~\cite{:2007jq} data. Hence, only the experimental errors were propagated,
which, in the absence of a model hypothesis, become large for the $t \to 0$
extrapolation.

\subsection{Potential of an electron-ion collider}

A high luminosity machine in the collider mode with polarized electron and proton or ion beams would be an ideal instrument to quantify QCD phenomena. It is expected that such a machine, combined with designated detectors, would allow for precise measurements of exclusive channels. Besides hard exclusive vector meson and photon electroproduction, one might address the behavior of parity-odd GPDs $\widetilde{\cal H}$ (related to polarized PDFs) and $\widetilde{\cal E}$ via the exclusive production of pions even in the small $x$ region. It is obvious from what was said above that an access of GPDs requires a large data set with small errors.  In the following we would like to illustrate the potential of such a machine for DVCS studies, where
we also address the GPD deconvolution problem.

Let us remind that already the isolation of CFFs is rather intricate. For a spin-1/2 target, we have four twist-two, four twist-three, and four gluon transversity-related complex valued CFFs. The photon helicity non-flip amplitudes are dominated by twist-two CFFs, the transverse--longitudinal flip amplitudes by
twist-three effects, and the transverse--transverse flip ones by gluon transversity. 
Hence, the first, second, and third harmonics w.r.t.~the azimuthal angle of the interference term are twist-two, twist-three, and gluon transversity dominated, respectively.
In an ideal experiment, assuming that transverse photon helicity flip effects are negligible, cross section measurements would allow to separate the sixteen quantities that are then given in terms of twist-two and twist-three CFFs.
The reader might find a more detailed discussion, based on a $1/Q$ expansion, in~\cite{Belitsky:2001ns}. We also note that the definition of CFFs is convention-dependent.

\begin{figure}[t]
\begin{center}
\includegraphics[width=15.0cm]{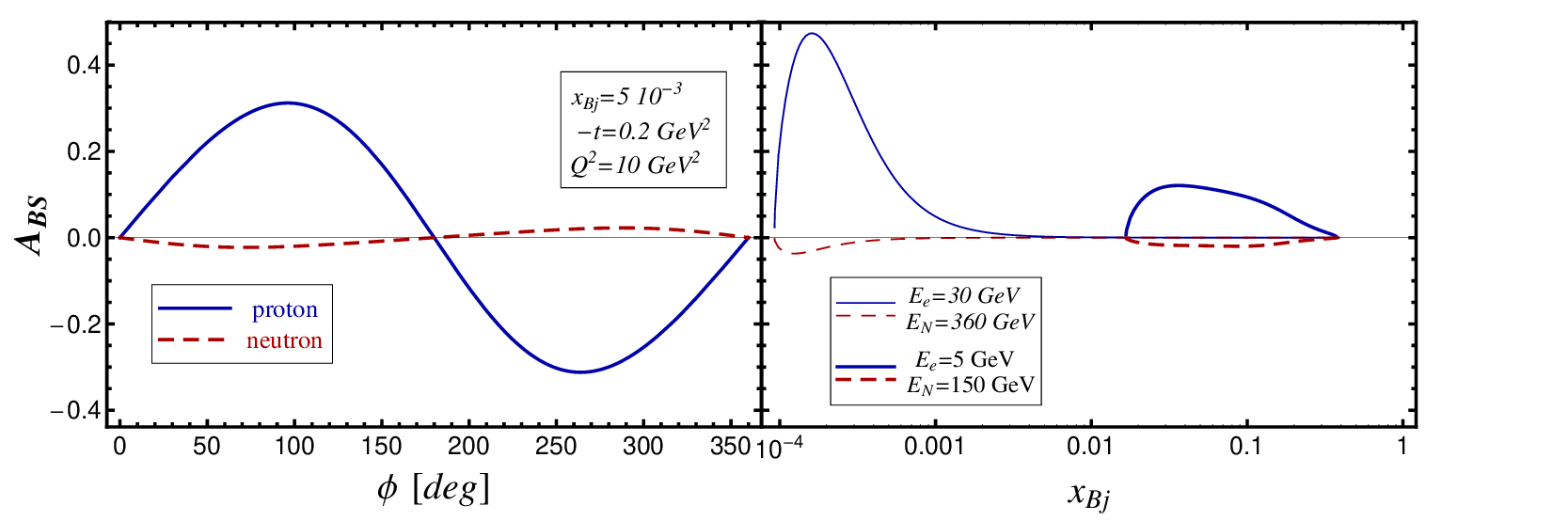}
\end{center}
\vspace{-0.7cm}
\caption{\small {\em KM10b} model estimate for the DVCS beam spin asymmetry 
with a proton (solid) and neutron (dashed) target.
Left panel: $A_{\rm BS}$ {\it vs.}~$\phi$ for $E_N=250~{\rm GeV}$, $E_e=5~{\rm GeV}$, $x_{B}=5\times 10^{-3}$, $Q^2 = 10~{\rm GeV}^2$, and $t=-0.2~{\rm GeV}^2$. Right panel: Amplitude $A_{\rm BS}^{(1)}$ of the first harmonic {\it vs.}~$x_{\rm Bj}$ at $t=-0.2~{\rm GeV}^2$ for small $x_{B}$ (thin)
[$E_e=30~{\rm GeV}$, $E_p=360~{\rm GeV}$, $Q^2=4~{\rm GeV}^2$] and large $x_{B}$ (thick)
[$E_e=5~{\rm GeV}$, $E_p=150~{\rm GeV}$, $Q^2=50~{\rm GeV}^2$]  kinematics.
\label{KLMSPM-Fig_BSA}}
\end{figure}
In a twist-two analyzes on unpolarized, longitudinally and
transversally polarized protons, one might be able to disentangle the four different twist-two CFFs via the measurement of single beam and target spin
asymmetries. In figure~\ref{KLMSPM-Fig_BSA}, we illustrate that
the beam spin asymmetry
for a proton target (solid curves),
\begin{eqnarray}
\label{KLMSPM-A_BS1}
A_{\rm BS}^{(1)}\propto  \frac{\sqrt{t_{\rm min} -t}}{2M}\,
y \left[F_1(t) H(\xi,\xi,t,Q^2)
-\frac{t}{4 M^2} F_2(t) E(\xi,\xi,t,Q^2)  + \cdots\right] \,,
\end{eqnarray}
might be rather sizeable over a large kinematical region in which the lepton energy loss $y$ is not too small.  Here the helicity conserved CFF ${\cal H}$  is the dominant contribution, while $\cal E$ appears with a kinematic suppression factor $t/4 M^2$, induced by the helicity flip.
For a neutron target, the ${\cal H}$ contribution is suppressed by the accompanying Dirac form factor $F^n_1$  ($F_1^n(t=0)=0$) and, hence, one becomes sensitive to the CFF ${\cal E}$. Unfortunately, one also has to worry about other non-dominant CFF contributions, indicated by the ellipsis. Note that the asymmetry for the neutron (dashed curves in figure~\ref{KLMSPM-Fig_BSA}) might be underestimated since we set in our model $E(x,x,t,Q^2)$ to zero.

\begin{figure}[t]
\begin{center}
\includegraphics[width=15cm]{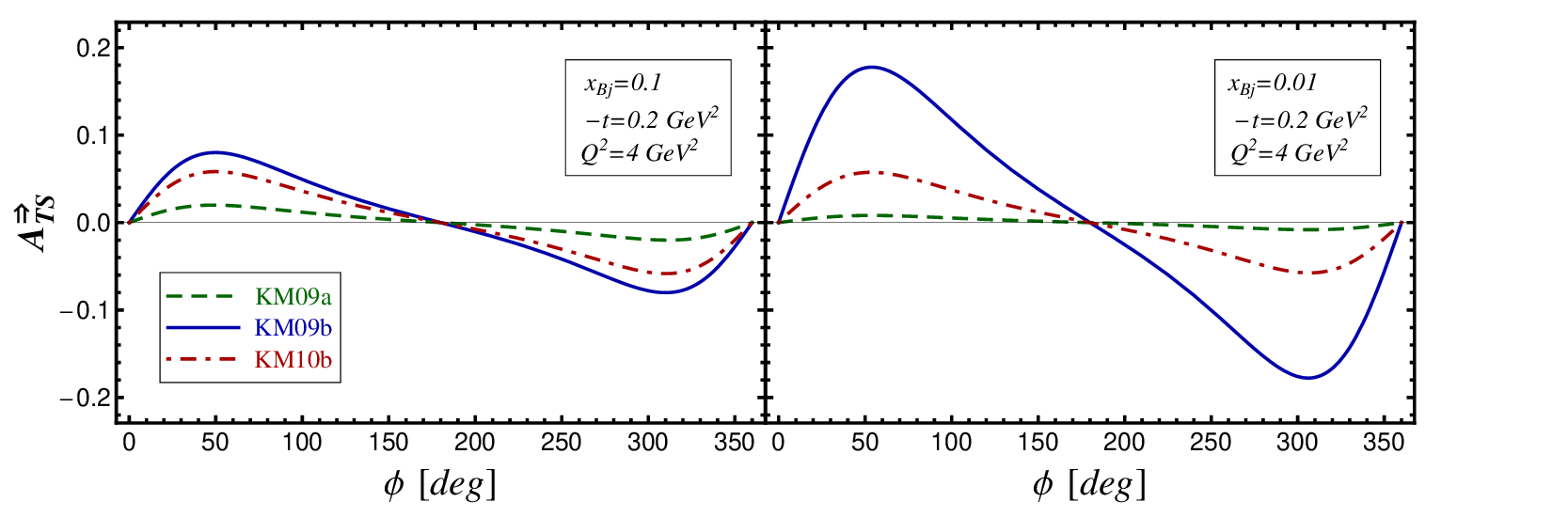}
\end{center}
\vspace{-0.7cm}
\caption{\small
DVCS longitudinal target spin asymmetry {\it vs.}~$\phi$ for {\em KM09a} (dashed), {\em KM09b} (solid), and  {\em KM10b}  hybrid (dash-dotted) models
at  $E_e=5~{\rm GeV}$, $t=-0.2~{\rm GeV}^2$, ${\cal Q}^2 = 4~{\rm GeV}^2$
within  $E_p=150~{\rm GeV}$, $x_{\rm Bj}=0.1$  (left)  and  $E_p=350~{\rm GeV}$,  $x_{\rm Bj}=0.01$ (right).
\label{KLMSPM-Fig_TSA}}
\end{figure}

For a longitudinally polarized target, the asymmetry
\begin{eqnarray}
\label{KLMSPM-A_LTS}
A^{\Rightarrow {(1)}}_{\rm TS}\propto \frac{\sqrt{t_{\rm min} -t}}{2M}
\left[F_1(t) \widetilde H(\xi,\xi,t,{\cal Q}^2) -
\frac{t}{4 M^2} F_2(t) \xi\widetilde E(\xi,\xi,t,{\cal Q}^2)
+\cdots \right]
\end{eqnarray}
is sensitive to the GPD $\widetilde H$, while $\xi\widetilde E$ and other GPDs might contribute to some extent. Naively, one would expect that
this asymmetry vanishes in the small $x_{B}$ region and might be sizeable at $x_{B}\sim 0.1$, see the left panel of figure~\ref{KLMSPM-Fig_TSA}. Not much
is known about the small $x$ behavior of $\widetilde H$ and it might be even accessible at smaller values of $x_{B}$, as illustrated by the {\em KM09b} model with its big $\widetilde H$ contribution (solid curve, the right panel of figure~\ref{KLMSPM-Fig_TSA}). For a
neutron target, the asymmetry becomes sensitive to the $\xi \widetilde E$  GPD. Note that here the factor $\xi$ is
annulled by a conventional $1/\xi$ factor in the definition of the $\widetilde E$  GPD.

Finally, we emphasize that a single spin asymmetry measurement with a transversally polarized target provides another handle on the
helicity-flip GPDs $E$ and $\widetilde E$. If the target spin is perpendicular to the reaction plane, the asymmetry
\begin{eqnarray}
\label{KLMSPM-A_TTS1}
A^{\Uparrow (1)}_{\rm TS}\propto \frac{t}{4 M^2}
\left[F_2(t) H(\xi,\xi,t,{\cal Q}^2)
- F_1(t) E(\xi,\xi,t,{\cal Q}^2) + \cdots\right] \,,
\end{eqnarray}
is dominated by a linear combination of the GPDs $H$ and $E$. In the case when the target spin is aligned with the reaction plane, the asymmetry
\begin{eqnarray}
\label{KLMSPM-A_TTS2}
A^{\Downarrow  (1)}_{\rm TS}\propto \frac{t}{4 M^2}
\left[F_2(t) \widetilde H(\xi,\xi,t,{\cal Q}^2)
- F_1(t) \xi\widetilde E(\xi,\xi,t,{\cal Q}^2)
+ \cdots \right]
\end{eqnarray}
Unfortunately, compared to the single beam spin (\ref{KLMSPM-A_BS1}) and longitudinal target
(\ref{KLMSPM-A_LTS}) asymmetries, the transversally ones are kinematically suppressed 
by an additional factor $\sim \sqrt{-t}/(2M)$ and, for a neutron target, in addition by the
Dirac form factor $F_1(t)$.

Although the given formulae (\ref{KLMSPM-A_BS1}--\ref{KLMSPM-A_TTS2}) are rather crude, they illustrate that a measurement of single spin asymmetries would
allow to access the imaginary part of the four twist-two related CFFs. 
However, the normalization of these asymmetries depends to some extent also on the real part of the twist-two related CFFs and the remaining eight ones. Measurements of cross section differences would allow one to eliminate the normalization uncertainty, and in combination with the harmonic analysis, one can separate to some extent twist-two, twist-three, and gluon transversity contributions. However, the  extracted harmonics might also be contaminated by DVCS cross section contributions which are bilinear in the CFFs.
To get rid of these admixtures, one needs cross section measurements with a positron beam.
Forming differences and sums of cross section
measurements with both kinds of leptons, allows one to extract the pure interference and DVCS squared terms and, thus,  might allow one to quantify twist-three effects.
Existing data indicate that these effects are small as expected based on
kinematic factors. However, even obtaining only an upper limit is
important for the determination of the systematic uncertainties
of twist-two CFFs.

\begin{figure}[t]
\begin{center}
\includegraphics[width=16.0cm]{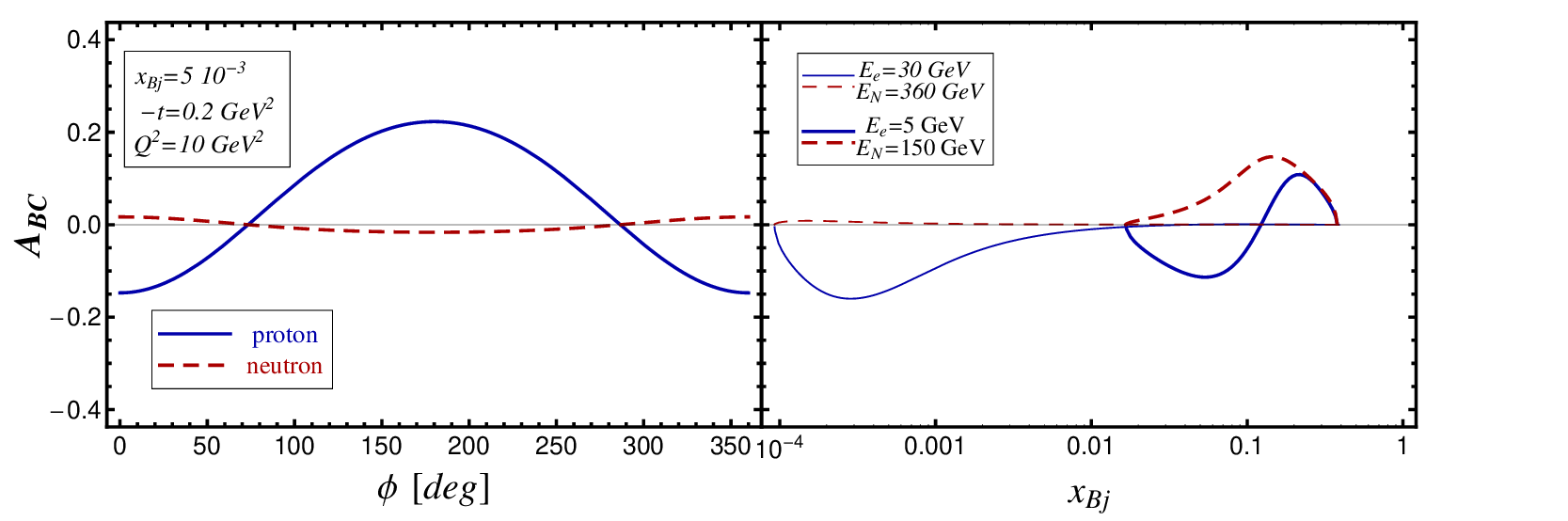}
\end{center}
\vspace{-0.7cm}
\caption{\small
{\em KM10b} model estimate for the DVCS beam charge asymmetry 
with a proton (solid) and neutron (dashed) target.
Left panel: $A_{\rm BC}$ {\it vs.}~$\phi$ for $E_N=250~{\rm GeV}$, $E_e=5~{\rm GeV}$, 
$x_{B}=5\times 10^{-3}$, $Q^2 = 10~{\rm GeV}^2$, and $t=-0.2~{\rm GeV}^2$. Right panel: Amplitude $A_{\rm BC}^{(1)}$ of the first harmonic {\it vs.}~$x_{B}$ at $t=-0.2~{\rm GeV}^2$ for small $x_{B}$ (thin)
[$E_e=30~{\rm GeV}$, $E_p=360~{\rm GeV}$, $Q^2=4~{\rm GeV}^2$] and large $x_{B}$ (thick)
[$E_e=5~{\rm GeV}$, $E_p=150~{\rm GeV}$, $Q^2=50~{\rm GeV}^2$]  kinematics.
\label{KLMSPM-Fig_BCA}}
\end{figure}
We also emphasize that having both kinds of lepton beams available allows one to measure the real part of CFFs. In figure~\ref{KLMSPM-Fig_BCA}, we show
the beam charge asymmetry,
\begin{eqnarray}
\label{KLMSPM-A_BC1}
A_{\rm BC}^{(1)}\propto
\Re{\rm e}\left[F_1(t) {\cal H}(x_{\rm Bj},t ,{\cal Q}^2)
-\frac{t}{4 M^2} F_2(t) {\cal E}(x_{\rm Bj},t ,{\cal Q}^2)
+ \cdots\right]\,,
\end{eqnarray}
for an unpolarized target, which is expected to be sizeable. For a proton target, this asymmetry  should possess a node in the transition from the valence to sea region(thick solid curve, right panel). In our parameterization, the real part of the $\cal E$ CFF is determined by the
$\cal D$ subtraction term, which induces a sizeable asymmetry (thick dashed curve, right panel), even for a neutron target.

The large kinematical coverage of the proposed high-luminosity EIC raises the question: Can one utilize evolution, even at moderate $x_{B}$ values, to access GPDs away from their cross-over line?
\begin{figure}[t]
\begin{center}
\includegraphics[width=15.0cm]{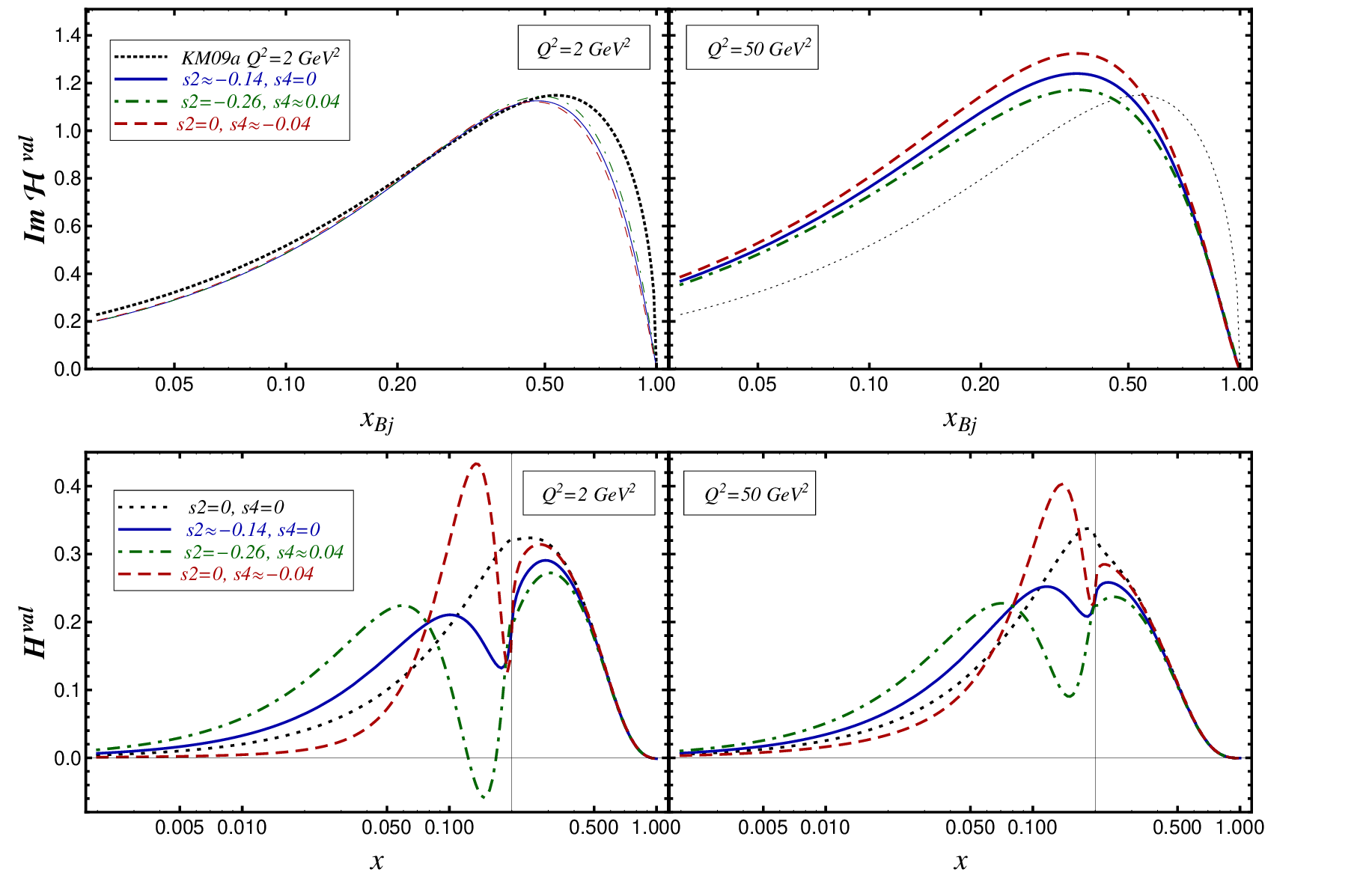}%
\vspace{-0.7cm}
\end{center}
\caption{\small Upper left panel: The valence-like contribution (\ref{KLMSPM-ansHval}) to the CFF ${\cal H}$ extracted with a ``dispersion-relation" fit {\em KM09a} from fixed target
measurements (dotted) at $t=-0.2~{\rm GeV}^2$ and $Q^2=2~{\rm GeV}^2$ {\it vs.}~$x_{B}$ 
together with various models. Lower left panel: The corresponding models of the GPD $x H(x,\eta,t,{\cal Q}^2)$ together with a minimalist GPD parameterization (dotted curve) 
{\it vs.}~$x$ at $\eta=0.2$, $t=-0.2~{\rm GeV}^2$, and $Q^2=2~{\rm GeV}^2$. 
The same quantities at $Q^2 = 50~{\rm GeV}^2$ are displayed in the right panels.
\label{KLMSPM-fig-CFF}
}
\end{figure}
Similarly to what has been done for the small $x_{B}$ region, 
we use the Mellin-Barnes integral technique to address the problem. Taking different
non-leading SO(3) partial waves in the ansatz for the conformal moments (\ref{KLMSPM-mod-nnlo},\ref{KLMSPM-mod-nnlo1}), we build three different GPD models for valence quarks that provide almost identical CFFs, see the upper left panel in figure~\ref{KLMSPM-fig-CFF}.  
They are compatible with 
(\ref{KLMSPM-ansHval}) from the "dispersion-relation" fit {\em KM09a} (dotted curves). 
We note that the different model behavior at large $x_{B}$ results only in a small 
discrepancy for the real part of the CFF in the kinematics of interest.
In the lower left panel of figure~\ref{KLMSPM-fig-CFF}, we illustrate that for fixed $\eta$, the $x$-shape of the three GPD models looks quite differently. Compared to the minimalist model (dotted curve), a model with a negative next-to-leading partial wave (solid) decreases the 
size of the GPD on the cross-over line $\eta=x$ and generates an oscillating behavior in the central region. The model with an alternating-sign SO(3) partial wave expansion (dash-dotted) possesses more pronounced oscillation effects in the central region or even nodes. In the third model (dashed curve), the reduction on the cross-over line is reached within a next-to-next leading SO(3) partial wave. Note that the GPDs in the region $\eta \ll x$ are governed by the $x$-behavior of the PDF analogues. In the right panels, we demonstrate that for a large lever arm in $Q^2$ (e.g., $Q^2=50~{\rm GeV}^2$), the evolution effects 
are important in the valence quark region. However, for CFFs (the upper right panel), the discriminating  power of evolution effects remains moderate even if the GPD shapes look rather different.

\subsection{Conclusions and summary}

With all the theoretical tools sketched above plus those which are presently
under development, it is clear that our understanding of hadron structure will be
revolutionized once most of the diverse asymmetries are measured with
percent or permille precision (depending on the observable).
At present, first steps have been undertaken to access GPDs from
experimental data in the small $x_{B}$ region and in the fixed target kinematics
providing us with some insight into the GPD $H$. 
In particular, for DVCS in the fixed target kinematics, LO model fits 
are compatible with least-square CFF fits and first results from neural networks 
(assuming $H$ dominance).  The large uncertainties in extracting CFFs
are mainly related to the lack of experimental data. Thus, not only the extraction
of the very desired $\cal E$ playing an important role
in the ''spin-puzzle'', but also of other CFFs,
requires a comprehensive measurement of all possible observables
in dedicated experiments.
A further comparison  
shows that while in the valence region the
extracted quark GPDs are somewhat different, 
they become compatible for small $x$. 
The main difference lies in the gluonic sector;  
a more appropriate
analysis 
requires the inclusion of radiative corrections in a
global fitting procedure, which is in progress. We should also mention here
that hard exclusive processes with
nuclei, which at present are not extensively studied, open a new window 
for the partonic view of nuclei. 

Imaging the partonic content of the nucleon and the
phenomenological  access to the proton spin sum rule from hard exclusive processes can only be reached through proper understanding of GPD models. 
We also point out that GPDs can also be formulated in terms of an
effective nucleon (light-cone) wave function, which links GPDs to transverse momentum dependent parton distributions.
The whole framework consisting of perturbative QCD, lattice simulations, and dynamical modeling 
is available to reveal GPDs  and access the nucleon wave function. 
Such a unifying description can be considered as the primary
goal in quantifying the partonic picture. While such a task looks rather straightforward, 
much effort is needed on the theoretical, phenomenological,
and experimental sides, with experimental data with small uncertainties playing the key role.
A high-luminosity EIC is an ideal machine that would cover a wide kinematical range and
complement the planned fixed target experiments at JLab@12 GeV.
Thus, besides new measurements, an EIC has a great potential to significantly improve existing data sets.


\noindent
{\it Acknowledgments}. We are grateful to P. Kroll for many fruitful discussions.


\section{Monte Carlo studies on DVCS with an EIC}
\label{sec:fazio}


\hspace{\parindent}\parbox{0.92\textwidth}{\slshape 
  Salvatore Fazio}
%


\index{Fazio, Salvatore}





\subsection{Exclusive processes with a dedicated EIC detector}

Our current knowledge of the role of gluons in hadronic matter comes mainly from DIS 
experiments of electrons off protons most notably from HERA at DESY. Although electrons 
only interact with electrically charged particles, and gluons carry only color charge, a high-energy 
electron beam can still be used as an excellent gluon microscope.  


The HERA physics program of $ep$ collisions surprisingly showed a large fraction
of diffractive events contributing $10-15~\%$ to the total DIS cross-section. 
One of the key signatures of
these ``diffractive'' events is an intact proton traveling at nearly beam energies, together with a gap
in rapidity before some final-state particles are produced at mid-rapidity. However, the detectors
(H1 and ZEUS) were not optimized for this important physics and were unable to measure the
scattered proton; this was only achievable after a program of upgrades.
In fact, to measure
diffractive physics events, it is desirable to have very forward detectors at small angles with respect to the
beam line, referred to as ``Roman Pots''.
Other requirements are that the detector should be able to measure all processes: 
inclusive ($ep\rightarrow e^{\prime}X$), semi-inclusive ($ep\rightarrow e^{\prime}X$+hadrons), 
and exclusive (e.g., $ep\rightarrow e^{\prime}p+J/\psi$) reactions. 
The requirements for $ep$ and $eA$ collisions are very similar, the only additional
complication in $eA$ collisions arises from the need to tag the struck nucleus, or to veto events with the nucleus break-up by detecting neutrons and other breakup products with high efficiencies.

Briefly, a possible EIC detector 
consists (in the barrel region at mid-rapidity) of a solenoidal field with Si tracking with full rapidity coverage
around the interaction point itself, followed by Cherenkov detectors for particle identification and
then by both electro-magnetic and hadronic calorimeters. There are further trackers and calorimeters
at forward rapidities, but this time, the magnetic field is a dipole. 
A Roman Pots spectrometer can be installed along the beam-pipe in the direction of the outgoing proton. 
This is just a starting point and other technologies are under active investigation.

\subsection{DVCS and GPDs: from HERA to an EIC}

Measurements of observables associated with hard exclusive processes at an $ep/eA$ collider requires substantially higher luminosities than traditional inclusive DIS because of the small cross sections
and the need for differential measurements. The detectors and the interaction region
have to be designed to permit full reconstruction of the final state. 
\begin{figure}[t]
\centering
\includegraphics[width=0.9\textwidth,angle=0]{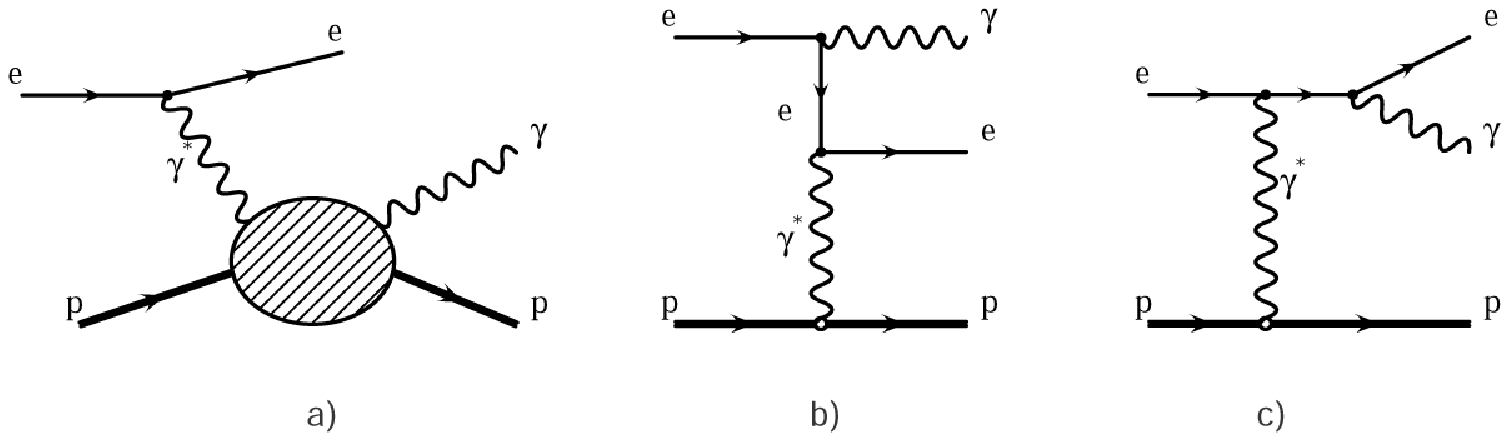}
\caption{\small Diagram of the DVCS (a) and BH processes for a photon emitted from the initial (b) and final (c) lepton line.}
\label{topo_dvcs_bh}
\end{figure}

In assessing the prospects for measurements of exclusive processes in $ep$ scattering at
collider energies, $W^2 \gg 10$~GeV$^2$, one needs to distinguish between ``diffractive'' (no
exchange of quantum numbers between the target and the projectile/produced
system) and ``non-diffractive'' processes (exchange of quantum numbers). In
diffractive channels, such as $J/\psi$, $\rho$, $\phi$ production and DVCS 
(production of a real photon), the
cross sections rapidly rise with the collision energy, $W$. At large $Q^2$, these processes
probe the gluon GPD and/or the singlet quark GPD. 
In non-diffractive channels, such as $\pi^\pm$, $\pi^0$, $\rho^+$, $K$ production, 
the cross sections 
decrease with energy. These processes at high $Q^2$ probe the flavor/charge/spin non-singlet quark GPDs describing the quark structure of the target.

The final state of a DVCS event, shown in figure~\ref{topo_dvcs_bh}a, contains one
track and two electromagnetic clusters together with a proton scattered at
a very small angle. The technique for measuring DVCS consists of first
extracting a data sample of events characterized by the DVCS topology.
Apart from the DVCS process, the data selection comprises also Bethe-Heitler (BH) events, 
a well known QED process, because DVCS and BH share the same final state.
The selected sample will contain a mixture of DVCS and BH contributions.
For not too large $y$ (see figure~\ref{bhfrac}), the BH contribution is not
much larger than DVCS and thus can be subtracted from the data
using
the MC predictions and control data samples containing BH only
since at high enough $Q^2$ the interference contribution
drops out to a good accuracy when  averaging over the angle $\phi$ between the production and scattering planes.

The differential cross section as a function of $|t|$ can be parameterized by an exponential: $d\sigma/dt \propto e^{b|t|}$. The H1 Collaboration~\cite{:2009vda} measured $|t|$ from the transverse momentum distribution of the photon and studied the $b$-slope in a few bins in $Q^2$ and $W$.
The slope $b$ seems to decrease with $Q^2$ up to the value expected for a hard process but it does not depend on $W$. The ZEUS Collaboration~\cite{Chekanov:2008vy} performed a direct measurement of the proton final state using a Roman Pots spectrometer: 
the resulting $b=4.5\pm 1.3\pm 0.4$~GeV$^{-2}$ at $Q^2=3.2$~GeV$^2$ and $W=104$~GeV is consistent, within the large uncertainties due to the low acceptance of the spectrometer, with the H1 result of $b=5.45\pm 0.19 \pm 0.34$~GeV$^{-2}$ at $Q^2=8$~GeV$^2$ and $W=82$~GeV~\cite{:2009vda}.

A comprehensive program of parton imaging in the nucleon would need precise measurements of $b$ for wide range of $x_{B}$ values, $10^{-4} < x_{B} < 10^{-1}$; this is currently beyond the possibilities of any experiment. Building an EIC with a properly designed detector could finally make it 
possible --- a preliminary feasibility study is reported in section~\ref{sec:bspace}.

The beam-charge asymmetry (BCA) provides an access to the real part of the DVCS amplitude through the interference between the DVCS and BH amplitudes (for illustration, we keep only the dominant $\cos(\phi)$ harmonic):
\begin{equation}
A_C= \frac{\frac{d\sigma^+}{d|t|}-\frac{d\sigma^-}{d|t|}}{\frac{d\sigma^+}{d|t|}+\frac{d\sigma^-}{d|t|}}
 = p_1 \cos(\phi) \propto 2{\cal A}_{\rm BH}\frac{\Re e ({\cal A}_{\rm DVCS})}{|{\cal A}_{\rm DVCS}|^2+|{\cal A}_{\rm BH}|^2} \cos(\phi) \,. 
\label{eq:fazio:AC}
\end{equation}
The measurement of $A_C$ is complementary to the measurement of the $|t|$ distribution.
%
The DVCS beam-charge asymmetry has been measured by the 
H1~\cite{:2009vda} and HERMES~\cite{Airapetian:2006zr,Airapetian:2009bi} experiments. 

The large rapidity acceptance and high precision tracker of the EIC detector together with its very accurate electromagnetic calorimeter and the high luminosity of the machine, make it an ideal tool for the measurement of both the DVCS cross section differential in $|t|$ and the DVCS+BH cross section asymmetries. Specifically for the BCA, a positron beam would be required.

\subsection{Monte Carlo simulations of DVCS at an EIC}

The Monte Carlo generator used for our studies is MILOU~\cite{Perez:2004ig}, which simulates both the DVCS and the BH processes together with their interference term. 
DVCS is simulated using the framework of GPDs at next-to-leading order (NLO) accuracy,
including the NLO evolution of GPDs~\cite{Belitsky:2001ns}. 
The $t$ dependence is introduced as an exponential $d\sigma /dt \propto e^{B(Q^2)|t|}$, where 
$B(Q^2)$ is either a constant or can have a weak logarithmic dependence on $Q^2$, $B(Q^2)\sim \mathrm{ln}(Q^2)$. In the present simulation, we used the former option, $B(Q^2)=5$ GeV$^{-2}$.
In addition, the proton dissociation background, $ep\rightarrow e\gamma Y$, has not been included.

The DVCS and BH processes have been simulated in the following kinematic range:
\begin{itemize}
\item$Q^2 \ge 1$~GeV$^2$;
\item $10^{-4}<x_{B}<10^{-1}$;
\item $0.01<y<0.85$;
\item $0.01<|t|<1$~GeV$^2$.
\end{itemize}
The $Q^2$ and $x_{B}$ ranges correspond to the phase space achievable with an 
EIC;
the lower $y$ limit is chosen according to the acceptance of the detector;  the interval in $|t|$ relates to the acceptance of a forward proton spectrometer. The energy configuration considered for the present study is a 5-20 GeV electron beam colliding with a  250~GeV proton beam.  
\begin{figure}[htbp]
\centering
\includegraphics[width=0.7\textwidth,angle=0]{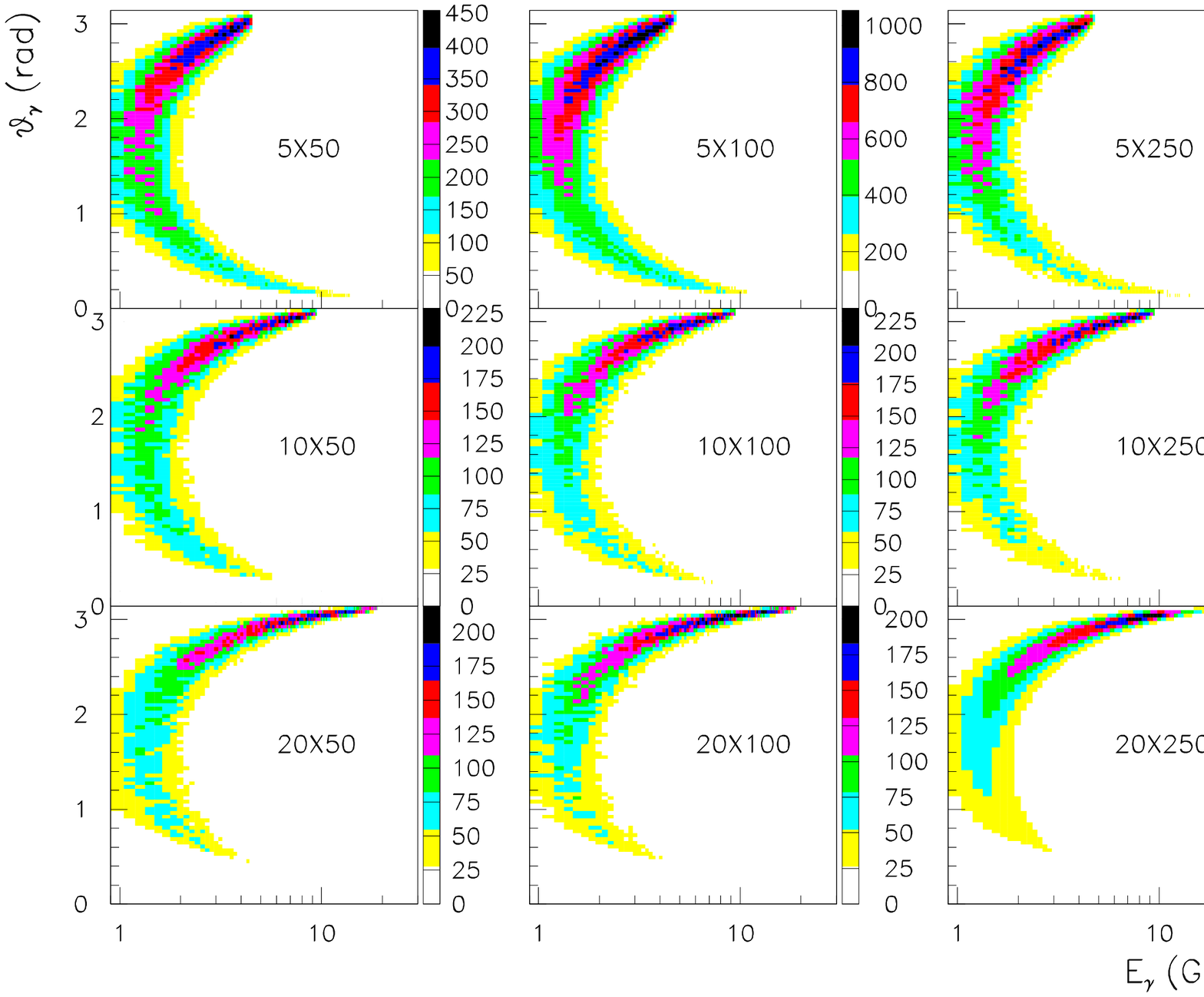}
\caption{\small The angle of the produced real photon in a DVCS event as a
  function of the photon energy for different EIC energy configurations.
  Each plot shows also the distribution of the photon and scattered
  electron energies.}
\label{angle_ene_gamma}
\end{figure}

\begin{figure}[htbp]
\centering
\includegraphics[width=0.95\textwidth,angle=0]{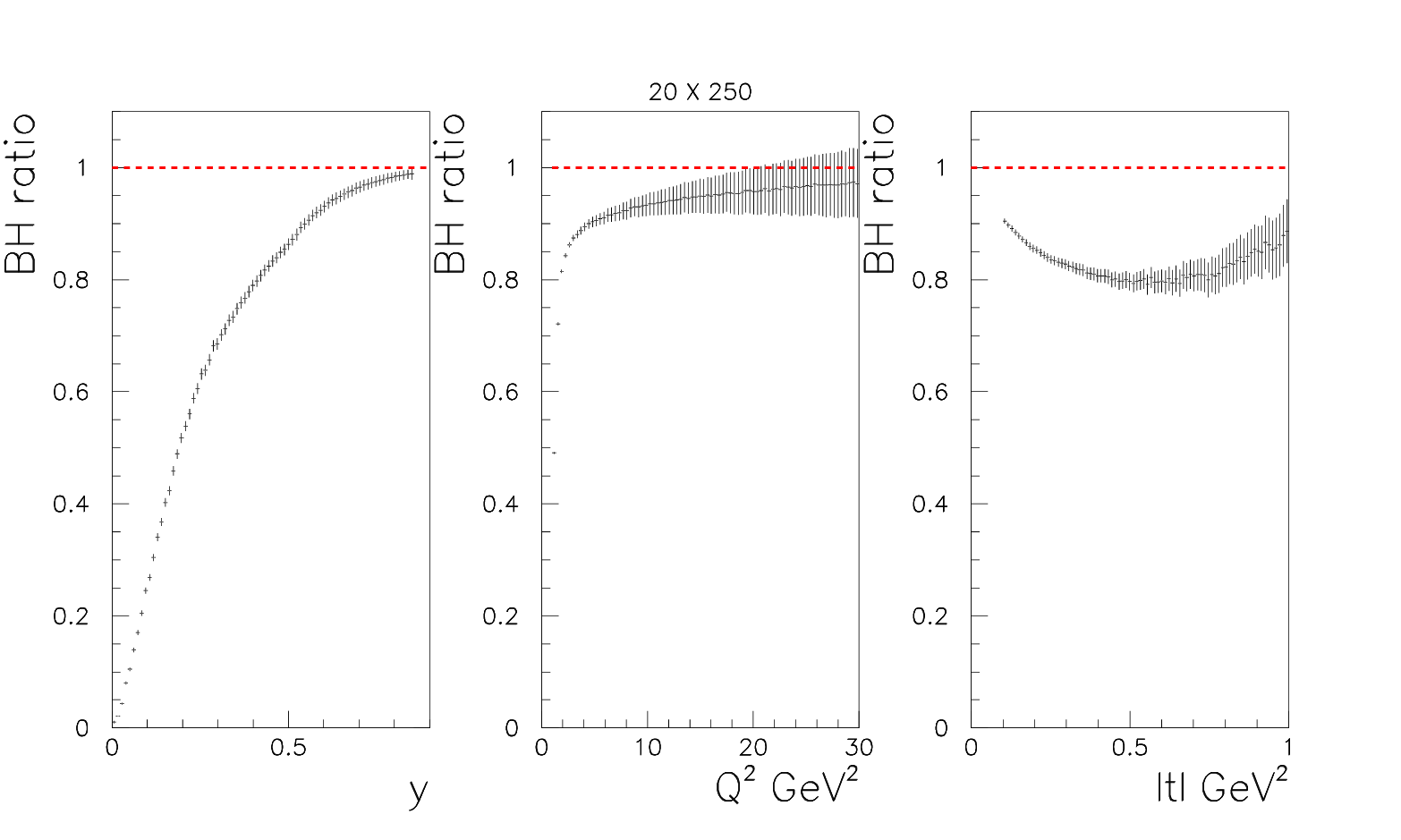}
\caption{\small The fraction of BH events in the $ep\rightarrow ep\gamma$
  sample as a function of $y$ (left), $Q^2$ (middle), and $|t|$ (right).}
\label{bhfrac}
\end{figure}

Figure~\ref{angle_ene_gamma} shows the correlation between the scattering angle of the real photon produced in the interaction and its energy 
for 
different EIC energy configurations.
 For the $20\times 250$ configuration, the photons with an energy greater than $\sim5$~GeV are produced backward at an angle larger than 2.7~rad, corresponding to the rear end-cap calorimeter in the detector. Since for the DVCS process the electron is always scattered backward, this can lead to problems in discriminating the photon and electron clusters and makes it crucial to have an 
electromagnetic calorimeter with high spatial resolution and a good tracker coverage at backward rapidity to measure the electron track. Indeed, this is extremely important for the $t$ resolution in the case of a measurement performed without a Roman Pots because the four-momentum $t$ must be reconstructed, using momentum conservation, from the transverse momenta of the electron and the photon.
\begin{figure}[htb]
\centering
\includegraphics[width=0.47\textwidth,angle=0]{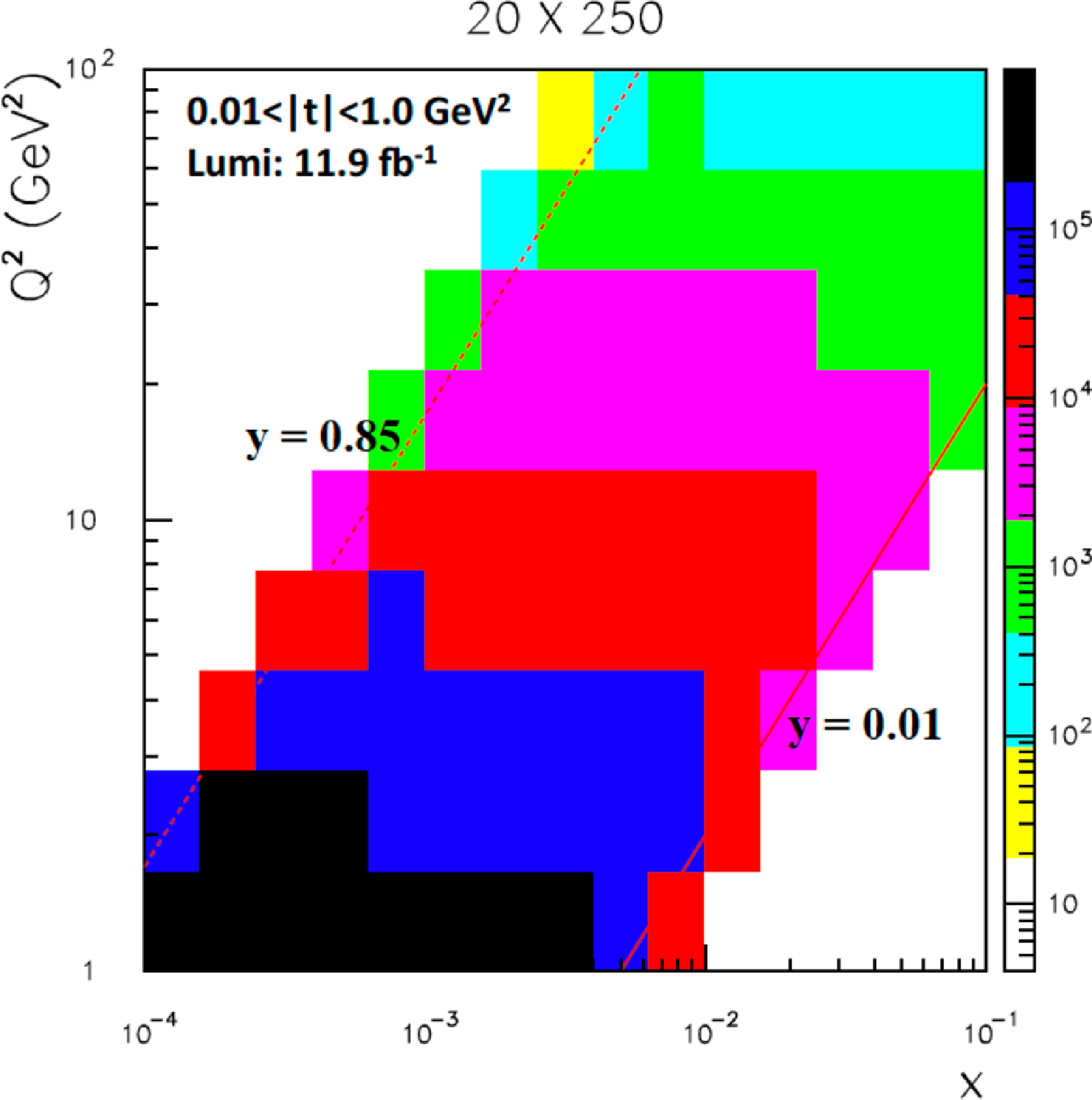}
\includegraphics[width=0.47\textwidth,angle=0]{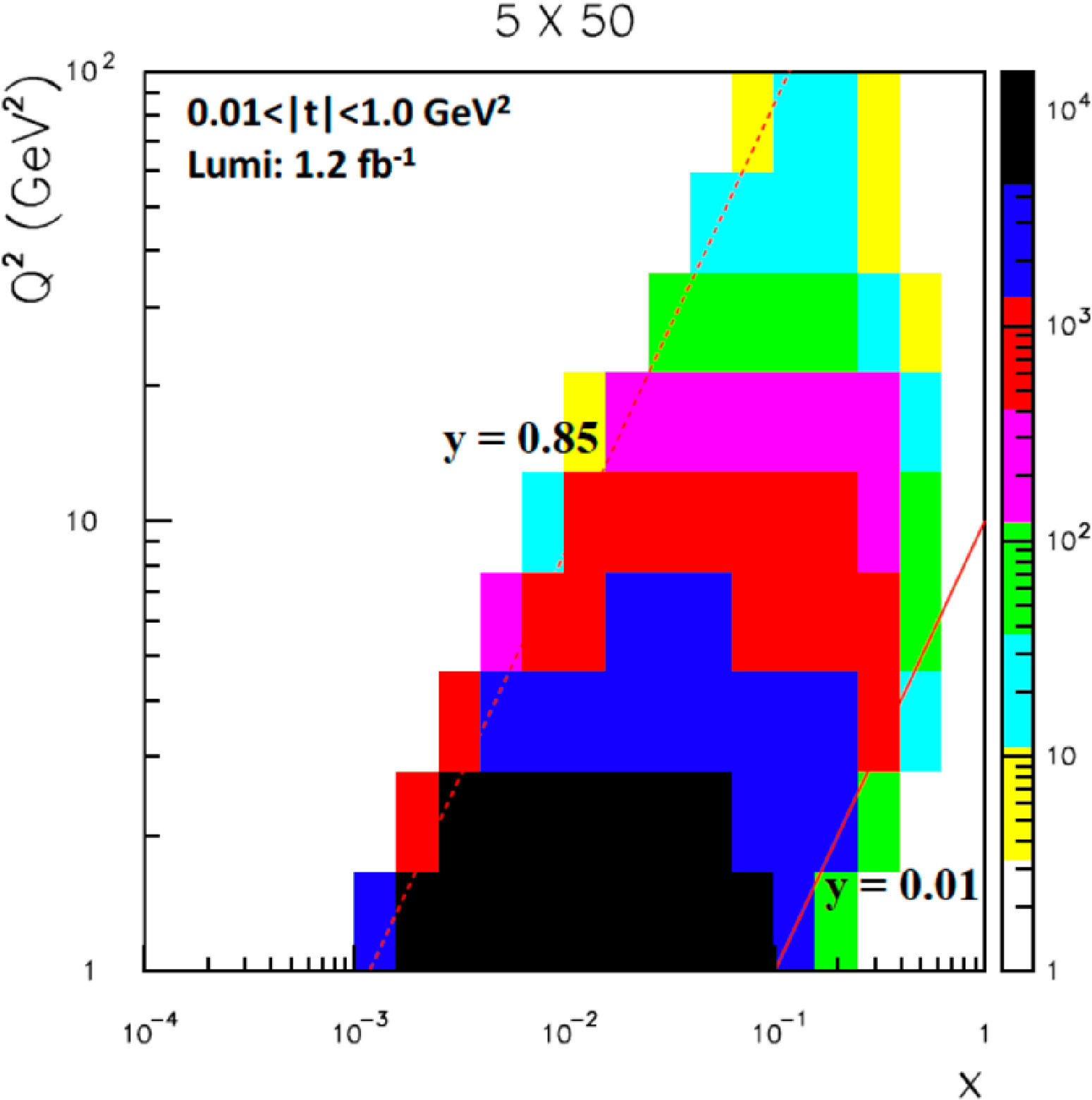}
\caption{\small The distribution of the DVCS events in a logarithmically
  binned phase space for the $20\times250$ (left) and $5\times50$ (right)
  EIC beam energy configurations.}
\label{phase_space}
\end{figure}

The fraction of BH events has been estimated using an MC sample containing both DVCS and BH processes.  The samples have been normalized to the luminosity. The fraction of BH events is calculated as follows:
\begin{equation}
F_{BH}=\frac{BH_{evt}}{BH_{evt}+DVCS_{evt}} \, .
 \label{eq:bhfrac}
\end{equation}
Figure~\ref{bhfrac} shows the fraction of BH events as a function of $y$, $Q^2$, and $|t|$. As expected, DVCS is dominant at low 
$y$ whereas BH dominates at higher 
$y$ with its fraction increasing up to $100\%$ for $y>0.85$. 

In the present study, the kinematic domain has been binned logarithmically in $1<Q^2<100$~GeV$^2$ and $10^{-4}<x_{B}<10^{-1}$. Figure~\ref{phase_space} shows the distribution of the statistics per bin
for the $20\times 250$ configuration (left panel) and  $5\times 50$ configuration (right panel).

As an example of the precision that could be achieved at 
an EIC,
figure~\ref{dvcs} shows the expectations for a measurement of the DVCS $ep$ cross section differential in $|t|$, 
$\mathrm{d}\sigma_{ep\rightarrow ep\gamma}/\mathrm{d}t$, for several bins of $x_{B}$ and $Q^2$.
The estimated luminosity for the  $20\times 250$ configuration is $10^{34}$~cm$^{-2}$s$^{-1}$.
The integrated luminosity of the simulated events is 11.9~fb$^{-1}$, corresponding to approximately one month of running at $20\times 250$ and 
assuming $50\%$ operational efficiency. The $b$ slope parameter with its uncertainty is extracted for each data set via a an exponential fit $\sim e^{-b|t|}$, and its value is reported in figure~\ref{dvcs} together with uncertainties. 

One can see that an excellent measurement (binning over a wide range in $Q^2$ and $x_{B}$) 
can already be obtained with a relatively modest beam time, allowing for numerous detailed studies of the reaction mechanism ($Q^2$-scaling
behavior, QCD evolution) and extraction of information about the nucleon GPDs and its change with $x_{B}$. 
The statistical uncertainty for the differential cross section can be, at small $|t|$ values, significantly below $1\%$, as well as the uncertainty on the extracted slope parameter, $b$. This implies that the measurement is actually limited by systematics. Thus the utilization of a high resolution spectrometer based on the Roman Pots technique becomes important for an EIC. 
For example,  the leading proton spectrometer based on 6 Roman Pots stations equipped with silicon micro-strips  detectors used at ZEUS for the DVCS $d\sigma/d|t|$ measurement, allowed to measure $P_t$ with a resolution of 5~MeV~\cite{Chekanov:2008vy} under test beam conditions which corresponded 
to $\Delta(P_t^2)=|\Delta t|=10^{-2}P_t$ for a $|t|$ measurement. A new properly designed 
Roman Pots spectrometer, potentially based on a radiation-hard silicon pixel technology, could reach a geometrical acceptance of about 60\%, with a better $P_t$ resolution.  Since for an EIC 
systematics are the challenge, it is worth sacrificing the acceptance and therefore increasing the beam time for a more accurate measurement.

Figure~\ref{dvcs_ht} shows the expectations for a DVCS measurement for large-$|t|$. 
The data have been simulated for $1 <|t|<2$ GeV$^2$ in several bins of $x_{B}$ and $Q^2$. 
The luminosity of the simulated sample is 151~fb$^{-1}$ corresponding to approximately 52 weeks of data taking in the $20\times 250$ configuration. One can see that even if the cross section drops drastically for large $|t|$ values, 
the EIC 
still allows for good binned measurements, 
but this requires years of data taking. (For a relevant discussion, see section~\ref{sec:bspace}.) 
In this regime, the main detector offers a much better acceptance then Roman Pots and can be used for measuring $|t|$.

\begin{figure}[htbp]
\centering
\includegraphics[width=0.8\textwidth,angle=0]{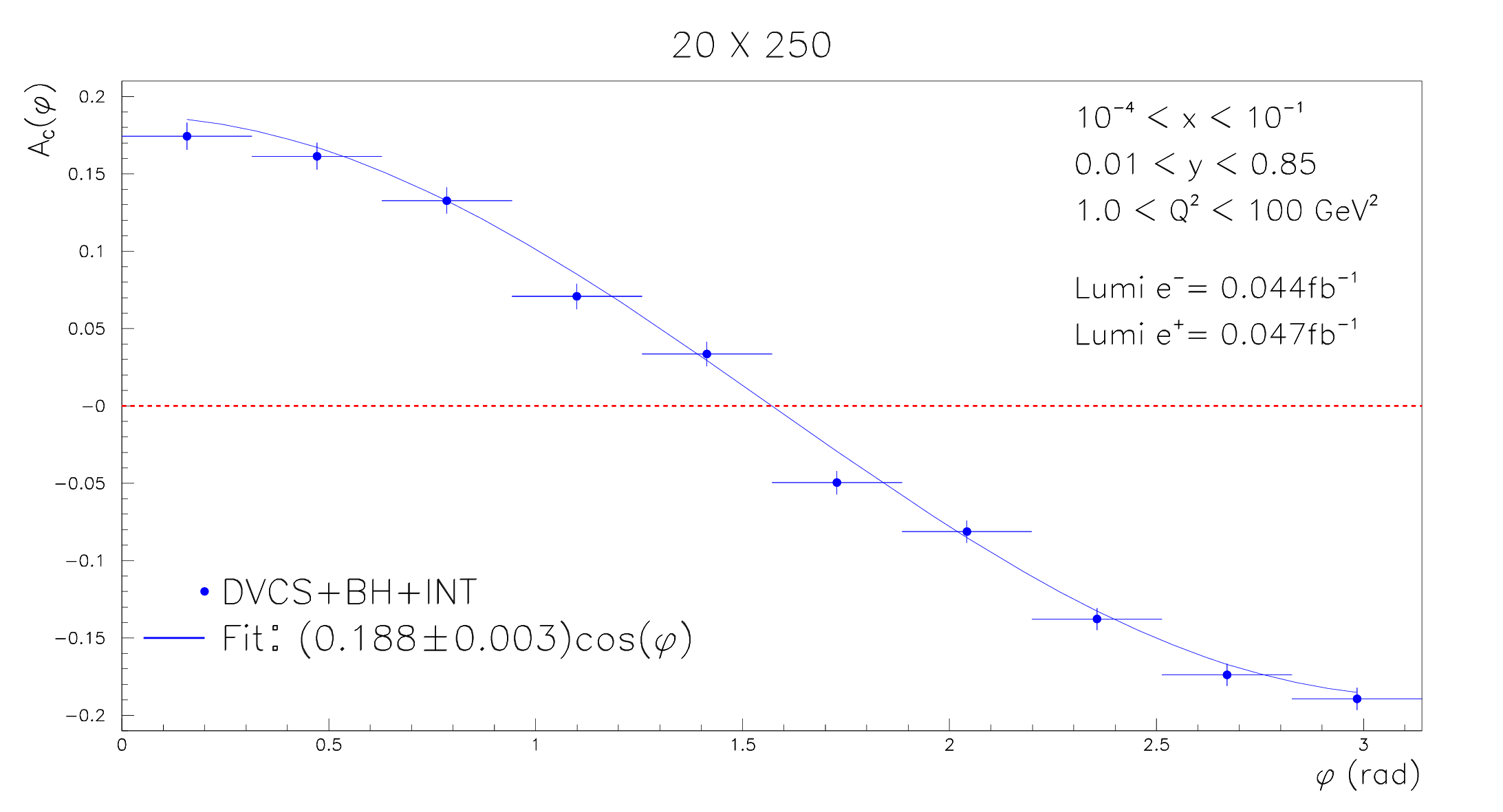}
\caption{\small The beam-charge asymmetry $A_C$ as a function of the azimuthal angle $\phi$ between the production and scattering planes.}
\label{bca}
\end{figure}

\begin{figure}[htbp]
\centering
\includegraphics[width=1.1\textwidth,angle=0]{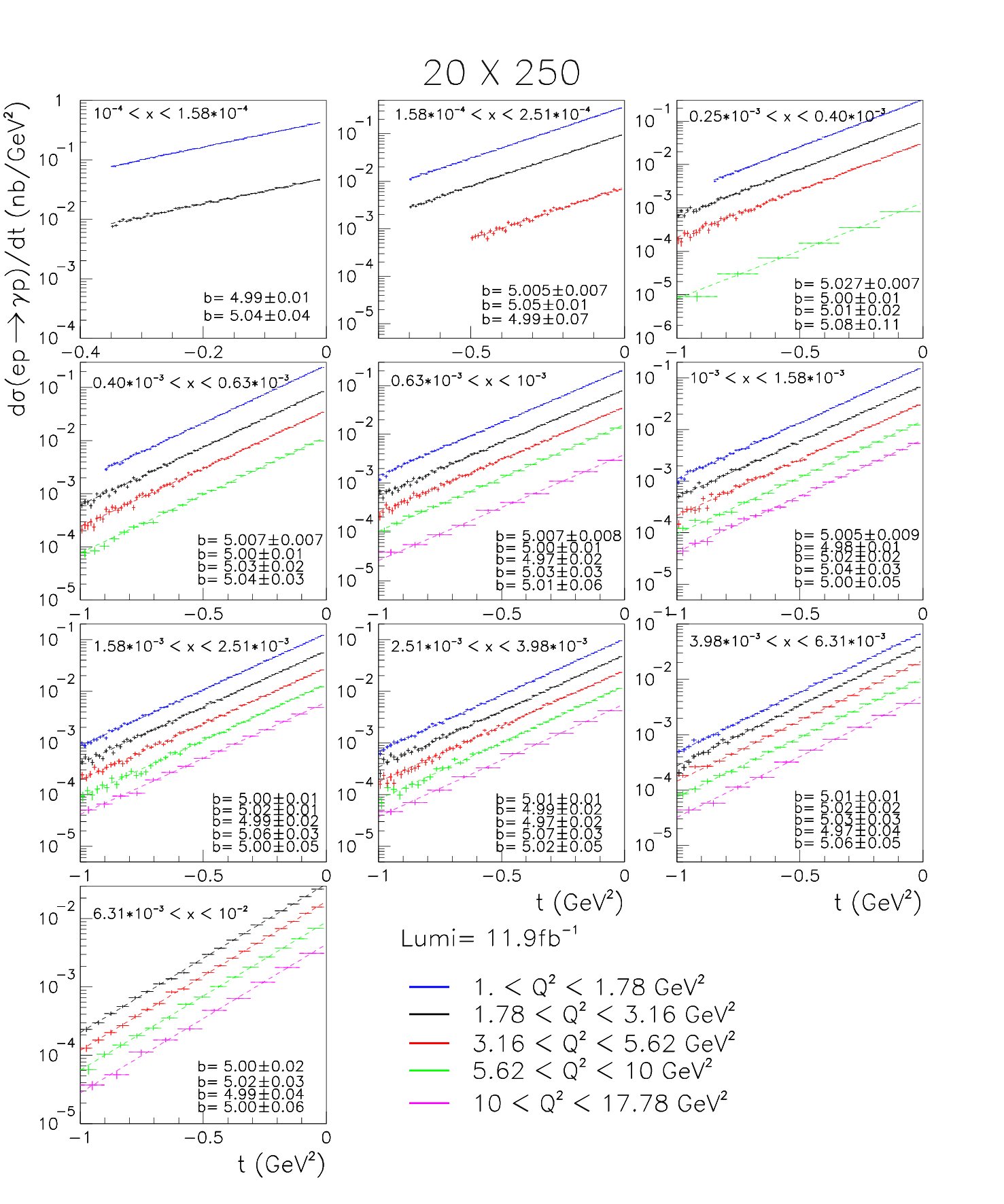}
\caption{\small {The DVCS cross section has been simulated in the range $1.0<Q^2<100~GeV^2$, $10^{-4}<x_{B}<0.1$ for the $20\times250~GeV$ energy configuration. The DVCS cross section is simulated in several bins of $x_{B}$ and $Q^2$ and is shown for small $|t|$ values.}}
\label{dvcs}
\end{figure}

\begin{figure}[htbp]
\centering
\includegraphics[width=1.1\textwidth,angle=0]{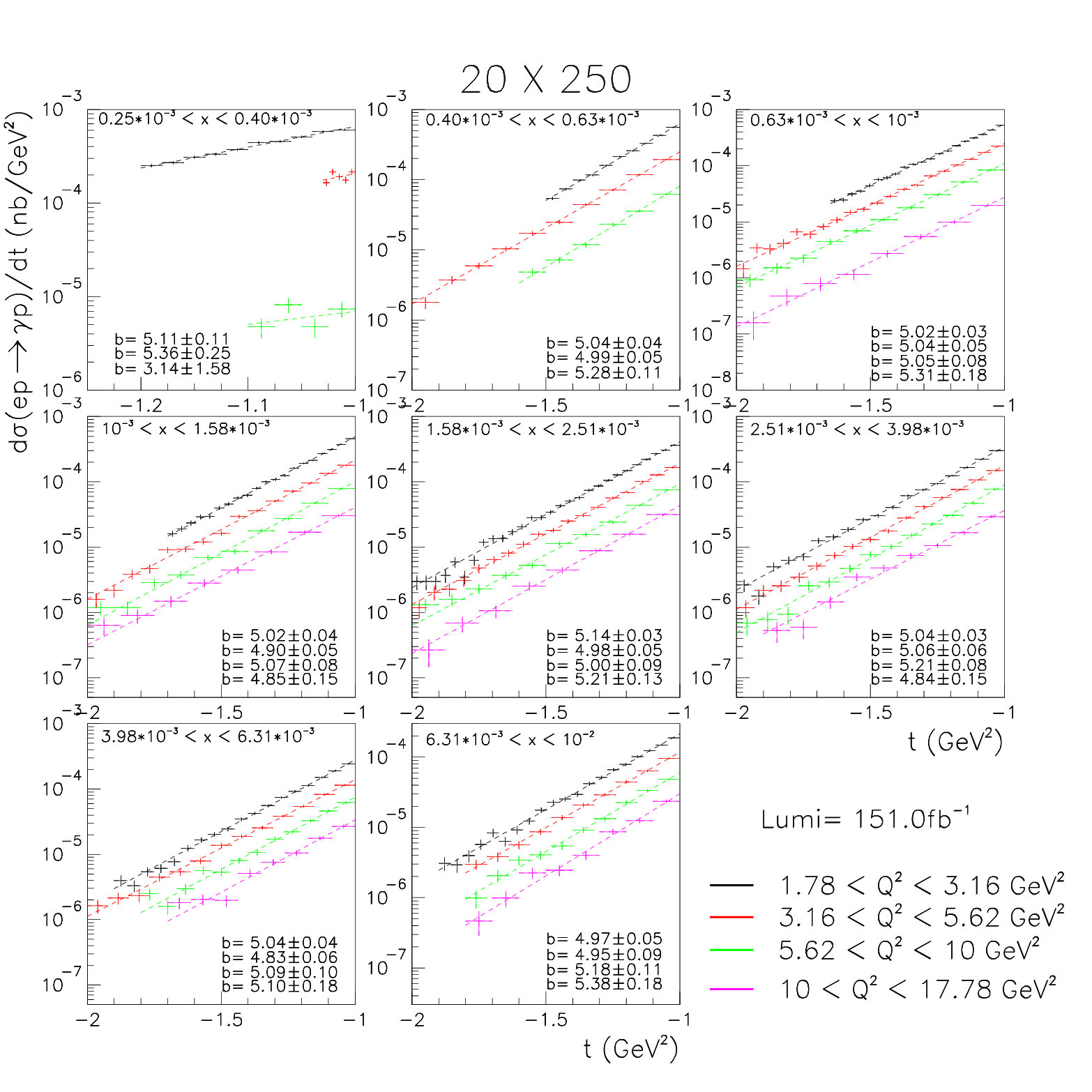}
\caption{\small {The DVCS cross section has been simulated in the range $1.0<Q^2<100~GeV^2$, $10^{-4}<x_{B}<0.1$ for the $20\times250~GeV$ energy configuration. The DVCS cross is section simulated in several bins of $x_{B}$ and $Q^2$  and is shown for large $|t|$ values.}}
\label{dvcs_ht}
\end{figure}

A data sample containing DVCS, BH and their interference term has been simulated considering 
separately an electron beam (luminosity is 44~pb$^{-1}$) and a positron beam (luminosity is 47~pb$^{-1}$) and used to calculate the beam-charge asymmetry, $A_C$. 
The result is shown in figure~\ref{bca} together with a fit in the form 
$A_C=p_1\cos(\phi)$~(\ref{eq:fazio:AC}), where 
$p_1$ is a free parameter. One can see that a fair accuracy for the BCA can be obtained at 
an EIC 
for a modest integrated luminosity.

\section{DVCS Beam Spin Asymmetries with an EIC}
\label{sec:dvcsbsa}


\hspace{\parindent}\parbox{0.92\textwidth}{\slshape 
R.~G\'eraud, H.~Moutarde, F.~Sabati\'e}
%

\index{Sabati\'e, Franck}
\index{Moutarde, Herv\'e}
\index{G\'eraud, R\'emi}





\subsection{Deeply Virtual Compton Scattering polarization observables}

The photon electroproduction $e p \to e p \gamma$ can either occur by radiation along
one of the electron lines (Bethe-Heitler or BH) or by emission of a real
photon by the nucleon (Deeply Virtual Compton Scattering or DVCS). The
total cross section as given by~\cite{Belitsky:2001ns} reads:
\begin{equation}
\frac{\mathrm d\sigma^{e p \rightarrow e p \gamma}}{\mathrm dx_B dy \,\mathrm d\Delta^2 \,\mathrm d\phi \,\mathrm d\varphi} =
\frac{\alpha^3 x_B y}{16 \pi^2 Q^2 \sqrt{1+\epsilon^2}}
\left | \frac{\mathcal T}{e^3} \right |^2 \, ,
\end{equation}

\begin{figure}[ht]
\centerline{\includegraphics[width=8.5cm]{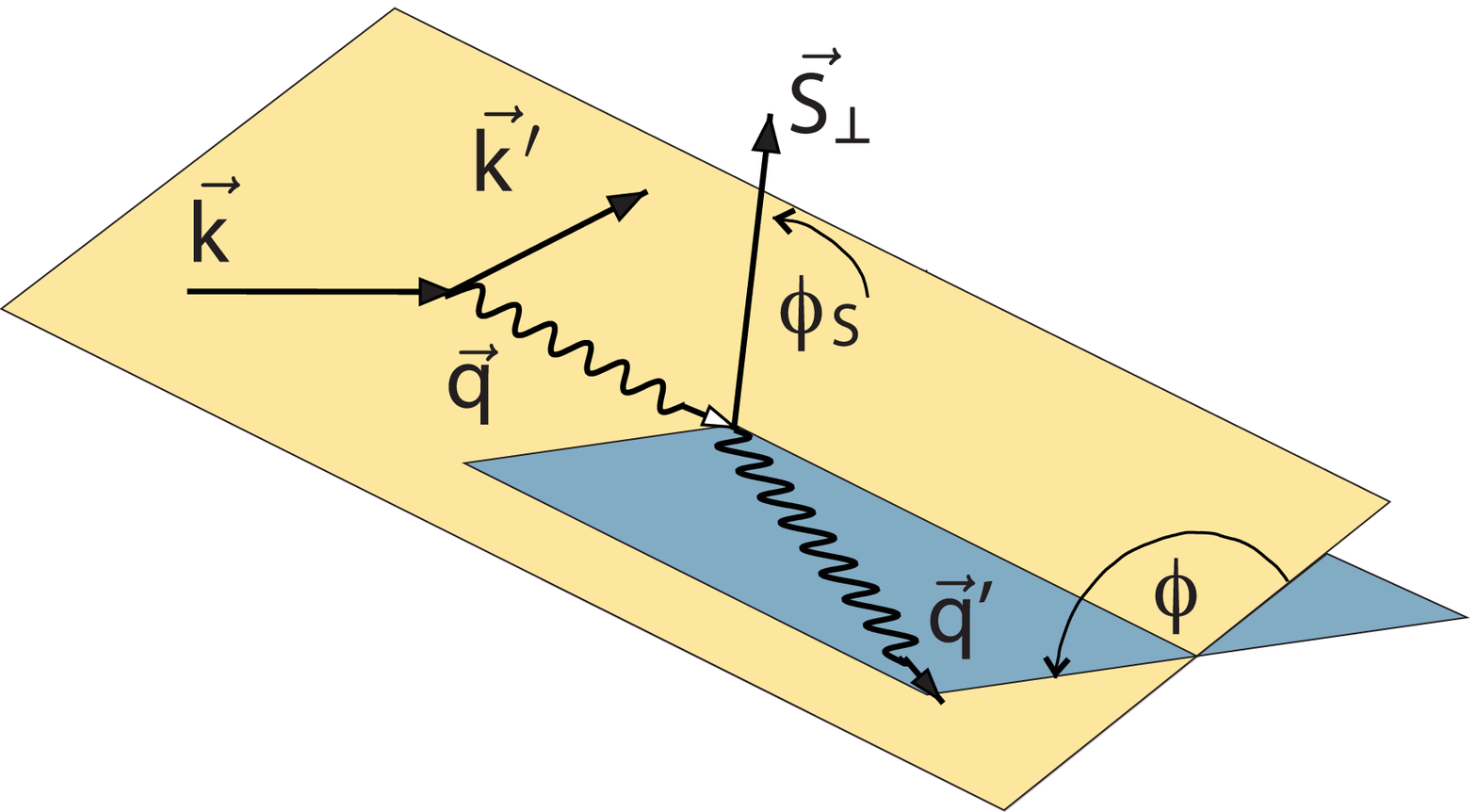}}
\caption{\small Kinematics of the photon leptoproduction in the target rest frame (the Trento notations). The incoming and outgoing leptons define the scattering plane,
and the outgoing photon and recoil protons define the hadronic plane. In this reference system, the azimuthal angle between the lepton and recoil proton planes is $\phi$.
The angle $\phi_S$ defines the orientation of the target spin in the case of a polarized target (it will not be used in the present contribution).
} 
\label{fig:angles}
\end{figure}

\vskip 0.3cm

\noindent where $\Delta$ is the 4-momentum transfer between the initial and final proton; 
$Q^2$ the virtuality of the exchanged photon; $x_B$ the usual Bjorken variable; $\epsilon=2 x_B M/Q$ ($M$ is the proton mass); $y$ is the fraction of the electron energy lost
in the nucleon rest frame; $\phi$ is the angle between the leptonic
plane $(e,e')$ and the photonic plane ($\gamma^{\ast},\gamma$) as shown in figure~\ref{fig:angles}. The angle $\varphi$ is defined as the difference between $\phi$ and $\phi_S$, the orientation of the target spin in the case of a polarized target, shown also in figure~\ref{fig:angles}.

The total amplitude $\mathcal T$ is the superposition of the BH and DVCS amplitudes:
\begin{eqnarray}
\left | \mathcal T \right |^2 & = & \left | \mathcal T_{\rm BH} \right |^2+\left |
\mathcal T_{\rm DVCS} \right |^2+ \mathcal I \,, \nonumber\\
\mathcal I & = & \mathcal T_{\rm DVCS}^{\ast} \mathcal T_{\rm BH} +
\mathcal T_{\rm DVCS} \mathcal T_{\rm BH}^{\ast} \, ,
\end{eqnarray}

\noindent where $\mathcal T_{\rm DVCS}$ and $\mathcal T_{\rm BH}$ are the amplitudes
for the DVCS and Bethe-Heitler processes, and $\mathcal I$ denotes the
interference between these amplitudes. The individual contributions to
the total $e p \to e p \gamma$ cross section can be written as (up to twist-3 contributions and corrections in $1/Q$)~\cite{Belitsky:2001ns}:
\begin{eqnarray}
\left| \mathcal T_{\rm BH} \right|^2 & = & \frac{\Gamma_{\rm BH}(x_B,Q^2,t)}{\mathcal P_1(\phi) P_2(\phi)}
\left\{ c_0^{\rm BH} + \sum_{n=1}^2 c_n^{\rm BH} \cos (n\phi) +s_1^{\rm BH} \sin\phi \right\}  \, , \label{eq:epepgBH} \\
\left| \mathcal T_{\rm DVCS} \right|^2 & = & \Gamma_{\rm DVCS}(x_B,Q^2,t)
\left\{ c_0^{\rm DVCS} + \sum_{n=1}^2 [ c_n^{\rm DVCS} \cos (n\phi) +s_n^{\rm DVCS} \sin(n\phi) ] \right\} \, , \label{eq:epepgDVCS} \\
\mathcal I & = & \frac{\Gamma_{I}(x_B,Q^2,t)}{\mathcal P_1(\phi) P_2(\phi)} ]
\left\{ c_0^{I} + \sum_{n=1}^3 [ c_n^{I} \cos (n\phi) +s_n^{I} \sin(n\phi) ] \right\} \, , \label{eq:epepgI}
\end{eqnarray}

\noindent where $\Gamma_{\rm BH}, \Gamma_{\rm DVCS}$ and $\Gamma_{I}$ are known kinematical prefactors. $\mathcal P_1(\phi)$ and $\mathcal P_2(\phi)$ come from the BH electron propagators and can be written as:
\begin{equation}
Q^2 \mathcal P_1 = Q^2 + 2 k\cdot\Delta \,, \quad
Q^2 \mathcal P_2 = - 2 k\cdot\Delta + \Delta^{2} \, ,
\label{eq:p1p2}
\end{equation}

\noindent where $k$ is the 4-momentum of the incoming lepton. 

In the case of scattering on unpolarized or longitudinally polarized targets, all $\sin (n\phi)$ coefficients in~(\ref{eq:epepgBH}-\ref{eq:epepgI}) depend either on the beam helicity $\lambda$ or on the target longitudinal polarization $\Lambda$; they disappear in the unpolarized cross section.

Using a polarized beam, two separate quantities can be extracted: the difference of cross section
with opposite beam helicities and the total cross section, which at leading twist can be written respectively as:
\begin{eqnarray}
d\sigma^\rightarrow-d\sigma^\leftarrow & = & 2 \cdot\mathcal T_{\rm BH} \cdot \Im m(\mathcal T_{\rm DVCS}) \,, \nonumber\\
d\sigma^\rightarrow+d\sigma^\leftarrow & = & \left|T_{\rm BH}\right|^2 + 2 \cdot\mathcal T_{\rm BH} \cdot \Re e(\mathcal T_{\rm DVCS}) + \left|T_{\rm DVCS}\right|^2 \, ,
\end{eqnarray}

\noindent where the arrows correspond to the beam helicity. 
At low $y$,
the interference term entering the total cross section is small compared to the DVCS and BH contributions, which contrasts with the 
case of intermediate or large $y$
where the DVCS contribution is small with respect to the interference, which itself is in general significantly smaller than the BH term. Note that the DVCS contribution to the difference of cross section only appears at higher twist.

From these two natural observables, one can write asymmetries which are experimentally easier to determine than cross sections:
\begin{equation}
A_{LU}=\frac{d\sigma^\leftarrow-d\sigma^\rightarrow}{d\sigma^\leftarrow+d\sigma^\rightarrow} \, ,
\end{equation}
\vskip 0.3cm

Beam spin asymmetries are mostly sensitive to the GPD $H$ and are complementary to unpolarized cross sections and beam charge asymmetry measurements presented in section~\ref{sec:fazio}.

\subsection{Monte Carlo}

The PROPHET package \cite{PROPHET} was used in its Monte Carlo configuration to generate photon electroproduction pseudo-data in the EIC kinematics. We relied on the Goloskokov-Kroll model for 
GPDs~\cite{Goloskokov:2008ib} evaluated at NLO, integrated over the LO hard kernel to obtain Compton Form Factors $\mathcal H$, $\widetilde{\mathcal H}$, $\mathcal E$ and $\widetilde{\mathcal E}$
which are the complex counterparts of GPDs and directly relate to the DVCS amplitude at the leading order of $\alpha_s$~\cite{Belitsky:2001ns}.
Note that $\widetilde{\mathcal E}$ only enters the unpolarized cross section for DVCS, but was neglected in this evaluation.

The photon electroproduction observables were evaluated using the GV package \cite{VG}, which in contrast with the usual BMK formalism~\cite{Belitsky:2001ns}, does not make approximations of the order of $1/Q$ in the treatment of the interference term. It was checked that the unpolarized cross sections generated by our Monte Carlo give the results simular to those shown in section~\ref{sec:fazio},
but with more realistic $b$-slopes since they were not taken as a constant but form a part of the Goloskokov-Kroll model.

\subsection{Projected results}

\begin{figure}[h]
\centering
\includegraphics[scale=0.32,angle=0]{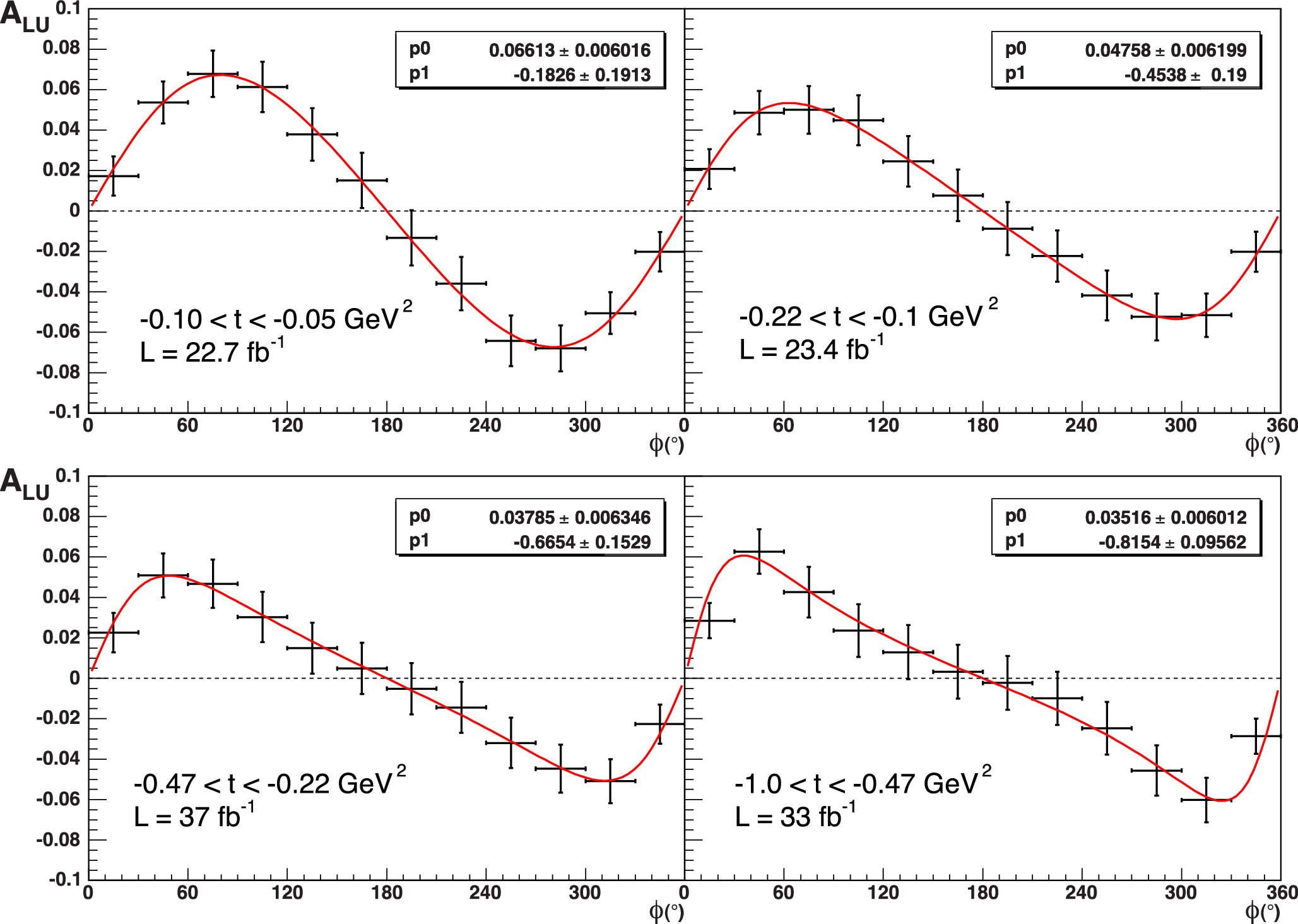}
\caption{\small Photon electroproduction Beam Spin Asymmetries for the $20\times 250$ EIC configuration, in the typical kinematic bin: $1.58 \cdot 10^{-3} < x_B < 2.51 \cdot 10^{-3}$, $3.16 < Q^2 < 5.61~\rm{GeV}^2$ for four different $t$-bins as shown on each plot. The Monte Carlo was set up as to generate 90k events for each $t$-bin and the corresponding integrated luminosity is shown on each plot. Up to about 3 months of beam time with 50\% efficiency is necessary to achieve 10 to 15\% accuracy on the extracted $\sin\phi$ coefficient $\tt p0$, sensitive to the imaginary part of CFF $\mathcal H$.}
\label{bsa_dvcs}
\end{figure}

The Beam Spin Asymmetry evaluated in a typical $(x_B,Q^2)$ bin is shown in figure~\ref{bsa_dvcs}. 
The $\sin\phi$ coefficient turns out 3 to 5 times lower than that at typical lower energy and higher $x_B$ kinematics. Therefore, this asks for a rather large integrated luminosity. 
For the considered $x_B$ range (and lower), about three months of EIC at $10^{34}$~cm$^{-2}$s$^{-1}$ luminosity assuming 50\% operational efficiency is necessary to achieve a 10 to 15\% accuracy on the extracted $\sin\phi$ coefficient, mostly linked to the imaginary part of CFF $\mathcal H$. For higher $x_B$ and $Q^2$ values, much larger integrated luminosities will be necessary to achieve similar statistical accuracy.

Using longitudinal and transverse target asymmetries, one will be able to obtain information on $\widetilde H$ and even the elusive $E$, essential for the evaluation of Ji's sum rule. Studies of these observables for the EIC are in progress using the same formalism.

\section{Hard exclusive photoproduction of Quarkonia}
\label{sec:quarkonia}

\hspace{\parindent}\parbox{0.92\textwidth}{\slshape 
      Peter Kroll}

\index{Kroll, Peter}

\vspace{1\baselineskip}

Photoproduction of quarkonia (e.g., $J/\Psi$ and $\Upsilon$) 
forms another class of hard exclusive processes which 
allow one to scrutinize the handbag approach and extract information 
on generalized parton distributions (GPDs). Neglecting intrinsic 
heavy quarks in the proton, only the gluonic subprocess
$\gamma^{\ast}g\to Mg$ (accompanied by the gluon GPDs) 
contributes to the scattering amplitude.
A particular feature of quarkonium production is the appearance of
the large mass, $m_Q$, of the heavy quark which provides a hard 
scale and allows one to treat photoproduction
within a QCD factorization approach. 
A first leading-order calculation of 
quarkonium production within the handbag approach has been carried 
out in \cite{Vanttinen:1998zd}. Recently this analysis has been 
improved by the inclusion of NLO corrections~\cite{Ivanov:2004vd}. 
In these studies, a non-relativistic scenario for the description of 
the quarkonium state has been adopted in which the quark and 
antiquark share the meson's momentum equally. Hence, the quarkonium 
wave function is proportional to $\delta(x-1/2)$ and the quarkonium 
mass is given by $M_Q\simeq 2m_Q$. There are other theoretical 
approaches to the process of interest, e.g., the dipole approach, 
the leading $\ln(1/x)$ approximation or the BFKL Pomeron. Due to 
limitation of space these approaches will not be discussed here.

In the kinematical range of large photon-proton center-of-mass energy, 
$\sqrt{s}\gg M_Q$, and small momentum transfer, $t$, the skewness is 
given by
\begin{equation}
\xi = M_Q^2/(2s)\,.
\label{eq:skewness}
\end{equation}
Neglecting terms of order of $\sqrt{-t}/m_Q$, $t/4m^2$ and $\xi$, one
finds the following expressions 
for the helicity amplitudes of the quarkonium photoproduction:
\begin{eqnarray}
{\cal M}_{\mu +,\mu +} &=&\frac{e_0}{2}\, e_Q\,
            \int_0^1 \frac{dx} {(x+\xi) (x-\xi + i{\varepsilon})} 
           \sum_\lambda \,{\cal H}_{\mu\lambda,\mu\lambda}\,
      \left[\,H^g +\lambda \widetilde{H}^g\,\right]\,,\nonumber\\
{\cal M}_{\mu -,\mu +} &=& -e_0 \,e_Q\,
             \frac{\sqrt{-t}}{4m} \; \int_0^1 \, \frac{d x}{(x+\xi) 
             (x-\xi+i{\varepsilon})}
        \sum_\lambda \,  {\cal H}_{\mu\lambda,\mu \lambda} E^g\,, \nonumber\\
{\cal M}_{- -,+ +} &=& e_0\, e_Q\,\frac{\sqrt{-t}}{2m} \;
            \int_0^1 \frac{d x} {(x+\xi) (x-\xi + i{\varepsilon})}\,       
            {\cal H}_{--,++}\,H_T^g\,.     
\label{eq:amplitudes}
\end{eqnarray}
Other helicity amplitudes are zero except for those related by parity 
conservation to the above ones. 
The helicity labels $\mu$ and $\mu^\prime$ refer to the initial photon and final meson, 
respectively; the labels $\lambda$ and $\lambda^\prime$ refer to the initial and final gluon,
respectively.
The explicit helicity labels of 
${\cal M}$ refer to the proton. 
In the non-relativistic scenario,
the LO subprocess amplitudes read
\begin{equation} 
{\cal H}_{\mu^\prime\lambda^\prime,\mu\lambda} =
                                \frac{8\pi\alpha_s(\mu_R)f_Q}{3m_Q}
                          \,\delta_{\mu^\prime\mu}\,\delta_{\lambda^\prime\lambda} \,,
\label{eq:subprocess}
\end{equation}
where $f_Q$ is the decay constant 
of the quarkonium; $\mu_R$ is an appropriate renormalization scale. 
Thus, at this level of accuracy, only the process amplitudes,
\begin{equation} 
{\cal M}_{\mu +,\mu +}=\phantom{-} e_0e_Q\,\frac{8\pi\alpha_s f_Q}{3m_Q}
       \langle H^g\rangle\,, \quad
{\cal M}_{\mu -,\mu +}= -e_0e_Q\,\frac{8\pi\alpha_s f_Q}{3m_Q}\,
\frac{\sqrt{-t}}{2m}\langle E^g\rangle \,,
\label{eq:ampl}
\end{equation}
are non-zero. The terms $\langle F\rangle$ denote the convolutions 
of the subprocess amplitudes and GPDs. One sees that the unpolarized 
cross section for quarkonium production at small $t$ is only fed 
by $H^g$, while the asymmetry measured with a transversally polarized 
target is given by an interference term of $E^g$ and $H^g$. Other 
GPDs, like the chiral-odd $H^g_T$, do not contribute at this 
level of accuracy. 

\begin{figure}[ht]
\begin{center}
\includegraphics[width=0.35\textwidth,bb=99 309 434 690,clip=true]
{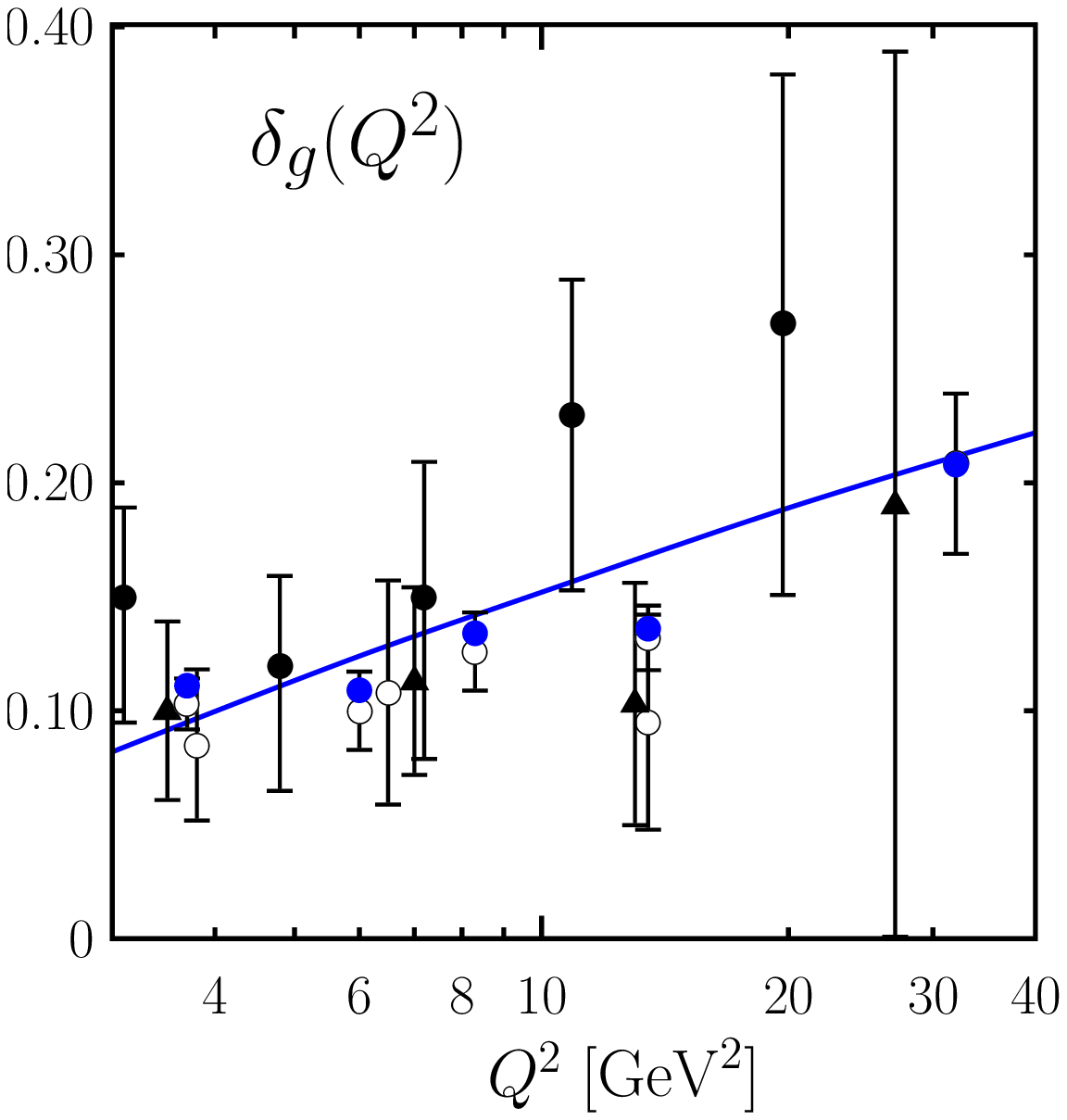}
\includegraphics[width=.5\textwidth,bb= 5 -35 515 350, clip=true]
{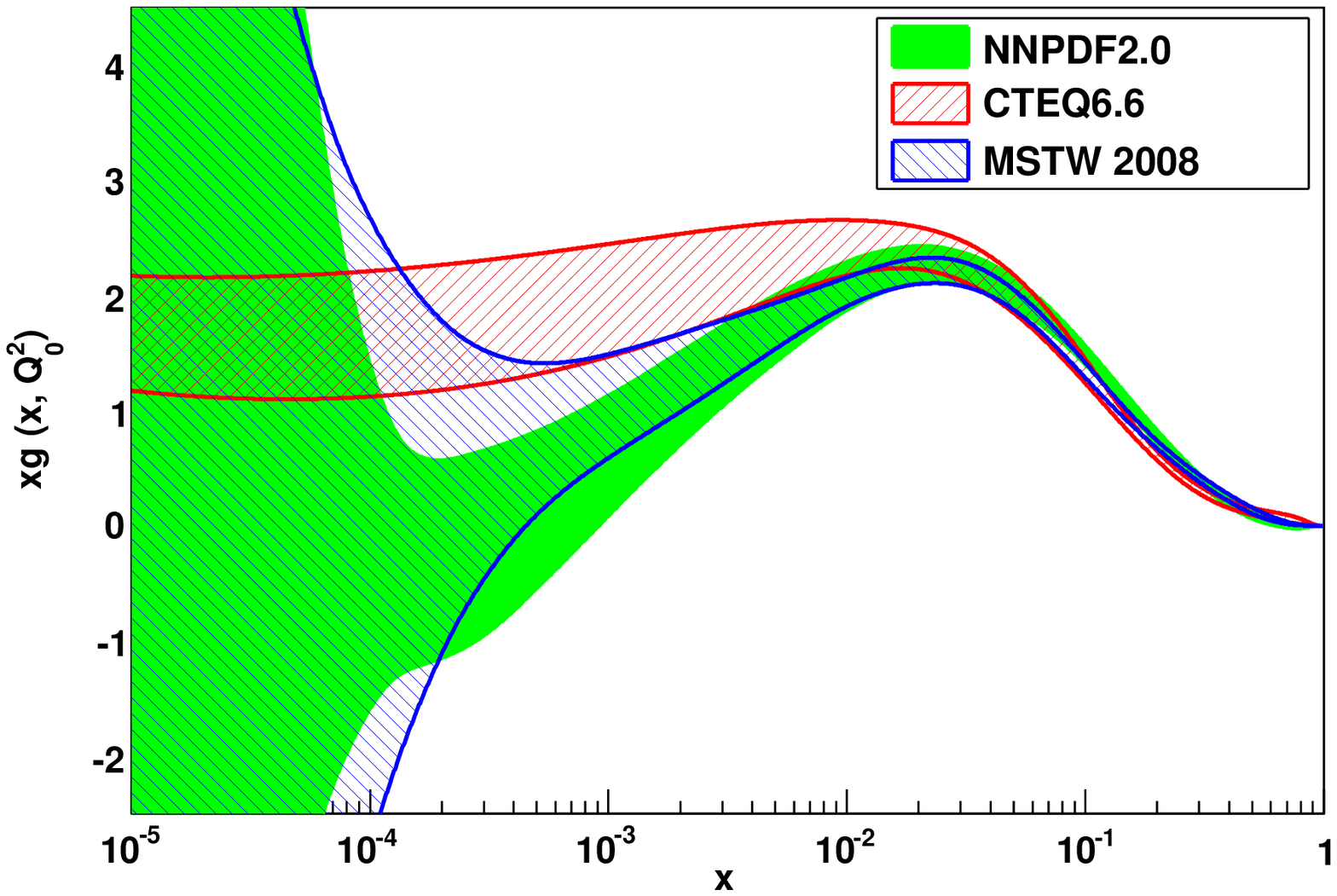}
\end{center}
\caption{\label{fig:delta} \small Left: The intercept of the gluonic trajectory
  shifted by one unit; the data points are from \cite{Aaron:2009xp}. Right: 
  Various NLO gluon PDFs. The figure is from \cite{Ball:2008by}.}
\end{figure}

For the EIC kinematics for which 
$\xi < 0.01$,
quarkonium production is a diffractive process, i.e., the amplitudes 
are dominantly imaginary. Thus, essentially the GPDs are only needed 
at the cross-over line $x=\xi$, while the small real part may be 
estimated with the help of analyticity.
Many methods for the construction of GPDs, e.g., 
the double distribution ansatz~\cite{Radyushkin:1998bz} 
or the Shuvaev transform~\cite{Martin:2009zzb}, lead to GPDs which
at $x=\xi$
are proportional to the usual parton distributions. 
In particular, for the reggeized double distribution ansatz used in 
\cite{Goloskokov:2007nt,Goloskokov:2005sd}, one has
\begin{equation}  
H^g(\xi,\xi,t) = c_h\big[2\xi g(2\xi)\big]\,(2\xi)^{-\alpha^\prime_h t}\,
      e^{b_ht}\,.
\label{eq:GPD}
\end{equation}
Assuming that $xg(x) = c x^{-\delta_h}$
at low $x$, the constant $c_h$ reads (with $b=2$ \cite{Radyushkin:1998bz}):
\begin{equation}
c_h = c \big[(1-\delta_h/5)(1-\delta_h/4)(1-\delta_h/3)\big]^{-1}\,.
\end{equation}
Since $\delta_h$ is positive, $c_h>c$, which implies that 
$H^g(\xi,\xi,t=0)> 2\xi g(2\xi)$ (this is termed the skewness effect). 
In \eqref{eq:GPD}, a linear gluonic ('Pomeron-like') Regge trajectory 
is assumed:
\begin{equation}
\alpha_h=1 +\delta_h + \alpha_h^\prime t\,.
\end{equation}
Its slope is taken from the $J/\Psi$ photoproduction data 
\cite{Chekanov:2002xi} ($\alpha_h^\prime=0.15\,{\rm GeV}^{-2}$), while 
$\delta_h$, the intercept minus 1, can be fixed from the data on the cross section 
for electroproduction of $\rho^0$ and $\phi$ mesons 
\cite{Aaron:2009xp} that behaves as $\sigma \propto s^{2\delta_h}$.
The data provide  a scale-dependent intercept (see figure~\ref{fig:delta}):
\begin{equation}
\delta_h(\mu) = 0.1 + 0.06 \ln(\mu^2/\mu^2_0) \,,
\label{eq:delta-exp}
\end{equation}
where $\mu_0=2\,{\rm GeV}$ and $\mu=Q$ for electroproduction. The scale 
dependence of $\delta_h$ is in agreement with evolution. 

The construction of the GPD along the lines described above requires 
the knowledge of the gluon PDF; in figure~\ref{fig:delta}, some recent results 
for it are shown. For $x$ smaller than about $10^{-3}$, the errors of
most PDFs become very large. The only exception is the CTEQ6 results 
\cite{Pumplin:2002vw} for which the error stays constant at the level 
of about $25\%$ for $x<10^{-2}$. Since the uncertainties of the PDF are 
conveyed to the GPDs and, hence, to the quarkonium cross section, any 
prediction of the latter will suffer from huge uncertainties rendering 
any comparison with experiment meaningless. In order to arrive at 
reasonable predictions, the following remedial measure has been 
proposed in~\cite{Goloskokov:2007nt,Goloskokov:2005sd}: The gluon PDF 
is expanded as the following series,
\begin{equation}
x g(x,\mu) = x^{-\delta_h(\mu)} (1-x)^5\sum_{i=0}^2 c_i(\mu) x^{i/2} \,,
\end{equation}
and the expansion parameters are fitted to a given PDF for intermediate 
values of $x$, say, $0.003 < x < 0.3$, and relevant scales. The 
power $\delta_h$ is fixed at the experimental value~\eqref{eq:delta-exp}. 
Applying this prescription to the NLO CTEQ6 PDF \cite{Pumplin:2002vw}, 
one can reproduce the CTEQ6 result for $x < 0.003$. Therefore, the 
CTEQ6 gluon PDF with its errors may be used for numerical predictions. 
(The same method applied to other current gluon PDFs leads, in 
most cases  and within uncertainties, to the results for cross sections that are in a reasonable agreement with 
those evaluated using the CTEQ6 PDF~\cite{Goloskokov:2007nt}.)
 
\begin{figure}[ht]
\begin{center}
\includegraphics[width=0.47\textwidth, clip=true]{Imaging/Imaging_Figures/Kroll_sj1.pstex}
\hspace*{0.04\textwidth}
\includegraphics[width=0.47\textwidth, bb= 98 364 596 718,clip=true]
{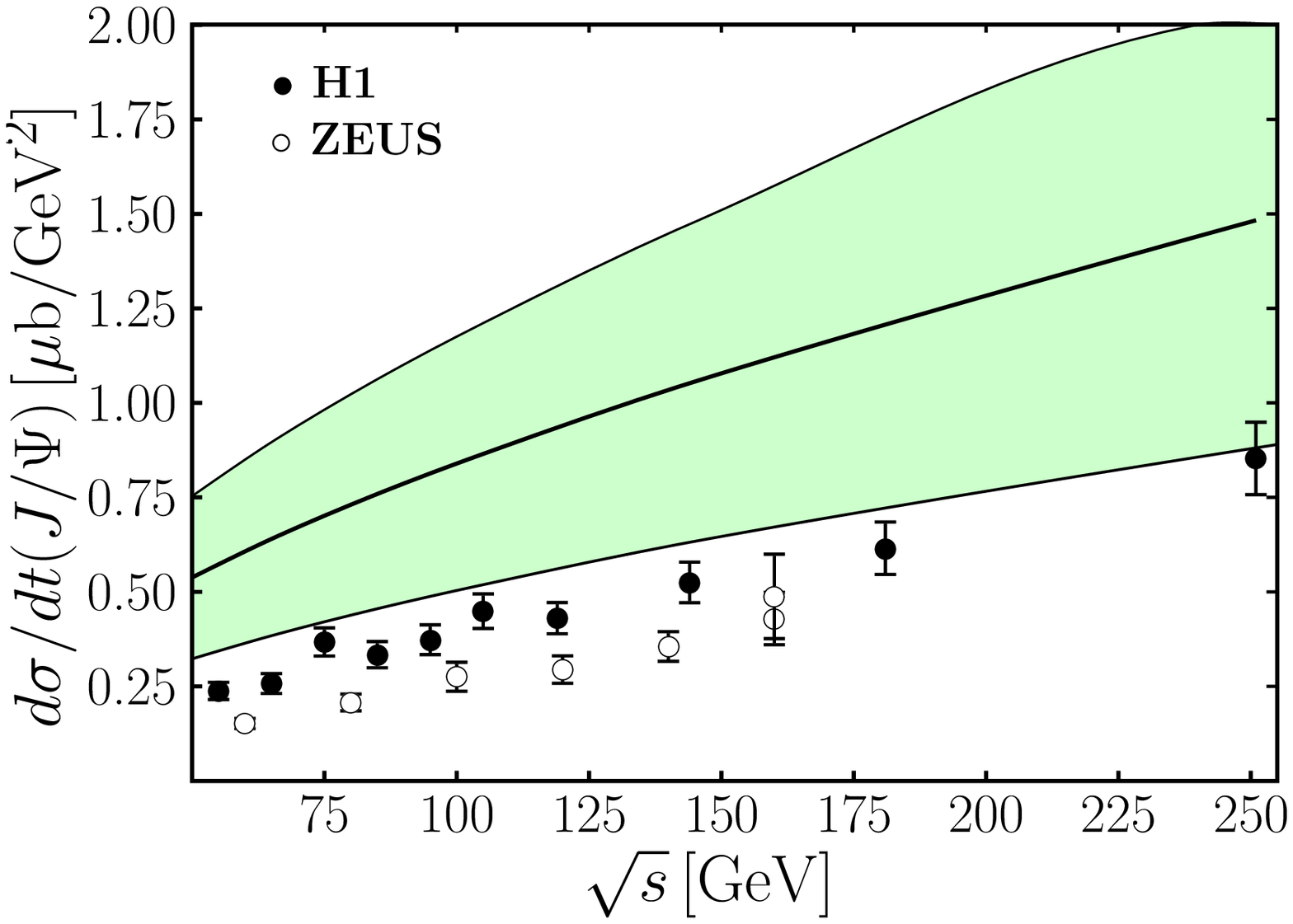}
\end{center}
\caption{\label{fig:cross} \small Left: LO and NLO predictions for the forward $J/\Psi$
photoproduction {\it vs.}~$\sqrt{s}$ 
  (the figure is from \cite{Ivanov:2004vd}). Right: LO prediction 
  using the GPD given in \cite{Goloskokov:2005sd,Goloskokov:2007nt}, 
  see the text. The green band indicates the uncertainty of the prediction; the 
  data points are from \cite{Aktas:2005xu,Chekanov:2002xi}.}
\end{figure}

The calculation of the quarkonium cross section is further complicated 
by large NLO corrections \cite{Ivanov:2004vd}, see figure~\ref{fig:cross}. 
The results shown in  figure~\ref{fig:cross} are evaluated from a GPD that is also generated 
from the NLO CTEQ6 PDF but under the assumption that, at the initial 
scale, $H^g$ is given by the PDF multiplied by an appropriate function 
of $t$. Evidently, the large NLO corrections necessitate a resummation 
of higher orders for a reliable prediction of the quarkonium cross section.

In order to examine the dependence of the predictions on the GPD used 
in the calculation, I have repeated the LO order calculation exploiting 
the GPD proposed in \cite{Goloskokov:2007nt,Goloskokov:2005sd} and 
whose construction is briefly described above. The result, obtained 
for a scale of $2m_Q$ which is chosen in concord with the construction 
of the GPD [see \eqref{eq:delta-exp}], overestimates the data but is 
in agreement with the experimental energy dependence. It however 
differs strongly from the LO result presented in \cite{Ivanov:2004vd}. 
Similar observations can be made for photoproduction of the $\Upsilon$.

The target asymmetry for quarkonium production may give access to the GPD
$E^g$, of which not much is known, see the discusion in section~\ref{sec:gpd-e}.
Indeed,
\begin{equation}
A_{UT} = -\frac{\sqrt{-t}}{m} \,\frac{|\langle E^g\rangle|}{|\langle
  H^g\rangle|}\,\sin{\phi}\,,
\end{equation}
where $\phi$ is the relative phase between $\langle H^g\rangle$ and 
$\langle E^g\rangle$. Since $E^g$ is expected to behave similarly to 
$H^g$ at small $\xi$ but with a Regge trajectory 
$\alpha_e= 1+\delta_e+\alpha_e^\prime t$, analyticity tells us that the 
relative phase is approximately given by
\begin{equation}
\phi(t) = \pi/2 (\alpha_e(t) - \alpha_h(t))\,.
\end{equation}
In a soft Pomeron scenario, one would have $\alpha_e = \alpha_h$ and, 
hence, $A_{UT}=0$. However, the QCD evolution of the GPDs may generate differences
in the trajectories and, therefore, $A_{UT}$ may be non-zero. To work this 
out, one needs a detailed study of the evolution of $E^g$, which is lacking at
present. There are examples of $\alpha_e$ in the literature that 
lead to tiny asymmetries of the 
order of $1 - 2\%$. In section~\ref{sec:gpd-e}, it has been proposed a parameterization 
of $E^g$ with a node at some intermediate value of $x$. In this case one 
may obtain a larger $A_{UT}$. However, this possibility has not yet been explored 
in detail. 

{\it Summary}: Comparing precise data of the cross section for 
photoproduction of quarkonia with theoretical calculations within the 
handbag approach may allow for an extraction of $H^g$ at the cross-over 
line and small values of $\xi$ since the real part of the amplitude 
provides only small corrections to the cross section of the order of 
$10\%$. This may lead to a useful constraint on $H^g$ and $g(x)$ at 
low $x$. However, in order to arrive at reliable results for $H^g$, the 
theoretical calculation should include resummed higher orders of 
perturbative QCD. Also, deviations from the non-relativistic scenario 
should be investigated and the strength of contributions from 
intrinsic heavy quarks estimated. Furthermore, a detailed comparison 
of various gluon GPDs should be made and their errors taken into 
account. For photoproduction of charmonium production in particular, one should be aware 
of possible substantial power corrections since the charm quark mass 
although being large enough to allow for a perturbative treatment of 
the subprocess, is not large enough to suppress power corrections 
decisively. On the other hand, for electroproduction of charmonium power corrections
are likely to be smaller.
In principle the target asymmetry gives an access to $E^g$. 
However, $A_{UT}$ will likely be very small except for the case when $E^g$ markedly 
differs from common parameterizations. 


\section{Simulations of non-diffractive exclusive processes at an EIC}
\label{sec:Horn}

\hspace{\parindent}\parbox{0.92\textwidth}{\slshape 
  T. Horn}
%

\index{Horn, Tanja}


\subsection{Introduction}
\label{sec:intro}

Exclusive processes in $ep$ scattering at collider energies can be either ``diffractive'' (no exchange of quantum numbers between the target and the projectile/produced system) or  ``non-diffractive'' (there is an exchange of quantum numbers). By measuring diffractive channels ($J/\Psi$, $\rho^0$, or $\phi$ production) at sufficiently high $Q^2$, one probes the gluon GPDs and/or the singlet quark GPDs. In particular, $J/\Psi$ production probes the gluon GPD in the nucleon, and its $t$-dependence reveals the transverse spatial distribution of the gluons. Measurements of DVCS and exclusive $\rho^0$ production at high $Q^2$ provide access to the singlet quark and gluon GPDs.

Non-diffractive channels like $\pi^+$, $\pi^0$, or $K^+$ production are sensitive to the flavor and spin structure of the nucleon at small $x_B$, which complements the information obtained from DVCS and meson production experiments in the valence region, e.g., HERMES and 6 GeV and 12 GeV JLab. 

For moderate values of $x_B$, the proposed electron-ion collider (EIC) could reach $Q^2 > 10$ GeV$^2$, where higher-twist contributions, which complicate the extraction of GPDs from the data, are expected to be small. Indeed, the comparison of different meson channels alone provides model-independent information about the ratio of quark spin and spatial distributions, and a comparison between, for instance, $\pi^+$ and $K^+$ production may allow for the studies of SU(3)
symmetry in parton distributions.
%

\subsection{Rate predictions}

Rate predictions were made for several exclusive reaction channels using a new exclusive Monte Carlo generator. Here, we will focus on the $\pi^+$ and $K^+$ channels. These are the simplest systems also allowing for comparisons of non-strange and strange distributions similarly to the comparative studies of singlet quarks and gluons with diffractive exclusive channels.

\begin{figure}[!htbp]
\begin{center}
{
\includegraphics[width=2.75in]{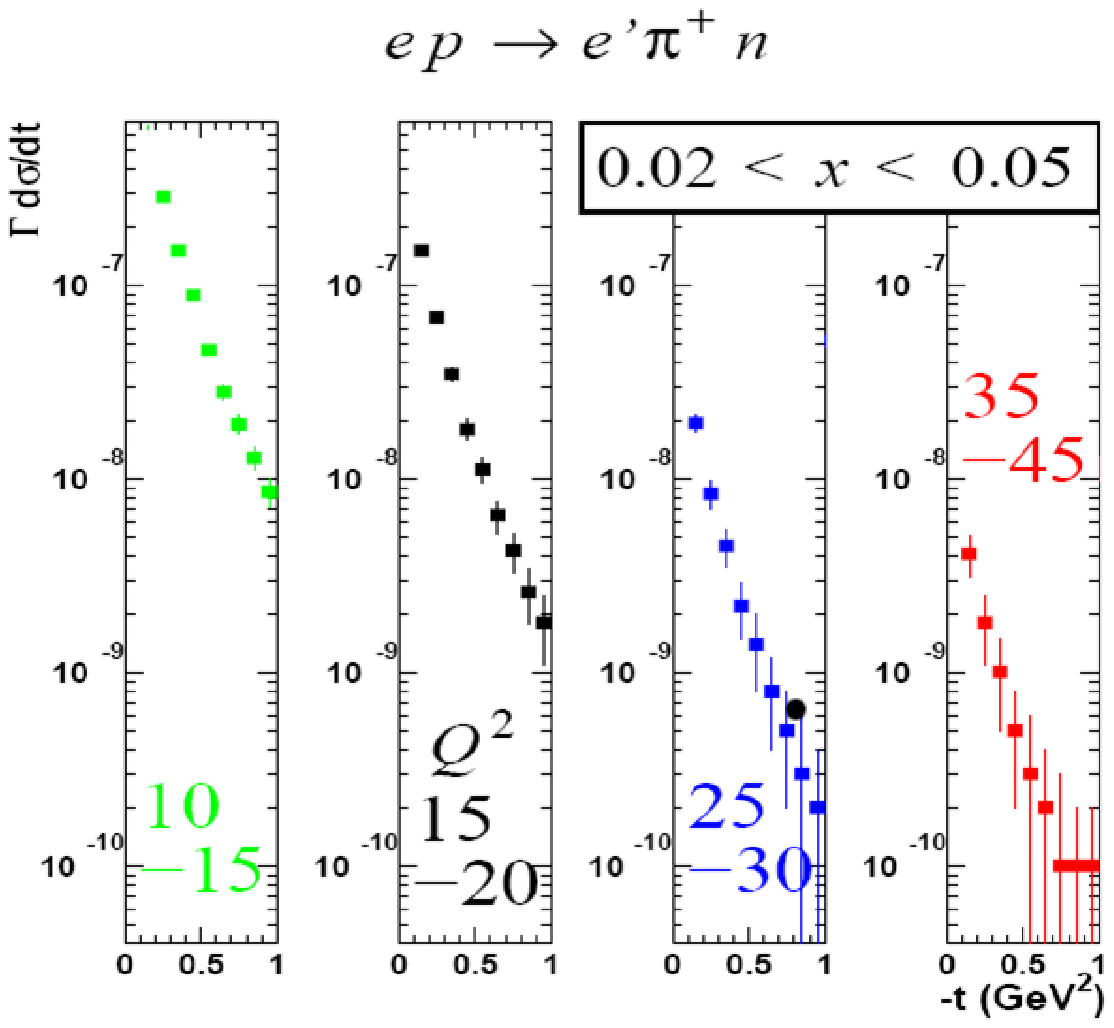}
}
{\includegraphics[width=2.75in]{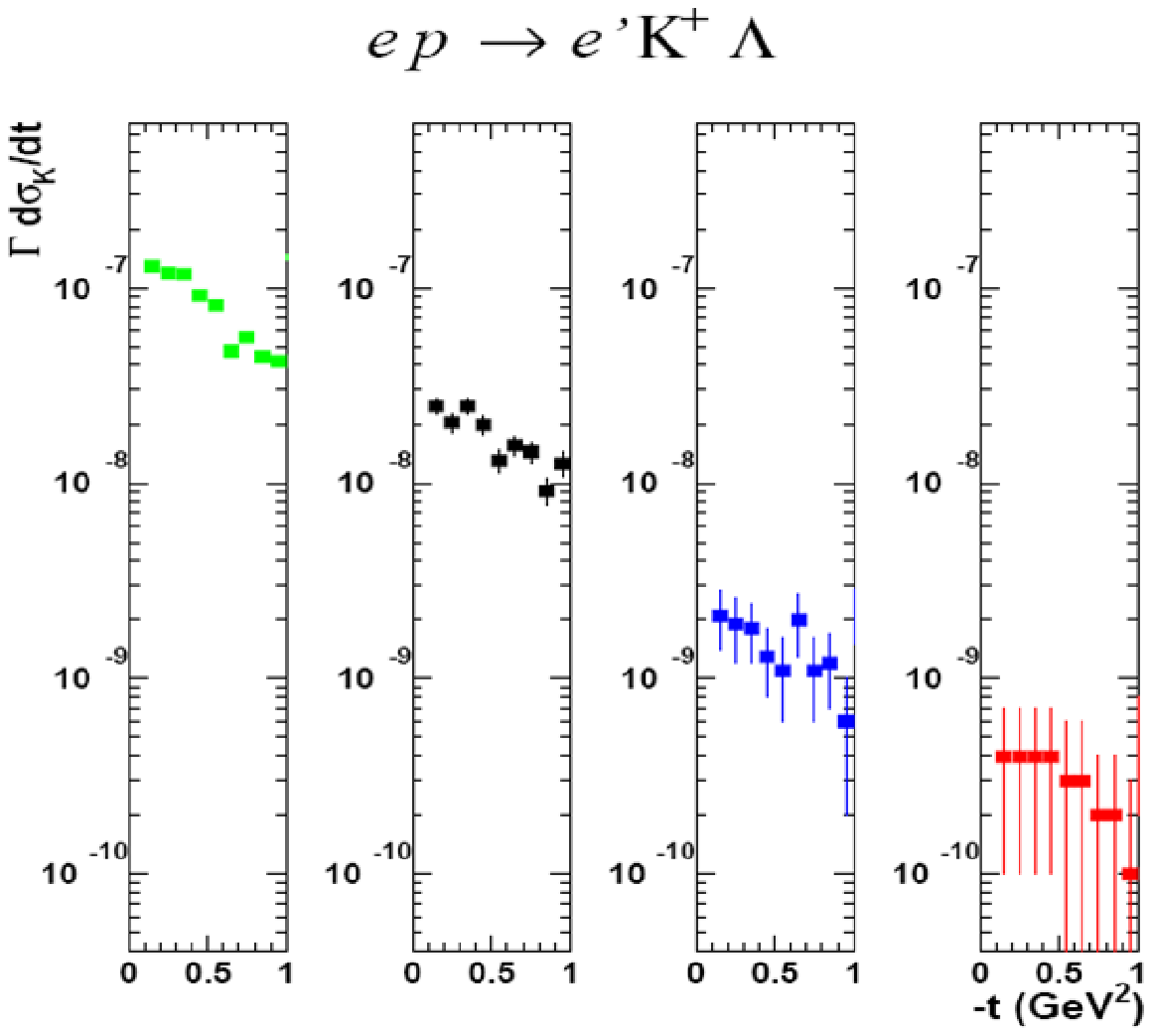}}
\caption{\small EIC simulations for the exclusive pion and kaon electroproduction cross sections at a electron beam energy of 5 GeV and a proton beam energy of 50 GeV ($\sqrt{s}$=31.6 GeV), 100 days running, and a luminosity of 10$^{34}$ cm$^{-2}$ s$^{-1}$. The points are shown for a bin in $x_B$ of 0.02 to 0.05. The four panels denote bins in $Q^2$ (from left to right) of 10-15 GeV$^2$, 15-20 GeV$^2$, 25-30 GeV$^2$, and 35-40 GeV$^2$.} 
\label{fig-xsec}
\end{center}
\end{figure}

Figure~\ref{fig-xsec} shows the simulated cross section for exclusive pion and kaon production in the $5 \times 50$ GeV configuration in $ep$ collisions ($\sqrt{s}$=31.6 GeV) at a luminosity of 10$^{34}$ cm$^{-2}$ s$^{-1}$~\footnote{These energies and luminosities correspond to those given as medium-energy collider design prior to the INT 10-3 program, e.g., $\sqrt{s}$=31.0 GeV.}, and data taking for 100 days. The simulated data shown here are divided into four $Q^2$ bins between 10 and 45 GeV$^2$ for a bin in $x_B$ between 0.02 and 0.05. Each $Q^2$ bin was divided into nine $-t$ bins. The simulated pion data cover a range in $-t$ up to 1 GeV$^2$ with acceptable rates for the assumed run time and luminosity for all $Q^2$ bins. The rates in each $Q^2$ bin are highest at small values 
of $-t$ 
and smallest at high values of $-t$. This makes sense as one of the features of pion production is the dominance of the "pole term" at low $-t$. Furthermore, the pion rate decreases rapidly with higher $Q^2$ bins as the cross sections decrease, which is a characteristic behavior of exclusive reactions. 
However, one should keep in mind that reaching high $Q^2$ is needed due to the factorization requirement for studying the transverse spatial structure of sea quarks. The kaon simulated data are presented in the same kinematic bins as the pion data. The kaon cross section is smaller than the pion one, although it does not fall off as rapidly with increasing $-t$ for each $Q^2$ bin, because the kaon pole is not as dominant as the pion pole. The kaon rates are generally lower than the pion rates, but the effect is most visible in the largest $Q^2$ bins. For a fully differential kaon measurement, it is thus essential to have luminosities of at least 10$^{34}$ cm$^{-2}$ s$^{-1}$. Given the design parameters of the medium-energy collider, this would correspond to a range in $\sqrt{s}$ between 31 and about 45 GeV.

The rate prediction depends to some extent on the cross section models included in the simulation. For pions, a Regge-based cross section model that describes existing data well was used~\cite{Weiss08}. The model dependence of the rate prediction  was estimated using a different cross section model based on an empirical parameterization of charged pion data~\cite{Blok:2008jy}. 
For kaons, the rate estimate was based on the empirical fits to world kaon production data. The resulting uncertainty in the simulated rates was about a factor of two.

\subsection{Kinematic considerations}

Measurements in exclusive reactions require, besides knowledge of the beam quantities, information of all particles in the exclusive reaction, i.e., the scattered electron, the scattered meson, and the recoil baryon. Below we will illustrate the kinematic features of exclusive reactions using the H($e,e^\prime \pi^+$)n reaction. However, the kinematic distributions shown are independent of the exclusive channel, and are thus generally applicable to all exclusive reactions (diffractive and non-diffractive).

\begin{figure}[!htbp]
\centering
\includegraphics[width=3.5in, angle=270]{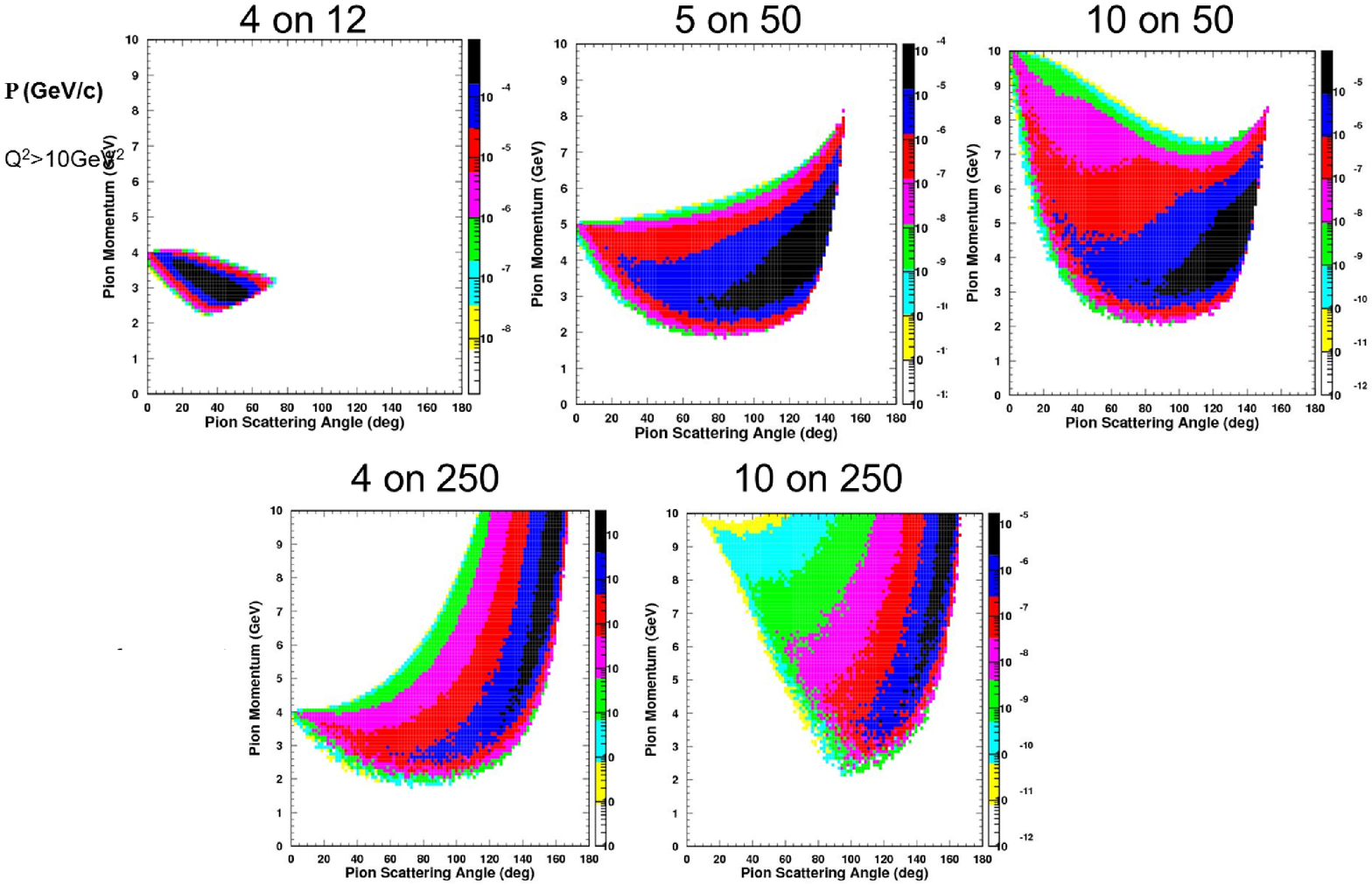}
\caption{\label{fig-meson-kinmat} \small The kinematic phase space for light mesons in deep exclusive reactions for different $ep$ collisions. A cut of $Q^2 >$ 10 GeV$^2$ is applied to focus on events needed for studies of transverse spatial distributions. The darker regions in the figures denote regions of the highest intensity. The center of mass energies for the medium energies are $\sqrt{s}$=13.8 GeV, 31.6 GeV, and 44.7 GeV, and for the high energies $\sqrt{s}$=63.2 GeV and 100 GeV. In this simulation, the direction of the electron beam is toward increasing angles.} 
\end{figure}

Figure~\ref{fig-meson-kinmat} shows the accessible phase space for exclusive reactions in $ep$ collisions for five center of mass energies~\footnote{The energies have been chosen to correspond to the preliminary values of the medium and high energy collider designs as given prior to the INT 10-3 program.}. A cut of $Q^2 >$ 10 GeV$^2$ was applied to focus on the region of interest for transverse spatial structure studies. At a value of $\sqrt{s}$=13.8 GeV, the meson distribution covers the angular range of about 30-40 degrees at relatively small momentum. Up to values of $\sqrt{s}$=44.7 GeV, which correspond to nearly symmetric collisions, the exclusive meson distribution spreads over a wide angular range, still at a moderate momentum. At even higher values of $\sqrt{s}$, the angular spread is reduced significantly. Indeed, the meson distribution is pushed into a relatively narrow forward cone. Furthermore, the events of interest in this narrow angular range also have very high momentum approaching the beam energy. Exclusive measurements at these large values of $\sqrt{s}$ would thus require the detection of a high energy meson over a very small angular range.

\begin{figure}[!htbp]
\centering
\includegraphics[width=4.5in, angle=0]{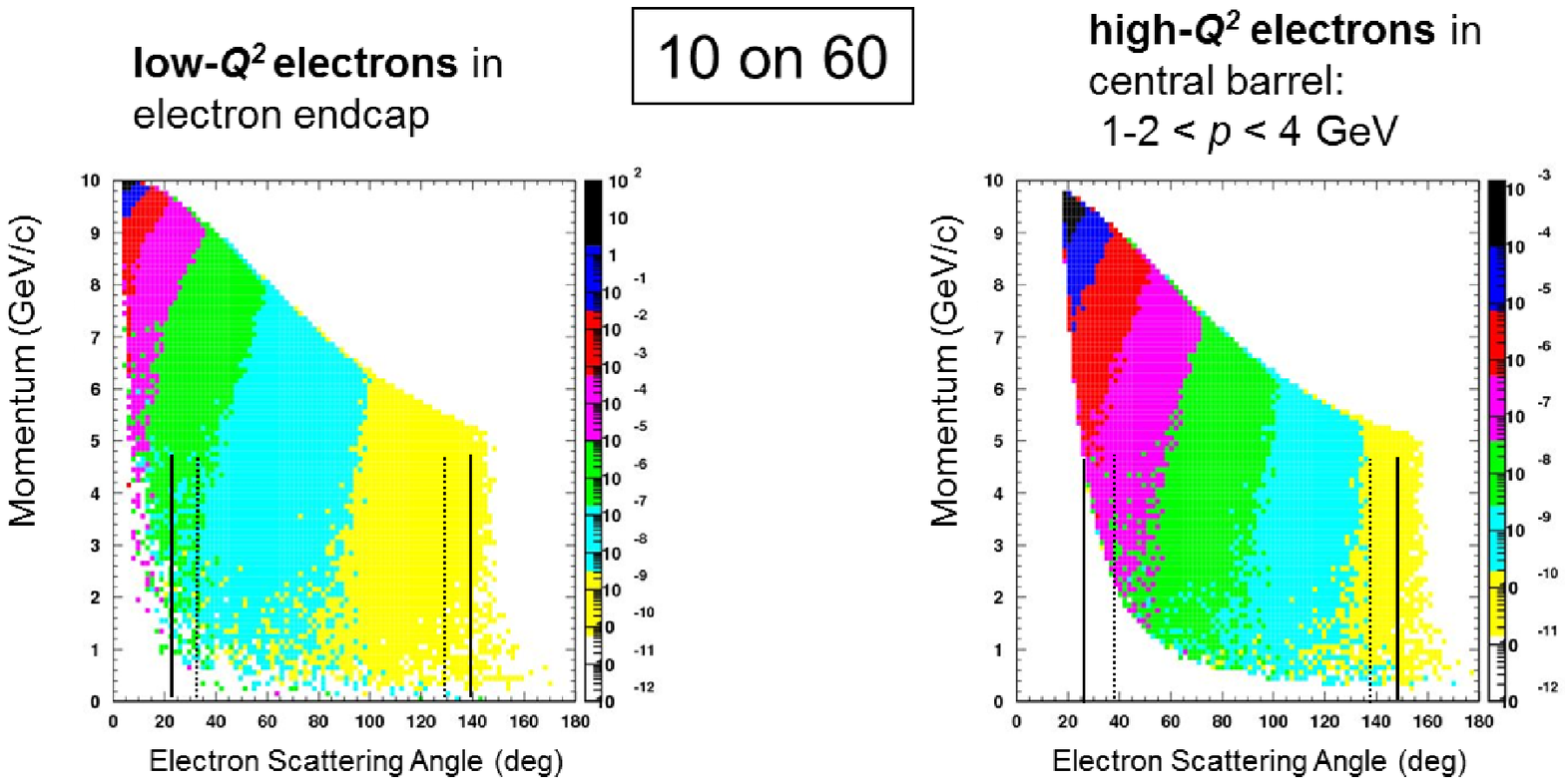}
\caption{\label{fig-electron-kinmat} \small The kinematic phase space for electrons in exclusive reactions at low and high $Q^2$ at a fixed value of $\sqrt{s}$=49.0 GeV.} 
\end{figure}

The momentum resolution ($dp/p$) to first order scales linearly with the momentum. The best resolution is thus achieved by keeping the laboratory momenta as low as possible for a given $\sqrt{s}$. This is achieved in symmetric, or nearly symmetric collisions. As illustrated in figures~\ref{fig-meson-kinmat} and \ref{fig-electron-kinmat} such kinematics also offer the advantage that the angular distribution of the outgoing electrons and mesons covers nearly $4\pi$, providing the best angular resolution.

Figure~\ref{fig-electron-kinmat} shows the scattered electron distribution in deep exclusive reactions. At modest electron energies (up to about 6 GeV) electrons predominantly scatter into the central and forward direction. Kinematically these correspond to high-$Q^2$ events, which are also the events of interest in studies of the transverse spatial structure of sea quarks. On the other hand, electrons at larger energies (up to the electron beam energy) scatter into the forward-electron direction. These events correspond to low-$Q^2$ events, which are of interest in photoproduction or heavy meson measurements. 

\begin{figure}[!htbp]
\centering
\includegraphics[width=3.5in, angle=270]{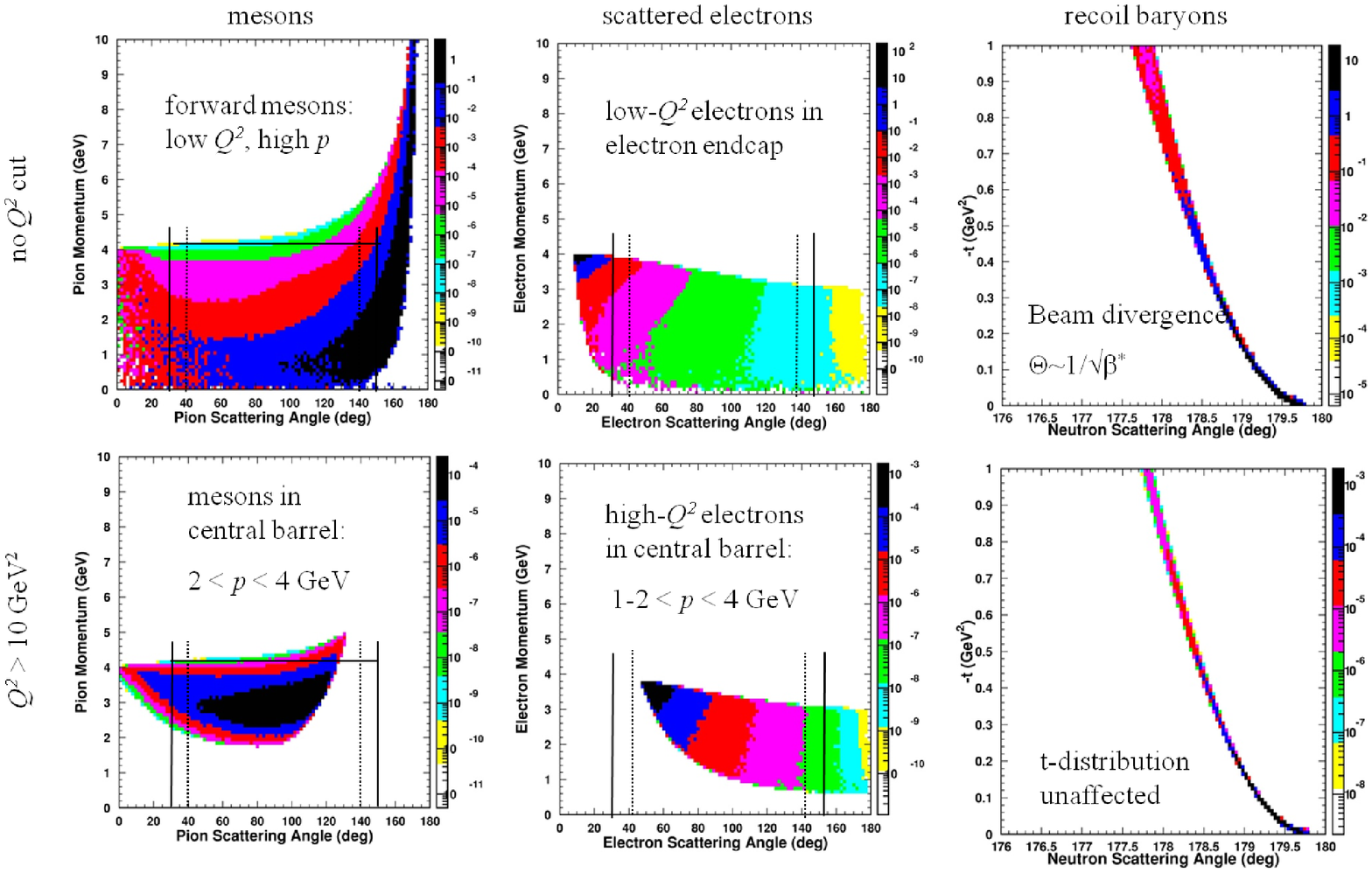}
\caption{\label{fig-meson-kinmat-compare} \small A comparison of the kinematic phase space for the scattered meson, electron, and the recoiling baryon in exclusive photo- and meson electroproduction at $\sqrt{s}$=21.9 GeV. The three upper panels are dominated by photoproduction; the three lower panels focus on light meson events for studies of transverse spatial distributions.} 
\end{figure}

The meson momentum distribution has a strong $Q^2$ dependence with the high momentum region dominated by low-$Q^2$ (photoproduction) events. This is illustrated in figure~\ref{fig-meson-kinmat-compare} with a comparison of photo- and electroproduction 
at fixed $\sqrt{s}$=22 GeV using a cut of $Q^2 >$ 10 GeV$^2$ to select the electroproduced light mesons. The forward scattered photoproduced mesons dominate the low $Q^2$ region populating a narrow angle cone with high momentum while the light mesons with $Q^2 >$10 GeV$^2$ are centered around central angles at momenta between 2 and 4 GeV. The scattered electron distribution shows the same general features as discussed above. The $t$ distribution of the recoil baryons does not change with selecting the low or high $Q^2$ region.

\begin{figure}[!htbp]
\centering
\includegraphics[width=2.1in, angle=270]{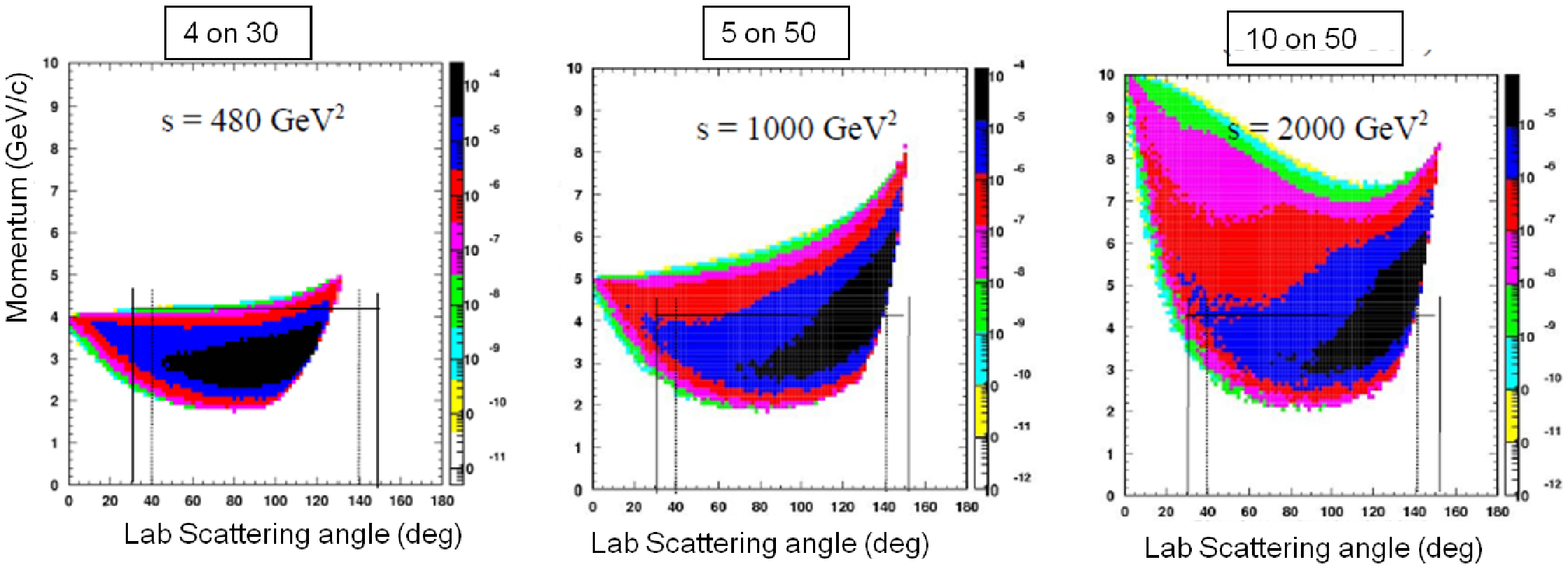}
\caption{\label{fig-meson-kinmat-higher} \small The meson kinematic phase space for higher energies for deep exclusive reactions for different combinations of the beam electron and proton energies. The first value in the labels denotes the electron beam energy. A cut of $Q^2 >$ 10 GeV$^2$ is applied to focus on events needed for studies of transverse spatial distributions. } 
\end{figure}

The EIC includes the option of using higher electron energies up to 11 GeV. Figure~\ref{fig-meson-kinmat-higher} shows a comparison of the meson distribution at fixed values of the ion beam energy for three values of $\sqrt{s}$=21.9, 31.6, 44.7 GeV. Here, we will focus on the distribution in the central region, which is indicated by the vertical lines. As mentioned above, the meson distribution is pushed into a narrow angular cone with an increasingly higher momentum as $\sqrt{s}$ increases. Furthermore, the average meson momentum in the central region between $\pm$ 30 deg increases from 4 GeV/c to about 8 GeV/c as the electron beam energy doubles from about 5 to 10 GeV at a fixed ion beam energy. Measurements of exclusive reactions at electron beam energies of about 10 GeV/c and fixed ion beam energies would thus require the detection of high momentum mesons in the central angle region ($\pm$ 30$^\circ$).

\begin{figure}[!htbp]
\centering
\includegraphics[width=3.5in, angle=270]{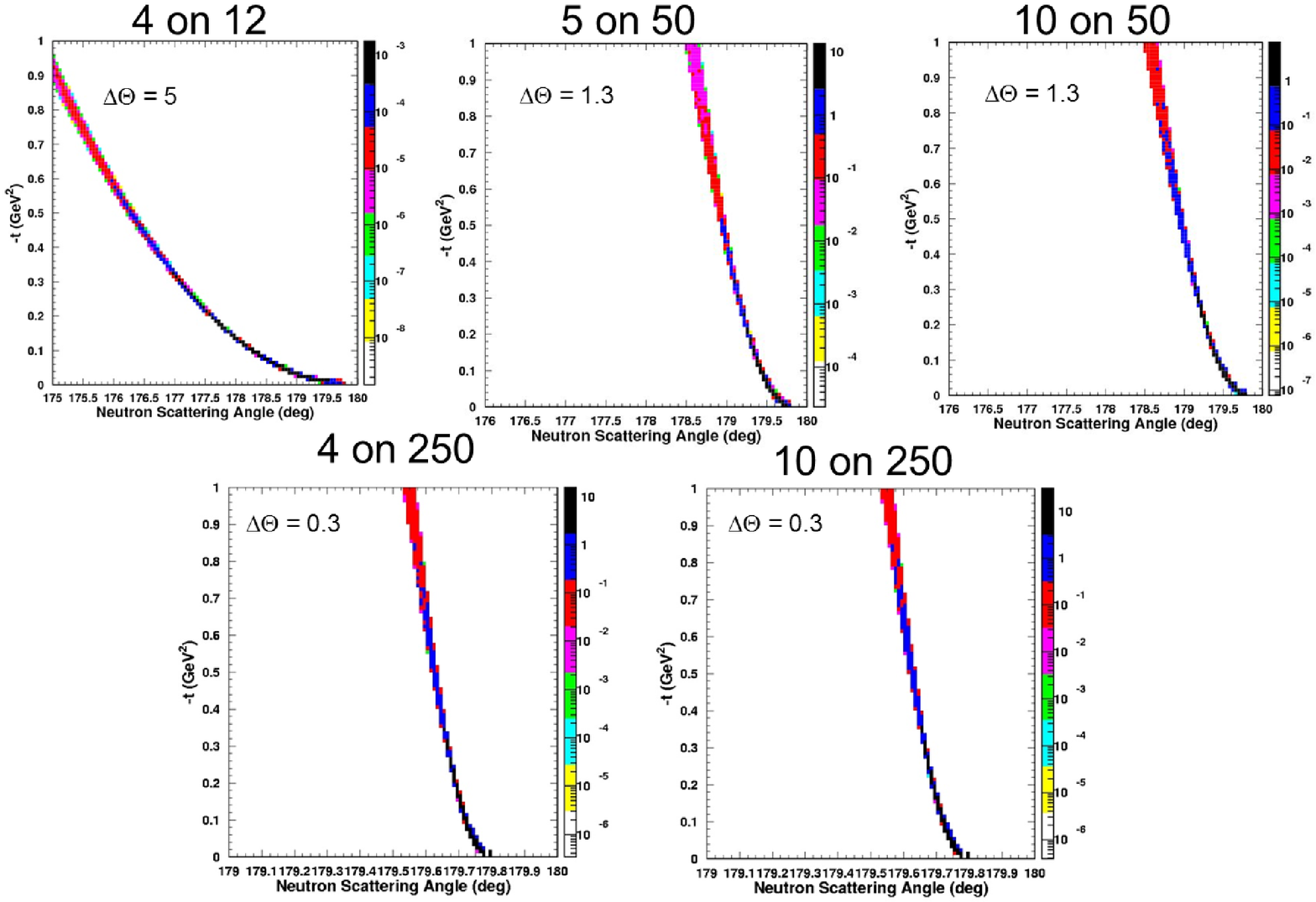}
\caption{\label{fig-recoil-kinmat} \small The kinematic phase space as $-t$ {\it vs.} the scattering angle of the recoil baryon in exclusive reactions for five values of $\sqrt{s}$. The first value in the labels denotes the electron beam energy. A cut of $Q^2 >$ 10 GeV$^2$ is applied to focus on events needed for studies of transverse spatial distributions. } 
\end{figure}

To access the physics of interest in exclusive reactions and extract information about the GPDs, one needs data binned over a sufficiently large range in $-t$:  
a range of at least $0<|t|<1$ GeV$^2$ is needed. In $ep$ collisions, the main challenge is that the outgoing baryons are scattered at relatively small angles, especially at low $-t$, as the resolution goes roughly as the inverse of the proton beam energy,
\begin{equation}
   \frac{\delta t}{t} \sim \frac{t}{E_p}\,.
\end{equation}
Figure~\ref{fig-recoil-kinmat} illustrates the deep exclusive recoil baryon $-t$ angular resolutions for values of $\sqrt{s}=(13.8, 31.6, 44.7, 63.2,100)$ GeV. 
The nearly symmetric collisions at lower proton beam energy provide the largest recoil baryon angular distributions of values of at least 1$^{\circ}$. For asymmetric collisions, the distribution rapidly decreases to the angular distributions of less than 0.3$^\circ$. To access the physics of interest, a better $-t$ resolution would thus be achieved with lower-energy and more symmetric kinematics.

\subsection{L/T separation}

\begin{figure}[!htbp]
\centering
\includegraphics[width=2.5in, angle=0]{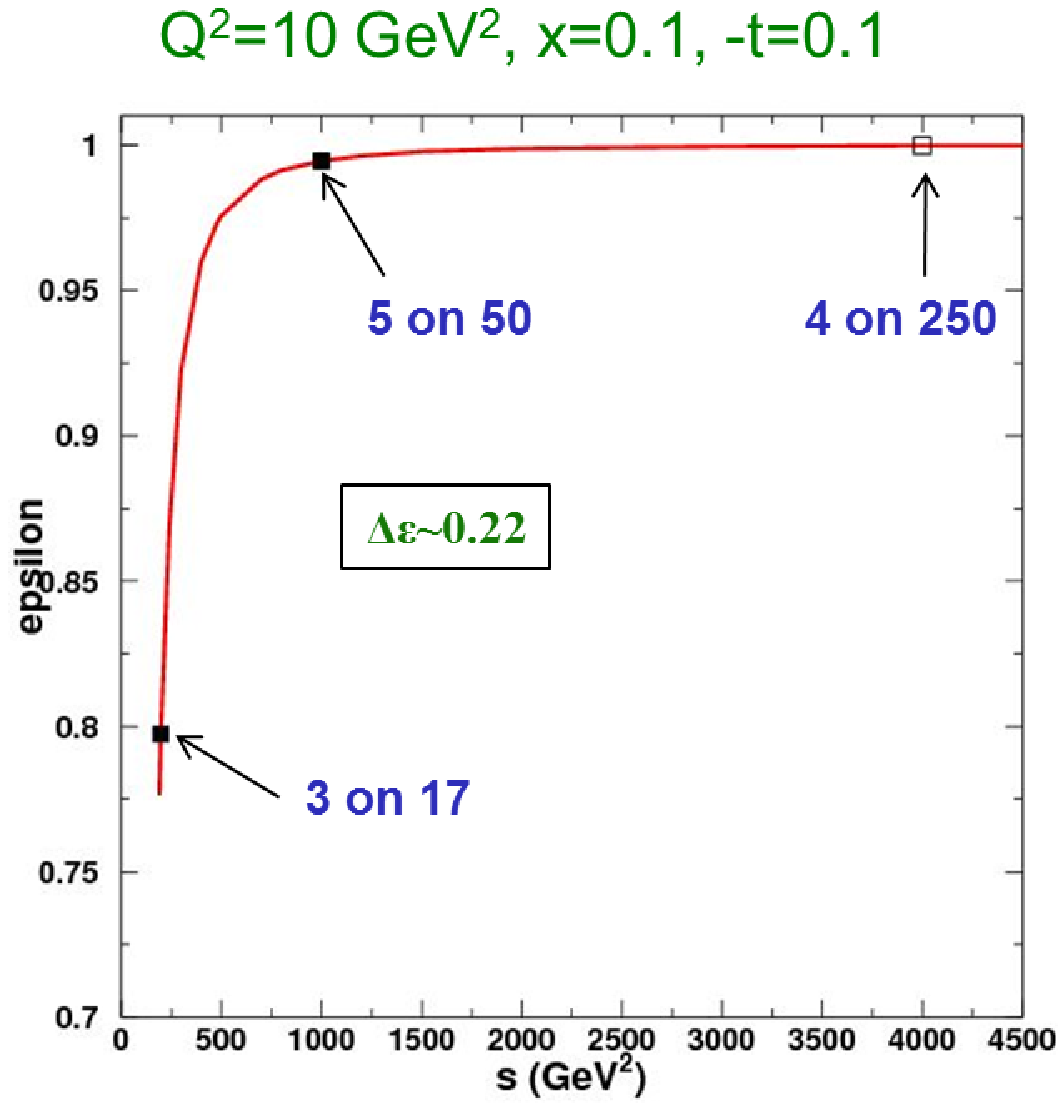}
\caption{\label{fig-epsilon-s} \small The virtual photon polarization, $\epsilon$ as a function of $s$ for different combinations of the electron and proton beam energies, at fixed $Q^2$=10 GeV$^2$, $x_B$=0.1, and $-t$=0.1~GeV$^2$. 
At high values of $s$, 
$\epsilon \to 1$
complicating the L/T separation.} 
\end{figure}
Beyond studies of transverse spatial structure of sea quarks, non-diffractive processes provide the opportunity for additional studies, for instance, the tests of hard-soft QCD factorization and measurements of the pion form factor. These measurements require isolating the longitudinal part of the electroproduction cross section using the L/T separations. This technique requires comparing data taken at two different beam energies with sufficiently large separation of the virtual photon polarization, $\Delta \epsilon$, to control systematic uncertainties. Based on previous L/T separations, a minimum acceptable value of $\Delta \epsilon$ is 0.1. 

Figure~\ref{fig-epsilon-s} shows the accessible values of $\epsilon$ as a function of $s$ at fixed values of $Q^2$, $x_B$, and $-t$. The lowest value of $\epsilon$ of about 0.8 is reached 
at $\sqrt{s}$=14.3 GeV, increasing to near unity as $s$ increases. Beyond $\sqrt{s}$=31.6 GeV, $\epsilon$ is effectively unity making the L/T distributions impossible.

\subsection{Summary of basic requirements for exclusive reactions}

Studies of exclusive non-diffractive processes provide important information on the transverse spatial distribution of non-perturbative sea quarks. These measurements require high luminosity for fully differential measurements in $x_B$, $-t$, and $Q^2$ as well as recoil detection for exclusivity. They require a kinematic reach in $t$ of at least up to 1 GeV$^2$ with good resolution. Our studies suggest that exclusive processes for values of $x_B>0.01$ have better prospects with lower-energy and more symmetric kinematics.

The following list summarizes the basic experimental requirements for studies of the transverse spatial structure of sea quarks through non-diffractive exclusive processes.\\
\underline{Energies}\\
 -- More symmetric energies favorable in exclusive non-diffractive reactions;\\
 -- Lower energies essential for a range in $\epsilon$ for the L/T separation.\\
\underline{Kinematic Reach}\\
 -- Need $Q^2>10$ GeV$^2$ (pointlike configurations);\\
 -- $x_B$ range between 0.001 and 0.1 overlapping with HERA and JLab 12 GeV;\\
 -- $s$ range between 200 and 1000 GeV$^2$.\\
\underline{Luminosity}\\
-- Exclusive non-diffractive processes require high luminosity for low rates for fully differential measurements;
-- Kaons push luminosity to $> 10^{34}$ cm$^{-2}$ s$^{-1}$.\\
\underline{Detection}\\
      -- Need recoil detection for exclusivity;\\
      -- Range in $-t$ and resolution.


\section{Partonic transverse spin in deep-inelastic exclusive experiments}
\label{sec:Liuti}


\hspace{\parindent}\parbox{0.92\textwidth}{\slshape 
  Gary R. Goldstein, Simonetta Liuti}
%

\index{Goldstein, Gary}
\index{Liuti, Simonetta}



\subsection{Introduction}

In this contribution, we suggest a class of deeply virtual exclusive reactions, namely pseudoscalar meson electroproduction, as a 
means to access chiral-odd distributions.
These are described by a set of four chiral-odd GPDs which enter the matrix elements for the various terms of the cross section.  
We conducted an analysis using a parameterization of the GPDs that is inspired by a physically motivated picture of the nucleon as a quark-diquark system with a Regge behavior. In the chiral-even sector a quantitative parameterization can be obtained from a global fit to PDFs, nucleon form factors, and DVCS data where the masses, couplings and Regge power behavior that set the scale for the dependence on the kinematic variables, $X, \zeta, t, Q^2$, are determined via a {\em recursive} procedure~\cite{Goldstein:2010gu}.

The extension of this parameterization scheme to the chiral-odd GPDs is critical for the phenomenology of deeply virtual meson electroproduction, which was begun particularly for the $\pi^0$ in~\cite{Ahmad:2008hp}. In the diquark spectator model,
chiral-even helicity amplitudes are simply related to their chiral-odd counterparts via parity transformations. 
For the $d$-quark case it is only the axial diquark relations that are involved, while the $u$-quark involves the scalar  contribution, as well. We thereby obtain the full set of four  chiral-odd GPDs, each being linearly related to helicity amplitudes. This allows us to predict the behavior of pseudoscalar electroproduction~\cite{GGL_progress}. 

It has now become particularly pressing to study the heavy quark components of the nucleon because of the advent of the LHC.  
For the types of precision measurements in the unprecedented multi-TeV CM energy regimes envisaged at the LHC it will be necessary to provide accurately determined QCD inputs.
The analyses in~\cite{Pumplin:2007wg}
have shown how the inclusion of non perturbative charm quarks could modify the outcome of  global PDF analyses. However, the situation is not clear-cut. 
We therefore extended our analysis to strange and charm pseudoscalar meson production \cite{Liuti:2010xy}. 
We proposed that in order to refine analyses such as the one in~\cite{Pumplin:2007wg}, new observables need to be identified from deeply virtual meson production and spin correlation measurements. 
We presented preliminary results involving the following electroproduction exclusive processes: 
(1) $\gamma^{\ast} p \rightarrow J/\psi \, p^\prime$; (2) $\gamma^{\ast} p \rightarrow D \, \overline{D} \, p^\prime$; (3) $\gamma^{\ast} p \rightarrow \overline{D} \, \Lambda_c$; 
(4) $\gamma^{\ast} p \rightarrow \eta_C \, p^\prime$.
These processes necessitate: {\it i)}  high luminosity because they are exclusive; {\it ii)}  high enough $Q^2$ to produce the various charmed mesons, and {\it iii)} a wide kinematical range in Bjorken $x$. 

Finally, a few questions have emerged concerning on one side  the applicability of dispersion relations to deeply virtual exclusive processes \cite{Goldstein:2009ks}, and on the other, the commonly assumed partonic picture of the ERBL region \cite{Goldstein:2010ce}. 
Newer deeply virtual exclusive cross section and asymmetry measurements in extended kinematical regimes will provide essential tests of the theory.

\subsection{Transverse spin from pseudoscalar meson production}
 
The basic definition of the quark-nucleon GPDs is through off-forward matrix elements of quark field correlators.
Contracting with the Dirac matrices, $\gamma^\mu$ or $ \gamma^\mu \gamma^5$ ($\sigma^{\mu \nu} \gamma^5$), and integrating over the internal quark momenta gives rise to the four chiral-even  GPDs, $H, E$ or $\widetilde{H}, \widetilde{E}$, and four chiral-even GPDs,
$H_T, E_T, \widetilde{H}_T, \widetilde{E}_T$~\cite{Diehl:2003ny}. 
The crucial connection of the eight GPDs to spin dependent observables in DVCS and DVMP is through the helicity decomposition~\cite{Diehl:2003ny}.
For example,
\begin{equation}
A_{++,++}(X,\xi,t)=\frac{\sqrt{1-\xi^2}}{2}(H^q+{\tilde H}^q-\frac{\xi^2}{1-\xi^2}(E^q+{\tilde E}^q))\,,
\label{Goldstein:chiraleven}
\end{equation}
\begin{equation}
A_{++,--}(X,\xi,t)=\sqrt{1-\xi^2}(H_T^q+\frac{t_0-t}{4M^2}{\tilde H}_T^q-\frac{\xi}{1-\xi^2}(\xi E_T^q+{\tilde E}_T^q)) \,.
\label{Goldstein:chiralodd}
\end{equation}

We have constructed a robust model for the GPDs, extending previous work~\cite{Ahmad:2007vw} that is based on the parameterization of diquark spectators and Regge behavior at small $X$. The GPD model parameters are constrained by their relations to PDFs (at $\zeta=0, t=0$)
and to nucleon form factors $F_1(t)$,  $F_2(t)$, $ g_A(t)$, and $g_P(t)$ through the first $x$ moments.
For the chiral-odd GPDs, there are fewer constraints. 
In particular, $H_T(X,0,0)=h_1(X)$ can be fit using the loose constraints 
in~\cite{Anselmino:2007fs} since the first moment of $H_T(X,\xi,t)$ is the ``tensor form factor'', called $g_T(t)$.  
It is conjectured that the first moment of $2{\tilde H}_T^q(X,0,0)+E_T^q(X,0,0)$ is a  ``transverse anomalous moment'', $\kappa_T^q$, 
defined in~\cite{Burkardt:2005hp}.

\begin{figure}[h] 
\begin{center}
\includegraphics[width=.40\textwidth]{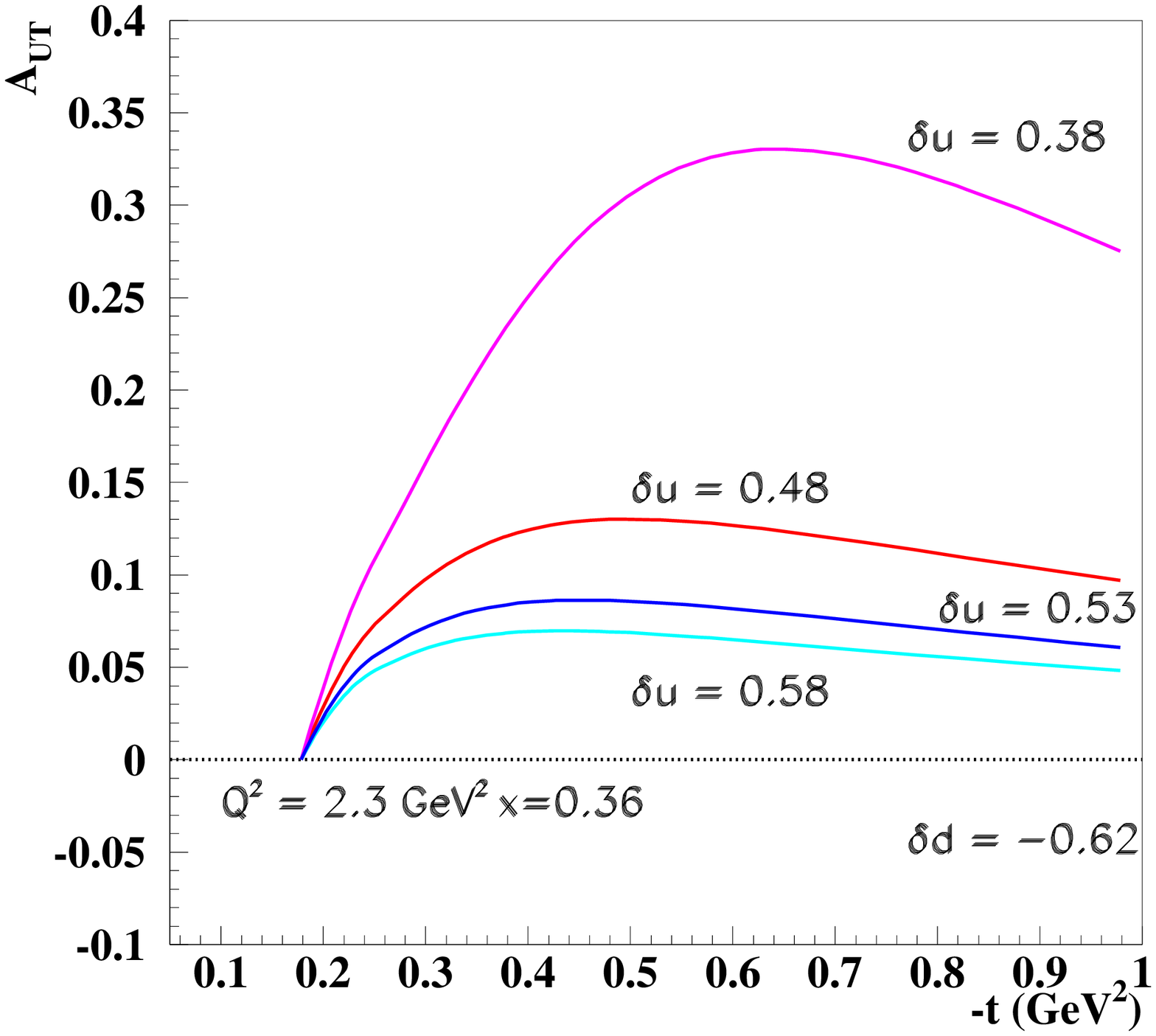} 
\hspace{0.5cm}
\includegraphics[width=.40\textwidth]{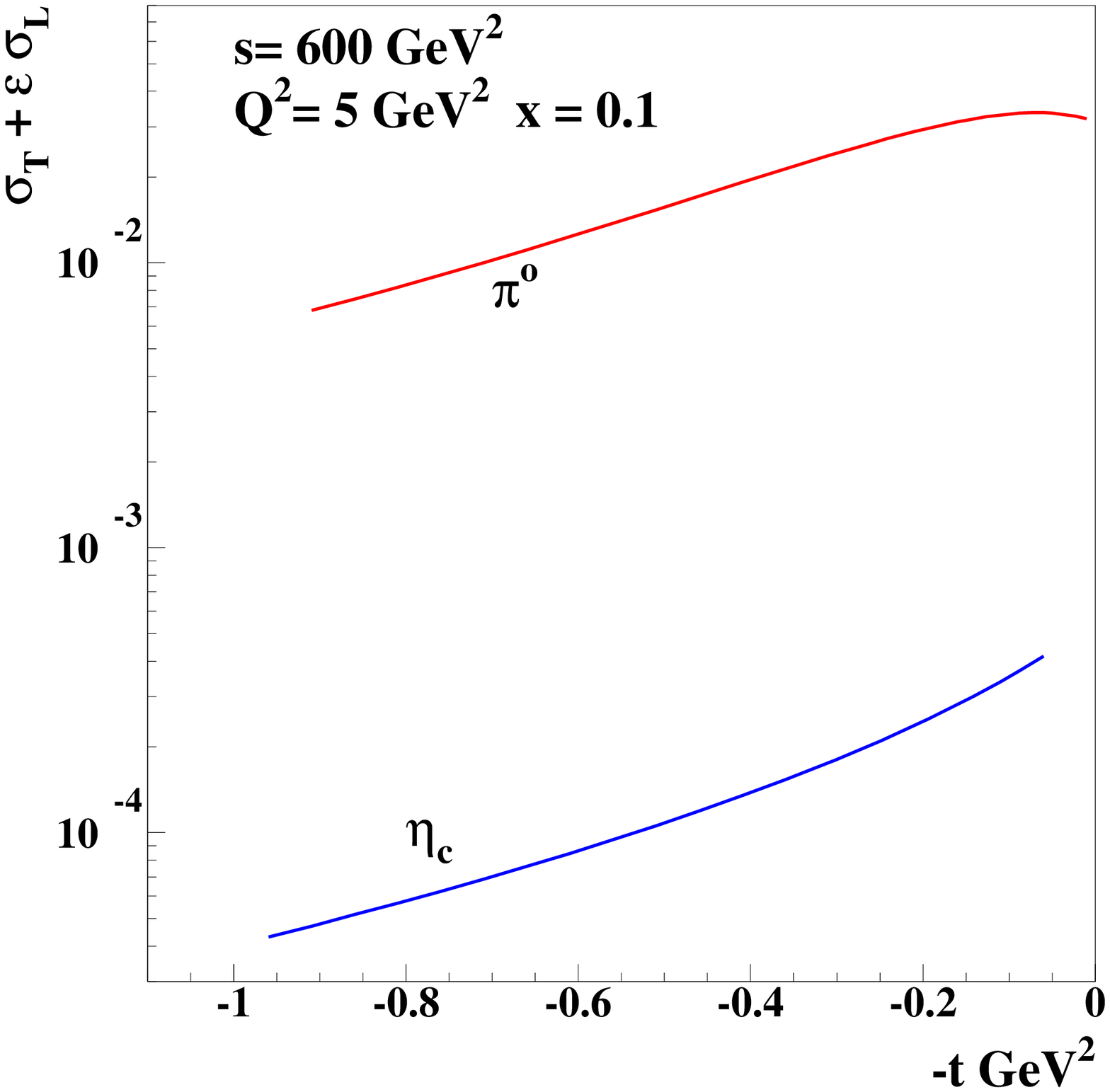} 
\end{center}
\caption{\label{fig1}\small  Left: Transverse spin asymmetry, $A_{UT}$, {\it vs.} $-t$ at 
$Q^2=2.3$ GeV$^2$ and $x_{B}=0.36$ for different
values of the tensor charge, $\delta u$, with fixed 
 $\delta d=-0.62$.
Right: Comparison of $\pi^0$ and $\eta_c$ cross sections. 
The range between the two lines gives an estimate of where the cross sections for the other processes
will lie.}
\end{figure}

With our ansatz, many observables can be determined in parallel with the corresponding Regge predictions. Since the initial work~\cite{Ahmad:2008hp}, we have undertaken a more extensive parameterization and  presented several new predictions~\cite{Goldstein:2010gu}. 
In figure~\ref{fig1}, we show an example corresponding to the transversely polarized proton
target.

The measured cross section for $\pi^0$ is sizable and has large transverse $\gamma^*$ contributions. This indicates that the main contributions should come from chiral-odd GPDs, for which the $t$-channel decomposition is richer. In particular, because these GPDs arise from the Dirac matrices $\sigma^{\mu \nu}$, there are two series of $J^{PC}$ values for each GPD~\cite{Hagler:2004yt} corresponding to space-space or time-space combinations  $1^{--}$ and $1^{+-}$. These series occur for three of the four chiral-odd GPDs, with exception of $\widetilde{E}_T$. We are thus led to the conclusion that chiral-odd GPDs will dominate the neutral pseudoscalar leptoproduction cross sections. 
This result has interesting consequences. First, in a factorized handbag picture, these GPDs will couple to the hard part, the $\gamma^{\ast}+\rm{quark} \rightarrow \pi^0 +\rm{quark}$, 
provided that $\pi^0$ couples through $\gamma^5$, which is naively twist-three, rather than 
the twist-two coupling $\gamma^+ \gamma^5$.  
Second, the vector  $1^{--}$ and axial-vector $1^{+-}$ in the $t$-channel, viewed as particles ($\rho^0, \omega \,\, \rm{and} \,\,b_1^0, h$), couple primarily to the transverse virtual photon. 
For Reggeons, the $1^{--}$ does not couple at all to the longitudinal photon, while the axial-vector $1^{+-}$ does so through helicity flip~\cite{Goldstein:1973xn}. Guided by these observations~\cite{Ahmad:2008hp}, we assume that the hard part depends on whether the exchange quantum numbers are in the vector or axial-vector series, thereby introducing orbital angular momentum into the model. We use $Q^2$ dependent electromagnetic ``transition'' form factors for vector or axial-vector quantum numbers going to a pion. We calculate these using pQCD for $q+\bar{q} + \gamma^{\ast}(Q^2) \rightarrow q+\bar{q}$ and a standard $z$-dependent pion wave function, convoluted in the impact parameter representation that allows orbital excitations to be easily implemented. 

With our model for the chiral-odd spin-dependent GPDs and these transition form factors, we can obtain the full range of cross sections and asymmetries in kinematic regimes that coincide with ongoing JLab experiments. (A similar emphasis on chiral-odd contributions for $\pi$ electroproduction has recently been proposed~\cite{Goloskokov:2009ia}, although the details of that model are quite different from ours.) We are able to predict the important transverse photon contributions to the observables~\cite{Ahmad:2008hp}.  In figure \ref{fig1} (left), we show one striking example of the predictions that depend on the values of the tensor charges, thereby providing a means to narrow down those important quantities. This program has been presented~\cite{Goldstein:2010gu} and  further details will soon appear, as the refinements of the chiral-odd parameterization are completed \cite{GGL_progress}. In figure \ref{fig1} (right), we show the cross section, $\sigma_T + \epsilon \sigma_L$, for charmed meson production and compare it to the one for $\pi^0$ production~\cite{Liuti:2010xy}.

In summary, through the use of physically motivated models and the new horizons provided by the EIC, a far reaching interpretation of the separate spin-dependent GPDs and thereby, a picture of the transverse structure of the nucleons will emerge. 
The connection of chiral-odd GPDs to the transversity structure of the nucleon is of great interest as a manifestation of quark and gluon orbital angular momentum.


%

\section{Ways to access  transversity GPDs at the EIC}
\label{sec: transversityGPDs}

\hspace{\parindent}\parbox{0.92\textwidth}{\slshape 
  B. Pire, L. Szymanowski, S. Wallon}
%

\index{Pire, Bernard}
\index{Szymanowski, Lech}
\index{Wallon, Samuel}





\subsection{Introduction}
Transversity quark distributions in the nucleon remain among the most unknown leading twist hadronic observables. This is mostly due to their chiral-odd character which enforces their decoupling in most hard amplitudes. Generalized parton distributions (GPDs) offer a new way to access the transversity dependent quark content of the nucleon. The factorization properties of exclusive amplitudes allow in principle to extract the four chiral-odd transversity 
GPDs~\cite{Diehl:2001pm}, 
$H_T$, $E_T$, $\tilde{H}_T$, $\tilde{E}_T$. However, one-photon or one-meson electroproduction leading twist amplitudes are insensitive to them~\cite{Diehl:1998pd,Collins:1999un}.  The strategy which we followed in~\cite{Ivanov:2002jj,Enberg:2006he} is to study the leading twist contribution to exclusive processes where more mesons are present in the final state. Note that, contrarily to transversity PDFs, transversity GPDs enter the formulae for exclusive cross sections even when considering {\em unpolarized} proton
target, provided one selects the polarization state of an outgoing meson.

\subsection{Diffractive photoproduction of two $\rho$ mesons}
\begin{figure}
\centerline{\includegraphics[width=7.cm]{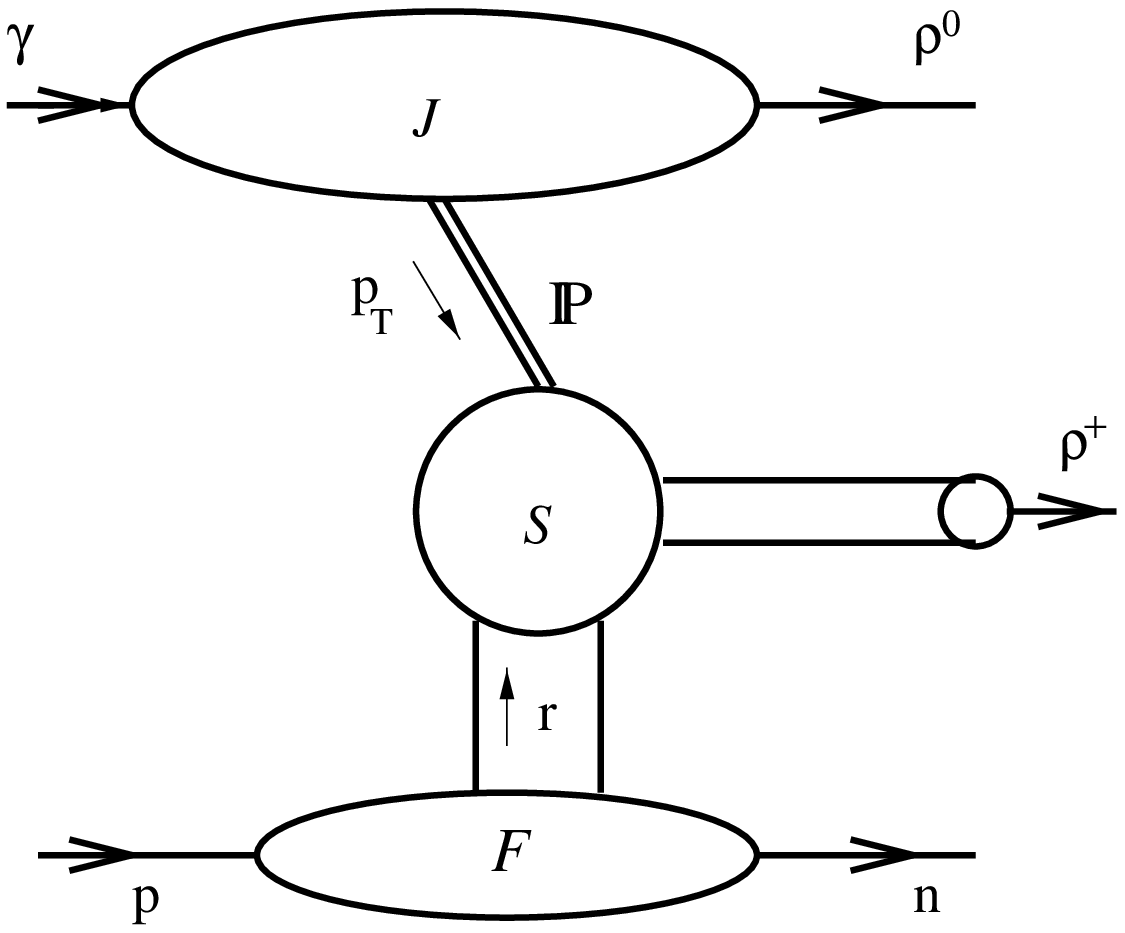}}
\caption{\small Factorization of the process $\gamma_{T} (q) + p (p_2) \to  \rho_{L,T}^0(q_\rho) + \rho_{T}^+(p_\rho)+
n(p_{2^{\prime}})$
in the asymmetric kinematics discussed in the text. $\pom$ is the hard Pomeron modeled by 
a two-gluon exchange.}
\label{figPro}
\end{figure}

We consider~\cite{Ivanov:2002jj,Enberg:2006he}, in analogy with the virtual photon exchange occurring in the deep inelastic electroproduction of a meson, the subprocess:
\begin{equation}
\pom (q_P)+ p (p_2) \to  \rho_{T}(p_\rho)+N^{\prime}(p_{2^{\prime}})\;,
\end{equation}
of almost forward scattering of a  virtual  Pomeron (the hard scale is the virtuality $- q_P^2$ of this Pomeron) on a nucleon. This subprocess is at work in the process
\begin{equation}
\label{2mesongen}
\gamma^{(\ast)}_{L/T} (q) + p (p_2) \to  \rho_{L,T}^0(q_\rho)+ \rho_{T}(p_\rho)+
N^{\prime}(p_{2^{\prime}})\;,
\end{equation}
where a real or  virtual  photon scatters on a proton
 $p$, which leads via a two-gluon exchange to the production
of  two vector mesons separated by a large
rapidity gap  and the scattered nucleon $N^{\prime}$, as shown on  figure~\ref{figPro}.   The final state may be either $\rho^0 \rho^0 p$ or  $\rho^0 \rho^+ n$. In both cases, 
the two-gluon exchange  with the nucleon line is forbidden by charge conjugation or charge conservation and the process is thus sensitive only to quark GPDs. 
We consider the kinematical region where the rapidity gap 
between $\rho (p_\rho)$ and
$N^{\prime}$ is much smaller than the one between
$\rho (q_\rho)$ and $\rho (p_\rho)$, i.e., the energy of the 
($\rho (p_\rho)+ N^{\prime}$) system is smaller
than that of the ($\rho + \rho$) system but is still large enough to
justify our approach  (in particular, it is much larger than baryonic resonance masses). 
Since quasi-real transverse photons are more abundant in electron-ion collisions and charged pions are most easily detected, one may specialize to the reaction:
\begin{equation}
\label{2mesonspec}
\gamma_{T} (q) + p (p_2) \to  \rho_{L,T}^0(q_\rho)+ \rho_{T}^0(p_\rho)+
p(p_{2^{\prime}})\;,
\end{equation}
where the initial quasi-real photon is treated as if it were real. 

In  this kinematical regime, the amplitude for this process is calculable consistently within
the collinear factorization method, as an
integral (over  the longitudinal momentum fractions of the quarks)  of
the product of two amplitudes: the first one (the {\em impact factor}) describes
the transition $\gamma^{(\ast)} \to \rho_{L,T}^0$ via a two-gluon exchange and
the second one  describes the subprocess
$\pom+p\to \rho^0_T+p$. The fact that the latter process  is
closely related to the electroproduction process 
$\gamma^{\ast}\,p \to\rho^0\,p$
allows us to separate  its long distance dynamics  expressed
through the GPDs from a perturbatively calculable coefficient function. 
The
skewness parameter $\xi$ is related in the usual way ($\xi \approx x_B/(2-x_B)$) to the  Bjorken variable
defined by the Pomeron momentum $x_B = -q_P^2/(2q_P \cdot p_2)$. 
The choice of a transversely
 polarized vector meson $\rho_{T}^0$  involves a chiral-odd distribution amplitude, which in turn selects the chiral-odd GPDs. 

The resulting scattering amplitude ${\cal M}^{\gamma^*\,p\,\to \rho_L^0\, \rho^0_T\,p}$ then receives contributions from the four chiral-odd GPDs $H_T, \tilde H_T,E_T $ and $\tilde E_T$,  but  only the first one does not vanish kinematically in the forward direction. Thus, assuming  that the Mandelstam variable $-t= -(p_2-p_{2'})^2$ is sufficiently small,
the transversity GPD $H_T$ contribution dominates the amplitude of process
(\ref{2mesonspec}) which reads :
\begin{eqnarray}
\label{CON}
&&{\cal M}^{\gamma \,p\,\to \rho_L^0\, \rho^0_T\,p}
= \sin \theta \;16\pi^2 s \alpha_s f_\rho^T \xi \sqrt\frac{1-\xi}{1+\xi}
\frac{C_F}{N\,(p_T^{\;2})^2}\nonumber \\
&&
\times\int\limits_0^1
\frac{\;du\;\phi_\perp(u)}{ \,u^2 \bar u^2 }
 J^{\gamma \to \rho^0_L}(up_T,\bar up_T)
 \frac{H^{u\,d}_T(\xi(2u-1),\xi,t)}{\sqrt 2} \,,
\end{eqnarray}
where $ H_T^{u\,d}= H_{T}^u- H_T^{d}$; $\phi_\perp(u)$ is the distribution
amplitude (DA) of the $\rho_T$ meson; $\theta$ is the angle between the transverse polarization vector of the target $\vec{n}$ and the polarization vector
$\vec{\epsilon}_T$ of the produced $\rho^0_T$ meson; $\vec{\varepsilon}$ is the polarization vector of the initial photon. The impact factor reads
\begin{equation}
\label{TL}
J^{\gamma \to \rho_L}(k_{T1},k_{T2}=p_T-k_{T1})=
-\frac{e\,\alpha_s\,\pi\,f_\rho^0}{\sqrt{2}\,N}\;\int\limits_0^1\,dz\,(2z-1)\,
\phi_\parallel(z)\,\left( \vec{\varepsilon}\,\cdot \vec Q_P \right) \,,
\end{equation}
with
\begin{eqnarray}
\label{Q}
&&\vec Q_P(k_{T1},k_{T2}=p_T-k_{T1})=
\frac{z\,\vec p_T}{z^2\,p_T^2+Q^2\,z\,\bar z +m_q^2}
- \frac{\bar z\,\vec p_T}{\bar z^2\,p_T^2+Q^2\,z\,\bar z +m_q^2} \nonumber
 \\
&&\hspace{1cm}+\frac{\vec k_{T1} - z\,\vec p_T}{(k_{T1} - z\,p_T)^2+Q^2\,z\,\bar z +m_q^2}
-\frac{\vec k_{T1} - \bar z\,\vec p_T}{(k_{T1} - \bar z\,p_T)^2+Q^2\,z\,\bar z +m_q^2}\;.
\nonumber
\end{eqnarray}
 The scattering amplitude~(\ref{CON}) receives a contribution only from the 
ERBL region.

\subsection{Cross section estimates}

To obtain an estimate of the differential cross section of this process, we need 
a model for the transversity GPD $H_T^q(x,\xi,t)$ ($q=u,\ d$). 
 We proposed a simple meson-pole approach
 starting with the effective interaction Lagrangian, 
\begin{equation}
{\cal L_{ANN}}= 
 \frac  {g_{A\, NN}}{2M}\bar N \sigma_{\mu\nu}\gamma_5
\partial^\nu A^\mu N \;,
\label{ANN}
\end{equation}
in which $g_{A NN}$ is the coupling constant determining  the strength of  the interaction of 
the axial meson $A$ with the nucleon $N$. This yields
\begin{equation}
\label{pole}
H^a_T(x,\xi)=\frac{g_{ANN}f_A^{a\perp} \left(\Delta\cdot S_T  \right)^2 }{2M_N\,m_A^2}\,
\frac{\phi_\perp(\frac{x+\xi}{2\xi})}{2\xi}\;,
\end{equation}
where $\Delta$ is the transverse part of the momentum transfer vector $r$;  $f^{a\perp}_A$ 
is related to the $A$ meson decay constant. Identifying the scalar product
$\left(\Delta\cdot S_T  \right)^2$ with the average of the 
intrinsic transverse  momentum of the quarks, $\left(\Delta\cdot S_T  \right)^2 
\to 1/2 \langle k_\perp^2\rangle$, and the axial meson $A$ with the $b_1$ meson,
$A=b_1(1235)$, we obtain our final expression for  $H^{ud}_T$: 
\begin{equation}
\label{Hpole}
H^{ud}_T(x,\xi,0)=\frac{g_{b_1 NN}f_{b_1}^{T}\langle k_\perp^2\rangle }{
2\sqrt{2}M_N\,m_{b_1}^2}\,
\frac{\phi^{b_1}_\perp(\frac{x+\xi}{2\xi})}{2\xi}\;,
\end{equation}
where
$f_{b_1}^T = \sqrt{2}f_{a_1}/m_{b_1}$ with
$f_{a_1}=(0.19\pm0.03)$ GeV$^2$;
$g_{b_1NN}=5/(3\sqrt{2})g_{a_1NN}$ with $g_{a_1NN}=7.49 \pm 1.0$;
$\langle k_\perp^2 \rangle = (0.58 - 1.0)$ GeV$^2$. The $t$ dependence of the chiral-odd GPDs may be parameterized in the following simple way:
\begin{equation}
\label{t-dep}
H^q_T(x,\xi,t) = H^q_T(x,\xi,t=0)\times  \frac{C^2}{(t  - C)^2} \,,
\end{equation}
with the standard dipole form factor with  $C=0.71~$GeV$^2$.

\begin{figure}
\begin{center}
\includegraphics[width=7.5cm]{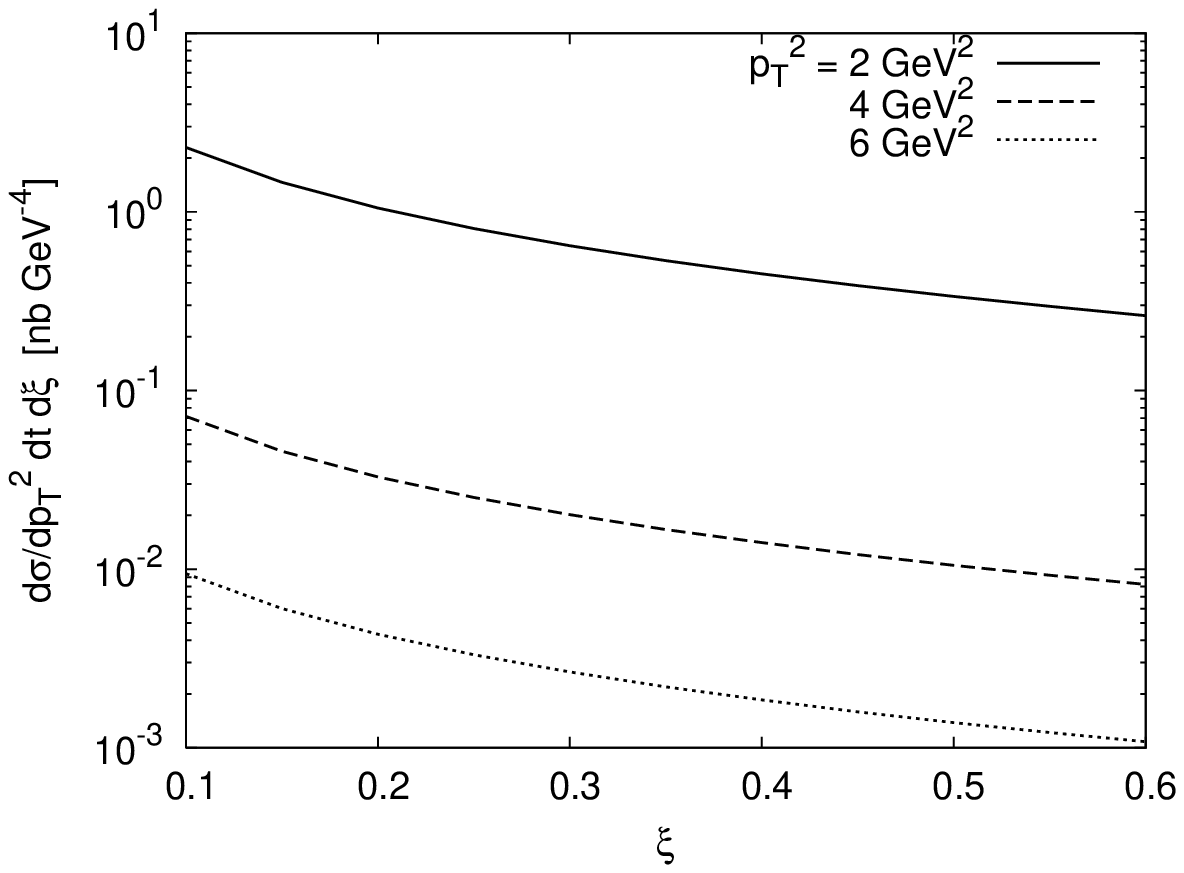}
\includegraphics[width=7.5cm]{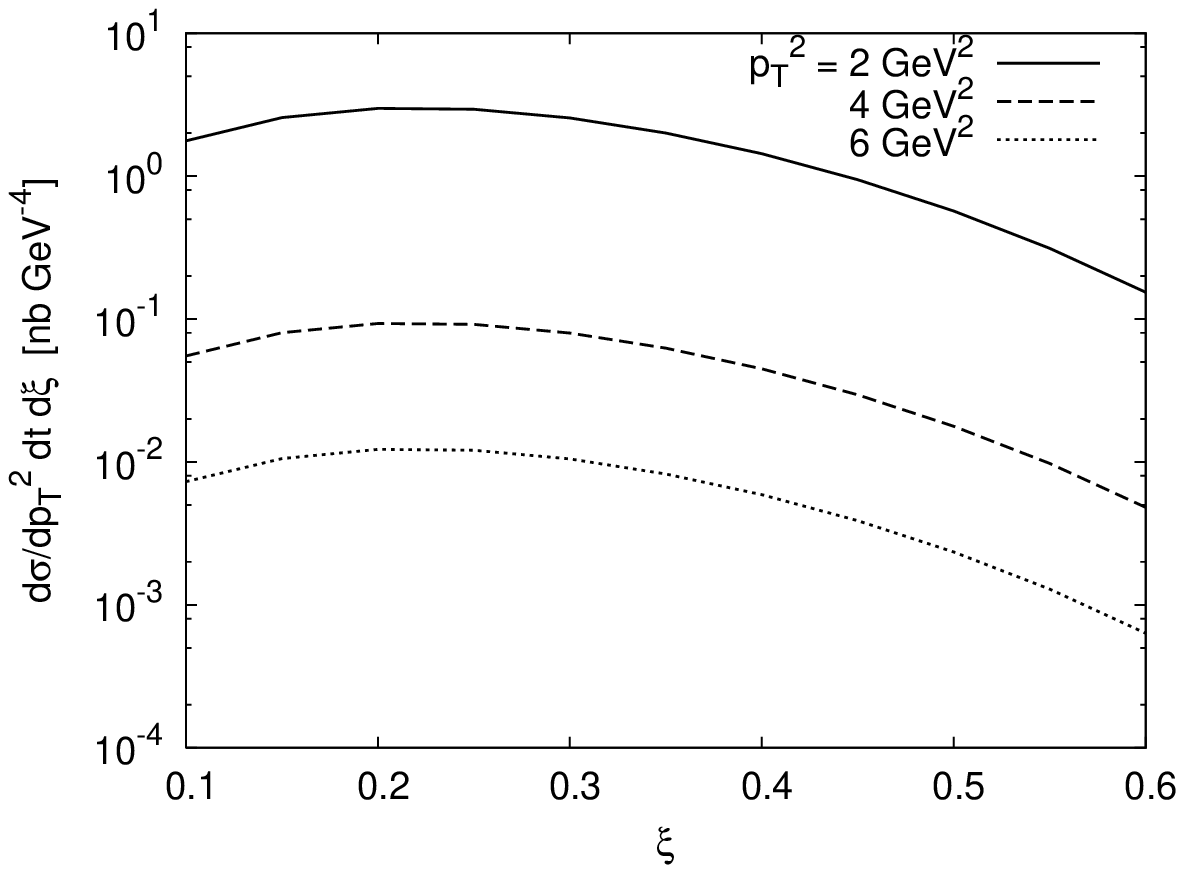}
\caption{\small
The differential cross section for the photoproduction of
$\rho^0_T$ and $\rho^+_T$ (left panel) and 
$\rho^0_L$ and $\rho^+_T$ (right panel) as a function of $\xi$ for $p_T^2 =
2,\,4,\,$ and $6$ GeV$^2$. The cross sections for the processes  with two 
neutral $\rho^0$ mesons in the final
state are two times smaller than those with $\rho^0~\rho^+$.
}
\label{fig2}
\end{center}
\end{figure}

In  figure~\ref{fig2}, we show our model estimates for the differential cross sections 
of photoproduction of two vector mesons~(\ref{2mesongen}), $\rho^0$ and transversely polarized $\rho^+$, 
with the unpolarized beam and target.
Note that these cross sections depend on the $\gamma$--nucleon energy only through 
the variable $\xi$.
 The cross sections for the processes  with two 
neutral $\rho^0$ mesons in the final
state are two times smaller than those with $\rho^0~\rho^+$.

\subsection{Photoproduction at lower photon energies }

Diffractive physics requires high photon energies. If low energy photon tagging may be performed at the EIC,  a QCD study based solely on the collinear factorization ({\em i.e.}, without any Pomeron exchange) approach may be followed, opening other interesting channels. 
In~\cite{Beiyad:2010cx,Beiyad:2009ju}, we considered the process:
\begin{equation}
\gamma(q) + p(p_1,\lambda) \rightarrow \pi^+(p_\pi) + \rho^0_T(p_\rho) + n(p_2,\lambda')\,,
\label{process2}
\end{equation}
on a  polarized or unpolarized proton target,  in the kinematical regime of large invariant mass $M_{\pi\rho}$ of the final meson pair (the hard factorization scale is now this invariant mass) and small momentum transfer $t =(p_1-p_2)^2$ between the initial and the final nucleons.   Roughly speaking, this kinematics means a moderate-to-large, and approximately opposite, transverse momentum of each  meson. The cross sections obtained are sizeable at values of $ s_{\gamma N}$ of the order $10 - 20$ GeV$^2$ but decrease quickly with the photon energies. This regime is more in the range of the JLab 12 program than of EIC.


 In conclusion, we stress that this approach only assumes leading twist factorization of non-perturbative  quantities, such as meson DAs and chiral-odd GPDs. 

\noindent
{\it Acknowledgments.}
We are grateful to  D.Yu.~Ivanov, R.~Enberg and O.V.~Teryaev for their contributions to the results presented here. 




\chapter{Input from lattice QCD}

\noindent
{\Large Chapter editors: \\[1em]
D. Hasch, F. Yuan}

\newpage




\section{Introduction}
\label{ch:lattice-sec:intro}

\hspace{\parindent}\parbox{0.92\textwidth}{\slshape
  Philipp H\"agler, Bernhard Musch, Andreas Sch\"afer}

\index{H\"agler, Philipp}
\index{Musch, Bernhard}
\index{Sch\"afer, Andreas}

\vspace{\baselineskip}

The focus of research at an EIC is a precise and comprehensive understanding
of the quark-gluon structure and dynamics of hadrons and nuclei within the
scope of traditional QCD, as well as beyond it, e.g., beyond the formalism
based on collinear parton distributions. This requires the combination of
input from many different fields, including lattice QCD (LQCD). In this
chapter, the status of LQCD as well as the prospects for the next decade
are sketched. The main tasks of LQCD is to increase precision and to
extend the scope of LQCD calculations. [Both depends also significantly on
progress in the understanding of perturbative QCD (pQCD).] Although
working out solutions for all technical details is a formidable task,
recent developments suggest that LQCD should have settled most of the open
theory issues by the time the EIC starts operating.

LQCD results are by now routinely used as input for phenomenology if direct experimental 
information is not available. This trend will intensify when in the future ever more subtle aspects are 
investigated. 
Therefore, the EIC and a dedicated effort in LQCD have to form a strong union.
If direct comparison with experiment has proven certain types of LQCD calculations to be reliable, 
LQCD can provide easily information which is hard to obtain experimentally, for example on 
moments of PDFs 
and GPDs and the flavour decomposition of structure functions.   
In this context it is, unfortunately, quite often not sufficiently appreciated that most
quantities of interest calculated on the lattice can only be linked to experiment 
by highly non-trivial input from pQCD. Thus all three elements, experiment, LQCD and pQCD have to 
be combined to reach optimal results.  

The two main sources of difficulty are:
\begin{itemize}
\item
The basis of LQCD is the observation that the analytic continuation to imaginary times 
$x^0\rightarrow {\rm i}x^4$ relates quantum field theory to statistics/thermodynamics. 
The latter allows for a purely numerical treatment by means of Monte Carlo techniques.
This analytic continuation is only simple for time-independent quantities. The
quantities of this type usually studied are matrix elements of local operators (which can be 
evaluated at $x^0=0=x^4$).
\begin{equation}
\langle h'(p') \Big| {\cal O}(x=0) \Big| h(p) \rangle\ .
\label{AS_Eq:1}
\end{equation}
Here $h$, $h'$ can be any hadronic state, including the QCD vacuum.  
One typically needs the continuum operator product expansion (OPE) to link such quantities 
to observables.  
\item 
Most QCD quantities of interest are scheme and scale dependent. Only in leading order (LO) 
this dependence can be neglected, but LO calculations are in most cases insufficient
for a high precision machine like the EIC. Thus LQCD results for matrix elements of
the type Eq.~(\ref{AS_Eq:1}) have to be matched to a specific pQCD setting, typically the 
$\overline{MS}$ scheme 
at a certain scale $\mu$. This requires also a matching of renormalization effects, 
which are quite different in the continuum and on the lattice due to the
loss of continuum symmetries (as discussed below).
The lattice discretization leads to different Feynman rules, in particular
the appearance of tadpole diagrams.
Another concrete, simple example is the modification of the fermion
propagator on the lattice, which typically might read
(depending on the specific lattice action)
\begin{equation}
D_{Lattice}(p) ~=~ \frac{m-{\rm i} a^{-1} \sum_{\mu} \gamma_{\mu} \sin(p_{\mu}a)}{m^2+ a^{-2} \sum_{\mu} \sin^2(p_{\mu}a)}
~~~\ .
\label{AS_Eq:3}
\end{equation}  
Thus, renormalization factors on the lattice and in the continuum differ by finite amounts, typically 
of the order of a few up to 30 percent. 
If one aims at an overall 
precision of order percent, the matching of the renormalization factors 
between non-perturbative (i.e. all order) lattice calculations and fixed order 
continuum calculations has to be achieved with high precision. 
To achieve this for all quantities of interest is clearly one of the major challenges for theory, both 
LQCD and pQCD, in the next decade. 
\end{itemize}

\begin{figure}
\begin{center}
\includegraphics[width=0.6\textwidth]{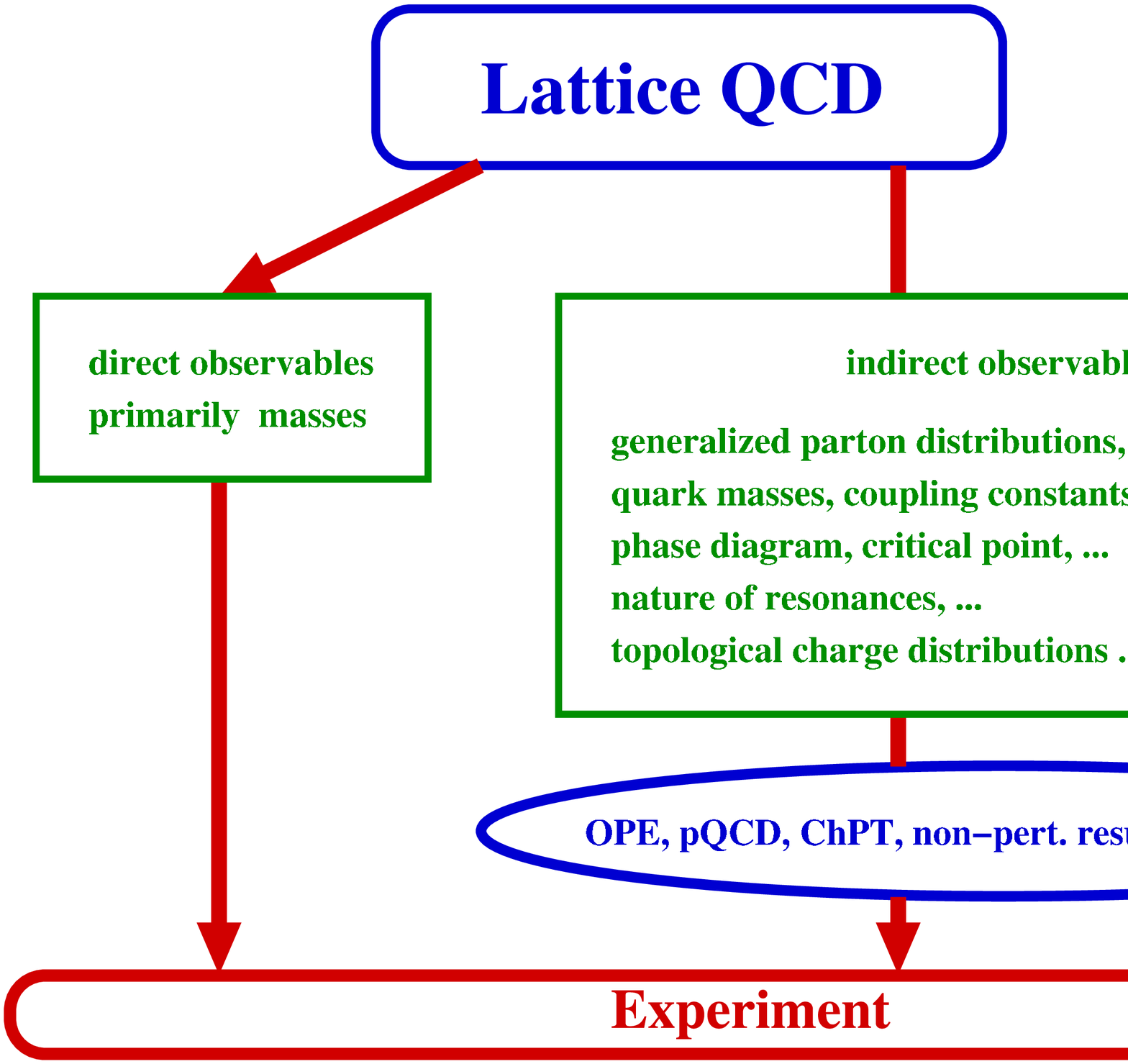}
\end{center}
\vspace*{-0.4cm}
\caption{\label{AS_Fig:1} Sketch of the different types of lattice observables.
For nearly all quantities of interest for an EIC, a combination of pQCD and LQCD 
is needed to make contact to experiment.}
\end{figure}

The points just discussed apply to 'indirect observables', as illustrated in the right column of 
Fig. \ref{AS_Fig:1}.
There do also exist some observables which can be 
compared directly, without the need for renormalization, especially 
hadron masses. However, these are well known experimentally, while the aim 
of LQCD is clearly to provide information on 
hitherto unknown correlators. The observables of
interest at the EIC, require nearly always non-perturbative renormalization
in the corresponding lattice studies.  

The calculation of matrix elements like (\ref{AS_Eq:1}) proceeds as follows: 
\begin{enumerate}
\item
 One generates a number of ensembles of gauge field configurations 
with the correct statistical weights. The parameters for these ensembles are 
chosen such that one has best control (for given computer resources) of the 
combined limit: lattice spacing $a\rightarrow 0$; physical lattice size 
$L\gg 1/\Lambda_{QCD}$; quark masses $m_q \rightarrow m_q(physical)$; large number 
of independent field configurations, typically $N\gg 100$.
\item
One generates hadronic states using
products of quark fields with the correct quantum numbers
(sources), e.g., one can use for a proton ($C=i\gamma^2\gamma^4$ is the charge 
conjugation matrix, $i,j,k$ run over the three color states):
\begin{equation}
\hat B_{\alpha}(t,\vec p) ~=~ \sum_{\vec x} e^{{\rm i}\vec p\cdot \vec x}
\epsilon_{ijk} \hat u^i_{\alpha}(x) ~\hat  u^j_{\beta}(x)
(C^{-1}\gamma_5)_{\beta\gamma}\hat  d^k_{\gamma}(x) ~~~~~.
\label{AS_Eq:4}
\end{equation}
Propagation in Euclidean time generates real exponentials 
rather than phases. Consequently, when expanding into a series in the correct physical 
multi-particle hadronic states, propagation in Euclidean time filters out the 
lowest mass state for large enough times,
\begin{eqnarray}
\hat B(0,\vec p)|0\rangle ~&=& ~c_0|N\rangle ~+~ c_1 |N'\rangle ~+~ c_2|N\pi\rangle ~+~...
\nonumber\\
\hat B(t,\vec p)|0\rangle &=& ~c_0 e^{-E_Nt}|N\rangle ~+~ 
c_1 e^{-E_{N'}t}|N'\rangle ~+~ c_2e^{-E_{N\pi}t}|N\pi\rangle ~+~...
\nonumber\\
& \sim & ~c_0 e^{-E_Nt}|N\rangle ~~~~~~ . 
\label{AS_Eq:5}
\end{eqnarray}
To improve signals and to investigate higher lying states one uses a set of 
sources and calculates a full correlation matrix.
\item One constructs ratios for quantities of interest in which the 
exponential factors cancel, e.g., 
\begin{equation}
\frac{\tilde \Gamma_{\alpha\beta}\langle B_{\beta}(t,\vec p) {{\cal O}} 
\bar B_{\alpha}(0,\vec p)\rangle}
{\Gamma_{\alpha\beta}\langle B_{\beta}(t,\vec p) 
\bar B_{\alpha}(0,\vec p)\rangle} \ .
\label{AS_Eq:6}
\end{equation}
\item
Finally, one determines the relevant renormalization factors
for the operator $\mathcal{O}$ 
non-perturbatively on 
the lattice and relates the lattice results to a specific pQCD scheme. 
\end{enumerate}

One has to appreciate that the efficient combination of experimental and LQCD results 
requires a good and efficient parametrization for the quantities of interest.
If there exists  e.g. an efficient parametrization of a specific GPD etc. in terms 
of just a few parameters, each result will constrain the acceptable parameter range .
Thus also high quality model building is necessary. 

Over the years it became clear that it is very non-trivial 
to derive realistic estimates, in particular of the systematic
uncertainties, from the highly
correlated quantities extracted from Lattice Monte Carlo data.
The ultimate test revealing potentially underestimated systematic uncertainties is the comparison 
of certain benchmark observables with experimental measurements, in this case, EIC data.  
The next best option is  
to compare results obtained with substantially 
different lattice formulations. In principle, each analysis should be repeated 
at least once with a different action. The latter is typically done in such a way that different
collaborations specialise on one specific action each. 

The most critical extrapolation is the continuum limit  $a\rightarrow 0$. 
Lattice actions violate basic symmetries of QCD (isotropy and homogeneity of space-time, 
chiral symmetry, isospin symmetry in the case of twisted-mass fermions  ...) for finite lattice spacing. 
Thus the  $a\rightarrow 0$ limit, which restores 
all symmetries, could be non-trivial. Unfortunately, $a$ can only be varied in very limited 
ranges because the needed 
CPU time is always proportional to a large power of $1/a$. Therefore, a variety of 
{\em improved} lattice actions was proposed in which lattice artifacts are not proportional to 
$a$ but e.g. $a^2$. Many variants exist, all of which are well motivated in one 
way or the other.
Substantial effort is invested to further improve such actions, and 
it would be very surprising if by the time an EIC starts operating also 
the systematic uncertainties due to the multiple extrapolation 
$a\rightarrow 0$, $L\rightarrow \infty$, $m_q \rightarrow m_q(physical)$.     
were not much better under control.\\
The purely statistical uncertainty will for sure become much smaller due to 
increased computer power. While the most powerful present day computers are of 
the Petaflop class, various initiatives aim already at Exaflop computing. 
In the next sections we will discuss in detail some of the physics quantities 
calculated on the lattice, which are especially important for the EIC.


\section{Generalized form factors}

Most correlators relevant for hadron structure which were determined on the lattice
are related to Generalized Parton Distributions (GPDs) or Distribution 
Amplitudes (DAs). For GPDs the hadronic states in Eq. (\ref{AS_Eq:1}) are equal, $h=h'$ 
but the momenta are usually different ($p\neq p'$). For the best known GPDs 
$H_q$ and $E_q$,
\begin{eqnarray}
&&\int \frac{d z^-}{2\pi}  e^{ix \bar P^+ z^-}
  \langle P_2|\, \bar{q}(-\frac{1}{2} z)\, \gamma^+ q(\frac{1}{2} z) 
  \,|P_1 \rangle \Big|_{z^+=0,\, z_{\perp}=0}
\nonumber \\
&=& \frac{1}{P^+} \left[
  {H_q(x,\xi,t)}\, \bar{N}(P_2) \gamma^+ N(P_1) +
  {E_q(x,\xi,t)}\, \bar{N}(P_2) 
                 \frac{i \sigma^{+\alpha} \Delta_\alpha}{2M} N(P_1)
  \, \right]  \ .
\label{AS_Eq:7}
\end{eqnarray}
The OPE gives moments in terms of {\em generalized form factors}  $A_{n,k}(t)$,
$B_{n,k}(t)$, $C_{n}(t)$,
\begin{eqnarray}
\int_{-1}^1dx\, x^{n-1}\, H(x,\xi,t) &=&
\sum_{\substack{k=0\\{\mathrm{even}}}}^{n-1}
(2\xi)^k\, A_{n,k}(t) + ({\rm n}+1~ {\rm mod} ~2)\,(2\xi)^{n} C_{n}(t)
\nonumber \\
\int_{-1}^1dx\, x^{n-1}\, E(x,\xi,t) &=&
\sum_{\substack{k=0\\{\mathrm{even}}}}^{n-1}
(2\xi)^k\, B_{n,k}(t) -  ({\rm n}+1~ {\rm mod} ~2)\,(2\xi)^{n} C_{n}(t) \,,
\label{AS_Eq:8}
\end{eqnarray}
which can be expressed in terms of local correlators 
by equations like
\begin{eqnarray}
\langle P'| ~ \bar q(0)  \gamma^{\{ \mu} {{\rm i} D}^{\mu_1}
... {{\rm i} D}^{\mu_n\}}q(0) |P \rangle &=&
\bar U (P') \left[ \sum_{i=0, even}^n \left\{ 
\gamma^{\{ \mu} {\Delta}^{\mu_1} ... {\Delta}^{\mu_i}
\bar P^{\mu_{i+1}}...\bar P^{\mu_n\}}
A_{n+1,i}(\Delta^2) \right.\right.
\nonumber \\
&-& {\rm i} \frac{\Delta_{\alpha}\sigma^{\alpha\{ \mu} \Delta^{\mu_1}... \Delta^{\mu_i} 
\bar P^{\mu_{i+1}}...\bar P^{\mu_n\}}}{2m} B_{n+1,i}(\Delta^2) 
\nonumber \\
&+& \left. \frac{\Delta^{\mu} ... \Delta^{\mu_n}}{m}~C_{n+1,0}(\Delta^2)\large|_{n~odd}
\right] U(P) \,,
\label{AS_Eq:9}
\end{eqnarray}
where $\{ \cdots \}$ denotes symmetrization and subtraction of trace terms.
The fact that the sum in Eq.(\ref{AS_Eq:8}) extends only up to $n-1$ 
is called {\em polynomiality}. The proton alone has eight independent quark 
GPDs for each quark flavour and typically one can calculate the 
leading three moments with satisfactory accuracy on the lattice. 
Adding the 
gluon GPDs and repeating the analysis for all octet and decuplet 
baryons and octet mesons one is already speaking about several
hundred quantities. In future one will also increasingly 
analyse hadron resonances and transition form factors, 
such that the lattice data base 
will become even richer. For each of these observables one has 
to analyse the renormalization properties, the quark/pion mass dependence  
and the finite 
volume dependence (within suitable versions of effective field
theory/chiral perturbation theory (ChPT)). Finally one has to compare 
results for different lattice actions and analyse the origin 
of discrepancies. Obviously it is impossible to review all of this 
here. Rather, we refer to the comprehensive paper 
\cite{Bratt:2010jn} for an example of a state of the art analysis. 
Fig. \ref{AS_Fig:2}, taken from this paper,  gives a typical example.  This figure shows a number of common aspects:

\begin{figure}
\begin{center}
\includegraphics[width=0.45\textwidth]{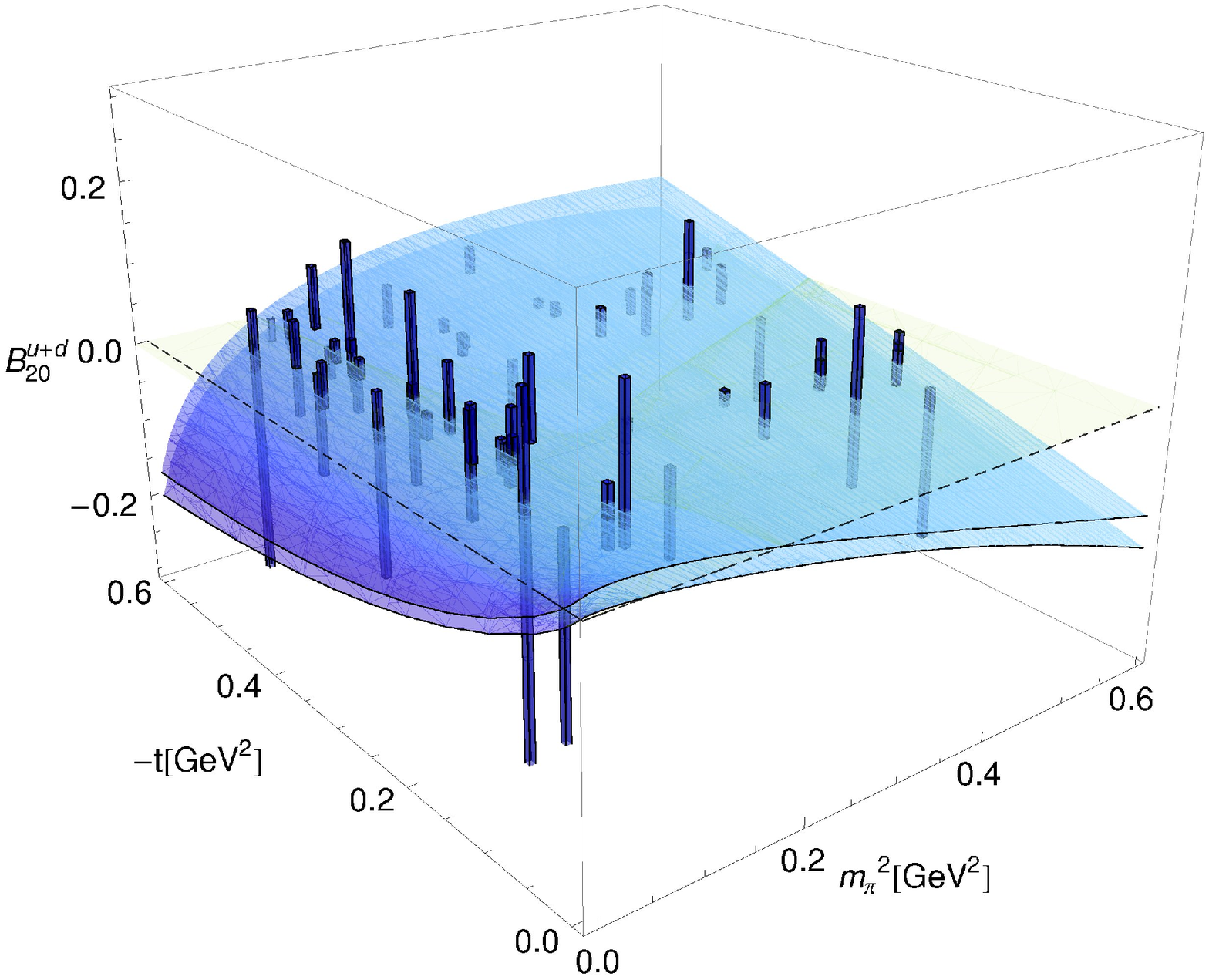}
\end{center}
\vspace{-0.5 cm}
\caption{\label{AS_Fig:2} The isosinglet moment $B_{20}^{u+d}(t)$ as a function
of simulated pion mass and $t$~\cite{Bratt:2010jn}.}
\vspace{0.5 cm}
\end{figure}

\begin{figure}
\begin{center}
\includegraphics[width=0.65\textwidth]{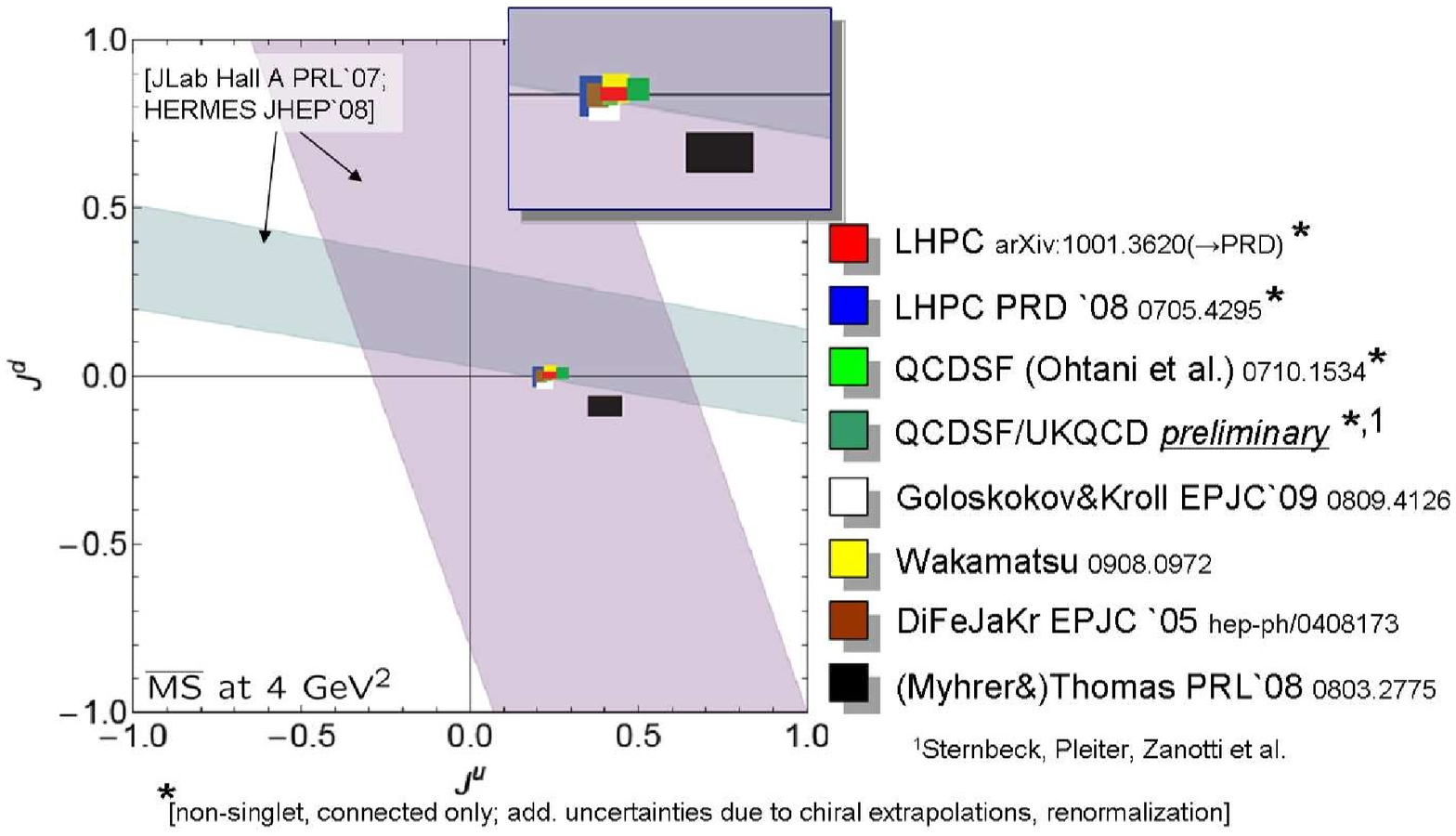}
\end{center}
\vspace{-0.5 cm}
\caption{\label{AS_Fig:2a} Lattice results for $J_u$ and $J_d$ compared with various models 
\cite{Diehl:2004cx,Wakamatsu:2009gx,Goloskokov:2008ib,Thomas:2008ga} and constraints derived 
from experiment (colored bands) }
\end{figure}

\begin{enumerate}
\item
Fluctuations are strongly suppressed for heavy quark/pion masses. This is 
why the statistical errors (dark blue bars) increase drastically for 
smalles pion masses.
\item
The difference between the two sheets gives the variation of HBChPT fits.
However, it would be safer to only use ensembles with squared pion masses below 
$m_{\pi}^2\leq 0.25{\rm GeV}^2$, where ChPT is rather well under control, which was 
obviously not possible with the ensembles available for this analysis.
\item
One is especially interested in the $t=0$ limit of $B_{20}$ in view of 
Ji's sum rule,
$$\Big\langle J_q^3 \Big\rangle~=~\frac{1}{2}[A_{2,0}^q(0)+B_{2,0}^q(0)]\ .$$
Already today lattice simulations give rather precise results for the total 
angular momentum carried by the different quark species in a nucleon, see Fig. \ref{AS_Fig:2a}. 
In future these results will further improve, e.g. due to the use of twisted 
boundary conditions to realize proton momenta different 
from the natural ones on a lattice, i.e. different from $p_j=\frac{2\pi}{L}n_j$.
\end{enumerate}

Thus, much has been done already, and much more will be done in future. 
Extrapolating the progress of recent years to the time an EIC will 
start operation it seems realistic to expect that by then pictures  
like Fig.~\ref{AS_Fig:2} will be  
numerically precise and will include reliable error bands.

\begin{figure}
\begin{center}
\includegraphics[width=0.45\textwidth]{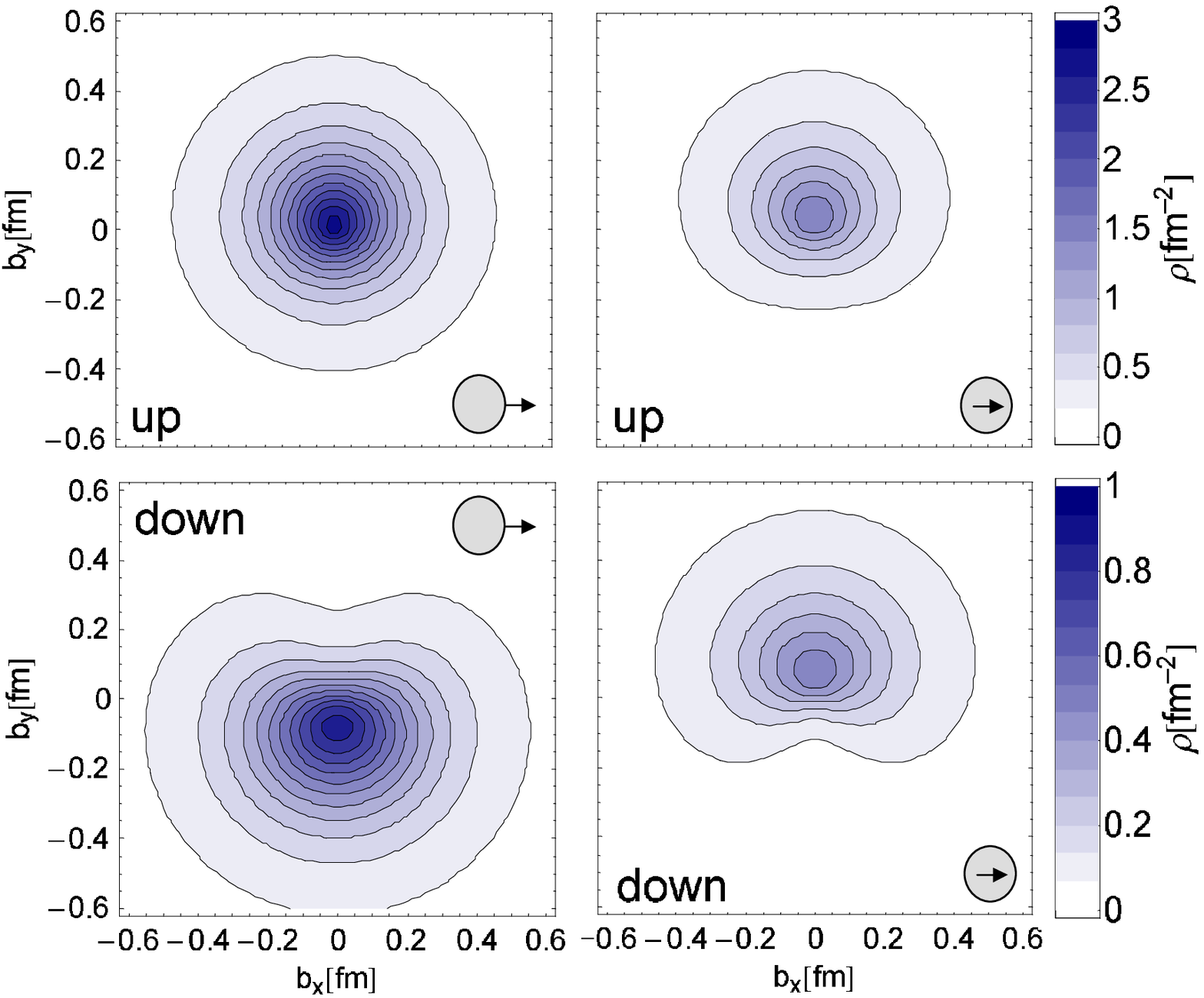}
\end{center}
\caption{\label{AS_Fig:3} An illustration for the transverse probability distribution of 
the nucleon quark distributions as a function of the transverse quark and nucleon spin direction. 
Figure taken from \cite{Gockeler:2006zu}.}
\end{figure}

Another important example are the quark density distributions in the
transverse plane plotted in Fig.~\ref{AS_Fig:3}.
Their form is mainly determined by the Fourier transformations of (moments
of) the GPDs.
\begin{eqnarray}
B^q_{n0}(x,0,b_{\perp}^2) & =& \frac{1}{(2\pi)^2}~\int_{-1}^1 dx x^{n-1}~\int d^2\Delta_{\perp} ~ 
e^{{\rm i}b_\perp\cdot \Delta_{\perp}} E(x,0,\Delta_{\perp}^2)
\nonumber \\
\bar B^q_{Tn0}(x,0,b_{\perp}^2) & =& \frac{1}{(2\pi)^2}~
~\int_{-1}^1 dx x^{n-1}~
\int d^2\Delta_{\perp} ~ 
e^{{\rm i}b_\perp \cdot \Delta_{\perp}}
\bar E_{T}(x,0,\Delta_{\perp}^2) ~~~.
\label{AS_Eq:10}
\end{eqnarray} 
The information on transverse structure contained in GPDs 
is, e.g., relevant in the following context: 
First LHC data show strong disagreement between observed interaction rates and 
predictions from event generators, see e.g.~\cite{Chatrchyan:2011id}
for the so-called ``underlying event'' which denotes the 
whole of all medium hard reaction channels, which are completely dominated by QCD. 
Part of the explanation might be related to multiple-hard interactions,
a class of reactions which was shown to be already relevant at the Tevatron,
see \cite{Abe:1997bp}. In these reactions multiple hard quark-gluon interactions occur 
in the same proton-proton collision, which are not described by the usual
inclusive factorization theorems. The correction terms have a complicated structure, 
see e.g. \cite{Diehl:2011tt} and references cited there, but can be partially related to GPD profiles 
in the transverse coordinate plane. By combining experimental results from an EIC with 
improved lattice calculations it should be possible to describe these effects much more 
precisely than currently. In this context, as always, experimental results are crucial, 
because it is very difficult to judge the reliability of lattice results without 
being able to compare with at least some experimental facts.  

A multitude of angular asymmetries and hadronic correlations, many of which include spin degrees of 
freedom, can be measured with a high luminosity EIC. For many of these, the microscopic 
reaction mechanism is not yet understood. Some of the proposals made depend crucially on the
transverse hadronic structure encoded in GPDs, see e.g. \cite{Burkardt:2002ks}. One of the 
main missions of an EIC is to clarify both the transverse structure and the reaction mechanisms. This is a demanding task which can only be mastered with input from LQCD.




\section{TMDs on the lattice}

The availability of methods to study GPDs on the lattice motivates us to develop similar techniques for the calculation of TMDs  \cite{Hagler:2009mb,Musch:2010ka}. 
In contrast to other, more mature areas of lattice QCD,  the present focus of TMD calculations on the lattice is on the development of methodology and on qualitative observations rather than precision. The ultimate goal is to obtain results from first principles only that can potentially be compared to experimental observations. The first step to reach this goal is to describe precisely which matrix elements need to be calculated, and how they can be regularized in the context of TMD factorization. Already at this step, the situation is much more challenging for TMDs than for moments of GPDs, where the matrix elements needed are well-known. These issues are not specific to lattice QCD, but they play a central role in the development of methods to calculate TMDs non-perturbatively. 

In its basic form, the correlator that needs to be calculated is that of eq.~(\ref{eq:corr}). 
For our purposes, we write the trace projections $\Phi^{[\Gamma]} = \frac{1}{2} \mathrm{Tr} (\Gamma\, \Phi)$ of this correlator  as 
\begin{align}
 \Phi^{[\Gamma]}(x,\BMkeiT) 
 = \frac{1}{P^+} \underbrace{
   \int \frac{d(\BMelll {\cdot} P)}{2\pi} e^{-i(\BMelll {\cdot} P)x}
 }_{
   \displaystyle \BMFTx }\ 
 \underbrace{
   \int \frac{d^2\BMelllT}{(2\pi)^2} e^{i \BMelllT{\cdot}\BMkeiT}
 }_{
   \displaystyle \BMFTT}\ 
   \underbrace{ 
   {\textstyle \frac{1}{2}} \langle  P, \BMSpin \vert\ \overline{\BMqfield}(\BMelll)\ \Gamma\ \mathcal{W}_\eta\ \BMqfield(0)\ \vert P, \BMSpin \rangle
 }_{
   \displaystyle \widetilde{\Phi}^{[\Gamma]}(\BMelll,P,\BMSpin)}  \Big \vert_{\BMelll^+ = 0}
  \label{eq:BMcorrel}
\end{align}
where $\Gamma$ is a Dirac matrix. 
The gauge link $\mathcal{W}_\eta$ is discussed in sec.~\ref{secIV:mulders-rogers},
and its geometry is depicted for the SIDIS process in figure~\ref{fig:BMlinks} a).
With the generalization of eq.~(\ref{eq:wilsonline}) 
it can be written as a concatenation of straight Wilson lines
$\mathcal{W}_\eta = V_{[\BMelll,\BMelll{+}\eta v]}\,\allowbreak V_{[\BMelll{+}\eta v,\eta v]}\,\allowbreak V_{[\eta v,0]}$. 
Here $v$ is a time-like vector normalized to $v^2 = 1$.
A staple shaped gauge link $\mathcal{W}_{\infty}$ extending to $\eta \rightarrow \infty$ corresponds to SIDIS,
while a staple $\mathcal{W}_{-\infty}$ directed in the opposite direction corresponds to the Drell-Yan process. 
Beyond tree level, eq.~\eqref{eq:BMcorrel} needs to be modified in order to take 
the collective effect of soft momentum gluons
into account and to subtract divergences. This can be achieved, e.g., by dividing $\tilde \Phi^{[\Gamma]}$ by appropriate vacuum expectation values (soft factors), see, e.g., \cite{Cherednikov:2008ua,Ji:2004wu,Ji:2004xq,Aybat:2011zv,Collins_book2011}.


First studies of transverse momentum dependence on the lattice follow the strategy to determine matrix elements of the form $\tilde \Phi^{[\Gamma]}$ in eq.~\eqref{eq:BMcorrel} directly from three-point functions. The idea of using a discrete representation of the non-local operator $\overline{\BMqfield}(\BMelll)\ \Gamma\ \mathcal{W}_\eta\ \BMqfield(0)$ is a novel technique and requires investigations about the properties of such extended operators on the lattice.
Considering this and the ambiguities about the precise operator geometry suitable for TMD extraction, it seems reasonable to begin with a simplified setup. 
The following two operator geometries are under investigation:
\begin{itemize}
  \item straight gauge link connecting the two quark fields directly, $\mathcal{W}_0 = V_{[\BMelll,0]} $.
    This simple setup yields high statistics results, but does not correspond to the situation in SIDIS or Drell-Yan. 
    For example, non-zero time-reversal odd TMDs such as the Sivers function $f_{1T}^\perp$ are forbidden by symmetry with this link geometry. 
    However, the qualitative features of the results are interesting, especially the spin-dependence. 
    A brief outline of findings obtained with straight links is given in sec.~\ref{sec:BMstraightlink}. 
  \item staple shaped gauge link of finite extent $\mathcal{W}_\eta$ for a spacelike choice of the direction $v$ as depicted in figure~\ref{fig:BMlinks} a). 
    Results for the SIDIS link $\mathcal{W}_{+\infty}$ and the Drell-Yan link $\mathcal{W}_{-\infty}$ can be read off if the lattice results converge to a constant for longer and longer staple extents $\eta$. 
    Ongoing studies with this operator geometry are discussed in sec.~\ref{sec:BMstaplelink}.
  \end{itemize}

\subsection{Straight gauge links}
\label{sec:BMstraightlink}

\begin{figure}
\begin{center}
a)\quad
\begin{minipage}[t]{0.45\textwidth}
\vspace{0pt}%
\includegraphics[width=\linewidth]{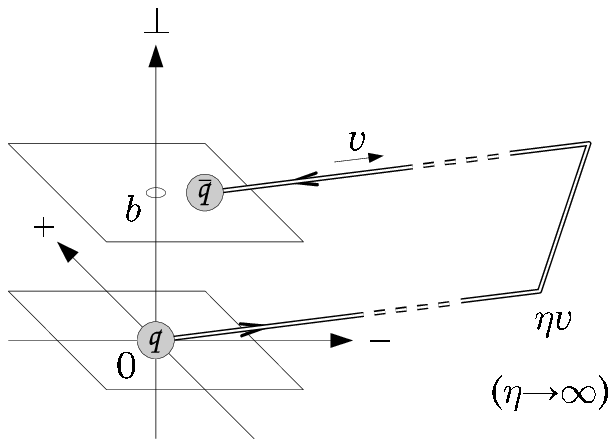}
\end{minipage}
\qquad b) \quad
\begin{minipage}[t]{0.25\textwidth}
\vspace{0pt}%
\includegraphics[width=\linewidth]{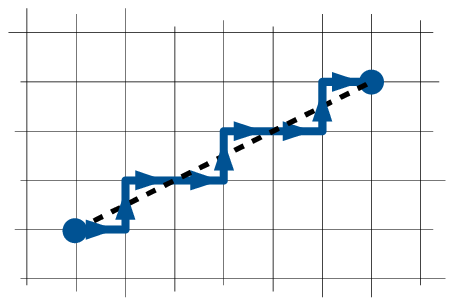}
\end{minipage}\par
\end{center}
\vspace*{-0.3cm}
\caption{\label{fig:BMlinks} a) Staple shaped Wilson line.
b) Representation of a straight Wilson line (dashed line) as a step-like product of link variables. }
\end{figure}

In lattice QCD, it is possible to determine matrix elements $\widetilde{\Phi}^{[\Gamma]}$ appearing in eq.~\eqref{eq:BMcorrel} directly from a ratio of three- and two-point functions, provided the operator has no extent in Minkowski-time. To do this, we employ the standard methods described in sec.~\ref{ch:lattice-sec:intro}. Only the operator we insert is specific to our method. Operators with straight gauge links can be approximated on the lattice by a step-like product of link variables, as depicted in figure~\ref{fig:BMlinks} b). At present, disconnected diagrams are neglected. Disconnected diagrams cancel in the isovector channel, i.e., in $u-d$ quark distributions.

The key element to relate the matrix elements $\widetilde{\Phi}^{[\Gamma]}$ determined on the lattice to the TMDs is a parametrization in terms of Lorentz-invariant amplitudes $\widetilde{A}_i(\BMelll^2,\BMelll {\cdot} P)$, similar to the parametrization in terms of amplitudes $A_i(k^2,k {\cdot} P)$ in ref.~\cite{Mulders:1995dh}. For straight gauge links one obtains
\begin{align}
  \widetilde{\Phi}^{[\gamma^\mu]} & =
    2\,P^\mu\,\tilde{A}_2 
    + 2i\,{M}^2\,\BMelll^\mu\,\widetilde{A}_3\,,
    \label{eq:BMdecompvect}\\
  \widetilde{\Phi}^{[\gamma^\mu \gamma^5]} & = 
    - 2\, M\, S^\mu\, \widetilde{A}_6
    - 2i\,M\,P^\mu (\BMelll \cdot S)\, \widetilde{A}_7 
    + 2\,{M}^3\,\BMelll^\mu (\BMelll \cdot S)\,\widetilde{A}_8 \ ,\label{eq:BMdecompaxvect}
    %
  \end{align}
To translate the amplitudes into TMDs, the Fourier transform in eq.~\eqref{eq:BMcorrel} must be carried out. For example
\begin{align}
   f_1(x,\BMkeiT^2) & =  2\, \BMFTT\,\BMFTx\, \widetilde{A}_2(\BMelll^2,\BMelll {\cdot} P) \,, \label{eq:BMf1}  \\
   g_{1T}(x,\BMkeiT^2) & =  4 M^2 \partial_{\BMkeiT^2} \, \BMFTT\,\BMFTx\, \widetilde{A}_7(\BMelll^2,\BMelll {\cdot} P)  \,, \label{eq:BMg1T}
\end{align}
where the two independent Fourier transforms $\BMFTx$ and $\BMFTT$ are defined in eq.~\eqref{eq:BMcorrel}. 
On the Euclidean lattice, the matrix elements can only be evaluated in the range 
\begin{align}
  \BMelll^2 & \leq 0 & |\BMelll {\cdot} P| &\leq |\mathbf{P}_{\text{lat}}|\,\sqrt{-\BMelll^2}
  \label{eq:BMlatconstraint}
\end{align}
where $\mathbf{P}_{\text{lat}}$ is the three-momentum of the nucleon chosen on the lattice.
As a result, data points are only available in a wedge shaped area.
The opening angle of the wedge can be potentially increased by using larger lattice nucleon momenta $\mathbf{P}_{\text{lat}}$, but full coverage of the $|\BMelll|,\BMelll{\cdot}P$-plane can never be achieved with this method. Thus the information required to reconstruct the $x$-dependence of the distributions is not fully available; the Fourier transform $\BMFTx$ in eqs.~\eqref{eq:BMf1}-\eqref{eq:BMg1T} cannot be carried out. However, model assumptions about the correlation of $x$- and $\BMkeiT$-dependence can be compared to the lattice data, see sec.~VI of ref.~\cite{Musch:2010ka}. Moreover, the lowest $x$-moment of TMDs can be calculated, since the required information is encoded in the data at $\BMelll {\cdot} P = 0$. For example,
\begin{align}
   f_1^{[1]}(\BMkeiT^2)  & \equiv \int dx\ f_1(x,\BMkeiT^2) =  \int_{0}^{1} dx\ \left( f_1(x,\BMkeiT^2) - \bar{f}_1(x,\BMkeiT^2) \right) \nonumber \\
   & = 2\, \BMFTT\, \widetilde{A}_2(\BMelll^2,0) \,, \label{eq:BMf1mom}  
\end{align}
where $\bar{f}_1$ is the unpolarized anti-quark distribution function. 
First results for the lowest $x$-moments $f_1^{[1]}$, $g_{1T}^{[1]}$ and $h_{1L}^{\perp [1]}$ using straight gauge links have been presented in Ref. \cite{Hagler:2009mb,Musch:2010ka}.
The calculations were carried out at a pion mass of about $500\, \mathrm{MeV}$, taking advantage of existing gauge configurations from the MILC collaboration \cite{Bernard:2001av} and propagators from the LHP collaboration \cite{Hagler:2007xi}. 
\begin{figure}
\begin{center}
\includegraphics[width=0.6\textwidth]{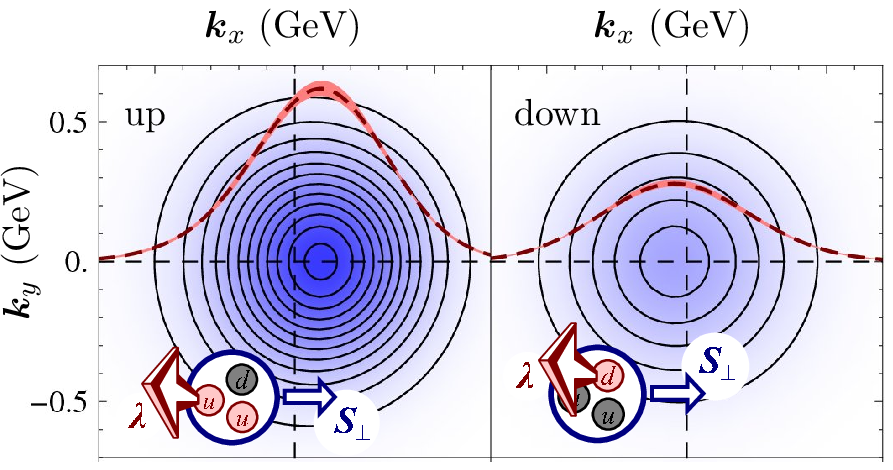}
\end{center}
\vspace*{-0.3cm}
\caption{\label{fig:BMdensities}$x$-integrated density of longitudinally polarized quarks inside a nucleon polarized in the transverse $x$-direction. These results have been obtained with straight gauge links at a pion mass $m_\pi \approx 500 \mev$ \cite{Hagler:2009mb}. The insets display the spin polarization of the quarks and of the nucleon.}
\end{figure}
The worm gear distribution $g_{1T}$ gives rise to dipole deformations in the $x$-integrated, $\BMkeiT$-dependent density of longitudinally polarized quarks inside a transversely polarized nucleon, as shown in figure~\ref{fig:BMdensities}. 
Due to the dipole deformation, this density is not axially symmetric. The peak is clearly shifted away from the center along the axis defined by the transverse spin vector. This shift is associated with a non-zero average transverse quark momentum $\langle \mathbf{\BMkei}_x \rangle_{TL}$, which can be expressed in terms of a ratio of amplitudes $ \widetilde{A}_7(0,0)/\widetilde{A}_2(0,0)$.
The lattice computations yield $\langle \mathbf{\BMkei}_x \rangle_{TL} = 67(5)\mev$ for down quarks and $\langle \mathbf{\BMkei}_x \rangle_{TL} = -30(5)\mev$ for up quarks  (errors statistical only). Reference \cite{Pasquini:2009eb} reveals that these results are of the same sign and of quite similar magnitudes as those obtained with a light-cone constituent quark model \cite{Pasquini:2008ax}, despite the unphysically large quark masses employed in the lattice calculation. 

We note that the straight link results discussed above depend on two additional important ingredients: a non-perturbative renormalization condition and a Gaussian parametrization to perform the Fourier transform, see ref. \cite{Musch:2010ka} for details. The renormalization condition is necessary to fix the length-dependent renormalization factor $\exp(-\delta m |\BMelll|)$ due to the self-energy of the spacelike Wilson line $\mathcal{V}_{[\BMelll,0]}$ \cite{Dorn:1986dt,Maiani:1991az,Martinelli:1995vj}. At the present level of statistical precision, the parametrization of the renormalized data as Gaussian functions is very successful and acts as a provisional regulator of contributions from large $\BMkeiT$. 
A better understanding of the operator in the transition from the short range (small $\sqrt{-\BMelll^2}$, corresponding to large $\BMkeiT$) to the long range behavior may lead to a an improved parametrization of the lattice data, beyond the Gaussian assumption, and may open the possibility to make contact with a perturbatively defined renormalization scheme.


\subsection{Staple-shaped links and the Sivers function}
\label{sec:BMstaplelink}



TMDs obtained with a straight gauge link as discussed in the previous section are not strictly identical to those relevant in, e.g., SIDIS or the Drell-Yan process. 
Instead, a link geometry as depicted in figure~\ref{fig:BMlinks} a) is required. 
In particular, naively time-reversal odd TMD such as the Sivers function can only be non-vanishing once the operator structure involves another direction $v$ related to final or initial state interactions.
Lattice QCD can profit from frameworks that avoid rapidity divergences by considering directions $v$ slightly off the lightlike $n^-$-direction \cite{Collins:1981uk,Collins:1981uw,Ji:2004wu,Ji:2004xq,Aybat:2011zv,Collins_book2011}. TMDs introduced this way follow an evolution equation in the rapidity cutoff parameter $\zeta \equiv (2 P \cdot v)^2/ |v^2|$ \cite{Collins:1981uk,Idilbi:2004vb}.
The restriction to operators $\overline{\BMqfield}(\BMelll)\ \Gamma\ \mathcal{W}_\eta\ \BMqfield(0)$ that have no extent in Euclidean time permits only the implementation of spacelike directions $v$ on the lattice, and furthermore limits the rapidity cutoff parameter to the range $0 \leq \zeta \leq 4 | \mathbf{P}_\text{lat} |^2$, where $\mathbf{P}_\text{lat}$ is the selected nucleon three-momentum on the lattice.
The dependence on $v$ leads to additional amplitudes $\tilde A_i$, $\tilde B_i$ in the decomposition of the correlator eq.~\eqref{eq:BMdecompvect}, compare also \cite{Goeke:2005hb}:
\begin{align}
\label{eq-phidecomp}
\widetilde \Phi^{[\gamma^\mu]} = \frac{2}{\widetilde{S}} \Bigl\{ & P^\mu\, \widetilde{A}_2 + i M^2 \BMelll^\mu\, \widetilde{A}_3 + i M \epsilon^{\mu \nu \alpha \beta} P_\nu \BMelll_\alpha \BMSpin_\beta\, \widetilde{A}_{12} + \frac{M^2}{(v {\cdot} P)} v^\mu\, \widetilde{B}_1 \nonumber \\ + & \frac{M}{v {\cdot} P} \epsilon^{\mu \nu \alpha \beta} P_\nu v_\alpha \BMSpin_\beta\, \widetilde{B}_7 + \frac{ i M^3}{v {\cdot} P} \epsilon^{\mu \nu \alpha \beta} \BMelll_\nu v_\alpha \BMSpin_\beta\, \widetilde{B}_8 \nonumber \\ - & \frac{M^3}{v {\cdot} P} (\BMelll {\cdot} S) \epsilon^{\mu \nu \alpha \beta} P_\nu \BMelll_\alpha v_\beta\, \widetilde{B}_9 + \frac{i M^3}{(v {\cdot} P)^2} (v {\cdot} \BMSpin) \epsilon^{\mu \nu \alpha \beta} P_\nu \BMelll_\alpha v_\beta \widetilde{B}_{10}\Bigr\}\ . 
\end{align}
Here $\widetilde{S}$ generically represents a soft factor modification, as needed, e.g., in the formalism of refs.~\cite{Ji:2004wu,Ji:2004xq,Aybat:2011zv,Collins_book2011}.
\begin{figure}
\begin{center}
\includegraphics[width=0.55\textwidth]{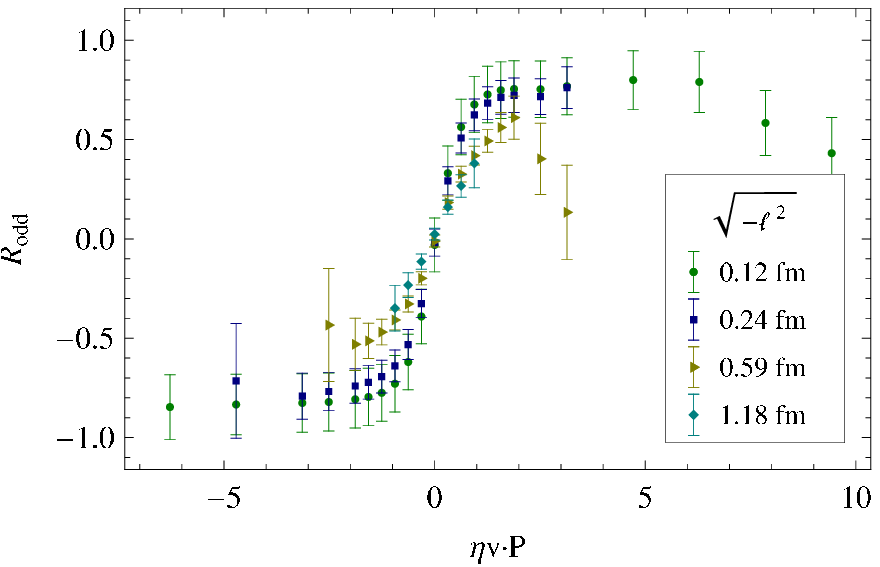}
\end{center}
\vspace*{-0.4cm}
\caption{\label{fig:BMstapletest}Test calculation of a T-odd ratio of amplitudes using staple shaped links at $m_\pi \approx 800\mev$ \cite{Musch:2009ku}.}
\end{figure}
First lattice studies are ongoing for ratios of amplitudes, in which renormalization factors and potential soft factors cancel \cite{Musch:2009ku,Musch:2010ka}, similar as in the asymmetries discussed in 
sec.~\ref{sec:TMD-weighted-asymmetries}.
Figure~\ref{fig:BMstapletest} shows results from a test calculation \cite{Musch:2009ku} of a ratio of time-reversal odd over time-reversal even amplitudes $R_\text{odd} \equiv ( \widetilde{A}_{12} - (M/|\mathbf{P}_\text{lat}|)^2 \widetilde{B}_8)/\widetilde{A}_2$, evaluated at $\BMelll {\cdot} P=0$, $|\mathbf{P}_\text{lat}| \approx 0.5\gev$, for selected values of $\BMelll^2$.  Note that $\widetilde{A}_{12}$ would correspond to the Sivers function $f_{1T}^\perp$ for lightlike $v$. In the test calculation, the operator has been evaluated with staple shaped links $\mathcal{W}_\eta$ for a large range of extents $\eta$. 
The result of the test calculation is, within statistics, an odd function of $\eta v {\cdot} P$, as expected for a time-reversal odd function. 
Moreover, we see the onset of a plateau at $|\eta v {\cdot} P| \gtrsim 2$.
The plateau at large positive $\eta$ correspond to the SIDIS result with a $\mathcal{W}_{\infty}$ link, while the plateau at large negative $\eta$ correspond to the Drell-Yan result with the $\mathcal{W}_{-\infty}$ link. This is a promising indication that lattice estimates could be feasible for, e.g., the average transverse momentum shift due to the Sivers function given by 
\begin{align}
	\langle \mathbf{\BMkei}_y \rangle_{TU} \equiv \left.\frac{ \int d^2\BMkeiT\, \mathbf{\BMkei}_y\, \Phi^{[\gamma^+]} }{ \int d^2\BMkeiT \Phi^{[\gamma^+]} }\right\vert_{\scriptstyle \BMSpinT=(1,0)} = 
	M \frac{\int dx\ f_{1T}^{\perp(1)}(x)}{\int dx\ f_{1}^{(0)}(x)},
\end{align}
see also \cite{Boer:2003cm,Burkardt:2003uw}. Here $\Phi^{[\gamma^+]}$ intuitively has an interpretation as the density of unpolarized quarks in a transversely polarized proton, and $f_{1T}^{\perp(1)}$ and $f_{1}^{(0)}(x)$ are $\BMkeiT$-moments defined as $f^{(n)}(x) \equiv \int d^2 \BMkeiT\ (\BMkeiT^2/2M^2)^n f(x,\BMkeiT^2)$. A generalized version of the above quantity can be formed directly from the amplitudes determined on the lattice, namely
\begin{align}
	\langle \mathbf{\BMkei}_y \rangle_{TU}(\mathcal{B}_\perp) \equiv  
	M \frac{\int dx\ \tilde f_{1T}^{\perp(1)}(x,\mathcal{B}_\perp^2)}{\int dx\ \tilde f_{1}(x,\mathcal{B}_\perp^2)} =
	-M \left. \frac{\widetilde{A}_{12} - R(\zeta) \widetilde{B}_8}{\widetilde{A}_{2} + R(\zeta) \widetilde{B}_1} \right\vert_{
	\shortstack[l]{$\scriptstyle \BMelll^2=-\mathcal{B}_\perp^2$\\$\scriptstyle \BMelll{\cdot}P=0$}}
\end{align}
with $R(\zeta) = 1-\sqrt{1 + 4M^2/\zeta}$ and where $\tilde f_{1}$ and $\tilde f_{1T}^{\perp(1)}$ are now $\BMkeiT$-Fourier-transformed TMDs as they appear in 
eq.~(\ref{eq:ssa_sivers_final}).
Keeping the length $\mathcal{B}_\perp$ sufficiently large compared to the lattice spacing and correspondingly assuming renormalization properties as in continuum field theory, one finds that multiplicative renormalization factors, including Wilson line self-energies, as well as potential soft factors cancel in the ratio of amplitudes above. The extrapolation to $\mathcal{B}_\perp=0$, where $\langle \mathbf{\BMkei}_y \rangle_{TU}(\mathcal{B}_\perp)$ is equal to $\langle \mathbf{\BMkei}_y \rangle_{TU}$, will require special attention to UV divergences and cutoff effects. However, already the generalized object $\langle \mathbf{\BMkei}_y \rangle_{TU}(\mathcal{B}_\perp)$ at nonzero $\mathcal{B}_\perp$ may offer opportunities to compare with phenomenology,  by means of an $x$-integrated version of the Bessel-weighted quantities introduced in 
eq.~(\ref{eq:ssa_sivers_final})
A possible difficulty for lattice computations will be to reach large enough values $\zeta$, in the regime where evolution equations \cite{Collins:1981uk,Idilbi:2004vb} can be applied.




\section{Spectroscopy and other physics topics}
\begin{figure}[h]
\begin{center}
\includegraphics[width=0.5\textwidth]{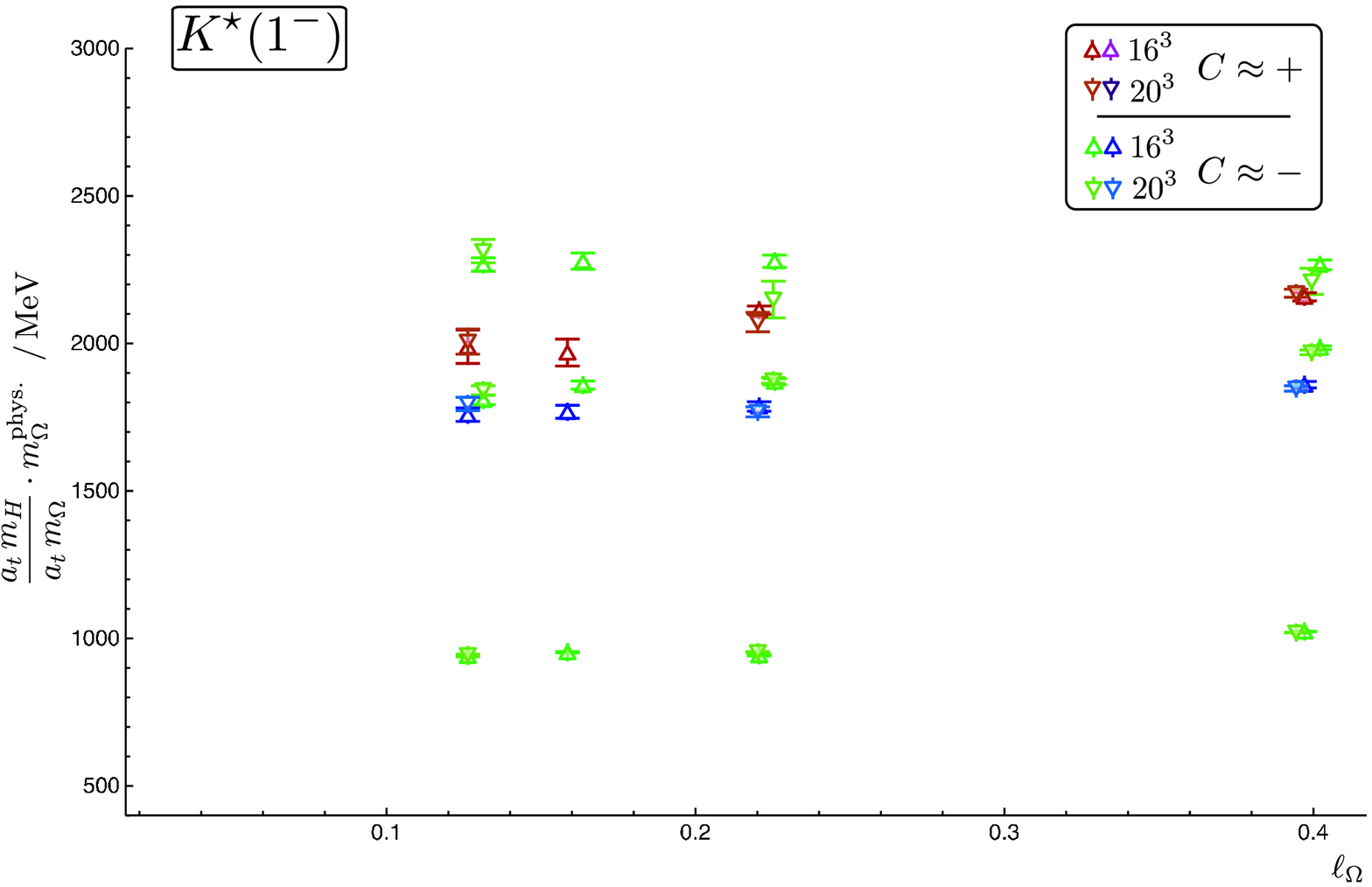}
\end{center}
\caption{\label{AS_Fig:5} A typical example from \cite{Dudek:2010wm}.
The $1^-$ kaon states are shown. Color coding indicates dominance of a particular
charge-conjugation eigenstate}
\end{figure}

Still another approach to elucidate hadron structure is offered by lattice spectroscopy. 
Spectroscopy has the great advantage that it allows to avoid the subtle renormalization 
issues mentioned above, but the disadvantage that the deduction of information on hadron
structure is less direct. A typical recent example is found in \cite{Dudek:2010wm}, see Fig.
\ref{AS_Fig:5}. The 
basic idea is that one uses a set of interpolating currents (sources) ${\cal O}_i$ with the same quantum 
numbers to calculate and analyse the correlation matrix as a function of separation of the 
time-hyper planes
\begin{equation}
C_{ij}(t)~=~ \Big\langle 0\Big|  {\cal O}_i(t) {\cal O}_j(0) \Big| 0 \Big\rangle
\end{equation}
and solves the generalized eigenvalue problem. One thus obtains not only the
eigenvalues (masses) but also the eigenvectors in terms of the different sources. 
If done with care the relative overlap of  the physical mass states with the different
sources allows to draw conclusions about their
structure. This provides information, which is often complementary to that obtained with the methods 
sketched above. Again for practical purposes this information is most sensitive to 
leading Fock-state components. While this is a very powerful method, its results must be 
interpreted with care. 
The eigenvectors of different mass states give the amplitudes with which each source  
contributes. All of these are forced to be orthonormal by construction. This constraint 
can lead to substantial artifacts if the chosen source functions span 
too small a function space. 
To avoid premature conclusions
one, therefore, has to compare results obtained for different lattice actions and many 
different choices of sources. Presently the lattice community is still in the process of 
optimizing this method, but it seems already clear by now that in a few years this 
approach will be 
a standard source of many detailed information about hadron structure.  

\subsection*{Acknowledgments}

Thanks are due to our collaborator John~Negele. We
are grateful to the LHP and MILC collaborations, for providing gauge
configurations and propagators. We also thank Gunnar Bali,  Alexei
Bazavov, Vladimir Braun, Markus Diehl, Robert Edwards, Meinulf G\"ockeler,
Alexei Prokudin, Dru Renner and David Richards for helpful
discussions. Our software uses the {\tt Chroma}-library
\cite{Edwards:2004sx}, and we use USQCD computing resources at Jefferson
Lab.




\chapter{QCD matter under extreme conditions}

\noindent
{\Large Convenors and chapter editors: \\[1em]
A. Accardi, M. Lamont, C. Marquet}

\newpage

\section{Chapter summary: overview and golden measurements}
\label{sec:overview}

\hspace{\parindent}\parbox{0.92\textwidth}{\slshape
 Alberto Accardi, Matthew Lamont and Cyrille Marquet}
\index{Accardi, Alberto}
\index{Lamont, Matthew}
\index{Marquet, Cyrille}


A basic quest of nuclear physics is the understanding of the structure of hadrons and 
nuclei (nucleon number $A > 1$) in terms of QCD Lagrangian degrees of
freedom, the quarks and gluons. 
Deviations of the nuclear quark and gluon densities from the sum of the 
free nucleon densities directly attest to binding effects and elucidate 
the QCD origin of the inter-nucleon interactions. Such deviations can 
arise through different mechanisms, such as a modification of the free 
nucleon structure, the presence of non--nucleonic degrees of freedom,
and quantum--mechanical interference of the quark/gluon fields of 
different nucleons at small parton fractional momentum $x$
(``shadowing''), creating a fascinating landscape of many-body QCD. 
At even smaller $x$, the gluon density increases to the point where
gluons become closely packed, leading to a strong field regime of
non-linear QCD evolution called saturation. This regime is argued to have 
universal properties for any hadronic system, ranging from pions, to protons
and nuclei, but its onset is enhanced in nuclear targets due to the
superposition of the gluon field of many nucleons.

A peculiar pattern of nuclear modifications was observed
in fixed-target experiments and caused much excitement; it shows
suppression for $0.2 < x < 0.8$ (``EMC  effect''), 
some signs of enhancement for $0.05 < x < 0.2$, and 
significant suppression (shadowing) at smaller $x$. 
However, such experiments were unable to reach deep into the shadowing 
region or probe gluons. The EIC will overcome these limitations,
extend measurements to very high scales of $Q^2$, and determine with high
precision the nuclear effects on gluon distributions. Full
reconstruction of the hadronic final state also opens up for the
first time the possibility of measuring charged current interactions
on nuclei, and to perform a full quark flavour separation based on
nuclear DIS data only.
Crucially, an EIC will access much lower values of $x<0.01$ and study the onset of the saturation regime, which has never been
directly probed experimentally, although tantalising (but
not unequivocal) signatures have been found at the Relativistic
Heavy-Ion Collider (RHIC). 

Another possibility offered by nuclear targets is the study of the
propagation of colour charges in nuclear matter and the
space-time evolution of hadronization. The unique feature of an EIC,
compared to previous fixed target experiments, is its large energy span.  This
 allows one to experimentally boost hadronization effects
completely out of the nucleus, in order to focus attention on the
propagation of fast quarks and gluons, and their accompanying parton
showers, through the nucleus.  Thus one can use the partons as coloured
probes of the soft components of the target nuclear wave
function, and conversely experimentally test QCD mechanisms of parton
energy loss in a known nuclear medium. At lower energies,
hadronization happens partially inside the nucleus, which can then be
used as a femtometer scale detector of the process. A good control of
energy loss mechanisms in the partonic phase will yield unambiguous 
insights into the dynamics of colour confinement whereby hadrons
emerge from coloured quarks and gluons. 

Novel observables will be available
thanks to the high energy reach, namely heavy flavours, charmonium and
bottomonium, and jets, greatly expanding the experimental toolbox and
sensitivity to nuclear effects, and thereby allowing a close connection to
first principles calculations in QCD. The collider mode will also make it feasible to 
study in detail target fragmentation and its correlation to current
fragmentation through multi-particle correlations, thereby expanding considerably the study of shower development and hadronization mechanisms.

\subsection{Gold and silver measurements}

One of the goals of the program at the INT was to identify a small number of
measurements whose ability to extract novel physics is beyond question and which are 
 feasible at an EIC. Such measurements are referred to as
``golden'' measurements. These are complemented by other ``silver''
measurements/observables, to form a broad, robust, and compelling
physics program. The gold and silver measurements are summarised in
Tables~\ref{tab:eA_golden} and \ref{tab:eA_silver}, where also their
feasibility in phase-I 
(medium energy) and phase-II (full energy) is indicated, 
and further discussed below.
Many more observables than can fit in this section will be available at
an EIC, contributing to a very rich physics program exploring the QCD
basis of nuclear physics. Many of these will be reviewed in detail in the rest
of this chapter. 

\begin{table}[htdp]
\centering

\noindent\makebox[\textwidth]{%
\footnotesize
\begin{tabular}{|c|c|c|c|c|}
\hline
Deliverables & Observables & What we learn & Phase-I & Phase-II \\
\hline
\hline
integrated gluon  &  $F_{2,L}$  & nuclear wave.fn.;    & gluons at   &  explore sat. \\
distributions     &            & saturation, $Q_s$ & $10^{-3}\lesssim x \lesssim 1$  &  regime \\
\hline
$k_T$-dep. gluons; &   di-hadron        &  non-linear QCD     &  onset of     &  RG evolution\\
gluon correlations    &  correlations     &  evolution/universality  &    saturation; $Q_s$ &   \\
\hline
transp. coefficients &  large-$x$ SIDIS;  &  parton energy loss,  & light flavours, charm&  precision rare \\
in cold matter       &  jets              &  shower evolution;    & bottom; jets        &   probes; \\
                     &                    &  energy loss mech.  &                    &  large-$x$ gluons \\
\hline

\end{tabular}}

\caption{\small Golden measurements in $e+A$ collisions at an EIC}
\label{tab:eA_golden}
\end{table}

\begin{table}[htdp]
\centering

\noindent\makebox[\textwidth]{%
\footnotesize
\begin{tabular}{|c|c|c|c|c|}
\hline
Deliverables & Observables & What we learn & Phase-I & Phase-II \\
\hline
\hline
integrated gluon & $F_{2,L}^c$, $F_{2,L}^D$ &  nuclear w.fn.;   & early sat. onset & saturation \\
distributions  &                        & saturation, $Q_s$ & challenge to measure    & regime \\
\hline
flavour separated      & charged current & EMC effect origin & full $q_i$ separation & larger $Q^2$, \\
nuclear PDFs &   \& $\gamma Z$ str. fns.   &                   & at $0.01\lesssim x \lesssim 1$ & smaller $x$ \\ 
\hline
$k_T$-dep. gluons &   SIDIS at       &  non-linear QCD     &  extract $Q_s$;     &  RG evol.; \\
                                &  small-$x$     &  evolution/universality   & multipole corr.  &  flavour sep.  \\
\hline
$b$-dep. gluons;              & DVCS;   & interplay between   & moderate $x$ with & smaller $x$, \\
 gluon correlations & diffractive $J/\Psi$,  & small-$x$ evolution & light, heavy nuclei   & saturation \\
     & \&  vector mesons                    & and confinement     &  & \\
\hline

\end{tabular}}

\caption{\small Silver measurements in $e+A$ collisions at an EIC}
\label{tab:eA_silver}
\end{table}

\subsection{QCD at high gluon density}

The fact that we do not know the dynamics of gluons in nuclei over
basically any $x$ range seems a compelling enough reason to build an
EIC. The non-Abelian nature of QCD is its most distinguishing feature and controls emergent phenomena such as colour confinement, chiral symmetry breaking and the generation of the vast bulk of the visible mass in the Universe. These are, however, non-perturbative phenomena which
are difficult to attack from first principles; where this is
possible, such as in the case of the hadron spectrum from lattice QCD,
only static aspects of the strong interactions are addressed. An EIC would allow one for the first time to experimentally probe at small $x$  dynamical non-Abelian aspects of a
fundamental force of nature in a controlled setting where weak coupling 
methods apply. The physics in this regime is the non-perturbative physics of strong colour fields; the important new feature is that the applicability of weak coupling methods allow for systematic comparisons of theory to experiment. 

In addition to the intrinsic interest in this novel many-body regime of QCD, experimentally establishing and refining an effective
field theory for the saturation regime -- such as the Colour Glass
Condensate (CGC) -- as well as precisely imaging
the distribution and correlations of small-$x$ partons in nuclei, besides being of intrinsic 
interest, would have wide-ranging applications. The universality of the
saturation regime implies that such a theory would provide a
microscopic basis for understanding and calculating total hadronic
cross sections, with important applications to, for example, ultra-high energy cosmic ray physics, where extrapolations
in energy of several orders of magnitude are required to compute their
spectrum and detect possible new physics effects.
In high-energy relativistic heavy-ion collisions, the release of
saturated low-$x$ partons represents the starting point of the
subsequent space-time evolution of the Quark-Gluon Plasma (QGP). Testing
and benchmarking the underlying theory opens the prospect of a controlled, 
and precise, first principles calculation of such an
initial state.  Thereby reducing one of the largest sources of uncertainty in the
interpretation of experimental observables, and the measurements of
the QGP properties: an EIC would offer to the RHIC and LHC heavy-ion
programs an important asset, as valuable as the one HERA provided to
the LHC p+p program.  

\begin{figure}[tb]
\centering
\includegraphics[height=7cm]{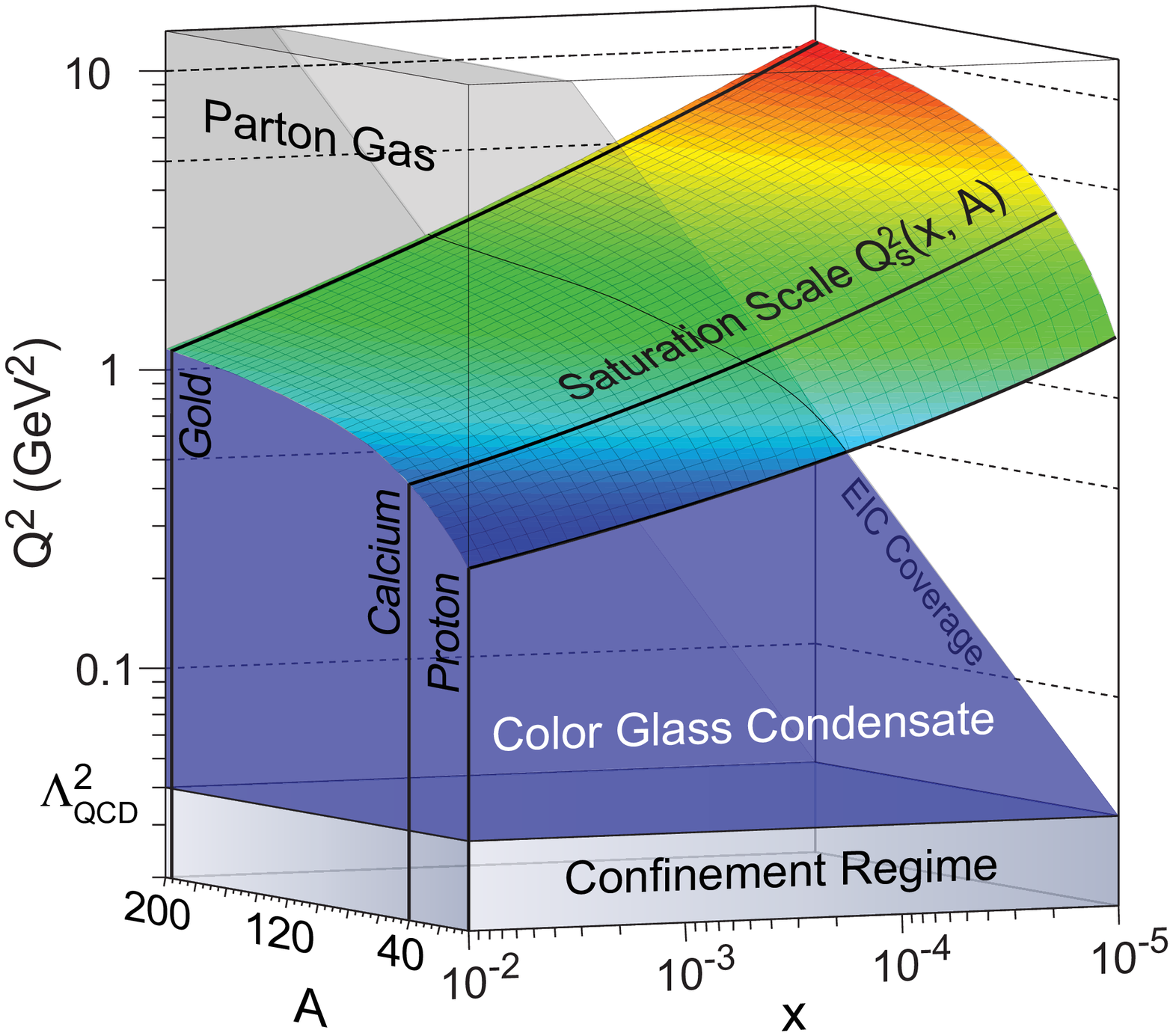}
\caption{\small
 The saturation scale, $Q\equiv Q_s$, and how it scales with $x$, and $A$.
}
\label{fig:eA_satscale}
\end{figure}

The onset of the saturation regime, when the gluon density becomes so large that
further growth with energy is tamed, is characterised by the 
saturation scale $Q_s(x)$; partons with momenta below this scale overlap
in transverse space, so that parton recombination and screening stops further growth in their number density. Given that parton
distributions grow as $x$ decreases, and dramatically so as discovered
at HERA, the saturation scale is clearly expected to grow as $x$ decreases.
It is further enhanced in nuclear targets because of the overlap
of the gluon fields originating from different nucleons. This is
illustrated in figure~\ref{fig:eA_satscale}.
In the saturated, dense regime at small $x$, non-linear QCD dynamics becomes
dominant but at the scale being set by a semi-hard $Q_s$ (of order 1 GeV),
calculations can be carried out by weak coupling techniques and
suitable effective field theories, of which the CGC is a prime example, can be derived from first principles. 

The dilute-dense separation of scales is more subtle  than just described. 
The larger the gluon's transverse momentum $k_T$, the smaller its
longitudinal energy fraction $x$ needs to be to enter the saturation
regime. In a scattering process, dilute partons (with $k_T\gg
Q_s(x)$) behave incoherently, whilst when the parton density is large
($k_T\lesssim Q_s(x)$), gluons scatter coherently. Therefore, 
transverse momentum dependent observables will be able to uncover more
details than inclusive observables, which can only access averaged
saturation effects. 
The interplay of saturation and the transverse spatial distribution of
gluons is also important; as $x$ decreases, gluon densities saturate
first in the centre of the nucleus.  To accommodate further growth,
gluons will be pushed more and more to the periphery, so that the
average gluon radius is expected to increase with decreasing $x$. 

For all these reasons regarding the small-$x$ physics program in e+A
collisions, 
the physics deliverables of an EIC have been classified in three main
categories giving access to the integrated, transverse-momentum-dependent,
and impact-parameter-dependent gluons. 
Here, by ``gluons" we mean not only
the conventional single-gluon distributions but also multi-gluon correlations. These have often been of secondary interest, but are now recognised as essential to a full
understanding of the low-$x$ regime. 
Indeed, except for the most inclusive observables which
are subject to cancellations, consistent QCD calculations in the
non-linear regime require the knowledge of multi-gluon
distributions. Integrated, transverse-momentum-dependent, and
impact-parameter-dependent gluon distributions and correlations in
nuclei are all unknown, and the processes we discuss below have never
been measured at small $x$. 

\subsubsection{Integrated gluons and sea-quarks}

As the most basic observables from both the theory and experimental
sides, the inclusive ($e$+A$\to e$+X) structure functions $F_2$ and $F_L$
stood out among other measurements, already well before the INT 
programme. They were the first potential golden measurements discussed; 
the pros and cons of these candidates to pin down the gluon and sea-quark 
distributions in nuclei were further reviewed during the program. 

$F_2$ is the most inclusive observable in deep inelastic scattering.
Its measurement presents no particular experimental challenge.  On the
theory side, it is the simplest process to calculate, along with
$F_L$, with the fewest input assumptions. For instance, $F_2$ and $F_L$ will be the first observables for which a full NLO calculation in QCD including non-linear effects will be
available.  (The existing phenomenology is still based on
leading-order ``impact factor" computations.)

Although it is harder to extract experimentally, $F_L$ is a golden measurement for high density QCD because
it is more directly related to the gluon distribution. Furthermore, it is more sensitive to
non-linear effects than $F_2$.  In the latter,  higher-twist contributions cancel each other out 
 delaying the onset of non-linear effects. 
The necessity of performing an energy scan to measure $F_L$
implies that the accessible $x$ range is a bit smaller than accessible with the
$F_2$ measurement.  However, the increased sensitivity to non-linear
effects more than compensates for this shortcoming. 
Deviations of DGLAP fits of the simultaneous ``singlet" dominated (at small $x$) evolution of $F_2$ and $F_L$ determined from EIC data should be able to quantitatively determine the onset of the saturation regime. The case for the low-energy EIC needs to be investigated more; in particular, the implementation of non-linear effects must be made more accurate, and
more detailed DGLAP fits of EIC pseudo-data should be performed before
establishing its sensitivity to saturation physics in the inclusive channel.
One caveat discussed is that QED  radiative corrections for nuclear targets can be large, and it remains to be proven that they can be controlled to the required precision. 
Computations addressing the role of these radiative corrections are underway and are discussed further in the detector studies section of this report.


The charm structure functions $F_{2,L}^c$ were considered as silver
measurements for the non-linear regime.
As in the case of  $F_L$, these observables give more direct
access to the gluon distribution relative to $F_2$; however due to the
mass of the charm quark, they also probe higher values of $x$ and are
therefore less sensitive to non-linear effects. In addition, QCD
calculations with non-zero charm mass are scheme dependent, which can 
absorb signals of non-linear effects if not appropriately handled.
However, since they can be measured precisely with a properly-designed
vertex detector, charm structure functions will be a very important
complementary measurement to pin down the nuclear gluon distribution
throughout the $(x,Q^2)$ plane.

Last but not least, silver measurements in the inclusive category are those  of the
diffractive structure functions $F_2^D$ and $F_{L,D}$. These are sensitive to the
square of the gluon distribution. As may therefore be anticipated, the strongest hints for manifestations of non-linear effects in e+p collisions at HERA come from diffractive measurements. A striking example is  the fact that the ratio of the diffractive to inclusive structure function is constant with energy, an observation not easily reconciled in a leading twist scenario. Furthermore, the leading-twist approximation does not explain the geometric scaling of the diffractive cross section. Finally, as a sign of enhanced sensitivity to non-linear effects, the DGLAP analysis of diffractive structure functions from HERA is problematic at larger values of $Q^2$ relative to the same for $F_2$ ( $\sim 8$ GeV$^2$ in the former compared to 2 GeV$^2$ in the latter).  However, measurements of diffractive structure functions are relatively more difficult and the additional kinematic variables make the analyses more involved than for $F_{2,L}$.

\subsubsection{Transverse momentum dependent gluons and sea-quarks}


The golden measurement here is that of di-hadron azimuthal correlations in $e$+A$\to e$+h$_1$+h$_2$+X processes. Di-hadron correlations are not only
sensitive to the $k_T$ dependence of the gluon distribution but also
to the $k_T$ dependence of gluon correlations. These correlations are sensitive to multi-gluon distributions 
for which first principles computations are only now becoming available.  Precise
measurements of these di-hadron correlations at an EIC would allow one to extract these multi-gluon correlations 
and study their non-linear evolution. Saturation effects in this channel correspond to a progressive disappearance 
with decreasing $x$ of the peak in the di-hadron azimuthal angle
difference around $\Delta\phi=\pi$. In a leading twist picture, where there is only one hard scattering, one expects, from momentum conservation, that the peak will persist. A comparison of the heights and widths of the di-hadron 
azimuthal distributions in $e$+A and $e$+p collisions respectively would clearly mark out such an
effect experimentally. An analogous phenomenon has already been observed for di-hadrons produced at forward
rapidity in d+Au and p+p collisions at RHIC. In that case, di-hadron production proceeds from valence quarks
in the deuteron (proton) scattering on small-$x$ gluons in the target Au nucleons (proton), $q_V$+Au(p)$\to$ h$_1$+h$_2$+X. Lacking direct experimental control over $x$, the onset of the
saturation regime is controlled by changing the centrality of the
collision, the di-hadron rapidity and the transverse momenta of the produced particles.
Experimentally, a striking flattening of the $\Delta\phi$ peak in d+Au
collisions is observed in central collisions, but the peak reappears
in peripheral collisions or for mid-rapidity di-hadrons. Directly using a point-like electron probe, as opposed to a quark bound in a proton or deuteron, is extremely beneficial.  It is experimentally much cleaner as there is no ``spectator" background to subtract and the
access to the exact kinematics of the process allows for more accurate 
extraction of the physics than is possible at RHIC or in the future with
p+A collisions at the LHC. Because there is such a clear correspondence between the physics of this particular final state in e+A collisions to the same in p+A collisions, this measurement is an excellent testing ground for quantitative studies of the universality of multi-gluon correlations in p+A and e+A collisions.

The simplest process to extract the transverse momentum
dependence of the gluon distribution is single inclusive DIS (SIDIS), $e$+A$\to$
$e$+h+X. One reason why these processes are especially
interesting is that by having two momentum scales at one's disposal, it is
possible to keep $Q^2$ large and access the saturation regime at
transverse momenta $p_T\lesssim Q_s$. This way, non-perturbative effects
and higher-twist contributions are suppressed, but one can nonetheless 
access non-linear QCD dynamics. Considering that $Q_s$ will not
exceed a few GeV at an EIC, this helps one  disentangle 
strong coupling effects, characterized by a fixed scale
($\Lambda_{QCD}$), from weak coupling non-linear effects more cleanly relative to  inclusive observables. Furthermore, in the large $Q^2$ and small $x$ limits, the relation between the transverse momentum of the produced hadron $p_T$ and that of the small-$x$ glue $k_T$ is quite direct, enabling a rather straightforward experimental probe of the gluon transverse momentum distribution. From the theoretical point of view, important connections have been  
established between the framework of Transverse Momentum Dependent (TMD) distributions discussed previously in this report and the CGC effective theory at small $x$. 
Thus SIDIS has all the pre-requisites to be considered a golden observable.  It is nonetheless classified as silver because di-hadron correlations are more directly sensitive to non-linear QCD 
evolution.

\subsubsection{Transverse position dependence of gluons and sea-quarks}

To pin down the transverse distribution and correlations of small-$x$
gluons, exclusive measurements are needed. The prototypical
observables discussed are diffractive vector meson production (DVMP)
and deeply virtual Compton scattering (DVCS).  Coherent
diffraction, where the nuclear target is intact, gives access to the transverse spatial distribution of the gluon density in a nucleus. Incoherent diffraction, where the nuclear target breaks up, but is separated by a rapidity gap from the projectile fragmentation region, allows one to extract, in addition, transverse plane correlations. These shed important light on the spatial picture of the partonic sub-structure of nuclei. In addition, both contribute crucial information 
necessary to understand the spatial gluon distributions that form the initial conditions for heavy ion collisions. $J/\Psi$ meson production off nuclei is the most widely considered exclusive channel; those of other vector mesons $\rho$, $\phi$) provide important complementary information.  DVCS, though luminosity hungry, is free of the uncertainty from incomplete knowledge of vector meson wave-functions. 

Coherent diffractive $J/\Psi$ production has been extensively
discussed as potentially the golden measurement in this
category. However, while the physics goals are golden, the technical challenges are formidable. Coherent diffraction dominates over incoherent diffraction only at
rather low values of $t$. It was determined that a rejection of the
target-dissociation background with at least 95\% efficiency is required in
order to measure the coherent cross section up to large enough
momentum transfers, and a 20 MeV resolution on the
momentum transfer is also needed in order to extract precise enough
information in impact parameter space. While this measurement is more feasible in light nuclei, it becomes more challenging in heavier nuclei. For light nuclei, coherent diffraction could shed important light on short range nuclear forces.
For larger nuclei, it is unclear at present whether what one learns is distinguishable from the distribution of gluons obtained from the  Woods-Saxon distribution of nucleons in the nucleus; 
 the ability for coherent diffraction to distinguish between different dynamical models for large nuclei is disputed. For these reasons, it is classed as a silver measurement.

Studies of the incoherent regime of diffractive vector meson production are slowly but surely emerging. This process is {\it a priori} more sensitive to high parton densities than is coherent diffraction. This is because
it is much easier to measure at large $t$, corresponding  to small values of b, nearer the center of the nucleus where the  gluon density is the largest. However, the amount of information that can be extracted from nuclear fragments is not clear, since the theoretical description of the nuclear break-up remains a challenge. The minimum requirement is to be able to identify if the nucleus breaks up into its constituent nucleons or if the nucleons themselves break-up, as
the corresponding calculations require different theoretical tools. Neither experimental or theoretical works on this process are mature enough to classify it as a golden measurement. However, the effects predicted by saturation models are large and unique enough that an observation of these will be convincing evidence of this physics. Therefore, both theoretical and experimental studies of this channel should be pursued vigorously.

%



\subsection{Parton substructure of nuclei}

Nuclear deep--inelastic scattering with an EIC will provide a unique measurement of gluon and sea quark densities in the ``dilute''
regime at $x\gtrsim 0.01$ in a range of nuclei. While the quark
densities in the region $0.05 \lesssim x \lesssim 0.6$ were 
studied in fixed target experiments and will be further explored at
JLab with 12 GeV electron beams, the behaviour of the gluon and sea quark densities in
this region is essentially unknown.  An EIC will have sufficient coverage in $Q^2$ 
to extract the nuclear gluon density through the $Q^2$ dependence
of the nuclear structure function $F_2^A$. Furthermore, direct access 
to gluons can be gained from the longitudinal structure 
function $F_{L}^A$ through measurements at different beam energies, or additionally, by  tagging charm production.

A reliable determination of the nuclear gluon
density in the dilute regime is essential for a
quantitative assessment of the onset of the new QCD regime of high
parton densities and non-linear gluon interactions, which will be more
widely accessible at a full-energy EIC.
 At $x \gtrsim 0.1$, an EIC will
also explore gluon anti-shadowing and EMC effects -- a step that might prove as revolutionary for our understanding of nuclei as the discovery of the quark EMC effect 30 years ago. For these reasons, inclusive $F_{2,L}$ structure function measurements at larger $x$ complement those discussed for the small-$x$ regime. 

As it turns out, the high luminosity envisaged for an EIC enables 
measurements of nuclear electromagnetic structure functions up to
$x\approx 1$ competitively with, or even surpassing, what has been
achieved to date in fixed target experiments. (Since the maximum $x$ in a nucleus is $A$, collider high luminosity measurements could uncover interesting physics in the Fermi regime where partons carry more more momenta than a bound nucleon.)
Furthermore, the large $Q^2$ range and hadronic event
reconstruction capabilities will also likely allow measurement of
charged current structure functions, and possibly of $\gamma-Z$
interference structure functions.  See the chapter on electroweak physics for further discussion. 
This will enable full quark flavour separation utilising only nuclear
DIS data, and offer, for example, new handles on the origin of the EMC effect
such as its flavour dependence. In this context, one should also mention the possibility of extracting information on particular twist four operators, which play an important role in parity violating DIS in the EMC region.
These measurements are
highly interesting, important, and in some cases unique to an EIC  
compared to previous facilities. However, more work is 
needed to establish to what extent full flavour separation can be effectively
carried out at an EIC; we therefore classify them as silver measurements. 

Much more information on the nuclear modification of the
quark/gluon structure of the proton and neutron can be gained from 
deep--inelastic measurements with detection of the spectator system
of ($A - 1$) nucleons in the final state. In particular, measurements 
on deuterium with a spectator proton can measure 
structure functions of the bound neutron ranging from nearly on-shell
to far off-shell, facilitating the extrapolation to an on-shell
neutron. Measurements with a spectator neutron, which are 
extremely difficult with a fixed target but feasible at a collider 
using a zero degree calorimeter, provide completely new information on
the off-shell proton structure functions, and constrain theoretical
models by comparison to the well known free proton wave
function. With heavier nuclear targets, one could explore the effects
of parton/nucleon embedding in a complex nuclear environment.
While no technical difficulty is foreseen, detailed studies of the
required detectors are needed to determine the feasibility and
precision of these measurements. 

\subsection
{Parton propagation and hadronization in nuclear matter}
\label{sec:eA_intro_ppf}

\begin{figure}[tb]
\centering
\includegraphics[height=5cm]{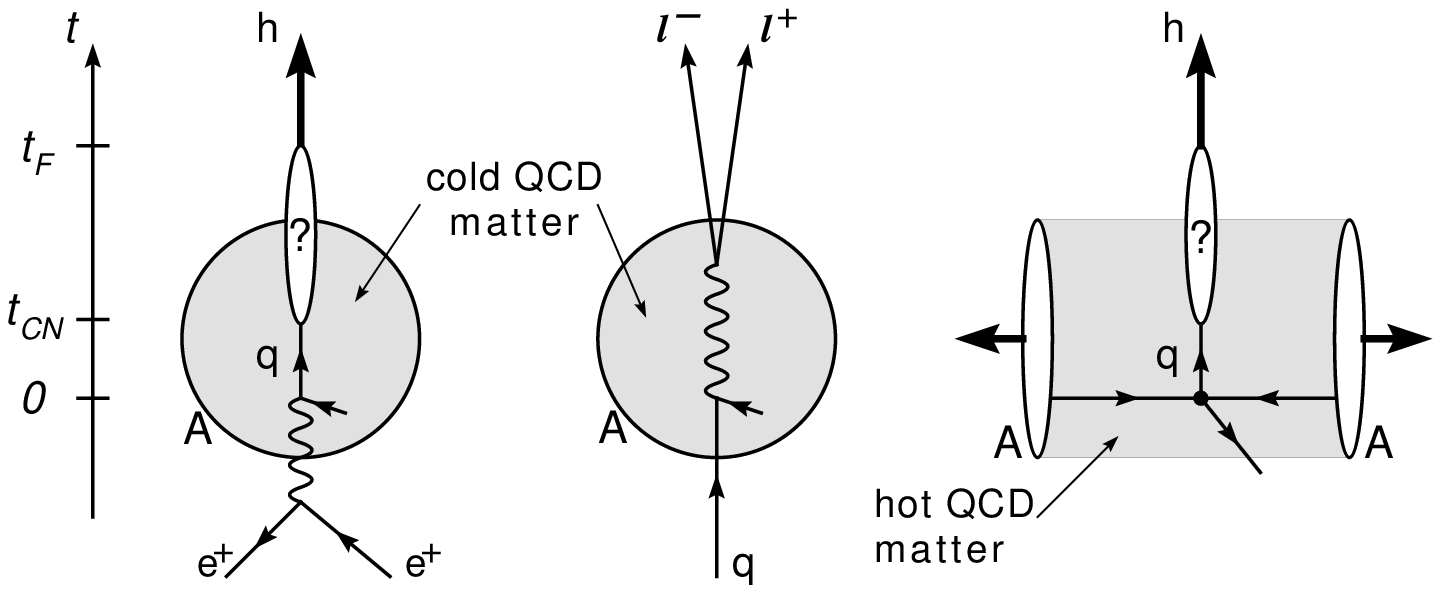}
\
\caption{\small
  Parton propagation and hadronization in cold and hot nuclear
  matter. A scenario of possibly distinct colour neutralization ($t_{CN}$) and hadron formation
  ($t_F$) time scales is illustrated on the vertical time axis.
}
\label{fig:intro_coldhot}
\end{figure}

The transition from coloured partons (quarks and gluons) to colourless
hadrons -- the so-called hadronization or fragmentation process -- exemplifies a fundamental process in QCD which still lacks a quantitative  understanding from first principles calculations.
Fragmentation functions, which encode the probability
that a parton fragments into a hadron, have been obtained
by fitting experimental data covering large kinematic ranges and numerous
hadron species. However, knowledge about the dynamics of the process 
remains limited and model dependent.
A particular model, see figure~\ref{fig:intro_coldhot}, posits a separation of scales between 
a short time scale for colour neutralization due to confinement generating a colourless ``pre-hadron" and a longer time scale (presumably controlled by chiral symmetry breaking), which governs the formation of hadrons. The dynamical consequences of such a model are distinguishable from other models where the separation of scales is reversed or is non-existent. 
Extracting these time scales would be an important  step towards understanding how hadrons emerge dynamically from partons, complementing the information on properties of colour confinement extracted from lattice measurements of ground state ``static" correlators. 

Nuclear deep inelastic scattering (nDIS) provides a known and stable
nuclear medium (``cold QCD matter'') and a final
state with strong experimental control on the kinematics of the hard
scattering. This permits one to use nuclei as femtometer-scale
detectors of the hadronization process, see
figure~\ref{fig:intro_coldhot}.  
In fact, both the energy loss due to medium-induced gluon {\em
  bremsstrahlung} off a 
quasi-free parton and the pre-hadron re-interaction with the surrounding
nucleons lead to attenuation and transverse momentum
broadening of hadron yields compared to proton targets, and allow
experimental access to the 
space-time evolution of hadronization. Theoretical models
of this process can be calibrated in nDIS and  
then applied, for example, to the study of the Quark-Gluon Plasma (``hot
QCD matter'') created in high-energy nucleus-nucleus collisions. 

The combination of high energy and luminosity offered by an EIC
promises a truly qualitative advance in this field, compared with
current and planned fixed target experiments. 
The large $Q^2$ range permits measurements in the fully calculable
perturbative regime with enough leverage 
to determine nuclear modifications in the QCD evolution of
fragmentation functions; the high-luminosity permits 
the multidimensional binning necessary for separating the
many competing effects and for detecting rare
hadrons. The large 
$\nu$ range ($\approx 10-1000$ GeV) allows one to
experimentally boost the hadronization process in 
and out of the nuclear medium, in order to isolate in-medium
parton propagation effects (large $\nu$) and cleanly extract colour
neutralization and hadron formation times (small $\nu$);
furthermore, using the quark flavour separated nuclear PDFs expected
from an EIC, one could analyze nuclear Drell-Yan data, which are free
from hadronization effects, and isolate initial state parton energy
loss from nuclear wave function effects, enabling a complete
experimental study of colour charge interactions in cold nuclear
matter. 
For the first time, one will be able to study hadronization of 
open charm and open bottom meson production in $e+A$ collisions, as well as the
in-medium propagation of the associated heavy quarks: these allow one to
fundamentally test high-energy QCD predictions for energy loss, and confront
puzzling measurements of heavy flavour suppression in the Quark-Gluon Plasma
at RHIC.
Within a collider environment, one would also be able to separate
target from current hadronization and cross-correlate these two,
adding a new dimension to hadronization studies.

The scattered quarks and gluons, from which the final-state hadrons
emerge, couple to other nuclear gluons.
Good control over the colour neutralization time scale will
allow one to use this internally created colour radiation to explore
the structure of 
nuclear matter in close analogy with the well-known exploration of
matter with electromagnetic radiation or electrically charged particles. 
Furthermore, an EIC with $\surd s \gtrsim 30$ GeV will permit for the first
time the measurement of jets and their substructure in $e$+A collisions,
furnishing a novel and extensive set of observables which directly access
quark energy loss and the as yet untested parton shower mechanism,
fundamentally described in QCD and pervasive in applications to
particle physics simulations.
Jet nuclear modifications can also be directly related to the propagation 
of the coloured partons shower in the nuclear medium, and used to measure
the cold nuclear matter transport coefficients which encode
basic information on the non-perturbative soft gluon structure of the
nuclei. 
These measurements are complementary to direct inclusive and diffractive
structure functions measurements at small $x$ in accessing the
high-density non-linear QCD regime, but are entirely feasible with a
low-energy EIC.

The outlined parton propagation and hadronization program can for the
most part be carried out in phase-I. In phase-II, we do not anticipate
any qualitative new lesson will be learned, while the increased energy
and $Q^2$ range may prove useful, for instance, for more refined studies
in the jet and heavy flavour sectors, and offering an increased reach
towards small $x$ for nuclear gluon measurements via 2+1 jet production.

In conclusion, due to the physics interest, theoretical
interpretability and feasibility in phase-I, this jet and heavy quark study program as a whole was
classified as a golden measurement for $e$+A collisions at an EIC, with light quark SIDIS classified as a 
silver measurement.


\section{Review of linear and non-linear approaches in QCD at small-$x$}  
\label{sec:introduction}
\label{sec:eA_smallx_part}
   
\subsubsection{Collinear factorization and DGLAP evolution}
\label{sec:dglap}
\hspace{\parindent}\parbox{0.92\textwidth}{\slshape
  Anna M. Sta\'sto}
\index{Sta\'sto, Anna M.}

\vspace{\baselineskip}

The evaluation of strong interaction cross-sections which involve hard scales is possible thanks to QCD  factorization theorems. The latter are derived from first principles in QCD \cite{Collins:1985ue,Collins:1988ig} and allow the factorization of cross sections into hard scattering coefficients (computed in a perturbative expansion in the strong coupling constant) and parton densities which contain information about nonperturbative dynamics. 
Parton densities, due to their intrinsically non-perturbative nature,  cannot be directly evaluated from first-principles lattice computations except perhaps in very limited kinematic windows.  Nevertheless, their evolution with hard scale can be calculated. This is done usually using the renormalization group  DGLAP equations,
\begin{equation}
\mu \frac{d}{d\mu}f_{j/h}(x,\mu) \; = \; \sum_k \int_x^1 \frac{dz}{z} \, P_{jk}(z,\alpha_s(\mu)) \, f_{k/h}(x/z,\mu) \; ,
\label{eq:dglap}
\end{equation}
with the splitting functions which  have perturbative expansion in powers of the strong coupling constant $\alpha_s$,
\begin{equation}
 P_{jk}(z,\alpha_s(\mu))=\sum_i (\alpha_s(\mu))^i P_{jk}^{(i)}(z) \; .
\label{eq:pjk}
\end{equation}

Coefficient functions and splitting functions are known up to NNLO accuracy \cite{Moch:2004pa,Moch:2004xu,Vogt:2004mw}. It has been found that at this order large corrections appear which are enhanced by the logarithmic terms in $1/x$. The collinear approach suffers also from other limitations. The kinematical approximations mostly suitable  for the evaluation of the  inclusive observables are not sufficient for exclusive processes and can lead to large discrepancies \cite{Collins:2005uv}. 

There are also other direct indications of the breakdown of the fixed order approach.
From the global fits \cite{Martin:2009iq,Nadolsky:2008zw}, it is known that the gluon density suffers from large uncertainties at the NLO level in the region of  small values of $x$,  and the gluon density even turns negative.
Even though the gluon density is not a directly observable quantity, the aforementioned  uncertainties propagate into  the observable longitudinal structure function $F_L$. The problem is concentrated in the low $Q$ and low-$x$ region, though the uncertainties remain even at larger values of $Q$ when $x$ is decreased. A systematic study of the  compatibility of the HERA deep inelastic data with DGLAP evolution has been performed in \cite{Caola:2009iy}. This analysis, originally based on the NNPDF1.2 analysis~\cite{Ball:2008by,Ball:2009mk}, was then extended  to the
global NNPDF2.0 set, which includes the very precise combined HERA-I dataset as well as all the relevant hadronic data.  A `safe' region was defined as the one in which the non-DGLAP effects are expected to be negligible, and it was defined by the cut on low-$x$ and $Q$ data. A fit was then performed to the data that pass the cut and only belong to the safe region and the structure functions evaluated at different scales.
It turned out that the prediction for the structure functions at low $Q^2$ obtained from the backward-evolution of the data above the cut
exhibits a systematic downward trend. Thus the precise HERA measurements indicate that the fixed order DGLAP evolution is incompatible with the data in the low $Q^2$ and low $x$ region.

\subsubsection{Small-$x$ re-summations}
\label{sec:re-sum}
\hspace{\parindent}\parbox{0.92\textwidth}{\slshape
 Anna M. Sta\'sto}
\index{Sta\'sto, Anna M.}

\vspace{\baselineskip}

Since the seminal works \cite{Kuraev:1977fs,Balitsky:1978ic}, it is well known that observables at small $x$ receive substantial corrections due to the large logarithms $\alpha_s \ln 1/x$ which need to be re-summed in this regime. The BFKL approach \cite{Kuraev:1977fs,Balitsky:1978ic} provides a framework for this summation and it is known up to next-to-leading logarithmic accuracy. 
The resulting evolution of the gluon Green's function provided by this framework is with respect to the $\ln 1/x$ or rapidity variable, with the transverse momenta of the gluons being summed over all possible configurations. 
The evolution has the following form
\begin{equation}
G(Y;{\bf k},{\bf k}_0) = \delta^{(2)}({\bf k}-{\bf k}_0) \; +\;  \int d^2 {\bf k}'  \; K({\bf k},{\bf k}') \;  G(Y;{\bf k}',{\bf k}_0) \; ,
\label{eq:bfkl}
\end{equation}
with the branching kernel having also the perturbative expansion
\begin{equation}
 K({\bf k},{\bf k}') =\sum_i (\alpha_s(\mu))^i K^{(i)}({\bf k},{\bf k}') \; .
\label{eq:kernel}
\end{equation}

\begin{wrapfigure}{r}{0.5\textwidth}
  \begin{center}
    \includegraphics[width=0.48\textwidth]{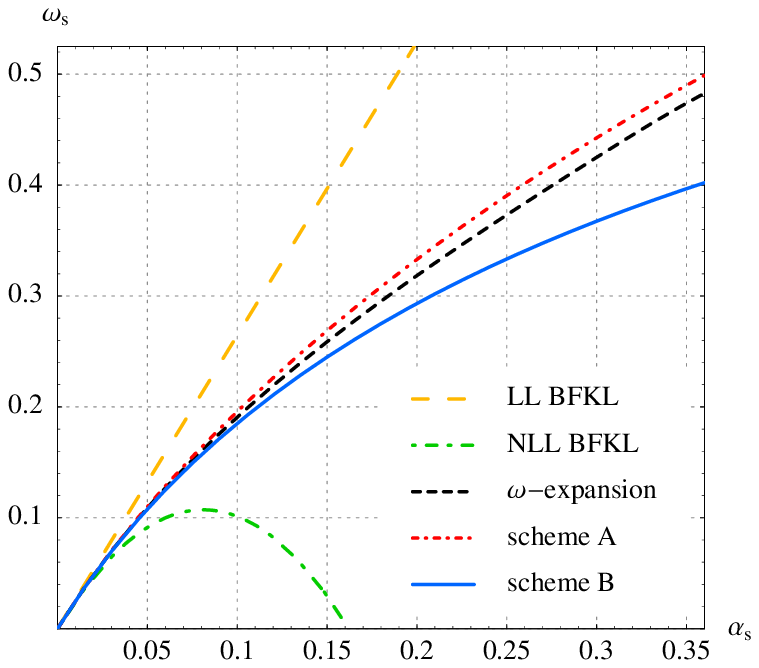}
  \end{center}
  \caption{\small The intercept of the hard Pomeron extracted from the BFKL equation with fixed strong coupling in LL, NLL and re-summed cases.}
\label{fig:intercept}
\end{wrapfigure} 
 
A solution for the gluon Green's function and therefore the resulting cross sections exhibit strong growth with the energy, the hard Pomeron, with the intercept being significantly larger than unity in the LO approximation, $\omega_P=1+\alpha_s N_c 4 \ln 2/\pi$. This growth turns out to be incompatible with both the hadronic data and the data on structure functions from deep-inelastic scattering. NLL (next-to-leading-log) corrections \cite{Fadin:1998py,Ciafaloni:1998gs} turned out to be rather large numerically and point to the need for the re-summation of subsequent powers of higher order corrections $\alpha_s ^k \ln 1/x$. The sizes of the various NLL corrections can be understood on physical grounds. Firstly, unlike in the DGLAP limit, the strong coupling constant is not naturally a small parameter. On top of that, the BFKL approach does not satisfy the momentum sum rule for the longitudinal momentum fractions (the transverse momenta are however conserved, unlike in the collinear approach). The kinematical approximations made in the BFKL limit cannot be efficiently recovered by the truncated higher orders of the perturbative expansion.




The strategy of re-summation at small $x$ has been developed in a series of works  \cite{Ciafaloni:2003rd,Ciafaloni:2003kd,Ciafaloni:2007gf,Altarelli:2005ni,Altarelli:2003hk}. It involves the construction of the appropriate re-summed kernel of the form given by Eq.~\ref{eq:kernel}, which includes at the same time known terms in the expansion of the splitting function, Eq.~\ref{eq:pjk}. Although the details of the various approaches differ, there are common fundamental ingredients. The evolution in rapidity is subjected to 
kinematical constraints which originate from the requirement of the consistency of the assumption about the Regge kinematics. The evolution is matched with the 
DGLAP  evolution by including the splitting function at LO and NLO.  The momentum sum rule is imposed onto the resulting re-summed splitting function. The running of the coupling is included into the evolution. Finally, matching to the NLL BFKL is performed with the  suitable subtractions in order to avoid double counting. The resulting Green's function and splitting function turn out to be very stable with minimal variations across the different re-summation schemes. \\


c
\noindent {\bf Gluon Green's function and the splitting function:}  In Fig.~\ref{fig:intercept}, we show the results on the intercept of the gluon Green's function in the case of the fixed strong coupling constant, obtained within the re-summation framework of \cite{Ciafaloni:2003rd}. The linear growth is given by the LO approximation. The NLO value of the intercept is significantly below the lowest order, and turns negative even for the intermediate values of $\alpha_s$.  The re-summed result is between the NLO and LO, it exhibits clear growth with increasing values of the coupling constant, albeit much reduced with respect to the LO value and much closer to the phenomenology.

\begin{figure}
  \centering 
  \parbox{0.4\linewidth}{
    \includegraphics[width=\linewidth]{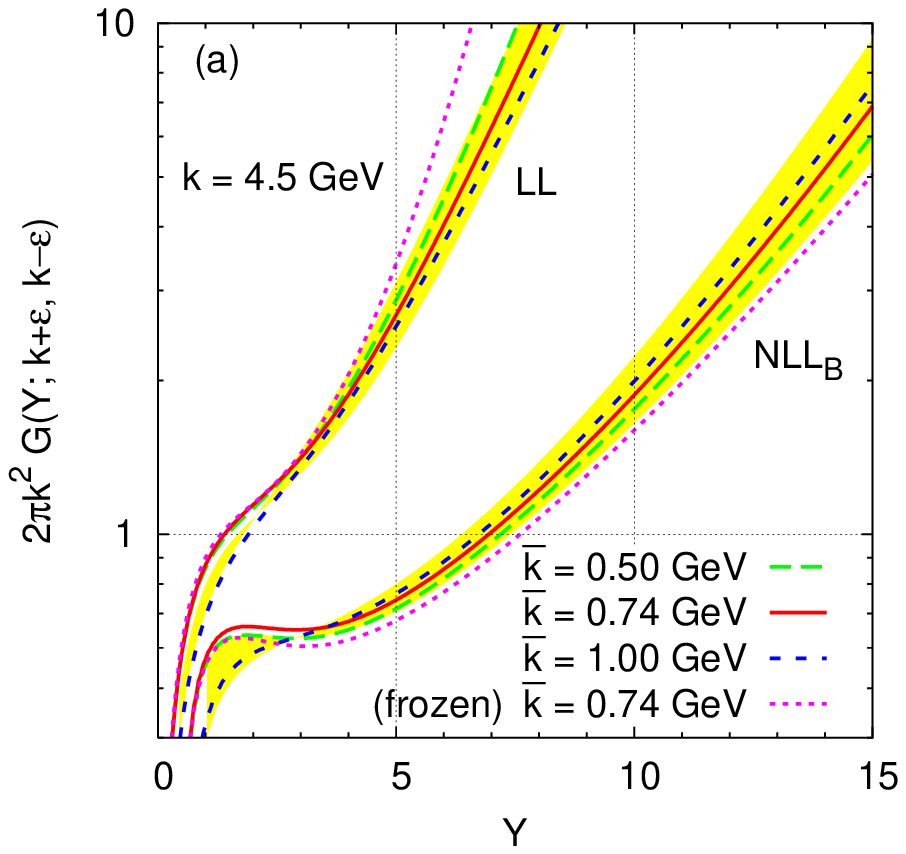}
  }
  \hspace*{0.5cm}
  \parbox{0.45\linewidth}{
    \includegraphics[width=\linewidth]{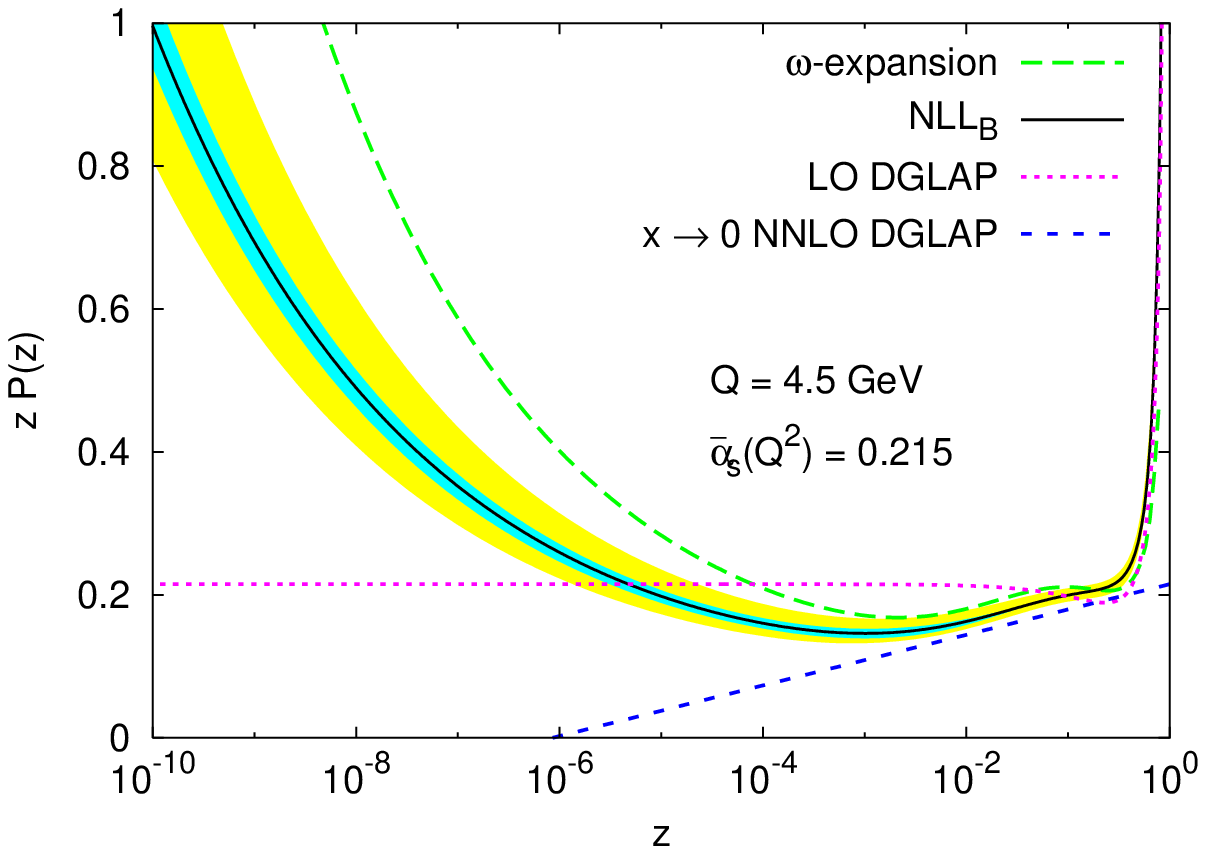}
  }
  \caption{\small 
    {\it Left:} The gluon Green's function extracted from the BFKL
    equation in LL and re-summed cases. The coupling is running in all
    computations, and $\bar{k}$ denotes the scale at which the strong
    coupling is regularized. 
    {\it Right:} The extracted effective splitting function from the
    re-summed approach: solid line. The scale was taken to be $Q= 4.5
    {\rm GeV}$.  Dotted pink line indicates the LO DGLAP splitting
    function and the blue dashed indicates small $x$ part of the NNLO
    DGLAP. The dashed green line corresponds to the re-summed
    spliitting function from the  $\omega$ expansion. The band
    correspond to the scale variation. 
} 
\label{fig:greensfunction}
\label{fig:splittingfunction}
\end{figure}

The rapidity dependence of the gluon Green's function is shown in
Fig.~\ref{fig:greensfunction} (left). The scale was chosen to be equal to $k=4.5 {\rm GeV}$. The reduction of the speed of growth is clear in the re-summed case. Also the scale variations are relatively small in this case.

By using the deconvolution of the integral equation, one can calculate the integrated gluon density.
As a result, it is possible to solve the re-summed splitting function numerically. In this way, the perturbative and non-perturbative contributions are factorized in $Q^2$. 
In Fig.~\ref{fig:splittingfunction} (right),  we show the results for the splitting function as a function of the momentum fraction for the re-summed case. It is compared with the results on the LO and NNLO (only small $x$ part) splitting functions. 
The results on the splitting function demonstrate that the small x growth is delayed to much smaller values of x (beyond HERA). The splitting function also has an interesting feature, namely that of the dip.
It turns out that this is a  universal feature, present also
 in other schemes of re-summation. In general, it was found that the  dip comes from the interplay
between NNLO order and the re-summation.

Thus far, re-summation  was demonstrated to give stable results for the gluon channel only.
For the complete description, however, one needs  to include quarks in the evolution.
A matrix approach was developed which was shown to be  consistent with the collinear matrix
factorization of the parton densities in the singlet evolution~\cite{Ciafaloni:2007gf}. This approached enables the calculation of the anomalous dimensions matrix, which
can be directly compared with the standard DGLAP matrix. It was shown that it is possible to incorporate NLLx BFKL + NLO DGLAP in this framework~\cite{Ciafaloni:2007gf}. \\

\noindent{\bf Conclusions and outlook:}  The small-$x$ regime requires a formalism which incorporates re-summation of the large terms  $\alpha_s \ln 1/x$. The BFKL formalism was extended to include re-summation to higher orders. This  formalism includes both  DGLAP NLO and BFKL NLL and higher
order terms. Stability of the results was demonstrated for scale changes and model
changes. There are certain universal and characteristic features which  come from the solutions to the evolution equations: the rapid growth with $x$ is delayed to smaller values of $x$, and  the splitting function has a minimum. A matrix model was developed which  gives consistent results on the gluon Green's function and the splitting functions. For the complete framework, one needs to include the re-summed coefficient functions. Detailed fits to the data need to be performed. In this regard, the EIC will generate very important information on parton densities at small $x$, and in distinguishing small $x$ re-summation effects from higher twist 
saturation effects.


\subsubsection{Parton Saturation}
\label{sec:partsat}
\hspace{\parindent}\parbox{0.92\textwidth}{\slshape
 Yuri V. Kovchegov and Cyrille Marquet}
\index{Kovchegov, Yuri V.}
\index{Marquet, Cyrille}

\vspace{\baselineskip}

The QCD description of hadrons in terms of quarks and gluons depends on the processes considered and on what part of the hadron wave function they are sensitive to. Consider a hadron moving at nearly the speed of light along the light cone direction $x^+$, with momentum $P^+$. Depending on their transverse momentum $k_T$ and longitudinal momentum $xP^+$, the virtual partons inside the hadron behave differently, reflecting the different regimes of the hadron wave function. Soft hadronic processes are mostly sensitive to the non-perturbative part of the wave function, they involve quantum fluctuations with transverse momenta of the order of $\Lambda_{QCD}\sim 200$ MeV. A hadron can then be thought of as a bound state of strongly-interacting partons, but a QCD description of the associated dynamics is still lacking. By contrast, hard processes in hadronic collisions are sensitive to the weakly-coupled part of the wave function and resolve the partonic structure of hadrons. They probe partons with $k_T\gg\Lambda_{QCD}$ whose QCD dynamics is better understood.

One can distinguish two weakly-coupling regimes in the wave function: a linear one called the hard regime, involving a small density of partons, typically with $x\lesssim 1$, in which the hadron looks like a dilute system of independent partons, and a non-linear one called the saturation regime, involving a large density of partons with $x\ll1$, in which the hadron looks like a dense system of nevertheless weakly-interacting partons, mainly gluons (called small-$x$ gluons). The dilute-dense separation is a bit subtler than that: the larger $k_T$ is, the smaller $x$ needs to be to enter the saturation regime. Indeed the separation between the two regimes is characterized by a momentum scale $Q_s(x)$, called the saturation scale, which increases as $x$ decreases. In a scattering process, dilute partons (with $k_T\gg Q_s(x)$) behave incoherently, while when the parton density is large ($k_T\lesssim Q_s(x)$), gluons scatter coherently. The dynamics of the dilute regime is well described by the leading-twist approximation of QCD, whose hallmark is collinear factorization. As explained in the previous section, when $x$ becomes small while not yet reaching the non-linear regime, so-called small-$x$ re-summations are also needed to improve the approximation.

To describe the small-$x$ non-linear part of hadronic/nuclear wave functions in QCD, the Color Glass Condensate (CGC) effective theory was proposed. Rather than using the standard Fock-state expansion which is ineffective in dealing with numerous small$-x$ gluons, the CGC approach employs collective degrees of freedom, static color sources at large $x$ and dynamical  classical color fields at small $x$. The traditional approach to saturation physics consists of two stages, corresponding to two different levels of approximations. The first level corresponds to the classical gluon field description of nuclear wave functions and scattering cross sections. It re-sums all multiple re-scatterings in the nucleus, but lacks energy dependence. The latter is included through quantum corrections, which are re-summed by non-linear evolution equations. This constitutes the second level of approximation. We will present both stages below.

\subsubsection{Classical gluon fields} 

\begin{figure}[tbh]
  \centering
  \parbox{0.45\linewidth}{
    \includegraphics[width=\linewidth]{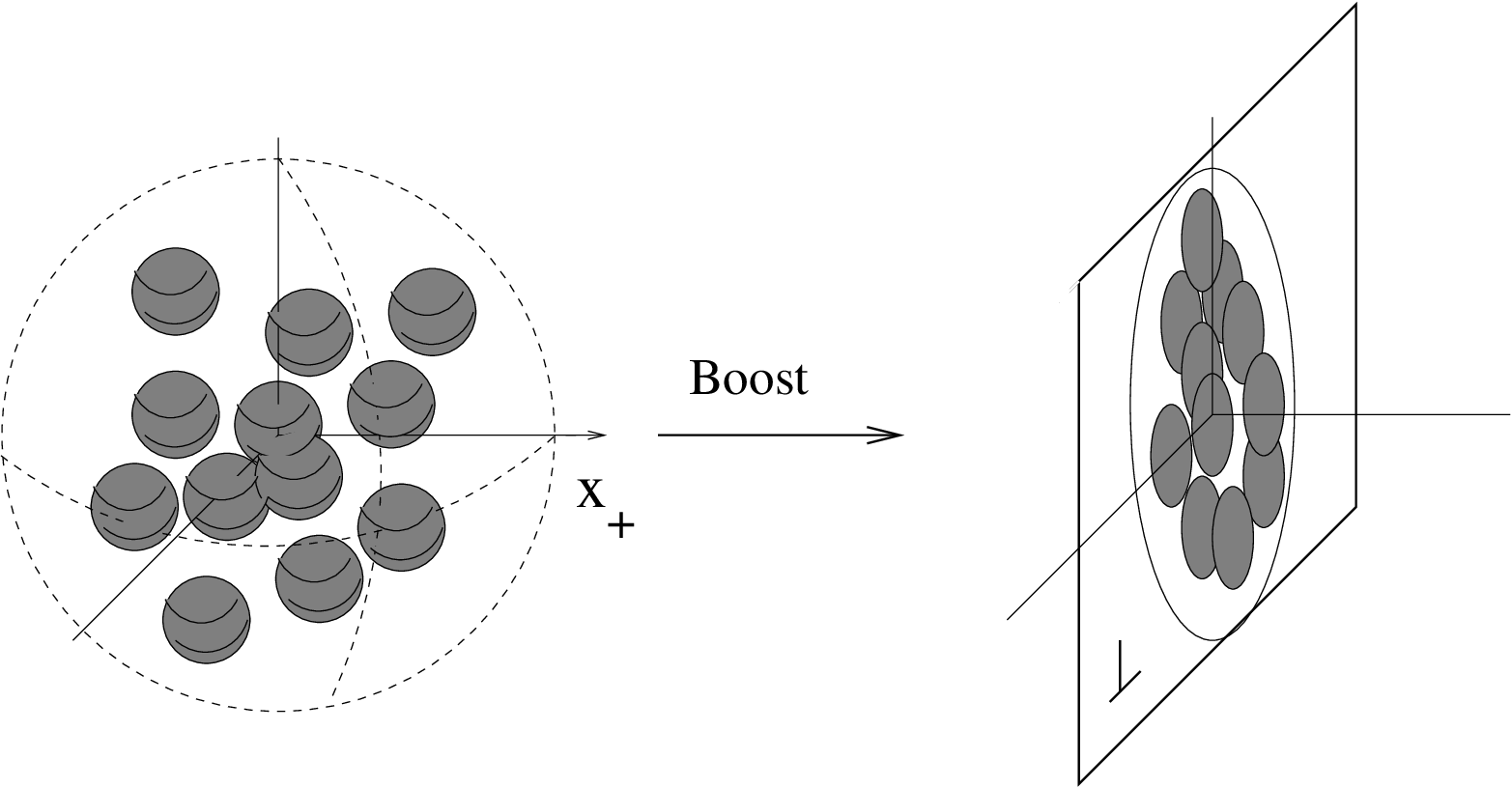}
  }
  \hfill
  \parbox{0.45\linewidth}{
    \includegraphics[width=\linewidth]{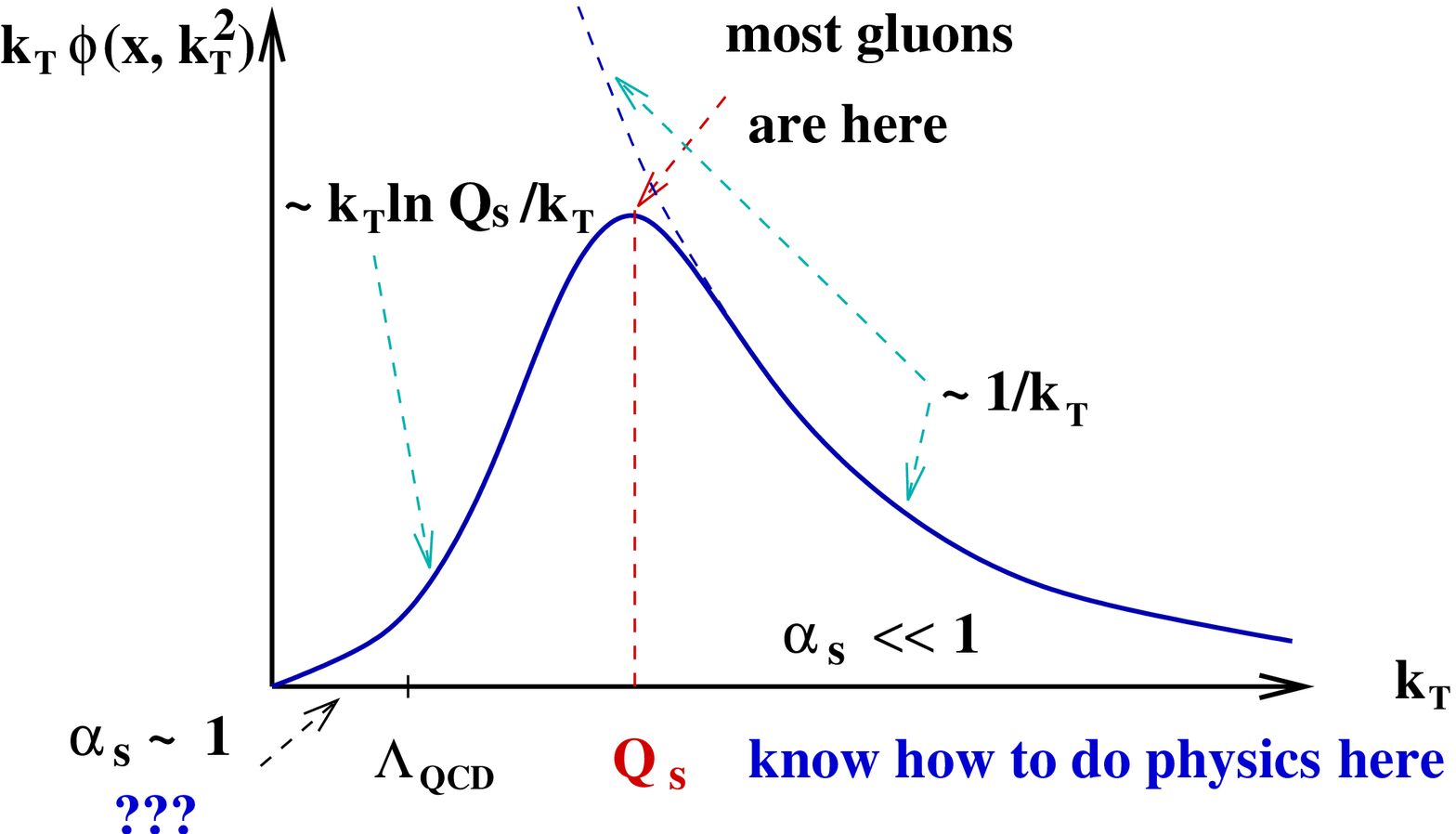}
  }
  \caption{\small 
    {\it Left:} Large nucleus before and after an ultrarelativistic boost.
    {\it Right:} Unintegrated gluon distribution $\phi (x, k_T^2)$ of a 
    large nucleus due to classical gluon fields (solid line). Dashed
    curve denotes the lowest-order perturbative result.
  }
\label{nucl_boost}
\label{mv2}
\end{figure}

\noindent{\bf McLerran--Venugopalan model:}  Imagine a single large nucleus, which was boosted to some
ultrarelativistic velocity, as shown in Fig.~\ref{nucl_boost} (left). We are
interested in the dynamics of small-$x$ gluons in the wave function of
this relativistic nucleus. The small-$x$ gluons interact with the
whole nucleus coherently in the longitudinal direction: therefore,
only the transverse plane distribution of nucleons is important for
the small-$x$ wave function. As one can see from
Fig.~\ref{nucl_boost}, after the boost, the nucleons, as ``seen'' by
the small-$x$ gluons, appear to overlap with each other in the
transverse plane, leading to high parton density. Large occupation
numbers of color charges (partons) lead to classical gluon fields
dominating the small-$x$ wave function of the nucleus. This is the
essence of the McLerran-Venugopalan (MV) model
\cite{McLerran:1993ni,McLerran:1994vd,McLerran:1993ka}. According to
the MV model, the dominant gluon field is given by the solution of the
classical Yang-Mills equations ${\cal D}_\mu \, F^{\mu\nu} \, = \, J^\nu$
where the classical color current $J^\nu$ is generated by the valence
quarks in the nucleons of the nucleus from Fig.~\ref{nucl_boost}.

The Yang-Mills equations were solved for a single nucleus exactly
\cite{Kovchegov:1996ty,Jalilian-Marian:1997xn} resulting in the
unintegrated gluon distribution $\phi (x, k_T^2)$ (multiplied by the
phase space factor of the gluon's transverse momentum $k_T$) shown in
Fig.~\ref{mv2} right as a function of $k_T$. (Note that in the MV model, $\phi (x, k_T^2)$ is independent of Bjorken-$x$.)
Fig.~\ref{mv2} demonstrates the emergence of the saturation scale
$Q_s$. As one can see from Fig.~\ref{mv2}, the majority of gluons in
this classical distribution have transverse momentum $k_T \approx
Q_s$. Since in this classical approximation $Q_s^2 \sim A^{1/3}$, for a
large enough nucleus, all of its small-$x$ gluons would have large
transverse momenta $k_T \approx Q_s \gg \Lambda_{QCD}$, justifying the applicability of the perturbative approach to the problem. Note that the
gluon distribution slows down its growth with decreasing $k_T$ for
$k_T < Q_s$ (from power-law of $k_T$ to a logarithm) and the distribution
{\sl saturates}.  \\

\begin{figure}
  \centering
  \includegraphics[height=3cm]{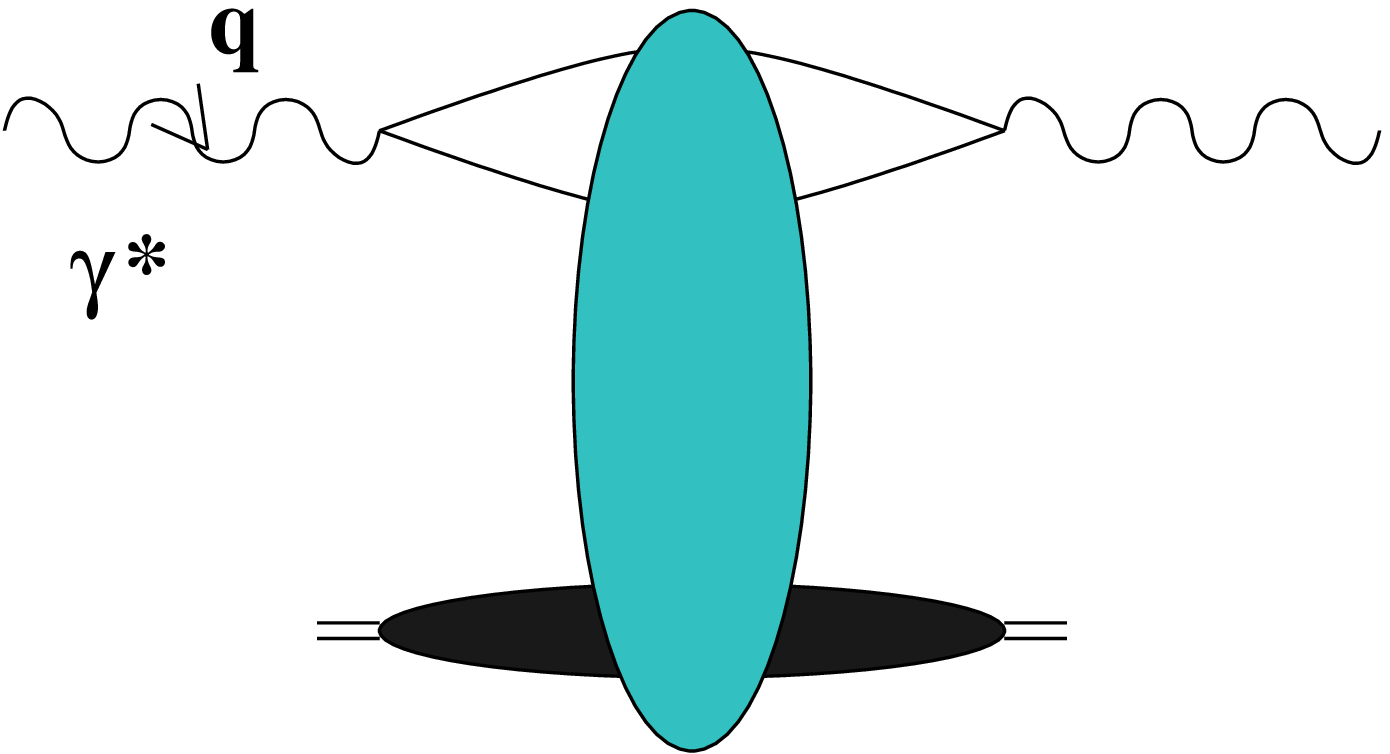}
  \hspace*{1cm}
  \includegraphics[height=3cm]{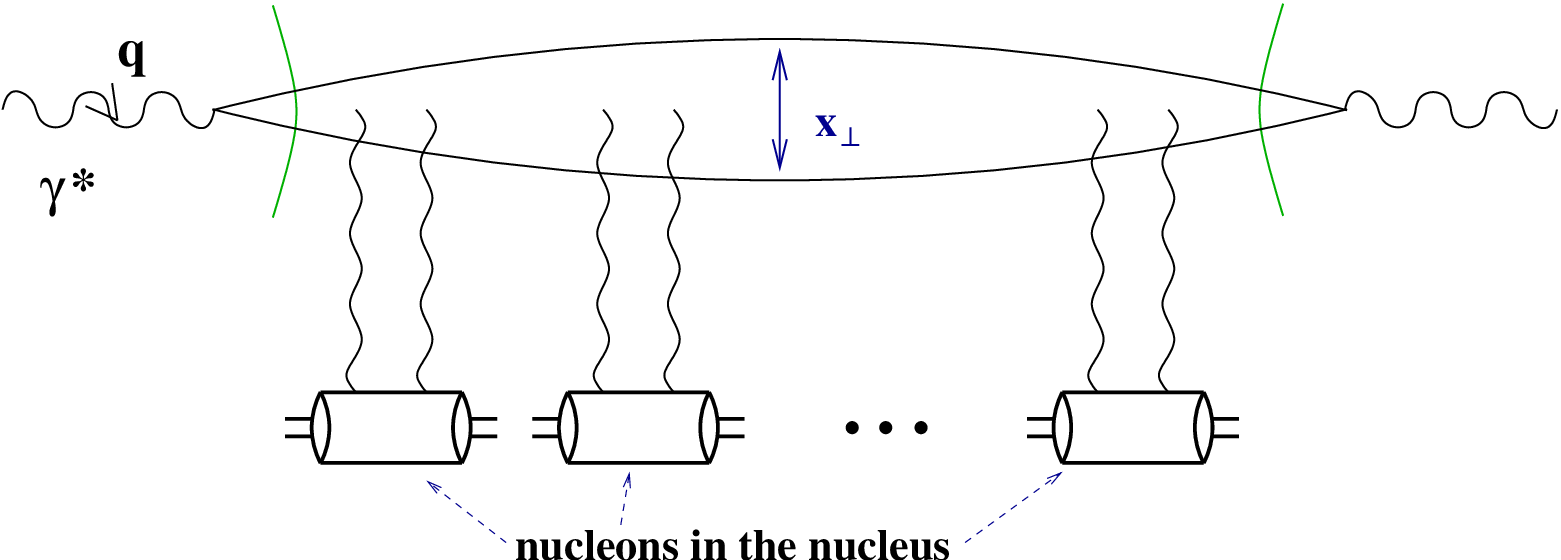}
  \caption{\small 
    {\it Left}: Deep inelastic scattering in the rest frame of the
    target.
    {\it Right}: Deep inelastic scattering in the quasi-classical 
    Glauber-Mueller approximation in $\partial_\mu A^\mu =0$ gauge.
    \label{qqbar}
 }
\label{kovchegov:dis}
\end{figure}

\noindent{\bf DIS at high energy: Glauber-Mueller formula:}  Let us consider deep inelastic scattering (DIS) on a large nucleus. In
DIS, the incoming electron emits a virtual photon, which in turn
interacts with the proton or nucleus. In the rest frame of the
nucleus, the interaction can be thought of as the virtual photon
splitting into a quark-antiquark pair, which then interacts with the
nucleus (see Fig.~\ref{kovchegov:dis}, left panel). Since the light cone lifetime of the $q\bar
q$ pair is much longer than the size of the target nucleus, the total
cross section for the virtual photon--nucleus scattering can be
written as a convolution of the virtual photon's light cone wave
function (the probability for it to split into a $q\bar q$ pair) with
the forward scattering amplitude of a $q\bar q$ pair interacting with
the nucleus
\begin{align}
  \label{sigN}
  \sigma_{tot}^{\gamma* A} (Q^2, x_{Bj}) \, = \, \int \frac{d^2 x \, d
    z }{2 \pi} \, [\Phi_T ({\underline x}, z) + \Phi_L ({\underline
    x}, z) ] \ d^2 b \ N({\underline x}, {\underline b} , Y)
\end{align}
with the help of the light-cone perturbation theory \cite{Lepage:1980fj}.
Here the incoming photon with virtuality $Q$ splits into a
quark--antiquark pair with the transverse separation ${\underline x}$ and the
impact parameter (transverse position of the center of mass of the
$q\bar q$ pair) ${\underline b}$. $Y$ is the rapidity variable given by $Y =
\ln (s \, x_T^2) \approx \ln 1/x_{Bj}$. The square of the light cone 
wave function of $q \overline{q}$ fluctuations of a virtual photon is
denoted by $\Phi_T ({\underline x}, z)$ and $\Phi_L ({\underline x}, z)$ for
transverse and longitudinal photons correspondingly, with $z$ being
the fraction of the photon's longitudinal momentum carried by the
quark. At the lowest order in electromagnetic coupling ($\alpha_{EM}$)
$\Phi_T ({\underline x}, z)$ and $\Phi_L ({\underline x}, z)$ are given by
\cite{Nikolaev:1990ja,Kovchegov:1999kx}
\begin{equation}
  \Phi_T ({\underline x}, z) = \frac{2 N_c}{\pi} \, \sum_f \alpha^f_{EM} \,
  \left\{ a_f^2 \ K_1^2 (x_\perp a_f) \ [z^2 + (1 - z)^2] + m_f^2 K_0 (x_\perp
  a_f)^2 \right\},
\label{eq:QED-T}
\end{equation}
\begin{equation}
  \Phi_L ({\underline x}, z) = \frac{2 N_c}{\pi} \, \sum_f \alpha^f_{EM} \, \
  4 \, Q^2 \, z^2 (1 - z)^2 \ K_0^2 (x_\perp a_f),
\label{eq:QED-L}
\end{equation}
with $a_f^2 = Q^2 z (1 - z) + m_f^2$, $x_\perp = |{\bf x}|$ and
$\sum_f$ denoting the sum over all relevant quark flavors with quark
masses denoted by $m_f$. $\alpha^f_{EM} = e_f^2 /4 \pi$ with $e_f$ the
electric charge of a quark with flavor $f$.

Our first goal is to calculate the forward scattering amplitude of a
quark--anti-quark dipole interacting with the nucleus, which is
denoted by $N({\underline x}, {\underline b} , Y)$ in
Eq.~(\ref{sigN}), including all multiple re-scatterings of the dipole
on the nucleons in the nucleus.To do this we need to construct a model
of the target nucleus. We assume that the nucleons are dilutely distributed in the nucleus~\cite{Mueller:1989st}.
There we can represent the dipole-nucleus interaction as a sequence of
successive dipole-nucleon interactions, as shown in Fig.~\ref{qqbar},
right panel.
Since each nucleon is a color singlet, the lowest order dipole-nucleon
interaction in the forward amplitude from Fig.~\ref{qqbar} is a
two-gluon exchange. The exchanged gluon lines in Fig.~\ref{qqbar} are
disconnected at the top: this denotes a summation over all possible
connections of these gluon lines either to the quark or to the
anti-quark lines in the incoming dipole.

Re-summation of the diagrams like the one in Fig.~\ref{qqbar} yields
\cite{Mueller:1989st}
\begin{align}
  \label{glaN2}
  N({\underline x}, {\underline b} , Y=0) \, = \, 1 - \exp \left\{ -
    \frac{x_\perp^2 \, Q_s^2 ({\underline b}) \, \ln (1/x_\perp \,
      \Lambda)}{4} \, \right\}
\end{align}
with the {\sl saturation} scale defined by
\begin{align}\label{qsmv}
  Q_s^2 ({\underline b}) \, \equiv \, \frac{4 \, \pi \, \alpha_s^2 \,
    C_F}{N_c} \, \rho \, T({\underline b}).
\end{align}

\begin{wrapfigure}{l}{0.5\textwidth}
  \begin{center}
    \includegraphics[width=0.48\textwidth]{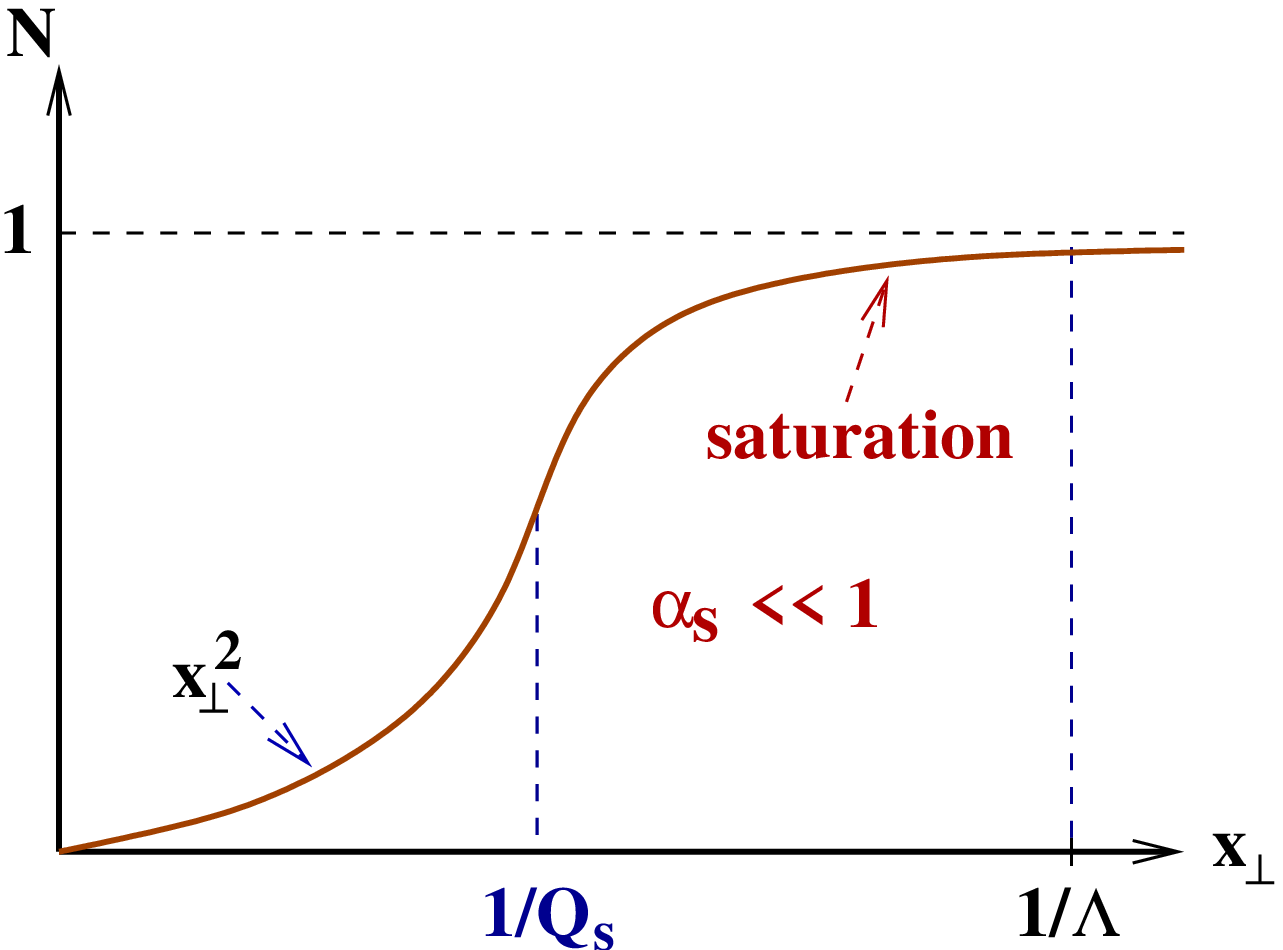}
  \end{center}
  \caption{\small The forward amplitude of the dipole--nucleus scattering $N$ ,plotted as a function of the transverse separation between the quark and the anti-quark in a dipole ($x_\perp$) using Eq.~(\ref{glaN2}).}
\label{glamue}
\end{wrapfigure} 

Here, $\rho$ is the density of nucleons in the nucleus ($\rho = A /
[(4/3) \pi R^3]$ for a spherical nucleus of radius $R$ with atomic
number $A$) and $T({\underline b})$ is the nuclear profile function
equal to the length of the nuclear medium at a given impact parameter
$\underline b$, such that $T({\underline b}) = 2 \, \sqrt{R^2 -
  {\underline b}^2}$ for a spherical nucleus. $\Lambda$ is an infrared
cutoff. We put $Y=0$ in the argument of $N$ in Eq.~(\ref{glaN2}) to
underline that this expression does not include any small-$x$
evolution which would bring in the rapidity dependence.

Eqs.~(\ref{glaN2}) and (\ref{qsmv}) allow us to determine the
parameter corresponding to the re-summation of the diagrams like the
one shown in Fig.~\ref{qqbar}. Noting that for large nuclei, the
profile function scales as $T({\underline b}) \sim A^{1/3}$ and the
nucleon density scales as $\rho \sim A^0$, we conclude that the
re-summation parameter of multiple re-scatterings is
\cite{Kovchegov:1997pc}: $\alpha_s^2 \, A^{1/3}$.
The physical meaning of the parameter $\alpha_s^2 \, A^{1/3}$ is rather
straightforward: at a given impact parameter the dipole interacts with
$\sim A^{1/3}$ nucleons exchanging two gluons with each. Since the
two-gluon exchange is parametrically of the order $\alpha_s^2$ we obtain
$\alpha_s^2 \, A^{1/3}$ as the re-summation parameter for the
quasi-classical approximation.


The dipole amplitude $N$, from Eq.~(\ref{glaN2}), is plotted
(schematically) in Fig.~\ref{glamue} as a function of $x_\perp$. One
can see that, at small $x_\perp$, $x_\perp \ll 1/Q_s$, we have $N \sim
x_\perp^2$ and the amplitude is a rising function of $x_\perp$.
However, at large dipole sizes $x_\perp \gtrsim 1/Q_s$, the growth
stops and the amplitude levels off ({\sl saturates}) at $N = 1$. This
regime corresponds to the black disk limit for the dipole-nucleus
scattering where, for large dipoles, the nucleus appears as a black disk. To
understand that the $N = 1$ regime corresponds to the black disk limit, let
us note that the total dipole-nucleus scattering cross section is
given by:
\begin{align}
  \label{sigtot}
  \sigma^{q{\bar q}A}_{tot} \, = \, 2 \, \int d^2 b \, N({\underline
    x}, {\underline b} , Y)
\end{align}
where the integration goes over the cross sectional area of the
nucleus. If $N=1$ at all impact parameters $\underline b$ inside the
nucleus,  for a spherical nucleus of radius $R$, Eq.~(\ref{sigtot}) becomes
$\sigma^{q{\bar q}A}_{tot} \, = \, 2 \, \pi \, R^2$,
which is a well-known formula for the cross section of a particle
scattering on a black sphere \cite{LL3}.

The transition between the $N \sim x_\perp^2$ to $N=1$ behavior in
Fig.~\ref{glamue} happens at around $x_\perp \sim 1/Q_s$. For dipole
sizes $x_\perp \gtrsim 1/Q_s$, the amplitude $N$ {\sl saturates} to a
constant.  This translates into the saturation of quark distribution
functions in the nucleus, as was shown in \cite{Mueller:1989st} (as
$xq + x\bar q \sim F_2 \sim \sigma_{tot}^{\gamma* A}$), and thus can
be identified with parton saturation, justifying the name of the {\sl
  saturation scale}.

Before we proceed, let us finally note that since $T({\underline b})
\sim A^{1/3}$, the saturation scale in Eq.~(\ref{qsmv}) scales as
$Q_s^2 \, \sim \, A^{1/3}$ with the nuclear atomic
number~\cite{McLerran:1994vd,McLerran:1993ka,McLerran:1993ni,Mueller:1989st}.
This implies that for a very large nucleus, the saturation scale would become
very large, much larger than $\Lambda_{QCD}$. If $Q_s \gg \Lambda_{QCD}$, the
transition to the black disk limit in Fig.~\ref{glamue} happens at
momentum scales (corresponding to inverse dipole sizes) where the
physics is perturbative and gluons are the correct degrees of freedom.


\subsubsection{Nonlinear evolution equations}


\noindent{\bf General picture:}  While the classical gluon fields of the MV model exhibit many correct
qualitative features of saturation physics, and give predictions about the
$A$-dependence of observables which may be compared to the data, they
do not lead to any rapidity/Bjorken-$x$ dependence of the
corresponding observables, which is essential in the data on nuclear
and hadronic collisions. To include rapidity dependence, one has to
calculate quantum corrections to the classical fields described above.

\begin{figure}[ht]
\centerline{\includegraphics[width=12cm]{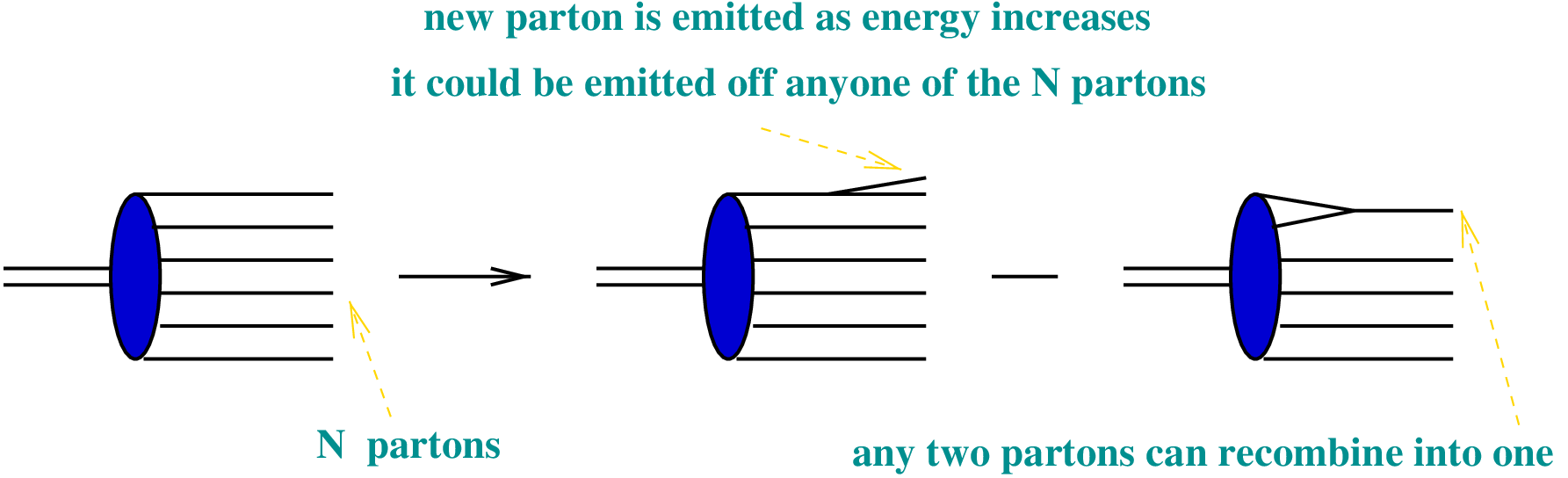}}
\caption{\small Nonlinear small-$x$ evolution of a hadronic or nuclear wave 
  functions. All partons (quarks and gluons) are denoted by straight
  solid lines for simplicity.}
\label{BKfig}
\end{figure}

The inclusion of quantum corrections is accomplished by the small-$x$
evolution equations. The first small-$x$ evolution equation was
constructed before the birth of saturation physics. This is the
Balitsky-Fadin-Kuraev-Lipatov (BFKL) evolution equation
\cite{Balitsky:1978ic,Kuraev:1977fs}. This is a linear evolution equation,
which is illustrated by the first term on the right hand side of
Fig.~\ref{BKfig}. Consider a wave function of a high-energy nucleus or
hadrons: it contains many partons, as shown on the left of
Fig.~\ref{BKfig}. As we make one step of evolution by boosting the
nucleus/hadron to higher energy, either one of the partons can split
into two partons, leading to an increase in the number of partons
proportional to the number of partons $N$ at the previous step,
\begin{align}\label{BFKL}
  \frac{\partial \, N (x, k_T^2)}{\partial \ln (1/x)} \, = \, \alpha_s \,
  K_{BFKL} \, \otimes \, N (x, k_T^2),
\end{align}
with $K_{BFKL}$ an integral kernel. Clearly the BFKL equation
(\ref{BFKL}) introduces a Bjorken-$x$/rapidity dependence in the
observables it describes.

The main problem with the BFKL evolution is that it leads to the
power-law growth of the total cross sections with energy,
$\sigma_{tot} \sim s^{\alpha_P -1}$, with the BFKL pomeron intercept
$\alpha_P -1 = (4 \, \alpha_s \, N_c \, \ln 2) /\pi >0$.  Such a power-law
cross section increase violates the Froissart bound, which states that
the total hadronic cross section can not grow faster than $\ln^2 s$ at
very high energies.  Moreover, the power-law growth of cross sections with
energy violate the black disk limit known from quantum mechanics: the high-energy total scattering
cross section $\sigma_{tot}$ of a particle on a sphere of radius $R$ is bounded by
$2 \, \pi \, R^2$ (note the factor of 2 which is due to quantum mechanics, this is not
simply a hard sphere from classical mechanics!).

We see that something has to modify Eq.~(\ref{BFKL}) at high energy.
The modification is illustrated on the far right of Fig.~\ref{BKfig}:
at very high energies, partons may start to recombine with each other
on top of the splitting. The recombination of two partons into one is
proportional to the number of pairs of partons, which, in turn, scales
as $N^2$. We end up with the following non-linear evolution equation:
\begin{align}\label{BK}
  \frac{\partial \, N (x, k_T^2)}{\partial \ln (1/x)} \, = \, \alpha_s \,
  K_{BFKL} \, \otimes \, N (x, k_T^2) - \alpha_s \, [N (x, k_T^2)]^2.
\end{align}
This is the Balitsky-Kovchegov (BK) evolution equation
\cite{Balitsky:1996ub,Kovchegov:1999yj}, which is valid for QCD in the
limit of large number of colors $N_c$. An equation of this type was
originally suggested by Gribov, Levin and Ryskin
\cite{Gribov:1984tu} and by Mueller and Qiu \cite{Mueller:1986wy},
though at the time it was assumed that the quadratic term is only the
first non-linear correction with higher order terms possibly appearing
as well: in \cite{Balitsky:1996ub,Kovchegov:1999yj} the exact form of
the equation was found, and it was shown that in the large-$N_c$ limit, Eq.~(\ref{BK}) does not have any higher-order terms in $N$.
Generalization of Eq.~(\ref{BK}) beyond the large-$N_c$ limit is
accomplished by the
Jalilian-Marian--Iancu--McLerran--Weigert--Leonidov--Kovner (JIMWLK)
\cite{Jalilian-Marian:1997gr,Iancu:2000hn} evolution equation, which
is a functional differential equation. Both the BK and JIMWLK
evolution equations will be discussed in more details later.

The physical impact of the quadratic term on the right of
Eq.~(\ref{BK}) is clear: it slows down the small-$x$ evolution,
leading to {\sl parton saturation} and to total cross sections
adhering to the black disk limit. The effect of
gluon mergers becomes important when the quadratic term in
Eq.~(\ref{BK}) becomes comparable to the linear term on the
right-hand-side. This gives rise to the saturation scale $Q_s$, which
now grows with energy (on top of its increase with $A$). \\



\noindent{\bf The Balitsky-Kovchegov equation:}  Let us now include the energy dependence in the dipole amplitude $N$
from Eq.~(\ref{glaN2}). Similar to the BFKL evolution equation
\cite{Kuraev:1977fs,Balitsky:1978ic}, we are interested in quantum evolution
in the leading longitudinal logarithmic approximation re-summing the
powers of $\alpha_s \, \ln \frac{1}{x_{Bj}} \, \sim \, \alpha_s \, Y$,
with $Y$ the rapidity variable. Again we will be working in the rest
frame of the nucleus, but this time we choose to work in the light
cone gauge of the projectile $A^+ = 0$ if the dipole is moving in the
light cone $+$ direction.

\begin{figure}[ht]
  \centering
  \includegraphics[width=6cm,bb=0 276 635 489,clip=true]{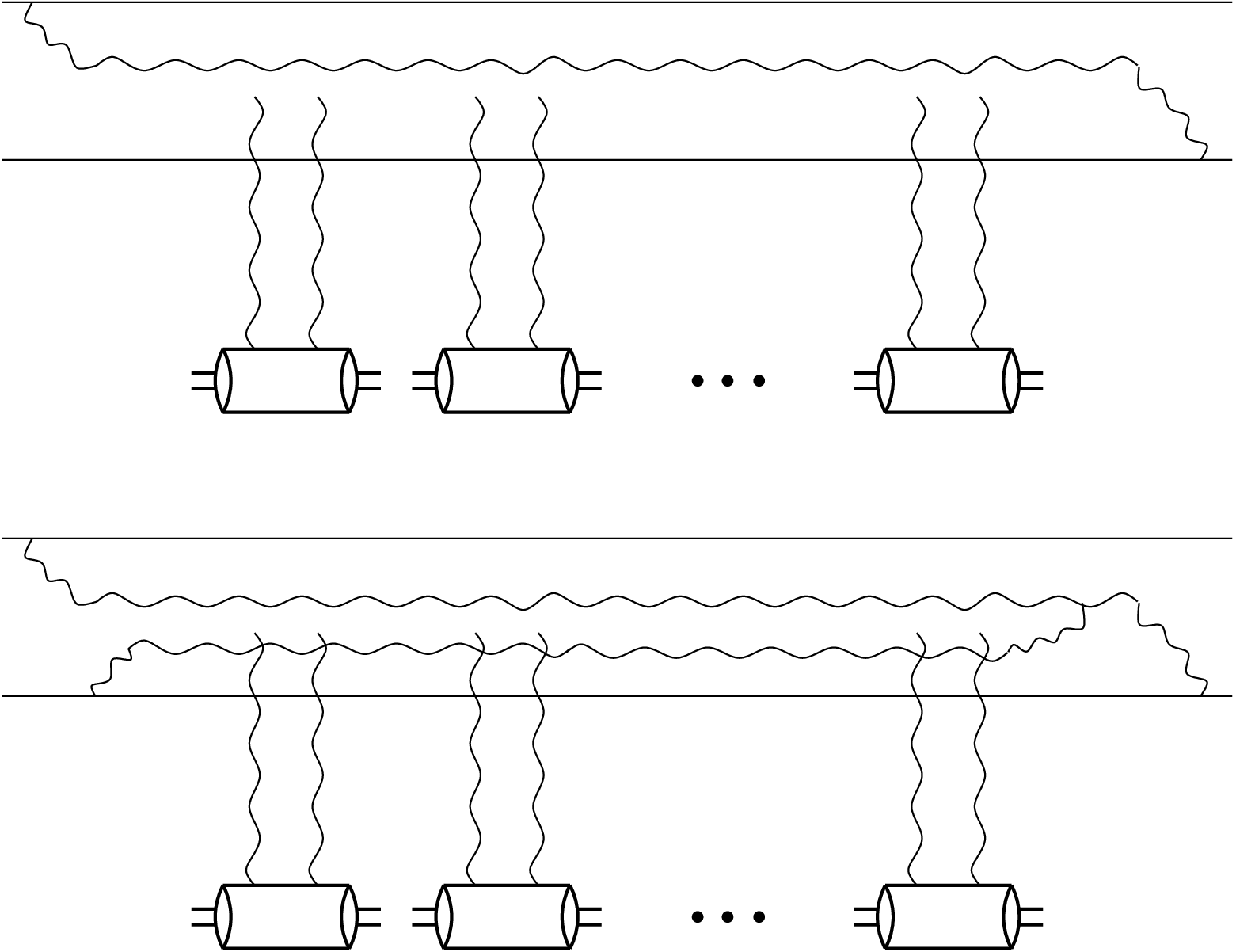}
  \hspace*{1cm}
  \includegraphics[width=6cm,bb=0   0 635 212,clip=true]{eA-final/Figs/fock.eps}
  \caption{\small Quantum corrections to dipole-nucleus scattering. }
  \label{fock}
\end{figure}

Leading logs in $x$ corrections appear in the
diagrams through emissions of long-lived $s$-channel gluons, as shown
in Fig.~\ref{fock}. These $s$-channel gluons interact with the target
nucleus through multiple re-scatterings.  In the large-$N_c$ limit of
QCD such diagrams can be re-summed by the BK evolution equation
\cite{Balitsky:1996ub,Balitsky:1997mk,Kovchegov:1999yj,Kovchegov:1999ua}:
\begin{align}
  \label{eqN}
  \frac{\partial N ({\underline x}_{0}, {\underline x}_1, Y)}{\partial Y} \, = \,
  \frac{\alpha_s \, C_F}{\pi^2} \, \int d^2 x_2 \, \frac{x_{01}^2}{x_{20}^2
    \, x_{21}^2} \, \bigg[ N ({\underline x}_{0}, {\underline x}_2, Y) + N ({\underline
    x}_{2}, {\underline x}_1 , Y) - N ({\underline x}_{0}, {\underline x}_1, Y) \notag \\
  - N ({\underline x}_{0}, {\underline x}_2 , Y) \, N ({\underline x}_{2}, {\underline x}_1, Y)
  \bigg],
\end{align}
where we have redefined the arguments of $N$ to depend on the
transverse coordinates of the quark and antiquark (instead of dipole
size and the impact parameter as was done in Eq.~(\ref{glaN2})). Here
$x_{ij} = |{\underline x}_{ij}|$ and ${\underline x}_{ij} =
{\underline x}_i - {\underline x}_j$.

\begin{figure}[ht]
\centerline{\includegraphics[width=12cm]{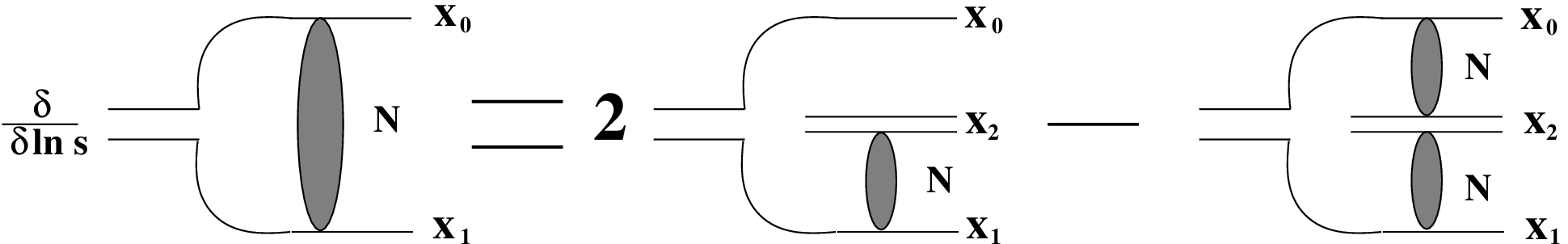}}
\caption{\small  Diagrammatic representation of the nonlinear evolution equation (\ref{eqN}).}
\label{dipeqn}
\end{figure}

In the large-$N_c$ limit, gluon cascades reduce to a cascade of color
dipoles. Summation of the dipole cascade is illustrated in
Fig.~\ref{dipeqn} where the dipole cascade and its interaction with
the target are denoted by a shaded oval. In one step of the evolution
in energy (or rapidity) a soft gluon is emitted in the dipole. If the
gluon is real, than the original dipole would be split into two
dipoles, as shown in Fig.~\ref{dipeqn}.  Either one of these dipoles
can interact with the nucleus with the other one not interacting,
which is shown by the first term on the right hand side of
Fig.~\ref{dipeqn} with the factor of $2$ accounting for the fact that
there are two dipoles in the wave function now.  Alternatively, both
dipoles may interact simultaneously, which is shown by the second term
on the right hand side of Fig.~\ref{dipeqn}. This term comes in with
the minus sign. The emitted gluon in one step of evolution may be a
virtual correction, which is not shown in Fig.~\ref{dipeqn}: in that
case, the original dipole would not split into two, it would remain the
same and would interact with the target. In the end, the evolved system
of dipoles interacts with the nucleus. In the large-$N_c$ limit, each
dipole does not interact with other dipoles during the evolution which
generates all the dipoles. For a large nucleus, the dipole-nucleus
interaction was given above in Eq.~(\ref{glaN2}). That result
re-sums powers of $\alpha_s^2 \, A^{1/3}$: hence the BK equation re-sums
powers of $\alpha_s \, Y$ and powers of $\alpha_s^2 \, A^{1/3}$.


\subsubsection{Map of high-energy QCD}

\begin{figure}[bt]
  \centering
  \includegraphics[height=6cm]{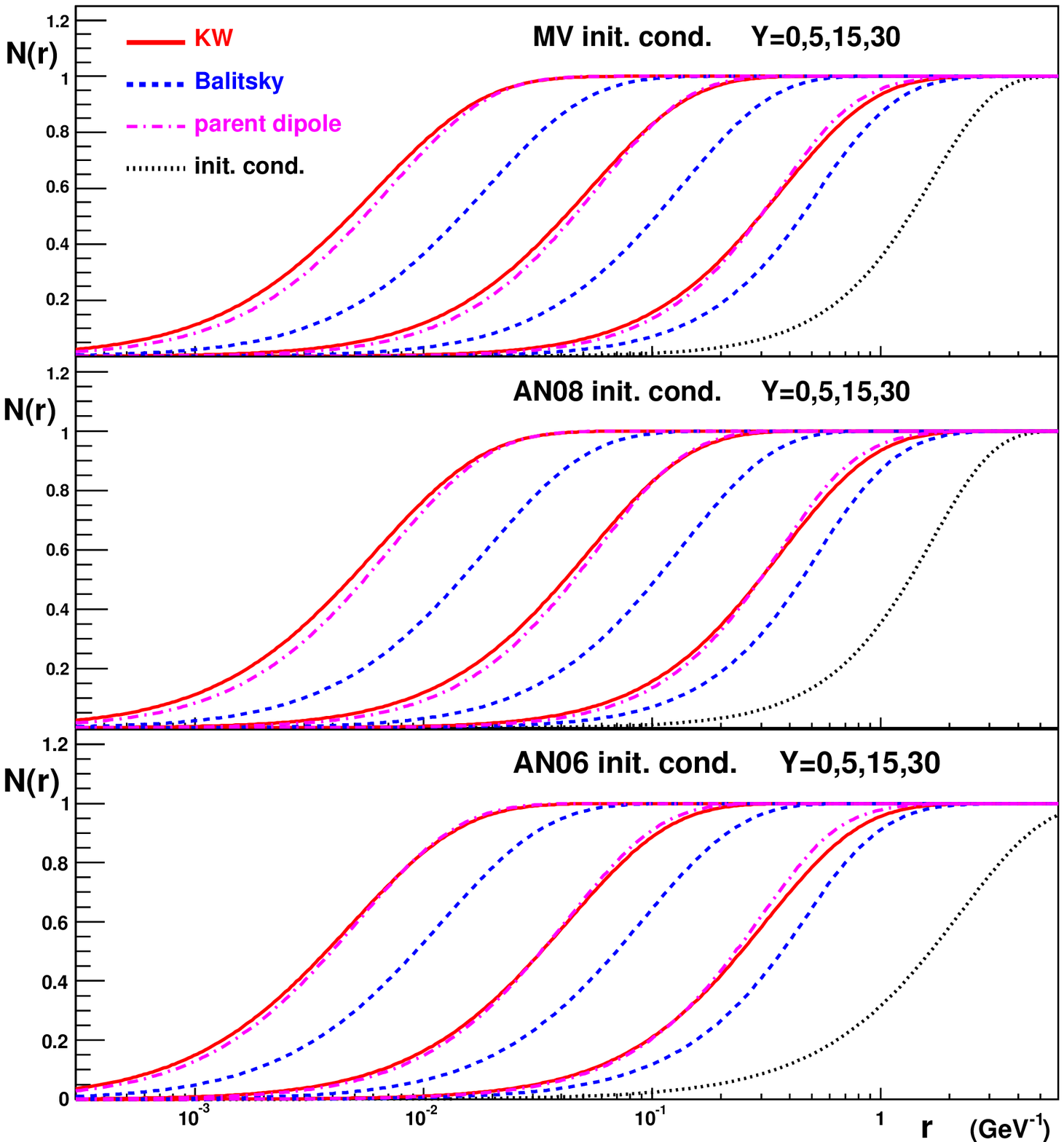}
  \hspace*{0.5cm}
  \includegraphics[height=6cm]{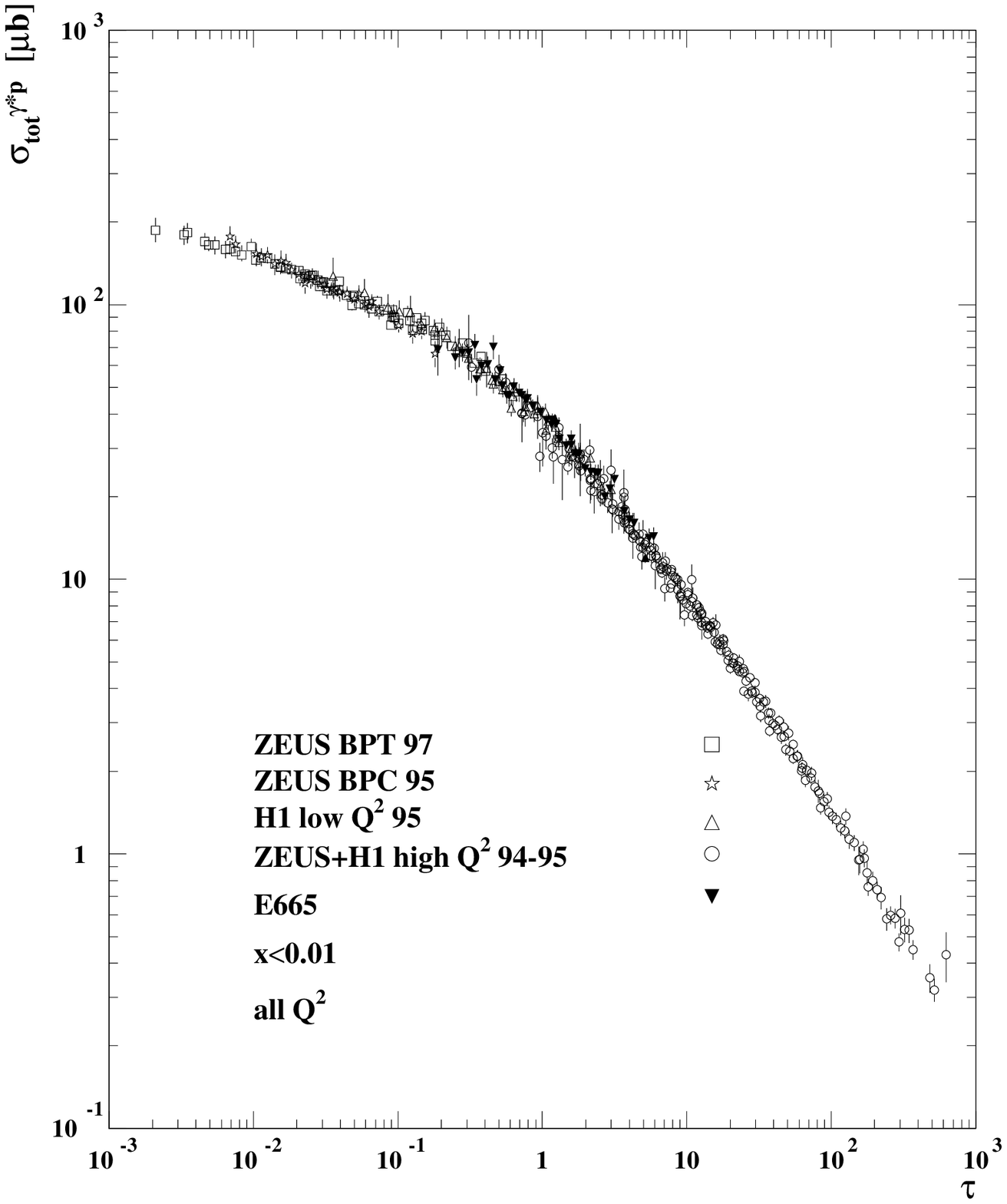}
  \caption{\small 
    {\it Left:} Solutions of the BK equation at rapidities Y=0, 5, 15 and 30
    (curves are labeled from right to left) for the three different
    running coupling schemes considered in \cite{Albacete:2007yr}
    {\it Right:} HERA data on the total DIS $\gamma^* p$ cross section
    plotted in \cite{Stasto:2000er} as a function of the scaling variable
    $\tau = Q^2/Q_s^2 (x_{Bj})$.  
  } 
\label{sols}
\label{gsdis}
\end{figure}

Solutions of the BK and JIMWLK evolution equations have been calculated numerically
\cite{Braun:2000wr,Golec-Biernat:2001if,Rummukainen:2003ns}, with
asymptotic limits studied analytically. The numerical solution (for the BK equation with running coupling,
which will be described later) is presented in Fig.~\ref{sols}~\cite{Albacete:2007yr}. These plots
are the same dipole amplitude $N$ plotted as a function of the dipole
size labeled $r$ as was done in Fig.~\ref{glamue}. In Fig.~\ref{sols}, different
curves correspond to different energies/rapidities $Y$. 
One can clearly see that the curves tend to drift to the left with
increasing energies, corresponding to increasing saturation scale with
the energy/rapidity. Therefore we see that the saturation scale
increases with rapidity, making the corresponding physics more perturbative.

\begin{figure}[t]
  \centering
  \includegraphics[width=10cm]{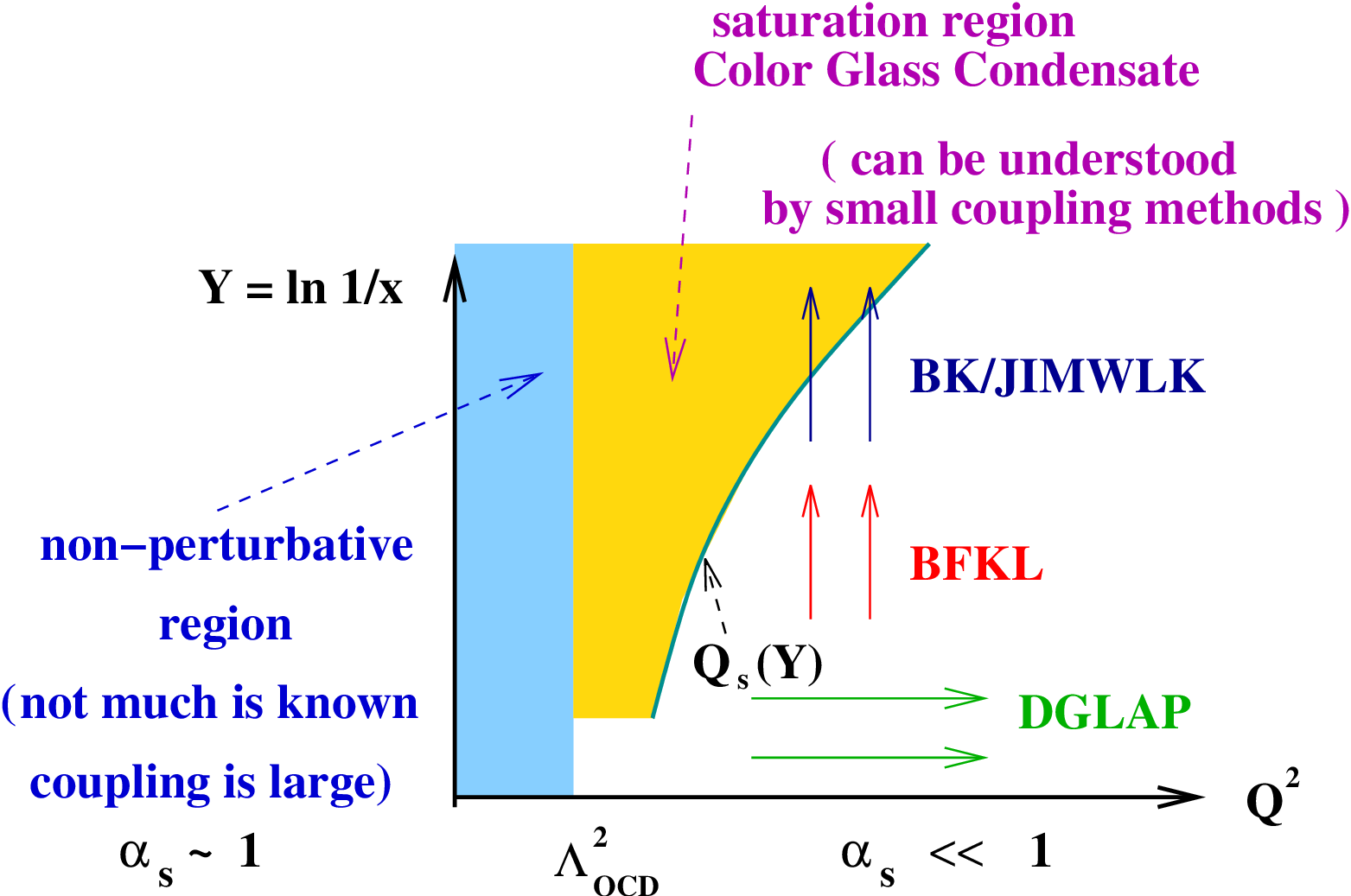}
  \caption{\small 
    Map of high energy QCD in the $(Q^2, Y=\ln 1/x)$ plane.
  }
\label{satbk}
\end{figure}

We summarize our knowledge of high energy QCD in Fig.~\ref{satbk}, in
which different regimes are plotted in the $(Q^2, Y=\ln 1/x)$ plane,
by analogy with DIS. For hadronic and nuclear collisions one can think
of typical transverse momentum $p_T^2$ of the produced particles
instead of $Q^2$. Also rapidity $Y$ and Bjorken-$x$ variable are
interchangeable.  On the left of Fig.~\ref{satbk} we see the region
with $Q^2 \le \Lambda_{QCD}^2$ in which the coupling is large,
$\alpha_s \sim 1$, and small-coupling approaches do not work.  In the
perturbative region, $Q^2 \gg \Lambda_{QCD}^2$, we see the standard
DGLAP evolution and the linear BFKL evolution. The BFKL equation
evolves gluon distributions toward small-$x$, where parton densities
becomes large and parton saturation sets in. The transition to saturation
is described by the non-linear BK and JIMWLK evolution equations. Most
importantly, this transition happens at $Q_s^2 \gg \Lambda_{QCD}^2$
where the small-coupling approach is valid.

One of the most important predictions of nonlinear small-$x$ evolution
is that, at high enough rapidity, the scattering amplitude $N$ (and,
consequently, DIS structure functions) would be a function of a single
variable $x_\perp \, Q_s (Y)$, such that $N(x_\perp, Y ) = N (x_\perp
\, Q_s (Y))$. This prediction is spectacularly confirmed by HERA data.
Geometric scaling has been demonstrated in an analysis of the HERA DIS
data~\cite{Stasto:2000er},
presenting one of the strongest arguments for the observation of
saturation phenomena at HERA. These results are shown here in
Fig.~\ref{gsdis} from \cite{Stasto:2000er}, where the authors combined HERA data on the total DIS
$\gamma^* p$ cross section $\sigma^{\gamma^* p}_{tot}$ for $x_{Bj} <
0.01$ as a function of the scaling variable $\tau = Q^2/Q_s^2
(x_{Bj})$. One can see that, amazingly enough, all the data falls on
the same curve, indicating that $\sigma^{\gamma^* p}_{tot}$ is a
function of a single variable $Q^2/Q_s^2 (x_{Bj})$!  This gives us the
best to date experimental proof of geometric scaling. (For a similar
analysis of DIS data on nuclear targets see \cite{Freund:2002ux}.)
The fact that geometric scaling is a property of the solution of the
BK equation was later demonstrated in
\cite{Levin:1999mw,Iancu:2002tr}.




\subsubsection{Universality aspects of the Color Glass Condensate}
\label{sec:universality}

\hspace{\parindent}\parbox{0.92\textwidth}{\slshape
 Fran\c{c}ois Gelis}
\index{Gelis, Fran\c{c}ois}

\vspace{\baselineskip}


The Color Glass Condensate (CGC) is an {\em effective field theory (EFT)} based on the
separation of the degrees of freedom into fast frozen color sources
and slow dynamical color fields~\cite{McLerran:1993ni,McLerran:1993ka,McLerran:1994vd}. A {\em
  renormalization group equation} --the JIMWLK equation
\cite{JalilianMarian:1997jx,JalilianMarian:1997gr,JalilianMarian:1998cb,JalilianMarian:1997dw,Kovner:2000pt,Weigert:2000gi,Iancu:2000hn,Iancu:2001ad,Ferreiro:2001qy}--
ensures the independence of physical quantities with respect to the
cutoff that separates the two kinds of degrees of freedom.

The fast gluons with longitudinal momentum $k^+>\Lambda^+$ are frozen
by Lorentz time dilation in configurations specified by a color
current $J^\mu_a \equiv \delta^{\mu +}\rho^a$, where
$\rho^a(x^-,x_\perp)$ is the corresponding color charge density. On
the other hand, slow gluons with $k^+<\Lambda^+$ are described by the
usual gauge fields $A^\mu$ of QCD. Because of the hierarchy in $k^+$
between these two types of degrees of freedom, they are coupled
eikonaly by a term $J_\mu A^\mu$.  The fast gluons thus act as
sources for the fields that represent the slow gluons. Although it is
frozen for the duration of a given collision, the color source density
$\rho^a$ varies randomly event by event. The CGC provides a gauge
invariant distribution $W_{\Lambda^+}[\rho]$, which gives the
probability of a configuration $\rho$. This encodes all the
correlations of the color charge density at the cutoff scale
$\Lambda^+$, separating the fast and slow degrees of freedom. Given
this statistical distribution, the expectation value of an operator
at the scale $\Lambda^+$ is given by
\begin{equation}
\left<{\cal O}\right>_{\Lambda^+}\equiv
\int\big[D\rho\big]\;W_{\Lambda^+}\big[\rho\big]\;{\cal O}\big[\rho\big]\; ,
\end{equation}
where ${\cal O}[\rho]$ is the expectation value of the operator for a
particular configuration $\rho$ of the color sources.

The power counting of the CGC EFT is such that in the saturated regime,
the sources $\rho$ are of order $g^{-1}$. Attaching an additional source to a given Feynman graph does not alter its order in $g$; the
vertex where this new source attaches to the graph is compensated by
the $g^{-1}$ of the source. Thus, computing an observable at a certain
order in $g^2$ requires the re-summation of all the contributions
obtained by adding extra sources to the relevant graphs.  The leading
order in $g^2$ is given by a sum of tree diagrams, which can be
expressed in terms of classical solutions of the Yang-Mills equations.
Moreover, for inclusive observables~\cite{Gelis:2006yv,Gelis:2006cr}, these
classical fields obey a simple boundary condition: they vanish when
$t\to -\infty$.

Next-to-leading order (NLO) computations in the CGC EFT involve a sum
of one-loop diagrams embedded in the above classical field. To prevent
double counting, momenta in loops are required to be below the cutoff
$\Lambda^+$. This leads to a logarithmic dependence in $\Lambda^+$ of
these loop corrections. These logarithms are large if $\Lambda^+$
is well above the typical longitudinal momentum scale of the
observable considered, and must be re-summed.

For inclusive observables, the leading logarithms are universal and
can be absorbed into a redefinition of the distribution
$W_{\Lambda^+}[\rho]$ of the hard sources. The evolution of
$W_{\Lambda^+}[\rho]$ with $\Lambda^+$ is governed by the functional
JIMWLK equation
\begin{eqnarray}
 \frac{\del\, W_{\Lambda^+}[\rho] }{{\del \ln(\Lambda^+)}}=
-{\cal H}\left[\rho,\frac{\delta}{ {\delta \rho}}\right]\,W_{\Lambda^+}[\rho]\;,
\end{eqnarray}
where ${\cal H}$ is known as the JIMWLK Hamiltonian. This operator
contains up to two derivatives $\del/\del\rho$, and arbitrary powers
in $\rho$. Its explicit expression can be found in
refs.~\cite{JalilianMarian:1997jx,JalilianMarian:1997gr,JalilianMarian:1998cb,JalilianMarian:1997dw,Kovner:2000pt,Weigert:2000gi,Iancu:2000hn,Iancu:2001ad,Ferreiro:2001qy,Iancu:2002xk,Iancu:2003xm}.
The derivation of the JIMWLK equation will be sketched below.

Numerical studies of JIMWLK evolution were performed in
\cite{Rummukainen:2003ns,Kovchegov:2008mk}.  An analytic, albeit formal, solution to the
JIMWLK equation was constructed in \cite{Blaizot:2002xy} in the form of a
path integral.  Alternatively, the evolution can can be expressed as
an infinite hierarchy of coupled non-linear equations for $n$-point
Wilson line correlators--often called the Balitsky
hierarchy~\cite{Balitsky:1995ub}.  In this framework, the BK equation is a mean
field approximation of the JIMWLK evolution, valid in the limit of a
large number of colors $N_c\to\infty$. Numerical studies of the JIMWLK
equation~\cite{Rummukainen:2003ns,Kovchegov:2008mk} have found only small differences
with the BK equation.

Let us finally comment on the initial condition for the JIMWLK
equation which is also important in understanding its derivation. The
evolution should start at some cutoff value in the longitudinal
momentum scale $\Lambda^+_0$ at which the saturation scale is already
a (semi)hard scale, say $Q_{s0}\gtrsim 1$~GeV, for perturbation theory
to be applicable.  The gluon distribution at the starting scale is in
general non--perturbative and requires a model.  A physically
motivated model for the gluon distribution in a large nucleus is the
McLerran-Venugopalan model~\cite{McLerran:1993ni,McLerran:1993ka,McLerran:1994vd}. In a large
nucleus, there is a window in rapidity where evolution effects are not
large but $x$ is still sufficiently small for a probe not to resolve
the longitudinal extent of the nucleus. In this case, the probe
``sees'' a large number of color charges, proportional to $A^{1/3}$.
These charges add up to form a higher dimensional representation of
the gauge group, and can therefore be treated as classical color
distributions~\cite{McLerran:1993ni,McLerran:1993ka,McLerran:1994vd,Jeon:2004rk}.  Further, the
color charge distribution $W_{\Lambda_0^+}[\rho]$ is a Gaussian
distribution\footnote{There is a additional term, corresponding to the
  cubic Casimir; which is parametrically suppressed for large
  nuclei~\cite{Jeon:2005cf}. This term generates Odderon excitations in the
  JIMWLK/BK evolution~\cite{Hatta:2005as,Kovchegov:2003dm}.} in $\rho$. The
variance of this distribution --the color charge squared per unit
area-- is proportional to $A^{1/3}$ and provides a semi-hard scale
that makes weak coupling computations feasible. In addition to its
role in motivating the EFT and serving as the initial condition in
JIMWLK evolution, the MV model allows for direct phenomenological
studies in p+A and A+A collisions in regimes where the values of $x$
are not so small as to require evolution. \\

\begin{figure*}[tbhp]
  \centering
  \parbox{0.4\linewidth}{
    \includegraphics[width=0.45\linewidth]{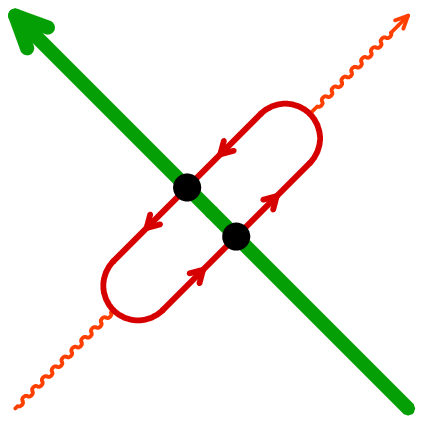}
    \hfill
    \includegraphics[width=0.45\linewidth]{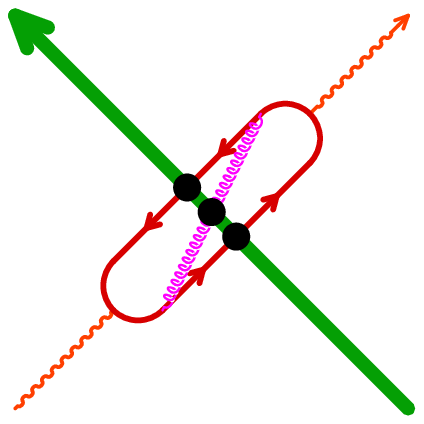}
  }
  \hspace*{1.5cm}
  \parbox{7cm}{
    \includegraphics[width=\linewidth]{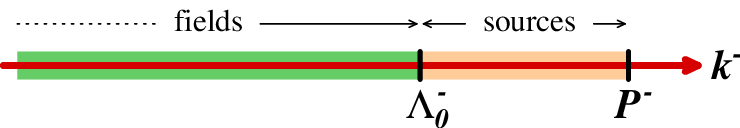}
    \vskip2mm
    \includegraphics[width=\linewidth]{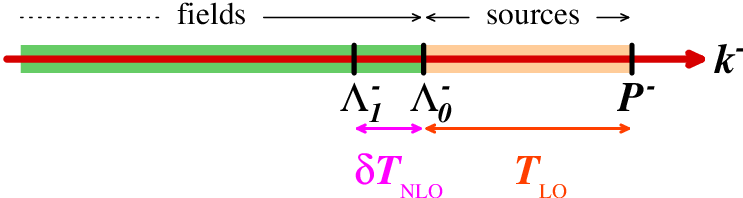}
  }
  \caption{\small 
    {\it Left:} LO and NLO contributions to DIS off the CGC. 
    {\it Top right:} sources and fields in the CGC
    effective theory. 
    {\it Bottom right:} NLO correction from a layer of field
    modes just below the cutoff.
  }
  \label{fig:DIS-LO}
  \label{fig:CGC-1}
\end{figure*}

\noindent{\bf The CGC in DIS at small $x$:}  We denote
$\sigma_{\rm dipole}(x,\r_\perp)$ the QCD ``dipole'' cross-section for the
quark-antiquark pair to scatter off the target. This process is shown
in fig.~\ref{fig:DIS-LO} left, where we have assumed that the target
moves in the $-z$ direction. In the leading order (LO) CGC description
of DIS, the target is described, as illustrated in
fig.~\ref{fig:CGC-1} right, as static sources with $k^->\Lambda_0^-$. The
field modes do not contribute at this order.

Employing the optical theorem, $\sigma_{\rm dipole}(x,\r_\perp)$ can
be expressed in terms of the forward scattering amplitude ${\bs
  T(\x_\perp,\y_\perp)}$ of the $q\bar{q}$ pair at LO as
\begin{equation}
\sigma_{\rm dipole}^{_{\rm LO}}(x,\r_\perp) 
= 
2 \int \; d^2{\bf b}\;\int \; [D\rho] W_{\Lambda_0^-}[\rho]\;
 {\bs T}_{_{\rm LO}}(\b+\frac{\r_\perp}{2}, \b - \frac{\r_\perp}{2})\; ,
\label{eq:opt-thm-LO}
\end{equation}
where, for a fixed configuration of the target color sources
\cite{McLerran:1998nk,Venugopalan:1999wu}
\begin{equation}
{ {\bs T}_{_{\rm LO}}(\x_\perp,\y_\perp)}
=
1-\frac{1}{N_c}\,{\rm tr}\,( U(\x_\perp)U^\dagger(\y_\perp))\; ,
\label{eq:Wilson-amp}
\end{equation}
with $U(\x_\perp)$ a Wilson line representing the interaction between
a quark and the color fields of the target, defined to be
\begin{equation}
U(\x_\perp)
=
 {\rm T}\,\exp i{g}\int^{1/x P^-} 
 dz^+\,{{\cal A}^-(z^+,\x_\perp)}\; .
\end{equation}
In this formula, ${\cal A}^-$ is the minus component of the gauge
field generated (in Lorentz gauge) by the sources of the target; it is
obtained by solving classical Yang-Mills equations with these sources.
The upper bound $xP^-$ (where $P^-$ is the target longitudinal
momentum) indicates that source modes with $k^-<xP^-$ do not
contribute to this scattering amplitude. Thus if the cutoff
$\Lambda_0^-$ of the CGC EFT is lower than $xP^-$, ${\bs T}_{_{\rm
    LO}}$ is independent of $\Lambda_0^-$.

However, when $\Lambda_0^-$ is larger than $xP^-$, the dipole
cross-section is in fact independent of $x$ (since the CGC EFT does
not have source modes near the upper bound $xP^-$) and depends on the
unphysical parameter $\Lambda_0^-$. As we shall see now, this is
related to the fact that eq.~(\ref{eq:opt-thm-LO}) is incomplete and
receives large corrections from higher order diagrams. Consider now
the NLO contributions (one of them is shown in the right panel in
figure \ref{fig:DIS-LO} left with gauge field modes in the slice
$\Lambda_1^-\le k^- \le \Lambda_0^-$ (see fig.~\ref{fig:CGC-1} right).
An explicit computation of the contribution of field modes in this
slice gives
\begin{equation}
    {\delta {\bs T}_{_{\rm NLO}}(\x_\perp,\y_\perp)}
    =
    \ln\left(\frac{\Lambda_0^-}{\Lambda_1^-}\right)\;
    {{\cal H}}\;
    {{\bs T}_{_{\rm LO}}(\x_\perp,\y_\perp)}\; ,
    \label{eq:DIS-NLO}
\end{equation}
where ${\cal H}$ is the JIMWLK Hamiltonian. All dependence on the
cutoff scales is in the logarithmic prefactor alone. This Hamiltonian
has two derivatives with respect to the classical field ${\cal
  A}\sim{\cal O}(1/g)$; ${\cal H}\,{\bs T}_{_{\rm LO}}$ is of order
$\alpha_s {\bs T}_{_{\rm LO}}$ and therefore clearly an NLO
contribution. However, if the new scale $\Lambda_1^-$ is such that
$\alpha_s\ln(\Lambda_0^-/\Lambda_1^-)\sim 1$, this NLO term becomes
comparable in magnitude to the LO contribution. Averaging the sum of
the LO and NLO contributions over the distribution of sources at the
scale $\Lambda_0^-$, one obtains
\begin{eqnarray}
    \int[D\rho]\;W_{\Lambda_0^-}[\rho]\;
    \left({\bs T}_{_{\rm LO}}+\delta{\bs T}_{_{\rm NLO}}\right)=
    \int[D\rho]\;W_{\Lambda_1^-}[\rho]\;{\bs T}_{_{\rm LO}}\; ,
\label{eq:int-by-parts}
\end{eqnarray}
where $W_{\Lambda_1^-}\equiv (1+\ln(\Lambda_0^-/\Lambda_1^-)\,{\cal
  H})\,W_{\Lambda_0^-}$. We have shown here that the NLO correction
from quantum modes in the slice $\Lambda_1^-\le k^- \le \Lambda_0^-$
can be absorbed in the LO term, provided we now use a CGC effective
theory at $\Lambda_1^-$ with the modified distribution of sources
shown in eq.~(\ref{eq:int-by-parts}).  In differential form, the
evolution equation of the source distribution
is the JIMWLK equation stated previously.

Repeating this elementary step, one progressively re-sums quantum
fluctuations down to the scale $k^-\sim xP^-$. Thanks to
eq.~(\ref{eq:int-by-parts}), the result of this re-summation for the
dipole cross-section is formally identical to
eq.~(\ref{eq:opt-thm-LO}), except that the source distribution is
$W_{xP^-}$ instead of $W_{\Lambda_0^-}$. Note that if one further
lowers the cutoff below $xP^-$, the dipole cross-section remains
unchanged. \\


\noindent{\bf The CGC in A+A collisions:}  Collisions between two nuclei (``dense-dense'' scattering) are
complicated to handle on the surface. However, in the CGC framework,
because the wave functions of the two nuclei are saturated, the
collision can be treated as the collision of classical fields coupled
to fast partons of each nucleus respectively described by the external
current $J^\mu=\delta^{\mu+}\rho_1+\delta^{\mu-}\rho_2$.  The source
densities of fast partons $\rho_{1,2}$ are both parametrically of
order $1/g$, which implies that graphs involving multiple sources from
both projectiles must be re-summed.

At leading order, inclusive observables\footnote{Exclusive observables
  may also be expressed in terms of solutions of the same Yang-Mills
  equations, but with more complicated boundary conditions than for
  inclusive observables.} depends on the retarded classical color
field ${\cal A}^\mu$, which solves the Yang-Mills equations $[{\cal
  D}_\mu,{\cal F}^{\mu\nu}]=J^\nu$ with the boundary condition
$\lim_{x^0\to -\infty}{\cal A}^\mu =0$. Among the observables to which
this result applies is the expectation value of the energy-momentum
tensor at early times after the collision. At leading order,
  \begin{equation}
T^{\mu\nu}_{_{\rm LO}}
=
\frac{1}{4}g^{\mu\nu}\,{{\cal F}^{\lambda\sigma}{\cal F}_{\lambda\sigma}}
-{{\cal F}^{\mu\lambda}{\cal F}^\nu{}_\lambda}\; ,
\end{equation}
where ${\cal F}^{\mu\nu}$ is the field strength of the classical field
${\cal A}^\mu$.

Although A+A collisions are more complicated than e+A or p+A
collisions, one can still factorize the leading higher order
corrections into the evolved distributions $W_{\Lambda^-}[\rho_1]$ and
$W_{\Lambda^+}[\rho_2]$. At the heart of this factorization is a
generalization of eq.~(\ref{eq:DIS-NLO}) to the case where the two
projectiles are described in the CGC framework
\cite{Gelis:2008rw,Gelis:2008ad,Gelis:2008sz}. When one integrates out the field
modes in the slices $\Lambda_1^\pm\le k^\pm\le \Lambda_0^\pm$, the
correction to the energy momentum tensor is
\begin{equation}
{\delta T^{\mu\nu}_{_{\rm NLO}}}
    =
    \Big[
    \ln\left(\frac{\Lambda_0^-}{\Lambda_1^-}\right)\,{{\cal H}_1}
    +
    \ln\left(\frac{\Lambda_0^+}{\Lambda_1^+}\right)\,{{\cal H}_2}
    \Big]\;{T^{\mu\nu}_{_{\rm LO}}}\; ,
\label{eq:AA-NLO}
\end{equation}
where ${\cal H}_{1,2}$ are the JIMWLK Hamiltonians of the two nuclei
respectively. What is crucial here is the absence of mixing between
the coefficients ${\cal H}_{1,2}$ of the logarithms of the two
projectiles; they depend only on $\rho_{1,2}$ respectively and not on
the sources of the other projectile.  Although the proof of this
expression is somewhat involved, the absence of mixing is deeply
rooted in causality.  The central point is that because the duration
of the collision (which scales as the inverse of the energy) is so
brief, soft radiation must occur before the two nuclei are in causal
contact. Thus logarithms associated with this radiation must have
coefficients that do not mix the sources of the two projectiles.

Following the same procedure for eq.~(\ref{eq:AA-NLO}), as for the e+A
and p+A cases, one obtains for the energy-momentum tensor in an A+A
collision the expression
\begin{equation}
\left<T^{\mu\nu}\right>_{_{\rm LLog}}
=
\int 
\big[D{\rho_{_1}}\,D{\rho_{_2}}\big]
\;
{ W_1\,[\rho_{_1}\big]}\;
{ W_2\big[\rho_{_2}\big]}
\;
T^{\mu\nu}_{_{\rm LO}}\; .
\label{eq:Tmunu}
\end{equation}
This result can be generalized to multi-point correlations of the
energy-mo\-men\-tum tensor,
\begin{equation}
\left<T^{\mu_1\nu_1}(x_1)\cdots T^{\mu_n\nu_n}(x_n)\right>_{_{\rm LLog}}
=\int \big[D{\rho_{_1}}\,D{\rho_{_2}}\big]\;{ W_1\,[\rho_{_1}\big]}\;
{ W_2\big[\rho_{_2}\big]}\ T^{\mu_1\nu_1}_{_{\rm LO}}(x_1)\cdots T^{\mu_n\nu_n}_{_{\rm LO}}(x_n)\; .
\label{eq:Tmunu2}
\end{equation}
In this expression, all the correlations between the energy-momentum
tensor at different points are from the distributions
$W_{1,2}[\rho_{1,2}]$. Thus, the leading correlations are already
built into the wavefunctions of the projectiles prior to the
collision. 

Note that the expressions in eqs.~(\ref{eq:Tmunu}) and
(\ref{eq:Tmunu2}) are valid for proper times $\tau\sim1/Q_s$ after the
heavy ion collision. Complicated final state effects, possibly driven
by instabilities, are expected to bring this non-equilibrium gluonic
matter into a quark-gluon plasma. Although this aspect of A+A
collisions is very different from what happens in DIS reactions, the
Color Glass Condensate provides a universal description of the
hadronic and nuclear wavefunctions prior to the collision in both
cases, and a powerful framework to show that the logarithms of the
collision energy are universal for inclusive enough observables.
Thanks to this universality, measurements at small $x$ in e+A
collisions can provide valuable constraints on the distributions
$W[\rho]$ for a nucleus, that can then be used in order to compute the
state of the system formed at early times in A+A collisions.


\subsubsection{Shadowing}   
\label{sec:shadowing}
\hspace{\parindent}\parbox{0.92\textwidth}{\slshape
 Boris Z. Kopeliovich}
\index{Kopeliovich, Boris Z.}

\vspace{\baselineskip}

In terms of the dipole formalism, nuclear shadowing is related to the interaction of different Fock components
of the projectile particle with the nuclear target. The lowest Fock states (i.e. $\gamma^*\to \bar qq$) are responsible for higher twist shadowing, while higher Fock components (i.e. $\gamma^*\to \bar qqg$) give rise to leading twist gluon shadowing. \\


\noindent{\bf Quark shadowing:}  The magnitude of higher twist shadowing is controlled by the interplay between
two fundamental quantities.

(i) The lifetime of photon fluctuations, or coherence time.
\beq
l_c=\frac{2\,\nu}{Q^2+M^2}= 
\frac{P}{x_{Bj}\,m_N} =
P\,l_c^{max}\ ,
\label{kopel-20}
\eeq
where $x_{Bj}=Q^2/2m_N\nu$,
$M$ is the effective mass of the fluctuation,
$P=(1+M^2/Q^2)^{-1}$, and $l_c^{max}=1/m_Nx_{Bj}$. 
The usual approximation is to assume that $M^2 \approx Q^2$ since $Q^2$ is the only large 
dimensional scale available. In this case, $P=1/2$ and the corresponding value of $l_c$ is called 
Ioffe length of time.

Shadowing is possible only if the coherence 
time exceeds the mean nucleon spacing in nuclei,
and shadowing saturates (for a given Fock component)
if the coherence time substantially exceeds the nuclear 
radius.

(ii) Equally important for shadowing is the transverse
separation of the $\bar qq$.  This controls the dipole-nucleon cross section $\sigma_{\bar qq}^N(r)$,
and correspondingly the total nuclear cross section~\cite{zkl,Nikolaev:1991et},
\beq
\Bigl(\sigma_{tot}^{\gamma^*A}
\Bigr)^{T,L}_{l_c\gg R_A}
= 2\, \int d\alpha\int d^2r
\left|\Psi_{\bar qq}^{T,L}\left(\varepsilon r\right)\right|^2\!
\int d^2b\left[1-\exp\left(-{1\over2}\,\sigma^N_{q\bar q}\left(r\right)
T_A(b)\right)\right]
\label{kopel-40}
\eeq
where the perturbative light-cone distribution function for the $\bar qq$
has the form~\cite{ks,bks},
\beq
\Psi^{T,L}_{\bar qq}(\vec r_T,\alpha)=
\frac{\sqrt{\alpha_{em}}}{2\,\pi}\,
\bar\chi\,\widehat O^{T,L}\,\chi\,K_0(\epsilon r_T);
\label{kopel-60}
\eeq
$\chi$ and $\bar\chi$ are the spinors of the quark
and antiquark respectively;
$K_0(\epsilon r_T)$ is the modified Bessel function; $\epsilon^2 = \alpha(1-\alpha)Q^2 + m_q^2$;
and the operators $\widehat O^{T,L}$ for transversely and longitudinally polarized photons have the form,
\beq
\widehat O^{T}=m_q\,\vec\sigma\cdot\vec e +
i(1-2\alpha)\,(\vec\sigma\cdot\vec n)\,
(\vec {e}\cdot \vec\nabla_{r})
+ (\vec\sigma\times\vec e)\cdot\vec\nabla_{r},
\label{kopel-80}
\eeq
\beq
\widehat O^{L}= 2\,Q\,\alpha(1-\alpha)\,\vec\sigma\cdot\vec n\ .
\label{kopel-100}
\eeq
Here
$\vec n=\vec p/p$ is a unit vector parallel to
the photon momentum; $\vec e$ is the polarization vector
of the photon;
$m_q$ and and $\alpha$ are the mass, and fractional 
light-cone momentum carried by the quark. See also eqs.~(\ref{eq:QED-T}) and (\ref{eq:QED-L}) discussed previously. 

In order to be shadowed, a 
$\bar qq$-fluctuation of the photon has to interact with
a large cross section. 
As a result of color transparency~\cite{zkl,bm}, small size
dipoles with $r^2\sim 1/Q^2$ interact only weakly and are therefore less shadowed. The dominant
contribution to shadowing comes from  the
aligned jet configurations ($\alpha\to 0,1$)~\cite{ajm} of $\bar qq$  pairs, which have large transverse separation, $\la r^2\ra\sim 1/[Q^2\alpha(1-\alpha)]$ according to (\ref{kopel-60}).
Although the weight of such configurations is small, $1/Q^2$, this is compensated by the large interaction cross section~\cite{Kopeliovich:1995ju}.

\begin{figure}[tbh]
\centerline{\includegraphics[width=0.45\textwidth]{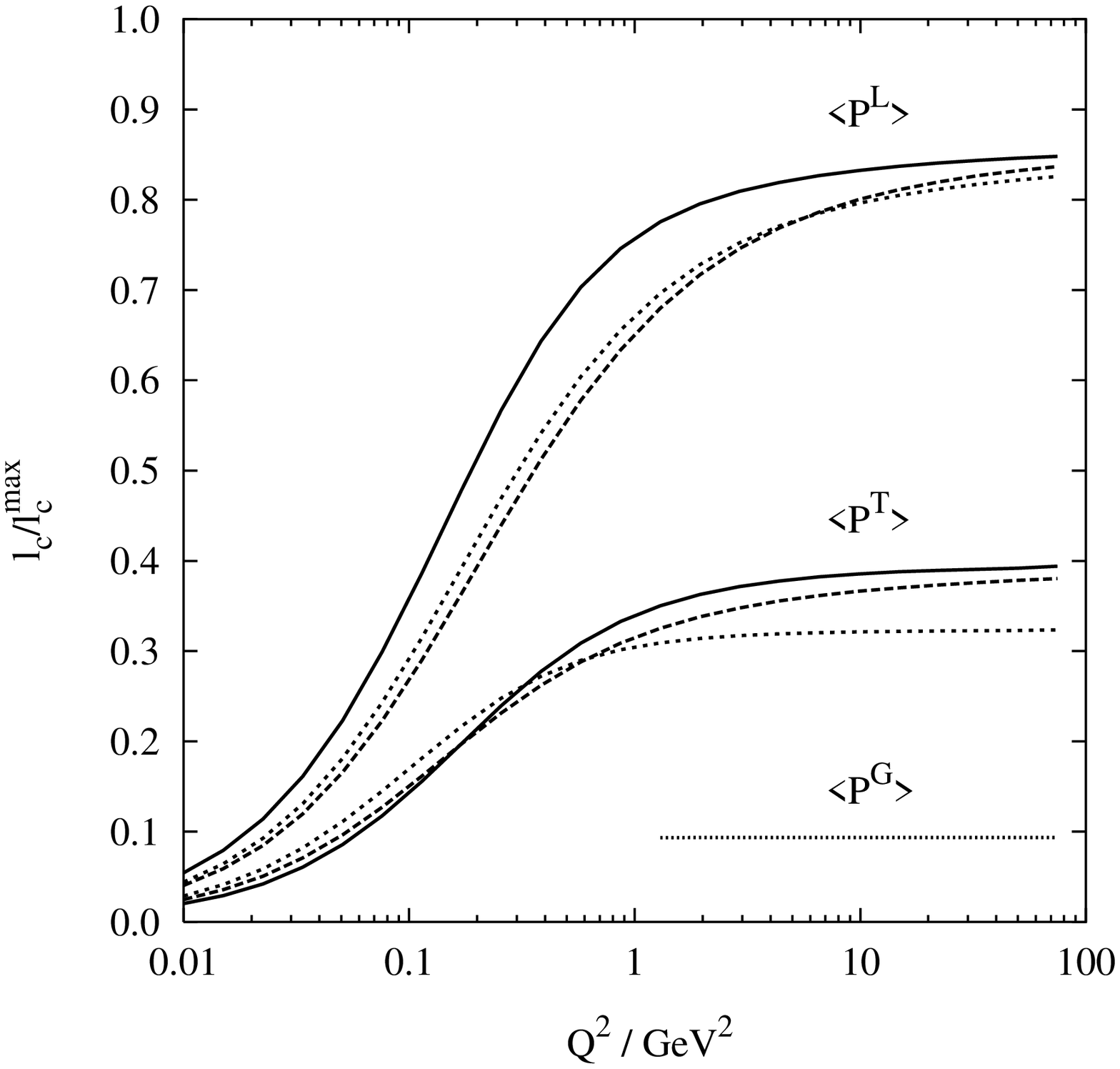}
\hfill\includegraphics[width=0.40\textwidth]{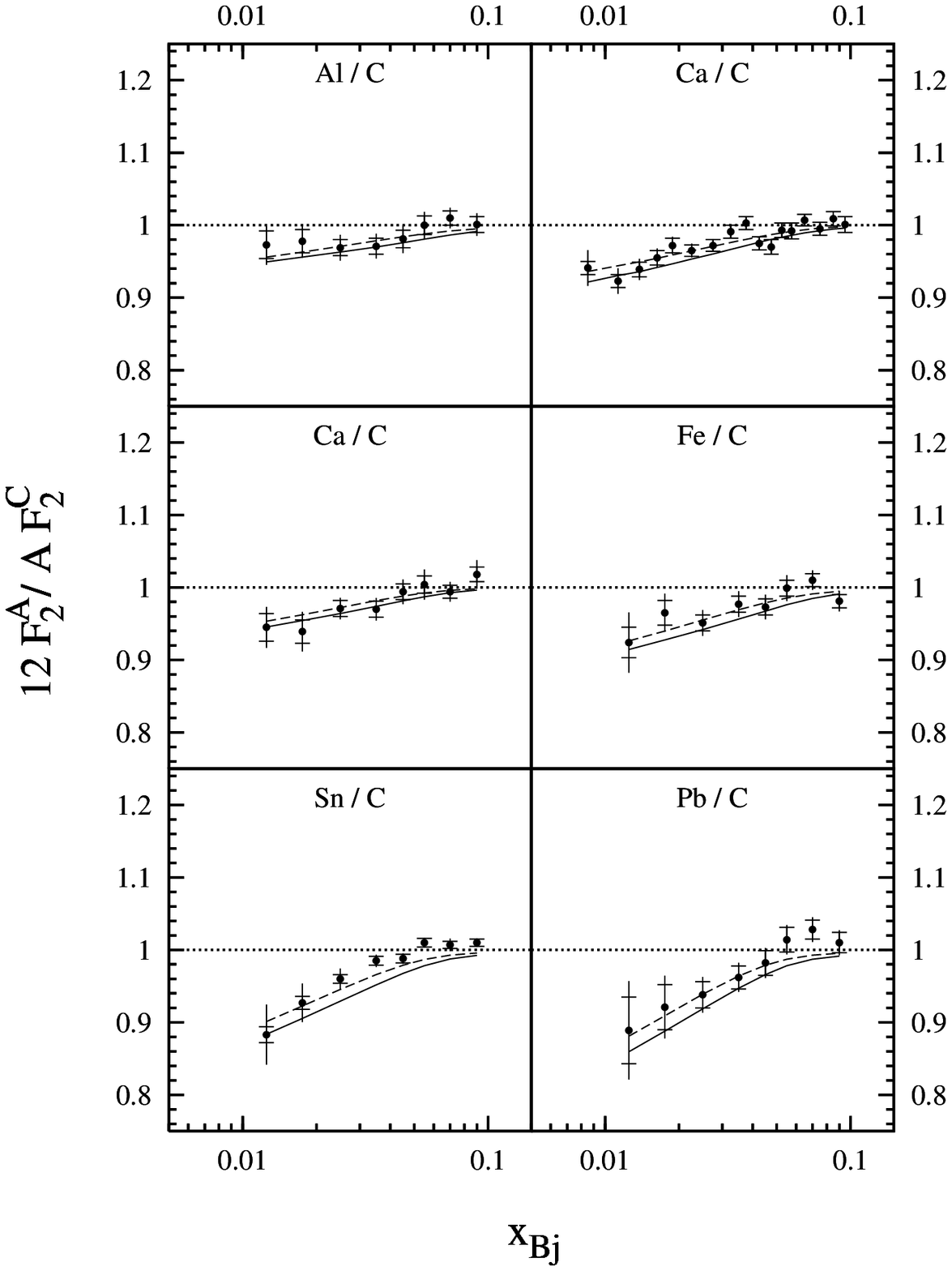}}
\caption{\small \label{fig:l-coh} {\it Left panel:}  Factor $\left\la P^{T,L}\right\ra$
and $\left\la P^{g}\right\ra$ 
defined in (\ref{kopel-20})  
for $\bar qq$ fluctuations of transverse and longitudinal
photons, and for $\bar qqg$ fluctuations, from the top to 
bottom. Calculations are done as a function of $Q^2$ at $x_{Bj}=0.01$. 
Dotted curves correspond to perturbative wave functions and an approximate dipole
cross section $\propto r_T^2$. Dashed curves rely on
the realistic parameterization for the dipole cross section~\cite{gbw}.
The solid curves show the most realistic case 
based on the nonperturbative wave functions.
{\it Right panel:} Comparison between calculations for quark shadowing and experimental data from NMC~\cite{nmc1,nmc2} for the structure functions of different nuclei relative to carbon as function of $x_{Bj}$. The $Q^2$ range covered by the data is approximately $3\gev^2 \leq Q^2 \leq 17\gev^2$ from the lowest to the highest $x_{Bj}$ bin.
Solid and dashed curves are calculated with and without the real part of the light-cone potential in (\ref{kopel-140}).}
\end{figure}

The coherence length (Eq.~(\ref{kopel-20})) averaged over interacting $|\bar qq\ra$ and $|\bar qqg\ra$ fluctuations calculated in~\cite{Kopeliovich:1998gv} is presented in Fig.~\ref{fig:l-coh}.
The mean values of the factor $P=l_c/l_c^{max}$ in (\ref{kopel-20}) are plotted for $\bar qq$ fluctuations of transverse and longitudinal photons, as well as for $\bar qqg$ fluctuations as a function of $Q^2$ at fixed $x_{Bj}$ (left panel). We see that $\bar qq$ fluctuations in a longitudinal photon live about twice as long as in a transverse one. Both are different from $P=1/2$ corresponding to the Ioffe time. The lifetime of the higher order Fock states containing gluons is about order of magnitude shorter. \\


\noindent{\bf Onset of shadowing:}  Eq.~(\ref{kopel-40}) describing quark shadowing is valid only in the limit of $l_c \gg R_A$, i.e. at very small $x_{Bj}$ where the magnitude of shadowing nearly saturates. However, all available data for DIS on nuclei are in the region of   shorter coherence length, and one needs theoretical tools to describe the onset of shadowing. 

\begin{figure}[tbh]
\centerline{\includegraphics[width=0.6\textwidth]{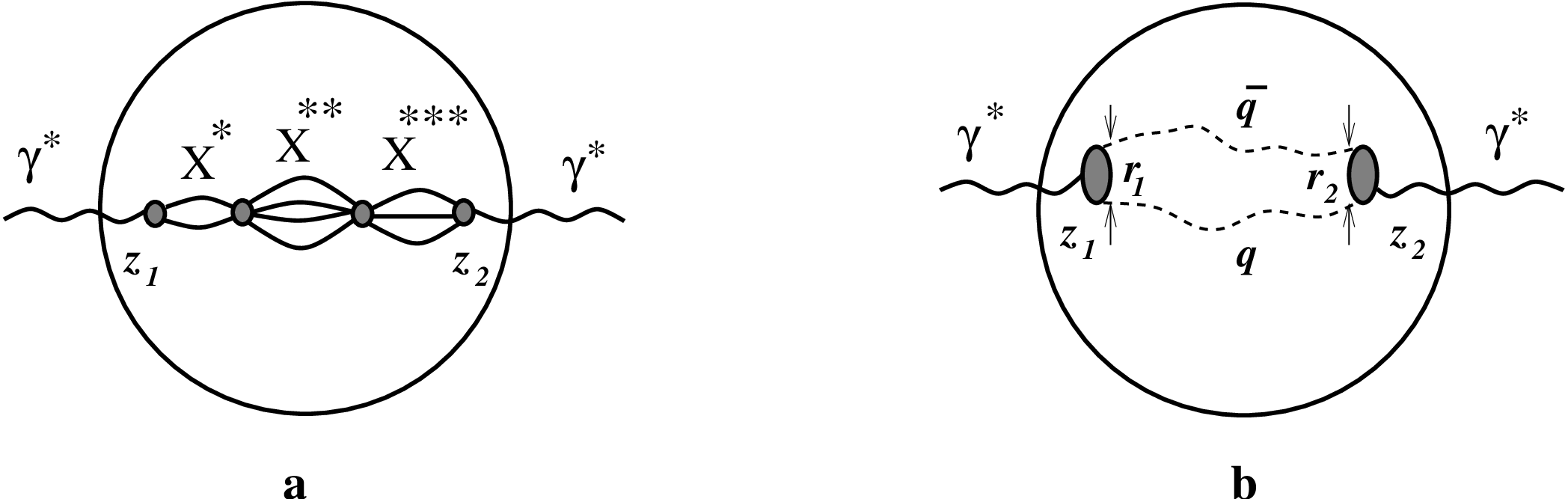}}
\caption{\small {\bf a:} A high order term in Gribov inelastic shadowing corrections to $F^A_2(x,Q^2)$;
{\bf b:} Dipole description based on the path integral technique, which sums up the Gribov corrections in all orders.}
\label{shad}
\end{figure}

The Gribov theory of inelastic shadowing~\cite{Gribov:1968jf} relates nuclear shadowing to the cross section 
of diffractive dissociation. In the case of a deuteron target, this approach provides a full and model independent 
description of shadowing. The onset of shadowing can be accurately calculated, since the phase shift $\Delta z/l_c$ between the impulse approximation term and the inelastic shadowing term is under control.
However, a description of shadowing for heavy nuclei is a challenge in this approach.
Indeed, only the lowest order of Gribov corrections can be calculated using data on diffraction. 
The higher order corrections, illustrated in Fig.~\ref{shad}a, need information unavailable from data, like the diffractive amplitudes between different excited states, $X^*$, $X^{**}$, the attenuation of these states in the nuclear medium, etc. 

An alternative description with the path integral technique was proposed in~\cite{Kopeliovich:1998gv}.
One should sum up over all possible trajectories of the quark and antiquark propagating through the nucleus, as is illustrated in Fig.~\ref{shad}b. This leads to the 2-dimensional Schr\"odinger equation for the Green function describing propagation of a dipole with initial (final) transverse separation $\vec r_1$ ($\vec r_2$) at longitudinal coordinate $z_1$ ($z_2$),
\beq
\left[i\frac{\partial}{\partial z_2}
+\frac{\Delta_\perp\left(r_2\right)-\varepsilon^2}
{2\nu\alpha\left(1-\alpha\right)}
+\frac{i}{2}\rho_A\left(b,z_2\right)
\sigma^N_{q\bar{q}}\left(r_2\right)
-\frac{a^4\left(\alpha\right)r_2^2}
{2\nu\alpha\left(1-\alpha\right)}
\right]
G\left(r_2,z_2\,|\,r_1,z_1\right)= 0
\label{kopel-120}
\eeq
The last two terms represent the imaginary and real parts of the light cone potential. The former describes the attenuation of the dipole in the nuclear medium, while the latter models the non-perturbative interactions inside the dipole.
Solving this equation, one can calculate the shadowing corrections as
\beqn
&&
\Bigl(\sigma^{\gamma^*A}_{tot}\Bigr)^{T,L} = 
A\,\Bigl(\sigma^{\gamma^*N}_{tot}\Bigr)^{T,L}
- \frac{1}{2} Re\int d^2b
\int\limits_0^1 d\alpha
\int\limits_{-\infty}^{\infty} dz_1 \int\limits_{z_1}^{\infty} dz_2
\int d^2r_1\int d^2r_2
\label{kopel-140}\\
&\times& 
\Bigl[\Psi^{T,L}_{\bar qq}\left(\varepsilon,
\lambda,r_2\right)\Bigr]^*
\rho_A\left(b,z_2\right)\sigma_{q\bar{q}}^N\left(s,r_2\right)
G\left(r_2,z_2\,|\,r_1,z_1\right)
\rho_A\left(b,z_1\right)\sigma_{q\bar{q}}^N\left(s,r_1\right)
\Psi^{T,L}_{\bar qq}\left(\varepsilon,\lambda,r_1\right)
\nonumber
\eeqn

At $l_c\ll1/\rho\sigma$, the second term vanishes.  For $l_c\gg R_A$, it saturates at the value given by Eq.~(\ref{kopel-40}).  The numerical results are compared with data from the NMC experiment~\cite{nmc1,nmc2} in the right panel of Fig.~\ref{fig:l-coh}.  The solid and dashed curves are calculated with and without the real part of the light-cone potential in (\ref{kopel-140}). It worth emphasizing that this is a parameter-free calculation, no adjustment to nuclear data has been done. The dipole cross section was fitted to DIS data on a proton.

Note that these calculations were performed for the lowest Fock component $|\bar qq\ra$ of the photon; they miss gluon shadowing related to the higher Fock states containing gluons. \\


\noindent{\bf Gluon shadowing:}\label{kopel-glue}  Gluon shadowing is related to specific channels of diffractive gluon radiation. In terms of Regge phenomenology, these processes correspond to the triple-Pomeron contribution, and can be seen in data as the large mass tail of the invariant mass distribution, $d\sigma_{diff}/dM_X^2\propto 1/M_X^2$. Such an $M_X^2$-dependence is the undebatable evidence of radiation of a vector particle, i.e. a gluon.

\begin{figure}[tbh]
\centerline{\includegraphics[width=0.52\textwidth]{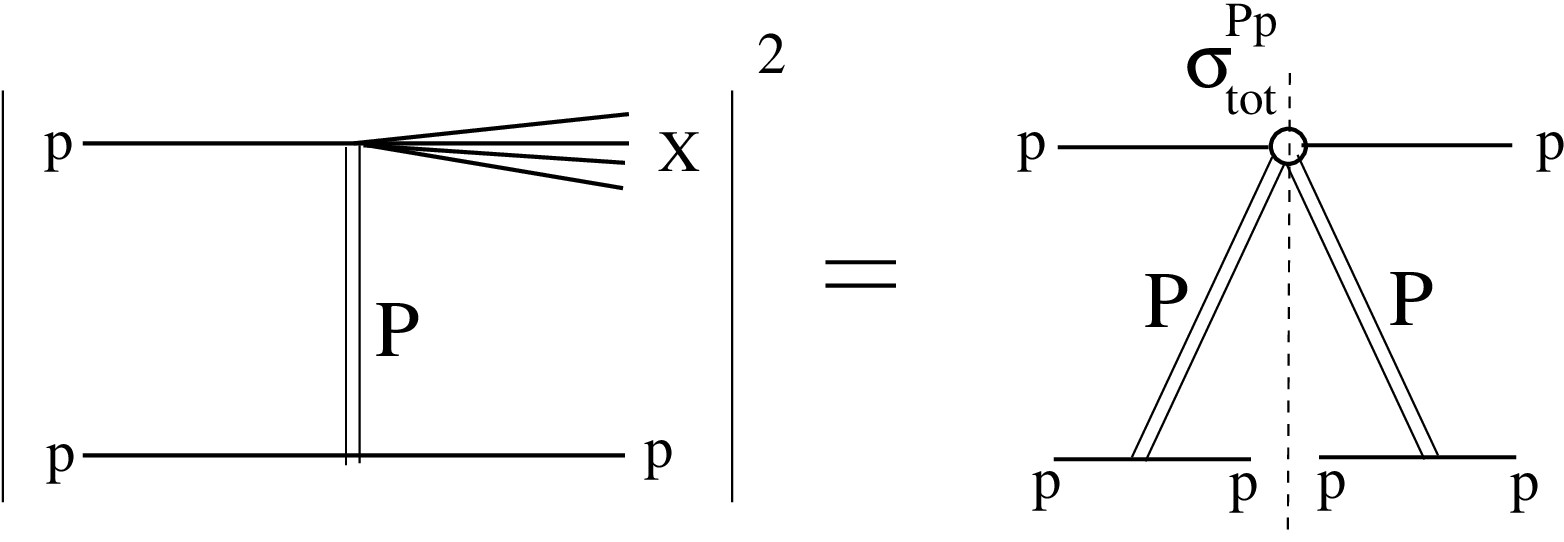}
\hfill\includegraphics[width=0.4\textwidth]{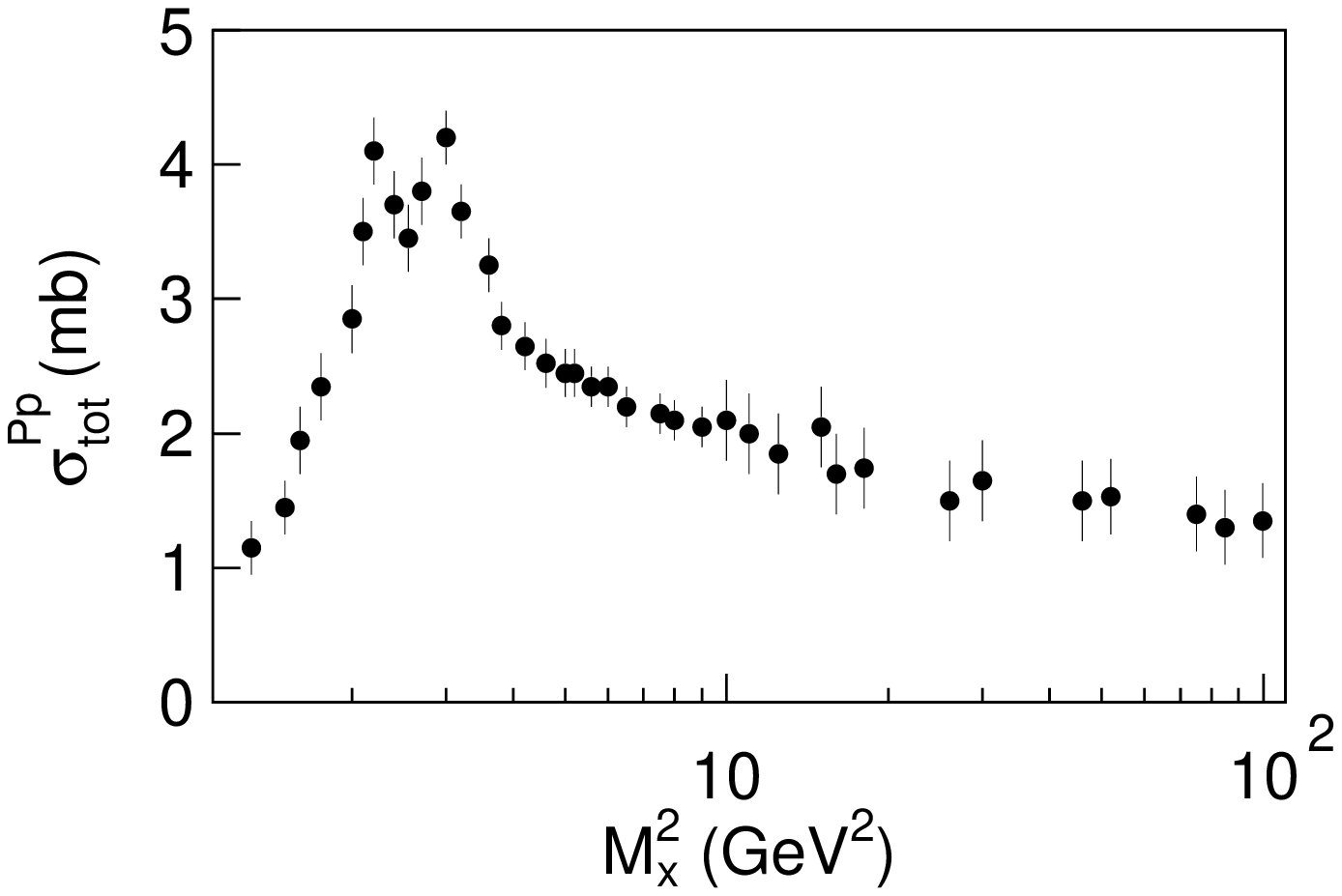}}
\caption{\small \label{fig:pom-p} {\it Left panel:} The amplitude squared of diffractive excitation of the projectile proton, summed over all the excitations with invariant mass $M_X$, is related via the optical theorem with the total Pomeron-proton cross section at c.m. energy $M_X$.
{\it Right panel:} The Pomeron-proton cross section extracted~\cite{kaidalov} from data on single diffraction $pp\to pX$ as function of $\Pom$-$p$ c.m. squared.}
\end{figure}

Data show that the magnitude of  diffractive gluon radiation is amazingly small. The way to see that is to express the single diffraction cross section in terms of the Pomeron-proton cross section as is illustrated in the left panel of Fig.~\ref{fig:pom-p}.
The Pomeron can be treated as a gluonic dipole and its cross section is expected to be about twice as big as for a $\bar qq$ dipole, i.e. $\sigma_{tot}^{\Pom p}\sim 50\mb$. However, data depicted in the right panel of Fig.~\ref{fig:pom-p}
show that $\sigma_{tot}^{\Pom p}< 2\mb$. Such a dramatic disagreement gives a clue that diffractive gluon radiation is strongly suppressed compared with the expectation based on pQCD. This problem has been known in the Regge phenomenology as smallness of the triple-Pomeron coupling~\cite{Kazarinov:1975kw}.

Gluon radiation can be described within the dipole approach via the propagation of a $\bar qqg$ dipole through the nuclear medium~\cite{kst1}. As the mean fractional momentum of the radiated gluon is very small, $\la \alpha_g\ra\sim
1/\ln(s)$, one can rely on eq.~(\ref{kopel-40}) for very small $x_{Bj}$, or eqs.~ (\ref{kopel-120})-(\ref{kopel-140}) for the onset of gluon shadowing, by replacing $\sigma_{\bar qq}(r)\Rightarrow \sigma_{gg}(r)$.
The only way to explain the observed suppression of gluon radiation is to reduce the mean size of the glue-glue dipole.
This can be achieved by introducing a specifically strong nonperturbative interaction within the glue-glue dipole, which
comes as the real part of the light-cone potential in eq.~(\ref{kopel-120}). Adjusting the strength of this interaction to data on diffractive gluon radiation (triple-Pomeron term) one arrives at the light-cone distribution functions in (\ref{kopel-40}) and (\ref{kopel-140}) with the mean glue-glue separation $r_0\approx 0.3\fm$~\cite{Kopeliovich:1999am}. This distance is smaller than the confinement radius $\sim1/\Lambda_{QCD}=1\fm$ and
is in accord with the lattice evaluations of the $gg$ correlation radius~\cite{pisa}, and the instanton radius~\cite{shuryak}. There is more experimental evidence supporting the existence of a semi-hard scale in hadrons~\cite{Kopeliovich:2007pq}.

\begin{figure}[tbh]
  \centering
  \includegraphics[width=0.3\textwidth]{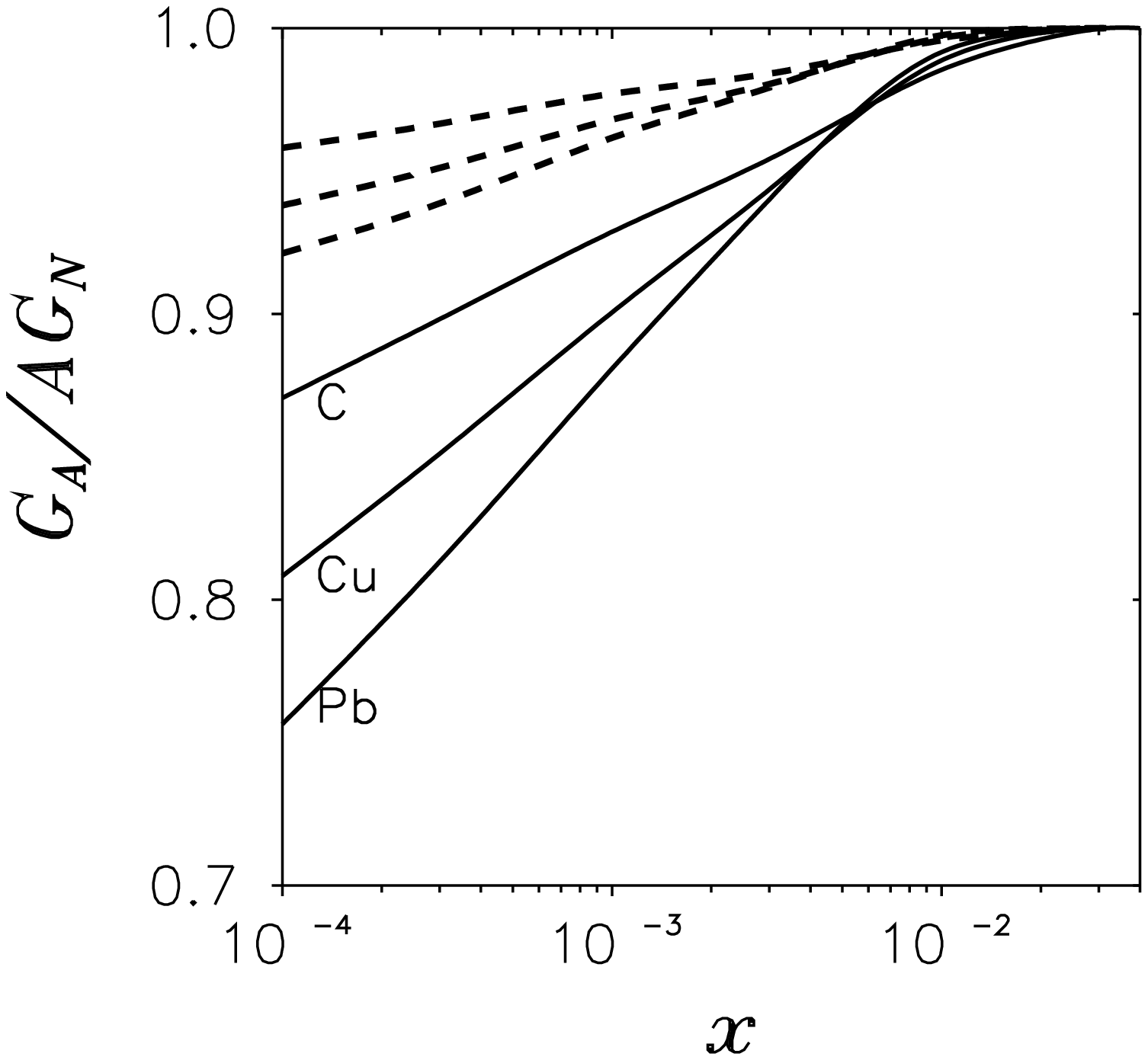}
  \hspace*{2cm}
  \includegraphics[width=0.35\textwidth]{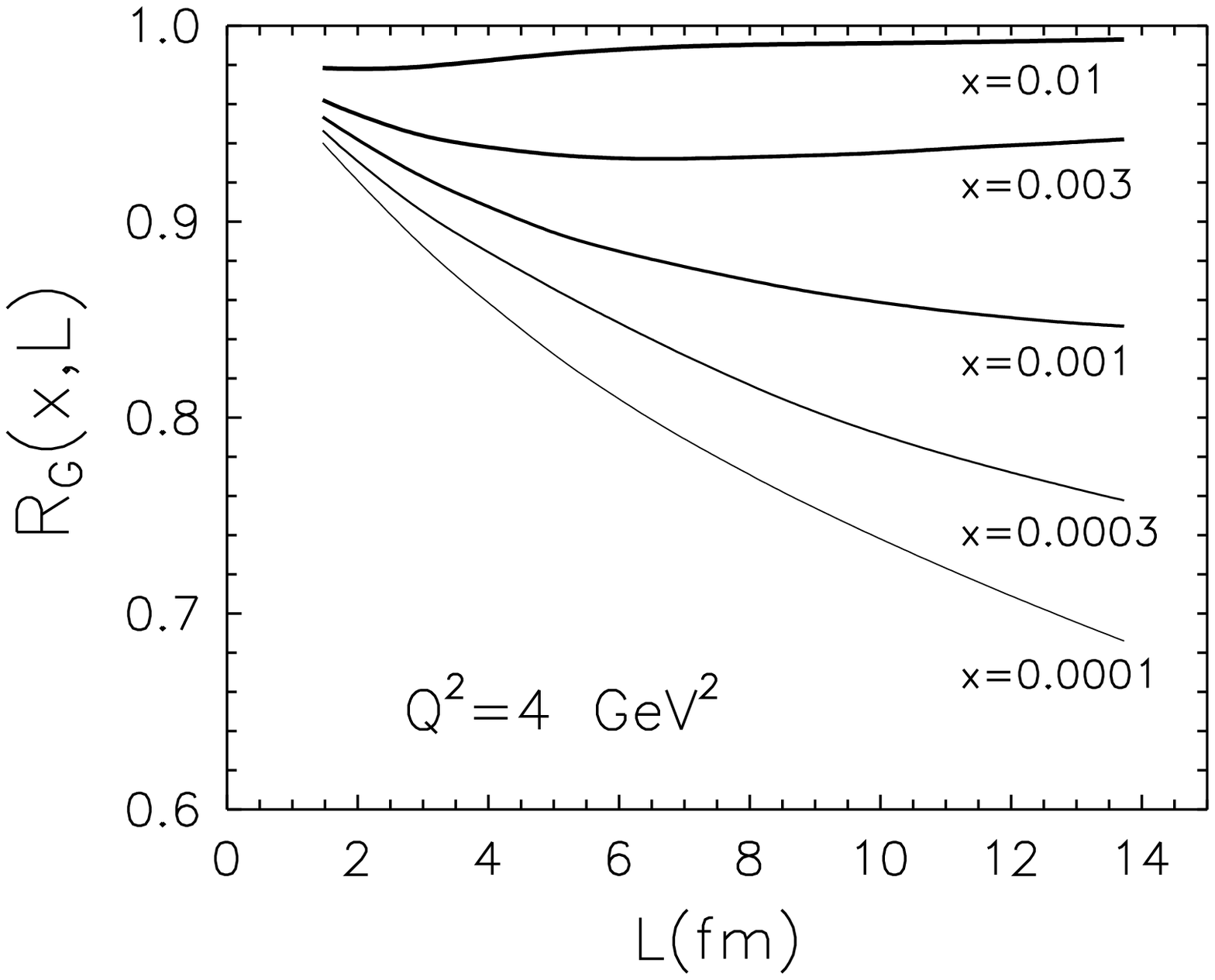}
  \caption{\small \label{glue-shad} 
    {\it Left panel:} Ratio of the gluon distribution functions
    in nuclei (carbon, copper and lead) and nucleons versus Bjorken $x$
    at $Q^2 = 4\,\gev^2$ (solid curves) and $40\,\gev^2$ (dashed
    curves)~\cite{Kopeliovich:1999am}. {\it Right panel:} Nuclear
    ratio $R_g=G_A(x,Q^2)/AG_N(x,Q^2)$ for gluons as function of path
    length in nuclear matter, calculated in~\cite{jan} at  
    $Q^2=4GeV^2$ for several fixed values of $x$.
  }
\end{figure}

Thus, the magnitude of gluon shadowing evaluated in~\cite{Kopeliovich:1999am,jan} is expected to be rather small, as is depicted in Fig.~\ref{glue-shad}.  The nuclear ratio $R_g=G_A(x,Q^2)/AG_N(x,Q^2)$ is plotted as a function of $x_{Bj}$ at $Q^2=4$ and $40\gev^2$ (left panel); and as a function of the path length in nuclear matter at $Q^2=4\gev^2$ and different
values of $x_{Bj}$.

The path-integral  approach is the most accurate method, which is valid in all regimes of gluon 
radiation, from incoherent to fully coherent. Nevertheless,  this is still
the lowest order calculation, which might be a reasonable approximation only for light nuclei, or 
for the onset of shadowing. The contribution of higher Fock components is still a challenge. This 
problem has been solved so far only in the unrealistic limit of long coherence lengths for all 
radiated gluons, described by the Balitsky-Kovchegov (BK) equation~\cite{Balitsky:1996ub,Kovchegov:1999yj}.
A numerical solution of this equation is quite complicated and includes lots of modelling~\cite{wiedemann}.
A much simpler bootstrap equation, which only requires modelling the shape of the saturated gluon distribution, was 
derived in~\cite{Kopeliovich:2010aa}. It includes the self-quenching effect for gluon shadowing,  and leads to a gluon distribution in nuclei which satisfies the 
unitarity bound~\cite{Kopeliovich:2008is}  The results are quite similar to the numerical solutions of the BK equation~\cite{wiedemann}. The magnitude of the self-quenched gluon shadowing found in~\cite{Kopeliovich:2010aa} is similar to the above results obtained in the leading order.



\subsubsection{Leading-twist nuclear shadowing}
\label{sec:LT_nuclear_shadowing}

\hspace{\parindent}\parbox{0.92\textwidth}{\slshape
 Vadim Guzey and Mark Strikman}
\index{Guzey, Vadim}
\index{Strikman, Mark}

\vspace{\baselineskip}

Nuclear shadowing in hadron (photon)-nucleus scattering
is the firmly established experimental phenomenon that 
at high energies the scattering cross
section on a nuclear target is smaller than the sum of the scattering
cross sections on the individual nucleons.
In the nucleus rest frame,
the theory of nuclear shadowing is based on the connection between nuclear shadowing and diffraction which has been established long time ago by Gribov~\cite{Gribov:1968jf}.
In the derivation, the key assumption is that nuclei can be described as dilute systems of nucleons.
The accuracy of the resulting theory for hadron-nucleus cross sections 
is very high, with the corrections at the level of a few \% which reflect the small admixture of non-nucleonic degrees of freedom in nuclei and the small off-shellness of the nucleons in nuclei as compared to the soft strong interaction scale.
Gribov's result can be understood~\cite{Frankfurt:1998ym} as a manifestation of unitarity
as reflected in the Abramovsky-Gribov-Kancheli (AGK) cutting rules~\cite{Abramovsky:1973fm}.   

The connection between shadowing and diffraction is also valid in deep inelastic scattering
(DIS) with nuclei; the approach based on this connection is called the leading twist theory
of nuclear shadowing~\cite{Frankfurt:1998ym,Frankfurt:2003zd,Guzey:2009jr,LT_shadowing_Phys_Rep}.
In this theory, parton distribution functions (PDFs)
in nuclei at small $x$ are calculated combining the unitarity relations for different 
cuts of the shadowing  diagrams corresponding to the diffractive and inelastic final states
(AGK cutting rules) with the QCD factorization theorem for hard diffraction~\cite{Collins:1997sr} (which provides a good description of the totality of the HERA hard diffractive data). 
The resulting multiple scattering series for the quark nuclear PDFs is presented in 
fig.~\ref{fig:gfs10_Master1_quarks}, where graphs $a$, $b$, and $c$ correspond to the 
interaction with one, two, and three nucleons of the target, respectively. 
Graph $a$ gives the impulse approximation;
graphs $b$ and $c$ contribute to the shadowing correction.
The interaction with $N > 3$ nucleons, though not shown, is taken into account in the 
final expression for nuclear PDFs.

At the level of the interaction with two nucleons of a nucleus with
the atomic mass number $A$, one can derive the
{\it model-independent} expression for the shadowing correction to the nuclear PDF of flavor $j$~\cite{Frankfurt:1998ym} (corresponding to graph $b$ of fig.~\ref{fig:gfs10_Master1_quarks}):
\begin{eqnarray}
x f_{j/A}^{(b)}(x,Q^2)&=&
-8 \pi A(A-1) \Re e \frac{(1-i\eta)^2}{1+\eta^2} \int^{0.1}_x d x_{\Pomeron}
\beta f_j^{D(4)}(\beta,Q^2,x_{\Pomeron},t_{{\rm min}}) \nonumber\\
&\times& \int d^2 \vec{b}  
\int^{\infty}_{-\infty}d z_1 \int^{\infty}_{z_1}d z_2 \,\rho_A(\vec{b},z_1) \rho_A(\vec{b},z_2) 
e^{i (z_1-z_2) x_{\Pomeron} m_N}   \,,
\label{eq:fgs10_eq1}
\end{eqnarray}
where $f_j^{D(4)}$ is the diffractive parton distribution of the nucleon; 
$\rho_A$ is the nuclear matter density; $\eta$ is the ratio of the real to imaginary
parts of the elementary diffractive amplitude, 
$\eta =\Re e A^{\rm diff}/ \Im m A^{\rm diff} \approx 0.17$. 
The diffractive PDF $f_j^{D(4)}$ depends on two light-cone fractions
$x_{\Pomeron}=(M_X^2+Q^2)/(W^2+Q^2)$ and $\beta=x/x_{\Pomeron}$ and the invariant
momentum transfer $t$, where $W$ is the invariant virtual photon-nucleon energy, 
$W^2=(q+p)^2$, and $M_X^2$ is the invariant mass squared of the produced intermediate
diffractive state denoted as ``X'' in fig.~\ref{fig:gfs10_Master1_quarks}.
The longitudinal (collinear with the direction of the photon momentum) 
coordinates $z_1$ and $z_2$ and the transverse coordinate (impact parameter) $\vec{b}$ 
refer to the two interacting nucleons; $m_N$ is the nucleon mass.
The $t$ dependence of $f_j^{D(4)}$ can be safely neglected as compared to the strong 
fall-off of the nuclear form-factor for $A> 4$
and, as a result, $f_j^{D(4)}$ enters 
eq.~(\ref{eq:fgs10_eq1}) at $t_{\rm min}\approx -x ^2 m_N^2(1+M_X^2/Q^2)^2$ and all 
nucleons enter with the same impact parameter $\vec{b}$.
Equation~(\ref{eq:fgs10_eq1}) satisfies the QCD evolution equations at all orders in 
the strong coupling constant $\alpha_s$.

\begin{figure}
\centerline{\includegraphics[width=0.6\textwidth]{eA-final/Figs/Master1b_2011.epsi}}
\caption{\small \label{fig:gfs10_Master1_quarks} Multiple scattering series for nuclear quark PDFs. 
Graphs $a$, $b$, and $c$ correspond to the interaction with one, two, and three
nucleons, respectively. 
Graph $a$ gives the impulse approximation;
graphs $b$ and $c$ contribute to the shadowing correction. }
\end{figure}

To evaluate the contribution to nuclear shadowing of the interactions with $N\ge 3$ nucleons in fig.~\ref{fig:gfs10_Master1_quarks}, one needs to invoke additional {\it model-dependent} considerations, since the interaction of a hard probe (virtual photon) with $N \ge 3$ nucleons is sensitive 
to fine details of the diffractive dynamics.
In particular, the hard probe can be viewed as a coherent superposition of configurations which
interact with the target nucleons with very different strengths. This effect of color (cross section)
fluctuations is analogous to the inelastic shadowing in hadron-nucleus scattering with the important
difference that the dispersion of the interaction strengths is much smaller in the hadron case than 
in DIS. 
However, the observation that $\alpha_{\Pomeron}(0)=1.11$ found in the analysis of 
hard diffraction at HERA~\cite{Aktas:2006hy} is very close to $\alpha_{\Pomeron}^{\rm soft}(0)=1.08$ in soft hadronic
interactions~\cite{Donnachie:1992ny} indicates that hard diffraction in DIS is dominated by large-size
hadron-like (aligned jet) configurations which evolve to large $Q^2$ via the 
DGLAP evolution.
(As to the point-like configurations, they give an important and increasing with $Q^2$
contribution to graph $a$ in  fig.~\ref{fig:gfs10_Master1_quarks}.)

This important observation reduces theoretical uncertainties in the treatment of 
the interactions with $N\ge 3$ nucleons and allows one to reliably parameterize the strength of the 
interaction with $N\ge 3$ nucleons by a single effective hadron-like cross section $\sigma_{\rm soft}^j$.
The final expression for the nuclear PDFs at a certain initial scale $Q_0^2$ reads~\cite{Guzey:2009jr,LT_shadowing_Phys_Rep}:
\begin{align}
  xf_{j/A}(x,Q_0^2)&=Axf_{j/N}(x,Q_0^2) \nonumber\\
  &-8 \pi A (A-1)\, \Re e \frac{(1-i \eta)^2}{1+\eta^2}
  \int^{0.1}_{x} d x_{\Pomeron} \beta f_j^{D(4)}(\beta,Q_0^2,x_{\Pomeron},t_{\rm min})\int d^2 b \int^{\infty}_{-\infty}d z_1
  \nonumber\\
  & \int^{\infty}_{z_1}d z_2\ \rho_A(\vec{b},z_1) \rho_A(\vec{b},z_2) e^{i (z_1-z_2) x_{\Pomeron} m_N}
  e^{-\frac{A}{2} (1-i\eta) \sigma_{\rm soft}^j(x,Q_0^2) \int_{z_1}^{z_2} dz^{\prime} \rho_A(\vec{b},z^{\prime})} \,.
  \label{eq:fgs10_eq2}
\end{align}
Due to the QCD factorization theorem, these nuclear PDFs $f_{j/A}(x,Q^2)$ can be used to
calculate many
different observables at small $x$ including the nuclear structure function $F_{2A}$ and the
longitudinal structure function $F_L^A$, the charmed contributions to these structure
functions $F_{2A}^{c}$ and $F_L^{A(c)}$, etc.;  $f_{j/A}(x,Q^2)$ can also be applied to the 
calculations of hard processes in heavy-ion collisions.

Removing the integration over $d^2b$ in right-hand side of
eq.~(\ref{eq:fgs10_eq2}), one obtains the impact parameter dependent
nuclear PDFs (nuclear GPDs in the $\xi=0$ limit in the impact
parameter representation)~\cite{LT_shadowing_Phys_Rep}, see
Section~\ref{subsec:nGPDS}.


\begin{wrapfigure}{l}{0.60\textwidth}
  \begin{center}
    \includegraphics[width=0.59\textwidth]{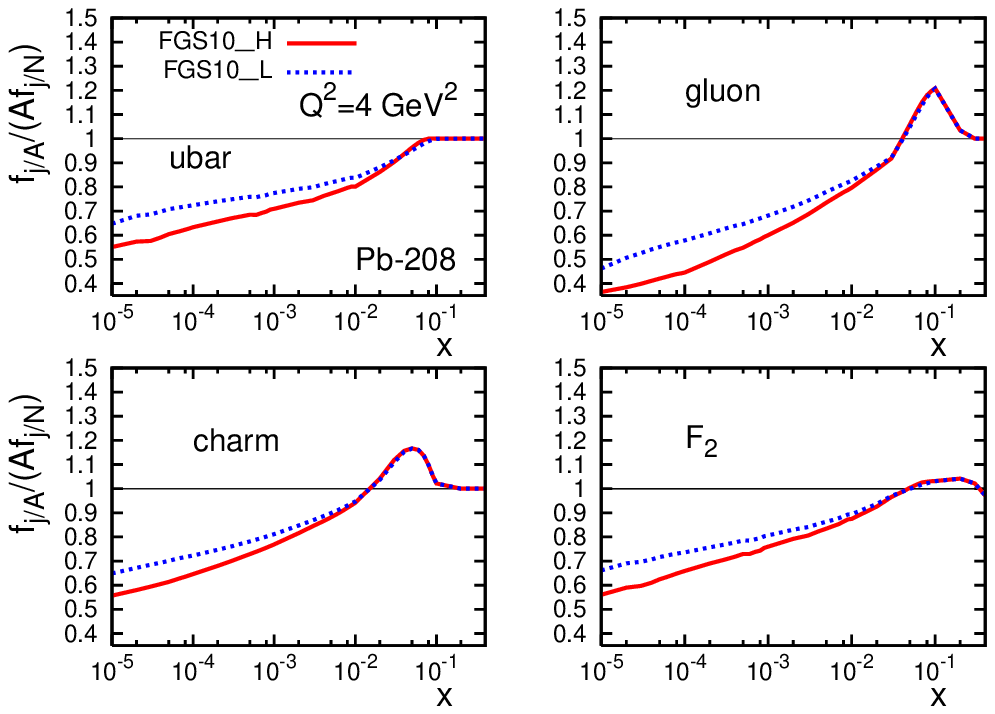}
  \end{center}
  \caption{\small The leading twist theory of nuclear shadowing predictions for $f_{j/A}/(A f_{j/N})$ ($\bar{u}$ and $c$ quarks and gluons) and $F_{2A}/(A F_{2N})$ as functions of $x$ at $Q_0^2=4$ GeV$^2$.  The two sets of curves correspond to the two extreme scenarios of nuclear shadowing (see the text).}
\label{fig:gfs10_LT2009_pb208_models12}
\end{wrapfigure} 

In our analysis~\cite{LT_shadowing_Phys_Rep}, we used two models for 
the color fluctuations in the virtual photon which correspond to two models 
for $\sigma_{\rm soft}^j$
which cover essentially all reasonable possibilities for the resulting
nuclear shadowing. An example of our predictions for the gluon, $\bar
u$-quark, $c$-quarks, and $F_{2A}$ structure functions is presented in
fig.~\ref{fig:gfs10_LT2009_pb208_models12} for the
two scenarios for $\sigma_{\rm soft}^j$ that we have mentioned above
(labeled FGS10\_H and FGS10\_L).
As one can see from fig.~\ref{fig:gfs10_LT2009_pb208_models12}, 
we predict large nuclear shadowing for 
each singlet parton flavor with
the characteristic feature that nuclear
shadowing in the gluon channel is larger than that in the quark channel.
The difference between the two extreme scenarios of color fluctuations
(the solid and dotted curves in fig.~\ref{fig:gfs10_LT2009_pb208_models12})
is  less than  20\% for $A\sim 200$ and much smaller for light nuclei.
The spread between the solid and dotted curves is the theoretical
uncertainty of our predictions.  
Note also that these results weakly depend on the choice of nucleon PDFs.

Accounting for the color fluctuations as done in eq.~(\ref{eq:fgs10_eq2}) 
tends to reduce the amount of nuclear shadowing as compared to the 
quasi-eikonal approximation used in the literature
\cite{Frankfurt:1998ym,Armesto:2010kr}. Also, the AGK technique allows
one to calculate other quantities such as nuclear diffractive PDFs and
fluctuations of multiplicity in non-diffractive
DIS~\cite{Frankfurt:1998ym,LT_shadowing_Phys_Rep,Frankfurt:2003gx}
both of which turn out to be  sensitive to the pattern of the color
fluctuations, see Section~\ref{sec:LT_Diffraction}.

Our approach to nuclear shadowing assumes the applicability of the linear (in parton densities) 
leading-twist DGLAP approximation. 
Numerical studies indicate that the dominant contribution to nuclear shadowing in eq.~(\ref{eq:fgs10_eq2}) comes from the region of relatively large $\beta=Q^2/(M^2+Q^2)$ corresponding to the rapidity intervals 
$\le 3$ for which the small-$x$ approximation used in the BFKL-type approaches is not applicable.  
These approaches predict $\alpha_{\Pomeron}(0) \sim 1.25$, 
while the HERA experiments find $\alpha_{\Pomeron}(0) \sim 1.11\approx 
\alpha_{\Pomeron}^{\rm soft}(0)$  consistent with the expectations of the QCD aligned jet approximation~\cite{Abramowicz:1995hb} that we effectively implemented in the 
derivation of eq.~(\ref{eq:fgs10_eq2}).

  
\subsubsection{Non-perturbative approaches}  
\label{pirner:approaches}  

\hspace{\parindent}\parbox{0.92\textwidth}{\slshape
 Hans J. Pirner}
\index{Pirner, Hans J.}

\vspace{\baselineskip}

One of the challenges in QCD is the  
description and understanding of  high-energy scattering
 on protons and nuclei.  
Even for high energies and large $Q^2$ in deep inelastic electron scattering   
a non-perturbative framework may be necessary.  
 For the transverse structure function, the $q \bar q$ dipole in the photon can be large  
 and the saturation scale $Q_s$  is small   
for the energies we discuss. In the following  
I will present the main features of such an approach, which of course will   
also include the perturbative aspects.

The most important phenomenon observed in high-energy scattering  
is the rise of the total cross sections with increasing c.m. energy.  
While the rise is slow in hadronic reactions of {\em large} particles  
such as protons, pions, kaons, or real photons, it  
is steep if one {\em small} particle is involved such as an  
incoming virtual photon or an outgoing charmonium.  
This energy behavior is best seen in the proton structure function  
$F_2(x,Q^2)$. With increasing photon virtuality $Q^2$, the increase of  
$F_2(x,Q^2)$ towards small Bjorken $x$ becomes significantly stronger.  
It is tempting to test the growth of the structure function with nuclei.  
In the following I will summarize my work with Shoshi, Steffen and Dosch
 which is   
published in two main papers~\cite{Shoshi:2002in,Shoshi:2002fq}. For    
references to other work please see these two papers.

In the two-pomeron model of Donnachie and  
Landshoff, the energy  
dependence of the cross sections at high energies results from the  
exchange of a soft and a hard pomeron.  The first dominates in  
hadron-hadron and $\gamma^* p$ reactions at low  
$Q^2$ and the second in $\gamma^* p$  
reactions at high $Q^2$. The two pomerons may be related to  
a glueball trajectory, which is inherently non-perturbative, and a gluon  
ladder \`a la BFKL, which includes the perturbative aspects.   
The two-pomeron model, however, does not contain parton saturation  
nor unitarity effects.  
A model  
motivated by the concept of parton saturation is the one of  
Golec-Biernat and W\"usthoff which  
allows very successful fits to $\gamma^* p$ data, but cannot be applied  
to hadron-hadron reactions.  A successful  
description of dipole nucleon scattering which  
 can be used for hadron-nucleon scattering and  
DIS with moderate $Q^2$ has been found~\cite{Kopeliovich:1999am}.

We have combined perturbative and  
non-perturbative QCD to compute high-energy reactions of hadrons and  
photons with special emphasis on saturation effects that manifest   
$S$-matrix unitarity~\cite{Shoshi:2002in}. We follow the {\em functional  
integral approach} to high-energy scattering of Nachtmann,  
in which the $S$-matrix element factorizes into the universal  
correlation of two light-like Wegner-Wilson loops $S_{DD}$.  
The light-like  
Wegner-Wilson loops describe color dipoles given by the quark and  
antiquark in the meson or photon projectile and  the   
quark and di-quark in the baryon target. This approach treats projectile and  
target symmetrically.    
$S$-matrix unitarity is respected as a  
consequence of a matrix cumulant expansion and the Gaussian  
approximation of the functional integrals. The resulting  
dipole cross sections do not show Glauber-like behavior  with  
the dipole size as in the Golec-Biernat model.  
The loop-loop correlation function $S_{DD}$ is expressed in terms of the  
gauge invariant bi-local gluon field strength correlator integrated  
over two connected minimal surfaces. Due to the symmetric treatment of  
the two dipoles this formalism can explicitly investigate the  
dependence on the impact parameter of the two scattering partners.

The gluon field strength correlator has a non-perturbative and a  
perturbative component. The {\em stochastic vacuum model}  
of Dosch and Simonov  is used for the non-perturbative low  
frequency background field and {\em perturbative BFKL gluon exchange}  
for the high frequency contributions.  This combination allows us to  
describe long and short distance correlations in agreement with  
Euclidean lattice calculations of the static quark-antiquark potential  
with color-Coulomb behavior at short distances and confining linear  
rise at long distances.   
We have tried to model both components in AdS/QCD,  
 but the long range loop-loop correlation  
cannot be established on a classical level, since the connecting  
surface in 5 dimensions breaks off at large distances~\cite{Nian:2009mw}.  
   
Energy dependence in the loop-loop correlation  
function, $S_{DD}$, is introduced by hand in order to describe  
 simultaneously the energy  
behavior in hadron-hadron, photon-hadron, and photon-photon reactions  
involving real and virtual photons as well.  Motivated by the  
two-Pomeron picture of Donnachie and  
Landshoff, we ascribe to the soft and hard  
component a weak and strong energy dependence, respectively. The parameter   
describing the energy dependence of the perturbative correlation  
 function is very large  
because we include  
{\em multiple gluonic interactions}.  
In ref.~\cite{Shoshi:2002in} we   
have considered not only the dependence of the dipole cross section   
on dipole size with increasing energy and the resulting $k_t$-saturation,   
but also the scattering amplitudes in impact parameter space, where  
the $S$-matrix unitarity imposes rigid limits on the impact parameter  
profiles such as the {\em black disc limit}. We  
present profile functions for longitudinal  
photon-proton scattering that provide an intuitive  
geometrical picture for the energy dependence of the cross sections.  
The profile function first becomes greyer, turns black and then  
 increases in transverse size.  
Using a leading-twist NLO DGLAP relation, we estimated the {\em impact parameter dependent gluon distribution} of  
the proton $xG(x,Q^2,|\vec{b}_{\perp}|)$ from the profile function for  
longitudinal photon-proton scattering.   
We have not found saturation of the profile function at HERA energies, but  
at higher energies, $xG(x,Q^2,|\vec{b}_{\perp}|)$ does saturate as a manifestation of  
 the $S$-matrix  
unitarity.   
  
In the same framework, we have studied the unintegrated gluon  
distribution $x G(x,k_t)$ as a function of transverse momentum $k_t$ for  
increasing energies~\cite{Shoshi:2002fq}. To obtain the   
unintegrated gluon  
distribution, one uses the possibility to rewrite  
the non-perturbative scattering of an artificial  
external dipole as a superposition of perturbative contributions.   
In other words the string of the projectile dipole  
can be decomposed mathematically in a superposition of dipoles of  
smaller sizes, from which $x G(x,k_t)$  can be extracted.   
  
The long range confining  
character of the non-perturbative field strength correlators determines  
the low $k_t$ behavior of the gluon structure function of the hadron  
as $x G(x,k_t) \propto 1/k_t$. In the low momentum limit, 
 $x G(x,k_t)\cdot k_t $ converges  
towards a constant independent of $x$, related to the size of the hadron.  
The cross-over from the non-perturbative region to the perturbative region  
occurs at around $k_t=1\gev$ at $x$-values $10^{-4}<x<10^{-2}$.   
  
On a more fundamental level, we   
 have analysed   
correlations of Wilson lines in vacuum as one approaches the light  
cone from space-like distances~\cite{Pirner:2002fe}.  The dominant terms of the near light cone  
Hamiltonian for the Wilson lines define a field theory in 2+1  
dimensions. In the limit of small x, the SU(3) QCD for Wilson lines  
reduces to a critical Z(3) theory with a diverging correlation length  
$\xi(x) \propto x^{-1/(2\lambda_2)}$ where the exponent  
$\lambda_2=2.52$ is obtained from the center group Z(3) of SU(3). We  
conjecture that the dipole wave function of the virtual photon behaves  
as the correlation function of Wilson lines in the vacuum. For  
transverse sizes smaller than the correlation size it scales like  
$\Psi \propto 1/(x_t)^{1+n}$ with $n=0.04$ and for distances larger than the  
correlation length it decays exponentially which makes this region
 negligible.   
For $F_2$ we  
integrate the square of the photon wave function weighted with   
a dipole proton cross section of fixed size $R_0$ independent of x. All  
the energy dependence is absorbed into the photon. Because of the approximate  
conformality of the dipole wave function $(n \approx 0)$, the result depends  
only on $R_0^2/\xi(x)^2 \propto R_0^2 x^{1/\lambda_2}$, i.e.  
 the saturation scale   
varies as    
as $ Q_s^2= Q(x_o)^2 (x0/x)^{1/\lambda_2}$.  
The critical index in this theory is a characteristic feature of Z(3)  
 theory i.e.  
the center group of SU(3) in an external field given by the light 
quarks.  
This is very different from the perturbative color glass
 condensate where $Q_s$  
depends on the running coupling similarly to the power behaviour of BFKL.   
  

\section{Inclusive DIS (F$_2$, F$_L$, F$_2^c$)}  
\label{sec:InclusiveDIS}  

\subsubsection{Estimates of higher twist in deep inelastic nucleon and nucleus scattering}
\label{sec:twists}

\hspace{\parindent}\parbox{0.92\textwidth}{\slshape
 Joachim Bartels, Krzysztof Golec-Biernat and Leszek Motyka}
\index{Bartels, Jochen}
\index{Golec-Biernat, Krzysztof}
\index{Motyka, Leszek}


A deeper understanding of the transition region at low $Q^2$ and small $x$ 
in deep inelastic electron proton scattering has been one of the 
central tasks of HERA physics. It will be one of the key questions to be 
addressed by a future Electron Ion Collider. Approaching this transition region from the 
perturbative side, one expects to see the onset of corrections to the 
successful DGLAP description, based upon leading twist operators in QCD.
The twist expansion defines a systematic approach to the short distance 
limit probed in deep inelastic scattering. The study of higher-twist 
corrections therefore provides an attractive route for  
investigating the region of validity of the leading twist DGLAP 
evolution equations.

The validity of the leading-twist QCD evolution equations is based upon the fact that,
for sufficiently large $Q^2$ and not too small $x$, the gluons 
inside the proton are dilute. The DGLAP evolution equations, however, 
predict that, at small $x$ and low $Q^2$, the gluon density grows. 
As a result, the gluons start to interact and the gluon density eventually 
saturates. The onset of saturation is encoded in the saturation scale, 
$Q_{\rm sat}^2(x)$. 

The investigation of saturation is of highest importance for our 
understanding of QCD. Saturation can be viewed as a first step of entering 
the strong interaction region. While the QCD coupling constant 
is still small, saturation phenomena probe nonlinear dynamics  
of the gluon sector which plays a crucial role in many areas of 
strong interactions. It is expected that saturation effects in deep inelastic 
scattering on a nucleus are enhanced in comparison with 
deep inelastic scattering on a proton. In the former case, the incoming photon `sees'
the gluons of many nucleons, whereas in the case of a single nucleon, 
one has to go to smaller $x$ values (higher energies) in order to 
reach the same gluon density.           

A brief discussion of the connection between saturation and 
the twist expansion has been given in \cite{Bartels:2009tu}. 
Whereas in the GBW model \cite{GolecBiernat:1998js,GolecBiernat:1999qd} there is a rather direct classification   
of eikonal-type exchanges of gluon ladders in terms of twist quantum numbers, 
in saturation models based upon the nonlinear BK-equation 
\cite{Albacete:2009fh,Albacete:2010sy} a twist decomposition is much 
less obvious. In the following we present some numerical estimates of
higher-twist contributions, using the improved version of the GBW
model \cite{Bartels:2002cj}.\\

\noindent{\bf The method:} The theory of higher-twist operators and their evolution equations has been outlined in \cite{Bukhvostov:1985rn}: in leading order, the higher-twist evolution equations are described by the 
nonforward DGLAP splitting functions, and there is a particular pattern of 
mixing between different operators of the same twist.
In the same way as for leading twist, a numerical analysis of higher twists 
requires initial conditions for the set of evolution equations, which 
have to be adjusted to data. In \cite{Bartels:2009tu} the magnitude of higher-twist 
corrections was evaluated in a slightly different way. Starting from the 
observation that within the GBW saturation model the multiple exchanges 
of leading-twist gluon ladders can be put into a one-to-one correspondence 
with contributions of definite twist quantum numbers, it is possible to 
arrive at quantitative estimates of the leading-twist contributions and  
corrections due to twist $\tau=4,\,6,\,...$. Details have been described in 
\cite{Bartels:2009tu} and will not be repeated here.

While the analysis in \cite{Bartels:2009tu} was performed for the case of $e+p$ scattering, 
it is straightforward to extend it to electron-nucleus scattering. Assuming a cylindrical nucleus with a 
characteristic size $R_A \approx A^{1/3} R_p$ (with $R_p$ being the proton radius),
we simply replace the dipole-proton cross section (eq.(42) in \cite{Bartels:2009tu})
\beq
\sigma_{\rm dipole-proton} = \sigma_0 \left( 1 - \exp( - \Omega(x,r^2)) \right)      
\eeq
by the dipole-nucleus cross section
\beq
\sigma_{\rm dipole-nucleus} = A^{2/3} \sigma_0 \left( 1 - \exp( - A^{1/3} \Omega(x,r^2)) \right),     
\eeq
where $\Omega(x,r^2)$ is the eikonal function given in \cite{Bartels:2009tu}.
With the parameters from \cite{Bartels:2009tu} we simply repeat the electron proton 
calculations for electron gold scattering, using the modified dipole 
cross section formula in (2). \\

\begin{figure}[t]
\centering
  \includegraphics[width=0.49\linewidth]{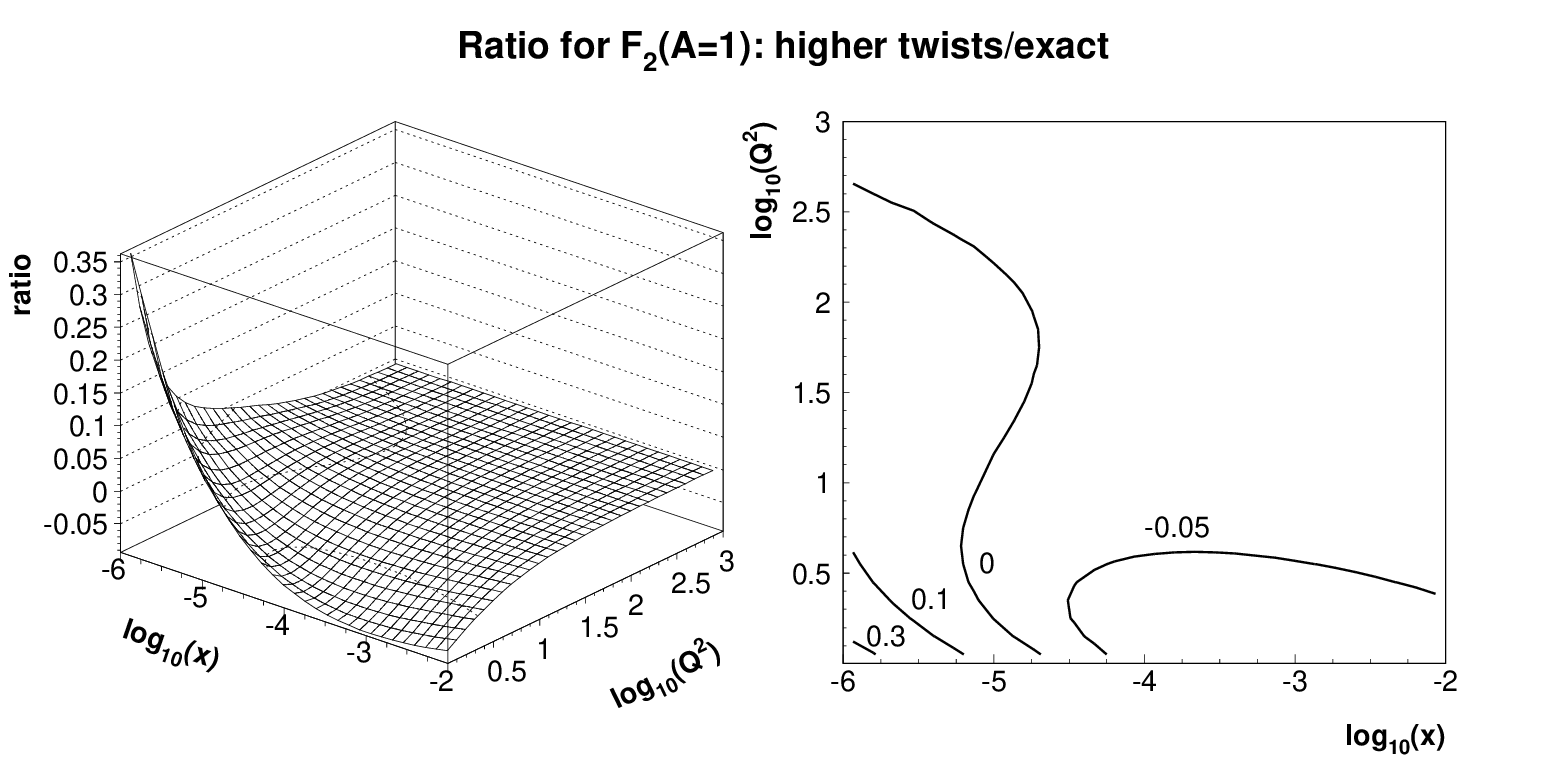}
  \includegraphics[width=0.49\linewidth]{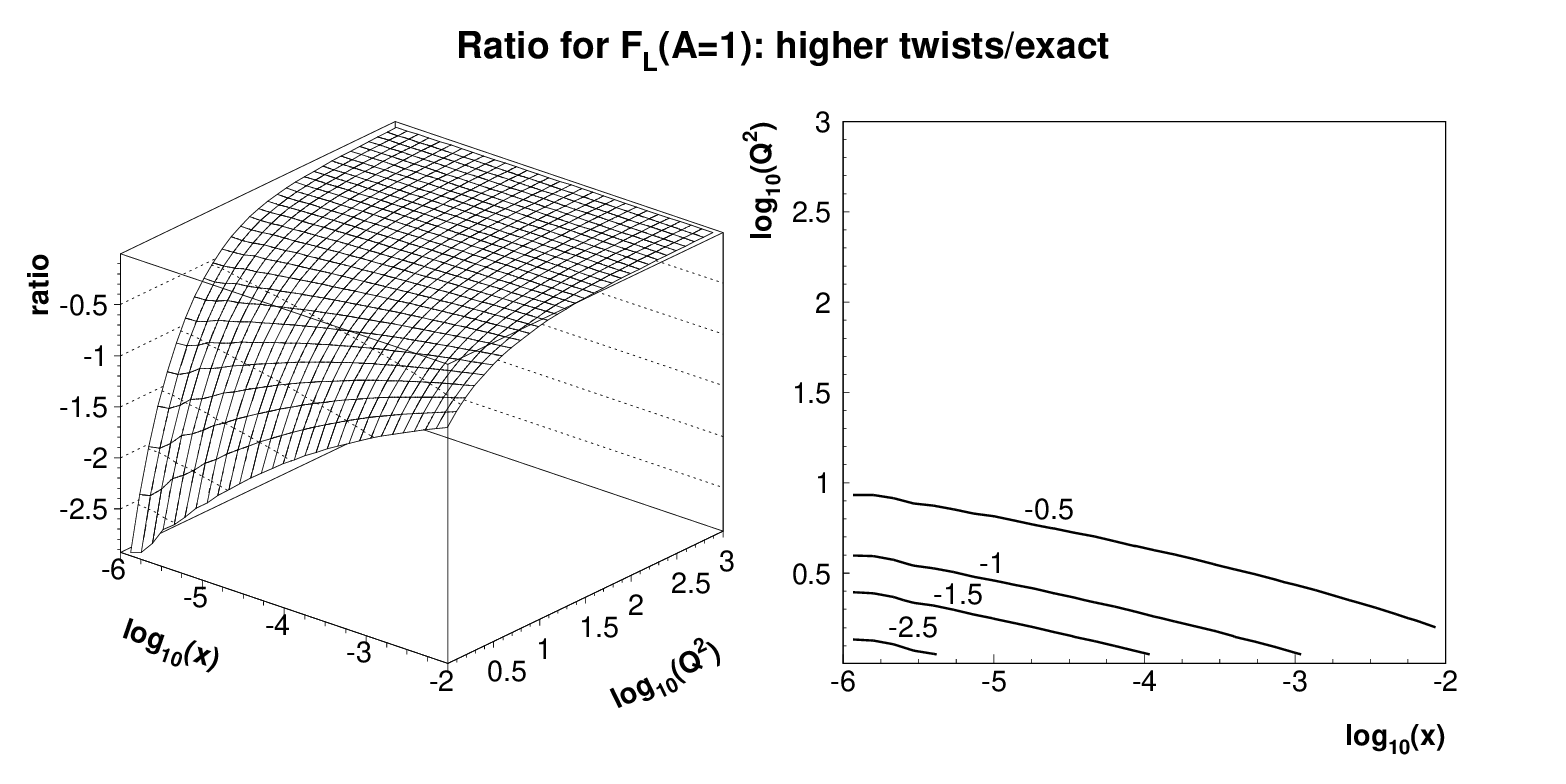}
  \caption{\small \label{fig:B_F2Lp}
    Higher-twist contribution estimate for $F_2$ (left) and $F_L$
    (right) of the proton.
  }
\end{figure}

\begin{figure}[h]
  \centering
  \includegraphics[width=0.49\linewidth]{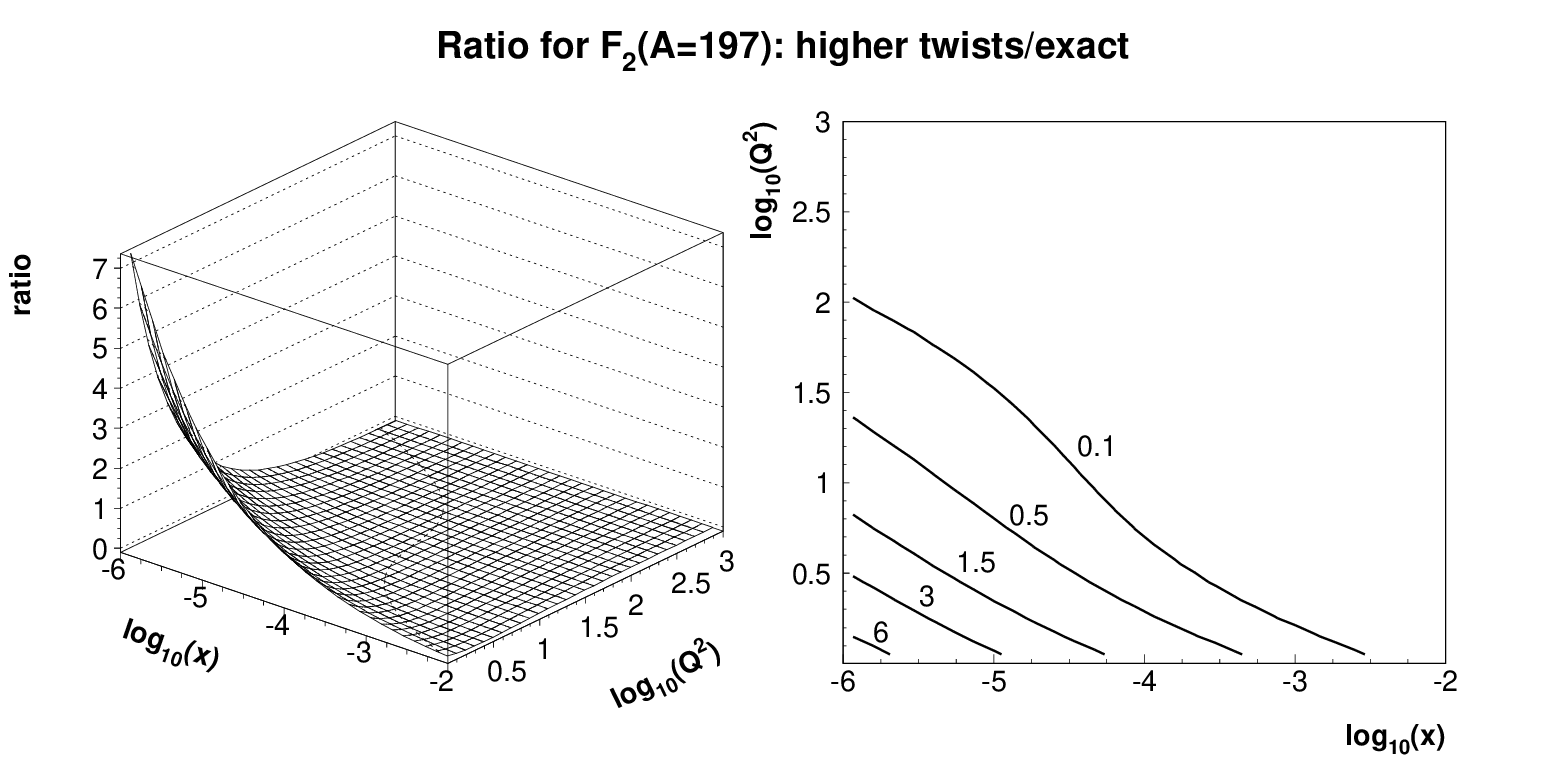}
  \includegraphics[width=0.49\linewidth]{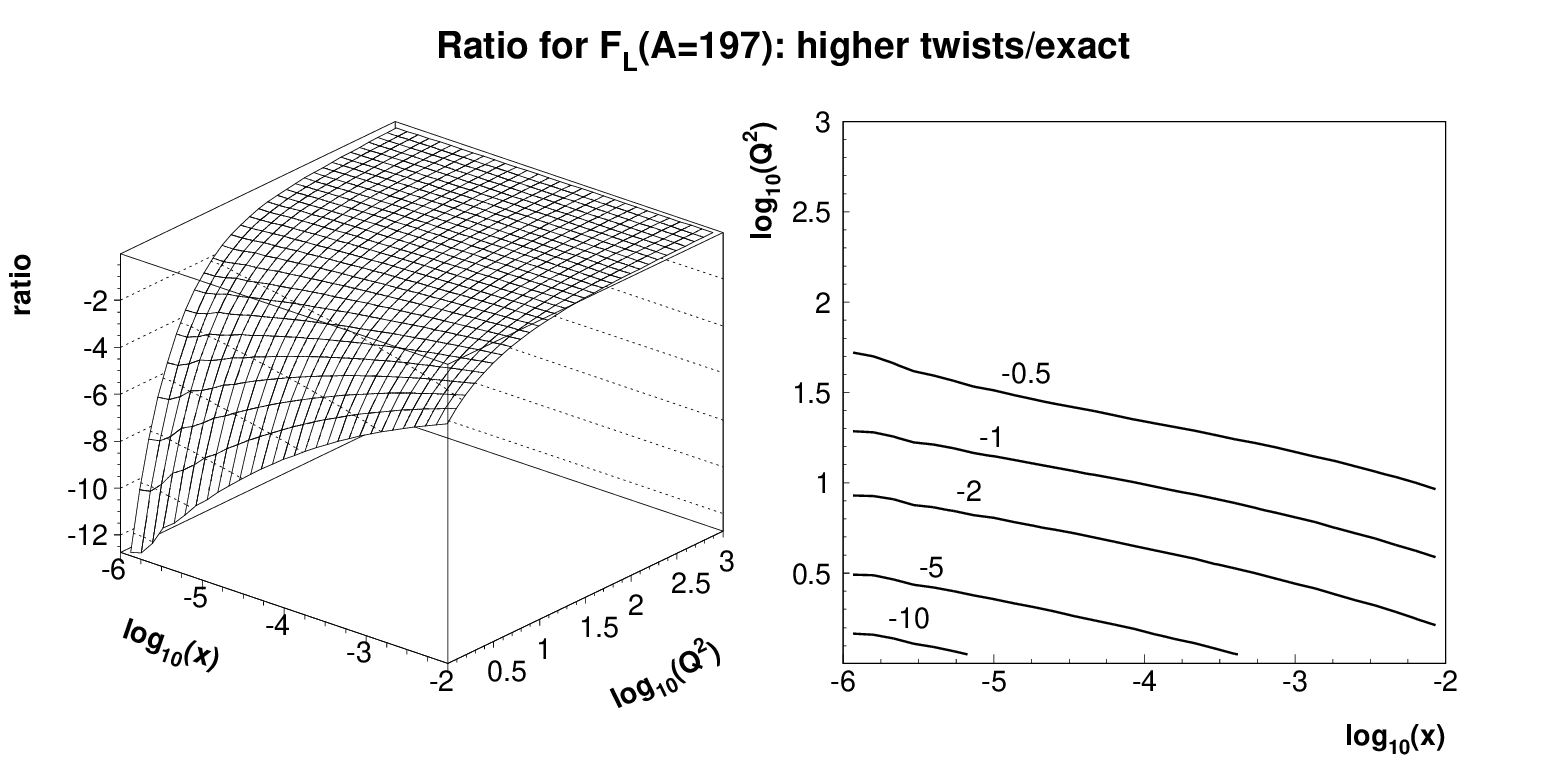}
  \caption{\small \label{fig:B_F2LAu}
    Higher-twist  contribution estimate for $F_2$ (left) and $F_L$
    (right) of the gold nucleus.
  }
\end{figure}

\noindent{\bf Numerical results:} The numerical results for $F_2$ and
$F_L$ are shown in Fig.~\ref{fig:B_F2Lp} for the
proton, and Fig.~\ref{fig:B_F2LAu} for the gold nucleus.
In each figure we show, on the l.h.s in a 3-dimensional view,
the ratio of the higher-twist corrections and the full structure
function as a fucntion of $x$ and $Q^2$,  
\begin{equation}
  \label{eq:ratio}
  {\rm ratio}=\frac{F^{\rm (total)}_{2,L} - F^{\rm (\tau=2)}_{2,L}}{F^{\rm (total)}_{2,L}}\,.   
\end{equation}
The r.h.s. shows the projection onto the $(\log x,\log Q^2$) plane: the lines 
belong to fixed values of the ratio (\ref{eq:ratio}). 
One recognizes the general trend: the corrections are getting larger 
when $x$ and $Q^2$ decrease (moving towards the lower left corner). For given $x$ and $Q^2$, the corrections for the 
longitudinal structure functions are larger than for $F_2$. This is a consequence of the 
sign structure of the corrections in $F_L$ and $F_T$: twist four corrections 
to $F_L$ and $F_T$ have opposite signs, and in the analysis \cite{Bartels:2009tu} of 
$F_2=F_T +F_L$ a strong cancellation has been found. This explains the small higher twist contribution for $F_2$.   

One also recognizes the general trend that for gold, all 
corrections are larger than for the proton. Finally, the corrections to 
$F_L$ are negative and those to $F_2$ are mostly positive. In the case of the proton $F_2$ there is a 
change in sign in the region of very small values of $Q^2$: this again 
is a consequence of the sign structure of the twist corrections 
to $F_T$ and $F_L$. \\

\noindent{\bf Conclusions:} Our numerical analysis confirms that, in general, the structure functions $F_L$ are more 
sensitive to higher-twist corrections than $F_2$ which, because of the sign structure 
in the twist 4 corrections, seems to much better ``protected'' against higher twist.
Also, nuclear targets are more sensitive to higher twist corrections relative to the proton.
In view of these results, it seems clear that in a future electron-ion 
collider, the measurement of $F_L$ is of vital importance in the search for saturation.   

\subsubsection{Strength of nonlinear effects in nucleons and nuclei}
\label{sec:lappi_nonlin}
\hspace{\parindent}\parbox{0.92\textwidth}{\slshape
 Tuomas Lappi}
\index{Lappi, Tuomas}

\vspace{\baselineskip}

\begin{wrapfigure}{r}{0.5\textwidth}
  \begin{center}
    \includegraphics[width=0.48\textwidth]{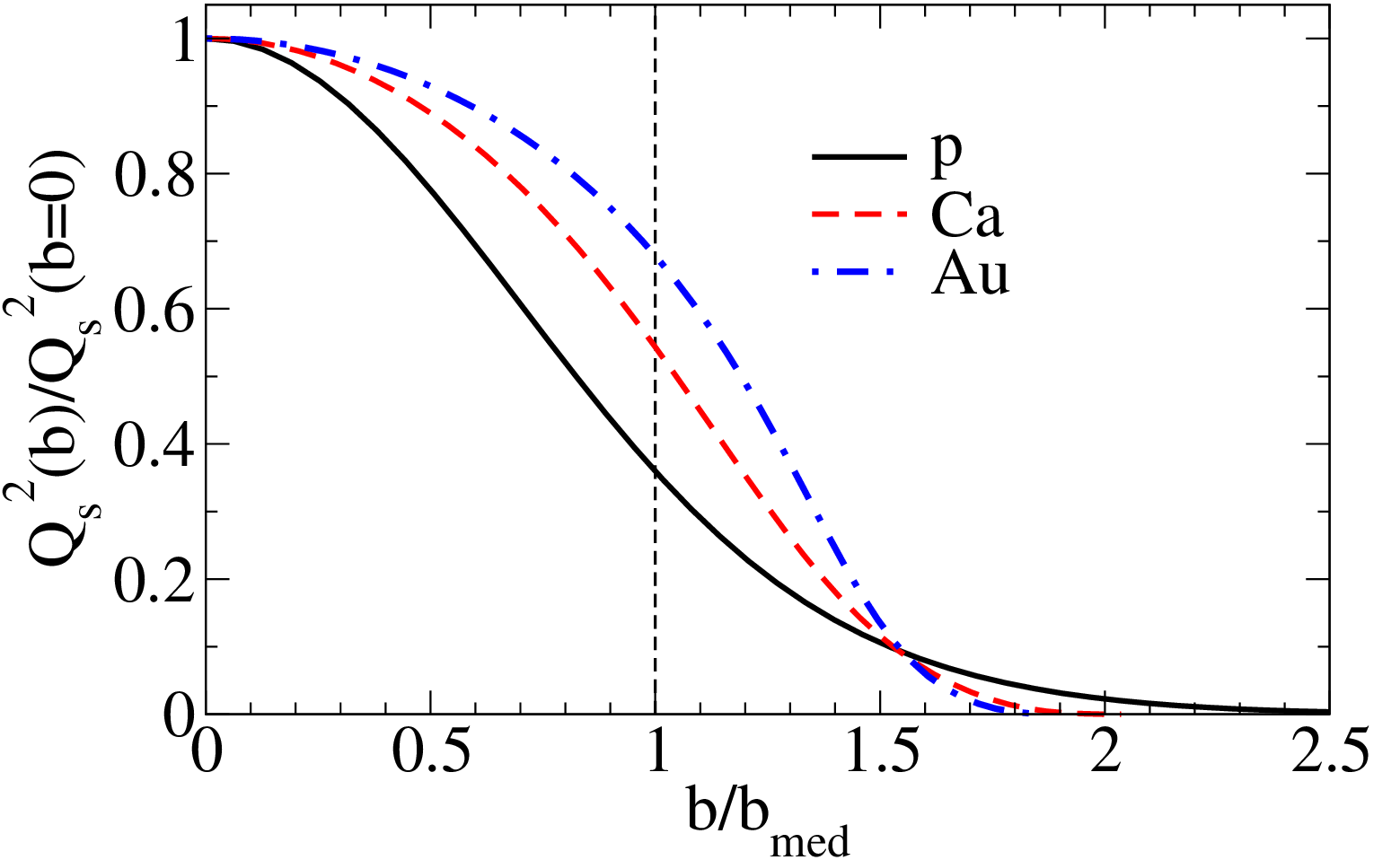}
  \end{center}
  \caption{\small \label{fig:qsabt} The saturation scale in a proton and Ca and Au nuclei as a function of $b/b_\mathrm{med},$ where $b_\mathrm{med}$ is the median impact parameter probed in inclusive DIS at $x=0.001$ and $Q^2=1 \gev^2$.}
\end{wrapfigure} 

The effects of nonlinearity and unitarity in small $x$
DIS are most clearly visible in the dipole framework. We denote ${\mathcal{N}}(x,{\mathbf{b}_T},{\mathbf{r}_T})$
the imaginary part of the scattering amplitude for a dipole of size
${\mathbf{r}_T}$ and rapidity $y= \ln(1/x)$ to scatter off the target at impact
parameter ${\mathbf{b}_T}$. The total dipole cross section is
given by twice the integral of ${\mathcal{N}}(x,{\mathbf{b}_T},{\mathbf{r}_T})$
over the impact parameter.
While the formal unitarity limit would be for ${\mathcal{N}}$ to lie between 
$0$ and $2$, in practice the reasonable physical area is between $0$ (no 
scattering) and $1$ (complete absorption or the black disk limit). 
The typical value of the dipole 
scattering amplitude therefore serves as a good measure of the 
degree of nonlinearity of the scattering process.

As the total cross section depends on the integral of the scattering amplitude
over the impact parameter, 
statements about the magnitude of the scattering amplitude
depend on the profile of the target in $\mathbf{b}_T$.
 The ${\mathbf{b}_T}$-depdendence for the scattering
amplitude on a nucleon is, however, very much constrained
by the $t$-dependence of exclusive vector meson production.
Using this information, in addition to the total cross section,
results in the two commonly used ${\mathbf{b}_T}$-dependent 
dipole amplitude parametrizations that we will use here, the
IPsat and bCGC models~\cite{Kowalski:2003hm,Kowalski:2006hc,Watt:2007nr}.
They have successfully been used to describe HERA data on the 
inclusive cross section, exclusive vector meson production and 
diffractive structure functions~\cite{Kowalski:2008sa}. \\



\noindent{\bf The saturation scale:}  To a first approximation the impact parameter dependence of the 
nuclear scattering amplitude can then be obtained by combining the nucleon one
 with  basic knowledge of nuclear geometry in a Glauber-like treatment
(see e.g. Refs.~\cite{Kowalski:2007rw,Lappi:2010dd} for details).
This yields a characteristic pattern of nuclear suppression (shadowing) 
of the inclusive cross section, a nuclear enhancement of diffraction to small
mass states and a suppression in diffraction to large masses 
(small $\beta$)~\cite{Kowalski:2008sa}. 

One way of quantifying the importance of nonlinear effects is to compare
the value of the ($\mathbf{b}_T$ and $x$-dependent) \emph{saturation scale}
$Q^2_\mathrm{s}$ to the virtuality $Q^2$ of the process. The saturation 
scale is defined as the inverse of the dipole size at which the scattering
amplitude $\mathcal{N}$ 
reaches some specific value defined by convention.
For $Q^2 \gg Q^2_\mathrm{s}$
one is in the dilute limit and for $Q^2 \sim Q^2_\mathrm{s}$ nonlinear
effects become important. A naive argument of the $A$-dependence of the 
saturation scale for nuclei would give $Q_{\mathrm{s}A}^2\sim A^{1/3}$.
The importance of a realistic
impact parameter dependence for nuclei was discussed in more detail
in Ref.~\cite{Kowalski:2007rw}, where it was found that this 
dependence is indeed true to a very good approximation, but the
picture is more intricate than that. For the center of a nucleus vs. the center 
of a proton the saturation scale is suppressed by a geometrical factor
$\sim 0.3 \approx R_p^2 A^{2/3}/R_A^2$. 
Both a nucleon and a nucleus have a dilute edge at large impact parameters.
The thickness of this edge is determined by confinement scale physics
and is thus of the same order for both. 
The proton is, however, a much smaller object and
therefore the dilute edge region is responsible for a much larger fraction of
the total cross section than in a nucleus.
One way to see this is to look at the saturation scale at the
\emph{median impact parameter} contributing to the inclusive cross DIS cross 
section. The value of $Q_\mathrm{s}^2(b_\mathrm{med})$
is $\sim 35\%$ of the value at $b=0$ for a proton, but $\sim 70\%$ for a 
gold nucleus (see Fig.~\ref{fig:qsabt}, \cite{Kowalski:2007rw}). \\

\begin{figure}[tbh]
\centerline{\includegraphics[width=0.49\textwidth]{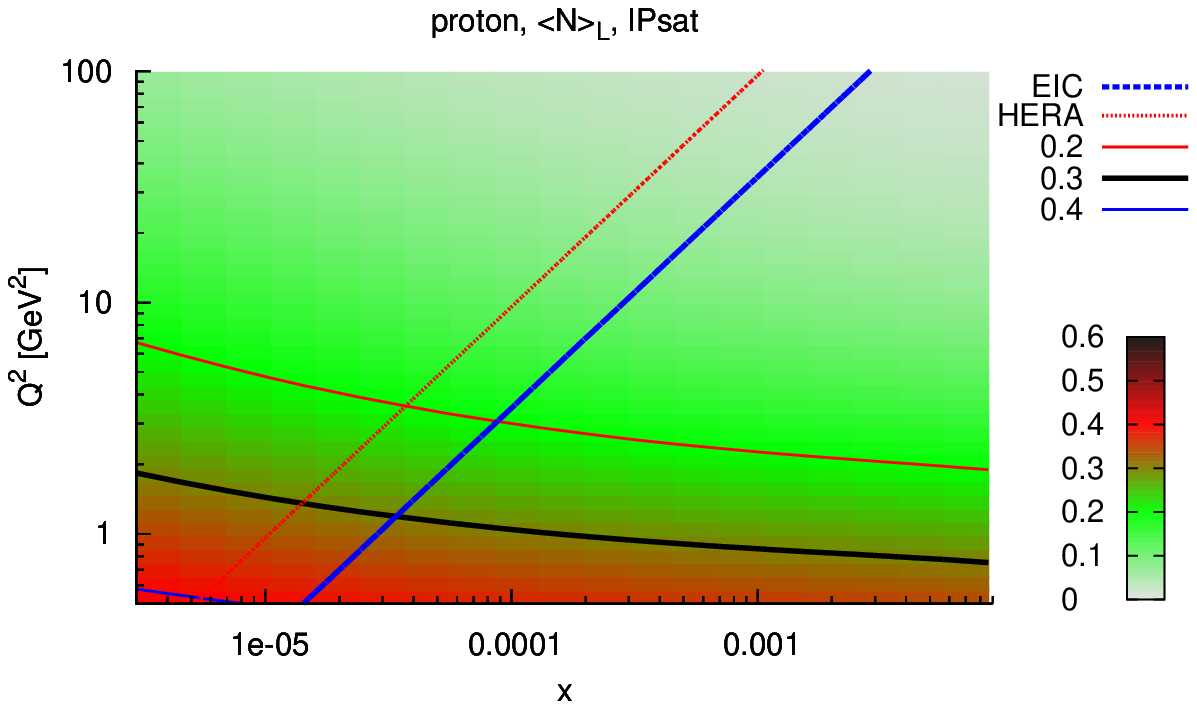}
\hfill\includegraphics[width=0.49\textwidth]{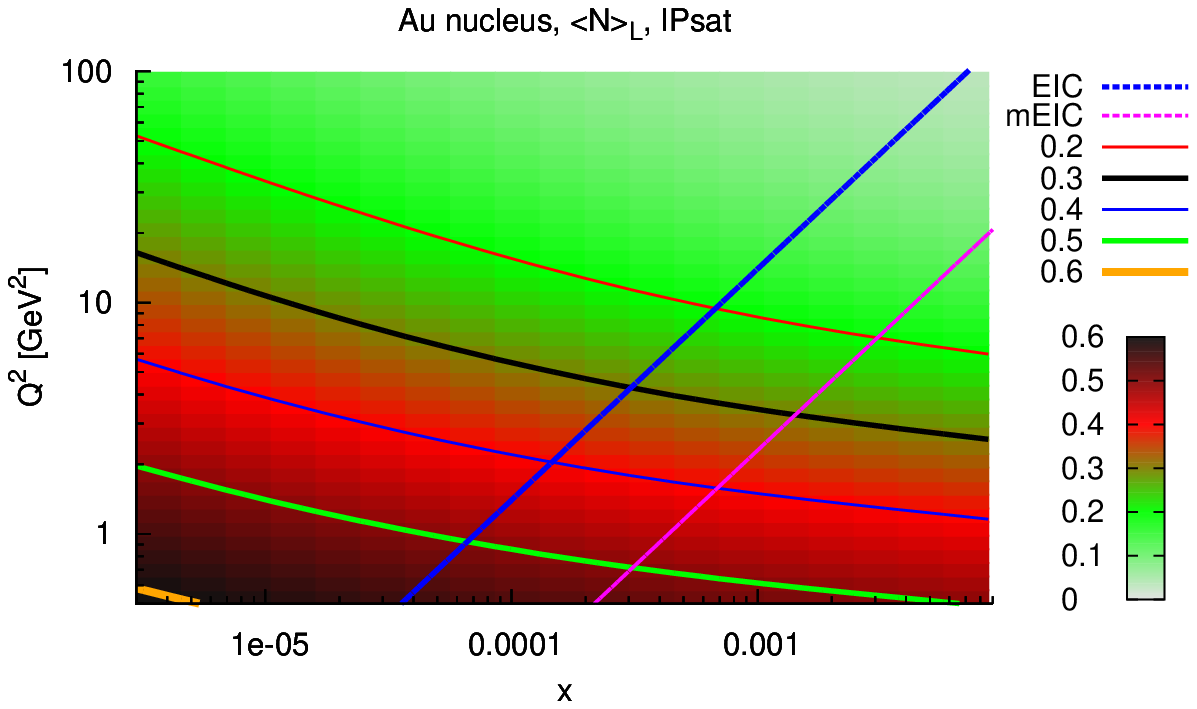}}
\vskip-0.8cm
\centerline{\includegraphics[width=0.49\textwidth]{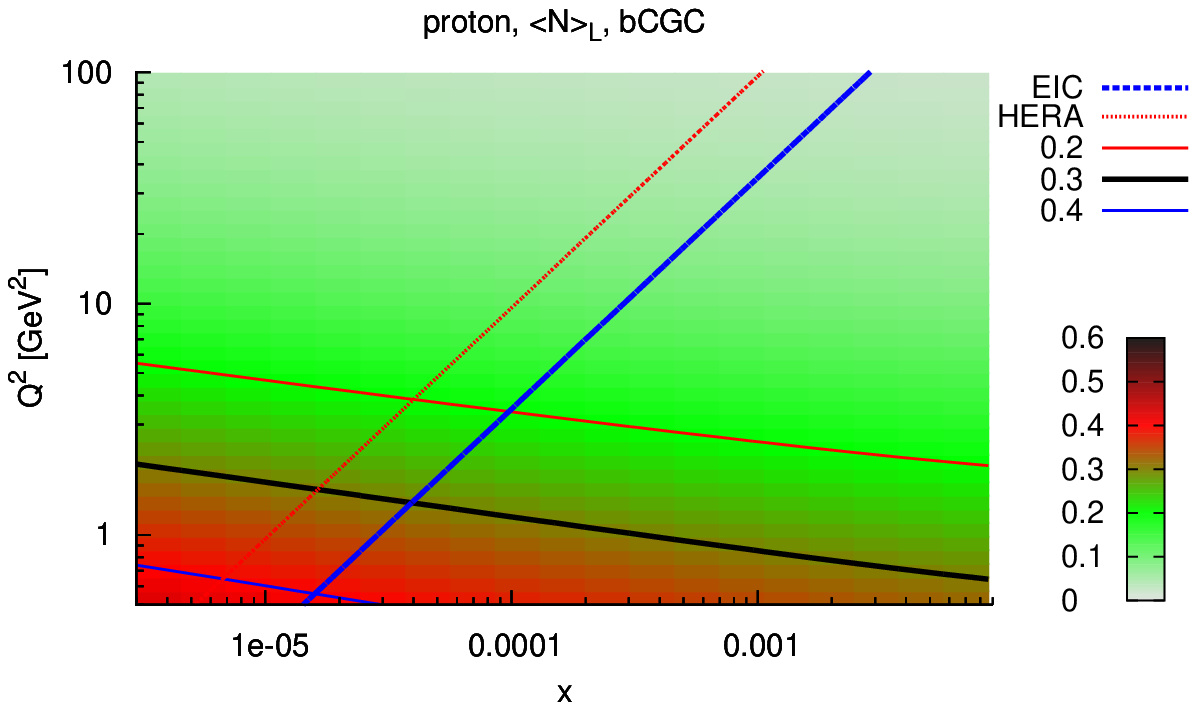}
\hfill\includegraphics[width=0.49\textwidth]{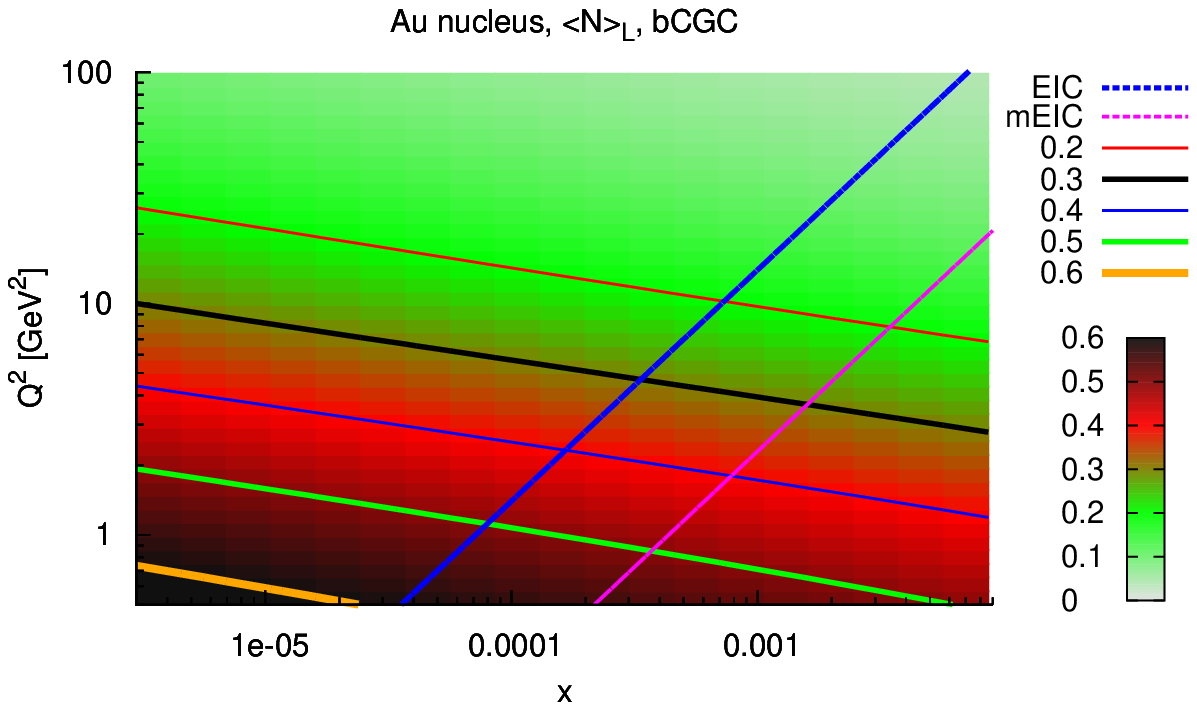}}
\vskip-0.6cm
\caption{\small Longitudinal mean scattering 
amplitude $\langle \mathcal{N} \rangle_L$ for a proton (left) and
a  a gold nucleus (right) with the IPsat parametrization (first row)
and bCGC parametrization (second row).}
\label{fig:meannL}
\label{fig:meannLb}
\end{figure}


\noindent{\bf The mean scattering amplitude:}  An alternative way of assessing the typical values of the scattering amplitude
is to calculate its expectation value weighted by the cross section of a 
particular process. We thus define the mean scattering amplitude as
\begin{equation}
\langle\mathcal{N} \rangle_{T,L}
= \frac{
 \int\! \mathrm{d}^2 {\mathbf{r}_T} \int_0^1 \! \mathrm{d} z \left| \Psi^{\gamma^*}_{L,T}
\right|^2
\int  \mathrm{d}^2 \mathbf{b}_T
 \mathcal{N}^2 (x,{\mathbf{b}_T},{\mathbf{r}_T})
}{
 \int\! \mathrm{d}^2 {\mathbf{r}_T} \int_0^1 \! \mathrm{d} z \left| \Psi^{\gamma^*}_{L,T}
\right|^2
\int  \mathrm{d}^2 \mathbf{b}_T
 {\mathcal{N}}(x,{\mathbf{b}_T},{\mathbf{r}_T})
} 
.
\end{equation}
This will yield a value between 0 and 1 for all points in the 
$Q^2,x$-plane. Note that although in 
principle $\langle \mathcal{N} \rangle$ varies between $0$ and $1$, the maximal 
value for a Gaussian $\mathbf{b}_T$-distribution, which describes the proton very 
well, is only $1/2$. The longitudinal and transverse structure functions probe 
a slightly different distribution of dipole sizes $r$, with the longitudinal 
structure function showing a stronger $Q^2$-dependence. 
The same quantities can  easily be computed also for charm quarks only.

\begin{figure}[tbh]
\centerline{\includegraphics[width=0.49\textwidth]{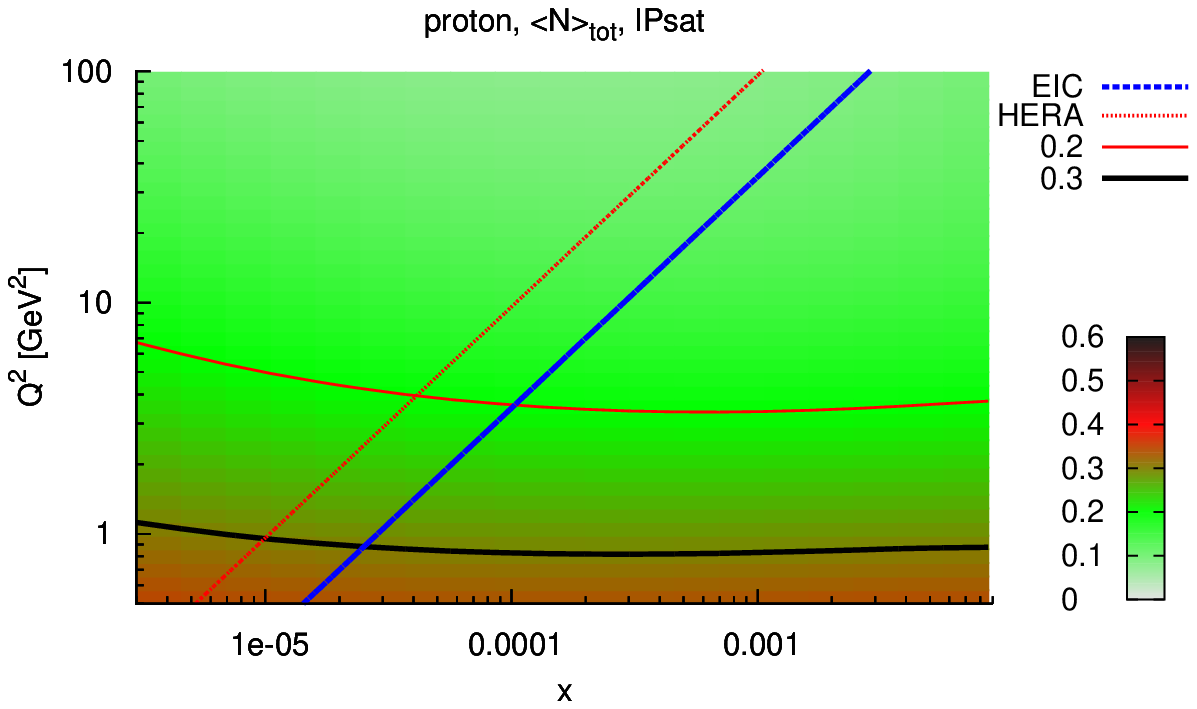}
\hfill\includegraphics[width=0.49\textwidth]{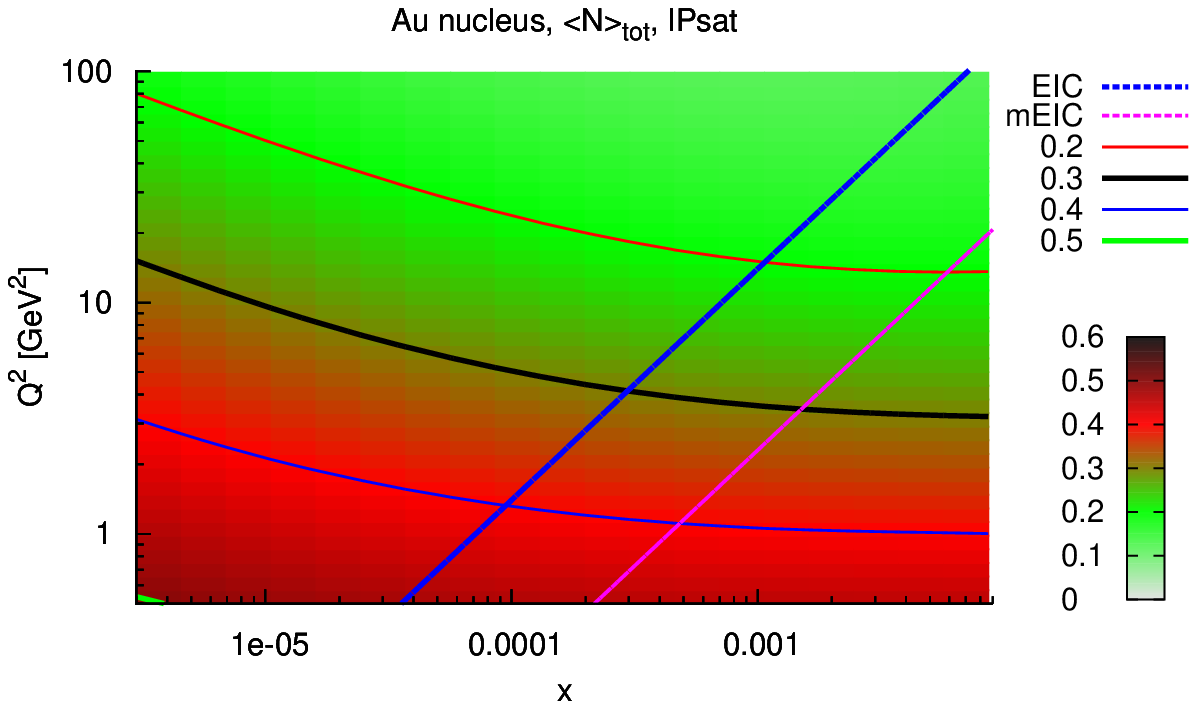}}
\caption{\small The mean scattering  amplitude $\langle \mathcal{N} \rangle_\mathrm{tot}$ 
for the total cross section for a proton (left) and
a  a gold nucleus (right) with the bCGC parametrization.}
\label{fig:meannt}
\end{figure}

Figure~\ref{fig:meannL} shows the mean scattering amplitude probed in the longitudinal 
total cross section in a proton and a gold nucleus in the IPsat model. 
 The characteristic
feature of the eikonalized DGLAP-evolved gluon distribution in this parametrization
is the fact that the $x$-dependence becomes faster at higher energies.
The same quantity for the bCGC 
cross section is plotted in fig.~\ref{fig:meannLb}.
 Here one sees
the characteristic constant energy dependence
$Q_\mathrm{s}^2 \sim x^{-\lambda}$ in the bCGC parametrization leading to 
straight lines of constant $\mathcal{N}$ in a log-log plot.
The amplitude weighted by the total cross section is shown in 
Fig.~\ref{fig:meannt} for the IPsat parametrization. It shows a slower 
$Q^2$-dependence than the longitudinal one,  connected with the
well-known fact that the longitudinal structure function is more
sensitive to higher-twist effects than the total one.

In all plots for protons we have shown the  kinematical limits 
for HERA and the EIC ($325\gev$ proton on $30\gev$ electron with $y< 0.9$)
and  in the nucleus plots for  the EIC 
($130A\gev$ nucleus on $30\gev$ electron with $y< 0.9$)
and lower energy mEIC option
($130A\gev$ nucleus on $5\gev$ electron with $y< 0.9$).
The comparison between nuclei and protons is striking. In the IPsat 
parametrization, as is typical of DGLAP evolution, the energy dependence
at the initial small $Q^2$-scale is very slow. Thus the lower energy of the EIC
compared to HERA would be insignificant in face of the effect of using nuclei.
A value of $\langle \mathcal{N}\rangle_\mathrm{tot}$ of $0.3$ could, for example,
be reached at $Q^2=4\gev^2$ at the EIC vs. $Q^2=1\gev^2$ at HERA; much more safely
in the weak coupling regime. With nuclei the 
EIC could, at $Q^2=1\gev$, reach values of $\langle \mathcal{N}\rangle_L \approx 0.5$
that are simply inaccessible in an ep collider at practically any energy
for an approximately Gaussian proton profile.


\ \\ \noindent{\it Acknowledgments:}
M.~Diehl came up with the idea of visualizing the strength of the nonlinear effects in the way presented here.

\subsubsection{Nuclear PDFs and deviations from DGLAP evolution}
\label{sec:AGR-DGLAPdev}

\hspace{\parindent}\parbox{0.92\textwidth}{\slshape
 Alberto Accardi, Vadim Guzey and Juan Rojo}
\index{Accardi, Alberto}
\index{Guzey, Vadim}
\index{Rojo, Juan}


In this contribution we present a preliminary
analysis which aims at determining the potential
of the EIC to measure gluon shadowing and
anti-shadowing and its sensitivity to
saturation dynamics. 


\begin{wrapfigure}{r}{0.45\columnwidth}
\centerline{\includegraphics[width=0.4\columnwidth]{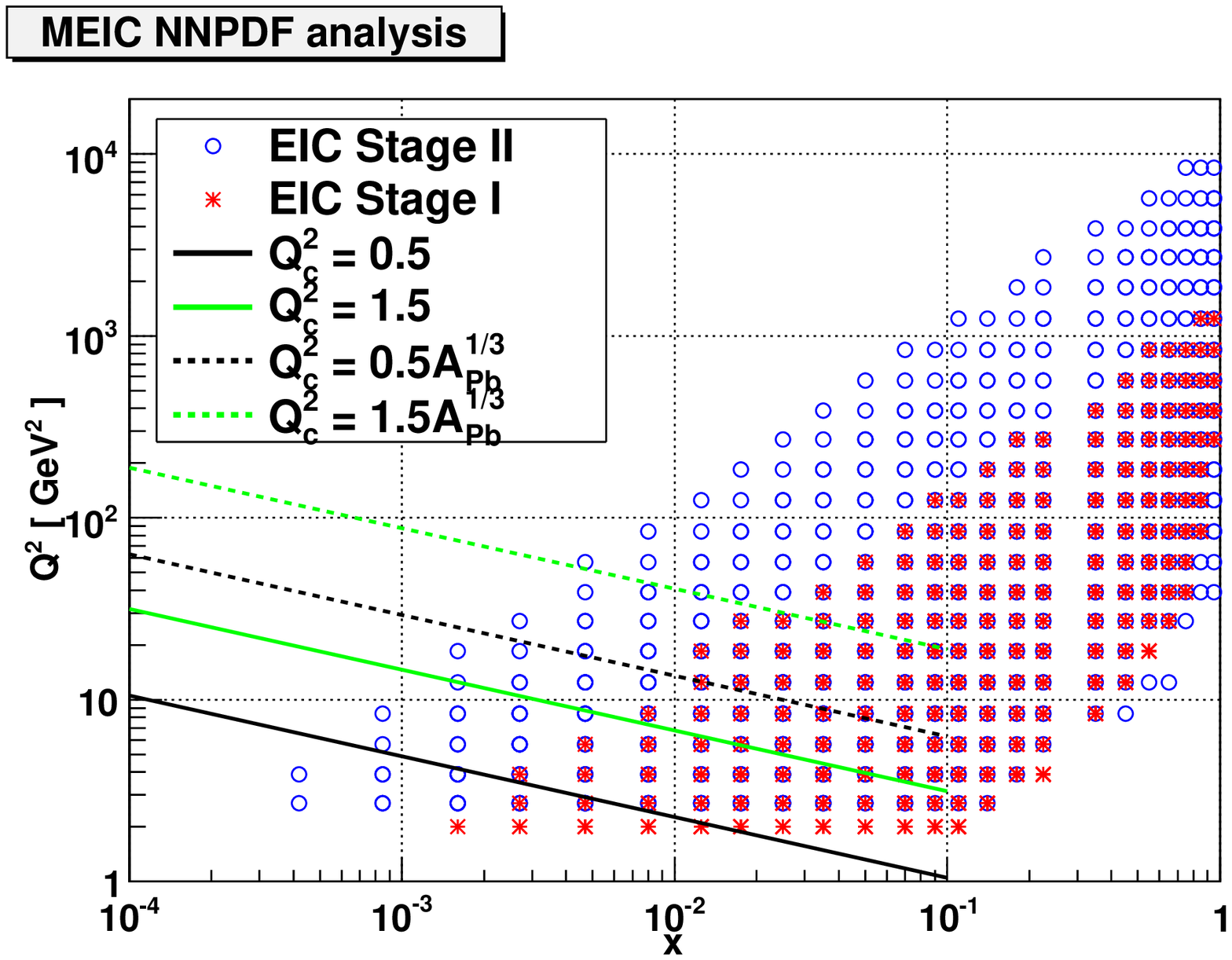}}
\caption{\small \label{fig:kin-meic} Kinematical coverage of the pseudo-data included in the NNPDF analysis of the EIC e+Pb cross-sections, both for stage I and for stage II.  Possible kinematical cuts relevant to the study of the onset of non--linear phenomena are also shown.}
\end{wrapfigure}

The input for this analysis is the EIC pseudo data for the inclusive DIS
cross section in two scenarios,
a medium  energy EIC ($\sqrt{s}=12,17,24,32,44$ GeV, denoted by
stage I) and a full energy EIC ($\sqrt{s}=63, 88, 124$ GeV, stage
II), with $0.004 < y < 0.8$ in either case.
The kinematic coverage is summarized in Fig.~\ref{fig:kin-meic}. 
The pseudo-data was generated starting from $e+p$ and $e+n$ cross
sections computed using the central values of the NNPDF2.0 parton
distributions~\cite{Ball:2010de}. An integrated 
luminosity of 4 fb$^{-1}$ was assumed for all energies, and the
pseudo-data has been
corrected for the expected statistical fluctuations. For most of
the $x$ range the resulting statistical errors are negligible compared
to the assumed 2\% systematic error.
Nuclear effects have been included in a $K$-factor approximation,
so that the longitudinal and transverse cross sections 
in $^{208}$Pb can be
expressed in terms of the proton cross sections as
\begin{equation}
  \sigma_{T,L}^{\rm Pb}\left( x,Q^2,y \right) =K^{\lambda}_{T,L}\left( x,Q^2,y \right)
  \sigma_{T,L}^{\rm p}\left( x,Q^2,y\right) \ ,
\label{eq:sigma_TL}
\end{equation}
where the label $\lambda$ sets the intensity of the assumed saturation
effects, and $\lambda=1$ corresponds to the nominal saturation in 
the IP Non-sat model~\cite{Kowalski:2003hm}.
In particular, the $K$-factor in Eq.~(\ref{eq:sigma_TL}) is given by the following piece-wise expression. For small $x$, $x \leq 0.01$, 
\begin{equation}
  K^{\lambda}_{T,L}=\frac{2}{\langle \sigma_{q \bar q}\rangle_{T,L}}
    \int d^2 b \left\langle 
    \left(1-e^{-\lambda \frac{1}{2} A \sigma_{q \bar q} T_A(b)}\right)
    \right\rangle_{T,L}  \, , 
\label{eq:K_factor}
\end{equation}


where $\sigma_{q \bar q}$ is the dipole cross section in the IP
Non-sat model (we assume for simplicity that in
the EIC kinematic range there is no saturation at the proton level,
and search for the nuclear medium-induced saturation); $T_A(b)=\int dz
\rho_A(b,z)$, where $\rho_A(b,z)$ is the nuclear density normalized to unity;
the brackets $\langle \dots \rangle_{T,L}$ stand for the integration
with the wave function squared of a virtual photon with
transverse or longitudinal polarization, respectively. 
In the $0.01 \leq x \leq 0.1$ interval, we assume that $K^{\lambda}_{T,L}$
increases linearly from the value given
by Eq.~(\ref{eq:K_factor}) at $x=0.01$ up to $K^{\lambda}_{T,L}=1$ at $x=0.1$.
For $x >0.1$, we assumed that $K^{\lambda}_{T,L}$ is equal to the ratio of the nuclear to
free nucleon structure functions, $F_{2A}(x,Q^2)/[AF_{2N}(x,Q^2)]$, which is given by the leading-order parameterization
of Ref.~\cite{Eskola:1998df}

Nuclear parton distributions are then determined by a Next-to-Leading
Order QCD fit of the pseudo-data within the NNPDF
framework~\cite{Ball:2008by,Ball:2010de}.
The kinematic cuts used to ensure the validity of DGLAP evolution are
$Q^2\ge 2$ GeV$^2$ and $W^2\ge$ 12.5 GeV$^2$.
In this preliminary study, we consider pseudo-data for Pb targets only, and
postpone discussion of the dependence of the nuclear PDFs on $A$ to a
future investigation. 
In the collinear factorization approximation, Pb structure
functions are related to Pb parton distributions in the same way as
in the proton case (see Section~\ref{sec:Sassot-nPDFstatus} and
Ref.~\cite{Forte:2010dt}).  
We also assumed for simplicity the Pb nucleus to be isoscalar, so
that the structure functions depend only on three independent nuclear
PDFs: the singlet quark PDF, $\Sigma^{\rm Pb}\left( x,Q^2\right)$, the
gluon PDF $g^{\rm Pb}(x,Q^2)$, and the strange PDF; the latter was
furthermore set to be a fixed fraction of the singlet PDF.

\begin{figure}[ht]
  \centering
  \includegraphics[width=0.8\textwidth]{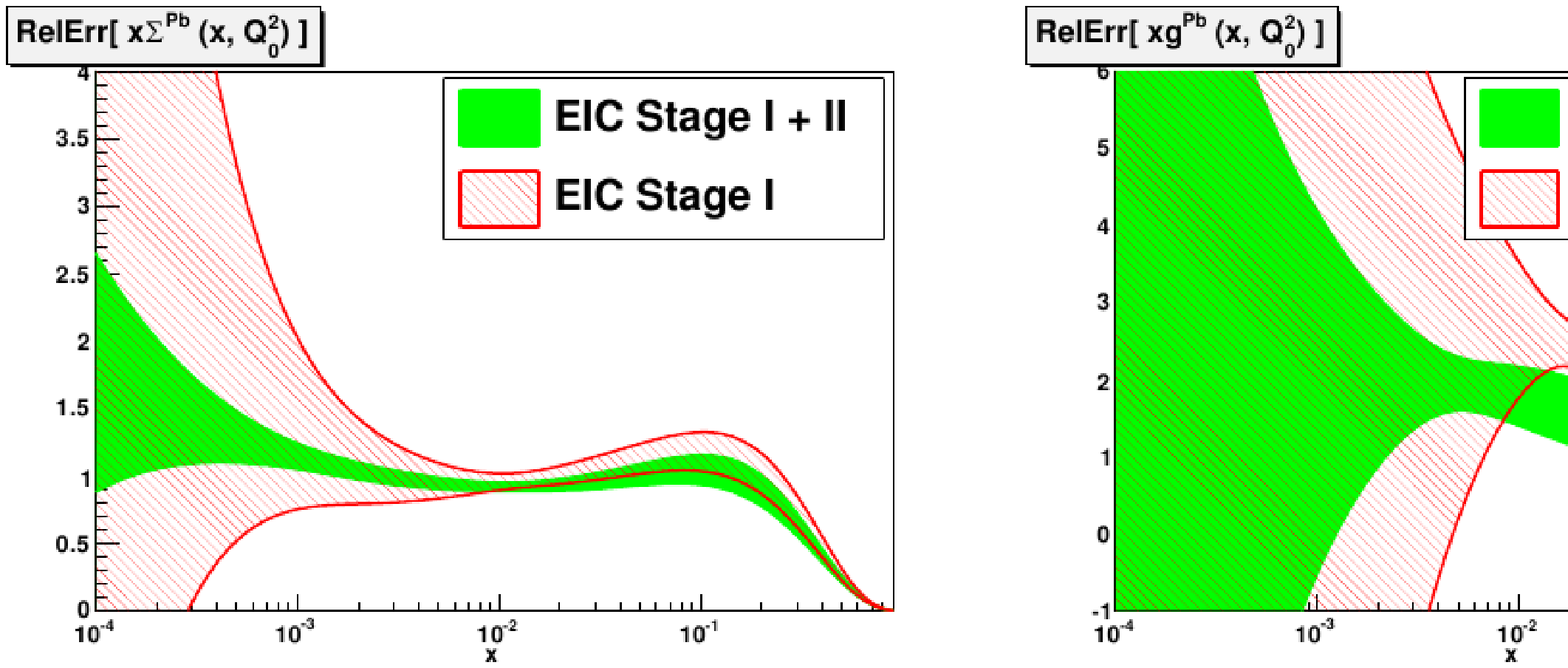}
  \caption{\small \label{fig:pb-pdfs} 
    The quark singlet (left plot) and the gluon PDFs (right plot) in Pb 
    at the initial evolution scale $Q_0^2=2$ GeV$^2$, for stage
    I and stage I+II.
  }
\end{figure}

\begin{figure}[t]
  \centering
  \includegraphics[width=0.37\textwidth]{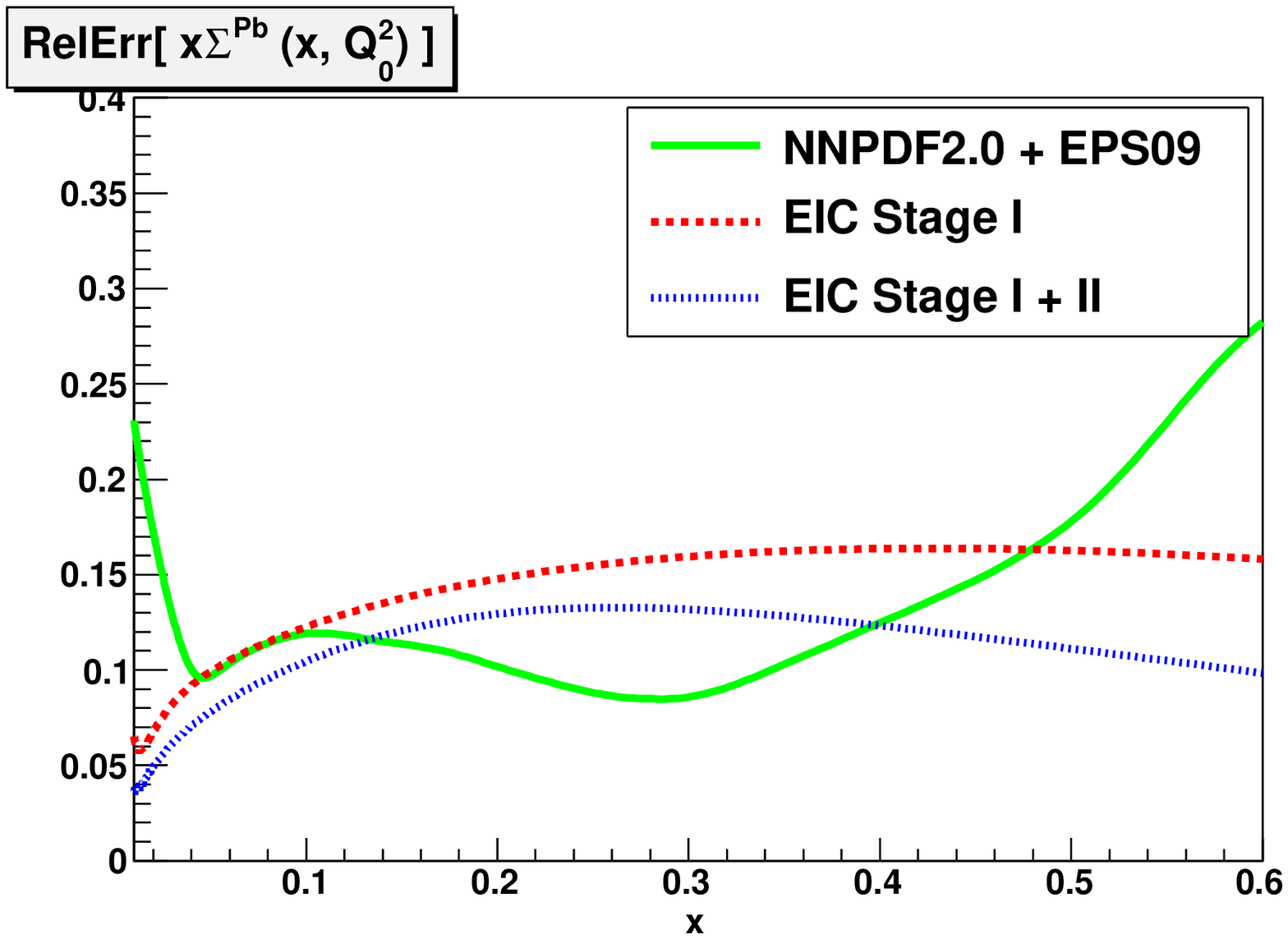}
  \includegraphics[width=0.37\textwidth]{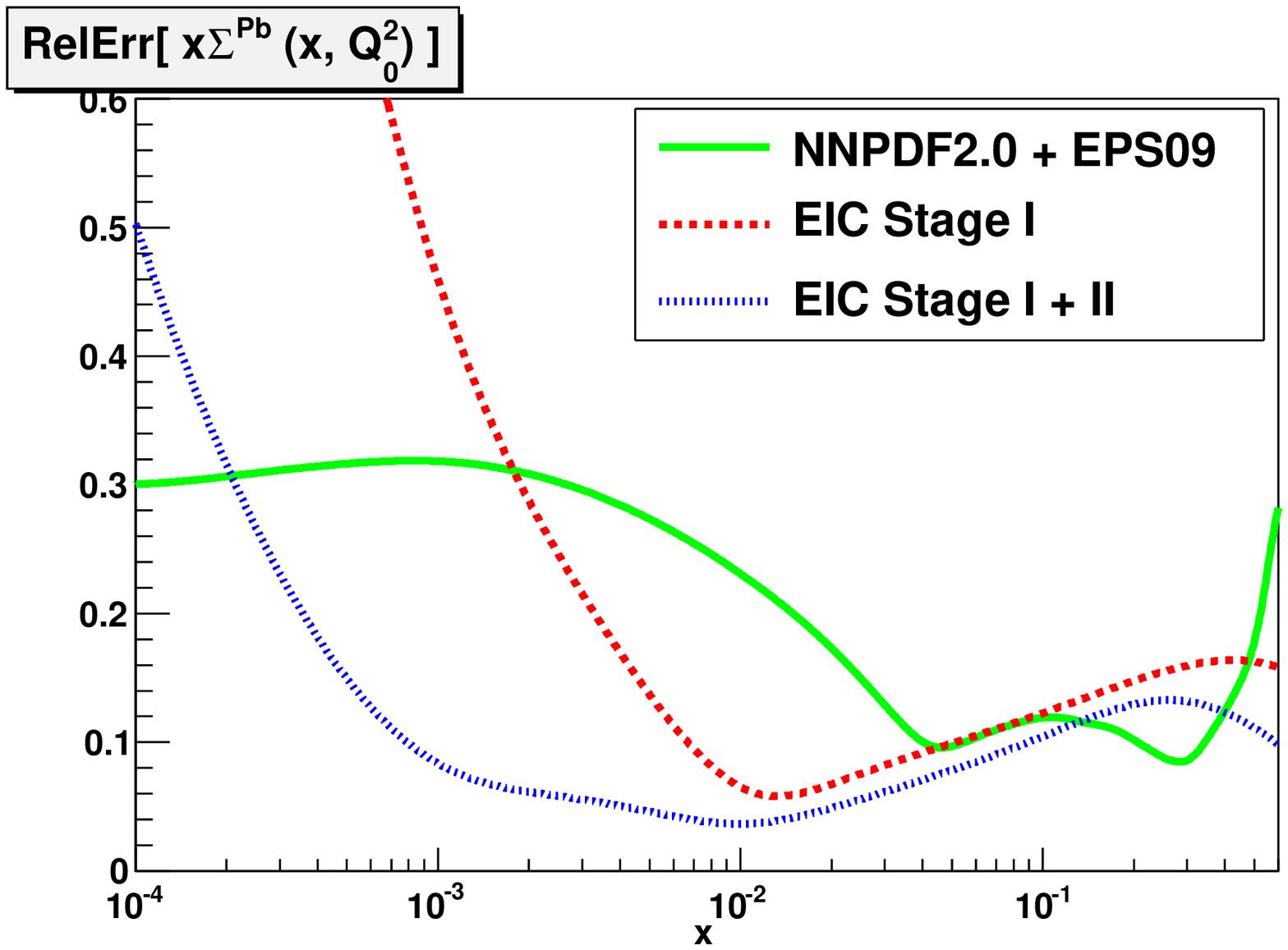}
  \includegraphics[width=0.37\textwidth]{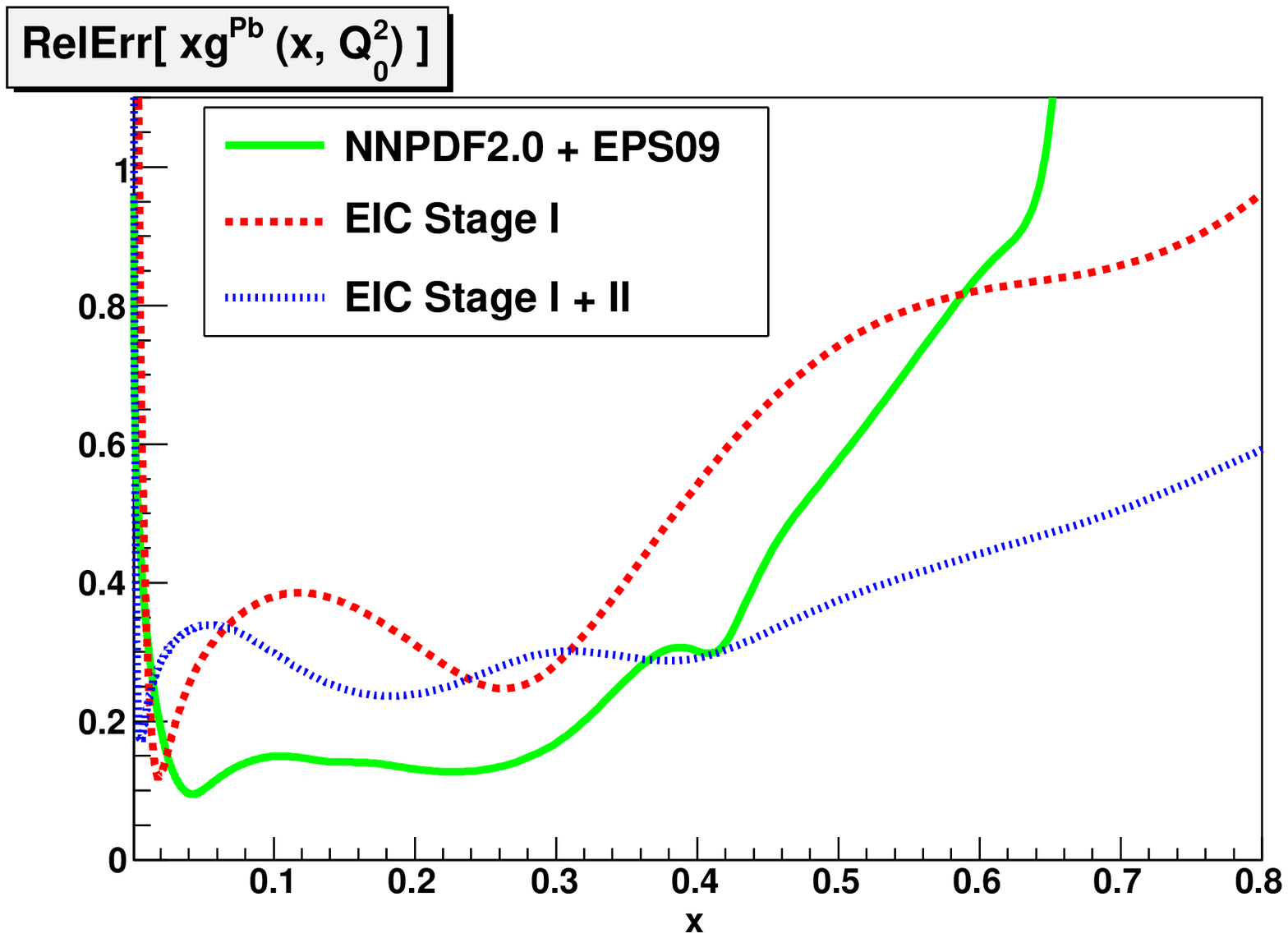}
  \includegraphics[width=0.37\textwidth]{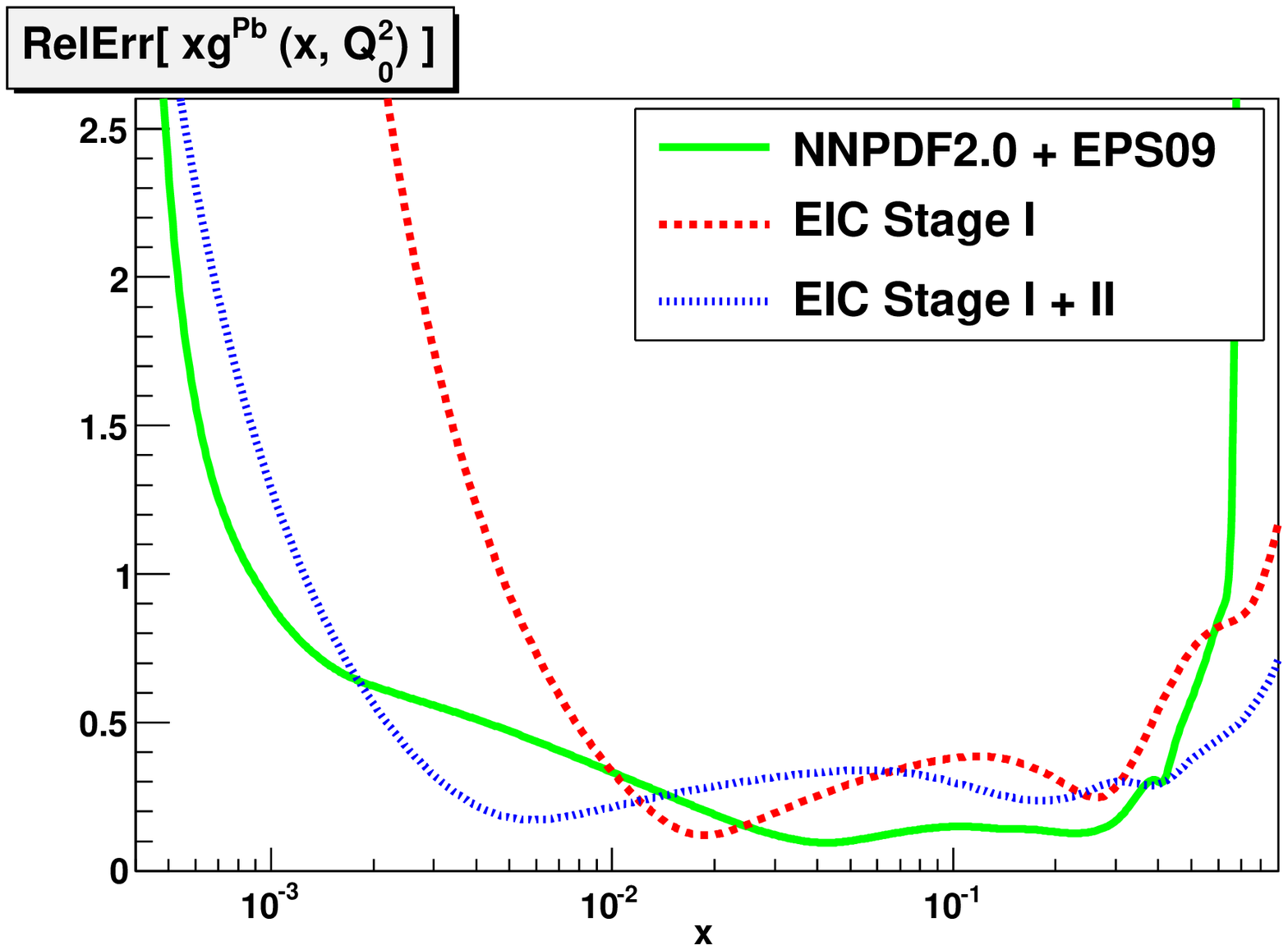}
  \caption{\small \label{fig:pb-pdfs-rel} 
    The relative uncertainty in the
    quark singlet (two upper panels) and the gluon PDFs in Pb 
    (two lower panels)
    at the initial evolution scale $Q_0^2=2$ GeV$^2$, with stage I and
    stage I+II data. Results are shown on linear
    (left plot) and logarithmic (right plot) scales. For reference,
    the analogous results for the Pb PDFs using NNPDF2.0+EPS09
    parametrizations are also shown.
  }
\end{figure}

Now we discuss some preliminary results of the nuclear
PDF fits. We show in Fig.~\ref{fig:pb-pdfs} 
the singlet and the gluon PDFs at the initial scale $Q^2=2$ GeV$^2$ 
obtained using only stage I data for e+Pb collisions, and then adding the stage II data.
To illustrate the accuracy that the EIC can reach in the determination
of nuclear PDFs we show in Fig.~\ref{fig:pb-pdfs-rel} 
their relative uncertainties alongside those of the proton's
NNPDF2.0~\cite{Ball:2010de} combined with those of 
the EPS09 nuclear modifications~\cite{Eskola:2009uj} 
for $^{208}{\rm Pb}$, which allows a comparison of the 
relative error bands. Since the restrictive EPS09 parametrization may
underestimate the nuclear uncertainties outside the region where
data is presently available, notably at $x\lesssim0.01$, we added the
relative NNPDF2.0 and EPS09 relative uncertainties linearly for a
conservative estimate of the total uncertainty.

The measurement of the nuclear modifications of the gluon
are one of the most important measurements at the EIC, as
this quantity is essentially unknown from present data.
Inclusive cross sections are sensitive to the gluon distribution 
both via scaling violations and, to a lesser extent,
through the longitudinal structure function accessed through the
proposed $\sqrt{s}=12-124$ GeV energy scan. From 
Fig.~\ref{fig:pb-pdfs} we see that one can determine,
with reasonable accuracy, the gluon shadowing down to $x\sim 10^{-3}$
in stage II and  down to $x\sim 10^{-2}$ in stage I.
The better capabilities of stage II stem both from its
greater lever arm in $Q^2$ and its coverage of smaller
values of $x$, see Fig.~\ref{fig:kin-meic}. In particular,
the precision of the Pb gluon in Stage II at small $x$
is comparable to estimates from
global proton fits. 
On top of this, at the EIC it will be possible to study gluon
anti-shadowing, EMC and Fermi motion effects with much better 
accuracy than afforded by current global nuclear fits (see 
Sections \ref{sec:Sassot-nPDFstatus} and \ref{sec:hkn-nPDFs}.
We can also see
that EIC will measure accurately the sea quark shadowing, and
that nuclear modifications of light quarks at large $x$ could be
measured a precision similar or even better than for the proton case.  

This analysis was based on the validity of collinear
factorization for nuclei, and the validity of linear DGLAP evolution in
$Q^2$. However, at small enough $x$ and $Q^2$, deviations
from linear fixed order DGLAP evolution are
expected to appear, e.g., due to small-$x$ re-summation 
effects~\cite{Altarelli:2008aj} or
gluon saturation, see Section \ref{sec:partsat}.
In heavy nuclei, the effects due to gluon saturation
are boosted to higher $Q^2$ and $x$ by the atomic number; one then
has the possibility of experimentally separating small-$x$ and
saturation effects, which is not be possible with HERA $e+p$ data.

In Refs.~\cite{Caola:2009iy,Caola:2010cy} a general 
strategy was presented to quantify potential deviations from NLO DGLAP
evolution, which was then applied to proton
HERA data.
In a global PDF fit, deviations from DGLAP in the data can be hidden
in a distortion of parton distributions; however, these can be singled
out by determining undistorted PDFs from data in regions where
such effects are 
expected to be
small.  In more detail, one can fit PDFs 
using data at large $x$ and $Q^2$, where DGLAP is likely to hold with
high accuracy, and then evolving them down in the $Q^2$  region 
where deviations are expected to arise. DGLAP deviations can then be
quantitatively determined by comparing calculations to data in this
region, which were not used in the PDF determination. 

This approach can be applied as well to the nuclear case. 
From simple theoretical arguments about the energy and $A$ dependence
of the saturation scale (see Section \ref{sec:partsat}), 
we expect deviations from linear evolution to appear when
$Q^2 \lesssim \bar Q^2 \left( A \bar x/x \right)^{\frac13}$, where $\bar x$ is a reference value, say $\bar x=10^{-3}$, and 
$\bar Q^2$ is the scale where DGLAP evolution at $\bar x$ would be
broken in the proton. 
Note however that the $A$-dependence of  the saturation scale may in fact be
tamed by the leading twist nuclear shadowing, see Section \ref{sec:LT_nuclear_shadowing}.
While saturation models may give an indication of the value of
$\bar Q^2$, we wish to determine this scale in a model independent way
as the scale at which deviations from DGLAP evolution can be
detected from EIC nuclear target (pseudo-)data. The unsafe region for DGLAP evolution
can also be written as $Q^2 \lesssim Q^2_c x^{-\frac13}$ with 
$Q^2_c$ some constant setting the strength of the deviations from
DGLAP. In Refs.\cite{Caola:2009iy,Caola:2010cy} the range $Q_c^2\in\left[
0.5,1.5\right]$ GeV$^2$ was considered for the proton case; in the nuclear
case this range should be rescaled by a factor $A_{\rm
  Pb}^{1/3}\approx 6$. Typical values of these kinematical cuts for
the Pb nucleus are shown in  Fig.~\ref{fig:kin-meic}. 

\begin{figure}[h]
  \centering
  \includegraphics[width=0.4\textwidth]{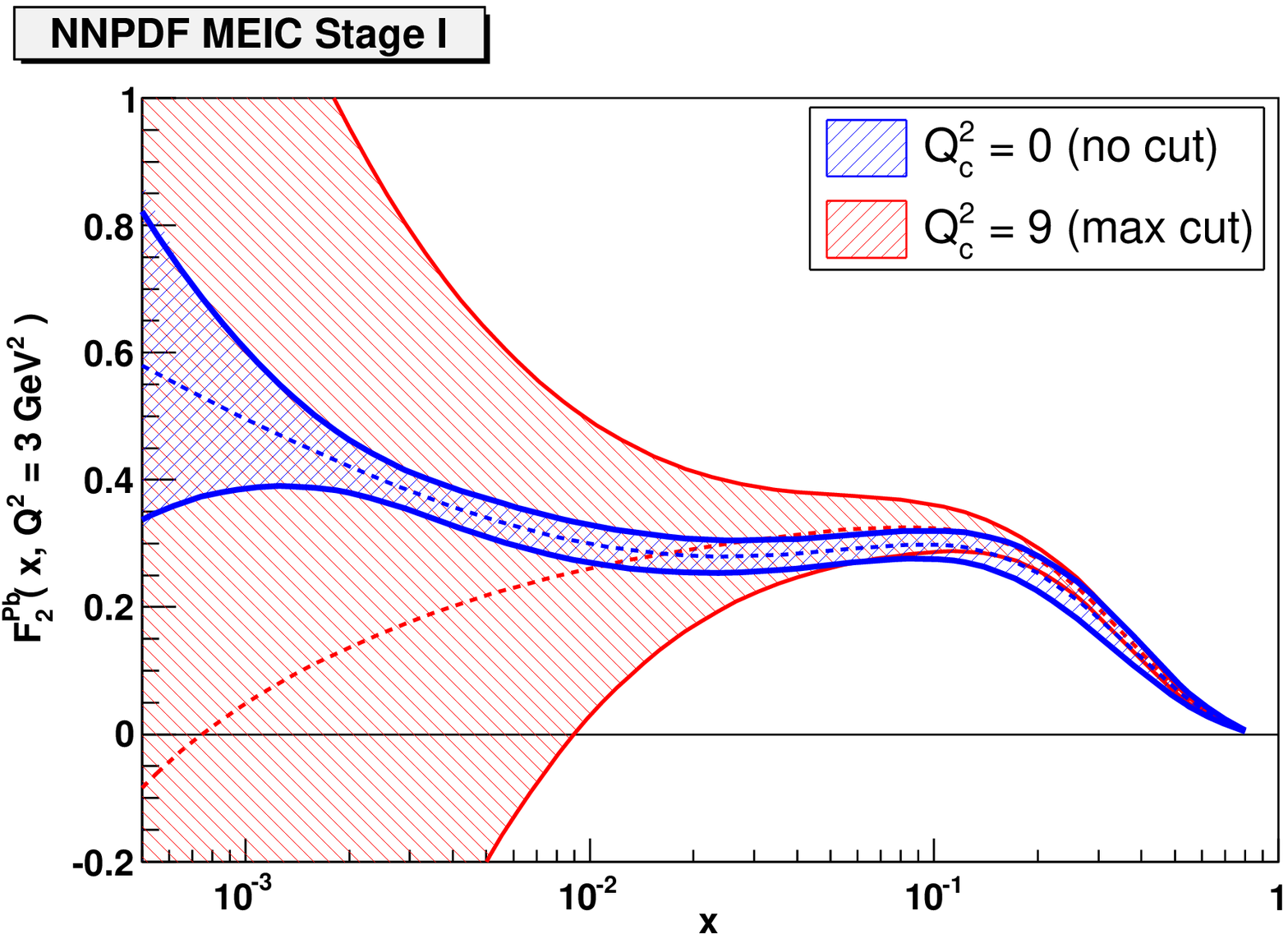}
  \includegraphics[width=0.4\textwidth]{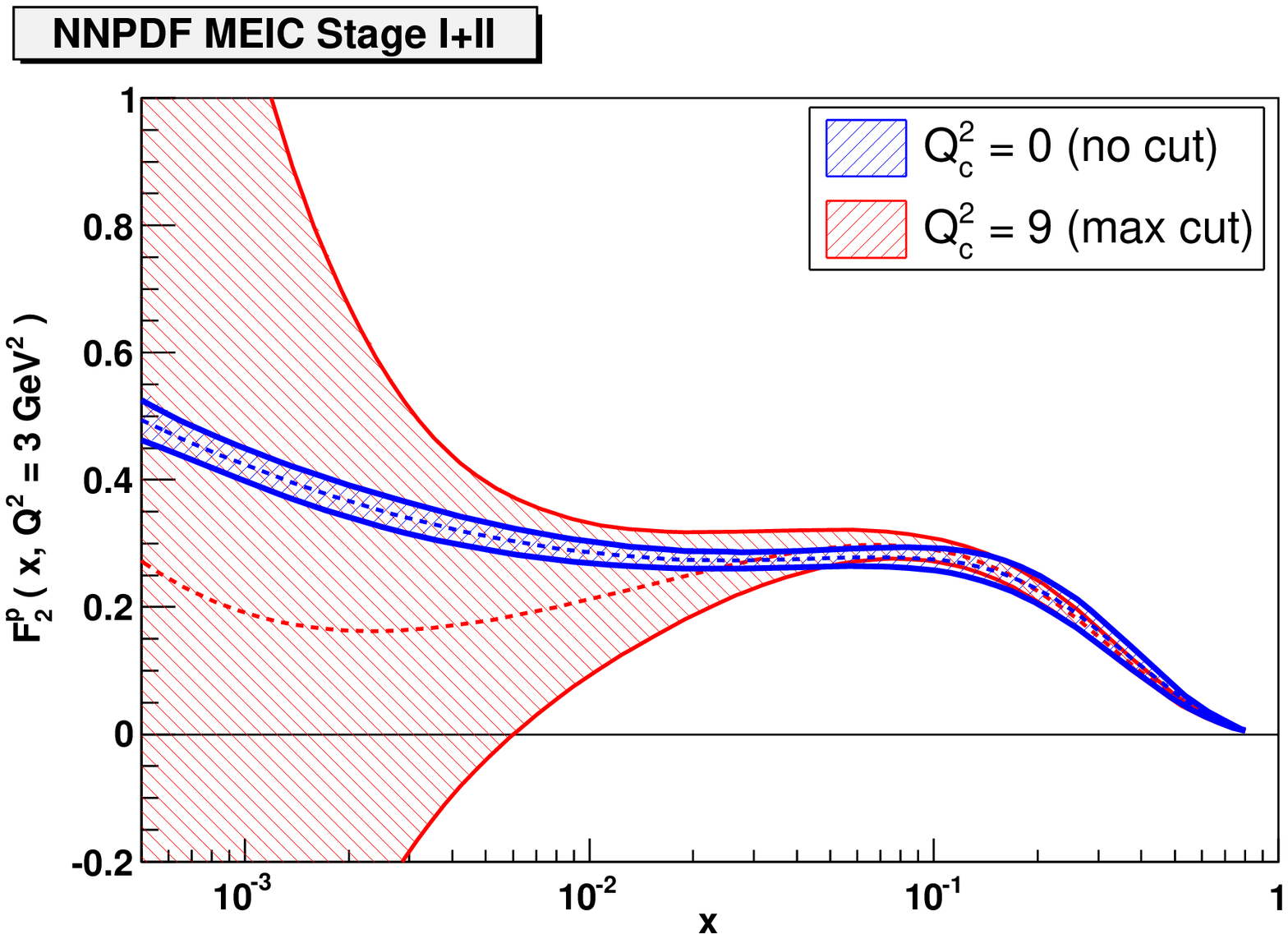}
  \caption{\small \label{fig:f2p-cut} 
    The Pb structure function $F_2^{\rm Pb}(x,Q^2)$ at $Q^2=3$ GeV$^2$
    from the analysis of the EIC stage I (left plot) and
    stage I+II (right plot) simulated data with $\lambda=1$, without
    kinematical cuts and with cuts using $Q_c^2=1.5A_{\rm Pb}^{1/3} \sim 9$.
  }
\end{figure}

We show in Fig.~\ref{fig:f2p-cut} a representative result
of the fits to the EIC pseudo-data after applying the cut with $\bar Q^2 =1.5A_{\rm Pb}^{1/3}\sim 9$,
compared to the reference uncut fits to stages I and I+II pseudo-data with $\lambda=1$.
As expected when data is removed the uncertainties in the
physical observables become much larger, but one can still
see a systematic downwards shift in the central value, which is the
signature of the departure from linear 
evolution~\cite{Caola:2009iy,Caola:2010cy}. Note that this signal is
already apparent with stage I data only, although its statistical
significance might be marginal.

We plan to systematically explore the sensitivity of the EIC
to non-linear dynamics using this technique, by 
optimizing the kinematical cuts for different values of the saturation
scale used to generate the pseudo-data, exploit the interplay between
the $F_2^{\text{Pb}}$ and $F_L^{\text{Pb}}$ structure functions, and
quantitatively measuring the statistical significance of the signal.
This will determine in a
fairly model-independent way the smallest saturation scale that can be
detected at the EIC in either stage I or stage II.


\ \\ \noindent{\it Acknowledgments:}
We thank F.~Caola, R.~Ent, S.~Forte and L.~Zhu for discussions and collaboration.

\subsubsection{Constraining the nuclear gluon distribution using inclusive observables}
\label{sec:constraining_victor}

\hspace{\parindent}\parbox{0.92\textwidth}{\slshape
 Victor P. Gon\c{c}alves}
\index{Gon\c{c}alves, Victor P.}

\vspace{\baselineskip}

\begin{wrapfigure}{r}{0.5\columnwidth}
\centerline{\includegraphics[width=0.48\columnwidth]{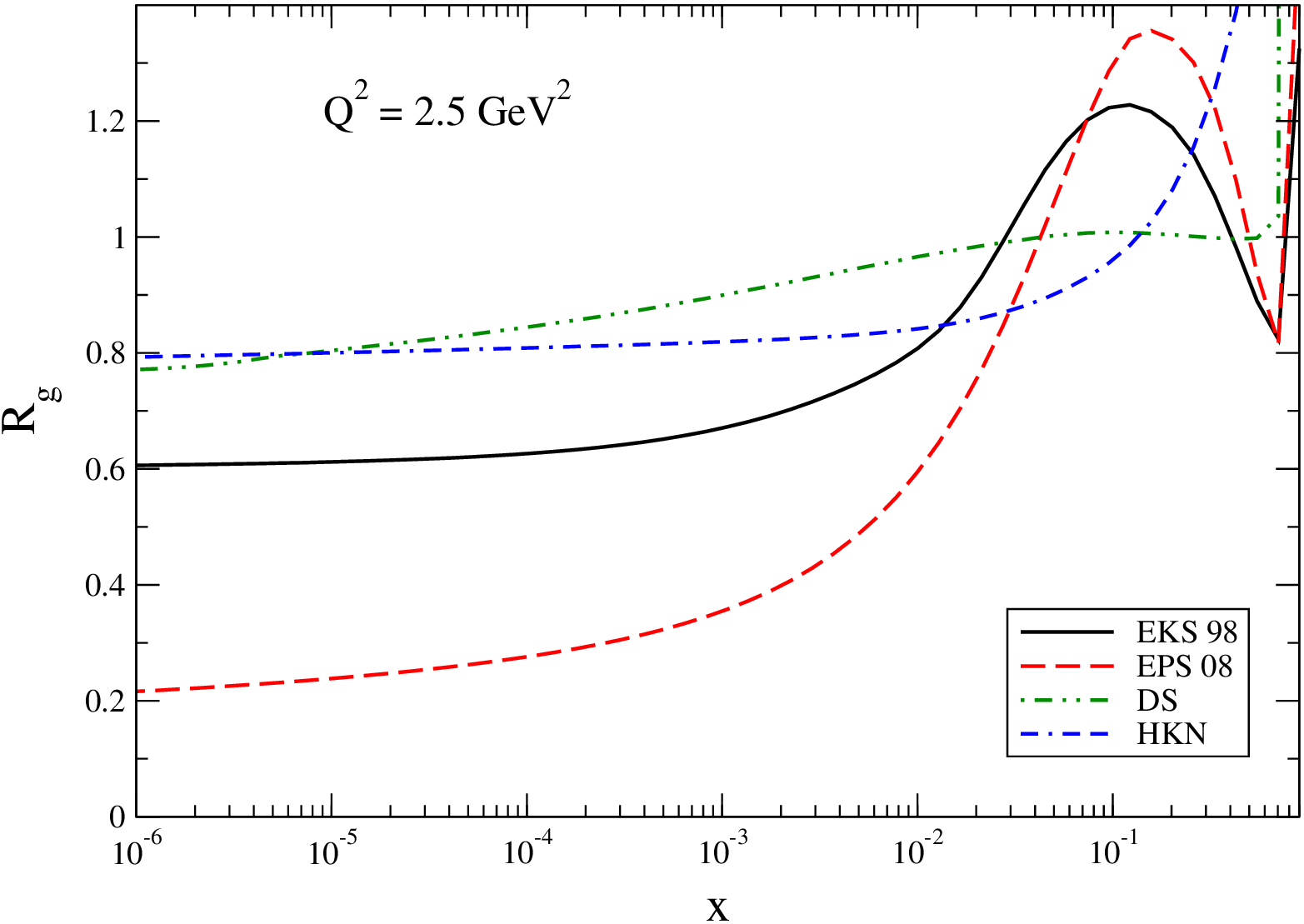}}
\caption{\small \label{fig:ratio_gluons}  The ratio $R_g = xg^A/A.xg^N$ predicted by the EKS, DS, HKN and EPS parametrizations for $A=208$ and $Q^2=2.5$ GeV$^2$.}
\end{wrapfigure}

Data from HERA allow for a  good determination of the gluon density of the proton.  A much harder task has been to determine the gluon  distribution of nucleons bound in a nucleus, the nuclear gluon distribution 
($xg^A (x,Q^2)$).  
Existing data, taken over a wide kinematic range $10^{-5}\,\le\,x\le\,0.1$ and  $0.05\,GeV^2\,\le\,Q^2\le\,100\,GeV^2$, show a systematic reduction of the nuclear structure function $F_2^A(x,Q^2)/A$ with respect to  the free nucleon structure  function $F_2^N(x,Q^2)$. This phenomenon is known as the {\it  nuclear shadowing effect} and is associated to  the modification of  the target parton distributions so that $xq^A(x,Q^2) \, < \,  Axq^N(x,Q^2)$, as expected from a superposition of $ep$ interactions. The modifications depend on the parton momentum fraction: for momentum fractions $x < 0.1$ (shadowing region)
and $0.3 < x < 0.7$ (EMC region), a depletion is observed in the nuclear structure functions. These two regions are
bridged by an enhancement known as antishadowing for $0.1 < x < 0.3$. The experimental data for the nuclear structure function  determine the behaviour of the nuclear quark distributions, while the behaviour of the  nuclear gluon distribution is indirectly determined using the momentum sum rule as a constraint and/or studying the $\log Q^2$ slope of the ratio $F_2^{Sn}/F_2^{C}$.  Currently, the behaviour of $xg^A (x,Q^2)$ at small $x$ (high energy) is completely uncertain as shown in Fig. \ref{fig:ratio_gluons}, where we present  the ratio $R_g = xg^A/(A.xg^N)$, for $A=208$, predicted by four different groups which realize a global analysis of the nuclear experimental data using the DGLAP evolution equations in order to determine the parton densities in nuclei.
In particular, the magnitude of shadowing and the presence or not of the antishadowing effect is completely undefined. 

%

In this contribution we study the behaviour of the nuclear longitudinal structure function  $F_L^A$ and the charm structure function $F_{2 }^{c,A}$ and analyse the possibility to constrain the nuclear effects present in $xg^A$ using these inclusive observables (For more details and references see Ref. \cite{Cazaroto:2008qh}). \\



\noindent{\bf $F_L^A$  and $F_{2 }^{c,A}$ in the collinear formalism:}  The longitudinal structure function in deep inelastic scattering is one of the observables from which the gluon distribution can be unfolded. In the collinear formalism, $F_L$ is described in terms of the Altarelli-Martinelli equation
\begin{eqnarray}
F_L(x,Q^2) = \frac{\alpha_s(Q^2)}{2\pi}\,x^2\, \int_x^1 \frac{dy}{y^3}[\frac{8}{3}\,F_2(y,Q^2) + 4\,\sum_q e_q^2 (1-\frac{x}{y})yg(y,Q^2)]\,\,.
\label{flalta}
\end{eqnarray}
 At small $x$, the second term with the gluon distribution is the dominant one. This  expression can be reasonably approximated  by  $F_L(x,Q^2) \approx 0.3\, \frac{4 \alpha_s}{3 \pi} xg(2.5x,Q^2)$, which  demonstrates  the close relation between the longitudinal structure function and the gluon distribution. Therefore, we expect  
the longitudinal structure function to be sensitive to  nuclear effects.

In order to estimate the charm contribution to the structure function we treat the  charm quark 
as a heavy quark and estimate its contribution  by fixed-order perturbation
theory. This involves the computation of the boson-gluon fusion process. A $c\overline{c}$ pair can be created  by boson-gluon fusion when  the squared  invariant mass  of the hadronic final state is $W^2 \ge 4m_c^2$. Since
$W^2 = \frac{Q^2(1-x)}{x} + M_N^2$, where $M_N$ is the nucleon mass, the charm production  can  occur well below the $Q^2$ threshold, $Q^2 \approx  4m_c^2$, at small $x$. The charm contribution to the proton/nucleus structure function, in leading order (LO), is given by 
\begin{eqnarray}
\frac{1}{x} F_2^c(x,Q^2,m_c^2) = 2 e_c^2 \frac{\alpha_s(\mu^{\prime 2})}{2\pi}
\int_{ax}^1 \frac{dy}{y}\, C_{g,2}^c(\frac{x}{y},\frac{m_c^2}{Q^2})\,g(y,\mu^{\prime 2})  \,\,,
\label{f2c}
\end{eqnarray}
where $a=1+\frac{4m_c^2}{Q^2}$ and the factorization scale $\mu^{\prime}$ is
assumed $\mu^{\prime 2}=4m_c^2$.  $C_{g,2}^c$ is the coefficient function
given by
\begin{eqnarray}
C_{g,2}^c(z, \frac{m_c^2}{Q^2})  & = & \frac{1}{2} \{ [z^2 + (1-z)^2 +z(1-3z)\frac{4m_c^2}{Q^2} - z^2 \frac{8m_c^4}{Q^4}]
ln \frac{1+\beta}{1-\beta} \nonumber \\ & + & \beta[-1 +8z(1-z) -z(1-z)\frac{4m_c^2}{Q^2}]\}\,\,,
\end{eqnarray}
where $\beta= 1 - \frac{4m_c^2 z}{Q^2 (1-z)}$  is the velocity of one of the charm quarks in the boson-gluon center-of-mass
frame.
Therefore, in leading order, ${\cal{O}}(\alpha_s)$, $F_2^c$ is directly sensitive only to the gluon density via the well-known Bethe-Heitler process
$\gamma^*g \rightarrow c\overline{c}$.
The dominant uncertainty in the QCD calculations arises from the uncertainty
in the  charm quark mass. In this contribution we assume $m_c = 1.5\,GeV$. \\


\noindent{\bf The nuclear ratios:}  Let us now study the behaviour of the nuclear longitudinal structure function  $F_L^A$ and the charm structure function $F_{2 }^{c,A}$ and analyze the possibility to constrain the nuclear effects present in $xg^A$ using these inclusive observables. We estimate the normalized ratios
\begin{eqnarray}
R_L(x,Q^2) = \frac{F_L^A(x,Q^2)}{A F_L^p(x,Q^2)} \,\,\,\,\mbox{and}\,\,\,\, R_C(x,Q^2) = \frac{F_{2 }^{c,A}(x,Q^2)}{A F_{2}^{c,p}(x,Q^2)}
\label{rat}
\end{eqnarray}
considering four distinct parametrizations for the nuclear gluon distributions and compare  their  behaviour with those predicted for the  ratio $R_g = xg^A/A xg^N$.

\begin{figure}
\centerline{\includegraphics[width=0.65\textwidth]{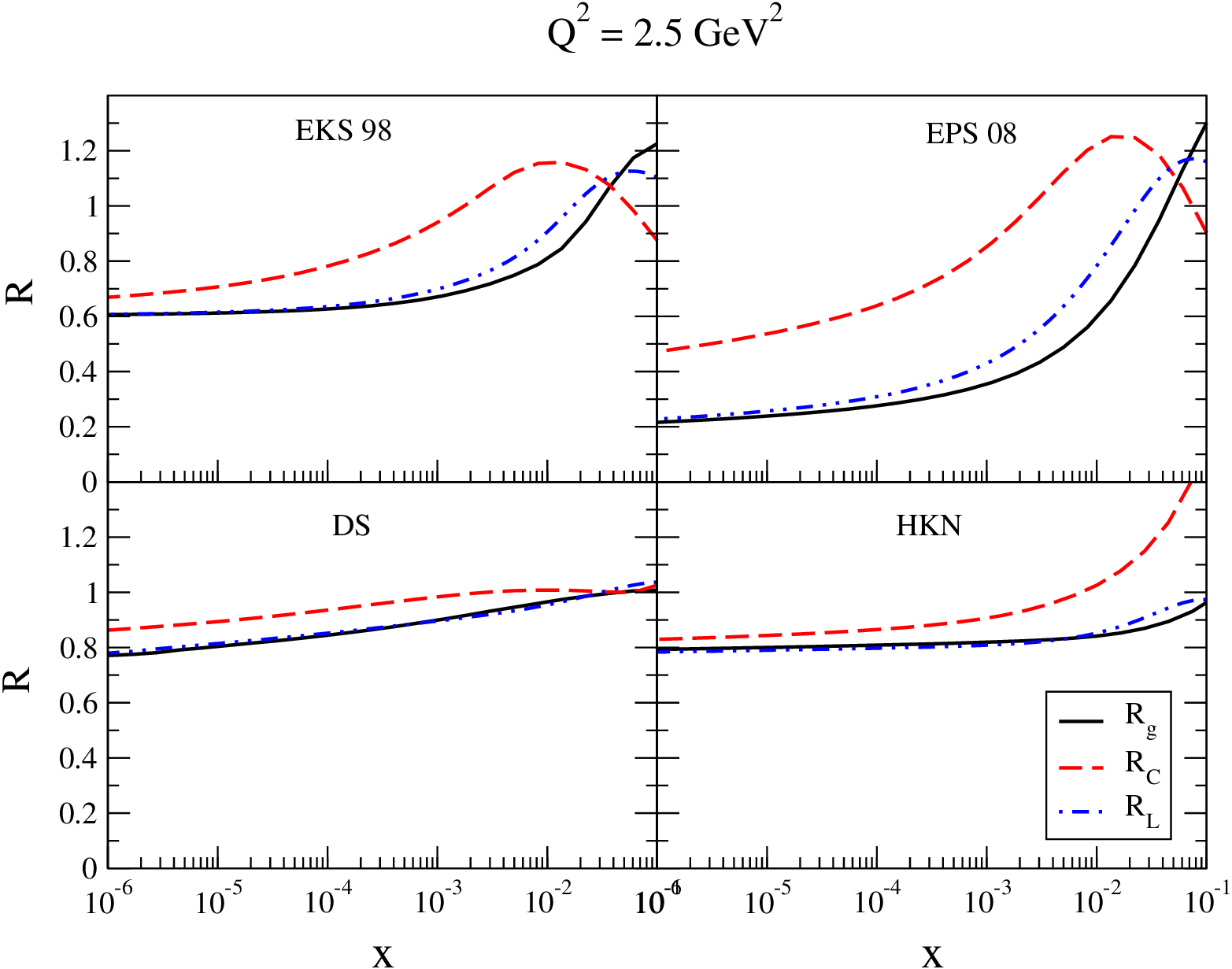}}
\caption{\small Ratios $R_g$,   $R_C$  and  $R_L$ for the
 four considered nuclear parametrizations,  $Q^2 = 2.5$ GeV$^2$ and $A = 208$.}
\label{fig:ratios}
\end{figure}

In Fig. \ref{fig:ratios}   we present our results. Firstly, let us  discuss the small-$x$ region, $x \le 10^{-3}$, determined by shadowing effects. We observe  that $R_L$ practically coincides with $R_g$ for all parametrizations and for the two values of $Q^2$ considered. This  suggests that  shadowing effects can be easily constrained  in an $eA$ collider by  measuring $F_L$. This conclusion is, to a good extent, model independent. On the other hand, the ratio $R_C$ gives  us an upper bound for the magnitude of the shadowing effects. For example, if it is found that $R_C$ is equal to $\approx 0.6$ at $x = 10^{-4}$ and $Q^2 = 2.5$  GeV$^2$ the  nuclear gluon distributions from 
DS and HKN parametrizations are very large and should be modified. 
Considering now the kinematical range of $x > 10^{-3}$ we can analyse the correlation between the behaviour of $R_L$ and $R_C$ and the antishadowing present or not in the nuclear gluon distribution. Similarly to what is observed at small values of $x$, the behaviour of $R_L$ is very close to the $R_g$ one in the large-$x$ range. In particular, the presence of antishadowing in $xg^A$ directly implies an  enhancement in $F_L^A$. It is almost 10\% smaller in magnitude that the enhancement predicted for $xg^A$ by the EKS and EPS parametrizations. Inversely, if we assume the non-existence of the antishadowing in the nuclear gluon distribution at $x < 10^{-1}$, as in the DS and HKN parametrizations, no enhancement will be present in $F_L^A$ in this kinematical region. Therefore,  
it suggests that also the antishadowing effects can be easily constrained in an $eA$ collider measuring $F_L$. On the other hand, in this kinematical range the behavior of $R_C$ is distinct of $R_g$ at a same $x$. However, we observe that the behavior of $R_C$ at $x = 10^{-2}$ is directly associated to $R_g$ at $x = 10^{-1}$. In other words, the antishadowing is shifted in $R_C$ by approximately one order of magnitude in $x$. For example, the large growth of $R_g$ predicted by the HKN parametrization at $x \ge 10^{-1}$ shown in Fig. \ref{fig:ratio_gluons} implies the  steep behavior of $R_C$ at $x \ge 10^{-2}$   observed in Fig. \ref{fig:ratios}.  Consequently, by measuring $F_2^c$ it is also possible  to constrain the existence and magnitude of the antishadowing effects.

\ \\ \noindent{\it Acknowledgments:}
The author thanks E.R. Cazaroto, F. Carvalho, and  F.S. Navarra for
collaboration. 


\subsubsection{DIS in the high-energy limit at next-to-leading order}
\label{sec:Chirilli}

\hspace{\parindent}\parbox{0.92\textwidth}{\slshape
 Giovanni A. Chirilli}
\index{Chirilli, Giovanni A.}

\vspace{\baselineskip}

Nowadays,it is widely accepted that non-linear dynamics effects dominate deep inelastic lepton hadrons scattering 
processes (DIS) at very high-energy (Regge limit), and non-linear equations have been derived in order to 
describe the evolution of the structure of hadronic matter at this regime. One of these equations is the 
Balitsky-Kovchegov equation (BK) derived by Balitsky \cite{Balitsky:1995ub} 
in the Wilson lines formalism, and by Kovchegov \cite{Kovchegov:1999yj,Kovchegov:1999ua} in the dipole frame.
The Wilson line formalism is an operator language based on the concept of factorization of the scattering amplitude in rapidity space
and on the extension of the application of the Operator Product Expansion (OPE) formalism to high-energy (Regge limit).
So far, the OPE formalism was known only in the Bjorken limit as an expansion in terms of local operators or in terms of light ray operators.

The relevance of the BK equation for future experiments at an Electron Ion Collider (EIC) or Large electron Hadron Collider (LeHC) 
can be determined by the running of the coupling constant and the evolution kernel at the next-to-leading-order (NLO) approximation
(NLO corrections in power of the strong coupling constant $\alpha_s$). 
The argument of the coupling constant has been obtained by the authors of ref. \cite{Balitsky:2006wa,Kovchegov:2006vj} 
where only the quark contribution has been calculated explicitly, while the gluonic part was obtained 
conjecturing that its contribution would follow the same pattern of the quark contribution. 
However, this result did not fully solve the problem of the argument of the running coupling constant due to an ambiguity of one term which
is not proportional to $b={11\over 3} N_c -{2\over 3}n_f$. 
The complete results of the NLO-BK kernel including the gluon contribution to the argument of the coupling constant has been obtained
in \cite{Balitsky:2008zza} where it was shown that the result agrees with the NLO Balitsky-Fadin-Kuraev-Lipatov (BFKL) kernel. 
The BFKL equation \cite{Fadin:1975cb,Balitsky:1978ic} can be obtained from the BK equation by dropping out the non linear terms. Indeed, a caveat of such a linear 
evolution equation is the violation at very high energy of the unitarity condition which is instead preserved by the BK equation.

Conformal symmetry is a symmetry violated in QCD by the running of the coupling constant. What one would then expect from the
calculation of the NLO BK-kernel is that the only source of violation of such symmetry come 
from the running of coupling while the rest of the kernel
preserves conformal (M\"obius) symmetry. 
However, although Wilson lines are formally conformal invariant, at one loop correction they are rapidity-divergent,
and since it is not known how to regulate them in a conformally invariant way, 
the NLO-BK kernel contains non-conformal terms (besides to the running coupling constant) 
as a remnant of the prescription used to cure such divergences. 
In order to study the source of the loss of conformal invariance, it is convenient to consider a conformally invariant theory like the 
$\cal N$=4 super-symmetric Yang-Mills (SYM) theory. The NLO evolution kernel obtained in this framework is also not conformally 
invariant~\cite{Balitsky:2009xg},
contrary to what one would expect from a conformal field theory. It was then shown in \cite{Balitsky:2009xg}, that 
suitable operators for the description of processes at high-energy (Regge) theory are composite 
conformal (Wilson line) operators constructed order by order in perturbation theory. These operators 
absorb the undesired non conformal terms in the same way as counterterms are added to renormalize local composite operators in 
order to restore the symmetry that the bare operator lost at the level of NLO (and higher) corrections. 
Indeed, the NLO evolution of such composite conformal operators in QCD resolve in a running coupling part and in a conformally invariant part.
In ref. \cite{Balitsky:2009xg,Balitsky:2008rc}, the conformal expression for the NLO BFKL has been obtained for the first time.

In order to obtain the full NLO amplitude for DIS at high energy, one needs to calculate the coefficient function (photon impact factor) 
at NLO and convolute it with the NLO evolution kernel of the relative operator (the NLO BK kernel). The NLO
impact factor has been calculated in ref. \cite{Balitsky:2010ze} where an analytic expression (in coordinate space) 
has been obtained for the first time. \\


\noindent{\bf High-energy operator product expansion:}  In the usual OPE, due to the presence of two different scales of the transverse momentum $k_\perp$, 
one introduces a factorization scale, usually denoted by $\mu$, which factorizes
the amplitude of DIS processes in pertubatively calculable contributions (hard part) and in a non-pertubatively calculable ones (soft part)
represented by matrix elements made of light-ray operators. The evolution of such
matrix elements with respect to the renormalization point $\mu$ is the DGLAP evolution equation.

At high-energy (Regge limit), all the transverse momenta 
are of the same order of magnitude.  Therefore, a suitable factorization scale
would be the rapidity scale: one introduces rapidity ($\eta$) which separates ``fast'' fields from ``slow'' fields. Thus,
the amplitude of the process can be represented
as a convolution of contributions coming from fields with rapidity $\eta<Y$ (fast fields) and contributions coming from fields
with rapidity $\eta>Y$ (slow fields). As in the case of the usual OPE, the integration over the fields with rapidity $\eta<Y$ 
gives us the coefficient functions while the integrations over fields with rapidity $\eta>Y$ are the matrix elements of the operators.
A general feature of high-energy scattering is that a fast particle moves along 
its straight-line classical trajectory and the only quantum effect is the eikonal phase 
factor acquired along this propagation path. In QCD, for the fast quark or gluon scattering off some target, 
this eikonal phase factor is a Wilson line - an infinite gauge link ordered along 
the straight line collinear to the particle's velocity $n^\mu$:
\begin{equation}
U^\eta(x_\perp)={\rm Pexp}\Big\{ig\int_{-\infty}^\infty\!\! du ~n_\mu 
~A^\mu(un+x_\perp)\Big\},~~~~
\label{defU}
\end{equation}
Here, $A_\mu$ is the gluon field of the target, $x_\perp$ is the transverse
position of the particle which remains unchanged throughout the collision, and the 
index $\eta$ labels the rapidity of the particle. Repeating the above argument for the target (moving fast in the spectator's frame) we see that 
particles with very different rapidities perceive each other as Wilson lines and
therefore Wilson-line operators are
the convenient effective degrees of freedom in high-energy QCD (for a review, see Ref. \cite{Balitsky:2001gj}). 
The expansion of the ${\rm T}$ product of two electromagnetic currents at high-energy (Regge limit) is then in terms of Wilson lines 
\begin{eqnarray*}
&&\hspace{-8mm}
T\{\hat{j}_\mu(x)\hat{j}_\nu(y)\}=\int\! d^2z_1d^2z_2~I^{\rm LO}_{\mu\nu}(x,y ; z_1,z_2)
\hat{\cal U}(z_1,z_2)\\
&&\hspace{-8mm}
+\int\! d^2z_1d^2z_2d^2z_3~I^{\rm NLO}_{\mu\nu}(x,y ; z_1,z_2,z_3)
[ \hat{\cal U}(z_1,z_3)+\hat{\cal U}(z_2,z_3)
-\hat{\cal U}(z_1,z_2)-\hat{\cal U}(z_1,z_3)\hat{\cal U}(z_3,z_2)]
\label{high-energy-ex}
\end{eqnarray*}
where 
\begin{equation}
\hat{\cal U}^\eta(x_\perp,y_\perp)=1-{1\over N_c}
{\rm Tr}\{\hat{U}^\eta(x_\perp)\hat{U}^{\dagger\eta}(y_\perp)\}
\label{chirilli-fla1}
\end{equation}
The evolution of the Wilson line operator in eq. (\ref{chirilli-fla1}) is given by the BK equation 
\cite{Balitsky:1995ub,Kovchegov:1999yj,Kovchegov:1999ua}
\begin{eqnarray}
&&\hspace{-1.2cm}
{d\over d\eta}~\hat{\cal U}(x,y)=
{\alpha_sN_c\over 2\pi^2}\!\int\!d^2z~ {(x-y)^2\over(x-z)^2(z-y)^2}
[\hat{\cal U}(x,z)+\hat{\cal U}(y,z)\nonumber\\
&&\hspace{5cm}
-\hat{\cal U}(x,y)-\hat{\cal U}(x,z)\hat{\cal U}(z,y)]
\label{bk}
\end{eqnarray}
The first three terms correspond to the linear BFKL evolution equation \cite{Fadin:1975cb,Balitsky:1978ic} 
and describe parton emission while the last term is responsible for parton annihilation. 
For sufficiently low $x_B$, parton emission balances parton annihilation so the partons reach the state of saturation 
\cite{Gribov:1984tu,Mueller:1985wy,Mueller1990115} with
the characteristic transverse momentum $Q_s$ growing with energy $1/x_B$.
The NLO evolution equation for composite Wilson line operator
(preserving conformal invariance as explained in the introduction) has
been calculated in \cite{Balitsky:2008zza}, where one can find the
full analytic expression.

In order to obtain the DIS amplitude at high-energy at the NLO, we now need the coefficient function (``impact factor'') at next to leading order.
Here, we present the NLO impact factor for the study of DIS in the linearized case (two gluon approximation) where the NLO BK equation 
reduces to the NLO BFKL equation. In this case the OPE at high energy for DIS reduces to 
\begin{eqnarray}
&&\hspace{-4mm}
{1\over N_c}(x-y)^4T\{\bar{\hat{\psi}}(x)\gamma^\mu \hat{\psi}(x)\bar{\hat{\psi}}(y)\gamma^\nu \hat{\psi}(y)\}~
\label{ifresult}\\
&&\hspace{-4mm}
=~{\partial\kappa^\alpha\over\partial x^\mu}{\partial\kappa^\beta\over\partial y^\nu}
\!\int\! {dz_1 dz_2\over z_{12}^4}~\hat{\cal U}_{a_0}(z_1,z_2)\big[{\cal I}_{\alpha\beta}^{\rm LO}\big(1+{\alpha_s\over\pi}\big)+ 
{\cal I}_{\alpha\beta}^{\rm NLO}\big]
\nonumber
\end{eqnarray}
where
\begin{equation}
{\cal I}^{\alpha\beta}_{\rm LO}(x,y;z_1,z_2)~=~ {\cal R}^2{g^{\alpha\beta}
(\zeta_1\cdot\zeta_2)-\zeta_1^\alpha\zeta_2^\beta-\zeta_2^\alpha\zeta_1^\beta\over \pi^6(\kappa\cdot\zeta_1)(\kappa\cdot\zeta_2)}
\label{loif1}
\end{equation}
is the LO impact factor and where we used the notation 
${\cal R}~\equiv~{\kappa^2(\zeta_1\cdot\zeta_2)\over 2(\kappa\cdot\zeta_1)(\kappa\cdot\zeta_2)}$,
and the conformal vectors $\kappa~=~{\sqrt{s}\over 2x_\ast}({p_1\over s}-x^2p_2+x_\perp)-{\sqrt{s}\over 2y_\ast}({p_1\over s}-y^2p_2+y_\perp)$, 
$\zeta_i~=~\big({p_1\over s}+z_{i\perp}^2 p_2+z_{i\perp}\big)$ with $x_*=p_2^\mu x_\mu = {\sqrt{2}\over s}x^+$ (s is the Mandelstam variable). 
The analytic expression of the NLO impact factor for DIS at high
energies can be found in Ref.~\cite{Balitsky:2010ze}.
Note that the NLO impact factor is conformally (M\"obius) invariant and is given by a linear combination of five conformal tensor structures as predicted in \cite{Cornalba:2009ax}. The next natural step would be the Fourier transformation of the result in ref. \cite{Balitsky:2010ze} (the NLO impact factor), which gives the momentum-space impact factor convenient for phenomenological applications (and available at present only as a combination of numerical and analytical expressions~\cite{Bartels:2004bi,Bartels:2002uz,Bartels:2001mv}). \\


\noindent{\bf Conclusions:}  We have briefly summarized the status of the NLO calculation of the structure function for DIS at high energy. The main ingredients for the full
amplitude, namely the NLO BK kernel and the NLO IF, have been calculated. The main result of this analysis is that the 
OPE for high energy (Regge limit) is at the same status as the usual OPE in the Bjorken limit. This means that the factorization in rapidity did 
not break down at NLO accuracy. As an application of the factorization in rapidity, the full NLO
analytic amplitude in ${\cal N}=4$ SYM was calculated, the NLO result for the Pomeron intercept at small $\alpha_s$ was confirmed, 
and for the first time the NLO Pomeron residue was obtained \cite{Balitsky:2009yp}.

The Wilson line formalism proved to be very successful, not only in obtaining in a more efficient way many results that in the usual pertubative QCD mechanism
(pQCD), were obtained after many years of calculations by several groups, but also to obtain some results that have not been 
obtained (not for lack of efforts) in the usual pQCD mechanism, like the NLO impact factor, the NLO conformal BFKL kernel and the NLO pomeron residue, and 
in addition to generalize these results to include the non linear 
effects dominant at high energies. Another example which proves the efficiency of this formalism is 
the calculation, in a very easy way, of the triple pomeron vertex for diffractive and non-diffractive (``fan diagrams'') 
processes, including the subleading $N_c$ contributions \cite{Chirilli:2010mw}.


\ \\ \noindent{\it Acknowledgments:}
The author is grateful to the organizer of the workshop, in particular
to Markus Diehl and Raju Venugopalan, and to the INT institute for the
warm hospitality.  

\subsubsection{Running Coupling in Small-$x$ Physics}
\label{sec:rcsmallx}

\hspace{\parindent}\parbox{0.92\textwidth}{\slshape
 Yuri V. Kovchegov}
\index{Kovchegov, Yuri V.}

\vspace{\baselineskip}

Running coupling corrections have been included into BFKL/BK/JIMWLK
evolution following the Brodsky-Lepage-Mackenzie (BLM) scale-setting procedure \cite{Brodsky:1982gc} in
\cite{Gardi:2006rp,Kovchegov:2006vj,Balitsky:2006wa,Kovchegov:2006wf,Albacete:2007yr}.
The BLM prescription requires one to first re-sum the contribution of
all quark bubble corrections giving powers of $\amu \, N_f$, with
$N_f$ the number of quark flavors and $\amu$ the physical coupling at
some arbitrary renormalization scale $\mu$. One then has to complete
$N_f$ to the full beta-function by replacing
$ 
N_f \rightarrow - 6 \, \pi \, \beta_2
$
in the obtained expression. Here, $\beta_2 = (11 N_c - 2 N_f)/(12\pi)$ is the one-loop QCD beta-function. After this, the powers of $\amu \,
\beta_2$ should combine into physical running couplings 
$  
  \as(Q^2) =  \amu / (1 + \amu \beta_2 \ln(Q^2/\mu^2))
$
at various momentum scales $Q$ which would follow from this
calculation. The running coupling below will be written in the
$\overline {\text MS}$ renormalization scheme.

Below we will concentrate on the case of running coupling corrections
to the BFKL and BK evolution equations. Running-coupling corrections
to the JIMWLK equation can be found in
\cite{Kovchegov:2006vj,Albacete:2007yr}. At the moment the running
coupling corrections to BK have been better explored numerically than
those for JIMWLK.\\


\noindent{\bf Analytic result:}  Let us briefly summarize the results of
\cite{Kovchegov:2006vj,Balitsky:2006wa,Albacete:2007yr}. The
Balitsky-Kovchegov evolution equation with the running coupling
corrections included (rcBK) reads
\begin{align}
  \frac{\partial S(\ud{x}_0,\ud{x}_1;Y)}{\partial Y} \, = \,
  \mathcal{R}\left[S\right]-\mathcal{S}\left[S\right]\,.
\label{frs}
\end{align}
Here we use the $S$-matrix notation, related to the forward dipole
amplitude by 
$
  S(\ud{x}_0,\ud{x}_1;Y) \, = \, 1 - N (\ud{x}_0,\ud{x}_1;Y).
$
The first term on the right hand side of \eq{frs} is referred to as
the running coupling contribution, while the second term on the right
hand side of \eq{frs} is referred to as the subtraction contribution.
Separation into the two parts is arbitrary, and was done differently
in \cite{Balitsky:2006wa} and \cite{Kovchegov:2006vj}, with the net
sum being the same \cite{Albacete:2007yr}.

The running coupling part was calculated independently in
\cite{Balitsky:2006wa} and in \cite{Kovchegov:2006vj}: the results of
those calculations are
\begin{align}\label{bal_run}
  \mathcal{R}^{\text{Bal}} \left[S\right] \, =& \, \int d^2 z \,
  \tilde{K}^{\text{Bal}} (\ud{x}_0,\ud{x}_1, {\un z})
  \left[S(\ud{x}_0,\ud{z};Y)\,S(\ud{z},\ud{x}_1;Y)-S(\ud{x}_0,\ud{x}_1;Y)\right]\\
\label{kw_run}
  \mathcal{R}^{\text{KW}} \left[S\right] \, =& \, \int d^2 z \,
  \tilde{K}^{\text{KW}} (\ud{x}_0,\ud{x}_1, {\un z})
  \left[S(\ud{x}_0,\ud{z};Y)\,S(\ud{z},\ud{x}_1;Y)-S(\ud{x}_0,\ud{x}_1;Y)\right].
\end{align}
The integral kernels in the two cases are given by 
\begin{align}
  \tilde{K}^{\text{Bal}}(\ud{r},\ud{r}_1,\ud{r}_2)=\frac{N_c\,\alpha_s(r^2)}{2\pi^2}
  \left[\frac{r^2}{r_1^2\,r_2^2}+
    \frac{1}{r_1^2}\left(\frac{\alpha_s(r_1^2)}{\alpha_s(r_2^2)}-1\right)+
    \frac{1}{r_2^2}\left(\frac{\alpha_s(r_2^2)}{\alpha_s(r_1^2)}-1\right)
  \right]
\label{kbal}
\end{align}
as found in \cite{Balitsky:2006wa} and by 
\begin{align}
  \tilde{K}^{\text{KW}}(\ud{r},\ud{r}_1,\ud{r}_2)=\frac{N_c}{2\pi^2}\left[
    \alpha_s(r_1^2)\frac{1}{r_1^2}-
    2\,\frac{\alpha_s(r_1^2)\,\alpha_s(r_2^2)}{\alpha_s(R^2)}\,\frac{
      \ud{r}_1\cdot \ud{r}_2}{r_1^2\,r_2^2}+
    \alpha_s(r_2^2)\frac{1}{r_2^2} \right]\,,
\label{kkw}
\end{align}
as found in \cite{Kovchegov:2006vj}, where 
\begin{align}
  R^2(\ud{r},\ud{r}_1,\ud{r}_2)=r_1\,r_2\left(\frac{r_2}{r_1}\right)^
  {\frac{r_1^2+r_2^2}{r_1^2-r_2^2}-2\,\frac{r_1^2\,r_2^2}{
      \ud{r}_1\cdot\ud{r}_2}\frac{1}{r_1^2-r_2^2}}\,.
\label{r}
\end{align}

One notices immediately that $\mathcal{R}^{\text{Bal}} \left[S\right]$
calculated in \cite{Balitsky:2006wa} is different from
$\mathcal{R}^{\text{KW}} \left[S\right]$ calculated in
\cite{Kovchegov:2006vj} due to the difference in the kernels
$\tilde{K}^{\text{Bal}}$ and $\tilde{K}^{\text{KW}}$ in Eqs.
(\ref{kbal}) and (\ref{kkw}). However that does not imply disagreement
between the calculations of \cite{Balitsky:2006wa} and
\cite{Kovchegov:2006vj}: after all, it is the full kernel on the right
of \eq{frs}, $\mathcal{R}\left[S\right]-\mathcal{S}\left[S\right]$,
that needs to be compared. To do that, one has to calculate the second
term on the right hand side of \eq{frs} (the subtraction
contribution). This was done in \cite{Albacete:2007yr}, yielding 
\begin{align}
  {\mathcal S} [S] \, = & \, \am^2 \, \int d^2 z_1 \, d^2 z_2 \, 
  K_{\oone} ({\un x}_0, {\un x}_1 ; {\un z}_1, {\un z}_2) \,  
  \nonumber \\
  &\times [ S ({\un
    x}_{0}, {\un w}, Y) \, S ({\un w}, {\un x}_1, Y) - S ({\un x}_{0},
  {\un z}_1, Y) \, S ({\un z}_{2}, {\un x}_1, Y)] 
  \label{sub_full} 
\end{align}
and the re-summed BK kernel $K_{\oone}$ can be found in the original
reference. 
Substituting ${\un w} = {\un z}_1$ (or, equivalently, 
${\un w} = {\un z}_2$) in \eq{sub_full} yields the subtraction term 
${\mathcal S}^{\text{Bal}} [S]$, 
which has to be subtracted from $\mathcal{R}^{\text{Bal}}
\left[S\right]$ calculated in \cite{Balitsky:2006wa} and given by
\eq{bal_run} to obtain the complete evolution equation re-summing all
orders of $\as \, N_f$ in the kernel. 
Substituting ${\un w} = {\un z} = \alpha \, {\un z}_1 + (1-\alpha) \,
{\un z}_2$ in \eq{sub_full} yields the term 
${\mathcal S}^{\text{KW}}[S]$,
which has to be subtracted from $\mathcal{R}^{\text{KW}}
\left[S\right]$ calculated in \cite{Kovchegov:2006vj} and given in
\eq{kw_run} again to obtain the complete evolution equation re-summing
all orders of $\as \, N_f$ in the kernel. \\



\begin{figure}[t]
\centerline{\includegraphics[height=5cm]{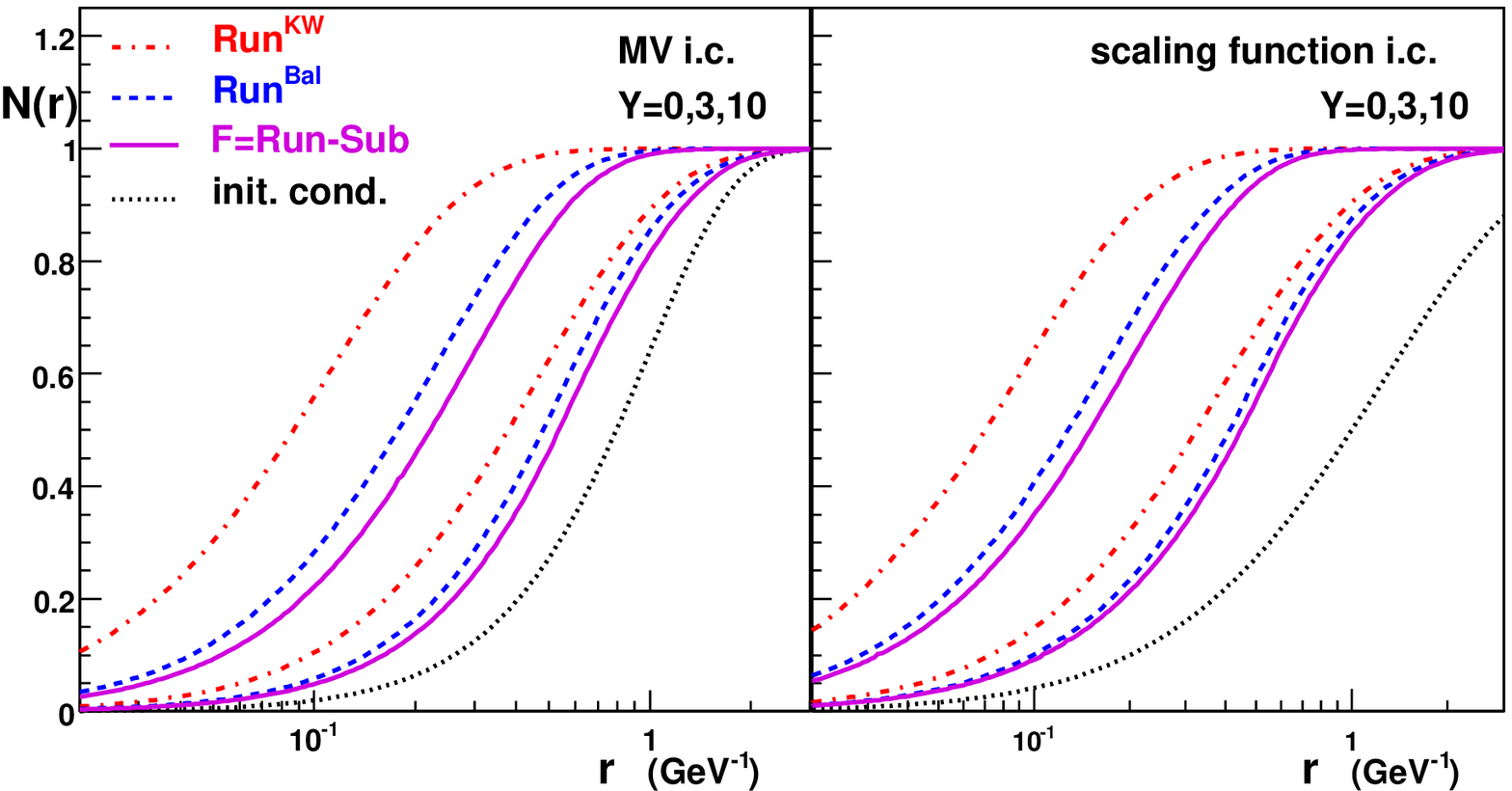}}
\caption{\small Solutions of the complete (all orders in $\as \, \beta_2$) 
  evolution equation given in \eq{frs} (solid lines), and of the
  equation with Balitsky's (dashed lines) and KW's (dashed-dotted)
  running coupling schemes at rapidities $Y=0$, $3$ and $10$.  Left
  plot uses quasi-classical McLerran-Venugopalan (MV) initial condition. The right plot
  employs the initial condition given by the dipole amplitude at
  rapidity $Y=35$ evolved using Balitsky's running coupling scheme and
  with $r$-dependence rescaled down such that $Q_s = Q'_s = 1$~GeV.}
\label{subev}
\end{figure}

\noindent{\bf Numerical Solution:}  The numerical solution of the
running-coupling BK (rcBK) evolution just presented was performed in
\cite{Albacete:2007yr} and plotted in \fig{subev}. One plots the
running-coupling parts from Eqs.~\peq{bal_run} and \peq{kw_run}
\cite{Kovchegov:2006vj,Balitsky:2006wa} (dashed and dash-dotted lines
correspondingly), along with the full solution (solid line). As one
can see the full solution is best approximated by the Balitsky's
running coupling scheme from \eq{bal_run} \cite{Balitsky:2006wa}.
Hence in most phenomenological applications one simply solves rcBK
with Balitsky's prescription \cite{Albacete:2009fh,Albacete:2007sm}. 
Note that the rcBK solution also exhibits the property of geometric
scaling \cite{Stasto:2000er}, as was shown in \cite{Albacete:2007yr}. \\

\noindent{\bf Running-coupling BFKL evolution:}  The running-coupling BFKL equation (rcBFKL) was constructed in
\cite{Kovchegov:2006wf} and reads
\begin{align}\label{rc_BFKL}
  \frac{\partial \phi (k, Y)}{\partial Y} \, = \, \frac{N_c}{2 \,
    \pi^2} \, \int d^2 q \, \Bigg\{ & \frac{2}{({\bm k} - {\bm q})^2} \,
  \as \left( ({\bm k} - {\bm q})^2 \, e^{-5/3} \right) \, \phi (q, Y)
  \nonumber \\ & - \frac{{\bm k}^2}{{\bm q}^2 \, ({\bm k} - {\bm q})^2}
  \, \frac{ \as \left({\bm q}^2 \, e^{-5/3} \right) \, \as \left(
      ({\bm k} - {\bm q})^2 \, e^{-5/3} \right)}{ \as \left({\bm k}^2
      \, e^{-5/3} \right)} \, \phi (k, Y) \Bigg\},
\end{align}
where the unintegrated gluon distribution $\phi (k, Y)$ is defined by
\begin{align}\label{FT}
  N (x_{01}, Y) \, = \, \int \frac{d^2 k}{(2 \pi)^2} \, \left(1 - e^{i
      {\bm k} \cdot {\bm x}_{01}} \right) \, {\tilde N} (k,Y)
\end{align}
with
\begin{align}\label{phi}
  \as (k^2) \, \phi (k,Y) \, = \, \frac{N_c \, S_\perp}{(2 \pi)^3} \, k^2 \,
  {\tilde N} (k,Y).
\end{align}
Here $S_\perp$ is the transverse area of the target. The
running-coupling BFKL equation \peq{rc_BFKL} was originally
conjectured in \cite{Braun:1994mw,Levin:1994di} by postulating the
validity of the bootstrap equation for running-coupling corrections.



\subsubsection{Running-coupling and higher-order effects on the saturation scale}
\label{sec:Beuf}

\hspace{\parindent}\parbox{0.92\textwidth}{\slshape
 Guillaume Beuf}
\index{Beuf, Guillaume}

\vspace{\baselineskip}

The DGLAP~\cite{Gribov:1972ri,Altarelli:1977zs,Dokshitzer:1977sg} and BFKL~\cite{Lipatov:1976zz,Kuraev:1977fs,Balitsky:1978ic} equations give the evolution with kinematics of the partonic content of hadrons and nuclei in the regime where these are dilute. As these equations are linear, they can be solved analytically by using a Mellin tranform. By contrast, when the phenomenon of gluon saturation~\cite{Gribov:1984tu} is taken taken into account, the relevant evolution equations - B-JIMWLK~\cite{Balitsky:1995ub,JalilianMarian:1997jx,JalilianMarian:1997gr,JalilianMarian:1997dw,Kovner:2000pt,Weigert:2000gi,Iancu:2000hn,Iancu:2001ad,Ferreiro:2001qy} or BK~\cite{Balitsky:1995ub,Kovchegov:1999yj,Kovchegov:1999ua} - are nonlinear, and thus cannot be solved analytically.

Nevertheless, the solutions of these nonlinear equations in the leading order (LO) approximation (where the coupling $\alpha_s$ is kept fixed) are well understood, by combining results from numerical simulations~\cite{GolecBiernat:2001if,Rummukainen:2003ns,Albacete:2004gw} and analytical asymptotic expansions~\cite{Iancu:2002tr,Mueller:2002zm,Munier:2003vc,Munier:2003sj,Munier:2004xu}. Indeed, the BK equation belongs to a well-studied class of nonlinear equations, whose solutions develop asymptotically a universal traveling wave-front structure~\cite{Bramson:1983,Ebert:2000}, which is independent of the initial condition\footnote{More precisely, in the QCD case, that asymptotic behavior in rapidity is reached from any initial condition compatible with perturbative QCD in the UV.}. In the context of QCD, that traveling wave-front structure of the solution implies the \emph{geometric scaling}~\cite{Stasto:2000er} property found in the DIS data at HERA: the total virtual photon - target cross sections $\sigma^{\gamma^*}_{T,L}(Y,Q^2)$ depend on $Y$ and $Q^2$ essentially only through the combination $Q^2/Q_s^2(Y)$, because the dipole-target amplitude solution of the BK equation depends only on $r^2 Q_s^2(Y)$ at large $Y$, $r$ being the dipole size. The evolution of the saturation scale $Q_s^2(Y)$ is obtained from the propagation of the wave-front. For the LO BK equation, one gets a large $Y$ expansion of the form
\begin{equation}
\label{beuf:eq:QsFC}
\log Q_s^2(Y) = a_{1} Y + a_{0} \log Y + \textrm{Const.} + a_{-1/2} Y^{-1/2} + {\cal O}(Y^{-1}),
\end{equation}
where $a_{1}$, $a_{0}$ and $a_{-1/2}$ are three known universal coefficients~\cite{Munier:2004xu}, whereas the constant term and all the ones of order $Y^{-1}$ or less do depend on the initial conditions, \emph{i.e.} on the nature of the target used for the DIS.  From geometric considerations, the initial $Q_s^2$ of a nucleus $A$ is enhanced by  a factor $A^{1/3}$ with respect to that of a proton. That nuclear enhancement of $Q_s^2(Y)$ is preserved by the LO high-energy evolution, in the constant term of the expansion \eqref{beuf:eq:QsFC}. Both from numerical simulations and from the expansion \eqref{beuf:eq:QsFC}, one learns that the evolution of $Q_s^2(Y)$ implied by the LO BK equation is too fast to be compatible with the data for DIS and other observables, which favor $\log Q_s^2(Y)\sim \lambda Y$, with $\lambda\simeq 0.2$ or $0.3$. We are thus forced to consider higher order corrections to the BK equation. \\


\noindent{\bf Running \emph{vs.} fixed coupling:}  As discussed in this section by Chirilli, the BK equation is now known at next-to-leading order (NLO)~\cite{Balitsky:2008zza,Balitsky:2009xg}. However, its solutions are much less understood than the ones of the LO equation. Indeed no numerical simulations of the full NLO BK equation have been performed yet, for technical reasons, but only simulations~\cite{GolecBiernat:2001if,Rummukainen:2003ns,Albacete:2004gw,Albacete:2007yr,Berger:2010sh} of the BK equation with LO kernel and running coupling $\alpha_s$, with various prescriptions used to set the scale in the coupling. By contrast, it is non-trivial to go from fixed coupling to running coupling in the analytical studies, since it leads to a different class of wave-front solutions, for which universality of the asymptotics is not fully established. The inclusion of other NLO corrections gives however no additional difficulty. Let us first discuss the effects of running coupling only.

$A$ $priori$, the running of the coupling brings the additional scale $\Lambda_{QCD}$ into the problem, which may spoil the geometric scaling property. Indeed, there is no interval where the solutions of the running coupling BK equation show exact geometric scaling, by contrast to fixed coupling solutions, but they satisfy an approximate geometric scaling in some range. Equivalently, the wave-front in the solutions is being slowly distorted during its propagation, instead of being uniformly translated as in the fixed coupling case.

Running coupling effects turn the asymptotic behavior of the saturation scale into $\log Q_s^2(Y)\propto \sqrt{Y}$, as found in early analytical studies~\cite{Gribov:1984tu,Iancu:2002tr,Mueller:2002zm,Munier:2003sj}.
More precisely its large $Y$ asymptotics writes
\begin{equation}
\label{beuf:eq:QsRC}
\log \left(Q_s^2(Y)/ \Lambda_{QCD}^2\right) = b_{1/2}\, \sqrt{Y} +b_{1/6}\, Y^{1/6}+ b_{0} +b_{-1/6}\, Y^{-1/6}  + b_{-1/3}\, Y^{-1/3} + {\cal O}\left(Y^{-1/2}\right)\, ,
\end{equation}
where the first five terms are universal and known\footnote{The  calculation of $b_{0}$, $b_{-1/6}$ and $b_{-1/3}$ has been performed recently in~\cite{Beuf:2010aw}.}, whereas the following ones of order $Y^{-1/2}$ or less are sensitive to the initial conditions.
The universality of the constant term $b_{0}$ in \eqref{beuf:eq:QsRC} implies that initial conditions effects such as the nuclear $A^{1/3}$ enhancement of $Q_s^2$ are washed-out at high rapidity when the coupling is running, as first predicted in~\cite{Mueller:2003bz}. Numerically, it has been found~\cite{Rummukainen:2003ns,Albacete:2004gw,Dusling:2009ni} that this effect happens at very high rapidity. Hence, the nuclear enhancement of $Q_s^2$, which is one of the motivations for doing nuclear DIS at the EIC, should still be present in the kinematical range accessible at the EIC.
Remarkably, the evolution of the saturation scale in the running coupling case is such that very good fits of DIS data can be performed with solutions of the running coupling BK equation~\cite{Albacete:2009fh,Albacete:2010sy}, by contrast to the fixed coupling case, without the inclusion of other NLO effects. \\


\noindent{\bf Other NLO effects:}  Apart from the contributions re-summed into the running of the coupling, there are large NLO corrections to the BK kernel, related to the large NLO corrections to the BFKL kernel~\cite{Fadin:1998py,Ciafaloni:1998gs}. 

In a conformal gauge field theory, terms of arbitrary N$^n$LO order from the kernel would contribute at each order of the expansion \eqref{beuf:eq:QsFC}. By contrast, the running of the coupling is dynamically quenching the effect on the solutions of higher order terms in the kernel. NLO contributions start to appear at order $Y^0$ in \eqref{beuf:eq:QsRC}, NNLO contributions at order $Y^{-1/2}$ and so on. Moreover, the coefficient $b_{-1/6}$ has been found to be NLO-independent~\cite{Beuf:2010aw}. Apart from the running of the coupling, NLO contributions thus affect mostly the normalization of $Q_s^2(Y)$ at large $Y$, via $b_{0}$, and only mildly the asymptotic $Y$-evolution of $Q_s^2(Y)$, via
$b_{-1/3} \, Y^{-1/3}$ and further subleading terms. That property is indeed seen in numerical simulations with running coupling and a subset of other NLO contributions included~\cite{Berger:2010sh}. That result shed some light on the spectacular success of the running coupling LO BK equation to describe DIS data. There is a degeneracy in \eqref{beuf:eq:QsRC} between the contribution of $\Lambda_{QCD}$ and $b_{0}$ to $Q_s^2(Y)$. Hence, treating $\Lambda_{QCD}$ as a free fit parameter as in Refs.~\cite{Albacete:2009fh,Albacete:2010sy} allows one to fit the bulk of NLO effects, without actually simulating the BK evolution with NLO kernel.

Several prescriptions~\cite{Balitsky:2006wa,Kovchegov:2006vj} have been proposed to split NLO corrections into contributions to the running coupling or to the kernel. Hence, BK equations with running coupling and LO kernel obtained following different prescriptions differ formally by terms of order NLO and beyond in the kernel. In numerical simulations of such running coupling LO BK equations~\cite{Albacete:2007yr}, solutions with different prescriptions differ at large $Y$ mostly by a constant rescaling of $Q_s^2(Y)$, in agreement with our previous discussion. \\


\noindent{\bf The problems brought by the impact-parameter dependence:}  Implicitly, we have discussed so far only results from studies of impact parameter independent solutions of the BK equation. The BK equation preserves unitarity at fixed impact parameter. However, its impact parameter dependent solutions violate unitarity since they violate the Froissart bound~\cite{Froissart:1961ux} on the cross-section~\cite{GolecBiernat:2003ym,Berger:2010sh}, due to the unphysical possibility of gluon emission at arbitrarily long range in the transverse plane.
The running of the coupling reveals another problem: there is a reappearance of the diffusion into the infrared~\cite{Berger:2010sh}, which was thought to be cured by gluon saturation, from studies of impact parameter independent solutions of the BK equation. Hence, the impact parameter dependent solutions of the BK solutions are very sensitive to strongly coupled infrared physics, which is not yet implemented in the formalism. This is the most challenging open theoretical problem with regard to gluon saturation. Therefore, it is not yet clear to what extent the results about impact parameter independent solutions presented in the previous sections are reliable for realistic proton or nuclear targets.

\section{Diffractive DIS (F$_2^D$, F$_L^D$, charm contribution)}  
\label{sec:DiffractiveDIS}  

\subsubsection{Diffraction in e+p and e+A collisions}   
\label{sec:diffintro}

\hspace{\parindent}\parbox{0.92\textwidth}{\slshape
 Cyrille Marquet}
\index{Marquet, Cyrille}

A non-negligible fraction of the events in DIS are diffractive, meaning that the hadronic target, of mass $M$, escapes the collision intact. As a colorless object has been exchanged in the t-channel, there is rapidity gap void of particles in the final state, between the outgoing target and the diffractive final state $X$, made up of all the other particles in the event. On top of $x$ and $Q^2$, two additional kinematic invariants are needed to characterize diffraction in DIS: the momentum transfer $t<0$ at the hadronic vertex, and the mass $M_X$ of the diffractive final state. In practice, the variable $M_X$ is sometimes traded for $\beta$ and the variable $x$ is traded for $\xpom$--these are defined as 
\begin{equation}
\beta=\frac{Q^2}{Q^2+M_X^2-t}\,\,;\,\, \xpom=\frac{x}{\beta}=\frac{Q^2+M_X^2-t}{Q^2+W^2-M^2}\, .
\end{equation}
Small values of $\beta$ refer to events with diffractive masses much bigger than the photon virtuality, while values of $\beta$ close to unity refer to the opposite situation. $\xpom$ is useful because it characterizes the size of the rapidity gap $\Delta\eta\simeq\ln(1/\xpom)$.

There are events in which the hadronic target, instead of staying intact, may dissociate into a low-mass excited state Y, while still leaving a rapidity gap in the final state. These events are also classified as diffractive, they occur only if the mass $M_Y$ of the excited state is close enough to the initial mass $M.$ Coherent diffraction is employed when the target scatters elastically (ep$\to$eXp), while incoherent diffraction refers to the more general case ep$\to$eXY which is a sum of coherent diffraction (Y=p) and target-dissociative diffraction (Y$\neq$p). The former dominates at low $|t|$ and the latter at large $|t|$.

While in the leading-twist approximation of QCD there is a collinear factorization theorem to compute diffractive structure functions in DIS at large $Q^2$, the description of hard diffraction in this framework is not as natural as for inclusive events. This is reflected in the fact that standard parton distribution functions (pdfs) are of no help to compute $F_2^D$, and one has to introduce a different set of parton distributions called diffractive pdfs (dpdfs). Therefore in the collinear factorization framework, the description of the parton content of the proton depends on whether or not the final state is diffractive. While this is successful - and should be since collinear factorization is a good approximation of QCD at large $Q^2$ - conceptually it is not so satisfactory as one would like to be able to describe any process with a single proton wave function.

No further conceptual advances are expected within the leading-twist approximation of QCD. There are some technical improvements that can be made, for instance it is nowadays practically impossible to extract dpdfs without assuming what is called Regge factorization: $\mbox{dpdf}(\xpom,t,\beta,Q^2)=f(\xpom,t)\ g(\beta,Q^2)$. This is not satisfactory, since such a factorization is not a property of QCD. However, there is little doubt that if one could bypass this practical problem - perhaps with a larger data sample in all four directions: $Q^2$, $\beta$, $\xpom$ and $t$ - this approach would succeed at large $Q^2$.

But in fact, the purpose of an electron-ion collider is not to check whether DGLAP evolution will work at large $Q^2$, the goal is rather to explore what we don't know as well: the non-linear regime of QCD where collinear factorization breaks down. To be more specific, we are interested in the regime
$Q^2<5$ GeV$^2$ and $x$ as small as possible. Interestingly enough, studying the non-linear {\it saturation} regime will be easier with diffractive than with inclusive measurements. This is so because at small $x$, diffractive processes are mostly sensitive to quantum fluctuations in the proton wave function that have a virtuality of order $Q_s^2$, instead of $Q^2$. As a result, power corrections (not the generic $\Lambda_{QCD}^2/Q^2$ corrections, but rather the sub-class of them of order $Q_s^2/Q^2$ important at small $x$) are expected to come into play starting from a higher value of $Q^2$ in diffractive DIS, compared to inclusive DIS. In fact, there is already a hint that this is happening at HERA: collinear factorization starts to fail below about 2 GeV$^2$ in the case of $F_2$, while already below about 8 GeV$^2$ in the case of $F_2^D$.

The QCD description of diffractive DIS in the small-$x$ limit turns out to be much more insightful than that of the large-$Q^2$ limit. It is so because at small $x$, DDIS can be expressed in the Good-Walker picture (which was originally imagined for soft diffraction in hadron-hadron collisions), with the benefit that, thanks to the point-like nature of the photon, the modeling part of the Good-Walker approach can be replaced by actual QCD computations. This remarkable realization of the Good-Walker picture in small-$x$ DIS is more commonly referred to as the dipole picture: dipoles are eigenstates of high-energy scattering in QCD, and it is known how to expand the photon wave function onto the dipole basis. At the end in this approach, the parton content of the proton - both in the linear and non-linear regimes - is parametrized through the dipole cross section. As a result, diffractive structure functions also feature geometric scaling \cite{Marquet:2006jb}. Another important fact is that at small $x$, diffraction can be entirely predicted, once the dipole cross section has been constrained with inclusive data.

In spite of the fact that this approach has been able to successfully predict $F_2^D$ at small $x$, there is still important conceptual progress to be made. For instance, the transverse impact parameter dependence of the dipole scattering amplitude is very poorly constrained. Indeed, one has been able to describe $F_2$ and correctly predict $F_2^D$ with two kinds of impact parameter dependences, neither of which is fully satisfactory. In a first class of dipole models, the impact parameter profile of the proton is independent of energy, yielding a dipole cross section bounded from above. In the other class of models, the black-disk regime of maximal scattering strength spreads too quickly in the transverse plane with increasing dipole size $r$, leading to a dipole cross section which diverges for large $r$. It is quite clear that the LHeC is needed to help us understand better this issue.

Finally, let us say a few important words on ep$\to$eXY diffractive events. In past experiments, events with $Y\neq p$ have mostly been regarded as background, and model-dependent subtractions have been applied to data, yielding large normalization uncertainties. Within the kinematic reach of HERA, it has been observed that the ratio $d\sigma^{ep\to eXY}/d\sigma^{ep\to eXp}$ is a constant independent of all kinematic variables other than $M_Y$ and $t$ (that ratio increases with $M_Y$ and $|t|$). Here we would like to emphazise that proton-dissociative events are also intrinsically interesting. For instance, at small $x$ the cross section difference $d\sigma^{ep\to eXY}-d\sigma^{ep\to eXp}$ is $1/N_c^2$ suppressed, meaning that if it were measured accurately, it would give access to details of the QCD dynamics which are untestable otherwise. The EIC provides such an opportunity.

After many fixed target experiments, it took a collider to discover diffractive events in e+p. Since no e+A collider has ever been built, diffraction in e+A has simply never been measured. That such a deficiency exists in our knowledge of nuclear structure is compelling enough to build the EIC. Everything we would learn about DDIS off nuclei at the EIC will be new, in any kinematical domain, implying a huge discovery potential. Nevertheless, we have expectations of what diffraction off nuclei should look like, based on our current understanding of QCD. For instance, the theory of nuclear shadowing allows the constuction of nuclear DPDFs for large $Q^2$ physics, while within the Color Glass Condensate framework, nuclear diffractive structure functions can be predicted at small $x$. Depending on these kinematics, different patterns of nuclear shadowing or antishadowing as a function of $\beta$ and $\xpom$ are expected. This is just one example out of many that should be checked with an e+A collider. Since the current predictions rely on rather simple models for impact parameter dependence, they need to be confronted to data, in order to, in return, improve our understanding.


Finally, there is one aspect of diffraction which is specific to nuclei that one should mention. The structure of incoherent diffraction eA$\to$eXY is more complex than with a proton target, and also can teach us a lot more. In the case of a target nucleus, we expect the following qualitative changes in the $t$ dependence. First, the low-$|t|$ regime in which the nucleus scatters elastically will be dominant up to a smaller value of $|t|$ (to about $|t|=0.05$ GeV$^2$) compared to the proton case, reflecting the larger size of the nucleus. Then, the nucleus-dissociative regime will comprise two parts: an intermediate regime in momentum transfer up to about $0.7$ GeV$^2$ where the nucleus will predominantly break up into its constituents nucleons, and a large$-|t|$ regime where the nucleons inside the nucleus will also break up, implying pion production in the $Y$ system for instance. These are only qualitative expectations, it is crucial to study this aspect of diffraction quantitatively in order to complete our understanding of the structure of nuclei.

\subsubsection{Expectations for e+A from the CGC}   
\label{sec:diffincgc}

\hspace{\parindent}\parbox{0.92\textwidth}{\slshape
 Cyrille Marquet}
\index{Marquet, Cyrille}

\vspace{\baselineskip}

In this work, hard diffraction in electron-nucleus (e+A) collisions is considered
within the IPsat model,\cite{Kowalski:2003hm} corresponding to the classical limit of the Color Glass Condensate approach. This effective theory of QCD at high partonic density is the most natural framework to describe the saturation phenomenon, and therefore to study e+A scattering at high energies, in particular diffractive observables. Here we shall focus on the nuclear diffractive structure function $F_{2,A}^D.$

Let us recall the kinematics of diffractive DIS: $\g^*A\!\rightarrow\!XA.$ With a momentum transfer $t\!\leq\!0,$ the proton/nucleus gets out of the $\g^*\!-\!A$ collision intact, and is separated by a rapidity gap from the other final-state particles whose invariant mass we denote $M_X.$ The photon virtuality is denoted $Q^2,$ and the $\g^*\!-\!A$ total energy $W.$ It is convenient to introduce the following variables: $x\!=\!Q^2/(Q^2\!+\!W^2),$
$\beta\!=\!Q^2/(Q^2\!+\!M_X^2)$ and $\xp\!=\!x/\beta.$ The size of the rapidity gap is
$\ln(1/\xp).$

The diffractive structure function is expressed as a function of $\beta,$ $\xp,$ $Q^2,$ and $t,$ and we will only consider the $t-$integrated structure function $F_2^{D,3}.$ While at large values of $\xp$ and $Q^2,$ the leading-twist collinear factorization is appropriate to describe hard diffraction off protons, this is not the case at small $\xp$ or off nuclei, as higher twists are enhanced by $\sim(A/\xp)^{0.3}.$ In this situation, the dipole picture is better suited to address the problem. It naturally incorporates the description of both inclusive and diffractive events into a common theoretical framework:\cite{Nikolaev:1991et,Bialas:1995bs,Bialas:1996tn} the same dipole-nucleus scattering amplitudes, which can be computed treating the nucleus as a CGC, enter in the formulation of the inclusive and diffractive cross-sections. \\


\noindent{\bf Diffractive structure functions in the dipole picture:}  In our approach, $F_2^D\!=\!F_T^{q\bar q}\!+\!F_L^{q\bar q}\!+\!F_T^{q\bar qg}$ where the different pieces correspond to transversely (T) or longitudinally (L) polarized photons dissociating into a $q\bar q$ or $q\bar qg$ final state. For instance, the $q\bar q$
contributions are
\bea
\xp F_T^{q\bar q}(\beta,\xp,Q^2)\!\!&\!\!=\!\!&\!\!
\f{N_c Q^4}{8\pi^3\beta}\sum_f e_f^2\int_0^1 dz\ \Theta(\kappa_f^2) 
z(1\!-\!z)\left[f_T(z)\varepsilon_f^2(z)I_1(\kappa_f,\epsilon_f)
\!+\!m_f^2I_0(\kappa_f,\epsilon_f)\right],\nonumber \\
\xp F_L^{q\bar q}(\beta,\xp,Q^2)\!\!&\!\!=\!\!&\!\!
\f{N_c Q^6}{8\pi^3\beta}\sum_f e_f^2\int_0^1 dz\ \Theta(\kappa_f^2)
z(1\!-\!z)f_L(z)I_0(\kappa_f,\epsilon_f)\ ,
\label{qq}\eea
with
\be
\varepsilon_f^2(z)\!=\!z(1\!-\!z)Q^2\!+\!m_f^2\ ,\hspace{0.2cm}
\kappa_f^2(z)\!=\!z(1\!-\!z)M_X^2\!-\!m_f^2\ ,\hspace{0.2cm}
f_T(z)\!=\!z^2\!+\!(1\!-\!z)^2\ ,\hspace{0.2cm}
f_L(z)\!=\!4z^2(1\!-\!z)^2\ .
\ee
The $\xp$ dependence comes in the functions $I_\lambda$ from $N_A(r,b,\xp),$ the $q\bar q$ dipole-nucleus scattering amplitude:
\be
I_\lambda(\kappa,\epsilon)\!=\!\int d^2b
\left[\int_0^\infty\!rdr J_\lambda(\kappa r)K_\lambda(\epsilon r)N_A(r,b,\xp)\right]^2
\label{ilambda}\ee
where $J_\lambda$ and $K_\lambda$ are Bessel functions. In equation (\ref{ilambda}), the integration variables $r$ and $b$ are the $q\bar q-$dipole transverse size and its impact parameter.

In principle, it is justified to neglect final states containing gluons, because these are suppressed by extra powers of $\alpha_s.$ However, for small values of $\beta$ or large values of $Q^2,$ the $q\bar q$ pair will emit soft or collinear gluons whose emissions are accomponied by large logarithms $\ln(1/\beta)$ or $\ln(Q^2)$ which compensate the factors of $\alpha_s.$ 
In those situations, multiple gluon emissions should be re-summed; in practice, including the
$q\bar qg$ final state is enough to describe the HERA data. In both the small$-\beta$ and
large$-Q^2$ limits, this can be done within the dipole picture. An implementation of the
$q\bar qg$ contribution $F_T^{q\bar qg}$ that correctly reproduces both limits was recently proposed~\cite{Marquet:2007nf}, while at large $\beta$ and small $Q^2,$ the $q\bar q$ contributions
in equation (\ref{qq}) dominate. The formulae that we shall use can be found in ref.~\cite{Marquet:2007nf}. \\


\noindent{\bf The dipole-nucleus scattering amplitude:}  We shall use the IPsat parametrization to describe the dipole-nucleus scattering amplitude:
\be
N_A(r,b,x)=1-e^{-r^2F(r,x)\sum_{i=1}^A T_p(b-b_i)}\ ,\quad
F(x,r^2)=\frac{\pi^2}{2N_c}\alpha_s\left(\mu_0^2\!+\!\frac{C}{r^2}\right)xg\left(x,\mu_0^2+\frac{C}{r^2}\right)\ .
\label{ipsat}\ee
This is a model of a nucleus whose nucleons interact independently. Indeed, $N_A$ is obtained from $A$ dipole-nucleon amplitudes $N_p\!=\!1\!-\!\exp[-r^2F(r,x)T_p(b)]$ assuming that the probability
$1\!-\!N_A$ for the dipole not to interact with the nucleus is the product of the probabilities
$1\!-\!N_p$ for the dipole not to interact with the nucleons. This assumption is not consistent with the CGC quantum evolution, which sums up nonlinear interactions between the nucleons. However, the classical limit of the dipole-CGC scattering amplitude can be thought of an initial condition (\ref{ipsat}). Note that in the small $r$ limit, one has $N_A=\sum_i N_p,$ and there is no leading twist shadowing.

In (\ref{ipsat}), $T_p(b)\!\propto\!\exp[-b^2/(2 B_G)]$ is the impact parameter profile function in the proton with $\int d^2b\ T_p(b)=1,$ and $F$ is proportional to the DGLAP evolved gluon distribution. The parameters $\mu_0,$ $C,$ and $B_{\rm G}$ (as well as two other parameters characterising the initial condition for the DGLAP evolution) are fit to
reproduce the HERA data on the inclusive proton structure function $F_2.$ The diffractive proton structure function $F_2^D$ is well reproduced \cite{Kowalski:2008sa} after adjusting $\alpha_s=0.14$ in the
$q\bar qg$ component. Vector-meson production at HERA is also well described.\cite{Kowalski:2006hc}

We introduced in (\ref{ipsat}) the coordinates of the individual nucleons $\{b_i\},$ they are distributed according to the Woods-Saxon distribution $T_A(b_i),$ which means that to compute
an observable, one has to perform the following average
\be
\langle\mathcal{O}\rangle_N \equiv \int \left(\prod_{i=1}^A d^2b_i 
T_A(b_i)\right)\mathcal{O}(\left\{b_i\right\})\ .
\label{wsavg}\ee
The Woods-Saxon parameters are measured from the electrical charge distribution, no additional parameters are introduced. The resulting dipole cross sections give a good agreement \cite{Kowalski:2007rw} with the small-$x$ NMC data on the nuclear structure function
$F_{2,A}$. We will use this parametrization of $N_A$ to predict the nuclear diffractive structure function $F_{2,A}^D$.

Note that performing the average at the level of the amplitude (\ref{wsavg}), meaning calculating
$\langle N_A\rangle_N^2$ in (\ref{qq}), imposes that the nucleus is intact in the final state. By contrast, when performing the average at the level of the cross-section, meaning calculating $\langle N_A^2\rangle_N$ in (\ref{qq}), one allows the nucleus to break up into individual nucleons, which will typically happen when the momentum transfer is bigger than the inverse nuclear radius. In what follows, we shall refer to those possibilities as ``non breakup'' and ``breakup'' cases. \\

\begin{figure}[t]
  \centering
  \includegraphics[width=6cm]{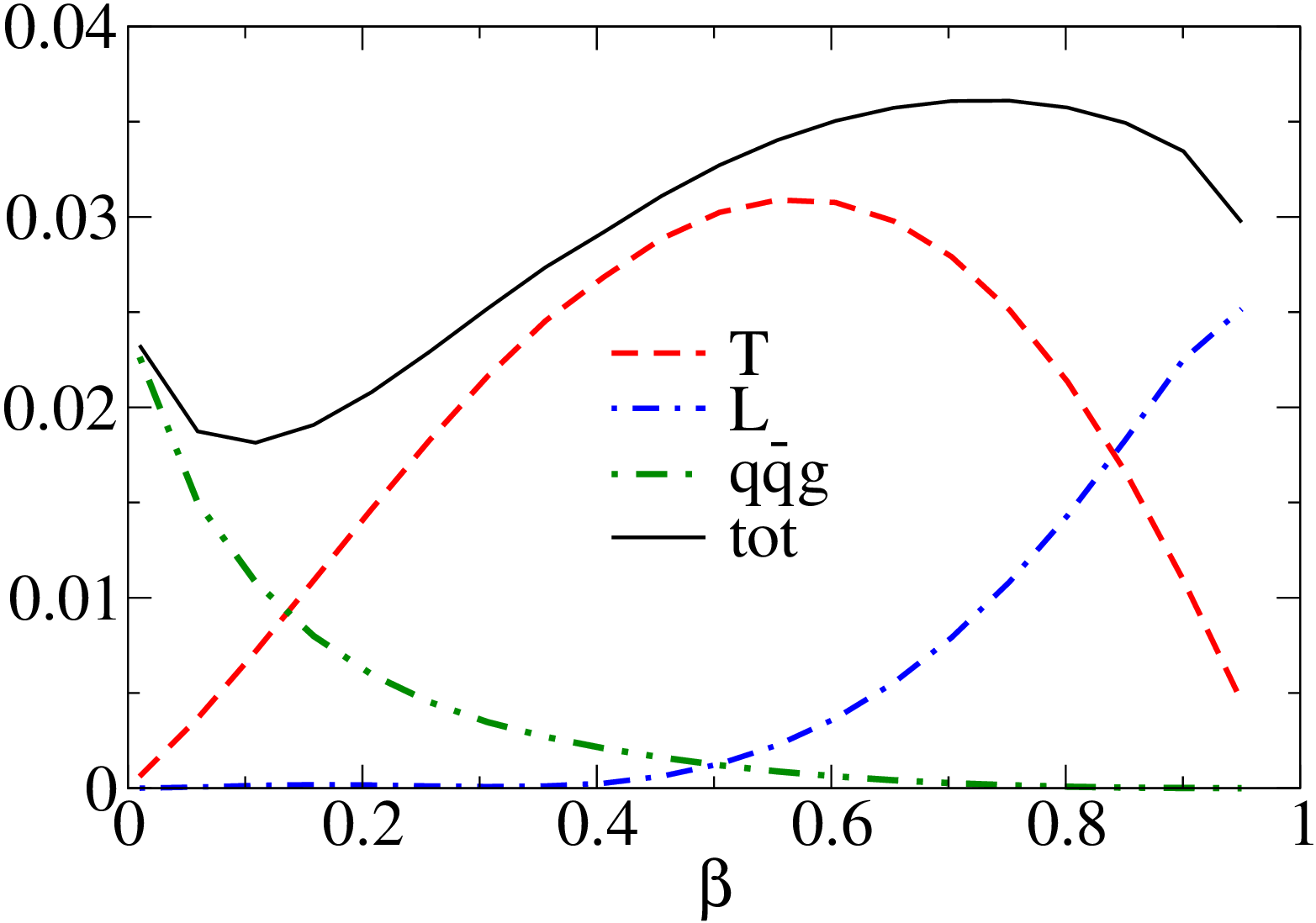}
  \hspace*{0.5cm}
  \includegraphics[width=6cm]{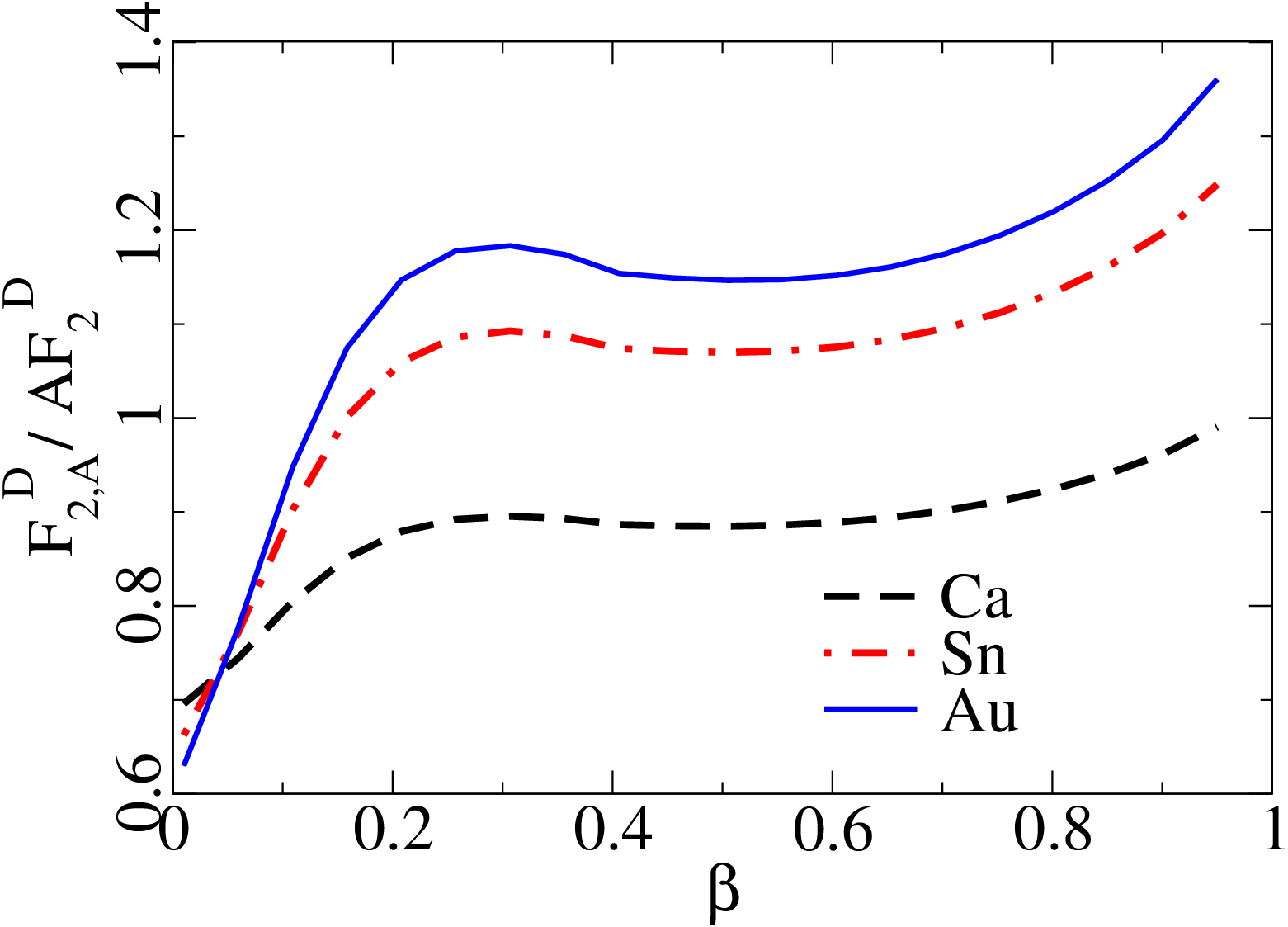}
  \caption{\small 
    Left plot: $\beta$-dependence of the different contributions to the proton
    diffractive structure function $F_{2,p}^D.$ Right plot: the ratio
    $F_{2,A}^D/(AF_{2,p}^D)$ as a function of $\beta$ for Ca, Sn and
    Au nuclei. In both cases, results are for the ``non breakup''
    case, and at $Q^2=5\ \mbox{GeV}^2$ and $\xp=0.001$. 
  }
\label{dipolehierarchy}
\end{figure}


\noindent{\bf Nuclear enhancement and suppression of $F_2^D$:}  In fig. \ref{dipolehierarchy}, the $\beta$ dependence of the diffractive structure function is displayed for
$Q^2=5\ \mbox{GeV}^2$ and $\xp=0.001.$
In the left plot, the hierarchy of the different contributions is analysed in the case of
$F_{2,p}^D.$ The dominant contribution is: the $q\bar q g$ component for
values of $\beta\!<\!0.1,$ the longitudinally polarized $q\bar q$ component for values of
$\beta\!>\!0.9,$ and the transversely polarized $q\bar q$ component for intermediate values.
In the case of $F_{2,A}^D,$ this separation is still true but the $q\bar q$ and $q\bar q g$ components behave differently as a function of $A.$ The $q\bar{q}$ components are enhanced compared to $A$ times the proton diffractive structure functions while the $q\bar{q}g$ component, on the contrary, is suppressed for nuclei compared to the proton (the $Q^2$ and $\xp$ dependence
of these effects will be discussed shortly).

This leads to a nuclear suppression of the diffractive structure function in the 
small $\beta$ region, and to an enhancement at large $\beta.$ This is illustrated
by the right plot of fig.~\ref{dipolehierarchy}, where the ratio $F_{2,A}^D/(AF_{2,p}^D)$ is shown as a function
of $\beta$ for different nuclei (for the ``non breakup'' case). The net result of the
different contributions is that $F_{2,A}^D/A$, for a large $\beta$ range down to 0.1, is close to
$F_{2,p}^D,$ and is increasing with $A.$

\begin{figure}[t]
  \centering
  \includegraphics[width=6cm]{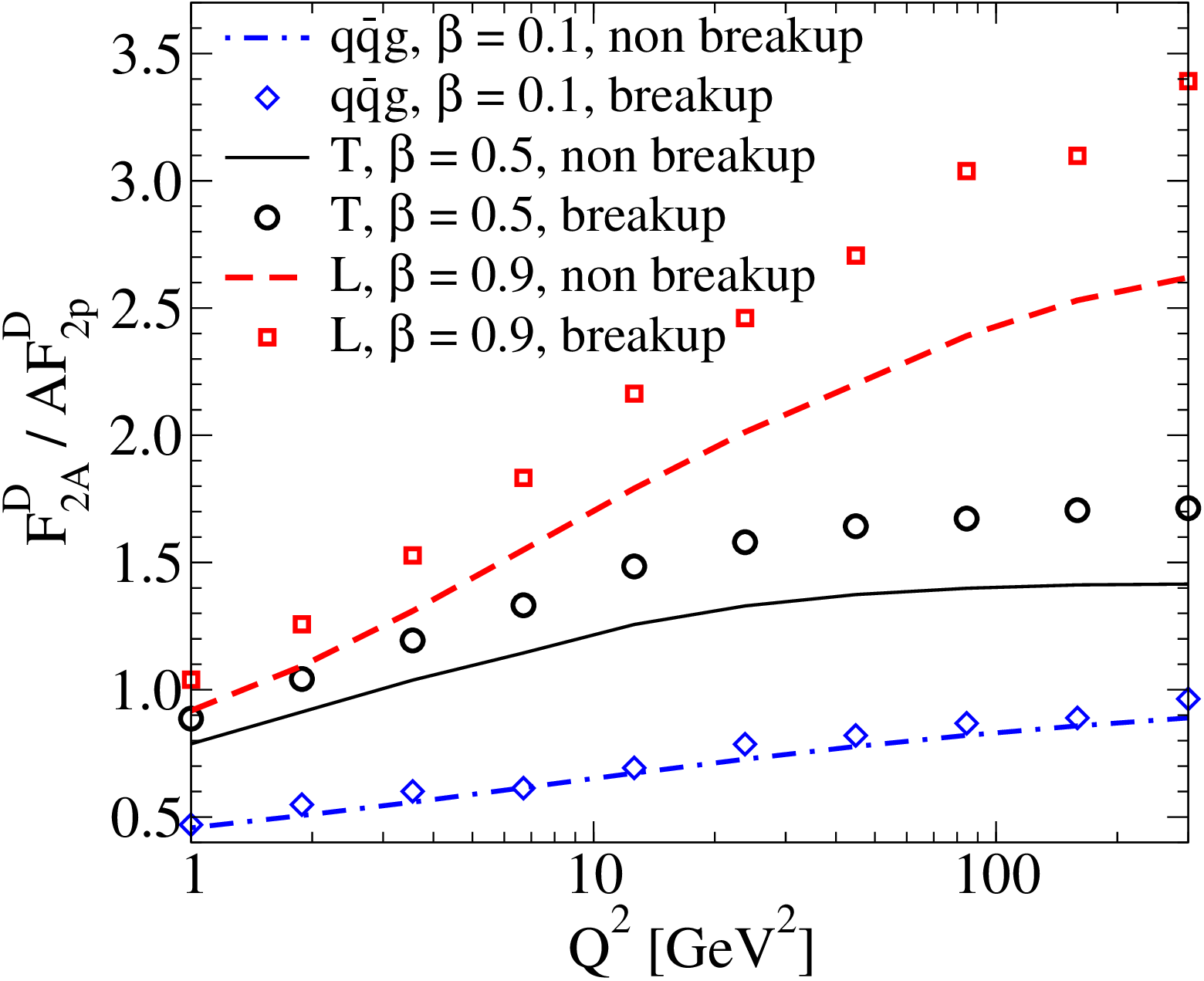}
  \hspace*{1cm}
  \includegraphics[width=6cm]{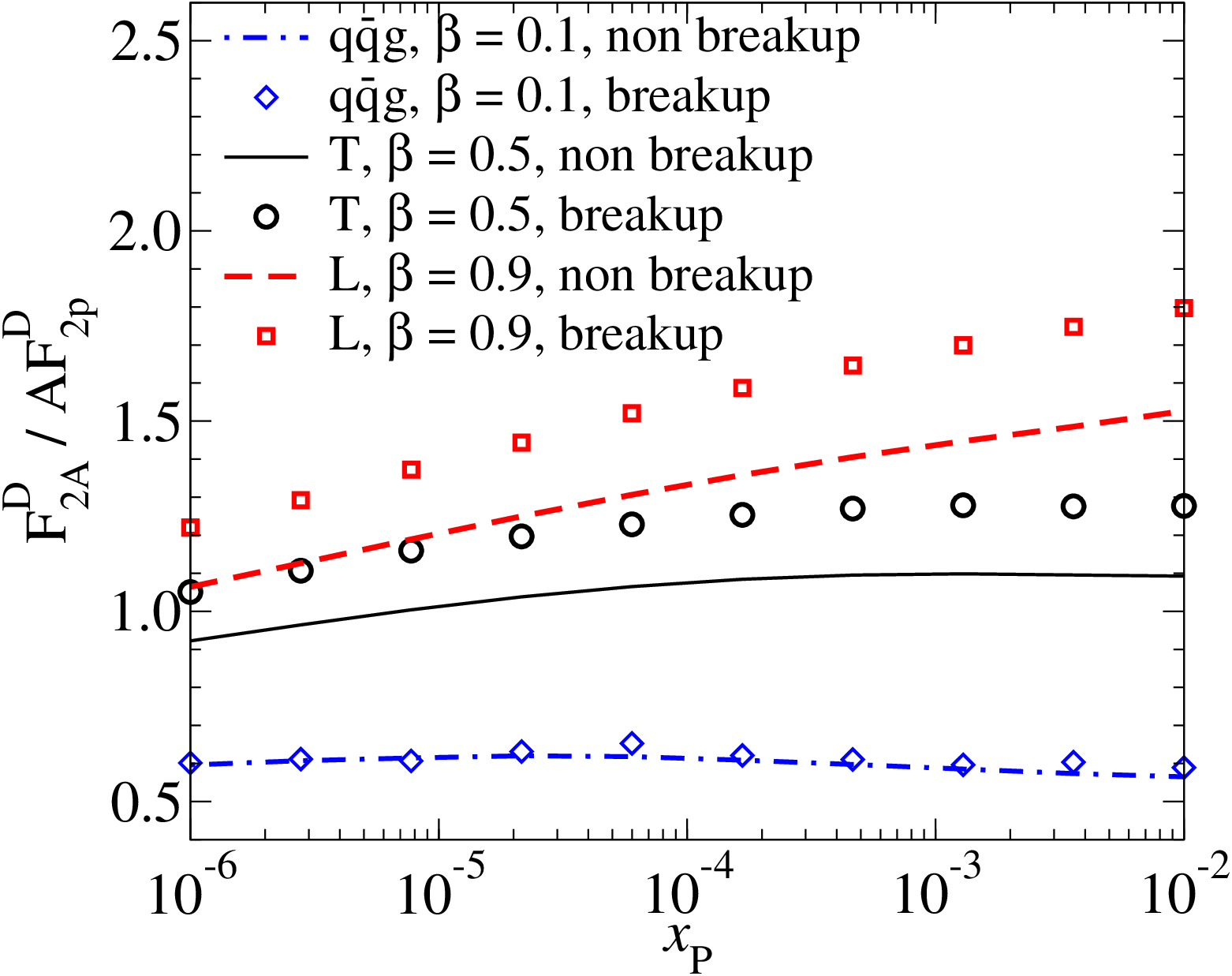}
\caption{\small The ratios $F_{2,A}^{D,x}/(AF_{2,p}^{D,x})$ of the different components
($x\!=\!q\bar qg,\ q\bar qT,\ q\bar qL$) of the diffractive structure function for both ``breakup''
and ``non breakup'' cases. Left plot: as a function of $Q^2$ for $\xp\!=\!0.001.$ 
Right plot: as a function of $\xp$ for $Q^2\!=\!5\ \mbox{GeV}^2.$
In both cases, results are for Au nuclei and the different components
are evaluated where they are dominant: at $\beta\!=\!0.1$ for $q\bar{q}g$,
$\beta\!=\!0.5$ for $q\bar qT$ and $\beta\!=\!0.9$ for $q\bar qL.$}
\label{eAdiffcgc}
\end{figure}

In fig. \ref{eAdiffcgc}, for the Au nucleus case, the ratios $F_{2,A}^D/(AF_{2,p}^D)$ of individual contributions are analyzed (for values of $\beta$ at which they are dominant).
Comparisons between the ``breakup'' and ``non breakup'' cases are made, as functions of $Q^2$ (left plot) and $\xp$ (right plot). For the $q\bar q g$ component, the nuclear suppression is almost constant (the suppression goes away slowly with $Q^2$). For the $q\bar q$ components, the enhancement becomes bigger with increasing $Q^2$ and $\xp.$ The result for the total diffractive cross-section in e+A scattering is that it decreases more slowly with increasing $Q^2$ or $\xp$ compared to the e+p case. Finally, cross sections in the ``non breakup'' case are about 15\%
lower than in the ``breakup'' case.

Comparing with other approaches, we obtain similar features. We notice one interesting difference with the results obtained using diffractive parton distributions modified by leading twist shadowing \cite{Frankfurt:2003gx}: even at large $\beta,$ it is found that $F_{2,A}^D/A$ is suppressed compared to $F^D_{2,p}$ as a function of $Q^2.$ This could be tested with measurements at a future EIC where diffraction will be an important part of a rich program. A typical nuclear enhancement of diffraction, for a Au nucleus, is a factor of $\sim\!1.2.$ Combining this with the typical nuclear suppression in the inclusive case ($\sim\!0.8,$ see \cite{Kowalski:2007rw}), we expect the fraction of diffractive events to be increased by a factor of $\sim\!1.5$ compared to the proton, meaning 25 to 35 \% at the EIC.

\subsubsection{Expectations for diffraction e+A DIS from LT shadowing}
\label{sec:LT_Diffraction}

\hspace{\parindent}\parbox{0.92\textwidth}{\slshape
 Vadim Guzey and Mark Strikman}
\index{Guzey, Vadim}
\index{Strikman, Mark}

\vspace{\baselineskip}

The leading twist theory of nuclear shadowing (see section~\ref{sec:LT_nuclear_shadowing}) 
that uses the connection between 
nuclear shadowing and diffraction~\cite{Gribov:1968jf} and allows one to 
predict nuclear parton distributions (PDFs) at small $x$~\cite{Frankfurt:1998ym,Frankfurt:2003zd,Guzey:2009jr,LT_shadowing_Phys_Rep} can
also be used to predict nuclear {\it diffractive} PDFs and {\it diffractive} 
structure functions~\cite{Frankfurt:2003gx}.
At small $x$ and in the nuclear target rest frame, the virtual photon interacts coherently 
with all nucleons of the nuclear target
and the $\gamma^{\ast}A \to X A$ scattering amplitude is given by the sum of the 
multiple scattering contributions presented in Fig.~\ref{fig:fgs10_coh_diff_multiple}.
Graphs $a$, $b$, and $c$ correspond to the coherent interaction with one, two, and three 
nucleons of the nuclear target, respectively:
 graph $a$ is the impulse approximation; graphs $b$ and $c$ 
contribute to the shadowing correction. Note that the interactions with four and more 
nucleons (at the amplitude level) are not shown, but 
they are implied.
The application of the Abramovsky-Gribov-Kancheli (AGK) cutting rules~\cite{Abramovsky:1973fm}
allows one to relate these diagrams to the corresponding diagrams for the total cross 
section in $\gamma^{\ast}A$ scattering.

\begin{figure}[htb]
\centerline{\includegraphics[width=0.6\textwidth]{eA-final/Figs/coh_diff_multiple_2011.epsi}}
\caption{\small 
  The multiple scattering series for the $\gamma^{\ast}A \to X A$
  diffractive scattering amplitude.
  Graph $a$ is the impulse approximation; graphs $b$ and $c$ 
  correspond to the interaction with two and three nucleons of
  the nuclear target, and contribute to the shadowing correction.
}
\label{fig:fgs10_coh_diff_multiple}
\end{figure}

Combining the Glauber-Gribov multiple scattering formalism for the
$\gamma^{\ast}A \to X A$ scattering amplitude 
with the QCD factorization
theorem~\cite{Collins:1997sr}, one can derive the nuclear 
diffractive parton distribution of flavor
$j$~\cite{LT_shadowing_Phys_Rep,Frankfurt:2003gx}: 
\begin{eqnarray}
  \beta f_{j/A}^{D(3)}(\beta,Q_0^2,x_{\Pomeron})&=&4 \pi A^2 \beta f_{j/N}^{D(4)}(\beta,Q_0^2,x_{\Pomeron},t_{\rm min}) \int d^2 b \nonumber\\
  &\times& \left| \int^{\infty}_{-\infty} dz e^{i x_{\Pomeron} m_N z}
    e^{-\frac{A}{2} (1-i\eta)\sigma_{\rm soft}^j(x,Q_0^2)
      \int_{z}^{\infty}dz^{\prime} \rho_A(b,z^{\prime})}
    \rho_A(b,z)\right|^2 \ ,
\label{eq:fgs10_diff1}
\end{eqnarray}
where the notation is the same as in eqs.~\eqref{eq:fgs10_eq1} and
\eqref{eq:fgs10_eq2}.

While at the level of the interaction with two nucleons 
(graphs $a$ and $b$ in fig.~\ref{fig:fgs10_coh_diff_multiple}) 
our predictions are model-independent, 
the contribution of the interaction with $N \geq 3$ nucleons requires additional
model-dependent considerations since these interactions probe the details
of the diffractive dynamics beyond what is encoded in the elementary 
diffractive distribution $f_j^{D(4)}$, as discussed in Section~\ref{sec:LT_nuclear_shadowing}.
Viewing the hard probe (virtual photon) as a coherent superposition of
the configurations that interact with the target nucleons with very different
strengths (from align-jet configurations to point-like configurations)
and which are present in the virtual photon 
with the probability $P(\sigma)$,
one immediately sees from fig.~\ref{fig:fgs10_coh_diff_multiple}
that diffractive scattering probes all moments of the cross section (color) 
fluctuations of the virtual photon, $\langle \sigma^n \rangle \equiv
\int d \sigma P(\sigma) \sigma^n$,
 up to the order $n=2A$. 
One should note that coherent diffraction probes these fluctuations differently
from inclusive scattering. For instance, while the shadowing correction to
the deuteron's usual parton distributions is proportional to $\langle \sigma^2 \rangle$
(i.e., it is unambiguously expressed in terms of the corresponding diffractive 
PDFs, see eq.~\ref{eq:fgs10_eq1} in section~\ref{sec:LT_nuclear_shadowing}), 
the shadowing correction to the deuteron's diffractive PDFs is proportional to
$\langle \sigma^3 \rangle$ (interference of graphs $a$ and $b$ in fig.~\ref{fig:fgs10_coh_diff_multiple}).
(Note that the square of graph $b$ in fig.~\ref{fig:fgs10_coh_diff_multiple}
proportional to $\langle \sigma^4 \rangle$ also contributes, but its contribution
is numerically very small.)
As the cross section fluctuations of the virtual photon 
($\langle \sigma^n \rangle$ moments) are rather weakly constrained 
by the present data, predictions of the leading twist theory of nuclear shadowing
contain unavoidable theoretical uncertainty
associated with modeling of $\langle \sigma^n \rangle$ with $n \geq 3$.
Precise measurements of the $t$ dependence of nuclear shadowing
in $eD$ diffraction at an EIC will dramatically reduce this uncertainty by determining 
exactly these  moments.


\begin{wrapfigure}{r}{0.65\columnwidth}
\centerline{\includegraphics[width=0.63\columnwidth]{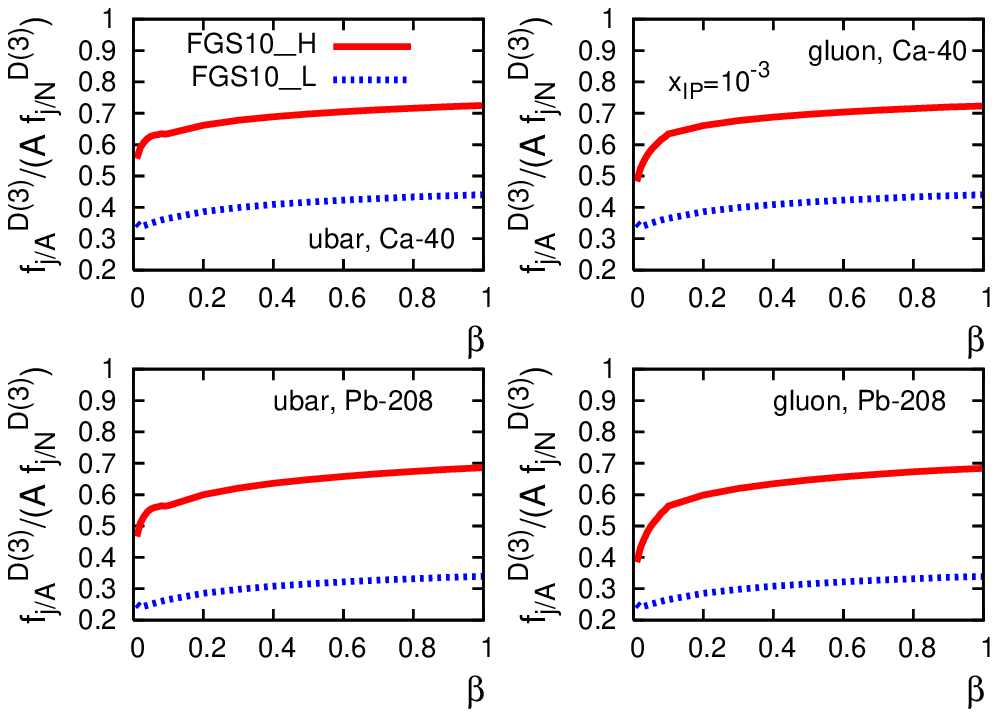}}
\caption{\small The leading twist theory of nuclear shadowing predictions for the ratio of nuclear to nucleon gluon and $\bar u$-quark diffractive PDFs, $f_{j/A}^{D(3)}/(Af_{j/N}^{D(3)})$, as a function of $\beta$ at $x_{\Pomeron}=10^{-3}$ and  $Q_0^2=4$ GeV$^2$.  The two sets of curves (labeled FGS10\_H and FGS10\_L) correspond to the two extreme scenarios of nuclear shadowing.}
\label{fig:fgs10_diffraction_ca40_beta_Jul10}
\end{wrapfigure}

Equation~(\ref{eq:fgs10_diff1}) determines nuclear diffractive PDFs at a certain initial
scale $Q_0^2$ ($Q_0^2=4$ GeV$^2$ in our case).
As a consequence of QCD factorization~\cite{Collins:1997sr},
the subsequent $Q^2$ evolution is given by the DGLAP evolution equations (at fixed
$x_{\Pomeron}$ and $t$). As another consequence of the
QCD factorization, the same nuclear diffractive
 PDFs $f_{j/A}^{D(3)}$ enter the perturbative QCD description of many processes and 
observables: the diffractive structure function $F_{2A}^{D(3)}$, the longitudinal 
diffractive structure function $F_{LA}^{D(3)}$, the charm structure functions
$F_{2A}^{D(3)(c)}$ and $F_{LA}^{D(3)(c)}$, and 
diffractive electroproduction of jets and heavy flavors.


As an example of our predictions for nuclear diffractive PDFs, 
in fig.~\ref{fig:fgs10_diffraction_ca40_beta_Jul10} we present 
the ratio of the nuclear ($^{40}$Ca and $^{208}$Pb)
to nucleon diffractive PDFs,
$f_{j/A}^{D(3)}/(Af_{j/N}^{D(3)})$, as a function of $\beta$ 
at fixed $x_{\Pomeron}=10^{-3}$ and $Q_0^2=4$ GeV$^2$. 
The left column of panels corresponds to the ${\bar u}$-quark distribution;
the right column corresponds to the gluon distribution.
The two sets of curves 
(labeled FGS10\_H and FGS10\_L) correspond to the two scenarios for
the effective cross section $\sigma_{\rm soft}^j$, which also
determines shadowing
effects as discussed in Section~\ref{sec:LT_nuclear_shadowing}.
As one can see from the comparison of fig.~\ref{fig:fgs10_diffraction_ca40_beta_Jul10}
to our predictions for the usual nuclear PDFs presented in fig.~\ref{fig:gfs10_LT2009_pb208_models12},
nuclear diffractive PDFs are much more sensitive to the effect of the color fluctuations
(the spread between the solid and dotted curves is much larger for $f_{j/A}^{D(3)}/(Af_{j/N}^{D(3)})$ than for $f_{j/A}(x,Q_0^2)/[Af_{j/N}(x,Q_0^2)]$).

\begin{table}[htb]
\centering
\begin{tabular}{|c|c|c|}
\hline
$A$/model & $F_{2A,{\rm incoh}}^{D(3)}/F_{2A}^{D(3)}$, $x_{\Pomeron}=10^{-3}$ 
& $F_{2A,{\rm incoh}}^{D(3)}/F_{2A}^{D(3)}$, $x_{\Pomeron}=10^{-2}$
\\
\hline
$^{40}$Ca, FGS10\_H & 0.35 & 0.33\\
$^{40}$Ca,  FGS10\_L & 0.43 & 0.38\\
$^{208}$Pb, FGS10\_H & 0.12 & 0.11\\
$^{208}$Pb, FGS10\_L & 0.20 & 0.16 \\
\hline
\end{tabular}
\caption{\small The leading twist theory of nuclear shadowing predictions for 
the ratio of the nuclear structure
functions measured in incoherent and coherent diffraction in $eA$ DIS,
$F_{2A,{\rm incoh}}^{D(3)}/F_{2A}^{D(3)}$,
at $x_{\Pomeron}=10^{-3}$ and $10^{-2}$ and $Q_0^2=4$ GeV$^2$. 
The ratio is approximately 
$\beta$-independent.}
\label{table:fgs10_F2diff}
\end{table}

The simplest observable to measure at an EIC is the 
diffractive structure function $F_{2A}^{D(3)}$. 
Our predictions for $F_{2A}^{D(3)}/(AF_{2N}^{D(3)})$ for $Q^2 \sim$ few GeV$^2$
are similar in shape
and close in the absolute value for $^{40}$Ca and model FGS10\_H 
to the corresponding predictions
made in the framework of the color dipole model, 
where the main contribution originates from the aligned-jet 
configurations~\cite{Kowalski:2008sa}.
(Note that at the level of the interaction with two nucleons, the expressions for the shadowing
correction in our leading twist approach and in the dipole formalism are essentially the same 
and are unambiguously expressed in terms of $\gamma^{\ast}$-nucleon diffraction.)
Hence, it appears that the $x_{\Pomeron}$ and $\beta$ dependence of
coherent inclusive diffraction in $eA$ DIS at $Q^2 \sim Q_0^2$ may not give 
unambiguous information on the onset of the non-linear regime of parton dynamics;
to distinguish between the non-saturation and saturation regimes one will 
need to study the $Q^2$ dependence of various diffractive observables.


In addition to inclusive coherent diffraction that we have discussed above,
the leading twist theory of nuclear shadowing makes predictions for incoherent 
diffraction (with nuclear break-up into its constituents) in $eA$ DIS, see~\cite{LT_shadowing_Phys_Rep} for details. 
An example of our predictions for the ratio of the nuclear structure
functions measured in incoherent and coherent diffraction in $eA$ DIS
at $x_{\Pomeron}=10^{-3}$ and $x_{\Pomeron}=10^{-2}$ and $Q_0^2=4$ GeV$^2$
is presented in table~\ref{table:fgs10_F2diff}.
To a good accuracy, the ratio is approximately 
$\beta$-independent.


\section{$k_T$-dependent gluons: SIDIS and jets}  
\label{sec:kTDependentGluons}  

\subsubsection{Dijet and Dihadron production at EIC}
\label{sec:dijet}

\hspace{\parindent}\parbox{0.92\textwidth}{\slshape
 Fabio Dominguez, Cyrille Marquet, Bowen Xiao and Feng Yuan}
\index{Dominguez, Fabio}
\index{Marquet, Cyrille}
\index{Xiao, Bo-Wen}
\index{Yuan, Feng}



\noindent{\bf Dijet production at an EIC:}  The operator definition of the Weizs\"{a}cker-Williams (WW)
 gluon distribution can be written as follows \cite{Bomhof:2006dp,Dominguez:2010xd}:
 \begin{equation}
xG^{(1)}(x,k_\perp)=2\int \frac{d\xi ^{-}d\xi _{\perp }}{(2\pi )^{3}P^{+}}%
e^{ixP^{+}\xi ^{-}-ik_{\perp }\cdot \xi _{\perp }}\ \langle P|\text{Tr}\left[F^{+i}(\xi ^{-},\xi _{\perp })\mathcal{U}%
^{[+]\dagger }F^{+i}(0)\mathcal{U}^{[+]}\right]|P\rangle \ ,\label{g1fund}
\end{equation}%
where the gauge link $\mathcal{U}_\xi^{[+]}=U^n\left[0,+\infty;0\right]U^n%
\left[+\infty, \xi^{-}; \xi_{\perp}\right]$ represents final state interactions with $U^n$ being the light-like
Wilson line in covariant gauge. By choosing the light-cone gauge with
certain boundary conditions for the gauge potential ($A_{\perp }(\zeta ^{-}=\infty)=0$ for the specific case above), we can drop out the
gauge link contribution in equation~\eqref{g1fund} and find that this gluon distribution has the
number density interpretation. Then, it can be calculated from the wave
functions or the WW field of the nucleus target
\cite{McLerran:1993ni,McLerran:1993ka,Kovchegov:1998bi}. At small-$x$ for a large nucleus, it
was found
\begin{equation}
xG^{(1)}(x,k_{\perp })=\frac{S_{\perp }}{\pi ^{2}\alpha _{s}}\frac{%
N_{c}^{2}-1}{N_{c}}\int \frac{d^{2}r_{\perp }}{(2\pi )^{2}}\frac{%
e^{-ik_{\perp }\cdot r_{\perp }}}{r_{\perp }^{2}}\left( 1-e^{-\frac{r_{\perp
}^{2}Q_{s}^{2}}{4}}\right) \ ,
\end{equation}%
where $N_c=3$ is the number of colors, $S_\perp$ is the transverse area of the target nucleus, and $Q^2_{s}=\frac{g^2N_c}{4\pi}\ln\frac{1}{r_{\perp}^2\lambda^2}\int dx^{-} \mu^2(x^{-})$ is the gluon saturation scale with $\mu^2$ the color charge density in a large nuclei.

\begin{figure}[tbp]
  \centering
  \hfill
  \parbox{0.49\linewidth}{
    \includegraphics[width=\linewidth]{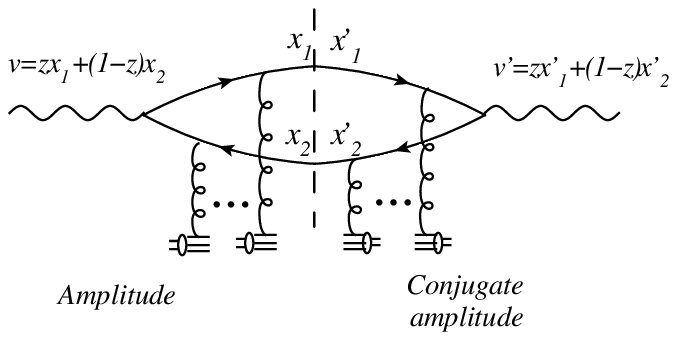}
  }
  \hfill
  \parbox{0.42\linewidth}{
    \includegraphics[width=\linewidth,bb=10 30 664 540,clip=true]{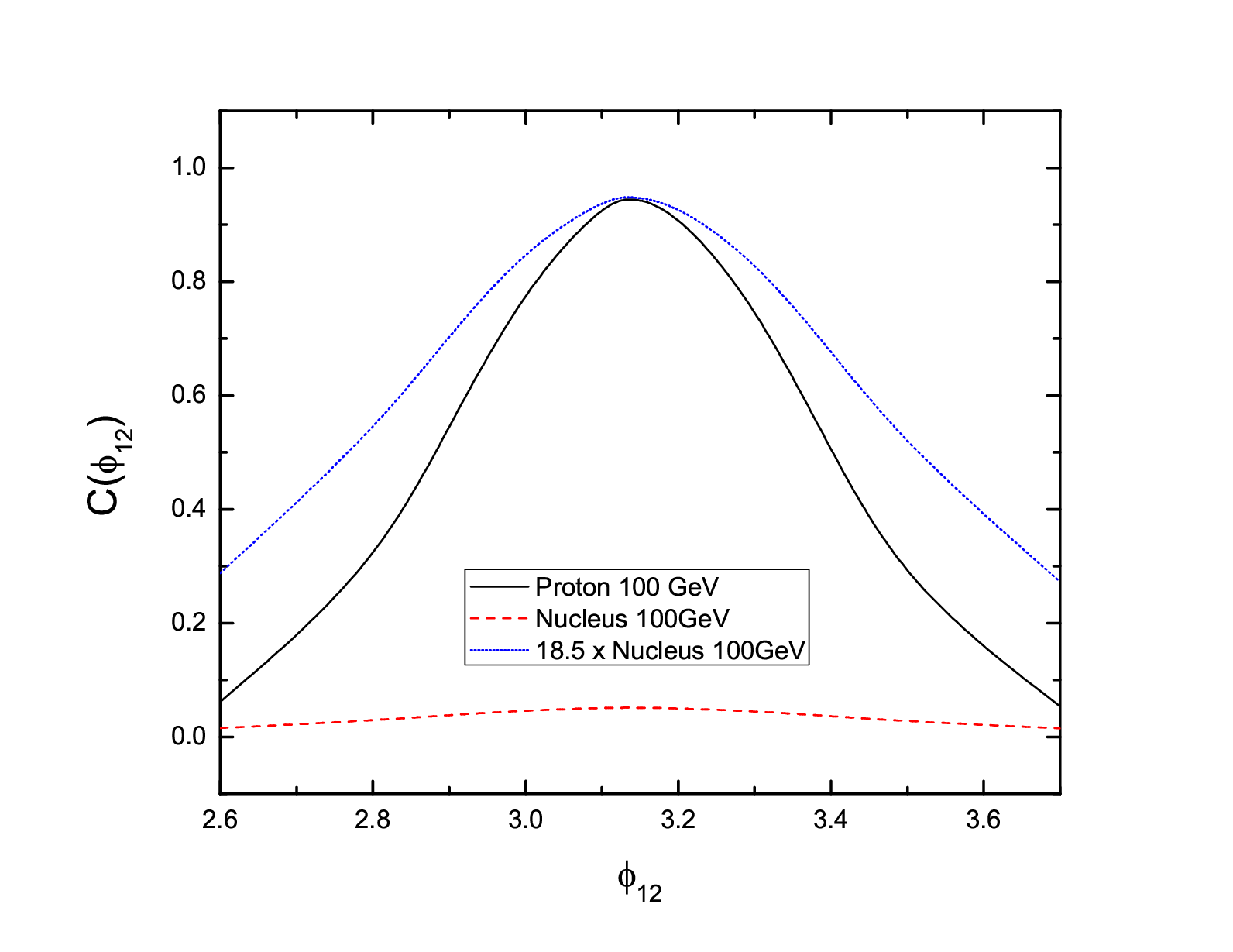}
  }
  \hfill
  \caption{\small 
    {\it Left:} Typical diagrams contributing to the cross section in
    the DIS at small-$x$ limit.
    {\it Right:} EIC dihadron correlation function
  } 
\label{discgc}
\label{norm}
\end{figure}

The second gluon distribution, the Fourier transform of the dipole cross
section, is defined in the fundamental representation
\begin{equation}
xG^{(2)}(x,k_{\perp }) =2\int \frac{d\xi ^{-}d\xi _{\perp}}{(2\pi)^{3}P^{+}}e^{ixP^{+}\xi ^{-}-ik_{\perp }\cdot \xi _{\perp }}\ \langle P|\text{%
Tr}\left[F^{+i}(\xi ^{-},\xi _{\perp })\mathcal{U}^{[-]\dagger }F^{+i}(0)%
\mathcal{U}^{[+]}\right]|P\rangle \ ,  \label{g2}
\end{equation}
where the gauge link $\mathcal{U}_\xi^{[-]}=U^n\left[0,-\infty;0\right]U^n\left[-\infty, \xi^{-}; \xi_{\perp}\right]$ stands for initial state interactions.
It has been shown in ref.~\cite{Dominguez:2010xd} that the Weizs\"{a}cker-Williams gluon distribution can be directly probed in the dijet production processes in DIS while the second gluon distribution enters in the total and semi-inclusive DIS cross sections.
The quark-antiquark dijet cross section in DIS can be calculated in both the CGC formalism and the TMD approach. In the CGC formalism,
the photon splits into a quark-antiquark pair which subsequently undergoes multiple interactions
with the nucleus (see figure~\ref{discgc} left).

After averaging over the photon's
polarization and summing over the quark and antiquark  colors and helicities in the splitting functions $\psi^{T, L\,\lambda}_{\alpha\beta}(p^+,z,r)$, we obtain,
\begin{eqnarray}
\frac{d\sigma ^{\gamma_{T,L}^{\ast }A\rightarrow q\bar{q}X}}{d^3k_1d^3k_2}
&=&N_{c}\alpha _{em}e_{q}^{2}\delta(p^+-k_1^+-k_2^+) \int
\frac{\text{d}^{2}x_1}{(2\pi)^{2}}\frac{\text{d}^{2}x_1^{\prime }}{(2\pi )^{2}}
\frac{\text{d}^{2}x_2}{(2\pi)^{2}}\frac{\text{d}^{2}x_2^{\prime }}{(2\pi )^{2}} \notag \\
&&\times e^{-ik_{1\perp }\cdot(x_1-x_1^{\prime })} e^{-ik_{2\perp }\cdot (x_2-x_2^{\prime })}
\sum_{\lambda\alpha\beta} \psi_{\alpha\beta}^{T, L \lambda}(x_1-x_2)
\psi_{\alpha\beta}^{T, L\lambda*}(x_1^{\prime }-x_2^{\prime })  \notag \\
&&\times \left[1+S^{(4)}_{x_g}(x_1,x_2;x_2^{\prime },x_1^{\prime})
-S^{(2)}_{x_g}(x_1,x_2)-S^{(2)}_{x_g}(x_2^{\prime },x_1^{\prime })\right] \ ,\label{xsdis}
\end{eqnarray}
where $k_1$ and $k_2$ are momenta for the final state quark and
antiquark, respectively. We further define $\vec P_\perp=\vec
k_{1\perp}-\vec k_{2\perp}$ and $\vec q_\perp=\vec
k_{1\perp}+\vec{k}_{2\perp}$. All transverse momenta are defined
in the center of mass frame of the virtual photon and the nucleus
target. The two- and four-point functions are defined as
\begin{equation} \small
S_{x_g}^{(2)}(x_1,x_2) = \frac{1}{N_c}\left\langle\text{Tr}\;
  U(x_1)U^\dagger(x_2)\right\rangle_{x_g} \!, \
S_{x_g}^{(4)}(x_1,x_2;x_2^{\prime},x_1^{\prime}) =
\frac{1}{N_c}\left\langle\text{Tr}\; 
  U(x_1)U^\dagger(x_1^{\prime})
U(x_2^{\prime})U^\dagger(x_2)\right\rangle_{x_g}\,.
\end{equation}
The notation $\langle\dots\rangle_{x_g}$ is used for the CGC average of the color
charges over the nuclear wave function where $x_g$ is the smallest fraction of
longitudinal momentum probed, and is determined by the kinematics.

In order to simplify the above result and obtain a factorized expression, we take the correlation limit of
equation~\eqref{xsdis}. For convenience, we introduce the transverse coordinate variables:
$u=x_1-x_2$ and $v=zx_1+(1-z)x_2$, and similarly for the primed coordinates.
The respective conjugate momenta are $\tilde P_\perp=(1-z)k_{1\perp}-zk_{2\perp}\approx P_\perp$
and $q_\perp$, and therefore the correlation limit ($\tilde P_\perp \gg q_\perp$)can be taken by assuming $u$ and $u'$
are small and then expanding the integrand with respect to these two variables before
performing the Fourier transform. Therefore, we can obtain the following expression which agrees perfectly with the TMD approach:
\begin{eqnarray}
\frac{d\sigma_{\textrm{TMD}}^{\gamma_{T,L}^{\ast }A\rightarrow q\bar{q}+X}}{d\mathcal{P.S.}}
=\delta (x_{\gamma ^{\ast }}-1) x_{g}G^{(1)}(x_{g},q_{\perp
})H_{\gamma_{T,L}^*g\to q\bar q}, \label{factdis}
\end{eqnarray}
where $x_g$ is the momentum fraction carried by the gluon and is determined
by the kinematics, $x_{\gamma^*}=z_q+z_{\bar q}$ with $z_q=z$ and $z_{\bar q}=1-z$
being the momentum fractions of the virtual photon carried by the quark and
antiquark, respectively. The phase space factor is defined as $d\mathcal{P.S.
}= dy_{1}dy_{2}d^2P_{\perp }d^{2}q_{\perp }$, and $y_1$ and $y_2$ are
rapidities of the two outgoing particles in the lab frame.
The leading order hard partonic cross section reads 
\begin{equation}
H_{\gamma_T^*g\to q\bar q}={\alpha_{s}\alpha _{em}e_{q}^{2}}\frac{\hat
s^2+Q^4}{(\hat s+Q^2)^4} \left(\frac{\hat{u}}{\hat{t}}+\frac{\hat{t}}{\hat{u}%
}\right)\ ,\quad H_{\gamma_L^*g\to q\bar q}={\alpha_{s}\alpha _{em}e_{q}^{2}}\frac{8\hat{s}%
Q^2}{(\hat s+Q^2)^4}\ ,
\end{equation}
with the usually defined partonic Mandelstam variables $\hat
s=P_\perp^2/(z(1-z))$, $\hat t=-(P_\perp^2+\epsilon_f^2)/(1-z)$,
and $\hat u=-(P_\perp^2+\epsilon_f^2)/z$ with
$\epsilon_f^2=z(1-z)Q^2$. \\


\noindent{\bf di-hadron correlations in DIS:}  By including the $k_t$ dependent fragmentation functions as proposed in ref.~\cite{Anselmino:2008sga}, one can compute the di-hadron production cross section and the correlation function $C(\phi_{12})$ which is defined as follows
\begin{equation}
C(\phi_{12}) =\frac{1}{\frac{d\sigma^{\gamma^{\ast }A\rightarrow h_1 X}_{\textrm{tot SIDIS}}}{dz_{h1}}} \frac{d\sigma_{\textrm{tot}}^{\gamma^{\ast }A\rightarrow h_1 h_2+X}}{dz_{h1}dz_{h2} d\phi_{12}},
\end{equation}
where $z_{h1}$ and $z_{h2}$ are the longitudinal momentum fractions of
two produced hadrons w.r.t. the photon momentum. $p_{1\perp}$ and
$p_{2\perp}$ are the transverse momenta of these two back-to-back
hadrons and $\phi_{12}$ is the azimuthal angle between them. Thus, it
is straightforward to numerically evaluate the correlation function
and plot it in figure~\ref{norm} right, where we fix $z_{h1}=z_{h2}=0.3$, $Q^2=4.0 \gev^2$, $\sqrt{s}=100\gev$. $p_{1\perp}$ and $p_{2\perp}$ are integrated in the range $[2, 3] \gev$ and $[1, 2] \gev$, respectively. For the gluon distribution in gold nuclei, we have used a parametrization inspired by $GBW$ model. From figure~\ref{norm}, one sees the suppression of the away-side peak in nuclei due to gluon saturation. \\


\noindent{\bf Conclusion:}  First of all, we would like to compare the dijet production process in DIS to the inclusive and semi-inclusive DIS. As shown above, we derive that the dijet production cross section in DIS is proportional to the WW gluon distribution in the correlation limit. On the other hand, it is well-known that inclusive and semi-inclusive DIS involves the dipole cross section instead~\cite{Marquet:2009ca}, which can be related to the second gluon distribution. This might look confusing at first sight, so let us take a closer look at equation~\eqref{xsdis}. If one integrates over one of the outgoing momenta, say $k_1$, one can easily see that the corresponding coordinates in the amplitude and conjugate amplitude are identified ($x_1=x_1^{\prime}$) and, therefore, the four-point function $S_{x_g}^{(4)}(x_1,x_2;x_2^{\prime},x_1^{\prime})$ collapses to a two-point function $S^{(2)}_{x_g}(x_2,x_2^{\prime})$. As a result, the SIDIS and inclusive DIS cross section only depend on two-point functions, thus they only involve the dipole gluon distribution.

Now we can see the unique feature of the dijet production process in DIS. By keeping the momenta of the quark and antiquark unintegrated, we can keep the full color structure of the four-point function which eventually leads to the WW gluon distribution in the correlation limit. Therefore, measuring the dijet production cross sections or dihadron correlations in DIS at future experimental facilities like EIC would give us a first direct and unique opportunity to probe and understand the Weizs\"acker-Williams gluon distribution. Last but not least, by measuring the SIDIS and inclusive DIS cross section at EIC, one can also probe and constrain the dipole gluon distribution.

\ \\ \noindent{\it Acknowledgments:}
We thank A.~Accardi, M.~Diehl, L.~McLerran, A.~Mueller, J.-W.~Qiu, A.~Stasto and R.~Venugopalan for stimulating discussion.  


\subsubsection{Heavy quark production in $e+A$ collisions}
\label{sec:heavy_victor}

\hspace{\parindent}\parbox{0.92\textwidth}{\slshape
 Victor P. Gon\c{c}alves}
\index{Gon\c{c}alves, Victor P.}

\vspace{\baselineskip}

In this contribution we calculate the  cross section  of heavy quark production  
using the dipole approach and a nuclear saturation model based on the physics of the  
Color Glass Condensate  (CGC)  (For more details and references see Ref. \cite{Goncalves:2010vq}). 
The main input of  our calculation  is the dipole-nucleus cross section, 
$\sigma_{dA} (x,r)$,  which is determined by the QCD dynamics at small $x$. In the eikonal 
approximation it is  given by twice the impact-parameter $b$ integral of ${\cal N}^A (x, r, b)$, the forward dipole-target scattering amplitude for a 
dipole with size $r$ which encodes all the information about the 
hadronic scattering, and thus about the nonlinear and quantum effects in the hadron wave 
function. In our calculations we will assume as before that the forward dipole-nucleus 
amplitude is given by 
\begin{eqnarray}
{\cal{N}}^A(x,r,b) = 1 - \exp \left[-\frac{1}{2}  \, \sigma_{dp}(x,r^2) 
\,T_A(b)\right] \,\,,
\label{enenuc}
\end{eqnarray}
where $\sigma_{dp}$ is the dipole-proton cross section and $T_A(b)$ is the nuclear profile 
function, which is obtained from a 3-parameter Fermi distribution for the nuclear
density normalized to $A$.
 It is important to emphasize that this model describes  the current 
experimental data on the nuclear structure function as well as includes the  impact parameter 
dependence in the dipole nucleus cross section. 
For the dipole-proton cross section we  will use 
the b-CGC model.

To estimate the magnitude of the saturation effects in  heavy quark production, let us compare  the CGC predictions  with those associated to linear QCD dynamics. As a model for the linear regime we consider the leading logarithmic approximation 
for the dipole-target cross section, where $\sigma_{dA}$ is directly 
related to the nuclear gluon distribution $xg_A$ as follows
\begin{eqnarray}
\sigma_{dA}(x,r^2) =  {\frac{\pi^2}{3}} r^2 \alpha_s xg_A(x, 10/r^2) \,\,.
\label{gluongrv}
\end{eqnarray}
The use of this cross section in the formulae given below will produce results which we
denote CT, from color transparency. In this limit we are disregarding multiple scatterings of the dipole with the nuclei and are assuming that the dipole interacts incoherently with the target. In what follows we consider two different models for 
the nuclear gluon distribution. In the first one we disregard the nuclear effects and 
assume that $xg_A (x,Q^2) = A.xg_N (x,Q^2)$, with $xg_N$ being the  gluon distribution in 
the proton and given by the GRV98 parameterization. We will refer to this 
model as CT. In the second model we take into account the nuclear effects in the nuclear 
gluon distribution as described by the EKS98 parameterization.  We will call this 
model CT + Shad. In our calculations the charm quark mass is $m_c = 1.5$ GeV and the  bottom 
quark  mass is $m_b = 4.5$ GeV. \\


\noindent{\bf Heavy quark production in the color dipole approach:}  Heavy quark production is usually estimated using the  collinear factorization approach, where all partons involved are assumed to be on mass shell, carrying only longitudinal momenta, and their transverse momenta are neglected in the QCD matrix elements.  On the other hand, in the large energy (small-$x$) limit, we have that the characteristic scale $\mu$ of the hard subprocess of parton scattering is much less than $\sqrt{s}$, but greater than the $\Lambda_{QCD}$ parameter. In this limit, the effects of the finite transverse momenta of the incoming partons become important, and the factorization must be generalized, implying that the cross sections are now $k_{\perp}$-factorized into an off-shell partonic cross section and a  $k_{\perp}$-unintegrated parton density function ${\cal{F}}(x,k_{\perp})$, characterizing the $k_{\perp}$-factorization  approach.  Recently, an alternative approach to calculating the heavy quark production at high energies was proposed, considering the quasi-multi-Regge-kinematics (QMRK)  framework. It  is based on an effective theory implemented with the non-Abelian gauge-invariant action.  The heavy quark production  can also be calculated using the  color dipole approach. This formalism can be obtained from the $k_{\perp}$-factorization approach  after the Fourier transformation from the space of quark transverse momenta into the space of transverse coordinates. It is important to emphasize that this equivalence is only valid in the leading logarithmic approximation, being violated if the exact gluon kinematics is considered. A detailed discussion of the equivalence or not  between the dipole  and the QMRK approaches  still is an open question.  The main advantage to use the color dipole formalism, is that it gives a simple unified picture of inclusive and diffractive processes and the saturation effects can be easily implemented in this approach.

\begin{figure}
  \centering
  \includegraphics[width=0.42\textwidth]{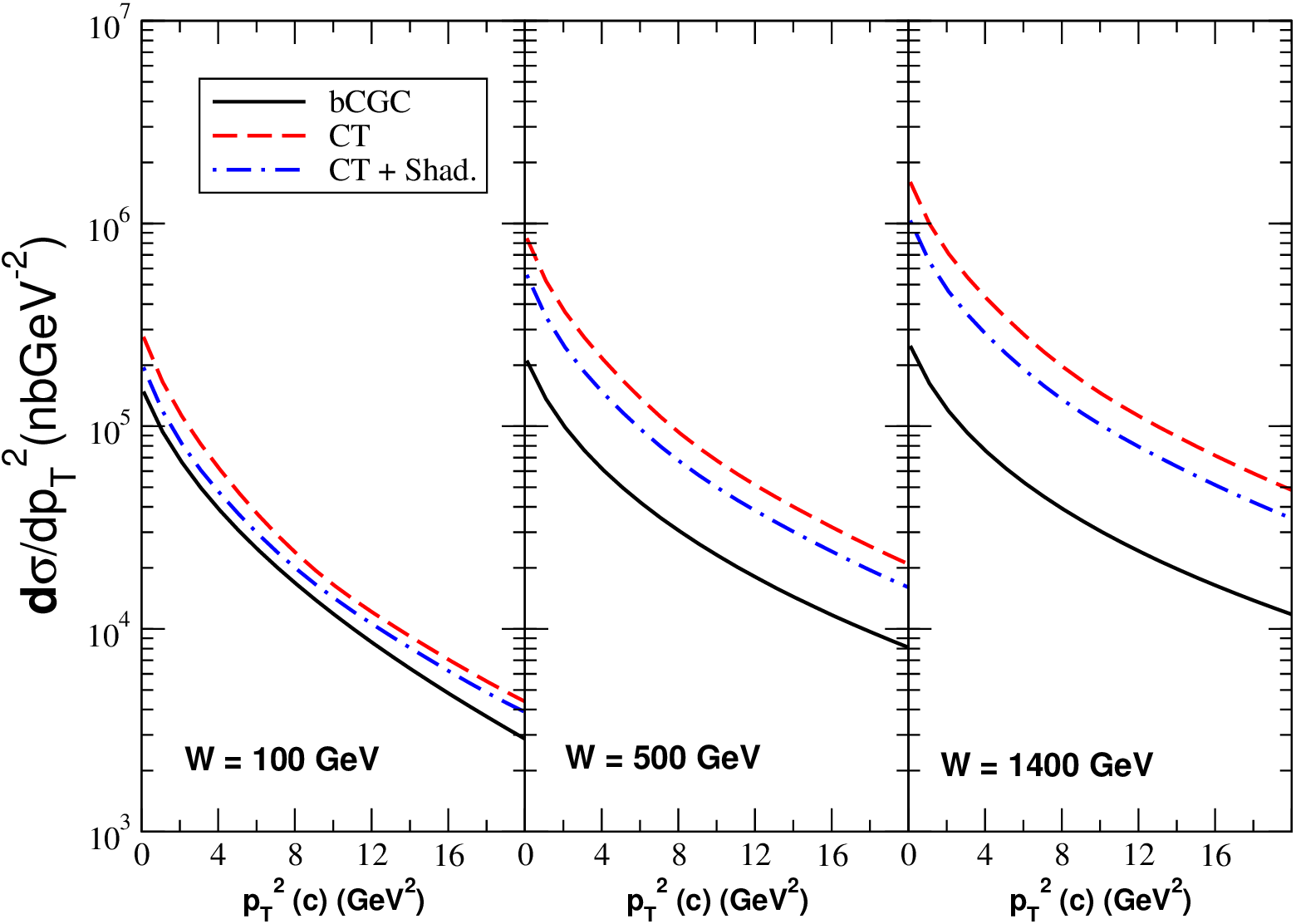}
  \hspace*{0.5cm}
  \includegraphics[width=0.42\textwidth,clip=true]{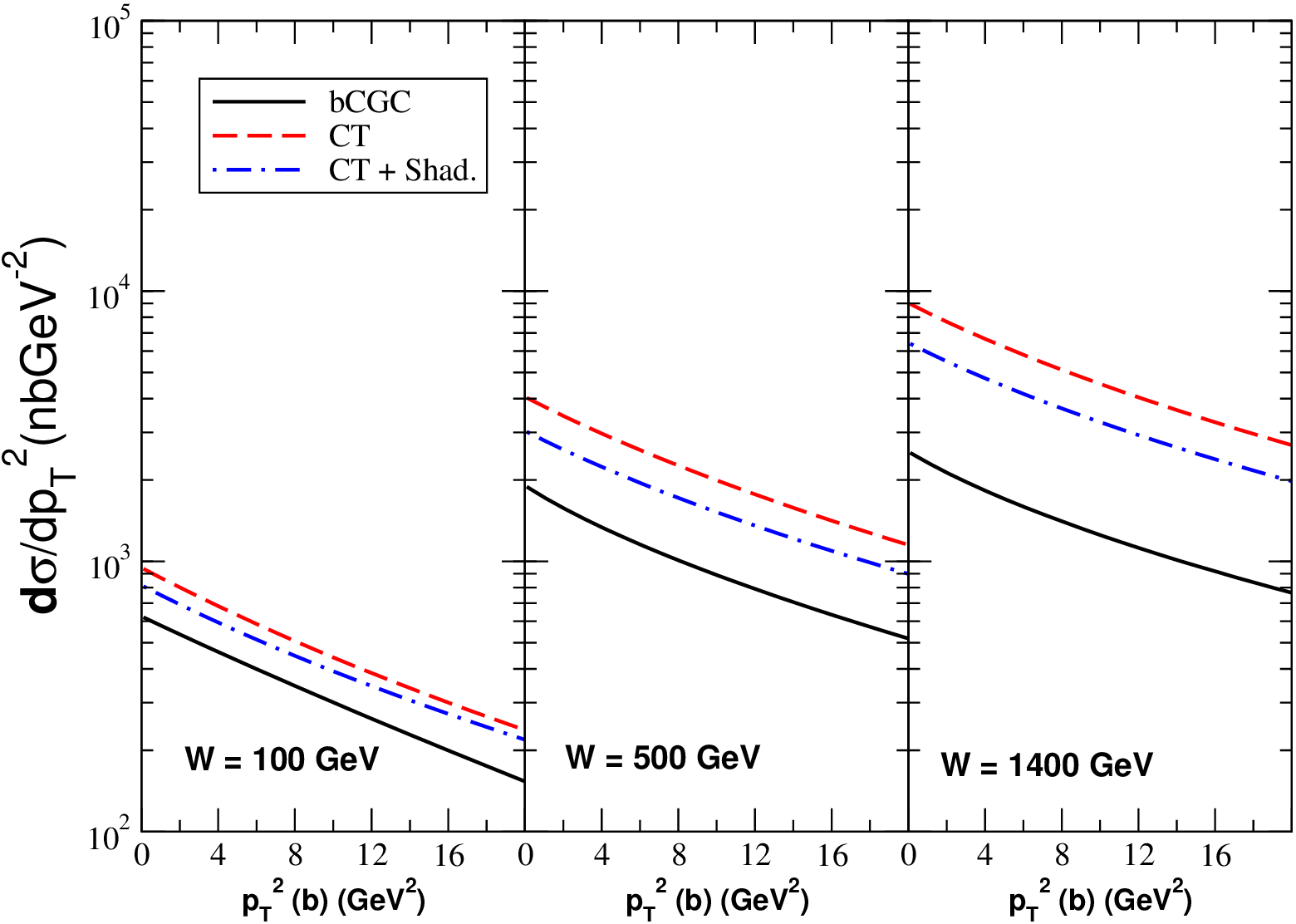}
  \caption{\small Transverse momentum charm spectrum (left) and bottom spectrum
    (right) for $Q^2 = 2$ GeV$^2$ and different energies.}
  \label{fig:charm}
  \label{fig:bottom}
\end{figure}

In the color dipole approach, the heavy quark production cross section is given by 
\begin{eqnarray}  
\frac{d\sigma(\gamma^{*}A\rightarrow Q\,X)}{d^{2}p_{Q}^{\bot}}  
&=&\frac{6e_{Q}^{2}\alpha_{em}}{(2\pi)^{2}}\int d\alpha
 \left\lbrace\left[\vphantom{\frac{1}{1}}m_{Q}^{2}+
4Q^{2}\alpha^{2}(1-\alpha)^{2}\right]  
\right.\left[\frac{I_{0}}{p_{Q}^{\bot 2}+\epsilon^{2}}-\frac{I_{2}}{4\epsilon}\right]
\nonumber\\  
&+&\left[\vphantom{\frac{1}{1}}\alpha^{2}+(1-\alpha)^{2}\right]\left.  
\left[\frac{p_{Q}^{\bot}\epsilon I_{1}}{p_{Q}^{\bot 2}+  
\epsilon^{2}}-\frac{I_{0}}{2}+\frac{\epsilon I_{2}}{4}\right]\right\rbrace  
\label{eqn:epI}  
\end{eqnarray}  
with   
\begin{eqnarray}  
I_{\lambda}=\int dr\,r\,J_{\lambda}(p_{Q}^{\bot}r)\,K_{\lambda}(\epsilon r)\,\sigma_{dA}(r)\,\,;\,\, 
I_{2}=\int dr\, r^{2}\, J_{0}(p_{Q}^{\bot}r)\,K_{1}(\epsilon r)\,\sigma_{dA}(r)\nonumber 
\label{eqn:I1-I3}  
\end{eqnarray}  
with $\lambda =0,1$, and $J_{0,1}$ and $K_{0,1}$ are Bessel functions, and $\epsilon^2 = \alpha (1 - \alpha) Q^2 + m^2$. \\


\begin{figure}[h]
\centerline{\includegraphics[width=0.45\textwidth,clip=true]{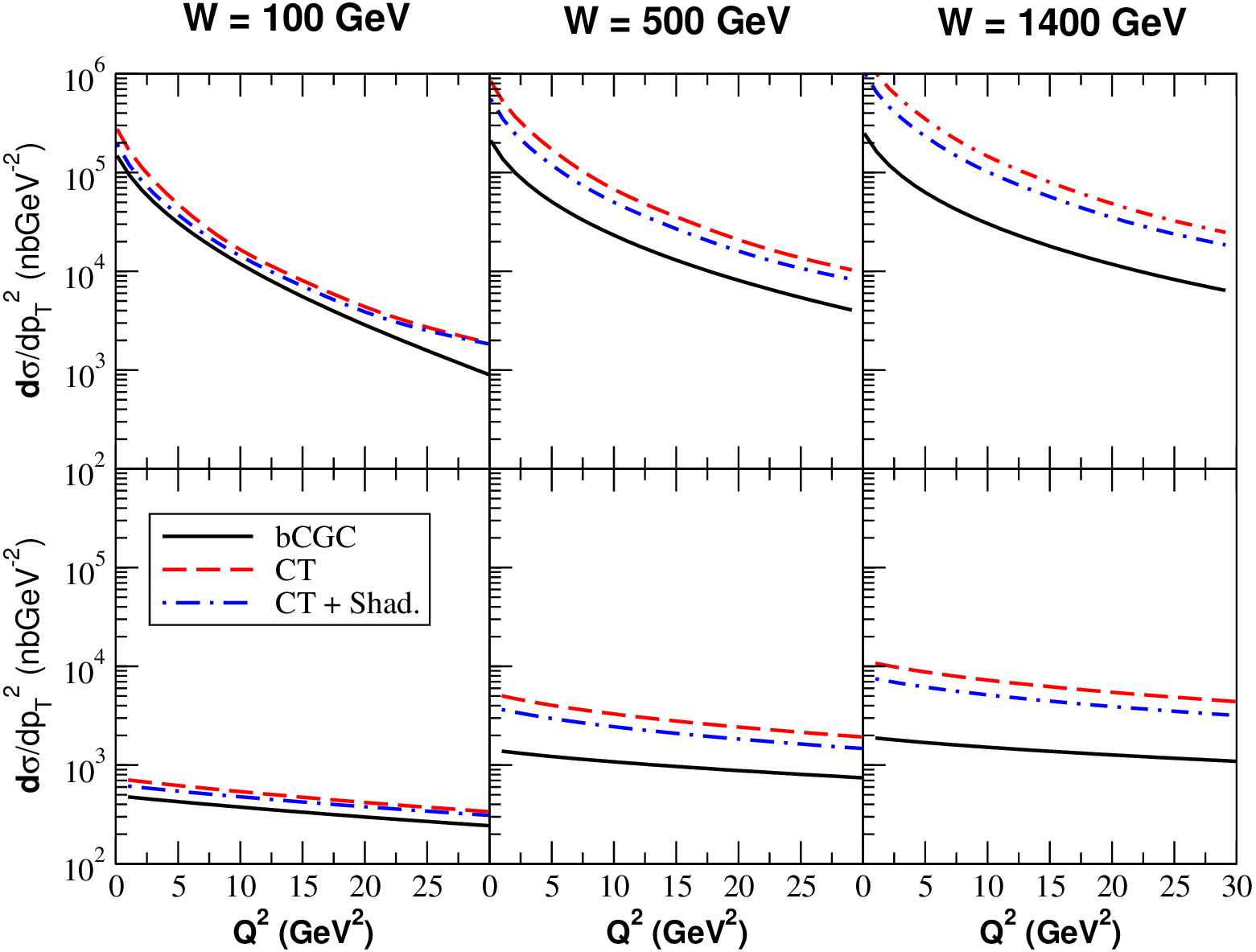}}
\caption{\small Dependence on the photon virtuality at $p_T^2 = 4$ GeV$^2$.}
\label{fig:q2} 
\end{figure}

\noindent{\bf Results:}  In Fig. \ref{fig:charm} we show the transverse momentum spectrum of charm quarks. The main 
purpose of this figure is to show that the predictions of the  linear physics (CT + Shad) 
differ from the total  (i.e. bCGC)  by a factor which increases with 
the energy $W$ and goes from  $1.5$ ($W=100$ GeV) to $4$ ($W=1400$ GeV). Moreover, this 
difference persists for a wide momentum window. At very large $p_{T}$ we enter the deep 
linear  regime and expect that the two curves coincide. 

In Fig. \ref{fig:bottom} we show the transverse momentum spectrum of bottom quarks. As expected, 
we observe the same features of the charm distribution, except that now the non-linear 
effects are weaker. Nevertheless they are still noticeable. 
In Fig. \ref{fig:q2} we show the $Q^2$ dependence of the $p_{T}$ distribution at a fixed 
value $p_{T} = 4$ GeV$^2$ for different energies. The upper and lower panels show the charm 
and bottom distributions respectively. Here again, we observe a remarkable strength and 
persistence up to large virtualities of the differences between CT + Shad and bCGC.

\ \\ \noindent{\it Acknowledgments:}
The author thanks  M. S. Kugeratski and  F.S. Navarra for collaboration.

\section{$b$-dependent gluons: exclusive VM, DVCS}  
\label{sec:bdep}  

\subsubsection{Gluon Density in e+A : KLN, CGC, DGLAP Glauber, or Neither?}
\label{sec:EVMPHorowitz}

\hspace{\parindent}\parbox{0.92\textwidth}{\slshape
 William A. Horowitz}
\index{Horowitz, William A.}

\vspace{\baselineskip}

Perturbative quantum chromodynamics (pQCD) predicts a nontrivial expansion in the size of the nuclear wavefunction at small $x$ as the perturbative power law tails of the gluon distribution near the edge of the nucleus become important compared to the exponential dropoff due to confinement effects \cite{Froissart:1961ux,Gribov:1984tu,Iancu:2002xk}.  Similarly, in order to not violate unitarity, the enormous growth in the gluon parton distribution function as $x$ becomes small found via na\"ive application of DGLAP evolution (see \cite{Nakamura:2010zzi} and references therein) must be tamed by perturbatively-calculable saturation effects \cite{Gribov:1984tu,Iancu:2002xk}.  However it is not yet clear from a theoretical standpoint at what values of $x$ these nontrivial changes in the dominant dynamics occur \cite{Iancu:2002xk}.   Additionally a quantitative theoretical understanding of experimental heavy ion data requires a quantitative understanding of the initial geometry of a heavy ion collision.  Certainly observables such as the azimuthal anisotropy of particles \cite{Hirano:2005xf,Luzum:2008cw,Jia:2010ee} are correlated with the anisotropy of the initial geometry; surprisingly the event-by-event fluctuations in the initial geometry also strongly affect these observables \cite{Schenke:2010rr,Jia:2011pi}.  In particular the viscosity to entropy ratio ($\eta/s$) of the quark-gluon plasma (QGP) found by comparing hydrodynamics simulations to heavy ion collision data is directly related to the eccentricity of the initial thermal quark-gluon plasma distribution that is evolved hydrodynamically.  Currently the uncertainty in the initial thermal distribution due to the uncertainty in the importance of saturation effects in the initial nuclear profiles is large enough that it is not clear whether the physics of the QGP is better described by leading order weakly-coupled perturbative quantum chromodynamics (LO pQCD) or by LO strongly-coupled anti-de-Sitter/conformal field theory (AdS/CFT) methods \cite{Luzum:2008cw}.  An experimental measurement of the spatial gluon distribution in a highly boosted nucleus, and hence the relevant physics in this kinematic range, would thus be a very interesting and important contribution to our understanding of QCD.  

\begin{figure}[!htb]
  \centering
  \parbox{2.3in}{
    \includegraphics[width=\linewidth,bb=0 -200 1182 650,clip=true]{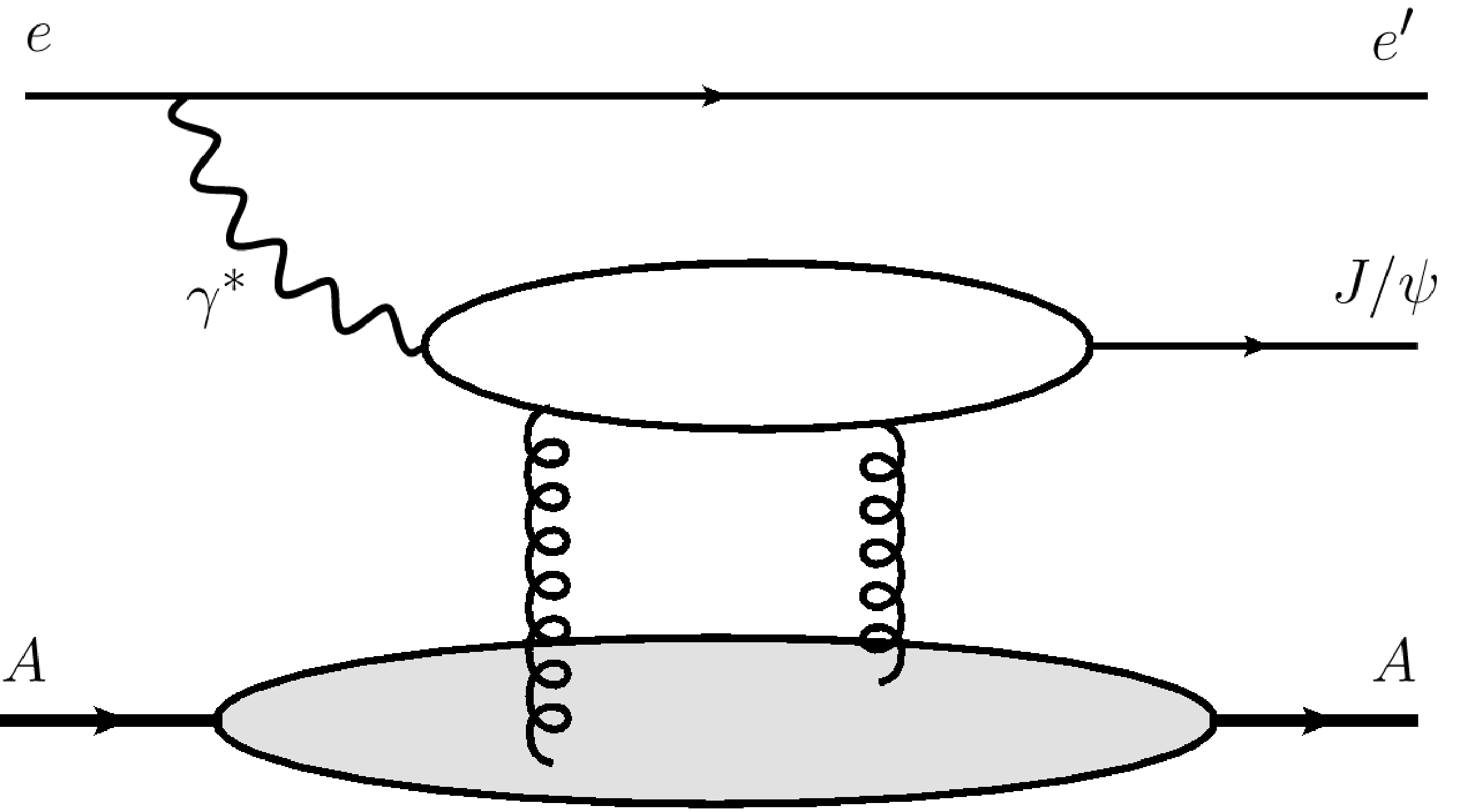}
  }  
  \caption{
    Leading order Feynman diagram for the exclusive vector
      meson production of a $\jpsi$  meson. 
    } 
\label{EVMPHU:handbag}
\end{figure}

Exclusive vector meson production (EVMP) in e + A collisions has been proposed as a channel for just such a measurement \cite{Munier:2001nr,Caldwell:2010zza,Lappi:2010dd}.  In this section we will focus on the production of heavy vector mesons, in particular $\jpsi$  mesons.  To leading order, EVMP of a $\jpsi$  meson occurs in an e + A collision when a photon emitted by the electron splits into a $c$-$\bar{c}$ pair which communicates with the gluon density in the highly boosted nucleus via a two gluon exchange and subsequently forms a $\jpsi$  meson and nothing else (we will be interested here in coherent EVMP, in which case the nucleus remains intact); see figure~\ref{EVMPHU:handbag} for a visualization of the process.  It is precisely this two gluon exchange which yields a diffractive measurement of the gluon density in a nucleus.  

Previous work \cite{Caldwell:2010zza} explored how modest changes in the Woods-Saxon distribution \cite{Woods:1954zz} of a nucleus might manifest themselves as changes in the diffractive peaks in EVMP if one assumes that the spatial distribution of gluons in a nucleus is proportional to the Glauber thickness function found from the Woods-Saxon distribution.  That these modest changes do result in a visually obvious modification of the diffraction pattern motivated our further study, in which we consider whether two very different physical pictures of the gluon distribution in a highly boosted nucleus can be experimentally distinguished via EVMP: in particular we wish to compare the diffraction patterns that emerge when the gluon distribution 1) has normalization dictated by DGLAP evolution and spatial distribution given by the Glauber thickness function and 2) is given by the KLN parameterization (see \cite{Hirano:2004en,Kuhlman:2006qp} and references therein) of the Color Glass Condensate (CGC) (see, e.g., \cite{Iancu:2002xk,JalilianMarian:2005jf} for a review).  We choose to investigate these two ans\"atze of the gluon distribution in nuclei as they have been the dominant models used in heavy ion physics calculations to estimate the uncertainty in the viscosity to entropy ratio of the QGP produced at RHIC due to the uncertainty of the currently poorly constrained initial conditions in heavy ion collisions \cite{Hirano:2005xf,Luzum:2008cw}.  

It is worth taking a moment to comment on some common---yet confusing---terminology in the EVMP field.  As mentioned above, to leading order the coherent production of a vector meson in an e + A collision involves a two-gluon exchange between the $q$-$\bar{q}$ pair and the nucleus.  If one assumes that all two-gluon exchanges occur independently, then one may exponentiate the single two-gluon exchange result.  Making this independence assumption is often referred to in the EVMP field as using ``saturation'' physics because the cross section is unitarized via the exponentiation process.  However this ``saturation'' \emph{does not refer to unitarizing the gluon distribution functions themselves}.  For instance in the ``IP-Sat'' \cite{Kowalski:2003hm} and ``b-Sat'' \cite{Kowalski:2006hc} models, where ``Sat'' is short for saturation, the $x$ evolution of the gluon PDF is effected through the use of the DGLAP equations.  On the other hand, the ``b-CGC'' model \cite{Kowalski:2006hc} incorporates both the exponentiation of the two-gluon exchange \emph{and} the CGC physics of the saturation of the gluon PDF.  We note that, in principle, small-$x$ evolution effects and exponentiation effects in the dipole cross section should become appreciable simultaneously \cite{Mueller:1989st}.  In order to (hopefully) make the presentation more clear, and to simplify some of the numerics, we will not exponentiate the two-gluon exchange; we will present results using only the leading order two-gluon exchange in which the gluon PDF is given either via DGLAP evolution or from the CGC.  Any subsequent reference to ``saturation'' in this paper will refer to the saturation of the gluon distribution function alone. \\


\noindent{\bf Formalism:}  Following \cite{Kowalski:2003hm,Caldwell:2010zza}, the diffractive production of a vector meson from a photon scattering off a target is
\begin{equation}
\label{EVMPHU:dsigdt}
\frac{d\sigma}{dt} = \frac{1}{16\pi}\left|\int d^2\boldsymbol{r}\int\frac{dz}{4\pi}\int d^2\boldsymbol{b} \, \langle V|\gamma\rangle_T \, e^{i\boldsymbol{b}\cdot\boldsymbol{\Delta}} \, \frac{d\sigma_{q\bar{q}}}{d^2\boldsymbol{b}}\right|^2,
\end{equation}
where $\langle V|\gamma\rangle_T$ is the overlap of the vector meson wavefunction and the transversely polarized virtual photon wavefunction---the contribution from the longitudinally polarized photon is zero as we are interested in $Q^2=0$ photoproduction---and we used the photon-meson overlap and Gauss-LC model for the $\jpsi$  wavefunction from \cite{Kowalski:2003hm}\footnote{Note that the normalization of the $\jpsi$  wavefunction in \cite{Kowalski:2003hm} is erroneously reported as a factor of 100 smaller than the correct value; one can readily see this by comparing with the normalization condition defined in \cite{Kowalski:2003hm} and with the results reported in \cite{Kowalski:2006hc}.  It is surprising that this error was not noted in \cite{Kowalski:2006hc}, in which the results found in \cite{Kowalski:2006hc} are explicitly compared to those in \cite{Kowalski:2003hm}.}, and $\boldsymbol{\Delta}^2 \, = \, -t$.  $d\sigma_{q\bar{q}}/d^2\boldsymbol{b}$ is the differential cross section for the interaction of the dipole with the target; its form depends on the physics assumptions we make for the nuclear gluon distribution, as we discuss in detail below. \\

\vspace{.1in}

\noindent{\bf DGLAP Evolution in $x$, Glauber Distribution of Gluons in $b$:}  If we assume that the two gluon exchange from the dipole to the nucleus occurs within an individual nucleon then 
\begin{equation}
\label{EVMPHU:DGLAP}
\frac{d\sigma_{q\bar{q}}}{d^2\boldsymbol{b}} = \frac{\pi^2}{N_c} \, r^2 \, \alpha_s(\mu^2) \, xg(x,\,\mu^2) \, T(b),
\end{equation}
where $r$ is the size of the dipole, $\mu \, = \, \surd(\mu_0+C/r^2)$ is the relevant momentum scale for the dipole, $xg$ is the gluon distribution function, and
\begin{equation}
T(b) = \frac{1}{2\pi B_G}e^{-b^2/2B_G}
\end{equation}
is the assumed spatial distribution of gluons in a nucleon.  We use the MSTW parameterization of the gluon PDF \cite{Martin:2009iq}.  As described in \cite{Bartels:2002cj}, $\mu_0$ and $C$ are free parameters; as in \cite{Bartels:2002cj,Kowalski:2003hm,Lappi:2010dd}, we take $\mu_0$ = 1 GeV$^2$ and $C$ = 4.  From HERA data \cite{Chekanov:2002xi} the measured slope of 
$d\sigma/dt$ yields $B_G \, \approx \, 4.25$ GeV$^{-2}$ \cite{Kowalski:2003hm}.  Then
\begin{equation}
\frac{d\sigma^{DGLAP}}{dt} = 4\pi \, \sigma_p^2 \, e^{-B_Gt} \, \left|\int db \, J_0(b\sqrt{t}) \, T_A(b) \right|^2,
\end{equation}
where $J_0$ is the usual Bessel function, 
$
  T_A(b) \equiv \int dz \, \rho_A\left(\sqrt{b^2+z^2}\right) 
$, 
with 
$
  \int d^2\boldsymbol{b} \, T_A(b) = A
$, is the usual thickness function, and $\rho_A$ is the density of the
nucleus (here taken as the Woods-Saxon distribution of $^{197}$Au with
the usual $R \, = \, 6.38$ fm and $a \, = \, 0.535$ fm
\cite{Hahn:1956zz}) and 
\begin{equation}
\sigma_p \equiv \frac{1}{4\pi}\int d^2\boldsymbol{r}\int dz \, \langle V|\gamma\rangle_T \, \frac{\pi^2}{N_c} \, r^2 \, \alpha_s(\mu^2) \, xg(x,\,\mu^2).
\end{equation}


\noindent{\bf CGC Distribution of Gluons in $x$ and $b$:}  Alternatively we may view the nucleus as a whole and that the gluon distribution is found from the CGC.  In this case
\begin{equation}
\label{EVMPHU:CGCdipole}
\frac{d\sigma_{q\bar{q}}}{d^2\boldsymbol{b}} = \frac{\pi^2}{N_C} \, r^2 \, \alpha_s(\mu^2) \, xg_A(\mu^2,\,Q_s^2),
\end{equation}
where $xg_A$ is the integrated gluon distribution function related to the unintegrated gluon distribution (UGD) $\phi_A$ by
\begin{align}
xg_A(\mu^2,\,Q_s^2)  = \int d^2\boldsymbol{k} \, \phi_A(k^2,\,Q_s^2) = \pi\,\int_0^{k_{max}^2\,=\,\mu^2}dk^2\,\phi_A(k^2,\,Q_s^2)
\end{align}
The $x$ and $b$ dependence of the two-gluon exchange dipole scattering formula, Eq.~\ref{EVMPHU:CGCdipole}, comes in implicitly through the $x$ and $b$ dependence of $Q_s^2$ \cite{Kuhlman:2006qp},
\begin{equation}
\label{EVMPHU:Qs}
Q_s^2 \equiv \frac{2\pi^2}{C_F} \, \alpha_s(Q_s^2) \, xg(x,\,Q_s^2) \, T_A(b),
\end{equation}
where $C_F \, \equiv \, (N_c^2-1)/2N_c$.

In principle one determines the UGD via the JIMWLK evolution equations or, in the large-$N_c$ limit, the BK evolution equations (see \cite{Iancu:2002xk,JalilianMarian:2005jf} and references therein).  However, instead of solving the full evolution equations many heavy ion physics calculations use instead the KLN prescription of the CGC (see, e.g., \cite{Hirano:2004en,Kuhlman:2006qp}), which attempts to capture the main feature of CGC physics; in particular, the KLN UGD becomes saturated at momenta on the scale of the saturation scale $Q_s$.  Because of its widespread use in heavy ion physics calculations and in order to simplify our own calculations we, too, will use the KLN UGD,
\begin{equation}
\label{EVMPHU:phiKLN}
\phi^{KLN}_A(k,\,Q_s^2) = \frac{\kappa \, C_F \, Q_s^2}{2\pi^3\,\alpha_s(Q_s^2)}\left\{ 
\begin{array}{ll}
\left(Q_s^2+\Lambda^2\right)^{-1}, & \,k^2 \, \le \, Q_s^2 \\ 
\left(k^2+\Lambda^2\right)^{-1}, & \,k^2 \, > Q_s^2,\end{array}\right.
\end{equation}
where $\kappa$ is an $O(1)$ parameter meant to represent higher order corrections to the UGD, and $\Lambda \, = \, 0.2$ GeV \cite{Kuhlman:2006qp}.  


In principle $\kappa$ is set by comparing to known experimental observables such as the measured multiplicity at midrapidity at RHIC \cite{Back:2002uc,Adler:2004zn,Alver:2008ck} or LHC \cite{Aamodt:2010pb,Aamodt:2010cz} or to the diffractive cross sections for protons measured at HERA \cite{Chekanov:2002xi}.  However we found that the results from the leading order multiplicity formula \cite{Hirano:2004en} are linearly dependent on the cutoff taken for $\alpha_s$, $\alpha_s^{max}$.  The KLN UGD itself, though, is not nearly as sensitive to $\alpha_s^{max}$, so the multiplicity prescription does not provide a robust way of setting $\kappa$.  We note in passing that the centrality dependence of the particles produced via the leading order CGC multiplicity formula using the KLN UGD's also depends on $\alpha_s^{max}$.  Perhaps the use of the next-to-leading order results in the UGD \cite{ALbacete:2010ad} and/or the production formula \cite{Horowitz:2010yg} will mitigate this dependence enough to make reasonable comparisons of CGC multiplicity to current data.  Currently, though, there does not appear to be any quantitative estimate of the size of the dependence of the predicted CGC multiplicity as a function of centrality on $\alpha_s^{max}$.  $\kappa$ also cannot be set by comparing to the proton diffractive cross section as the currently available data does not probe regions of $x$ small enough such that $Q_s^2$ is a perturbative scale (at least when using the LO MSTW PDFs).  In our calculations we will set $\kappa \, = \, 1$.

It is important to contrast the interaction of the dipole in the KLN CGC approach taken here, in which the $q$-$\bar{q}$ pair interacts with the entire nucleus, and the Glauber approach, in which the pair interacts with individual nucleons.  By interacting with individual nucleons the diffractive cross section for the DGLAP Glauber model picks up an extra exponential suppression in $t$ proportional to the square of the width of the nucleon, $B_G$.  \\


\noindent{\bf Results:}  
The saturation physics of the CGC has resulted in a wider and flatter gluon distribution than that from the Glauber treatment; the DGLAP growth of the small-$x$ gluon distribution---tamed by the saturation physics of the KLN CGC---leads to a significant enhancement in the cross section at $x \, = \, 10^{-5}$ compared to that found using the KLN CGC gluon distribution.  It is worth noting that the KLN prescription for the CGC satisfies the black disk limit.  

We attempt to quantify the changes in both the nuclear gluonic width and density as a function of $x$ and note that even out to extremely small values of $x\,\sim\,10^{-13}$, $b_{1/2}$ from the KLN CGC continues to rise sublinearly with $\log(s)$; thus the implementation of the KLN CGC used here, with the MSTW gluon PDF, satisfies the Froissart bound \cite{Froissart:1961ux}.  Intriguingly this sublinear (as opposed to linear) growth in radius as a function of $\log s$ is a surprise compared to other CGC parameterizations \cite{GolecBiernat:2003ym}.  Note the enormous growth of the dipole cross section as $x$ decreases for the LO DGLAP-evolved gluonic density.  This unitarity-violating enhancement is clearly reduced tremendously with the saturation physics of the KLN CGC.

The drastically faster increase in the gluon density from the DGLAP evolved PDF results in a cross section that increases much faster as a function of $x$ than for the KLN CGC case.  As was shown in \cite{Lappi:2010dd}\footnote{fig. 8 in \cite{Caldwell:2010zza} also shows that the incoherent process quickly dominates the coherent one as a function of $t$, although we note that there was an error in the calculation of the figure and that the curves plotted do not correspond to the equations in the text of the paper.} the incoherent cross section, in which the nucleus breaks up, begins to dominate the total diffractive cross section by $t \, \sim \, 0.02$ GeV$^{-2}$.  It is likely that the $t$ dependence of the incoherent EVMP of the two models will be different, although we do not provide a quantitative estimate here: the decrease in cross section as a function of $t$ for the DGLAP Glauber model will be enhanced by $\exp(-B_G\,t)$ due to the assumption that the heavy quark dipole interacts with individual nucleons.  And in the case of coherent scattering one can discern a stronger $t$ dependence in the DGLAP Glauber results due precisely to the extra $\exp(-B_G\,t)$ factor that results from treating the nucleus as a collection of individual nucleons.  More importantly, the much larger gluon density yields a particularly noticeable difference at $t \, = \, 0$, where possible nuclear breakup effects are negligible.
Even with the very large PDF uncertainties as $x$ decreases, there is a clear increase in the coherent diffractive cross section for the DGLAP Glauber dipole compared to the KLN CGC dipole.  


\noindent{\bf Conclusions and Discussion:}  An enormous wealth of information on the gluonic structure of highly relativistic nuclei can be found using exclusive vector meson production.  In particular we investigated the experimental signatures of the coherent scattering of a $c\bar{c}$ dipole onto a nucleus that results in an intact nucleus and a $\jpsi$  meson in e + A collisions at eRHIC energies.  We found that the diffractive cross section will readily experimentally differentiate between the two common initial highly boosted nucleus prescriptions used in heavy ion physics phenomenology: 1) the gluon density is found using DGLAP evolution and its spatial distribution is assumed to be proportional to the at-rest Glauber nuclear thickness function and 2) the gluon density and distribution is given by the KLN parameterization of the CGC.  In particular there is the exciting possibility of literally watching a nucleus grow with center of mass energy as the positions in $t$ of the minima and maxima in the diffractive cross section for the saturation physics calculation depend quite strongly on $\log(x)$.  On the other hand the DGLAP Glauber model yields a nucleus of constant size as a function of $x$; the positions in $t$ of the diffractive minima and maxima do not change as a function of $x$.  At the same time one is determining the width of a nucleus in e + A collisions, one will also measure the $x$ dependence of the normalization of $d\sigma/dt$.  Due to the explosion of small-$x$ gluons the DGLAP Glauber approach yields a normalization that rapidly increases as a function of $x$; additionally the $t$ dependence of the DGLAP Glauber $d\sigma/dt$ is also quite strong as it is proportional to $\exp(-B_G \, t)$ due to the assumption that the $q$-$\bar{q}$ dipole interacts with individual nucleons.  Conversely the KLN CGC dipole description does not have a strong $x$ dependence in its normalization due to its inclusion of saturation effects; similarly, the interaction of the dipole with the whole nuclear gluonic wavefunction yields a weaker $t$ dependence than is displayed by the DGLAP Glauber results.  

It is clear that, at the very least, the striking difference between the $x$ dependence of the peaks and minima from the DGLAP Glauber model and the KLN CGC model are robust: these differences will persist should we use even more sophisticated models of these two physical pictures; the $x$ dependence of the peaks and minima will persist should we attempt to approximate multiple scattering within the nucleus by exponentiating the dipole cross section, should we use a less approximate CGC calculation such as is found in \cite{ALbacete:2010ad}, or should we examine the results from other vector mesons such as the $\phi$ or $\rho$.  We regrettably leave the quantification of the diffractive cross section for these more sophisticated physical models and additional vector mesons for future work.  Exponentiating the two-gluon exchange cross section will reduce the enormous growth in the diffractive cross section in the DGLAP Glauber picture compared to the CGC case; we suspect this reduction will not be too large, although we also leave the quantification of this reduction to future work.


\ \\ \noindent{\it Acknowledgments:}
The author wishes to thank E.~Aschenauer, M.~Diehl, Y.~Kovchegov,
H.~Kowalski, T.~Lappi, C.~Marquet, T.~Ullrich, and R.~Venugopalan for invaluable
discussions and the INT for its hospitality and support.  The author
wishes to especially thank Y.~Kovchegov for reading and commenting on
the manuscript.  


\subsubsection{Coherent vs incoherent diffraction}
 \label{sec:cohvsincoh}

\hspace{\parindent}\parbox{0.92\textwidth}{\slshape
 Tuomas Lappi and Cyrille Marquet}
 \index{Lappi, Tuomas}
 \index{Marquet, Cyrille}
 
\vspace{\baselineskip}

The purpose of this section is to investigate incoherent diffraction in a simpler context than with inclusive diffraction $\gamma^* A\to XY$, mainly using diffractive vector meson production
$\gamma^* A \to V Y$, where the diffractive final state X consists of a vector meson and nothing else, $A$ stands for the target nucleus and $Y$ for the final state it may dissociate into. At high energies, the $q\bar q$ dipole that the virtual photon has fluctuated into, scatters off the gluonic field of the nucleus before recombining into the vector meson. While this scattering involves a color-singlet exchange, leaving a rapidity gap in the final state, the nucleus can still interact elastically ($Y=A,$ this is called coherent diffraction) or inelastically ({\it i.e.} break up, called incoherent diffraction). In this process, the momentum transfer $t$ can be determined from the meson regardless of the fate of the target, and elastic and inelastic interactions of the target can be experimentally distinguished.

Kinematically, a low invariant mass of the system $Y$ corresponds to a large rapidity gap in the final state between that system and the vector meson, and implies that the longitudinal momentum of the meson is close to that of the incoming photon. In this case, the eikonal approximation can be assumed to compute the dipole-nucleus scattering. At small values of $x=(Q^2+M_V^2)/(Q^2+W^2)$ where $Q^2$ is the photon virtuality, $M_V$ the vector meson mass, and $W$ the energy of the
$\gamma^*-A$ collision, a target proton can also be considered. Indeed in that case, since partons with an energy fraction as small as $x$ are probed in the target wave function, the dipole will
scatter off large gluon densities generated by the QCD evolution.

In e+p collisions, the cross-section is maximal at minimum momentum transfer with exclusive production (or coherent diffraction) dominating. As the transfer of momentum gets larger, the role of incoherent diffraction increases and eventually it becomes dominant, typically for momenta larger that the inverse target size; the elastic contribution decreases exponentially while the inelastic contribution decreases only as a power law. It is known that saturation models describe well the exclusive cross section \cite{Kowalski:2003hm,Forshaw:2003ki,Kowalski:2006hc,Marquet:2007qa}, while the BFKL Pomeron exchange approach works well for the target-dissociation cross-section \cite{Enberg:2003jw,Poludniowski:2003yk}.
In the section on proton breakup, we show that, within the Color Glass Condensate (CGC) picture of the small$-x$ part of the hadronic wave function, both coherent and incoherent diffraction can be described in the same framework. We also explicitly calculate both contributions to the diffractive vector meson production cross-section using the McLerran-Venugopalan (MV) model for the CGC wave function, and discuss phenomenological consequences in the context of a future electron-ion collider \cite{Deshpande:2005wd}.

Diffractive dissociation is an aspect of diffraction that changes qualitatively with nuclear targets.
Indeed, the structure of incoherent diffraction eA$\to$eXY is more complex than with a proton target,
and also can teach us a lot more. In the case of a target nucleus, we expect the following
qualitative changes in the $t$ dependence. First, the low-$|t|$ regime in which the nucleus
scatters elastically will be dominant up to a smaller value of $|t|$ (to about $|t|=0.05$
GeV$^2$) compared to the proton case, reflecting the bigger size of the nucleus. Then, the
nucleus-dissociative regime will be made of two parts: an intermediate regime in momentum
transfer up to about $0.7$ GeV$^2$ where the nucleus will predominantly break up into its
constituents nucleons, and a large$-|t|$ regime where the nucleons inside the nucleus will
also break up, implying pion production in the $Y$ system for instance. These are only
qualitative expectations, it is crucial to study this aspect of diffraction quantitatively
in order to complete our understanding of the structure of nuclei. The transition from the coherent to the intermediate regime is studied in the nuclear breakup section, following Ref.~\cite{Lappi:2010dd}. 


\noindent{\bf Proton breakup:}  In diffractive vector meson production, the relevant quantity is (the photon is a right mover, the CGC a left mover, and the gauge is ${\cal A}^+=0$):
\begin{equation}
T_{\textbf{xy}}[{\cal A}^-]=1-\frac{1}{N_c}
\mbox{Tr}\left(U^\dagger_{\textbf{y}}U_{\textbf{x}}\right)\ ,\quad\mbox{with }
U_{\textbf{x}}[{\cal A}^-]={\cal P}
\exp\left(ig_S\int dz^+ T^c {\cal A}_c^-(z^+,\textbf{x})\right)\ .
\label{wline}\end{equation}
In terms of this object, the differential cross sections for a transversely (T) or longitudinally (L) polarized photon are given by (with $t=-q_\perp^2$ the momentum transfer squared)
\begin{equation}
\frac{d\sigma_{T,L}}{dt}=\frac{1}{4\pi}\left\langle\left|
\int dz d^2x d^2y e^{iq_\perp.(z\textbf{x}\!+\!(1\!-\!z)\textbf{y})}
\Psi_{T,L}(z,\textbf{x}\!-\!\textbf{y})T_{\textbf{xy}}\right|^2\right\rangle_x\ ,
\label{diffcs}\end{equation}
where $2\Psi_T=\Psi_{V|\gamma}^{++}+\Psi_{V|\gamma}^{--}$ and $\Psi_L=\Psi_{V|\gamma}^{00}$ 
with
\begin{equation}
\Psi_{V|\gamma}^{\lambda'\lambda}(z,\textbf{r})=\sum_{h\bar{h}}
[\phi_{\lambda'}^{h\bar{h}}(z,\textbf{r})]^*\phi_{\lambda}^{h\bar{h}}(z,\textbf{r})\ ,
\label{overlap}\end{equation}
the overlap between the photon and meson wave functions. $\lambda$ and $h$ denote polarizations and helicities while $z$ is the longitudinal momentum fraction of the photon carried by the quark and
$\textbf{x}$ and $\textbf{y}$ are the quark and antiquark positions in the transverse plane.

The target average $\langle\ .\ \rangle_x$ is done with the CGC wave function squared
$|\Phi_x[{\cal A}^-]|^2:$
\begin{equation}
\langle f\rangle_x=\int DA^- |\Phi_x[A^-]|^2 f[A^-]\ .
\end{equation}
If one had imposed elastic scattering on the target side to describe the exclusive process
$\gamma^*A\!\to\!VA,$ the CGC average would be at the level of the amplitude, and the two-point
function $\langle T_{\textbf{xy}}\rangle_x$ inside the $|\ .\ |^2$ in (\ref{diffcs}), recovering the formula often used with dipole models.

Instead, when also including the target-dissociative part, the diffractive cross section
involves the 4-point correlator $\langle T_{\textbf{xy}}T_{\textbf{uv}}\rangle_x.$ In order to compute it, we must specify more about the CGC wave function. We shall use the McLerran-Venugopalan (MV) model \cite{McLerran:1993ni,McLerran:1994ka,McLerran:1994vd}, which is a Gaussian distribution for the color charges which generate the field ${\cal A}:$
\begin{equation} |\Phi_x[A^-]|^2=\exp\left(-\int d^2xd^2y dz^+
\frac{\rho_c(z^+,\textbf{x})\rho_c(z^+,\textbf{y})}{2\mu^2(z^+)}\right)\ ,\label{MV}\end{equation}
where the color charge $\rho_c$ and the field ${\cal A}_c^-$ obey the Yang-Mills equation
$-\nabla^2{\cal A}_c^-(z^+,\textbf{x})=g_S\rho_c(z^+,\textbf{x}).$ The variance of the distribution is the transverse color charge density squared along the projectile's path $\mu^2(z^+),$ with
\begin{equation}
\langle\rho_c(z^+,\textbf{x})\rho_d(z'^+,\textbf{y})\rangle=\delta_{cd}\delta(z^+-z'^+)
\delta^{(2)}(\textbf{x}-\textbf{y})\mu^2(z^+)\ .
\end{equation}

The only parameter is the saturation momentum $Q_s,$ with $Q_s^2$ proportional to the integrated color density squared. Note that there is no $x$ dependence in the MV model, it should be considered as an initial condition to the small$-x$ evolution.

The MV distribution is a Gaussian distribution, therefore one can compute any target average by expanding
the Wilson lines in powers of $g_S{\cal A}_c^-$ (see (\ref{wline})), and then use Wick's theorem \cite{Fujii:2002vh,Blaizot:2004wv}.
The results for the 4-point function $\langle T_{\textbf{xy}}T_{\textbf{uv}}\rangle$ are given in \cite{Dominguez:2008aa}. We note that, in the large$-N_c$ limit, one has $\langle T_{\textbf{xy}}T_{\textbf{uv}}\rangle=\langle T_{\textbf{xy}}\rangle\langle T_{\textbf{uv}}\rangle,$ which means that at small$-x,$ the target-dissociative part of the diffractive cross-section in suppressed at large $N_c,$ compared to the exclusive part. 

\begin{wrapfigure}{r}{0.4\columnwidth}
\centerline{\includegraphics[width=0.39\columnwidth]{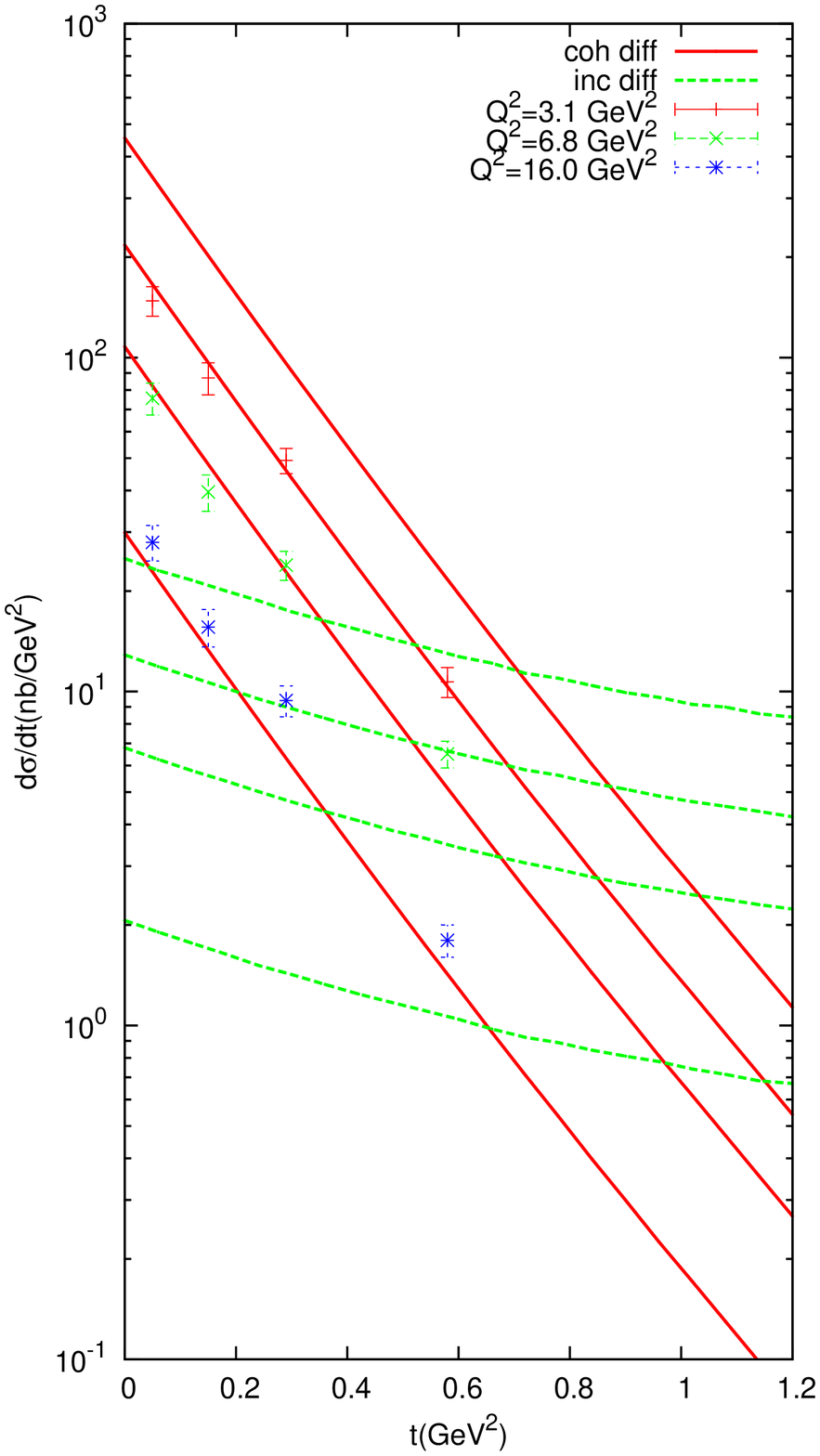}}
\caption{\small Diffractive J/$\Psi$ production in DIS at HERA for $W=90\ \mbox{GeV}$ and different $Q^2$ values. 
}
\label{Fig:Cyrille1}
\end{wrapfigure}

The numerical results presented below are obtained with the $x$ evolution of the saturation scale modeled as in \cite{GolecBiernat:1998js}:
$Q_s(x)=\left(x_0/x\right)^{\lambda/2}\ \mbox{GeV}$, with $\lambda=0.277$ and $x_0=4.1\ 10^{-5}$ for the case of a target proton. The collinear logarithm of $Q_s$ is neglected, which corresponds to exact geometric scaling \cite{Stasto:2000er,Marquet:2005qu,Marquet:2006jb}: $F(x,\textbf{r})=F[\textbf{r}^2Q^2_s(x)].$ As an illustration, the resulting cross-section for diffractive J/$\Psi$ production is displayed in Fig.~\ref{Fig:Cyrille1}, and separated into its coherent and incoherent contributions. The light-cone Gaussian J/$\Psi$ wave function \cite{Dosch:1996ss,Kulzinger:1998hw} has been used in (\ref{overlap}). At small values of $|t|$ where coherent diffraction dominates, our results are in agreement with HERA data \cite{Chekanov:2004mw} (one can get a better agreement with more realistic saturation models \cite{Kowalski:2003hm,Forshaw:2003ki,Kowalski:2006hc,Marquet:2007qa}, but this is not our point). Our model indicates that for $|t|>0.7\ \mbox{GeV}^2$ or so (this value slightly decreases when $Q^2$ increases), incoherent diffraction starts to dominate.  This may be the reason why the data on exclusive production stop: there is too much proton-dissociative `background'. We observe that this part of the cross-section decreases as a power law with $|t|,$ rather than exponentially as the exclusive part does.

The model discussed in this work is well adapted to describe the low- and large$-|t|$ regimes in the case of scattering off a nucleus, but not the intermediate regime since the constituent nucleons are absent from the description. This problem has been addressed in a complementary setup in the case of inclusive diffraction off nuclei \cite{Kowalski:2007rw,Kowalski:2008sa}, and the coherent diffraction regime was found to be dominant up to about $|t|=0.05\ \mbox{GeV}^2.$ The vector meson production case will be addressed next. While in the proton case, both exclusive and diffractive processes can be measured, it is likely that at a future electron-ion collider, the exclusive cross section cannot be extracted: when the momentum transfer is small enough for the nucleus to stay intact, then it will escape too close to the beam to be detectable. Therefore the diffractive physics program will rely on our understanding of incoherent diffraction.



\noindent{\bf Nuclear breakup into its constituent nucleons:}  To simplify our calculation, we will here use a factorized 
impact parameter profile for the dipole cross section in a proton
\begin{equation}\label{eq:factbt}
{\frac{\mathrm{d} \sigma^\textrm{p}_\textrm{dip}}{\mathrm{d}^2 {\mathbf{b}_T}}}({\mathbf{b}_T},{\mathbf{r}_T},x) 
=  2 \left( 1 - S_p({\mathbf{r}_T},{\mathbf{b}_T},x)\right)
= 2 \,T_p({\mathbf{b}_T}) {\mathcal{N}}(r,x),
\end{equation}
where $T_p$ is a Gaussian profile $T_p({\mathbf{b}_T}) = \exp\left(-b^2/2 B_p\right)$.
In the following we shall consider two dipole cross section parametrizations, 
the IIM model~\cite{Iancu:2003ge,Soyez:2007kg,Marquet:2007nf}, for which we take
take $B_p=5.59\gev^{-2}$, and a factorized approximation of the IPsat 
parametrization~\cite{Kowalski:2003hm,Kowalski:2006hc}, for which
$B_p=4.0\gev^2$. See \cite{Lappi:2010dd}
for a discussion of the origin of these values in different fits.

To extend the dipole cross section from protons to nuclei,
 we will take the independent
scattering approximation that is usually used in Glauber theory 
and write the $S$-matrix as
\begin{equation}\label{eq:sfact}
S_A({\mathbf{r}_T},{\mathbf{b}_T},x) = \prod_{i=1}^A S_p({\mathbf{r}_T},{\mathbf{b}_T}-{\mathbf{b}_T}_i,x).
\end{equation}

\begin{wrapfigure}{r}{0.5\columnwidth}
\centerline{\includegraphics[width=0.49\columnwidth]{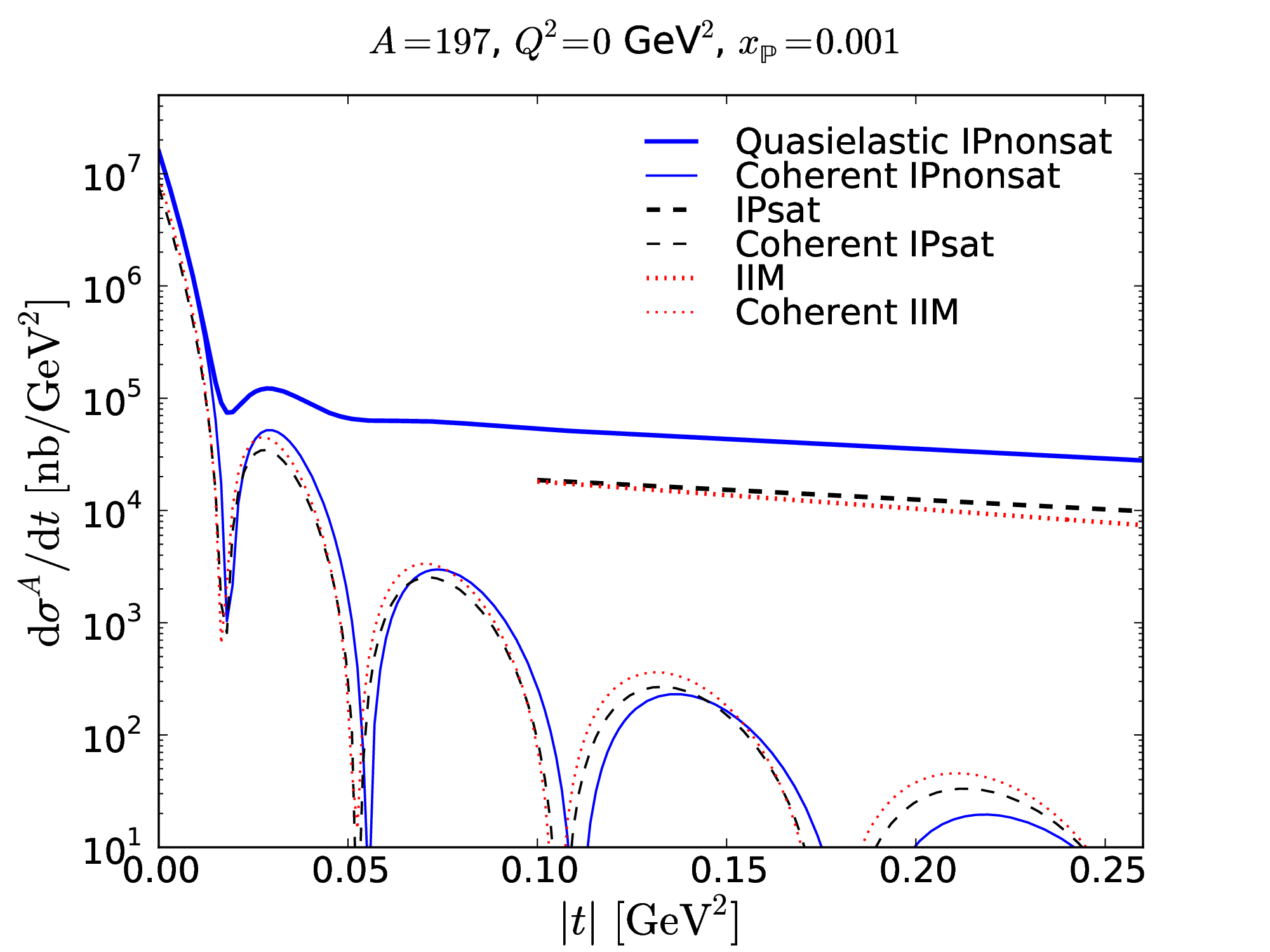}}
\caption{\small The quasielastic and coherent diffractive $J/\Psi$ cross sections in gold nuclei  at  $Q^2= 0$ and  $x_\mathbb{P} = 0.001$.  The IPsat and IIM parametrizations are shown. We also show the result for the linearized  ``IPnonsat'' version (used e.g. in Ref.~\cite{Caldwell:2009ke}) where the incoherent cross section is explicitly $A$ times that of the proton.  Our approximation (\ref{eq:amplisqn1}) is not valid for small $|t|$ and has been left out of the plot.}
\label{fig:dsigmavst}
\end{wrapfigure}

Here ${\mathbf{b}_T}_i$ are the nucleon coordinates.
This independent scattering assumption is natural in IPsat-like parametrizations or the MV-model, 
where $r = |{\mathbf{r}_T}|,$ $S({\mathbf{r}_T}) \sim e^{-r^2 {Q_\mathrm{s}}^2/4}$ with a  saturation scale 
${Q_\mathrm{s}}^2$ proportional to the nuclear thickness $T_A(b)$.
High energy evolution, however, introduces an anomalous dimension that leads,
in the nuclear case, to what could be called leading twist shadowing.
With an anomalous dimension
$S\sim e^{-({Q_\mathrm{s}} r)^{2\gamma}}$ with $\gamma \neq 1$, a proportionality
${Q_\mathrm{s}}^2 \sim T_A(b)$ is not equivalent to Eq.~(\ref{eq:sfact}). 
A solution to this 
problem (see also the more detailed discussion in~\cite{Kowalski:2008sa}) 
would require a realistic impact parameter dependent solution to the 
BK~\cite{Balitsky:1995ub,Kovchegov:1999yj,Kovchegov:1999ua} 
equation which is not yet available.
We point the reader to Ref.~\cite{GolecBiernat:2003ym}, for example, for a 
discussion of the  difficulties. These are related to the long distance 
Coulomb tails that, physically, are regulated at the confinement length
scale that is not enforced in a first principles weak coupling calculation.

The average over the positions of the nucleon in the nucleus was given in eq.~(\ref{wsavg}). 
The expectation valuedefined there is equivalent to the average
over nucleon configurations in a Monte Carlo Glauber calculation.
We are assuming that the positions ${\mathbf{b}_T}_i$ are independent, i.e. 
neglecting nuclear correlations that would be a subject of
interest in their own right (see e.g.~\cite{Alvioli:2009ab}).
The coherent cross section is obtained by averaging the amplitude
before squaring it, $|\left\langle \mathcal{A} \right\rangle_\mathrm{N} |^2$, 
and the incoherent one is
the variance $\left\langle |\mathcal{A}|^2\right\rangle_\mathrm{N} -  |\left\langle \mathcal{A} \right\rangle_\mathrm{N}|^2$ that measures the fluctuations
of the gluon density inside the nucleus. Because
$\left\langle \mathcal{A} \right\rangle_\mathrm{N}$
 is a very smooth function of ${\mathbf{b}_T}$, its Fourier transform 
vanishes rapidly for $\Delta \gtrsim 1/R_A$. Therefore, at large $\Delta$,
the quasielastic cross section is almost purely incoherent.

The cross section for quasielastic vector meson production is now expressed
in terms of the dipole scattering amplitude as
\begin{multline} \label{eq:incxsec}
\frac{\mathrm{d} \sigma^{\gamma^* A \to V A^* }}{\mathrm{d} t} 
= 
\frac{R_g^2(1+\beta^2)}{16\pi}  
\int 
\frac{\mathrm{d} z}{4 \pi}  \frac{\mathrm{d} z'}{4 \pi}
\mathrm{d}^2 {\mathbf{r}_T} \mathrm{d}^2 {\mathbf{r}_T}'
\\ \times
\left[ \Psi^*_V \Psi \right] (r,z,Q)
\, \left[ \Psi^*_V  \Psi \right](r',z',Q)
\left\langle \left| \mathcal{A}_{q\bar{q}}\right|^2({x_\mathbb{P}},r,r',\boldsymbol{\Delta}_T) \right\rangle_\mathrm{N},
\end{multline}
where we have applied corrections for the skewedness, $R_g$, and the real 
part of the scattering amplitude (see e.g. \cite{Watt:2007nr})
We now average the square of the dipole scattering amplitude over the 
nucleon coordinates, using the assumptions of
Eqs.~(\ref{eq:sfact}) and~(\ref{eq:factbt}) and taking the large $A$ limit.
We are additionally assuming that $T_A$ is a smooth function on the 
distance scale defined by $B_p$.
Averaging the square of the amplitude gives the total quasi-elastic 
contribution.

Note that  Eqs.~(\ref{eq:sfact}) and~(\ref{eq:factbt}) have enabled us to
write the leading contributions as proportional to the
(Gaussian) proton impact parameter profile, which can then be 
Fourier-transformed analytically. Giving up either of these approximations
would force us to numerically Fourier-transform the ``lumpy'' 
$b$-dependence corresponding to a fixed configuration
of the nucleon positions. 
Keeping only the terms 
that contribute at large $|t| \gg 1/R_A^2$
leaves us with the expression
\begin{multline}\label{eq:amplisqn1}
\left| \mathcal{A}_{q\bar{q}}\right|^2({x_\mathbb{P}},r,r',\boldsymbol{\Delta}_T)  
=
16 \pi^2 B_p^2 A \int \mathrm{d}^2 {\mathbf{b}_T} 
\\ \times 
e^{-B_p \boldsymbol{\Delta}_T^2}
e^{-2 \pi B_p (A-1) T_A(b) 
\left[ {\mathcal{N}}(r) + {\mathcal{N}}(r') \right] }
\mathcal{N}(r){\mathcal{N}}(r') T_A(b).
\end{multline}
Equation (\ref{eq:amplisqn1}) has a very clear interpretation. The 
squared amplitude is proportional to $A$ times the squared amplitude
for scattering off a proton, corresponding to the dipole scattering 
independently off the nucleons in a nucleus. This sum of independent
scatterings is then multiplied by a nuclear attenuation factor
which accounts for the requirement that the dipole must \emph{not}
scatter inelastically off the other $A-1$ nucleons in the target (otherwise the
interaction would not be diffractive). 
Note that factor 
$4 \pi B_p {\mathcal{N}}(r,{x_\mathbb{P}})=\sigma_p(r,{x_\mathbb{P}})$ is the proton-dipole cross section for  a
dipole of size $r$. Thus this attenuation corresponds to the
probability of a dipole with a cross section which is the average 
of dipoles with $r$ and $r'$ to pass though the nucleus.
A similar expression 
can be found in Ref.~\cite{Kopeliovich:2001xj} for example.

The coherent cross section in our approximation is given by
\begin{equation} \label{eq:coh}
\frac{\mathrm{d} \sigma^{\gamma^* A \to V A }}{\mathrm{d} t} 
=\frac{R_g^2(1+\beta^2)}{16\pi} \left| \left\langle \mathcal{A}( x_\mathbb{P} ,Q^2,\boldsymbol{\Delta}_T) \right\rangle_\mathrm{N} \right|^2,
\end{equation}
where in the large $A$ and smooth nucleus limit the amplitude is
\begin{multline}\label{eq:cohampli}
\left\langle \mathcal{A}({x_\mathbb{P}},Q^2,\boldsymbol{\Delta}_T ) \right\rangle_\mathrm{N}
= \int \frac{\mathrm{d} z}{4\pi} \mathrm{d}^2 {\mathbf{r}_T}  \mathrm{d}^2 {\mathbf{b}_T} e^{-i {\mathbf{b}_T} \cdot  \boldsymbol{\Delta}_T }  
[\Psi_V^*\Psi](r,Q^2)
 2 \left[ 1-e^{ - 2 \pi B_p A T_A(b) {\mathcal{N}}(r,{x_\mathbb{P}}) } \right].
\end{multline}

Figure \ref{fig:dsigmavst} summarizes the $t$-dependence of the quasielastic and coherent
cross sections. Also shown is the approximation used in \cite{Caldwell:2009ke} where 
nonlinear effects are left out. The most striking result is the large 
suppression by a factor of $\sim 3$ 
of the incoherent cross section due to nonlinear effects. The 
incoherent and coherent curves cross saround $|t| \approx 0.05 \gev^2,$ as anticipated.
With a very good detection of the nuclear breakup events, the first, even the second,
diffractive dips in the coherent cross section could be measurable at the EIC,
providing detailed information about the average spatial distribution of gluons inside
the nucleus. For understanding the initial conditions of ultra-relativistic heavy-ion 
collisions what has turned out to be equally important are the fluctuations in the
gluon density, which are directly measured by the incoherent part of the spectrum.


\subsubsection{Electroproduction of $J/\Psi$}
\label{sec:dvmp-boris} 

\hspace{\parindent}\parbox{0.92\textwidth}{\slshape
 Boris Z. Kopeliovich}
\index{Kopeliovich, Boris Z.}

\vspace{\baselineskip}


\noindent{\bf Proton target:}  The diffractive electro-production of charmonia and the charmonium-nucleon elastic scattering are closely related. The amplitudes of diffractive electro-production of a charmonium and elastic charmonium-proton scattering in the dipole approach have the form,
\beqn
  {\cal M}_{\gamma^* p}(s,Q^2) &=& \sum_{\mu,\bar\mu}
   \,\int\limits_0^1 \!\!d\alpha \int d^2r_T
   \,\Phi^{\!*(\mu,\bar\mu)}_{\Psi}(\alpha,\vec r_T) \,\sigma_{q\bar q}(r_T,s)
   \,\Phi^{(\mu,\bar\mu)}_{\gamma^*}(\alpha,\vec r_T,Q^2);
  \label{kopel-1020}\\
  {\cal M}_{\Psi\,p}(s)    &=& \sum_{\mu,\bar\mu}
    \,\int\limits_0^1 \!\!d\alpha \int d^2r_T 
    \,\Phi^{\!*(\mu,\bar\mu)}_{\Psi}(\alpha,\vec r_T) \,\sigma_{q\bar q}(r_T,s)
    \,\Phi^{(\mu,\bar\mu)}_{\Psi}(\alpha,\vec r_T) \ .
\label{kopel-1040}
\eeqn
Here, $\mu$ and $\bar\mu$ are the spin indices of the $c$ and
$\bar c$ quarks, $Q^2$ is the photon virtuality, $\Phi_{\gamma^*}(\alpha,
r_T,Q^2)$ is the light-cone distribution function of the photon for a
$c\bar c$ fluctuation of separation $r_T$ and relative fraction $\alpha$ of
the photon light-cone momentum carried by $c$ or $\bar c$. Correspondingly,
$\Phi_{\Psi}(\alpha,\vec r_T)$ is the light-cone wave function of $\jpsi$, or
$\Psi'$, or $\chi$. 

The wave functions of charmonia are calculated in~\cite{Hufner:2000jb} solving the Schr\"odinger equation with 
four realistic potentials, which are labelled as COR~\cite{Eichten:1979ms}, BT~\cite{Buchmuller:1980su}, LOG~\cite{Quigg:1977dd}, and POW~\cite{Martin:1980jx}.  Then one should make a Lorentz boost from the charmonium rest frame to the infinite momentum frame, and to switch from 3-dimensional coordinates to the light-cone variables, $p_T$ and $\alpha$, which are the $c$-quark transverse and fractional longitudinal momenta respectively. This was done in~\cite{Hufner:2000jb} using the popular prescription~\cite{terentev}.

The important ingredient of the calculations performed in~\cite{Hufner:2000jb} (compare with~\cite{hoyer})
is the Melosh spin rotation~\cite{melosh} which relates
the 2-dimensional spinors $\chi_c$ and $\chi_{\bar c}$, describing $c$
and $\bar c$ in the infinite momentum frame,  to the spinors 
$\bar\chi_c$ and $\bar\chi_{\bar c}$ in the rest frame:
\beqn
  \bf\overline{\chi}_c        =\widehat R(  \alpha, \vec p_T)\,\chi_c\ ,\quad\bf\overline{\chi}_{\bar c} = \widehat R(1-\alpha,-\vec p_T)\,\chi_{\bar c}\ ,
  \label{kopel-1060}
\eeqn
where the matrix $R(\alpha,\vec p_T)$ has the form:
\beq
  \widehat R(\alpha,\vec p_T) = 
    \frac{  m_c+\alpha\,M - i\,[\vec\sigma \times \vec n]\,\vec p_T}
    {\sqrt{(m_c+\alpha\,M)^2+p_T^2}} \ .
\label{kopel-1080}
\eeq

Since the $c\bar c$ pair is in $S$-wave, the spatial and spin 
dependences in the wave function factorize, and one arrives at the following
light cone wave function of the $c\bar c$ in the infinite momentum frame
\beq
 \Phi^{(\mu,\bar\mu)}_\psi(\alpha,\vec p_T) =
     U^{(\mu,\bar\mu)}(\alpha,\vec p_T)\cdot\Phi_\psi(\alpha,\vec p_T)\ ,
\label{1100}
\eeq
where 
\beq
  U^{(\mu,\bar\mu)}(\alpha,\vec p_T) = 
    \chi_{c}^{\mu\dagger}\,\widehat R^{\dagger}(\alpha,\vec p_T)
    \,\vec\sigma\cdot\vec e_\psi\,\sigma_y
    \,\widehat R^*(1-\alpha,-\vec p_T)
    \,\sigma_y^{-1}\,\widetilde\chi_{\bar c}^{\bar\mu}.
\eeq

Now we can determine the light-cone wave function in the mixed
longitudinal momentum - transverse coordinate representation:
\beq
 \Phi^{(\mu,\bar\mu)}_\psi(\alpha,\vec r_T) =
    \frac{1}{2\,\pi} 
    \int d^2p_T\,e^{-i\vec p_T\vec r_T}\,
    \Phi^{(\mu,\bar\mu)}_\psi(\alpha,\vec p_T)\ .  
    \label{kopel-1120}
\eeq

\begin{wrapfigure}{r}{0.52\columnwidth}
\centerline{\includegraphics[width=0.47\columnwidth]{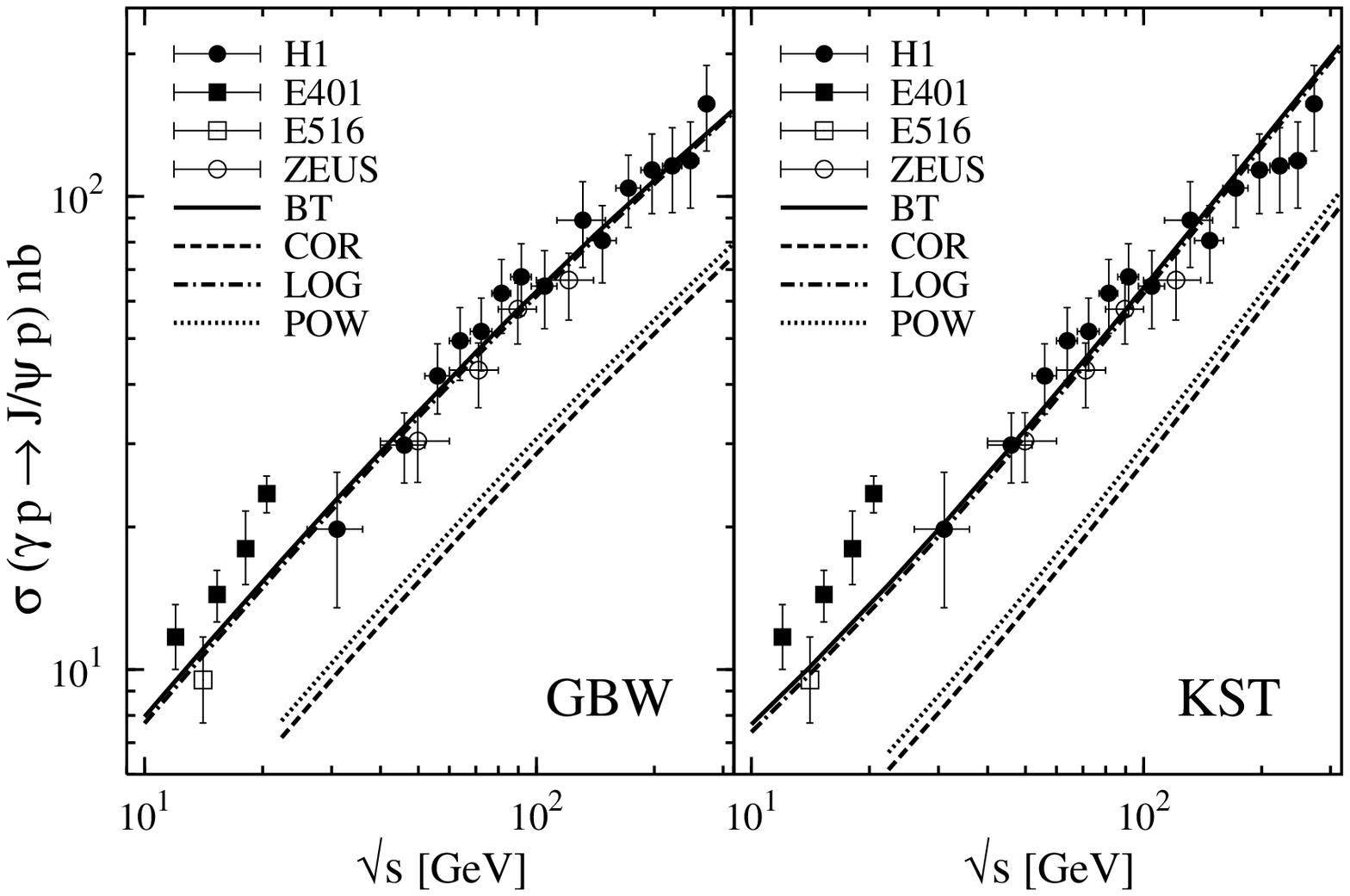}}
\caption{\small Integrated cross section for elastic photoproduction $\gamma\,p%
      \rightarrow \jpsi\,p$ with real photons ($Q^2=0$) as a function
      of the energy calculated with GBW and KST dipole cross sections
      and for four potentials to generate $\jpsi$ wave functions.
      }
\label{fig:gam-psi-s}
\end{wrapfigure}

With this wave function and with the standard distribution functions of the photon one can calculate the amplitudes in (\ref{kopel-1020})-(\ref{kopel-1040}) and predict the cross section of $\jpsi$ photoproduction on a proton.
The results for the energy dependence are compared with HERA data (see references in~\cite{Hufner:2000jb}) in Fig.~\ref{fig:gam-psi-s}. The calculation was performed in~\cite{Hufner:2000jb} with two parametrizations of the dipole cross section labelled as GBW~\cite{gbw} and KST~\cite{Kopeliovich:1999am}.

\begin{figure}[tbh]
  \centering
  \includegraphics[width=0.4\textwidth]{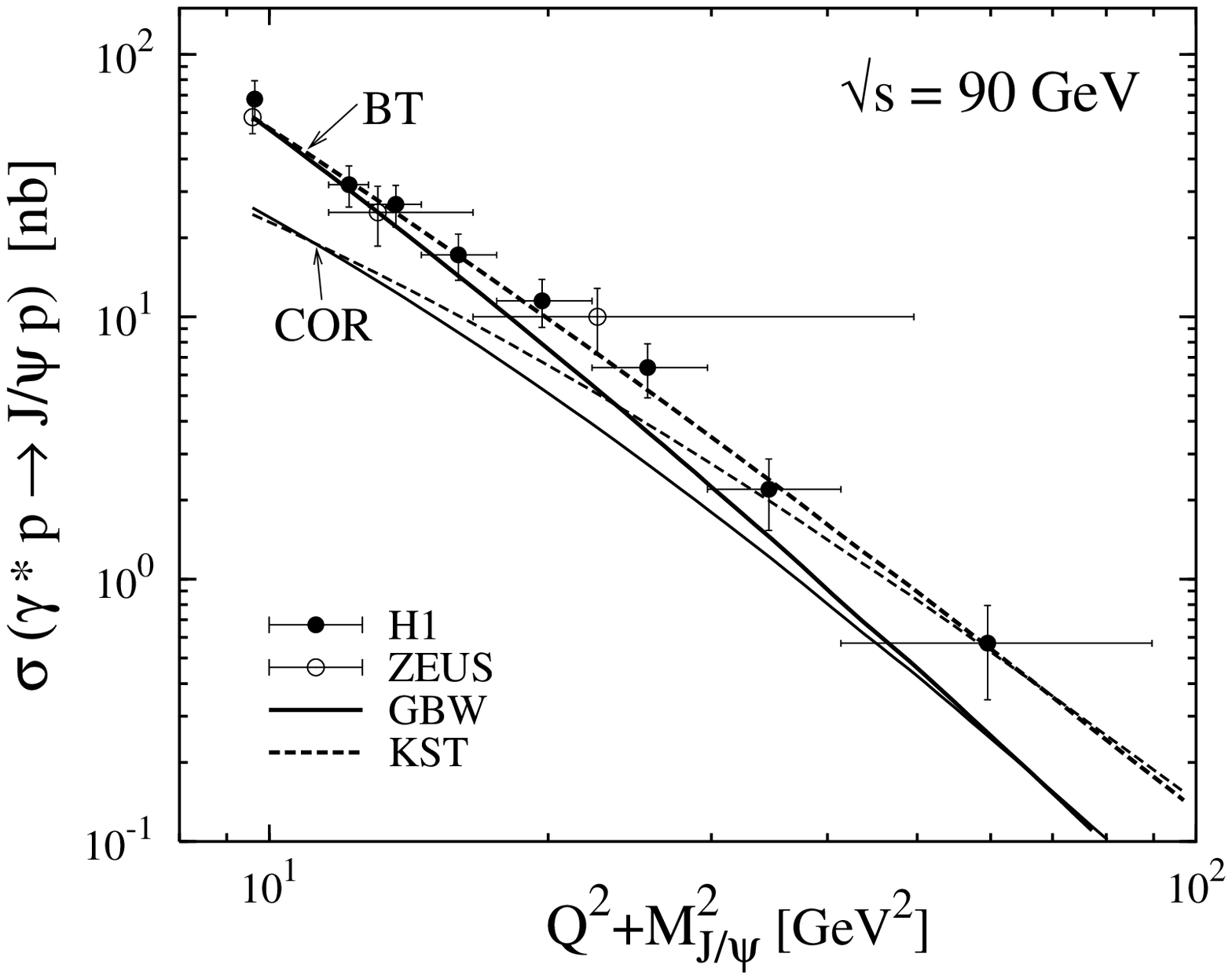}
  \hspace*{1cm}
  \includegraphics[width=0.5\textwidth]{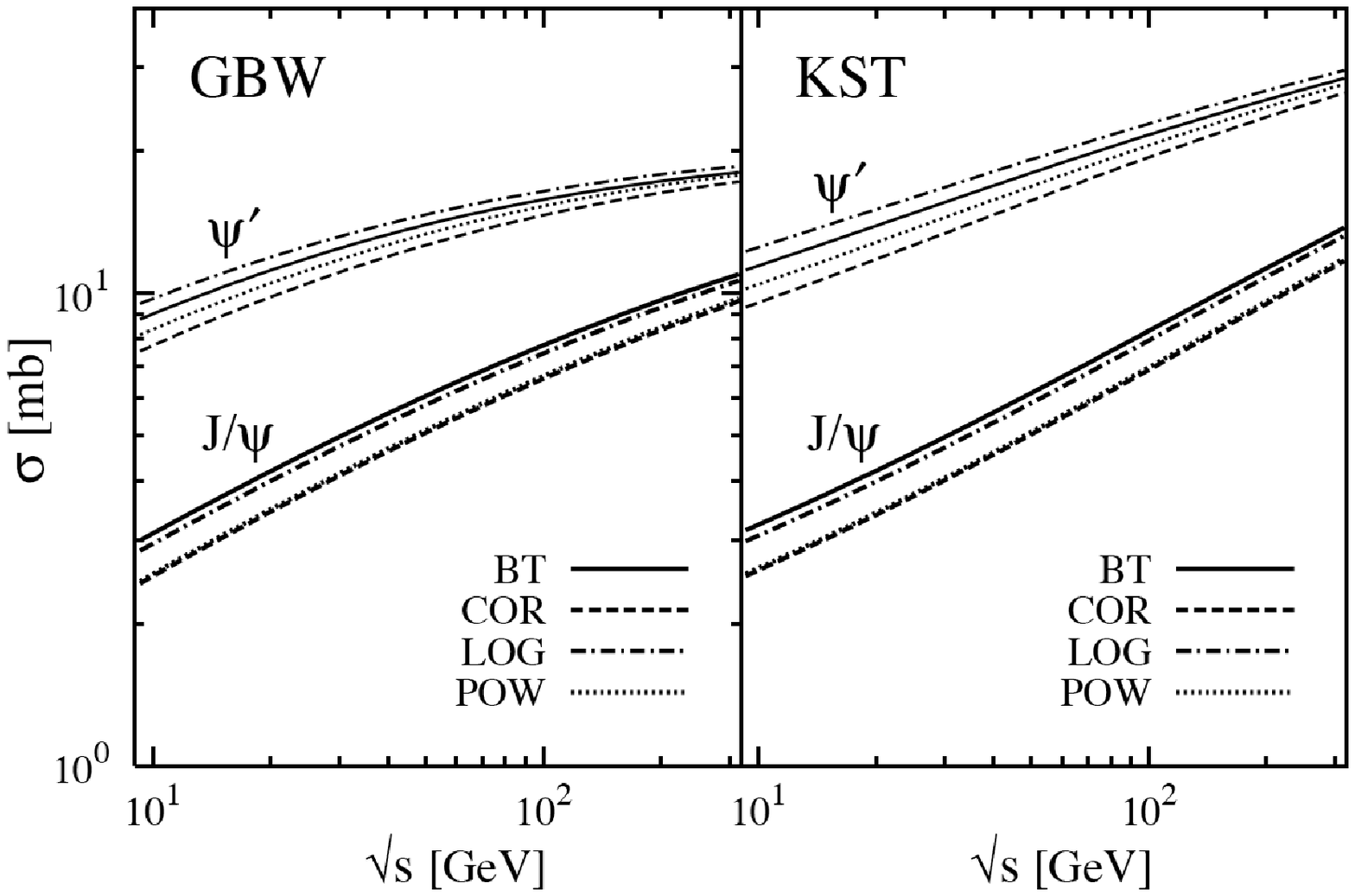}
  \caption{\small 
    {\it Left:} Integrated cross section for elastic photo production
    as a function of the photon virtuality $Q^2+M_{\jpsi}$ at energy
    $\sqrt{s}=90\gev$.
    {\it Right:} Total $\jpsi$-$p$ (thick curves) and $\Psi'$-$p$
    (thin curves) cross sections with the GBW and KST
    parameterizations for the dipole cross section. 
  } 
\label{fig:gam-psi-q}
\label{fig:psi-psi} 
\end{figure}

We see that only BP and LOG potentials describe the data well, which, however, are not sensitive to the choice of the phenomenological dipole cross section.  The $Q^2$ dependence of the cross section is compared to HERA data
(see references in~\cite{Hufner:2000jb}) in Fig.~\ref{fig:gam-psi-q} (left) for the LOG and BT potentials.  It turns out that the effects
of Melosh spin rotation have a gross impact on the cross section of elastic
photoproduction $\gamma\,p \to \jpsi(\psi)p\,$. It increases the photoproduction cross section by about $50\%$.
These effects have even more dramatic impact on the $\psi'$, increasing the photoproduction cross section by a factor of 2-3 and eliminating the large discrepancy with data observed previously~\cite{hoyer}.

Eventually, we are in a position to predict the charmonium-proton total
cross section, which is impossible to extract directly from
photoproduction data, either on protons, or nuclear targets. Indeed,
neither vector dominance~\cite{Hufner:1997jg}, nor Glauber
model~\cite{Kopeliovich:1991pu} can be used for data analysis. We
believe that the only way is to predict the charmonium cross section
within a model, which successfully describe data on photoproduction in
a parameter free way. Our predictions for the energy dependent
charmonium-proton total cross section are depicted in
Fig.~\ref{fig:psi-psi} (right) for $\jpsi$ and $\Psi'$. \\

\noindent{\bf Nuclear targets:}  Charmonium photoproduction on nuclei is controlled by two length scales. 
\beq
  l_c = \frac{2\,\nu}{M_{c\bar c}^2+Q^2} \approx 
        \frac{2\,\nu}{M_{\jpsi}^2+Q^2}\,\,;\,\, l_f = \frac{2\,\nu}{M_{\Psi'}^2 - M_{\jpsi}^2}\, ,
  \label{kopel-1140}
\eeq
The first one is called coherence length can be interpreted as the lifetime of a $\bar cc$ fluctuation in the projectile photon in the nuclear rest frame. When $l_c$ is short compared to the mean nucleon spacing, one can treat $\bar cc$ production as instantaneous, with following propagation of the $\bar cc$ dipole through the nucleus. In the opposite limit of $l_c\gg R_A$ the $\bar cc$ dipole propagates and attenuates through the whole nucleus.
The second scale$l_f$  is the formation length, which characterizes the formation of the charmonium wave function. Indeed, the produced $\bar cc$ dipole has a certain size and interaction cross section, but does not have any certain mass. It might be the $\jpsi$, or its radial excitation. To disentangle between them, takes time according to the uncertainty principle. 

The cross section of charmonium photo-production on nuclei is easiest to write in the limit of long $l_c\gg R_A$.
In this case, the size of the $\bar cc$ dipoles ``frozen'' by Lorentz time dilation for propagation of the dipole through the nucleus. The cross sections of incoherent (the nucleus break up to fragments) and coherent (the nucleus remains intact) production have the form~\cite{Kopeliovich:1991pu,Ivanov:2002kc},
\beqn
 \sigma^{\gamma_{T,L}^*A}_{inc}(s,Q^2) = 
  \int d^2b\,T_A(b)\,
  \left|\left\la\Psi\left|\sigma_{\bar cc}(r_T,s)\,
  \exp\left[ -{1\over2}\, \sigma_{\bar cc}(r_T,s)\,T_A(b)\right]
  \right|\Psi^{T,L}_{c\bar c}
  \right\ra\right|^2
  \label{kopel-1180}\\
  \sigma^{\gamma_{T,L}^*A}_{coh}(s,Q^2) =
  \int d^2b\,\left|\left\la\Psi\left|1\,-\,
  \exp\left[-{1\over2}\,\sigma_{\bar cc}(r_T,s)\,T_A(b)\right]
  \right|\Psi^{T,L}_{c\bar c}
  \right\ra\right|^2,
  \label{kopel-1200}
\eeqn
where $\Psi_{\bar cc}^{T,L}$ are the photon wave functions given by Eq.~(\ref{kopel-60}); $\Psi(\vec r_T,\alpha)$ is the charmonium light-cone wave function calculated in the previous section.
These expressions are significantly different from the Glauber model~\cite{Hufner:1996dr} and effectively include the Gribov corrections in all orders.

\begin{figure}[tbh]
  \centering
  \includegraphics[width=0.24\textwidth]{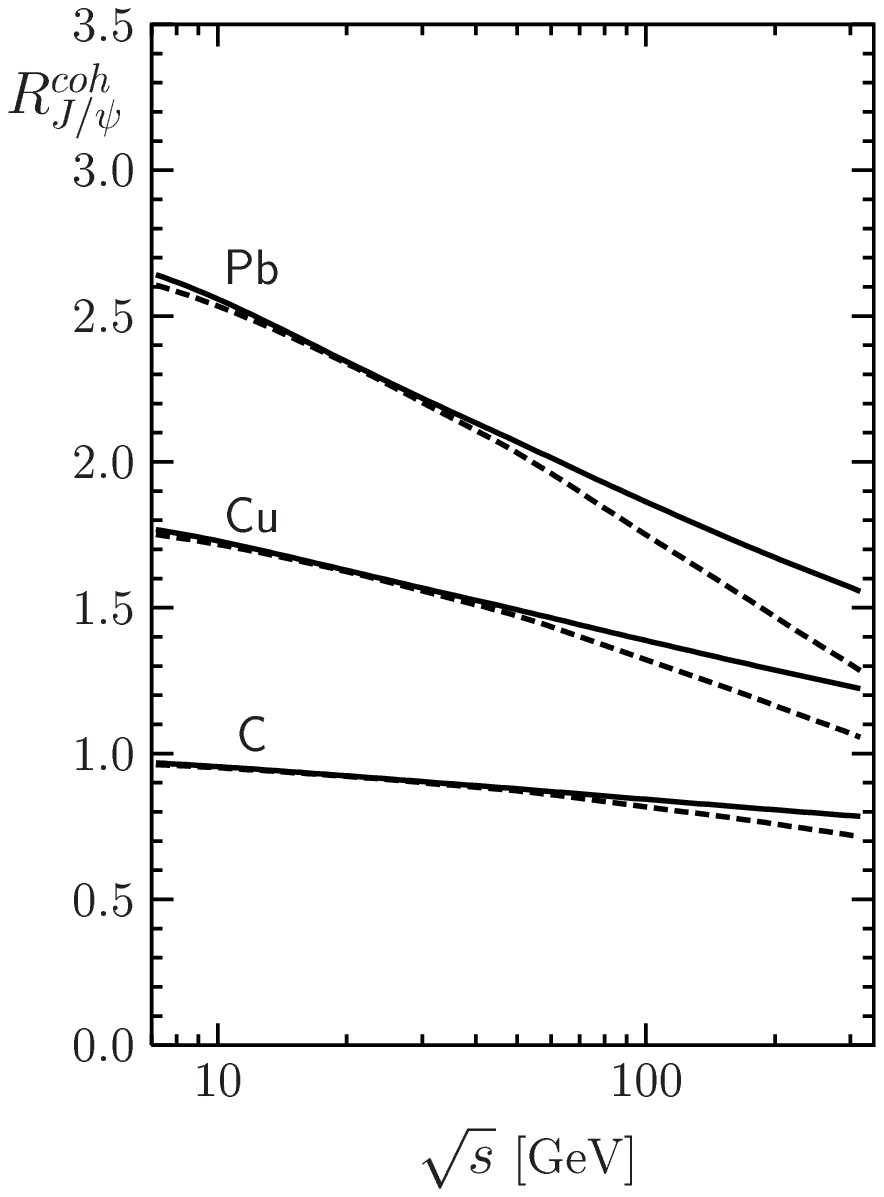}
  \includegraphics[width=0.24\textwidth]{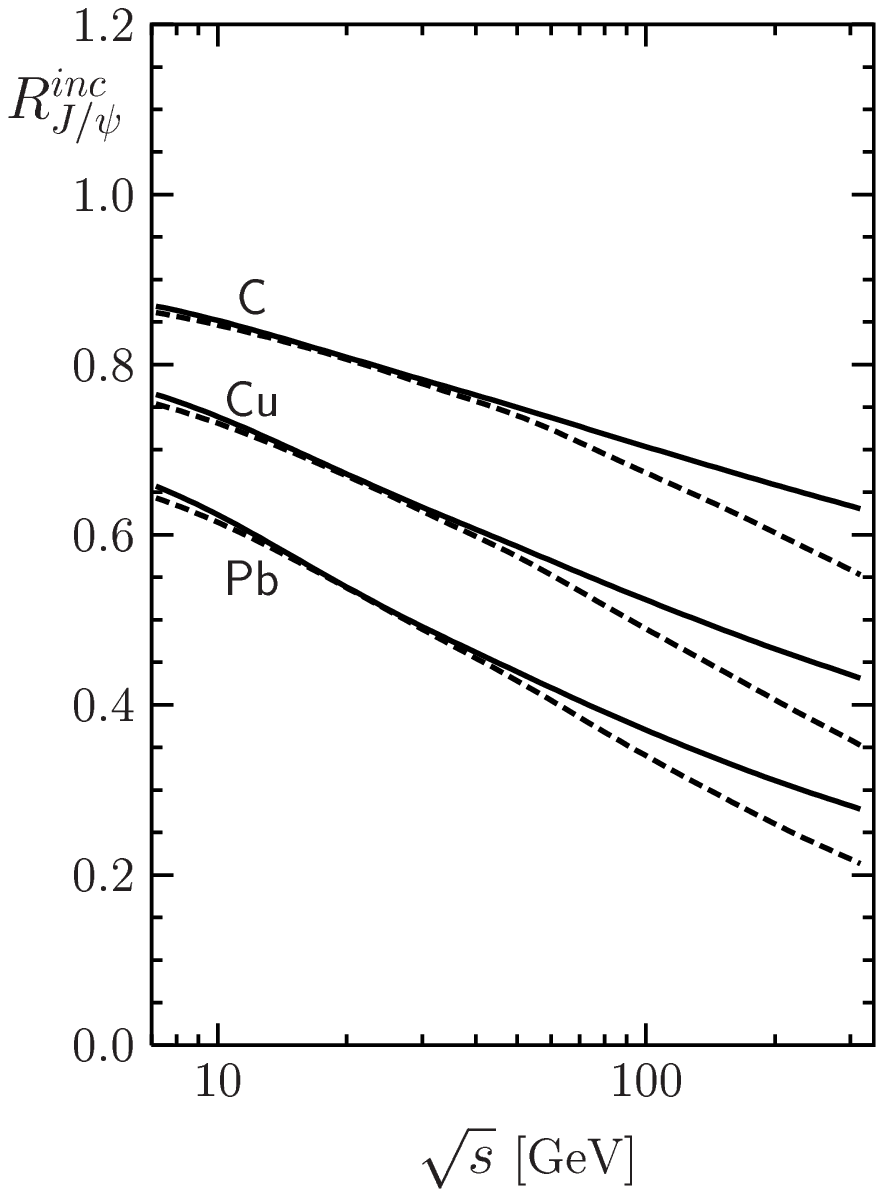}
  \includegraphics[width=0.24\textwidth]{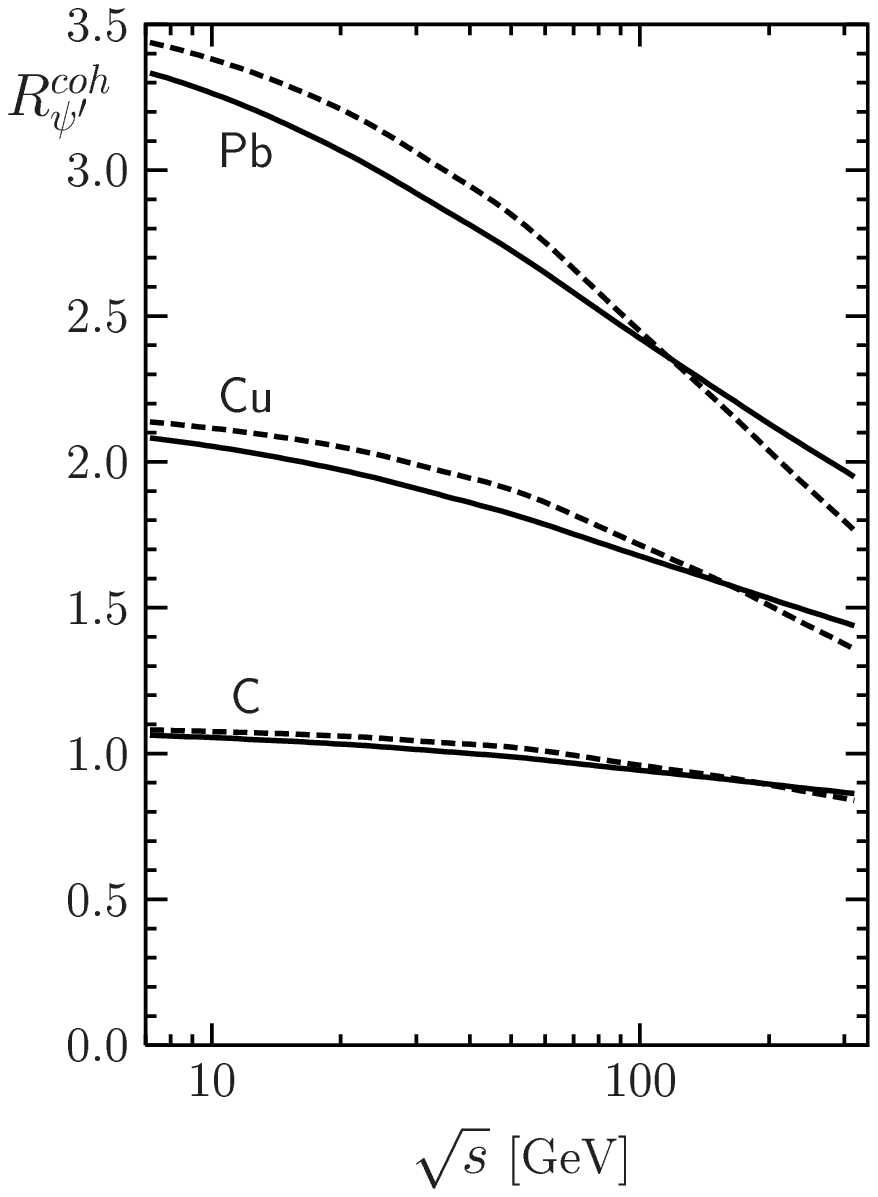}
  \includegraphics[width=0.24\textwidth]{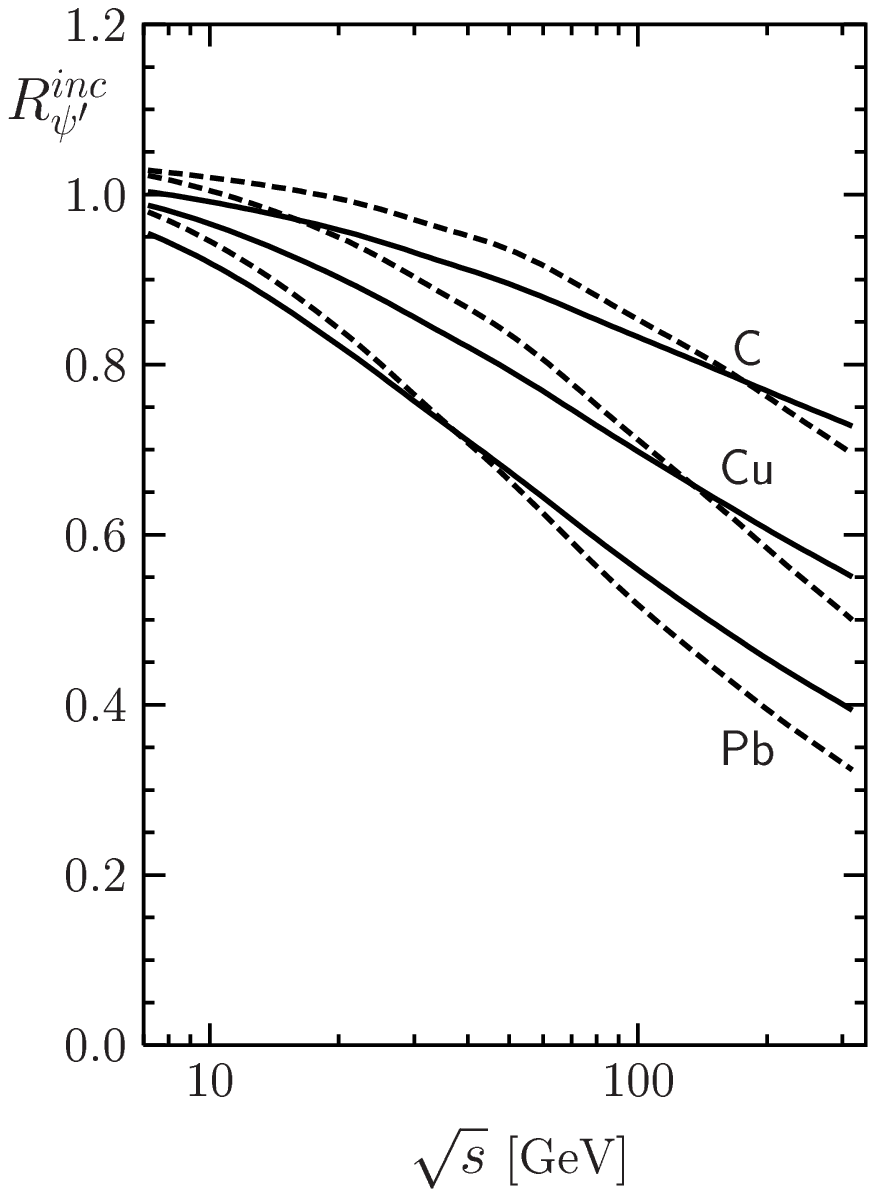}
  \caption{\small Ratios $R^{coh}_{\jpsi}$, $R^{inc}_{\jpsi}$, $R^{coh}_{\Psi'}$ and $R^{coh}_{\Psi'}$ for coherent and incoherent production on carbon, copper and Pbas function of $\sqrt s$ and at $Q^2=0$. The solid and dashed curves refer to the GBW and KST parameterizations respectively.}
\label{fig:inc-frozen}
\end{figure}

We define the nuclear ratios for coherent and incoherent reactions as,
\beq
 R^{coh}_\Psi(s,Q^2) =
  \frac{\sigma_{coh}^{\gamma^*A}(s,Q^2)}{A\,\sigma^{\gamma^*N}(s,Q^2)}\ ,\quad
   R^{inc}_\Psi(s,Q^2) =
  \frac{\sigma_{inc}^{\gamma^*A}(s,Q^2)}{A\,\sigma^{\gamma^*N}(s,Q^2)}\ .
  \label{kopel-1220}
\eeq
These ratios, calculated with Eqs.~(\ref{kopel-1180})-(\ref{kopel-1200}) for real photoproduction of $\jpsi$ and $\Psi'$, are depicted as a function of energy in Fig.~\ref{fig:inc-frozen}. For coherent production, the cross section rises with $A$ nearly as $A^{4/3}$, so the ratio may reach a large magnitude.

One can also predict the dependence on the momentum transfer $\vec k_T$
for the charmonium electro-production on nuclei. In the case of incoherent
production, this dependence is the same as for production on free nucleons.
However, in coherent production, the nuclear form factor comes into play
and one has
\beq
  \frac{d\sigma^{\gamma_{T,L}^*A}_{coh}(s,Q^2)}{d^2k_T} =
  \left| \int d^2b \,\, e^{i\vec k_T\cdot\vec b}\,
  \left\la \Psi\left|1\,-\,\exp\left[-{1\over2}\,
  \sigma_{\bar qq}(r_T,s)\,T_A(b)\right]\right|\Psi^{T,L}_{c\bar c}
  \right\ra\right|^2\ .
  \label{kopel-1240}
\eeq

We introduce the ratios the sum
of $T$ and $L$ components of Eq.~(\ref{kopel-1240}) 
to the cross section at $Q^2=0$ and $k_T=0$,
\beq
  {\cal R}(s,Q^2,k_T) =
  \frac{d\sigma^{\gamma^*A}_{coh}(s,Q^2)}{d^2k_T} \left/\left.
  \frac{d\sigma^{\gamma^*A}_{coh}(s,Q^2=0)}{d^2k_T}
  \right\vert_{k_T=0}\right.
  \label{kopel-1260}
\eeq

\begin{wrapfigure}{r}{0.7\columnwidth}
\begin{minipage}[b]{0.34\columnwidth}
\centerline{\includegraphics[width=0.75\columnwidth]{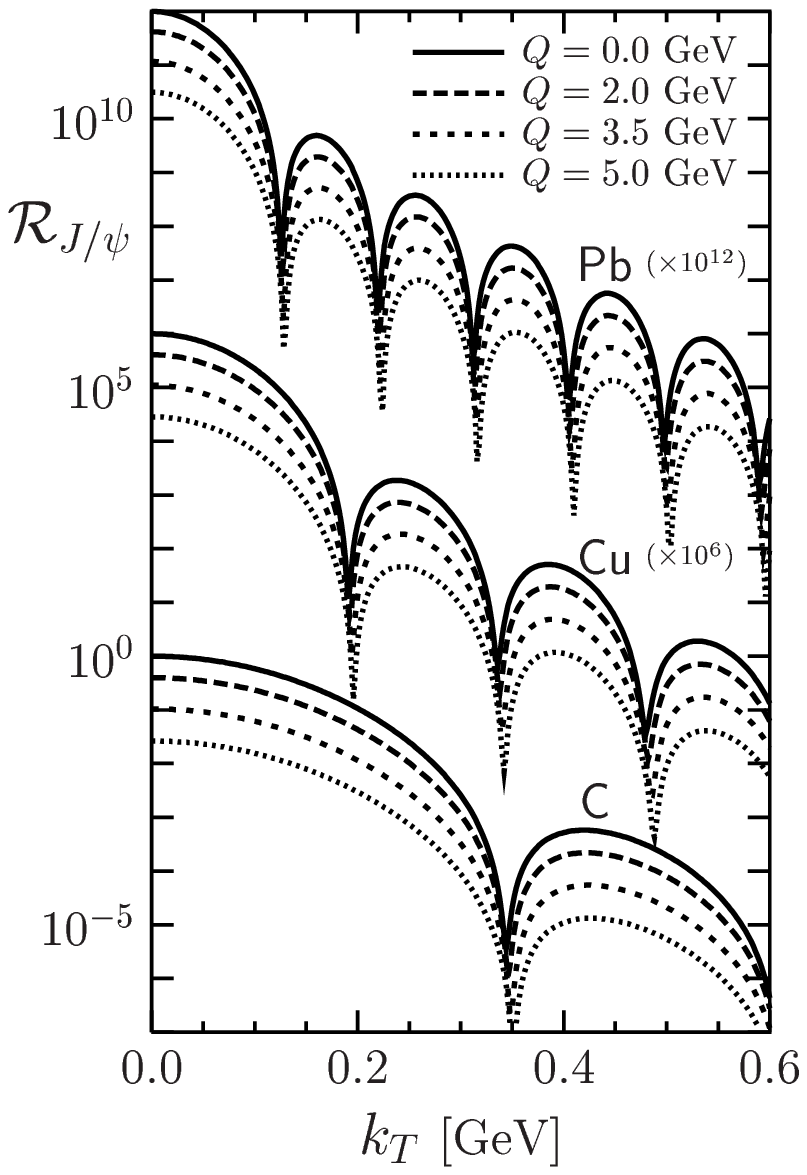}}
\end{minipage}
\begin{minipage}[b]{0.34\columnwidth}
\centerline{\includegraphics[width=0.75\columnwidth]{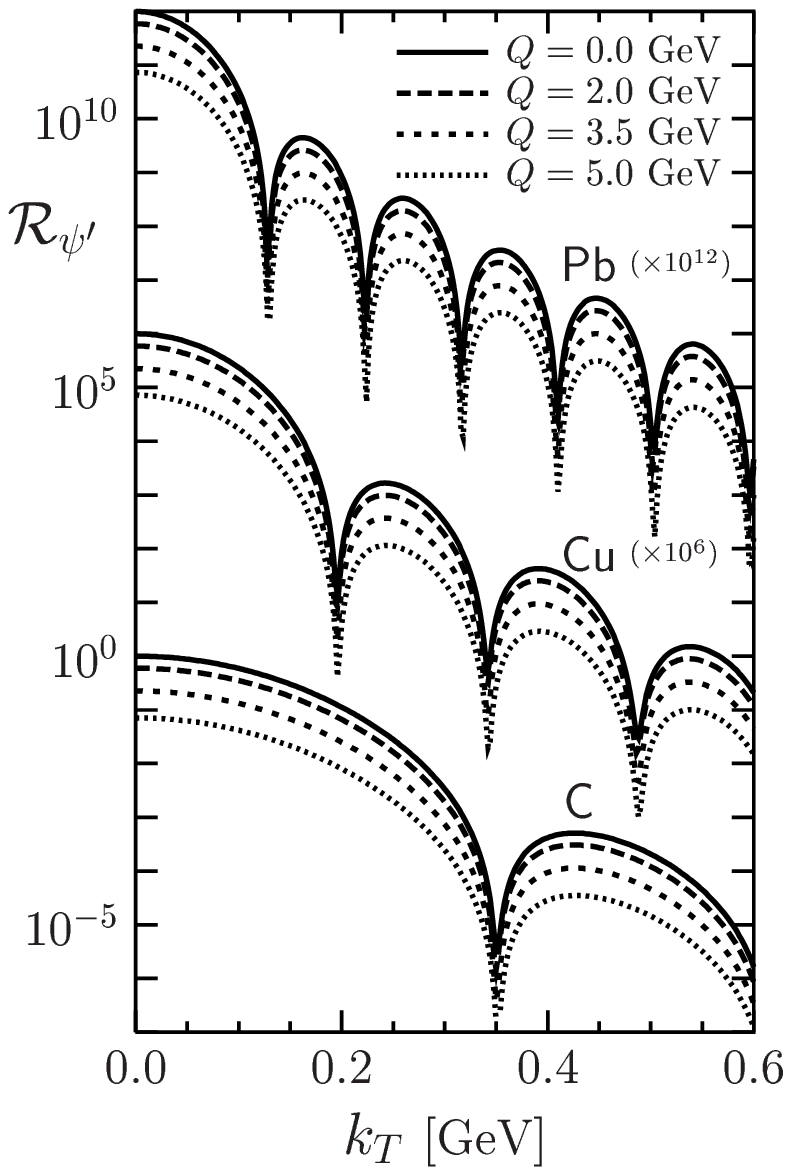}}
\end{minipage}
\caption{\small Ratios ${\cal R}_{\jpsi}$ and ${\cal R}_{\Psi'}$ as
      functions of $k_T$ at $s=4000\gev^2$ for different values of $Q$.
      All curves are calculated with the GBW parameterization of the dipole
      cross section.}
\label{fig:kt-dep}
\end{wrapfigure}

This ratio is plotted in Fig.~\ref{fig:kt-dep}
as a function of $k_T$ at $s=4000\gev^2$ for different virtualities of the photon.
We see that the $k_T$ dependences are rather similar for $\jpsi$ and
$\Psi'$. The shape of the distribution is determined mainly by the nuclear
geometry (and not by the size of the (small) charmonium). The calculated
curves show the familiar diffraction pattern known from elastic scattering
on nuclei.


 
\begin{wrapfigure}{r}{0.7\columnwidth}
\begin{minipage}[b]{0.34\columnwidth}
\centerline{\includegraphics[width=0.75\columnwidth]{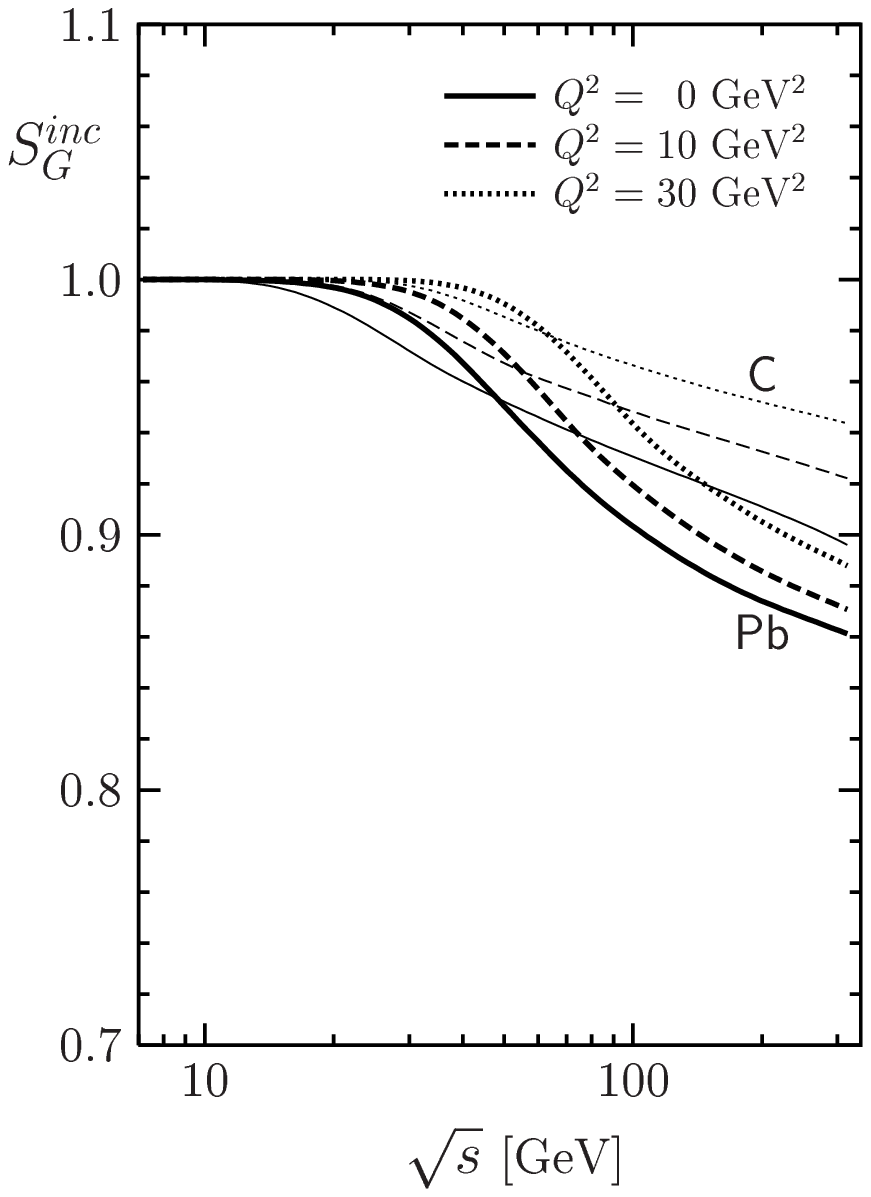}}
\end{minipage}
\begin{minipage}[b]{0.34\columnwidth}
\centerline{\includegraphics[width=0.75\columnwidth]{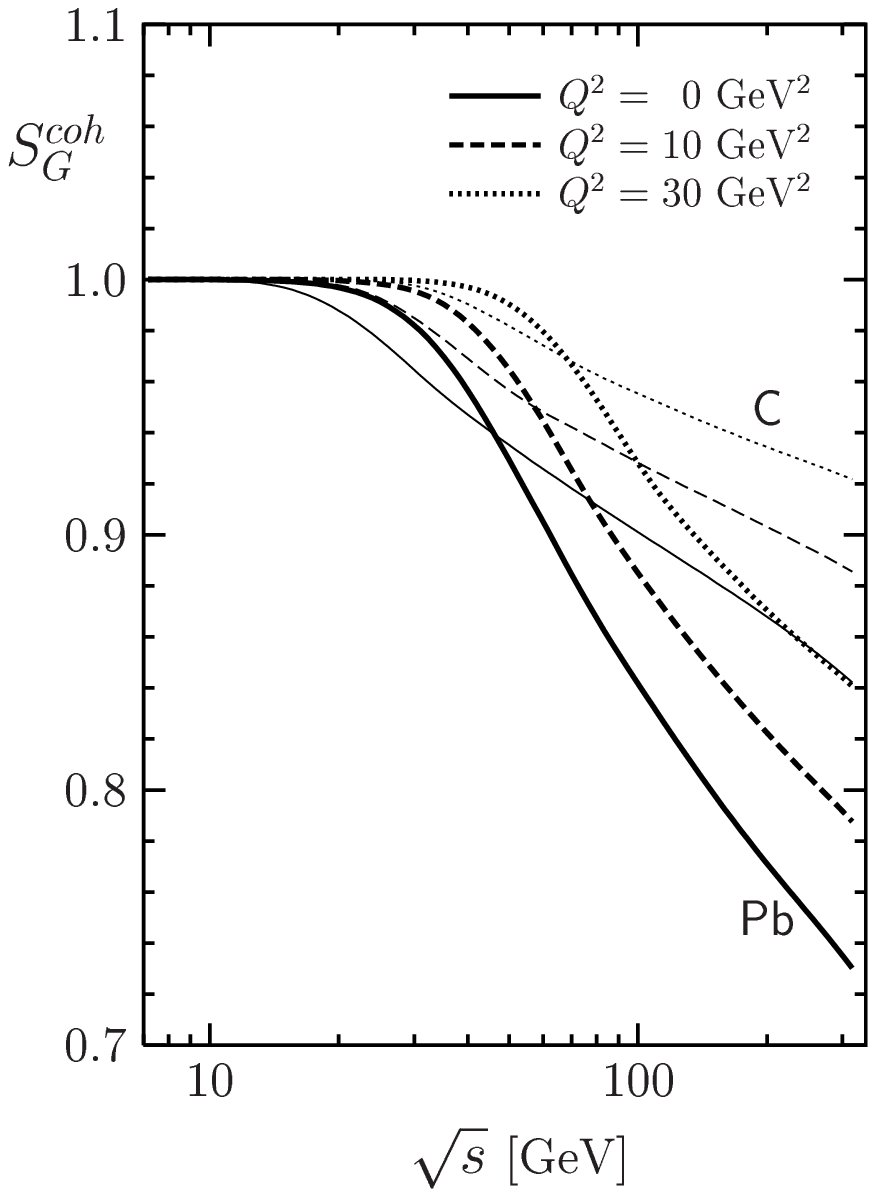}}
\end{minipage}
\caption{\small Ratios $S_g(s,Q^2)$, defined in (\ref{kopel-1280}), of
      cross sections calculated with and without 
      gluon shadowing for incoherent and coherent $\jpsi$ production..}
\label{fig:glue-shad2}
\end{wrapfigure}

It is interesting that the effects of gluon shadowing, calculated in~\cite{Ivanov:2002kc}, do not affect much the 
shape and position of the minima in $k_T$ dependence of the coherent cross section. However, the cross section integrated over $k_T$ may be significantly affected by gluon shadowing. To see the magnitude of gluon shadowing, we introduce the ratio of the cross sections calculated with
and without gluon shadowing,
\beq
  S_g(s,Q^2) = \frac{\sigma^{\gamma^*A}_g(s,Q^2)}
                    {\sigma^{\gamma^*A}(s,Q^2)}.
\label{kopel-1280}
\eeq
for incoherent and coherent exclusive charmonium electroproduction.
The predicted effects of gluon shadowing are depicted in Fig.~\ref{fig:glue-shad2}.

We only plot ratios for $\jpsi$ production, because ratios for
  $\Psi'$ are practically the same. All curves are
  calculated with the GBW parameterization of the dipole cross
  section. We see that the onset of gluon shadowing happens at a c.m. energy of a few
tens of GeV. This is controlled by the longitudinal nuclear form factor 
 \beq
F_A(q^g_c,b) = \frac{1}{T_A(b)}
\int\limits_{-\infty}^{\infty} dz\,
\rho_A(b,z)\,e^{iq_c z}\,
\label{kopel-1300}
 \eeq
 where the longitudinal momentum transfer $q^g_c=1/l^g_c$.
For the onset of gluon shadowing, $q^g_c\,R_A \gg 1$, one can keep only
the double scattering shadowing correction,
 \beq
S_g \approx 1-{1\over4}\,\sigma_{eff}
\int d^2b\, T^2_A(b)\,F_A^2(q^g_c,b)\ ,
\label{kopel-1320}
 \eeq
 where $\sigma_{eff}$ is the effective cross section which depends on the dynamics
of interaction of the $\bar qqg$ fluctuation with a nucleon.

It was found in~\cite{Kopeliovich:1998gv} that the coherence length
for gluon shadowing is rather short, $l^g_c \approx (10\,x\,m_N)^{-1}$,  
where Bjorken $x$ in our case should be an effective one,
$x=(Q^2+M_\Psi^2)/2m_N\nu$.  The onset of shadowing according to
(\ref{kopel-1300}) and
(\ref{kopel-1320}) should be expected at $q_c^2\sim 3/(R_A^{ch})^2$
corresponding to 
$
  s_g \sim 10 m_N R_A^{ch}(Q^2+M_\Psi^2)/\sqrt{3}
$,
where $(R_A^{ch})^2$ is the mean square of the nuclear charge radius.
This estimate is in a good agreement with Fig.~\ref{fig:glue-shad2}. Remarkably,
the onset of shadowing is delayed with rising nuclear radii and $Q^2$.
This follows directly from Eq.~(\ref{kopel-1320}) and the fact that the formfactor
is a steeper falling function of $R_A$ for heavy than for light nuclei, provided
that $q_c^GR_A\gg1$.

At medium energies, the effects of finite coherence length, $l_c\sim R_A$ become important. They increase the incoherent and suppress coherent cross sections of charmonium electroproduction. One can find the details of the corresponding calculations in~\cite{Ivanov:2002kc}.


\subsubsection{Exclusive processes in $e+A$ collisions}
\label{sec:exclusive_victor}

\hspace{\parindent}\parbox{0.92\textwidth}{\slshape
 Victor P. Gon\c{c}alves}
\index{Gon\c{c}alves, Victor P.}

\vspace{\baselineskip}

Exclusive processes in deep inelastic scattering (DIS) have appeared as key reactions to 
trigger the generic mechanism of diffractive scattering. In particular, diffractive vector meson production and deeply virtual Compton 
scattering (DVCS) have been extensively studied at HERA and provide a valuable  probe of the  
QCD dynamics at high energies.  The cross sections for exclusive processes in DIS are proportional 
to the square of the scattering amplitude, which makes them  strongly sensitive to the 
underlying QCD dynamics.

In this contribution, we present our estimate for the coherent and 
incoherent cross sections for  exclusive $\rho$, $J/\Psi$, and  $\phi$ production as well as for nuclear DVCS, making use of the numerical solution of the  
Balitsky-Kovchegov equation  including running coupling corrections in order to estimate the contribution of the saturation 
physics to exclusive processes 
(For more details and references see Refs. \cite{Goncalves:2009za,Cazaroto:2010yc}). \\


\noindent{\bf Exclusive production:} In the color dipole approach, exclusive  production $\gamma^* A \rightarrow EY$  ($E = \rho, \phi, J/\Psi$ or $\gamma$) in electron-nucleus interactions at high energies ($l_c \gg R_A$) is given by 
\begin{eqnarray}
\sigma^{coh}\, (\gamma^* A \rightarrow EA)  =  \int d^2b \left\langle 
\mathcal{N}^A(x,r,b) \right\rangle^2
\label{totalcscoe}
\end{eqnarray}
where
\begin{eqnarray}
\left\langle \mathcal{N} \right\rangle = \int d^2r
 \int dz  \Psi_E^*(r,z) \, \mathcal{N}^A(x,r,b)\, \Psi_{\gamma^*}(r,z,Q^2)
 \label{totalcscoe1}
\end{eqnarray}
and $ {\cal N}^A (x, r, b)$, defined in eq.~(\ref{enenuc}), is the forward dipole-target scattering amplitude for a 
dipole with size $r$ and impact parameter $b$. We will assume that $\sigma_{dp}$ in eq.~(\ref{enenuc}) is given by the bCGC saturation model or the 
solution of the running coupling BK equation.

On the other hand, if the nucleus scatters inelastically, i.e. breaks up ($Y = X$),   
the process is called incoherent production. 
 In this case, one sums over all final states of the target nucleus, 
except those that contain particle production. The $t$ slope is the same as in the case of a 
nucleon target. Therefore we have
\begin{eqnarray}
\sigma^{inc}\, (\gamma^* A \rightarrow EX) = \frac{|{\cal I}m \, 
{\cal A}(s,\,t=0)|^2}{16\pi\,B_E} \;
\label{totalcsinc}
\end{eqnarray}
where at high energies ($l_c \gg R_A$) :
\begin{eqnarray}
|{\cal I}m \, {\cal A}|^2  =  \int d^2b \, T_A(b)   |  \Psi_E^*(r,z) \sigma_{dp} \, 
\exp[- \frac{1}{2} \, \sigma_{dp} \, T_A(b)]\Psi_{\gamma^*}(r,z,Q^2)|^2 
\label{totalcsinc1}
\end{eqnarray} 
and $ \sigma_{dp}$ is the dipole-proton cross section. In the incoherent case, the $q\bar{q}$ pair attenuates with a constant absorption cross 
section, as in the Glauber model, except that the whole exponential is averaged 
rather than just the cross section in the exponent.  
The coherent and incoherent cross sections depend  
differently on  $t$.  
At small-$t$ ($-t\,R_A^2/3 \ll 1$) coherent 
production dominates, with the signature being a sharp 
forward diffraction peak. On the other hand, incoherent production will dominate  
at large-$t$ ($-t\,R_A^2/3 \gg 1$), with the $t$-dependence being to a good 
accuracy the same as in the production off  free nucleons.  

In Eqs. (\ref{totalcscoe1}) and (\ref{totalcsinc1}) the functions 
$\Psi^{\gamma}(z,\,r)$ and $\Psi^{E}(z,\,r)$  
are the light-cone wavefunctions  of the photon and the exclusive final state, respectively.  The 
variable $r$ defines the relative transverse
separation of the pair (dipole) and $z$ $(1-z)$ is the
longitudinal momentum fraction of the quark (antiquark). 
In the dipole formalism, the light-cone
 wavefunctions $\Psi(z,\,r)$ in the mixed
 representation $(r,z)$ are obtained through a two dimensional Fourier
 transform of the momentum space light-cone wavefunctions
 $\Psi(z,\,k)$. The photon wavefunctions  are well known in the literature. For the meson 
wavefunction, we considered the Gauss-LC  model. In the DVCS case, as one has a 
real photon in the final state, only the transversely polarized overlap function contributes 
to the cross section.  Summed over the quark helicities, for a given quark flavour $f$, it is 
given by,
\begin{align}
  (\Psi_{\gamma}^*\Psi)_{T}^f & =  \frac{N_c\,\alpha_{\mathrm{em}}
e_f^2}{2\pi^2}\left\{\left[z^2+\bar{z}^2\right]\varepsilon_1 K_1(\varepsilon_1 r) 
\varepsilon_2 K_1(\varepsilon_2 r) 
 +     m_f^2 K_0(\varepsilon_1 r) K_0(\varepsilon_2 r)\right\},
  \label{eq:overlap_dvcs}
\end{align}
where we have defined the quantities $\varepsilon_{1,2}^2 = z\bar{z}\,Q_{1,2}^2+m_f^2$ and 
$\bar{z}=(1-z)$. Accordingly, the photon virtualities are $Q_1^2=Q^2$ (incoming virtual 
photon) and $Q_2^2=0$ (outgoing real photon).


\begin{figure}
  \centering
  \includegraphics[width=0.49\textwidth,clip=true]{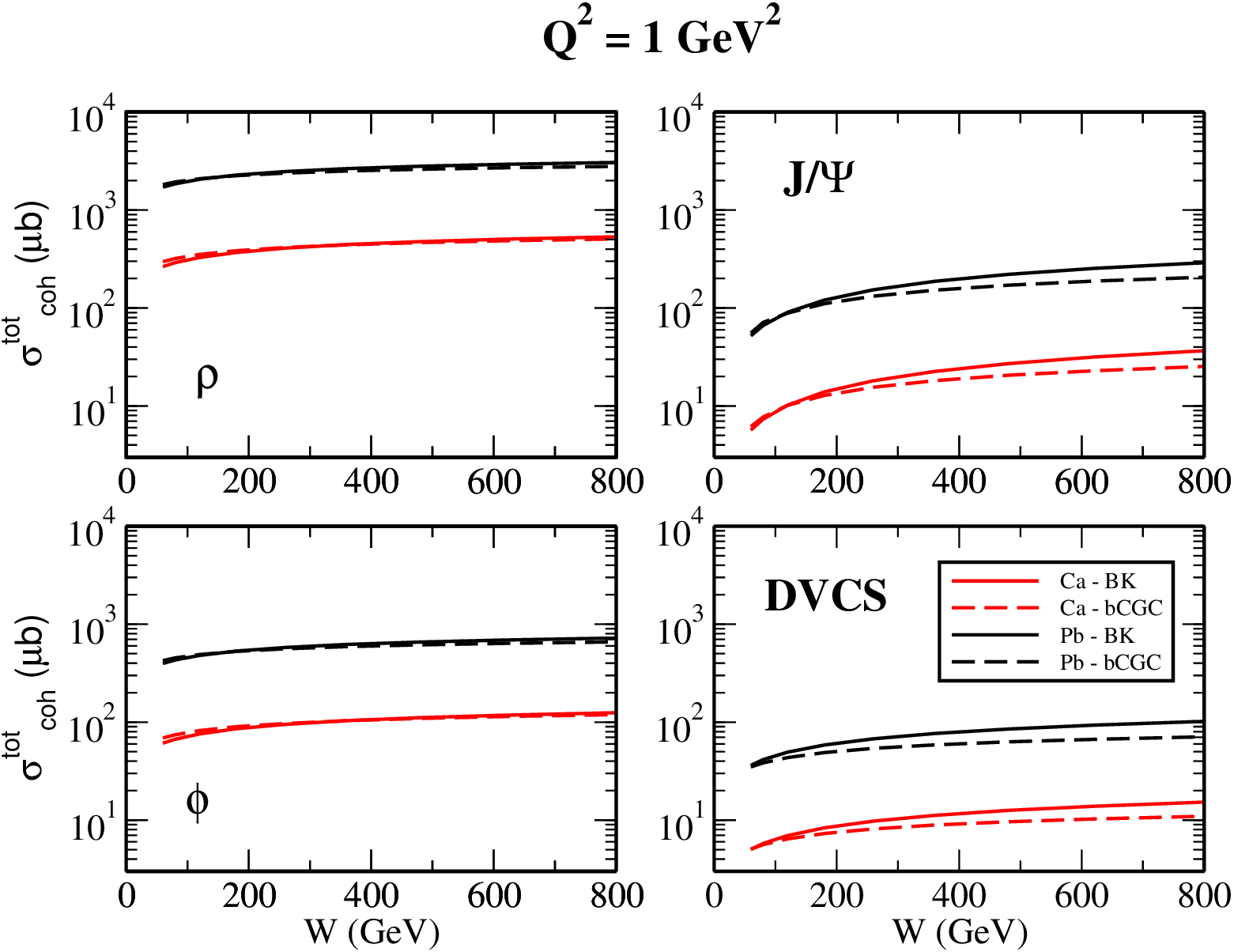}
  \hfill
  \includegraphics[width=0.49\textwidth,clip=true]{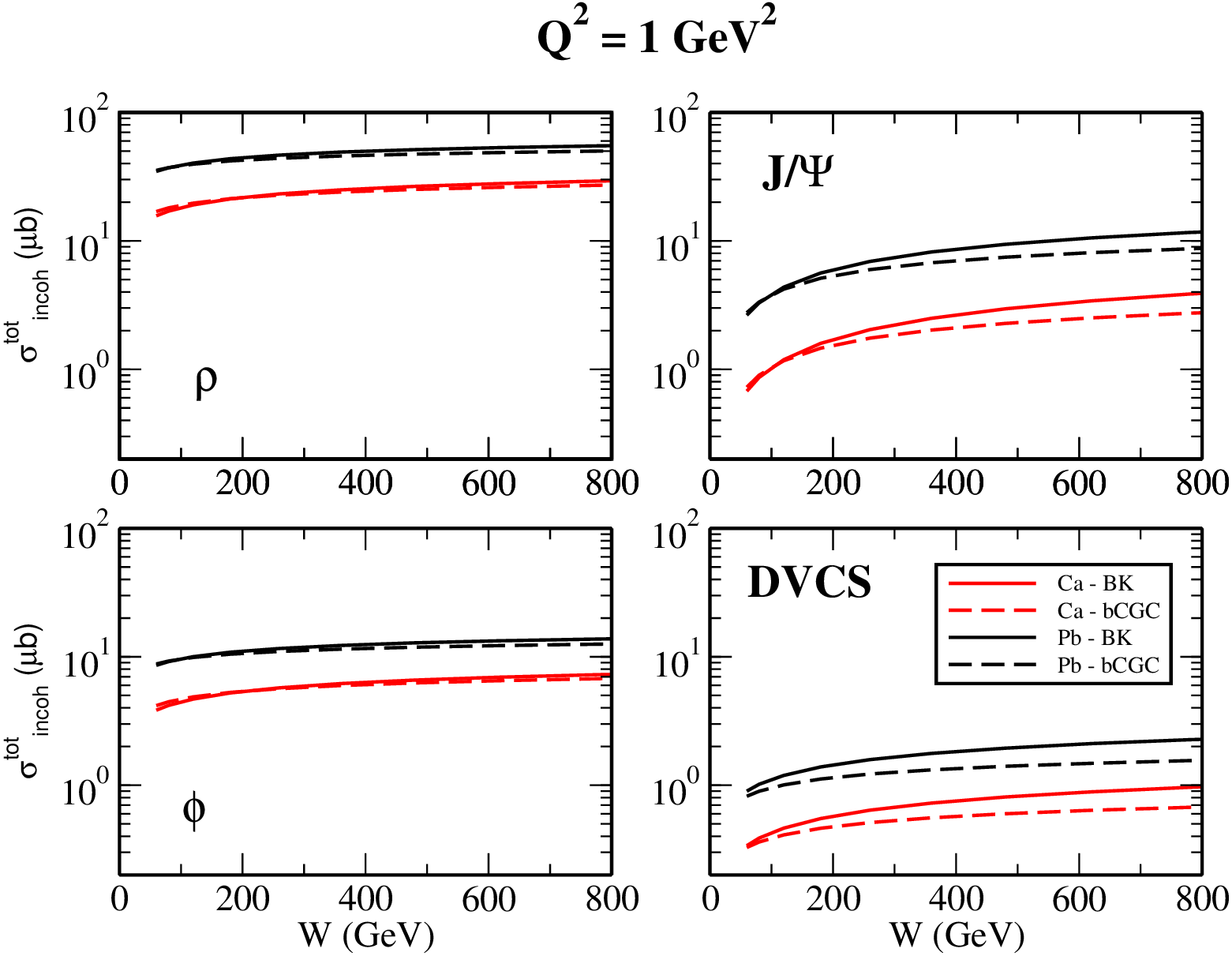}
  \caption{\small 
    Energy dependence of the coherent (left) and incoherent
    (right) cross sections for different final states and $Q^2 = 1$
    GeV$^2$.
  } 
\label{fig:energy_coh}
\label{fig:energy_inc}
\end{figure}


\noindent{\bf Results:}  In Fig. \ref{fig:energy_coh} left, we show the coherent production cross section as a function of the photon-target 
c.m.s energy, $W$, for a fixed photon virtuality $Q^2=1$
GeV$^2$. Fig. \ref{fig:energy_inc} right is the exact analogue  for the corresponding incoherent 
cross sections. Each one of the panels 
shows  the results obtained for one specific  final state. In each figure, the two upper (lower) curves show the results for a Pb (Ca) target.  
In  all figures, the dashed (solid) lines are obtained with the bCGC (rcBK) dipole-proton cross section. 
At  low $W$, the  bCGC and rcBK  production cross sections are indistinguishable 
from one another because the  dipole cross sections tend to coincide. These latter have been 
tuned to fit DIS data,  which are taken in this kinematical region. Another expected feature 
is the observed decrease of the cross sections with increasing  
vector meson masses, which comes from the wave functions.  
Differences are expected to appear at higher energies, where we enter  
the  lower $x$ (extrapolation)  region. In all cases we see that  the  results obtained 
with the rcBK cross section are larger  than those obtained with the bCGC one. This is 
related to the fact that the numerical solutions of the BK equation tend to reach the unitarity limit later.  Due to this fact, the
results obtained with the rcBK dipole cross section grow faster with energy than those 
obtained with the bCGC one.  Another feature is that the differences between bCGC and rcBK are 
larger for heavier vector mesons. 
Comparing the results shown in Fig.~\ref{fig:energy_coh} we verify the  dominance of 
the coherent production  with a small contribution coming from incoherent processes. 

\ \\ \noindent{\it Acknowledgments:}
The author thank E.R. Cazaroto, F. Carvalho, M. S. Kugeratski,
M.V.T. Machado, and  F.S. Navarra by collaboration. 

\subsubsection{Constraining the $\rho$ wavefunction}
\label{sec:sandapen}

\hspace{\parindent}\parbox{0.92\textwidth}{\slshape
 Jeffrey R. Forshaw and Ruben Sandapen}
\index{Forshaw, Jeffrey R.}
\index{Sandapen, Ruben}

\vspace{\baselineskip}

In the dipole model~\cite{Nikolaev:1990ja,Mueller:1994jq}, the 
imaginary part of the amplitude  for diffractive $\rho$ production is written 
as~\cite{Marquet:2007qa}
\begin{equation}
\Im \mbox{m} \mathcal{A}_{\lambda} (s,t;Q^2) =
\sum_{h,\bar{h}} \int \d^2  \mathbf{r} \d z
\Psi^{\gamma^*,\lambda}_{h,\bar{h}} (r,z;Q^2)
\Psi^{\rho,\lambda}_{h,\bar{h}}(r,z)^{*}
e^{-iz\mathbf{r}.\mathbf{\Delta}} \mathcal{N}(x,\mathbf{r},\mathbf{\Delta}) 
\label{non-forward-amplitude}
\end{equation}
where $t=-|\mathbf{\Delta}|^2$. In a standard
notation~\cite{Forshaw:2010py,Marquet:2007qa,Forshaw:2003ki},
$\Psi^{\gamma^*,\lambda}_{h,\bar{h}}$ and $\Psi^{\rho,\lambda}_{h,\bar{h}}$ are
the light-cone wavefunctions of the photon and the $\rho$ meson respectively
while $\mathcal{N}(x,\mathbf{r},\mathbf{\Delta})$ is the imaginary part of the
dipole-proton elastic scattering amplitude. The energy dependence of the latter
is via the dimensionless variable $x$, taken here to
be $x = (Q^2 + 4m_f^2)/(Q^2+ s)$ where $m_f$ is a phenomenological light quark mass.\footnote{We shall take
$m_f=0.14~\mbox{GeV}$, the value used when extracting the dipole
cross section from $F_2$ data.} Setting $t=0$ in equation
\eqref{non-forward-amplitude}, we obtain the forward
amplitude used in reference~\cite{Forshaw:2010py}:

\begin{equation}
\left.\Im \mbox{m} \, \mathcal{A}_\lambda(s,t;Q^2)\right|_{t=0} = s \sum_{h,
\bar{h}}
\int \d^2 {\mathbf r} \; \d z \; \Psi^{\gamma,\lambda}_{h, \bar{h}}(r,z;Q^2)
\hat{\sigma}(x,r) \Psi^{\rho,\lambda}_{h, \bar{h}}(r,z)^*
\label{forward-amplitude}
\end{equation}
where we have used the optical theorem to introduce the dipole cross-section
$\hat{\sigma}(x,r)=\mathcal{N}(x,r,\mathbf{0})/s$.
Note that since the momentum transfer $\mathbf{\Delta}$ is Fourier conjugate to
the impact parameter $\mathbf{b}$, the dipole cross-section at a given energy
is simply the $b$-integrated dipole-proton scattering amplitude:

\begin{equation}
\hat{\sigma}(x,r)=\frac{1}{s}\int
\d^2 \mathbf{b}\ \mathcal{N}(x,r,\mathbf{b}) \;.
\label{dipole-xsec-b-int}
\end{equation}
This dipole
cross-section can be extracted from
the $F_2$ data since
\begin{equation}
F_2(x,Q^2)  \propto \int \d^2 \mathbf{r}\ \d z\ |\Psi_{\gamma^*}(r,z;Q^2)|^2
\hat{\sigma}(x,r)
\end{equation}
and the photon's light-cone wavefunctions are known in QED, at least for large
$Q^2$. The
$F_2$-constrained dipole cross-section can then be used
to predict the imaginary part of the forward amplitude 
for diffractive $\rho$ production and thus the forward differential
cross-section, 
\begin{equation}
\left. \frac{d\sigma_{\lambda}}{dt} \right|_{t=0} =\frac
{1}{16\pi} (\Im\mathrm{m} \mathcal{A}_\lambda(s,0))^2 \; (1 + \beta_\lambda^2)~,
\label{gammap-xsec}
\end{equation}
where $\beta_\lambda$ is the ratio of real to imaginary parts of the
amplitude and is computed as in reference~\cite{Forshaw:2010py}. The
$t$-dependence
can be assumed to be the
exponential dependence as suggested by experiment~\cite{Chekanov:2007zr}:
\begin{equation}
\frac{d\sigma_{\lambda}}{dt}= \left. \frac{d\sigma_{\lambda}}{dt} \right|_{t=0}
\times
\exp(-B|t|)\ ,\quad
B=N\left(
  14.0 \left(\frac{1~\mathrm{GeV}^2}{Q^2 + M_{\rho}^2}\right)^{0.2}+1\right)
\label{Bslope}
\end{equation}
with $N=0.55$ GeV$^{-2}$. After integrating over $t$, we can compute the total
cross-section $\sigma=\sigma_L + \epsilon \sigma_T$ 
which is measured at HERA.\footnote{To compare with the HERA data, we take
$\epsilon=0.98$.} 

Presently, several dipole
models~\cite{Forshaw:2004vv,Watt:2007nr,Soyez:2007kg,Bartels:2002cj,Kowalski:2006hc} 
are able to fit the current HERA $F_2$ data and 
there is evidence that the data prefer those incorporating
some form of saturation~\cite{Motyka:2008jk}. We can use the $F_2$-constrained
dipole cross-section in order to
extract the $\rho$ light-cone wavefunction using the current precise HERA data
~\cite{Chekanov:2007zr,Aaron:2009xp}. This
has recently been performed in reference~\cite{Forshaw:2010py} using the
Regge-inspired  FSSat dipole
model~\cite{Forshaw:2004vv} and we shall report the results of
this work here. In addition, we repeat the analysis 
using two alternative models~\cite{Soyez:2007kg,Watt:2007nr,Kowalski:2006hc}
both
based on the original Colour Glass Condensate (CGC) model~\cite{Iancu:2003ge}.
They differ from the original CGC model by including the
contribution of charm quarks when fitting to the $F_2$
data. Furthermore in one of them~\cite{Soyez:2007kg,Watt:2007nr}, the anomalous
dimension  $\gamma_s$ is treated as an additional free parameter instead of
being fixed to its LO BFKL value of $0.63$. We shall refer to
these models as CGC[$0.74$] and CGC[$0.63$] models  where the number in the
square brackets stands for the fitted and fixed value of the anomalous
dimension respectively. For both models, we use the set of fitted parameters
given in reference~\cite{Watt:2007nr}. All
three models, i.e FSSat, CGC[$0.63$] and CGC[$0.74$] account for saturation
although in a $b$- (or equivalently $t$-) independent way. Indeed, at
a given energy, the dipole cross-section is equal to the forward
dipole-proton amplitude or to the $b$-integrated dipole proton amplitude given by equation \eqref{dipole-xsec-b-int}. Finally, all three dipole models we consider here give a good description of
the diffractive structure function data~\cite{Marquet:2007nf,Forshaw:2006np}. \\


\begin{table}[h]
  \centering
  \parbox{6cm}{\centering
    \textbf{Boosted Gaussian predictions}
    \[
    \begin{array} 
      [c]{|c|c|}\hline
      \mbox{Dipole model} & \chi^2/\mbox{data point}\\ \hline
      \mbox{FSSat}& 310/75 \\ \hline
      \mbox{CGC}[0.74]& 262/75 \\ \hline
      \mbox{CGC}[0.63]&  401/75 \\ \hline
    \end{array}
    \]
  }\hspace{-0.5cm}
  \parbox{5cm}{\centering
    \textbf{BG fits}
    \[
    \begin{array} 
      [c]{|c|c|}\hline
      \mbox{Model} & \chi^2/\mbox{d.o.f}  \\ \hline
      \mbox{FSSat~\cite{Forshaw:2010py}}&  82/72 \\ \hline
      \mbox{CGC}[0.74]& 64/72 \\ \hline
      \mbox{CGC}[0.63] & 83/72 \\ \hline
    \end{array}
    \]
  }\hspace{-0.5cm}
  \parbox{5cm}{\centering
    \textbf{Improved fits}
    \[
    \begin{array} 
      [c]{|c|c|}\hline
      \mbox{Model} & \chi^2/\mbox{d.o.f} \\ \hline
      \mbox{FSSat~\cite{Forshaw:2010py}}&  68/70 \\ \hline
      \mbox{CGC}[0.63] & 67/70  \\ \hline
    \end{array}
    \]
  }
  \caption {
    {\it Left:} Predictions of the $\chi^2/\mbox{data point}$ using the BG
    wavefunction.
    {\it Center:} $\chi^2/\mbox{d.o.f}$ obtained when fitting $R_{\lambda}$ and
    $b_{\lambda}$ to the leptonic decay width and HERA data.
    {\it Right:} $\chi^2/\mbox{d.o.f}$ obtained when fitting $b_{\lambda}$,
    $R_{\lambda}$ $c_{T}$, $d_{T}$ the leptonic decay width and HERA data.
  }
\label{tab:BG-chisqs}
\label{tab:BG-Fits-chisqs}
\label{tab:Enhancement-fits-chisqs}
\end{table}

\noindent{\bf Fitting the HERA data:}  Previous work~\cite{Forshaw:2003ki,Marquet:2007qa,Watt:2007nr}
has shown that a reasonable assumption for the
scalar part of the light-cone wavefunction for the $\rho$ is of the form
\begin{eqnarray}
\phi^{{\mathrm{BG}}}_\lambda(r,z) &=&
\mathcal{N}_\lambda \;  4[z(1-z)]^{b_{\lambda}} \sqrt{2\pi R_{\lambda}^{2}} \;
\exp \left(\frac{m_f^{2}R_{\lambda}^{2}}{2}\right)
\exp \left(-\frac{m_f^{2}R_{\lambda}^{2}}{8[z(1-z)]^{b_{\lambda}}}\right) \\
\nonumber
& &\times \exp \left(-\frac{2[z(1-z)]^{b_\lambda}
r^{2}}{R_{\lambda}^{2}}\right) 
\label{boosted-gaussian} 
\end{eqnarray}
and is referred to as the `Boosted Gaussian' (BG). This wavefunction is a
simplified version of that proposed originally
by Nemchik, Nikolaev, Predazzi and Zakharov~\cite{Nemchik:1996cw}. In the
original BG wavefunction, $b_\lambda=1$ while the parameters $R_{\lambda}$ and
$\mathcal{N}_{\lambda}$ are fixed by the leptonic decay width constraint and
the wavefunction normalization conditions~\cite{Forshaw:2010py}.
However, when the BG wavefunction is used in conjunction with either the FSSat
model or any of the CGC models, none of them is able to give a good quantitative
agreement with the current HERA $\rho$-production data.
This is illustrated by the large $\chi^2$ values in table \ref{tab:BG-chisqs}, the
situation is considerably improved by fitting $R_\lambda$ and $b_\lambda$ to
the leptonic decay width and HERA data (we fit to the same data set and with the same cuts as in reference~\cite{Forshaw:2010py}). 

For the FSSat and CGC[$0.63$] models, we can further improve the quality of the fit
by allowing for
additional end-point enhancement in the transverse wave-function, i.e. using a
scalar
wave-function of the form
\begin{equation}
\phi_T (r,z)= \phi^{{\mathrm{BG}}}_T (r,z) \times [1+ c_{T}
\xi^2 + d_{T} \xi^4]
\label{EG} 
\end{equation}
where $\xi=2z-1$. The results are shown in table
\ref{tab:Enhancement-fits-chisqs}.


\begin{table}[t]
\centering
\textbf{Best fit parameters}
\[
\begin{array} 
[c]{|c|c|c|c|c|c|c|}\hline
                 & R_L^2 & R_T^2 & b_L & b_T & c_T & d_T \\ \hline
\mbox{FSSat~\cite{Forshaw:2010py}}     &26.76  &27.52 &0.5665 &0.7468 &0.3317
&1.310  \\ \hline
\mbox{CGC}[0.63] &27.31  &31.92 &0.5522 &0.7289 &1.6927 &2.1457  \\ \hline
\mbox{CGC}[0.74] &26.67  &21.30 &0.5697 &0.7929  &0&0 \\ \hline
\end{array}
\]
\caption {Best fit parameters for each dipole model.}
\label{tab:Best-fit-params}
\end{table}

\begin{figure}[htb]
  \centering
  \includegraphics[width=0.48\textwidth]{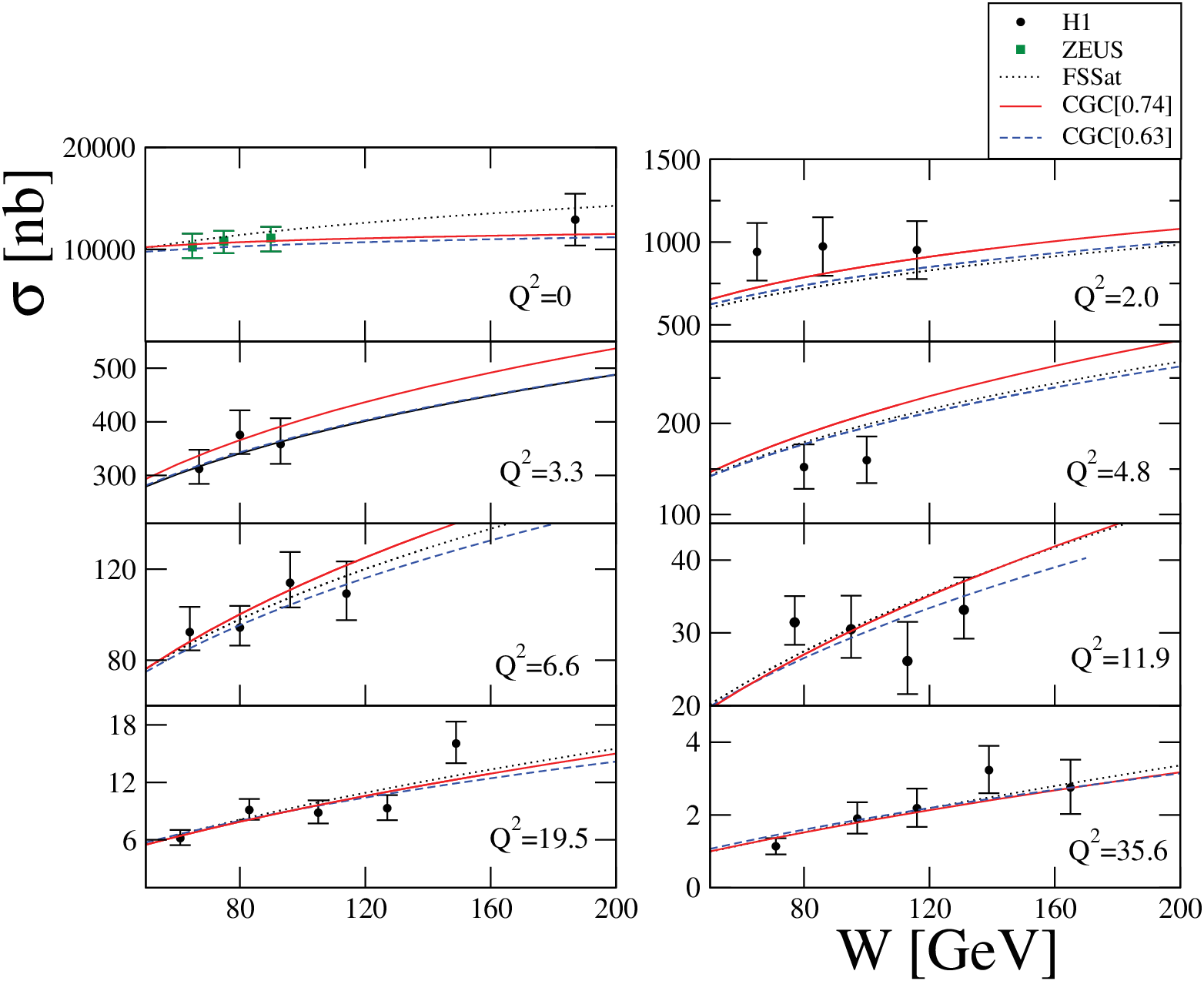}
  \hspace*{0.2cm}
  \includegraphics[width=0.48\textwidth]{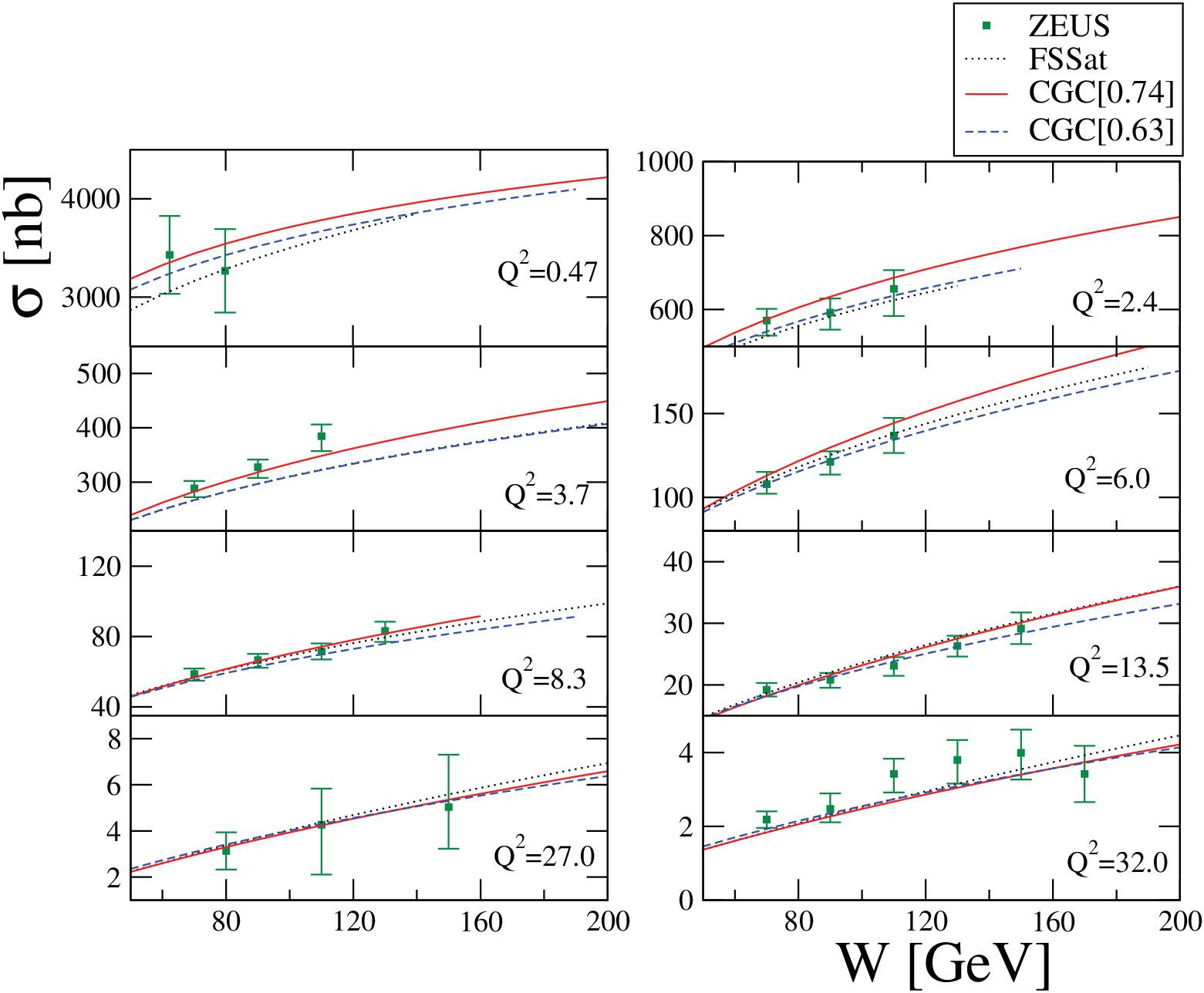}
  \caption{\small Best fits to the HERA (left) and ZEUS (right) total
    cross-section data. CGC[$0.74$]: solid; FSSat: dotted;
    CGC[$0.63$]: dashed.} 
  \label{fig:ZEUS-xsec} 
  \label{fig:H1-xsec}
\end{figure}

\begin{figure}[htb]
  \centering
  \includegraphics[width=0.35\textwidth]{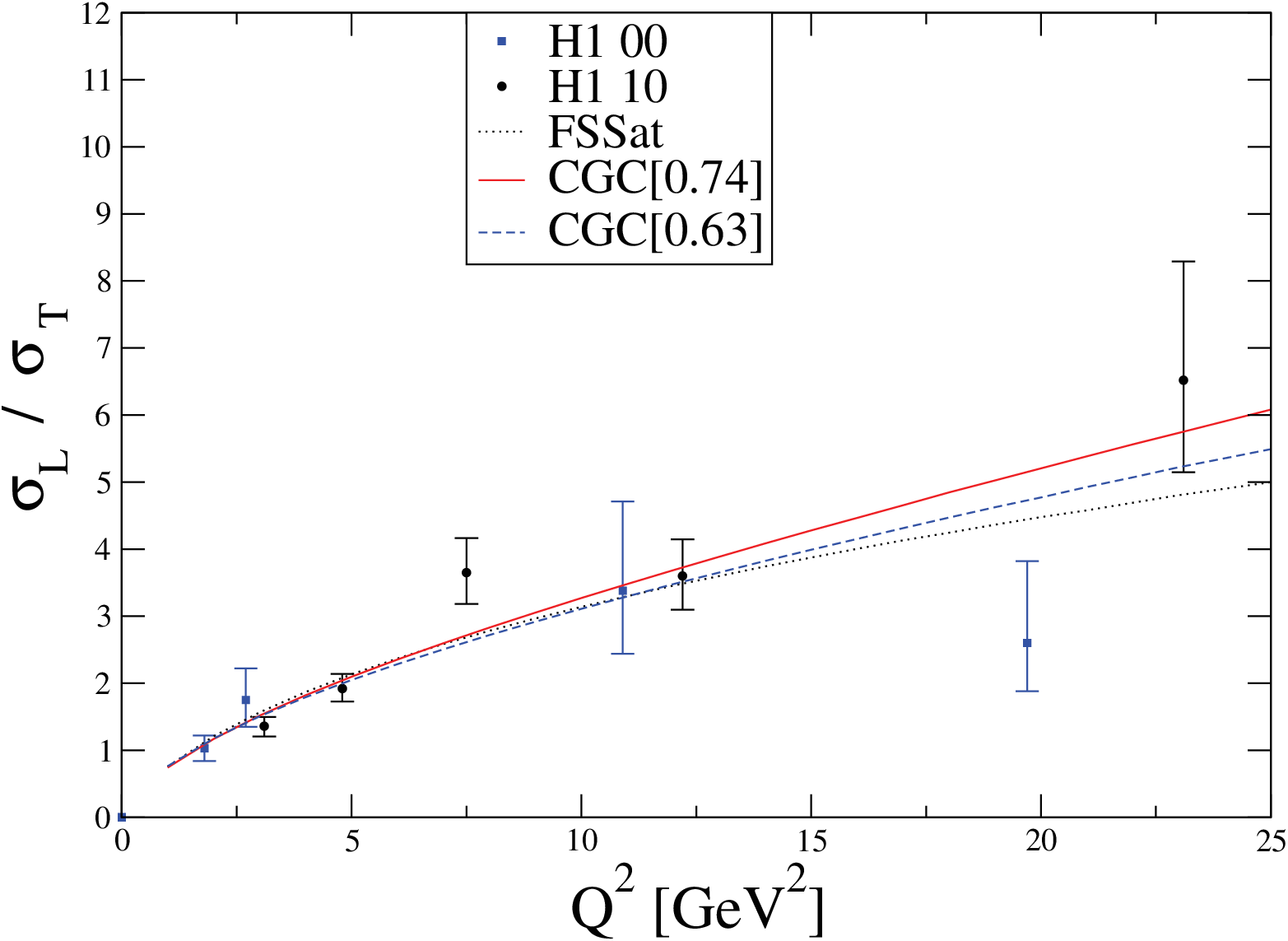}
  \hspace*{1cm}
  \includegraphics[width=0.35\textwidth]{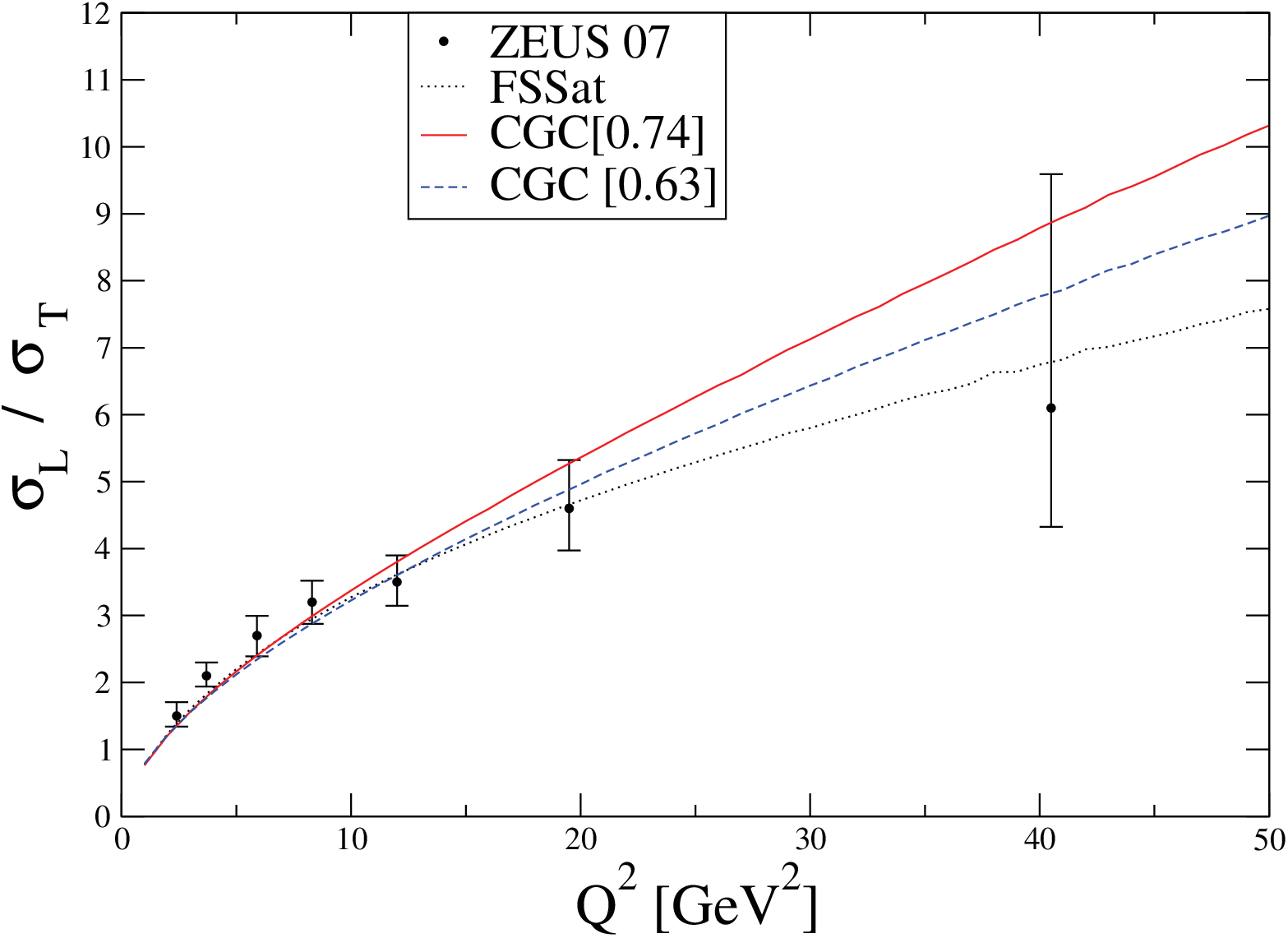}
  \caption{\small Best fits to the $\sigma_L/\sigma_T$ data. The H1 data are
    at $W=75$~GeV while the ZEUS data are at $W=90$~GeV. 
    CGC[$0.74$]: solid; FSSat: dotted; CGC[$0.63$]: dashed.}
\label{fig:HERA-ratio} 
\end{figure}

\begin{figure}
  \centering
  \includegraphics[width=0.32\textwidth]{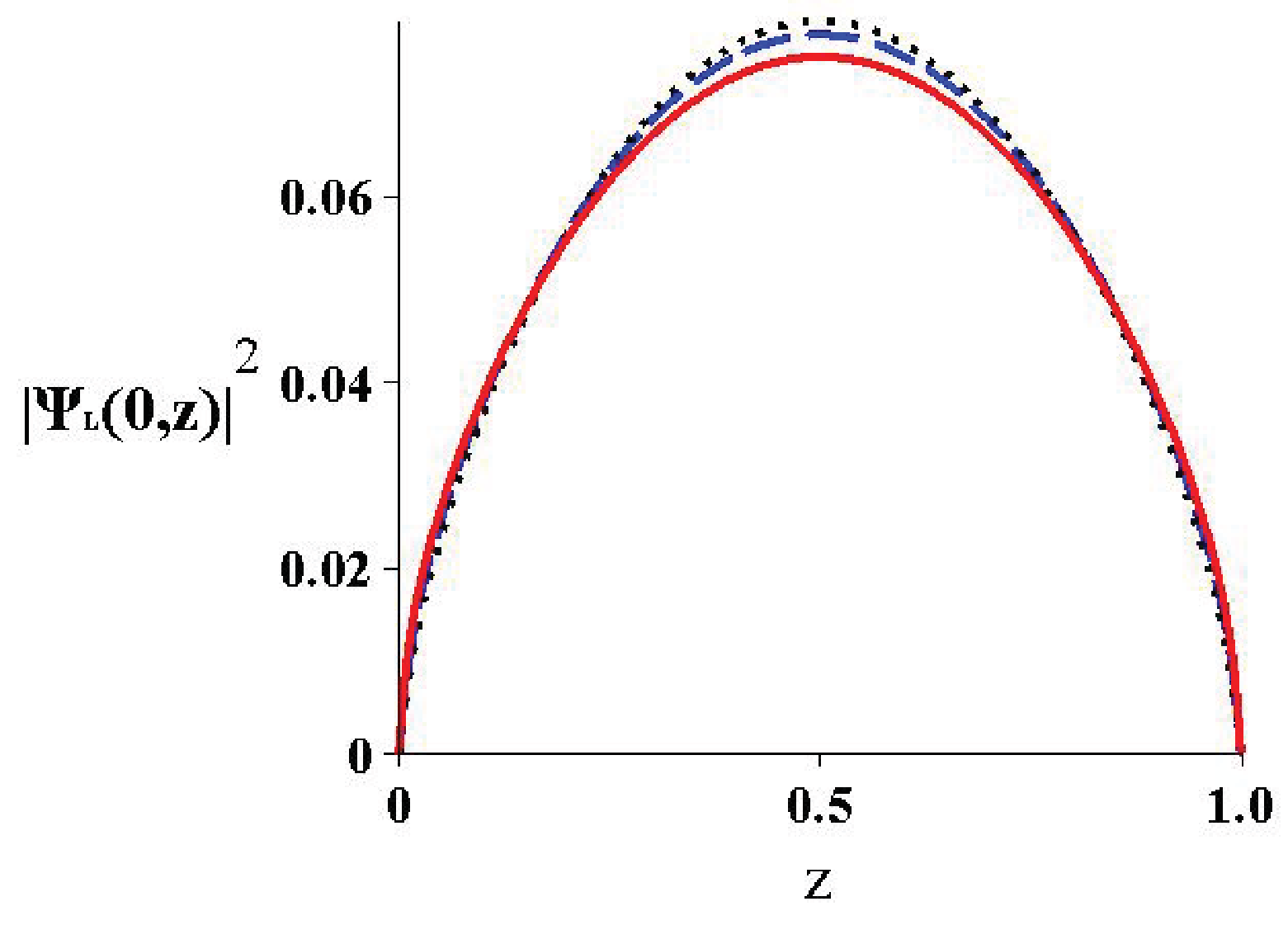}
  \hfill
  \includegraphics[width=0.32\textwidth]{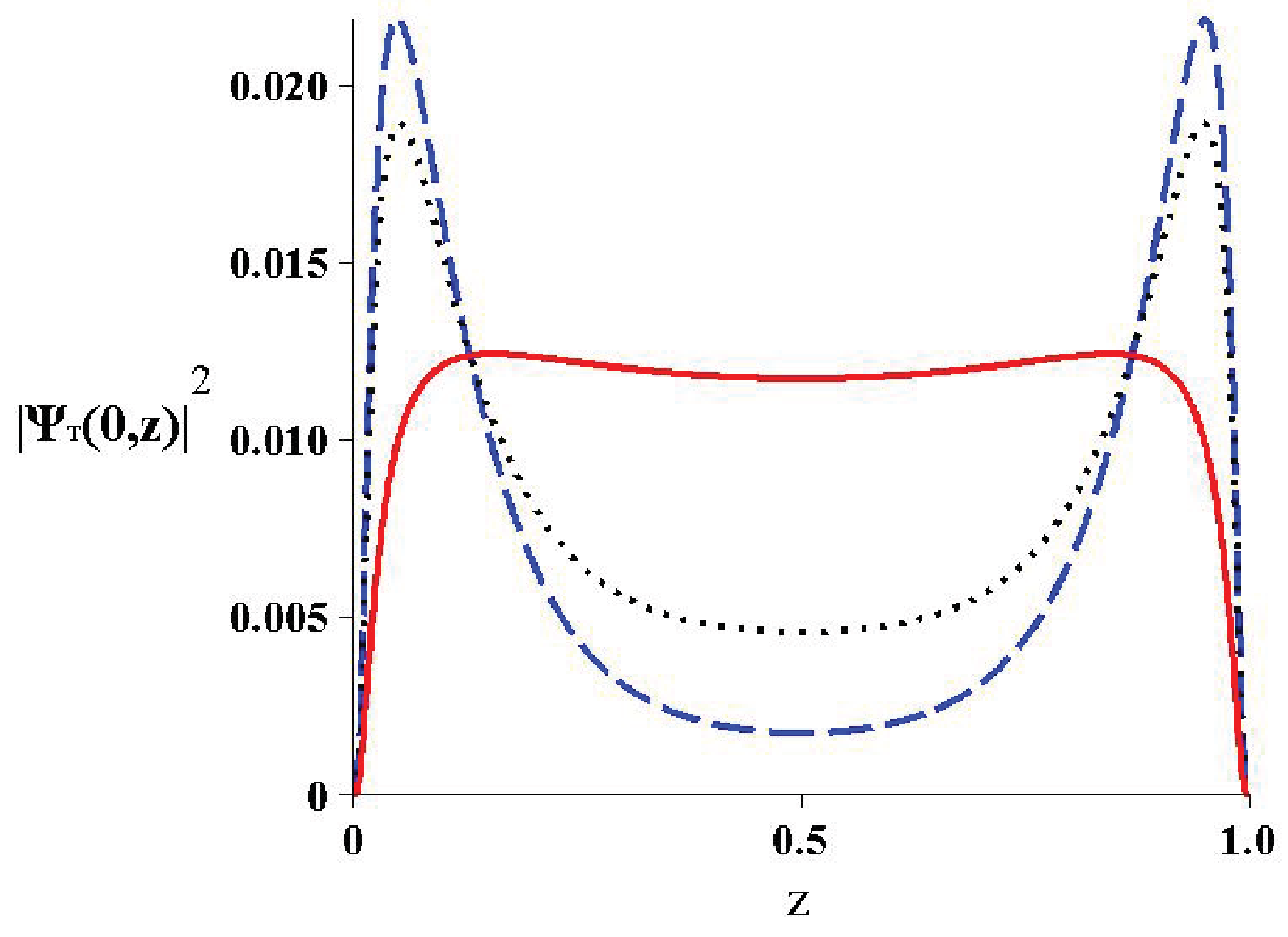}
  \hfill
  \includegraphics[width=0.32\textwidth]{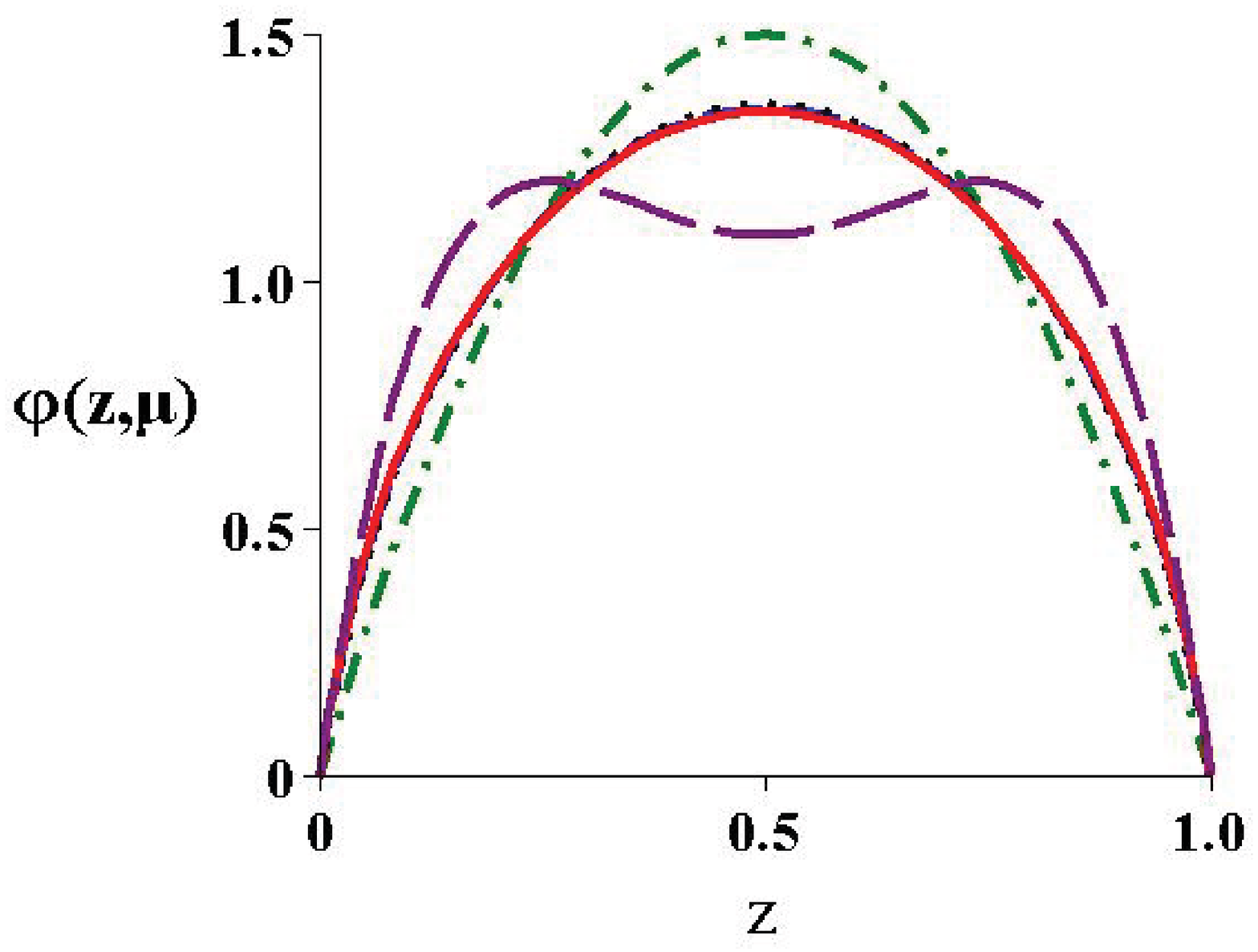}
  \caption{\small 
    {\it Left and center:} The longitudinal and transverse
    light-cone wavefunctions squared at $r=0$. 
    (CGC[$0.74$]: solid; FSSat: dotted; CGC[$0.63$]: dashed.)
    {\it Right:} The extracted leading twist-$2$ DAs at $\mu=1$ GeV 
    compared to the DA of reference  \cite{Ball:1996tb} also at $1$
    GeV (long-dashed) and the asymptotic DA (dot-dashed). 
  }
\label{fig:Psisq_r0}
\label{fig:DA_1GeV}
\end{figure}

The best fits obtained with each dipole model are compared to the HERA data in
figures \ref{fig:H1-xsec} and \ref{fig:HERA-ratio}. 
The corresponding fitted parameters are given in table
\ref{tab:Best-fit-params}.
Note that we achieve a lower $\chi^2/\mbox{d.o.f}=0.89$ with CGC[$0.74$]
than with CGC[$0.63$] and FSSat for which we obtain $\chi^2/\mbox{d.o.f}=0.96$
and $\chi^2/\mbox{d.o.f}=0.97$ respectively. Compared to
the FSSat and CGC[$0.63$] fits, note that no
additional enhancement in the transverse wavefunction is required in the
CGC[$0.74$] fit. Nevertheless the
extracted wavefunction still exhibits enhancement compared to the old BG
wavefunction. The extracted light-cone wavefunctions are shown in figure
\ref{fig:Psisq_r0} left. \\

\noindent{\bf Distribution Amplitudes:}  The leading twist-$2$ Distribution Amplitude (DA) reads \cite{Forshaw:2010py}:

\begin{equation}
\varphi(z,\mu) \sim \left( 1 - \mathrm{e}^{-\mu^2/\Delta(z)^2}\right)
  \mathrm{e}^{-m_f^{2}/\Delta(z)^2} [z(1-z)]^{b_L}~,
\end{equation}
where $\Delta(z)^2 = 8[z(1-z)]^{b_L}/R_L^2$. This leading twist DA is only
sensitive to the longitudinal wavefunction and, as illustrated in figure
\ref{fig:DA_1GeV} right, we expect little variation in the predictions using the
different dipole models. 
To compare with existing theoretical predictions for the DA, we
compute moments:

\begin{equation}
\langle \xi^n \rangle_{\mu} = \int_0^1 \d z \; \xi^n \varphi(z,\mu)~. 
\end{equation}
where by convention~\cite{Forshaw:2010py} 
$
\int_0^1 \d z ~\varphi(z,\mu) = 1
$.
In reference~\cite{Forshaw:2010py}, we noted that our DA is very slowly varying
with
$\mu$ for $\mu > 1$~GeV, i.e  our parameterization neglects the
perturbatively known
$\mu$-dependence of the DA. This statement remains true
if we use the CGC[$0.63$] or CGC[$0.74$] instead of the FSSat model. 

Our results are compared with the existing predictions in table
\ref{tab:moments-mu}. The moments
obtained with our best fit, i.e with the CGC[$0.74$] model, are very similar to
those obtained with FSSat model or the CGC[$0.63$]. In all cases, the results
are in very good agreement with expectations based on QCD sum rules and the
lattice.
Finally, in figure \ref{fig:DA_1GeV} right, we compare our DAs with that
predicted by Ball and Braun~\cite{Ball:1996tb}, at a scale $\mu=1$ GeV. 
The agreement is reasonable given that in reference~\cite{Ball:1996tb}, the
expansion
in Gegenbauer polynomials is truncated at low order, which is presumably
responsible for the local minimum at $z=1/2$. Certainly, all $4$ distributions are broader than the asymptotic prediction $\sim 6z(1-z)$. \\

\begin{table}[bt]
\centering
\textbf{Moments of the leading twist DA at the scale $\mu$}
\small
\[
\begin{array} 
[c]{|c|c|c|c|c|c|c|c|c}\hline
\mbox{Reference} & \mbox{Approach} & \mbox{Scale}~\mu &\langle \xi^2
\rangle_{\mu}&\langle \xi^4 \rangle_{\mu}&\langle \xi^6 \rangle_{\mu}&\langle
\xi^8 \rangle_{\mu}&\langle \xi^{10} \rangle_{\mu} \\ \hline
\mbox{(This paper)} & \mbox{CGC[$0.74$] fit}&\sim 1~\mbox{GeV} &0.227 &
0.105& 0.062 &0.041 &0.029\\ \hline
\mbox{(This paper)} & \mbox{CGC[$0.63$] fit}&\sim 1~\mbox{GeV} &0.229 &
0.107& 0.063 &0.042 &0.030 \\ \hline
\mbox{\cite{Forshaw:2010py}} & \mbox{FSSat fit}&\sim 1~\mbox{GeV} &0.227&
0.105&0.062&0.041&0.029\\ \hline
\mbox{(This paper)} & \mbox{Old BG prediction}&\sim 1~\mbox{GeV} &0.181&
0.071&0.036&0.021&0.014\\ \hline
\mbox{\cite{Bakulev:1998pf}}&
\mbox{GenSR}&1~\mbox{GeV}&0.227(7)&0.095(5)&0.051(4)&0.030(2)&0.020(5) \\ \hline
\mbox{\cite{Chernyak:1983ej}}& \mbox{SR}&1~\mbox{GeV} &0.26&0.15 & & &  \\
\hline
\mbox{\cite{Ball:1996tb}} & \mbox{SR}&1~\mbox{GeV} &0.26(4)& & & &   \\ \hline
\mbox{\cite{Ball:2007zt}}&\mbox{SR}&1~\mbox{GeV} &0.254& & & &  \\ \hline
\mbox{\cite{Ball:2004ye}}&\mbox{SR}&1~\mbox{GeV} &0.23\pm^{0.03}_{0.02}&
0.11\pm_{0.02}^{0.03}& & &  \\ \hline
\mbox{\cite{Boyle:2008nj}}&\mbox{Lattice} &2~\mbox{GeV} &0.24(4)& & & &  \\
\hline 
 & 6z(1-z)
&\infty&0.2&0.086&0.048&0.030&0.021 \\ \hline
\end{array}
\]
\caption {Our extracted values for  $\langle \xi^n \rangle_{\mu}$, compared to
predictions based on the QCD sum rules (SR),
Generalised QCD Sum Rules (GenSR) or lattice QCD.}
\label{tab:moments-mu}
\end{table}


\noindent{\bf Conclusions:}  We have used the current HERA data on diffractive $\rho$ production to extract
information on the $\rho$ light-cone wavefunction. We find that the
corresponding leading twist-$2$ DA is broader than the asymptotic shape
and agrees very well with the expectations of QCD sum
rules and the lattice. We also find that the data
prefer a transverse wavefunction with end-point enhancement although the degree
of such an enhancement is model-dependent. 


\ \\ \noindent{\it Acknowledgments:}
We thank H.~Kowalski and C.~Marquet for useful discussions.
R.S. also thanks the organisers for their invitation and for
making this workshop most enjoyable.



\section{Nuclear effects across the $x$-$Q^2$ plane: quarks and gluons}

\subsubsection{Introduction}
\label{sec:eA_across_xQ_intro}

\hspace{\parindent}\parbox{0.92\textwidth}{\slshape 
  Rodolfo Sassot, Marco Stratmann, Pia Zurita 
}
\index{Sassot, Rodolfo}
\index{Stratmann, Marco}
\index{Zurita, Pia}

In spite of the remarkable phenomenological success of QCD as the
theory of strong interactions, a detailed understanding of the role of
quark and gluon degrees of freedom in nuclear matter is still lacking
and poses great challenges for the theory. 
Ever since the discovery that quark and gluons in bound nucleons exhibit 
momentum distributions different from those measured in
free or loosely bound nucleons~\cite{Aubert:1983xm}, the precise determination of nuclear parton distribution
functions (nPDF) has attracted growing attention, driving both
increasingly accurate and comprehensive nuclear structure functions
measurements \cite{Arneodo:1992wf} and a more refined theoretical
understanding of the underlying physics. 

The precise knowledge of nPDFs is not only required for a deeper
understanding of the mechanisms associated with nuclear binding from a
QCD improved parton model perspective, but is also a crucial input for
the theoretical interpretation and analyses of a wide variety of
ongoing and future high energy physics experiments, such as, for
instance, heavy ion collisions at BNL-RHIC \cite{Adler:2003kg},
proton-nucleus collisions to be performed at the CERN-LHC
\cite{Accardi:2004be}, or neutrino-nucleus interactions in long
baseline neutrino experiments \cite{Paschos:2001np}. Consequently, the
kinematic range and the accuracy at which nPDFs are known has evolved
into a key issue in many areas of hadronic and particle physics. 

The standard description of DIS processes off nuclear targets is
customarily done in terms of the hard scale $Q$ set by the virtuality
of the exchanged photon and a scaling variable $x_A\equiv
Q^2/(2p_A\cdot q)$, analogous to the Bjorken variable used in DIS off nucleons. 
Here, $p_A$ is the target nucleus momentum, and, consequently, $x_A$
is kinematically restricted to $0<x_A<1$, just like the standard
Bjorken variable. Alternatively, one can define another scaling
variable $x_B\equiv A\,x_A$, where $A$ is the number of the nucleons
in the nucleus.  Under the assumption that the nucleus momentum $p_A$
is evenly distributed between the nucleons $p_N=p_A/A$, this variable
resembles the Bjorken variable corresponding to the scattering off
free nucleons,  $x_B\equiv Q^2/(2p_N\cdot q)$.  
However, in the context of nuclear scattering, it spans the interval
$0<x_B<A$ by definition, reflecting the fact that a parton may in
principle carry more than the average nucleon momentum. 

In a naive picture, parton distributions in a nucleus are
simply given by the incoherent sum of the parton
distributions in the $Z$ protons and $(A-Z)$ neutrons that constitute
the nucleus. In that case, the ratios between the structure functions
or cross sections of two iso-scalar nuclei (with the same proportion of
protons and neutrons, such as carbon and deuteron) should be just
proportional to the ratio of their respective number of nucleons (or
to unity if we normalize the structure functions by the number of
nucleons $A$). 

\begin{figure}
  \centering
  \parbox{0.45\linewidth}{
    \includegraphics[width=0.9\linewidth]{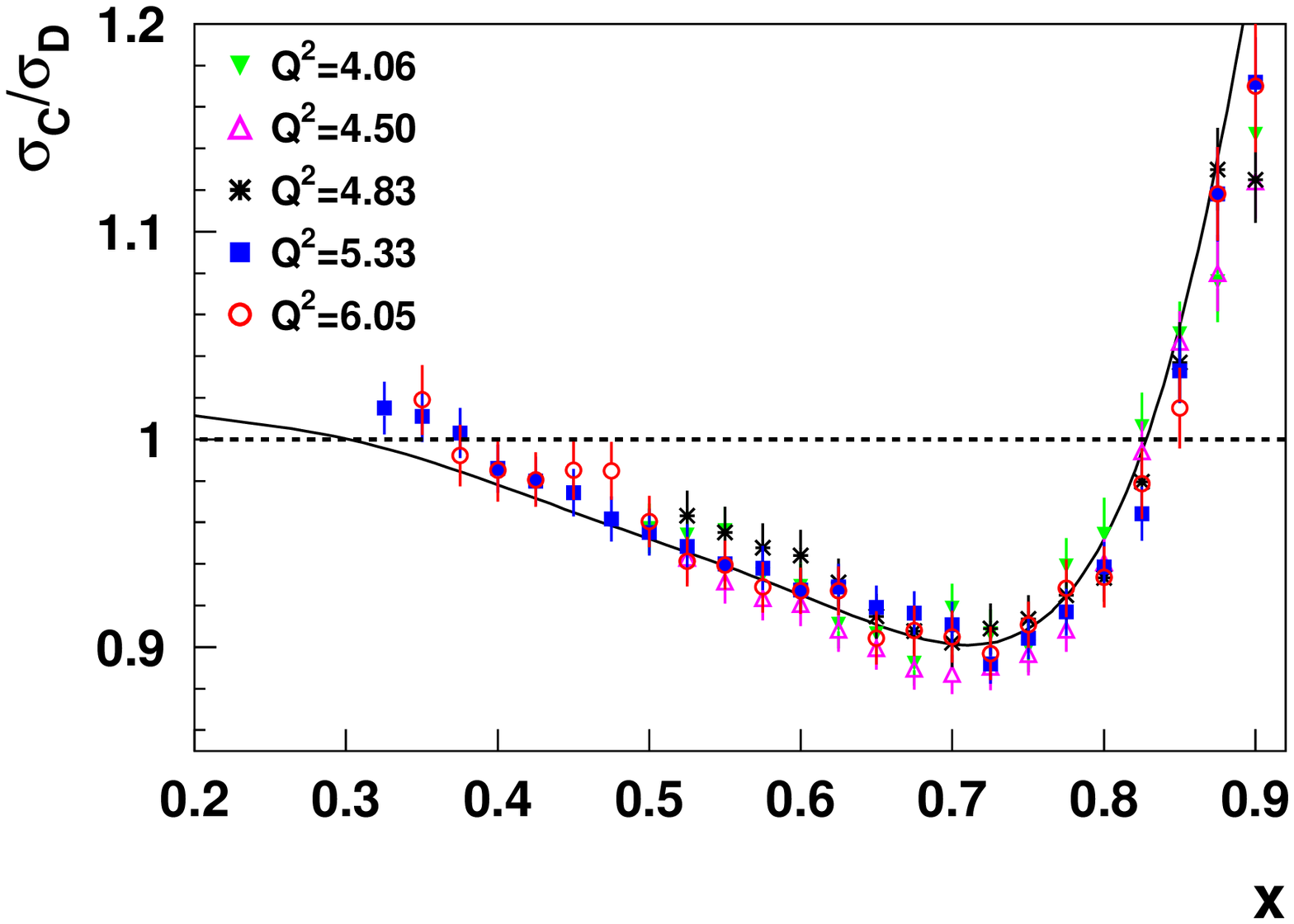}
  }
  \parbox{0.45\linewidth}{
    \includegraphics[width=0.9\linewidth]{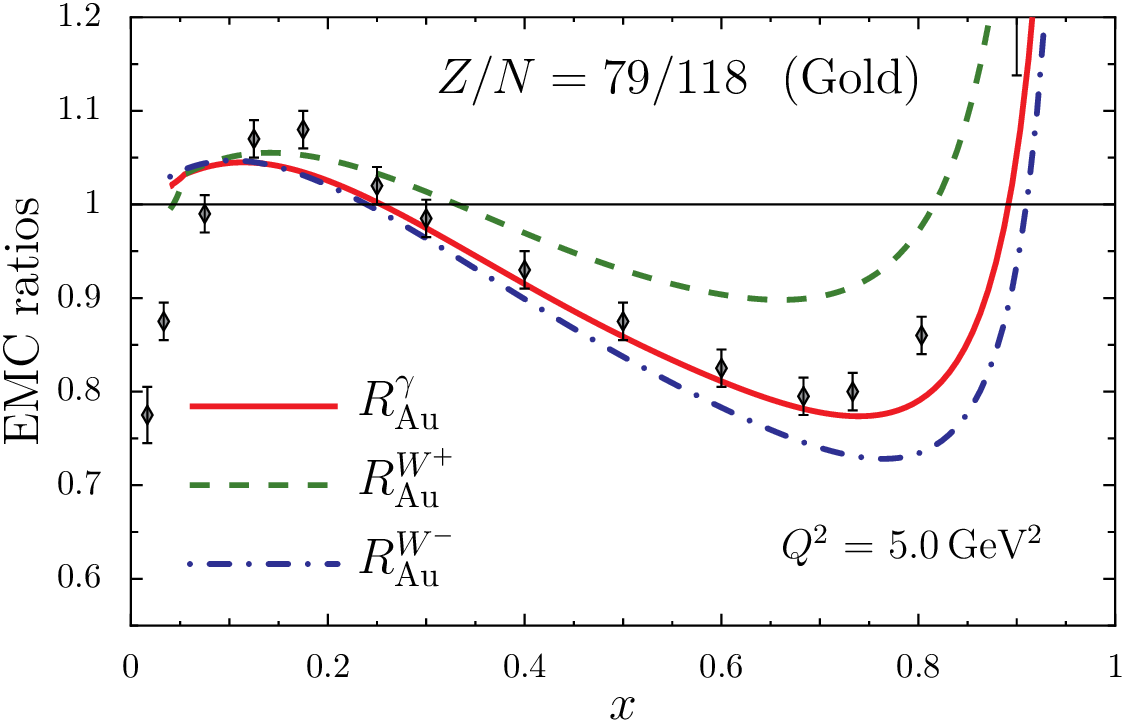}
  }
  \caption{\small 
    {\it Left:} 
    EMC effect and Fermi motion as measured at JLab \cite{Seely:2009gt}.
    {\it Right:} 
    The experimental results of Ref.~\cite{Gomez:1993ri} are
    illustrated for a gold target.  The solid line is the result from
    Ref.~\cite{Cloet:2009qs} for the usual EMC effect and the dashed
    and dashed--dotted lines are respectively the EMC effects in the
    $F_2^{W^+}$ and $F_2^{W^-}$ structure functions. In both panels,
    $x\equiv x_B$ as defined in the text.
  }
\label{fig:EMC} 
\label{fig:emc_gomez}
\end{figure}

If we take into account Fermi motion effects, one would expect that in
the larger nuclei, the cross section extends up to larger $x_B$, so
the rates should typically grow to larger than unity at high
$x_B$. What the EMC experiment found was that in addition to this
motion effect, there was a significant and quite unexpected drop in
the rates between approximately $x_B \approx 0.3$ and $x_B \approx
0.7$. In fig, \ref{fig:EMC}, we show a precise measurement that illustrates both effects. which was recently performed at JLab \cite{Seely:2009gt} 
Later on, it was found that the situation was even worse for the naive
picture outlined above, because at lower $x_B$ values, the rates
showed non-trivial patterns of suppression and
enhancement. These effects are called shadowing and anti-shadowing,
respectively. The phenomenon has been measured at different $Q^2$ and
persists at higher $Q^2$ but with a dependence specific for each $x_B$ region. 

After more than 30 years of experimental and theoretical
studies, a standard picture of nuclear modifications of structure
functions and parton densities has not yet emerged. This is a clear
target for detailed studies at the EIC, which have a large potential
to qualitatively improve the current situation.

\subsubsection{The EMC effect at an EIC}

\hspace{\parindent}\parbox{0.92\textwidth}{\slshape 
  Ian C. Clo\"et
}
\index{Clo\"et, Ian C.}

\vspace{\baselineskip}

The EMC effect has an immediate parton model interpretation, which is
that the valence 
quarks in nuclei carry a smaller momentum fraction than the valence quarks in a 
free nucleon. There have been numerous attempts to explain the EMC
effect, for example 
nuclear structure~\cite{Bickerstaff:1989ch}, nuclear pion
enhancement~\cite{Berger:1984na}, dynamical rescaling and
inter-nucleon color
conductivity~\cite{Close:1983tn,Nachtmann:1983py,Pirner:2010fw}, 
point like configurations~\cite{Frankfurt:1985cv} and the medium modifications to the bound 
nucleons~\cite{Miller:2001tg,Smith:2002ci,Cloet:2005rt,Cloet:2006bq}. 
However, after more than a quarter of
a century since the original EMC experiment, there is still no universally accepted 
explanation of the EMC effect. Therefore, it appears likely that to gain a deeper insight
into the origins of the EMC effect we require new experimental information that is not
accessed in traditional DIS. 

An electron ion collier (EIC) provides excellent opportunities to access different aspects
of the EMC effect, which are not as accessible with traditional fixed target experiments.
A standout example is $W$--production via the DIS processes 
\begin{align*}
\ell^- + A &\longrightarrow W^- + \nu_\ell + A \longrightarrow \nu_\ell + X,  \\
\ell^+ + A &\longrightarrow W^+ + \bar{\nu}_\ell + A \longrightarrow \bar{\nu}_\ell + X.   
\end{align*}
The extraction of the target structure functions from these reactions is possible at an EIC
because of the unique ability to reconstruct the final state and therefore avoid the need
to directly determine the outgoing momentum of the neutrino or anti--neutrino. The 
parton model expressions for
the $F_2$ structure functions that characterize these processes are~\cite{Amsler:2008zzb}
\begin{align}
F_{2A}^{W^+}(x) &= \bar{u}_A(x) + d_A(x) + s_A(x) + \bar{c}_A(x), \\
F_{2A}^{W^-}(x) &= u_A(x) + \bar{d}_A(x) + \bar{s}_A(x) + c_A(x),
\end{align}
where $u_A(x),\,\bar{u}_A,\,\ldots$ are the various quark distributions of the target. 
In the valence quark 
region these $W^{\pm}$ structure functions are completely dominated by quark distributions
of a single flavour, and hence a measurement of these structure functions provides direct
access to the flavour decomposition of the nuclear parton distributions functions in this
region. The flavour dependence of the EMC effect can then be determined, which will provide
extremely important new information on the nature of this important phenomena.

The EMC effect ratio can be defined as
\begin{align}
R^i = \frac{F_{2A}^i}{Z\,F_{2p}^i + N\,F_{2n}^i}, \quad \text{where} \quad i \in \g,\,W^{\pm},
\end{align}
and $F_{2p}^i$, $F_{2n}^i$, $F_{2A}^i$  are respectively the proton, neutron and
nuclear structure functions. The atomic number
of the nucleus is labelled by $Z$, $N$ is the neutron number. Using the nuclear
quark distribution results from Ref.~\cite{Cloet:2009qs}, we can construct the 
usual EMC effect associated with
the exchange of a virtual photon and also the EMC effect in the $W^{\pm}$ structure
functions. These results are illustrate in Fig.~\ref{fig:emc_gomez} for an Au nucleus.

Therefore, measurements of $F_{2A}^{W^\pm}(x)$ for various nuclei, for example C, Fe,
Au and Pb would provide important new information on the flavour dependence of the 
EMC effect, which in ref.~\cite{Cloet:2009qs} is predicted to be large for nuclei 
like Pb and Au. It is also claimed that a significant part of the NuTeV anomaly
may also be explained by this isovector EMC effect~\cite{Cloet:2009qs}.
Therefore, these measurements present an excellent opportunity for an EIC
and will undoubtedly help us understand the origins of the EMC effect, which
is essential if we are to ever have a QCD based description of nuclei.

\subsubsection{Nuclear  gluons}  
\label{pirner:gluons}

  
\hspace{\parindent}\parbox{0.92\textwidth}{\slshape   
  Hans J. Pirner    
}  
\index{Pirner, Hans J.}  
  
\vspace{\baselineskip}  
  
Historically, the very accurate NMC measurements of DIS  
on Tin and Carbon nuclei has allowed one to extract the gluon distribution  
from the scaling violation in $F_2(A)$. This has been done by Gousset  
and myself~\cite{Gousset:1996xt} for the first time. That analysis  
shows an enhancement of $10 \%$ i.e. antishadowing for $x \approx 0.1$  
and the same amount of shadowing, namely also $10 \%$ at $x \approx  
0.01$. A high experimental accuracy is demanded, therefore   
only a trend could be established. 
The asymptotic calculation of heavy  
charmonium production on nuclei is often proposed as another method to  
extract the nuclear gluon distribution based on the gluon-gluon fusion  
process. As shown in various papers by Kopeliovich this production is  
more complicated, especially for $J/\Psi$, because of initial and  
final state effects.  Measurements of the gluon distribution would  
give an experimental window on the importance of gluonic effects in  
nuclear binding. Very little is known about the role of gauge fields  
in nuclei.

To gain insight on gluons in bound nucleons system,    	  
we have studied an abelian QED model~\cite{Kaluza:1993ps} where the  
nucleon is replaced by an atom and the nucleus by a molecule, {\it
  i.e.}, we  
have analysed the structure function of the photon in the  
$H_2$-molecule and compared it with the structure function in the  
$H$-atom. The electron orbits of the hydrogen atoms in the molecule  
are polarized and modified by the electron exchange interaction  
leading to a suppression of photons at small $x$. At the momentum  
corresponding to the relative distance of the two protons, a small  
antishadowing peak is visible~\cite{Kaluza:1993ps}. In analogy, gluon  
antishadowing in the region $x=0.1$ may indicate the distance $\Delta r  
\approx 2 \fm$ between the centers of the nucleons which act as color  
sources of common gluon fields between nucleons.  A covalent binding  
of quarks may manifest itself as a density dependent lack of long  
range gluons at $x<0.1$ similarily to the deformation of the photon  
cloud in the hydrogen molecule. In addition, in non-Abelian QCD, one  
expects at small $x$ that the gluons from different nucleons overlap  
and merge. Both of these effects have also an interpretation in the  
nuclear rest frame in terms of the absorption of various partonic  
components in the wave function of the photon.

During the last ten years, the available data have been used to extract nuclear parton
distributions and evolve them to high $Q^2$, as reviewed below.  
In a careful analysis one  
has to respect the large errors of the starting distribution at low  
$Q^2$ for the nuclear gluon distributions and also the larger  
$x$ region has to be included correctly - at least the fact that the  
nuclear gluon distribution~\cite{Merabet:1993du} is more strongly  
affected by Fermi-motion of the nucleons than the quark distribution,  
since it has a stronger decrease at large $x$. Enhancement of the  
nuclear gluon distribution sets in already at $x=0.5$ which may be of  
importance for charmonium production at JLab~\cite{Merabet:1993du}. 


\subsubsection{Global fits of nuclear PDFs: current  status}
\label{sec:Sassot-nPDFstatus}

\hspace{\parindent}\parbox{0.92\textwidth}{\slshape 
  Rodolfo Sassot, Marco Stratmann, Pia Zurita 
}
\index{Sassot, Rodolfo}
\index{Stratmann, Marco}
\index{Zurita, Pia}

From the point of view of perturbative QCD (pQCD), the extraction of nPDF can be 
performed in close analogy to what is routinely done for free nucleons: 
they are considered as non-perturbative inputs, to be inferred from data, whose relation to the
measured observables and their energy scale dependence can be computed order by order 
in perturbation theory. 
Although one cannot discard potentially larger higher-twist or 
power corrections than in the case of free nucleons, or non-linear 
nuclear recombination effects, standard QCD factorization and universality of nPDFs 
are found to hold to a very good approximation in the
kinematical range covered by present experiments.   

At variance with PDFs for free nucleons, which, driven by the demand for 
increasingly precise predictions of the standard model, obtained an impressive degree of
accuracy and refinement, extractions of nPDFs are done at a
considerably lower level of sophistication. Not only the number,
variety, kinematical coverage, and precision of nuclear data are much
more limited, but the precise parameterization of nPDFs is also much
more involved as it depends not only on the energy scale $Q$ and the
parton's momentum fraction $x$, but also on the size of the nucleus
characterized by the atomic number $A$. In the following, we present a
brief summary of the current status of nPDFs and outline limitations
in the analyses imposed by the data available so far. 

Thanks to its variable beam energy, the possibility to run with
different nuclei, and the envisioned large luminosities, an EIC will
add invaluable novel information on 
nPDFs from studies of the inclusive structure functions $F_{2,L}$. It will
extend the kinematic range toward lower values of $x$ as well as
higher scales of $Q$, allowing precise determination of the gluon distribution from scaling
violations of $F_2$, permit the flavour separation of the quark sea and
the study the onset of non-linear saturation effects at small $x$ (see 
Section~\ref{sec:AGR-DGLAPdev}), which eventually spoil the factorized
pQCD approach.\\ 


\noindent {\bf Status of Nuclear Parton Densities.}
From the point of view of pQCD and a factorized approach, the description 
of nuclear DIS can be viewed as follows. In a DIS processes off a
nuclear target, we also have a hard momentum scale $Q$ that allows one
to factorize the measured cross section into a point-like partonic
cross section and non-perturbative parton densities, characteristic of
partons seen in that nucleus. These ``effective'' parton densities
factorize and encode all the non-perturbative information, including
the details about the nuclear structure, and every mechanism,
interaction, or effect we can imagine. Since the hard partonic cross
sections are just the same as those appearing in the factorization for
free nucleons, the nuclear parton densities will evolve with scale in
the same way as ordinary parton densities. For similar reasons, the
approach could be extended to higher orders. 
What is clearly not obvious within this line of reasoning is why, or how, 
one could split the non-perturbative effective nuclear parton density
into a piece containing only the effects related to quarks and gluons
belonging to single nucleon from those related to the nucleons bound
in the nuclei. No field theoretical tool gives us a precise
prescription of how to achieve this.  It is important to keep in mind
that even in lepton-nucleon scattering standard PDFs are not just
naive probability densities; they are non-trivial, though perfectly
well defined, objects which depend on the choice of factorization scheme
and contain other ingredients such as gauge links. 

What can be done, of course, is to follow a program of global QCD analyses 
completely analogous to the one carried out for PDFs, i.e.,
to extract the nPDFs and their $A$ dependence from data.
In doing so one can explore if the basic properties 
of factorization and universality still hold in a nuclear environment. 
The first QCD extractions of nPDFs defined in this way were done at the end of the 90's 
by two pioneering groups who performed leading order (LO) analyses of nuclear DIS data (EKS98, HKM01) 
\cite{Eskola:1998iy,Eskola:1998df,Hirai:2001np}.

\begin{figure}
  \centering
  \includegraphics[height=9cm]{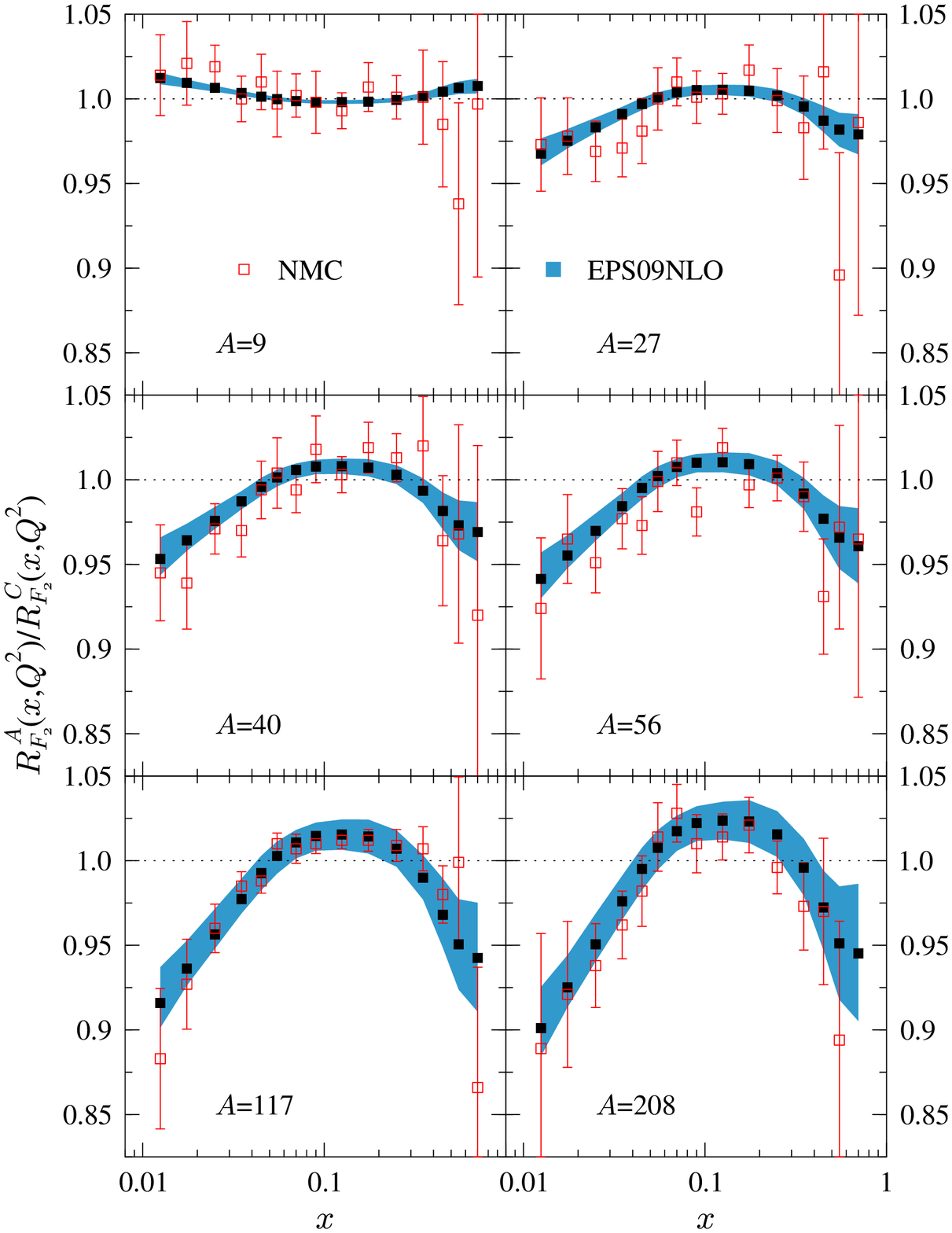}
  \includegraphics[height=9cm]{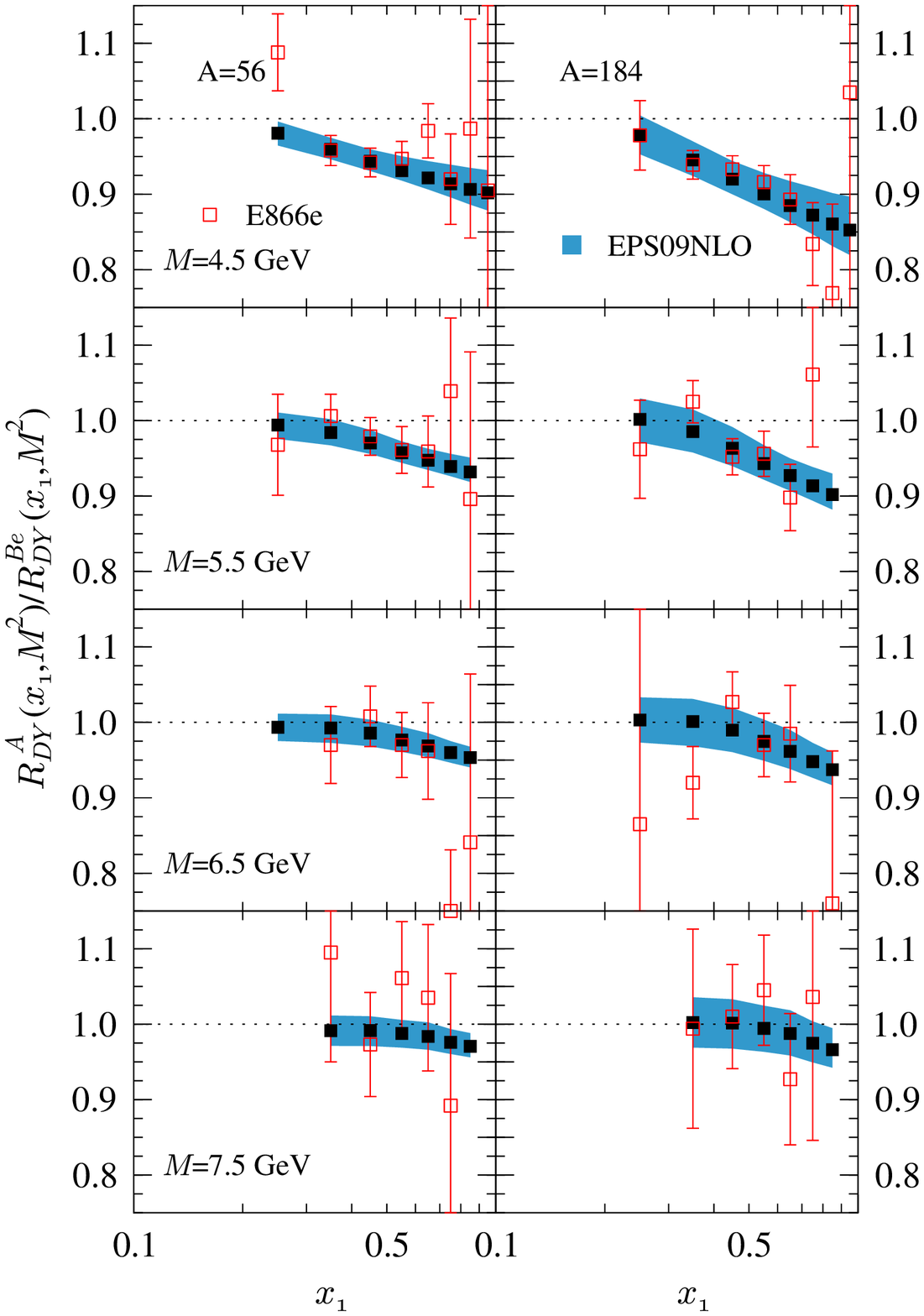}
  \caption{\small \label{fig:dis&DY} Quality of the fit to nuclear DIS and Drell-Yan data;
    taken from Ref.~\cite{Eskola:2009uj}.}
\end{figure}

When introducing nPDFs, the usual approach was to propose a very 
simple relation between the parton distribution of a proton bound in the nucleus, 
$f_i^A$, and those for free protons $f_i$,
\begin{equation}
\label{eq:f1}
f_i^A(x_B,Q_0^2) = R_i(x_B,Q_0^2,A,Z)\, f_i(x_B,Q_0^2),
\end{equation}
in terms of a multiplicative nuclear correction factor $R_i(x_B,Q^2,A,Z)$, specific
to a given nucleus $(A,Z)$, parton flavor $i$, and initial energy scale 
$Q^2_0$. Such a description is convenient since the ratio $R_i(x_B,Q^2,A,Z)$ 
compares directly the parton densities with and without nuclear effects, and
is closely related to the most common nuclear DIS observables, which are the 
ratios between the nuclear and deuterium structure functions. 
In Ref.~\cite{deFlorian:2003qf} the alternative to 
relate nPDFs to standard PDFs by means of a convolution was introduced. 
The convolution approach implements straightforwardly 
effects related to rescalings or shifts in the parton's momentum fraction due to interactions 
with the nuclear medium.  
In addition, convolution integrals are the most natural language for parton dynamics 
beyond the LO and allow for the straightforward application of the Mellin transform techniques, 
convenient for a numerical fast and accurate computation of the scale dependence of PDFs
and relevant cross section estimates.

Following the developments for standard PDFs, nPDFs analyses subsequently 
incorporated various improvements such as a consistent next-to-leading
order (NLO) framework (nDS) 
\cite{deFlorian:2003qf}, a thorough uncertainty analysis (HKN04 LO) \cite{Hirai:2004wq}, 
and periodical updates of the different sets in order to incorporate new data (EKPS07 LO) 
\cite{Eskola:2007my}, up to NLO accuracy  (HKN07 NLO, EPS09 NLO) 
\cite{Hirai:2007sx,Eskola:2009uj}. In the latest sets
\cite{Eskola:2009uj,Schienbein:2009kk} 
particular attention has been paid to the possible impact of d+Au collision data
from RHIC and neutrino DIS data on the global fits.
A typical comparison to nuclear DIS and Drell-Yan data is shown in Fig.~\ref{fig:dis&DY}.

It is worth noticing that the inclusion of $d+Au$ data in nPDF fits,
although neglecting any nuclear modifications in the hadronization
process,  leads to significantly larger gluon shadowing and  
antishadowing, as has been pointed out in \cite{Eskola:2009uj}. The
same data, however, can be described with much more moderate nuclear
gluon PDFs, but including medium modified nFFs~\cite{Sassot:2009sh},
see Section~\ref{sec:Sassot-FF}.
\begin{figure}[htb]
  \centering
  \includegraphics[width=0.95\textwidth,bb= 0 25 512 250,clip=true]
                  {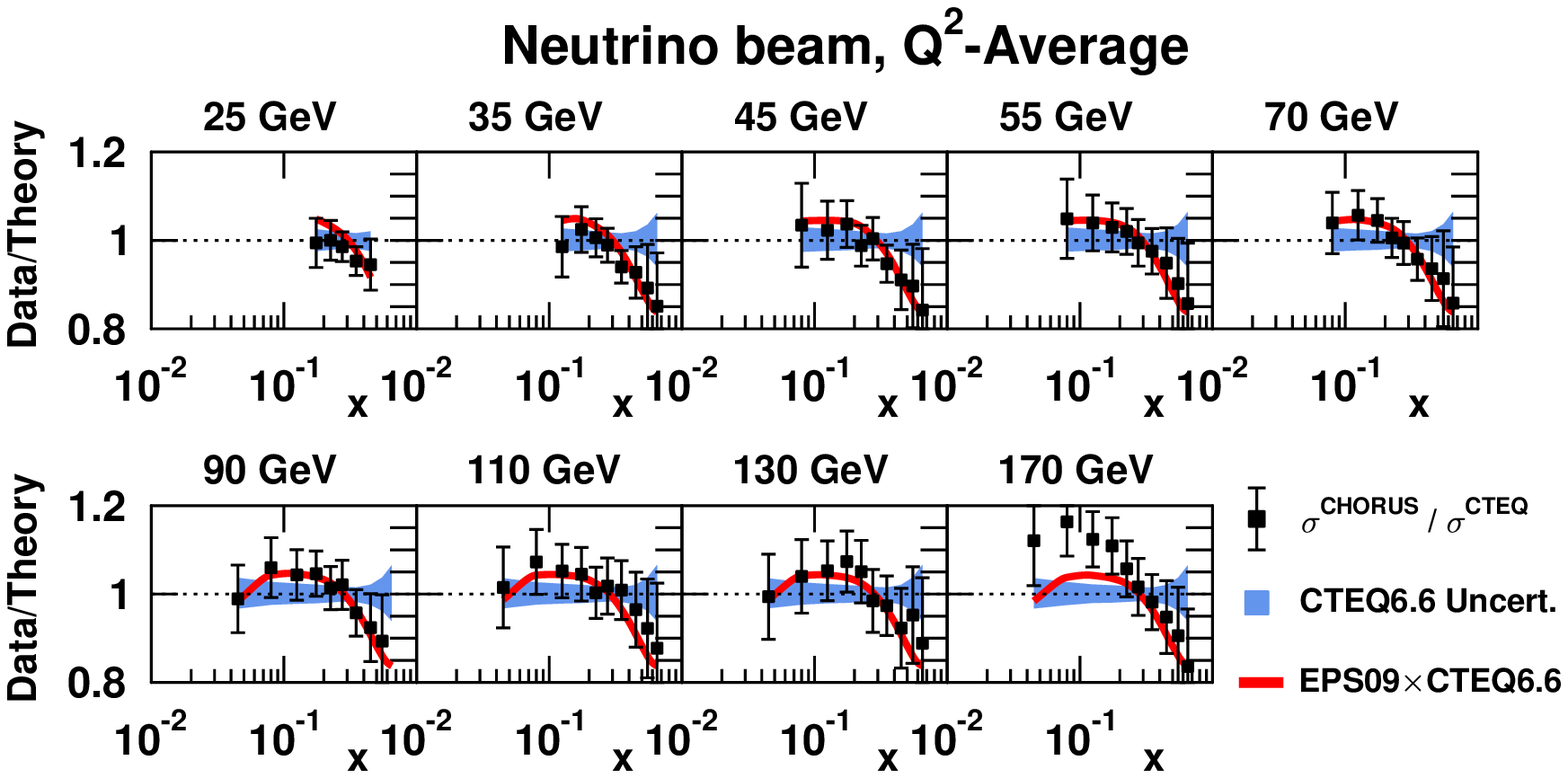}
  \caption{\small \label{fig:neutrino} Comparison to neutrino data; taken from Ref.~\cite{Paukkunen:2010qj}. }
\end{figure}

\begin{figure}[htb]
  \centering
  \includegraphics[width=0.6\textwidth,bb=0 10 625 435,clip=true]{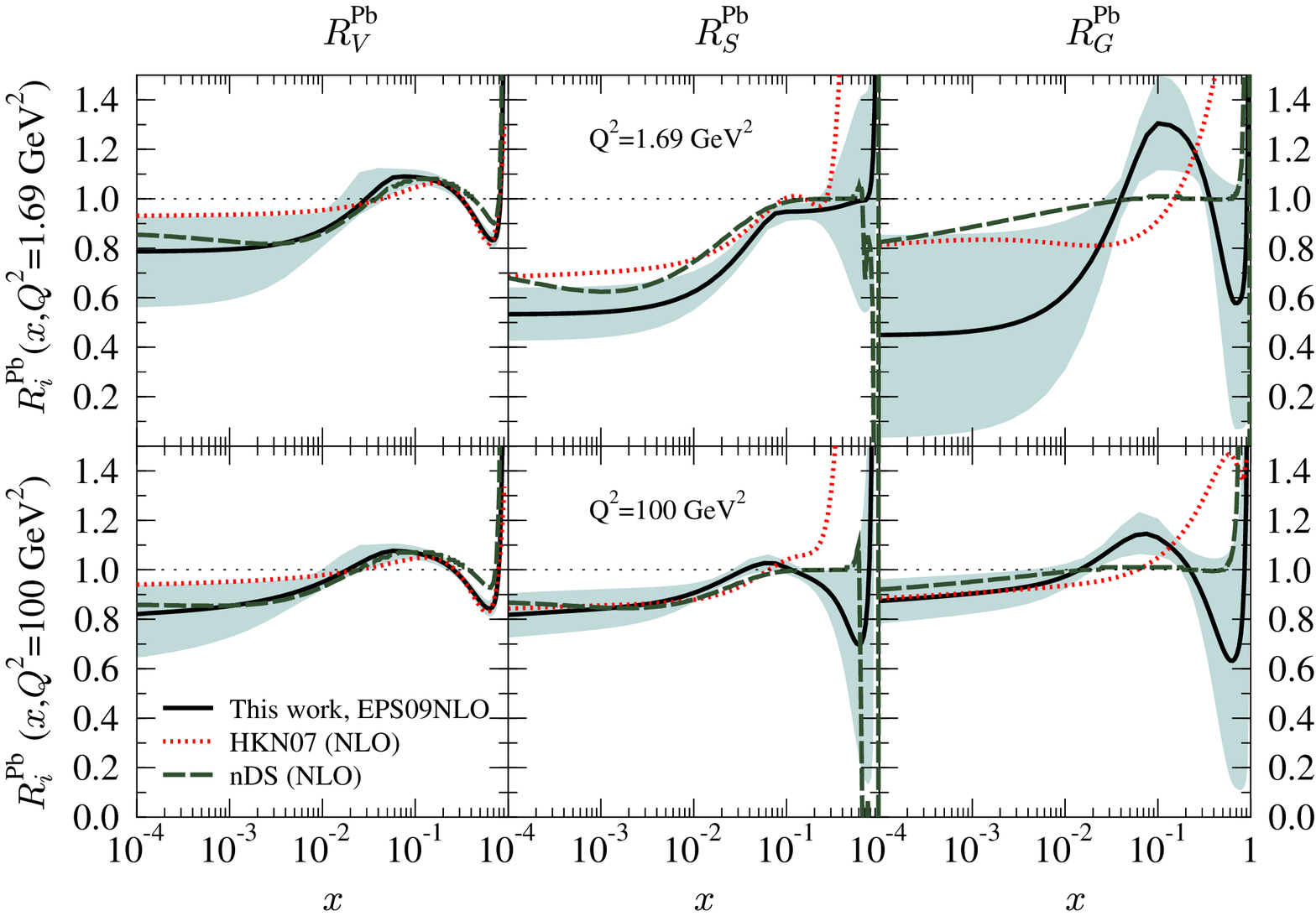}
  \caption{\small \label{fig:nPDFs} Comparison between different sets of nPDFs from \cite{Paukkunen:2010qi}. }
\end{figure}

Regarding the impact of neutrino data, Schienbein et al. \cite{Schienbein:2009kk} claim that 
within their analysis it is not
possible to reproduce simultaneously the trend of the data coming from electromagnetic nuclear DIS
and some observables derived from neutrino DIS measurements.  Of course, these 
conclusions are reached under rather stringent assumptions such as a very specific 
parameterization for nuclear effects and those implicit in the derivation of the neutrino DIS rates
to deuteron, which have not been actually measured yet. On the other hand, using the EPS09 
analysis and neutrino DIS data, Paukkunen and Salgado \cite{Paukkunen:2010qj} find no traces 
of such tension, besides some energy dependent fluctuations in the NuTeV data.
A typical comparison to neutrino data is given in Fig.~\ref{fig:neutrino}.

Different recent extractions of nPDFs are shown in Fig.~\ref{fig:nPDFs}.
A general shortcoming of all present fits is that independent nuclear
modification factors can be determined only for gluons, valence, and sea quarks 
without distinguishing different quark flavors.
Also, present fixed-target data do not constrain nPDFs below about $x_B\simeq 0.01$,
and the curves shown at smaller values of $x_B$ are mere extrapolations.
Uncertainties on nPDFs are large, in particular for the nuclear gluon distribution.
There is clearly a need for more precise data covering also the small
$x_B$ region. \\


\noindent {\bf Conclusions.} 
In the last few years, our knowledge of the way that both parton
densities and fragmentation probabilities are modified in a nuclear
environment have improved significantly. Different studies performed
so far have clearly demonstrated that pQCD factorization and
universality are extremely good approximations within the precision
and kinematic range of the available data. Although the uncertainties
and differences between different QCD global analysis are still large,
the availability of more data for different processes, and their
subsequent inclusion in the analyses will certainly help to reduce
them further. 
Ultimately, the EIC will be required for precise quantitative studies and 
to explore the small $x_B$ regime where novel non-linear recombination
and saturation phenomena are expected. 
A preliminary study of the capabilities of the EIC in these respects
has been presented in Section~\ref{sec:AGR-DGLAPdev}: the EIC has the
potential to determine gluon and quark nPDFs to a precision comparable
to the nucleon PDFs down to $x\sim10^{-3}$, and indeed to detect
saturation effects as a deviation from DGLAP linear evolution.

\subsubsection{HKN nuclear parton distribution functions}
\label{sec:hkn-nPDFs}

\hspace{\parindent}\parbox{0.92\textwidth}{\slshape 
  Shunzo Kumano 
}
\index{Kumano, Shunzo}

\vspace{\baselineskip}



The Hirai, Kumano and Nagai (HKN) nuclear PDFs
\cite{Hirai:2004wq,Hirai:2007sx}
are determined 
by a global analysis of world data on charged-lepton DIS and
Drell-Yan processes with nuclear targets. 
Since the PDFs of the nucleon are relatively well determined,
it is appropriate to parametrize the nPDFs 
at the initial $Q^2_0=1$ GeV$^2$ using Eq.~\eqref{eq:f1} and  
\begin{eqnarray}
 R_i(x_B,Q_0^2,A,Z)=1+\Bigl( 1-\frac{1}{A^\alpha} \Bigr )
  \frac{a_i+b_ix +c_i x^2 +d_i x^3}{(1-x)^{\beta_i}},
\end{eqnarray}

\begin{wrapfigure}{r}{0.4\textwidth}
   \centering
       \includegraphics[width=0.35\textwidth]{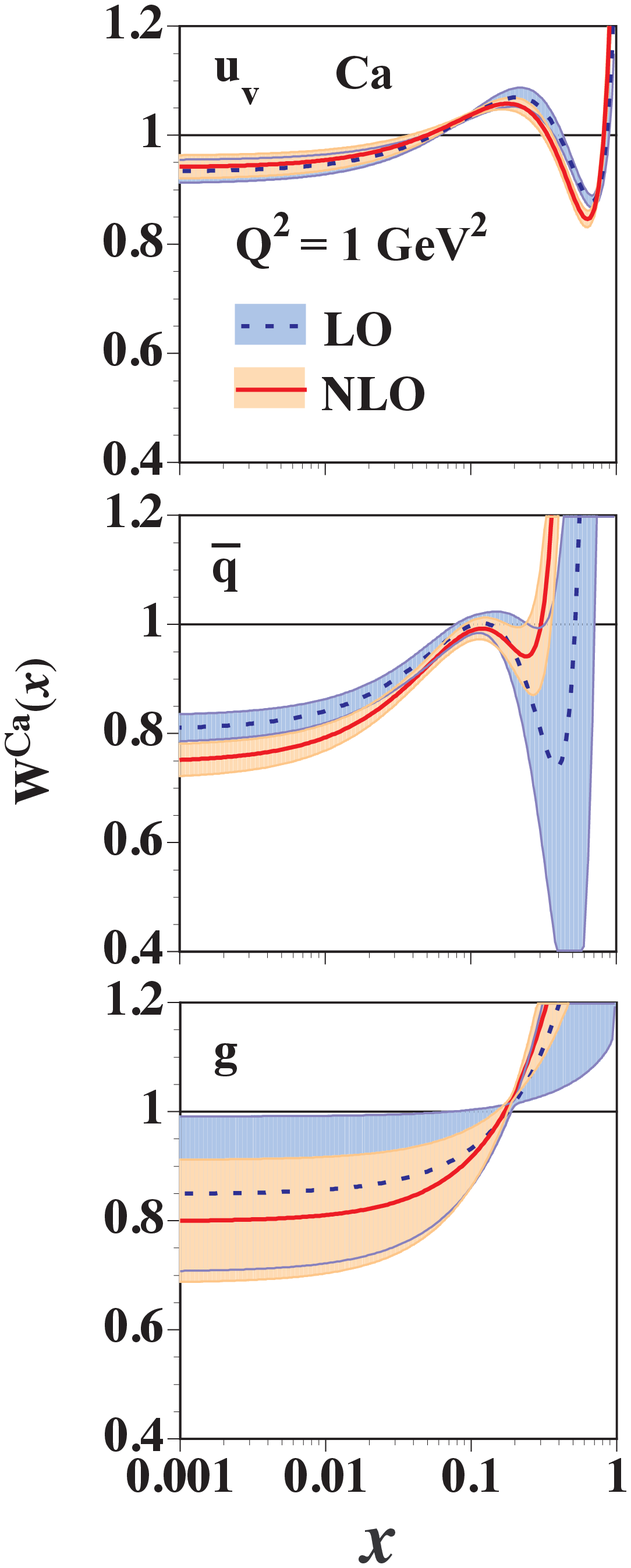} 
   \caption{\small \label{fig:w-ca} Determined nuclear modifications in Ca
   \cite{Hirai:2007sx}.}
\end{wrapfigure}


The determined $u_v$, $\bar q$, and $g$ nPDFs from the HKN07 analysis
\cite{Hirai:2007sx} are shown for the
calcium nucleus in Fig. \ref{fig:w-ca} at $Q^2$=1 GeV$^2$. LO and NLO results
are shown with uncertainty bands, showing that nPDFs are determined
more accurately at NLO. We obtain $\chi_{min}^2/$d.o.f.=1.35 and
1.21 for the LO and NLO fits, respectively.

The valence-quark modifications are well
determined because of accurate measurements on the $F_2$ ratios
at medium $x$. The small-$x$ region is fixed by the baryon-number
and charge conservations together with the modifications in 
the medium- and large-$x$ regions.
The antiquark modifications are also determined well at small $x$
due to measurements on $F_2$ shadowing, and they are also 
fixed at $x \sim 0.1$ because of Fermilab Drell-Yan measurements.
However, the region at $x>0.2$ is not determined at all. 
The E906/SeaQuest collaboration is currently measuring this medium-$x$
region, and there is also a possibility to measure
this region with an experiment at J-PARC. 
In the near future, the uncertainty bands should be significantly
reduced for the antiquark. 

The gluon distribution has the largest uncertainties since
it contributes to the $F_2$ and Drell-Yan ratios only
as higher-order effects, and
the $Q^2$ dependence of $F_2^A/F_2^{A'}$
is not measured accurately on nuclear targets, which makes it difficult to
pin down the gluon modifications measured by scaling violations of
$F_2$. The small-$x$ nPDFs are 
dominated by huge gluon distributions, so that it is essential
to determine them accurately for new discoveries by high-energy
heavy-ion experiments. Therefore, it is important to measure the $Q^2$
dependence of $F_2^A/F_2^{A'}$ at EIC for determining nuclear gluon
distributions. 

In HKN07, the nPDFs are also investigated for the deuteron.
In obtaining the ``nucleonic'' PDFs, deuteron data are used  
after crude nuclear corrections. Since the current PDFs 
could possibly contain nuclear effects, appropriate nuclear
corrections should be applied in future for excluding
such effects. 
Our codes for calculating the nPDFs and their uncertainties
are available at the web site \cite{HKN:npdfweb}.
The technical details are explained in Refs. 
\cite{Hirai:2004wq,Hirai:2007sx} and within the subroutine. 


\section{Color transparency}

\subsubsection{Color transparency phenomena}
\hspace{\parindent}\parbox{0.92\textwidth}{\slshape 
  B.~Z.~Kopeliovich 
}
\index{Kopeliovich, Boris Z.}

\vspace{\baselineskip}

The nuclear medium is more transparent for colorless hadronic wave
packets than predicted by the Glauber model. One can treat this
phenomenon either in the hadronic basis as a result of Gribov's
inelastic corrections~\cite{Gribov:1968jf}, or in QCD as a result of color
screening~\cite{zkl}, an effect called color transparency (CT). Although the two approaches are
complementary, the latter interpretation is more intuitive and
straightforward. Indeed, a point-like colorless object cannot interact
with external color fields, therefore its cross section vanishes
as $\sigma(r)\propto r^2$ when $r \to 0$~\cite{zkl}.  When a colorless
wave packet propagates through a 
nucleus, the fluctuations with small size have an enhanced survival
probability which leads to a non-exponential attenuation $\propto 1/L$~\cite{zkl}, where $L$ is the path length in nuclear matter.

Diffractive electro-production of vector mesons off nuclei is affected by
shadowing and absorption which are different phenomena. Final state
absorption of the produced meson exists even in the classical
probabilistic approach which relates nuclear suppression to the
survival probability $W(z,b)$ of the vector meson produced at the
point with longitudinal coordinate $z$ and impact parameter $\vec b$,
\begin{equation}
  W(z,b) = exp\Bigg[-\,\sigma_{in}^{VN}\,
    \int\limits_{z}^{\infty}
    dz^\prime\,\rho_A(b,z^\prime)\Bigg]\ ,
\label{kopel-1370}
\end{equation}
where $\rho_A(b,z)$ is the nuclear density and $\sigma_{in}^{VN}$ is
the inelastic $VN$ cross section.
Shadowing, is also known to cause nuclear
suppression. In contrast to final state absorption, it is a pure
quantum-mechanical effect which results from destructive interference of
the amplitudes for which the interaction takes place on different bound
nucleons. It can be interpreted as a competition between the different
nucleons participating in the reaction: since the total probability
cannot exceed one, each participating nucleon diminishes the chances of
others to contribute to the process.
The interplay between absorption and shadowing is controlled by the
two time scales introduced for the case of charmonium in
eq.~(\ref{kopel-1140}). They are defined similarly
for other hadrons.

In the low-energy limit of short $l_c<l_f\ll R_A$ (shorter than the mean
nucleon spacing $\sim 2\fm$) only final state
absorption matters. The ratio of the quasielastic 
$\gamma^*\, A \to V\,X$ and $\gamma^*\, N \to V\,X$ cross sections reads,
\begin{align}
  R_{inc}\Bigr|_{l_c,l_f\ll R_A} &\equiv
  \frac{\sigma_V^{\gamma^*A}}
  {A\,\sigma_V^{\gamma^*N}} =
  \frac{1}{A}
  \,\int d^2b\,
  \int\limits_{-\infty}^{\infty}
  dz\,\rho_A(b,z)\,
  \exp\left[-\sigma^{VN}_{in}
    \int\limits_z^{\infty} dz'\,
    \rho_A(b,z')\right]
  \label{kopel-1380}
\end{align}
In the limit of long $l_c\gg R_A$, it takes a different form; in the Glauber approximation,
\begin{align}
  R_{inc}\Bigr|_{l_c\gg R_A} = 
  \int d^2b\,T_A(b)\,
  \exp\left[-\sigma^{VN}_{in}\,T_A(b)\right]\ ,
  \label{kopel-1400}
\end{align} 
One can see that the $V$ meson attenuates along the whole nucleus
thickness in Eq.~(\ref{kopel-1400}), but only along roughly half of that length
in Eq.~(\ref{kopel-1380}).  
The exact expression beyond VDM which interpolates
between the two regimes (\ref{kopel-1380}) and (\ref{kopel-1400}) can
be found in~\cite{Hufner:1996dr}.

The effects of color transparency lead to deviation from this
expression. These effects, which can be understood as Gribov inelastic
corrections lead to equation~(\ref{kopel-1180}), which
should be used to study the effects of color transparency.    \\


\noindent {\bf Light-cone distribution functions for the photons and vector mesons.} In what follows, we rely on the dipole description and need to know the distribution functions for the photon and vector mesons. To be self-consistent, we should use the same light-cone potential for describing both.
In equation (\ref{kopel-120}) for the Green function, we chose the
real part of the potential of the $\bar q q$ dipole as in
Refs.~\cite{Kopeliovich:1999am,Kopeliovich:2001xj}. 
Solving Eq.~(\ref{kopel-120}) for the Green function with this
potential 
and assuming similar spin structures for the vector mesons and
photons, one can obtain an explicit formula for the vector meson
light-cone wave function ~\cite{Kopeliovich:2001xj}, 
depending on a ``width'' and a ``quark mass'' phenomenological
parameters that were fitted to data in \cite{Nemchik:1996cw}.

\noindent {\bf Cross section on a proton.} 
Now we are in a position to calculate the forward electroproduction diffractive amplitudes, which have the following form,
The forward scattering amplitude $ {\cal M}_{\gamma^{*}N\rightarrow V\,N}^{T,L}(s,Q^{2})\Bigr|_{t=0}$ can be extracted from eq.~(\ref{kopel-1020}) discussed previously. 
 These amplitudes are normalized as ${|{\cal M}^{T,L}|^{2}}=
\left.16\pi\,{d\sigma_{N}^{T,L}/ dt}\right|_{t=0}$. In what follows we calculate the cross sections
$\sigma = \sigma^T + \epsilon\,\sigma^L$ assuming that the photon
polarization is $\epsilon=1$.


For HERA data, the normalization of the cross section and its energy and $Q^2$
dependence are remarkably well reproduced, see
\cite{Kopeliovich:2001xj}.  This is important, since the absolute
normalization is usually much more difficult to reproduce than nuclear
effects, which we switch to in the nest section.

As a cross-check for the choice of the $\rho^{0}$ wave function, we also calculated the total
$\rho^0$-nucleon cross section, which is usually expected to be roughly
similar to the pion-nucleon cross section $\sigma_{tot}^{\pi N}\sim 25\mb$. 
For the dipole cross section, we adopt the KST
parameterization~\cite{Kopeliovich:1999am}, which has been used above,
and is designed to describe low-$Q^2$ data. Then, at $\nu = 100$ GeV,
we obtain $\sigma_{tot}^{\rho N} = 27\mb$ which is quite a reasonable
number. 

\begin{figure}[tb]
  \centering
  \includegraphics[width=0.35\textwidth]{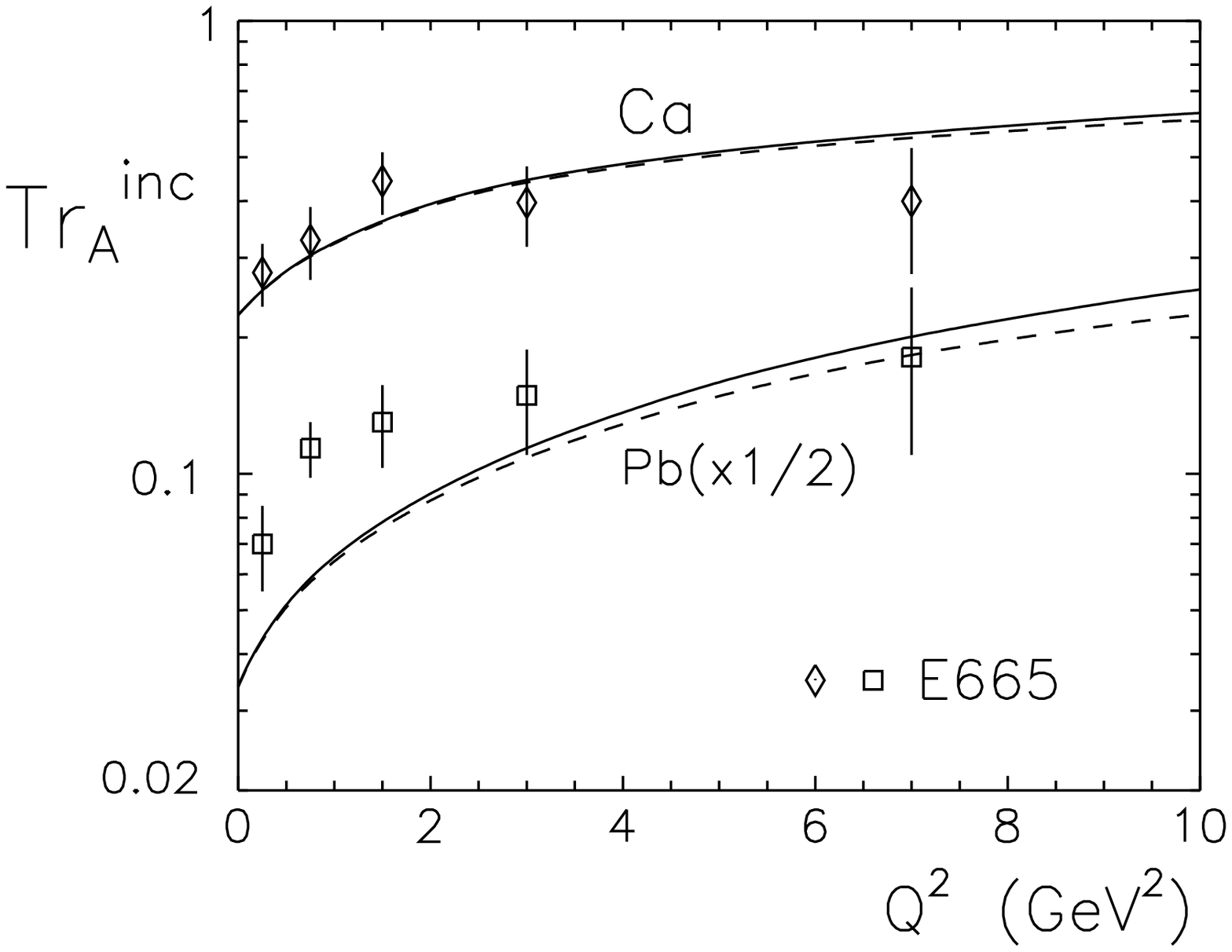}
  \hspace*{0.2cm}
  \includegraphics[width=0.35\textwidth]{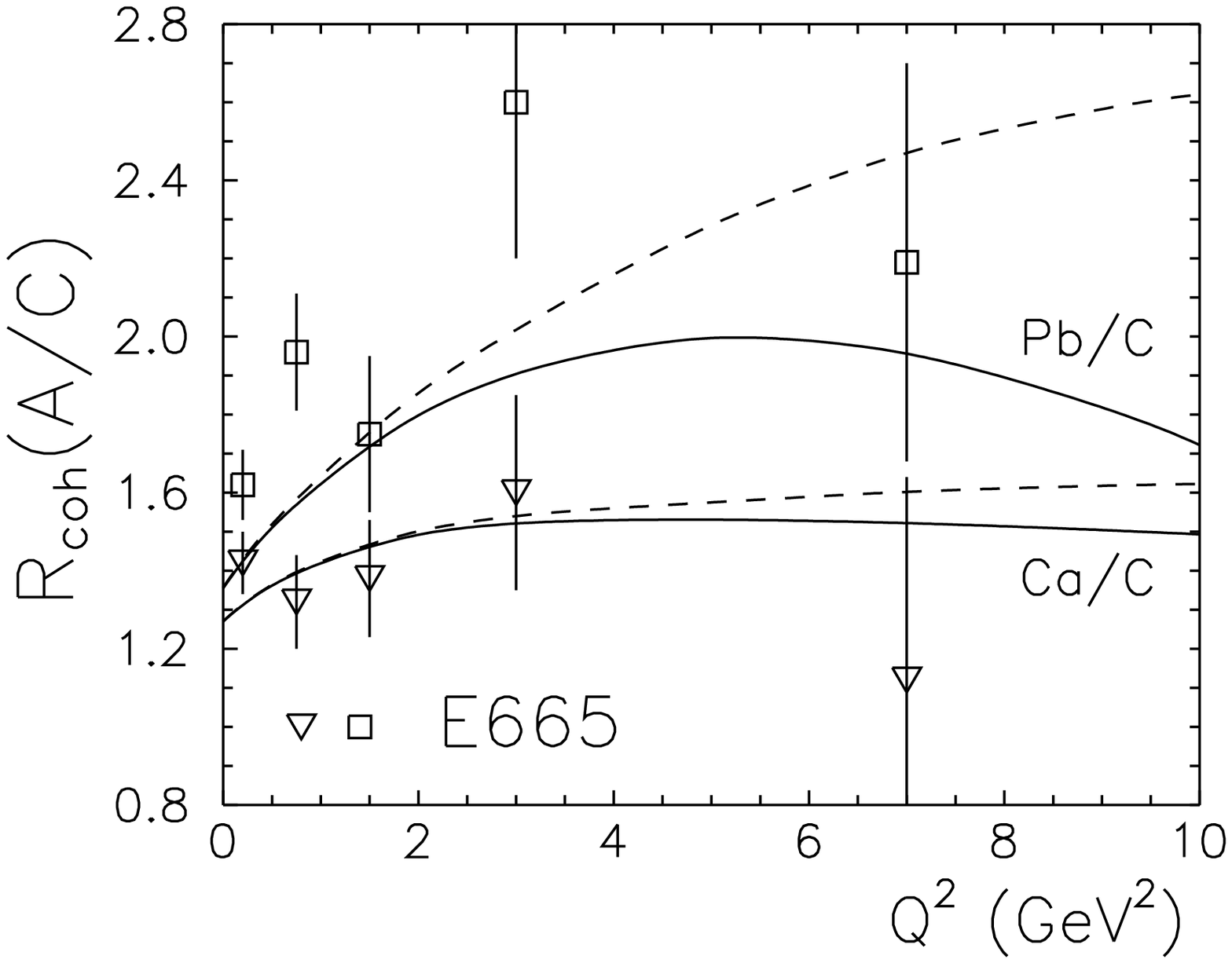}
  \caption{\small \label{fig:e665} 
    Comparison of the dipolea pproach with E665
    data~\cite{Adams:1994bw} for nuclear effect in electroproduction
    of $\rho$-mesons. 
    {\it Left panel:} $Q^2$- dependence of nuclear transparency for
    lead and calcium. Solid and dashed curves show the results of
    using the Green function approach and the ``frozen'' approximation
    respectively. {\it Right panel:} $Q^2$-dependence of the total
    cross section ratio $R_{coh}(A/C) = 12 \sigma_{coh}/A\sigma_{coh}$. 
  } 
\end{figure}


\noindent {\bf Diffractive electroproduction on nuclei.}
In the high energy regime of $l_c\gg R_A$ one can rely on Eq.~(\ref{kopel-1240})  for incoherent electroproduction of $\rho$-mesons (with different quark mass and meson wave function).
As a manifestation of color transparency, 
the nuclear ratio, also called nuclear transparency, $Tr_A^{inc}\equiv R_{inc}$ defined in (\ref{kopel-1220}),
was predicted in~\cite{Kopeliovich:1993gk} to rise as function of $Q^2$. Indeed, the mean size of the $\bar qq$ component of the virtual photon decreases qith $Q^2$, so the nucleus becomes more transparent.
The results of the E665 experiment at Fermilab~\cite{Adams:1994bw} depicted in Fig.~\ref{fig:e665} are in a good accord with the predicted behavior.
The calculations performed in the "frozen" approximation ($l_c\gg R_A$) are presented with dashed curves.
The more realistic results including finiteness of $l_c$ and $l_f$ are plotted by solid curves.
While the "frozen" approximation is rather accurate for incoherent production, the deviation from its expectation for coherent process at the energy of the E665 experiment is significant.

\begin{figure}[tb]
  \centering
  \includegraphics[width=0.49\textwidth]{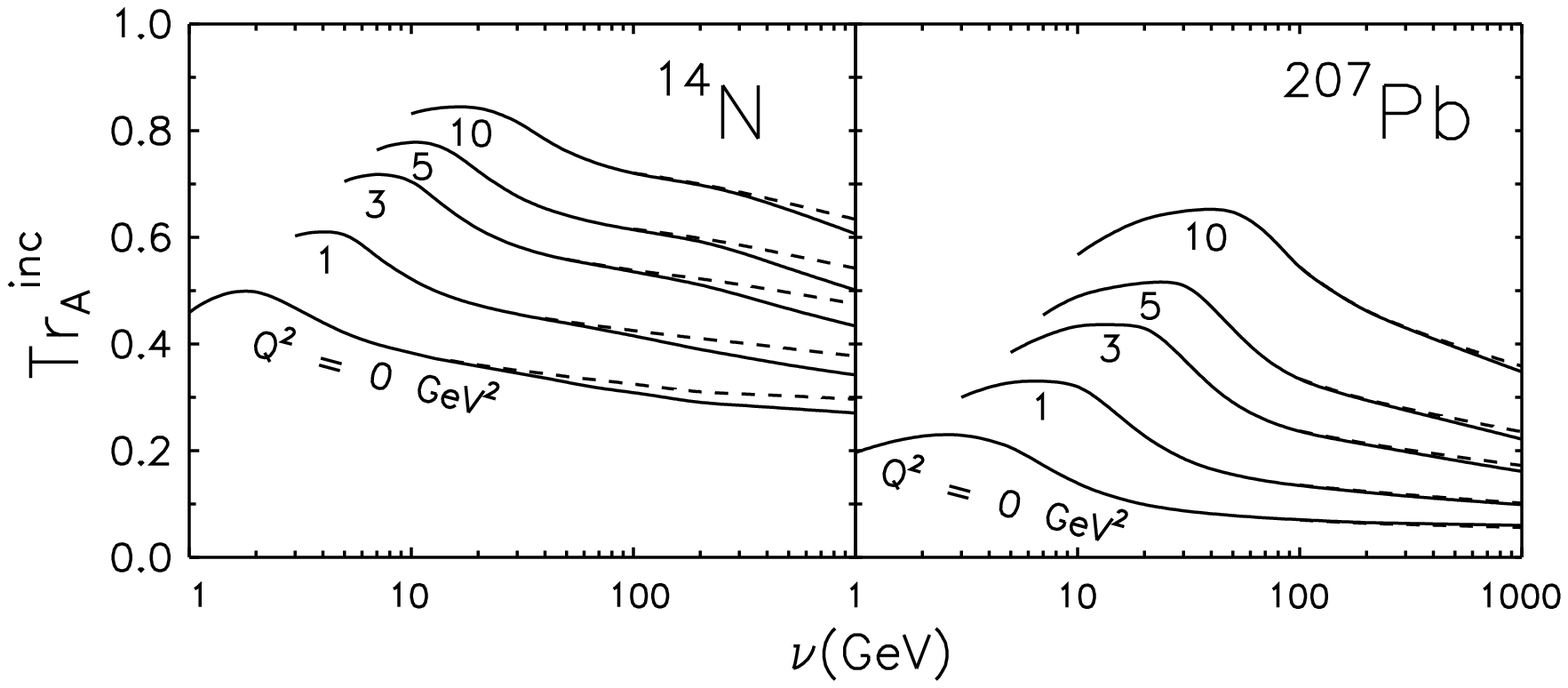}
  \includegraphics[width=0.49\textwidth]{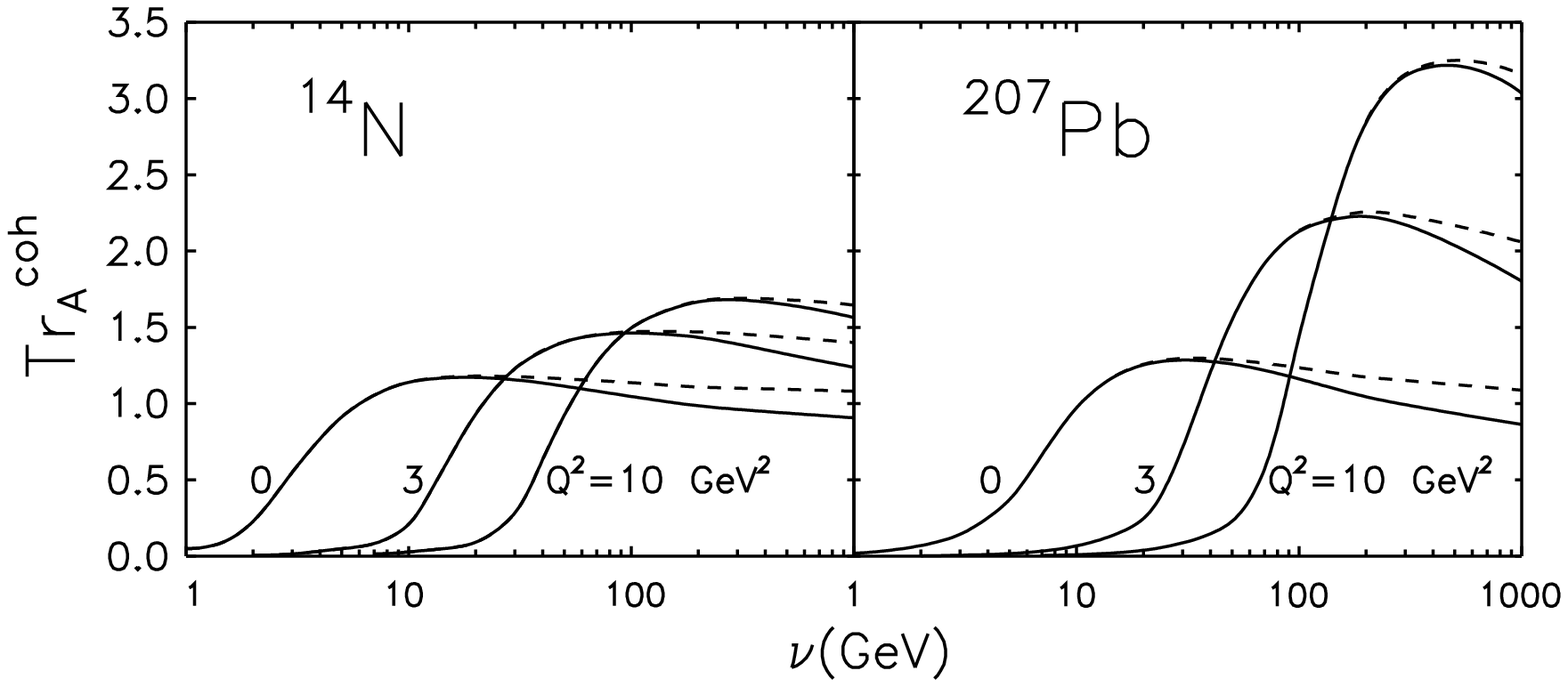}
  \caption{\small \label{fig:e-dep} 
    Nuclear transparency for incoherent and coherent electroproduction
    of $\rho^0$ on nitrogen and lead as function of energy. Solid and
    dashed curves correspond to calculations with and without gluon
    shadowing, respectively. {\it Left two panels:} Incoherent production 
    at $Q^2=0$, 1, 3, 5, 10 GeV$^2$. {\it Right two panels:} Coherent
    production at $Q^2=0$, 3, 10 GeV$^2$. 
  }
\end{figure}

The predicted energy dependence of the nuclear ratios in incoherent and coherent $\rho$ production on nitrogen and lead are depicted in Fig.~\ref{fig:e-dep}.
As was expected, the nucleus becomes more opaque with energy for incoherent production. This happens because when the hadronic fluctuations of the virtual photon live longer, they propagate through the whole nucleus and attenuate more. 
On the other hand, in incoherent production the phase shifts between the amplitudes of $\rho$ production on different nucleons must me very small in order the nucleus remained intact. This is why the nuclear ratio depicted in the bottom part of Fig.~\ref{fig:e-dep} is so suppressed at low energies.

At high energies, such as at an EIC, gluon shadowing causes an
additional nuclear suppression of $\rho$ production. 
This correction is calculated as was described in Sect.~\ref{kopel-glue} and the final results are plotted in Fig.~\ref{fig:e-dep} by solid curves. As was expected, the effect of gluon shadowing is not significant.

\subsubsection{From color transparency to color opacity}
\label{sec:Strikman-ct}
\hspace{\parindent}\parbox{0.92\textwidth}{\slshape 
  Mark Strikman 
}
\index{Strikman, Mark}

\vspace{\baselineskip}

Color transparency (CT) phenomena play  several roles. They 
probe both the high energy dynamics of the strong interaction  and  the
minimal small size  components of the hadrons. In  the case when some
of the produced particles have energies smaller than 10 GeV in the nucleus
rest frame, these processes could be also used to study the space-time
evolution of small wave packets - a question  
relevant for interpretation of heavy-ion collisions.  They also
provide an important link to the  hard QCD black disk regime - the
regime of strong absorption for the processes which at lower energies
exhibit the CT regime, and determine the kinematics where
factorization can be applied to generalized parton distribution studies.


The basic feature of CT is the suppression of the interaction of small
size color singlet configurations: for a dipole of transverse
size $d$, perturbative QCD gives  
 \begin{equation}
\sigma(d,x_N)= {\pi^2\over 3} \alpha_s(Q^2_{eff}) d^2\left[x_NG_N(x_N,Q^2_{eff})  +2/3  x_NS_N(x_N, Q^2_{eff})\right], 
\label{pdip}
\end{equation}
where $Q^2_{eff} \propto 1/d^2, x_N= Q^2_{eff}/W^2$, and the second term 
is due to the contribution of quark exchanges which is important for
intermediate energies \cite{Frankfurt:2000jm}. 
There are two critical requirements for CT phenomena: {\it squeezing},
 the selection of small size configurations, and {\it freezing}, the selection of high enough 
energies to allow the squeezed configuration to live long enough. 

At high energies, one can select CT processes by selecting special final
states: for example, the diffraction of a pion into two high $p_t$  jets, or a small
initial state $\gamma_L^*$ such as in the exclusive production of mesons. 
QCD factorization theorems \cite{Collins:1996fb,Frankfurt:2000jm}
were proven  for these processes  based on the CT property of QCD.
The space time picture of these processes in the nucleus rest frame is
as follows: long before the target, the projectile pion or virtual
photon fluctuates into a $q\bar q$ configuration with transverse
separation $d$, which elastically scatters off the target with an
amplitude which for $t=0$ is given by Eq.~\eqref{pdip} (up to  
small corrections due to different off shellness of the $q\bar q$ pair
in the initial and final states), followed by the transformation of
the pair into two jets or a vector meson. With a slight
simplification, the amplitude for dijet diffractive production can be
written as 
\begin{equation}
  A(\pi\,  N \to \, 2\,{\rm  jets}\, + \,N) (z,p_t,t=0) 
    \propto \int d^2d\;\psi^{q\bar q}_{\pi}(d,z) 
    \sigma_{q\bar q - N(A)}(d,s)e^{ip_td}, 
\label{dijet}
\end{equation}
where $z$ is the light-cone fraction of the pion momentum carried by a
quark, and $\psi^{q\bar q}_{\pi}(z,d)\propto z(1-z)_{d\to 0}$ is the
quark-antiquark Fock component of the meson light cone wave
function. The presence of the plane-wave factor in the final state leads
to an expectation of an earlier onset  of  scaling than in the case of
the vector meson production, where the vector meson wave function
appears instead. CT was observed  in the pion diffraction into two jets 
\cite{Aitala:2000hc}, confirming predictions in~\cite{Frankfurt:1993it}.
The HERA data on exclusive vector meson production are also well
described. 

\subsubsection*{Investigations at an EIC}

Studies at an EIC will require investigations of different exclusive
meson production channels as a function of $x,Q^2$. In the CT limit
and $-t \ge \mbox{0.1 GeV}^2$, where  coherence effects are negligible,
one expects  
\begin{equation}
\sigma^{incoh}_{\gamma^*_{L} A \to "meson" A^*}(t)=Z \sigma_{\gamma^*_{L} p \to "meson" N}(t) + 
N \sigma_{\gamma^*_{L} n \to "meson" N}(t)
\end{equation}
In EIC kinematics, the coherence length is $\gg 2 R_A$ so deviations from the CT prediction could be due to leading twist effects - leading twist shadowing, and higher twist effects of multiple interactions of  the $q\bar q$ pair with the target nucleus. 
There are two distinctive regimes: $x\ge 0.03$ where nuclear PDFs are practically linear in A, and $x \le 0.01$ where a significant  LT shadowing of nPDFs is expected (see discussion in section \ref{sec:LT_nuclear_shadowing}). 

\noindent 
{\bf The $ x \ge 0.03$ region.} Multiple interactions should reduce
the cross section. At an EIC, it would be possible to perform a scan
as a function of $Q^2$. For low $Q^2$ and especially for $\sigma_T$,
one expects a hadron-like regime in which absorption is strong and
$\sigma^{incoh}_{\gamma^*_{L} A \to "meson" A^*}(t) \propto
A^{1/3}$. With an increase of $Q^2$, one expects a  transition from
soft dynamics with Gribov-Glauber type screening to the CT regime
without significant  LT gluon shadowing.  
In the case of $J/\psi$ production, one expects the CT regime already at
low $Q^2$ while for the light mesons, the onset of CT can be much
slower as essential transverse sizes of the $q\bar q$ pair decrease rather
slowly with $Q^2$ as manifested in the slow convergence of the t-slope of $\rho$-meson production to the t-slope of $J/\psi$ production with increasing $Q^2$~\cite{Frankfurt:1995jw}.

\begin{figure}[tb]
  \centering
  \includegraphics[width=0.8\textwidth]{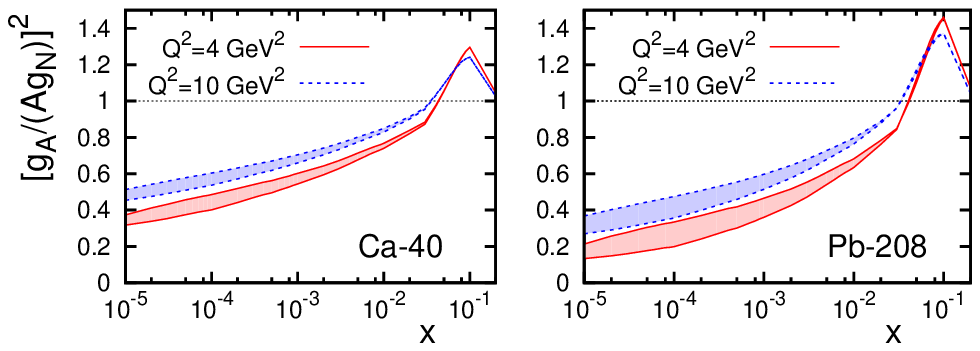}
  \caption{\small Leading twist shadowing effect for coherent vector meson production off Ca and Pb. Bands reflect the range  of predictions given by the  FGS10$_{-}$L and FGS10$_{-}$H parametrizations of the gluon LT shadowing. }
\label{fig:shadcoh} 
\end{figure}

\noindent 
{\bf  The $x \le 0.01$ region.} In this regime, one expects large
shadowing due to the LT mechanisms both for the incoherent
and coherent contribution in which
case~\cite{Brodsky:1994kf}, perturbative color opacity is given by
\begin{equation}
{d\sigma_{\gamma_L A\to V\; A}\over dt}/  {d\sigma_{\gamma_L N\to V\; N}\over dt}={G_A^2(x,Q_{eff}^2)/G_N^2(x,Q_{eff}^2)} \cdot F_A^2(t),
\label{shadcoh}
\end{equation}
where $F_A(t)$ is the nucleus  form factor. Typical results for the expected suppression effect are given in Fig. \ref{fig:shadcoh}.  Note here that effective $Q^2$, which enters in Eq.\ref{shadcoh}, is much smaller than $Q^2$ in the electro-production of light vector mesons. For example, in the case of the $\rho$ meson,  $Q_{eff}^2\sim 3$ GeV$^2$ for $Q^2\sim 10$ GeV$^2$. For  $J/\psi$
 photo-production, $Q_{eff}^2\sim 3 \div 4$ GeV$^2$ and grows slowly with $Q^2$ \cite{Frankfurt:1997fj}. 
Hence, for the top EIC energies, one expects a reduction in the coherent $J/\psi$ photo/electro production of at least  a factor of two.
 Numerically, the LT  shadowing mechanism leads to a larger screening effect for the interaction of the small dipoles than the HT  dipole eikonal models (cf. \cite{Frankfurt:1995jw}).

The incoherent cross section, $\sigma_{incoh}$, is shadowed somewhat
more strongly than  
the coherent cross section, $\sigma_{coh}$. The effect grows with the increasing strength of the elementary interaction.   As a result, the ratio  
$B_{\gamma^* N\to "V"N}^{-1}\cdot \sigma_{incoh}/\sigma_{coh}$
of incoherent and coherent cross sections integrated over $t$ and
divided by the slope of the elementary cross section is  expected to
decrease slowly  with decreasing $x$ at fixed $Q^2$ (cf. Fig.43  in
\cite{Baltz:2007kq}). For example, for $B= 4$ GeV$^{-2}$,
$R\equiv \sigma_{incoh}/\sigma_{coh}$ changes from $R\approx 0.3$ in
the impulse approximation limit to $R\approx  0.18$ in the regime of
strong absorption (strength of dipole interaction of the order
$\sigma_{tot}(\pi N)$). Simultaneous measurements of coherent and
incoherent diffraction will allow the testing of the underlying dynamics in
greater detail.  

Note that it will be feasible to measure the coherent cross section at
$t\sim 0$ due to the very steep $t$ dependence of coherent peak and the 
ability to kill most of the incoherent diffraction experimentally. At
the same time, measurements of the $t$ dependence of coherent  
diffraction beyond the first minimum are unlikely (except for the lightest nuclei like $^4He$)
due to the dominance of processes of the nuclear excitations for $-t
\ge -t_1$. (Measurements of very soft photons at rather large
opening angles are required\cite{White:2010tu}.) 
Note that the cross section of inelastic diffraction with production of
hadrons in the nucleus fragmentation region is comparable to that of
quasi-elastic diffraction. Studies of the
t-dependence of the meson production and/or hadron production in the
nucleus fragmentation regionare required to separate these two processes. 

\noindent 
{\bf Testing the onset of the black disk regime.}
The study of vector meson production provides a  fine probe  to
test the onset of high density color opacity regime where the LT
approximation breaks down - the black disk regime in which interactions
of  small dipoles with heavy nuclei become completely absorptive. In
this limit, one can derive a model independent
prediction for the cross section of the vector meson production~\cite{Frankfurt:2001nt}:
\begin{equation}
  {d\sigma^{\gamma^{\ast}_{T} +A\to V+A} \over dt} 
  =
  {M_V^2\over Q^2} {d\sigma^{\gamma^{\ast}_{L} +A\to V+A} \over dt}= 
  {(2\pi R_A^2)^2\over 16\pi}{3 \Gamma_V M_V^3 \over \alpha
    (M_V^2 + Q^2)^2 } \frac{4~\left|J_1(\sqrt{-t}R_A)\right|^2}{-tR_A^2}\, ,
\label{Strikman-vm}
\end{equation}
where $\Gamma_V$ is the electronic decay width $V\to e^+e^-$,
$\alpha$ is the fine-structure constant. Eq.~\eqref{Strikman-vm} corresponds
to a drastically different result: a factor of $Q^4$ slower $Q^2$ 
dependence of the cross section than the LT result. 

\begin{figure}[t]
  \centering
  \includegraphics[width=0.7\textwidth,clip=true]{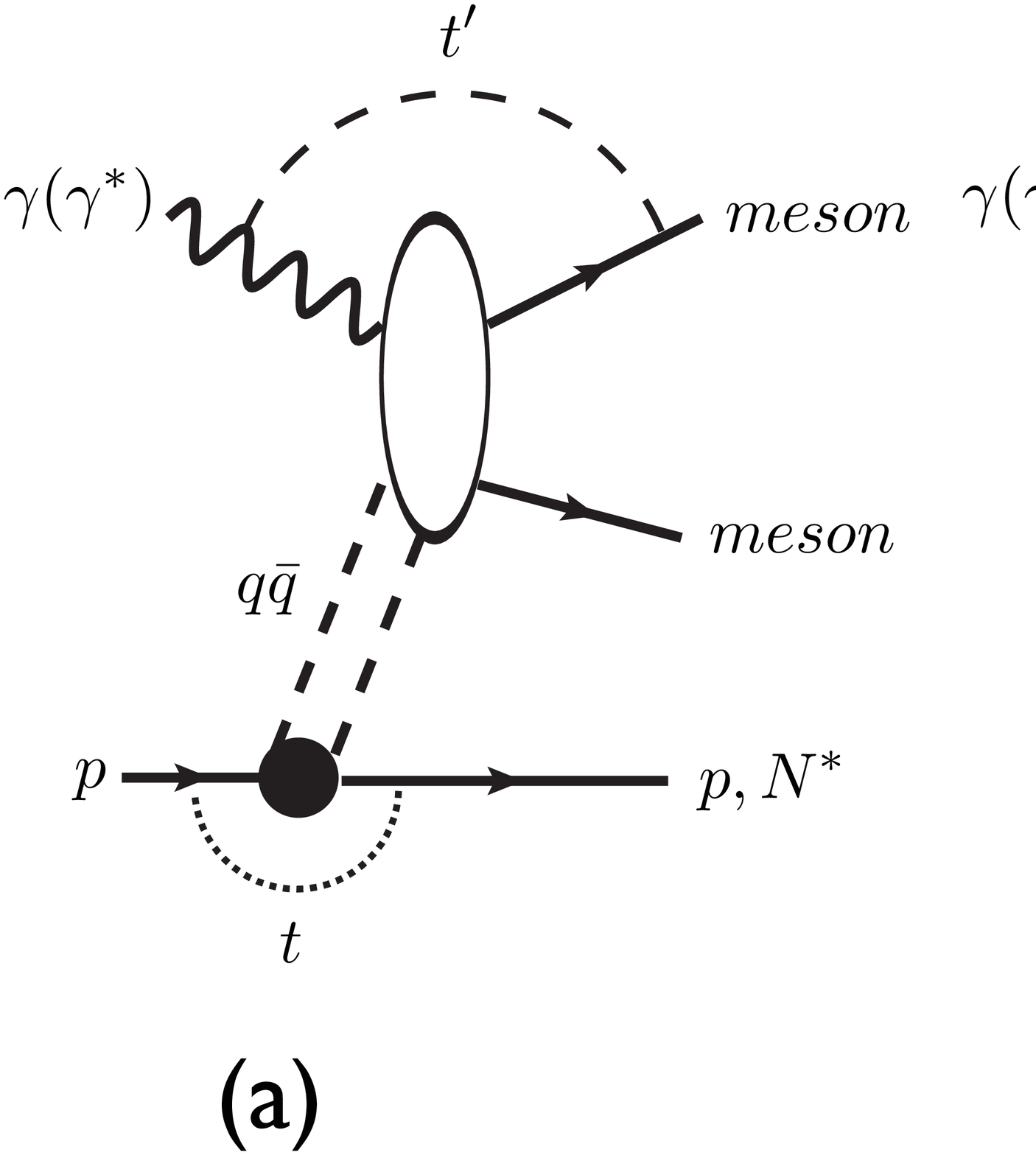}
  \caption{\small  
    Examples of $2\to 3$ processes probing (a) nucleon GPDs and large
    angle $\gamma^* (q\bar q) \to \pi \pi (\pi \rho) $ scattering; (b)
    photon GPDs and large angle $nucleon  (q\bar q) \to \pi N $
    scattering, (c) nucleon $\to$ meson GPD and 
    large angle $\gamma^* (qq q) \to \to \pi N $ scattering.} 
\label{fig:2to3}
\end{figure}
\noindent 


\noindent {\bf Other directions of studies.} 
Recently, a number of novel processes were suggested to check the interplay
between CT and color opacity phenomena as well as to use CT to
understand the dynamics of various elementary processes. 
\begin{enumerate}
\item
It was demonstrated that it is possible to trace small dipoles through
the center of the nucleus by selecting large $t$ VM production with
rapidity gap $\gamma^* A\to V + gap +Y$ for $x_g= -t /(-t+M_Y^2)$
\cite{Frankfurt:2008et}. 
\item
It was suggested that amplitudes of high energy $2 \to 3$ branching processes: 
$a+b \to c+ d + e$, where $t=(p_b-p_e)^2$ is small,  $t'= (p_a-p_c)^2$, $s'=(p_c+p_ed)^2$ are  large, and   $t'/s' = const$ can be written in  a factorized form as a convolution of different  nucleon quark GPDs and hard $2\to 2$ amplitudes  \cite{Kumano:2009he}. Several examples of such processes are depicted in Fig.~\ref{fig:2to3}. In the case of the $ep$ collider one would be able to study both the nucleon GPDs and GPDs of the real (virtual photon). Also, it will be possible to study large angle $\gamma(\gamma^*) +(q\bar q) \to meson_1 + meson_2$ and 
$\gamma(\gamma^*) +(qqq) \to meson + baryon$ reactions.
\item 
Embedding these processes in nuclei, for example by studying the process
$\gamma +A \to \pi^+\pi^+ A^*$, will make it possible to determine at what $p_T $ of the pions CT sets in and hence determine minimal $p_T$ for which these processes could be used to study various  quark GPDs.  The nuclear transparency for these processes   is very sensitive to the size of the meson $\bar q$ configurations
 \cite{Kumano:2009ky}. Hence it may be possible   to determine the characteristic  transverse size of the $q\bar q$ dipole involved in the hard process   using Eq.~\eqref{pdip}. Also, by studying the transparency as a function of $s$ for fixed $s',t$ and $t'$, one could measure in great detail the rate and the pattern of the space time evolution of small $q\bar q$ wave packets.
\end{enumerate}



 

\section{Nuclear GPDs and TMDs}

\subsection{Nuclear quark and gluon GPDs}
\label{subsec:nGPDS}

\hspace{\parindent}\parbox{0.92\textwidth}{\slshape 
  Vadim Guzey, Mark Strikman 
}
\index{Guzey, Vadim}
\index{Strikman, Mark}

\vspace{\baselineskip}

Generalized parton distributions (GPDs) parameterize the response of hadronic targets
(nucleon, nucleus) when probed by hard probes in exclusive reactions.
The QCD factorization theorems state that GPDs are universal distributions that
can be accessed in a wide range of hard exclusive processes: 
deeply virtual Compton scattering (DVCS)~\cite{Collins:1998be}, 
electro-production of mesons by 
longitudinal virtual photons~\cite{Collins:1996fb}, time-like Compton
scattering, etc. GPDs are fundamental and rigorously-defined
quantities that encode information on: 
(i) the distributions and correlations of partons in hadrons that is much richer than that
contained in usual diagonal parton distributions and elastic form factors 
(in a certain sense, GPDs provide three-dimensional parton imaging), 
(ii) parton total angular momentum (thus, GPDs are believed to help resolve the so-called proton spin crisis), etc.
For the detailed discussion of GPDs, see section \ref{sec:imaging_executive_summary} on ``Imaging QCD Matter" 

While what has been said above holds true for any hadronic target, 
nuclear GPDs are also interesting in their own right:
\begin{itemize}
\item[(i)]
Nuclear GPDs give access to both proton and neutron 
GPDs~\cite{Berger:2001zb,Cano:2003ju,Kirchner:2003wt,Guzey:2003jh,Guzey:2008th}. 
Incoherent reactions
(with nuclear break-up) can be used to study
quasi-free neutron GPDs~\cite{Mazouz:2007vj}.
\item[(ii)]
Traditional nuclear effects---off-diagonal EMC effect~\cite{Scopetta:2004kj,Liuti:2005gi}, 
nuclear shadowing and antishadowing~\cite{Freund:2003ix,Goeke:2009tu,LT_shadowing_Phys_Rep}---have been predicted to be more prominent 
than in the diagonal case.
\item[(iii)]
Nuclear GPDs may be a good tool to study not well-established/controversial and novel nuclear effects such as the medium modifications of bound nucleon 
GPDs~\cite{Liuti:2005gi,Guzey:2008fe} and 
presence of non-nucleonic degrees of freedom~\cite{Guzey:2005ba}.
\end{itemize}
 
\subsubsection{Medium $x_B > 0.05$}

The cleanest way to study GPDs is deeply virtual Compton scattering
(DVCS), $\gamma^{\ast}+A \to \gamma+A^{\prime}$. Nuclear DVCS is more
complex and versatile than that with the free proton because the
nuclear target, $A$, can have various spins (the number of GPDs
increases with the spin of the target)  and many different final
states, $A^{\prime}$, can be produced ($A^{\prime}=A, A^{\ast}, A+\pi,
A-1+N$, etc.).  
In the situation when the final nuclear state cannot be detected, one
can sum over all final states $A^{\prime}$ assuming their completeness
and obtain for the nuclear DVCS cross section~\cite{Guzey:2003jh}:
\begin{equation}
\sigma_{\rm DVCS}=A(A-1)\sigma_{\rm DVCS}^{\rm coh}+A  \sigma_{\rm DVCS}^N \,.
\label{eq:sigma_dvcs_summed}
\end{equation}
In this expression, the first term is the coherent-dominated 
contribution (without nuclear break-up or excitation) 
which is proportional to the nuclear form factor squared, $F_A^2$, and
significant only at the small momentum transfer $t$.
The second term is the incoherent contribution whose $t$ dependence is governed by that of
the nucleon GPDs; this term dominates at large $t$.

Similarly to Eq.~(\ref{eq:sigma_dvcs_summed}), the expressions interpolating between the coherent and incoherent regimes can also be derived
for the interference between DVCS and Bethe-Heitler (BH) amplitudes and BH cross section.
For instance, the coherent-dominated contribution to the interference between DVCS and BH amplitudes scales as $Z(A-1)$ and that to the BH cross section scales as $Z(Z-1)$
($Z$ is the nuclear charge).
Therefore, one immediately and model-independently predicts the enhancement of
the ratio of 
the nuclear to free proton DVCS beam-spin asymmetries at small $t$,
$A_{\rm LU}^A/A_{\rm LU}^p \sim (A-1)/(Z-1)$ \cite{Kirchner:2003wt,Guzey:2003jh}.
At large $t$, the cross section is dominated by the incoherent
contribution, no nuclear enhancement is expected, and $A_{\rm
LU}^A/A_{\rm LU}^p \sim 1$ (in fact, the neutron contribution somewhat
suppresses the ratio and makes $A_{\rm LU}^A/A_{\rm LU}^p <
1$ \cite{Guzey:2008th}). 
While the HERMES analysis of nuclear DVCS with
$^4$He, $^{14}$N, $^{20}$Ne, $^{84}$Kr, and $^{132}$Xe targets supports
that $A_{\rm LU}^A/A_{\rm LU}^p \sim 1$ at large $t$ and $A$-independent at all $t$,
it finds that at small $t$, $A_{\rm LU}^A/A_{\rm LU}^p=0.91 \pm 0.19$~\cite{Airapetian:2009bi}.

Quark nuclear GPDs in the kinematic region of the off-diagonal EMC effect,
$0.1 < x_B <0.3$,
 will
be constrained with high precision by 
the analysis CLAS data on DVCS
on $^4$He~\cite{He4-JLab}. 
The experiment measured purely coherent nuclear DVCS (the recoiled
nucleus was detected using the BoNuS spectator tagger) and also
DVCS on a quasi-free proton. The latter will probe possible  
nuclear medium modifications of the bound proton quark
GPDs~\cite{Guzey:2008fe}. Gluon GPDs in nuclei can be accessed best in
hard exclusive production of heavy vector 
mesons. For instance, coherent $J/\psi$ production for $x_B >0.1$  can
be used to learn about the off-diagonal EMC effect in the gluon
channel. The incoherent production of $J/\psi$ can be used to probe
medium modifications of the gluon GPD of the bound nucleon. 

The EIC will be the only other accelerator beside JLab 12 GeV to study
GPDs in $e+A$ collisions, and will contribute considerably to their
knowledge. In particular, it will access sea quark and gluon
distributions, which are hard to measure at 6 GeV due to the limited
$x$ and $Q^2$ range, and open dedicated channels like $J/\Psi$
diffreactive production.

\subsubsection{Small $x_B < 0.05$: leading twist
shadowing and exclusive diffraction}

The EIC will open the way to experimental measurements of nuclear GPDs
at small $x_B$, where nuclear shadowing is known to occur for PDFs.
The leading twist theory of nuclear shadowing (see section~\ref{sec:LT_nuclear_shadowing}) 
allows one also to predict the impact parameter dependence of nuclear PDFs~\cite{LT_shadowing_Phys_Rep,Frankfurt:1998ym,Frankfurt:2003zd,Guzey:2009jr}.
The resulting impact parameter dependent nuclear PDFs, $f_{j/A}(x,Q^2,b)$ 
are  the corresponding nuclear generalized parton distributions (GPDs)
in the $\xi \to 0$ limit and in impact parameter space~\cite{Goeke:2009tu},  
$f_{j/A}(x,Q^2,b)=H^j_A(x,\xi=0,b,Q^2)$, where the latter GPD depends in general on two light-cone fractions $x$ and $\xi$; $\xi$ is fixed by the external
kinematics,  $\xi=x_B/(2-x_B)$, where $x_B$ is the standard Bjorken $x$.
The number of GPDs depends on the spin of the target; we shall consider only spinless
targets characterized by one twist-two chirally-even GPD $H^j$ ($j$ is the parton 
flavor).

Using the predictions of the leading twist theory of nuclear shadowing
for the impact parameter dependence of nuclear PDFs
(Eq.~\eqref{eq:fgs10_eq2}) and the connection of these to GPDs, 
one can obtain the nuclear GPD $H_A^j$ at small $x$ in the $\xi=0$ limit. 
The final result for the GPDs in the momentum space is
\begin{align}
  & H_{A}^{j}(x,\xi=0,t,Q^2) =A F_A(t) f_{j/N}(x,Q^2) \nonumber\\
  & \qquad\quad - \frac{A(A-1)}{2} 
  \, 16 \pi  \, \Re e \Bigg\{\frac{(1-i \eta)^2}{1+\eta^2}
  \int d^2 b\, e^{i \vec{\Delta}_{\perp} \cdot \vec{b}}
  \int^{\infty}_{\infty} dz_1 \int^{\infty}_{z_1} dz_2
  \int_{x}^{0.1} dx_{\Pomeron} \rho_A(b,z_1) 
  \nonumber\\
  & \qquad\quad \times  \rho_A(b,z_2) e^{i m_N x_{\Pomeron}(z_1-z_2)}
  e^{-\frac{A}{2} (1-i \eta) \sigma_{\rm soft}^j(x,Q^2) \int^{z_2}_{z_1}dz^{\prime} \rho_A(b,z^{\prime})}
  \frac{1}{x_{\Pomeron}} f_{j}^{D(4)}(\beta,Q^2,x_{\Pomeron},t_{\rm min})
  \Bigg\} \ ,
  \label{eq:xiAlimit}
\end{align}
where the notation is the same as in eqs.~\eqref{eq:fgs10_eq1} and
\eqref{eq:fgs10_eq2}.

\begin{figure}
  \centering
  \includegraphics[width=0.75\textwidth]{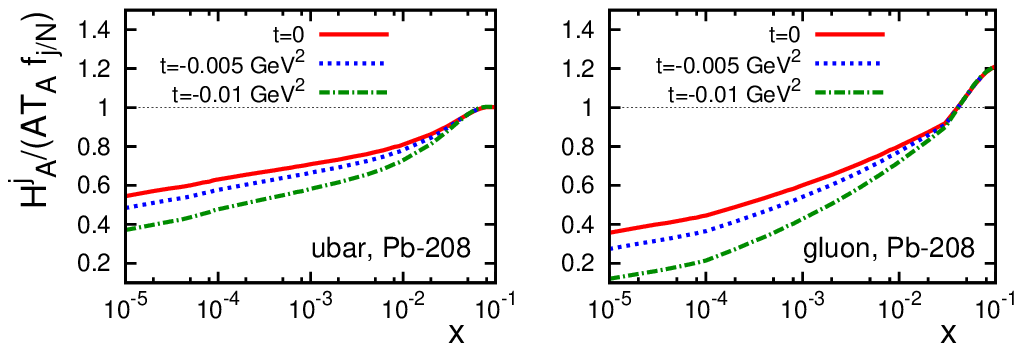}
  \caption{\small 
    The ratio of the gluon and $\bar u$-quark
    $H_{A}^{j}(x,\xi=0,t)/[A F_A(t) f_{j/N}(x,Q^2)]$ for $^{208}$Pb 
    as a function of $x$ for different values
    of $t$. All curves correspond to $Q^2=4$ GeV$^2$ and model FGS10\_H. 
  }
  \label{fig:IMP_pb208_2011_writeup} 
\end{figure}

Fig.~\ref{fig:IMP_pb208_2011_writeup} presents our predictions for the ratio $H_{A}^{j}(x,\xi=0,t,Q^2)/[A F_A(t) f_{j/N}(x,Q^2)]$ for $^{208}$Pb 
as a function of $x$ for different values of $t$. 
The left panel corresponds to the ratio of the ${\bar u}$-quark distributions;
the right panel corresponds to the gluon distributions.
All curves correspond to $Q^2=4$ GeV$^2$ and model FGS10\_H (see details in 
section~\ref{sec:LT_nuclear_shadowing}).
Since the $t$ dependence of the shadowing correction to $H_{A}^{j}(x,\xi=0,t)$
(second term in Eq.~(\ref{eq:xiAlimit})) is somewhat slower than that
of the impulse approximation (the first term), the effect of nuclear shadowing
increases as $|t|$ is increased, as expected.

Experimental observables measured in hard exclusive processes
such as, {\it e.g.}, $\gamma^{\ast} +A \to \gamma (J/\Psi, \rho,
\dots) + A$,
 probe the GPD $H_{A}^j(x,\xi,t,Q^2)$ integrated over the entire region of
the light-cone variable $x$, $0 \leq x \leq 1$. However, 
at high energies (small $\xi$ or $x_B$), the situation simplifies: 
the predominantly imaginary $\gamma^{\ast} +A \to \gamma (J/\Psi, \rho, \dots) +A$ scattering
amplitudes are expressed solely in terms of the GPDs at the $x=\xi$  cross-over line,
$H_{A}^j(\xi,\xi,t,Q^2)$ (to the leading order in the strong coupling constant $\alpha_s$).
In addition, it was shown in~\cite{Frankfurt:1997ha} that at high energies
and in the leading logarithmic approximation (LLA), GPDs at an 
input scale $Q_0^2 \sim $ few GeV$^2$ can be approximated well by the usual parton distributions,
i.e., it is safe to neglect the effect of the skewness $\xi$.
Therefore, for instance, for the imaginary part of the coherent nuclear deeply virtual
Compton scattering (DVCS) amplitude 
($\gamma^{\ast} +A \to \gamma +A$), we have at the leading order in $\alpha_s$:
\begin{eqnarray}
\Im m {\cal A}_{\rm DVCS}(\xi,t,Q^2)&=&-\pi \sum_q e_q^2 \left[H_A^q(\xi,\xi,t,Q^2)+H_A^{\bar q}(\xi,\xi,t,Q^2)\right] \nonumber\\
& \approx & -\pi \sum_q e_q^2 \left[H_A^q(\xi,\xi=0,t,Q^2)+H_A^{\bar q}(\xi,\xi=0,t,Q^2)\right] \;,
\label{eq:sigma_dvcs2}
\end{eqnarray}
where $e_q$ are the quark charges; $H_A^q(\xi,\xi=0,t,Q^2)$ are given by 
Eq.~(\ref{eq:xiAlimit}). 

The cleanest way to access GPDs is via DVCS.
At the photon level, the $\gamma^{\ast}+A \to \gamma +A$
cross section reads, (see, e.g.,~\cite{Belitsky:2001ns}):
\begin{equation}
\frac{d \sigma_{\rm DVCS}}{dt}=\frac{\pi \alpha_{\rm em}^2 x^2(1-\xi^2)}{Q^4 \sqrt{1+\epsilon^2}}|\Im m {\cal A}_{\rm DVCS}(\xi,t,Q^2)|^2 \,,
\label{eq:sigma_dvcs}
\end{equation}
where $\alpha_{\rm em}$ is the fine-structure constant; 
$\epsilon^2=4 x^2 m_N^2/Q^2$;
$\Im m {\cal A}_{\rm DVCS}$ is given by Eq.~(\ref{eq:sigma_dvcs2}).

The DVCS process interferes and competes with the purely electromagnetic
Bethe-Heitler (BH) process.
The BH cross section at the photon level can be written in the following
form~\cite{Belitsky:2001ns}:
\begin{equation}
\frac{d \sigma_{\rm BH}}{dt}=\frac{\pi \alpha_{\rm em}^2}{4 Q^2 t (1+\epsilon)^{5/2} (1-y-y^2/2)}
\int^{2 \pi}_{0} \frac{d \phi}{2 \pi} \frac{1}{{\cal P}_1(\phi) {\cal P}_2(\phi)}
|{\cal A}_{\rm BH}(\xi,t,Q^2,\phi)|^2 \,,
\label{eq:sigma_bh}
\end{equation}
where $y=(q \cdot P_A)/(k \cdot P_A)=Q^2/(x s)$ ($k$ is the incoming lepton momentum,
$q$ is the momentum of the virtual photon,
$P_A$ is the momentum of the incoming nucleus, $s$ is the total invariant energy
squared); 
$\phi$ is the angle between the lepton and hadron scattering planes; 
 ${\cal P}_1(\phi)$ and  ${\cal P}_2(\phi)$ are proportional to the lepton
propagators; $|{\cal A}_{\rm BH}(\xi,t,Q^2)|^2$ is the BH amplitude squared.
The expressions for ${\cal P}_{1,2}(\phi)$ and $|{\cal A}_{\rm BH}(\xi,t,Q^2)|^2$ 
can be found in~\cite{Belitsky:2001ns}.
Note that $|{\cal A}_{\rm BH}(\xi,t,Q^2)|^2$ is proportional to the nuclear
electric form factor squared ($|F_A(t)|^2$) and the nucleus charge
squared ($Z^2$).

Integrating the differential cross sections in Eqs.~(\ref{eq:sigma_dvcs}) and 
(\ref{eq:sigma_bh}) over $t$, one obtains the corresponding $t$-integrated cross
section $\sigma_{\rm DVCS\, (BH)}$ between $t_{\rm min} \approx -x^2 m_N^2$ and $t_{\rm max}=-1$ GeV$^2$:

\begin{figure}[tb]
  \centering
  \includegraphics[height=3.45cm]{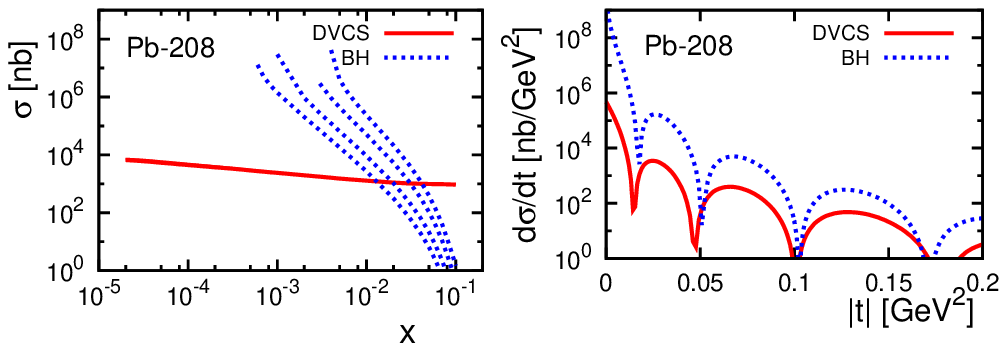}
  \includegraphics[height=3.45cm,trim=145 0 0 0,clip=true]{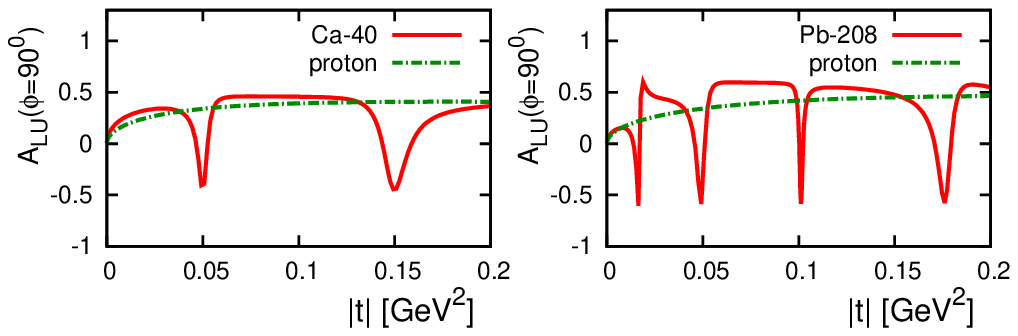}
  \vspace*{-0.2cm}
  \caption{\small   \label{fig:DVCS_2011}
    The DVCS (solid curves) and Bethe-Heitler (dot-dashed curves)
    cross sections for $^{208}$Pb at $Q^2=4$ GeV$^2$. 
    {\it Left panel:} $t$-integrated cross sections vs. $x$. 
    The four BH curves correspond, from left to right, to 
    $\sqrt{s}=(32, 44, 66, 90)$ GeV. 
    {\it Middle panel:} differential cross sections vs. $|t|$ at fixed
    $x=5 \times 10^{-3}$; the BH curve corresponds to $\sqrt{s}=32$ GeV. 
   {\it Right panel:} beam-spin asymmetry $A_{LU}$.
}
\end{figure}

In fig,~\ref{fig:DVCS_2011} we present our predictions for a
$^{208}$Pb target: in the left plot, the DVCS and BH cross sections at
$Q^2=4$ GeV$^2$, in the middle plot the differential cross sections as
a function of $|t|$ at fixed $x=5 \times 10^{-3}$, and in the right
plot the $A_{LU}$ asymmetry. 

In the considered kinematics, the $t$-integrated BH cross section is
much larger than the DVCS cross section for $x <10^{-2}$ due to
the dramatic enhancement of the BH cross section
at small $t\approx t_{\min}$ by the factor $1/t$, see Eq.~(\ref{eq:sigma_bh}).
Therefore, in order to extract a small DVCS signal on the background
of the dominant BH contribution for such $x$, one needs to consider
the observable differential in $t$. 
The $t$ dependence of the
DVCS and BH differential cross sections has the characteristic shape 
of the nuclear form factor squared, with distinct minima and maxima.
However, the minima of the DVCS cross section are slightly shifted
towards smaller $t$: 
this is the effect of the leading twist nuclear shadowing in quark nuclear GPDs.
The small shift of the minima toward smaller $t$ can be interpreted as
an increase of the transverse size of the distributions of quarks in
nuclei. One can enhance the effect by
using lighter nuclei (e.g., $^{4}$He and $^{12}$C)
or by considering observables sensitive to the interference between the 
BH and DVCS amplitudes. 
For instance, the DVCS beam-spin asymmetry at $A_{\rm LU}(\phi=90^{0})$,
dramatically oscillates as a function of $|t|$ \cite{Goeke:2009tu},
and the sole reason for these oscillations is the leading twist
nuclear shadowing.  

Another possibility to study nuclear shadowing in DVCS is offered by processes with nuclear 
break-up. In this case, the nuclear modification (suppression due to shadowing)
 of the DVCS break-up  cross section (as compared to the impulse approximation) 
is as large---or even bigger---as that for the coherent case.  
At the same time, in the impulse approximation, the relative contribution of the DVCS and BH
cross sections is enhanced by $A/Z$ compared to the $ep$ case. It allows one to observe the
DVCS signal on the large BH background down to much smaller $x$ than
in the $ep$ case, 
see the discussion in section \ref{sec:imaging_executive_summary}.

The leading twist theory of nuclear shadowing allows one also to make predictions for
certain observables in exclusive electroproduction of heavy vector mesons ($J/\psi$, $\Upsilon$)
with nuclear targets which probe the nuclear gluon distribution, with a pattern
similar to that discussed for small-$x$ nuclear DVCS~\cite{LT_shadowing_Phys_Rep}.
See  the discussion by M.Strikman in Section~\ref{sec:Strikman-ct}.

\subsection{Nuclear TMDs}
\label{sec:nuclearTMDs}
\hspace{\parindent}\parbox{0.92\textwidth}{\slshape
  Jian-Hua Gao, Zuo-tang Liang, Xin-Nian Wang, Jian Zhou
}
\index{Gao, Jian-Hua}
\index{Liang, Zuo-tang}
\index{Wang, Xin-Nian}
\index{Zhou, Jian}



Transverse momentum dependent distributions (TMDs) were discussed extensively for nucleons earlier in this report. 
%
TMDs play an important role in studying  final/initial state multiple re-scattering
effects in nuclei. Indeed, the leading power nuclear effect comes from the gauge link appearing in the nuclear
TMDs, in which the re-scattering effect is encoded.

The extraction of the TMDs from high energy scattering data
relies on TMD factorization theorems, established in
the $e^+e^-$ annihilation process~\cite{Collins:1981uw} and
semi-inclusive deep-inelastic (SIDIS) lepton-nucleon
scattering~\cite{Ji:2004wu}. It is not so clear whether TMD
factorization still holds in SIDIS off a large nucleus target.
In our recent work~\cite{Liang:2008vz}, we simply assume that 
it does. Correspondingly, one can introduce
leading power unpolarized nuclear TMDs. For simplicity, we restrict
our discussion to the light cone gauge,
$A^+=0$ \cite{Belitsky:2002sm}, where
\begin{eqnarray}
  f_q^A(x,\vec k_\perp) = \int \frac{dy^-}{2\pi}
  \frac{d^2y_\perp}{(2\pi)^2} e^{ixp^+y^- -i\vec k_\perp\cdot \vec
    y_\perp} \langle A \mid \bar\psi(0,\vec 0_\perp)\frac{\gamma^+}{2}
  {\cal L}_\perp(0,y) \psi(y^-,\vec y_\perp) \mid A \rangle ,
  \label{tmd0}
\end{eqnarray}
and the transverse gauge link is 
$
  {\cal L}_\perp \equiv
  P\exp\left[-ig\int_{\vec 0_\perp}^{\vec y_\perp} d\vec\xi_\perp\cdot
  \vec A_\perp(\infty,\vec\xi_\perp)\right]
$.
This gauge link is not only crucial to ensure the gauge invariance
of the TMD parton distribution functions, but
also leads to physical consequences such as single-spin asymmetries in
SIDIS and the Drell-Yan process in $e+p$ collisions
\cite{Sivers:1989cc,Brodsky:2002cx,Collins:2002kn}. 
For DIS off a 
nucleus target, it should also contain information on the quark transverse
momentum broadening due to multiple scattering
inside the nucleus \cite{Liang:2008vz}.

In the study of either cold or hot nuclear matter, parton transverse
momentum broadening plays a crucial role in unraveling the medium properties.
 One important parameter that controls parton energy loss is the
 parton transport parameter $\hat q$, {\it i.e.}, the transverse 
momentum broadening squared per unit of propagation length~\cite{Baier:1996sk}.
Therefore, the calculation and measurement of the jet transport parameter is an important step toward understanding
the intrinsic properties of the QCD medium. Much effort has been
devoted to the study of transverse momentum broadening in high energy collisions within different approaches
~\cite{Baier:1996sk,Bodwin:1988fs,Luo:1992fz,Guo:1998rd,Wiedemann:2000za,Fries:2002mu,Majumder:2007hx,D'Eramo:2010ak}.

In this contribution, we start from the matrix element definition of
the nuclear TMD and  identify the 
gauge link as the main source of leading nuclear effects. The
broadened distribution has a Gaussian form, as found in earlier
studies~\cite{Majumder:2007hx}, and
suppresses the azimuthal asymmetry in SIDIS off nuclear targets.
This in turns gives direct experimental access to the cold nuclear
matter transport coefficient $\hat q$, 
and offers a way to determine the relative magnitude of the intrinsic
transverse momentum in various {\it nucleon} TMDs.

\noindent{\bf Nuclear TMDs and nucleon TMDs.}
The effect of final state interactions that lead to transverse
momentum broadening can be encoded in the gauge link. In fact, the 
nuclear dependent part of the quark TMD can be isolated from the
gauge link so that the nuclear TMD can be expressed as a convolution
of the Gaussian broadening and the nucleon TMD.
Assuming a weakly bound nucleon, neglecting the correlation between
different nucleons, and keeping only the matrix elements
with nuclear enhancement 
one obtains the nuclear TMD,
\begin{equation}
f_q^A(x,\vec k_\perp)=\frac{A}{\pi \Delta_{2F}} \int d^2\ell_\perp
e^{-(\vec k_\perp
-\vec\ell_\perp)^2/\Delta_{2F}}f_q^N(x,\vec\ell_\perp) \ ,
\label{tmd5}
\end{equation}
as a convolution of the nucleon TMD and a
Gaussian with a width $\Delta_{2F}$ given by the total transverse
momentum broadening squared,
\begin{equation}
\Delta_{2F}= \frac{1}{Af_q^N(x)} \int d^2 k_\perp  k_\perp^2
\left [ f_q^A(x,\vec k_\perp)-f_q^N(x,\vec k_\perp)\right ] =\int d\xi^-_N \hat q_F(\xi_N) \label{totbrd}
\ .
\end{equation}
where the quark transport parameter $\hat q_F(\xi_N)$ is defined as
\begin{equation}
\hat q_F(\xi_N)=-\frac{g^2}{2N_c}\rho_N^A(\xi_N)\int
\frac{d\xi^-}{2p^+} \langle N \mid F_{+\sigma}(0)F_+^{\sigma}(\xi^-)
\mid N \rangle
=\frac{2\pi^2\alpha_s}{N_c}\rho_N^A(\xi_N)[xf_N^g(x)]_{x=0},
\label{qhat1}
\end{equation}
with $\rho_N^A(\xi_N)$ is the spatial nucleon density inside the
nucleus and $f^N_g(x)$ is the gluon distribution function in a
nucleon. Eq.~\eqref{tmd5} is our main result.\\

\begin{figure} [htb]
  \centering
  \includegraphics[width=0.45\linewidth]{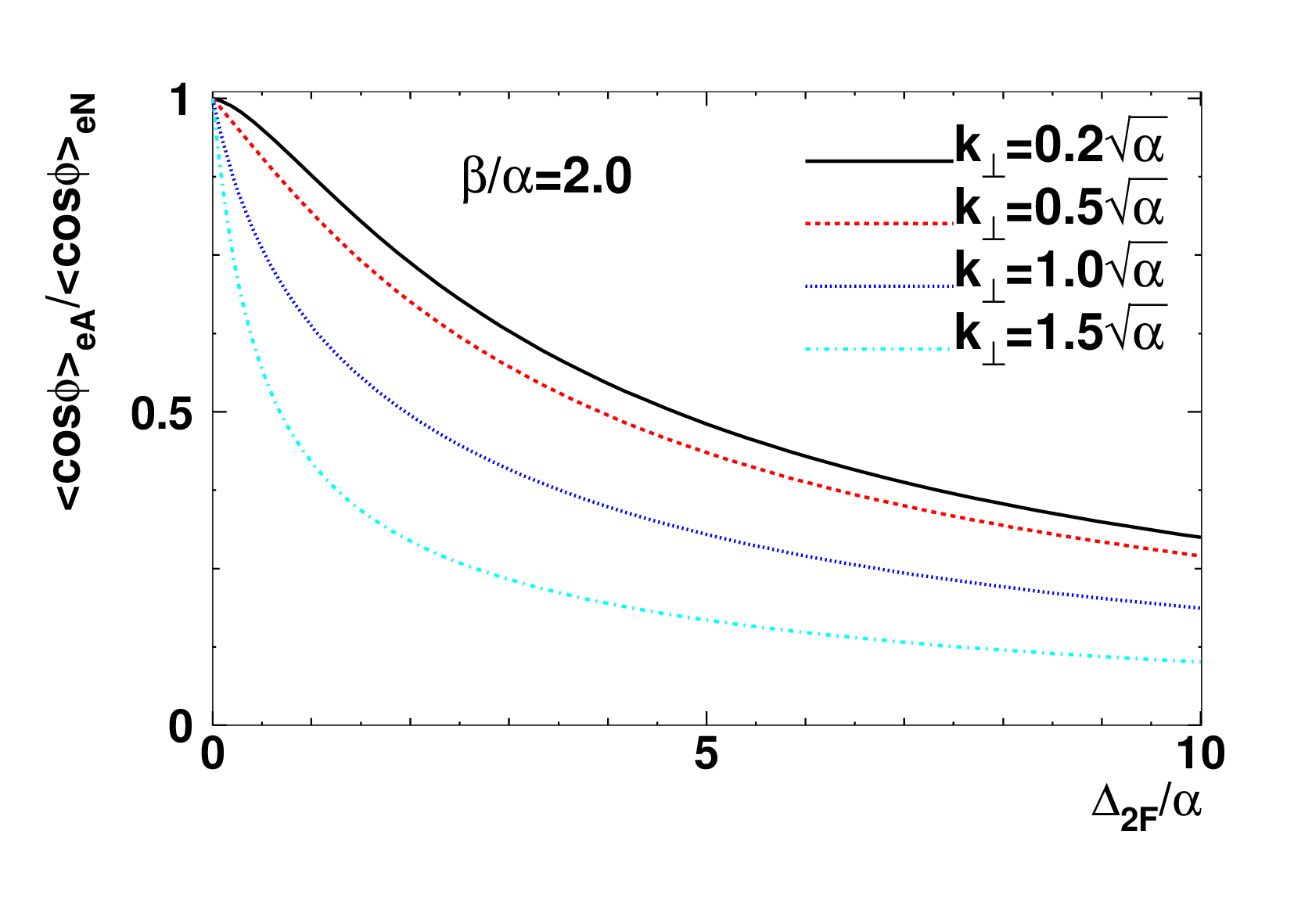}
  \includegraphics[width=0.45\linewidth]{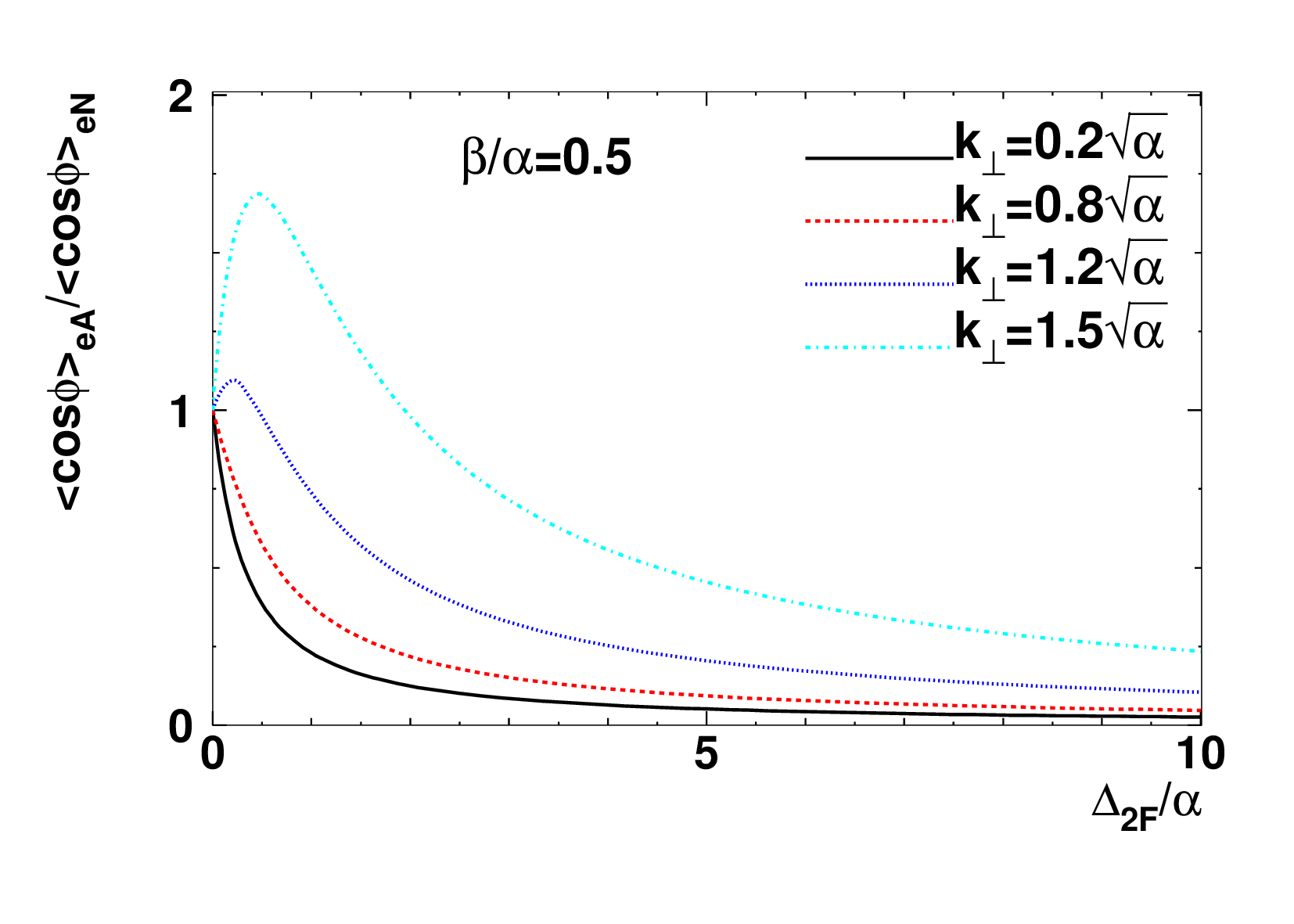}
  \vskip-.6cm
  \caption{\small 
    Ratio $\frac{\langle \cos\phi \rangle_{eA}}{\langle
    \cos\phi\rangle_{eN}}$ as a function of $\Delta_{2F}$ for
    different $k_\perp$ and the relative width $\beta/\alpha$. 
  } 
\label{fig:AfH}      
\end{figure}

\noindent{\bf Nuclear dependence of azimuthal asymmetry in SIDIS.} 
One can generalize the above approach to the nuclear modification of higher
twist TMD parton distributions. The case of twist-3 and twist-4
TMDs~\cite{Mulders:1995dh,Bacchetta:2006tn,Liang:2006wp}, which
account for the $\cos \phi$ and $\cos 2 \phi$ azimuthal asymmetries
in SIDIS, has been recently investigated in
Ref.~\cite{Gao:2010mj,Song:2010pf}. Here we review the 
nuclear dependent $\cos \phi$ azimuthal asymmetry in the two kinematic
regions: at small transverse 
momentum $P_{h\perp} \sim \Lambda_{QCD}$ and intermediate transverse
momentum $\Lambda_{QCD} \ll P_{h\perp} \ll Q$, where $Q$ is the virtual photon
momentum.  The central ingredient of the treatment in
Ref.~\cite{Gao:2010mj} is the relation between the nucleon twist-3
TMDs and nuclear ones. If we look at jet production in SIDIS,
the azimuthal asymmetry is solely determined by one twist-3 TMD
distribution $f^\perp(x,k_\perp)$.  The ratio of the asymmetry between
SIDIS off nucleons and nuclei is,
\begin{equation}
  \frac{\langle \cos\phi \rangle_{eA}}{\langle\cos
    \phi\rangle_{eN}}=\frac{f_\perp^A(x,k_\perp)/f^A(x,k_\perp)}
    {f_\perp^N(x,k_\perp)/f^N(x,k_\perp)} 
\label{nTMD:ratio}
\end{equation}
The ratio depends on how the twist-3 TMD distributions $f_\perp^A$
is enhanced/suppressed due to the stronger final state interaction
taking place inside a nucleus. Following the same approach applied
to the twist-2 TMD distribution, we relate the function $f_\perp^A$
to $f_\perp^N$,
\begin{eqnarray}
f_{\perp}^A(x,k_\perp)& \approx & \frac{A}{\pi \Delta_{2F}} \int
d^2\ell_\perp \frac{(\vec k_\perp\cdot\vec\ell_\perp)}{\vec
k_\perp^{2}} e^{-(\vec k_\perp
-\vec\ell_\perp)^2/\Delta_{2F}}f_{\perp}^N(x,\ell_\perp)
\label{eq:fqperp2}
\end{eqnarray}
Given the  TMDs $f^N(x,k_\perp)$ and $f_\perp^N(x,k_\perp)$, one will
be able to calculate the ratio \eqref{eq:fqperp2}. To illustrate the nuclear
dependence of the  asymmetry qualitatively, we consider an ansatz of
the Gaussian distributions in $k_\perp$ for both TMDs,
\begin{equation}
f^N(x,k_\perp)=\frac{1}{\pi\alpha}f_q^N(x)e^{- k_\perp^2/\alpha}\ ,
\qquad
f_{\perp}^N(x,k_\perp)=\frac{1}{\pi\beta}f_{q\perp}^N(x)e^{-
k_\perp^2/\beta}.
\end{equation}
As shown in Fig.~\ref{fig:AfH}, the azimuthal asymmetry is suppressed
in $e+A$ SIDIS as compared to that in $e+N$ SIDIS.
Note also that the suppression pattern as a function of $k_\perp$ is
sensitive to the relative magnitude of the intrinsic transverse
momentum in the {\it nucleon} TMDs.

Now let us discuss the asymmetry at intermediate transverse
momentum. The fact that TMDs are perturbatively calculable when
$p_{\perp} \gg \Lambda_{QCD}$ or $k_\perp\gg\Lambda_{QCD}$ allows us
to reduce the theoretical uncertainty, since the twist-3 
TMDs are poorly known so far.
In the parton model, the azimuthal asymmetry for hadron production in
SIDIS can be expressed as a convolution of a few TMD distributions and
TMD fragmentation functions \cite{Mulders:1995dh,Bacchetta:2006tn}.
It turns out that fragmentation
functions  $H_1^\perp$ and $\tilde H$ are power suppressed compared to 
$\tilde D_\perp$ and $D$ at large $p_\perp$
\cite{Ji:2006ub,Bacchetta:2008xw,Yuan:2009dw}. Therefore, at
intermediate transverse momentum, the 
leading power terms are proportional to $f_1  \tilde D_\perp$ and 
$f_\perp D$. In the current fragmentation region, where $p_\perp$ is large,
we make a collinear expansion around $p_\perp=q_\perp$ in terms of the
power $k_\perp/q_\perp$ and keep the quadratic terms
$k_\perp^2/q_\perp^2$ in order to extract the nuclear dependent
contributions. After carrying out the integrals over $p_\perp$, we
find the nuclear dependent azimuthal asymmetry is related to the term
$D(z) \int \frac{k_\perp^2}{q_\perp^2} f_1(x, k_\perp) d^2 k_\perp$.
Therefore, the difference of the $\cos \phi_h$ azimuthal asymmetry is
proportional to the transverse momentum broadening.
\begin{eqnarray}
\langle\cos \phi_h\rangle_{eA}-\langle\cos \phi_h\rangle_{eN} \propto \int
\frac{k_\perp^2}{q_\perp^2} \left  [f_1^A(x, k_\perp)-f_1^N(x,
k_\perp) \right ] =\frac{\Delta_{2F}}{q_\perp^2}
\end{eqnarray}

\vskip1ex\noindent{\bf Conclusions.}
In summary, we can get a direct handle on the crucial transport
parameter $\hat q$, which descrcibed the properties of the QCD medium,
by measuring the nuclear dependent azimuthal asymmetry at intermediate
transverse momentum. 
Conversely, the target nucleus can be used as a filter to study nucleon TMDs,
{\it e.g.}, to determine the relative magnitude of the intrinsic
transverse momentum of $f^N$ and $f_{\perp}^N$.




\ \\ \noindent{\it Acknowledgments:}
J.Z. thanks A.~Metz and M.~Diehl for helpful discussion.


\section{Current fragmentation}
\label{sec:PartonProp}

\subsubsection{Introduction and the role of $e+A$ collisions}
\label{sec:eA_ppf_intro}

\hspace{\parindent}\parbox{0.92\textwidth}{\slshape 
Rapha\"el Dupr\'e and Alberto Accardi 
}

\index{Dupr\'e, Rapha\"el}
\index{Accardi, Alberto}

\vspace{\baselineskip}

\begin{figure}[htbp]
\centering
\includegraphics[width=10cm]{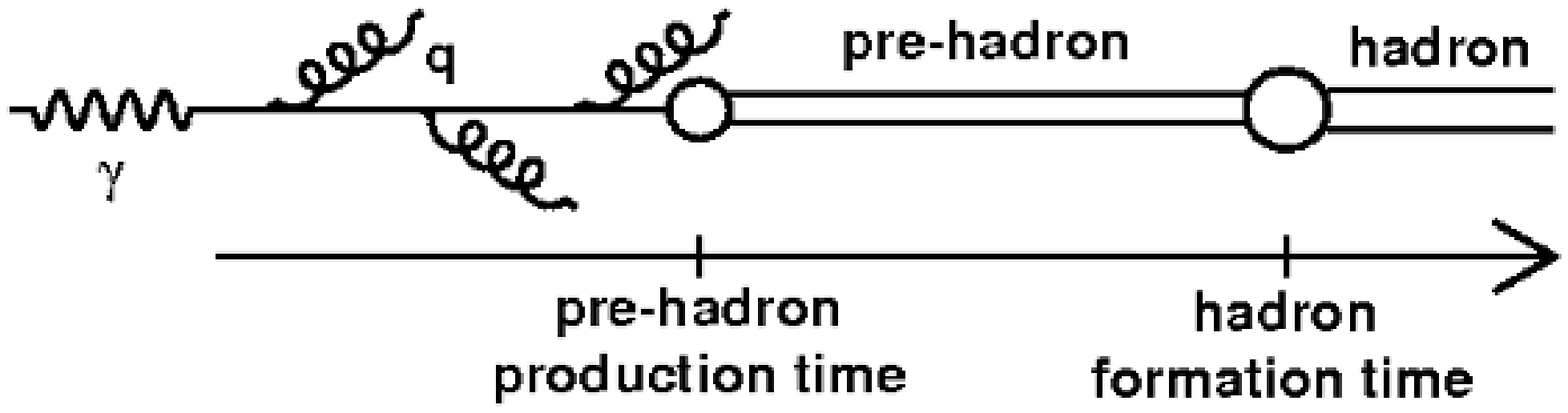} 
\caption {A model sketch of the hadronization process.}
\label{fig-AccDup:hadro}
\end{figure}

The fragmentation process, by which hard partons turn into hadrons, is only
partly known due to its non perturbative nature. 
Fragmentation functions, which encode
the probability that a parton fragments into a hadron, have been
obtained by fitting experimental data covering large kinematic ranges
and numerous hadron species, see Section \ref{sec:Sassot-FF}.
However, knowledge about the dynamics of hadronization
remains fragmentary: this process has been studied in a number of
model calculations, but lacks a first-principles description in QCD. 
One possible scenario for the hadronization process is sketched in figure~\ref{fig-AccDup:hadro} 
as an example for DIS. At LO the virtual photon strikes a quark,
which then propagates quasi-freely emitting gluons; after a time
called {\it  production time}, the quark neutralizes its color and
gluon emission stops. The quark becomes a pre-hadron, which will
eventually form a hadron at the {\it formation time}. In fact, a color
string connects the struck quark to its nucleon, and hadrons can be
formed all along this string, but we focus our attention on the hadron
that contains the struck parton.
In nuclear DIS, the hadronization process happens at
least in part in the target nucleus (cold nuclear matter). Thus the
quark is subject to 
energy loss by medium-induced gluon {\it brehmsstrahlung}, and the
prehadron (as well as the hadron) can have inelastic interactions with the
surrounding nucleons, leading to attenuation and
broadening of the produced particle spectra. The relative weight of
one mechanism compared to the other is determined by the magnitude of
the color neutralization time. For full reviews, see
Refs.~\cite{Accardi:2009qv,Majumder:2010qh,Albino:2008gy}.  Alternative scenarios are also feasible and final states in
 nuclear DIS (nDIS) can help untangle these from the scenario outlined here to provide genuine insight into the hadronization process.

These nuclear effects are both an opportunity for a first principles study  
hadronization and nuclear properties as well as important benchmarks for reducing existing  
uncertainties in many nuclear measurements. For example, in neutrino
experiments, nuclei are used to maximize the cross section and the
kinematics are reconstructed from the hadronic final state.  Therefore,
a poor knowledge of hadron attenuation leads to a tangible systematic error. 
In heavy-ion collisions,
hadrons are produced in hot and expanding nuclear matter, whose
properties can be measured, among other methods, by the modifications
of high-energy particle spectra compared to proton-proton and
proton-nucleus collisions. It is clear that the details and the time
scales of the hadronization process can profoundly modify the
interpration of the data, see Fig.~\ref{fig:intro_coldhot}.
\\

\noindent {\bf The role of $e+A$ collisions.} 
Nuclear deep inelatic scattering provides a known and stable
cold nuclear medium and a low-multiplicity final state with strong
experimental control on the kinematics of the hard scattering. This
permits one to use nuclei as femtometer-scale detectors and study the
time scales of the hadronization process and 
calibrate theoretical models for parton energy loss and prehadronic
scattering, that can then be applied, for instance, to the study of the QGP,
see Figure~\ref{fig:intro_coldhot}. Initial state parton energy
loss can furthermore be studied in isolation from hadronization in Drell-Yan
lepton pair production in $p+A$ collisions, where however it can be
masked by nuclear modification of the target wave function such as
antishadowing and the EMC effect. So, an interplay of nuclear DIS and nuclear
Drell-Yan can help isolate hadronization effects on one hand, and on
the other to clarify the differences in quark and anti-quark antishadowing. 
Perhaps more interestingly, the study of hadronization in nuclear DIS
can give direct information about the gluon structure of the nuclei. For
example, one can link energy loss and transverse momentum 
broadening to the gluon density \cite{Arleo:2002kh} or more directly to the 
saturation scale \cite{Kopeliovich:2010aa}. In models like GiBUU \cite{Gallmeister:2007an}, focussing on 
hadron absorption, access to the pre-hadron evolution and its color transparency
evolution is possible. All these physical interpretation
of the data are model dependent and based on very different 
assumptions about the relative importance of the interaction
mechanisms, therefore they are fragile and need to be carefully
validated and calibrated with precise data.

The typical observables used to explore hadronization in nuclear DIS are the 
multiplicity ratio and the transverse momentum broadening, in both cases they
are comparison of deuterium with heavier nuclei. The multiplicity ratio,
representing the production rate of a hadron $h$ in a nuclear
target~$A$ compared to Deuterium, is defined as
\begin{equation}
R_A^h (Q^2,\nu,z_h,p_T^2) = {{N_A^h (Q^2,\nu,z_h,p_T^2) / N_A^e (Q^2,\nu)} 
                       \over {N_D^h (Q^2,\nu,z_h,p_T^2) / N_D^e (Q^2,\nu)}}
\end{equation}
with $N^e_t$ and $N_t^h$ respectively the number of electrons and the number
of semi-inclusive hadrons $h$. $1 - R_A^h$ is the 
attenuation of hadron production in a nucleus of atomic
mass $A$. This ratio minimizes the influence of nuclear PDF
modifications, which have been shown to cancel to a large degree up to
NLO. The hadron transverse momentum broadening, representing the
increase of transverse momentum in a nuclear target A compared to
Deuterium, is defined by $\Delta \langle p_T^2 \rangle = \langle p_T^2 \rangle_A - \langle p_T^2 \rangle_D$, 
with $\langle p_T^2 \rangle_t$ the average hadron transverse momentum
measured in a nucleus. When integrated over a large kinematic range,
these observables  they are dominated by the geometry of the nuclei
and do not discriminate well between the models. One needs to
also consider more differential observables, including a
multi-dimensional analysis of $R_M$ and $\Delta \langle p_T^2
\rangle$, and hadron-hadron and photon-hadron correlations. 

Another possibility is to use experimental settings in which we can
isolate the involved processes. In the 
case of EIC, the high energy boost imparted to the struck quark in
events with large $\nu$ can increase dramatically the production length,
which leads to pre-hadron production far outside the nuclei and an
experimental isolation of pure parton energy loss effects. Since the
pre-hadron production time is expected to roughly be inversely
proportional to the mass squared of the hadron, measuring
attenuation and $p_T$-broadening of many meson and baryon species,
together with the large $\nu$ leverage afforded by the EIC, will give
another important handle in the exploration of the hadronization mechanism. New features will be availabe at the EIC, the high rate for heavy
flavor production ($D$ and $B$ mesons) will allow
the measurement of heavy quark energy loss. Finally, jet production, 
will open the possibility to study the dynamics of parton showers and
the detailed transport properties of cold nuclear matter using
specific jet observables.

\noindent{\bf Overview of theoretical models.}
Three processes are typically included in theoretical
descriptions of in-medium  hadronization: quark energy loss, pre-hadron
absorption and modified fragmentation functions. The models in the
literature are usually 
based on one or two of those and neglect the others. In this
section we will discuss a few examples to give an idea  
of the abundant existing literature; for a detailed review, including 
models specific to heavy-ion collision experiments, see
Ref.~\cite{Accardi:2009qv}.

Pure quark energy loss models assume a very long production time and
are typically used to describe hadron suppression in the hot nuclear
matter produced in heavy-ion collision. In a few cases they have been
applied to nDIS data as well \cite{Wang:2002ri,Accardi:2006ea} permitting a 
common interpretation of hadron suppression in cold and hot nuclear matter.
In these models, hadron suppression is due to the lower energy of the quark 
when it fragments, so that hadrons are produced in lower
number and at lower energy.
The differences in the models depend on the way calculations of
medium-induced gluon radiation are performed, on the modeling of the
medium, and on assumptions about its coupling to the hard parton.

Typically, parton energy loss is determined by the transport
coefficient $\hat q$, which is defined as the transverse momentum
square transfered to a quark after propagating through a length of
nuclear matter and is a characteristic property of that matter. It is expected
to be much larger in a Quark-Gluon Plasma than in the nucleus
of a nDIS experiment, which is what is observed from the analysis of
experimental data from RHIC and HERMES
\cite{Wang:2002ri,Accardi:2006ea,Adare:2008cg,Bass:2008rv}.
The $\hat q$ transport coefficient is directly related to the observed
broadening of the $p_T$ distribution of hadrons in nDIS; it  
follows that the main challenge for these pure energy loss models is to make a 
coherent picture of both multiplicity ratios and hadron $p_T$
broadening. In particular, for some of the models, the $\hat q$
extracted from multiplicity ratios is larger by an order of magnitude
than what one would estimate from the hadron transverse momentum
broadening. This has led some authors \cite{Kopeliovich:2003py} 
to the conclusion that quark energy loss is not enough to explain 
the observed nuclear effects; nevertheless, the variation between theoretical 
models is still too big for a definitive statement.

The GiBUU model \cite{Gallmeister:2007an} is an absorption model based
on Boltzmann equation including only hadronic and pre-hadronic
interactions, see Section~\ref{sec:e+A_with_GiBUU}. It assumes
short productions times obtained from the Lund string model and
neglects gluon brehmstralung from the partonic stage.
It can describe very well well most of the hadron multiplicity
ratios measured at HERMES and EMC using
a linear growth of the pre-hadron cross section between production time and 
formation time. Other pure absorption models
\cite{Accardi:2002tv,Accardi:2005jd,Accardi:2005hk} are also successful in 
describing hadron attenuation. However, the transverse momentum
broadening remains a challenge for this kind of models; some progress
within GiBUU has been presented during the meeting by Kai
Gallmeister.

To resolve the problems of the previous ``pure'' models, Kopeliovich 
{\it et al.} \cite{Kopeliovich:2003py} describe hadronization
including both quark energy  
loss and hadron absorption. In their model, the transverse momentum broadening 
is linked to quark energy loss and the multiplicity ratio suppression is 
explained by hadron absorption, therefore the two processes can be 
independently quantified. This model describes HERMES data to a large
extent, and highlights the fact several processes are involved and
need to be disentangled. 

Recently, HERMES data have also been described by assuming
factorization and universality to
hold at the nuclear level not only for parton distributions but also
for fragmentation functions, and a set of nuclear Fragmentation
Functions have been fitted to experimental data using both $e+A$
interactions and $d+Au$ collisions at RHIC. In this case,  no
dynamical assumption is made of the physical mechanism for
nuclear modifications of hadron production; this information is 
subsumed into the non-perturbative nuclear FFs--see Section \ref{sec:Sassot-FF}.

A number of other models exist using different variants of the
discussed mechanisms, and most of them are 
able to describe the existing data to a good extent: no 
consensus is reached yet on which mechanisms are dominant, and indeed
this is the main motivation for future precise measurements of hadronization at
Jefferson Lab \cite{JLab-PR12-06-117},  which will be completed by the time EIC starts its operations, and will help
settle some of the issues related to early time color dynamics and
interaction in cold nuclear matter.

\noindent{\bf Previous mesurements and open questions.}
Unidentified charged hadron multiplicity ratios in nuclei were
measured in numerous lepton facilities, the earlier results were by 
Osborne {\it et al.} \cite{Osborne:1978ai} at SLAC, Hand {\it et al.}
\cite{Hand:1978tx} and the E665 collaboration \cite{Adams:1994ri} at
FNAL, and the 
European Muon Collaboration \cite{Arvidson:1984fz,Ashman:1991cx} at CERN. Those 
measurements revealed a general picture: hadron suppression is
stronger at low $\nu$ and high $z$. On the opposite side, at low $z$,
an increase of the number of hadron is observed.

\begin{figure}[tb]
\centering
\includegraphics[width=0.48\linewidth]{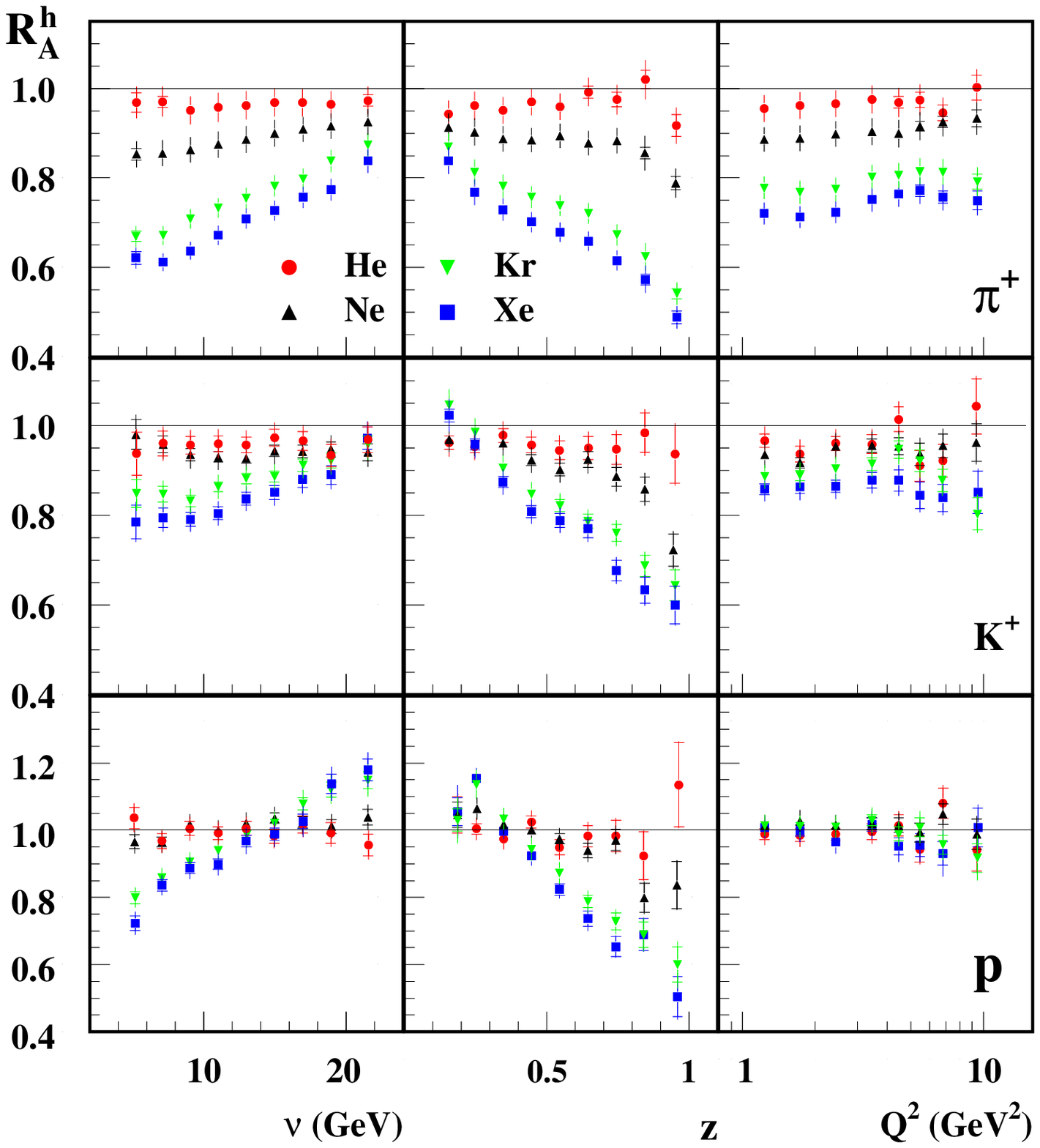} 
\hspace*{0.3cm}
\includegraphics[width=0.48\linewidth]{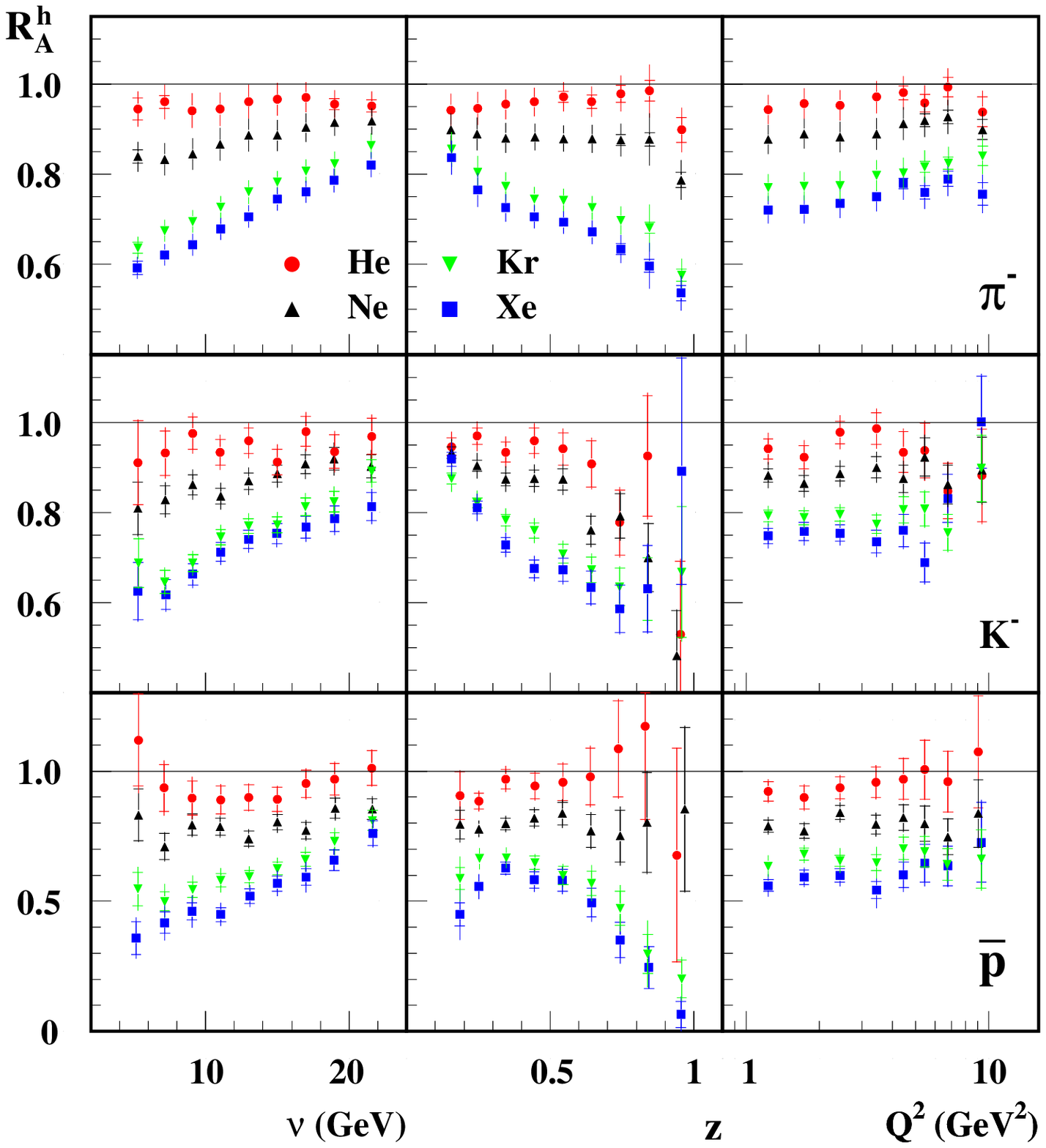} 
\caption {
Multiplicity ratio of positively charged hadrons (left) and
  negatively charged hadrons (right) from the HERMES experiment
  \cite{Airapetian:2007vu}
} 
\label{fig-AccDup:her1}
\end{figure}

In the more recent data from the HERMES collaboration
\cite{Airapetian:2007vu,Airapetian:2009jy}  
several hadrons are studied individually (Fig.~\ref{fig-AccDup:her1}), and new 
observables such as transverse momentum broadening (figures 
\ref{fig-AccDup:ptz} and \ref{fig-AccDup:rQ2evo}) and two hadrons multiplicity 
ratios \cite{Airapetian:2005yh} are measured.  Because of their
improved precision and the  
large number of hadron species, these data provide us today with a much more 
detailed picture, which leads to new questions.  The behavior of the kaons, 
for example, is very interesting: K$^+$ are less suppressed than pions, but 
K$^-$ have the same behavior as pions (figure \ref{fig-AccDup:her1}). This 
difference is not reproduced by existing models, showing
that the relatively simple phenomenological models utilized so far cannot 
fully describe the data. Furthermore, the introduction by HERMES of precise 
and flavor dependent $\Delta p_T^2$ measurement
\cite{Airapetian:2009jy} has revealed another strange 
behavior: the $p_T$ broadening of $K^+$ is larger than for the pion 
(Figure~\ref{fig-AccDup:ptz} right). This seems  to indicate more
interaction for  
kaons, and yet they are less suppressed. To solve this apparent incongruity, 
one may have to consider models based on different processes involved at 
different stages of hadronization, like in Reference \cite{Kopeliovich:2003py}, reinforcing 
the indications coming from kaon suppression. Furthemore, no model is able to 
describe the $z$ dependence of the $p_T$ broadening, highlighting once again 
the need for a more detailed theoretical understanding of hadronization. 
Finally, proton observables are very different from anti-protons (figure 
\ref{fig-AccDup:her1}), and no model  is yet able to reproduce them 
correctly, although few attempts have been made
\cite{Gallmeister:2007an,CiofidegliAtti:2004pv}.  
At the low energies of HERMES, part of the problem may be due to protons coming 
from the target fragmentation region, which is interesting in its own right. 
The collider geometry and the large 
energy range of EIC will permit to experimentally separate clearly target and 
current fragmentation, allowing to address hadronization in either
region. Indeed developing a consistent picture within a given model for both
current and target fragmentation would be a great theoretical progress.

To complete the review of existing data, we should mention the preliminary 
results on pion and kaon production from the CLAS collaboration at
Jefferson Lab, where electrons up to 5 GeV scatter
on fixed targets ranging from Carbon to Lead
\cite{Hicks:2009wf,Brooks:2009xg}.

\subsubsection{Studying hadronization at an EIC}
\label{sec:AccardiDupre_EIChadronization}
\hspace{\parindent}\parbox{0.92\textwidth}{\slshape 
Rapha\"el Dupr\'e and Alberto Accardi 
}
\index{Dupr\'e, Rapha\"el}
\index{Accardi, Alberto}

\vspace{\baselineskip}

The experimental study of the hadronization process using nDIS is well 
established; however the high energy available at the EIC creates novel 
opportunities. The main interest in going at higher energy is to ensure that 
hadron formation occurs outside of the nuclei, in order to isolate in-medium 
parton interactions and energy loss. Furthemore, an EIC will permit, for the first 
time in $e+A$ collisions, the study of hadronization of the open charm and 
eventually open bottom mesons. Recent results from RHIC
\cite{Adare:2006nq,Abelev:2006db} are 
showing unexpected results for open charm and bottom suppression in $A+A$ 
collisions, and several contrasting explainations have already been suggested, 
with more detailed experiments planned at RHIC. However, due to the intricated 
interplay of the many variables in $A+A$ collisions and to the poorly known 
nature of the Quark-Gluon Plasma partons, the $e+A$ input seems necessary to 
confirm any interpretation. Also, the considerable energy leverage offered by 
an EIC is a chance to map precisely the $Q^2$ evolution of parton energy 
loss, and determine possible nuclear modifications of DGLAP evolution. The 
high luminosity will also facilitate the study of two particle correlations (such 
as hadron-hadron or photon-hadron) over a wide energy range, largely improving 
recent HERMES measurements, and complementing the low-energy measurements 
planned at CLAS. Finally, high energy permits access to jets, which give an 
opportunity to use new observables with improved sensitivity to quark energy 
loss and the medium modification of fragmentation functions, see
Section~\ref{sec:eAjets}. They also 
facilitate a detailed determination of the cold nuclear matter transport 
coefficients, which encode basic information on the non perturbative gluonic 
structure of the nuclei and can be calculated from first 
principles, e.g., in lattice QCD \cite{Majumder-INT}.

To illustrate the possibilities offered by EIC, we show projections done using
the PYTHIA Monte-Carlo generator to evaluate cross sections at 
$s = 200$ or $1000$~GeV$^2$, and $L = 200$~fb$^{-1}$. We
apply a series of cuts on the generated events to ensure the DIS nature of the 
interaction ($Q^2 > 1$ GeV$^2$ and $W^2 > 4$ GeV$^2$), to limit radiative
corrections ($y < 0.85$), to insure that we can detect the scattered
electron ($y > 0.1$) and to limit di-parton production in the hard  
scattering of the virtual photon ($x_{Bj} > 0.1$). Finally we assume
an acceptance of 50\% for pions, eta meson and kaons, and, an
acceptance of 2\% for heavy mesons. The acceptance is set low for
heavy mesons to account for the small number of decay channels that
can be effectively detected. EIC observables are plotted on arbitrary
vertical scales, and include statistical errors only.

\begin{figure}[tb]
\centering
\includegraphics[height=5.5cm]{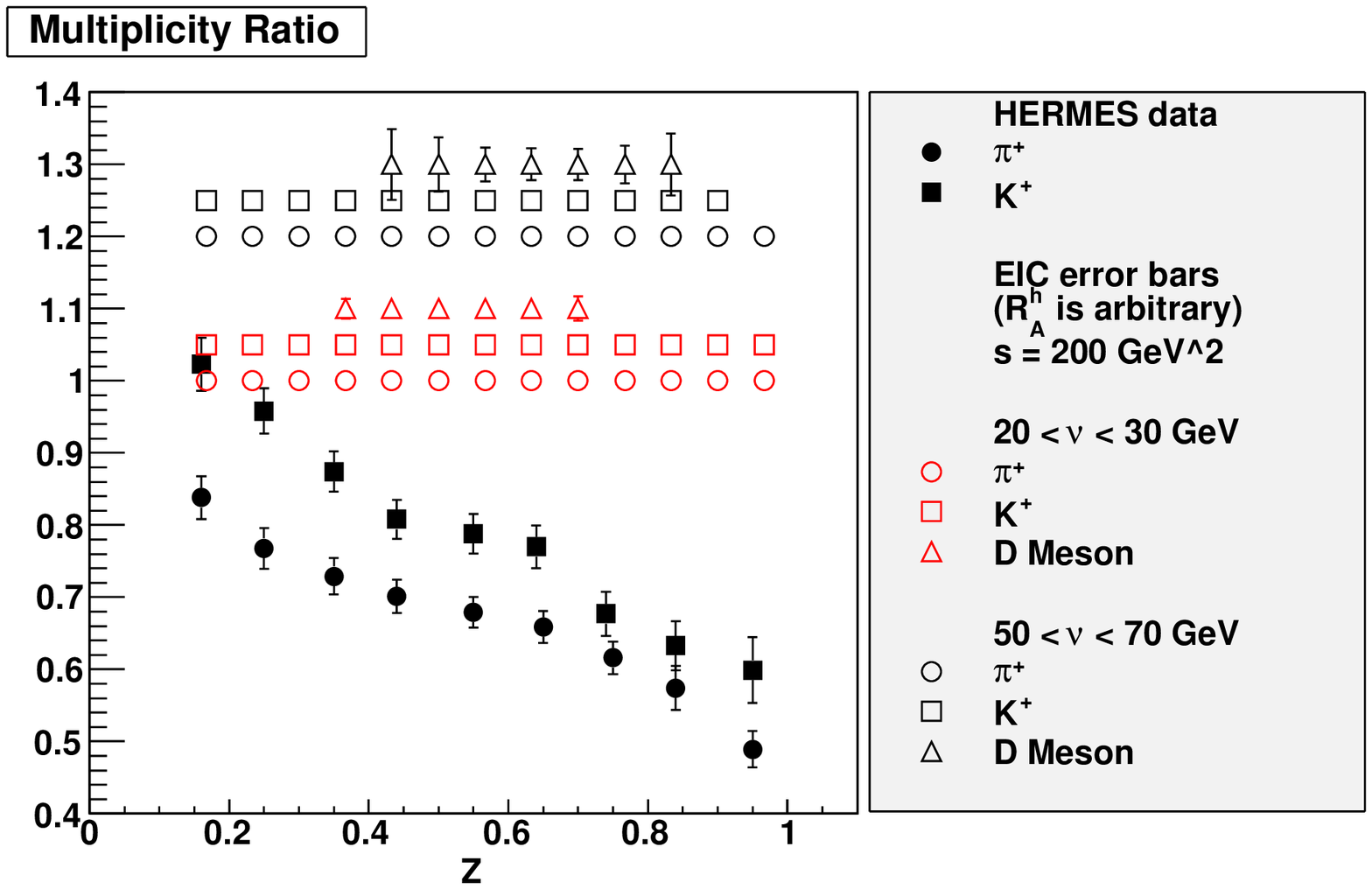} 
\hspace*{0.3cm}
\includegraphics[height=5.5cm]{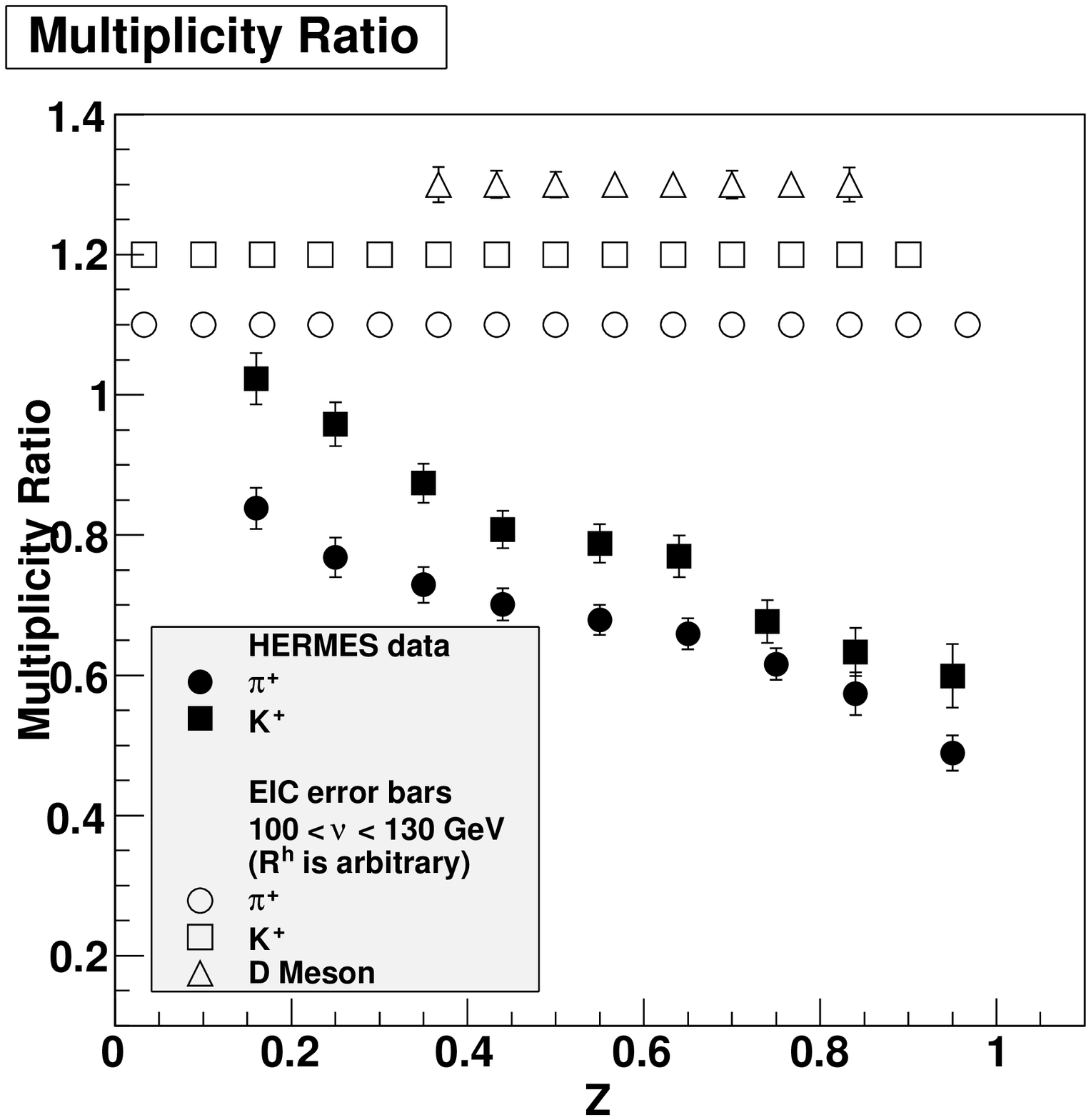} 
\caption {
  Multiplicity ratio in function of z for various $\nu$ bins. 
  Full points are data from HERMES \cite{Airapetian:2007vu}, empty are projections
  for statistical errors at the EIC, at arbitrary vertical position.
  The left panel shows EIC measurements at $s = 200$GeV$^2$, for 2
  different $\nu$ cuts ($20 < \nu < 30$~GeV and $50 < \nu < 70$~GeV); 
  the right panel at $s = 1000$GeV$^2$ with $100 < \nu < 130$~GeV.
}
\label{fig-AccDup:rlowe}
\end{figure}

\begin{figure}[tb]
\centering
\includegraphics[height=5.5cm,bb=60 430 580 720,clip=true]{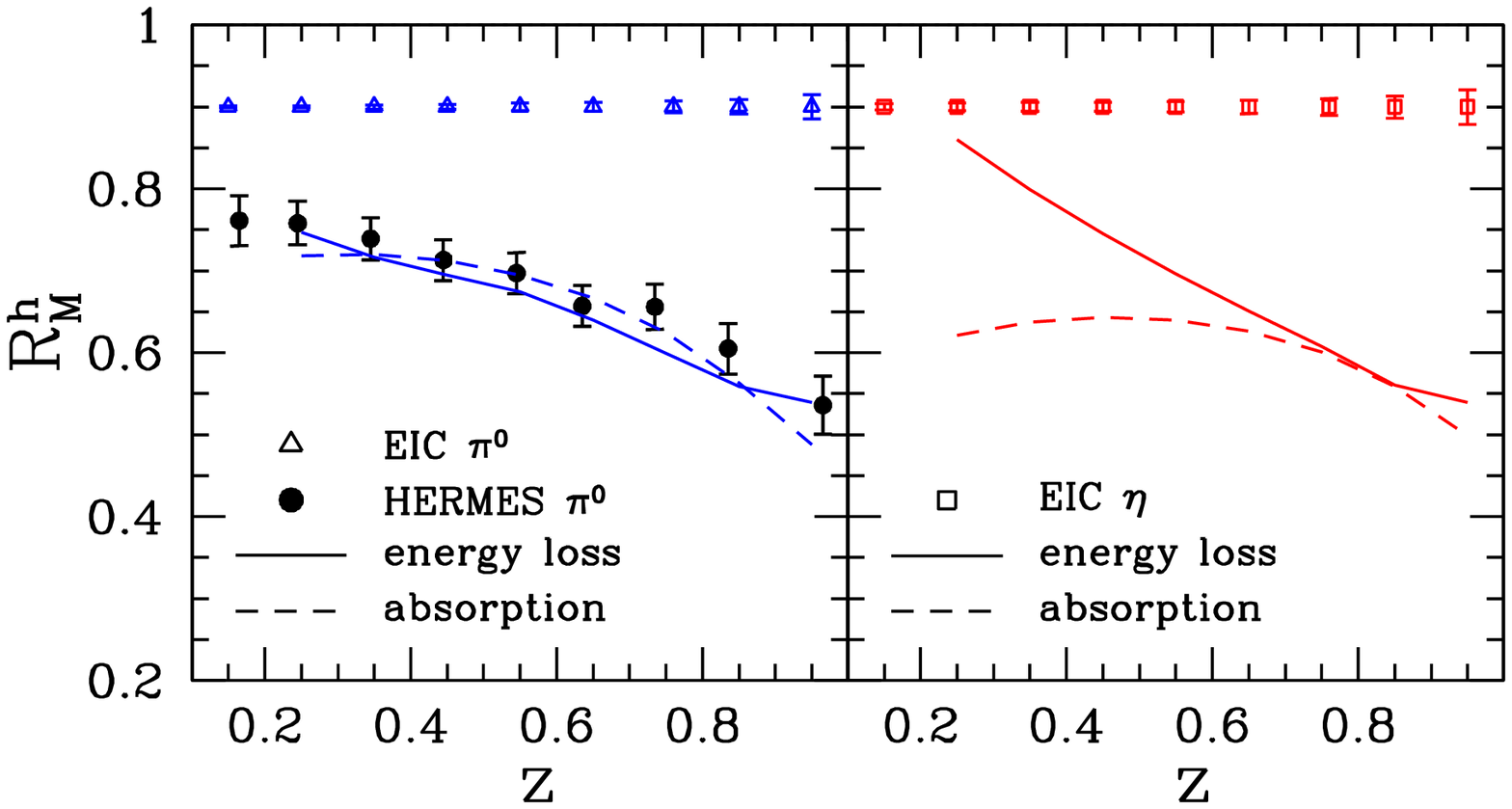}
\caption {
  Multiplicity ratio for $\pi^0$ and $\eta$ mesons compared to pure energy
  loss and pure prehadron absorption computations.
}
\label{fig-AccDup:etapi0}
\end{figure}

An EIC is the perfect tool for precise measurement of quark energy 
loss and transverse momentum broadening. One may object that at 
the higher EIC energies, because of the large $\nu\gtrsim 150$
GeV, the relative effect on the quark momentum is too little to
produce an appreciable hadron attenuation. This is true at least for 
the pions, as shown by EMC data. However, attenuation may in fact
disappear at a yet higher value of $\nu$ for large $z$ or
for heavier particles, because of reduced production times, or for
large $Q^2$, because of a faster evolution in virtuality as discussed
in Section~\ref{sec:eA-AbhijitMC}. 
Anyway, because of the EIC kinematic flexibility, interesting 
multiplicity ratios can be measured. For example,
Figure~\ref{fig-AccDup:rlowe} shows projections 
for light and heavy flavors, which would shed light on the heavy
quarks at RHIC, where they unexpectedly display a similar suppression
compared to their light counterparts. It is also interesting to
compare mesons of different 
mass but the same valence quark contents, such as $\pi^0$ vs. $\eta$, and
$K^0$ vs $\Phi$. Figure~\ref{fig-AccDup:etapi0} shows projections for
the former case compared to calculations in a pure energy loss or pure
prehadron absorption scenario. The sensitivity of such measurement to the
hadronization time scales is obvious.

\begin{figure}[tbp]
\centering
\includegraphics[height=5.5cm]{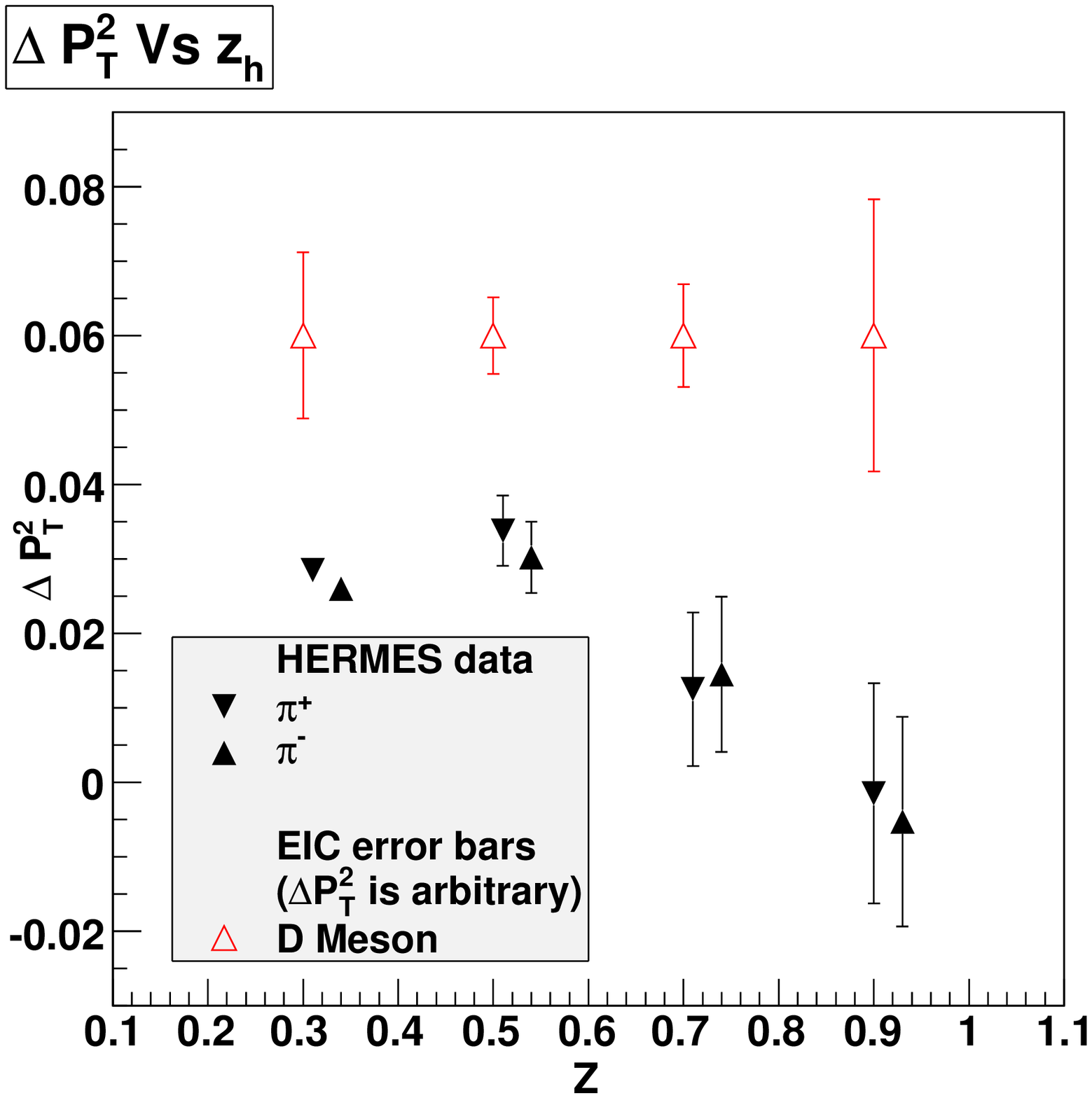} 
\hspace*{0.3cm}
\includegraphics[height=5.5cm]{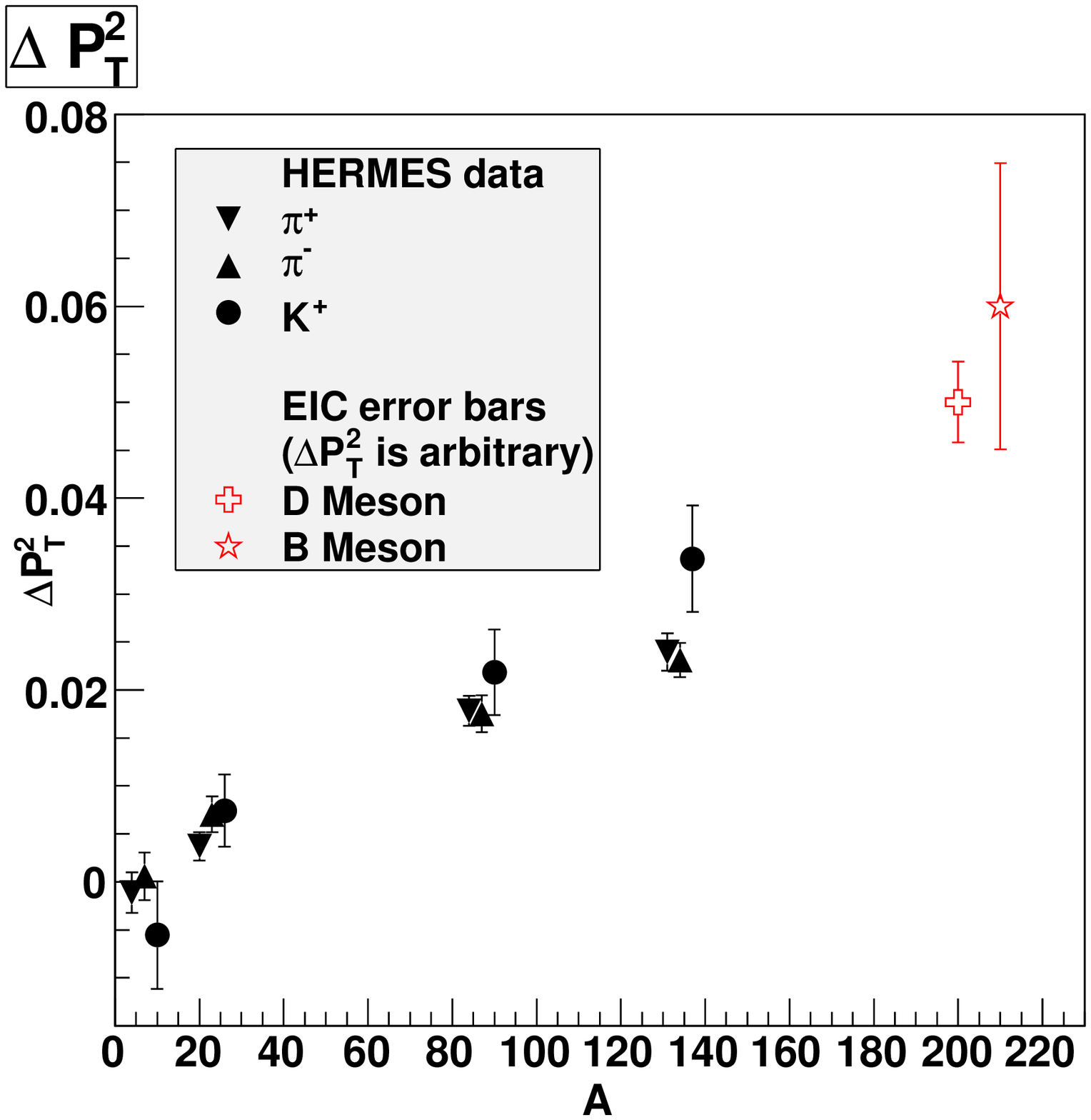} 
\caption {
  Transverse momentum broadening in function of z (left) and A (right),
  empty triangles and star are projections for EIC at $s = 1000$GeV$^2$, 
  full points are HERMES data.
}
\label{fig-AccDup:ptz}
\end{figure}

Changing observables, measurements of the hadron transverse momentum
broadening permit getting around the small values of hadron attenuation at
large energies. Indeed the $p_T$ broadening to first approximation is
independent of $\nu$, and even very little effects can be
experimentally observed; moreover, the induced transverse momentum has
a theoretical interpretation in terms of transport
coefficients. However, one should keep in mind that $\Delta \langle
p_T^2 \rangle$ of pions or other hadrons is not a direct measurement
of $\hat q$, which is the parton transverse
momentum broadening, and that it is essential to use 
dependences in $\nu$ and $z$ to make a model independent extraction of
$\hat q$. One may also access $\hat q$ through nuclear modifications
of hadron azimuthal asymmetries, see Section~\ref{sec:nuclearTMDs}.
The importance of this topic, especially in the scope of 
other EIC measurements, is enhanced by the connection between $\hat q$ and
the saturation scale \cite{Kopeliovich:2010aa}, enabling an
independent large-$x$ measurement of the latter, complementary to the
more traditional small-$x$ measurements discussed in
Section~\ref{sec:eA_smallx_part}. 
An EIC will not only allow one to make those measurements with pions
but also, and uniquely compared to previous e+A facilities, with heavy
mesons (see figures \ref{fig-AccDup:ptz}).

\begin{figure}[tbp]
\centering
\includegraphics[height=5.5cm]{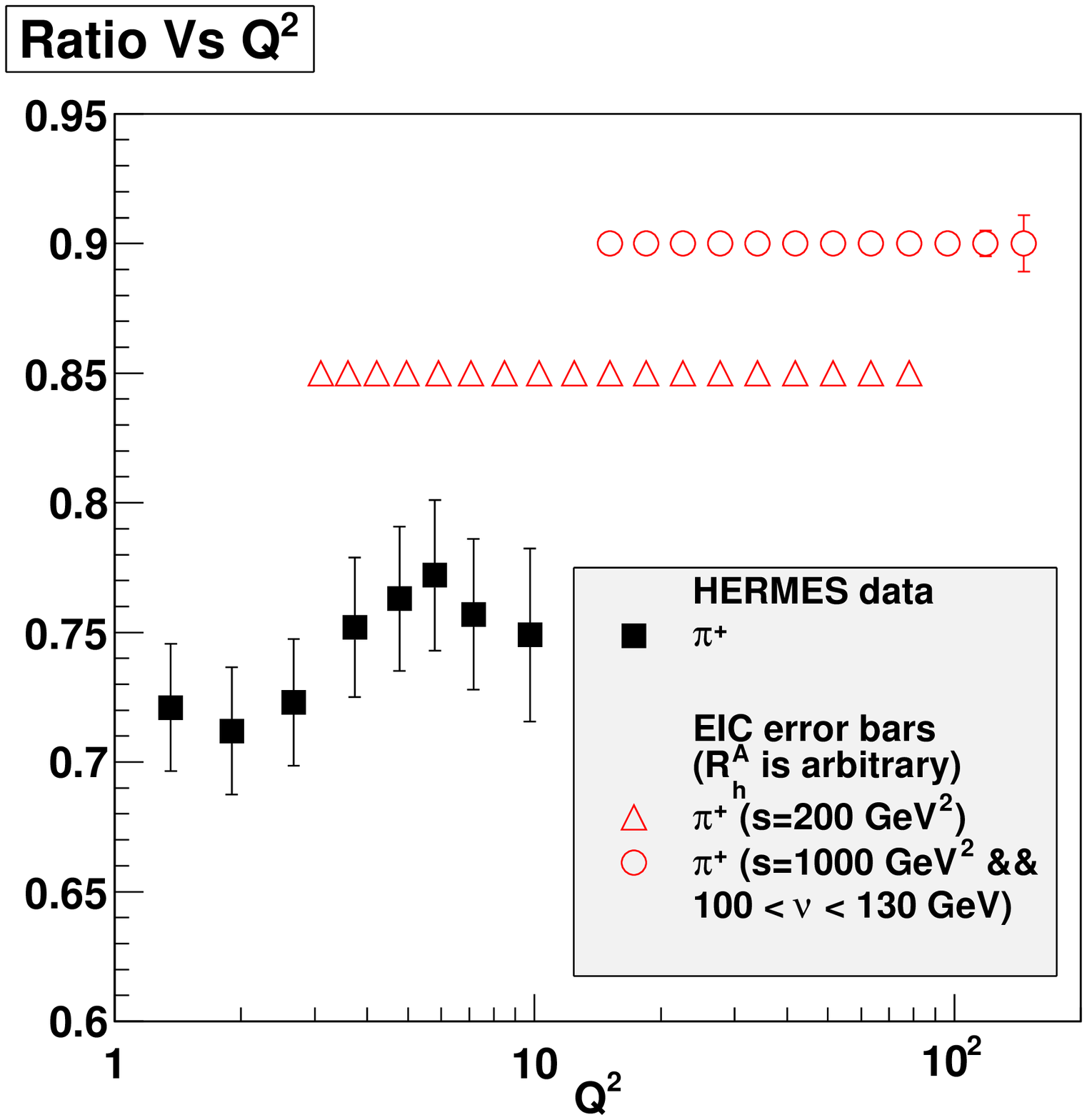} 
\hspace*{0.7cm}
\includegraphics[height=5.5cm]{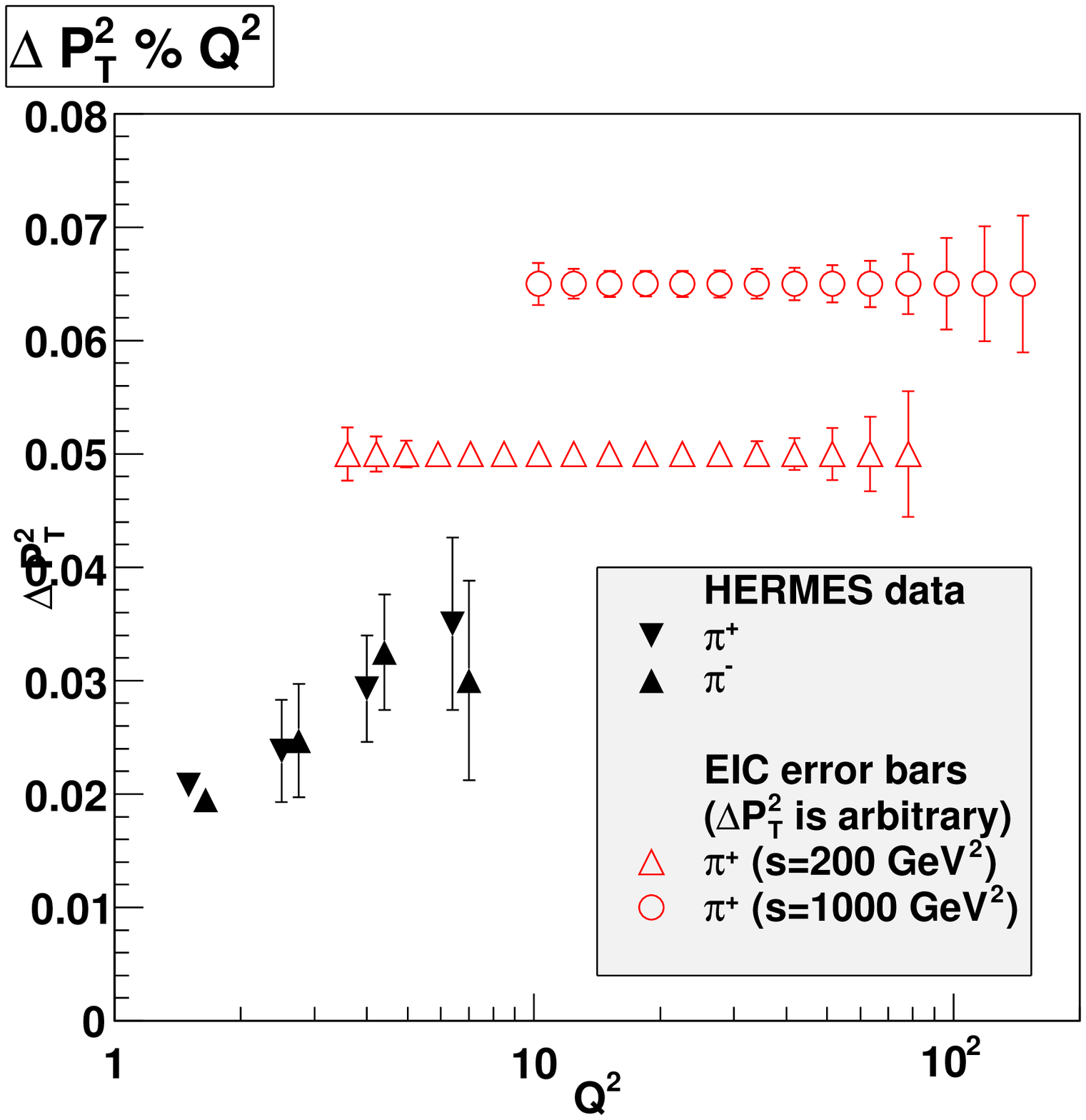} 
\caption {
  Multiplicity ratio (left) and transverse momentum broadening (right) 
  in function of $Q^2$, empty markers are projections for EIC at $s = 200$GeV$^2$ 
  (triangles) and at $s = 1000$GeV$^2$ (circles), full markers are HERMES data.
}
\label{fig-AccDup:rQ2evo}
\end{figure}

The $Q^2$ evolution of hadron attenuation is not clearly understood:
HERMES data indicate a small rise of the transverse momentum
broadening, but the $Q^2$ coverage is not large enough to make a
definite statement. An EIC can do a far better job as shown in figure
\ref{fig-AccDup:rQ2evo} and provide a unique probe to detect any
modification of the DGLAP evolution in nuclear medium. 

The scaling of the hadronization times and the quark energy loss 
with the mass of quarks is an important question that can be used to reveal 
pQCD effects in parton energy loss and
non perturbative effects in hadronization
\cite{Vitev:2008jh,Horowitz:2008ig}. Many measurements to 
explore this at the EIC are 
possible, as the figures in this section illustrate.

To achieve the discussed measurement the key experimental requirement
are good particle ID in general; for heavy flavors one needs in
particular a very good vertex detector resolution, which needs to be
of the order of few tens of micrometer, and high luminosity to reach
a statistical precision allowing unambigous theoretical
interpretations. Having a $\nu$ range covering low values for
studies of hadronization and large values for studies of parton
propagation and energy loss will require energies spanning
$s=200-1000$ GeV$^2$. The lowest required energy can be
increased provided measurements of $y<0.1$ can be achieved for SIDIS
observables. 

Finally, the high energy of an EIC provides the chance, for the first
time in $e+A$ collisions, to study hadronization through jet observables. Jets are a
new and independent way to access transport coefficient $\hat q$ and
confirm other measurements, to explore in detail the medium induced
gluon radiation and transport properties of cold nuclear matter, and
to study the conversion of the parton shower into hadrons, see
Section~\ref{sec:eAjets}.

\subsubsection{Hadronization in $\bm e+A$ collisions within GiBUU}
\label{sec:e+A_with_GiBUU}

\hspace{\parindent}\parbox{0.92\textwidth}{\slshape 
Kai Gallmeister and Ulrich Mosel 
}
\index{Gallmeister, Kai}
\index{Mosel, Ulrich}

\vspace{\baselineskip}

The study of the interaction of hadrons, produced by elementary probes
in a nucleus, with the surrounding nuclear medium can help to investigate
important topics, such as color transparency and hadronization
time scales. We investigate this by means of the
semiclassical GiBUU transport code \cite{GiBUU}, which not only allows
for the absorption of newly formed hadrons, but also for elastic and
inelastic scattering as well as for side feeding through coupled
channel effects.  A study of parton interactions in cold, ordinary
nuclear matter of known properties is important to disentangle effects
of the interaction of partons from those of the medium in which they
move.

\begin{figure}
  \centering
  \includegraphics[width=0.8\textwidth]{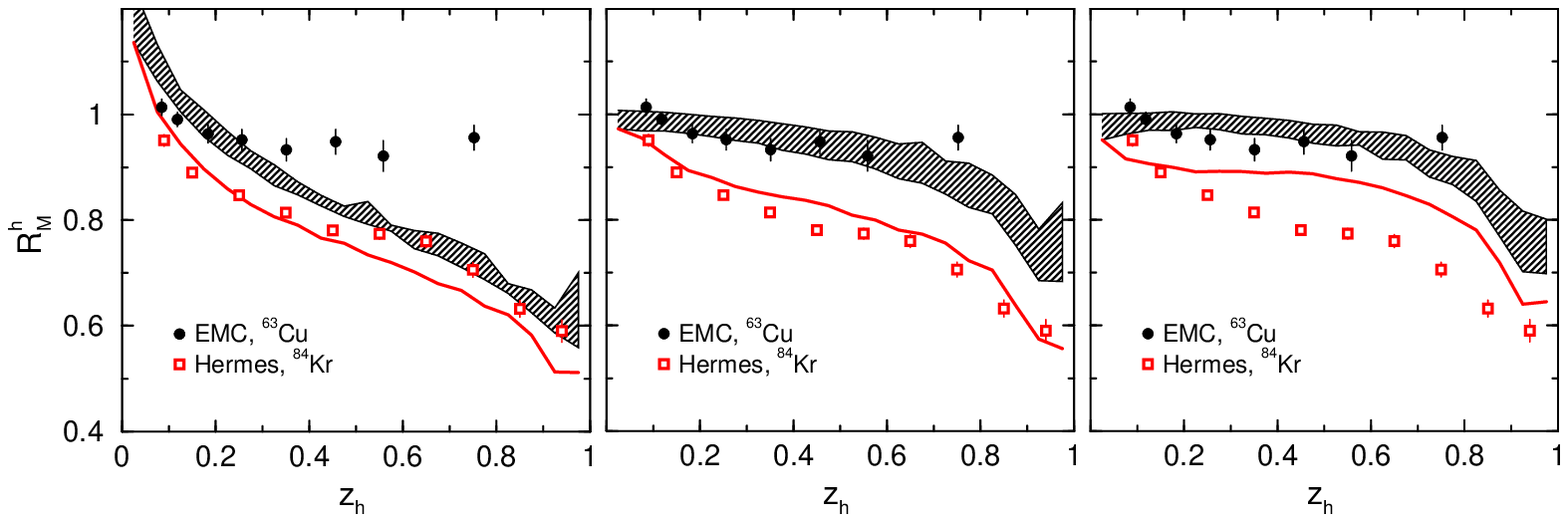}
  \caption{\small Nuclear modification factor for charged hadrons. 
    Experimental data are shown for HERMES at 27 GeV and 
    for EMC at 100-280 GeV.  The cross section scenarios are (from
    left to right): constant, linear and quadratic increase with time
    after production.} 
  \label{fig:GiBUU_att}
\end{figure}

We summarize here the main features of our model, for details see
\cite{Gallmeister:2007an}. The model relies on a factorization of
hadron production into the 
primary interaction process of the lepton with a nucleon, essentially
taken to be the free one, followed by an interaction of the produced
hadrons with nucleons. We have modeled the prehadronic interactions
such that the description is applicable at all energy regimes and
describes the transition from high to low energies correctly.
For the first step, we use the PYTHIA model that has been proven to
very successfully describe hadron production, also at the low values
of $Q^2$ and $\nu$ treated in our studies. This model contains not
only string fragmentation but also direct interaction processes such
as diffraction and vector-meson dominance. In this first step, we take
nuclear effects such as Fermi motion, Pauli blocking and nuclear
shadowing into account \cite{FalterPHD}. The relevant production and
formation times \cite{Gallmeister:2007an} are obtained directly from PYTHIA
\cite{Gallmeister:2005ad}. In the second step we introduce
prehadronic interactions between the production and the formation time
and the full hadronic interactions after the hadron has been formed.

The actual time dependence of the prehadronic interactions presents an
interesting problem in QCD.  Dokshitzer et
al.~\cite{Dokshitzer:1991wu} have pointed out that QCD and quantum
mechanics lead to a time-dependence somewhere between linear and
quadratic. We also note that a linear behavior has been used by Farrar
et al.~\cite{Farrar:1988me} in their study of quasi-exclusive
processes. In our calculations, we work with different time-dependence
scenarios, among them a constant, lowered pre-hadronic cross section, a
linearly rising one, and a quadratically rising one. In addition, we
study a variant of the latter two, where the cross section for leading
hadrons, i.e., hadrons that contain quarks of the original target
nucleon, starts from a pedestal value $\sim 1/Q^2$, thus taking into
account possible effects of color transparency.

\begin{wrapfigure}{r}{0.66\textwidth}
  \begin{center}
    \includegraphics[width=0.65\textwidth]{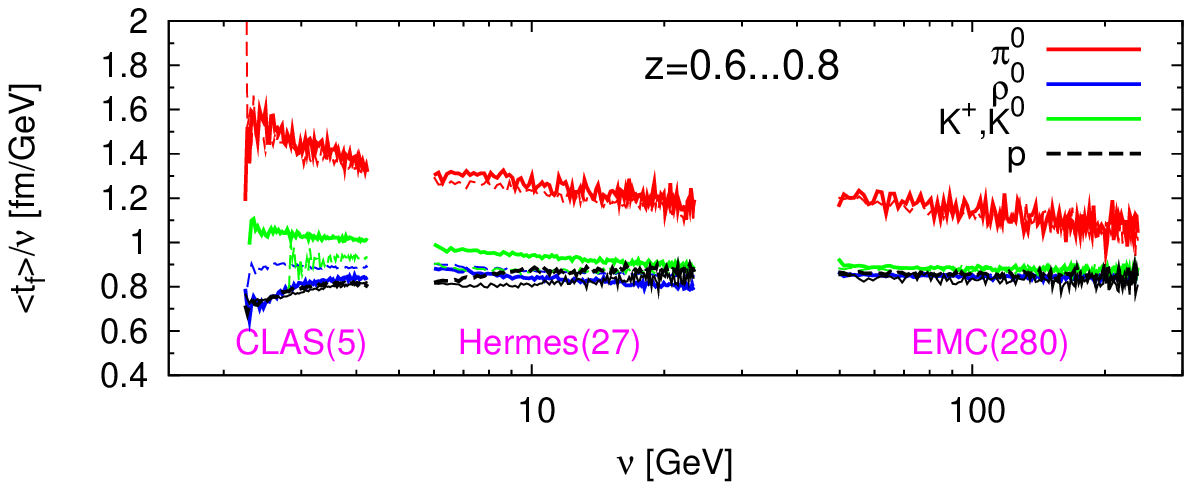}
  \end{center}
  \caption{\small The average formation time of different particles divided by $\nu$ as a function of $\nu$ for several experimental setups.}
\label{fig:GiBUU_AveTimes}
\end{wrapfigure} 

Fig.~\ref{fig:GiBUU_att} shows a comparison of these various model
assumptions to HERMES and EMC data on unidentified charged hadron
attenuation.  
A good description of both data sets simultanousely is obtained only
with a linear time dependence of the cross sections. Furthermore,
a nearly perfect agreement is observed in HERMES data for pions, kaons, and protons, which
give the attenuation $R_M$ as a function of energy transfer $\nu$,
relative energy $z_h = E_h/\nu$, momentum transfer $Q^2$ and the
squared transverse momentum $p_T^2$ \cite{Airapetian:2007vu}. 
The rise of $R_M$ with $\nu$ is mainly
an acceptance effect, as we have shown in \cite{FalterPHD}, whereas
the weaker rise of $R_M$ with $Q^2$ reflects the pedestal value $\sim
1/Q^2$ of the pre-hadronic cross sections.

In Fig.~\ref{fig:GiBUU_AveTimes} we show the average formation time for
different particle species as a function of the boson energy $\nu$.
One realizes a smooth transition from CLAS at $5\gev$ up to EMC at
$280\gev$ for all particle species.
One observes a somehow larger formation time for pions than for the heavier particles. Nevertheless, this effect, being somewhere
on a 50\% level, is much smaller than mass ratios would suggest:
$m_N/m_\pi\sim 7$. Thus, recalling the basic boost relation, 
$t_h=\gamma_h \tau_h=(E_h/m_h) \tau_h$, 
the factor $\tau_h$ and the factor $m_h$ in
the nominator/denominator cancel each other.
We therefore conclude that, within our model, the formation time of a
hadron in its rest frame is proportional to its mass, $\tau_{f} \propto
m_H$, contrary to common assumptions of a constant formation time for
all hadron species, which can also be obtained 
from uncertainty principle considerations
\cite{Vitev:2008jh,Accardi:2009qv}.


\begin{figure}[htb]
  \centering
  \includegraphics[width=0.55\textwidth]{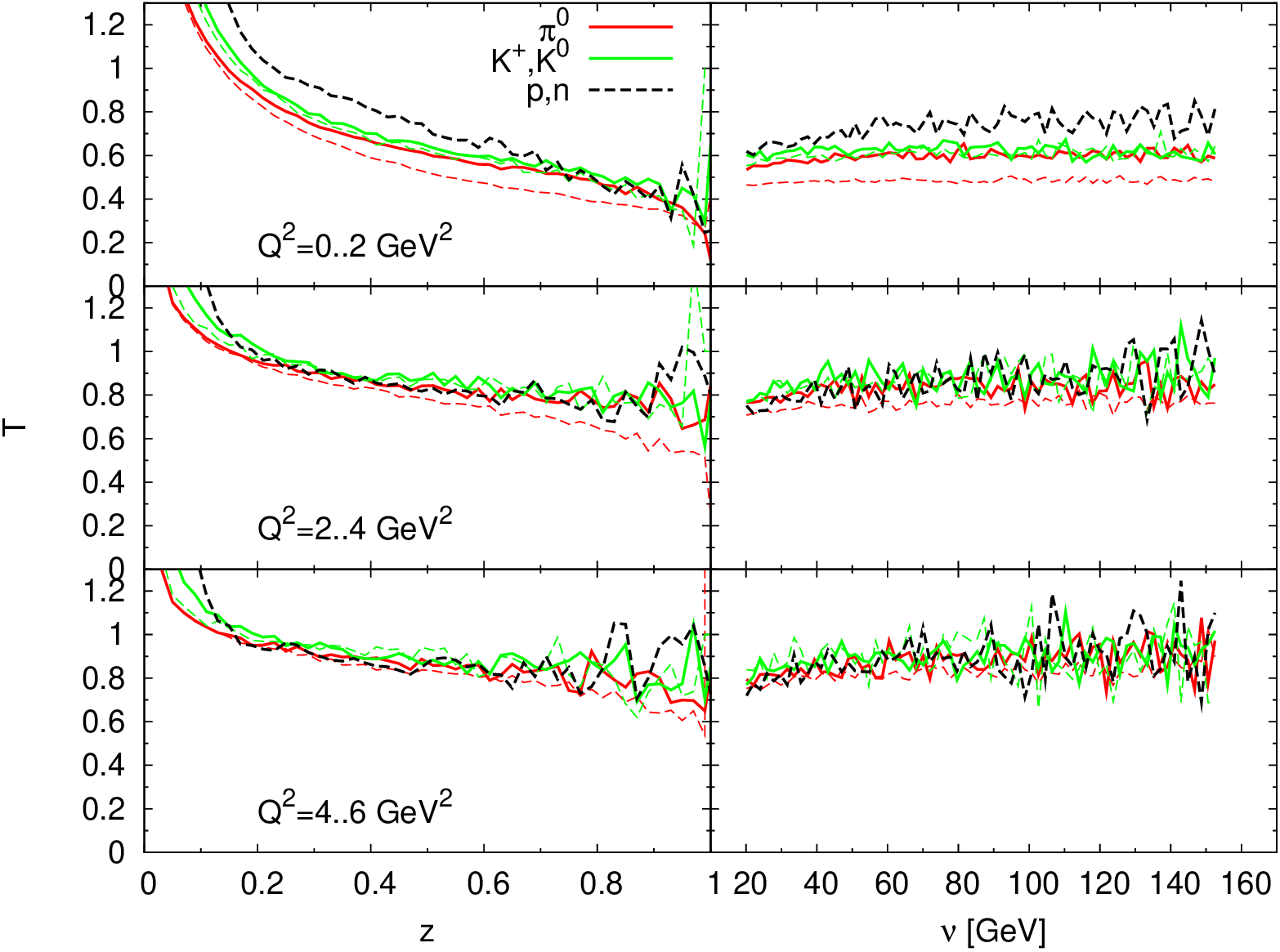}
  \includegraphics[width=0.39\textwidth]{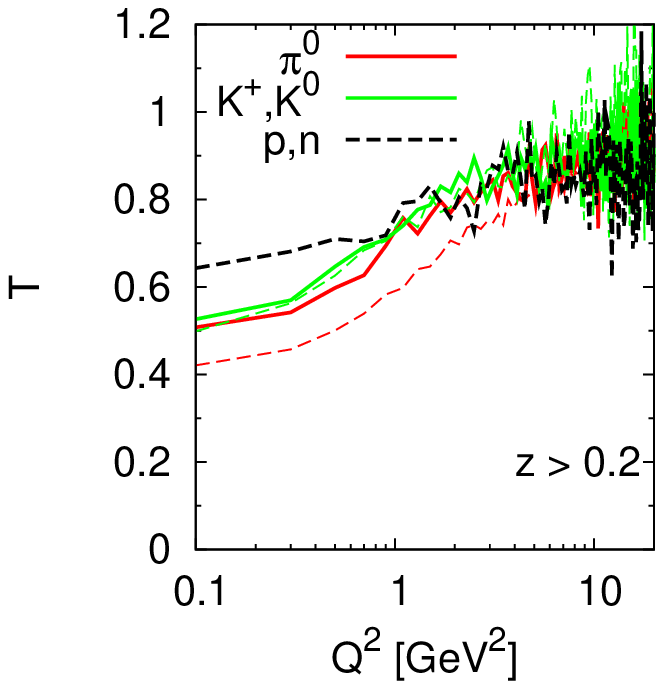}
  \caption{\small 
    The hadron attenuation for different hadron species within
    several $Q^2$ bins as function of $z$ (left panel) and $\nu$ (for
    $z>0.2$, right panel) for a collider setup $(3+30)\gev$.
  }
  \label{fig:GiBUU_EICatt}
\end{figure}

\noindent {\bf Hadron Attenuation at an EIC: Strong $\bm Q^2$ Dependence.} 
One may now look at hadron attenuation at an EIC.
Fig.~\ref{fig:GiBUU_EICatt} shows the expected attenuation for
different hadron species within several $Q^2$ bins as functions of
$\nu$ and $z$ for a very low energy collider setup $(3+30)\gev$, which is close to
former EMC conditions. 
One observes a large $Q^2$ dependence: while for low $Q^2$ values, the
attenuation of all hadron species decreases to approx.~0.5 at $z\to
1$, the attenuation is only approx.~0.8 for $Q^2>4$ GeV$^2$. This is
also shown in Fig.~\ref{fig:GiBUU_EICatt}, where the same
attenuation is shown, but now as a function of $Q^2$ and integrated
over all $\nu$ and $z>0.2$ values.
It is worthwhile mentioning that there is nearly no $\nu$ dependence
for all $Q^2$ bins visible in our calculations. 

\begin{figure}
  \centering
  \includegraphics[width=0.47\textwidth]{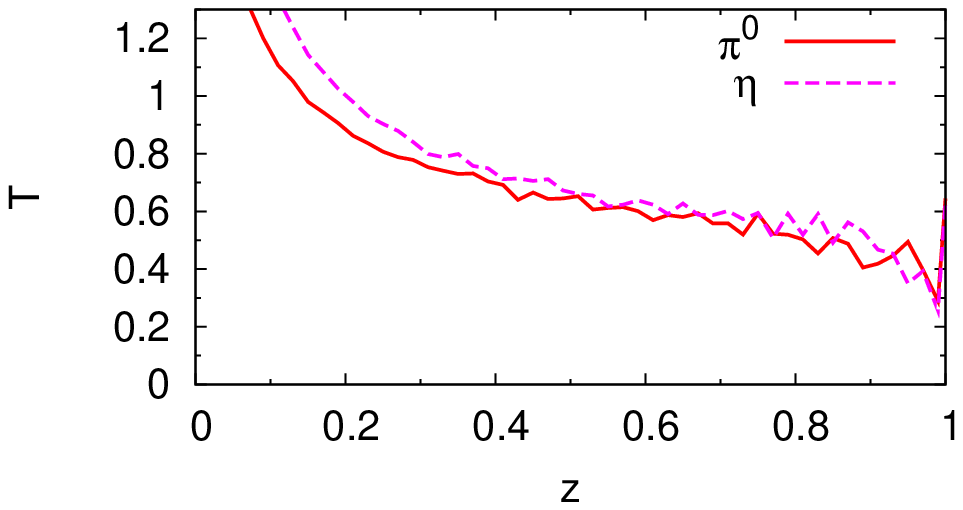}
  \hspace*{.5cm}
  \includegraphics[width=0.35\textwidth]{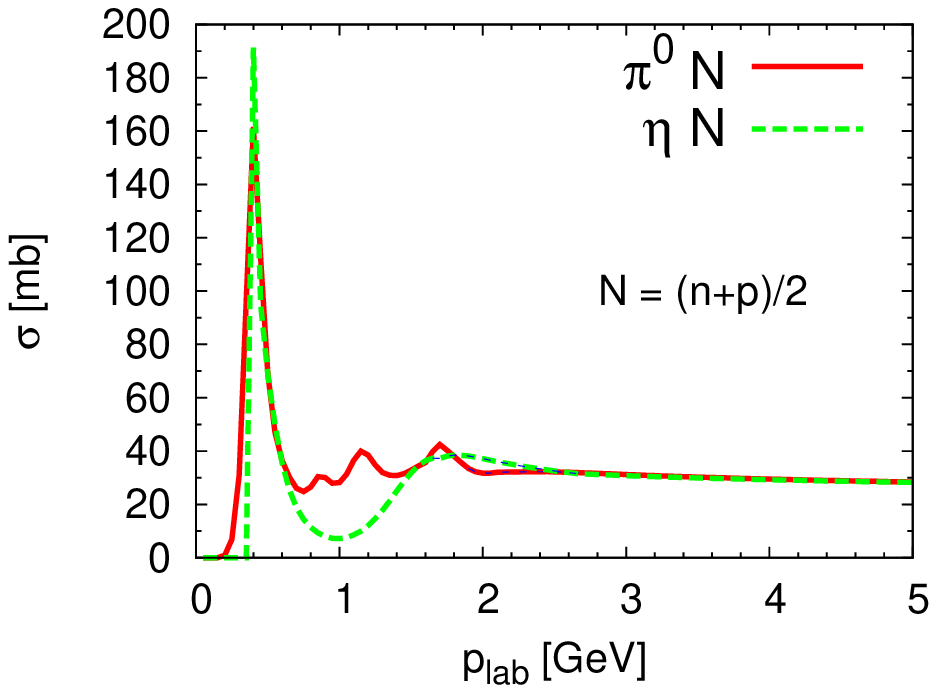}
  \caption{\small Left panel: The hadron attenuation for $\pi^0$ and $\eta$
    mesons for a collider setup of 3+30 GeV. Right panel: The
    hadronic interaction cross section of $\pi^0$ and $\eta$ mesons
    with nucleons at rest as a function of the meson momentum.}
  \label{fig:GiBUU_EICattPieta}
\end{figure}

\noindent {\bf Hadron Attenuation at an EIC: $\bm \pi^0$ vs. $\bm \eta$.}
As already shown in Fig.~\ref{fig:GiBUU_EICatt}, some differences in
the resulting attenuation ratio show up for different hadron
species. In Section~\ref{sec:AccardiDupre_EIChadronization}, it has
been suggested that a comparison of $\eta$ and 
$\pi^0$ attenuation ratios will distinguish between energy-loss models
and absorption models. In Fig.~\ref{fig:GiBUU_EICattPieta} we show
our results for the attenuation of these two particle species.
Both attenuation signals are close to each other, but show stronger absorption for $\pi^0$ than for $\eta$ mesons, in
which case the discriminatory power would weaken. 
In Fig.~\ref{fig:GiBUU_EICattPieta} we also show the hadronic
interaction cross section of pions and eta mesons with nucleons. For
laboratory momenta larger than 2 GeV, these are nearly
identical. Thus differences in the attenuation are due to formation
time effects.

\subsubsection{A global fit of nuclear fragmentation functions}
\label{sec:Sassot-FF}

\hspace{\parindent}\parbox{0.92\textwidth}{\slshape 
  Rodolfo Sassot, Marco Stratmann, Pia Zurita 
}
\index{Sassot, Rodolfo}
\index{Stratmann, Marco}
\index{Zurita, Pia}

Similarly to modifications of PDFs in nuclei, the production of
hadrons in the final-state 
is known to be affected when occurring in a nuclear environment.
For example, semi-inclusive deep-inelastic scattering (SIDIS)
off large nuclear targets shows significant differences as compared to
hadron production off light nuclei or proton targets, as reviewd in
Section~\ref{sec:eA_ppf_intro}. 

The past few years have seen a significant improvement in the pQCD description 
of hadron production processes, and, more specifically, in the precise determination of 
vacuum fragmentation functions (FFs), including estimates of their uncertainties
\cite{deFlorian:2007aj}.
FFs carry the details of the non-perturbative hadronization process, factorized from
the hard scattering cross section in the same way as for PDFs.
The most important result of these studies is that the standard pQCD framework 
not only reproduces data on electron-positron annihilation into hadrons, but it describes 
with remarkable precision also other processes like semi-inclusive deep-inelastic scattering and hadron production in 
proton-proton collisions.
It is then quite natural to ask if pQCD factorization can be also generalized to 
final-state nuclear effects, i.e., to introduce medium modified or nuclear 
fragmentation functions (nFFs), and to assess how good such an approximation
works or to determine where and why it breaks down.
From theoretical considerations alone, the answer is, however, not obvious since on the one hand, 
interactions with the nuclear medium may spoil the requirements of the factorization theorems, 
but, on the other hand, any estimates of possible
factorization breaking effects are strongly model dependent. 

Within the factorization ansatz,
nFFs should contain (factorize) all the non-perturbative 
details related to hadronization in a nuclear environment,
would be exchangeable from one process to another (universal), 
and would allow for QCD estimates at any given order in perturbation theory in a well defined 
and unified framework. 
These features can be explicitly tested using data from an increasing 
but still limited number of experiments 
that have performed precise measurements of hadron production off nuclear targets, for instance,
in SIDIS by HERMES \cite{Airapetian:2007vu} or in 
deuteron-gold collisions studied at RHIC \cite{Adler:2006wg,Adams:2006nd}. Both type of processes 
are compatible with a universal nuclear modification of the hadronization mechanism
in the currently accessible kinematic regime.
The inclusion of next-to-leading order QCD corrections and the possibility to use 
different observables have been proven to be crucial for an accurate parametrization of nFFs
\cite{Sassot:2009sh}.

In addition to the primary goal of testing the factorization properties of nFFs
and to constrain them from different data sets in a consistent theoretical framework 
(for further comparison with the different model estimates), a thorough analysis of nFFs 
also serves as a baseline for ongoing studies of hadron production processes in heavy-ion collisions 
performed at RHIC and the LHC \cite{Armesto:2009ug}.
In the following, we present a brief summary of the first global fit of nFFs 
and outline limitations in the analysis imposed by the data available so far.


\noindent {\bf Medium Modified Fragmentation Functions.}  Even though nuclear effects in the hadronization process have been known to be 
significant for quite some time, only recent experiments have become precise 
enough and selective from a kinematical point of view to allow for more detailed
and quantitative studies. 
Specifically, the HERMES collaboration has performed a series of
measurements of pion, kaon and proton attenuation on  
different nuclear targets as a function of the 
hadron momentum fraction $z$ and the photon virtuality $Q^2$, which both are 
used to characterize fragmentation functions, as well as the virtual
photon energy $\nu$, that can be related to the nucleon momentum
fraction $x$ carried by initial-state parton, see
Fig.~\ref{fig-AccDup:her1}. 


\begin{figure}[tb]
  \centering
  \includegraphics[width=0.85\textwidth]{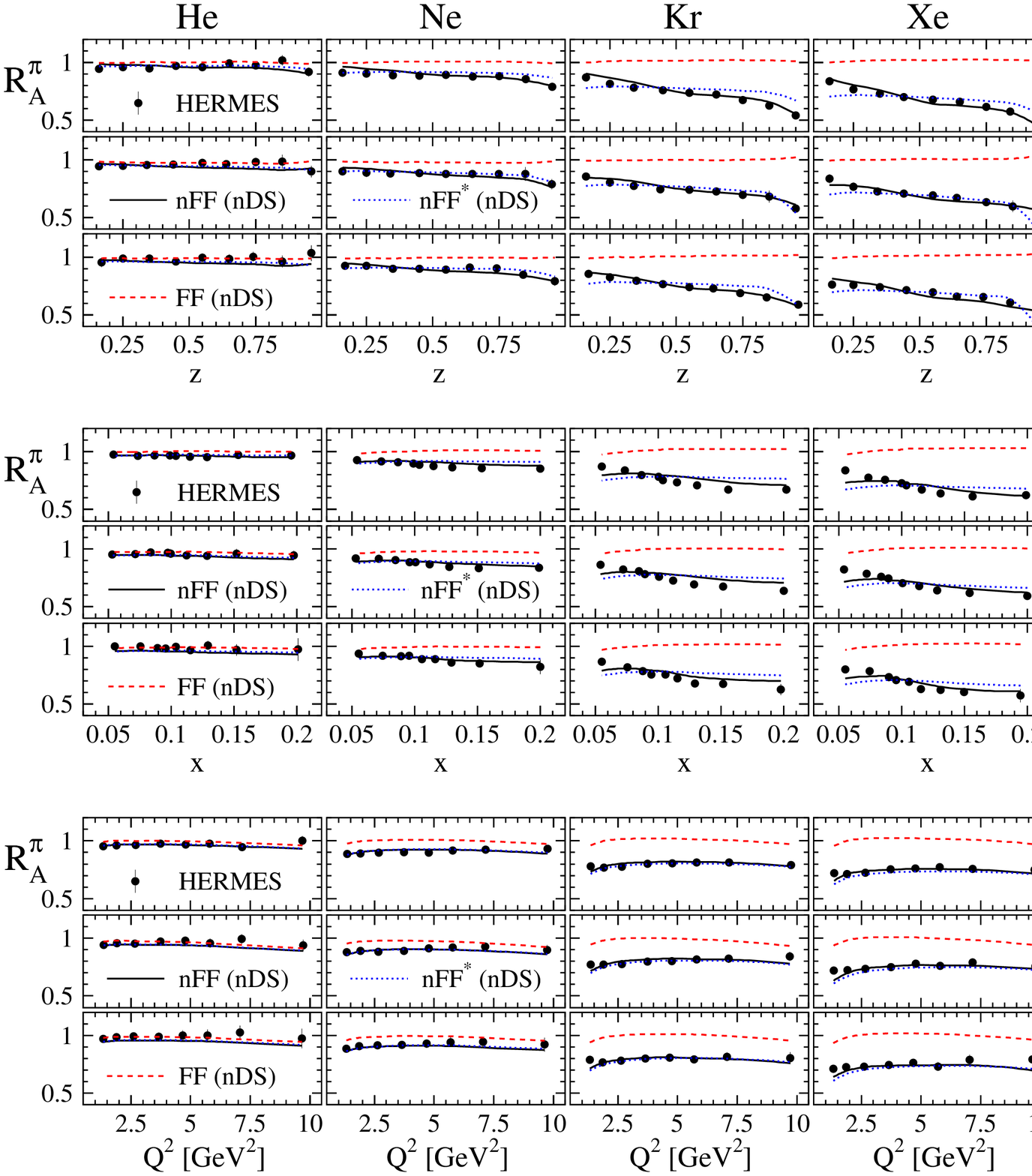}
  \caption{\small \label{fig:nsidis} 
    Quality of the nFF fit to nuclear SIDIS data from HERMES.
  }
\end{figure}

Single-inclusive identified hadron yields obtained in $d+Au$
collisions at mid-rapidity at BNL-RHIC, which show a characteristic
nuclear suppression and enhancement pattern as a function of the
hadron transverse momentum $p_T$, are another source of information on
nuclear modification effects in the hadronization process.
These measurements are often seen as ``control experiments''
associated with the heavy-ion program 
at RHIC to explore the properties of nuclear matter under extreme conditions. 
However, in view of the evidence for strong medium induced effects in
the fragmentation process found in SIDIS, $d+Au$ data are also of
particular relevance for extracting nFFs and testing the assumed
factorization and universality properties. 


To perform global nFF fits, it was proposed in
Ref.~\cite{Sassot:2009sh} to relate the medium modified fragmentations
to the standard ones in a convolution approach with a very simple
ansatz for the weight functions. The fits gives a very good
description of the full kinematic dependence of  
the HERMES data as can be seen in Fig.~\ref{fig:nsidis} while an
approach which ignores all final-state nuclear effects clearly fails.
The same set of nFFs that account for nuclear modification in SIDIS
also reproduce the main features of the $d+Au$ data from RHIC. The
peculiar $p_T$ dependence of the effects is found to come from an
interplay between quark and gluon fragmentation as a function of $p_T$  
in the hadron production cross section.
It is interesting to notice that there seems to be no visible conflict
between the standard $Q^2$ dependence assumed for the nFFs and the data.
In this respect, there have been many interesting suggestions and
model dependent calculations at the LO level, motivating the use of
medium modified evolution equations. However, in the range of $Q^2$
covered by present SIDIS and $d+Au$ data, there is no evidence  
for any significant departure from standard time-like evolution
equations 
\cite{Guo:2000nz,Majumder:2005jy,Armesto:2007dt,Albino:2009hu}.

\begin{figure}[tb]
  \centering
  \includegraphics[width=0.55\textwidth]{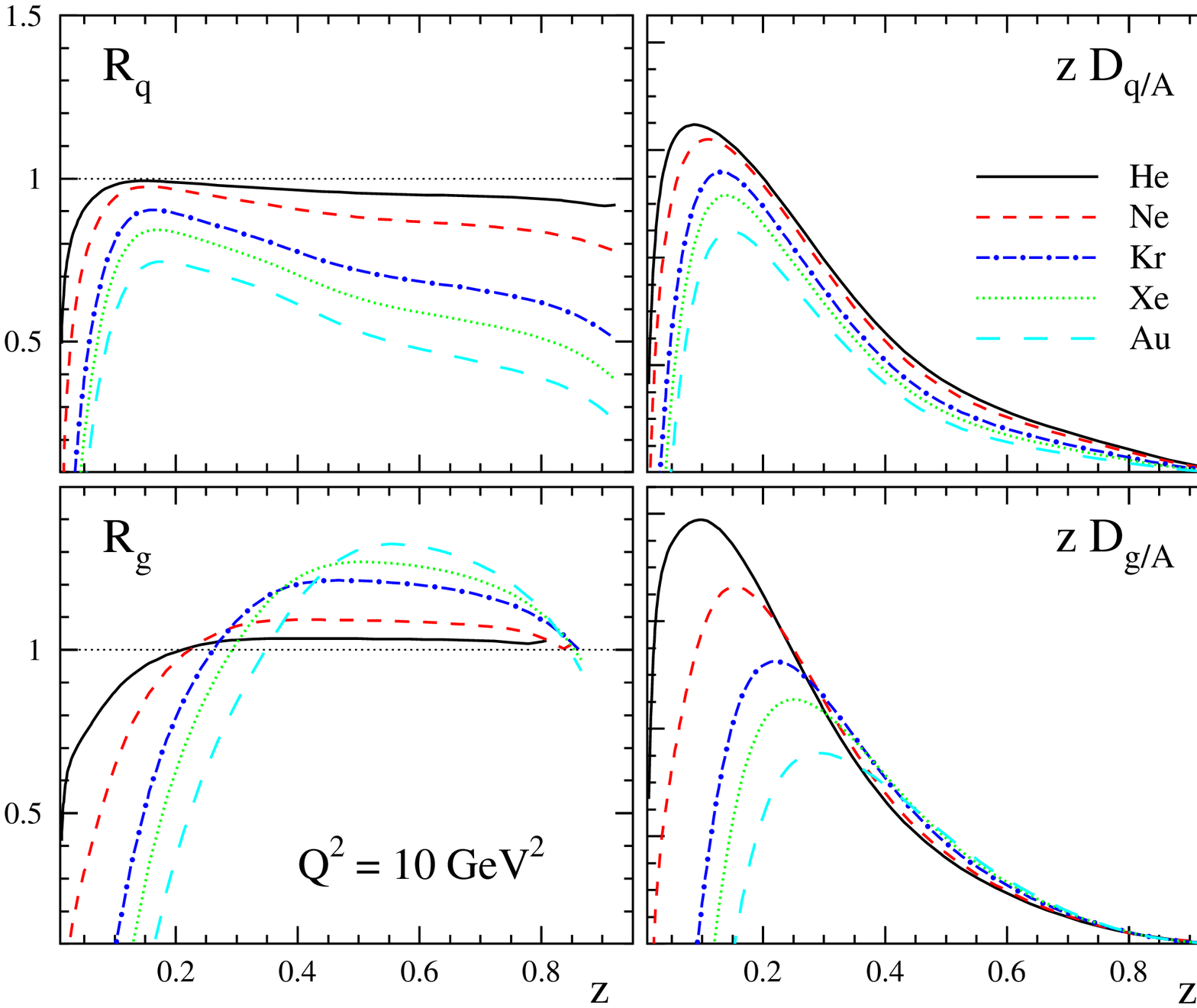}
\vspace{-1em}
  \caption{\small \label{fig:nFFs} Comparison of medium modified and
    standard FFs. } 
\end{figure}

The pattern of medium induced modifications is rather different for quarks and 
for gluons, see Fig.~\ref{fig:nFFs}. The dominant role of quark
fragmentation in SIDIS leads to a suppression, i.e., $R_q^{\pi}<1$,
increasing with nuclear size $A$ as dictated by the pattern of hadron  
attenuation found experimentally. The enhancement of hadrons observed
in $d+Au$ collisions for $p_T\approx 10\,\mathrm{GeV}$, along with the
dominant role of gluon fragmentation at low values of $p_T$ explains
that $R_g^{\pi}>1$ for $z\rightarrow 0.2$. Below $z\simeq 0.2$, where
all the data used in the fit have very limited or no constraining
power, both quark and gluon nFFs drop rapidly. For the time being, the
behavior in this region could easily be an artifact of the currently
assumed functional form for the parameterization.   The
extended $Q^2$ range of EIC  will allow
one to accurately test the factorization assumption for nFFs, which is
at the basis of the presented approach to nuclear modfications of
hadron production.\\

\subsubsection{Heavy quarks and quarkonia in a nuclear environment}

\hspace{\parindent}\parbox{0.92\textwidth}{\slshape 
  B.~Z.~Kopeliovich 
}
\index{Kopeliovich, Boris Z.}

\vspace{\baselineskip}

\noindent {\bf Time dependence of vacuum radiation.}  
The color field of a quark originating from a hard reaction
(DIS, high-$p_T$, $e^+e^-$, etc.) is stripped off, {\it i.e.}, such a quark is
lacking a color field up to transverse frequencies $q\lsim Q$, and
starts regenerating its field by radiating  gluons, i.e.,
forming a jet. This can be described by means of an expansion
of the initial ``bare'' quark over the Fock states containing a physical
quark and different number of physical gluons with different
momenta. Originally, this is a coherent
wave packet equivalent to a single bare quark $|q\ra$. However,
different components have different invariant masses and they start
gaining relative phase shifts as a function of time. As a result,
the wave packet is losing coherence and gluons are radiated in
accordance with their coherence times. The required time
is to the jet energy, since the radiation time (or length) depends on
the gluon energy and transverse momentum $k$ (relative to the jet axis),
 \begin{align}
l_c=\frac{2E}{M_{qg}^2-m_q^2}= \frac{2Ex(1-x)}{k^2+x^2\,m_q^2}.
\label{kopel-160}
 \end{align}
 Here, $x$ is the fractional light-cone momentum of the radiated gluon; $m_q$ is the quark mass;
$M_{qg}^2=m_q^2/(1-x)+k^2/x(1-x)$ is the invariant mass squared of
the quark and radiated gluon.

One can trace how much energy is radiated over the path length $L$ by the gluons which have lost 
coherence during this time interval~\cite{Kopeliovich:1995jt,Kopeliovich:2003py,Kopeliovich:2006xy,Kopeliovich:2007yv,Kopeliovich:2010aia},
 \begin{align}
\Delta E(L) =
E\int\limits_{\Lambda^2}^{Q^2}
dk^2\int\limits_0^1 dx\,x\,
\frac{dn_g}{dx\,dk^2}
\Theta(L-l_c),
\label{kopel-180}
 \end{align}
 where $Q\sim p_T$ is the initial quark virtuality; the infra-red cutoff is fixed at 
$\Lambda=0.2\gev$.
 The radiation spectrum reads
 \begin{align}
\frac{dn_g}{dx\,dk^2} =
\frac{2\alpha_s(k^2)}{3\pi\,x}\,
\frac{k^2[1+(1-x)^2]}{[k^2+x^2m_q^2]^2},
\label{kopel-200}
 \end{align}
 where $\alpha_s(k^2)$ is the running QCD coupling, which is regularized at low scale by the substitution:
$k^2\Rightarrow k^2+k_0^2$ with $k_0^2=0.5 \gev ^2$.
In the case of heavy quark the $k$-distribution Eq.~(\ref{kopel-200}) peaks at $k^2\approx x^2\,m_q^2$, 
corresponding to the polar angle (in the small angle approximation) $\theta=k/xE=m_q/E$. This is 
known as the dead cone effect~\cite{Dokshitzer:1991fc,Dokshitzer:2001zm}.

\begin{wrapfigure}{r}{0.76\columnwidth}
\begin{minipage}[b]{0.38\columnwidth}
\centerline{\includegraphics[width=0.75\columnwidth]{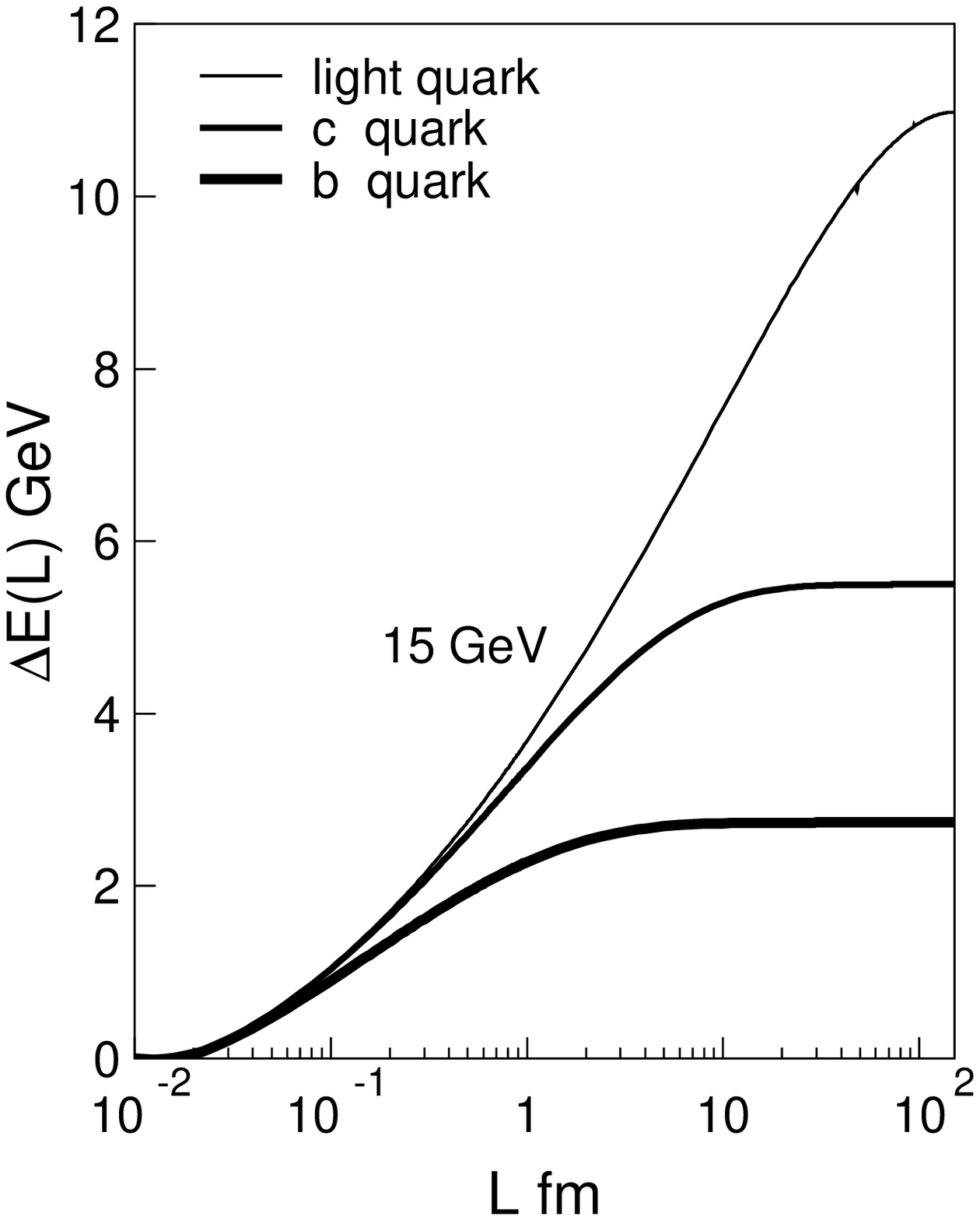}}
\end{minipage}
\begin{minipage}[b]{0.38\columnwidth}
\centerline{\includegraphics[width=0.75\columnwidth]{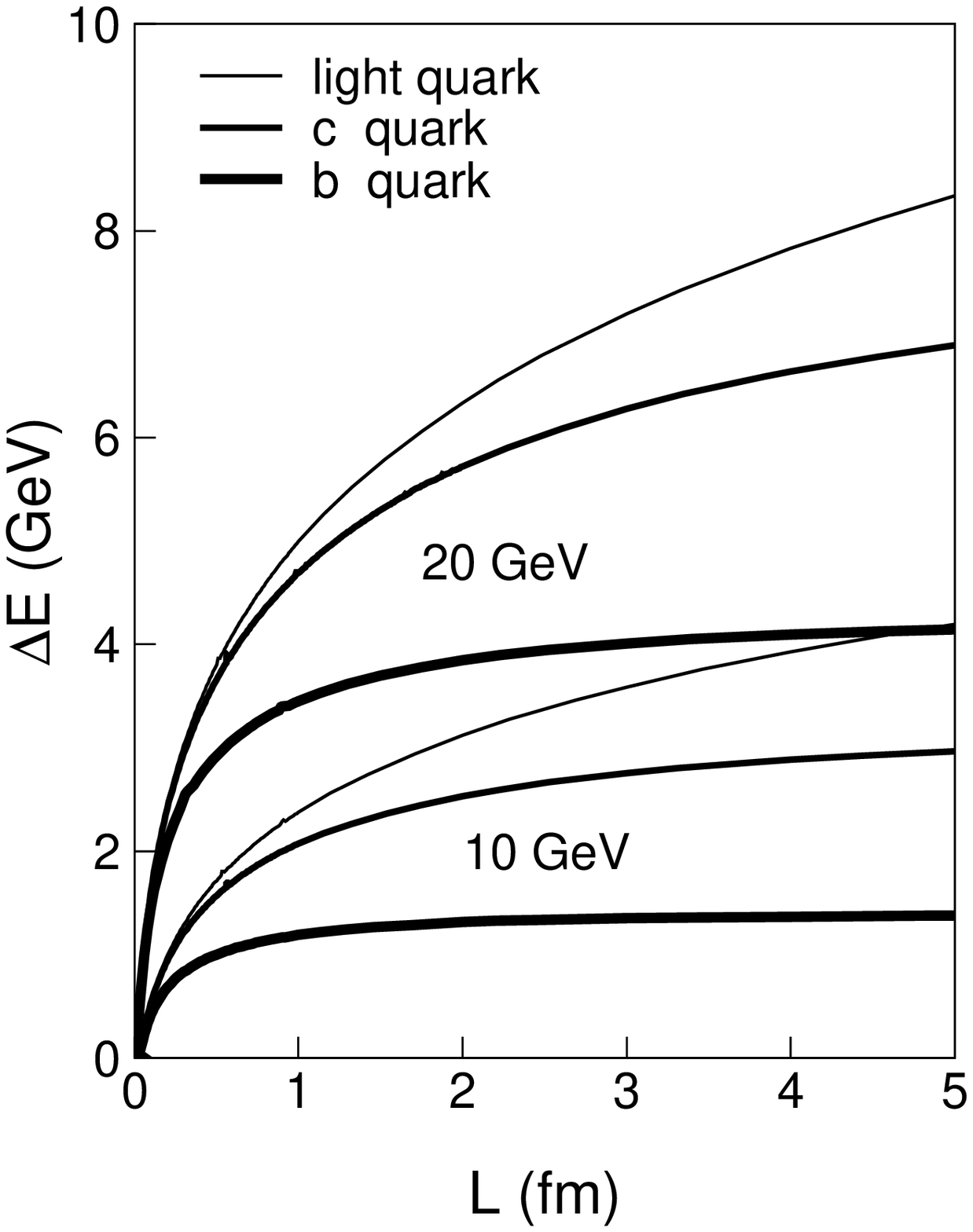}}
\end{minipage}
\caption{\small {\it Left panel:} Vacuum energy loss by light ($m_q=0$), charm ($m_c=1.5 \gev $) and bottom ($m_b=4.5 \gev $) quarks with $E=15\gev $ and virtuality $Q\sim E$ as function of path length.  {\it Right panel:} The same, but for energies $20$ (three upper curves) and $10\gev$ (three bottom curves), and zoomed in at short path lengths.\vspace*{0cm}}
\label{l-dep}
\end{wrapfigure}

The step function in Eq.~(\ref{kopel-180}) creates another dead cone~\cite{Kopeliovich:2010aia}: 
since the quark is lacking a gluon field, no
gluon can be radiated unless its transverse momentum is sufficiently high, 
$ k^2>{2Ex(1-x)}/{L}-x^2m_q^2$. This bound relaxes with the rise of
$L$ until it reaches $k^2\sim x^2m_q^2$, characterizing the heavy
quark dead cone at $L_q= {E(1-x)}/{xm_q^2}$. 
The radiation of such a ``naked'' quark has its own dead cone controlled
by its virtuality {$ Q^2\gg m_q^2$}, and is much wider than the one
related to the quark mass. Therefore, there is no mass dependence of the  
radiation until the quark virtuality cools down to {$ Q^2\Rightarrow
  Q^2(L)\sim m_q^2$}.  At the early stage of hadronization,
when {$ Q^2(L)\gg m_q^2$}, all quarks radiate equally, and the
results of~\cite{Dokshitzer:2001zm} for a reduced energy loss of heavy
quarks should be applied with a precaution.  
The numerical results demonstrating this behavior are depicted in
Fig.~\ref{l-dep}.


 %
%

One can see that a substantial difference between the radiation of energy
by charm and light quarks onsets at rather large distances, above
$10\fm$.
However the $b$-quark radiation is suppressed already
at a short distance, less than one fermi.
Moreover, it completely regenerates the color field already at a distance of the order of $1\fm$
and does not radiate any more. Of course, this $b$-quark still may have a medium induced radiation,
which is very weak according to~\cite{Dokshitzer:2001zm}. Notice that the interference between vacuum and induced 
radiations is absent because they occur on different 
time scales. 

\begin{figure}[htb]
\centering
 \includegraphics[width=10cm]{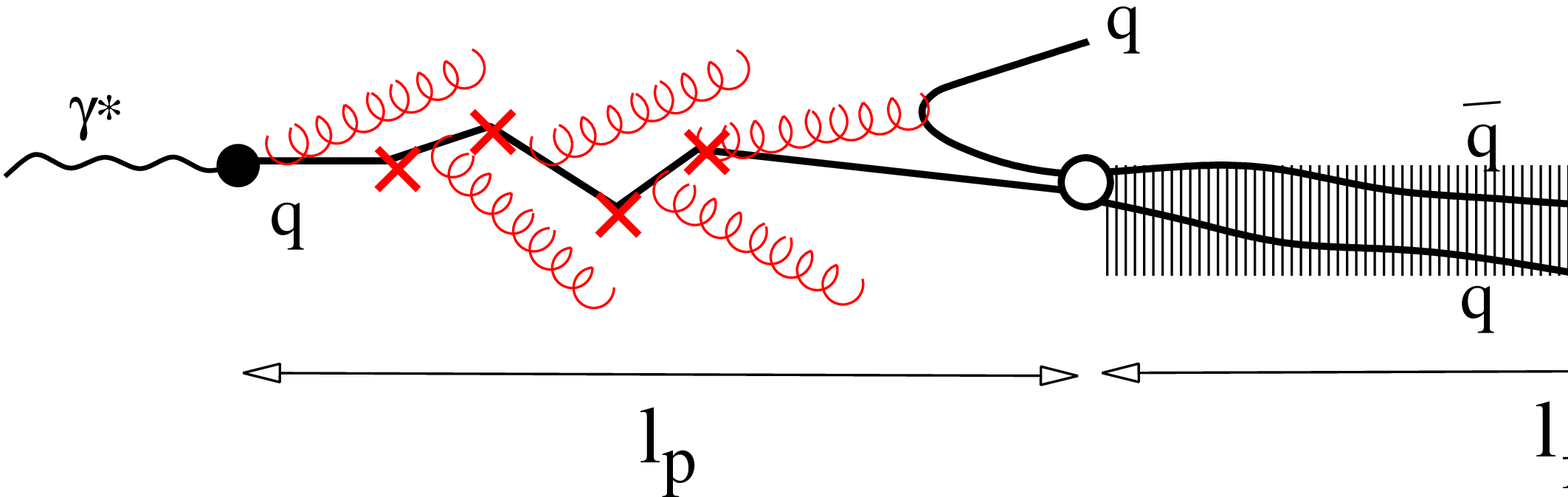}
 \caption{\small \label{2-step} The two-step process of leading hadron production. On the production
length $l_p$ the quark is hadronizing experiencing multiple interactions
broadening its transverse momentum and inducing an extra energy loss.
Eventually, the quark color is neutralized by picking up an antiquark. The
produced color dipole (pre-hadron) is attenuating in the medium and
developing the hadron wave function over the formation path length $l_f$.}
 \end{figure}

\noindent {\bf Production and formation length.}
One should clearly distinguish between the production time scales for a colorless dipole (pre-hadron) and the final hadron.
The former signals color neutralization, which stops the intensive energy loss caused by vacuum radiation following
the hard process, while the latter is a much longer time taken by the
dipole to gain the needed hadronic mass, i.e. to develop the hadron
wave function. While the former is proportional to $1-z_h$ and
contracts at large fractional momentum $z_h$ of the hadron, the latter
keeps rising proportionally to $z_h$. These two time scales are frequently mixed
up. The shortness of the production lengths at large $z_h$ is dictated
by energy conservation. Indeed, a parton originating from a hard
reaction intensively radiates, losing energy.  This cannot
last long, otherwise the parton energy will drop below the energy of
the detected hadron. Only the creation of a colorless pre-hadron, which does not radiate gluons any more, can stop the dissipation 
of energy. Energy conservation thus imposes a restriction on the color
neutralization time~\cite{Kopeliovich:1984ur},  $l_p\leq \frac{E_q}{\la dE/dz\ra}\,(1-z_h)$, which must vanish at $z_h\to1$.
One should also distinguish between the mean hadronization time of a jet, whose energy is shared between many hadrons, and specific events containing  a leading hadron with $z_h\to1$. The production of such a hadron in a jet is a small probability fluctuation, usually associated with large rapidity gap events. The space-time development of such an unusual jet is different from the usual averaged jet.
It is illustrated in Fig.~\ref{2-step}.
Notice that one should not mix up the production time with the time
scale evaluated in~\cite{Markert:2008jc}, Eq.~(2), 
which  is just the well known coherence time.
This is not the time of duration of hadronization which we are interested in.
If hadronization were lasting as long as the coherence time, energy
conservation would be broken. Besides, a pre-hadron does not have any
certain mass, since according to the uncertainty relation it  
takes time, called formation time, to resolve between the ground and
excited states, which have certain masses. Therefore, one cannot
evaluate the production time of a pre-hadron relying on the mass of
the hadron. 

Since the produced pre-hadron strongly attenuates in the
nuclear medium, the position of the color neutralization point is crucial
for the resulting nuclear suppression. Notice that such a picture of
space-time development of hadronization is  classical. In quantum
mechanics one cannot say with certainly whether the pre-hadron is
produced inside or outside the medium: the inside-outside interference
term is significant~\cite{Kopeliovich:2008uy}. 



\noindent {\bf Heavy flavored hadrons.} 
The production length distribution calculated for light quarks \cite{Kopeliovich:1995jt,Kopeliovich:2003py,Kopeliovich:2007yv} should be similar to that  for charm quarks, which
have a similar vacuum radiation during the first several fermi. However, a bottom quark, according to 
Fig.~\ref{l-dep}, dissipates considerably less energy, moreover, its vacuum radiation ceases at the 
distance of about $1\fm$, because the quark completely restores its color field.  Of course, 
confinement does not allow a colored quark, even with a restored field, to propagate freely. It 
keeps losing energy via nonperturbative mechanisms~\cite{Kopeliovich:2007yv}, like in the string (flux tube) model. 
Surprisingly, nonperturbative dynamics is more involved into hadronization of heavy compared with light quarks. 
However, one should remember that this is true only for jets which end up producing leading hadrons with $z_h\to1$.

A high-energy heavy quark always escapes from the medium
and produces an open flavor hadron with no suppression. 
Therefore, a break-up of a light-heavy dipole propagating in a medium should not lead to a suppression,
unless the fractional momentum $z_h$ of the detected hadron is fixed at a large value. In such a 
case, break up of the dipole ignites continuation of vacuum energy loss, which slows down the quark 
to smaller values of $z_h$. This is why a quark should stop radiating at a distance $l\sim l_p$ and 
produce a colorless dipole, which then survives through the medium. 

It is interesting that the produced heavy-light, $c$-$q$ or $b$-$q$ dipoles expand their sizes faster than  a 
light $\bar qq$ dipole. This happens because of the very asymmetric sharing of the longitudinal 
momentum in such dipoles. Minimizing the energy denominator one gets the fractional momentum carried 
by the light quark,
$
  \alpha\sim\frac{m_q}{m_Q},
$
which indeed is very small, about $0.1$ for charm and $0.03$ for bottom.
Then according to~\cite{Kopeliovich:2010aia,Kopeliovich:2010uz}, the dipole size is evolving with time as
$r_T^2(t)=\frac{2t}{\alpha(1-\alpha)\,E}+r_0^2$, 
where $r_0$ is the initial dipole separation:
the $b$-$q$ dipole is expanding much faster than $\bar qq$. 

{\bf Conclusions.} The hadronization of charm and bottom quarks ends
up at a short distance $l_p$ with production of a colorless
dipole
which is strongly absorbed by the
medium. This may explain why both of them are strongly suppressed in
$A+A$ collisions. Studies of light {\it vs.} heavy meson productions at
the EIC will clearly be able to validate the discussed effects.

\section{Jets}
\label{sec:eAjets}

\subsubsection{Jets, in-medium parton propagation and nuclear gluons}


\hspace{\parindent}\parbox{0.92\textwidth}{\slshape 
  Alberto Accardi, Matthew A. C. Lamont, Gregory Soyez 
}
\index{Accardi, Alberto}
\index{Lamont, Matthew}
\index{Soyez, Gregory}

Preliminary results from the SLAC E665 fixed target experiment have
demonstrated jet production in $e+A$ collisions at $s \approx 1000$
GeV$^2$ \cite{Salgado:1993mm,Melanson:1993ep}. Thus, the start of the
jet study programme should be feasible in a Phase-I EIC.  This can be
confirmed by further simulations, required to study the capabilities
in a collider experiment as opposed to a fixed-target experiment like
E665.

As will be discussed in detail in the next 2 contributions, the 
nuclear modification of 1+1 jet production, i.e., 1 jet
from current fragmentation and 1 from target fragmentation, is of
great interest to study parton propagation through cold nuclear
matter, in order to extract cold nuclear transport coefficients,
and probing soft gluons in nuclei. In addition, the nucleus can be
used as a femtometer-scale detector of the evolution of parton
showers, allowing to test their perturbative descriptions ({\it e.g.},
$k_T$-ordering {\it vs.} rapidity ordering) and Monte-Carlo
implementations, which are used pervasively in all fields of
high-energy physics to analyze experimental data. 

The case of 2+1 jets is also interesting. Indeed, the cross section
for this prcess reads
\begin{equation}\label{eq:basic_xsect}
\frac{{\rm d}^2\sigma_{2+1}}{{\rm d}x_p\,{\rm d}Q^2}
   =   A_q(x_p,Q^2)\,q_A(x_p,Q^2)
 \,+\, A_g(x_p,Q^2)\,g_A(x_p,Q^2),
\end{equation}
where the two terms correspond to the quark-initiated and
gluon-initiated processes respectively, and the coefficients $A_q$ and
$A_g$ are matrix elements that can be computed at given order in
perturbation theory. Unlike the 1+1 case which is dominated by quark
initiated processes, the 2+1 cross section is now also sensitive to
nuclear gluons, and offers yet another way to measure them.

\begin{figure}
\centering
\parbox{0.35\linewidth}{
  \centering
  \includegraphics
    [height=2cm,bb=85 452 160 529,clip=true]
    {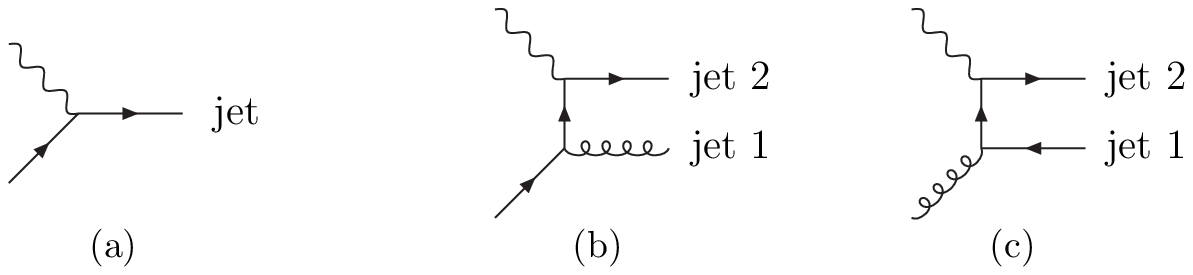} \\
  \includegraphics
    [height=2cm,bb=225 452 428 529,clip=true]
    {eA-final/Figs/pert_jets.eps}
}
\parbox{0.31\linewidth}{
  \includegraphics
    [height=\linewidth,angle=270,bb=-40 70 140 270,clip=true]
    {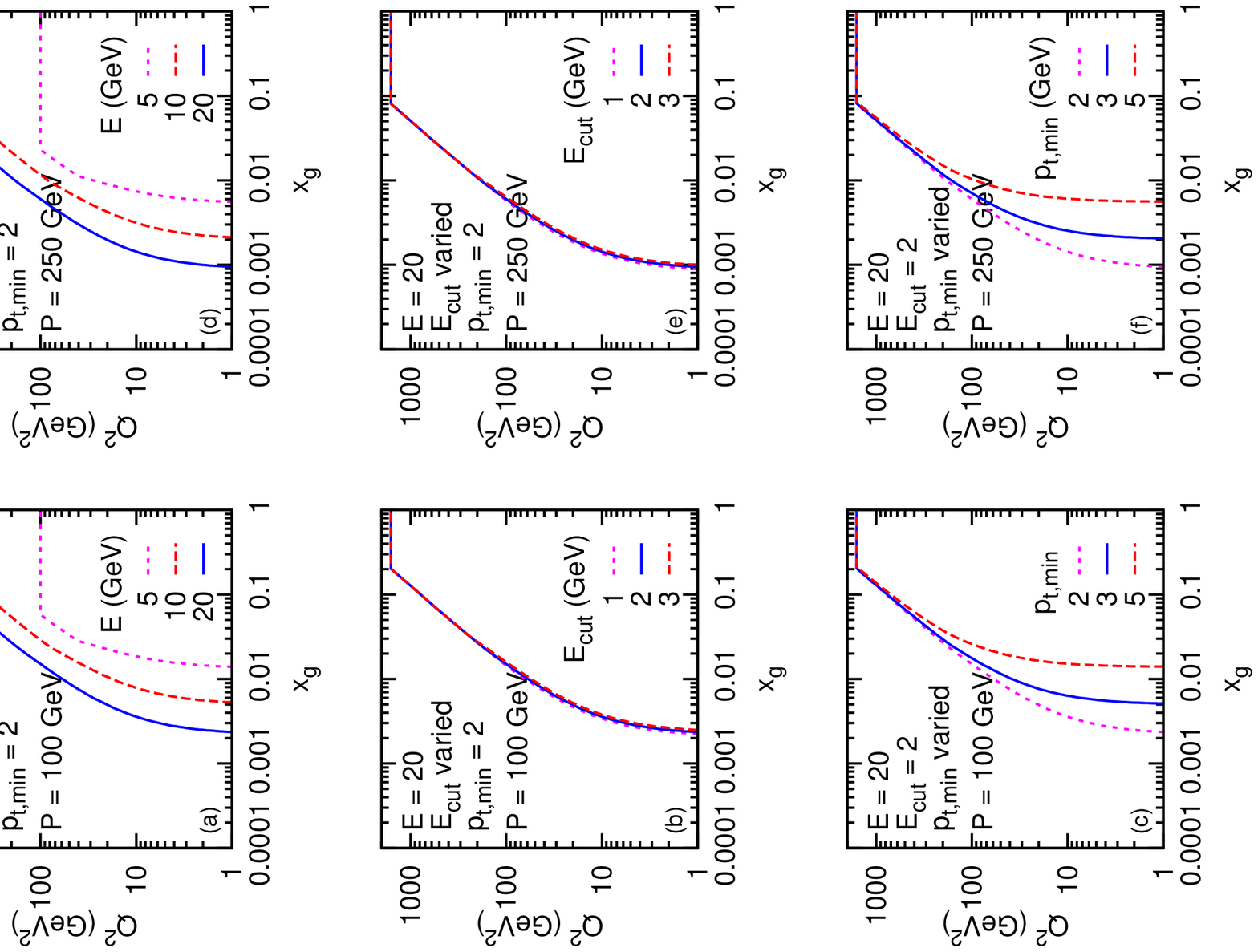}
}
\parbox{0.31\linewidth}{
  \includegraphics
    [height=\linewidth,angle=270,bb=363 70 543 270,clip=true]
    {eA-final/Figs/kinem_note.eps}
}
\caption{\small 
  {\it Left:} Parton-level processes that contribute (a) to the 1+1 and
(b,c) 2+1 jet cross-section.
  {\it Middle and Center:} Accessible kinematic range in $x_p$ and $Q^2$ for the 2+1
  jets scenario. The accessible region is plotted for different
  energies $E$ of the electron beam and hadron beam energy $E_p=100$
  GeV, corresponding to a phase-I and
  phase-II EIC, for different jet transverse momentum cuts
  $p_T>p_{T,min}$ at fixed jet energy cut $E_{cut}$.
}
\label{fig:2+1jets_eA}
\end{figure}

Since the outgoing jets have to travel in the medium, the coefficients
$A_q$ and $A_g$ will be affected by in-medium propagation. We shall
assume here that the measurements of 1+1 jet cross-sections allow to
control in medium quark jets, hence $A_q$. Then, by tagging or vetoing
gluon jets in 2+1 events one can study, respectively, gluon jets
in-medium propagation and the nuclear gluon distributions.
In Fig.~\ref{fig:2+1jets_eA}, we show the expected kinematic reach of
the gluon measurements for a phase-I and phase-II EIC, and for various
cuts on the jet transverse momentum $p_T$. Details can be found in
\cite{Soyez-eAnote}.  Detailed simulations are planned to study the feasibility and physics
reach of these jet studies.

\subsubsection{Monte-Carlo for hard jets in e+A collisions}
\label{sec:eA-AbhijitMC}  

\hspace{\parindent}\parbox{0.92\textwidth}{\slshape 
A. Majumder 
}
\index{Majumder, Abhijit}

\vspace{\baselineskip}

The production and modification of hard jets produced in lepton nucleus collisions is considered. 
The assumption of factorization of the hard scattering cross section from the structure functions and final fragmentation 
function allow one to compute the final medium modified fragmentation function in both cold nuclear matter and 
in a hot Quark-Gluon-Plasma (QGP) in an identical formalism. This allows for both a cross check of the 
basic energy loss formalism used in these reactions, and a comparative study of the partonic sub-structure of 
these different phases of QCD matter. 
Detailed descriptions are provided via a Monte-Carlo 
simulation of such calculations. We compare the results of analytical calculations in these two regimes and 
present preliminary Monte-Carlo simulations for jets produced in
deep-inelastic collisions.


\noindent{\bf Introduction to in-medium DGLAP.}
Collision processes which involve a hard scale can be factorized into separate probabilities of hard and 
soft processes which are convoluted via a single dimensionless variable~\cite{Collins:1985ue}. 
For example, for the case of single hadron  
inclusive production in deep-inelastic-scattering (DIS), the differential cross section may be expressed as,
\begin{eqnarray}
\frac{d \sigma}{ d z }  = \int dx G(x,Q^{2}) \otimes \frac{d \hat{\sigma}}{d Q^{2}} \otimes D(z,Q^{2}), \label{majumder_fact_1}
\end{eqnarray}
where, $G(x,Q^{2})$ represents the parton distribution function, $\frac{d \hat{\sigma}}{d Q^{2}}$  represents the 
electron quark scattering cross section via single photon exchange. $D(z,Q^{2})$ represents the fragmentation 
function to produce a hadron with a momentum $z \nu$ from the
fragmentation of the outgoing quark jet. The structure functions and
fragmentation functions are defined and  factorized from the hard
cross sections at a given scale $\mu^{2}$ which, in this case, is chosen
to be equal to the hard scale of the process $Q^{2}$.
They only need to be measured at a single scale, and the change of
these functions with scale is given by the DGLAP evolution
equations~\cite{Altarelli:1977zs}. 
For fragmentation functions, these equations read
\begin{eqnarray}
\frac{\partial D(z,Q^{2})}{\partial \ln Q^{2}} = \frac{\alpha_{S}}{2\pi} \int \frac{dy}{y} P(y) D \left(  \frac{z}{y}, Q^{2}\right) ,
\end{eqnarray}
where, $P(y)$ is the gluon splitting function and represents  the probability for a quark to radiate a gluon and retain a fraction $y$ of its light cone momentum.

In the case of DIS on a large nucleus, one may simply include the entire effect of the medium by including a length dependent 
multiplicative factor to the gluon splitting function~\cite{Majumder:2009zu}, 
which accounts for the fact that the radiated gluon will scatter in the medium 
influencing its radiation amplitude, i.e., $P(y) \rightarrow P(y) K(y,q^{-},L^{-},Q^{2})$. The medium dependent factor given 
as~\cite{Majumder:2009ge},
\begin{eqnarray}
K(y,q^{-},L^{-},Q^{2}) = \int_{0}^{L^{-}} d \zeta^{-} \frac{\hat{q}}{Q^{2}} 
\left[  2 - 2 \cos \left(   \frac{Q^{2} \zeta^{-} }{2 p^{+}q^{-} y (1-y)} \right)\right]  \label{in-medium-kernel}
\end{eqnarray}
In the equation above, $L^{-}$ is the maximum possible length traversed in the medium in the course of one emission. In an evolution equation, 
the formation time of the final radiation is chosen to be larger than the maximum medium length. This restricts the length to be no larger than 
$q^{-}/Q_{min}^{2}$, where $Q_{min}$ is the minimum allowed virtuality on exit from the medium. In an analytic solution to the DGLAP equation, 
one requires an input fragmentation function. The most unambiguous input is to use the known vacuum fragmentation function at the scale 
$Q_{min}^{2}$ where we have stipulated that the jet has emerged from the medium. This is then evolved in $Q^{2}$ up to the hard scale of the 
process using the medium modified evolution equation which includes the kernel of Eq.~\eqref{in-medium-kernel}.

\begin{figure}
\centering
\parbox{0.35\linewidth}{
  \includegraphics[width=\linewidth,trim=40 10 80 50,clip=true]{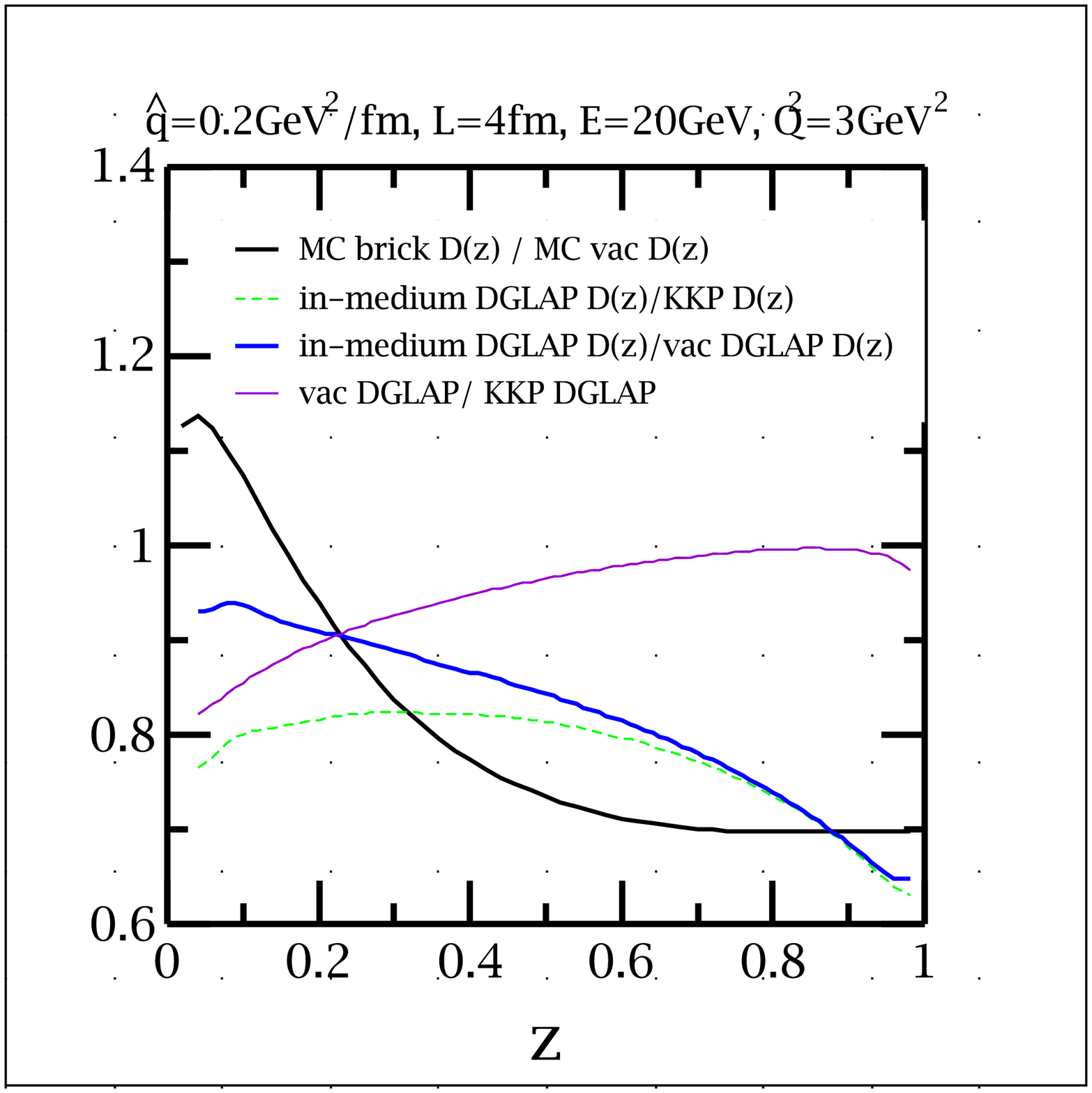}
}
\hspace*{0.4cm}
\parbox{0.44\linewidth}{
 \vspace*{-0.1cm}
  \includegraphics[width=\linewidth]{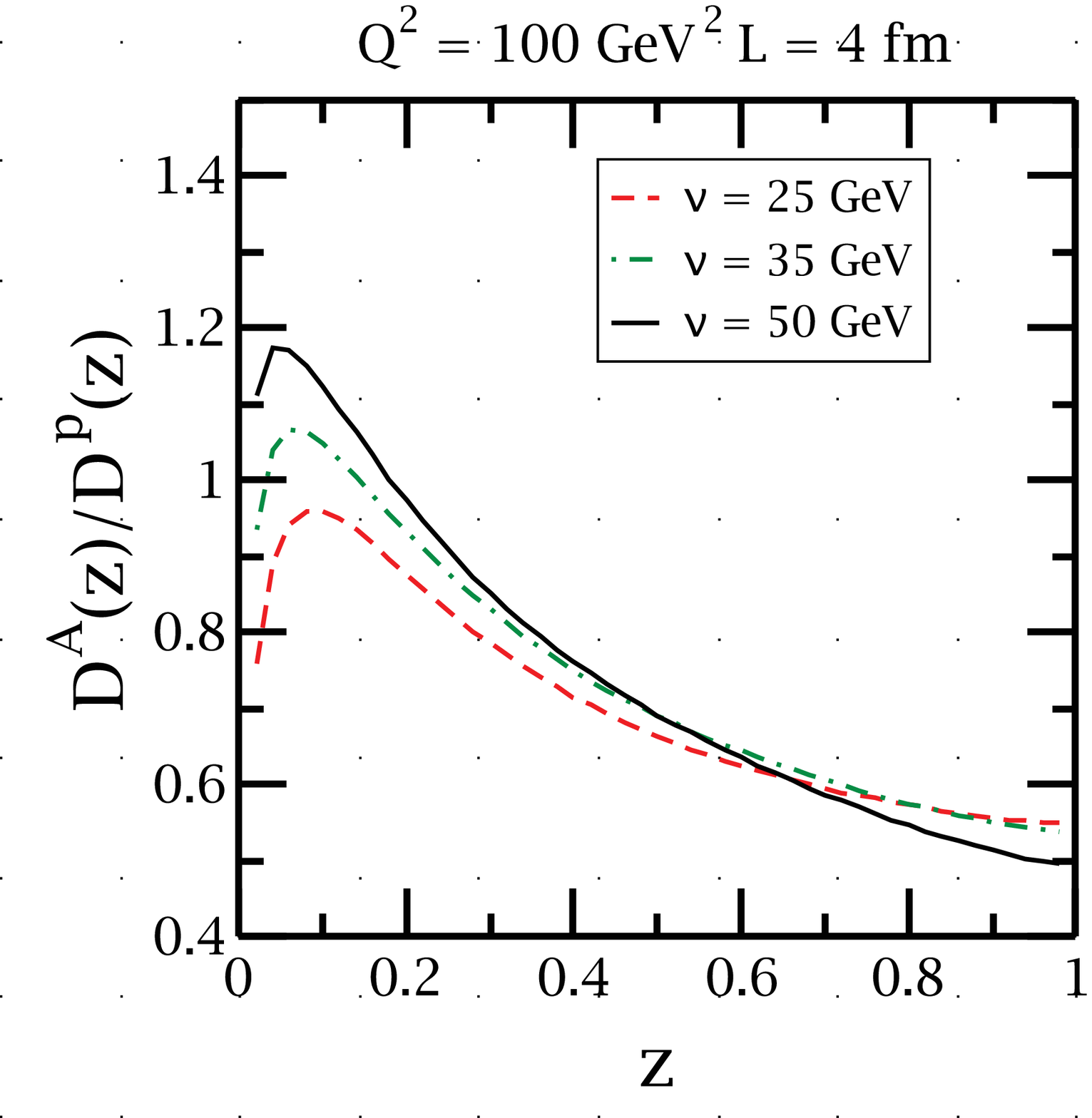}
}
\caption{\small 
  {\it Left:} A comparison of the results of an analytic DGLAP
  evolution calculation and a Monte-Carlo shower calculation for the
  same choice of input parameters.
  {\it Right:} Results of a set of Monte-Carlo simulations of a jet
  propagating through a 4 fm brick.
} 
\label{fig:DGLAP-MC-compare}
\label{fig:EIC-MC} 
\end{figure}

Results from such an in-medium DGLAP evolution are plotted in Fig.~\ref{fig:DGLAP-MC-compare} . The input distribution in 
vacuum is taken from KKP at an input scale of $\mu_{in}^{2} = 1$
GeV$^{2}$ and evolved up to $Q^{2}$. Its ratio to the
KKP fragmentation at the scale $Q^{2}$ is plotted as the green dashed
line in Fig.~\ref{fig:DGLAP-MC-compare}. 
Note that our numerical implementation of the DGLAP equation is different from that of KKP and so for comparison, we 
plot the ratio of the vacuum evolved fragmentation function in our implementation versus that in the KKP where both 
calculations start from the same input distribution i.e. the KKP function at the scale $\mu_{in}^{2}$, and are compared
at the higher scale of $Q^{2}$. The ratio is plotted as the magenta curve in Fig.~\ref{fig:DGLAP-MC-compare}. 
While over the range of $z$ considered, the curve is close to unity, it may deviate by up to 20\% at lower values of $z$.

The solid blue line in Fig.~\ref{fig:DGLAP-MC-compare} represents the ratio of the medium modified fragmentation 
function to the vacuum fragmentation function, where both numerator and denominator are calculated using the 
same numerical routine (for the vacuum FF we simply use $\hat{q} = 0$).
This ratio can be approximately compared to the
ratio of hadron yields in DIS experiments.  
It should be pointed out that in all the calculations reported in this article, the medium is assumed to be static 
and uniform with a fixed length. This fixed length is travelled by each jet. 
Realistic geometries will be considered in the future.

\noindent{\bf Monte-Carlo implementation.}
In any realistic calculation of jet modification in an extended medium a variety of approximations need to be made. 
For example, in the in-medium DGLAP evolution equations reported in the previous sections, we assumed that the 
entirety of the parton shower exits the medium and fragments in vacuum. This is obviously not the case. In reality, 
a large portion of the shower is trapped in the medium and does not undergo vacuum fragmentation. Such effects 
cannot be treated in a DGLAP setup where the input is the final vacuum
fragmentation. Note that such effects may be included 
with a position dependent input fragmentation function. However, such input is always ambiguous and the computation of the 
evolution of  a position, energy and obviously $z$-dependent 
fragmentation functions are prohibitively numerically intensive. 

The obvious solution to this is to use a Monte-Carlo jet routine. Unlike analytic in-medium DGLAP calculations which 
evolve upwards, numerical Monte-Carlo routines evolve downwards in virtuality. As such, they are a more 
natural calculation which reconstructs the shower forwards in time.  One starts with the original produced hard virtual 
parton and then constructs the Sudakov factor 
\begin{eqnarray}
\Delta(Q^{2},\mu^{2}) = \exp \left[  - \frac{\alpha_{S}}{2\pi} \int_{\mu^{2}}^{Q^{2}} \frac{d q^{2}}{q^{2}} 
\int dy P(y)\left\{ 1 +  K(y^{-},q^{-},L^{-}, q^{2})   \right\} \right],
\end{eqnarray}
which yields the probability of no resolvable emission between $Q^{2}$ and $ \mu^{2}$ and uses this to numerically estimate the 
probability of the initial parton being produced with a maximum virtuality of $\mu^{2}$. One then samples the splitting function to 
estimate the probability that the produced partons have fractions $y$ and $1-y$ of the parent parton. Unlike the 
case of the vacuum Sudakov factor, the equation above also contains in addition the medium dependent kernel $K$ defined in 
Eq.~\eqref{in-medium-kernel}. This means that at each point, the shower may undergo either a vacuum split or a 
medium induced split. It also clearly demonstrates how the probability of splitting increases in the medium. At each 
point, we estimate the location of the parton based on its formation time, which may be easily obtained from its 
virtuality and its energy. 

This showering routine is repeated to obtain partons with lower and lower virtuality. We terminate the shower when the 
virtuality of the parton reaches $\Lambda_{0} = 1$ GeV. If at this point the parton is found outside the medium, then it is 
convoluted with a vacuum fragmentation function. If it is found inside the medium then it is removed from the final shower. 
We point out again that the medium in all these calculations is not a real nucleus, but rather a static brick. Once the shower 
is calculated in the medium, it is then repeated in vacuum. Thus, both numerator and denominator of the ratio of 
fragmentation functions are calculated by an identical routine.

Using this implementation we may repeat our calculations in the HERMES-like systematics of Fig.~\ref{fig:DGLAP-MC-compare}.
The results of the Monte-Carlo is represented by the solid black line. We should mention in passing that the fragmentation 
function used in the Monte-Carlo calculation is BKK while that in the DGLAP is KKP. We note that the ratio of fragmentation 
functions are rather similar. The Monte-Carlo results are for the most part below the DGLAP calculation. This is because of the 
mechanism by which we can systematically remove the partons which fragment in the medium, which can only be done in 
the MC calculation. The excess at lower $z$ is partially due to the use of a different fragmentation function and partially due to 
some of these partons having a long formation time.

Having tested the Monte-Carlo calculation in HERMES-like systematics ($E=20$ GeV and $Q^{2} = 3 $ GeV$^{2}$),  we 
apply the MC calculation to the EIC-like systematics ($E=25, 35, 50$ GeV and $Q^{2} = 100$ GeV$^{2}$).
First off, we note that even with the larger energies there is a considerable amount of suppression. This is due to the 
larger $Q^{2}$ of the produced jet. Such jets tend to shower a lot and thus end up being considerably affected 
by the medium. This goes beyond what is known at HERMES that increasing the energy reduces the observed 
suppression. We also find a kind of universal suppression at large $z$ which is independent of energy. This kind of 
universal suppression was also noted in the DGLAP calculations performed for comparison with the HERMES data. 
In the earlier DGLAP calculations, the reason for the scaling was due to the vanishing of the real part of the evolution 
equation, leaving the same virtual corrections for different energies.
It is difficult to state at this point if the scaling observed in the Monte-Carlo calculations is due to a similar reason, i.e., 
the vanishing of the real part of the equivalent DGLAP calculation.

If the results reported here are verified by a future EIC, this would represent an interesting observation: to find an almost  50\% suppression 
in the large $z$ yield even for 50 GeV jets. Such high $Q^{2}$ jets should be describable using perturbation theory 
over a large part of their lifetime and would thus yield deep probes of the medium through which they propagate. 
This would allow for a much clearer understanding of the gluonic structure of nucleons inside nuclei. It would also 
greatly facilitate our understanding of how jets are modified in a dense extended environment, which would allow for 
more refined probes of matter produced in heavy-ion collisions.

  
\subsubsection{Jet evolution in hot and cold matter }  
\label{pirner:jetevolution}  

\hspace{\parindent}\parbox{0.92\textwidth}{\slshape   
  Hans J. Pirner   
}  
\index{Pirner, Hans J.}  

\vspace{\baselineskip}
  
We will discuss jet propagation in hot matter first before addressing jet propagation in the ``cold'' matter of 
electron-nucleus collisions. 
A common interpretation of the large pion attenuation in Au+Au
collisions at RHIC is parton  
energy loss, where hadronization occurs outside of the hot zone  
and is not affected by the medium.  There is no doubt that gluon  
radiation plays an important role for the energy loss and the parton  
evolution at RHIC and the LHC. The respective virtualities of partons  
are around $Q=20\gev $ and $Q=100\gev$. In our modeling of jet  
evolution~\cite{Domdey:2010id,Domdey:2008gp} the parton shower is  
treated together with the propagation of the parton in the medium  
which is more realistic because of the relevant time  
scales.  A typical shower at RHIC lasts about $\tau_{evo}=2 \fm$.  The  
non-perturbative part of hadronization involves the decay of the  
resonances at the pre-confinement scale $Q_0=1-2\gev $ into 3-4 pions.  
The lifetime of the plasma can be estimated at $\tau_c= 3.3 \fm$.  
Comparing the two time estimates, we see that at the end of the  
evolution at RHIC, resonances interact with hadronic resonance  
matter. This process can be described by a hadronic theory with cross  
sections slightly larger than hadronic cross sections in  
vacuum. Because of these large cross sections, absorptive effects   
play a decisive role in the observed suppression of hadrons in RHIC  
experiments.  We have advocated two scenarios. Scenario 1 uses the  
conservative radiative energy loss obtained from QCD   
and includes pre-hadron formation and resonance  
absorption. Scenario 2 neglects the resonance phase but tunes up the
energy loss parameter to fit the data.   
  
In more detail, our model~\cite{Domdey:2010id} works as follows:   
The parton produced in a hard process radiates  
successively to reduce its virtuality and become on mass-shell. This  
parton shower is modified by scattering in  
the medium. As both terms enter the same equation, one cannot separate  
scattering and radiation. This equation includes truly   
radiative energy loss, but without coherence.   
Quark fragmentation at RHIC and gluon fragmentation at the LHC   
should give the essential results. The indices on the fragmentation
functions and  the splitting functions can then be dropped and the
formalism becomes simpler.  
For the in-medium fragmentation function  $D^m(x,Q^2)$   
we include into the DGLAP evolution the scattering term $S(x,Q^2)$.  

\begin{equation}\label{mod-ev}  
\frac{ \partial\, D^m(x,Q^2)}{\partial\, \ln{Q^2}}  
=\frac{\alpha_s(Q^2)}{2 \pi} \int_{x}^1 \frac{\mbox{d}z}{z} P(z)  
D^m\left(\frac{x}{z},Q^2\right)+ S(x,Q^2)  
\end{equation}  
with  
\begin{equation}  
S(x, Q^2) \simeq  
  f\frac{ n_g\sigma \langle q_\perp^2 \rangle}{2 m_s Q^2}  
  \left(D(x,Q^2)+x \frac{\partial D}{\partial x}(x,Q^2)\right).  
\end{equation}  
The quantity appearing in the scattering term is the jet  
transport parameter   
$  
\hat q \simeq \bar n \bar \sigma \langle q_\perp^2 \rangle,  
$  
which describes the mean acquired transverse momentum of the parton per unit  
length.   
  
To allow a direct   
fit of experimental data with only parton energy  
loss, we introduce a possible enhancement  factor $f$ in the scattering term.  
The scattering term is most relevant at small virtualities $Q\simeq  
Q_0$ and consequently we have used the scale $Q_0$ in $\alpha_s$ to  
arrive at an upper boundary for $\hat q$. More explicitly, these  
expressions give $\hat q=0.5\gev^2/\fm$ for a temperature of $T=0.3$  
GeV for RHIC and $\hat q=5.2\gev^2/\fm$ for $T=0.5\gev$   
corresponding to the LHC.   
As shown in ref.~\cite{Domdey:2010id} we can fit the RHIC data including   
pre-hadron absorption in the final state resonance gas. The prediction  
for LHC gives $R_{AA} \approx 0.4$. If we use an enhancement factor  
$f=8$ which is beyond any higher order QCD correction, the measurement  
of hadrons with high transverse momentum would be totally suppressed at the LHC.


Let us now discuss jets in cold matter resulting from   
DIS on nuclei. Electron scattering on a target at  
intermediate Bjorken $x$ can be treated along similar lines as the  
DGLAP evolution of the quark jet in the cold medium, whereas  
electron-nucleus scattering at low $x$, in principle  
necessitates the evolution of the quark and antiquark produced from  
photon-gluon fusion. It is not clear whether the cascades from the two  
reaction products behave independently when they propagate through the  
target. In the Ariadne model, two strings result from the quark and  
antiquark produced by photon-gluon fusion. The first string connects  
the antiquark with the quark which emitted the gluon. The second  
string combines the quark with the remnant di-quark of the proton.  
Due to the aligned jet configuration, one of the two strings only
contains a few low momentum particles and perhaps may be neglected in the
first approximation.  The evolution equation outlined above can then be  
applied to jet propagation in cold matter, and applications to an EIC
are planned. Scattering partners of the quark are  
nucleons and the quantity $<\sigma q_{\bot}^2>$ can be derived from  
the dipole cross section on nucleons. The resulting transport  
parameter at HERMES energies is very small $\hat q= 0.035 \gev^2/\fm$  
and has been tested in hadronic broadening of the produced  
hadrons~\cite{Domdey:2008aq}.  For a high energy machine with an  
electron-nucleon energy $E_{cm}=100 \gev$ the transport parameter will  
be larger due to the increasing dipole cross section, we estimate that  
the transport parameter will increase to about $\hat q =0.1  
\gev^2/\fm$. So effects should well be observable, but smaller than at  
RHIC.  
   
\section{Target fragmentation}

\subsubsection{Fragmentation of nuclei - a critical tool for novel QCD
  phenomena }
\label{sec:strikman-targetfrag}


\hspace{\parindent}\parbox{0.92\textwidth}{\slshape 
  Mark Strikman 
}
\index{Strikman, Mark}

\vspace{\baselineskip}

\begin{figure}[b]
\centering
\includegraphics[width=0.8\textwidth]{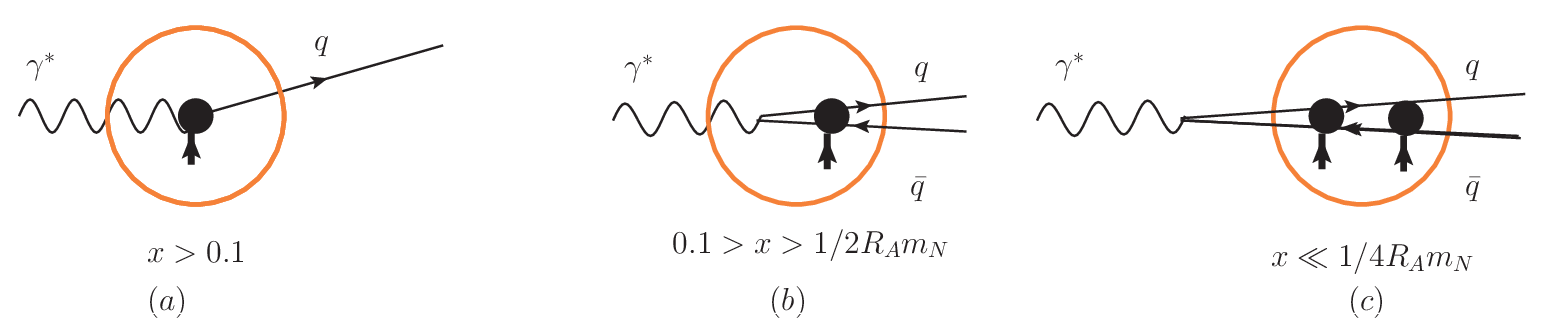}
\caption{\small \label{fig:distance} Space-time picture of DIS in the nucleus rest frame for different x}
\end{figure}

The main focus of the discussions on quark propagation through the
nucleus has been on current fragmentation processes, {\it e.g.}, the
suppression of the leading hadron spectrum, $p_t$ broadening and jet
propagation in nuclear matter. So far, very little
attention has been paid to nuclear fragmentation in DIS. 
To some extent, this is due to the lack of experimental data as such
measurements are very challenging. However, while nuclear effects in the
current fragmentation region decrease with increasing $Q^2$ at fixed
$x$, the nuclear effects in the fragmentation region persist in this
limit, and are likely to depend on $x$. They  may help address a
number of important questions: 
\begin{itemize}
\item 
Are color tubes formed in propagation of quarks through nuclear media?
\item 
How  different are the propagations of gluons and quarks through the nuclear media?
\item
How different are the propagations of quark and dipole?
\end{itemize}
To visualize these questions, it is 
convenient to consider the process in the nuclear rest frame and
distinguish three kinematic regions: (a) For $x\ge 0.1$,  a quark is
knocked out (or a gluon if we consider for example a leading di-jet or
charm production in DIS), (b)  for $0.1 > x\ge 1/(2R_Am_N)$ the  
virtual photon converts to a $q\bar q$ pair inside a heavy nucleus,
and (c) for $x < 1/2R_Am_N$, $\gamma^* \to q\bar q$ transition occurs
predominantly before the target, see Fig.~\ref{fig:distance}

\begin{figure}[t]
  \centering
  \parbox{0.37\linewidth}{
    \includegraphics[width=\linewidth]{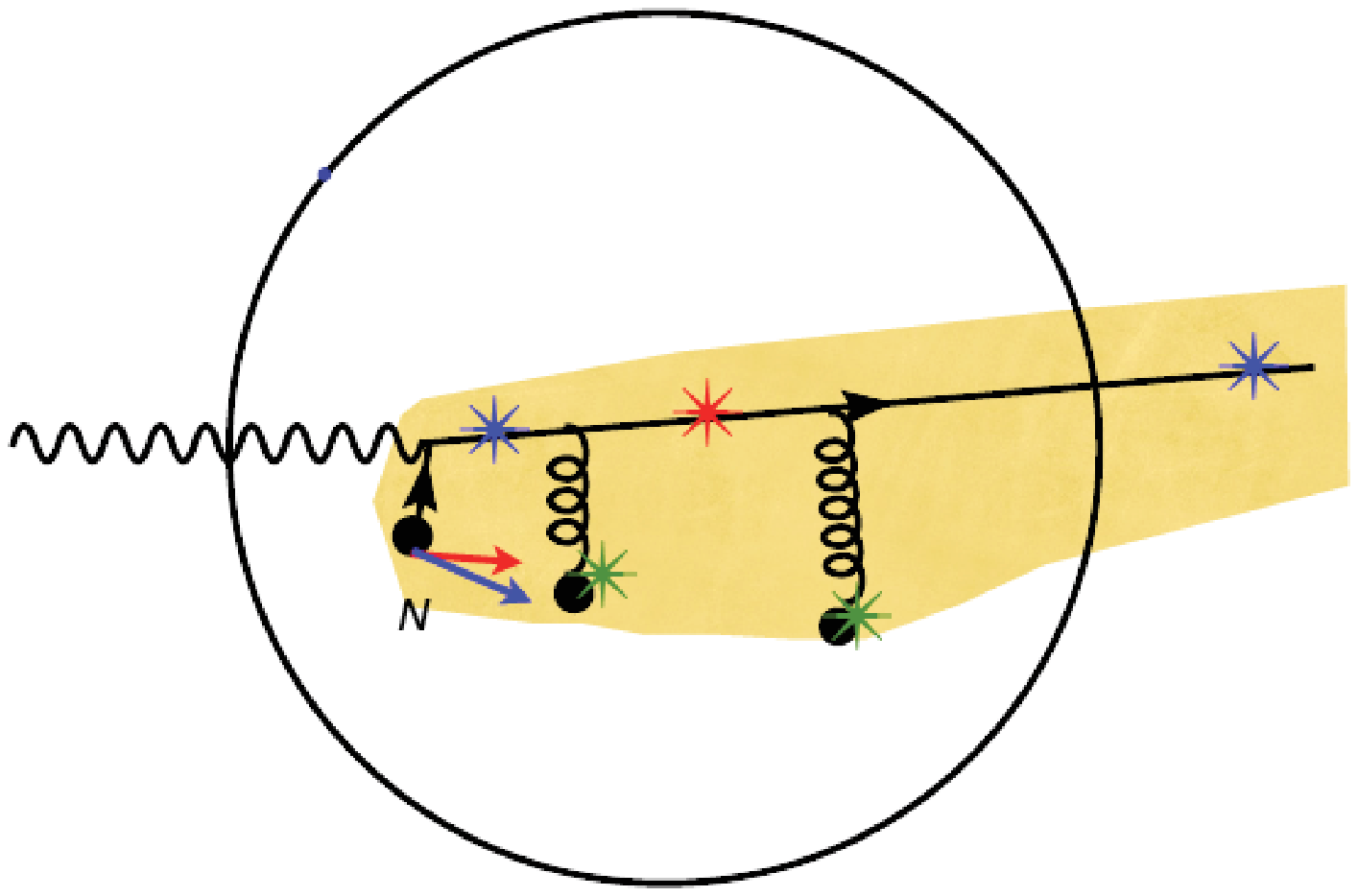}
  }
  \hspace*{1cm}
  \parbox{0.35\linewidth}{
    \includegraphics[width=\linewidth]{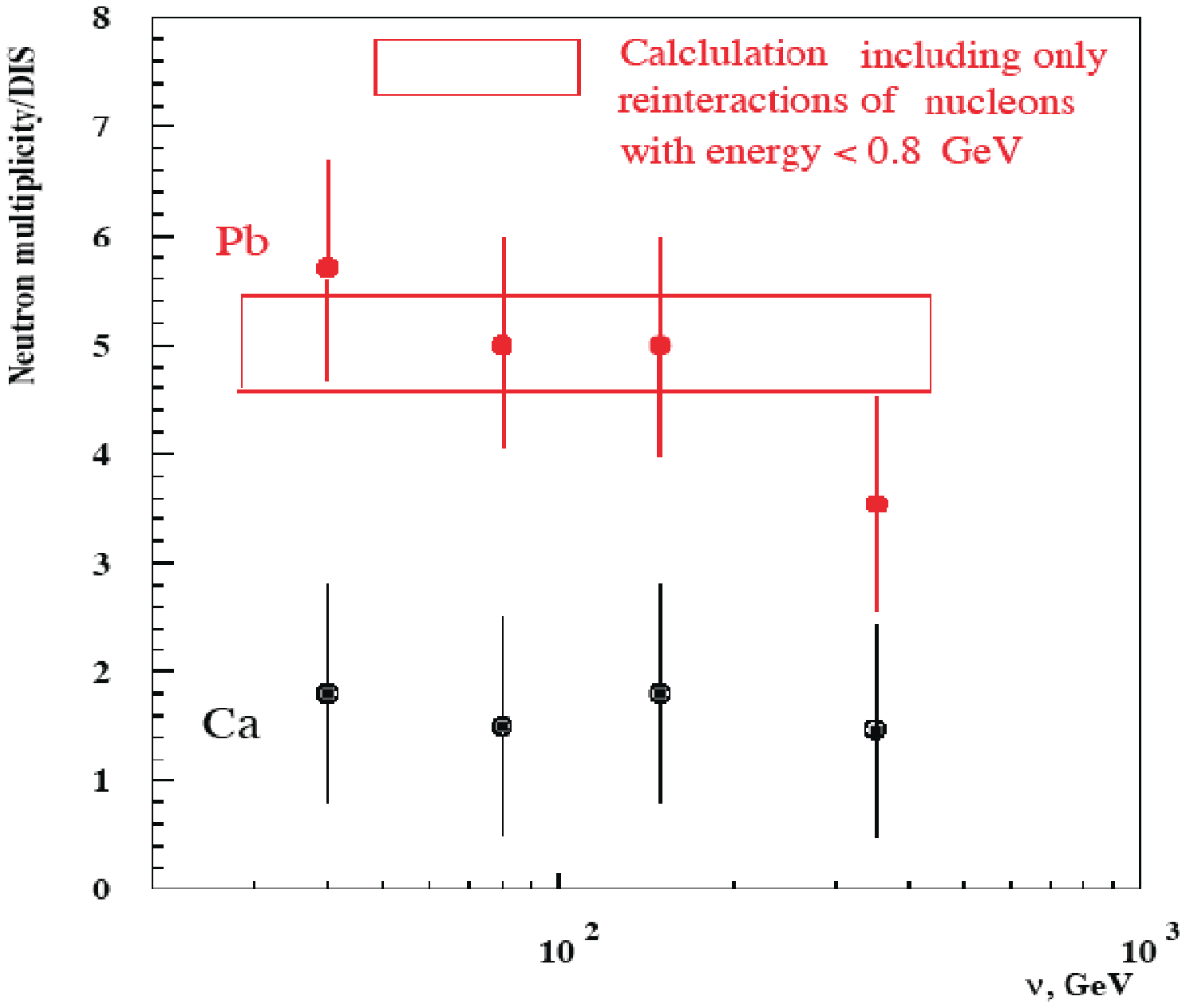}
  }
  \caption{\small 
    {\it Left:} Coulomb exchanges may lead to formation of extended
    spatial regions where color is not screened. 
    {\it Right:} The E665 data \cite{Adams:1995nu} for the soft
    neutron multiplicity compared with the calculation of
    \cite{Strikman:1998cc}.  
  } 
  \label{fig:neutron}
  \label{fig:largex}
\end{figure}

In the case of of $x> 0.1$ and large $Q^2$ corresponding to the knock out
of a quark, a color triplet $qq$ system is left inside the nucleus and
it is typically moving along the virtual photon momentum direction
with a relatively small velocity.  The knocked out quark fragments
into partons/hadrons at longitudinal distances $ y \ge 2p_q/\Delta
m^2 \gg R_A$, where $\Delta m^2\sim 1$ GeV$^2$ can be estimated based on
the current DIS data \cite{Gallmeister:2007an}. It is  
similar to that for color transparency processes. 
As a result, the leading hadron spectrum at large $Q^2$  approaches the
universal limit given by pQCD. This pattern is consistent with the
experimental data. Different to the naive expectations of the parton
model, an A-dependent  $p_t$ broadening in present in this limit.
Naively the hadrons produced in the fragmentation of the quark are formed at
distances given by   $y \ge 2p_h/\Delta m^2$, so that there should
be a depletion in the spectrum  at $p_h^{crit} \sim \Delta m^2 R_A/2 $
followed by an enhancement at rapidities close to the nuclear
rapidity (hadron pileup). Since for heavy nuclei $p_h^{crit} \sim
10\div 20$  GeV/c, one would expect a strong deformation of the hadron
spectrum with a large increase of multiplicity for  $|y-y_A| \le 2
\div 3 $ for $A \sim 200$. In particular, it would be manifested in
the strong break up of the heavy nuclei which is associated with
emission of many soft neutrons. 
One should also expect an increase of the multiplicity of soft
neutrons with an increase of $p_t$ of the leading hadron, since large
$p_t$ selects events with extra Coulomb exchanges which are more
likely for longer quark paths inside the nucleus and should result
in a larger number of wounded nucleons. These may also lead to the 
creation of large unscreened color regions in the nucleus - see
Fig.\ref{fig:largex}. 
 An open question is how these expectations could be  affected by  
  a high degree of coherence in the emission of the partons in pQCD. Such a coherence may lead to   strong screening effects in the formation of the final state and in particular a reduction of $\Delta m^2$ away from the current fragmentation region. 
  Also,  if the color tube is very narrow, a chance that the tube intersects with other nucleons maybe significantly reduced.

For intermediate $x\sim 0.05$, the virtual photon also penetrates any
point in the nucleus but it can hit either quark or antiquark, so in
principle, by studying the properties of the leading hadron one can
compare the structure of  the final state interaction for the removal
of quark and antiquark which maybe different, for example since $\bar
q$ can belong to a color singlet $q\bar q$ cluster.

For small $x  \le 0.03$, the virtual photon predominantly transforms 
into a $q\bar q$ pair  before the target nucleus. In the aligned jet
model one would expect that the number of wounded nucleons would be
given by $A\sigma(eN) /\sigma(ep)$ with the hadrons formed at the
similar distances as in the large $x$ case. 
Hence naively one would expect that  many nucleons will be wounded in
a heavy nucleus, leading to a strong excitation of the nucleus which
is known to be associated with multiple neutron emission, and emission
of protons with momenta of $\ge$ 300 MeV/c, see also
Section~\ref{sec:gallmeister-slown} .
 
The process of neutron emission in DIS off Pb was studied by the
E665 collaboration at FNAL for average $x \sim 0.05$ and $Q^2\sim $
few GeV$^2$ \cite{Adams:1995nu}. The results of the measurement are
compared the theoretical calculation of \cite{Strikman:1998cc} in
Fig.~\ref{fig:neutron}.
Calculations using a Monte Carlo event generator tuned to reproduce
the neutron emission in the proton-nucleus scattering reproduces both
the neutron multiplicity and  the neutron  momentum distribution,
provided only recoil nucleons with energy smaller than 1 GeV are
allowed to interact in the nucleus.  
Taken at face value, this suggests a very strong reduction of the
final state interactions at large energies which is consistent with
the trend of the E665 data to have a smaller neutron multiplicity for
larger $\nu$. 
  
At very small $x$ and moderate $Q^2$, one may reach the black disk
regime. In this regime, the leading hadron spectrum is reduced and the
pQCD factorization for the parton fragmentation breaks down in a gross
way \cite{Frankfurt:2001nt}, see also Section~\ref{sec:Strikman-ct}. 
In this limit, the selection of events with enhanced activity in the
nuclear fragmentation region should lead to reduction of the  
forward spectrum: this would provide a clear signal for a new regime, since
no such correlation is possible in the leading-twist pQCD regime. 

In summary, hadron production in the nuclear fragmentation region is
very sensitive to the dynamics of space-time evolution of the triplet
and octet color tubes as well as of color dipoles.  
This is one of the  unexplored frontiers where the collider kinematics
will allow a qualitative improvements in the data, and likely lead
to the discovery of a series of new regularities. This may include  a much
higher degree of coherence in the fragmentation (hinted at by the E665
data) than suggested by the current models. 
Understanding of the fragmentation dynamics will be also of great help
for understanding the dynamics in the nuclear
fragmentation region in heavy ion collisions, where high density
quark-gluon systems may be produced.

\subsubsection{In-medium hadronization and EMC effects in nuclear SIDIS}
\label{sec:eA-CCK}
\hspace{\parindent}\parbox{0.92\textwidth}{\slshape
  C. Ciofi degli Atti, L. P. Kaptari, B. Z. Kopeliovich,
  and C. B. Mezzetti 
}
\index{Ciofi degli Atti, Claudio}
\index{Kopeliovich, Boris Z.}
\index{Kaptari, Leonid P.}
\index{Mezzetti, Chiara Benedetta}


The SIDIS  process $A(e,e'(A-1))X$  in which, instead of the
leading hadron, a nucleus $(A-1)$ in the ground or in low excitation states is
detected in coincidence with the scattered electron, can provide new
information about the mechanism of hadronization and the origin of the
EMC effect \cite{CiofidegliAtti:2002as,CiofidegliAtti:1999kp,CiofidegliAtti:2003pb,Palli:2009it,Atti:2010yf}.
Two main advantages of the new SIDIS process over the classical SIDIS
\cite{Accardi:2009qv} and inclusive $A(e,e')X$ scattering
\cite{Norton:2003cb} are worth mentioning here. Firstly, it can
provide a new insight into the space-time development of
hadronization at the early stage, which can be probed only by
placing additional scattering centers at microscopic distances,
i.e. by using nuclear targets. By detecting a jet produced on a
nuclear target, one can get information about its time
development, but in a rather indirect and complicated way, since
cascading inside the nuclear medium essentially modifies the
observables.
Measuring the recoil nucleus supplies additional and cleaner
information about the dynamics of hadronization; in particular,
this process is free of the uncertainties caused by cascading, and
the survival probability of the recoil nucleus is extremely
sensitive to the multiparticle components of the jet
\cite{CiofidegliAtti:2002as}. Secondly,  a proper
ratio of the cross sections on a nucleus $A$ taken at different values
of the Bjorken scaling variable $x_{Bj}$ provides information on
the nucleon structure functions in the medium, $F_2^{N/A}$. Several
experimental projects to investigate the new process at $12\,\,
\gev$ have been proposed thanks to the
development of proper recoil detectors \cite{CETAL-proposals}, and the
experiment on Deuteron targets has already been performed
\cite{Klimenko:2005zz}.

\begin{figure}[tb]
  \centering
  \includegraphics[width=0.32\textwidth]{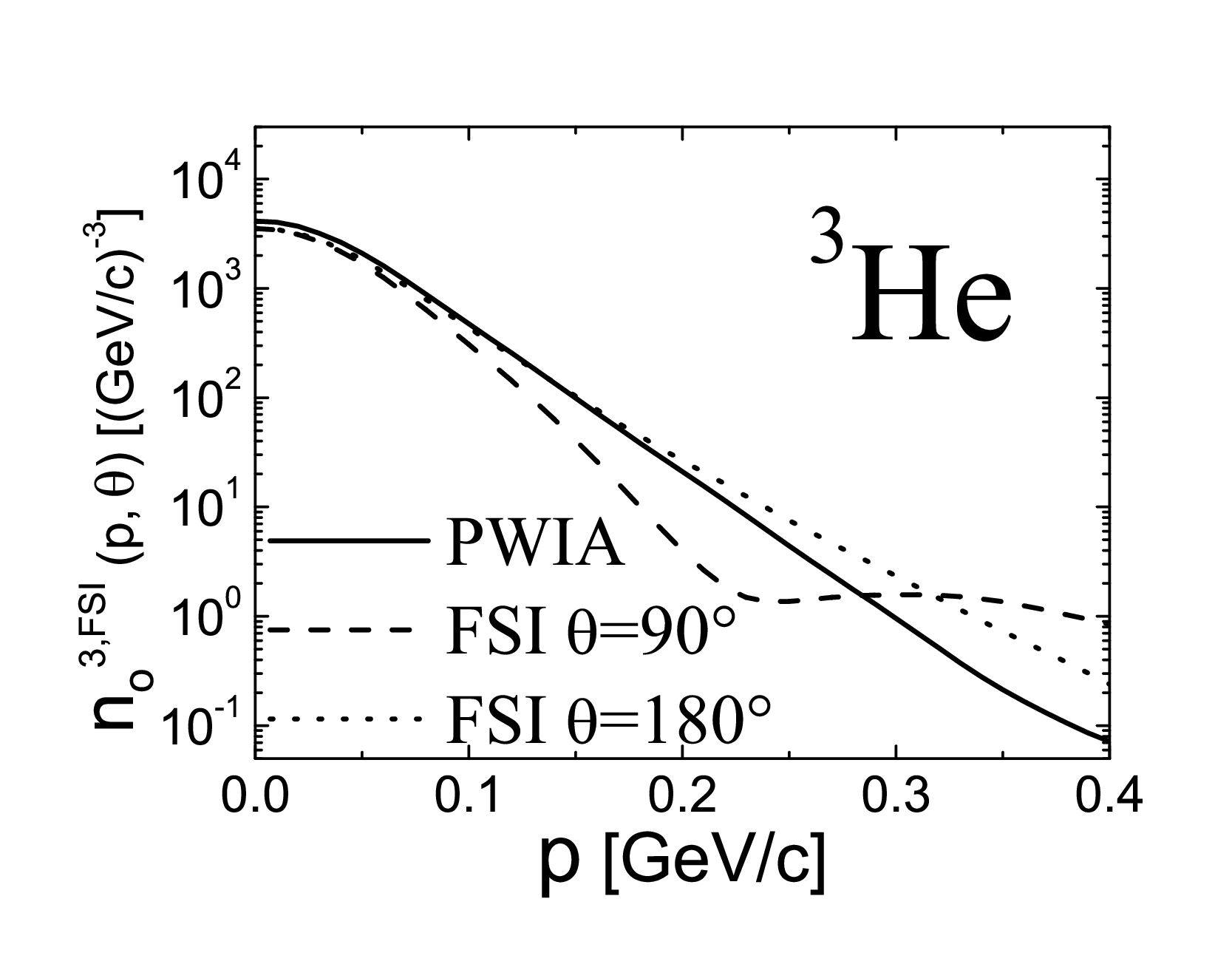}
  \includegraphics[width=0.32\textwidth]{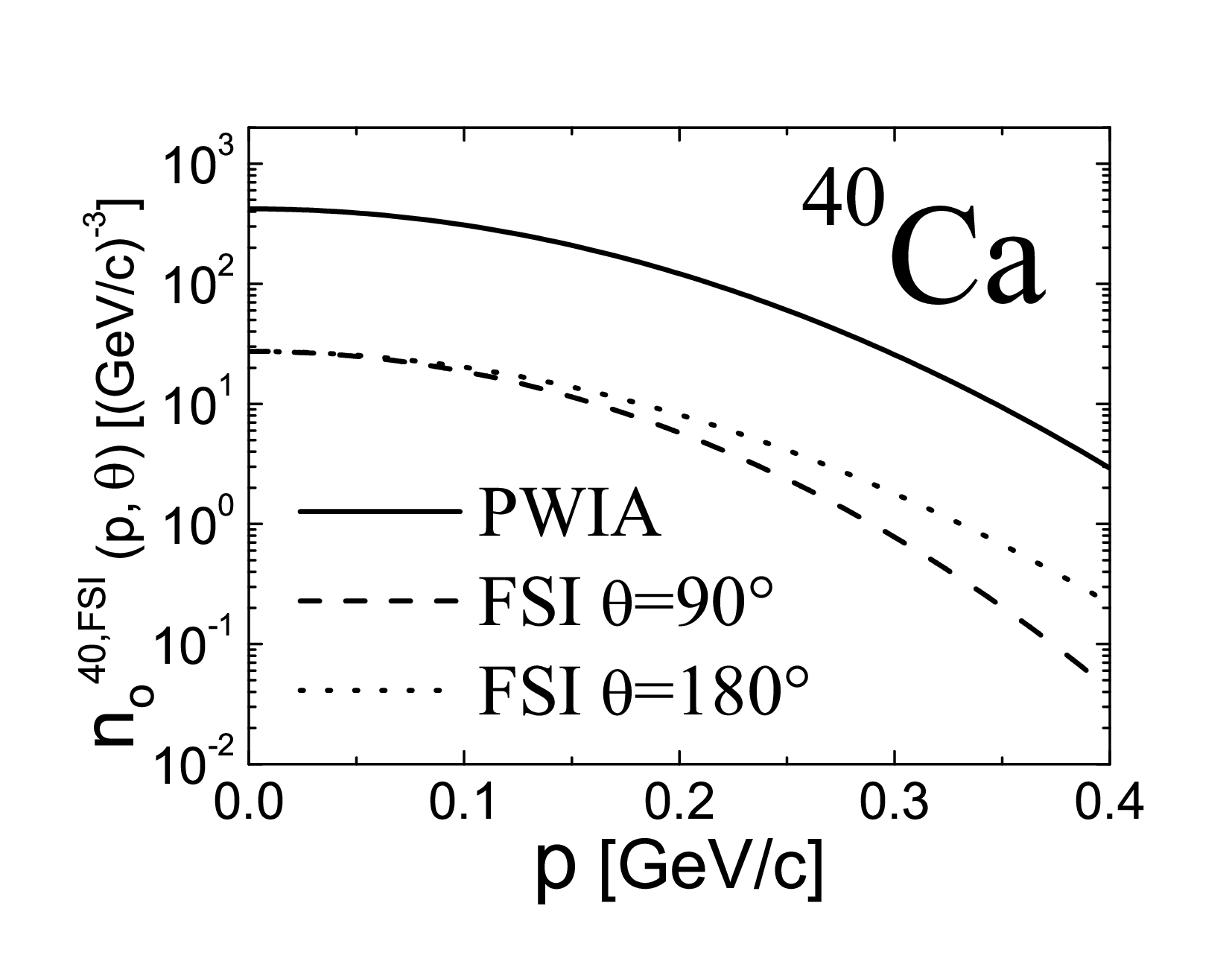}
  \vskip-0.5cm
  \includegraphics[width=0.32\textwidth]{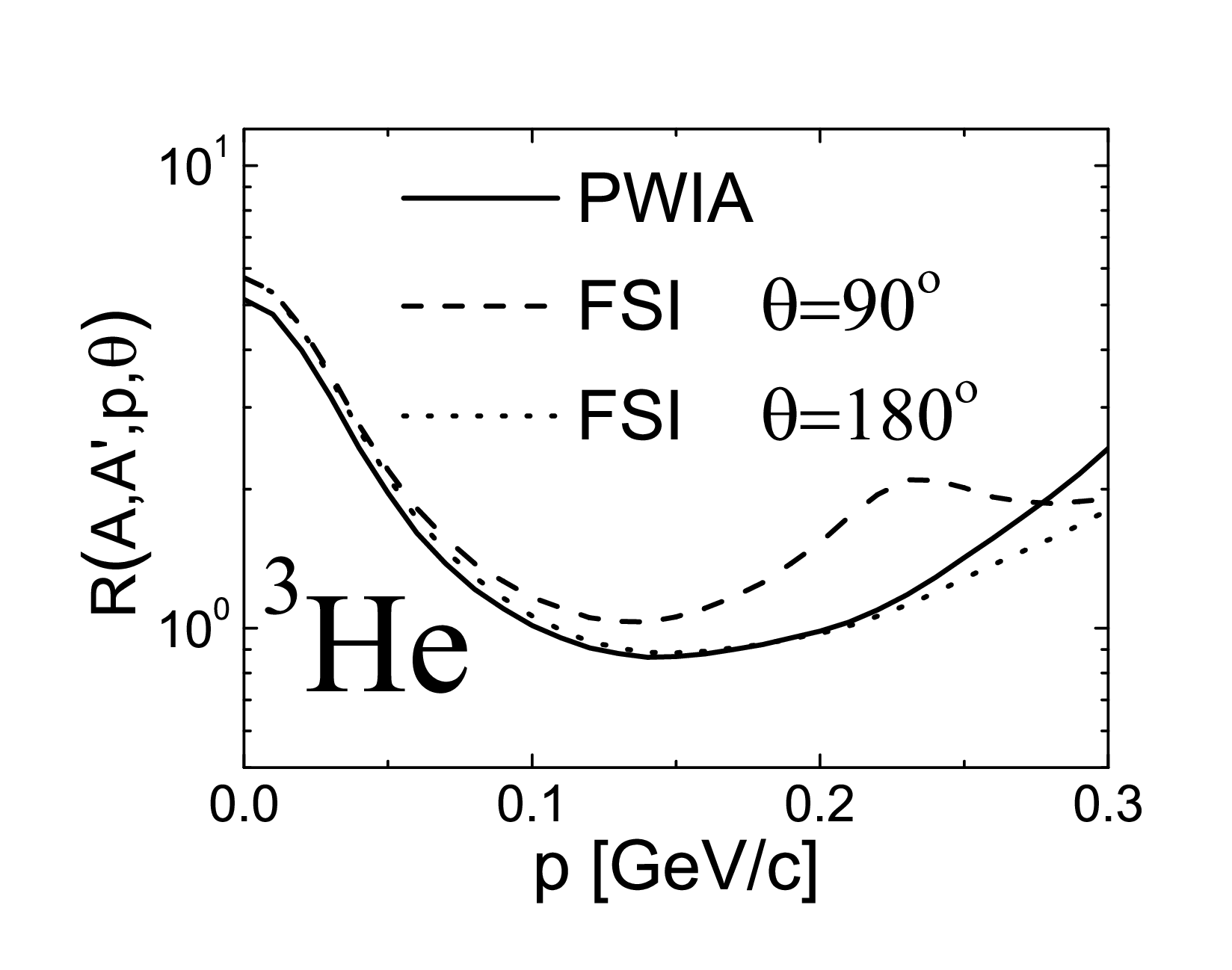}
  \includegraphics[width=0.32\textwidth]{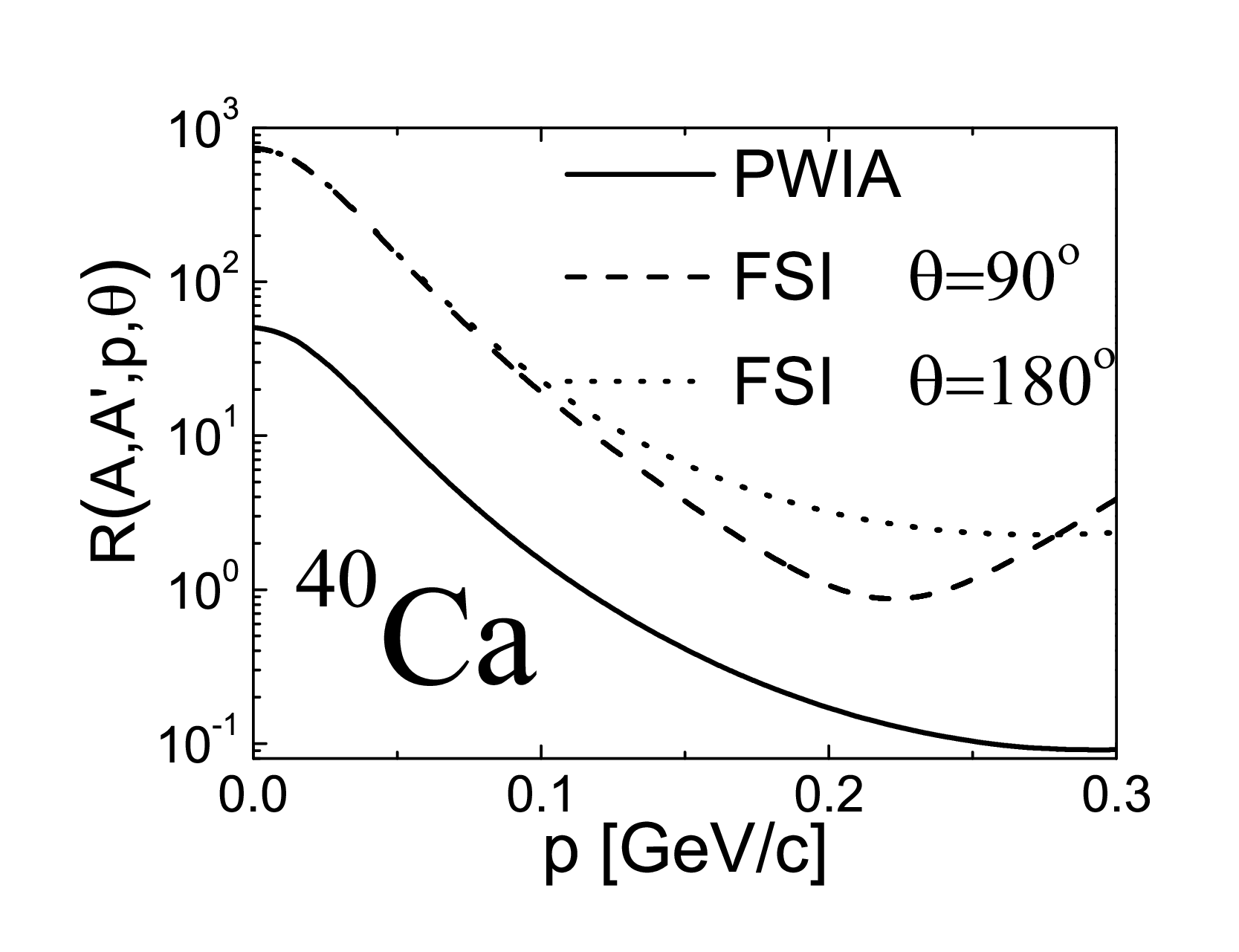}
  \vskip-0.5cm 
  \caption{\small 
    {\it Top panels:} 
    the distorted momentum distributions $n_0$ with 
    $\theta=\theta_{\widehat{ {\bf P}_{A-1}{\bf q}}}$ and 
    $ p\equiv |\textbf{P}_{A-1}|$ for $^3He$ and $^{40}Ca$. 
    {\it Bottom panels:} 
    The ratio $R(A,A')$ of Eq.~\ref{ratioa-1}
    with  $A =2$,and $A'=^3He$ or $^{40}Ca$.
  }
  \label{fig2_cck} 
  \label{fig3_cck}
\end{figure}

The basic ingredients of the theoretical calculation are the nuclear momentum
distributions, the nucleon structure function $F_2^{N/A}$ in the
medium, and the
effective cross section of  interaction between the hadronizing
nucleon debris and  the spectator nucleons. This last reads
\cite{CiofidegliAtti:2002as}
\begin{equation}
\sigma_{eff}(z,x_{Bj},Q^2)\,\equiv
\sigma_{eff}(z)\,=\,\sigma_{tot}^{NN}\,+\,\sigma_{tot}^{\pi
N}\,\big[\,n_{M}(z)\,+\,n_{G}(z)\,\big] \label{sigmaeff}
\end{equation}
where $\sigma_{tot}^{NN}$ and $\sigma_{tot}^{\pi N}$ are the total
nucleon-nucleon ($NN$) and pion-nucleon ($\pi N$) cross sections,
and the $Q^2$- and $x_{Bj}$-dependent quantities $n_{M}(z)$ and
$n_{G}(z)$ denote the pion multiplicities due to the breaking of
the color string and to gluon radiation, respectively. Their
explicit form directly follows from the
hadronization mechanism proposed in Ref.~\cite{Kopeliovich:2004kq},
leading to a satisfactory description of the grey track production in
DIS off nuclei \cite{CiofidegliAtti:2004pv}.

The cross section of the $A(e,e'(A-1))X$ process
\cite{CiofidegliAtti:2002as,CiofidegliAtti:2003pb} schematically reads
\begin{eqnarray}
  \frac{d\sigma^{A,FSI}}{d x_{Bj} d Q^2  d  \textbf{P}_{A-1}} =
  F_2^{N/A}(x_A,Q^2,k^2) \otimes n_0^{A,FSI}(\textbf{P}_{A-1})
  \label{crossdist}
\end{eqnarray}
where $x_A = x_{Bj}/z_1^{(A)}$, $z_1^{(A)} = (M_A k \cdot q)/(m_N
P_A\cdot q)$, $k$ is the
four-momentum of the bound nucleon and $P_A$ of the target nucleus. In
this equation, $n_0^{A,FSI}(\textbf{P}_{A-1})$ is the distorted momentum
distribution of the bound nucleon after final state interaction (FSI) 
with the debris nucleon (${\bf k}_1 = - {\bf
P}_{A-1}$ in Plane Wave Impulse Approximation):
\begin{eqnarray}
  n_0^{A,FSI}(\textbf{P}_{A-1})
  =\frac{1}{2J_A+1}
  \sum_{{\cal M}_A,{\cal M}_{A-1}} \left | \int\, d {\bf r}_1^{\prime}
    e^{i {\bf P}_{A-1} {\bf r}_1^{\prime}} \langle \Psi_{J_{A-1}, {\cal M}_{A-1}}^{0}
    |S_{FSI}^{XN}|
    \Psi_{J_{A}, {\cal M}_{A}}^{0} \rangle
  \right |^2\!\!\!\!
  \label{dismomfsi}
\end{eqnarray}
where
$S_{FSI}^{XN}$ is  the debris-nucleon eikonal scattering $S$-matrix
which differs from the Glauber form because of the $z$
dependence of $\sigma_{eff}$~\cite{glau2}. 
The results of some calculations are presented in what follows, using
for Deuteron and $^3He$ realistic wave functions \cite{Kievsky:1999ma}
corresponding to the AV18 interaction \cite{Wiringa:1994wb}, and for
heavy nuclei single particle mean field wave functions. 
A good agreement between our parameter-free calculation \cite{Atti:2010yf} and
the experimental data for $^{}2H(e,e'p)X$ around $\theta \simeq
90^o$ is exhibited.

\begin{figure}[tb]
  \centering
  \includegraphics[width=0.35\textwidth]{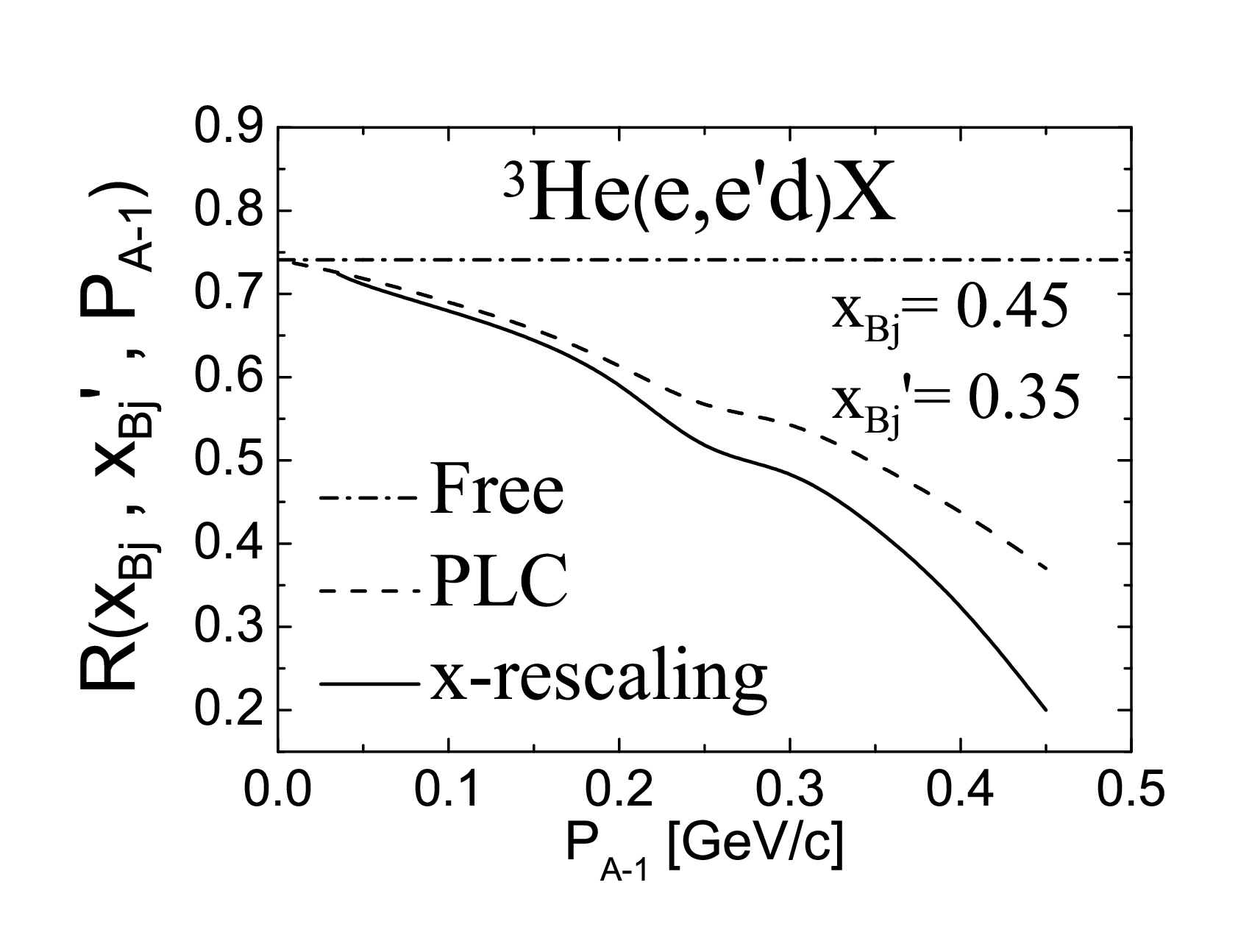}
  \includegraphics[width=0.35\textwidth]{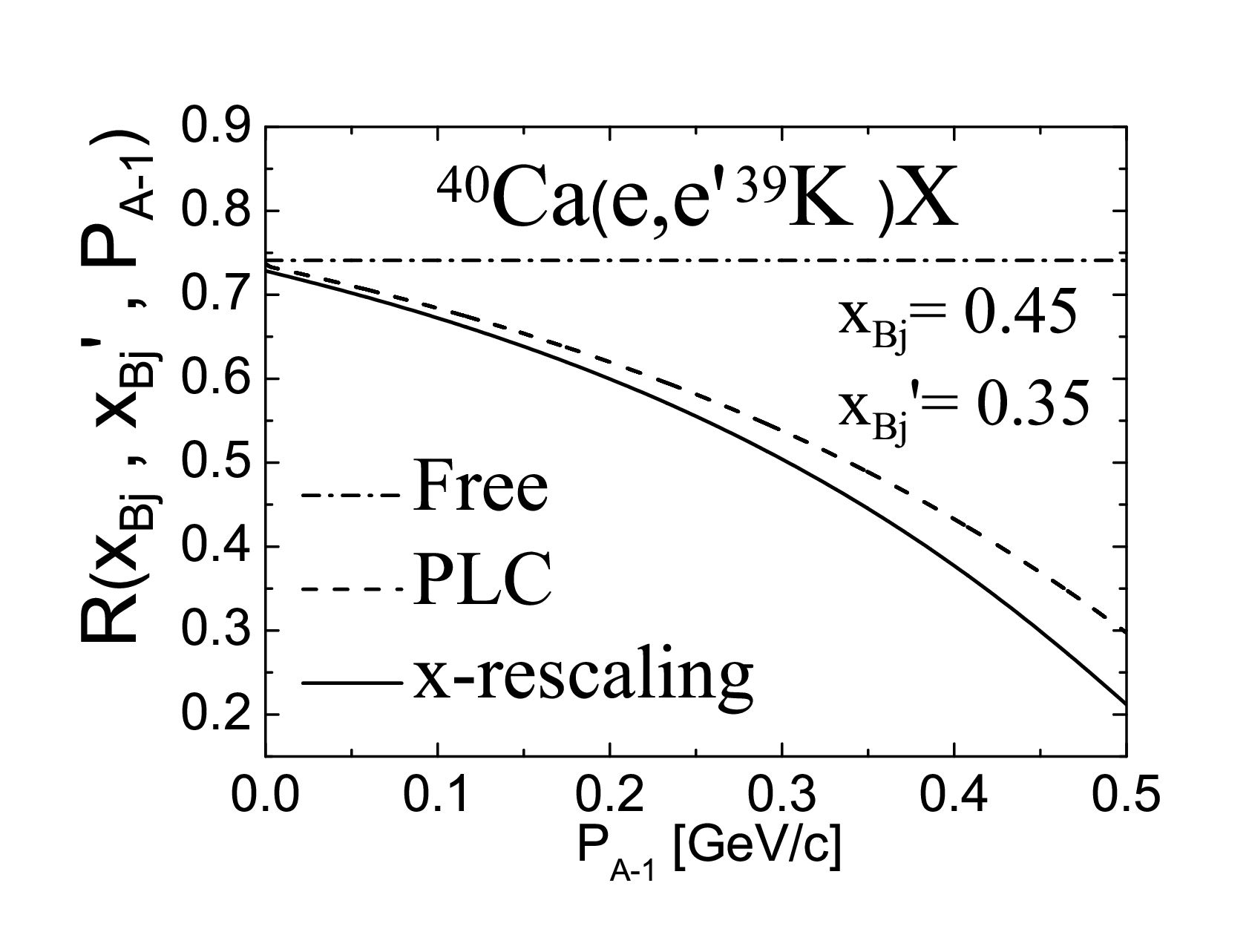}
  \vskip-0.5cm 
  \caption{\small \label{fig4_cck}
    The ratio $R(x_{Bj},x_{Bj})$ of Eq.~\eqref{ratioa} for the process
    $^3He(e,e'd)X$ and $^{40}Ca(e,e'\,^{39}K)X$ calculated with different
    nucleon structure functions: i)
    free structure function; ii) off mass-shell
    (x-rescaling) structure function; iii)
    with suppression of point-like configurations (PLC) in the medium
    depending upon the nucleon virtuality
    \cite{CiofidegliAtti:2007vx} ($P_{A-1}\equiv |\textbf{P}_{A-1}|$).}
\end{figure}

The distorted momentum 
distributions of $^3He$ and $^{40}Ca$ at kinematics more
appropriate for an EIC are shown in Fig. \ref{fig2_cck}.  As already
pointed out, the FSI is governed by
 the details of $\sigma_{eff}$ and strongly affects
the survival probability of $(A-1)$, as it can be seen by
comparing the results for $^3He$ and $^{40}Ca$. Let us denote the
cross section (\ref{crossdist}) by $\sigma^{A,FSI}$. Then,
if our description is correct, the ratio of cross sections on
different nuclei, 
\begin{eqnarray}
  R(A,A', \textbf{P}_{A-1}) = \frac{\sigma^{A,exp} ( x_{Bj},Q^2,|\textbf{P}_{A-1}|,z_1^{(A)},y_A ) }
  {\sigma^{A',exp} ( x_{Bj},Q^2,|\textbf{P}_{A-1}|,z_1^{(A')},y_{A'} ) }
  \rightarrow \frac{{n_0^{(A,FSI)}(\textbf{P}_{A-1})}}{{n_0^{(A',FSI)} ( \textbf{P}_{A-1})}}
   \label{ratioa-1}
\end{eqnarray}
should be  governed only by the FSI, as shown in Fig.~\ref{fig3_cck}.

In order to tag bound nucleon structure functions, whose nuclear
modification is one of the causes of the EMC effect, one has to get rid
of  the distorted nucleon momentum 
distributions and other nuclear structure effects. This can be
achieved by considering the ratio of the cross sections on a nucleus
$A$ measured at two different values of the Bjorken scaling
variable, $x_{Bj}$ and $x_{Bj}^{\prime}$, leaving unchanged all
other quantities in the two cross sections, i.e., the ratio
\begin{eqnarray}
R(x_{Bj}, x_{Bj}^{\prime}, |\textbf{P}_{A-1}|) =\frac{\sigma^{A,exp} ( x_{Bj},Q^2,|\textbf{P}_{A-1}|,z_1^{(A)},y_A ) }
             {\sigma^{A,exp} (  x_{Bj}^{\prime},Q^2,|\textbf{P}_{A-1}|,z_1^{(A)},y_{A} ) }
\approx \frac{F_2^{N/A}(x_A,Q^2,k^2)}
{F_2^{N/A}(x_{A}^{\prime},Q^2,k^2)}
\label{ratioa}
\end{eqnarray}

\begin{wrapfigure}{r}{0.45\textwidth}
  \begin{center}
    \includegraphics[width=0.44\textwidth]{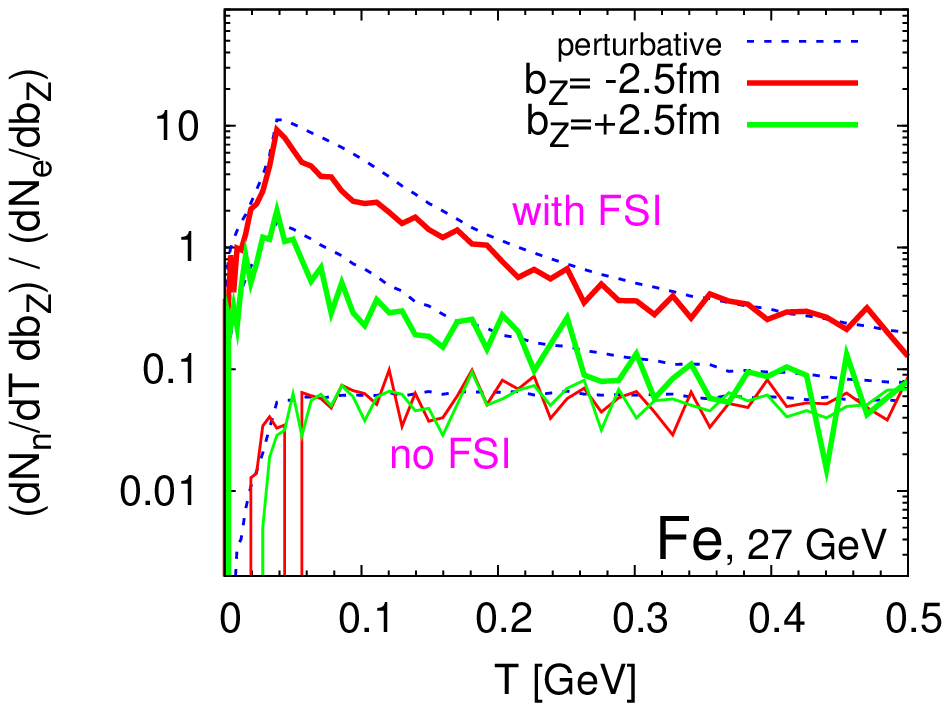}
  \end{center}
  \caption{\small The production cross section of neutrons with low momenta for different longitudinal production points, normalized to the corresponding number of events. (The calculations are preliminary.)}
\label{fig:GiBUU_Momenta}
\end{wrapfigure} 

which depends only upon the nucleon structure function $F_2^{N/A}$. Calculations
of the ratio (\ref{ratioa}) have been performed \cite{Atti:2010yf} using
three different  structure functions, namely, the free one, giving
no EMC effect, and  two medium dependent structure functions,
yielding only a few percent difference in the inclusive cross
section.
It can be seen from Fig. \ref{fig4_cck} that the
discrimination of  different models of the medium dependence
of $F_2^{N/A}(x_{A},Q^2,k^2)$ can indeed be achieved, especially at
large $P_{A-1}\equiv|{\bf P}_{A-1}|$.

In conclusion, from what shown here and in the original papers
\cite{CiofidegliAtti:1999kp,CiofidegliAtti:2002as,CiofidegliAtti:2003pb,Palli:2009it,Atti:2010yf}
it appears  that the SIDIS process $A(e,e'(A-1))X$,
with detection of a complex nucleus $(A-1)$,  would be extremely useful
to clarify the origin of the EMC effect and
to study the early stage of hadronization at short formation times.
At EIC kinematics (large $Q^2$ and $W_X^2$), the theoretical
assumptions underlying Eqs.\eqref{sigmaeff}-\eqref{dismomfsi}
are expected to be of higher validity than at
lower energy. The problem remains as to whether  experiments of the kind we are
discussing, i.e. the detection of low-momentum light nuclei at
specific angles, could be performed at an Electron Ion Collider.
We have calculated the process $^3He(e,e'd)X$ at various EIC
kinematics and found that, e.g. at $Q^2 \,\simeq 30\,\, \gev^2$
and $x_{Bj} \simeq 0.7$, when the Deuteron is emitted at about
$90^{0}$ in the target rest frame, this corresponds to about
$1^{0}$ in the direction of the incident nucleus in the collider
CM frame.

\subsubsection{Proving the microscopic  origin of nuclear forces}
\label{sec:strikman-nuclearforces}
\hspace{\parindent}\parbox{0.92\textwidth}{\slshape 
  Mark Strikman 
}
\index{Strikman, Mark}

\vspace{\baselineskip}

An important task for the EIC is to probe nuclear forces on the microscopic level using hard probes. 
Before describing some of the possible avenues for EIC research, it is worth summarising what is already known from the analyses of the experimental studies of the nuclear pdfs.

\begin{itemize}
\item The quark distributions at large $x$ are suppressed as compared to the naive expectations based on the picture of the nucleus built of nucleons with internal parton distributions coinciding with the free nucleon pdfs, the so called EMC effect -  for a review see e.g.~\cite{Arneodo:1992wf}.
 However, the EMC  effect modification of the nucleon pdfs remains small - $\le 2\%$ for $x\le 0.5$ after one takes into account the Coulomb field contribution into the wave function of the heavy nuclei and uses the proper scaling variable $x_A=AQ^2 /2q_0 M_A$ for the comparison of the nuclear cross sections \cite{Frankfurt:2010cb}. The modification of the nucleon pdfs strongly grows with $x$ at $x > 0.5$ reaching $\sim$10\% at $x$ = 0.6.
 \item The $A$-dependence of the EMC effect at large $x$ indicates that the main contribution to the EMC effect is due to scattering off the short-range correlations (SRC)  in nuclei.
 
 \item  Experiments at JLab confirm approximate $A$-independence  of the momentum distribution of nucleons in the short-range correlations, though the absolute probability is a factor of $\sim 5$ larger in heavy nuclei than in the deuteron, for a recent review see \cite{Arrington:2011xs}. 
 
 \item The measurements of the antiquark distributions in nuclei were performed using the Drell-Yan process. No enhancement of the $\bar q _A/ \bar q _N $ ratio was observed for  $x \sim 0.1$ where  the models of nuclear forces with dynamic pion fields predicted $10\div 20 \%$.
 
\item Application of the baryon and momentum sum rules indicate that the valence quarks and gluons are enhanced in nuclei at $x\sim 0.1$ \cite{Frankfurt:1990xz,Eskola:1998df}.
\end{itemize}

The region of $x\sim 0.1$ is especially interesting for the purposes of studying the QCD origin of the nuclear forces since it corresponds to the Ioffe distances $\sim 1/ 2xm_N \sim 1$ fm, characteristic of more medium and short-distance nuclear forces.  The regularities listed above suggest that meson exchanges which lead to the enhancement of the sea quark distributions are less important than it is suggested in the meson models of the nuclear forces, while quark and gluon interchanges between nearby nucleons play a significant role. The inclusive measurements at the EIC  will directly  measure 
$V_A/V_N, G_A/G_N$ for $x\sim 0.1$.

A new tool which will be available at the EIC is exclusive hard processes for which the QCD factorization theorem has been proven for the processes $\gamma_L + T \to VM + T'$ for the Bjorken limit and the mass of the final system $T'$ being fixed ~\cite{Collins:1996fb}. We will focus on the processes with deuteron target since in this case it is easier to select scattering off the compact proton - neutron configurations and measure a complete final state. 
The rational here is that the structure of the SRC in nuclei is approximately the same while using a heavier target, say $^4$He would increase the impulse approximation interaction rate by a factor of $\sim 3$ only (due to 
  a higher probability of SRCs in $^4$He).  However this apparent gain  will be compensated to a large extent by the final state absorption/distortions and multistep processes significantly complicating the interpretation of the observations.
 
 The first question one can address is whether the quark and gluon transverse distributions in bound nucleon are the same.
 The simplest possible processes are break up of the deuteron
 \begin{equation}
 \gamma + {}^2H \to J/\psi  + p + n.
 \label{vm1}
\end{equation}
and
\begin{equation}
 \gamma^*_L + {}^2H \to \rho^-  + p + p.
 \label{vm2}
\end{equation}
which probe gluon and quark transverse distributions.

The exclusivity of the process could be tested by measuring 
$p_t(N_1)+p_t(N_2)+ p_t(VM)=0$. 
To avoid an ambiguity which of the nucleons was interacting via the hard process $\gamma^* + N \to VM +N$, one needs to select transverse momenta of the vector meson  $\ge 600 \div 700 MeV/c$ with momentum of the nucleon $N_1$ in approximately the opposite direction.  For the spectator to belong to the SRC one needs to ensure that it has a large momentum in the deuteron rest frame $\ge 0.3 GeV/c$. It could be either mostly longitudinal or have a transverse component sufficiently deviating from the direction opposite to $p_t(VM)$.

The measurement involves studying the dependence of the ratio of the cross
section of the reaction \eqref{vm1}, \eqref{vm2} and elementary reaction 
$$R(p_sp)={ {d \sigma (\gamma^*+{}^2H \to NN +VM) \over dW_{\gamma N} dQ^2
    dt, d p_{sp} }\over {d \sigma (\gamma^*+ N \to N +VM) \over dW_{\gamma N} dQ^2 dt} }$$
on the momentum transfer to the vector meson - $t\approx -p_t(VM)^2$ for fixed values of $p_{sp}$. Deviations of the t-dependence from a constant (which can be calculated in the two nucleon approximation) would signal the change of the size of the bound nucleon. The theoretical expectation is that such effects are proportional to the nucleon ``off-shellness'' so they should rapidly increase with increasing $p_{sp}$, roughly $\propto p_{sp}^2$~\cite{Frankfurt:1988nt,CiofidegliAtti:2007vx}. It is worth noting that nucleon deformation along and transverse to the direction between the nucleons may differ (like in the case of polarization of the atoms in the molecules). Hence a nontrivial dependence of $R(p_sp)$ on the angle between $p_{sp} $ and $p_t(VM)$ is possible (such a dependence is absent if the deformation depends only on the virtuality). It would be possible to study the dependence of $R(p_sp)$ on $x$ (for photoproduction of $J/\psi$ on $m_{J/\psi}^2/W^2$ probing how the nucleon deformation depends on $x $ of the gluon in the bound nucleon wave function.

Another possible direction for studies is probing directly the pion exchange mechanism using exclusive hard processes - for example $\gamma + {}^2H \to J/\psi + \pi^- + pp$ with 
transverse momenta of $J/\psi$  and $ \pi^-$ back to back 
and large deuteron rest frame momenta of both protons (to ensure that the process occurs off the SRC).

The discussed class of the reactions is well suited also for looking for non-nucleonic baryonic components in the SRCs
(six quarks, $\Delta\Delta$, ...). For example one can study the process 
$\gamma + {}^2H \to J/\psi + \Delta^{++} + \Delta^- $ where transverse momenta of $J/\psi $ and one of $\Delta$'s are back to back.  The advantage of this reaction as compared to medium energy processes  is the absence of a non-vacuum exchange in t-channel. 

\subsubsection{Slow neutrons and final-state interaction length}
\label{sec:gallmeister-slown}
\hspace{\parindent}\parbox{0.92\textwidth}{\slshape 
  Kai Gallmeister, Ulrich Mosel 
}
\index{Gallmeister, Kai}
\index{Mosel, Ulrich}

\vspace{\baselineskip}

With collider kinematics, it is very instructive to look at
``slow'' nucleons of energy less than 10 GeV, considered slow with
respect to the (fast) target nucleon \cite{Strikman:1998cc}, see also
Section~\ref{sec:strikman-targetfrag}.
Performing some exploratory simulations within the GiBUU 
framework (see Section~\ref{sec:e+A_with_GiBUU}) we are confronted
with a lot of complications. In Fig.~\ref{fig:GiBUU_Momenta} we show
some distributions of slow neutrons as 
a function of energy for different production points in the
longitudinal axis, normalized to the corresponding number of scattered
electrons.
This result is to be considered as preliminary, since we learned that
we need a more accurate treatment of Pauli-blocking and binding
effects in the few MeV region. In addition, we need to take into
account the production of slow nucleons via evaporation and
fragmentation. This work is currently in progress by inclusion of a
multi-fragmentation framework (SMM) \cite{Bondorf:1995ua} and
correcting for effects of the large energy gap between initial
interaction and fragmenting nucleons.

It has been proposed by Ciofi degli Atti and coworkers in many papers
(see Section~\ref{sec:eA-CCK}) that the interaction cross section of
the jet particles within a SIDIS 
event with the debris of the target nucleus shows interesting
formation length dependencies. We see a large potential for our
GiBUU model to study all these questions.

\section{Bose-Einstein correlations at an electron-ion collider}

\hspace{\parindent}\parbox{0.92\textwidth}{\slshape
Gerald~P.~Gilfoyle 
}
\index{Gilfoyle, Gerald P.}

\vspace{\baselineskip}

QCD directs the formation of hadrons from quarks and gluons in hard scattering.
However, our understanding of this process is {\it ad hoc}; there is no full, QCD-based theory
to explain hadronization and fragmentation.
To probe these processes, we propose to take advantage of an iconic quantum mechanical effect, 
the symmetrization 
of the wave function required
for bosons.
Particles formed near one another will have overlapping wave functions and the interference 
of the wave functions
produces correlations in the intensity and momentum dependence of the final particles.
These Bose-Einstein Correlations (BEC) (or the Hanbury-Brown Twiss effect) are examples of 
intensity interferometry
and can be used to study the space-time extent of the source of the particles and/or learn 
about the dynamics 
of their formation.
They have been used to investigate hot nuclear matter, but there are only a few cases where
$e+A$ interactions have been studied.
That work revealed that BECs can be used to study the QCD string in hard scattering
and our simulations show we will be able to make precise measurements
of the BEC source size at an EIC.

Bose-Einstein Correlations arise when two identical bosons are detected and 
their joint
wave function $|p_1 p_2 \rangle$
($p_i$ is the particle 4-momentum) must be symmetric under particle exchange.
In other words, when the two bosons are detected from different points in space-time, the observer cannot 
distinguish the origin
of each particle so their amplitudes must add.
This requirement gives rise to interference terms in the intensity
that do not exist for non-identical particles.
In fact, for identical fermions there would an anti-correlation between the particles.
The BEC in energy-momentum space is related to the extent of the source in its spatial dimensions
and the correlation function can be written as 
\begin{equation}\label{GPG:eq1}
R(Q_{12}) = \frac{dN/dQ_{12}}{dN_{ref}/dQ_{12}}
\end{equation}
where $Q_{12} = \sqrt{-(p_1-p_2)^2}$ is the Lorentz-invariant momentum 
difference between the identical bosons and $N_{ref}$ is a reference spectrum constructed with no BECs.
The correlation function is often parameterized as
\begin{equation}\label{GPG:eq2}
R(Q_{12}) = \alpha \left ( 1 + \lambda \Omega(Q_{12} r_{12}) \right ) \left ( 1 + \beta Q_{12} \right ).
\end{equation}
In static models of particle sources, $\Omega(Q_{12} r_{12})$ can be interpreted as the Fourier 
transform of the spatial distribution
of the emission region of bosons with overlapping wave functions and is characterized by the 
size parameter $r_{12}$ of the source.
It is typically treated as a Gaussian ($e^{-Q_{12}^2 r_{12}^2}$) or an exponential ($e^{-Q_{12}r_{12}}$).
The parameter $\lambda$ measures the coherence of the source, $\alpha$ is a normalization factor, 
and $\beta$ accounts for long range
correlations. 

\begin{figure}
 \centering
 \includegraphics[height=2.3in]{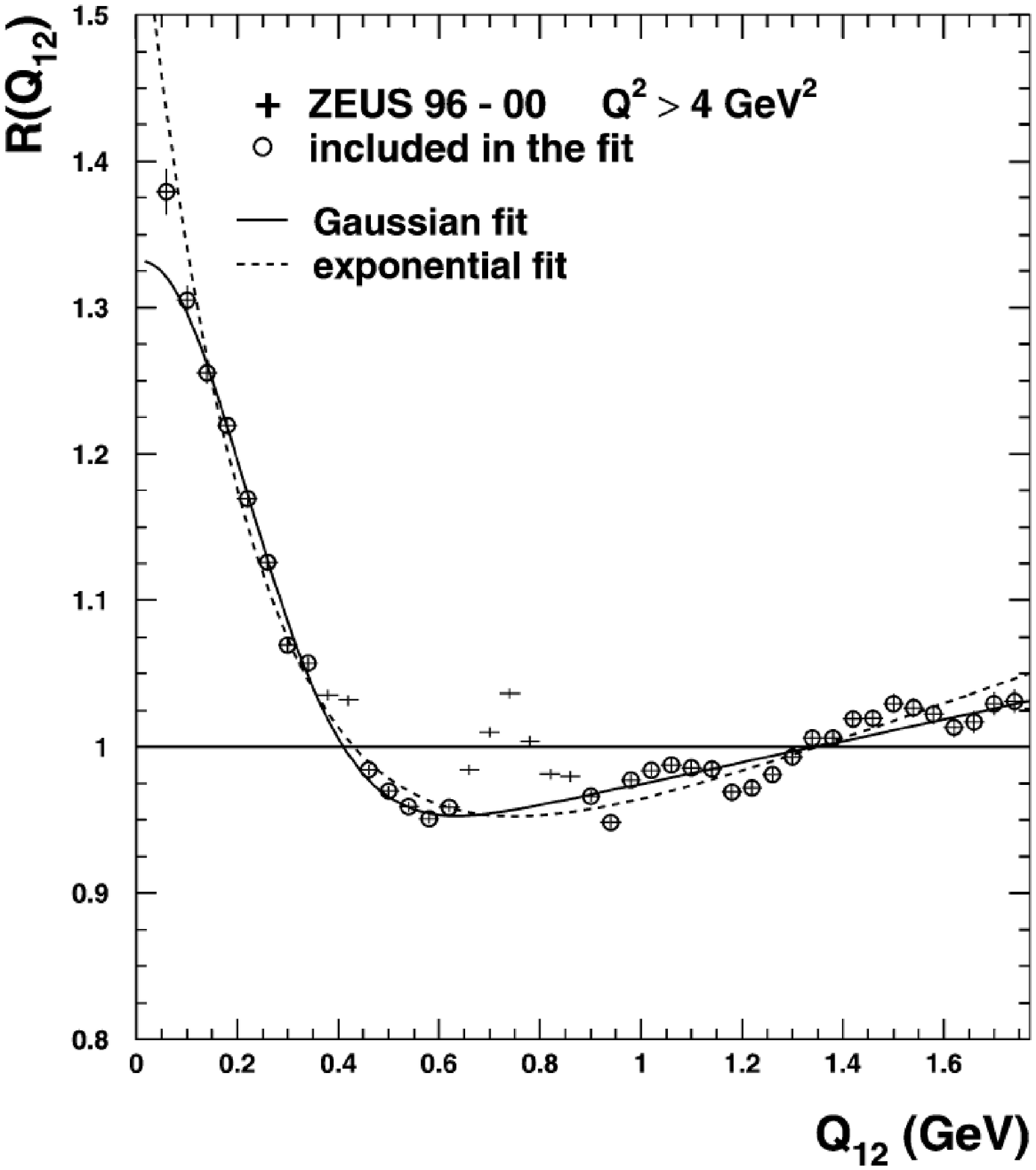}
 \hspace*{.5cm}
 \includegraphics[height=2.4in]{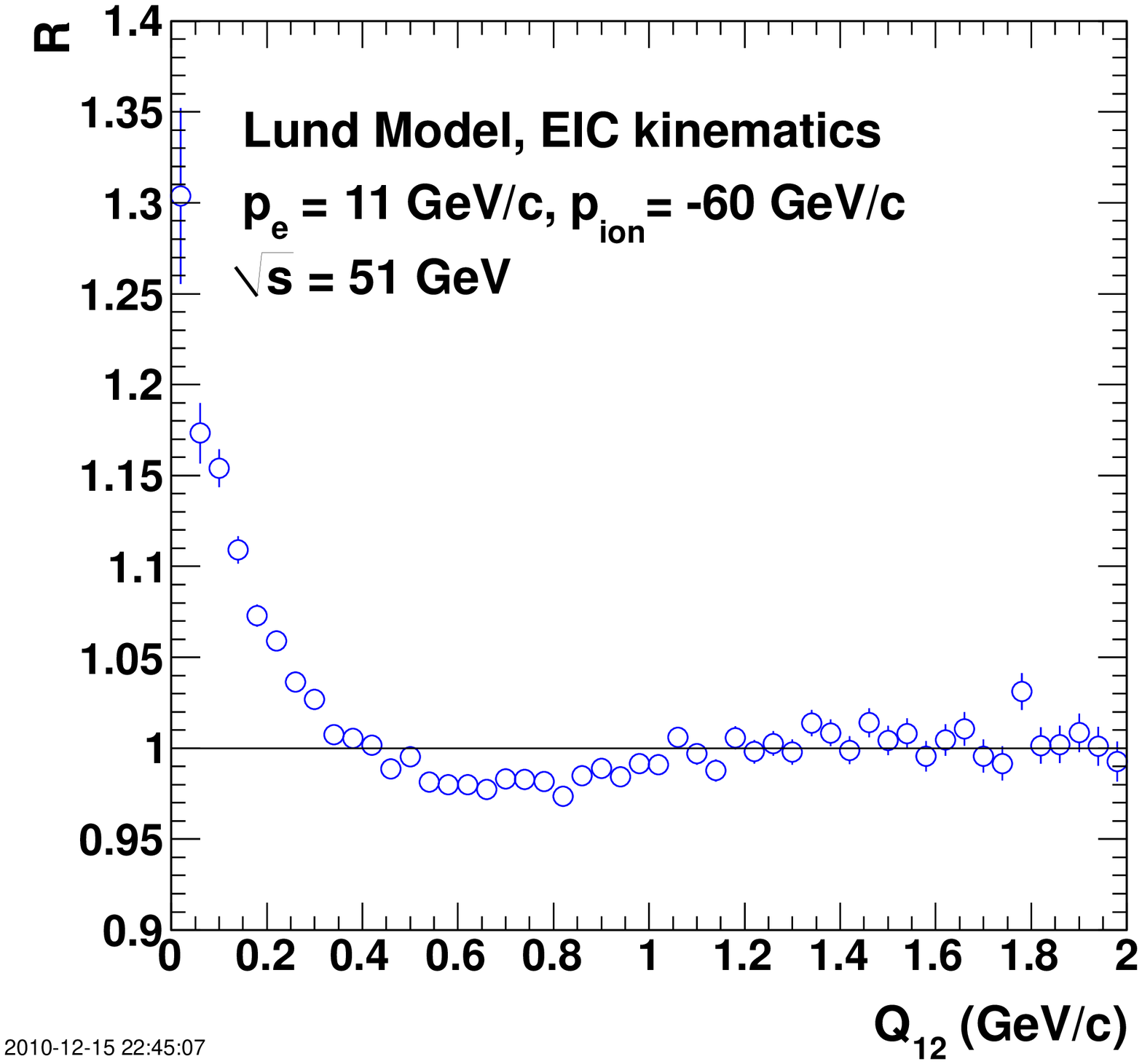}
 \caption{\small 
   {\it Left:} the measured Bose-Einstein correlation function,
   $R(Q_{12})$, together with Gaussian and  exponential fits
   \cite{Chekanov:2003gf}. The error bars show the 
   statistical uncertainties. The data points included in the fit are
   marked with the circles. The other points are excluded
   from the fit because the correlation is dominated by resonance
   effects.
   {\it Right:} Pythia simulation of $\pi^+\pi^+$ Bose-Einstein
   correlations (BEC) at Electron-Ion Collider kinematics. The BEC
   parameters were taken from Ref. \cite{Chekanov:2003gf}. The Lund
   fragmentation model was used. 
}
\label{GPG:fig1}
\label{GPG:fig3}
\end{figure}

\noindent{\bf Existing measurements.}
There is a long history of the study of BECs in particle and nuclear physics going back to 1960 when 
two-pion correlations
were measured in $p \bar p$ collisions \cite{Goldhaber:1960sf}.
They have been used to study geometric properties in $e+p$ reactions
\cite{Chekanov:2003gf}, the space-time extent of 
hot nuclear matter in $Au+Au$ collisions \cite{Adler:2004rq,Khachatryan:2010un},
and the dynamical properties of hadrons extracted from $Au+Au$
collisions \cite{Csorgo:2009pa}. Figure \ref{GPG:fig1} shows the
two-pion correlation function from Ref.~\cite{Chekanov:2003gf} 
for  $e+p$ reactions  measured at the DESY collider for an electron
momentum $p_e = 27.6~\rm GeV$ and proton momenta 
$p_p = - 820~\rm  GeV$ and $p_p = - 920~\rm  GeV$.
It shows several of the important features seen in many correlation functions.
There is a clear correlation that is maximal at $Q_{12}=0$ and drops rapidly to unity and below
with increasing momentum difference.
The height of the correlation function at $Q_{12} = 0$ measures the coherence in the source.
At moderate $Q_{12}$ the correlation drops below one, reflecting the usual practice of requiring the integral
of the entire correlation function to go to one.
There is a steady rise in $R$ at larger $Q_{12}$ due to long-range effects.
Recall the denominator Eq.~\eqref{GPG:eq1};
It should be free of the correlations arising from Bose-Einstein
statistics, but will not be free of all correlations:
momentum conservation will push $R$ up at large $Q_{12}$.
The width of the peak at $Q_{12}=0$ reflects the size of the source of the two bosons, {\it i.e.} 
large width in momentum space 
implies a small spatial source.
The width of $R$ in Fig.~\ref{GPG:fig1} corresponds to $r_{12}\approx 0.9~\rm fm$ for an exponential fit 
and is largely independent 
of $Q^2$, the square of the four-momentum transfer.

\begin{figure}
\begin{center}
  \includegraphics[height=2in]{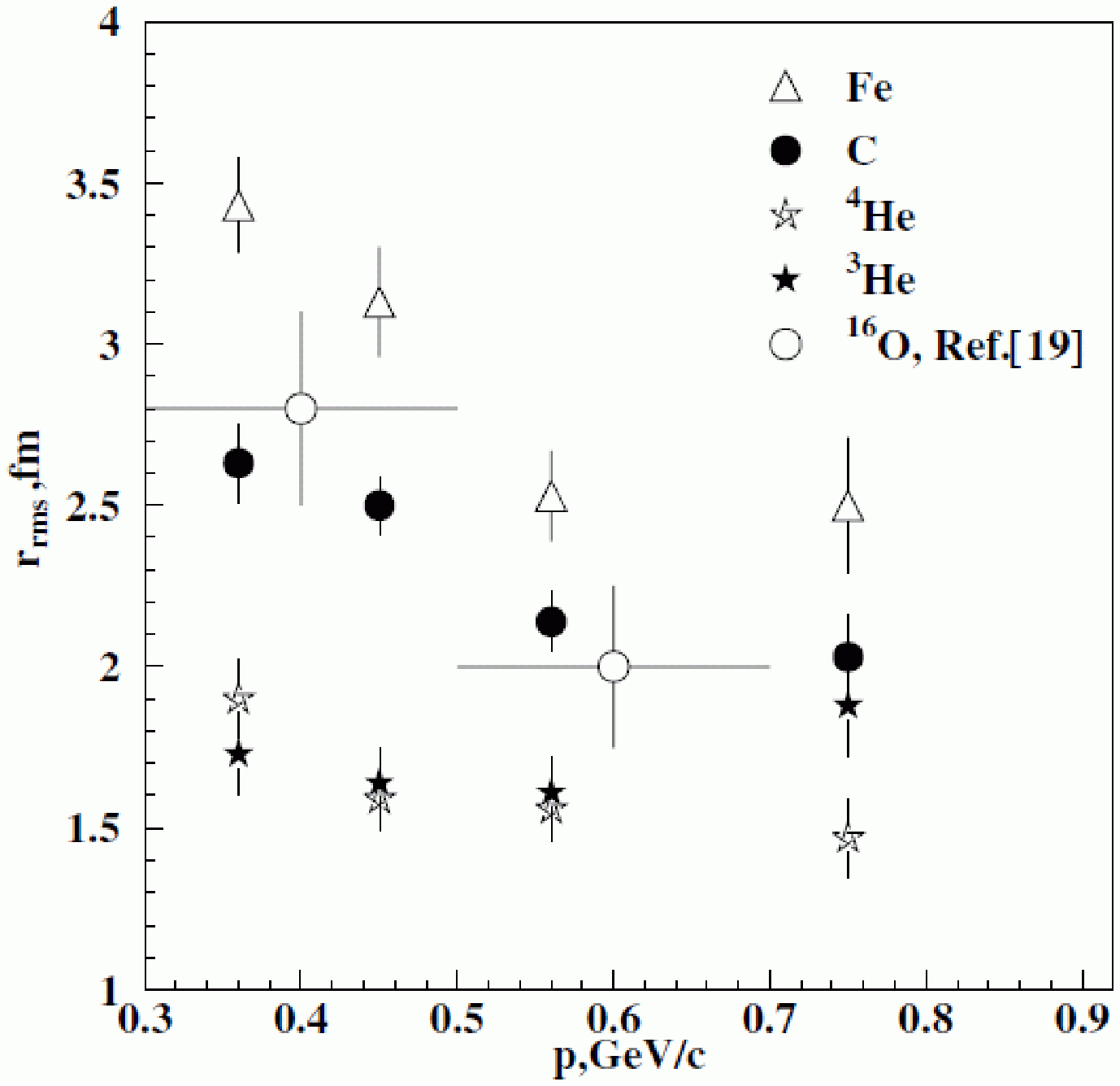}
  \hspace*{0.5cm}
  \includegraphics[height=2in]{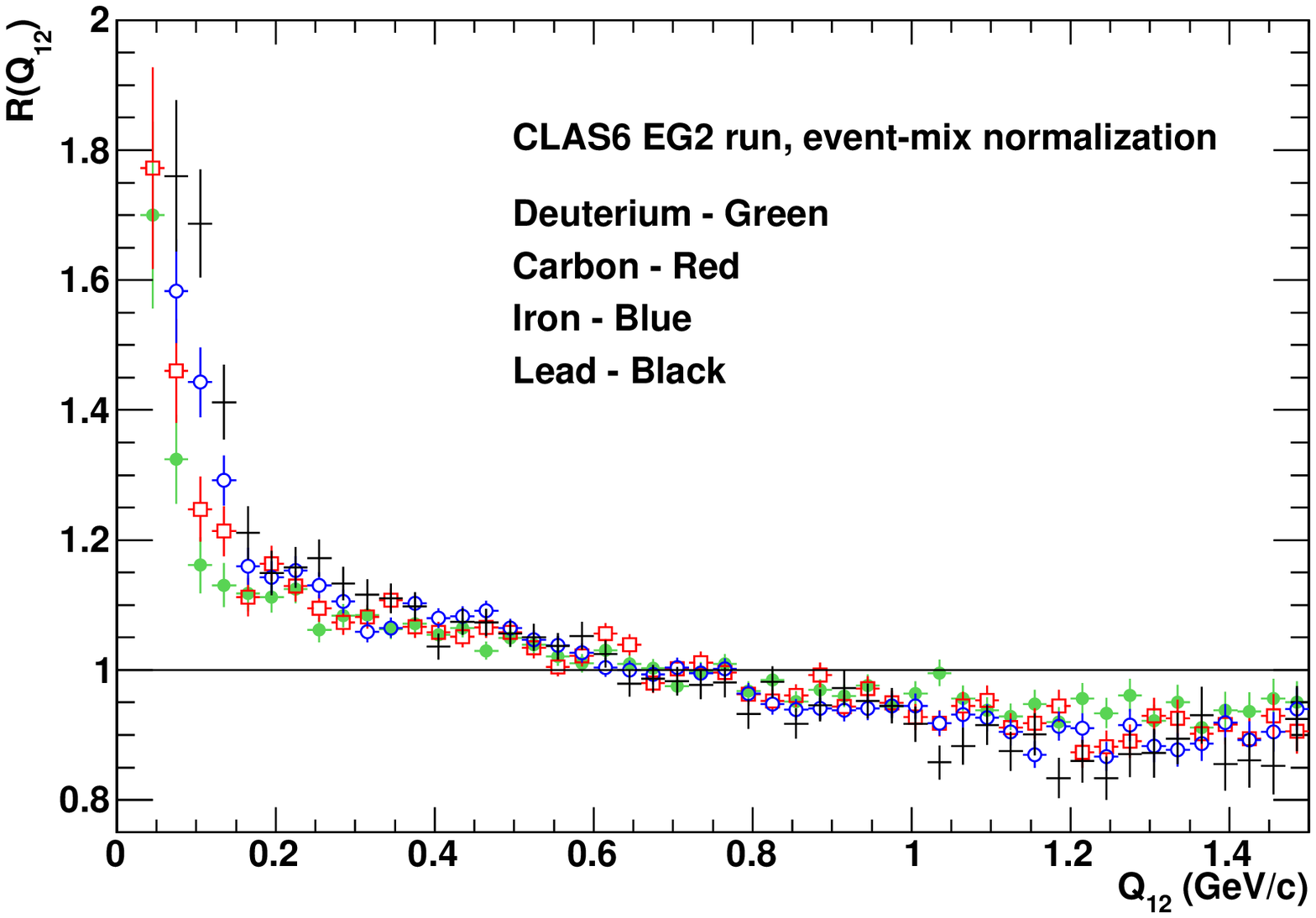}
\caption{\small 
  Left panel: The size parameter $r_{rms}$ as a function of the mean
  pair momentum $p=|\vec p_1 + \vec p_2|/2$ is shown for different
  nuclear targets \cite{Stavinsky:2004ky}. Data from Refs
  \cite{Degtyarenko:1989hy,Degtyarenko:1993kz,jlabcorrelation4} are
  shown which correspond to $e-{\rm ^{16}O}$ interactions at initial
  energy of $\rm 5~ GeV$ and $Q^2 < 0.1 \rm (GeV/c)^2$ are shown for
  comparison. Right panel: Preliminary correlation functions for
  $\pi^+\pi^+$ from the CLAS detector at Jefferson Lab
  \cite{GPG:hayk1}.
}
\label{GPG:fig2}
\end{center}
\end{figure}

\noindent{\bf BECs at an EIC.} Measurements with the CLAS detector  of a different type of correlations ({\it i.e.}, two protons)
have been performed 
on nuclear targets.
Some of the results are shown in the left-hand panel of Fig.~\ref{GPG:fig2} \cite{Stavinsky:2004ky}.
The figure shows the effects on the source size $r_{rms}$ (extracted from the correlation)
of the average pair momentum ($p=|\vec p_1 + \vec p_2|/2$) 
and the nuclear size on the correlation function.
At low average pair momentum $r_{rms}$ increases for the heavier nuclei and 
approaches the nuclear size;
implying the possible dominance of proton rescattering.
The density of the source was extracted in Ref. \cite{Stavinsky:2004ky} and found to be 
about 2-3 times the nuclear 
density in helium. In the right-hand panel of Fig.~\ref{GPG:fig2} we show preliminary results on $\pi^+\pi^+$ pairs 
on several nuclear targets \cite{GPG:hayk1}.
Below $Q_{12}\approx 0.15~\rm GeV/c$ the correlations from all nuclei rise to a 
large positive correlation. 
Above $Q_{12}\approx 0.15~\rm GeV/c$ the correlation functions overlap one another within 
the statistical uncertainty.

Measurement of Bose-Einstein correlations at an EIC will provide a new portal to studies of
cold, high-density nuclear matter and the process of hadronization.
The ground-state properties of nuclei are now well understood.
{\it Ab initio} calculations of the nuclear ground state are successful for nuclei up
to $A=8$ and higher \cite{McCutchan:2009th,Pieper:2001mp} and lattice QCD calculations continue to make progress toward
a fundamental understanding
of the nucleon \cite{Jansen:2008vs}.
However, the high-momentum components of the nuclear ground state are only now being
revealed.
These high-momentum nucleons are often paired with another, nearby neutron or proton forming regions of cold, dense nuclear matter.
Short-range correlations have shown the importance of high-density components and the influence of the
tensor force \cite{Egiyan:2005hs,Baghdasaryan:2010nv}.
The results of Ref. \cite{Stavinsky:2004ky} (left-hand panel of Fig.~\ref{GPG:fig2}) demonstrated the use of correlations to extract density 
information.
Measurements at an EIC could also help us to 
understand neutron stars \cite{Frankfurt:2008ke} and the EMC effect \cite{Sargsian:2002wc}.


\noindent{\bf Simulations.} 
We have simulated Bose-Einstein correlations for $\pi^+\pi^+$ pairs at
the kinematics of an Electron-Ion Collider
to investigate the feasibility of measuring BECs at an EIC.
For our starting point we used the results for $\pi^+\pi^+$ correlations from $ep$ reactions 
at DESY that are shown in 
Fig.~\ref{GPG:fig1} \cite{Chekanov:2003gf}.
That measurement covered the range $Q^2 = 4-8000~\rm (GeV/c)^2$ and there was limited $Q^2$ dependence in
the BEC parameters they extracted.
It is reasonable to believe those parameters may also apply to the EIC kinematics.
We chose the $\pi^+\pi^+$ channel because we expect them to be abundant and there is data 
from other experiments that enable us
to make comparisons.
We took advantage of several existing tools to perform the simulations.
The Pythia program \cite{Sjostrand:2006za} was used to generate events
with either Lund string model or independent fragmentation.
The code also includes a feature to simulate Bose-Einstein correlations \cite{Lonnblad:1995mr,Lonnblad:1997kk}.
The algorithm for the BECs starts with the usual fragmentation simulation and then
pairs of identical particles ({\it i.e.} $\pi^+\pi^+$) are selected.
For these pairs the relative 4-momentum $Q_{12}$ is modified according
to the desired parameterization (see discussion of Eq.~\eqref{GPG:eq2} above) with the constraint 
that the total 3-momentum
of the pair remains the same in the center-of-mass (CM).
The overall effect of applying the algorithm is to preserve momentum conservation, but reduce the energy.
To compensate for the energy reduction, the CM momentum vectors are then rescaled.

As a consistency check, we compared the simulated correlation function
$R$ for $\pi^+\pi^+$ pairs with the measurements from DESY shown in
Fig.~\ref{GPG:fig1}. 
The simulated correlation was weaker than the measured one, 
$R(Q_{12}=0)=1.2$ (simulated) versus $R(Q_{12}=0)=1.38$ (measured),
and not as wide, but still experimentally significant.
Since we are studying the possibility of observing BECs, the
parameters from Ref. \cite{Chekanov:2003gf}  
will provide a more conservative (and safer) test.
We also simulated the BECs at the same kinematics  as the preliminary
results shown in the right-hand panel
of Fig.~\ref{GPG:fig2} ($p_e = 5~\rm GeV$ and fixed target).
Here we found the simulated correlation disappeared entirely.
The multiplicity of the events generated by Pythia dropped
significantly at these kinematics reflecting the limitations of the code
at these lower energies.

At EIC kinematics ($p_e =
11~ \rm GeV/c$, $p_{ion} = -60~\rm GeV/c$, $\sqrt s = 51~\rm GeV$), 
we used the BEC parameters from ZEUS \cite{Chekanov:2003gf}. 
Since the EIC will run at energies lower than at HERA, but above the
current ones at Jefferson Lab, our estimates of the BECs are again
conservative ones. 
Our simulation of $R$ at EIC kinematics is shown in the right panel of
Fig.~\ref{GPG:fig3}.
There is, like in the Ref.~\cite{Chekanov:2003gf} data, a sizable
correlation at $Q_{12}=0$, a decrease in $R$ with  
width $\approx 0.2~\rm GeV/c$,
a dip below unity (recall discussion of Fig.~\ref{GPG:fig1}) and then the data approach 
one at high $Q_{12}$.
The Lund model was used here for the fragmentation and a calculation using the independent fragmentation 
model in Pythia yielded
similar results.
This result shows we can expect sizable correlation functions at the EIC.

\begin{figure}
\centering
\includegraphics[width=0.95\linewidth]{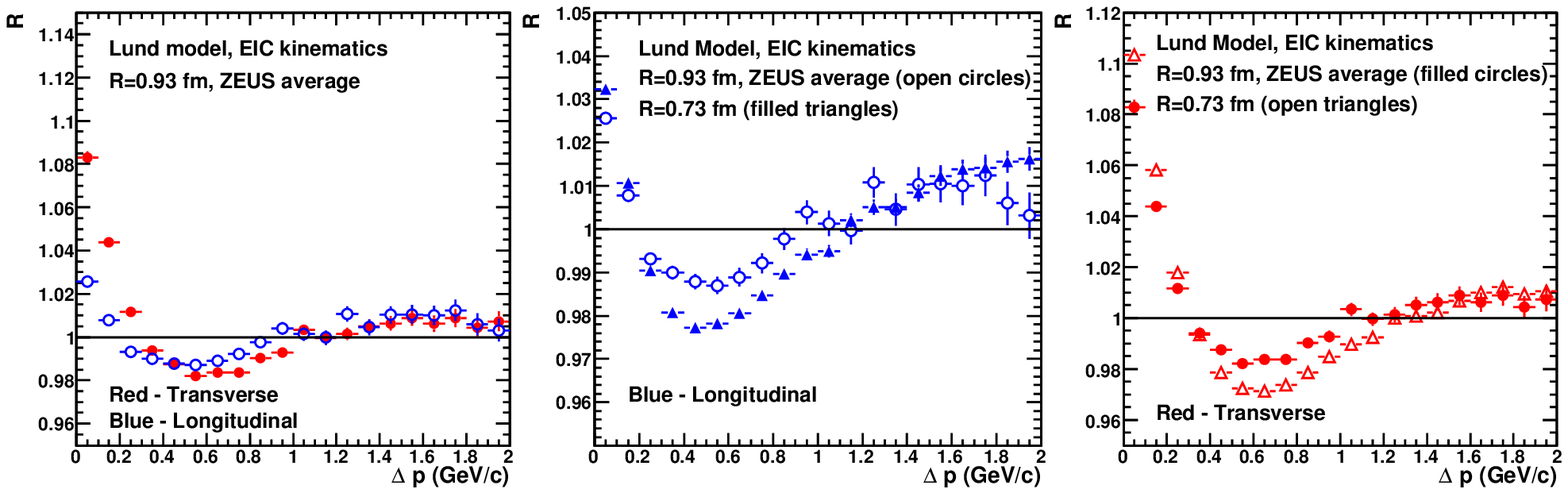}
\caption{\small Longitudinal and transverse ($LT$) correlation functions calculated with Pythia. 
The left-hand panel shows the
correlations functions using the Ref. \cite{Chekanov:2003gf} parameters. The other two panels show 
a comparison between those results and 
ones from a calculation with a smaller source size $r_{12}$.}
\label{GPG:fig4}
\end{figure}

One of the possible effects we may see at an EIC is the stretching of the QCD color string at high $Q^2$ 
and/or changes in
the string tension (recall Ref. \cite{Andersson:1985qn}).
The fragmentation region may not be spherical as observed in Ref. \cite{Andersson:1985qn}, but may have different 
sizes in the longitudinal and transverse
directions.
Such a difference was measured in Ref. \cite{Chekanov:2003gf} where
the longitudinal radius was $0.26\pm0.03$ fm  
bigger than the transverse one.
To search for such an effect in our simulation requires a different approach to extracting $R$.
We worked in the longitudinal Center-of-Mass System (LCMS), where the longitudinal components
of the pair momentum add to zero and extracted the transverse and longitudinal 3-momentum 
differences $\Delta p$.
Our initial results are shown in the left-hand panel of Fig.~\ref{GPG:fig4}.
The transverse (red, filled circles) and longitudinal (blue, open circles) produce the characteristic 
shapes seen above for $R$,
but with significant quantitative differences between the two.
The transverse correlation is about twice the longitudinal one at $Q_{12}=0$ and the widths are similar.
The large difference between the correlation functions suggests this may be a useful tool 
for studying space-time 
properties of the emission source.
To delve deeper into this question, we considered the sensitivity of the $LT$ distributions 
to changes in the
size parameter in the BEC parameterization.
The middle and right-hand panels in Fig.~\ref{GPG:fig4} show a comparison of the same $LT$ 
correlation functions shown in
the left-hand panel with ones calculated with a smaller size parameter
($r_{12} = 0.73$ fm 
versus $r_{12}=0.93$ fm from Ref.~\cite{Chekanov:2003gf}).
The smaller radius amplifies the shape of the correlation functions (the maximum at $Q_{12}=0$ 
increases and the dip
at $Q_{12}\approx 0.6~\rm GeV/c$ is deeper.
We can clearly separate the two distributions within the Monte Carlo statistics shown here.
We expect the statistical uncertainties for an EIC measurement to be better than the Monte Carlo statistical 
uncertainties shown here.
The cross sections for these reactions (from Pythia) multiplied by the EIC luminosity suggest 
a production rate of $10^5$ Hz.
We also fitted the correlation functions with Eq.~\eqref{GPG:eq2} and obtained uncertainties on the size parameter
$r_{12}$ less than 0.15 fm which is comparable to the precision of the results in Ref. \cite{Chekanov:2003gf}.
Thus, we will be able to discriminate between different size
parameters at least at the 0.2 fm level. 

\noindent{\bf Conclusions.} 
Bose-Einstein correlations will be an important tool at an Electron-Ion Collider for studying
high-density nuclear matter, the dynamics of the QCD string in hard scattering, and to gain a deeper
understanding of fragmentation and hadronization.
Our simulations have shown us that we can expect large ($20\%$) effects in the correlation function at small 
$Q_{12}$.
The longitudinal-transverse correlations are sensitive to the size parameter to a fraction of a $fm$.
Finally, the large $\pi^+\pi^+$ BECs observed at JLab that are not reproduced in our simulations hold the
promise of new physics to be uncovered with the EIC.


\section{e+A Monte Carlo simulation tools \label{sec:ea-monte}}

\subsection{A Monte Carlo Generator for Diffractive Events in e+A Collisions}


\hspace{\parindent}\parbox{0.92\textwidth}{\slshape 
  Tobias Toll and Thomas Ullrich
}
\index{Toll, Tobias}
\index{Ullrich, Thomas}

\vspace{\baselineskip}

While there is a rich
set of Monte Carlo (MC) event generators for $e+p$ collisions available ({\it e.g.}
PYTHIA6 \cite{Sjostrand:2000wi}, HERWIG++ \cite{Bahr:2008pv}, LEPTO
\cite{Ingelman:1996mq}, PEPSI \cite{Mankiewicz:1991dp}, RAPGAP
\cite{Jung:1993gf}, ARIADNE \cite{Lonnblad:1992tz}, CASCADE
\cite{Jung:2001hx}, SHERPA \cite{Gleisberg:2008ta}), the situation for
$e$A collisions is less favorable. The exception is DPMJET
\cite{Roesler:2000he} which attempts to describe deep-inelastic
$e$A events but does not include the rich physics accessible
via diffractive events.

In strong interactions, diffractive events can be interpreted as
resulting from scattering via the exchange of a pomeron that carries
the quantum numbers of the vacuum, as discussed in \ref{sec:diffincgc}. 
It was a surprise to see that a large fraction (approximately 15\%) of all $e$+$p$ events at
HERA were diffractive.  Calculations predict this fraction to be even
larger in $e$+A collisions at EIC where the large nuclei
remain intact $\sim$ 25-30\% of the time ({\it e.g.}
\cite{Kowalski:2007rw,Kowalski:2008sa}).
In fact diffractive events are considered the most
sensitive means of studying saturation since the dipole scattering
amplitude is proportional to the square of the gluon momentum
distribution $xg(x, Q^2)$.  Another fascinating aspect of the study
of diffractive events at an EIC is that that it would allow us to
measure the intensity and the {\em spatial} distribution of the strong
field that binds the nucleus together \cite{Caldwell:2010zza}.

For all the above measurements the most important process to study is
the production of exclusive diffractive vector mesons, such as
$\jpsi$, $\phi$, and $\rho$ mesons, as well as Deeply Virtual Compton
Scattering (DVCS) photons.  These processes give very clean final
states, consisting of the scattered electron and nucleus and one
extra particle: a vector meson or a real photon. This is a process
which is dominated by small momentum fractions $x<10^{-2}$.  $\jpsi$
production is particularly well suited for studies of the spatial
gluon distribution inside nuclei due to its well known wave function,
narrow decay width, and its large branching ratio for
electro\-magnetic decays $\jpsi\rightarrow e^++e^- ({\rm
    or~}\mu^++\mu^-)$.

The measurement of exclusive vector meson production in diffractive
events will be one of the key measurements at an EIC.  Therefore these
processes has been the starting point in our efforts to realise a new
multi-purpose MC generator. 


\noindent{\bf The Dipole Model:}  The dipole model is an important tool in investigations of
diffractive processes and for the purpose of
applying it to $e$+A collisions, we needed an impact
parameter dependent model as starting-point. Two known models fulfil
this requirement: bSat (or IPSat)\cite{Kowalski:2006hc} and bCGC
\cite{Kowalski:2006hc,Watt:2007nr}.  They are the underlying building blocks
used in the generator. In what follows, we will concentrate on the bSat model and 
not discuss the technical details of the generator but focus on how the dipole models
are applied with emphasis on the extension to $e$+A collisions.

The parameters of the dipole models described below have been tuned to
inclusive HERA data, and they describe a wide variety of HERA
measurements exceptionally well \cite{Kowalski:2006hc,Watt:2007nr}. \\


\noindent{\bf The Dipole Model in $e+p$:}  The production of exclusive vector mesons and DVCS photons at small
$x$ for $ep$ collisions, $e + p \rightarrow e^\prime + p^\prime +
V/\gamma$, in the dipole model has been extensively studied
\cite{Kowalski:2006hc,Watt:2007nr}.  Here the virtual photon splits into a quark-antiquark dipole which
interacts with the target diffractively via one or many two-gluon
pomeron exchanges (see Fig.~\ref{fig:dipole}).  The amplitude for this
process is
\begin{eqnarray}
    \mathcal{A}_{T,L}^{\gamma^*p\rightarrow Vp}(x,Q,\boldsymbol{\Delta}) = 
    i\int \mathrm{d}^2 \mathbf{r} \int\frac{{\rm d} z}{4\pi}\int{\rm d}^2{\bf b}\left(\Psi^*_V\Psi\right)(r, z) 
    e^{-i{\bf b}\cdot{\bf \Delta}}\frac{{\rm d}\sigma_{q\bar q}^{(p)}}{{\rm d}^2{\bf b}}(x, r, {\bf b})
    \label{eq:ttepamplitude}
\end{eqnarray}
Here $T$ and $L$ represent the transverse and longitudinal
polarizations of the virtual photon, $r$ is the size of the dipole,
$z$ the energy fraction of the photon taken by the quark,
$\Delta=\sqrt{-t}$ is the transverse part of the four-momentum
difference of the outgoing and incoming proton, and ${\bf b}$ is the
impact parameter of the dipole.  The wave function of the produced
vector meson or real photon is $\Psi_V$ while that of the incoming photon
that splits into the dipole is $\Psi$.  

In the bSat model the dipole cross-section in terms of the dipole scattering amplitude $\mathcal{N}^{(p)}(x, r, {\bf b})$ is
\begin{eqnarray}
    \frac{{\rm d}\sigma_{q\bar q}^{(p)}}{{\rm d}^2{\bf b}}\equiv 2\mathcal{N}^{(p)}(x, r, {\bf b}) =
    2\left[1-\exp\left(-\frac{\pi^2}{2N_C}r^2\alpha_S(\mu^2)xg(x,\mu^2)T(b)\right)\right]\,,
    \label{eq:bsatep}
\end{eqnarray}
where $\mu^2=4/r^2+\mu_0^2$ and $\mu_0^2$ is a cut-off scale in the
DGLAP evolution of the gluons $g(x, \mu^2)$. The nucleon shape
function $T^{(p)}(b)=1/(2\pi B_G)\exp(-b^2/(2B_G))$.  The parameter
$B_G$ is determined through fits to HERA data \cite{Kowalski:2006hc}.
We use $B_G=4~{\rm GeV}^{-2}$.  It should be noted that bSat is a
model of multiple two-gluon exchanges and does not contain any
gluon-gluon recombinations. It is however, by construction, a model that obeys unitarity,
so in this respect it is a saturation model.
\begin{figure}
    \begin{center}
        \includegraphics[width=\columnwidth]{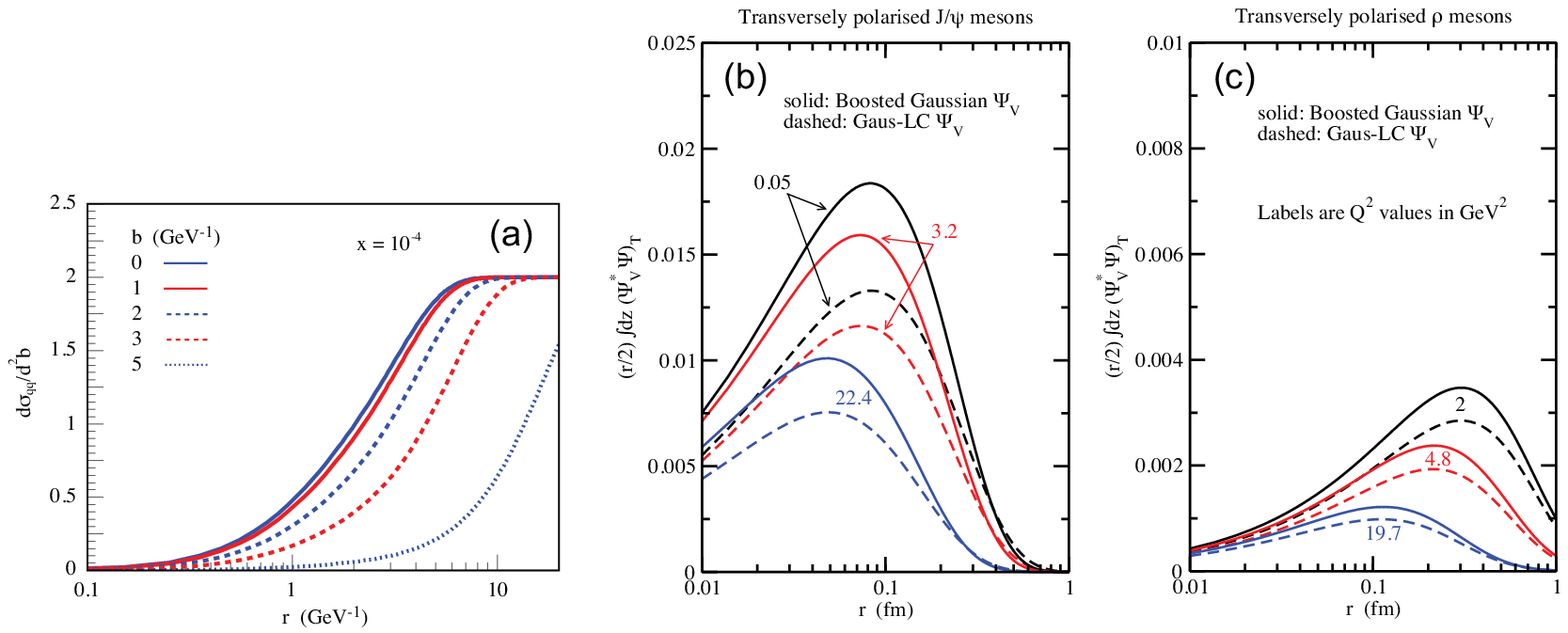}
    \caption{\small (a) shows the dipole cross-section for various impact
        parameters as a function of dipole size in the bSat model. (b)
        and (c) depict the wave overlap functions for $\jpsi$ and
        $\rho$ mesons respectively as a function of $r$ for various
        $Q^2$ for transversely polarized photons (from
        \cite{Kowalski:2006hc}).}
    \label{fig:mesons}
    \end{center}
\vspace{-1em}
\end{figure}

Figure \ref{fig:mesons}(a) shows the dipole cross-section as a
function of $r$ for different impact parameters.  Figure \ref{fig:mesons}(b) and (c) depict the wave overlap,
$(\Psi^*_V\Psi)(r,z)$, for $\jpsi$ (b) and $\rho$ mesons (c)
\cite{Kowalski:2006hc} used in Eq.~\ref{eq:ttepamplitude}.  It should
be noted that the $J/\psi$ is not necessarily the best suited
vector meson for probing saturation effects. Studying saturation
implies probing large dipole radii $r\gsim 2~{\rm GeV}^{-1}$ (0.4 fm).
However, the wave overlap with the $\jpsi$ vanishes almost entirely
for these dipole sizes. The lighter vector mesons $\rho$ and $\phi$
certainly appear more suited in this case.  Unfortunately the wave
functions of the lighter vector mesons are less well known than that
of the $\jpsi$ increasing the uncertainties in model-data comparisons.
This can be overcome in the future by improving our knowledge of the
light vector meson wave functions. 


\noindent{\bf Phenomenological Corrections to the Cross-Section:}  In the derivation of the dipole amplitude (eq.~\eqref{eq:ttepamplitude})
only the real part of the $S$-matrix is taken into account, making the
amplitude purely imaginary. The real part of the amplitude can be
included by multiplying the cross-section by a factor $(1+\beta^2)$,
where $\beta$ is the ratio of real to imaginary part of the amplitude.
It is calculated using
\begin{eqnarray}
    \beta=\tan\left(\lambda\frac{\pi}{2}\right),~~~{\rm where}~~
    \lambda\equiv\frac{\partial\ln\left(\mathcal{A}_{T,L}^{\gamma*p\rightarrow Vp}(x,Q,\Delta)\right)}{\partial\ln(1/x)}
\end{eqnarray}
Also, the two gluons interacting in each event do not carry the same
momentum fraction $x$. In the leading $\ln(1/x)$ limit, this skewedness
effect disappears, but can still be accounted for by a factor
$R_g(\lambda)$, where $R_g(\lambda)=2^{2\lambda+3}\Gamma(\lambda+5/2)/\Gamma(\lambda+4)/\sqrt{\pi}$. 

$R_g$ is multiplied to the gluon distribution $xg(x,\mu^2)$ and $\lambda$ is
defined as the derivative of $\ln(xg(x,\mu^2))$ with respect to
$\ln(1/x)$. It should be noted that while the correction of the real
part of the amplitude is on firm theoretical footing, the skewedness
correction should be viewed as a purely phenomenological correction.
Also, the correction variable $\lambda$ is only well behaving for
small values of $x<10^{-2}$. The combined magnitude of both
corrections is $x$ dependent and is typically of the order of
$10-60\%$. \\


\noindent{\bf Extending the Dipole Model from $e+p$ to $e$+A:}  When going from $+ep$ to $e$+A scattering we will use the independent scattering approximation (see also eq.~(\ref{eq:sfact})), 
\begin{eqnarray}
    1-\mathcal{N}^{(A)}=\prod_{i=1}^A
    \left(1-\mathcal{N}^{(p)}(x, r, |{\bf b}-{\bf b}_i|)\right)
    \label{eq:eptoea}
\end{eqnarray}
where ${\bf b}_i$ is the position of each nucleon in the nucleus. Here, 
these positions are generated according to the Wood-Saxon potential.
Combining equations \eqref{eq:bsatep} and
\eqref{eq:eptoea} the bSat dipole cross-section for $e$+A becomes:
\begin{eqnarray}
    \frac{{\rm d}\sigma_{q\bar q}^{(A)}}{{\rm d}^2{\bf b}}(x, r, {\bf b}, \Omega)=
    2\left[1-\exp\left(-\frac{\pi^2}{2N_C}r^2\alpha_S(\mu^2)xg(x,\mu^2)
            \sum_{i=1}^AT^{(p)}({\bf b}-{\bf b}_i)\right)\right]
\end{eqnarray}

At small gluon momentum fractions, $x<10^{-2}$, the dipole interacts
coherently with large volumes of the nucleus. Therefore the
configuration of nucleons in the nucleus is not an observable. To
obtain the total cross-section, these nucleon configurations have
to be averaged over:
\begin{eqnarray}
    \frac{{\rm d}\sigma_{\rm total}}{{\rm d} t}=
    \frac{1}{16\pi}\left<\left|\mathcal{A}(x, Q^2, t, \Omega)\right|^2\right>_\Omega
\end{eqnarray}
where $\Omega$ denotes nucleon configurations.

One defines two different kinds of diffractive events in $e$A:
coherent and incoherent. In incoherent diffractive processes the
nucleus breaks up into two or more color neutral fragments, something
not possible in diffractive $ep$.  If the nucleus stays intact the
diffractive processes are coherent.  In the Good-Walker picture
\cite{Good:1960ba} (also found in \cite{Caldwell:2010zza}) the
incoherent cross-section is proportional to the variance of the
amplitude with respect to the initial nucleon configurations $\Omega$
of the nucleus:
\begin{eqnarray}
    \frac{{\rm d}\sigma_{\rm incoherent}}{{\rm d} t}=
    \frac{1}{16\pi}\left(\left<\left|\mathcal{A}(x, Q^2, t, \Omega)\right|^2\right>_\Omega
        -\left|\left<\mathcal{A}(x, Q^2, t, \Omega)\right>_\Omega\right|^2\right)
\end{eqnarray}
where the first term on the R.H.S is the total cross-section and the
second term is the coherent part of the cross-section. 


\noindent{\bf The Generator:}  The Monte Carlo event generator is implemented in C++ through a set of
modular classes.  A rich set of input parameters let the user select
beam energy and species (A), wave function model, dipole model,
kinematic range and the final state particle to study: $\rho$,
$\pi$, $\jpsi$, or $\gamma$ (DVCS). Internally, the variables $t$,
$Q^2$, and $W^2$ are generated following a probability density
function (pdf). From these three variables, the complete final state
consisting of the scattered electron, the scattered proton or nucleus,
and the produced vector meson or photon can be unambiguously
calculated. \\


\noindent{\bf Generating Events for $ep$:}  The variables are generated from a probability density function
which for $ep$ is
\begin{eqnarray}
    {\rm pdf}(Q^2, W^2, t)=\frac{\partial^3\sigma_{\rm tot}}{\partial Q^2\partial W^2\partial t}=
    \frac{1}{16\pi}\sum_{T,L}f^{\gamma^*}_{T, L}(Q^2, W^2)
    \left|\mathcal{A}_{T,L}^{\gamma^*p\rightarrow Vp}(W^2, Q^2, t)\right|^2
\end{eqnarray}
where $f^{\gamma^*}_{T, L}$ is the photon flux for transversely and
longitudinally polarized photons.  The user may also choose to include
the corrections for the real part of the amplitude and/or the
skewedness effect as described above. 

\noindent{\bf Generating Events for $e$A, the MC-Glauber Approach:}  For $e$A the pdf is
\begin{eqnarray}
    \frac{\partial^3\sigma_{\rm total}}{\partial Q^2\partial W^2\partial t}(Q^2, W^2, t) = 
    \frac{1}{16\pi}\sum_{T,L}f^{\gamma^*}_{T, L}(Q^2, W^2)
    \left<\left|\mathcal{A}^{\gamma^*A\rightarrow VA}_{T,L}(Q^2, W^2, t, \Omega)\right|^2\right>_\Omega
\end{eqnarray}
Here the average of an observable $\mathcal{O}$ with respect to the
initial nucleon configurations $\Omega$ is defined as  $\left<\mathcal{O}\right>_\Omega\equiv\frac{1}{C_{\rm max}}\sum_{j=1}^{C_{\rm max}}\mathcal{O}(\Omega_j)$, 
where a number of $C_{\rm max}$ configurations $\Omega_j$ are
generated and summed over.  This sum will converge to the true average
for large $C_{\rm max}$. We call this way of performing the
average the MC-Glauber approach.  It should be noticed that this
method of averaging the initial nucleon configurations is different
than in previous publications, {\it e.g.} in \cite{Kowalski:2007rw}
and \cite{Caldwell:2010zza}.

For each event, the coherent part of the cross-section is calculated
simultaneously with the total cross-section, by averaging the
amplitude before squaring it. It is then decided probabilistically
that the nucleus breaks up if
\begin{eqnarray}
    \left( \frac{\partial^3\sigma_{\rm total}}{\partial Q^2\partial
        W^2\partial t}- 
        \frac{\partial^3\sigma_{\rm coherent}}{\partial Q^2\partial
          W^2\partial t} \right) \bigg / 
    \frac{\partial^3\sigma_{\rm total}}{\partial Q^2\partial
      W^2\partial t} 
    > R
\end{eqnarray}
where $R$ is a random number from a uniform distribution on $[0-1]$.
When this happens, the final state does not contain a scattered nucleus
but rather the decay products resulting from the break-up of the
nucleus. 


\begin{figure}
\begin{center}
  \includegraphics[width=0.42\columnwidth]{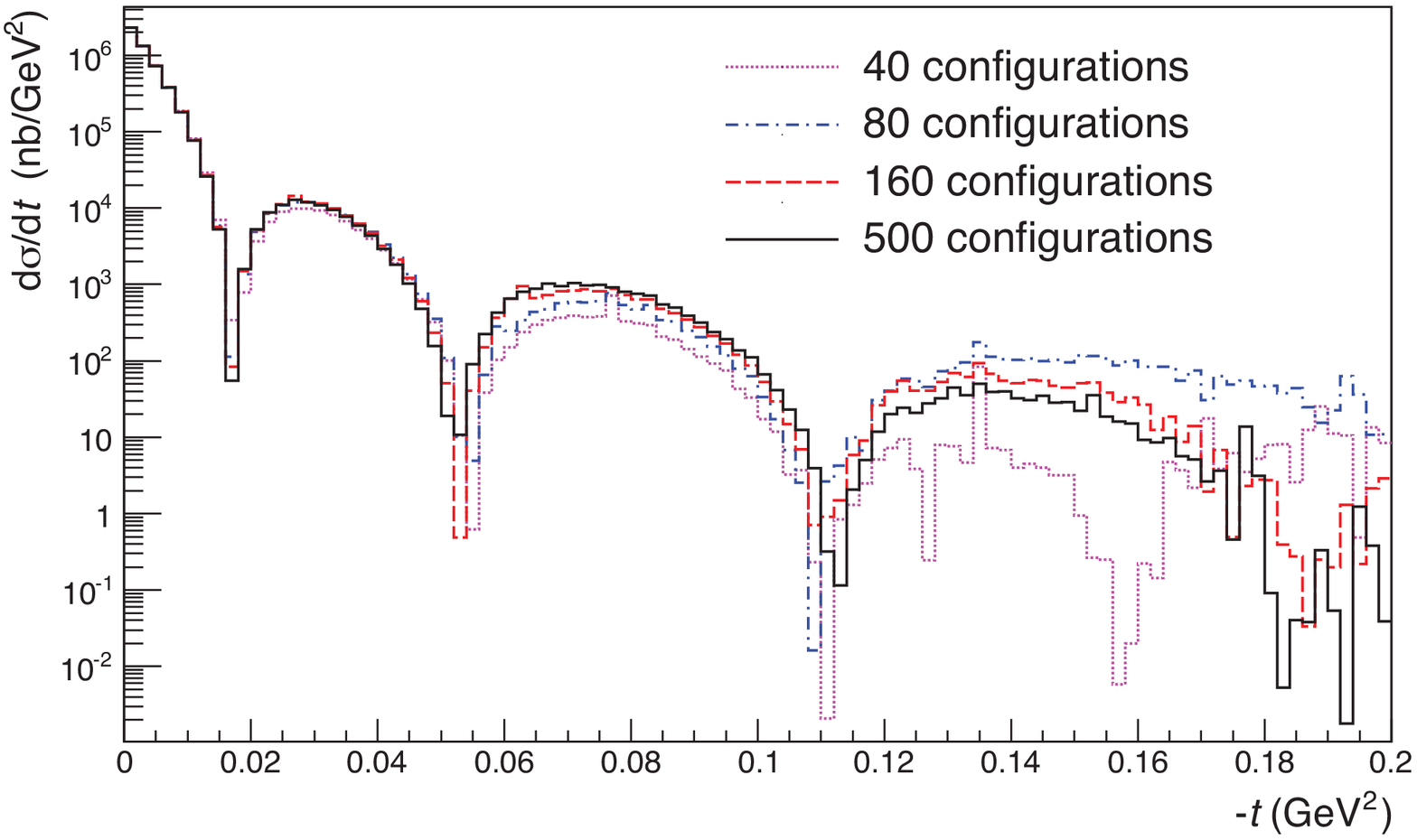}
  \
  \includegraphics[width=0.42\columnwidth]{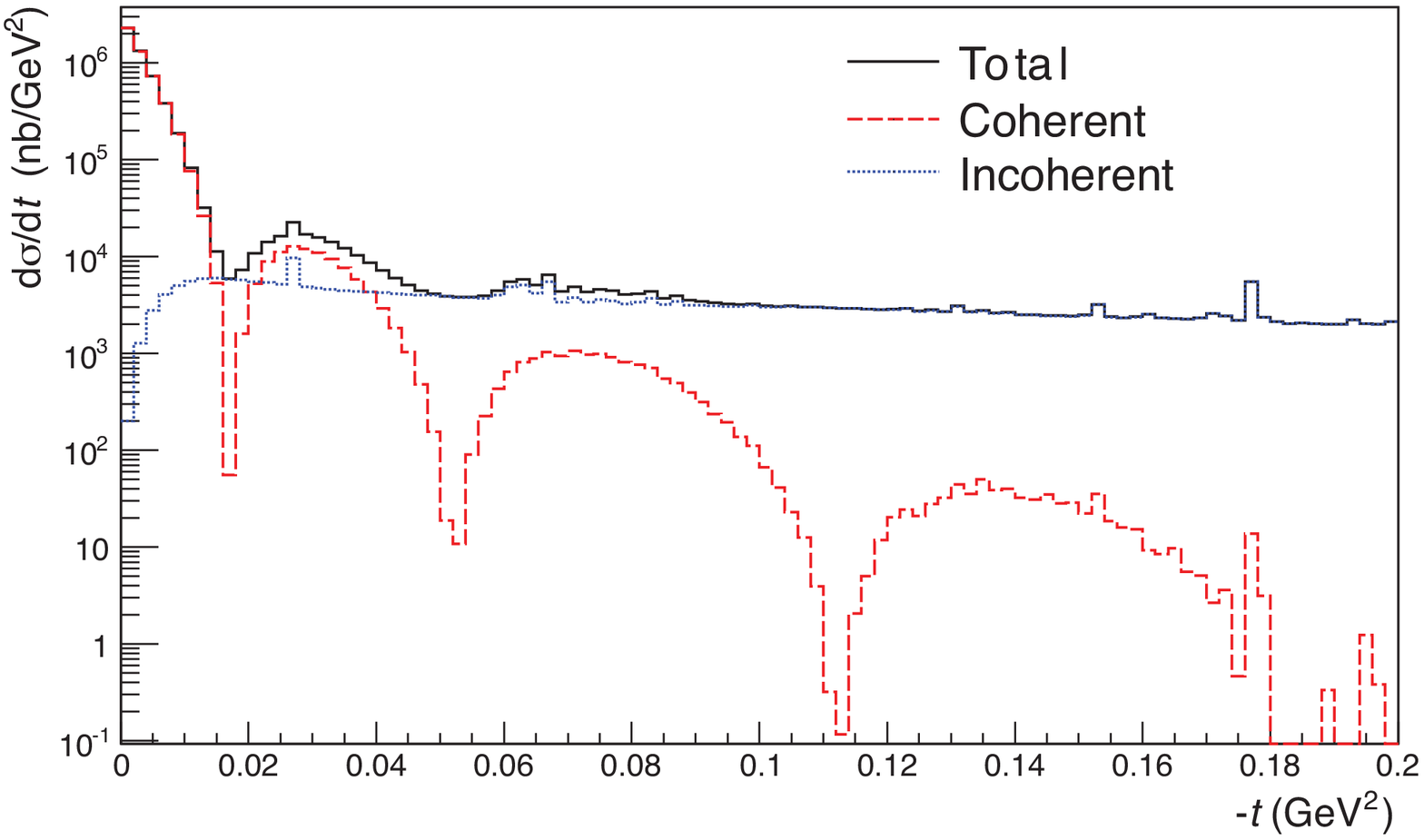}
  \caption{\small
    {\it Left plot.} The coherent part of
    the cross-section 
    as a function of $|t|$ for electron-gold scattering at
    $Q^2=10^{-4}~{\rm GeV}^2$ and $x_p=0.006$ averaged over 40,
    80, 160 and 500 configurations respectively.
    {\it Right plot.}
    The total, coherent, and incoherent cross-sections as a
    function of $|t|$ for $e$Au scattering at $Q^2=10^{-4}~{\rm
    GeV}^2$ and $x_p=0.006$ averaged over 500 configurations.
  }
\label{fig:cmax}
\label{fig:cs} 
\end{center}
\end{figure}

\noindent{\bf Generating Events for $e$A, the Optical Approach:}  A simpler and faster way of doing the average over the initial nucleon
configurations is what we call the optical approach. Here the average
is done implicitly in the dipole cross-section which becomes
\cite{Kowalski:2007rw}
\begin{eqnarray}
    \left<\frac{{\rm d}\sigma_{q\bar q}^{A}}{{\rm d}^2{\bf b}}\right>_{\Omega, {\rm Optical}}=
    2\left[1-\left(1-
            \frac{T_A({\bf b})}{2}\sigma_{q\bar q}(x, r)\right)^A\right] .
\end{eqnarray}
For processing speed reasons we approximate the integrated dipole
cross-section using the GBW model \cite{GolecBiernat:1999qd}:
\begin{eqnarray}
    \sigma^{\rm GBW}_{q\bar q}(x, r)=\sigma_0\left(1-\exp\left(-\frac{r^2Q^2_s(x)}{4}\right)\right)
\end{eqnarray}
where $Q_s^2(x)=(x_0/x)^\lambda$. Here, $\sigma_0=23.9~{\rm mb}$,
$\lambda=0.287$ and $x_0=1.1\cdot 10^{-4}$ \cite{Kowalski:2006hc}.
$T_A$ is the projection of the Woods-Saxon potential in the transverse
plane. This approximation is valid for large nuclei. In the optical
approach, {\it only} the coherent part of the cross-section can be
calculated, since it gives the average of the amplitude, but not of
the amplitude squared.  It is implemented in the program as a fast
alternative to the more accurate but CPU-time intensive MC-Glauber
approach. 

\begin{wrapfigure}{r}{0.5\textwidth}
  \begin{center}
    \includegraphics[width=0.48\textwidth]{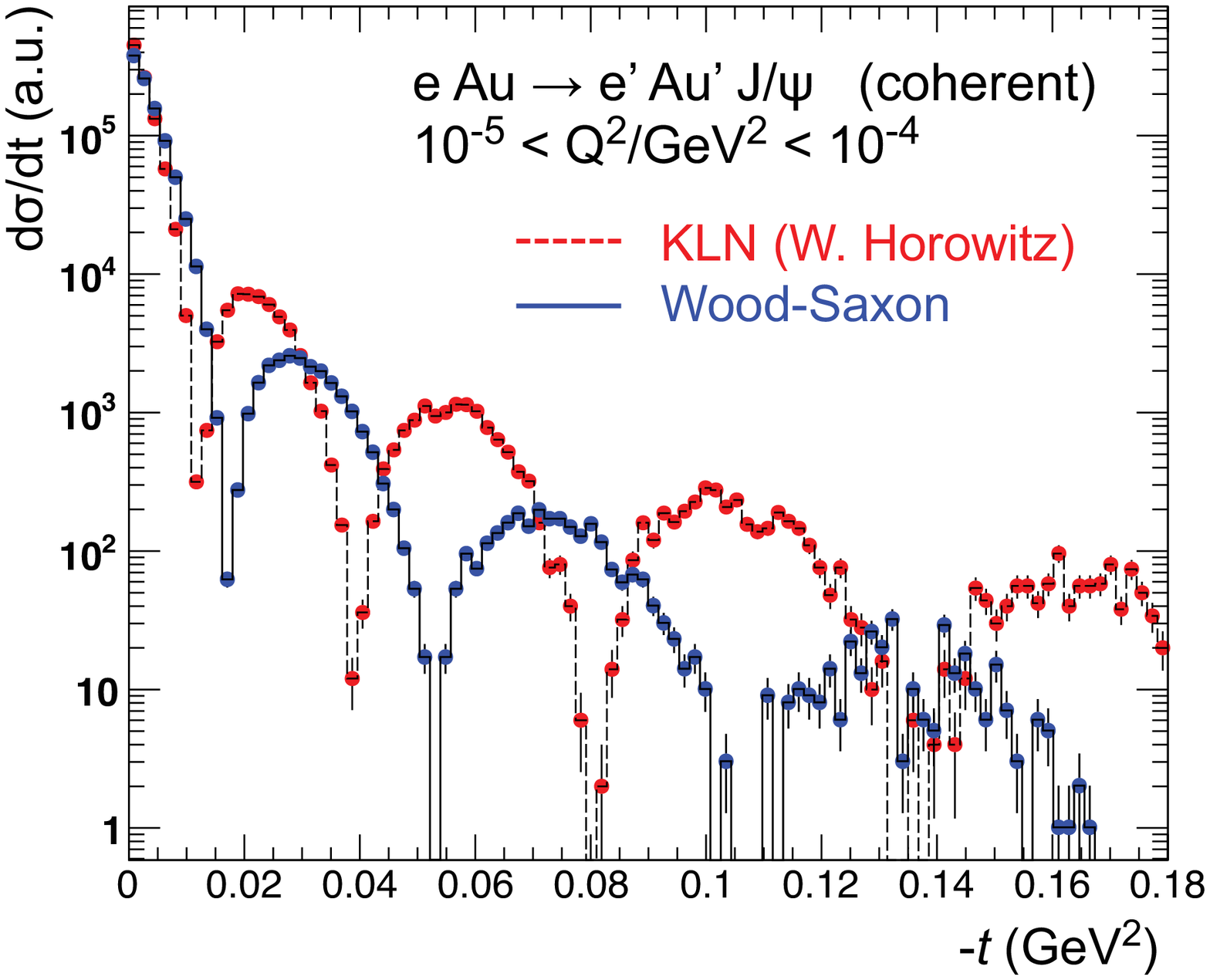}
  \end{center}
  \caption{\small The coherent part of the cross-section of
      $e+$A$\rightarrow e'+$A$'+\jpsi$ for two different distributions
      of the initial nucleon configurations: Woods-Saxon
      and KLN (from \cite{Will:2011}).}
\label{fig:KLN}
\end{wrapfigure} 

\noindent{\bf Results:}  In the following we only show results from the $e$+A part of the
generator.  In Figure \ref{fig:cmax}, the coherent part of the
cross-section for $e+\mathrm{A}\Rightarrow e'+ \mathrm{A}'+J/\Psi$ is
shown as a function of $|t|$, with $Q^2=10^{-4}~{\rm GeV}^2$ and
$x_p=0.006$.  The nucleus used is gold with A=197.  The cross-section
is calculated for different numbers of averaged nucleon configurations
$C_{\rm max}$.  The target is probed by the dipole at a scale $\Delta$
which means that at large $|t|$ the cross-section is much more
sensitive to smaller variations in the positions of the nucleons than
it is for small $|t|$.  Therefore, for small $|t|$, the sum over
configurations converges quickly, while for larger $|t|$, more
configurations are needed for the sum to converge.
As indicated in Fig.~\ref{fig:cmax} approximately 100 configurations are
needed to describe $e$A scatterings up to $|t|\approx 0.2~{\rm GeV}^2$.
In Figure \ref{fig:cs} the total cross-section and the incoherent
part of the cross-section are shown as averaged over 500 nuclear
configurations.  The $t$-slope of the incoherent cross-section is
close to $6~{\rm GeV}^{-2}$. This is a bit steeper than is found in
\cite{Lappi:2010dd}, where the impact parameter dependence was factorized 
out in the dipole cross-section and therefore the 
$t{\rm -slope}=B_G=4~{\rm GeV}^{-2}$.

In order to measure the spatial distribution of gluons inside the
nucleus, the coherent cross-section has to be well measured as a
function of $t$. The inverse Fourier transformation of this will then
give the transverse spatial dependence of the amplitude. To do this
the position of the several coherent maxima in the $t$-distribution
have to be measured accurately.

Experimentally, this requires the suppression of the large incoherent
fraction, which is of course also of great interest in itself
\cite{Kowalski:2008sa}.  Coherent and incoherent processes
can be separated by detecting the nuclear-breakup, i.e., detecting the
nuclear fragments. While this is experimentally straight forward in
fixed target experiments it is rather challenging at an EIC since the
charged fragments are transported along the ion beam line.  The most
promising approach is the measurement of emitted neutrons via
zero-degree-calorimeters, a technique used extensively at RHIC.
Preliminary studies using de-excitation models (e.g. Gemini++
\cite{Charity:2011} and SMM \cite{Botvina:2011}) and a realistic
layout of an EIC interaction region showed that rejection factors
of larger than $10^5$ can be achieved.

In Fig.~\ref{fig:cs}, the nucleon configurations have been explicitly
generated according to the Woods-Saxon configuration.
Fig.~\ref{fig:KLN} shows the same Woods-Saxon distribution in the
optical approach compared with a KLN distribution motivated by the CGC
as discussed in \cite{Will:2011}.  It can be seen that the difference
when using different initial nucleon distributions within the nucleus
is considerable and easily measurable by an EIC. It also demonstrates
the flexibility of the generator in adapting different models at all
stages of the generation process. 


\noindent{\bf Summary and Outlook:}  A new event generator for the generation of diffractive events in $ep$
and $e$A collisions has been implemented based on an impact parameter
dependent dipole model. It describes the coherent and incoherent
contributions to the cross-sections.  In its current version it is
limited to the production of exclusive vector mesons and DVCS photons.
We intend to include more general diffractive processes in the
same framework, $e+p/$A$\rightarrow e'+p'/$A$' + X$ where $X$ is a
general final state consisting of two or more hadrons. When completed
it will be the first diffractive event generator for $e$A collisions with a
broad range of processes relevant for the physics of a future EIC.


\ \\ \noindent{\it Acknowledgments:} 
The authors would like to
thank the INT for their hospitality and support. Also many thanks to
G.~Beuf, 
M.~Diehl, 
A.~Dumitru, 
W. Horowitz,
H.~Kowalski, 
T.~Lappi, 
and R.~Venugopalan for many helpful discussions.


\subsection{Parton propagation and hadronization simulations:
  overview}   
\label{sec:EnergyLossSims}  

\hspace{\parindent}\parbox{0.92\textwidth}{\slshape   
  Alberto Accardi 
}
\index{Accardi, Alberto}

\vspace{\baselineskip}

The ``Parton Propagation and Fragmentation'' working group is
currently working on several Monte Carlo simulations to address
hadronization in the cold nuclear medium.
More information, references and links are
available on the PPF working group wiki \cite{ppfwg_wiki}.

\begin{itemize}
\item  {\bf PyQM.} The ``Pythia Quenching Model''
  is an energy-loss simulation based on Pythia, see
  Section~\ref{sec:eA-PQM}. The 
  partons created in the hard scattering are allowed to lose energy
  according to the Salgado-Wiedemann quenching weights, and then fed
  into the Lund string fragmentation Pythia module. The goal is to
  determine if the Lund string fragmentation leads to observable
  differences compared to using Fragmentation Functions to describe
  leading hadron attenuation (as implemented e.g. in PQM, see below),
  and to provide a simulation for a broader range of hadron flavors. 
\item {\bf Q-Pythia extension to DIS.} Q-Pythia is an energy loss
  simulation by Armesto, Cunqueiro and Salgado
  based on medium-modified DGLAP evolution equations. 
  Currently, only energy loss in the QGP is implemented, and we are
  working on implementing energy loss in the cold nuclear target. 
  Pursuing this simulation is likely to have a very big pay-off: 
  it will allow to study jet nuclear modifications, the effects of
  medium modified DGLAP evolution on hadron observables, and compare
  this to the BDMPS energy loss formalism in the integrated PQM
  simulation, and the implementation of the Higher-Twist
  energy loss formalism. Comparison to simulations done with Q-Herwig,
  would also allow one to gauge the effects of cluster vs. Lund string
  hadronization.
  \begin{itemize}
  \item {\bf PQM.} The ``Parton Quenching Model'' is a simulation by 
    Dainese, Loizides and Paic, which uses Pythia as a parton level
    generator, and then applies the Salgado-Wiedemann quenching
    weights to determine the parton energy loss before using
    Fragmentation Functions to determine single hadron attenuation. 
    It has been integrated in Q-Pythia by C.~Loizides.
  \item {\bf PyQM integration.} 
    It will be interesting to integrate PyQM in Q-Pythia, to provide a
    direct comparison between hadronization performed according to the
    Lund string model and using Fragmentation Functions.
  \end{itemize}
\item {\bf Higher-Twist energy loss.} 
  The Higher-Twist energy loss formalism has
  recently been extended to include a resummation of all higher-twist
  contributions, and inmplemented in a Monte-Carlo simulation, see
  Section~\ref{sec:eA-AbhijitMC}.
\item {\bf GiBUU.} This is (among other things) a simulation of 
  nuclear modifications of hadron production in DIS based on the Lund
  string model and BUU coupled-channel transport equation for the
  (pre)hadrons, and completely neglects energy loss, see
  Section~\ref{sec:e+A_with_GiBUU}. It has been extensively tested on
  HERMES and EMC data, and is ready to use at the EIC energy. It will
  be interesting to implement the few variations in the space-time
  prehadron production schemes available on the market and investigate
  possible observable differences. 
  Inclusion of target fragmentation is currently in
  progress in the multi-fragmentation framework (SMM)
  \cite{Bondorf:1995ua} and correcting for effects of the large energy
  gap between initial interaction and fragmenting nucleons.
\end{itemize}

\subsection{PyQM: a pure energy loss Monte-Carlo simulation}
\label{sec-AccDup:pyqm}
\label{sec:eA-PQM}   
\hspace{\parindent}\parbox{0.92\textwidth}{\slshape 
Rapha\"el Dupr\'e and Alberto Accardi 
}
\index{Dupr\'e, Rapha\"el}
\index{Accardi, Alberto}

\vspace{\baselineskip}

Pure quark energy loss models are widely used to describe jet quenching in 
relativistic heavy ion collisions (RHIC), however most of the calculations 
were never applied to the nDIS experiments, which are usually at lower energy,
making any comparison difficult. EIC is the chance to have data of both 
processes at similar kinematic, in this context it is natural to develop PyQM, 
a pure energy-loss Monte-Carlo simulation for nDIS based on the
Salgado-Wiedemann quenching weights formalism \cite{Salgado:2002cd} widely used
to analyze RHIC data. This simulation will be utilized 
as a tool to evaluate the future EIC capabilities concerning quark
energy loss measurement; since it also provides rate estimates and
the kinematics of particles to detect, this information will be used
to discuss the relevance and interest of various observables and the
accelerator and detector requirements to access them. 

\begin{figure}[h]
  \centering
    \includegraphics[width=0.8\linewidth]{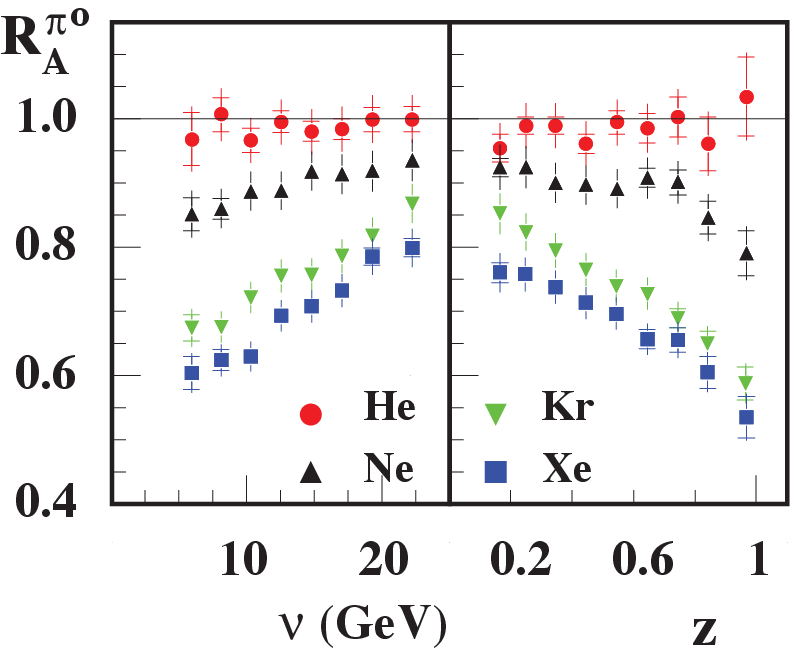} 
    \caption{\small \label{fig-AccDup:simherm}
      Multiplicity ratio of positive pions from
      HERMES\cite{Airapetian:2007vu} (points) compared to 
      the PyQM pure energy loss simulation (lines).
  }
\end{figure}

The PyQM Monte-Carlo simulation is based on PYTHIA 
\cite{Sjostrand:2006za} for the DIS interaction and the fragmentation process,
which is described by the Lund string model.
Between the intial hard scattering and string fragmentation,
we apply quark energy loss on the struck 
parton, using a nuclear density profile \cite{DeJager:1987qc} to estimate the quantity of 
matter the quark has to go through, and the Salgado-Wiedmann quenching
weights \cite{Salgado:2003gb} to calculate the energy loss itself. To account for
the geometry of the nuclei, we follow Ref.~\cite{Accardi:2007in}, and
pick randomly the interaction point 
according to the nuclear density distribution; the thickness of the nuclear 
matter seen by the parton is then given by  $R = { { 2 \bar \omega_C^2 (\vec b, y) } \over 
      { \int_y^\infty dz \hat q_A(\vec b, z)} }$
with $y$ the position along the propagation direction with its origin at 
the interaction point, and $\vec b$ the transverse position of the y
axis relative to the center of the nucleus. The characteristic energy
$\omega_C$, and the local transport coefficient $\hat q$ are given by 
\begin{equation}
\omega_C (\vec b, y) = \int _y^\infty dz (z-y) \hat q_A(\vec b, z)
\qquad
\hat q_A(\vec b, y) = {\hat q_0 \over \rho_0} \rho_A(\vec b, y)
\end{equation}
Then the only free parameter for the quenching weights, and indeed for
the whole simulation, is $\hat q_0$, the transport coefficient at the
center of the nucleus. This is found to be $\hat q_0 = 0.6$ GeV$^2$
fm$^{-1}$ from a fit of the HERMES data \cite{Airapetian:2007vu} (figure
\ref{fig-AccDup:simherm}), in agreement with the analytic calculations of
\cite{Accardi:2007in,Accardi:2009qv}. A full description of  
the results of this simulation compared to HERMES would be beyond the scope of 
this presentation; here we note that its results are satisfactory for the 
multiplicity ratio, but require a seemingly too large $\hat q$
compared to HERMES data on pion $p_T$-broadening. We are currently
working on an implementation of $p_T$-broadening in our simulation,
which, puzzingly, appears instead to produce the right amount of
integrated $p_T$ broadening as a function of $A$.
This issue is directly linked to the quenching weight calculation and
work is in progress to better understand it.

\section{Connections to other fields}
\label{sec:Connections}


\subsection{Gluon Tomography in Nuclei - The Heavy Ion Collision Initial State}
\label{sec:eAAA}
\hspace{\parindent}\parbox{0.92\textwidth}{\slshape
 William A. Horowitz}
\index{Horowitz, William A.}

\vspace{\baselineskip}

The main purpose of colliding large nuclei is the creation and study of the quark-gluon plasma (QGP), the deconfined state of QCD matter at high temperatures ($T\,\sim\,200$ MeV) and low baryon chemical potential ($\mu\,\sim\,0$).  Measuring the properties of the QGP is interesting as it is a known phase of the strong force, one of the  four known forces in Nature.  
The QGP is fascinating from a theoretical standpoint as there exists the possibility of experimentally measuring the emergent many-body properties of the non-linear, non-Abelian QCD field theory.  It was hoped that the collision of large, ultra-relativistic nuclei 
in a heavy-ion collision (HIC) might provide an experimental window with which to observe the 
properties of the QGP, and it appears that just such a novel phase of matter has been created 
at RHIC \cite{Adcox:2004mh,Adams:2005dq,Back:2004je,Arsene:2004fa,Gyulassy:2004zy}.  

\begin{figure}[!htb]
\centering
\includegraphics[width=4.5 in]{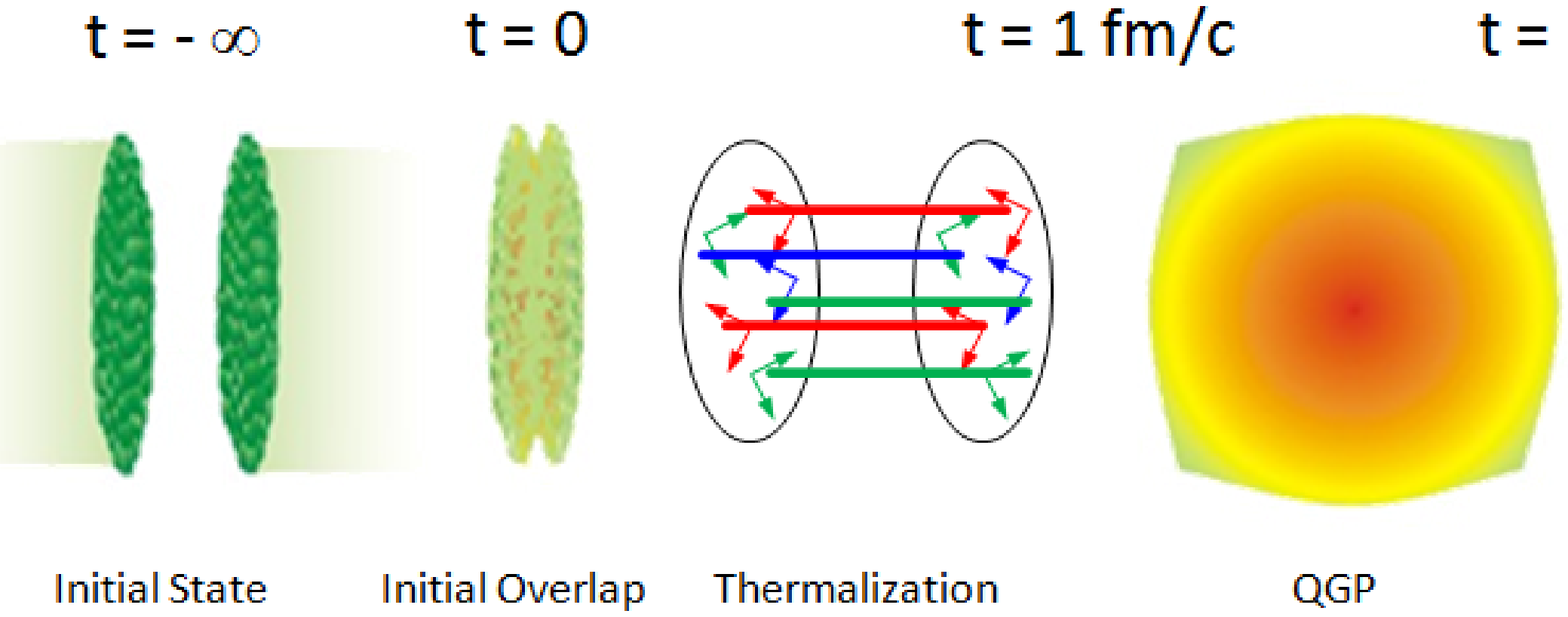}\vspace{-.1in}
\caption{\small \label{AAHorowitz:HICStages}
Cartoon of the stages of a heavy ion collision.  Timescales are approximate. 
}
\end{figure}

But what are the properties of this QGP that has been created?  Qualitatively: is the medium 
strongly or weakly coupled; what are its relevant degrees of freedom; does viscous 
relativistic hydrodynamics describe the bulk physics of the QGP; does either pQCD or the phenomenological string theory methods of the AdS/CFT correspondence 
or neither describe the physics of either the bulk medium or the high momentum probes of 
the medium?  Quantitatively: what is the viscosity of the medium; what are the values of 
its transport coefficients?  Is the QGP at RHIC the most perfect fluid created by mankind?  The difficulty faced when trying to answer these questions is 
that a heavy-ion collision is an incredibly complex event.  It is useful to think about a HIC, 
as currently best understood, as a series of separate stages: 1) $t = -\infty$, the time 
before overlap, when the nuclei are boosted to 200 GeV per nucleon at RHIC; 2) $t=0$, the 
actual collision of the nuclei and creation of large chromodynamic fields; 3) 
$0\lesssim t\lesssim 1$ fm/c, the initial large chromodynamic fields rapidly thermalize; 
4) $1\lesssim t\lesssim 3$ fm/c, evolution as a QGP; 5) $3\lesssim t \lesssim 4$ fm/c, 
hadronization; 6) 4 fm/c $\lesssim t \rightarrow \infty$, evolution as a hadron gas.  
A cartoon of a typical central heavy ion collision is shown in figure~\ref{AAHorowitz:HICStages}.  

As one can see, the system is in the QGP phase for only a brief period of its entire spacetime 
evolution!  
These times are important to understand not only because they are interesting in their own 
right---What are the non-linear evolution effects on the color charge density of highly boosted 
nuclei?  How do very large chromodynamic fields thermalize so rapidly?  How does hadronization 
occur?---but also because the interpretation of experimental observables associated with 
the QGP is sensitive to the details of the physics of the other stages of a HIC.  Any new 
means of experimentally extending our understanding of these other stages would provide a 
qualitative leap forward in our understanding of the QGP created at RHIC.  Of particular 
importance are the initial conditions of a heavy-ion collision, from 
$t = -\infty$ to $t\sim 1$ fm/c, from the time before the collision up through 
thermalization.  An electron-ion collider that probes gluons at $x\sim10^{-3}$ could 
provide precisely this qualitatively new physics understanding of the initial conditions.

The two most striking discoveries of the RHIC heavy-ion program so far are perfect 
fluidity and jet suppression.  The naive interpretation of the measured flow of low momentum particles is that the QGP is a strongly coupled fluid whose properties are described by AdS/CFT; the naive interpretation of the measured jet suppression is that the QGP is a weakly coupled plasma whose properties are described by pQCD.  These two interpretations are both 
mutually exclusive and highly dependent upon the initial conditions of HIC. 

\begin{figure}[tb]
\centering
\includegraphics[width=\textwidth]{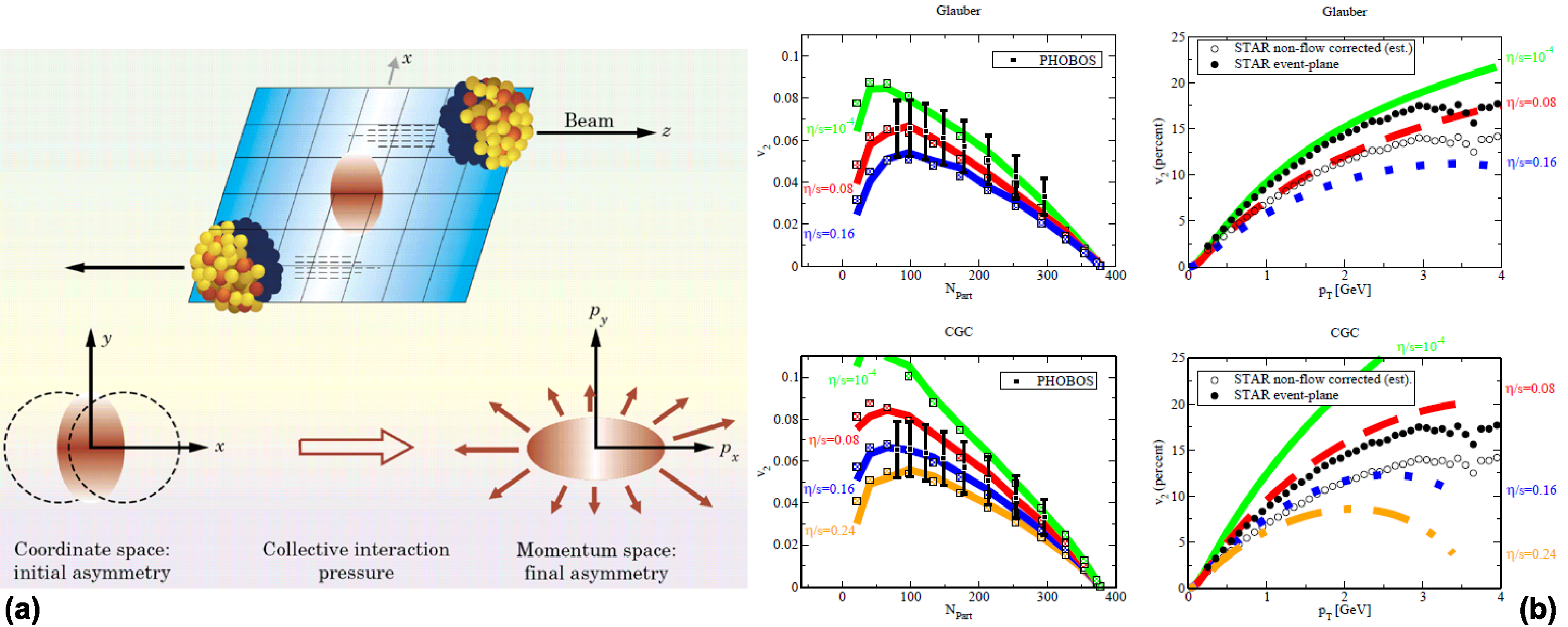}\vspace{-.1in}
\caption{\small \label{AAHorowitz:hydropic}
(a) Initial spatial anisotropy evolves into momentum anisotropy in non-central heavy ion collisions.  
Hydrodynamics aims to quantitatively model this process to gain information on the 
medium and its properties.  (b) Comparison of data and theoretical predictions using viscous 
relativistic hydrodynamics for $v_2^h(N_{part})$ (left) and $v_2^h(p_T)$ (right).  Viscous 
hydrodynamics predictions use Glauber-like initial conditions (top) or a simplified 
implementation of color glass condensate (CGC) physics (bottom).  Note the 100\% difference 
in extracted $\eta/s$ from the two naive geometry models.  figures adapted from 
\cite{Ludlam:2003rh,Luzum:2008cw}.}
\vspace{-0.5em}
\end{figure}

\noindent {\bf Hydrodynamics.} The stunning success of ideal relativistic hydrodynamics at RHIC as compared to its 
failure in lower energy machines \cite{Kolb:2000fha,Alt:2003ab,Hirano:2005xf}, led to 
the proclamation of the creation of a perfect fluid at RHIC 
\cite{Riordan:2006df,BNLRelease,Jacak:2010zz}.  
In HIC particle spectra are often conveniently reported as
\begin{equation}
\label{AAHorowitz:hydrovtwo}
\frac{dN^h}{dp_T}(p_T, \, \phi, \, N_{part}) = \frac{dN^h}{dp_T}(p_T, \, N_{part}) \left( 1 + 2 v^h_2(p_T, \, N_{part}) \cos 2\phi + \ldots \right),
\end{equation}
where $\phi$ is the angle of the observed particle with respect to the semiminor axis of the overlap region; see figure~\ref{AAHorowitz:hydropic} (a).  As pictured 
in figure~\ref{AAHorowitz:hydropic} (a) the $v_2^h$ develops from pressure gradients that build up as a result 
of the spatial anisotropy created in the initial overlap of the two nuclei.  

The nearly ideal fluid flow as surmised from hydrodynamics is exciting because the extracted 
value of $\eta/s$, the shear viscosity to entropy ratio, is smaller than for any other 
known substance \cite{Kapusta:2008vb}.  From a theoretical standpoint, this nearly ideal flow is a huge success for string 
phenomenology: the lower bound for $\eta/s$ in a strongly-coupled liquid as computed 
using the AdS/CFT correspondence is $1/4\pi$, in natural units.  This value of 
$1/4\pi\simeq0.1$ should be compared to the naive estimate from pQCD, $\eta/s\sim1$.  
Conservative estimates of the extracted value of $\eta/s$ from comparison between 
theoretical calculations and experimental data yield $\eta/s\sim 0.1-0.5$ \cite{Kapusta:2008vb}.  
Hydrodynamics 
is a set of partial differential equations: initial conditions, for which hydrodynamics 
can tell us nothing, must be supplied. Figure~\ref{AAHorowitz:hydropic} (b), in which a 2+1$D$ viscous hydrodynamics calculation is compared with data, 
shows the at least factor of 2 uncertainty in the extracted value of $\eta/s$ that arises from 
the poorly constrained mean value of the initial geometry.  
The uncertainty from fluctuations \cite{Takahashi:2009na}, in which hot and cold 
spots appear in the initial conditions, might also be very large \cite{Gyulassy:1996br}.  
This very large range of $\eta/s$ means that one cannot definitively claim that the medium is 
better understood as strongly coupled and near the lower bound set by AdS/CFT or weakly 
coupled, with pQCD providing a good physical description. 

\begin{figure}[tb]
\centering
\includegraphics[width=\textwidth,bb=0 163 746 387,clip=true]{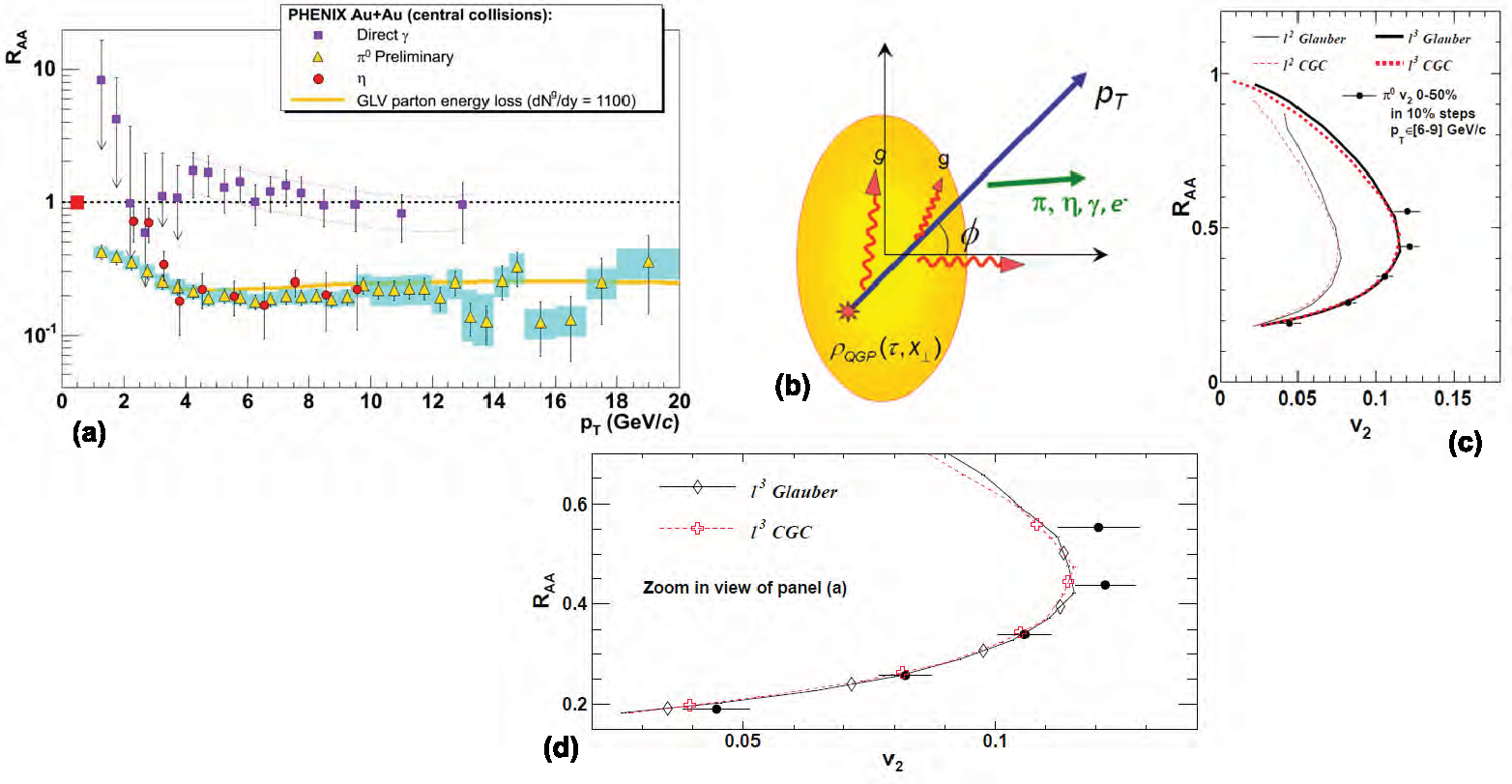}
\caption{\small \label{AAHorowitz:pqcdsuccess}(a) A plot of the early success
  of pQCD energy loss calculations in describing $R_{AA}(p_T)$,
  Eq.~\ref{AAHorowitz:raa}.  (b) Cartoon of the energy loss from a
  high-$p_T$ parton in the QGP medium. The longer the pathlength $L$
  the greater the energy loss: the spatial anisotropy manifests as a
  suppression anisotropy, which is represented by $v_2^h$. (c) 
  pQCD ($\Delta E~\sim L^2$) energy loss significantly underpredicts
  the anisotropy while AdS/CFT ($\Delta E\sim L^3$) loss is
  consistent.  The simultaneous description of $R_{AA}$ and $v_2$
  seems to require both $L^3$ energy loss and a CGC-like initial
  state. figures adapted from \cite{Akiba:2005bs,Jia:2011pi}.
}
\vspace{-0.5em}
\end{figure}


\noindent {\bf High-pT Physics.} Originally, high-$p_T$ particles were hoped to provide a tomographic probe of the QGP medium 
produced at RHIC.  Jet tomography, then, would provide a means, independent of 
hydrodynamics, for determining many medium properties; most important, jet tomography could 
be a tool to investigate the initial geometry of the HIC.  While early work showed great promise, see figure~\ref{AAHorowitz:pqcdsuccess} (a), there are several observables for which the perturbative energy loss calculations 
do not provide a good description of the data (see, e.g., figure \ref{AAHorowitz:pqcdsuccess}(c)).  There is currently not much theoretical control over the in-medium energy loss experienced 
by high-$p_T$ partons: different assumptions about the best physics approximations have yielded 
very different energy loss calculations (see, e.g., \cite{Horowitz:2007su,Majumder:2010qh}), and all these calculations suffer from large, 
mostly unquantified uncertainties due to simplifying mathematical approximations \cite{Horowitz:2009eb}.  Nevertheless, qualitatively fascinating discoveries 
can be made from high-$p_T$ observables.  In particular, one may compare the results of strong 
coupling calculations derived using the AdS/CFT correspondence to those derived using 
traditional pQCD methods; in this way, energy loss holds out the possibility of rigorously investigating, independent of hydrodynamics, 
whether RHIC creates a strongly-coupled perfect fluid or a weakly-coupled plasma.  

High-energy particle spectra are often reported as normalized by the $p+p$ spectrum multiplied 
by $N_{coll}(N_{part})$, where $N_{coll}(N_{part})$ is the expected number of $p+p$-like 
hard collisions in an $A+A$ collision with a given number of participants:
\begin{equation}
\label{AAHorowitz:raa}
R_{AA}^h(p_T, \, \phi, \, N_{part}) = \frac{dN_{AA}^h}{dp_T}(p_T, \, \phi)\Big/N_{coll}(N_{part})\;\frac{dN_{pp}^h}{dp_T}(p_T),
\end{equation}
where $h$ is the measured hadron species and $\phi$ is the same angle
as was defined in the discussion of hydrodynamics.  This ratio is also often reported as a Fourier expansion, with $v_2^h$ again representing twice the first Fourier coefficient (the same $v_2^h$ as in hydrodynamics).  However the physical understanding of the origin of the high-$p_T$ $v_2^h$ is very different from the hydrodynamics physics which dominates the generation of the low-$p_T$ $v_2^h$.  For high-$p_T$ observables, $v_2^h$ comes from high-$p_T$ partons traversing a medium asymmetric from the initial 
geometry: less energy loss occurs for partons traveling the short direction of the 
almond-shaped overlap region compared to those partons that travel the long direction.  
A cartoon of this physical picture is shown in figure~\ref{AAHorowitz:pqcdsuccess} (b).  The size of $v_2^h$ 
is then an entangled measure of the geometry of the medium and the pathlength dependence of 
the energy loss mechanism: perturbative elastic energy loss, which goes as $L^1$, produces 
less $v_2^h$ for a given geometry than perturbative inelastic energy loss, which goes as $L^2$, 
which produces less $v_2^h$ than strong-coupling energy loss, which, for light partons, 
goes as $L^3$ and as $\exp(-L)$ for heavy partons.  $v_2^h$ is of particular interest because 
it was recently measured out to $p_T\sim13$ GeV, well beyond momentum scales where 
hadronization effects might be important.  That 
the observed $v_2^h$ is significantly larger than that predicted by perturbative methods, shown 
in figure~\ref{AAHorowitz:pqcdsuccess} (c), is perhaps the best high-$p_T$ experimental evidence that AdS/CFT, 
as opposed to pQCD, is the best approximation to the relevant physics at RHIC.

As the theoretical prediction of high-$p_T$ $v_2^h$ comes directly from the azimuthal anisotropy 
of the QGP medium, knowledge and constraint of the initial geometry is crucially important 
for a rigorous scientific conclusion to be made: the sharper the produced medium the larger 
the $v_2^h$, regardless of energy loss mechanism. As one can see from figure~\ref{AAHorowitz:pqcdsuccess}, there are reasonable initial conditions for which no known energy loss calculation describes the data.  And just as in hydrodynamics, 
fluctuations may play an important, even outsized, role. 

\noindent{\bf Measuring the Initial State.}
From the above discussion it is clear that knowledge of the initial conditions at RHIC is crucial for interpreting the experimental data.  The density of the charged and neutral matter density of nuclei at rest is well understood from 
diffraction pattern experiments (see, e.g., \cite{Blanpied:1977bp}).  Knowledge of the 
rest frame density of protons and neutrons in nuclei has been used extensively in 
estimating the initial matter density created in HIC.  Matter production in HIC, though, 
depends on the distribution of quarks and, especially, gluons in the nuclear wavefunction.  
Below some value of Bjorken $x$ that is not yet precisely known, non-Abelian, non-linear 
QCD evolution effects become important.  The (mostly) gluonic initial state medium at midrapidity 
at RHIC consists of particles of $x\sim p_T/\surd s\sim10^{-3}$, which is at the order 
of magnitude for which small-$x$ physics likely becomes relevant.  Unfortunately the 
theory of small-$x$ physics in $A+A$ collisions is very complicated, and current knowledge is 
incomplete.  Additionally, the aforementioned theoretical calculations of $v_2^h$ are in 
fact most sensitive to the the quantitative shape of the edge of the initial nuclear overlap 
in HIC; it is just in this region that many of the theoretical tools developed for small-$x$ 
physics study break down.  
It turns out, though, that through careful measurements, diffraction patterns may be 
measured at an electron-ion collider using deeply-virtual Compton scattering and 
vector meson production.  These diffraction patterns, in turn, may be inverted to 
constrain the initial gluon and quark densities of the highly boosted nuclei.  
Fortuitously, these experimental measurements give the most sensitive determination 
of these densities at the edge of the nucleus, the region of the overlap which hydrodynamics and energy loss calculations are most sensitive to.

\subsection{Constraining initial conditions in A+A collisions}
\label{sec:adum}
\hspace{\parindent}\parbox{0.92\textwidth}{\slshape
 Adrian Dumitru}
\index{Dumitru, Adrian}

\vspace{\baselineskip}

Understanding small-$x$ gluon production in the \emph{initial} state
of relativistic A+A collisions constrains the amount of additional
entropy produced via ``final-state'' interactions such as parton
thermalization / QGP formation~\cite{Baier:2002bt} and its subsequent
hydrodynamic expansion. If these processes provide a significant
contribution, then that should presumably show in the centrality
dependence of the multiplicity in the final state: final state
interactions should be much more prevalent for a head-on collision of
two large nuclei than for a grazing shot or p+A or (minimum bias) p+p
collisions. It is therefore very important to test models for initial
particle production over a broad range of centralities -- perhaps down
to the level of p+p collisions -- in order to constrain entropy
production due to thermalization and viscous hydrodynamic
expansion~\cite{Dumitru:2007qr}.

To compute the number of small-$x$ gluons released from the
wavefunctions of the colliding nuclei, one frequently employs the
$k_\perp$-factorization formalism~\cite{Gribov:1984tu,Kharzeev:2001gp},
\begin{eqnarray}
  \frac{dN}{d^2 \mathbf{r}_{\perp}dy}&=& {\cal N}
  \frac{N_c}{N_c^2-1} \int \frac{d^2p_\perp}{p^2_\perp}
  \int^{p_\perp} \!\! {d^2 k_\perp} \;\alpha_s(Q^2)\nonumber\\ 
& & \hspace{-1.5cm} \times 
  \; \phi_A(x_1, \frac{(\mathbf{p}_\perp+\mathbf{k}_\perp)^2}{4};
  \mathbf{r}_\perp)
  \; \phi_B(x_2,\frac{(\mathbf{p}_\perp{-}\mathbf{k}_\perp)^2}{4}; 
  \mathbf{r}_\perp)~,
 \label{dumitru_eq:ktfac}
\end{eqnarray}
where $N_c = 3$ is the number of colors, and $p_{\perp}$, $y$ are
the transverse momentum and the rapidity of the produced gluons,
respectively. $x_{1,2} = p_{\perp} \exp(\pm y)/\sqrt{s_{NN}}$ denote
the light-cone momentum fractions of the colliding gluon ladders,
$\sqrt{s_{NN}}$ is the collision energy, and typically one chooses
$Q^2 = \mathrm{max}((\mathbf{p}_\perp+\mathbf{k}_\perp)^2, 
(\mathbf{p}_\perp{-}\mathbf{k}_\perp)^2)/4$.  The normalization factor
${\cal N}$ can be fixed from peripheral collisions, where final-state
interactions should be suppressed. It effectively also absorbs NLO
corrections and the contribution from sea (anti-)quarks.  The
unintegrated gluon distribution $\phi$ is related to the dipole
scattering amplitude in the adjoint representation, $N_G$, through a
Fourier transform~\cite{Kovchegov:2001sc}:
\begin{equation}
\phi(x,k_\perp^2;{\mathbf{r}}_\perp) = 
\frac{C_F}{\alpha_s(k_\perp)\,(2\pi)^3}\int d^2{\mathbf s}_\perp\
e^{-i{\mathbf k}_\perp\cdot{\mathbf s}_\perp}\,\nabla^2_{{\mathbf
  s}_\perp}\,N_G(x,s_\perp;{\mathbf r}_\perp)\,.
\label{dumitru_eq:phi}
\end{equation}
\vskip1ex

\begin{figure}
\begin{center}
\includegraphics[width=0.45\textwidth]{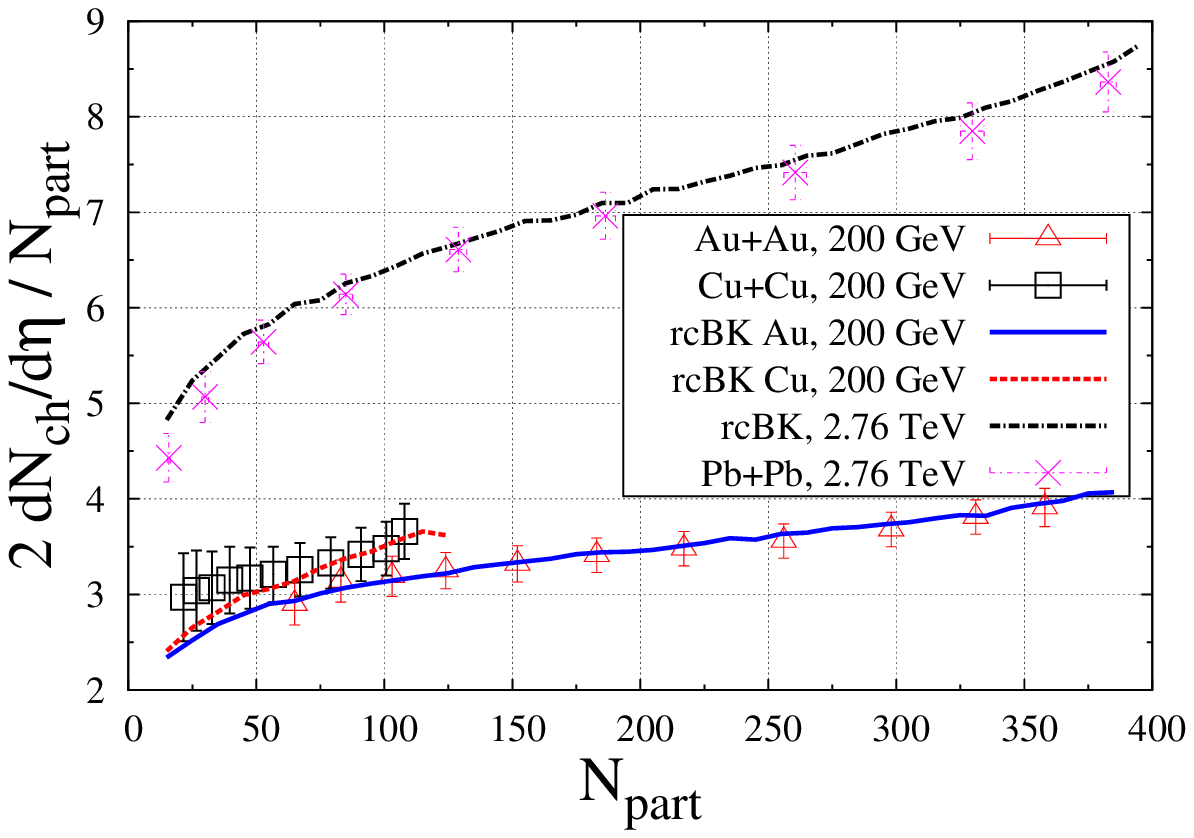}
\includegraphics[width=0.45\textwidth]{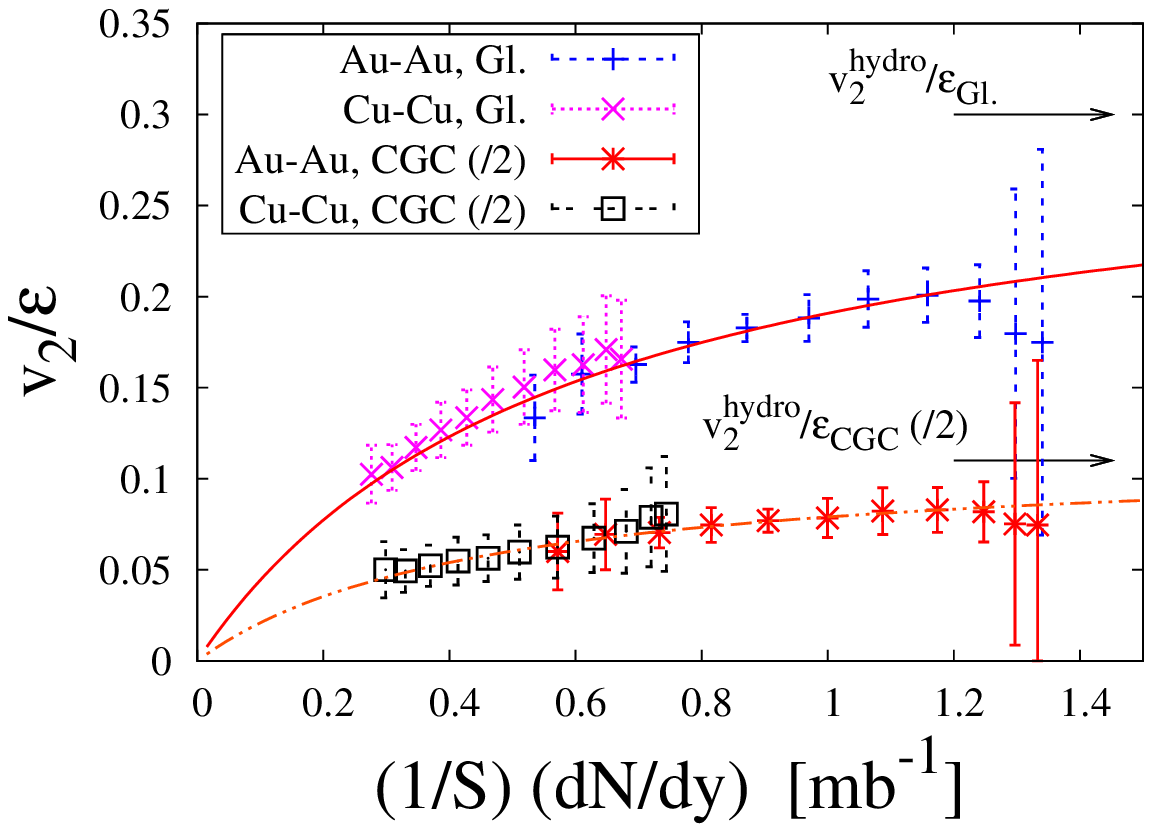}
\end{center}
\vspace*{-0.8cm}
\caption{\small Left: Centrality dependence of the
  multiplicity at $200\gev$ and $2760\gev$, respectively, from
  $k_\perp$ factorization with running-coupling BK unintegrated gluon
  distributions (see~\cite{ALbacete:2010ad} for details). PHOBOS data:
  \cite{Back:2002uc} (Au+Au), \cite{Roland:2005ei} (Cu+Cu); ALICE
  data from ref.~\cite{Aamodt:2010pb}.
Right: $v_2/\varepsilon$ versus the transverse particle
      density~\cite{Drescher:2007cd}; $v_2/\varepsilon_{\rm CGC}$ has
      been scaled by 1/2 for better visibility.
}
\label{dumitru_fig:f1}
\end{figure}


\noindent {\bf The multiplicity in heavy-ion collisions.} 
Figure~\ref{dumitru_fig:f1} (left) shows the centrality dependence of
particle production for heavy-ion collisions at $200\gev$ and
$2760\gev$, respectively, obtained by integrating
eq.~\eqref{dumitru_eq:ktfac} over the transverse overlap of the
colliding nuclei. The unintegrated gluon distributions are solutions
of the \emph{local} (impact parameter independent) Balitsky-Kovchegov
(BK) equation with running-coupling corrections according to the
Balitsky prescription~\cite{Albacete:2007yr}. The impact parameter
dependence is due entirely to the initial condition where it has been
assumed that essentially $Q_s^2(x_0;\mathbf{r}_\perp)=Q_0^2 \;
\sigma_0\; T_A(\mathbf{r}_\perp)$ increases in proportion to the
thickness of the nucleus ($Q_0$ and $\sigma_0$ denote constant scales;
for details see ref.~\cite{ALbacete:2010ad}). Neglecting the impact
parameter dependence of the dipole scattering amplitude $N_G$ \emph{in
  a nucleon} relies on the scale separation $R_A\gg R_N \gg Q_s^{-1}$
where $R_A$ is the size of the overlap region in the
collision~\cite{Dumitru:2010kb}.

Apparently, the model calculation describes both the centrality and
the energy dependence of particle production fairly well. If so, this
constrains final-state entropy production and correlates the
thermalization time and the shear viscosity to entropy density ratio:
extremely rapid thermalization and/or $\eta/s\gsim 0.3$ would be
excluded by stringent entropy production bounds~\cite{Dumitru:2007qr}.

Several caveats remain. As already mentioned above, the absolute
normalization of the gluon density at small $x$ (alternatively, the
factor ${\cal N}$ in the $k_\perp$ factorization formula) can be fixed
in practice only from very peripheral A+A or p+p
collisions\footnote{Small-$x$ partons do not contribute significantly
  to the momentum sum rule and a precise matching to the parton
  distributions at large $x$ and low $Q^2$ is lacking.}. For p+p
collisions, however, the impact parameter dependence of the dipole
scattering amplitude over distance scales $\sim R_N$ can not be
neglected, see for example ref.~\cite{Tribedy:2010ab}.

Furthermore, it may be important to consider in more detail the
structure of running coupling corrections to the
$k_\perp$-factorization
formula~\eqref{dumitru_eq:ktfac}~\cite{Horowitz:2010yg} and the effect
of a full NLO treatment of BK evolution. Indeed, if such corrections
modify the centrality dependence of particle production in A+A
collisions then they will also affect entropy production constraints
and thus the fundamental understanding of the thermalization processes
and time scales as well as estimates of the shear viscosity of thermal QCD.



\noindent {\bf The eccentricity in heavy-ion collisions.}
Other quantities of relevance for the interpretation of heavy-ion
collisions exhibit even greater sensitivity to the actual
\emph{distribution} of produced gluons in the transverse
$\mathbf{r}_\perp$ plane than its integral $dN/dy$. A
collision of two heavy ions at non-zero impact parameter, neglecting
fluctuations of the local density of participant nucleons, leads to a momentum
asymmetry called ``elliptic flow'', $v_2 \sim \langle\cos
2\phi\rangle$, as described in section \ref{sec:eAAA}. In the absence of any scales (such as the freeze-out
temperature $T_f$, the phase transition temperature $T_c$, or a
non-vanishing mean free path $\lambda$), hydrodynamics predicts that
$v_2$ is proportional to the eccentricity $\varepsilon$ of the overlap
area~\cite{Ollitrault:1992bk}, $\varepsilon = \langle
y^{\,2}{-}x^{\,2}\rangle/ \langle y^{\,2}{+}x^{\,2}\rangle$. The
average is taken with respect to the distribution of produced gluons
in the transverse $x$-$y$ plane. Clearly, $\varepsilon$ involves large
cancellations of the contributions of gluons produced near the center
$r_\perp \sim0$ of the overlap zone and so is more sensitive to
particle production in the periphery.

A simple geometry based initial condition assumes that by analogy to
the Glauber model for soft particle production $dN/dyd^2r_\perp \sim
\rho_{\rm part}^{\rm ave}(r_\perp) \equiv (\rho_{\rm part}^A(r_\perp)
+ \rho_{\rm part}^B(r_\perp))/2$, where $\rho_{\rm part}^i$ is the
density of participants of nucleus $i$ per unit transverse area.
High-density QCD (the ``Color-Glass Condensate'') predicts a somewhat
different distribution of gluons in the transverse plane,
corresponding to a higher eccentricity $\varepsilon$. In particular,
in the ``p+A limit'' when one of the nuclei is very dense while the
other is dilute, the number of produced particles is proportional only
to the density of the dilute collision partner, whose partons add up
linearly. Hence, \emph{in the reaction plane}, $dN/dyd^2r_\perp \sim
{\rm min}(Q_{s,A}^2,Q_{s,B}^2) \sim {\rm min}(\rho_{\rm part}^A ,
\rho_{\rm part}^B)$ drops more rapidly as one moves towards the edge
of the overlap zone than $dN/dyd^2r_\perp \sim \rho_{\rm part}^{\rm
  ave}$~\cite{Drescher:2006pi}. Thus, a higher eccentricity is a
generic effect due to a dense target or projectile. Specific numerical
estimates rely on an accurate determination of the unintegrated gluon
distribution, however. Ref.~\cite{ALbacete:2010ad} finds that the
energy dependence of $\varepsilon$ from RHIC to LHC is very weak.

Figure~\ref{dumitru_fig:f1} (right) shows the elliptic flow $v_2$
measured in heavy-ion collisions at RHIC scaled by the eccentricity
$\varepsilon$ of the overlap zone~\cite{Drescher:2007cd}. As already
mentioned above, in the absence of any scales such as a non-zero mean
free path, $v_2/\varepsilon$ would be independent of the transverse
density of particles. Indeed, if the $v_2$ data is scaled by the
eccentricity obtained from a CGC model implementation then the
required breaking of scale invariance is lower than for purely
geometry based (Glauber-like) initial conditions. Actual solutions of
viscous hydrodynamics (for $v_2$) appear to confirm this simple
observation in that the {\em slope} of $v_2/\varepsilon$ versus
transverse density is sensitive to the distribution of produced
particles~\cite{Heinz:2009cv}.

More recent studies attempt to understand also the relation of higher
moments of anisotropic flow $v_n$ to corresponding moments of the
initial eccentricity $\varepsilon_n$ -- such as the
``triangularity''~\cite{Alver:2010gr,Alver:2010dn,Qin:2010pf,Schenke:2010rr,Teaney:2010vd},
which is non-zero because of fluctuations of the large-$x$ sources in
the transverse impact parameter plane before the collision. A
quantitative interpretation of the ``response'' $v_n$ of the
Quark-Gluon Plasma medium to the initial geometry will also rely on a
good understanding of particle production in high-energy QCD.




\subsection{Particle production at low-$x$ and gluon saturation: from $p+A$ to $e+A$}
\label{sec:tuchin_pAeA}
\hspace{\parindent}\parbox{0.92\textwidth}{\slshape
 Kirill Tuchin}
\index{Tuchin, Kirill}

\vspace{\baselineskip}

In the beginning of the RHIC era, the $p(d)+A$ program was perceived as merely a useful baseline reference for the heavy-ion program. It very soon turned out that  due to a wise choice of colliding energy, 
RHIC probes the transition region to a new QCD regime of gluon saturation. While the first hints of gluon saturation were observed in DIS experiments at HERA, it is fair to say that  gluon saturation was discovered in $dA$ collisions at RHIC. At present, as we are heading toward the era of EIC, it is important to review what we have learned at RHIC and how it can be used to optimize the EIC program. The purpose of this section is to review phenomenological studies of gluon saturation at RHIC. 

The reason why $pA$ and $eA$ high energy physics programs are closely related is provided by the Pomerantchuk theorem, which states that all high energy scattering processes are mediated by the exchange of a collective gluon state -- known as a Pomeron -- that has vacuum quantum numbers.  For this reason, inclusive processes in both programs share many similarities in the low $x$ region. The main distinction arises from the difference in the characteristic scales of the projectile: in protons it is a soft scale $\Lambda$, while in virtual photons, it is the photon virtuality $Q^2$, which depends on the electron kinematics. A possibility to control the  $Q^2$ is a  great advantage of DIS. In particular, it allows one to study the total cross sections/structure functions. However, in practice, the requirement to keep $x$ low significantly restricts the range of $Q^2$'s available for low $x$ studies. 

The relation between $pA$ and $eA$ scattering at low-$x$ becomes particularly apparent in the framework of the dipole model \cite{Mueller:1989st}. In the dipole model,  the cross section for $eA\to X$ or $pA\to X$ scattering, where $X$ is an arbitrary final state,  can be represented as 
\begin{align}\label{tuchin:1}
d\sigma_{p(\gamma^*)+A\to X} =\int d^2r_\bot\,\Phi^ {p(\gamma^*)}(r_\bot)\, d\sigma_{\mathfrak{d}+A\to X}( r_\bot)\,,
\end{align}
where $\mathfrak{d}$ stands for color dipole (letter $d$ is reserved for deuteron) of size $r_\bot$ in the transverse  plane. Eq.~(\ref{tuchin:1}) is based on the separation  of scales: the interaction length $\ell_i\sim R_A$ (in the target rest frame) is much smaller than the coherence length $\ell_c= \gamma/M_N$, where $\gamma\gg 1$ is the Lorentz factor and $M_N$ is the nucleon mass.    $\Phi^ {p(\gamma^*)}(r_\bot)$ is the light-cone ``wave function" describing the Fourier decomposition of a projectile into dipoles; it can be  calculated in QED (for $\gamma^*$), or modeled (for proton), see e.g. \cite{Kovchegov:2001ni,Li:2008se}. The main theoretical concern in low $x$ $pA$/$eA$ scattering is  calculation of the dipole-nucleus cross section, which is universal for both processes. With this observation in mind, we   are going to consider some of the $pA$ processes at RHIC that are of relevance for low-$x$ physics at EIC. 


\noindent {\bf Inclusive hadron production: $p+A\to h+X$.}
The cornerstone for phenomenological applications of the Color Glass Condensate (CGC), which is the theory of gluon saturation, is the factorization theorem proved in \cite{Kovchegov:2001sc}, where the cross section was derived that  re-sums all leading logarithmic   contributions $\alpha_s\ln(1/x)\sim 1$ (LLA)  for a heavy nucleus  in the quasi-classical limit $\alpha_s^2A^{1/3}\sim 1$.  A similar result was reported in \cite{Braun:2000bh,Blaizot:2004wu,Braun:2010qsa}. One does not expect that any of the hard perturbative QCD (hpQCD) factorizations  apply in this case because higher twist interactions  of valence quarks and gluons give contributions of order unity. Nevertheless, despite the fact that individual diagrams break factorization in covariant and light-cone gauges, the final re-summed expression can be cast in the $k_T$-factorized form. Unlike in hpQCD, the physical quantity that is factorized -- the unintegrated gluon distribution $\varphi(x,Q^2)$ --  can be calculated perturbatively owing to the existence of a hard scale $Q_s\gg \Lambda_\mathrm{QCD}$.  Another surprising fact is that contrary to naive expectations, $\varphi(x,Q^2)$ is related not to the momentum space Fourier-image of the nucleus gluon-field correlation function $\langle A_\bot( 0_\bot)\cdot  A_\bot( x_\bot)\rangle$, but rather to the Fourier-image of $\nabla^2_r N(r_\bot,  b_\bot,y)$, where $N(r_\bot,  b_\bot,y)$ is the imaginary part of the forward elastic scattering amplitude of a color dipole of size $ r_\bot$ at impact parameter $b_\bot$ and rapidity $y=\ln (1/x)$ in the heavy nucleus. Although the inclusive gluon production in $pA$ collisions is the only known case were  $k_T$-factorization holds, factorization of the multipoles in the transverse coordinate space is a general feature of the low-$x$ cross sections. It must be stressed that this multipole factorization does not imply hpQCD factorizations ($k_T$ or collinear ones) and neither opposite is generally true. 

The $k_T$-factorization formula derived  in \cite{Kovchegov:2001sc} led to successful phenomenology of inclusive hadron production in $dA$ collisions at RHIC, where the suppression of hadrons at forward rapidities and  Cronin enhancement at mid-rapidity were qualitatively predicted \cite{Kharzeev:2003wz,Baier:2003hr} and then quantitatively described in the CGC framework \cite{Kharzeev:2004yx,Dumitru:2005kb,Albacete:2010bs}. The production of valence quarks in the forward direction gives an important contribution to inclusive hadron production at large-$x$ of the proton and was discussed in \cite{Dumitru:2002qt,Gelis:2001da,Gelis:2002nn}.

By integrating  the gluon spectrum over $p_\bot$, one arrives at the total hadron yield as a function of rapidity $y$. It is rather weakly dependent on the details of the gluon distributions. Therefore, a simple model suggested in \cite{Kharzeev:2002ei} is able to describe inclusive hadron yield with remarkable accuracy.


\noindent {\bf Open charm (beauty) production: $\bm {p+A\to D+X}$.} 
The production of heavy quarks in $pA$ collisions at low-$x$ was calculated in \cite{Tuchin:2004rb,Blaizot:2004wv,Kovchegov:2006qn}. One expects that  the hpQCD factorization is  applicable if the saturation momentum is much smaller than the quark mass $m$ \cite{Kharzeev:2003sk}. At RHIC, $Q_s\sim m$ for charm and bottom, hence factorization is broken in both cases. Indeed, analysis of \cite{Fujii:2005vj} indicates that  semi-classical calculations of \cite{Blaizot:2004wv} disagree with  $k_T$-factorization by about 10\% at the $t$-channel gluon transverse momenta around $m$. hpQCD factorization is restored in the kinematic region where the operator product expansion  is applicable, i.e.\ at transverse momenta much higher than the saturation momentum.

The phenomenology of open heavy quark production at RHIC was developed in \cite{Tuchin:2007pf}, where it was found that the production pattern of heavy quarks is qualitatively similar to that of light quarks and gluons, although the magnitude of nuclear effects (Cronin and suppression) slowly decrease with increasing quark mass. These qualitative features are in good agreement with  preliminary data. 

\noindent {\bf Inclusive production of $\bm{\jpsi}$: $\bm{p+A\to \jpsi+X}$.}
In addition to the scales $\ell_i$ and $\ell_c$ mentioned earlier, the production of a charmonium state is characterized by another scale: formation length $\ell_f= \gamma/\Delta M$, where  $\Delta M$ is its binding energy. The key  theory  observation is strong ordering of the scales at high energies: 
$\ell_i\ll \ell_c\ll\ell_f$ \cite{Brodsky:1994kf,Hufner:1996jw}.  Consequently, we can distinguish three stages of $\jpsi$ production. (i) $g^*\to c\bar c$ described by the light-cone amplitude $\psi^g( k_\bot, z)$ often referred to as gluon's light-cone wave function, (ii) interaction of the gluon or the $c\bar c$ with the target depending on whether the splitting has occurred after or before the interaction, and (iii) formation of charmonium wave function. Unlike stages (i) and (ii), which can be described using perturbation theory owing to the weakness of the strong interaction at the $\jpsi$-mass scale, stage (iii) is non-perturbative because $\Delta M\ll M$. This, however, does not preclude us from using perturbation theory for calculating the $\jpsi$ production cross section, since the fragmentation process is independent of energy and atomic weight  ($\ell_f\gg R_A$). In other words, fragmentation happens in the vacuum long after any interaction with the target. 

Thus, the problem of calculating the $\jpsi$ production cross-section reduces to the calculation of the cross section of  $\mathfrak{d}+A\to [c\bar c(1^{--})]+X$ dipole-nucleus scattering. This calculation was done in \cite{Kharzeev:2005zr,Kharzeev:2008cv}. Note, that interaction depends on the quantum state of the $c\bar c$ pair, which must be in the $1^{--}$ color singlet state. Therefore, only those higher twist contributions may be taken into account that lead to this quantum state, and which are also enhanced by $\alpha_s^2A^{1/3}\sim 1$.  At the lowest order in $\alpha_s$, the projectile gluon in the proton wave-function has two interaction possibilities:
(i) leading twist processes $g+g\to \jpsi +g$, which is of order $\mathcal{O}(\alpha_s^5 A^{1/3})$ and  (ii) higher twist process $g+g+g\to J/\psi$ (initial gluons come from different nucleons), which is of the order  $\mathcal{O}(\alpha_s^6 A^{2/3})$. Since $\alpha_s^2A^{1/3}\sim 1$, the higher twist mechanism (ii) is parametrically enhanced. 
Notice, that this leading contribution explicitly breaks
$k_T$-factorization as it is proportional to
$xG(x_1)[xG(x_2)]^2$. Results reported in
\cite{Kharzeev:2005zr,Kharzeev:2008cv} show strong coherence effects
consistent with expectations of CGC theory.


\noindent {\bf Electromagnetic probes.}
The main  advantage of electromagnetic probes, such as photons and dileptons, is that they are directly observable without an intermediate hadronization process, in contrast to quarks and gluons. Therefore, they are a cleaner probe of low-$x$ nuclear matter. Their disadvantage is a low production rate due to the smallness of electromagnetic coupling.  
Prompt photon production in $pA$ collisions was considered in \cite{JalilianMarian:2005zw} through the process $qA\to \gamma qX$. The production of di-leptons in a similar process   $qA\to l^+l^- qX$ was addressed in  \cite{Gelis:2002fw,Baier:2004tj,JalilianMarian:2004er}. At higher energies, gluons become much more abundant than quarks in the central rapidity region which implies that photon  (dilepton) production will go via the process $g^*A\to q\bar qX\gamma(l^+l^-)$. It is suppressed by $\alpha_s$ but enhanced by a positive power of energy. There have been no detailed phenomenological studies of electromagnetic probes in $pA$ collisions at RHIC. 


\noindent {\bf Double inclusive hadron production and correlations.}
Azimuthal correlations are an important tool to investigate properties of QCD at low $x$.
In \cite{Kharzeev:2004bw} it was proposed that azimuthal correlations  of hadrons produced at large rapidity separation ($\Delta y\gg 1$) may be depleted due to a quasi-classical nature of the saturated gluon fields. Unfortunately, accurate theoretical calculations in the region of large but finite $\Delta y$ are challenging as they must involve complicated NLO BFKL effects. Important progress has been made in the investigation of azimuthal correlations at smaller $\Delta y$. 

 It has been suggested that correlations at small $\Delta y$ in the forward direction can  be effectively used to study gluon saturation~\cite{Marquet:2007vb}, where the forward direction corresponds to low-$x$ of the nucleus where saturation effects are strongest. Theory predicts that back-to-back correlations are suppressed due to  gluon saturation. Phenomenological models based on the CGC were suggested in  \cite{Marquet:2007vb,Albacete:2010pg} and \cite{Tuchin:2009nf} and rely on different  approximations.  An approach of \cite{Marquet:2007vb,Albacete:2010pg}  is based  on the dipole model \cite{Mueller:1989st} in which  double inclusive gluon \cite{JalilianMarian:2004da}, quark--anti-quark \cite{Tuchin:2004rb,Blaizot:2004wv,Kovchegov:2006qn} and valence quark--gluon  \cite{Marquet:2007vb} cross sections were calculated. Another approach  \cite{Tuchin:2009nf} is based on an approximate $k_T$-factorization and relies on calculating double-inclusive production based on  NLO   BFKL \cite{Fadin:1996zv,Leonidov:1999nc}. 
  
Both models give a reasonable quantitative description of experimental data. However, in order to use azimuthal correlations to study low-$x$ physics in the most effective way, work remains to be done to reconcile the existing approaches and reduce model-dependencies in calculations. Measurements of forward azimuthal correlations in $eA$ will have a clear advantage over  that  in $pA$ due to much better theoretical control of the projectile current.


\noindent {\bf Diffraction.}
One of the most sensitive probes of low-$x$ QCD is diffraction. This is because scattering in the high energy limit of QCD is mediated by the same collective gluon state (Pomeron) as the diffractive scattering.
Saturation effects on diffractive processes in $pA$ collisions were investigated in \cite{Kovchegov:2001ni,Kovner:2001vi,Li:2008bm,Li:2008se,Tuchin:2008np} where the main focus was on diffractive hadron production. (In \cite{GolecBiernat:2005fe,Tuchin:2011dq} this work was extended to DIS).

In diffraction on nuclear targets, it is important to distinguish two processes: coherent and incoherent diffraction, depending on the final state of the target.  Coherent diffractive hadron production in $pA$ collisions is a process $p+A\to X+h+[LRG]+A$, where $[LRG]$ stands for Large Rapidity Gap. Coherent diffractive production exhibits a much stronger dependence on energy and atomic number  than the corresponding inclusive process. Indeed, the diffractive amplitude is proportional to the square of the inelastic one. At asymptotically high energies, coherent diffractive events are expected to constitute up to a half of the total cross section, the other half being all inelastic processes. Therefore, coherent diffraction  is a powerful tool for studying the low-$x$ dynamics of QCD. 

In all phenomenological applications of the CGC formalism, one usually relies on the mean-field approximations in which only the lowest order  Green's functions are relevant. Although corrections to the mean-field approximation, i.e.\ quantum fluctuations about the classical solution,  are assumed to be small in $pA$ collisions at RHIC, their detailed phenomenological study is absent. An observable that is directly sensitive to quantum fluctuations is incoherent diffraction: $p+A\to X+h+[LRG]+A^*$,
 where $A^*$ denotes  excited nucleus that subsequently decays into a system of colorless protons, neutrons and nuclei debris.  Incoherent diffraction  measures fluctuations of the nuclear color field.  Calculations show that unlike the nuclear modification factor for coherent diffractive gluon production, the nuclear modification factor for incoherent diffraction is not expected to exhibit a significant rapidity and energy dependence~\cite{Tuchin:2008np}.  Therefore, the two diffractive processes can in principle be experimentally distinguished and yield unique information about low-$x$ QCD. 
Unfortunately, the study of diffraction in $pA$  collisions at RHIC is a virgin subject in part due to technical difficulties associated with measurements at very small forward angles.


\noindent {\bf Instead of a summary.}
Studying particle  production in DIS at low $x$ has two main advantages: (i) one has much better theoretical understanding of the forward kinematic region owing to the weakness of the QED coupling and (ii) new kinematic regions open up for investigation depending on values of momentum scales $Q^2$, $k_\bot^2$ and $Q_s^2$, where $Q^2$ is photon virtuality, $Q_s^2$ is saturation scale and $k_\bot$ is transverse momentum of produced hadron.

\subsection{Small-$x$ dynamics  in ultraperipheral heavy ion collisions at the LHC}
\label{sec:upc}
\hspace{\parindent}\parbox{0.92\textwidth}{\slshape
 Mark Strikman}
\index{Strikman, Mark}

\vspace{\baselineskip}

\begin{wrapfigure}{r}{0.50\textwidth}
  \begin{center}
    \includegraphics[width=0.49\textwidth]{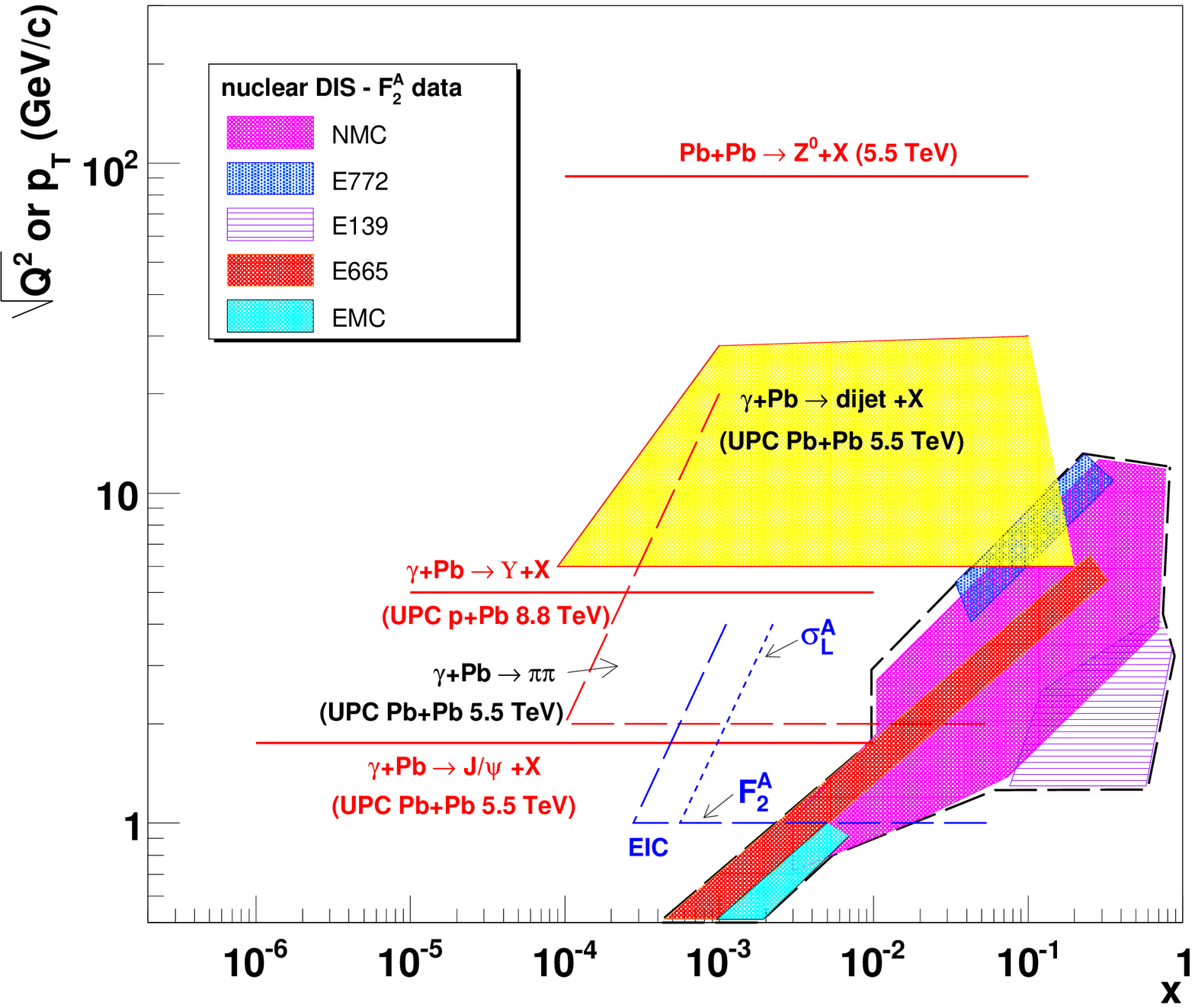}
  \end{center}
  \caption{\small The kinematic range in which UPCs at the LHC can 
      probe gluons in protons and nuclei in quarkonium,
      di-jet and di-hadron  production. For comparison, the kinematic 
      ranges for $J/\psi$ at RHIC, $F_2^A$ and $\sigma_L^A$ at eRHIC and $Z^0$
      hadroproduction at the LHC are also shown.}
\label{fig:upc}
\end{wrapfigure} 
Experiments at HERA have demonstrated that reactions with quasi real photons provide an effective tool of probing pQCD which complements studies of DIS processes. In the near future it will be possible to extend these studies to ultra-high energy photon - nucleus collisions via the study of ultra-periperal collisions (UPCs)  of heavy ions (protons and ions) at the LHC. The feasibility and the possible reach of these investigations was explored in  a five year long  study undertaken by the collaboration of theorists and experimentalists. The results of the study were published   as a volume of  Physics Reports \cite{Baltz:2007kq}. 
 Due to the high energy of the colliding nuclei and very good acceptance of the CMS and ATLAS detectors at large  rapidities, UPCs at the LHC allows to study a wide range of the processes sensitive to the small-x dynamics  
 for $W_{\gamma N} \le   \mbox{1}-\mbox{2 TeV}$. This would  extend the $x$ range probed at HERA down by at least by a factor of ten. A further advantage for the search for non-linear effects will be the use of the  nuclear targets.

 The kinematic range for which studies  of several processes of interest will be feasible is  presented  in figure~\ref{fig:upc} (taken from \cite{Baltz:2007kq}) as a function of $x$ and  $Q$ which is the typical gluon virtuality which, as the transverse momentum of the jet or leading pion, sets the scale for dijet 
and $\pi \pi$ production respectively.  The typical gluon virtuality scale for 
exclusive quarkonium photoproduction is shown for $J/\psi$ and $\Upsilon$.  
   Below we list some of the directions of the planned studies. 
  

\noindent {\bf Dijet production.}
Dijet production in the discussed kinematic range is dominated by photon - gluon fusion. Estimates of the counting rates including cuts due to the acceptance of the CMS detector were  performed in \cite{Strikman:2005yv}. It was found that
  measurements of the nuclear gluon  pdfs will be feasible down to $x\sim 10^{-4}$ via study of several channels: dijet, charm, beauty jets,  providing a number of cross checks. Use of the zero degree calorimeters (ZDCs) will also allow the separation of diffractive events and hence measure the nuclear gluon  diffractive  pdfs in the same kinematics. Hence, it will be possible to test a prediction of the leading twist theory of nuclear shadowing that the probability of the gluon induced  coherent diffraction at large $p_T$ and small $x$ should  be of the order $10-15$\%~\cite{Frankfurt:2003gx}.

The cutoff $p_t(jet) \ge 6-8 $ GeV/c  (necessary for selecting dijet production) reduces non-linear effects in dijet production.   The parameter which governs non-linear effects is $R_{NL}= C_F^2 \alpha_s(Q)xG_T(x,Q^2) / \pi r_T^2 Q^2$, where  $C_F^2 $ is the Casimir operator, equal to $4/3$ for $q\bar q$ and $3$  for 
$gg$, and $r_T$ is the transverse area of the target. For the smallest $x, p_T$ corner, $R_{NL} $ for the UPC processes $R_{NL} $ is about the same as for  $F_{2A}(x,Q^2 \sim 2-4 GeV^2)$ for the lowest $x$ which could be reached at the EIC.  
 
 It will be also possible to reach larger  $R_{NL} $ at smaller virtualities and $x\sim 10^{-4}$ using leading pion production in the central detectors $|y|\le 2.4$ - see dashed area in figure~\ref{fig:upc}.  This is a kinematics similar to the production of two forward pions in d+Au collisions at RHIC. Within the mechanism of fractional energy losses \cite{Frankfurt:2001nt,Frankfurt:2007rn},
 one expects a strong suppression of the two pion yield as compared to the single pion yield which would allow one to perform   clean tests of the onset of the black disk regime (BDR).
 
 Another sensitive probe  of the onset of BDR  would be exclusive
 diffractive production of two jets in the process $\gamma$ +A$ \to $
 2 jets  +A. In the case of light quark jets, this process is strongly
 suppressed in the pQCD regime, while it is a dominant contribution to
 the diffraction mechanism in the  BDR \cite{Frankfurt:2001nt}. \\
 

\noindent {\bf The interaction of  small dipoles  with nuclear media.}
In the leading twist approximation, the suppression of onium coherent production is given by the square of the ratio of the gluon densities in the nucleus and the proton gluon pdfs. It will 
 be feasible to investigate the suppression of coherent $J/\psi, \Upsilon$  production in nucleus-nucleus collisions down to $x \sim m_{onium}/2(E_A/A)$ corresponding to production at  the central rapidities.  At rapidities away from zero, photons of smaller energies dominate in the production of $J/\psi$, making it very difficult to probe  smaller x for virtualities $ \sim$ 3 GeV$^2$ characteristic for $J/\psi$ coherent photoproduction.  However, the use of incoherent diffractive onium production appears to solve this problem as one can use production of soft neutrons to  determine which of the nuclei emitted a photon and which was involved in the strong interaction \cite{Baltz:2007kq}. As a result, there is a potential for probing $J/\psi$ production down to $x \sim 10^{-6}$, see figure~\ref{fig:upc}.

A complementary method of tracking a small dipole through the nuclear media 
will be provided by the $J/\psi $ production in the $-t \ge $ few GeV$^2$  process $\gamma + A\to J/\psi + \mbox{rapidity gap + Y}$ \cite{Frankfurt:2008et}. It is possible in this case to  select the kinematics where $x_g$ of the gluon involved in the hard process is $ x_g\ge 0.01$. In this case, scattering 
at central impact parameters dominates and one  can  probe the propagation of a small dipole  through $\sim $ 10 fm of the nuclear media up to  $W_{\gamma N} \sim$ 1 TeV. 

In conclusion, it appears that UPC studies to be performed at the LHC in the next few years  will allow for the search of  several  signals of  the onset of the BDR. However, it will not be possible to perform a precision scan of the range of moderate $Q^2$ sensitive to the transition between non-linear and linear regimes in the $x$ range to be covered by the EIC. Hence the UPC - LHC and EIC  programs will nicely complement each other.

 








\chapter{Electroweak physics}

\noindent
{\Large Convenors and chapter editors: \\[1em]
K. Kumar, Y. Li, W. Marciano}

\newpage

\section{Electroweak physics at the EIC}
\label{sec:EWatEIC}


\hspace{\parindent}\parbox{0.92\textwidth}{\slshape
 Krishna Kumar, Yingchuan Li, William J. Marciano}

\index{Kumar, Krishna} \index{Li, Yingchuan} \index{Marciano, William J.}



\subsection{Introduction}

The $SU(3)_C \times SU(2)_L \times U(1)_Y$ standard model of particle physics has been extremely successful in describing strong and electroweak interactions. Its unbroken gauge symmetry, $SU(3)_C$ or Quantum Chromodynamics (QCD), taken on its own, represents a ``perfect theory''' with no arbitrary free parameter. Nevertheless, it beautifully encompasses all the basics of strong interactions: quark confinement, chiral symmetry breaking, asymptotic freedom, etc.  The electroweak sector is potentially much more mutable. In addition to its, as yet, undiscovered Higgs scalar remnant of $SU(2)_L \times U(1)_Y$ symmetry breaking, it contains many arbitrary free masses, couplings and mixing angles. They are accommodated but not understood at a deep level. Questions such as: why parity violation, why 3 generations of quarks and leptons? etc suggest that simplifying principles must await future new discoveries. However, precision measurements and searches for rare phenomena still have important roles to play. They have the capability of indirectly probing scales of physics beyond collider facilities and expanding the horizons of electroweak physics.

The EIC is being proposed mainly for the study of strong interactions but also has a unique ability to measure parity violating structure functions involving $W^{\pm}$ and $Z$ boson mediated interactions. The high energy and luminosity combined with polarized electrons and protons as well as a variety of heavy ion targets will provide a wealth of data in an area never explored before.

Two EIC capabilities for electroweak measurements,  outlined in table~\ref{tab:EW-sciencematrix}, are: 1) Precision measurements of the weak mixing angle over a broad range in $Q^2$ and 2) Searches for $e \rightarrow \tau$ flavor changing conversion. For the former, we show how parity violating, right-left, deep-inelastic polarized $ep$ and $ed$ asymmetries can be used to precisely determine the running ${\rm sin}^2 \theta_W(Q)$ as a function of $Q^2$. The comparison of those measurements with precision values obtained from other lower energy or Z-pole studies can be used to find hints of ``new physics''. Alternatively, the overall World average of ${\rm sin}^2 \theta_W$ can be compared with precisely determined quantities such as $\alpha_{EM}$, $G_F$, $m_Z$, and $m_W$ to test the SM at the quantum loop level and probe ``new physics'' effects. In the case of $e-\tau$ conversion, the $ep \rightarrow \tau X$ reaction is examined, including isolation cuts and $\tau$ identification. First estimates suggest that backgrounds are under control and the high luminosity goals of the EIC allow the search of reactions well beyond HERA sensitivities.

\begin{table}[hb]
 \small
\begin{center}
\begin{tabular}{|c|c|c|c|c|} \hline
Deliverables & Observables & What we learn & Phase I & Phase II\\
\hline\hline
Weak mixing &  Parity violating & physics behind EW & good precision & high precision \\
angle & asymmetries in & symmetry breaking & over limited & over wide range \\
  &  $ep$- and $ed$-DIS & $\&$ BSM physics & range of scales & of scales \\
\hline
 & & flavour violation & &  \\
e-$\tau$ conversion  & ep $\rightarrow$ $\tau$,X  & induced by BSM & challenging & very promising \\
  &  & physics &  &  \\
\hline
\end{tabular}
\end{center}
\caption{\label{tab:EW-sciencematrix}
\small Science Matrix for Electroweak physics at an EIC.
}
\end{table}



\section{The weak mixing angle via polarized electron scattering asymmetries}
\label{sec:weakmixing}


\hspace{\parindent}\parbox{0.92\textwidth}{\slshape
 Krishna Kumar, Yingchuan Li, William J. Marciano, Seamus Riordan}

\index{Kumar, Krishna} \index{Li, Yingchuan} \index{Marciano, William J.}
\index{Riordan, Seamus}



\subsection{Introduction}

The nature of spontaneous gauge symmetry breaking implies that the
masses and couplings of weak gauge bosons $W$ and $Z$ are
related by natural lowest order relations ${\rm sin}^2\theta^0_W=e^2_0/g^2_0=1-{m^0_W}^2/{m^0_Z}^2$. 
The weak mixing angle plays a central role in those
correlations. In the context of the Standard Model (SM) as a complete stand alone
theory, the renormalized weak mixing angle is related to the other precisely
measured quantities
\begin{eqnarray}
\alpha^{-1} &=& 137.03599959(40)  \\
G_{\mu} &=& 1.1663788(7)\times 10^{-5} ~ {\rm GeV}^{-2} \nonumber \\
m_Z &=& 91.1871(21) ~ {\rm GeV} \nonumber
\end{eqnarray}
via
\begin{equation}
\label{eq:msbar}
{\rm sin}^2 2 \theta_W(m_Z)_{{\overline {MS} }} = \frac{4\pi \alpha}{\sqrt{2} G_{\mu}m^2_Z[1-\Delta \hat{r}(m_t,m_H)]},
\end{equation}
where $\Delta \hat{r}$ denotes loop corrections that depend on the top quark and Higgs masses while the renormalized weak mixing angle is defined by modified minimal subtraction $\overline {MS}$ \cite{Marciano:1979yg,Marciano:1980be}. The value of ${\rm sin}^2 2 \theta_W(m_Z)_{{\overline {MS} }}$ can be determined from parity violating asymmetries and other weak interaction measurements. A comparison of the parameters in equation \ref{eq:msbar} at a high level of precision was used in the past to constrain the top quark mass (before its discovery) and more recently, to provide an upper bound on the Higgs boson mass, the missing particle of the SM. After the Higgs mass is directly measured, equation \ref{eq:msbar} will be used to probe for ``new physics" effects at the tree or loop level. 

Incorporating $m_W=80.398(25)$ GeV via
\begin{equation}
{\rm sin}^2 \theta_W(m_Z)_{{\overline {MS} }} = \frac{\pi \alpha}{\sqrt{2} G_{\mu}m^2_W[1-\Delta r(m_Z)_{\overline {MS}}-0.0085 S-{\cal O}(1)m^2_W/m^2_{W^*}]},
\end{equation}
with $\Delta r(m_Z)_{\overline {MS}}=0.0696(2)$ representing loop corrections insensitive to $m_t$ and $m_H$, 
one has another handle on ``new physics" parameters such as $S$ \cite{Peskin:1991sw,Marciano:1990dp}, a measure of possible new heavy chiral doublets such as a 4th generation, or $m_{W^*}$, the scale of possible Kaluza-Klein excitations.

The most precise determinations of ${\rm sin}^2 \theta_W$ come from two measurements at SLAC \cite{Abe:2000dq} and CERN
\cite{LEP2000}
\begin{eqnarray}
{\rm sin}^2 \theta_W(m_Z)_{{\overline {MS} }} &=& 0.23070(26) ~~~ ({\rm SLAC}) \\
{\rm sin}^2 \theta_W(m_Z)_{{\overline {MS} }} &=& 0.23193(29) ~~~
({\rm CERN}), \nonumber
\end{eqnarray}
with both extracting ${\rm sin}^2 \theta_W$ at the Z pole and
carrying an error of roughly $0.1\%$ level. Unfortunately, 
they disagree by about 3 sigma and therefore, individually provide completely
different implications for the Higgs mass and possible ``new physics". For
example, the SLAC left-right asymmetry result weighs heavily in the leptonic Z pole
average which indicates a Higgs mass of
\begin{equation}
m_H \approx 50^{+34}_{-23} ~ {\rm GeV}
\end{equation}
with the center value significantly below the LEP II direct search limit \cite{Nakamura:2010zzi}
\begin{equation}
m_H > 114 ~ {\rm GeV} ~(95\% {\rm C.L.}).
\end{equation}
On the other hand, the LEP $Z\rightarrow b {\bar b}$
forward-backward asymmetry weights heavily in the hadronic Z pole
average which implies a rather heavy Higgs \cite{Marciano:2006zu}
\begin{equation}
m_H \approx 480^{+350}_{-230} ~ {\rm GeV}.
\end{equation}
The often quoted bound $m_H < 150$ GeV results mainly from the Z-pole world average ${\rm sin}^2 \theta_W (m_Z)_{\overline {MS}}=0.23125(16)$. Is the world average correct? We may have to wait and see what the LHC tells us.

In addition to experiments at the Z pole, several precision measurements of ${\rm sin}^2 \theta_W$
have been carried out at lower $Q^2$, including atomic parity violation \cite{Wood:1997zq}, 
polarized Moller scattering \cite{Anthony:2005pm}, and deep-inelastic neutrino scattering \cite{Zeller:2001hh}, 
but with uncertainties about an order of magnitude larger, i.e. ${\cal O}(1\%)$. 
Together all such measurements play an important role in constraining ``new physics" appendages to the SM, such as heavy $Z'$ bosons of ${\cal O}$(1 TeV) and are useful for demonstrating the running of ${\rm sin}^2 \theta_W(Q)$, due to $\gamma-Z$ loop mixing, at about the 6 sigma level.

\begin{figure}
\begin{center}
\includegraphics[width=0.8\textwidth]{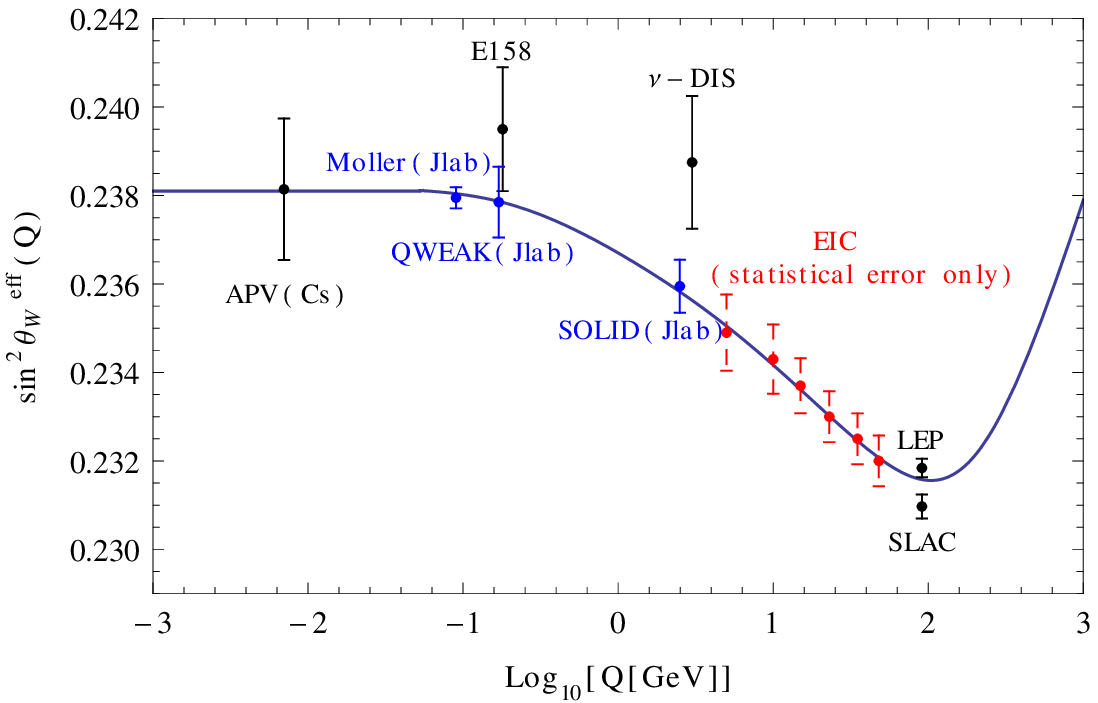}
\end{center}
\caption{\label{fig:running} The past, currently running, and future experiments
on extracting ${\rm sin}^2 \theta_W(Q)$.}
\end{figure}

It is highly desirable to have other experimental extractions of ${\rm
sin}^2 \theta_W$ with a precision roughly comparable to Z pole
measurements, given the 3 $\sigma$ discrepancy between the two best
values. Fortunately, several new measurements are in progress or planned at Jefferson lab, 
including $Q_{weak}$ using elastic $ep$ scattering \cite{Gericke:2009zz} ($\pm0.0008$), polarized 
Moller scattering \cite{JLabmoller} ($\pm0.00025$), and SOLID using polarized ed-DIS ($\pm0.0006$), which aim to extract ${\rm
sin}^2 \theta_W$ at low $Q^2$ in very high luminosity fixed target experiments. Their projected uncertainties are shown in figure \ref{fig:running}.

Here, we focus on the feasibility of measuring ${\rm
sin}^2 \theta_W$ at high $Q^2$ using an Electron-Ion Collider (EIC).
Since the center of mass energy of the EIC is expected to be much higher than fixed target experiments, parity violating asymmetries are larger and, therefore, potentially more sensitive to weak interaction effects. In addition, the EIC enables one to extract ${\rm
sin}^2 \theta_W(Q)$ and demonstrate its evolution over a wide range of $Q^2$.
Those measurements will test the predicted running of ${\rm sin}^2 \theta_W(Q^2)$, improve the world average 
${\rm sin}^2 \theta_W (m_Z)_{\overline {MS}}$, and test for ``new physics" such as $Z'$ bosons via comparison 
with Z-pole and low $Q^2$ results. We demonstrate those capabilities for an EIC
with integrated luminosity of 200 fb$^{-1}$, $\sqrt{s}\approx140$
GeV and electron (as well as perhaps hadron) polarization. A
statistical determination of ${\rm sin}^2 \theta_W(Q^2)$ to about
$\pm 0.25\%$ is found for a range of $Q^2$ with overall precision
roughly equal to the best Z-pole and proposed polarized $e^-e^-$ measurements.

\subsection{Extracting ${\rm sin}^2 \theta_W$ from parity-violating right-left polarization asymmetries}

Various parity violating asymmetries in $ep$- and $ed$(deuteron)-DIS can be obtained from ratios of
differences and sums of cross-sections with opposite polarizations
\begin{eqnarray} \label{eq:difference}
&& d \bar{\sigma}(P_e,P_{p,d}) - d \bar{\sigma}(-P_e,-P_{p,d}) \propto    \\
&& \frac{1}{2} \Sigma_i f_i(x) \left\{(P_e + \tilde{f}_i(x)
P_{p,d})(d \sigma^i_{RR} - d\sigma^i_{LL}) \right. \nonumber \\
&& \left. ~~~~~~~~~~ + (P_e - \tilde{f}_i(x) P_{p,d})(d
\sigma^i_{RL} - d\sigma^i_{LR}) \right\} \nonumber
\end{eqnarray}
and
\begin{eqnarray} \label{eq:sum}
&& d \bar{\sigma}(P_e,P_{p,d}) + d \bar{\sigma}(-P_e,-P_{p,d}) \propto   \\
&& \frac{1}{2} \Sigma_i f_i(x) \left\{(1 + \tilde{f}_i(x) P_e
P_{p,d})(d \sigma^i_{RR} + d\sigma^i_{LL}) \right. \nonumber \\
&& \left. ~~~~~~~~~~ + (1 - \tilde{f}_i(x) P_e P_{p,d})(d
\sigma^i_{RL} + d\sigma^i_{LR}) \right\},  \nonumber
\end{eqnarray}
where $P_e$ and $P_{p,d}$ are longitudinal polarizations of the electron and proton
(deuteron) beams. The $\alpha$ and $\beta$ in
$d\sigma^i_{\alpha \beta}$($\alpha,\beta=R,L$) label polarizations of
the electron and quark of type $i$, respectively. The $f_i(x)$ is the
unpolarized parton distribution function and
\begin{equation}
\tilde{f}_i(x) \equiv \Delta f_i(x)/f_i(x)
\end{equation}
is the ratio of polarized and unpolarized parton distribution
function. The quantity $\tilde{f}_i(x) P_{p,d}$ can be viewed as the
effective quark longitudinal polarization in a polarized proton (deuteron).

The polarized electron-quark cross-sections are proportional to \cite{Cahn:1977uu}
\begin{eqnarray}
d\sigma^i_{RR} &&\propto \left( \frac{Q^{\gamma}_{Re} Q^{\gamma}_{Ri}}{Q^2} + \frac{Q^{Z}_{Re} Q^{Z}_{Ri}}{Q^2+M^2_Z} \right)^2 \nonumber \\
d\sigma^i_{LL} &&\propto \left( \frac{Q^{\gamma}_{Le} Q^{\gamma}_{Li}}{Q^2} + \frac{Q^{Z}_{Le} Q^{Z}_{Li}}{Q^2+M^2_Z} \right)^2 \nonumber \\
d\sigma^i_{RL} &&\propto \left( \frac{Q^{\gamma}_{Re} Q^{\gamma}_{Li}}{Q^2} + \frac{Q^{Z}_{Re} Q^{Z}_{Li}}{Q^2+M^2_Z} \right)^2 (1-y)^2 \nonumber \\
d\sigma^i_{LR} &&\propto \left( \frac{Q^{\gamma}_{Le}
Q^{\gamma}_{Ri}}{Q^2} + \frac{Q^{Z}_{Le} Q^{Z}_{Ri}}{Q^2+M^2_Z}
\right)^2 (1-y)^2. \label{xsec_e_q}
\end{eqnarray}

Left-handed and right-handed couplings of electrons and quarks to the photon are the same
\begin{eqnarray}
Q^{\gamma}_{L}=Q^{\gamma}_{R}\equiv Q^{\gamma}
\end{eqnarray}
while those to the $Z$ are different (giving rise to parity violation)
\begin{eqnarray}
Q^{Z}_{L} &=& \frac{e}{{\rm sin} \theta_W {\rm cos}\theta_W} (T_{3L}-Q^{\gamma}{\rm sin}^2 \theta_W) \nonumber \\
Q^{Z}_{R} &=& \frac{e}{{\rm sin} \theta_W {\rm cos}\theta_W}
(-Q^{\gamma}{\rm sin}^2 \theta_W),
\end{eqnarray}
with
\begin{eqnarray}
&& Q^{\gamma}_u = \frac{2}{3}, ~ Q^{\gamma}_d = -\frac{1}{3}, Q^{\gamma}_e = -1, \nonumber \\
&& T^{u}_{3L} = \frac{1}{2}, ~ T^{d}_{3L} = T^{e}_{3L} = -\frac{1}{2}.
\end{eqnarray}

For an $ep$ collider, there are two single-polarization parity violating right-left asymmetries
\begin{eqnarray}
A^e_{ep} & \equiv & \frac{ d \bar{\sigma}(P_e,P_{p}=0) - d
\bar{\sigma}(-P_e,P_{p}=0)}{d \bar{\sigma}(P_e,P_{p}=0) + d
\bar{\sigma}(-P_e,P_{p}=0)}  \\
&=& P_e \frac{\Sigma_i f_i(x) \left[ (d \sigma^i_{RR} -
d\sigma^i_{LL}) + (d \sigma^i_{RL} - d\sigma^i_{LR})
\right]}{\Sigma_i f_i(x) \left[ (d \sigma^i_{RR} + d\sigma^i_{LL}) +
(d \sigma^i_{RL} + d\sigma^i_{LR}) \right]},  \nonumber
\end{eqnarray}
and
\begin{eqnarray}
A^{p}_{ep} & \equiv & \frac{ d \bar{\sigma}(P_e=0,P_{p}) - d
\bar{\sigma}(P_e=0,-P_{p})}{d \bar{\sigma}(P_e=0,P_{p}) + d
\bar{\sigma}(P_e=0,-P_{p})}  \\
&=& P_p \frac{\Sigma_i \Delta f_i(x) \left[ (d \sigma^i_{RR} -
d\sigma^i_{LL}) - (d \sigma^i_{RL} - d\sigma^i_{LR})
\right]}{\Sigma_i f_i(x) \left[ (d \sigma^i_{RR} + d\sigma^i_{LL}) +
(d \sigma^i_{RL} + d\sigma^i_{LR}) \right]},  \nonumber
\end{eqnarray}
with electron and proton separately polarized.

These asymmetries are simplified for an $ed$ collider since the deuteron is
an iso-singlet. Restricting to the large $x$ region ($x>0.2$), the
anti-quark contributions can be neglected. To first approximation, the parton distributions
of $u$ and $d$ quark are the same (up to charge symmetry violation
effects) in the deuteron and can thus be factored out of the sum over quark
flavors. They then cancel in the asymmetries
\begin{eqnarray}  \label{eq:ed_asy}
A^e_{ed}|_{x>0.2} & \equiv & \frac{ d \bar{\sigma}(P_e,P_{d}=0) - d
\bar{\sigma}(-P_e,P_{d}=0)}{d \bar{\sigma}(P_e,P_{d}=0) + d
\bar{\sigma}(-P_e,P_{d}=0)}  \\
&=& P_e \frac{\Sigma_i \left[ (d \sigma^i_{RR} - d\sigma^i_{LL}) +
(d \sigma^i_{RL} - d\sigma^i_{LR}) \right]}{\Sigma_i \left[ (d
\sigma^i_{RR} + d\sigma^i_{LL}) + (d \sigma^i_{RL} + d\sigma^i_{LR})
\right]} \nonumber
\end{eqnarray}
and
\begin{eqnarray}
A^{d}_{ed}|_{x>0.2} & \equiv & \frac{ d \bar{\sigma}(P_e=0,P_{d}) -
d \bar{\sigma}(P_e=0,-P_{d})}{d \bar{\sigma}(P_e=0,P_{d}) + d
\bar{\sigma}(P_e=0,-P_{d})}  \\
&=& \tilde{f}^D P_d \frac{\Sigma_i \left[ (d \sigma^i_{RR} -
d\sigma^i_{LL}) - (d \sigma^i_{RL} - d\sigma^i_{LR})
\right]}{\Sigma_i \left[ (d \sigma^i_{RR} + d\sigma^i_{LL}) + (d
\sigma^i_{RL} + d\sigma^i_{LR}) \right]}.  \nonumber
\end{eqnarray}
This leads to some simplification in the case of the single-polarization asymmetry
$A^e_{ed}(x)$ for the $ed$ collider over $A^e_{ep}(x)$ for the $ep$
collider. Both $A^e_{ep}(x)$ and $A^e_{ed}(x)$ are proportional to
electron polarization $P_e$ and thus carry smaller uncertainties than
asymmetries $A^p_{ep}(x)$ and $A^d_{ed}(x)$ which are proportional
to the hadron polarization $P_{p,d}$ which has a larger uncertainty. In fact, the
single-polarization asymmetries $A^{p}_{ep}(x)$ and $A^{d}_{ed}(x)$
with a hadron beam polarized would hardly play any role for the
purpose of measuring ${\rm sin}^2 \theta_W$ to high precision due to
the large uncertainty in $P_{p,d}$ expected to be ${\cal O}(\pm5\%)$. Instead, hadron polarization (or quark polarization) may be precisely determined from the asymmetries. 

The double-polarization asymmetries
\begin{eqnarray}
&& A^{ep,ed}_{ep,ed} \equiv \frac{ d \bar{\sigma}(P_e,P_{p,d}) - d
\bar{\sigma}(-P_e,-P_{p,d})}{ d \bar{\sigma}(P_e,P_{p,d})
+ d \bar{\sigma}(-P_e,-P_{p,d})} \\
&=& \frac{\Sigma_i f_i(x) \left\{(P_e + \tilde{f}_i(x) P_{p,d})(d
\sigma^i_{RR} - d\sigma^i_{LL})
 + (P_e - \tilde{f}_i(x) P_{p,d})(d \sigma^i_{RL} - d\sigma^i_{LR}) \right\}}{\Sigma_i f_i(x) \left\{(1 + \tilde{f}_i(x) P_e P_{p,d})(d \sigma^i_{RR} + d\sigma^i_{LL})
 + (1 - \tilde{f}_i(x) P_e P_{p,d})(d \sigma^i_{RL} + d\sigma^i_{LR}) \right\}} \nonumber
\end{eqnarray}
for both $ep$ and $ed$ collider running depend on hadron polarization; however, there are circumstances for which the
asymmetry can be simplified and carry a reduced uncertainty. First,
the $d\sigma_{RL,LR}$ are proportional to $(1-y)^2$ and thus suppressed
in the kinematic region $y \rightarrow 1$. Second, the
double-polarization asymmetry can be further simplified for a $ed$
collider at large $x$. As a result, the asymmetry
\begin{equation}
A^{ed}_{ed}|_{y\rightarrow 1, x>0.2} \approx P_{{\rm eff.}}
\frac{\Sigma_i (d \sigma^i_{RR} - d\sigma^i_{LL}) }{\Sigma_i (d
\sigma^i_{RR} + d\sigma^i_{LL}) }
\end{equation}
is proportional to the effective polarization
\begin{equation}
P_{{\rm eff.}} \equiv \frac{P_e + \tilde{f}(x) P_{d}}{1 +
\tilde{f}(x) P_e P_{d}}
\end{equation}
which carries a reduced fractional uncertainty.

\begin{figure}
\begin{center}
\includegraphics[width=0.45\textwidth]{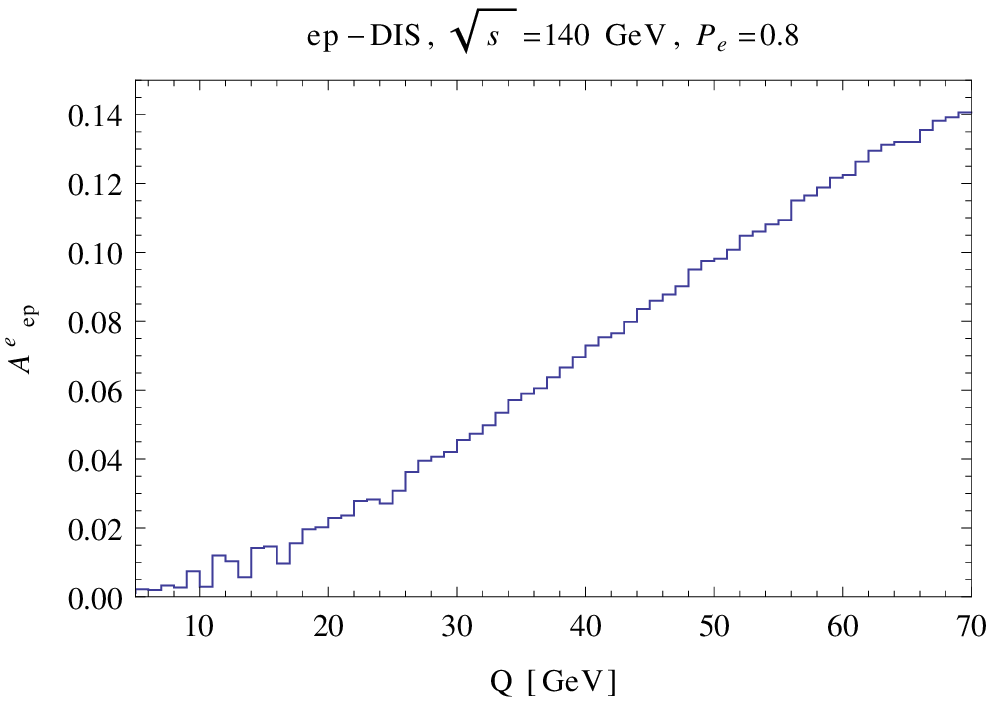}
\includegraphics[width=0.45\textwidth]{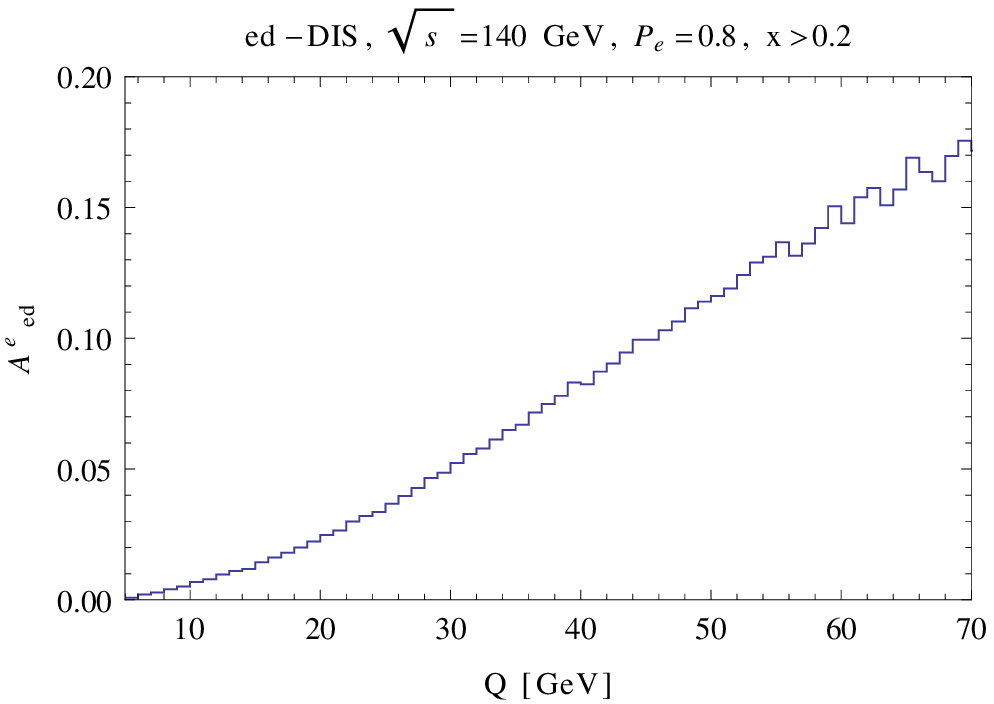}
\end{center}
\caption{\label{fig:asy} The right-left asymmetries $A^e_{ep}$ and $A^e_{ed}$
as functions of $Q$ for $ep$- and $ed$- DIS at $\sqrt{s} = 140$ GeV
with polarized electron ($P_e=0.8$).}
\end{figure}

\begin{figure}
\begin{center}
\includegraphics[width=0.45\textwidth]{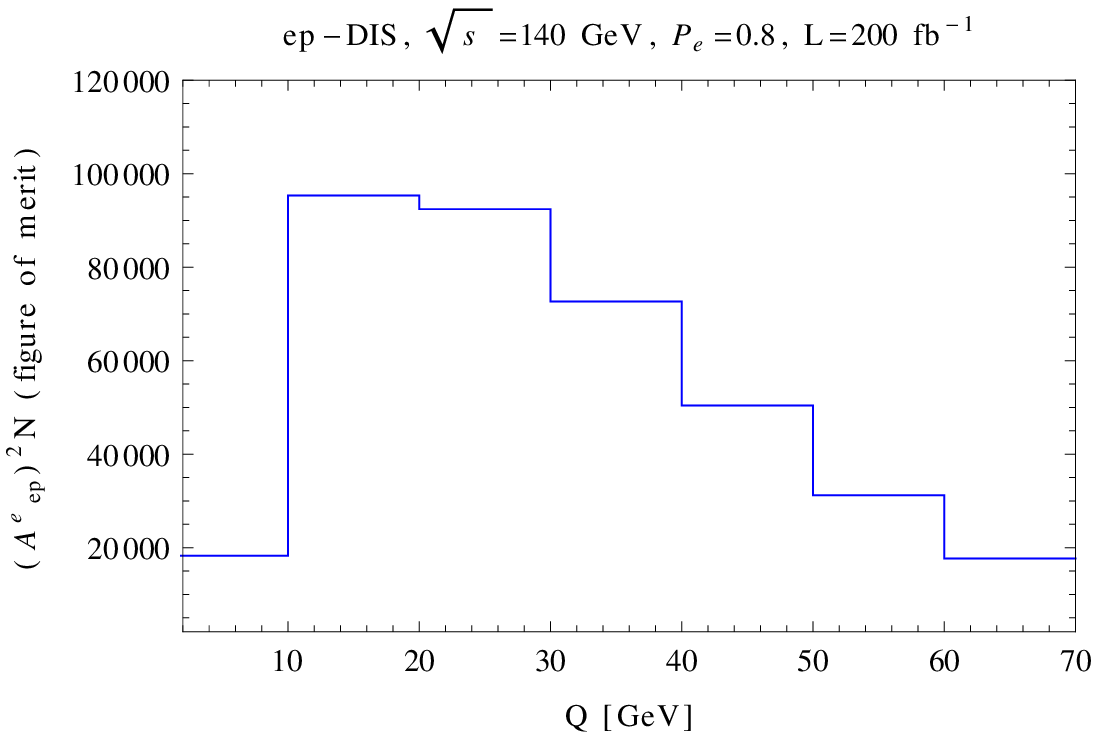}
\includegraphics[width=0.45\textwidth]{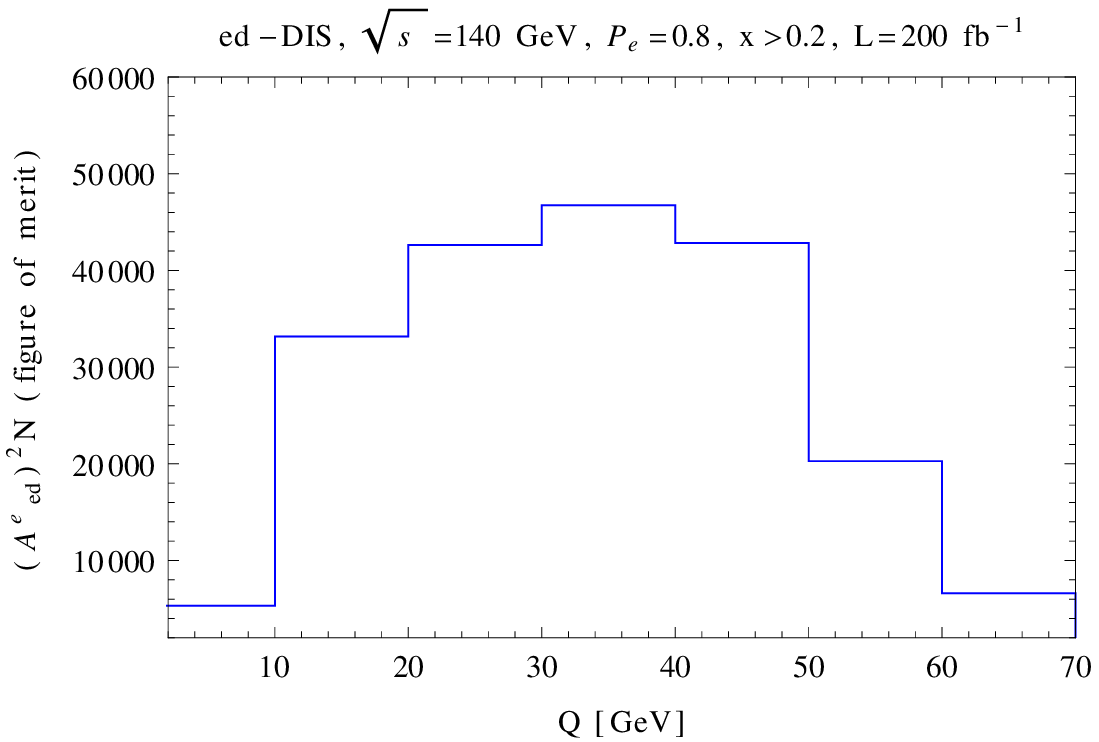}
\end{center}
\caption{\label{fig:fom} The figure of merit of measuring the
asymmetries $A^e_{ep}$ and $A^e_{ed}$ at an $ep$ and $ed$ collider with 
$\sqrt{s} = 140$ GeV and polarized electron ($P_e=0.8$), integrated luminosity of 200 fb$^{-1}$, for bin size of 10 GeV. A cut of
$x>0.2$ is imposed for $ed$ collisions.}
\end{figure}

From the above discussion, it is clear that with regard to precision ${\rm sin}^2 \theta_W$ measurements, the most useful asymmetries
are the two single-polarization asymmetries $A^{e}_{ep}$ and
$A^{e}_{ed}$ with only the electron polarized for $ep$ and $ed$ collisions,
respectively, and the double-polarization asymmetry $A^{ed}_{ed}$
for the $ed$ collision, since they carry the smallest systematic polarization
errors.

In general, the high energy EIC gains some advantage over
experiments at low energy. For example, the error from higher $1/Q^2$ twist
effects should be negligible at high $Q$. In addition, since the
uncertainty from parton distributions largely cancel in
$A^e_{ed}$, the major source of systematic error comes from the
polarization of electron beam $P_e$ which is expected to carry an uncertainty of
roughly $\pm0.5\%$. This leads to an uncertainty of $\pm$0.5$\%$ in the
single-polarization asymmetries $A^e_{ep,ed}$ and roughly $\pm$0.25$\%$ in ${\rm
sin}^2 \theta_W$. One possible way to obtain some leverage on
extracting ${\rm sin}^2 \theta_W$ with further reduced systematic
error is to make use of the $y$ dependence to extract the term
proportional to the vector coupling $g^e_V \propto 1-4 {\rm sin}^2
\theta_W$ of electrons to the $Z$ boson. It is well known that this coupling is
very sensitive to ${\rm sin}^2 \theta_W$. An accuracy of 1$\%$ of
the asymmetry proportional to this coupling determines ${\rm sin}^2
\theta_W$ at the 0.1$\%$ level. This may help with the systematic
precision but unlikely with the statistical one since the latter would
decrease in extracting various pieces from the $y$ dependence.  To assess the statistic error in measuring ${\rm sin}^2 \theta_W$,
we carry out a Monte Carlo simulation for polarized $ep$- and
$ed$- DIS at $\sqrt{s}=140$ GeV as an example. We use the
parton distribution functions of CTEQ6L \cite{Pumplin:2002vw}. 
We have included $u$ and $d$ quark and anti-quark contributions.
For $ed$-DIS, a cut of $x>0.2$ is imposed to suppress the anti-quark contribution as
needed to simplify the asymmetry in equation \ref{eq:ed_asy}. We
show the asymmetries $A^e_{ep}$ and $A^e_{ed}$ for $ep$ and $ed$
collider with polarized electron ($P_e=0.8$) in figure
\ref{fig:asy}. The asymmetries grow with $Q$ and reach 14$\%$ and 17$\%$ for
$Q \approx$70 GeV, for $ep$ and $ed$ collisions, respectively.

\begin{figure}
\begin{center}
\includegraphics[width=0.45\textwidth]{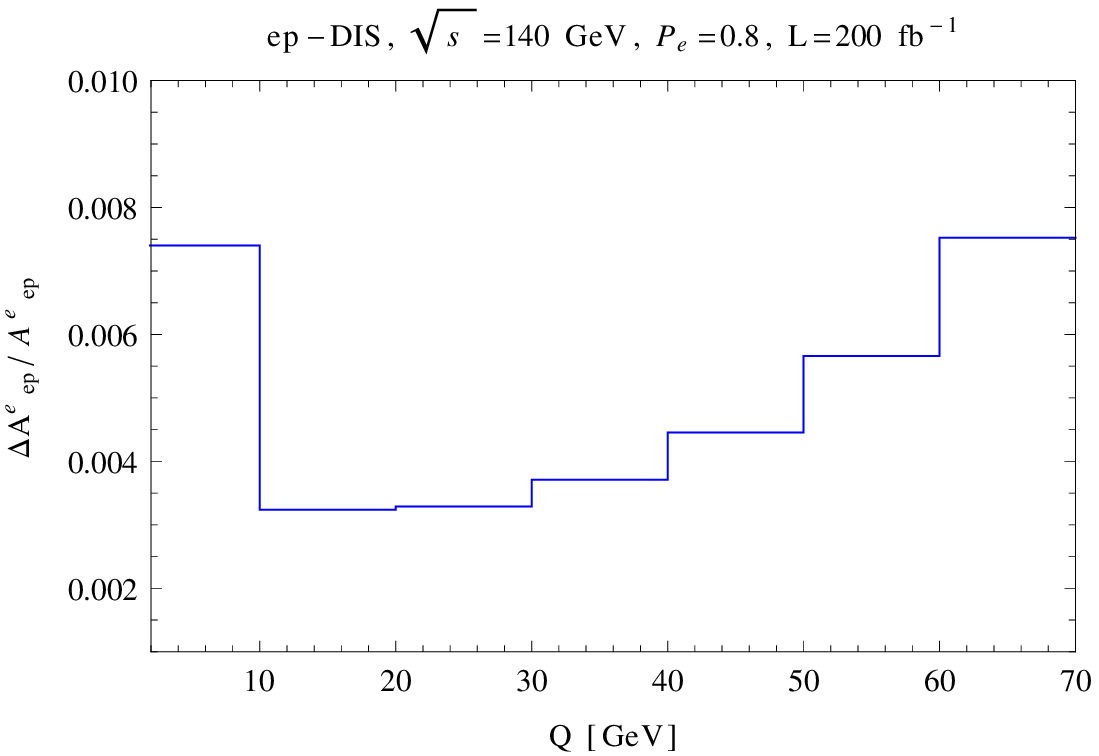}
\includegraphics[width=0.45\textwidth]{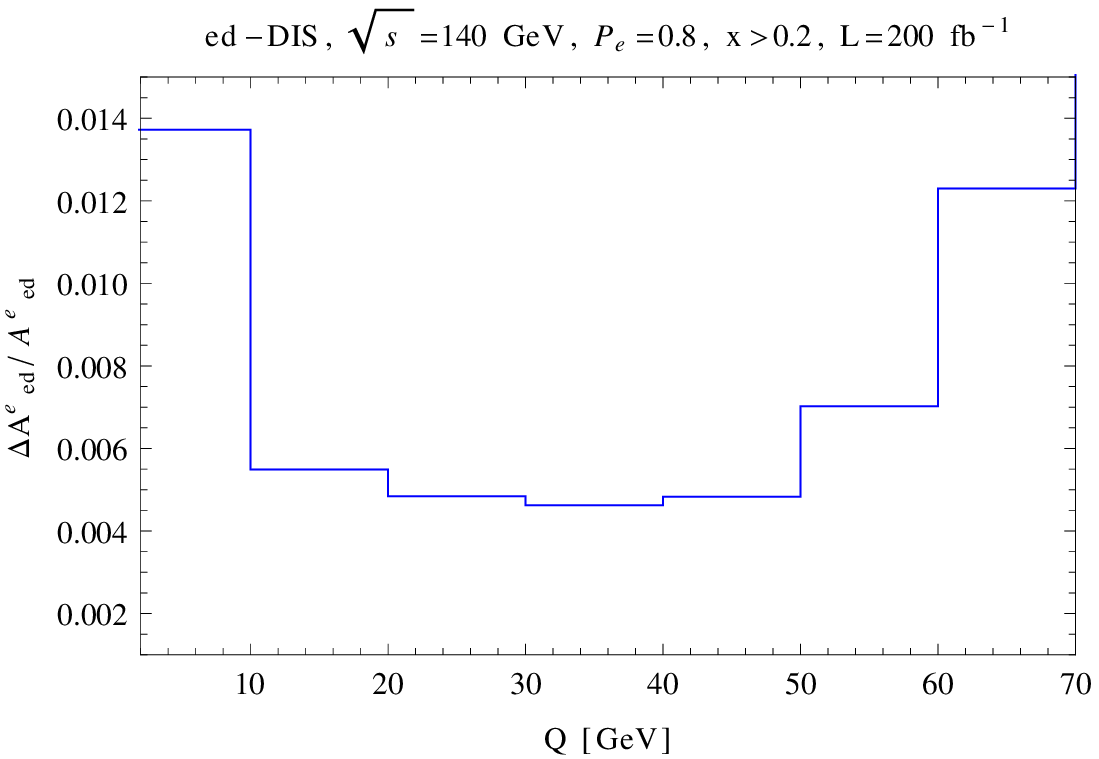}
\end{center}
\caption{\label{fig:error}The statistical error expected for the
asymmetries $A^e_{ep}$ and $A^e_{ed}$ for $ep$ and $ed$ collider at
$\sqrt{s} = 140$ GeV with polarized electron ($P_e=0.8$), and with
luminosity of 200 fb$^{-1}$, for bin size of 10 GeV. A cut of
$x>0.2$ is imposed for $ed$ collider. The statistical error for ${\rm sin}^2 \theta_W(Q)$ 
is roughly 1/2 the percentage error on $A^e_{ep}$ or $A^e_{ed}$.}
\end{figure}

Based on these polarized cross-sections, one can further obtain the statistical
figure of merit (F.O.M.) $A^2N/(1-A^2)\approx A^2 N$ for measuring the asymmetry and the
statistical errors for a given luminosity. In figure \ref{fig:fom}, we
show the figure of merit for $ep$ and $ed$ collisions with integrated luminosity
of 200 fb$^{-1}$ as function of $Q$, with bin size of 10 GeV.

The corresponding statistical errors, $\Delta A/A \approx (A^2N)^{-1/2}$, are
shown in figure \ref{fig:error} for $ep$ and $ed$ colliders. For
an $ed$ collider, the energy of the deuteron beam is shared by the proton and
neutron, thus effectively the CM energy for e-nucleon is reduced
from 140 GeV to roughly 100 GeV. For both $ep$ and $ed$ collider, with 10 GeV bin, the
statistical error is about $\pm0.5\%$ for $Q$ between 10 and 50 GeV.
For $Q>50$ and $Q<10$ GeV region, the statistical error 
is significantly higher. However, a smaller error is achievable for $Q>50$ GeV
region if a larger bin is used. Overall, the error in
extracting ${\rm sin}^2 \theta_W$ is roughly half of the error in
the asymmetry. Therefore, the statistical error in extracting ${\rm
sin}^2 \theta_W$ for most of the $Q$ region between a few
GeV and Z-pole is below $\pm0.25\%$ level.

\subsection{Conclusions}

The advantage of measuring ${\rm sin}^2 \theta_W$ at a polarized EIC
lies in its high $Q^2$, which enhances the parity violating asymmetry, reduces some of
the uncertainty from higher twist effects, and most
importantly enables one to extract ${\rm sin}^2 \theta_W$ over a wide
range of $Q$ from a few GeV to $Q\approx m_Z$. We demonstrated the
capability of measuring ${\rm sin}^2 \theta_W$ for an EIC with
integrated luminosity of 200 fb$^{-1}$, $\sqrt{s}\approx140$ GeV and
electron (as well as perhaps hadron) polarization. A statistical
determination of ${\rm sin}^2 \theta_W(Q^2)$ to about $\pm 0.25\%$
is found for most of the region of $Q$ with overall precision
roughly equal to the best Z-pole measurements. In figure
\ref{fig:running}, we have plotted values of ${\rm sin}^2 \theta_W(Q)$ obtained from past, ongoing and planned
as well a possible EIC measurements. 
The running of ${\rm sin}^2 \theta_W(Q)$ is based on ref. \cite{Czarnecki:1995fw,Ferroglia:2003wa}.
The error bar for EIC
measurements only represents the statistical error based on figure
\ref{fig:error}. A combination of all the measurements of ${\rm
sin}^2 \theta_W$ at various scales will play very important roles in
revealing the physics behind EWSB and other ``new physics''.

\subsection*{Acknowledgments} We thank T. Han for providing the Fortran
code HANLIB that is used in the Monte Carlo simulations.


\section{Electron-to-Tau conversion}
\label{sec:etau}


\hspace{\parindent}\parbox{0.92\textwidth}{\slshape
 Abhay Deshpande, Cyrus Faroughy, Matthew Gonderinger, Krishna Kumar,
 Swadhin Taneja} 

\index{Deshpande, Abhay}
\index{Faroughy, Cyrus}
\index{Gonderinger, Matthew}
\index{Kumar, Krishna}
\index{Taneja, Swadhin}




\subsection{Introduction and Motivation}\label{etau_sec:intro}

Every conservation law in the Standard Model (SM) is anticipated to have a symmetry associated with it. We 
have no knowledge of a symmetry that asserts Lepton Flavor Conservation in the Standard 
Model (SM) of particle physics and  yet its (direct) violation has never been seen. Although discovery of neutrino oscillations~\cite{Ahmad:2002jz,Fukuda:1998mi} 
indicates 
that charged Lepton Flavor Violation (LFV) processes such as $\mu \rightarrow e\gamma$ should be allowed (within the SM), 
its rate is expected to be very small  (BR$(\mu \rightarrow e\gamma) < 10^{-54}$) due to the very small values of the neutrino masses.
This level of sensitivity is beyond the reach of any present or planned experiment.
However, many models of physics Beyond the SM (BSM) predict rates of charged lepton flavor violation significantly higher than 
those within the SM, some of them even within the reach of present or planned experiments. 
LFV hence becomes a very attractive process for experimental discovery of physics beyond the Standard Model.

Many searches for specific reactions which violate lepton flavor have been performed. The most sensitive include 
searches for $\mu + N \rightarrow e + N$ using low energy muons (from the SINDRUM II collaboration~\cite{Bertl:2006up}), the muon decay 
$\mu \rightarrow e \gamma$ (MEGA collaboration \cite{Brooks:1999pu,Ahmed:2001eh}), and decays of kaons (\cite{Paradisi:2006zza}). The limits from these 
processes, though extremely precise, are all sensitive to $e \leftrightarrow \mu$ transitions (abbreviated LFV(1,2)) and not to 
$e \leftrightarrow \tau$  transitions (LFV(1,3)). Also, each of these processes involve specific quark flavors: in some, only 
the 1st generation quarks participate; in others the same quark flavor must couple to the initial and final leptons, or strange 
quarks must participate. These stringent bounds are related to the opportunities for such searches afforded by specific experimental apparatuses.
None of these searches involved the $\tau$ lepton either in the initial or in the final state. 
Since a general model with lepton flavor violation may involve a $\tau$ lepton and also initial and final state quarks of different 
flavors (not necessarily including strange quarks), the above measurements would be blind to such LFV mechanisms.   
Existing best limits on $e \leftrightarrow \tau$ conversion come from the BaBar Collaboration ($\tau \rightarrow e \gamma$)~\cite{Aubert:2009tk}
and the BELLE Collaboration ($\tau \rightarrow 3e$)~\cite{Miyazaki:2007zw}. These are notably worse than the limits on $e \leftrightarrow \mu$
by several orders of magnitude. LFV searches at proposed future experiments would further improve limits on $e\leftrightarrow\mu$ transitions.

The search for LFV involving $\tau$ leptons has been performed by the high energy lepton - hadron collider 
experiments H1 and ZEUS. The LFV process could proceed via exchange of a leptoquark (LQ), a color triplet boson -- scalar or vector -- with both lepton and baryon quantum numbers which appears naturally in many extensions of the SM such as GUTs, 
supersymmetry, compositeness, and technicolor (for a concise review of LFV in several such models, see~\cite{Albright:2008ke}). The most recent limits on the search for $e p \rightarrow \mu X$ and $e p \rightarrow \tau X$ were set by the H1 collaboration using HERA collisions at $320$ GeV center-of-mass energy
and an integrated luminosity of 0.5 fb$^{-1}$. They did not find any evidence for lepton flavor violation~\cite{Aaron:2011zz,Aktas:2007ji}, and in turn they put limits on the mass and couplings of the leptoquarks in the Buchm\"{u}ller-R\"{u}ckl-Wyler (BRW) effective model~\cite{Buchmuller:1986zs}.
  
A high energy, high luminosity electron-proton/ion collider (EIC) is being considered by the US nuclear science community with a variable center-of-mass
energy of $50 \rightarrow 160$ GeV and with $100-1000$ times the
accumulated luminosity of HERA over a comparable operation time,
%
%
see sections \ref{sec:eRHIC-design} and \ref{sec:MEIC_design}.
In a recent study~\cite{Gonderinger:2010yn} it has been argued that a $90$ GeV center-of-mass e-p collider with 10 fb${}^{-1}$ of integrated luminosity
could set a limit on leptoquark coupling-over-mass ratios that would surpass the current best limits from HERA experiments. The study also
shows that the proposed EIC could compete or surpass the updated leptoquark limits from $\tau \rightarrow e\gamma$ for a subset of 
quark flavor diagonal couplings. Lastly, the authors found that although $e\rightarrow \tau$ LFV is indeed severely suppressed,  $e \rightarrow \tau$ transition could still exist within the reach of the EIC, under certain situations~\cite{Gonderinger:2010yn}.  The present study of search for leptoquarks at the EIC was motivated by these exciting possibilities. 

For completeness, we studied leptoquark couplings with first, second and third generation leptons ($e \rightarrow e, \mu, \tau$) in our simulations, although 
the main focus of this study was the $e \rightarrow \tau$ transition. We comment here on all three.
\begin{enumerate}
\item Leptoquark decays to first generation leptons lead to final states similar to those in SM deep inelastic scattering (DIS) neutral current (NC, $ep \rightarrow eX$) and charged current (CC, $ep \rightarrow \nu X$) interactions. These processes contribute as backgrounds by mimicking the final state signature of the signal events, and hence are indistinguishable. Other SM backgrounds arise from photo-production $\gamma p \rightarrow X$, lepton-pair production 
($e p \rightarrow e l^{+}l^{-}X$), and W production ($e p \rightarrow eWX$). We simulate them and study the angular correlations of the final states and
the missing momentum spectra in cases where neutrinos are involved in the final state. Possibilities of misidentification of events due to 
detector inefficiencies will be commented upon in section~\ref{etau_sec:exp_concl}. 
\item Leptoquark decays with a $\mu$ in the final state give a back-to-back muon and hadronic system event characteristic in the
transverse plane. Since muons typically deposit a very small fraction of their energy in a calorimeter, in real experiments, such events are characterized
by a large missing calorimetric transverse momentum. Additionally, such muons are typically required to be isolated, well separated from the hadronic
jets or tracks in such an event. Such selections strongly suppress the NC component of the SM backgrounds, which mainly arise from muon-pair production
and muonic decays of W bosons. See details in \cite{Aaron:2011zz,Aktas:2007ji}.
\item The 1$\rightarrow$3 transition, $ep \rightarrow \tau X$, is the principle focus of this study; it is studied using three $\tau$ decay channels: electronic, muonic and hadronic. Electronic decays $\tau \rightarrow e \nu_{e} \nu_{\tau}$ have a topology similar to high $Q^{2}$ NC events, except for missing transverse momentum due to the escaping neutrinos, which can be exploited to reduce this background. Muonic decays $\tau \rightarrow \mu \nu_{\mu} \nu_{\tau}$ result in similar final states as the electronic decay of $\tau$ and hence a similar criteria for their selection is used. Hadronic decays of $\tau$ lead to a high transverse momentum, narrow jet resulting in a signal topology of a di-jet event with no leptons. These events can be selected using various well known algorithms to identify and separate the $\tau$-jet from other hadronic jets in NC DIS and photoproduction events.
\end{enumerate}

Many of the above mentioned strategies require detailed detector simulation of the response. This is not done in the present study. However,
we studied the event topologies of the SM processes and the leptoquark events through simulations with beam energies and detector acceptance guidelines suggested on the INT website~\cite{EICparams}. The differences in event topologies generated by $p_T^{miss}$ (the missing transverse momentum) and the angle $\phi$ (between the $\tau$-jet and the missing transverse momentum vector) present in SM and LQ events with final state neutrinos were studied. We ask in this study: are they different enough to be distinguishable from one another at the EIC energies, and for what range of leptoquark couplings and masses could the LFV LQ events be differentiated from a SM event at the EIC.

This report proceeds as follows.  In section~\ref{etau_sec:LQ_frame}, the leptoquark framework is introduced and the findings of~\cite{Gonderinger:2010yn} are summarized and updated to reflect recent developments regarding higher EIC integrated luminosities and the proposed reach of Super-B experimental searches for $\tau\rightarrow e\gamma$.  Section~\ref{etau_sec:eff_ops} discusses the possibility of $e\rightarrow\tau$ searches at the EIC in the broader context of an effective operator framework.  Concluding remarks for the theoretical analysis are presented in section~\ref{etau_sec:th_concl}.  An experimental analysis begins with section~\ref{etau_sec:SM_bg} in which we present the SM process generation and its study for the above correlations. In section~\ref{etau_sec:LQ_sim} we detail the MC generator study for
the leptoquark and study some of its parameters (leptoquark mass dependence and the coupling strength dependence) on the observable missing
$p_{T}$ and $\phi$ spectra. In section~\ref{etau_sec:exp_concl} we compare some of the selected spectra from SM and the leptoquark and show potentially how leptoquarks may be identified at a future EIC. We then conclude with a comment on the limitations of this study and a brief plan for the near future.

\subsection{Theory I: Leptoquark Framework}\label{etau_sec:LQ_frame}

We begin our study of $e\rightarrow\tau$ conversion at the EIC by assuming a leptoquark framework.  Leptoquarks (abbreviated LQs) are particles coupling to leptons and quarks which arise in models such as Pati-Salam color-$SU(4)$ and $SU(5)$ GUTs.  Leptoquarks provide a useful framework for an initial analysis of $e\rightarrow\tau$ conversion because they allow for the conversion process to occur at tree level, as described further below, and so larger cross sections may be expected relative to other models which induce LFV through loop effects.  Additionally, searches for leptoquark-induced $e\rightarrow\tau$ were performed at HERA, and so direct comparisons can be made between limits from HERA and potential limits from the EIC.

The class of particles which may be described as ``leptoquarks'' have a variety of properties: spin 0 or 1; fermion number $F=3B+L=$ 0 or $\pm 2$; $SU(2)_L$ singlet, doublet, or triplet representations; and chiral couplings to $L$- or $R$-handed leptons.  We use the Buchm\"{u}ller-R\"{u}ckl-Wyler (BRW) parameterization of LQs~\cite{Buchmuller:1986zs}.  In this parameterization, there are 14 different LQs encompassing all allowed combinations of the listed properties; their interactions with quarks and leptons are given by the renormalizable SM gauge-invariant Lagrangian in equation \eqref{eq:lq_lagrang}.

\begin{equation}\label{eq:lq_lagrang}
\begin{aligned}
&\mathcal{L}_{LQ} & &= & &\mathcal{L}_{F=0} + \mathcal{L}_{|F|=2}\\
&\mathcal{L}_{F=0} & &= & &h_{1/2}^L\bar{u}_R\ell_L S_{1/2}^L + h_{1/2}^R\bar{q}_L\epsilon e_R S_{1/2}^R + \tilde{h}_{1/2}^L\bar{d}_R\ell_L \tilde{S}_{1/2}^L\\ 
& & & & &\quad + h_0^L\bar{q}_L\gamma_\mu\ell_L {V_0^L}^\mu + h_0^R\bar{d}_R\gamma_\mu e_R V_0^{R\mu} + \tilde{h}_0^R\bar{u}_R\gamma_\mu e_R\tilde{V}_0^{R\mu}\\
& & & & &\quad + h_1^L\bar{q}_L\gamma_\mu\vec{\tau}\ell_L\vec{V}_1^{L\mu} + \mathrm{h.c.}\\
&\mathcal{L}_{|F|=2} & &= & &g_0^L\bar{q}_L^c\epsilon\ell_L S_0^L + g_0^R\bar{u}_R^c e_R S_0^R + \tilde{g}_0^R\bar{d}_R^c e_R \tilde{S}_0^R + g_1^L\bar{q}_L^c\epsilon\vec{\tau}\ell_L\vec{S}_1^L\\ 
& & & & &\quad + g_{1/2}^L\bar{d}_R^c\gamma_\mu \ell_L V_{1/2}^{L\mu} + g_{1/2}^R\bar{q}_L^c\gamma_\mu e_R V_{1/2}^{R\mu}\\
& & & & &\quad + \tilde{g}_{1/2}^L\bar{u}_R^c\gamma_\mu\ell_L\tilde{V}_{1/2}^{L\mu} + \mathrm{h.c.}
\end{aligned}
\end{equation}

In equation \eqref{eq:lq_lagrang}, $q_L$ and $\ell_L$ are the $SU(2)_L$ doublet quarks and leptons, $u_R,\ d_R,\ e_R$ are the $SU(2)_L$ singlet quarks and charged lepton, $\epsilon$ is the $SU(2)_L$ antisymmetric tensor ($\epsilon_{12} = -\epsilon_{21} = +1$), $\vec{\tau} = \left(\tau_1, \tau_2, \tau_3\right)$ are the Pauli matrices, and the charge conjugated fermion is defined as $\psi^c \equiv C\bar{\psi}^T = i\gamma_2\gamma_0\bar{\psi}^T$ in the Dirac basis for the $\gamma$ matrices.  Color, $SU(2)_L$, and flavor (generation) indices have been suppressed.  We follow the notation used in the recent literature where spin-0 leptoquarks are $S$ and spin-1 are $V$, the subscript indicates the $SU(2)_L$ quantum number (0 for a singlet, 1/2 for a doublet, 1 for a triplet), the superscript $L,R$ indicates the chirality of the lepton coupling to the leptoquark, and a tilde ($\tilde{\, }$) is used to distinguish between leptoquarks which have different hypercharges but are otherwise identical.  The dimensionless coupling constants $g$ and $h$ (which we assume to be real) carry the same lepton chirality and $SU(2)_L$ labels as their associated leptoquarks.  Lepton flavor violating processes mediated by LQs arise if the couplings --- which are matrices in flavor space --- have non-zero off-diagonal elements.

\begin{figure}
\begin{center}
\includegraphics[width=0.5\textwidth]{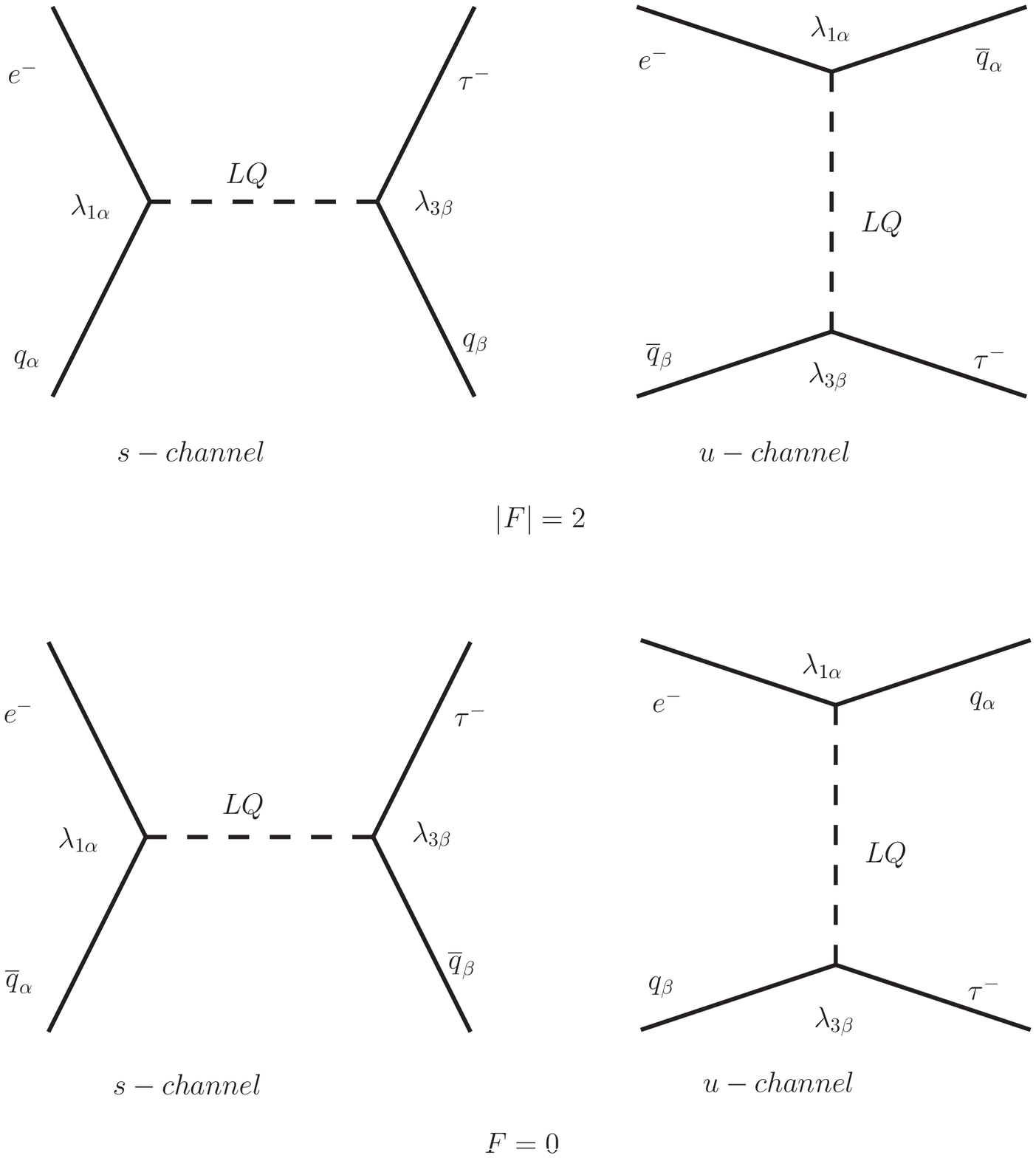}
\end{center}
\caption{\label{fig:e_tau_conv} Feynman diagrams showing the leptoquark-mediated $e\rightarrow\tau$ conversion process.  $\alpha$ and $\beta$ are the quark generation indices.}
\end{figure}

The $e\rightarrow\tau$ conversion process mediated by LQs is shown at the partonic level in the Feynman diagrams in figure \ref{fig:e_tau_conv}.  For simplicity, the couplings $g$ and $h$ in equation \eqref{eq:lq_lagrang} have been replaced by $\lambda_{ij}$ where the first index corresponds to the lepton generation and the second index the quark generation.  The cross section for the deep inelastic scattering conversion process $e^-+p\rightarrow\tau^-+X$ mediated by a single leptoquark is calculated using the Feynman rules derived from the Lagrangian of equation \eqref{eq:lq_lagrang} and convoluting the partonic subprocess with the appropriate parton distribution functions for the initial state quark or antiquark.  In the high mass approximation, where the LQ mass is much larger than the center-of-mass energy and all fermion masses are neglected, the momentum dependence of the LQ propagator can be neglected, effectively shrinking the propagator to a four fermion contact interaction.  The cross section is then given by~\cite{Chekanov:2002xz}
\begin{equation}\label{eq:e_tau_cxn}
\begin{aligned}
&\sigma_{F=0} & &= & &\sum_{\alpha,\beta}\frac{s}{32\pi}\left[\frac{\lambda_{1\alpha}\lambda_{3\beta}}{M_{LQ}^2}\right]^2 \left\{ \int dxdy\, x\bar{q}_\alpha\left(x,xs\right)f\left(y\right)\right.\\
& & & & &\qquad\qquad \left. + \int dxdy\, xq_\beta\left(x,-u\right)g\left(y\right)\right\}\, \, ,\\
&\sigma_{|F|=2} & &= & &\sum_{\alpha,\beta}\frac{s}{32\pi}\left[\frac{\lambda_{1\alpha}\lambda_{3\beta}}{M_{LQ}^2}\right]^2 \left\{ \int dxdy\, xq_\alpha\left(x,xs\right)f\left(y\right)\right.\\
& & & & &\qquad\qquad \left. + \int dxdy\, x\bar{q}_\beta\left(x,-u\right)g\left(y\right)\right\}\, \, .
\end{aligned}
\end{equation}
The functions $f$ and $g$ are defined differently for scalar and vector leptoquarks:
\begin{equation}\label{eq:f_and_g}
f\left(y\right) = \left\{\begin{array}{cc} 1/2 & \mathrm{(scalar)}\\2\left(1-y\right)^2 & \mathrm{(vector)}\end{array}\right. \, \, , \, \, 
g\left(y\right) = \left\{\begin{array}{cc} \left(1-y\right)^2/2 & \mathrm{(scalar)}\\2 & \mathrm{(vector)}\end{array}\right. \, \, .
\end{equation}
The parton distribution functions for the quarks and antiquarks are $q\left(x,Q^2\right)$ and $\bar{q}\left(x,Q^2\right)$, respectively, evaluated at momentum fraction $x$ and energy scale $Q^2$.  Also, $u = xs\left(y-1\right)$ and both $x$ and $y$ are integrated from 0 to 1.  As equation \eqref{eq:e_tau_cxn} shows, in the high mass approximation the unknown leptoquark couplings and masses appear in the cross section as the ratio $\lambda_{1\alpha}\lambda_{3\beta}/M_{LQ}^2$.

In the $e^\pm p$ collisions at HERA, no $e\rightarrow\tau$ conversion events were observed.  Limits on the LQ ratios $\lambda_{1\alpha}\lambda_{3\beta}/M_{LQ}^2$ were set by both the ZEUS~\cite{Chekanov:2005au} and H1~\cite{Aktas:2007ji} collaborations.  In our analysis, we determine how the EIC might improve on these limits set by ZEUS and H1 by answering the question, to what values of the ratios $\lambda_{1\alpha}\lambda_{3\beta}/M_{LQ}^2$ would the EIC be sensitive?  As with the ZEUS and H1 analyses, we consider all combinations of the quark generations $\alpha$ and $\beta$ (excluding the top quark) for all 14 BRW leptoquarks.  It is assumed that one of the BRW LQs dominates the cross section and the LQs in $SU(2)_L$ multiplets are degenerate in mass.  Full results of this analysis can be found in~\cite{Gonderinger:2010yn}; in this report, we summarize the results and discuss a few representative examples.  

\begin{figure}
\begin{center}
\includegraphics[width=0.6\textwidth]{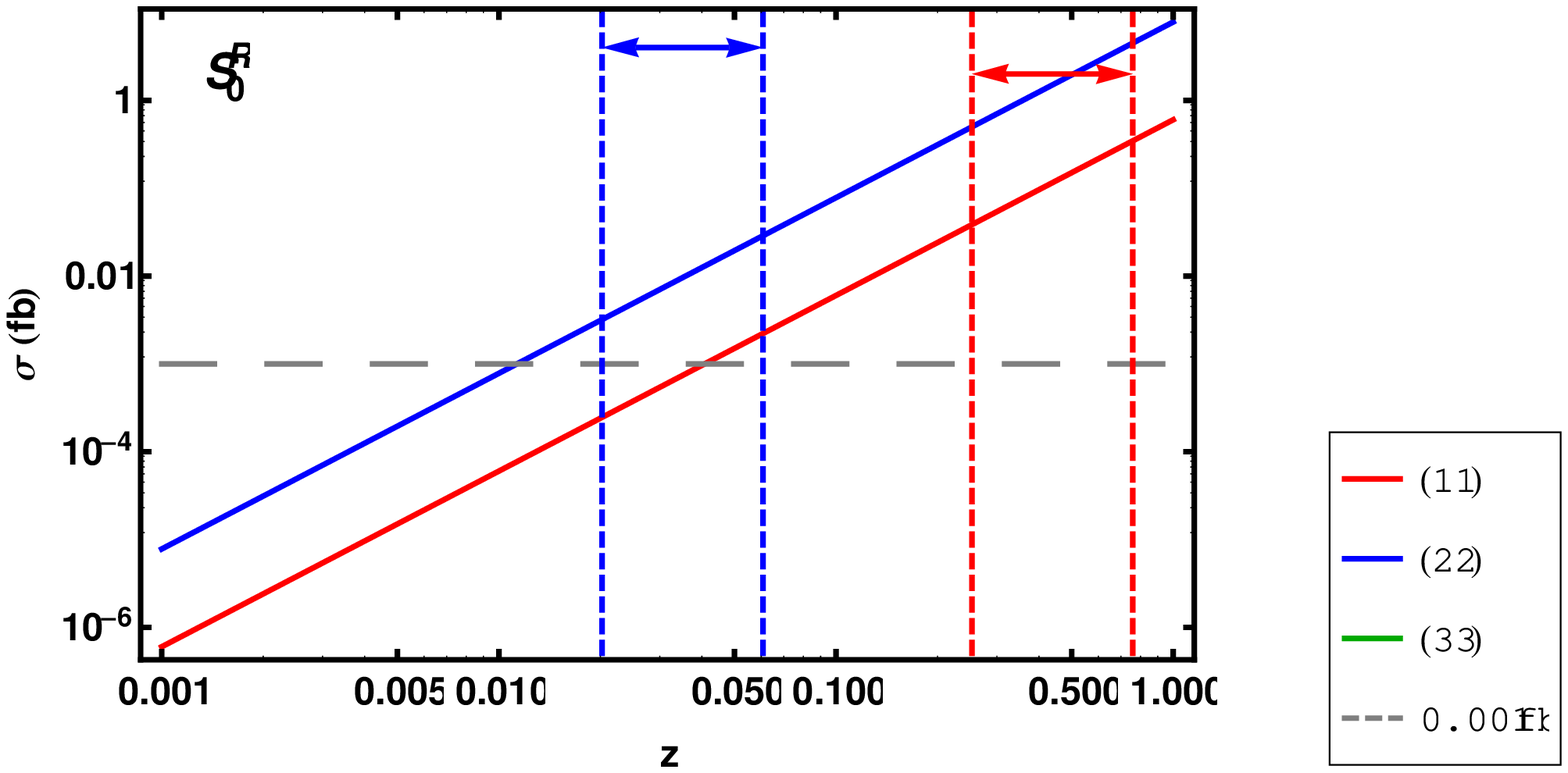}
\end{center}
\caption{\label{fig:cxn_diag} The $e\rightarrow\tau$ cross section for the leptoquark $S_0^R$ plotted as a function of $z$, defined to be the ratio $\lambda_{1\alpha}\lambda_{3\beta}/M_{LQ}^2$ scaled by the HERA limit.  A cross section of 0.001~$\fb$, corresponding to order 1 events with 1000~$\fb^{-1}$ integrated luminosity, is indicated with a gray dashed line.  The cross section is plotted for the different quark generation combinations, $\left(\alpha\beta\right)$.  Shown here are the quark flavor-diagonal contributions with $\alpha=\beta$.  The vertical dashed lines indicate the range of these ratios to which the Super-B experiments may be maximally sensitive from $\tau\rightarrow e\gamma$ searches.}
\end{figure}

\begin{figure}
\begin{center}
\includegraphics[width=0.6\textwidth]{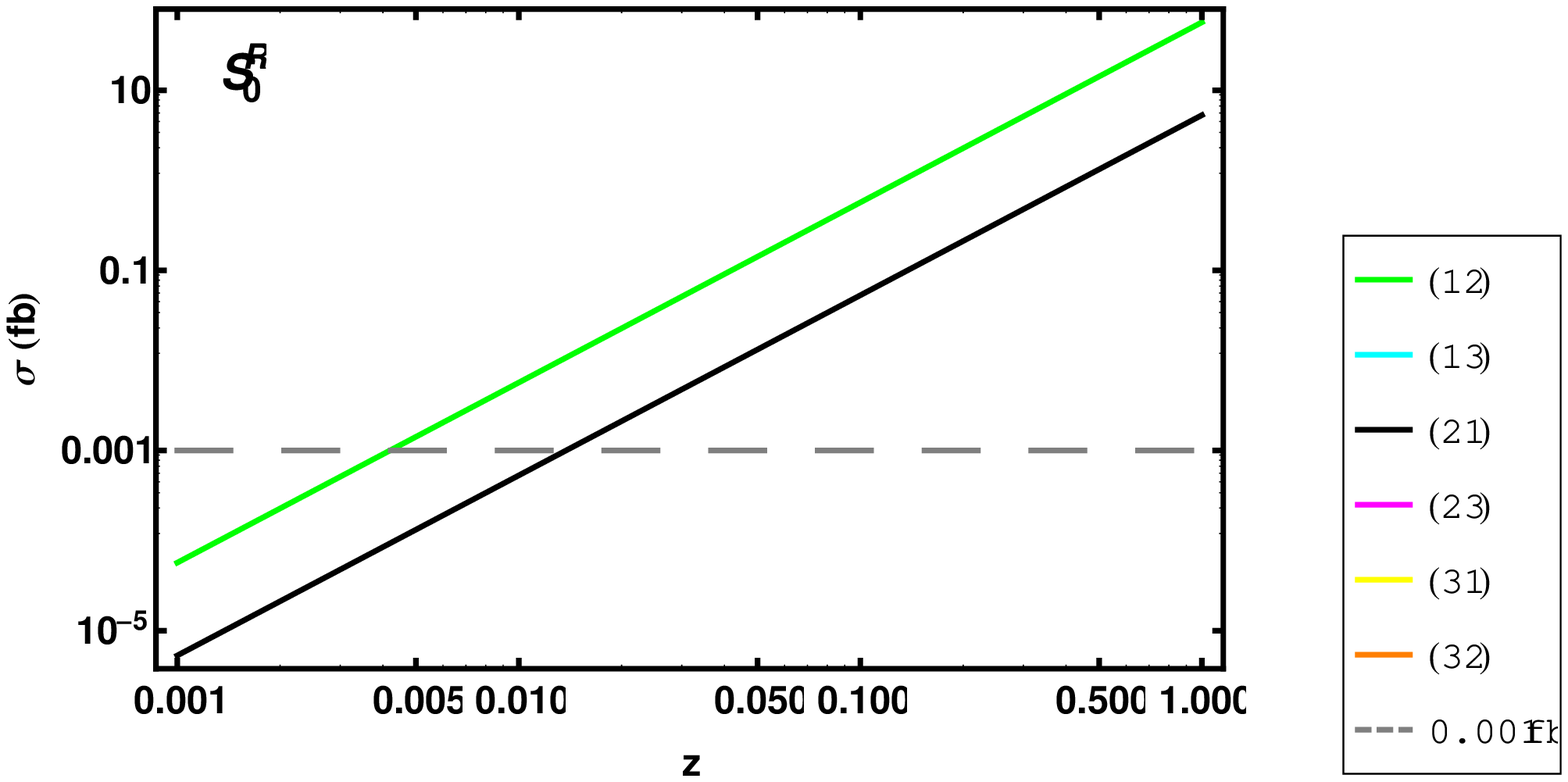}
\end{center}
\caption{\label{fig:cxn_offdiag} As for figure~\ref{fig:cxn_diag}, but shown here are the quark flavor-off-diagonal contributions with $\alpha\neq\beta$.  No $\tau\rightarrow e\gamma$ limits exist in this case.}
\end{figure}

With 1000~$\fb^{-1}$ of integrated luminosity (attainable within a reasonable length of time at a high luminosity machine such as the EIC), the EIC would in principle be sensitive to $e\rightarrow\tau$ conversion cross sections at a level of 0.001~$\fb$.\footnote{Reference~\cite{Gonderinger:2010yn} focused on a lower integrated luminosity and a larger cross section.}  This would yield on the order of one $e\rightarrow\tau$ conversion events (not accounting for backgrounds, $\tau$ reconstruction efficiency, \emph{etc}.).  Using this number for the cross section, and assuming a center-of-mass energy $\sqrt{s}=90~\gev$, the LQ ratios $\lambda_{1\alpha}\lambda_{3\beta}/M_{LQ}^2$ can be calculated from equation \eqref{eq:e_tau_cxn}.  Generally, for nearly all leptoquarks and combinations of quark generations $\alpha$ and $\beta$, the EIC could probe values of the ratios $\lambda_{1\alpha}\lambda_{3\beta}/M_{LQ}^2$ that are smaller than the HERA limits by a factor between 10 and 200.  This is demonstrated for the LQ $S_0^R$ in figures \ref{fig:cxn_diag} and \ref{fig:cxn_offdiag} where the cross sections for the different quark generation combinations $\left(\alpha\beta\right)$ are plotted as a function of the number $z$, defined to be the LQ ratio $\lambda_{1\alpha}\lambda_{3\beta}/M_{LQ}^2$ scaled by the corresponding HERA limit.  For example, the cross section for first generation initial and final state quarks (the red line in figure \ref{fig:cxn_diag}) is equal to 0.001~$\fb$ at $z\simeq 0.05$.  This means that the EIC could improve the HERA limit on the ratio $\lambda_{11}\lambda_{31}/M_{LQ}^2$ for the leptoquark $S_0^R$ by as much as a factor of 20; or, if such a leptoquark exists and has properties such that $\lambda_{11}\lambda_{31}/M_{LQ}^2$ is between 0.05 and 1 times the HERA limit, this LQ could induce a number of $e\rightarrow\tau$ conversion events sufficiently large enough to be observed at the EIC.


\begin{figure}
\begin{center}
\includegraphics[width=0.8\textwidth]{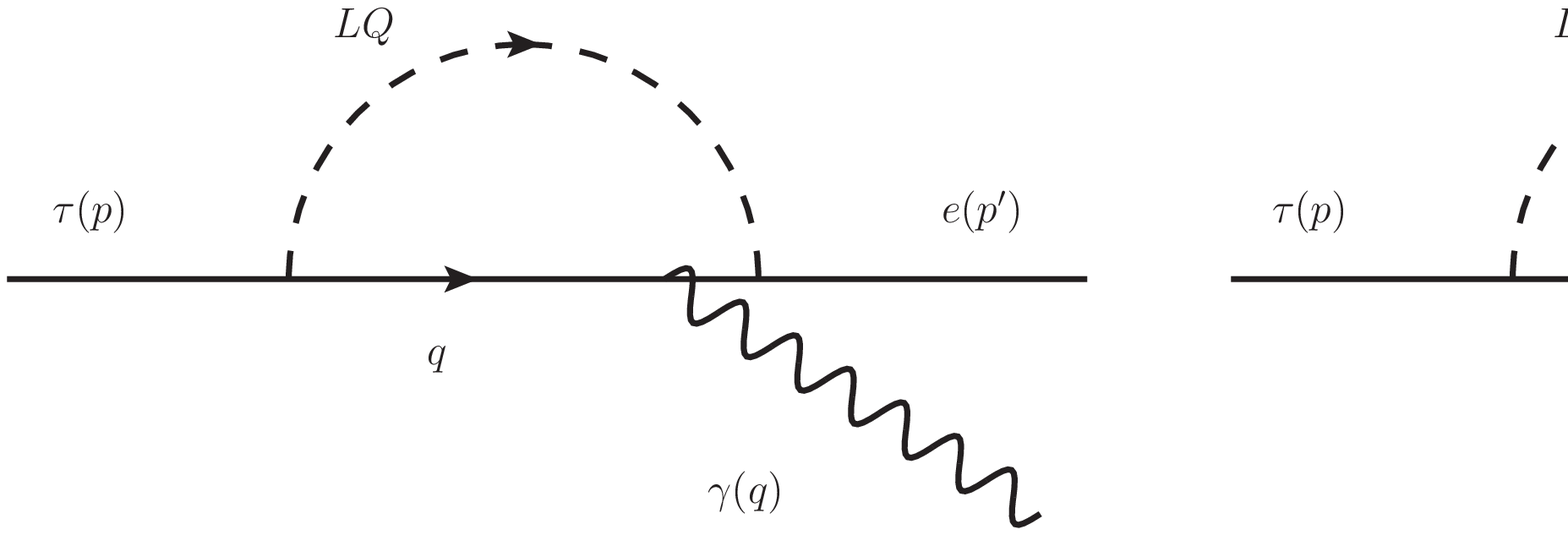}
\end{center}
\caption{\label{fig:tau_e_gamma} Feynman diagrams showing the leptoquark loops contributing to the $\tau\rightarrow e\gamma^*$ process.}
\end{figure}

Also shown in figure \ref{fig:cxn_diag} are the values of the LQ ratios $\lambda_{1\alpha}\lambda_{3\beta}/M_{LQ}^2$ (again scaled by the HERA limits) to which future Super-B experiments may be sensitive\footnote{Reference~\cite{Gonderinger:2010yn} used only the current $\tau\rightarrow e\gamma$ limit.}; these are indicated with vertical dashed lines in the figure.  The scalar leptoquarks can contribute to the $\tau\rightarrow e\gamma$ decay through loop diagrams shown in figure \ref{fig:tau_e_gamma}.\footnote{The contribution of the vector leptoquarks is less clear, for reasons explored in~\cite{Gonderinger:2010yn}, so we restrict our discussion to the scalar LQs.}  Limits on the LQ ratios are derived as follows.  The amplitude for the process $\tau\rightarrow e\gamma^*$ has the general form~\cite{RamseyMusolf:2006vr}
\begin{equation}\label{eq:tau_e_gamma_amp}
\begin{aligned}
\mathcal{M}_{\tau\rightarrow e\gamma^*} =& e {\epsilon^*}^\nu \overline{u}_e\left(p'\right)\left[\left(q^2\gamma_\nu - q_\nu\left(q\cdot\gamma\right)\right)\left(A_1^L P_L + A_1^R P_R\right)\right.\\
&\qquad\qquad\qquad\quad \left. + im_\tau q^\alpha\sigma_{\nu\alpha}\left(A_2^L P_L + A_2^R P_R\right)\right] u_\tau\left(p\right)\, \, ,
\end{aligned}
\end{equation}
where the $A_1$s and $A_2$s are model-dependent factors.  For a real photon, $q^2=0$, and so $\left|\mathcal{M}\right|^2$ depends only on the factors $A_2^{L,R}$.  Then, the $\tau$ decay rate ratio is given by
\begin{equation}\label{eq:br_tau_e_gamma}
R\left(\tau\rightarrow e\gamma\right) \equiv \frac{\Gamma\left(\tau^-\rightarrow e^-\gamma\right)}{\Gamma\left(\tau^-\rightarrow e^-\bar{\nu}_e\nu_\tau\right)}
	= \frac{48\pi^3\alpha_{EM}}{G_\mu^2} \left(\left|A_2^L\right|^2 + \left|A_2^R\right|^2\right)\, \, .
\end{equation}
Recent work by the BABAR collaboration~\cite{Aubert:2009tk} has set a 90\% C.L. limit $\Gamma\left(\tau\rightarrow e\gamma\right)/\Gamma_{total}\leq 3.3\times 10^{-8}$.  The current consensus is that future Super-B experiments will be able to improve this limit by a single order of magnitude, and so we take $R\leq 1.85\times 10^{-8}$.  The coefficients $A_2^L$ and $A_2^R$ can be determined for each scalar leptoquark by computing the amplitude for the $\tau\rightarrow e\gamma$ loop diagrams and picking out terms proportional to the magnetic moment operator $q^\alpha \sigma_{\nu\alpha}$.  When neglecting the lepton masses and expanding in powers of $m_q^2/M_{LQ}^2$, at zeroth order the $A_2$s will depend on a sum over $\alpha$ of the ratios $\lambda_{1\alpha}\lambda_{3\alpha}/M_{LQ}^2$ (here, $\alpha=\beta$ since there is only one quark present in the loop).  Thus, the experimental limit on $R$ determines a range of upper limits on the LQ ratios: the stronger upper limit is set by assuming all three quark generations contribute to the $A_2$ coefficients equally, while the weaker upper limit assumes only a single quark generation contributes to the $A_2$ coefficients.  Both upper limits on $\lambda_{1\alpha}\lambda_{3\beta}/M_{LQ}^2$ (again scaled by the HERA limit) are indicated by vertical dashed lines for each quark generation in figure \ref{fig:cxn_diag}.  

As figure \ref{fig:cxn_diag} shows for the $S_0^R$ LQ, the EIC could potentially surpass upper limits on the LQ ratios derived from an improved Super-B factory $\tau\rightarrow e\gamma$ limit.  For the other scalar leptoquarks, it is generally true that the EIC would be competitive with or surpass the future limits from Super-B factories.  The EIC also has two additional advantages over $\tau\rightarrow e\gamma$ searches.  First, $\tau\rightarrow e\gamma$ only constrains those leptoquark ratios $\lambda_{1\alpha}\lambda_{3\beta}/M_{LQ}^2$ for which $\alpha = \beta$, while the EIC can probe all combinations of quark generations.  Second, it is possible for the LQ-induced $\tau\rightarrow e\gamma$ to be suppressed relative to $e\rightarrow\tau$ conversion: the first non-zero contribution to the $A_2$ coefficient may be proportional to $m_q^2/M_{LQ}^2\ll 1$ because of a cancellation of electric charges in the zeroth order term.  Under these circumstances, the $\tau\rightarrow e\gamma$ yields relatively weak upper bounds on the LQ ratios.  This occurs for the scalar leptoquark $\tilde{S}_{1/2}^L$.

Finally, we discuss the impact of LFV(1,2) searches on the leptoquark limits.  As in the case of the effective operators below, \emph{a priori} there is nothing in the BRW leptoquark parameterization that relates the LQ couplings to second generation leptons to LQ couplings to third generation leptons.  Therefore, experimental limits on $\mu\rightarrow e$ conversion, $\mu\rightarrow e\gamma$, and $\mu\rightarrow 3e$ do not necessarily affect the expected size of the cross sections expected for leptoquark-mediated $e\rightarrow\tau$ conversion at the EIC.  Only by considering a specific model with an additional symmetry does a connection between LQ-induced LFV(1,2) and LFV(1,3) exist.  An example is the $SU(5)$ GUT studied in~\cite{Dorsner:2005fq,FileviezPerez:2008dw}.  The leptoquark present in this model has the same spin and gauge group quantum numbers as the BRW leptoquark $\tilde{S}_{1/2}^L$.  As mentioned above, this particular LQ evades limits from $\tau\rightarrow e\gamma$ as well as $\mu\rightarrow e\gamma$ for the same reason.  Additionally, the $SU(5)$ symmetry implies that the leptoquark couplings are proportional to the neutrino mixing angles and squared mass differences, and the stringent experimental bounds on $\mu\rightarrow e$ conversion further constrain the LQ couplings.  Imposing all of these limits, the LQ couplings can still yield an $e\rightarrow \tau$ conversion cross section within reach of the EIC with 1000~$\fb^{-1}$ integrated luminosity (for details, see~\cite{FileviezPerez:2008dw,Gonderinger:2010yn}).  In particular, the $e^-+p\rightarrow\tau^-+X$ cross section is dominated by the partonic subprocess $e+d\rightarrow \tau+b$, implying that a $\tau$ plus a $b$-jet may be a unique experimental signatature of this particular $SU(5)$ GUT at the EIC.

\subsection{Theory II: Effective Operators}\label{etau_sec:eff_ops}

\begin{figure}
\begin{center}
\includegraphics[width=\textwidth]{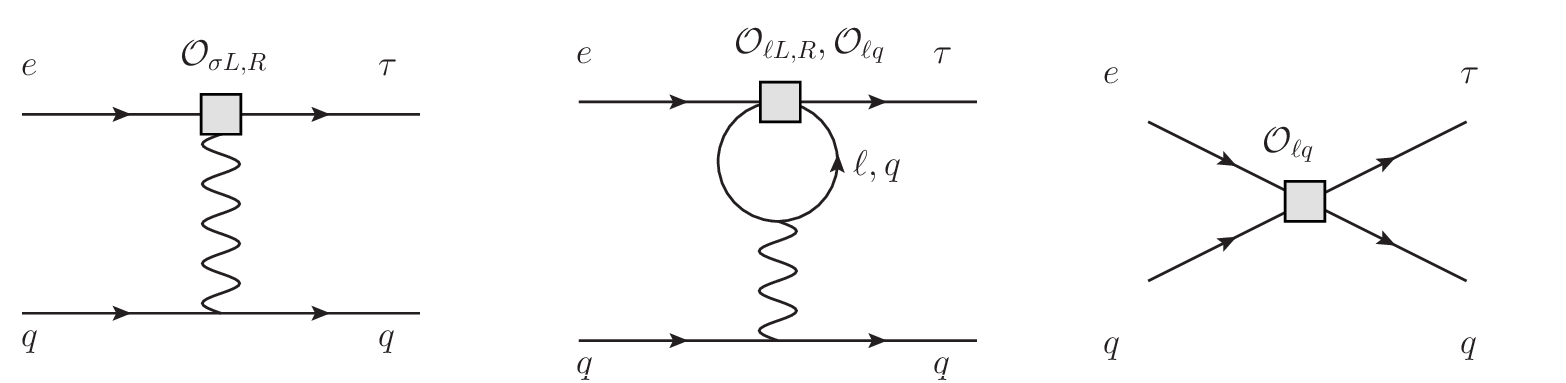}
\end{center}
\caption{\label{fig:e_tau_ops} Feynman diagrams showing the contributions of the magnetic moment ($\mathcal{O}_{\sigma L,R}$), four lepton ($\mathcal{O}_{\ell L,R}$), and four fermion ($\mathcal{O}_{\ell q}$) operators to the $e\rightarrow\tau$ conversion process.}
\end{figure}

We now examine the $e\rightarrow\tau$ conversion process at the EIC from the perspective of model-independent effective operators.  A complete list of $SU(3)_C\times SU(2)_L\times U(1)_Y$ gauge-invariant dimension-5 and -6 operators built from the SM field content can be found in~\cite{Buchmuller:1985jz} (for an updated list, see~\cite{Grzadkowski:2010es}).  There are three classes of operators which can contribute to $e\rightarrow\tau$ conversion and are of particular interest~\cite{Raidal:1997hq,Cirigliano:2004tc}:
\begin{enumerate}
\item magnetic moment operators (written here after electroweak symmetry breaking) 
\begin{equation}\label{eq:magnetic_moment_op}
\mathcal{O}_{\sigma L}=im_j\bar{\ell}_{Li} \sigma^{\mu\nu}\ell_{Rj}F_{\mu\nu} + h.c.\,\,;
\end{equation}
\item four lepton operators
\begin{equation}\label{eq:four_lepton_op}
\mathcal{O}_{\ell L} = \bar{\ell}_{Li}\ell^C_{Lj}\bar{\ell}^C_{Lk}\ell_{Lm}\,\,;
\end{equation}
\item four fermion (two quark, two lepton) operators
\begin{equation}\label{eq:four_fermion_op}
\mathcal{O}_{\ell q} = \bar{\ell}_i\Gamma_\ell \ell_j \bar{q}\Gamma_q q
\end{equation}
\end{enumerate}
We use indices $i,j,k,l$ to indicate the lepton generations and suppress the gauge group indices; the superscript ${}^C$ indicates charge conjugation.  Note that analogous operators with right-handed fields can also be constructed.  These operators can contribute to the deep inelastic electron-to-tau conversion process, as shown in figure \ref{fig:e_tau_ops}.

The leptonic current for the photon exchange diagrams in figure \ref{fig:e_tau_ops} (left and middle) has a general parameterization similar to the $\tau\rightarrow e\gamma$ amplitude in equation \eqref{eq:tau_e_gamma_amp}:
\begin{equation}\label{eq:lep_current}
\begin{aligned}
&j^\mu & &= & &\bar{u}_e\left[\left(q^2\gamma^\mu - q^\mu\left(q\cdot\gamma\right)\right)\left(A_1^L P_L + A_1^R P_R\right)\right.\\
& & & & &\qquad \left.+ im_\tau q_\nu\sigma^{\mu\nu}\left(A_2^L P_L + A_2^R P_R\right)\right]u_\tau\,\,.
\end{aligned}
\end{equation}
The two structures in this current are the charge radius term, $q^2\gamma^\mu - q^\mu\left(q\cdot\gamma\right)$, with $A_1^{L,R}$ coefficients, and the magnetic moment term, $im_\tau q_\nu\sigma^{\mu\nu}$, with $A_2^{L,R}$ coefficients.  The loop diagram (middle) in figure \ref{fig:e_tau_ops} which contains the four fermion and four lepton operators contributes to the charge radius term of the leptonic current, and so the Wilson coefficients of the operators $\mathcal{O}_{\ell L,R}$ and $\mathcal{O}_{\ell q}$ appear in the coefficients $A_1^{L,R}$.  These contributions are loop-suppressed but receive potentially large logarithmic enhancements which go like $\ln{\left(\Lambda_{LFV}^2/m^2\right)}$ (where $m$ is the mass of the quark or lepton in the loop and $\Lambda_{LFV}$ is the scale at which new degrees of freedom that induce LFV are no longer integrated out of the theory)~\cite{Raidal:1997hq}.  The left photon exchange diagram in figure \ref{fig:e_tau_ops} containing the magnetic moment operator implies that the $A_2^{L,R}$ factors in the leptonic current depend on the Wilson coefficients of the $\mathcal{O}_{\sigma L,R}$ operators.  These effective operators also are loop suppressed since they appear in the effective theory when heavy particles in loop diagrams (e.g., the leptoquarks in figure \ref{fig:tau_e_gamma}) are integrated out of the full theory.

The four fermion operator in figure \ref{fig:e_tau_ops} (right), similar to the Fermi theory for weak interactions, is a contact interaction that arises when a massive propagator is integrated out of the full theory at external momentum scales much smaller than the propagator's mass.  This operator contributes to $e\rightarrow\tau$ conversion at tree level.  Its cross section is expected to be larger than the cross sections from the other diagrams and operators discussed above (assuming the Wilson coefficients for all of the operators in equations \eqref{eq:magnetic_moment_op}-\eqref{eq:four_fermion_op} are all roughly the same order).  

\begin{table}
\begin{center}
\begin{tabular}[tb]{|l|l|}
\hline 
$\tau\rightarrow 3e$ & $\Gamma\left(\tau^-\rightarrow e^-e^+e^-\right)/\Gamma_{total} < 3.6\times 10^{-8}$\\\hline 
$\tau\rightarrow e\mu\mu$ & $\Gamma\left(\tau^-\rightarrow e^-\mu^+\mu^-\right)/\Gamma_{total} < 3.7\times 10^{-8}$\\\hline 
$\tau\rightarrow \mu ee$ & $\Gamma\left(\tau^-\rightarrow \mu^-e^+e^-\right)/\Gamma_{total} < 2.7\times 10^{-8}$\\\hline 
$\tau\rightarrow 3\mu$ & $\Gamma\left(\tau^-\rightarrow\mu^-\mu^+\mu^-\right)/\Gamma_{total} < 3.2\times 10^{-8}$\\\hline 
$\tau\rightarrow e\gamma$ & $\Gamma\left(\tau^-\rightarrow e^-\gamma\right)/\Gamma_{total} < 3.3\times 10^{-8}$\\\hline 
$\tau\rightarrow \mu\gamma$ & $\Gamma\left(\tau^-\rightarrow \mu^-\gamma\right)/\Gamma_{total} < 4.4\times 10^{-8}$\\\hline 
\end{tabular}
\end{center}
\caption{\label{tab:tau_decay_limits} All limits are taken from \cite{Nakamura:2010zzi} and are at a 90\% C.L.}
\end{table}

Limits on the magnetic moment and four lepton operator coefficients can be determined directly from relevant $\tau$ decay limits, some of which are listed in table \ref{tab:tau_decay_limits}.  The smallness of the limits on these $\tau$ decays, in conjunction with loop suppression factors, ensures that the contributions of the $\mathcal{O}_{\sigma L,R}, \mathcal{O}_{\ell L,R}$ coefficients to the leptonic current in equation \eqref{eq:lep_current} are negligible.  Therefore, as stated previously, it is expected that the greatest contributions to $e\rightarrow\tau$ conversion will come from the four fermion operators $\mathcal{O}_{\ell q}$, while photon exchange contributions will be negligibly small.  Limits on the four fermion operators' coefficients can be determined from the limits on the leptoquark ratios $\lambda_{1\alpha}\lambda_{3\beta}/M_{LQ}^2$.  The 14 leptoquarks in the BRW parameterization correspond to 7 of the four fermion operators listed in \cite{Buchmuller:1985jz}, as shown in table \ref{tab:lq_eff_ops} (though the correspondence is not one-to-one).  Hence the leptoquark limits set by direct $e\rightarrow\tau$ searches at HERA as well as the rare process searches cited by the HERA analyses~\cite{Chekanov:2005au}, such as $\tau\rightarrow\pi e$ and decays of $B$ and $K$ mesons, allow limits to be set on the four fermion operator coefficients.

\begin{table}
\begin{center}
\begin{tabular}[tb]{|l|l|l|}
\hline 
$\mathcal{O}_{\ell q}^{(1)}$ & $\bar{\ell}_L\gamma_\mu\ell_L\bar{q}_L\gamma^\mu q_L$ & $\left(S_0^L,\vec{S}_1^L\right);\left(V_0^L,\vec{V}_1^L\right)$ \\\hline 
$\mathcal{O}_{\ell q}^{(3)}$ & $\bar{\ell}_L\gamma_\mu \tau^a \ell_L \bar{q}_L\gamma^\mu \tau^a q_L$ & $\left(S_0^L,\vec{S}_1^L\right);\left(V_0^L,\vec{V}_1^L\right)$ \\\hline 
$\mathcal{O}_{eu}$ & $\bar{e}_R\gamma_\mu e_R\bar{u}_R\gamma^\mu u_R$ & $S_0^R;\tilde{V}_0^R$ \\\hline 
$\mathcal{O}_{ed}$ & $\bar{e}_R\gamma_\mu e_R\bar{d}_R\gamma^\mu d_R$ & $\tilde{S}_0^R;V_0^R$ \\\hline 
$\mathcal{O}_{\ell u}$ & $\bar{\ell}_L u_R \bar{u}_R \ell_L$ & $S_{1/2}^L;\tilde{V}_{1/2}^L$ \\\hline 
$\mathcal{O}_{\ell d}$ & $\bar{\ell}_L d_R \bar{d}_R \ell_L$ & $\tilde{S}_{1/2}^L;V_{1/2}^L$ \\\hline 
$\mathcal{O}_{qe}$ & $\bar{q}_L e_R \bar{e}_R q_L$ & $S_{1/2}^R;V_{1/2}^R$ \\\hline 
$\mathcal{O}_{qde}$ & $\bar{\ell}_L e_R \bar{d}_R q_L$ & \\\hline 
$\mathcal{O}_{\ell q}$ & $\bar{\ell}_L e_R \epsilon \bar{q}_L u_R$ & \\\hline 
\end{tabular}
\end{center}
\caption{\label{tab:lq_eff_ops} List of four fermion operators.  For the operator names, we follow the notation of~\cite{Buchmuller:1985jz}.  In the middle column, we maintain the same notation as in equation \eqref{eq:lq_lagrang}.  The right column lists the leptoquarks from which these operators are obtained upon integrating out the LQs.  Some operators are a linear combination of different LQs which are enclosed in parentheses.}
\end{table}

We conclude our discussion of the effective operators by noting that searches for $\mu\rightarrow e$ conversion, $\mu\rightarrow e\gamma$, and $\mu\rightarrow 3e$ bound the coefficients of the operators in \eqref{eq:magnetic_moment_op}-\eqref{eq:four_fermion_op} which mix first and second generation leptons.  However, \emph{a priori}, limits on such LFV(1,2) operators do not constrain the LFV(1,3) operators relevant for $e\rightarrow\tau$ conversion.  Only by assuming the existence of an additional symmetry or a particular underlying model can the two sets of operators be related.  One example of such an additional symmetry is the theory of minimal flavor violation (MFV) in the lepton sector~\cite{Cirigliano:2005ck}.  Under the assumptions of MFV, the breaking of the lepton flavor symmetry group $SU(3)_L\times SU(3)_E$ (for the left-handed doublets and the right-handed charged leptons) arises solely from the charged lepton and neutrino mass matrices.\footnote{We limit our discussion here to the scenario of ``minimal field content'' described in~\cite{Cirigliano:2005ck}.}  As a result, all higher-dimensional lepton flavor violating operators constructed from the lepton bilinears $\bar{\ell}_L^i\Gamma\ell_L^j, \bar{e}_R^i\Gamma\ell_L^j$, and $\bar{e}_R^i\Gamma e_R^j$ are suppressed by one or more powers of lepton masses and/or neutrino mixing parameters.  This is true even of the four fermion type operators.  Under the MFV hypothesis, the $e\rightarrow\tau$ conversion cross section is unobservably small; it is probable that any observation of $e\rightarrow\tau$ conversion at the EIC would therefore rule out the MFV hypothesis.

\subsection{Theory III: Conclusions and Future Work}\label{etau_sec:th_concl}

The theoretical analysis of the $e\rightarrow\tau$ DIS process presented in~\cite{Gonderinger:2010yn} and section~\ref{etau_sec:LQ_frame} shows that leptoquarks provide a framework in which $e\rightarrow\tau$ conversion searches at the EIC are feasible.  Present leptoquark limits are not prohibitive, and the EIC would be competitive with future Super-B experiments ($\tau\rightarrow e\gamma$ searches) on similar time scales, for several reasons: the EIC would have high luminosity and be sensitive to small cross sections; the EIC like HERA can set limits for all combinations of quark generations while $\tau\rightarrow e\gamma$ is more limited in this region; and the EIC could probe leptoquarks which may evade $\tau\rightarrow e\gamma$ searches.

Limits from LFV(1,2) searches may or may not be relevant for leptoquarks.  While the BRW framework implies no connection between LFV(1,2) limits and LFV(1,3) processes, in general it is presumed that leptoquarks will arise from physics at the high scale which does in fact constrain LFV(1,3) processes given the current stronger limits on LFV(1,2) processes.  However, at least one model, the $SU(5)$ GUT discussed above, exists in which limits from LFV(1,2) searches ($\mu\rightarrow e$ conversion, $\mu\rightarrow e\gamma$), $\tau\rightarrow e\gamma$ searches, and the neutrino sector can be implemented and still allow for observable $e\rightarrow \tau$ conversion cross sections at the EIC.

An estimation of $e\rightarrow\tau$ cross sections using model-independent effective operators and present limits on $LFV$ processes suggests that the best hope for observing $e\rightarrow \tau$ conversion at the EIC is with models which give rise to four fermion operators through tree level processes at low energies.  Four lepton and magnetic moment operators are generally too suppressed to give rise to large enough $e\rightarrow\tau$ cross sections via photon exchange, especially when the relevant limits (\emph{e.g.}, $\tau\rightarrow 3e, \tau\rightarrow e\gamma$) are imposed on the operator coefficients.  Limits from additional LFV(1,2) searches like $\mu\rightarrow e$ conversion and $\mu\rightarrow e\gamma$ can be applied to the effective operator analysis if an additional symmetry such as MFV is imposed.  MFV results in a suppression of all the LFV operators, including the four fermion operators, and hence negligibly small $e\rightarrow\tau$ cross sections.

There are several theoretical topics worthy of further attention for the $e\rightarrow\tau$ EIC search.  First is the study of leptoquarks and LFV(1,3) flavor structure at LHC.  While studies of LQ searches at the LHC have been performed in the past (as an example, see~\cite{Belyaev:2005ew}), such work has focused only on first generation fermions coupling to leptoquarks and has not considered LFV leptoquark final states.  Further work is required to determine the extent to which the EIC and the LHC may provide complementary probes of the leptoquark flavor violating parameter space.  

An additional topic which merits further study is a broader analysis of model-dependent $e\rightarrow\tau$ searches at the EIC.  Non-leptoquark models or symmetries may give promising results for the $e\rightarrow\tau$ conversion process.  For example, R-parity violating supersymmetry allows for tree level $e\rightarrow\tau$ conversion mediated by squarks; this suggests that large cross sections perhaps may be expected.  Furthermore, depending on the models which give rise to the effective operators discussed above, there may be large log enhancements in the charge radius contribution to photon exchange $e\rightarrow\tau$ which could overcome the limits on the four lepton operators.

Finally, we observe that many experiments have over many years placed limits on a wide variety of flavor-violating processes.  Many of these experiments constrain the leptoquark parameter space, as analyzed in~\cite{Davidson:1993qk}.  Updated limits from experimental searches for other flavor-violating processes may exist and still need to be considered in analyzing the potential of the EIC (and LHC) to search for LQ-mediated LFV events.  Such limits may also be relevant for non-LQ scenarios.  Improved limits from ongoing and future experiments searching for LFV(1,2) processes also need to be included, depending on the context for the $e\rightarrow\tau$ analysis.

This concludes the discussion of the theoretical analysis of leptoquark-induced $e\rightarrow\tau$ conversion in deep-inelastic scattering at the EIC.  The analysis so far has been optimistic and disregarded important experimental considerations that would impact a search for $e\rightarrow\tau$ events.  The next several sections address the questions of SM backgrounds and $\tau$ detection.

\subsection{Experiment I: Standard Model Backgrounds \& the Analysis Strategy}\label{etau_sec:SM_bg}

\begin{figure}[h]
\center
\includegraphics[width=\textwidth]{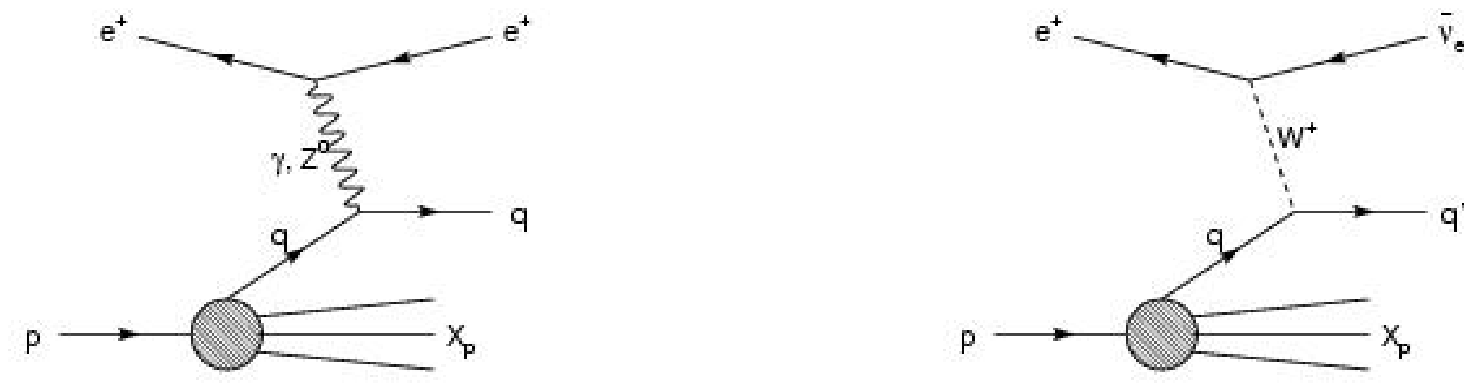}
\caption{ NC and CC DIS diagrams.}
\label{fig:Diagrams}
\end{figure}

In this section, we discuss the main SM processes that could mimic the LFV(1,3) signal at the EIC. In the SM, $ep$ scattering is caused by the exchange of an electroweak gauge boson between the electron and a quark inside the proton. Photon exchange dominates when the momentum transfer $Q$ is low, but the amplitude of weak gauge bosons becomes more important as $|Q^{2}|$ approaches $M^{2}_{W^{\pm}}$ and $M^{2}_{Z^{0}}$. Standard Model NC and CC DIS processes are shown in figure~\ref{fig:Diagrams}. The EIC acceptance from the beampipe ($0.1^{\circ} <\theta < 179.9^{\circ}$) restricts all EIC kinematics to $Q^{2}>0.01$ GeV$^{2}$~\cite{EICwiki}. This cut was implemented in the SM simulations at low momentum transfer. However, we focused our SM background analysis on events with very high $Q^{2}$ since a cut of $Q^{2}>1000$ GeV$^{2}$ was used in all simulations involving leptoquarks (given the range of LQ masses chosen, see table~\ref{table_ratio}).

Ignoring rare processes, the final state $\tau$ in the LFV(1,3) event can decay in three different ways: electronic channel, muonic channel, and hadronic channel. Like previous searches done at HERA \cite{Aaron:2011zz,Aktas:2007ji}, we consider five different SM events whose final states could be misidentified with a decaying $\tau$: NC DIS, CC DIS, photoproduction, lepton-pair production, and real W boson production.

SM processes lead to final state particles that could be misidentified as our candidate $\tau$. In other words, they produce particles that leave tracks in the detector that look like the leptons or hadrons produced in a $\tau$ decay. However, the geometry of the SM events and the LFV(1,3) events do differ. Indeed, the identified $\tau$ lepton in an $ep\rightarrow \tau X$ conversion must be back-to-back in azimuth with the hadronic sector $X$. In addition, the angular distribution in the $\theta$ direction of the decay products of scalar LQs, vector LQs, and SM DIS background will differ because their corresponding cross sections have a different $y$ dependence.\footnote{The Bjorken scattering variable is given by $y=Q^{2}/s_{ep}x=1/2(1-\cos \hat{\theta})$ where $\hat{\theta}$ is the decay polar angle of the lepton relative to the incident proton in the center-of-mass frame.} By inspecting equations \eqref{eq:e_tau_cxn} and \eqref{eq:f_and_g} we can see that in the s-channel (see figure \ref{fig:e_tau_conv}) vector LQs are distributed according to $d\sigma/dy\propto(1-y)^{2}$, whereas scalar LQs decay isotropically (flat $d\sigma/dy$ distribution) in their rest frame (and vice-versa in the u-channel). In contrast, NC DIS events have $d\sigma/dy\propto y^{-1/2}$, and this difference between LQ and SM $y$ spectra could be exploited to identify background DIS events~\cite{Adloff:1999tp}.

At the detector level, the events in all channels at the EIC must be accepted by a trigger for a large imbalance in the transverse energy flow. The energy flow summation runs over all energy deposits in the calorimeters and missing transverse momentum is associated to the neutrinos that escape the detector without any energy deposit. In this initial analysis, the missing transverse momentum $p_{T}^{miss}$ is defined as:
\begin{equation}
p_{T}^{miss} = \sqrt{(\sum P_{x,i})^{2} + (\sum P_{y,i})^{2}}
\end{equation}
where $i$ runs over all final state particles in an event, excluding all neutrinos.
\\
\indent PYTHIA 6.4.23 is used to generate all SM events with the CTEQ 5L parametrization of the parton distribution functions of the proton \cite{Lai:1999wy}.
Initial and final state radiation are included. No GEANT simulation of the EIC detector has been used. Two different energies are chosen for the $ep$ MC 
simulations: 20$\times 325$ GeV with $\sqrt{s}=161.25$ GeV 
and 10$\times 250$ GeV with $\sqrt{s}=100.01$ GeV. Although not directly relevant for the conclusions of this topological study, we allowed ourselves 
the possibility of gathering a total integrated luminosity  of $1000$ fb$^{-1}$ as suggested on this workshop's web page \cite{EICwiki}.

\subsubsection{Standard Model Event Generation}
\subsubsection*{NC DIS: ($ep\rightarrow eX$)}
NC DIS events are mediated by a photon or a $Z^{0}$ boson, and the final state includes an electron. The final state event topology of the tau electronic decay ($\tau\rightarrow e\nu_{e}\nu_{\tau}$) is therefore very similar to that of high $Q^{2}$ NC DIS. By energy-momentum conservation the $\sum(E-P_{z})$ distribution for NC DIS events is peaked at $2E_{0}$, where $E_{0}$ is the electron beam energy (10 or 20 GeV). We can also select NC events by implementing an upper and lower cut to the quantity $\sum(E-P_{z})$ measured. In contrast, the $\tau$ decay exhibits a large missing transverse momentum due to the neutrinos in the decay.

\subsubsection*{CC DIS: ($ep\rightarrow \nu X$)}
CC DIS events are mediated by a $W^{\pm}$ boson and are characterized by high missing transverse momentum $p_{T}^{miss}$ and higher $Q^2$.

\subsubsection*{Photoproduction: ($\gamma p\rightarrow X$)}
Events from photoproduction processes occur in the low $Q^{2}$ limit and may contribute to the final selection if a narrow hadronic jet fakes the tau signature or is misidentified as an electron. For $\gamma p$ events simulated with PYTHIA, the photon can be either direct (point-like) or resolved (VMD and GVMD/anomalous). A photon is assumed to be direct (point-like) when it can only interact in processes which explicitly contain the incoming photon \cite{Sjostrand:2006za}, such as $f_{i} \gamma\rightarrow f_{i}g$. A photon is considered to be resolved when it interacts through its constituent quarks and gluons. Each photoproduction subprocess leads to a different event structure and has a cross section that depends strongly on the virtuality of the photon. For high virtualities (high $Q^{2}$), DIS events dominate, and the photon is very virtual ($\gamma^{*}$). For very low $Q^{2}$, however, the photon can be treated as real and can have a partonic structure that can interact in different ways with the proton's quark (\emph{e.g.}, resolved photoproduction).
\\
\indent However, the LFV processes were simulated with a $Q^{2}>1000$ GeV$^{2}$ cut and hence the SM photoproduction background will automatically be reduced to zero. As table~\ref{table:photoproduction} below suggests, most of the background that concerns us is therefore in the DIS region where the photon is very virtual.
\\
\begin{table}[h]
\center
\begin{tabular}{|c|c|c|}
\hline
\textbf{Subprocess} &  \textbf{\% of tot. events, $Q^{2}>0.01$}& \textbf{\% of tot. events, $Q^{2}>1000$}\\ \hline
VMD& 61.56 & 0\\ \hline
Direct& 11.28 & 0\\ \hline
Anomalous& 9.05 & 0\\ \hline
DIS ($\gamma^{*}q\rightarrow q$)& 18.11 & 100\\
\hline
\hline
\end{tabular}
\caption{Event statistics for photoproduction/DIS subprocesses simulated at 20$\times$325 GeV with $Q^{2}>0.01$ GeV$^{2}$ and $Q^{2}>1000$ GeV$^{2}$.}
\label{table:photoproduction}
\end{table}
\\
\subsubsection*{Lepton-pair Production: ($ep\rightarrow el^{+}l^{-}X$) }
Lepton-pair production events contribute to the background because they may lead to high momentum leptons in the final state. An analysis of the event geometry is required to avoid misidentifying the three pencil-like tracks in the tau decay with the tracks left by $l^{+}$, $l^{-}$ and $X$ when the scattered electron is missed. The background samples include $e^{+}e^{-}$, $\tau^{+}\tau^{-}$ and $\mu^{+}\mu^{-}$ production. The simulation of these processes was not included in this analysis due to its very low cross section given the chosen EIC energy range. However, they can be included in the future using a better suited generator with improved efficiency compared to PYTHIA.

\subsubsection*{W Production: ($ep\rightarrow eWX$) }
Real W boson production leads to final states with isolated leptons with  high transverse momentum. The simulated W production samples include hadronic W decays (which can fake a tau decay) and leptonic ($l\bar{\nu_{l}})$ decays that contribute to the missing transverse momentum and could potentially produce a non-LFV $\tau$. However, the cross section of this process ($2.449\times10^{-13}$ mb for $10\times250$ and  $5.343\times10^{-11}$ mb for $20\times325$) at EIC energies and luminosities is negligible. 

\subsubsection{Results}
\begin{figure}[h]
\centerline{\includegraphics[angle=90, scale=0.4]{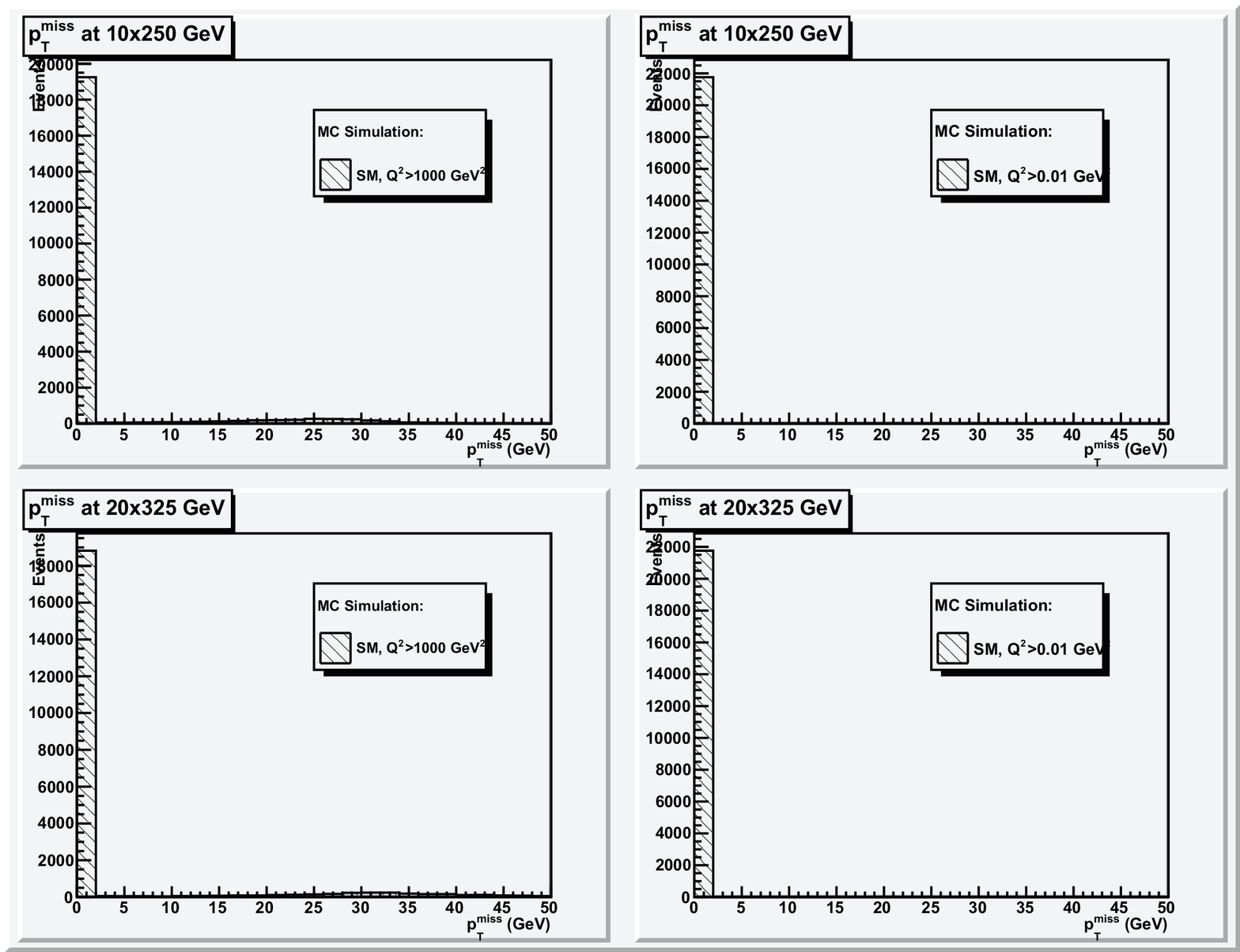}}
\caption{ $p_{T}^{miss}$ at 10$\times$250 and 20$\times$325 GeV with low $Q^{2}$ and $Q^{2}>1000$ GeV$^{2}$.}
\label{fig:Ptmiss}
\end{figure}

\begin{figure}[h]
\centerline{\includegraphics[angle=90, scale=0.4]{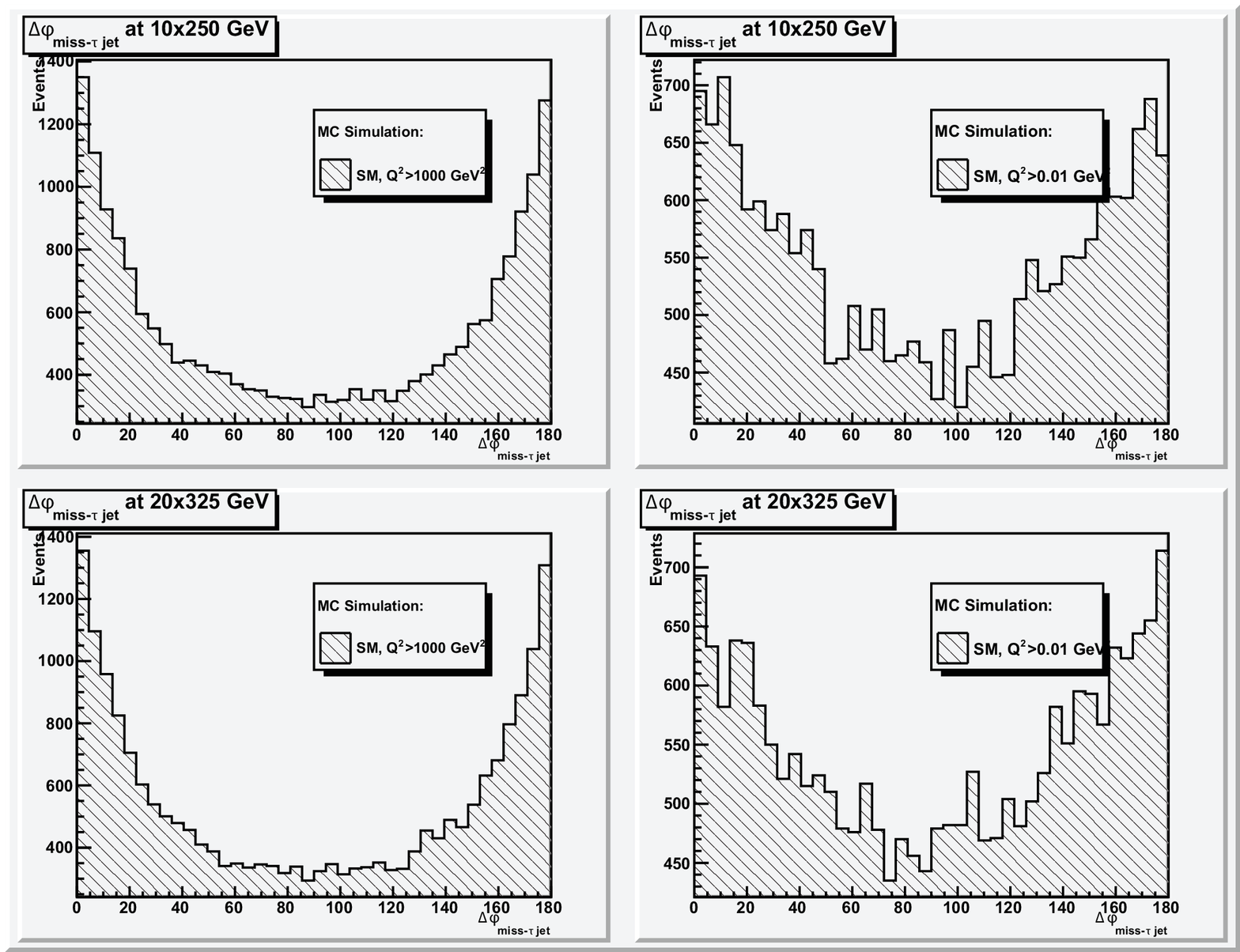}}
\caption{ Acoplanarity $\Delta \phi_{miss-\tau jet}$ at 10$\times$250 and 20$\times$325 GeV with low $Q^{2}$ and $Q^{2}>1000$ GeV$^{2}$ .}
\label{fig:acoplanarity}
\end{figure}

Figures~\ref{fig:Ptmiss} and ~\ref{fig:acoplanarity} include all SM processes. 
Shown are the $p_{T}^{miss}$ and acoplanarity $\Delta \phi_{miss-\tau jet}$ found in events due to 
missing neutrinos. 
The plots are made for two different beam energy combinations (top and bottom).  It is apparent that beam energies do not matter, the plots are very similar.
Two different $Q^{2}$ conditions were studied: left and right, which isolate predominantly high and  low $Q^{2}$ events, respectively.
With no $Q^{2}$ cut, the event sample is dominated by low $Q^{2}$ photo-production background. If a cut of $Q^{2} > 1000$ GeV$^{2}$ is made the EW-physics (W) 
events become apparent. 
Figure~\ref{fig:acoplanarity} shows the acoplanarity $\Delta \phi_{miss-\tau jet}$, the angle between the reconstructed $\tau$-jet direction and the missing momentum direction (presumably
the neutrinos in the primary collision) for the two different energies and virtualities. 
The figures also reveal that the shapes of the curves are very similar for the two different center-of-mass energies.


The particles in table~\ref{table:stat} are primarily produced from the decays of hadrons in the hadronic sector $X$; \emph{e.g.}, $\tau$s can be produced
from $D_s$ meson decay, but also include leptons from the processes mentioned above. The low $P_T$ suggests that these background particles
can be partially avoided by restricting the kinematics phase space to high $Q^2$ and high transverse momentum.

\begin{table}[h]
\center
\begin{tabular}{|c|c|c|c|c|}
\hline
\textbf{Particle ID} &  \textbf{$N_1$} & \textbf{$N_2$} & \textbf{10$\times$250} & \textbf{20$\times$325} \\ \hline
$\tau^{-}$ & 531 & 316 & $p_{T}<4$, $\theta_{p}\sim 2$ & $p_{T}<4$, $\theta_{p}\sim 2$\\ \hline
$\tau^{+}$ & 512 & 385 & $p_{T}<4$, $\theta_{p}\sim$ 1 & $p_{T}<4$, $\theta_{p}\sim 1$\\ \hline
$\mu^{-}$ & 38771 & 27849 & $p_{T}<2$, $\theta_{p}\sim$ 4 & $p_{T}<2$, $\theta_{p}\sim 4$\\ \hline
$\mu^{+}$ & 38691 & 27523 & $p_{T}<2$, $\theta_{p}\sim$ 4 & $p_{T}<2$, $\theta_{p}\sim 4$\\ \hline
$\nu_{\tau}$ & 1043 & 701 & $p_{T}<1$, $\theta_{p}\sim 4$ & $p_{T}<1$, $\theta_{p}\sim 4$\\ \hline
$\nu_{\mu}$ & 37200 & 26170 & $p_{T}<2$, $\theta_{p}\sim 4$ & $p_{T}<2$, $\theta_{p}\sim 4$\\ \hline
$\nu_{e}$ & 38343 & 27255 & $p_{T}<2$, $\theta_{p}\sim 4$ & $p_{T}<2$, $\theta_{p}\sim 4$\\ \hline
$\bar{\nu_{\tau}}$ & 1043 & 701 & $p_{T}<1$, $\theta_{p}\sim 4$ & $p_{T}<1.5$, $\theta_{p}\sim 4$\\ \hline
$\bar{\nu_{\mu}}$ & 37280 & 26496 & $p_{T}<2$, $\theta_{p}\sim 4$ & $p_{T}<2$, $\theta_{p}\sim 4$\\ \hline
$\bar{\nu_{e}}$ & 38836 & 28004 & $p_{T}<2$, $\theta_{p}\sim 3$ & $p_{T}<2$, $\theta_{p}\sim 4$\\
\hline
\hline
\end{tabular}
\label{table:stat}
\caption{Statistics of selected SM background particles for 10 million $e^{-}p$ collisions generated with PYTHIA. $N_1$ and $N_2$ are the number of times the particle is produced out of the 10 million events at energies of 20$\times325$ GeV and 10$\times250$ GeV respectively. $\theta_{p}$ is the peak of the particle's polar angle distribution in degrees with a FWHM$\sim11^{\circ}$ and $p_T$ is the transverse momentum in GeV. All $\phi$ distributions are flat.}
\end{table}
\vskip .3cm

\subsection{Experiment II: Leptoquark Simulation Study}\label{etau_sec:LQ_sim}

\subsubsection{Type of Leptoquark Studied: Parameter Space}
In this work we present the distribution of $p_{T}^{miss}$ generated from a LFV signal Monte Carlo sample of the leptoquark $\tilde{S}_{1/2}^{L}$.\footnote{See section~\ref{etau_sec:LQ_frame} for a description of the notation.  This leptoquark's interactions are given by the Lagrangian terms $\lambda \bar{d}_R\ell_L \tilde{S}_{1/2}^L + h.c.$  Also note that this leptoquark evades limits from $\tau\rightarrow e\gamma$ as explained in section~\ref{etau_sec:LQ_frame}.} The mass of the leptoquark is determined from the ratios $z \equiv \frac{\lambda_i\lambda_j}{M^2}$. The smallest value of the these ratios \cite{Matt:2010} which the EIC will potentially probe are listed in table \ref{table_ratio}. The LFV signal Monte Carlo events were generated using a LQ generator called ``LQGENEP" \cite{Bellagamba:2001fk}. LQGENEP is a LQ generator for electron/positron-proton scattering which simulates processes involving LQ production and exchange using the BRW \cite{Buchmuller:1986zs} effective model. The generator is interfaced with the PYTHIA
event generator. The value of $\lambda_{i} = \lambda_{j}=0.3$ is taken throughout this study. The values of $\lambda$ are correlated with $z$ through their relation to the mass of the leptoquarks, $M_{LQ}$. 
\begin{table}[h]
\center
\begin{tabular}{|c|c|c|}
\hline
$(q_i q_j)$  & $z (TeV^{-2})$ &  $Mass(GeV)$  \\ \hline
$11$  &  $0.024 $ &  $1936.5 $ \\ \hline
$13$  &  $0.03 $ & $1732.0$  \\ \hline
$22$  &  $0.039 $ & $1519.1 $ \\ \hline
$23$  &  $0.047 $ & $1383.8 $ \\ \hline
$31$  &  $0.03 $ & $1732.0 $ \\ \hline
$32$  &  $0.06 $ & $1224.7 $ \\ \hline
$33$  &  $0.084 $ & $1035.1 $ \\ 

\hline
\hline
\end{tabular}
\caption{The initial and final quark flavors $(q_i q_j)$ in the subprocess $e q_i \rightarrow \tau q_j$, the ratio 
$z$ and the mass of the LQ for $\lambda_i=\lambda_j=0.3$.}
\label{table_ratio}
\end{table}

\subsubsection{Leptoquark Event Characterization}

Electron-to-tau events were generated using the LFV generator LQGENEP for two EIC energies, namely, 10x250 GeV and 20x325 GeV, in $ep$ scattering. These events were restricted to sub-processes with a specific intermediary BRW LQ, $~\tilde{S}_{1/2}^{L}$. The kinematic region was restricted to $Q^2 > 1000 GeV^2$ and $y > 0.1$. 

\subsubsection*{Electronic \& muonic $\tau$ decays}

\begin{figure}
\begin{center}
\begin{minipage}[b]{0.49\columnwidth}
\centerline{\includegraphics[angle=90, scale=0.38]{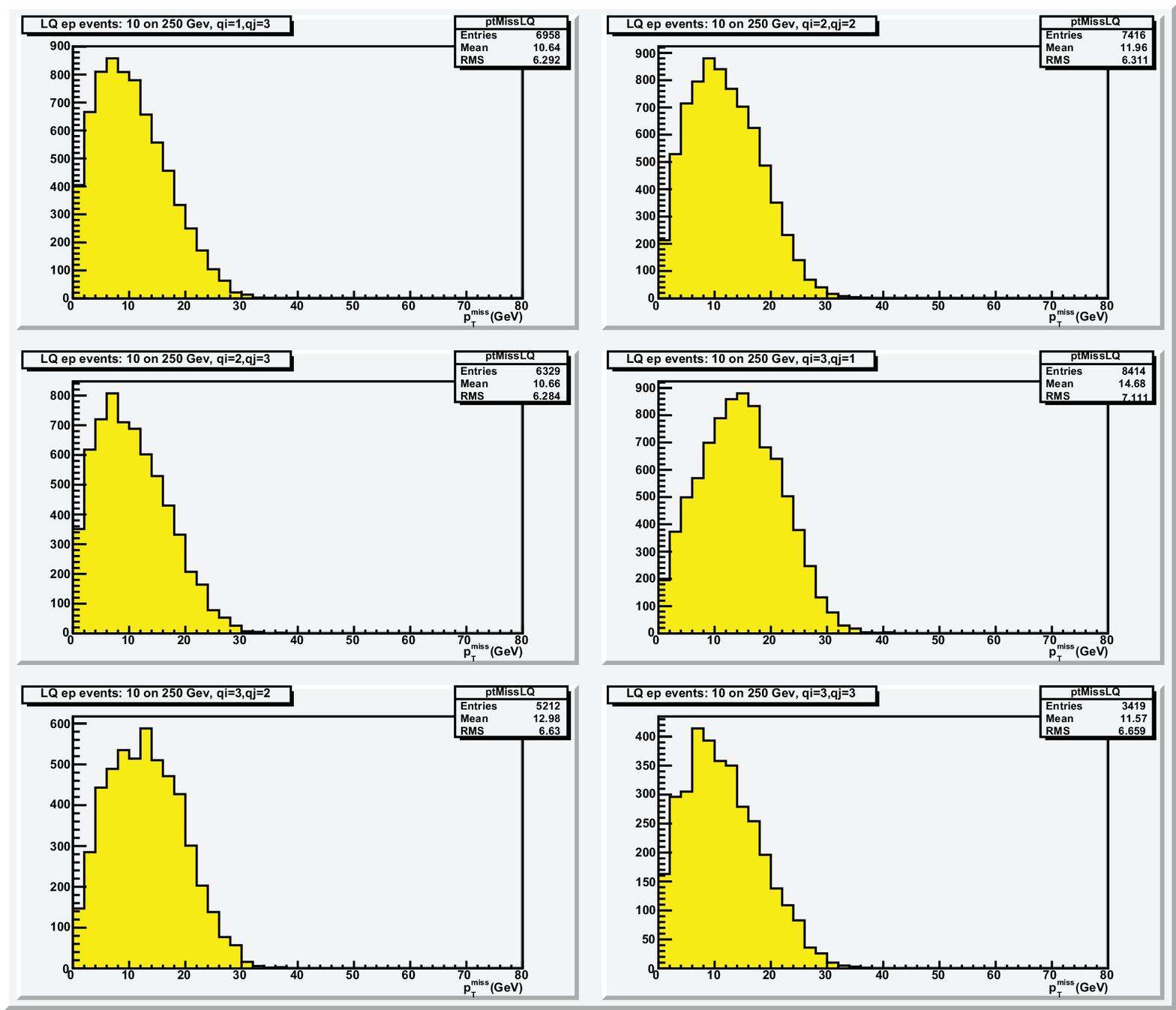}}
\end{minipage}
\begin{minipage}[b]{0.49\columnwidth}
\centerline{\includegraphics[angle=90, scale=0.38]{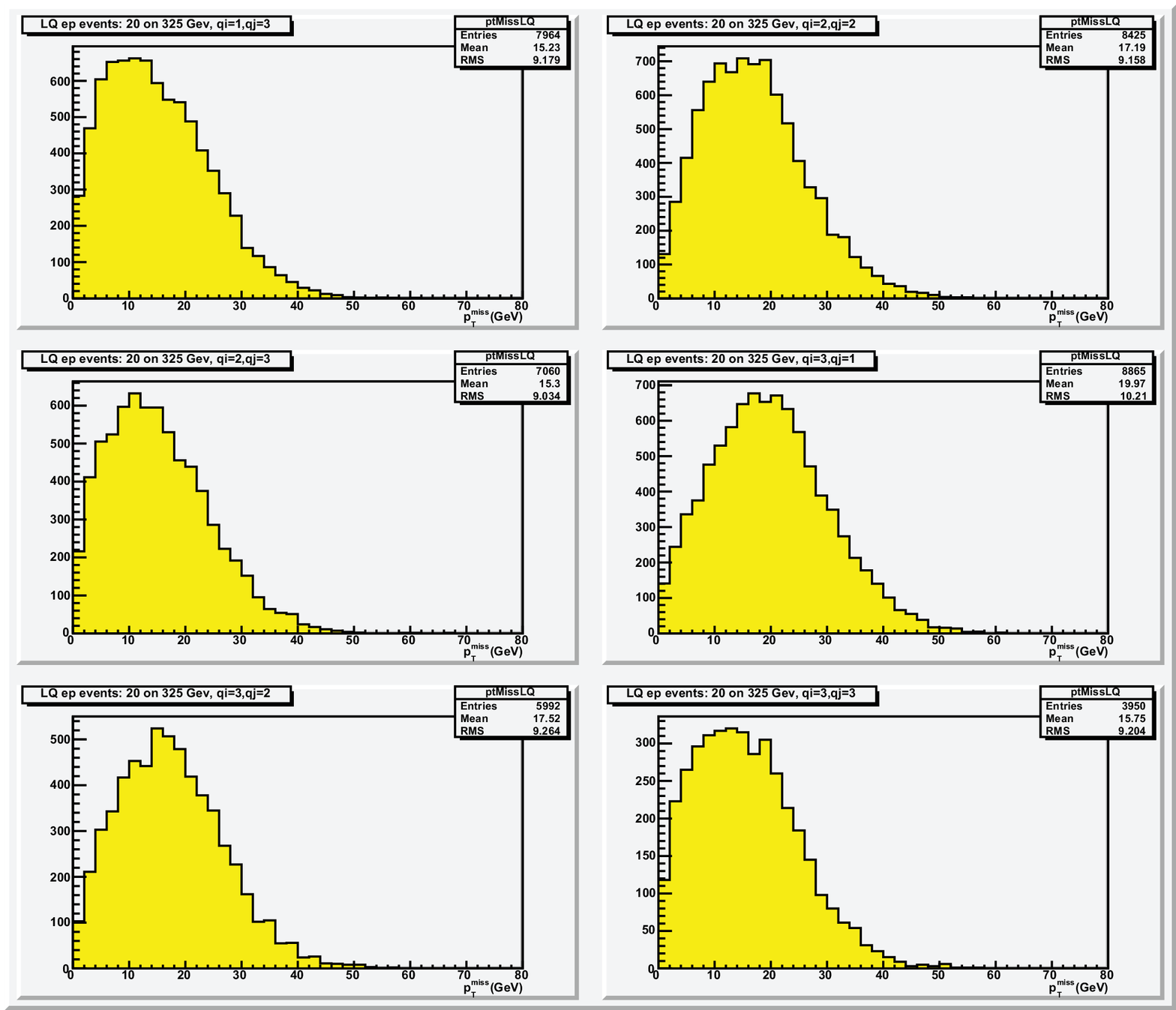}}
\end{minipage}
\caption{Missing transverse momentum in the electronic $\tau$ decay channel in the $ep$ scattering with 10x250 (left) and 20x325 (right) energies and the ratios, $z$, from table \ref{table_ratio} for $(q_i q_j)\equiv$ 13,22,23,31,32 and 33. The
lepton-quark couplings are $\lambda_i=\lambda_j=0.3$.}
\end{center}
\label{fig:EDPT}
\end{figure}

The leptonic decays of the tau, $\tau \rightarrow e\overline{\nu}_e\nu_\tau, \mu\overline{\nu}_\mu\nu_\tau$, were studied. Background for these events is present from SM neutral current events in $ep$ DIS. The $p_{T}^{miss}$ distribution for 10x250 and 20x325 are shown in figure \ref{fig:EDPT} (left) and (right), respectively. The plots shown are for electron final states. The muon final state plots are identical. The different panels indicate the $p_{T}^{miss}$ spectrum for each combination of quarks $i,j$ involved. The $p_{T}^{miss}$ spectrum is wider at higher center-of-mass energies, but otherwise the spectra are generally similar.

\subsubsection*{Hadronic $\tau$ decays}

The hadronic decay of high-$P_T$ $\tau$ leptons results in a characteristic narrow ``pencil-like'' jet with three pions. The event $ep \rightarrow \tau X$ 
would look like a a di-jet event with one narrow and one wide/high multiplicity jet. The jet 
associated with the $\tau$ decay is narrow. Thus one narrow and one wide jet in a di-jet event is a 
potential candidate for the signal. Various standard algorithms are used to identify such events \cite{Aktas:2007ji}. We did not simulate the detector response -- this is a topic for a future detector study -- but we studied the event characteristics and topology for such events.

\begin{figure}
\begin{center}
\begin{minipage}[b]{0.49\columnwidth}
\centerline{\includegraphics[angle=90, scale=0.38]{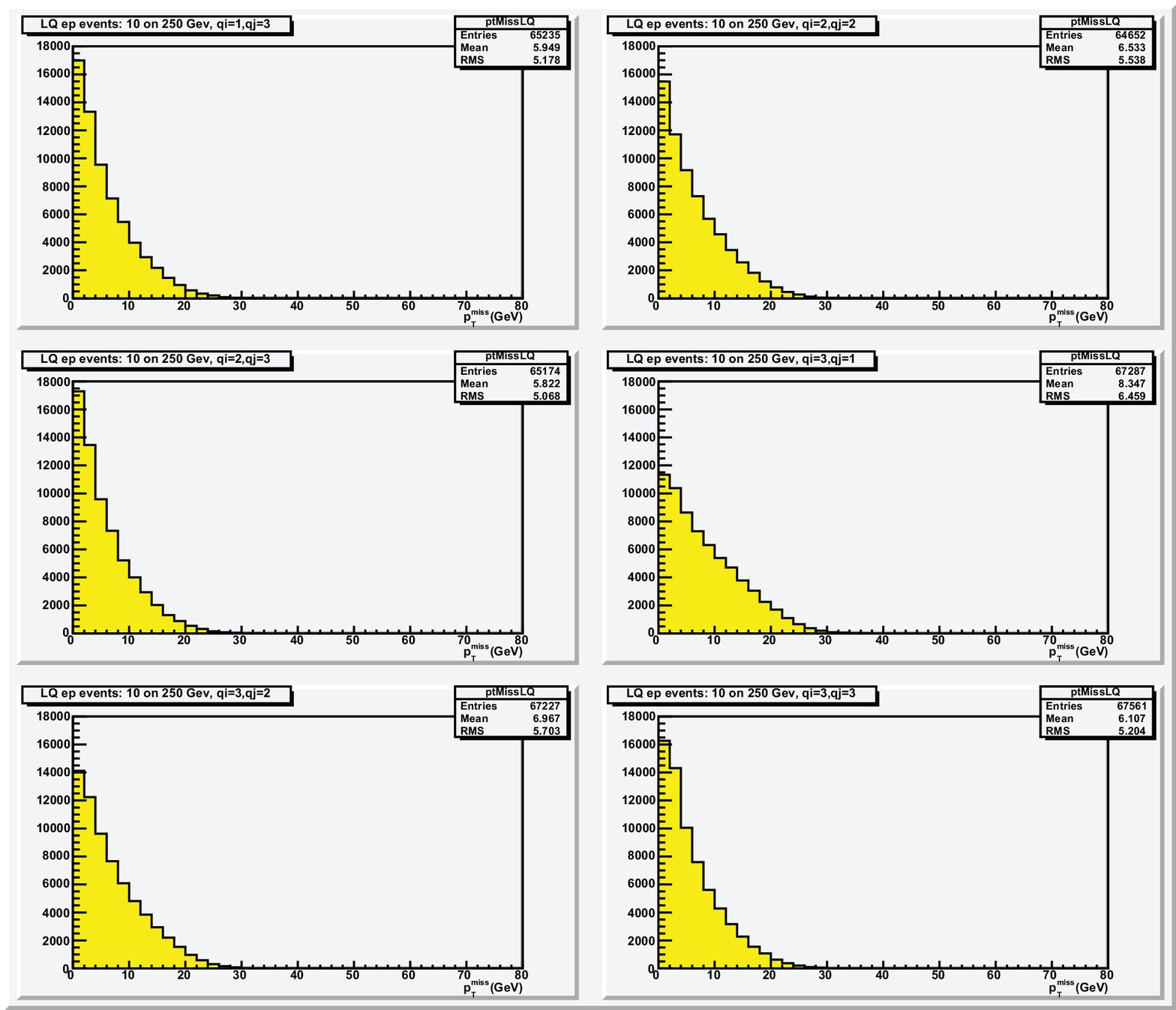}}
\end{minipage}
\begin{minipage}[b]{0.49\columnwidth}
\centerline{\includegraphics[angle=90, scale=0.38]{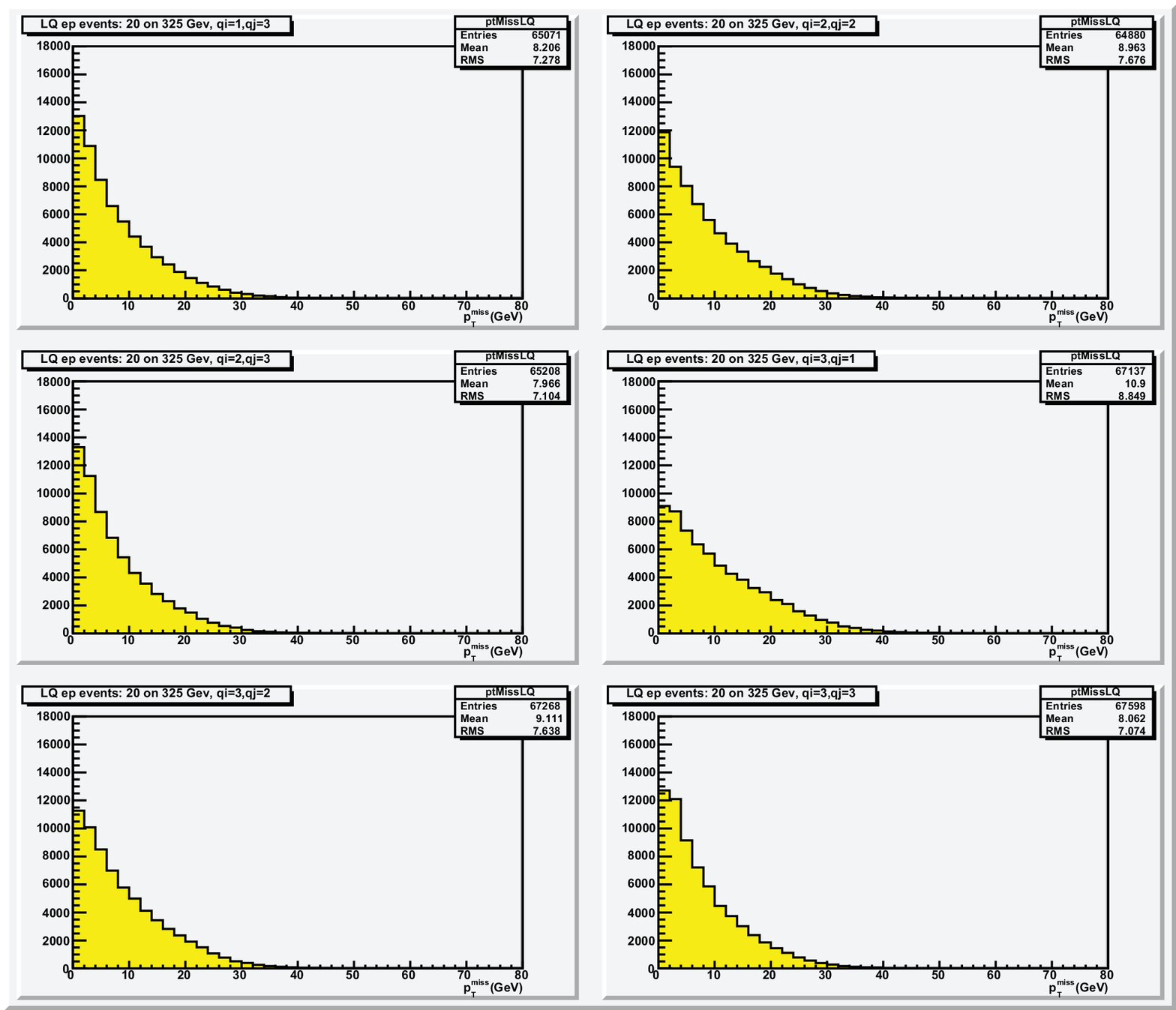}}
\end{minipage}
\caption{Missing transverse momentum in the hadronic $\tau$ decay channel in the $ep$ scattering with 10x250 (left) and 20x325 (right) energies and the ratios, $z$, from table \ref{table_ratio} for $(q_i q_j)\equiv$ 13,22,23,31,32 and 33. The
lepton-quark couplings are $\lambda_i=\lambda_j=0.3$.}
\end{center}
\label{fig:HDPT}
\end{figure}

\begin{figure}
\begin{center}
\begin{minipage}[b]{0.49\columnwidth}
\centerline{\includegraphics[angle=90, scale=0.38]{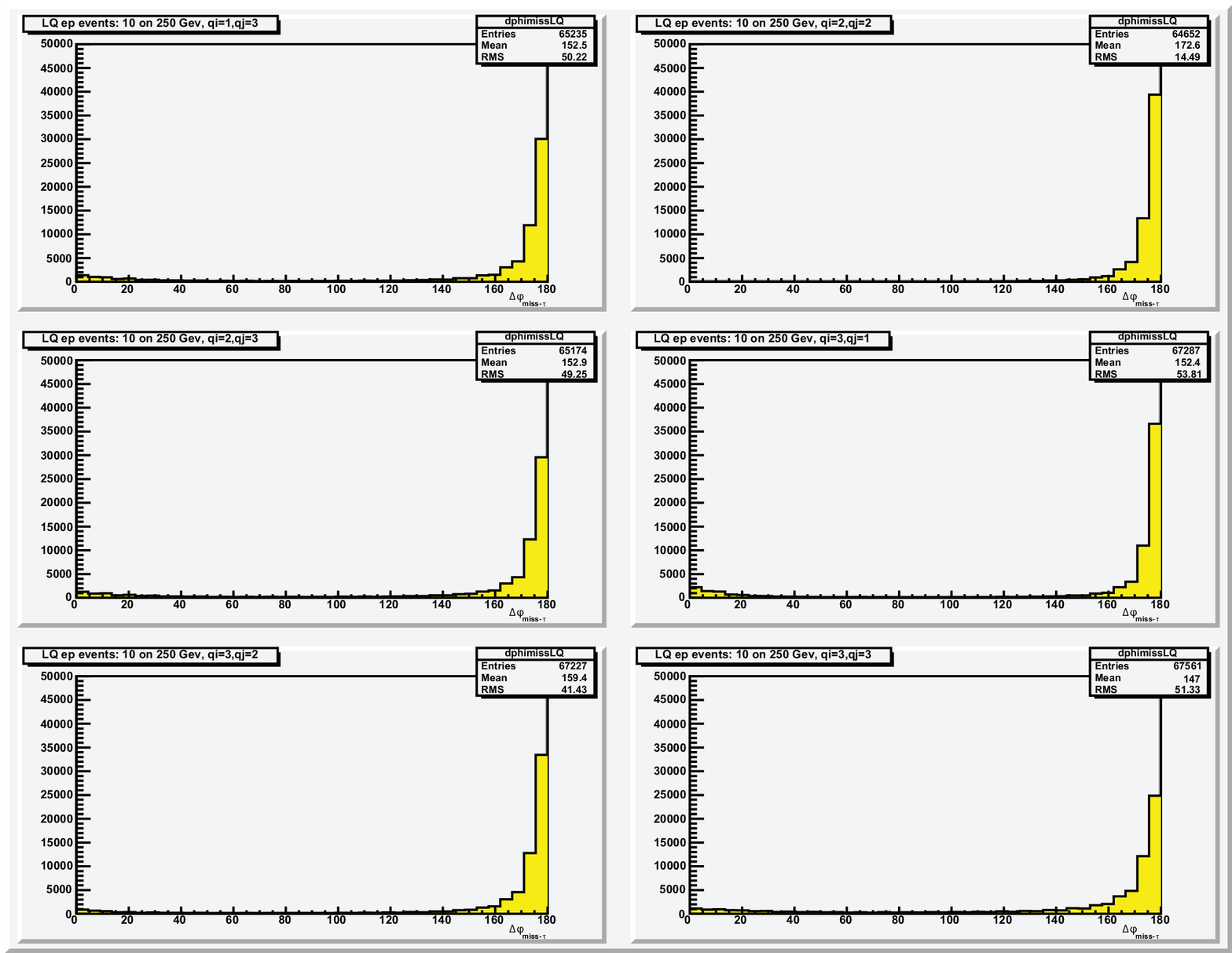}}
\end{minipage}
\begin{minipage}[b]{0.49\columnwidth}
\centerline{\includegraphics[angle=90, scale=0.38]{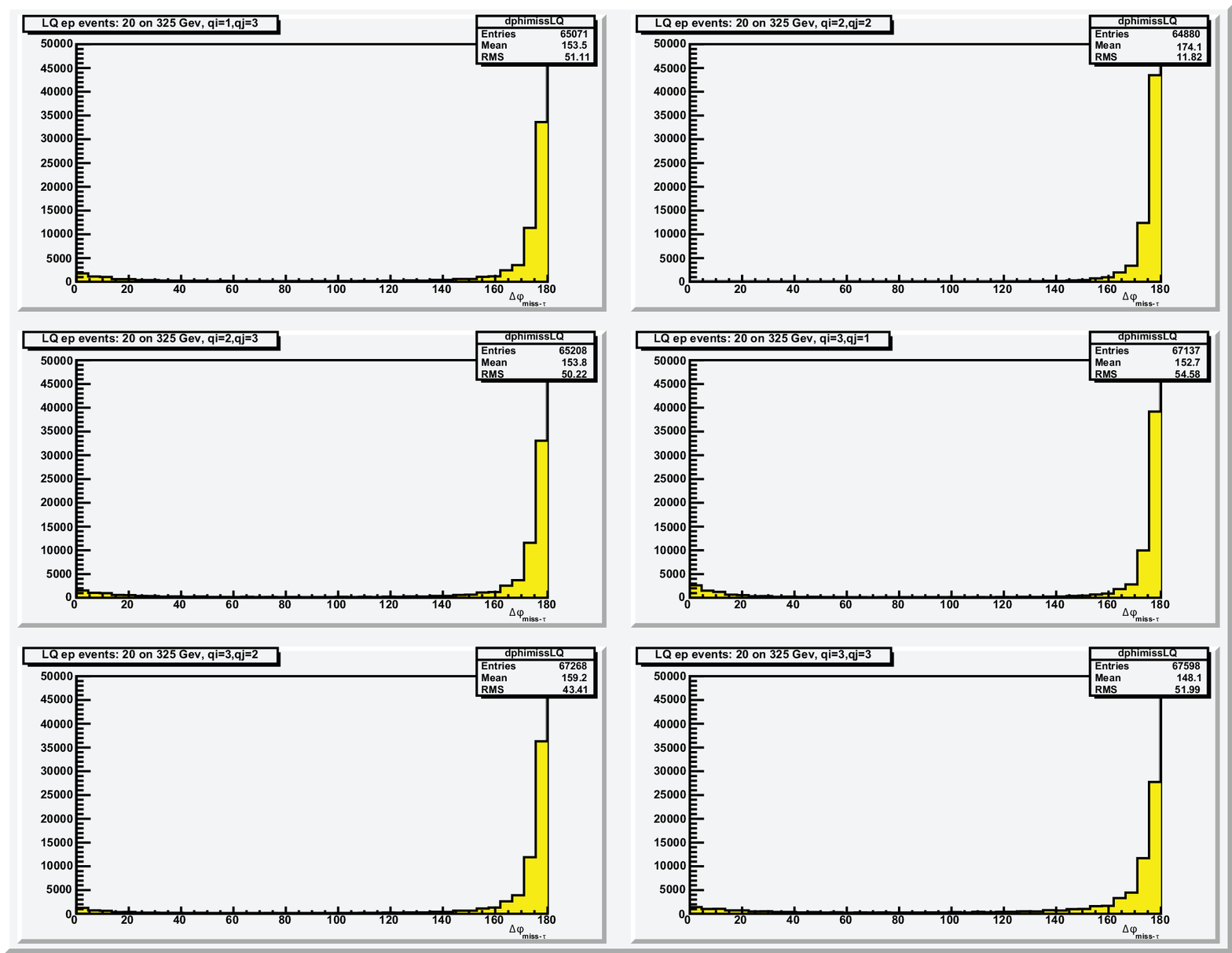}}
\end{minipage}
\caption{The acoplanarity,$~\Delta \phi_{miss-\tau jet}$, between the $\tau-jet$ and the missing transverse momentum in the hadronic $\tau$ decay channel in the $ep$ scattering with 10x250 (left) and 20x325 (right) energies respectively and the ratios, $z$, from table \ref{table_ratio} for $(q_i q_j)\equiv$ 13,22,23,31,32 and 33.The lepton-quark couplings are $\lambda_i=\lambda_j=0.3$.}
\end{center}
\label{fig:DPH}
\end{figure}

\begin{figure}
\begin{center}
\begin{minipage}[b]{0.49\columnwidth}
\centerline{\includegraphics[angle=90, scale=0.38]{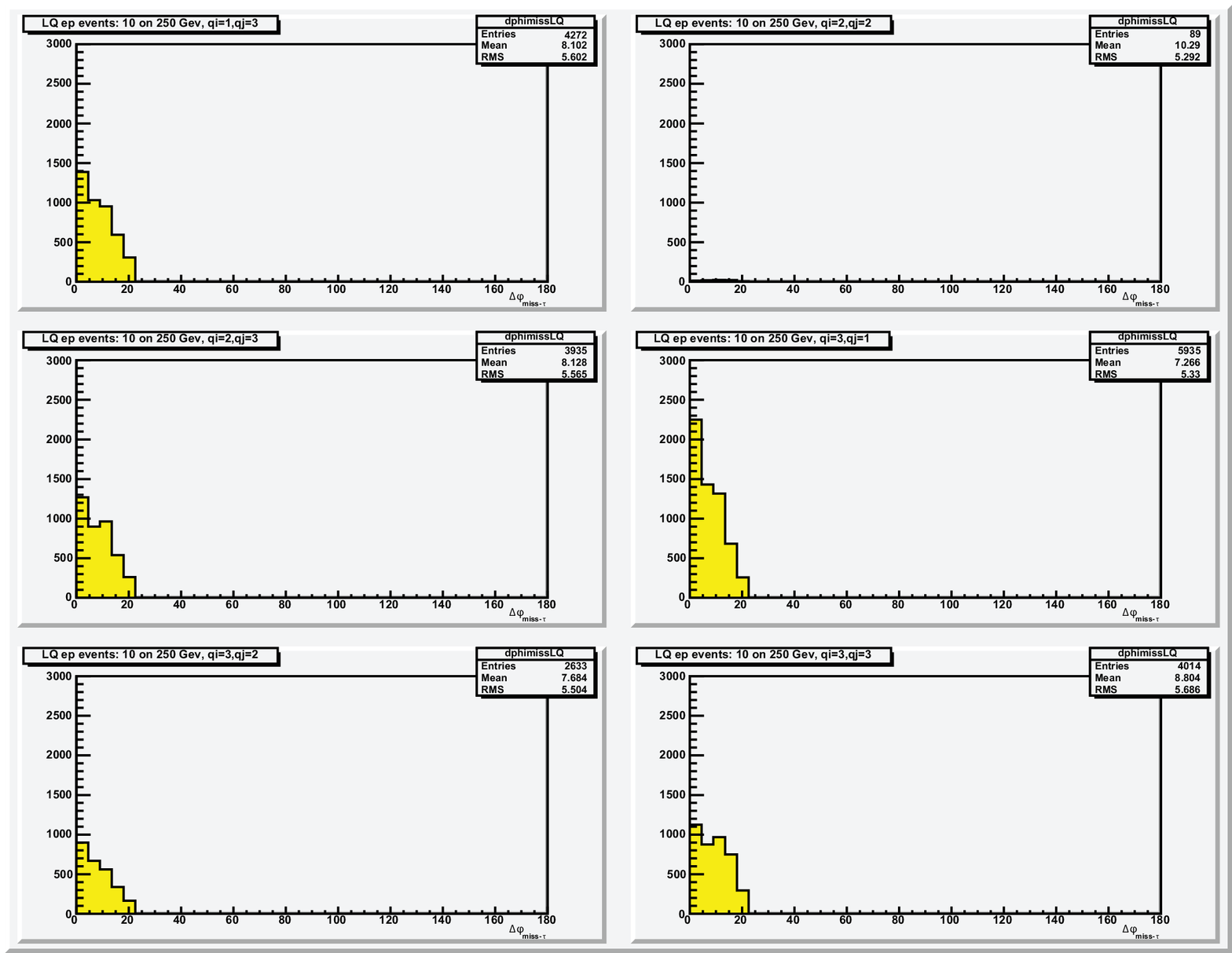}}
\end{minipage}
\begin{minipage}[b]{0.49\columnwidth}
\centerline{\includegraphics[angle=90, scale=0.38]{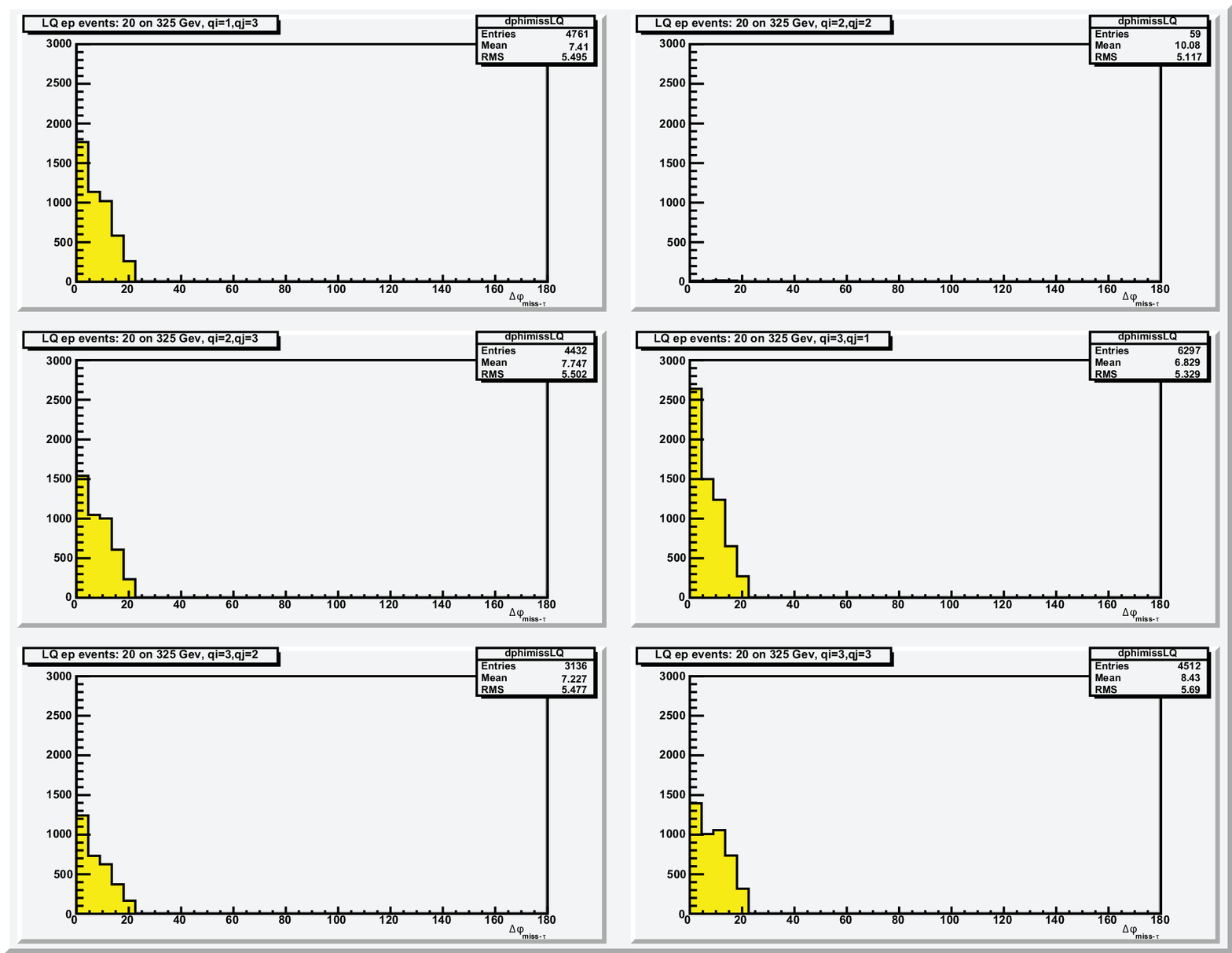}}
\end{minipage}
\caption{The acoplanarity,$~\Delta \phi_{miss-\tau jet}$, between the $\tau-jet$ and the missing transverse momentum in the hadronic $\tau$ decay channel in the $ep$ scattering with 10x250 (left) and 20x325 (right) energies respectively and the ratios, $z$, from table \ref{table_ratio} for $(q_i q_j)\equiv$ 13,22,23,31,32 and 33.  The lepton-quark couplings are $\lambda_i=\lambda_j=0.3$.  $~\Delta \phi_{miss-\tau jet}$ is required to be below $20^{\circ}$.}
\end{center}
\label{fig:DPHC}
\end{figure}

The $p_T^{miss}$ distribution for 10x250 and 20x325 are shown in figure~\ref{fig:HDPT} (left) and (right), respectively. Also plotted is the acoplanarity, $\Delta \phi_{miss-\tau jet}$, 
between the $\tau$-jet and the missing transverse momentum. $\Delta \phi_{miss-\tau jet}$ for the EIC energies 10x250 and 20x325, shown in figure~\ref{fig:DPH} (left) and (right) respectively. Figure~\ref{fig:DPHC} shows the same results as previous two figures but with an additional requirement of $\Delta \phi_{miss-\tau jet}$ below $20^{\circ}$.  A small $\Delta \phi_{miss-\tau jet}$ requirement means that the missing transverse momentum in the event, in the form of a $\tau$ neutrino, is aligned with the $\tau$ jet.  These
should be the events in which the $\tau$ decayed with neutrinos in the final state.   

\subsection{Experiment III: Concluding Remarks}\label{etau_sec:exp_concl}

We have studied the topological differences between events in the SM and a BRW-leptoquark extension of the SM. Leptoquark searches in electron-hadron machines are sensitive to the ratio of the product of coupling constant to the square of the leptoquark mass. Motivated by recent theoretical expectations first presented in~\cite{Gonderinger:2010yn} and summarized above in section~\ref{etau_sec:LQ_frame}, we have studied this for a range of leptoquark masses.  While we studied 
the topologies of leptoquark-mediated transitions between the electron and all three generations of charged leptons, we limit our comments to the LFV(1,3) transition, for now. 

\subsubsection{Observations}

The topological features of a SM event that produces a $\tau$ lepton which decays and an event in which a $\tau$ is created as a decay of a leptoquark are distinct in two different variables routinely studied in colliders. They are: 1)  $p_{T}^{miss}$ spectrum, transverse missing momentum in such an event, and 2) $\Delta\phi_{miss-\tau jet}$, defined as the transverse angle $\phi$ between the $\tau$ created in the event and the vector direction of missing momentum. 

\begin{figure}
\centerline{\includegraphics[angle=90,scale=0.4]{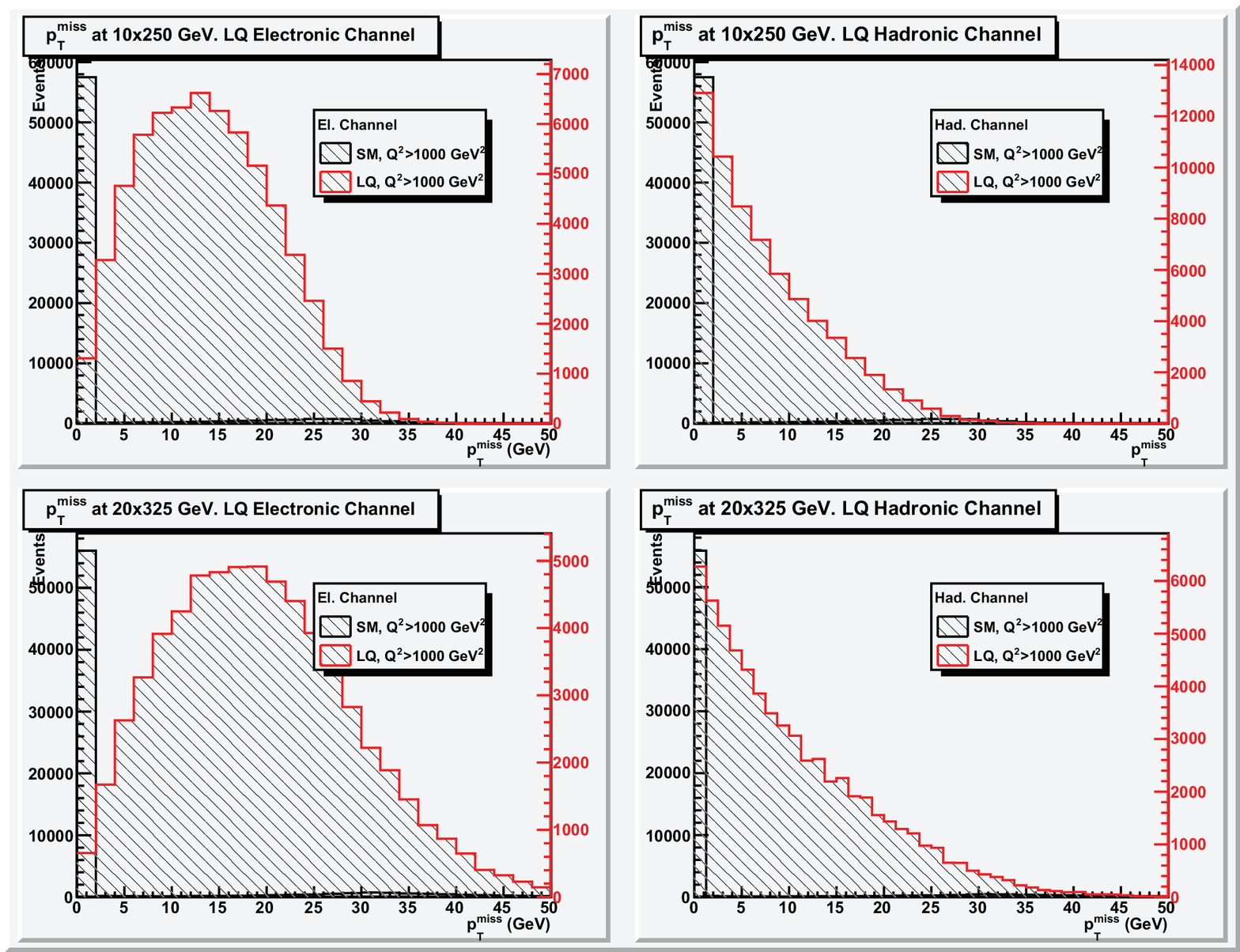}}
\caption{SM and LQ (both hadronic and electronic channels) $p_{T}^{miss}$ at 10$\times$250 and 20$\times$325 GeV with $Q^{2}>1000$ GeV$^{2}$. These plots are not normalized amongst
themselves.}
\label{fig:Overlap_PT}
\end{figure}

As shown in figure~\ref{fig:Overlap_PT}, the $p_{T}^{miss}$ spectrum is extremely narrow for SM events with final state leptonic or hadronic decays of the $\tau$ created in the collisions. This implies that the neutrinos released in the decay of the SM-produced $\tau$ are boosted in the direction of the $\tau$ and hardly any noticeable transverse momentum is lost. In contrast,
the $\tau$ produced in a leptoquark decay tends to have a larger spread in the $p_{T}^{miss}$ spectrum. These general features of the $p_{T}^{miss}$ spectra do not depend on the center-of-mass of the collision: the top histograms in figure~\ref{fig:Overlap_PT} correspond to 100 GeV center-of-mass energy, while on the bottom they correspond to 160 GeV. Note that these plots are not normalized amongst themselves.

\begin{figure}
\centerline{\includegraphics[angle=90, scale=0.4]{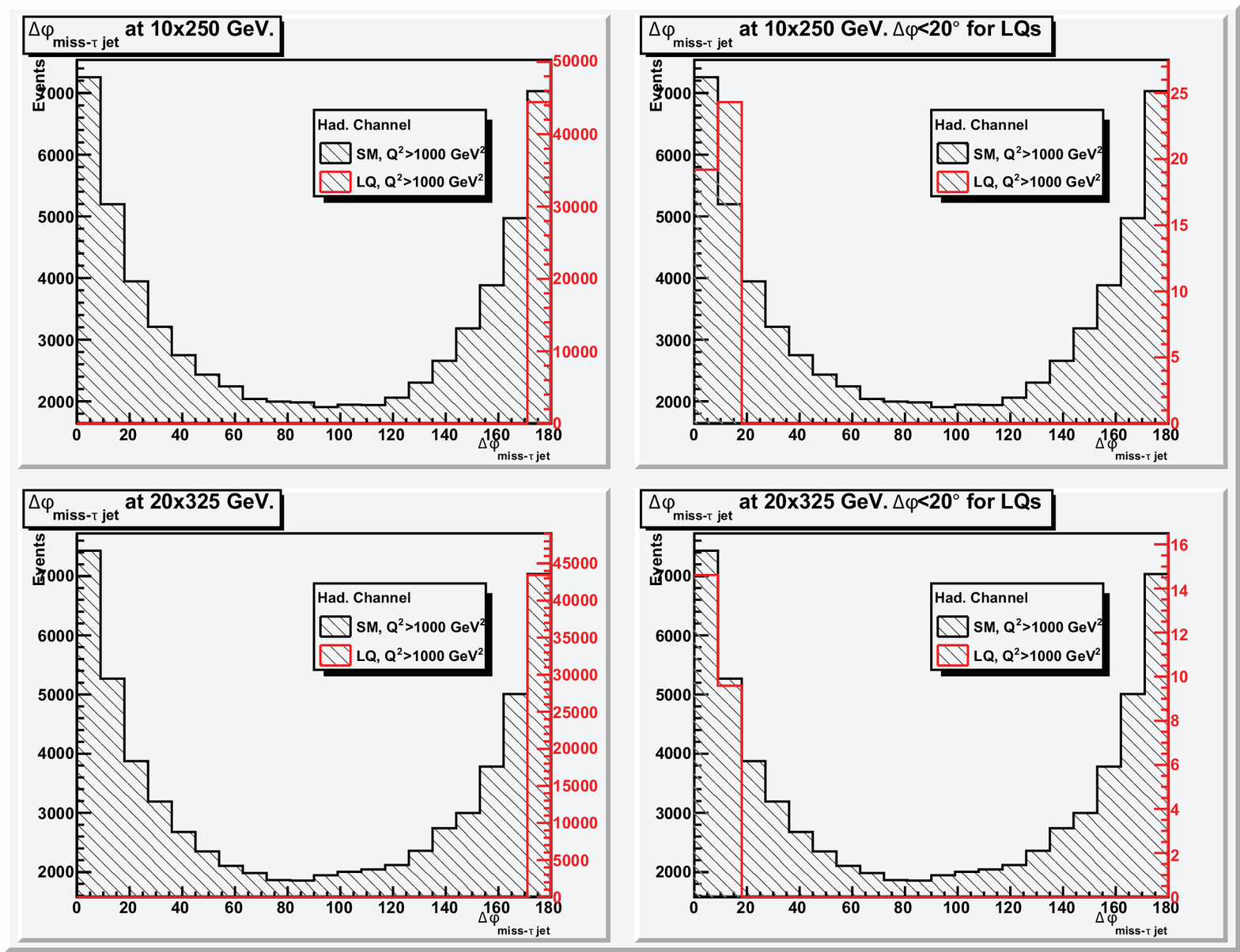}}
\caption{Acoplanarity $\Delta \phi_{miss,\tau jet}$ at 10$\times$250 and 20$\times$325 GeV with $Q^{2}>1000$ GeV$^{2}$. The two plots in the right side have a $\Delta\phi<20^{o}$ cut implemented on the LQ simulation. In this case, the LQ preserves the quark flavor: ($q_{i}=1$ $q_{j}=1$). These plots are not normalized amongst
themselves.}
\label{fig:Overlap_Phi}
\end{figure}

Figure~\ref{fig:Overlap_Phi} shows the acoplanarity plots, the angle between the $\tau$-jet in the event and the missing momentum vector reconstructed in the transverse plane using all other observable hadronic and leptonic activity (whether part of a jet or not). The left plots at each center-of-mass energy are unnormalized acoplanarity distributions showing that the SM events are distributed widely over the $\phi$ range, while the leptoquark-produced $\tau$ jets are narrowly peaked at $180^{o}$. If a proximity requirement cut of $20^{o}$ is made -- meaning that the
missing neutrinos were very close in $\phi$ angle with the direction of the generated $\tau$ -- the distribution
switches sides, indicating two categories of such events (the plots on the right of figure).  Again, the top two histograms are for 100 GeV 
center-of-mass energy and the bottom are for 160 GeV. They show no differences based on these center-of-mass energies. This feature by itself will be less deterministic of the event topology of the leptoquarks, but we expect (as was done in previous searches \cite{Aaron:2011zz}) the $p_{T}^{miss}$ spectrum and the acoplanarity distributions together will be utilized in a maximum likelihood or neural network analysis to search for
excesses seen in future EIC events.

\subsubsection{Outlook}

The study we performed is only a beginning. The estimates made in the theoretical motivation in section~\ref{etau_sec:LQ_frame} and \cite{Gonderinger:2010yn} assume a 100\% efficiency of final state leptonic and hadronic decay reconstruction of the $\tau$ created in the final state. The experimental publications \cite{Aaron:2011zz,Aktas:2007ji} indicate that their detailed simulation of the H1 detector resulted in a range of 7\% to 15\% in the reconstruction efficiency of the $\tau$ in those final states. BELLE and BaBar detectors have reported higher efficiencies reaching about 20\%. It is reasonable to assume that a future EIC detector may be able to achieve at least that. Assuming this we note
that the luminosity requirements stated in section~\ref{etau_sec:LQ_frame} and \cite{Gonderinger:2010yn} for the EIC to probe $e\rightarrow\tau$ cross sections at the level stated are
an underestimate by about 10-to-5 times. This means at 90 GeV center-of-mass, the stated 10 fb$^{-1}$ could be
as high as 100 or as low as 50 fb$^{-1}$.

The studies we performed were based on HERA studies in which the collision energies were about 300 GeV in 
the center-of-mass frame. The efficiencies of some cuts and selection criteria would certainly be better at those 
energies than at the EIC 100 and 160 GeV center-of-mass energies, so the cut efficiencies may not transfer 
exactly as it has been assumed in these estimates. However, it is not unreasonable to assume that a similar
but equally (if not more) efficient set of cuts and analysis techniques may be eventually found for this 
search at the future EIC.

The group now formed hopes to continue these studies with detailed detector simulation as it will become available in near future.

\subsection*{Acknowledgments}

The authors would like to thank William Marciano and Yingchuan Li for
useful discussions, and helpful suggestions.



\chapter{Experimental aspects}

\noindent
{\Large Chapter editors: \\[1em]
E. C. Aschenauer, R. Ent}

\newpage
\section{High-energy high-luminosity electron-ion collider eRHIC}
\label{sec:eRHIC-design}


\hspace{\parindent}\parbox{0.92\textwidth}{\slshape 
Vladimir N. Litvinenko, Joanne Beebe-Wang, Sergei
Belomestnykh, Ilan Ben-Zvi, Michael M. Blaskiewicz,
Rama Calaga, Xiangyun Chang, Alexei Fedotov, David
Gassner, Harald Hahn, Lee Hammons, Yue
Hao, Ping He, William Jackson, Animesh Jain,
Elliott C. Johnson, Dmitry Kayran, J\"org Kewisch,
Yun Luo, George Mahler, Gary McIntyre, Wuzheng
Meng, Michiko Minty, Brett Parker, Alexander
Pikin, Eduard Pozdeyev, Vadim Ptitsyn, Triveni
Rao, Thomas Roser, Brian Sheehy, John Skaritka,
Steven Tepikian, Yatming Than, Dejan Trbojevic, Evgeni
Tsentalovich, Nicholaos Tsoupas, Joseph Tuozzolo,
Gang Wang, Stephen Webb, Qiong Wu,  Wencan Xu,
Anatoly Zelenski}
%
%

\index{Litvinenko, Vladimir N.}
\index{Beebe-Wang, Joanne}
\index{Belomestnykh, Sergei}
\index{Ben-Zvi, Ilan}
\index{Blaskiewicz, Michael M.}
\index{Calaga, Rama}
\index{Chang, Xiangyun}
\index{Fedotov, Alexei}
\index{Gassner, David}
\index{Hammons, Lee}
\index{Hahn, Harald}
\index{Hao, Yue}
\index{He, Ping}
\index{Jackson, William}
\index{Jain, Animesh}
\index{Johnson, Elliott C.}
\index{Kayran, Dmitry}
\index{Kewisch, J\"org}
\index{Luo, Yun}
\index{Mahler, George}
\index{McIntyre, Gary}
\index{Meng, Wuzheng}
\index{Minty, Michiko}
\index{Parker, Brett}
\index{Pikin, Alexander}
\index{Pozdeyev, Eduard}
\index{Ptitsyn, Vadim}
\index{Rao, Triveni}
\index{Roser, Thomas}
\index{Skaritka, John}
\index{Sheehy, Brian}
\index{Tepikian, Steven}
\index{Than, Yatming}
\index{Trbojevic, Dejan}
\index{Tsentalovich, Evgeni}
\index{Tsoupas, Nicholaos}
\index{Tuozzolo, Joseph}
\index{Wang, Gang}
\index{Webb, Stephen}
\index{Wu, Qiong}
\index{Xu, Wencan}
\index{Zelenski, Anatoly}





\subsection{Introduction}

In this paper, we describe a future electron-ion collider (EIC), based on the existing Relativistic Heavy Ion Collider (RHIC) hadron facility, with two intersecting superconducting rings, each 3.8 km in circumference~\cite{VL:ref1}. The replacement cost of the RHIC facility is about two billion US dollars, and the eRHIC will fully take advantage and utilize this investment. We plan adding a polarized 5-30 $\gev$ electron beam to collide with variety of species in the existing RHIC accelerator complex, from polarized protons with a top energy of 325 $\gev$, to heavy fully-stripped ions with energies up to 130 $\gev /u$. 

BrookhavenÕs innovative design, (figure 1), is based on one of RHICÕs hadron rings and a multi-pass energy-recovery linac (ERL). Using the ERL as the electron accelerator assures high luminosity in the $10^{33}$-$10^{34} cm^{-2} sec^{-1}$ range, and for the natural staging of eRHIC, with the ERL located inside the RHIC tunnel.  eRHIC will provide electron-hadron collisions in up to three interaction regions. We detail eRHICÕs performance in subsection~\ref{VL:subsct2}.

Since the first paper on eRHIC in 2000, its design has undergone several iterations. Initially, the main eRHIC option (the so-called ring-ring, RR, design) was based on an electron ring, with the linac-ring (LR) option as a backup. In 2004, we published the detailed ``eRHIC 0th-Order Design Report"  including a cost-estimate for the RR design~\cite{VL:ref2}. After detailed studies, we found that an LR eRHIC has about a 10-fold higher luminosity than the RR. Since 2007, the LR, with its natural staging strategy and full transparency for polarized electrons, became the main choice for eRHIC. In 2009, we completed technical studies of the design and dynamics for MeRHIC with 3-pass 4-$\gev$ ERL. We learned much from this evaluation, completed a bottom-up cost estimate for this \$350M machine, but then shelved the design.

\begin{figure}[htb]
\centering
(a) \hspace{7.5cm} (b)
\includegraphics[width=0.48\textwidth]{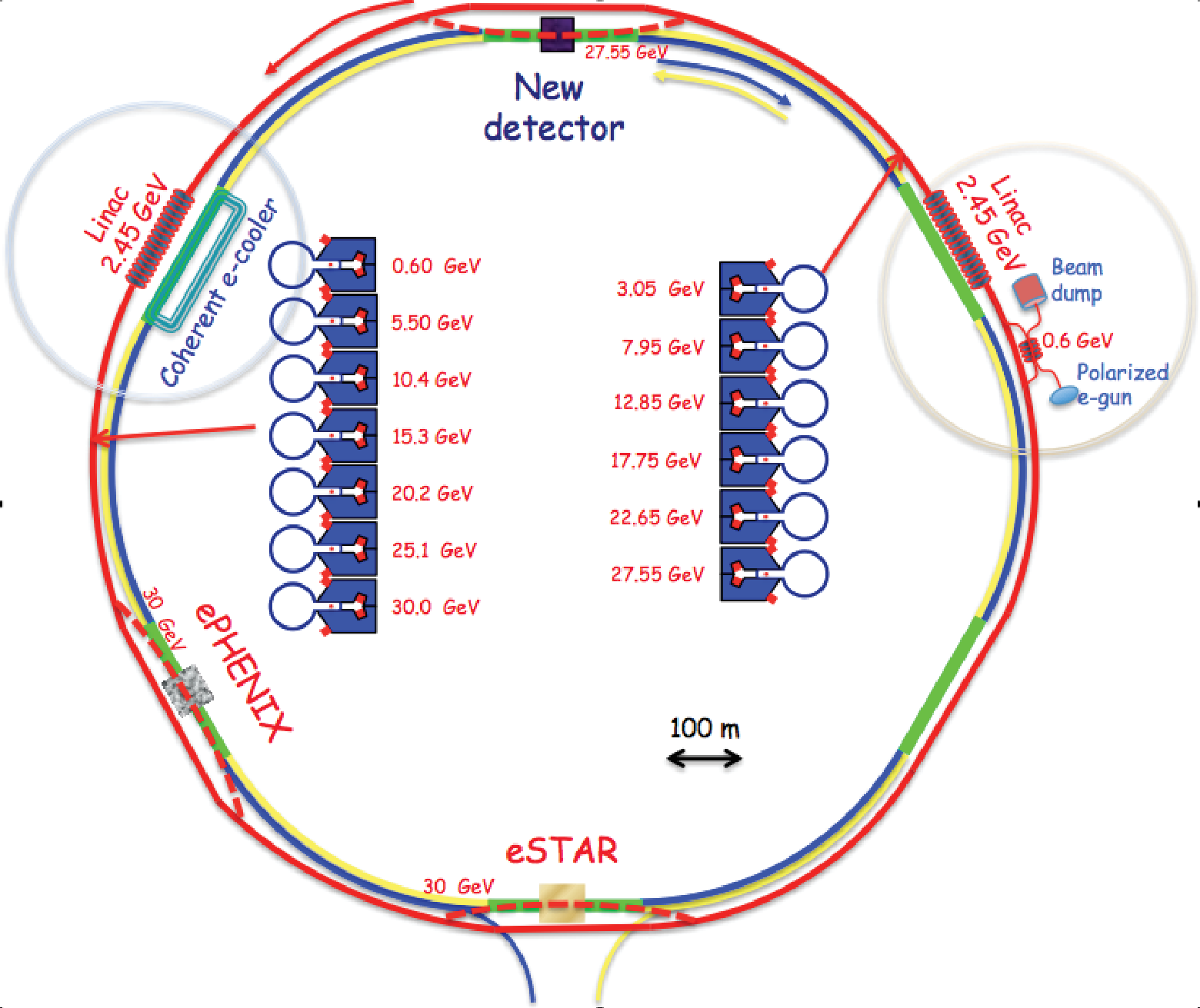}
\hfill
\includegraphics[width=0.48\textwidth]{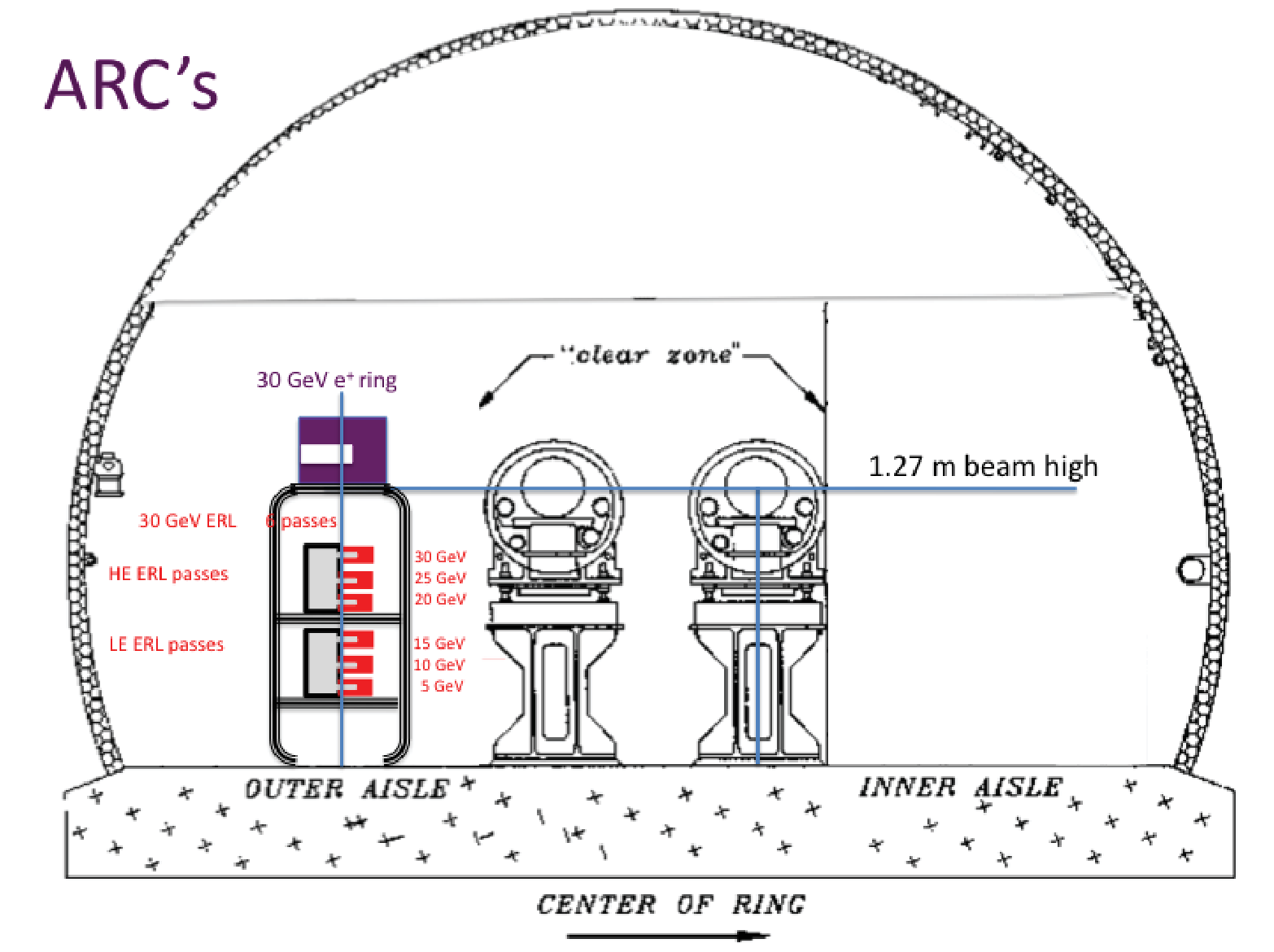}

\caption{\small (a) Layout of the ERL-based, all-in-RHIC-tunnel, 30 GeV x 325 GeV high-energy high-luminosity eRHIC. (b) Location of eRHICÕs six recirculation arcs in the RHIC tunnel.}
\label{VL:fig1}

\end{figure}

In the same year, we turned again to considering the cost-effective, all-in-tunnel six-pass ERL for our design of the high-luminosity eRHIC (figure \ref{VL:fig1}). In it, electrons from the polarized pre-injector will be accelerated to their top energy by passing six times through two SRF linacs. After colliding with the hadron beam in up to three detectors, the e-beam will be decelerated by the same linacs and dumped. The six-pass magnetic system with small-gap magnets~\cite{Hao:2010zzd} will be installed from the start. We will stage the electron energy from 5 $\gev$ to 30 $\gev$ stepwise by increasing the lengths of the SRF linacs. 

We considered several IR designs for eRHIC. The latest one, with a 10 mrad crossing angle and $\beta^*= 5 cm$, takes advantage of newly commissioned $Nb_3Sn$ quadrupoles~\cite{Caspi:2010zz}. Subsection \ref{VL:subsct4} details the eRHIC lattice and the IR layout.

The current eRHIC design focuses on electron-hadron collisions. If justified by the EIC physics, we will add a 30 $\gev$ polarized positron ring with full energy-injection from the eRHIC ERL. This addition to the eRHIC facility provides for positron-hadron collisions, but at a significantly lower luminosity than those attainable in the electron-hadron mode.

 As a novel high-luminosity EIC, eRHIC faces many technical challenges, such as generating 50~mA of polarized electron current. eRHIC also will employ coherent electron cooling (CeC)~\cite{Litvinenko:2009zz}  for the hadron beams. Staff at BNL, JLab, and MIT are pursuing vigorously an R\&D program for resolving addressing these obstacles. In collaboration with Jlab, BNL plans experimentally to demonstrate CeC at the RHIC. We discuss the structure and the status of the eRHIC R\&D in subsection \ref{VL:subsct5} .

\subsection{Main eRHIC parameters} \label{VL:subsct2}

eRHIC is designed to collide electron beams with energies from 5- to 30-GeV\footnote{ 
There is no accelerator problem with using lower energy electron beams.. According 
to statements from EIC physicists, using electron energies below 5 GeV would not 
contribute significantly to the physics goals. } with hadrons, viz., either with 
heavy ions with energies from 50- to 130-GeV per nucleon, or with polarized protons 
with energies between 100- and 325-GeV. Accordingly, eRHIC will cover the C.M. 
energy range from 44.7- to 197.5-GeV for polarized e-p, and from 31.6- to 125-GeV 
for electron heavy-ion-collisions.

Several physics and practical considerations influenced our choice of beam parameters 
for eRHIC. Some of these limitations, such as the intensity of the hadron beam, the space charge and beam-beam tune shift limits for hadrons, come from experimental 
observations at RHIC or other hadron colliders. Some of them, for example \ensuremath{\beta}* 
= 5 cm for hadrons, are at the limits of current accelerator technology, while 
others are derived either from practical or cost considerations. For example, from considering the operational costs, we limit the electron beam's power loss for synchrotron radiation to about 7 MW, corresponding to a 50 mA beam 
current at 20 GeV. Above 20 GeV, the electron beam's current will decrease in inverse 
proportion to the fourth power of energy, and will be restricted to about 10 mA 
at an energy of 30 GeV. It means that the luminosity of eRHIC operating with 30 GeV 
electrons will be a 1/5th of that with 20 GeV. 

Since the ERL provides fresh electron bunches at every collision, the electron beam 
can be strongly abused, i.e., it can be heavily distorted during a collision. The only 
known effect that might cause a serious problem is the so-called kink instability. 
The ways of suppressing it within range of parameters accessible by eRHIC is well-understood~\cite{Hao:2010zzc} 
and it no longer presents a problem. 

We list below some of our assumed limits and parameters: 

{
\parindent=20pt
1. Bunch-intensity limits:

{\parindent=45pt
a. For protons: 2~10\textsuperscript{11}

b. For Au ions: 1.2~10\textsuperscript{9}
}

2. Electron-current limits:

{\parindent=45pt
a. Polarized current: 50~mA

b. Un-polarized current: 250~mA
}

3. Minimum \ensuremath{\beta}* = 5 cm for all species

4. Space-charge tune shift for hadrons: $\leq$0.035

5. Proton (ion) beam-beam parameter: $\leq$0.015

6. Bunch length (with coherent electron-cooling):

{\parindent=45pt
a. Protons: 8.3 cm at energies below 250~GeV, 4.9 cm at 325~GeV 

b. Au ions: 8.3 cm in all energy ranges
}

7. Synchrotron radiation intensity limit is defined as that of a 50~mA beam at 
20~GeV

8. Collision rep-rate $\leq$ 50~MHz.
}

The limitations on luminosity resulting from various considerations are involved. 
The main trend is that eRHIC's luminosity \emph{does not depend on the electron 
beam's energy (below 20~GeV)} and reaches its maximum at the hadron beam's highest 
energy. We mentioned the exception for energies of electrons above 20~GeV. The top eRHIC performance for various species is shown in table \ref{VL:tbl_lumin_table}. 

\begin{table}
\centering
\begin{tabular}{|p{5.7cm}|c|c|c|c|c|}
\hline
  & { e } & { p} &  \textsuperscript{{ 2}}{ He}\textsuperscript{{ 3}}  & \textsuperscript{{ 79}}{ Au}\textsuperscript{{ 197}}  & \textsuperscript{{ 92}}{ U}\textsuperscript{{ 238}}  \\ \hline
 Energy, GeV  & { 5-20} & { 325} & { 215} & { 130} & { 130} \\  \hline
CM energy, GeV & ~ & { 80-161} & { 131} & { 102} & { 102} \\ \hline
Number of bunches or distance between bunches & { 74 nsec } & { 166} & { 166} & 
{ 166} & { 166}\\ \hline
Bunch intensity (nucleons), $10^{11}$  & 0.24 & 2 & 3 & 3 & 3.15 \\ \hline
Bunch charge, nC &  { 3.8 } & { 32} & { 30} & { 19} & { 20} \\ \hline
Beam current, mA & { 50 } & { 420} & { 390} & { 250} & { 260} \\ \hline
Normalized emittance of hadrons 95\% , $mm\cdot mrad$ & ~ & { 1.2} &{ 1.2} &{ 1.2} &{ 1.2}\\ \hline
Normalized emittance of electrons, rms, $mm\cdot mrad$ & {  } & { 5.8-23 } & { 7-35 } &{ 12-57} & { 12-57} \\ \hline
Polarization, \% & { 80} & { 70} & { 70} & { none} & { none}\\ \hline
RMS bunch length, cm & { 0.2} &{ 4.9} & { 8.3} &{ 8.3} & { 8.3}\\\hline
$\beta$*, cm & { 5} & { 5} &{ 5} &{ 5} &{ 5}\\ \hline
Luminosity per nucleon, $10^{34}$ cm\textsuperscript{-2}s\textsuperscript{-1}  & {  } & 1.46 & 1.39 & 0.86 &  0.92 \\ \hline
\end{tabular}
\caption{Projected eRHIC luminosity for various hadron beams at top energy.}
\label{VL:tbl_lumin_table}
\end{table}

Table \ref{VL:tbl_pol_ep} lists the luminosity of a polarized electron-proton collision for a set 
of electron- and proton-energies.  Table \ref{VL:tbl_pol_eAu} contains this information for a polarized electron beam colliding with Au ions, 
while tables \ref{VL:tbl_unpol_ep} and \ref{VL:tbl_unpol_eAu} provide data for the case of unpolarized electrons.

\begin{table}
\centering
\begin{tabular}{|c|c|c|c|c|}
\hline
\backslashbox{Electrons}{Protons} & 100 GeV & 130 GeV & 250 GeV & 325 GeV \\ \hline
5 GeV & 0.62\textsuperscript{.}10\textsuperscript{33} & 1.4\textsuperscript{.}10\textsuperscript{33} & 9.7\textsuperscript{.}10\textsuperscript{33} & 15\textsuperscript{.}10\textsuperscript{33} \\ \hline
10 GeV & 0.62\textsuperscript{.}10\textsuperscript{33} & 1.4\textsuperscript{.}10\textsuperscript{33} & 9.7\textsuperscript{.}10\textsuperscript{33} & 15\textsuperscript{.}10\textsuperscript{33}\\ \hline
20 GeV & 0.62\textsuperscript{.}10\textsuperscript{33} & 1.4\textsuperscript{.}10\textsuperscript{33} &9.7\textsuperscript{.}10\textsuperscript{33} & 1.5\textsuperscript{.}10\textsuperscript{33}\\ \hline
30 GeV & 0.12\textsuperscript{.}10\textsuperscript{33} &  0.3\textsuperscript{.}10\textsuperscript{33} & 1.9\textsuperscript{.}10\textsuperscript{33} & 3\textsuperscript{.}10\textsuperscript{33}\\ \hline
\end{tabular}
\caption{Projected eRHIC luminosity (in cm\textsuperscript{-2} sec\textsuperscript{-1}) 
for \emph{polarized electron-proton collisions}. }
\label{VL:tbl_pol_ep}
\end{table}

\begin{table}
\centering
\begin{tabular}{|c|c|c|c|c|}
\hline
\backslashbox{Electrons}{Au ions} & 50 GeV/u & 75 GeV/u & 100 GeV/u & 130 GeV/u\\ \hline
5 GeV & 0.49\textsuperscript{.}10\textsuperscript{33} & 1.7\textsuperscript{.}10\textsuperscript{33} & 3.9\textsuperscript{.}10\textsuperscript{33} & 8.6\textsuperscript{.}10\textsuperscript{33}\\ \hline
10 GeV & 0.49\textsuperscript{.}10\textsuperscript{33} & 1.7\textsuperscript{.}10\textsuperscript{33} & 3.9\textsuperscript{.}10\textsuperscript{33} & 8.610\textsuperscript{33} \\ \hline
20 GeV & 0.49\textsuperscript{.}10\textsuperscript{33} & 1.710\textsuperscript{33} & 3.9\textsuperscript{.}10\textsuperscript{33} & 8.6\textsuperscript{.}10\textsuperscript{33}\\ \hline
30 GeV & 0.1\textsuperscript{.}10\textsuperscript{33} & 0.34\textsuperscript{.}10\textsuperscript{33} & 0.8\textsuperscript{.}10\textsuperscript{33} & 1.7\textsuperscript{.}10\textsuperscript{33}\\ \hline
\end{tabular}
\caption{Projected eRHIC luminosity (in cm\textsuperscript{-2} sec\textsuperscript{-1}) for \emph{polarized electrons and Au ions}. }
\label{VL:tbl_pol_eAu}
\end{table}

\begin{table}
\centering
\begin{tabular}{|c|c|c|c|c|}
\hline
\backslashbox{Electrons}{Protons}  & 100 GeV & 130 GeV & 250 GeV & 325 GeV\\ \hline
5 GeV & 3.1\textsuperscript{.}10\textsuperscript{33} & 5\textsuperscript{.}10\textsuperscript{33} & 9.7\textsuperscript{.}10\textsuperscript{33} & 15\textsuperscript{.}10\textsuperscript{33}\\ \hline
10 GeV & 3.1\textsuperscript{.}10\textsuperscript{33} & 5\textsuperscript{.}10\textsuperscript{33} & 9.7\textsuperscript{.}10\textsuperscript{33} & 15\textsuperscript{.}10\textsuperscript{33}\\ \hline
20 GeV & 0.62\textsuperscript{.}10\textsuperscript{33} & 1.4\textsuperscript{.}10\textsuperscript{33} & 9.7\textsuperscript{.}10\textsuperscript{33} & 15\textsuperscript{.}10\textsuperscript{33}\\ \hline
30 GeV & 0.12\textsuperscript{.}10\textsuperscript{33} & 0.3\textsuperscript{.}10\textsuperscript{33} & 1.9\textsuperscript{.}10\textsuperscript{33} & 3\textsuperscript{.}10\textsuperscript{33}\\ \hline
\end{tabular}
\caption{Projected eRHIC luminosity (in cm\textsuperscript{-2} sec\textsuperscript{-1}) for  \emph{polarized protons and unpolarized electrons}. }
\label{VL:tbl_unpol_ep}
\end{table}

An additional major parameter describing eRHIC's overall performance is its expected 
average luminosity. Since the plans for eRHIC are to use coherent electron cooling 
to control the parameters of hadron beam, its lifetime will be affected only by 
scattering on residual gas, and by burn-off in collisions with electrons. Hence, 
the hadron beam's luminosity lifetime could be as long as a few days, and, in the 
most likely scenario, the average delivered luminosity will be determine by the 
reliability of RHIC systems. Hence we anticipate that the average luminosity will 
be $\sim 70\% $ of that listed in the tables.

\begin{table}
\centering
\begin{tabular}{|c|c|c|c|c|}
\hline
\backslashbox{Electrons}{Au ions} & 50 GeV/u & 75 GeV/u & 100 GeV/u & 130 GeV/u\\ \hline
5 GeV & 2.5\textsuperscript{.}10\textsuperscript{33} & 8.3\textsuperscript{.}10\textsuperscript{33} & 11.4\textsuperscript{.}10\textsuperscript{33} & 18\textsuperscript{.}10\textsuperscript{33}\\ \hline
10 GeV & 2.5\textsuperscript{.}10\textsuperscript{33} & 8.3\textsuperscript{.}10\textsuperscript{33} & 11.4\textsuperscript{.}10\textsuperscript{33} & 18\textsuperscript{.}10\textsuperscript{33}\\ \hline
20 GeV & 0.49\textsuperscript{.}10\textsuperscript{33} & 1.710\textsuperscript{33} & 3.9\textsuperscript{.}10\textsuperscript{33} & 8.6\textsuperscript{.}10\textsuperscript{33}\\ \hline
30 GeV & 0.1\textsuperscript{.}10\textsuperscript{33} & 0.34\textsuperscript{.}10\textsuperscript{33} & 0.8\textsuperscript{.}10\textsuperscript{33} & 1.7\textsuperscript{.}10\textsuperscript{33}\\ \hline
\end{tabular}
\caption{Projected eRHIC luminosity (in cm\textsuperscript{-2} sec\textsuperscript{-1}) 
for \emph{unpolarized electrons and Au ions}. }
\label{VL:tbl_unpol_eAu}
\end{table}

\subsection{The eRHIC interaction region}  \label{VL:subsct4}

\begin{figure}[htbp]
\centering
\includegraphics[width=0.85\textwidth]{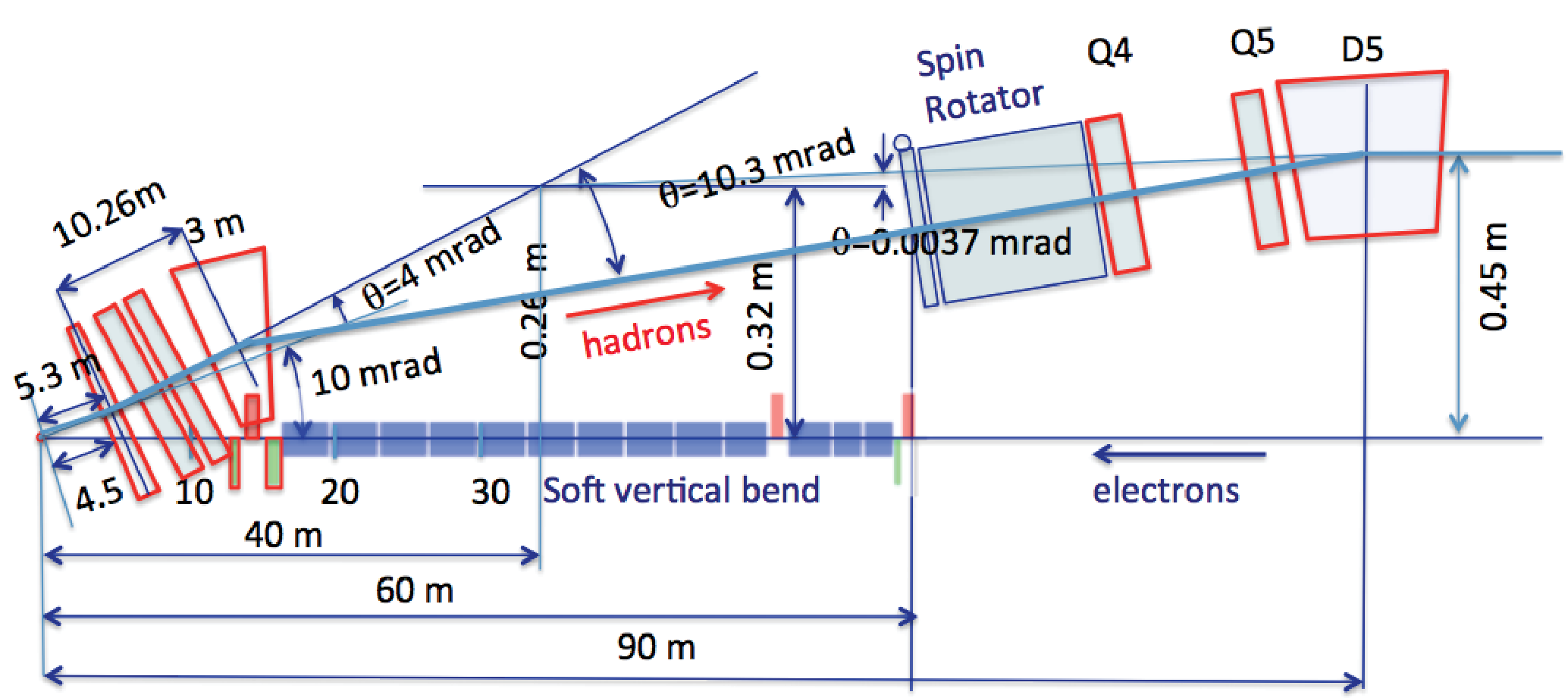}
\caption{\small Layout of the right side of eRHIC IR from the IP to the RHIC arc. The 
spin rotator is the first element of existing RHIC lattice remaining in place in 
this IR design.}
\label{VL:fig_IR_layout}

\end{figure}

The current high-luminosity eRHIC IR design incorporates a 10~mrad crab-crossing scheme; 
thus, hadrons traverse the detector at a 10~mrad horizontal angle, while electrons 
go straight through. Figure \ref{VL:fig_IR_layout} plots this scheme. The hadron beam is focused to $\beta$*=5cm 
by a special triplet wherein the first magnet is a combined function magnet (1.6~m 
long with 2.23~T magnetic fields and a -109~T/m gradient). It has two functions; 
it focuses the hadron beam while bending it 4~mrad. Two other quadrupoles do not 
bend the hadron beam but serve only for focusing. Importantly, all three magnets 
provide zero magnetic fields along the electron beam's trajectory. Quadrupoles 
for this IR require very high gradients, and can be built only with modern superconducting 
technology \cite{Caspi:2010zz,VL:ref15}

\begin{figure}[h]
\centering
(a) \hspace{7.5cm} (b)
\includegraphics[width=0.48\textwidth]{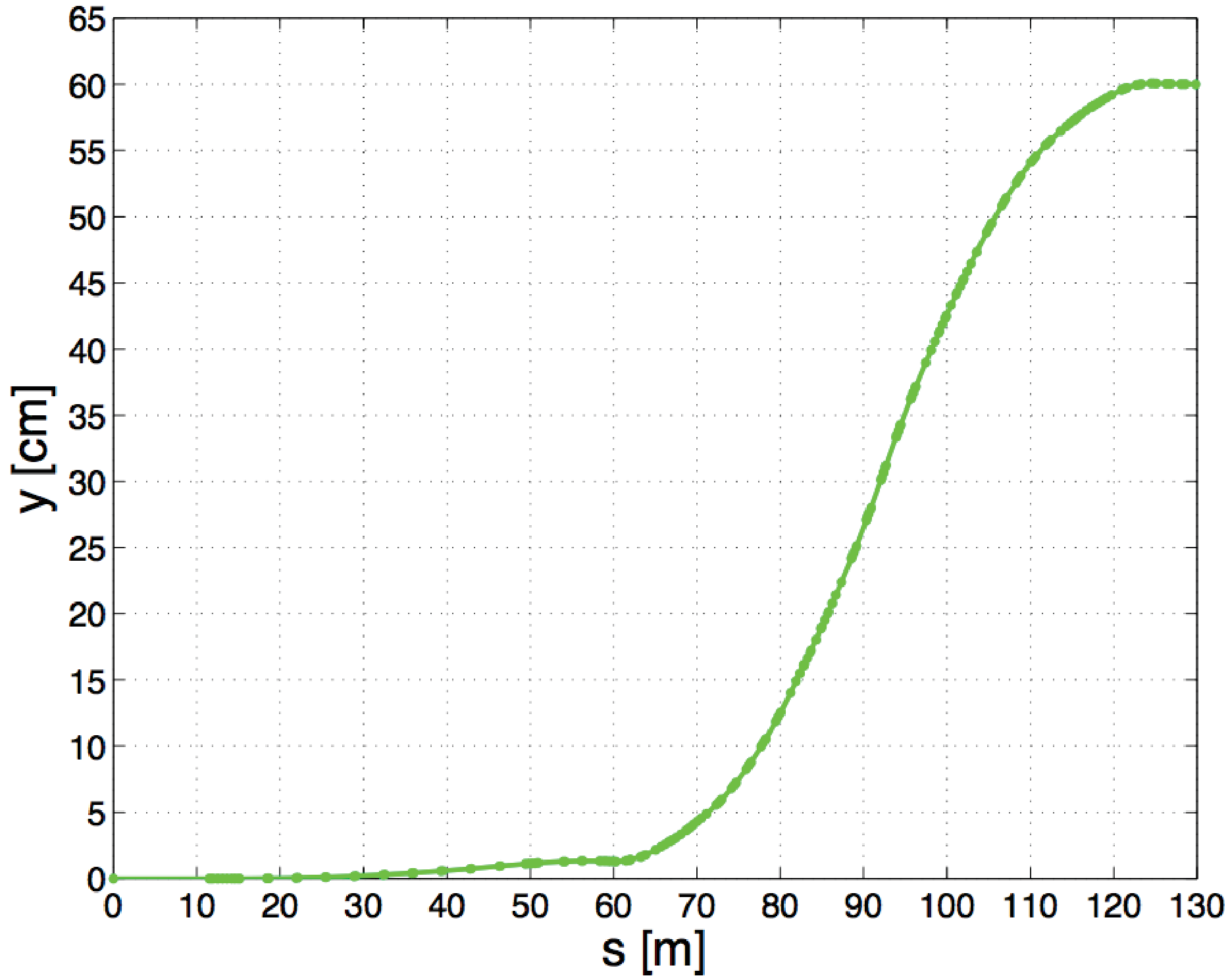}
\hfill
\includegraphics[width=0.48\textwidth]{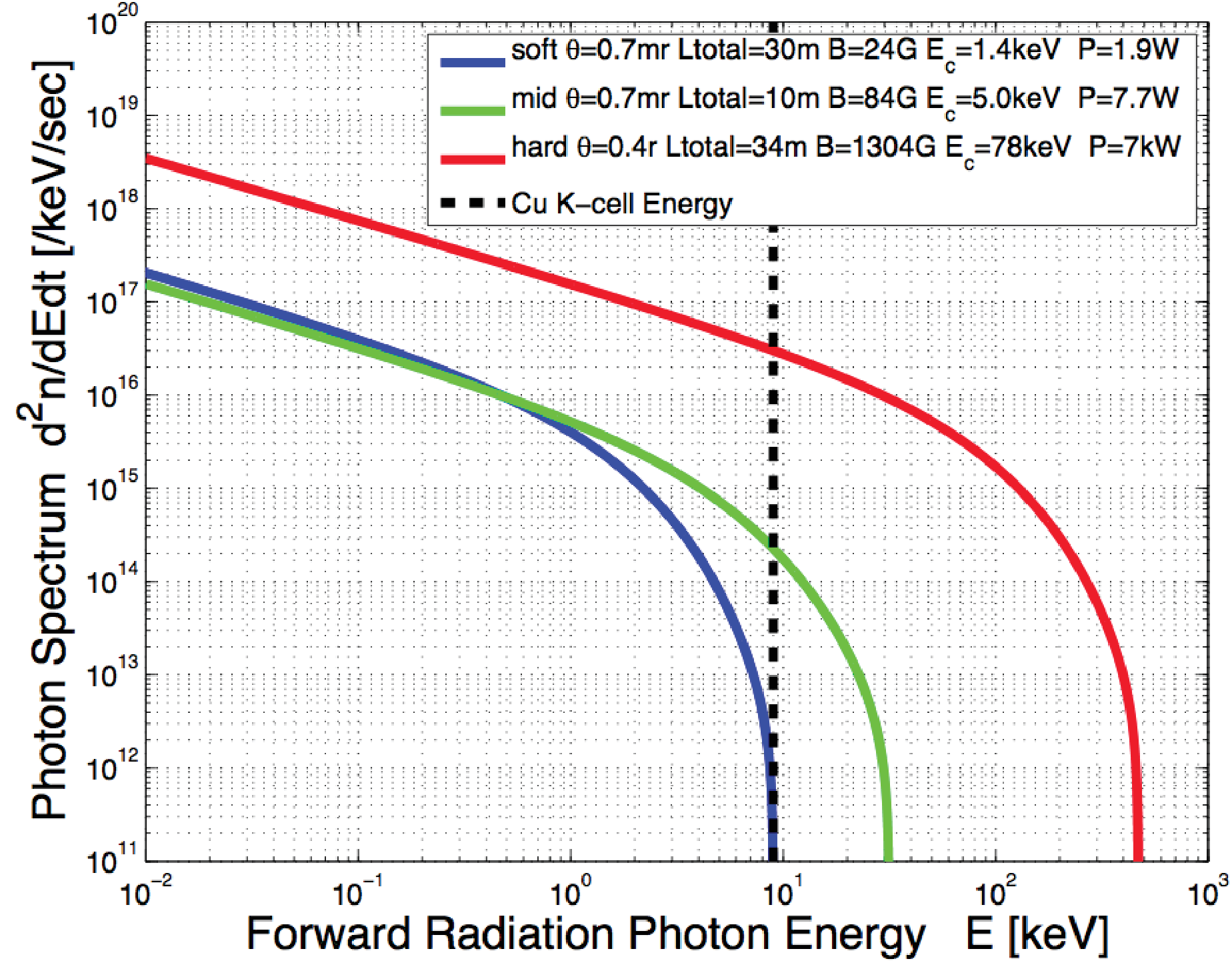}

\caption{\small (a) Vertical trajectory of 30 GeV electron beam merging over 130 meters 
into the IP. (b) Spectra of the radiation from various part of the merger. Only 
1.9 W of soft X-ray radiation will propagate through the detector; the absorbers 
intercept the rest of it. }
\label{VL:fig_IR_merging}

\end{figure}

This configuration guaranties the absence of harmful high-energy X-rays from 
synchrotron radiation. Furthermore, the electron beam is brought into the collision 
via a 130~meter long merging system (figure  \ref{VL:fig_IR_merging}). The radiation from regular bending 
magnets would be absorbed. The last 60~meters of the merging system use only soft 
bends: downwards magnets have strength of 84~Gs ( for 30~GeV beam ) and the final 
part of the bend used only 24~Gs magnetic field. Only $1.9$~W of soft radiation from 
the later magnets would propagate through the detector.

\begin{figure}[h]
\centering
(a) \hspace{7.5cm} (b)
\includegraphics[width=0.44\textwidth]{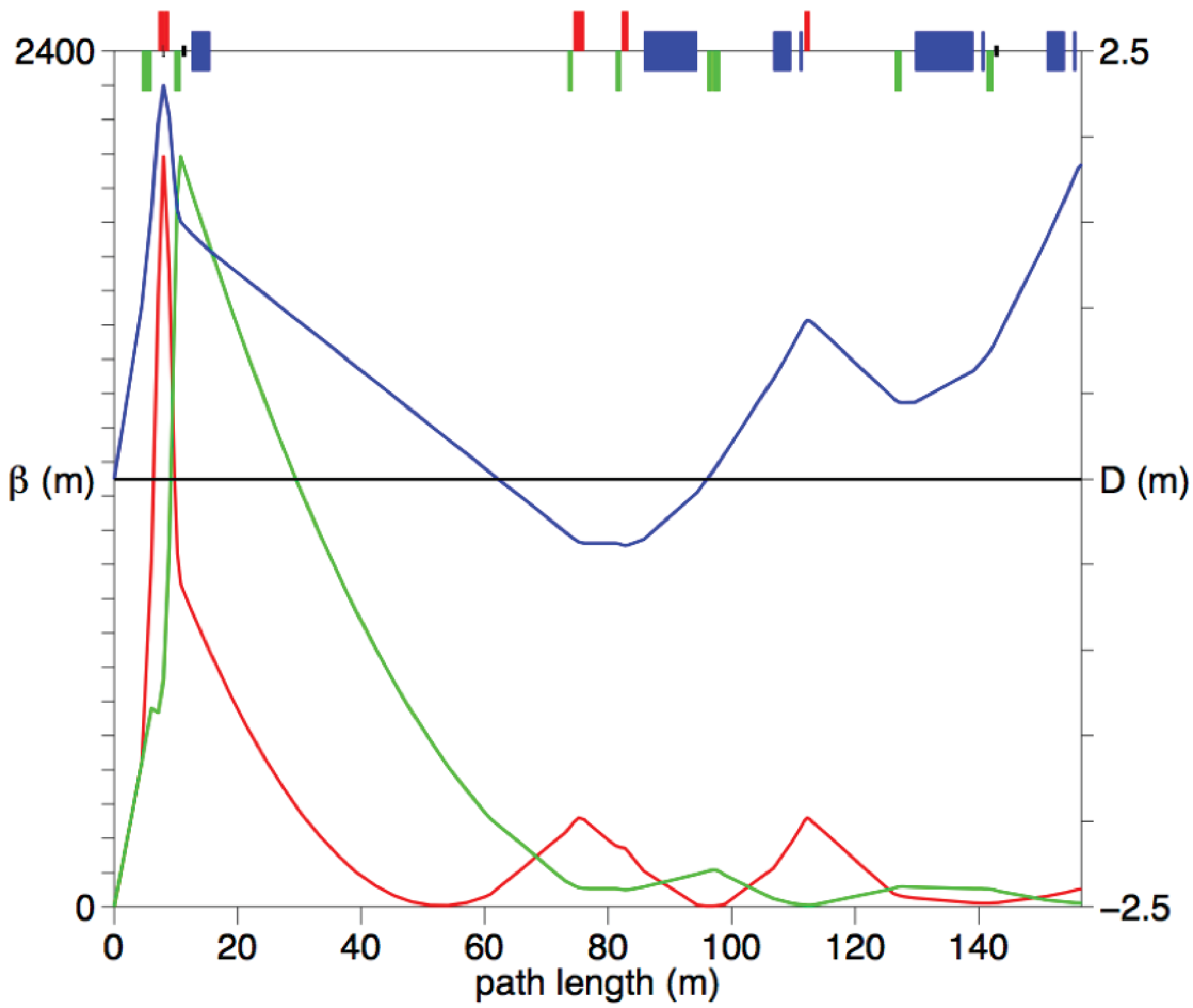}
\hfill
\includegraphics[width=0.33\textwidth]{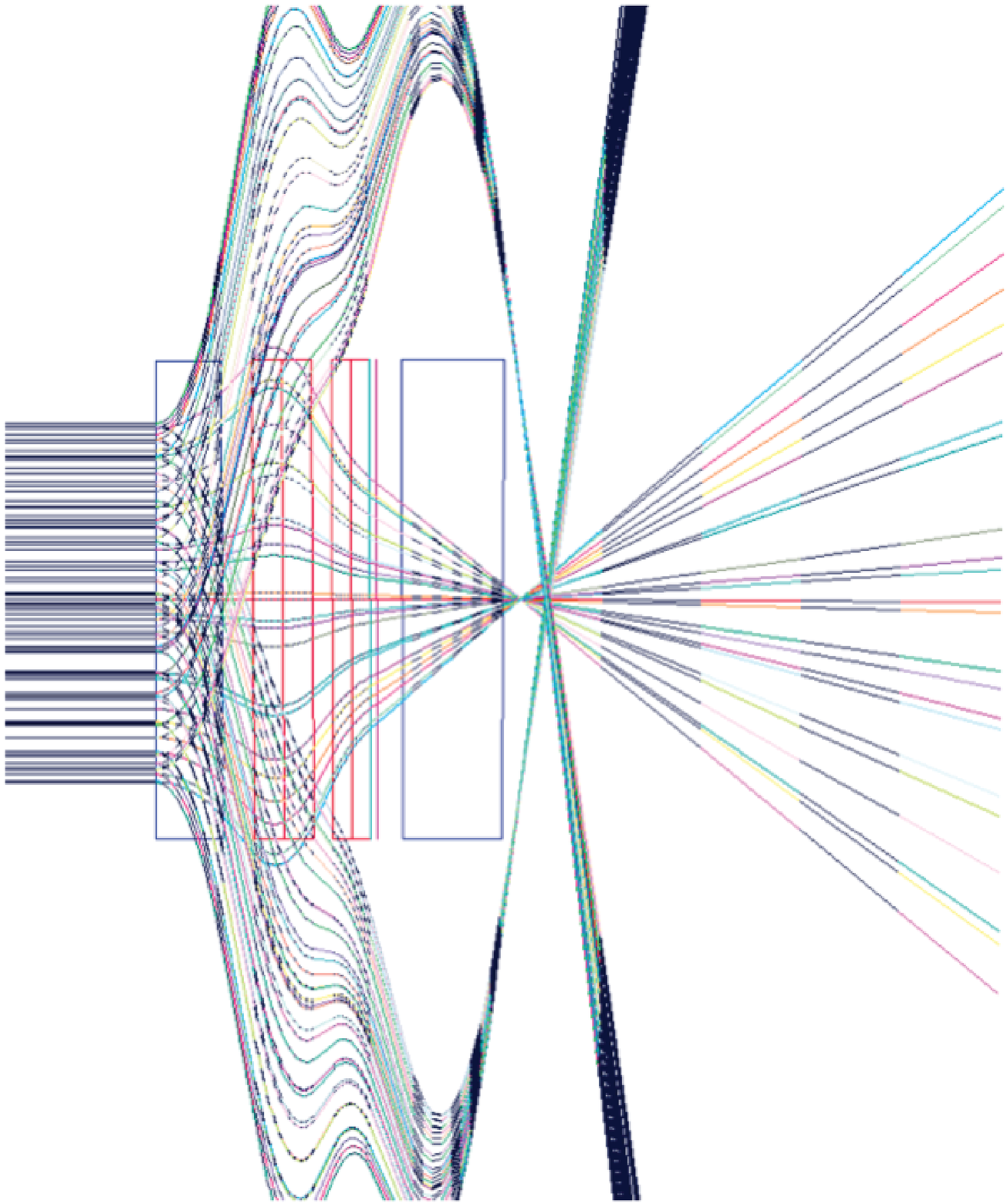}

\caption{\small (a) Hadron beam's optics at the eRHIC IR. The 5~cm $\beta$* is matched 
into the RHIC's arc lattice that starts about 60~m from the IR. (b) Tracking of 
hadrons with an energy deviation of $\pm  0.1\%$ through the first four magnets at 
the IR. }
\label{VL:fig_IR_optics}

\end{figure}

One important factor in the IR design with low $\beta$*=5 cm is that the chromatism 
of the hadron optics in the IR should be controlled, which is reflected in the 
maximum $\beta$-function of the final focusing quadrupoles. Figure \ref{VL:fig_IR_optics}a  shows the 
designed $\beta$- and dispersion-functions for the hadron beam. The values of the $\beta$-function 
are kept under 2~km, and the chromaticity held at the level typical for RHIC operations 
with $\beta^* \sim 1$ m. We are starting full-fledged tracking of hadron 
beams in RHIC, including characterizing beam-beam effects and all known nonlinearities 
of RHIC magnets: we do not anticipate any serious chromatic effects originating 
from our IR design.

Furthermore, we introduced the bending field in the first quadrupole for the hadrons 
thereby to separate the hadrons from the neutrons. Physicists considering processes 
of interest for EIC science requested our installing this configuration.  Since the electrons are used only once, the optics for them is much less constrained.  Hence, it does not present any technical- or scientific-challenges, and so we omit 
its description here. 

\begin{figure}[htbp]
\centering
(a) \hspace{7.5cm} (b)
\includegraphics[width=0.48\textwidth]{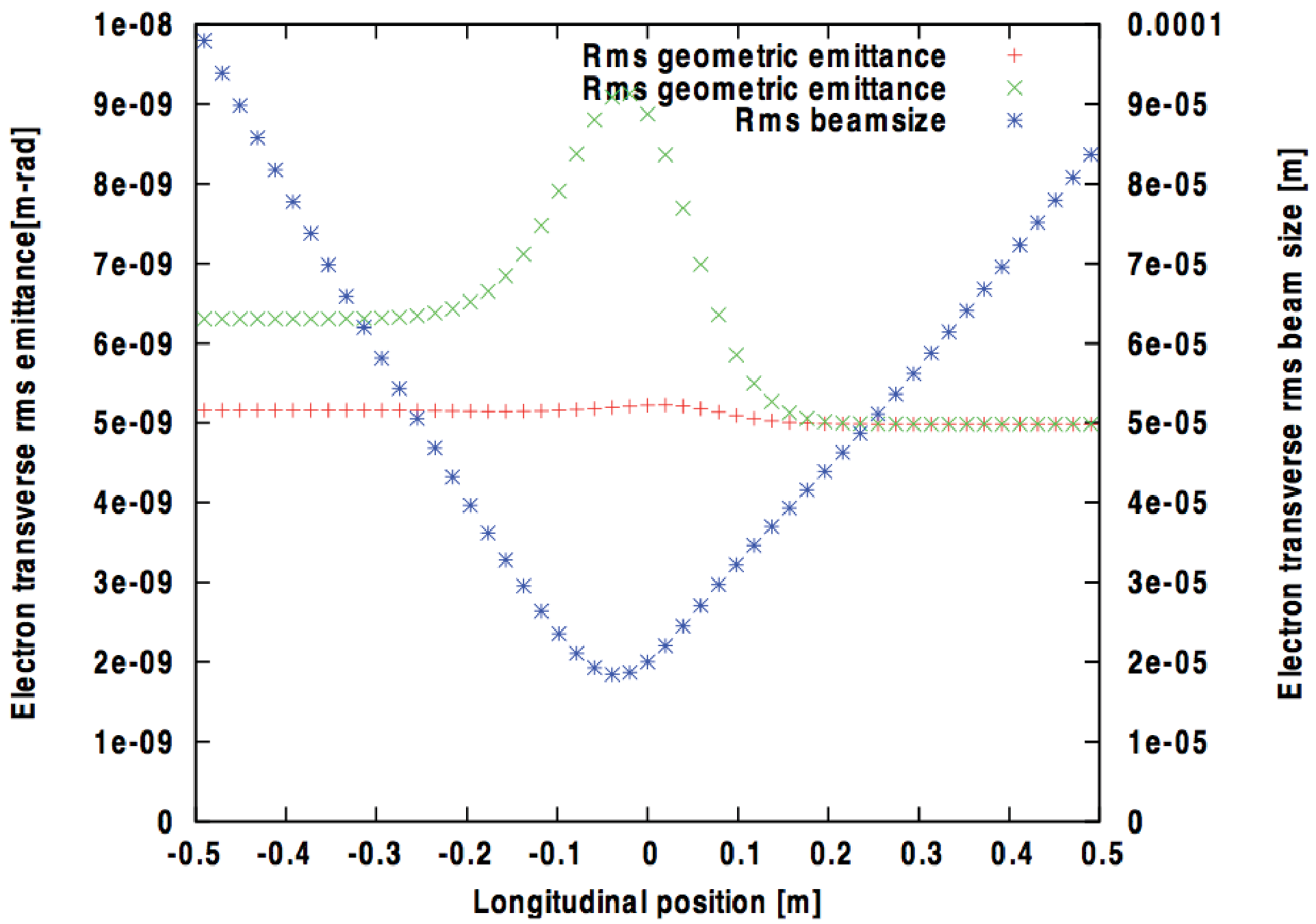}
\hfill
\includegraphics[width=0.48\textwidth]{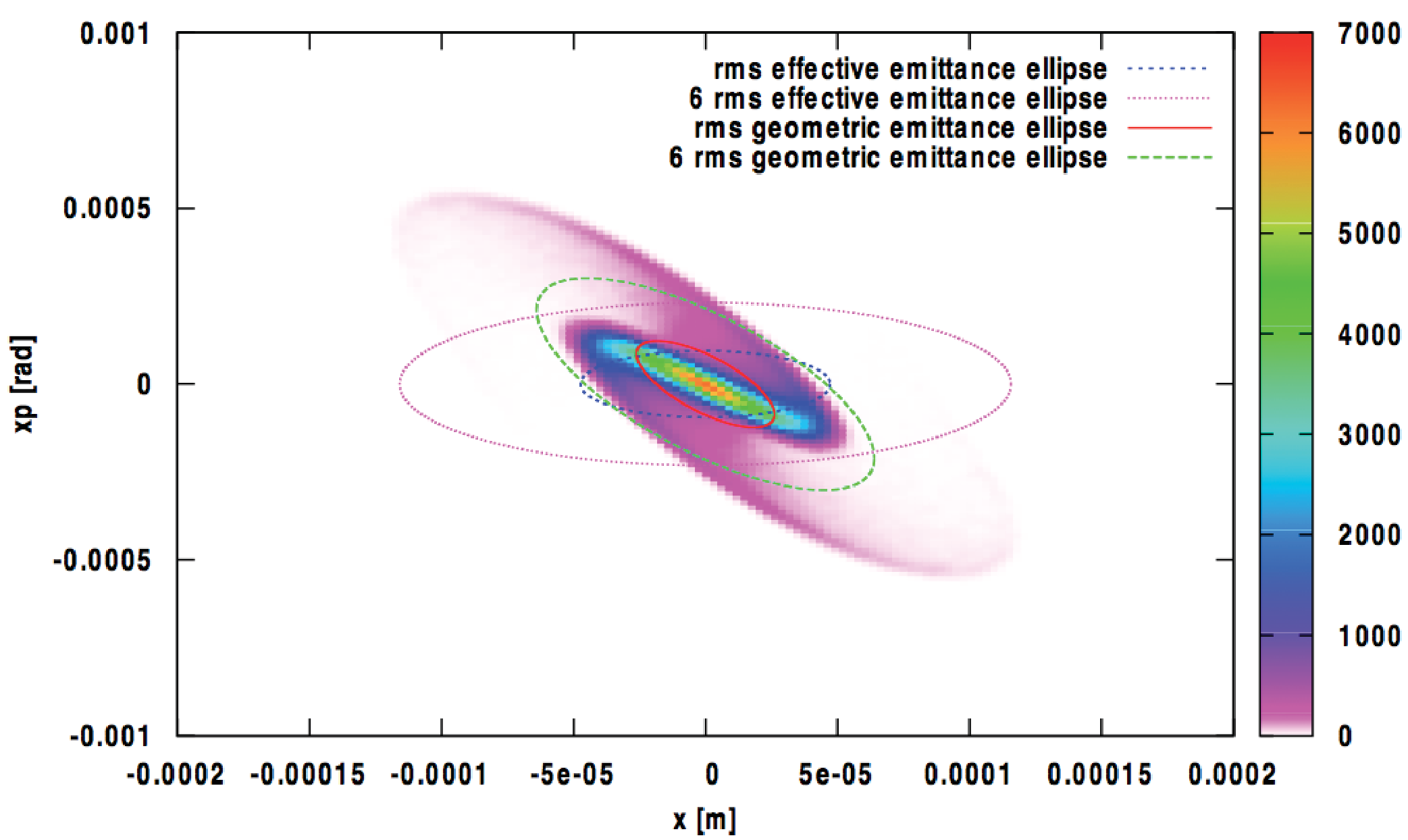}

\caption{\small (a) The optimized e-beam envelope during collision with the hadron beam 
in eRHIC; (b) Distribution of electrons after colliding with the hadron beam in 
eRHIC. }
\label{VL:fig_IR_disruption}
\end{figure}

Finally, beam-beam effects play important roles in eRHIC's performance. While 
we will control these effects on the hadron beam, i.e., we will limit the total 
tune shift for hadrons to about $0.015$, the electron beam is used only once and 
it will be strongly disrupted during its single collision with the hadron beam. 
Consequently, the electrons are strongly focused by the hadron beam (pinch effects), 
and the e-beam emittance grows by about a factor of two (disruption) during the 
collision. These effects, illustrated in figure \ref{VL:fig_IR_disruption}, do not represent a serious problem, 
but will be carefully studied and taken into account in designing the optics and 
the aperture . 

More details on the lattice and IR design are given in reference \cite{Trbojevic:2010zzb}.

\subsection{eRHIC R\&D}  \label{VL:subsct5}

The list of the needed accelerator R\&D on eRHIC is quite extensive, ranging from 
the 50~mA CW polarized source \cite{Tsentalovich:2008zz,VL:ref8,VL:ref9} to Coherent Electron Cooling \cite{Litvinenko:2009zz}. It includes designing 
and testing multiple aspects of SRF ERL technology in BNL's R\&D ERL \cite{VL:ref18}.

\begin{figure}[hp]
\centering
\includegraphics[width=0.9\textwidth]{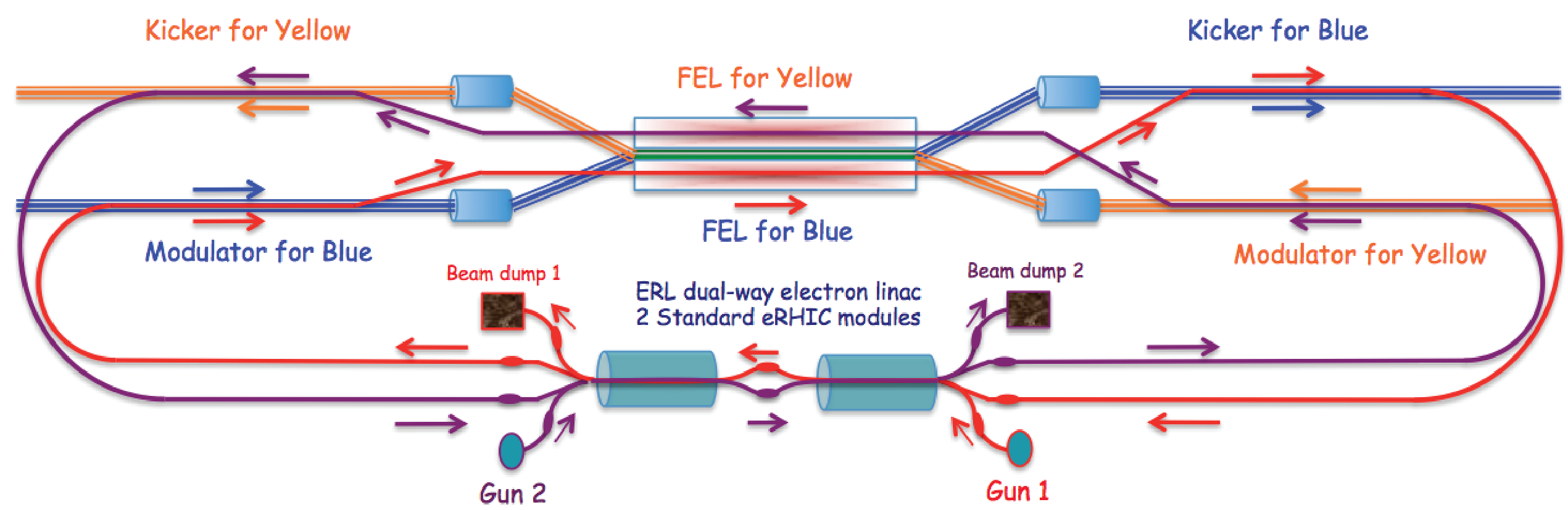}
\caption{\small Possible layout of RHIC CeC system cooling for both the yellow and blue 
beams.}
\label{VL:fig_CEC}

\end{figure}

Coherent Electron Cooling (figure \ref{VL:fig_CEC}) promises to cool both ion beams by 
an order of magnitude (both transversely and longitudinally) in under half an hour. Traditional stochastic or electron cooling techniques could not satisfy this demand.  Being a novel unverified technique, CeC will be tested in a proof-of-principle 
experiment at RHIC in a collaboration between scientists from BNL, JLab, and TechX \cite{VL:ref17}.

\begin{figure}[htbp]
\centering
(a) \hspace{7.5cm} (b)
\vfill
\includegraphics[width=0.28\textwidth]{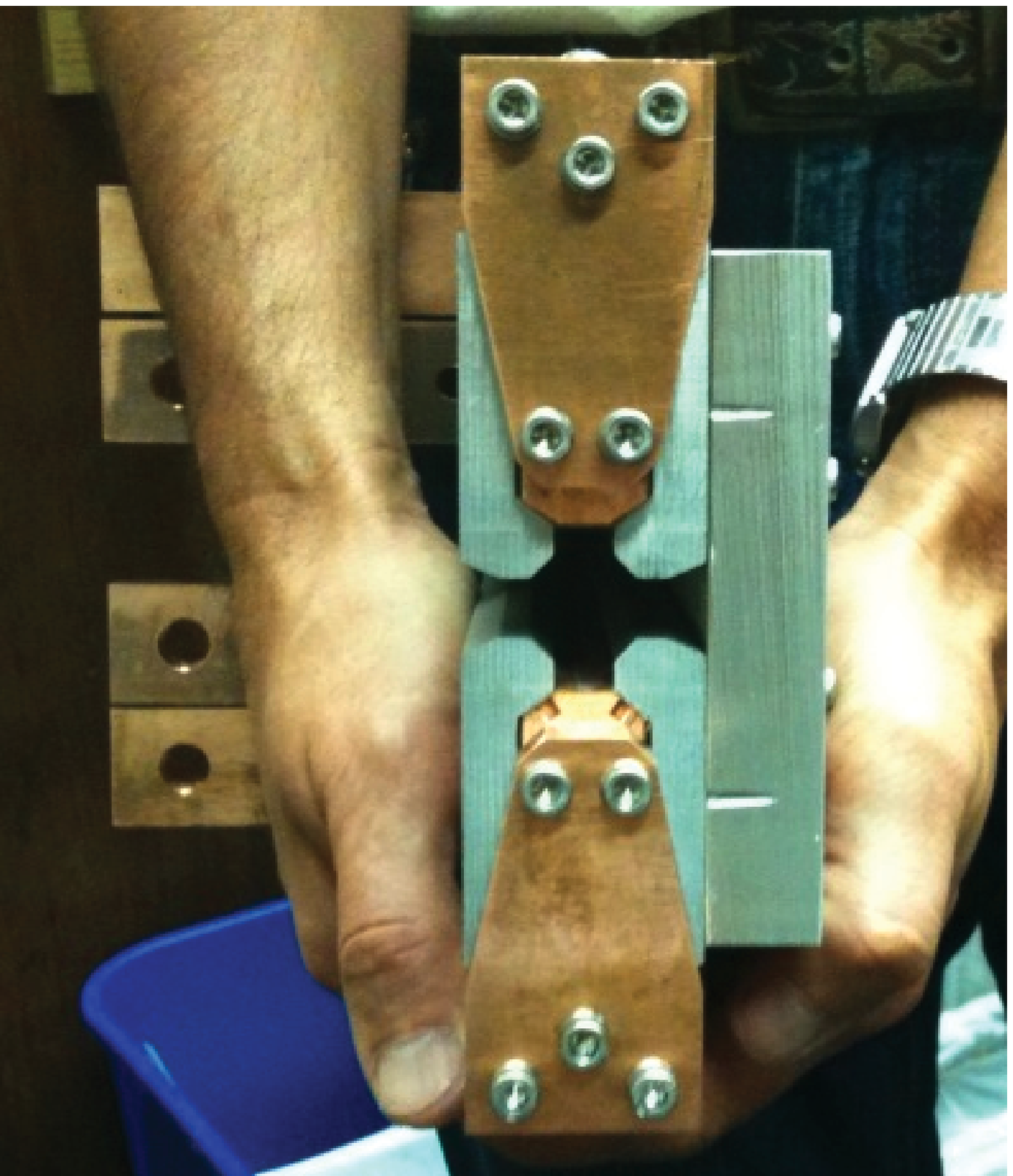}
\hspace*{0.5cm}
\includegraphics[width=0.60\textwidth]{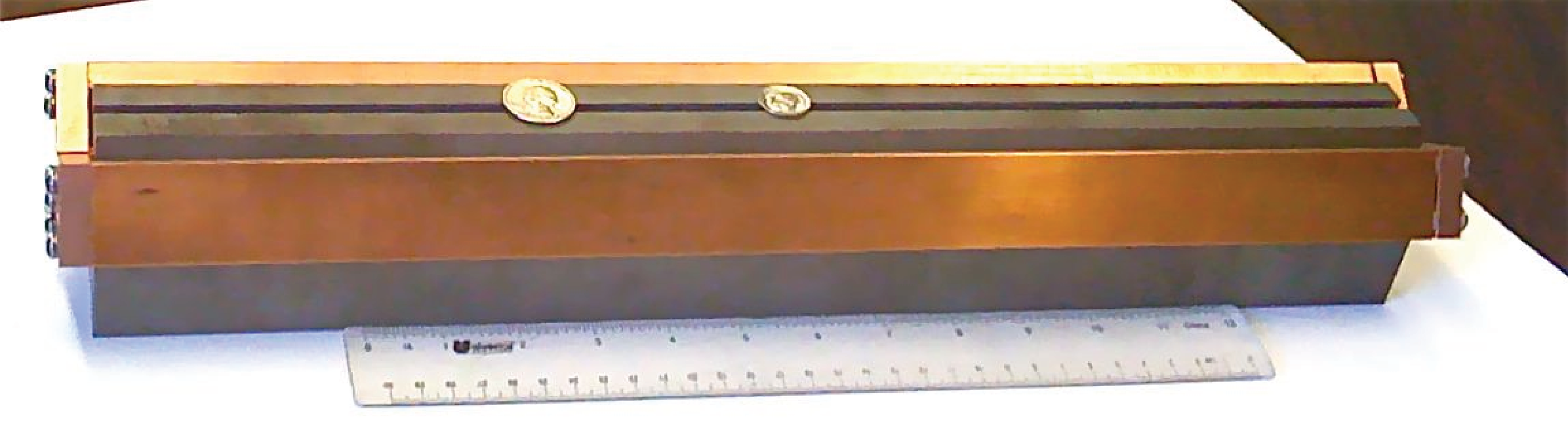}

\caption{\small (a) A prototype of eRHIC quadrupole with 1 cm gap; (b) Assembled prototype 
of eRHIC dipole magnet with 5 mm gap.}
\label{VL:fig_prototypes}

\end{figure}

Another important R\&D effort, supported by an LDRD grant, focuses on designing and 
prototyping small-gap magnets and vacuum chamber for cost-effective eRHIC arcs \cite{Hao:2010zzd}. 
In addition to their energy efficiency and inexpensiveness, small-gap magnets 
assure a very high gradient as room-temperature quadrupole magnets. Figure \ref{VL:fig_prototypes} shows 
two such prototypes; they were carefully tested and their fields were mapped using 
high-precision magnetic measurements. While the quality of their dipole field is 
close to satisfying our requirements, the quadrupole prototype was not manufactured 
to our specifications. We will continue this study, making new prototypes using 
various manufacturers and techniques. 

Another part of our R\&D encompasses testing the RHIC in the various modes that 
will be required for eRHIC's operation.

\subsection{Conclusions and Acknowledgements}

We are making steady progress in designing the high-energy, high-luminosity electron-ion 
collider eRHIC and plan to continue our R\&D projects and studies of various effects 
and processes. So far, we have not encountered a problem in our proposal that we 
cannot resolve. Being an ERL-based collider, eRHIC offers a natural staging of the 
electron beam's energy from 5-6 to 30 GeV. During this year, we will complete our 
cost estimate of all eRHIC stages. 

The authors would like to acknowledge contributions and advice from E.-C.Aschenauer, 
D.~Bruhwiler, G.~Bell, A.~Cadwell, A.~Deshpande, R.~Ent, W.~Gurin, A.~Hutton, H.~Kowalski,
G.~Krafft, M.~Lamont, T.~W.~Ludlam, R.~Milner, M.~Poelker, R.~Rimmer, B.~Surrow, B.~Schwartz, 
T.~Ullrich, S.~Vigdor, R.~Venugopalan, and W.~Vogelsang.

\section[A polarized medium-energy electron-ion collider at JLab]{A
  Polarized Medium-Energy Electron-Ion Collider at \linebreak JLab}
\label{sec:MEIC_design}


\hspace{\parindent}\parbox{0.92\textwidth}{\slshape 
B. Terzi{\'c}, Y. Zhang, S. Abeyratne, S. Ahmed,
A. Bogacz, P. Chevtsov, Ya. Derbenev, B. Erdelyi,
A. Hutton, A. Kondratenko, 
G. Krafft, R. Li, S. Manikonda, F. Marhauser,
V. Morozov, P. N. Ostrumov, F. Pilat,  
R. Rimmer, T. Satogata, H. Sayed,
M. Spata, M. Sullivan, H. Wang, B. Yunn}
%
%

\index{Terzi\'c, Bal\v{s}a}
\index{Zhang, Yuhong}
\index{Abeyratne, S.}
\index{Ahmed, S.}
\index{Bogacz, A.}
\index{Chevtsov, P.}
\index{Derbenev, Ya.}
\index{Erdelyi, B.}
\index{Hutton, A.}
\index{Kondratenko, A.}
\index{Krafft, G.}
\index{Li, R.}
\index{Marhauser, F.}
\index{Manikonda, S.}
\index{Morozov, V}
\index{Ostrumov, P. N.}
\index{Pilat, F.}
\index{Rimmer, R.}
\index{Sayed, H.}
\index{Satogata, T.}
\index{Sullivan, M.}
\index{Spata, M.}
\index{Wang, H.}
\index{Yunn, B.}

\vspace{2\baselineskip}

The conceptual design of MEIC, a polarized ring-ring electron-ion collider based on CEBAF, has been continuously optimized for best supporting the nuclear science program. MEIC covers a medium CM energy region up to 65 GeV (for 6 T superconducting dipole magnets) and achieves a luminosity of above $10^{34} {\rm cm}^{-2} {\rm s}^{-1}$ for one high-luminosity and one full-acceptance detectors. The unique compact figure-8 shaped collider rings, designed to accommodate 3 to 11 GeV electrons and up to 96 GeV protons or 48 (38) GeV/u  for light (heavy) ions (128 GeV protons or 64 (51) GeV/u  light (heavy) ions for 8 T superconducting magnet), provide a great advantage for delivering and preserving high polarization of ion beams (including polarized deuterons) for collisions at multiple interaction points. The design is upgradable to accommodate 20 GeV electrons and about 250 GeV proton energies at a late stage, with luminosities up to $10^{35} {\rm cm}^{-2}{\rm s}^{-1}$. The present focus of the Jefferson Lab accelerator team is to develop a coherent machine design that integrates all of the design features that have been explored over recent years, based upon state-of-the-art performance criteria. Various collider components including ion linac and boosters, spin rotators, and interaction regions have been designed and integrated into a unified design. These advances will be discussed in detail.   


\subsection{Introduction}  

Over the last decade, Jefferson Lab has been developing a conceptual design of an electron-ion collider for future nuclear physics research. This facility, fully utilizing the 12 GeV upgraded CEBAF, will provide collisions between polarized electrons and polarized light ions or unpolarized light to heavy ions up to lead over a wide CM energy range at multiple interaction points (IPs). Requirements of the science programs drive the design efforts to focus on achieving ultra-high luminosity ($10^{34} {\rm cm}^{-2}{\rm s}^{-1}$ or above) per detector, and high polarization (over 80\%) for both electron and light ion beams. 

Our primary design focus at the present time is a Medium-energy Electron-Ion Collider (MEIC), with a CM energy up to 65 GeV, which covers electron energy up to 11 GeV, proton energy up to 96 GeV and ion energy up to 48 GeV per nucleon. It is considered as an optimal compromise between science, technology and project cost. We also maintain a well-defined upgrade capability to higher energies, ELIC, which can reach up to 20 GeV electron energy, and 250 GeV proton energy or 100 GeV/u {\it heavy} ion energies (typically, for heavy ions the proton number is about 40\% of the atomic mass number). In both instances, high luminosity and high polarization remain the main design drivers. 

The present MEIC design features a traditional ring-ring collider with a high luminosity at a level of $10^{34} \rm{cm}^{-2}{\rm s}^{-1}$ per detector, over up to three IPs, by taking full advantage of an electron beam from the upgraded 12 GeV CEBAF recirculated SRF linac. As a design concept, the high luminosity of MEIC is attained by utilizing high bunch repetition rate, crab-crossing colliding electron and ion beams with short bunch length and small transverse emittances, and strong final focusing at collision points. Our choice of this luminosity concept was motivated by the remarkable success of two electron-positron colliders at KEK and SLAC B-factories, which had reached luminosities over $2\times10^{34} \rm{cm}^{-2}{\rm s}^{-1}$. In a way, Jefferson Lab is poised to replicate the same success in a collider involving {\it hadron} beams. The new concept requires the colliding ion beams of MEIC to be very different from all existing or previously operated hadron colliders in terms of bunch intensity (very small), bunch length (very short), transverse emittances (very small) and repetition frequency (very high), while, at the same time, it pushes the final focusing parameter $\beta^{*}$ to be much smaller than what has been achieved in hadron colliders. To support such a conceptual design, extensive R\&D programs have been established at Jefferson Lab, supplemented by several external collaborations.   

As a design strategy, we are taking a conservative technical position by limiting many MEIC design parameters within or close to the present state-of-the-art in order to minimize technical uncertainty. This conservative technical design will form a baseline for future design optimization guided by the evolution of the science program, technology innovations and R\&D advances.

\subsection{Baseline Design}

\begin{figure}[htp]
      \includegraphics[width=0.48\textwidth]{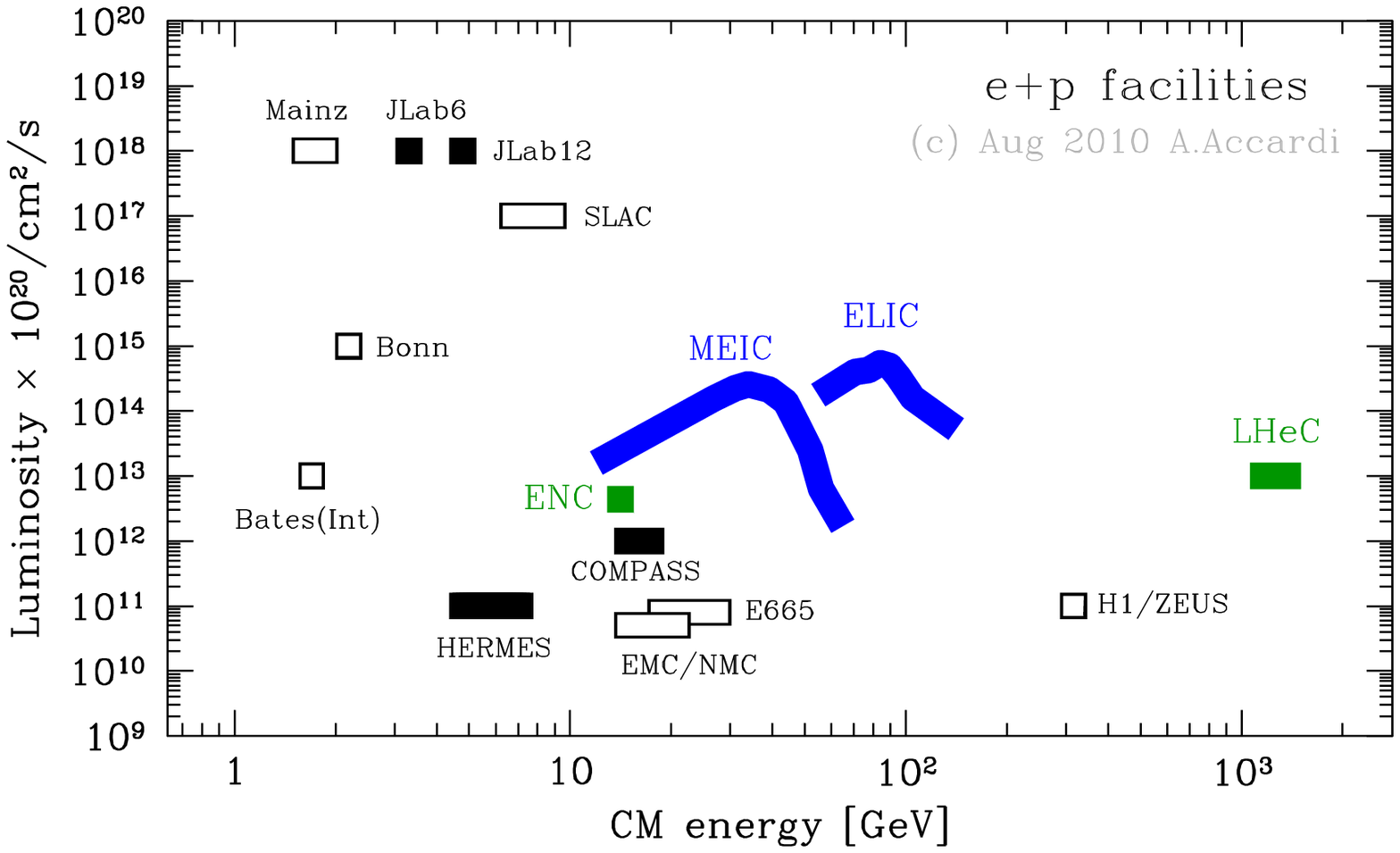}  
      \includegraphics[width=0.48\textwidth]{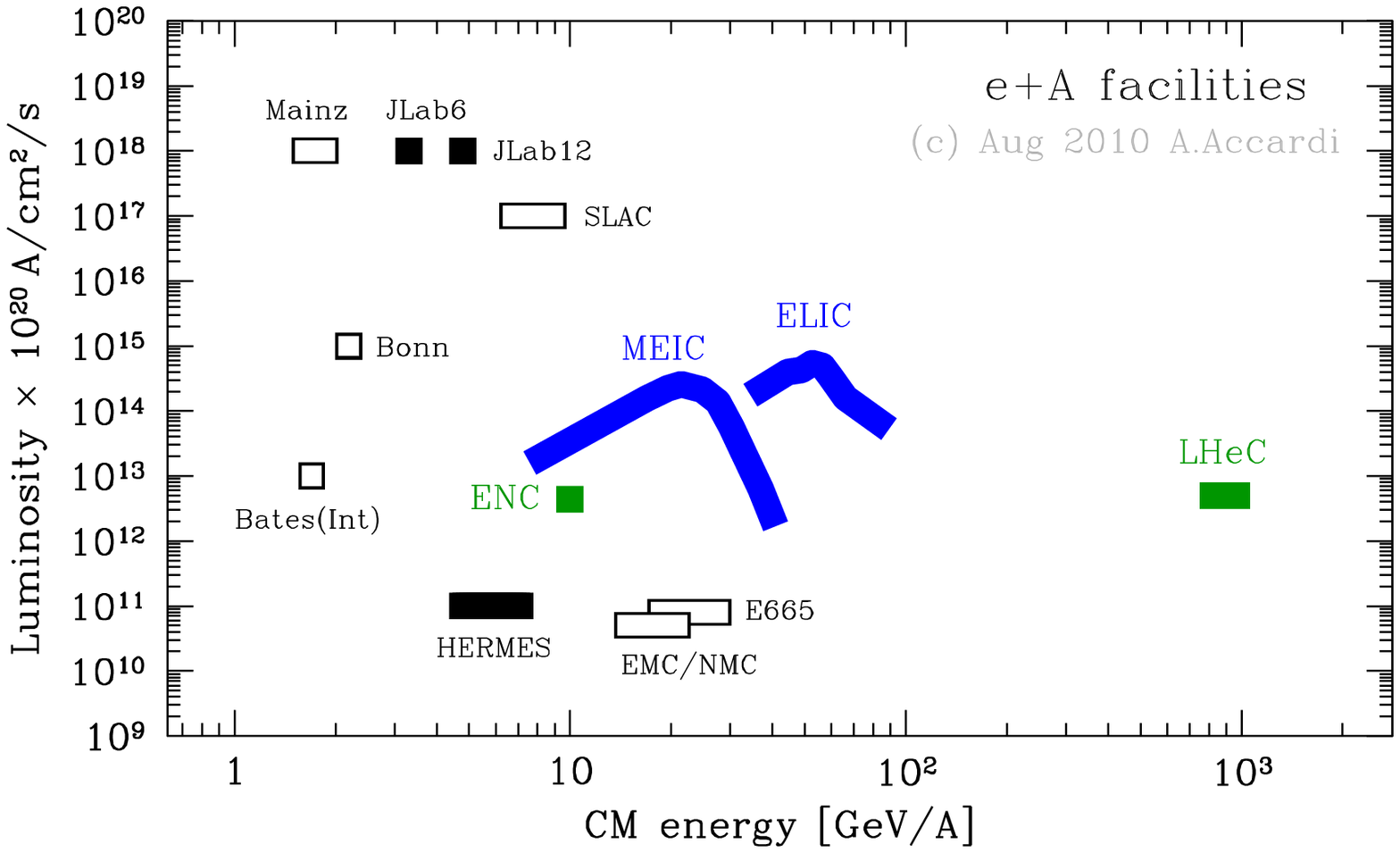}  
      \vspace*{-3cm}
\caption{\small{MEIC and its high-energy upgrade ELIC in CM energy-luminosity space.}}
\label{MEIC:fig_1}
\end{figure}

The MEIC main parameters are summarized in table~\ref{tab:MEIC1} for a design point of 60 GeV proton and 5 GeV electron. Figure~\ref{MEIC:fig_1} presents luminosities as a function of CM energy for both proton and ions.  In deriving this set of design parameters, we have imposed certain limits on several key machine or beam parameters in order to reduce technical risk and the accelerator R\&D challenges and to improve robustness of the design. These limits, based on largely previous lepton and hadron collider experiences and state-of-art of accelerator technologies, are:
\begin{itemize}
\item Average current of the stored beams are up to 1 A for protons/ions and 3 A for electrons,
\item Electron synchrotron radiation power density is less than 20 kW/m,
\item Peak bending field of ion superconducting dipole is no larger than 6 T,
\item The maximum betatron value at the beam extension area near an IP is no larger than 2.5 km,
\item Frequency of accelerating RF cavity in the electron ring is less than 1 GHz.
\end{itemize}
Also, different nuclear programs usually require different detector acceptances and arrangement of interaction regions (IR). While such detector requirements are still in a formation stage, we have considered two different types of IR designs, one for a full-acceptance detector (with 0.5 to 179.5 degree solid angular acceptance before the ion final focusing magnets, and the apertures of the latter sufficient to allow particles with angles up to 0.5 degrees to go through the bore of the magnet for downstream detection), the other for a high-luminosity detector. The key difference of the IR designs is a space between the collision point and the location of the first final focusing quad, and values of these distances for the two detectors are 7 m and 4.5 m respectively for ion beams, while the space for electron beams can be as low as 3 m for both cases. The relatively short distance of 4.5 m enables a further reduction of the final focusing $\beta^{*}$ to 8 mm, thus resulting in a more than a factor two increase of luminosity for that detector configuration as shown in table~\ref{tab:MEIC1}.  

\begin{wrapfigure}{l}{0.55\columnwidth}
\begin{minipage}[b]{0.55\columnwidth}
\centerline{\includegraphics[width=0.56\columnwidth]{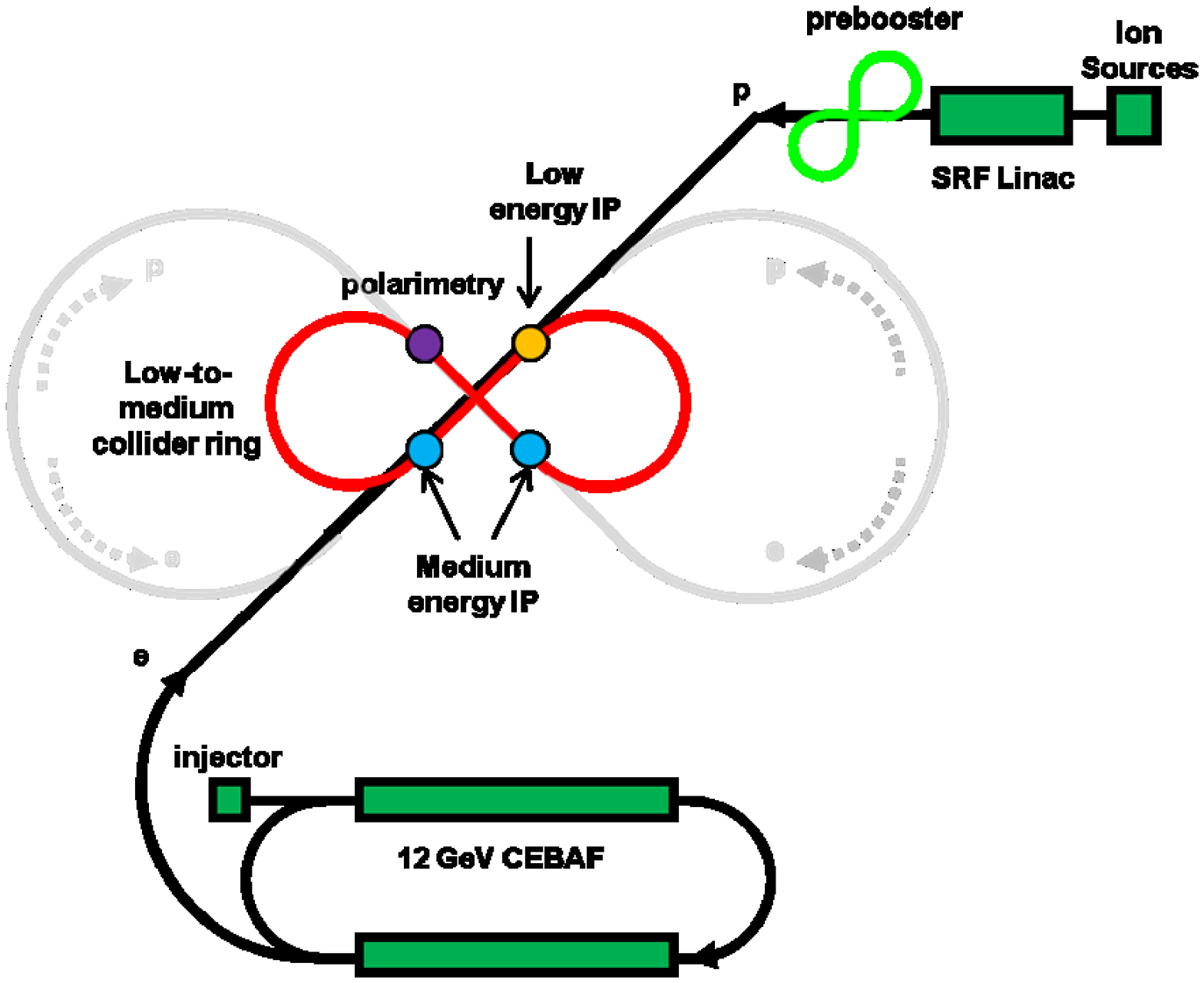}}
\end{minipage}
\begin{minipage}[b]{0.65\columnwidth}
\hspace*{0.3cm}
\includegraphics[width=0.65\columnwidth]{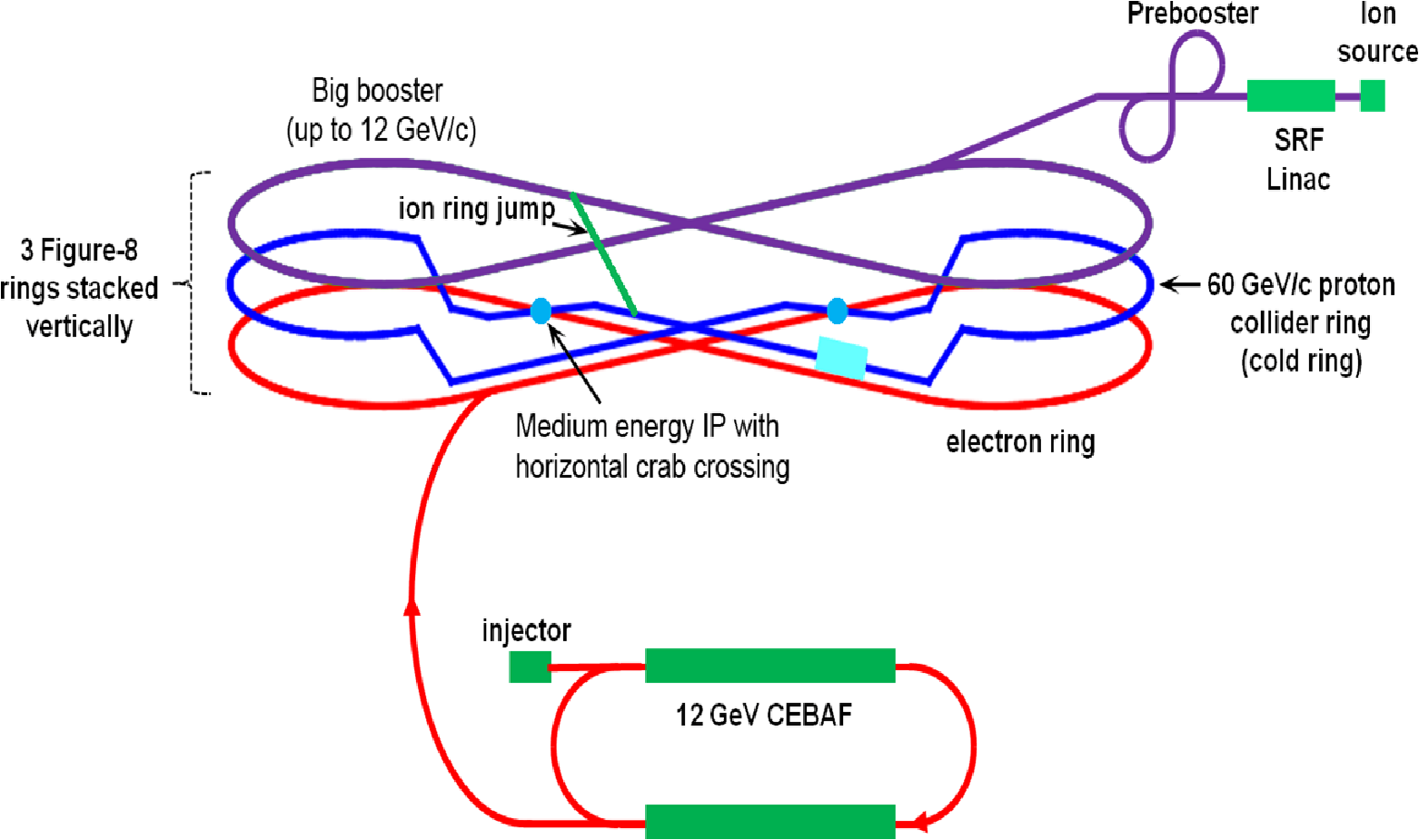}
\end{minipage}
\begin{minipage}[b]{0.65\columnwidth}
\hspace*{1.3cm}
\includegraphics[width=0.4\columnwidth]{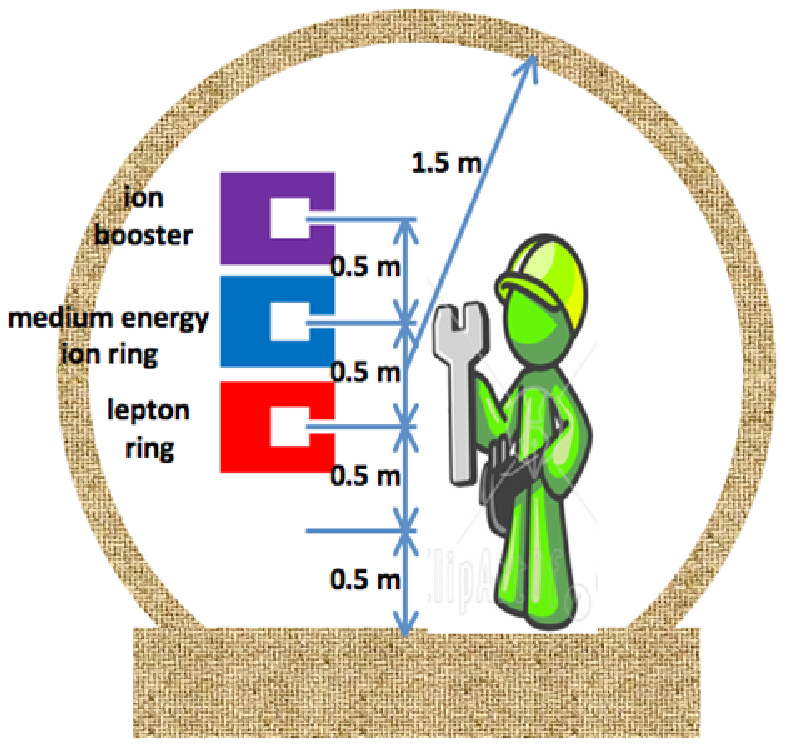}
\end{minipage}
\caption{Layout and side view of the MEIC and cross-section of tunnel.}
\label{MEIC:fig_2}
\end{wrapfigure}

The MEIC design calls for the construction of a green-field ion accelerator complex and two collider rings, one for electrons and the other for medium energy ions, as shown in figure~\ref{MEIC:fig_2}. There are four crossing points of these figure-8 collider rings which will accommodate three detectors, at least two of which are available for medium-energy collisions, and the other for low-energy collisions with ions stored in a large booster. As presently envisaged, the two collider rings of identical circumferences are vertically stacked and the ion beams are transported into the plane of the electron ring via a vertical chicane, where horizontal crab crossings were used to collide the two beams at the collision points.

\begin{table}
\centering
\noindent\makebox[\textwidth]{%
\footnotesize
    \begin{tabular}{lccc}
        \textbf{Quantity} & \textbf{Unit}               & \textbf{$p^{-}$ beam}  & \textbf{$e$ beam} \\ 
         Beam energy &     GeV    &    60     &     5    \\
         Collision frequency &     MHz    &     \multicolumn{2}{c}{749}    \\
         Particles per bunch &     $10^{10}$   &     0.416 &   2.5      \\       
         Beam current &     A   &     0.5    &    3    \\      
         Polarization & \% & $>70$ & $\sim 80$    \\
         Energy spread &   $10^{-3}$  &   0.3     &    0.71     \\  
         RMS bunch length &     mm    &   10  &  7.5     \\   
         Horiz. emit. (norm.) &     $\mu$m   &      0.35  &  53.5       \\ 
         Vertical emit. (norm.) &     $\mu$m       &    0.07    &     10.7  \\       
         Horizontal $\beta^{*}$ &     cm   &    10 (4)   &   10 (4)   \\    
         Vertical $\beta^{*}$ &     cm   &    2 (0.8)  &   2 (0.8) \\
         Vertical beam-beam tuneshift  &         &     0.015     &   0.03    \\
         Laslett tuneshift  &         &     0.06     &   small    \\
         Distance from IP to $1^{\rm st}$ final focusing quad  &  m   &    7 (4.5)     &   3.5    \\         
         Luminosity per IP&   $10^{33} {\rm cm}^{-2}{\rm s}^{-1}$  & \multicolumn{2}{c}{$5.6 ~(14.2)$}  \\        
    \end{tabular}}
    \label{tab:MEIC1}
     \caption{MEIC design parameters for the full-acceptance detector. Values for the high-luminosity detector are given in parentheses. }
\end{table}
%

The ion complex consists of ion sources, a 200 MeV SRF linac, a 3 GeV pre-booster and a large booster with energy up to 20 GeV. The ion beams are formed and accelerated in multiple stages in the low-energy ion complex, and are then filled into the collider ring for further acceleration to the colliding energy and stored for collision. A large figure-8 ring, also drawn in Fig.~\ref{MEIC:fig_2} (in grey), accommodates high-energy ion beams in a future energy upgrade. In that case, the compact medium-energy collider ring will act as another large booster. On the electron side, a 12 GeV upgraded CEBAF SRF linac will serve as a full-energy injector into the electron collider ring, which could also be operated in a top-off mode in order to maintain high beam current. It is possible to continue the fixed target program for the CEBAF whenever there is a need, since each filling of the electron ring is very short.

%


The MEIC�s figure-8 shape in all rings is an optimal solution to preserve full polarization of light ion beams by avoiding spin resonances during acceleration in multiple booster and collider rings. It is also the only practical way to accelerate polarized deuterons and to arrange for longitudinal spin polarization at IP. The figure-8 layout allows for energy independence of the spin tune, as well as the transverse polarization of deuterons.

An essential component of every version of Jefferson Lab electron-ion collider design is an electron cooling facility, which is required for reducing ion beam transverse emittance and along with strong RF bunching, shortening the bunch length to 1 cm.

\subsection{Ion Complex} 

Being primarily a lepton lab, Jefferson Lab does not have an ion complex at this time. This is usually considered a disadvantage to the Jefferson Lab electron-ion collider design effort, since a new ion complex is usually more expensive than a new electron complex. However, a green-field ion complex provides an excellent opportunity for applying new concepts and accelerator technologies which have been developed, tested and perfected over the last half century. Therefore, the MEIC�s ion complex could, in principle, be built to be far superior to the existing or legacy hadron facilities, thereby offsetting the relatively high project cost.

The main design goal of MEIC ion complex is to create and accelerate polarized or un-polarized ion beams with appropriate time, spatial and phase space structure matching the electron beam in order to implement the new luminosity concept. It is important to note that, while, on the one hand, the MEIC design requires bunch length and transverse emittances order of magnitude smaller than that of the conventional ion beams, on the other hand, due to a high bunch repetition rate, the MEIC ion bunch intensity is unusually low (at $4\times 10^9$), approximately 50 times smaller than RHIC ion beam, thereby drastically easing the process of forming such ion beams and intensity dependent collective beam instabilities. 

The MEIC ion complex is shown in figure~\ref{MEIC:fig_2}. Its layout also characterizes the scheme of ion beam acceleration and formation. The ions, coming out from the polarized or unpolarized sources, will be accelerated step-by-step to the colliding energy in the following major machine components: a 200 MeV SRF linac, a 3 GeV pre-booster, a 20 GeV large booster and finally a medium-energy collider ring of 20 to 60 GeV.  All rings are in figure-8 shape for the benefit of ion polarization. We will present a brief description on each component in the rest of this section. The pre-booster is also an accumulator ring, accepting and stacking ions (0.5 to 1 A average current) from a SRF linac in a multi-turn injection with assistance of a conventional DC electron cooling (except the case of H$^-$/D$^-$ for which a phase space paint technique will be used). The accumulated ion beam in the pre-booster will become a coasting beam and will be re-bunched later in the medium-energy collider ring in order to decouple the RF frequencies in the linac and collider rings, as well as to suppress space charge tune-shift at low-energy stage.   

\subsubsection{Ion Sources}
The MEIC ion sources will rely on existing and mature technologies. We will have an Atomic Beam Polarized Ion Source (ABPIS) with Resonant Charge Exchange ionization for producing polarized light ions H$^+$/D$^+$ and $^{3}$He$^{++}$. For unpolarized light to heavy ions, we will utilize Electron-Beam Ion Source (EBIS) which is current in operation at BNL. It is a realistic extrapolation, given future R\&D, that an ABPIS should be able to deliver 10 mA polarized H$^+$/D$^+$ pulses at 5 Hz repetition frequency, over a 0.5 ms pulse length with a polarization better than 90\%. An EBIS, on the other, hand is expected to generate unpolarized $^{208}$Pb$^{30+}$ pulses also at 5 Hz repetition rate and about 1.6 mA averaged current over a much shorter pulse length of 10 to 40 $\mu$s. Alternatively, an Electron Cyclotron Resonance Source (ECR) can generate heavy ion beams with similar averaged currents, but 10 to 50 times longer pulse lengths, resulting in a factor of 10 or more pulse charges. In all of these instances, the present ion source technologies should be able to meet the requirement of the MEIC.                 

\subsubsection{Ion Linac}
\begin{figure}[htp]
\centering
      \includegraphics[width=0.7\textwidth]{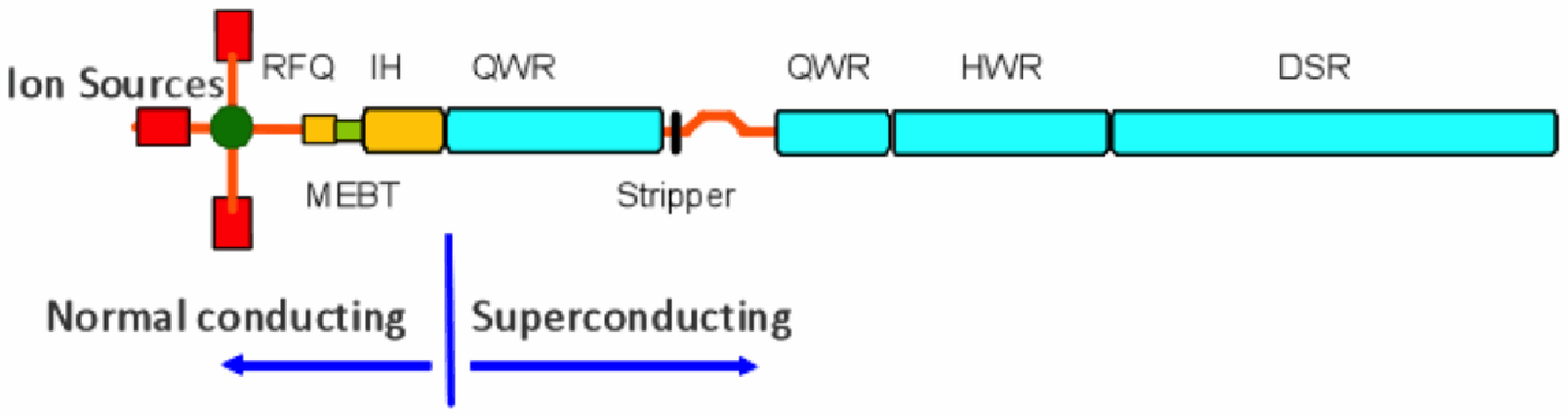}  
\caption{\small{Schematic drawing of MEIC SRF linac conceptual design.}}
\label{MEIC:fig_3}
\end{figure}

A technical design of an advanced SRF ion linac, originally developed at Argonne National Laboratory as a heavy-ion driver accelerator for Rare Isotope Beam Facility, has been adopted for the MEIC proposal. This 150-m-long linac, as shown in figure~\ref{MEIC:fig_3}, is very effective in accelerating a wide variety of polarized and unpolarized ions from H� (285 MeV) to $^{208}$Pb$^{67+}$ (100 MeV/u). Economic acceleration of lead ions up to 100 MeV/u requires a stripper with an optimal stripping energy of 13 MeV/u. The stripping efficiency of $^{208}$Pb ion beam to the most abundant charge state 67+ is 21\%.

\subsubsection{Pre-Booster/Accumulator Ring}

\begin{figure}[htp]
\begin{center}
      \includegraphics[width=0.7\textwidth]{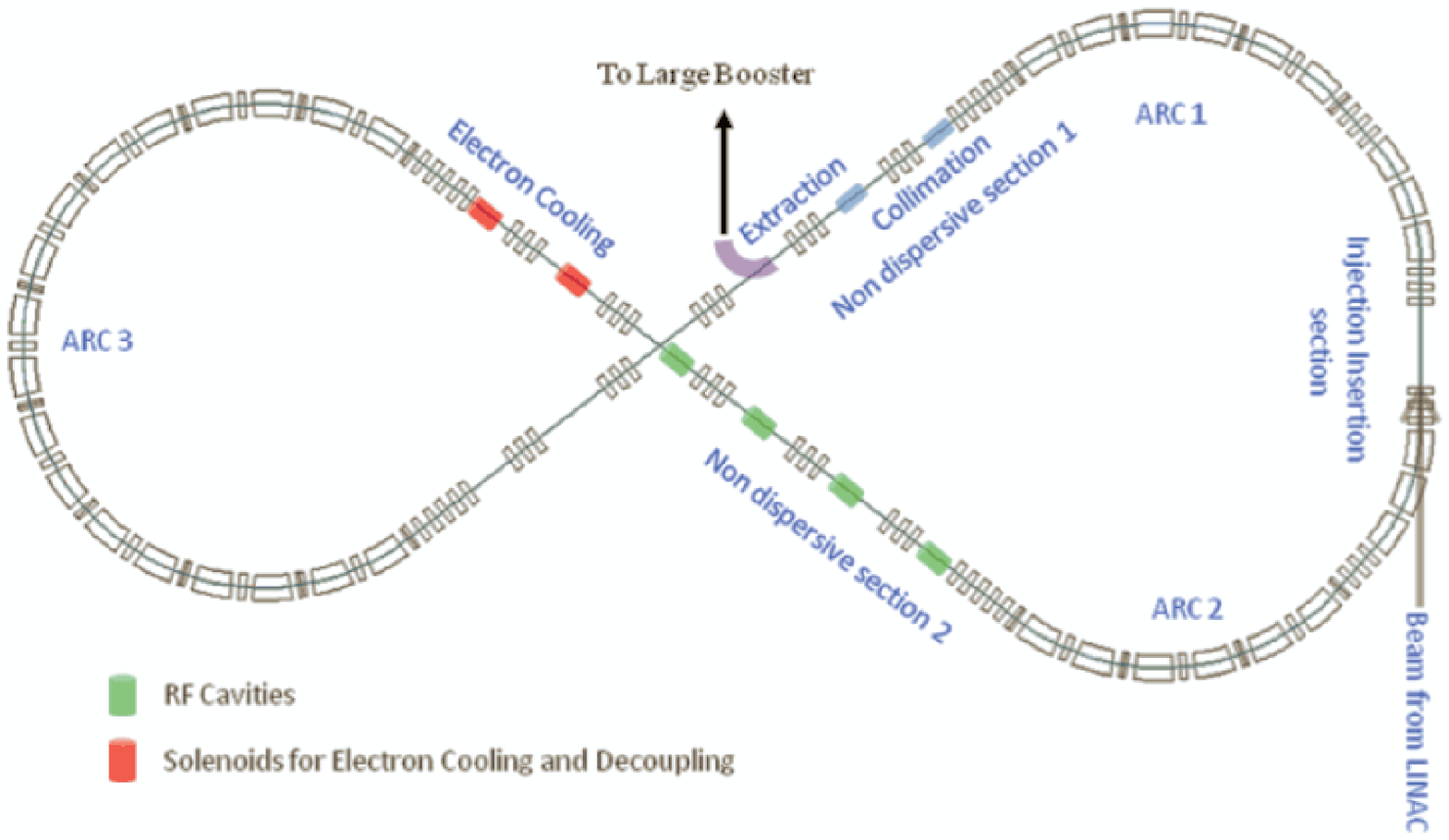}  
\end{center}
      \vskip-10pt
\caption{\small{A figure-8 shaped pre-booster ring.}}
\label{MEIC:fig_4}
\end{figure}

The pre-booster ring, as shown in figure~\ref{MEIC:fig_4}, is an essential component of the ion accelerator complex, which accepts beam pulses of any ions from the ion linac and, after accumulation and/or acceleration, transfers the beam to the subsequent large booster for further acceleration. The exact mechanisms of pre-booster operation depend on the ion species, relying on either combined longitudinal and transverse paint technique for H$^-$/D$^-$ or conventional DC electron cooling for lead or other heavy metals during multi-turn injection from the SRF linac. One important design consideration of the pre-booster is sufficiently high transition gamma, such that the ions never cross the transition energy during acceleration in order to prevent associated particle loss. In addition, the betatron motion working point should be carefully chosen such that the tune footprint does not cross low-order resonances.

\begin{table}[htdp]
\centering

\noindent\makebox[\textwidth]{%
\footnotesize
\begin{tabular}{|c|c|c|c|c|c|}
\hline
Length & Crossing angle & Max. beam size & $\gamma$ for 3 GeV particles & Transition $\gamma$ & Mom. compaction \\
\hline
\hline
234 m  & 75 deg  & 2.3 cm & 4.22 & 5 & 0.04 \\
\hline

\end{tabular}}

\caption{Parameters for the pre-booster ring.}
\label{tab:MEIC2}
\end{table}

\subsubsection{Large Booster}

The MEIC large booster shares the same tunnel as the electron and ion collider ring. It accelerates protons from 3 GeV to 12 to 20 GeV before sending them to the medium-energy collider ring. The extraction energy will be determined in the further design optimization. The boosters can handle all ion species. However, the energy will be affected by the ratio of charge and mass of the ion species. In principle, higher extraction energy is preferred. The key design requirements are that the ring must be also a figure-8 shape, its magnetic lattice should be built by warm magnets and a crossing of the transition energy must be avoided. It should be pointed out that the large booster will be also used as a low-energy collider ring for collision energies in the region of 5 to 20 GeV per nucleon over only one detector. Special insertions such as an interaction region and an electron cooling facility must be inserted or added to meet such physics demand.

\subsection{Collider Rings}

\begin{figure}[htp]
\begin{center}
      \includegraphics[width=0.8\textwidth]{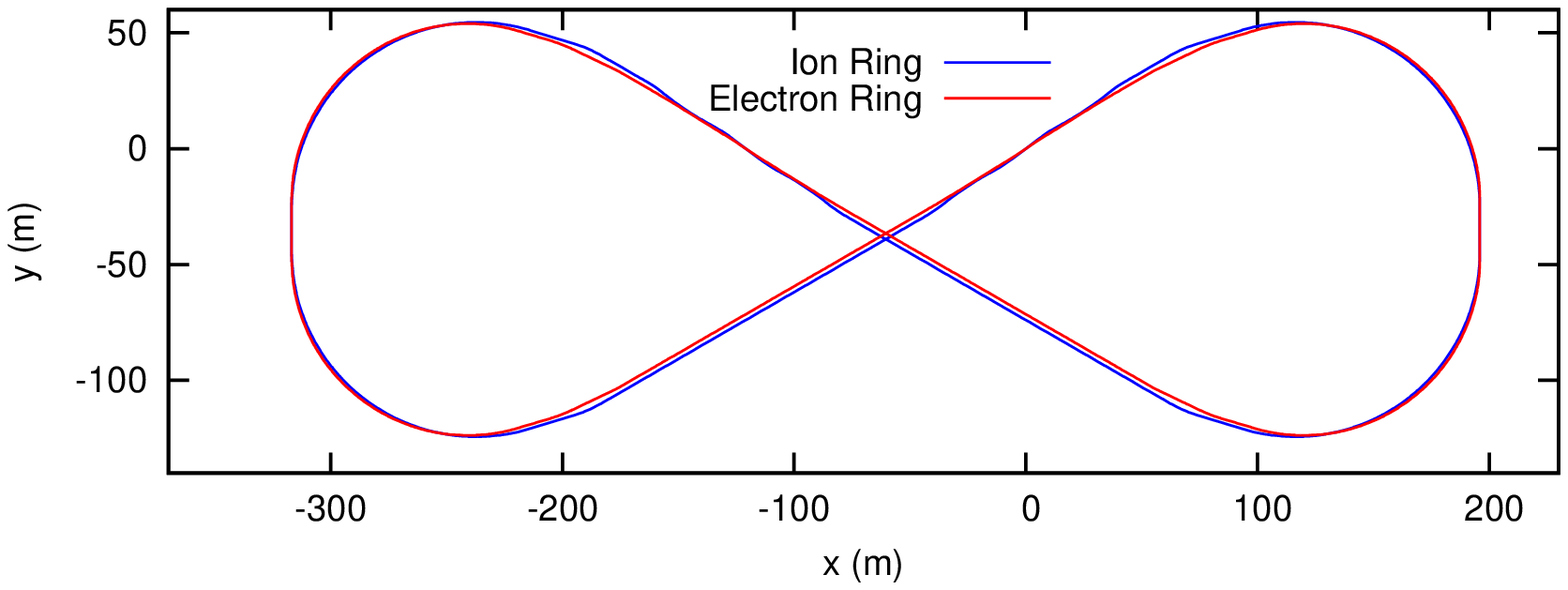}  
\end{center}
\caption{\small{Layout of the electron and ion collider rings.}}
\label{MEIC:fig_5}
\end{figure}

Figure~\ref{MEIC:fig_5} shows a scaled layout of the electron and ion collider rings. When designing the optics of the electron and ion collider rings, the following geometric constraints must be taken into account: 
\begin{itemize}
\item Figure-8 shape, and four intersection points in two straights,
\item Two short (20 m) straights in the middle of the two arcs for two Siberian snakes,
\item A 60 m Universal Spin Rotator consisting of two solenoids and two sets of arc bending dipoles on each end of two electron arcs,
\item Layouts of the interaction regions must match for both electron and ion rings,
\item Footprints of the two collider rings must be very close so that they can be housed in one tunnel,
\item The ion ring circumference must be equal to the big booster�s length and be an integer multiple of the pre-booster�s length for the purpose of RF matching.
\end{itemize}
The circumferences of the electron and ion rings are about 1340 m. The circumferences can be modified by adjusting length of straights, crossing angle and arc radii. The figure-8 crossing angle is 60$^{o}$. The electron and ion rings� IPs coincide. The maximum separation of the electron and ion beam lines in the design is less than 4 m, which can be further reduced to be within the required limits once the optics design concept is finalized. The parameters of the electron ring and optics design are summarized in table~\ref{tab:MEIC3}.

\begin{table}[htdp]
\centering

\noindent\makebox[\textwidth]{%
\footnotesize
\begin{tabular}{|c|c|c|c|c|c|}
\hline
Arc's net bend & Crossing angle & Arc length & Arc avg. radius & straight length & circumference \\
\hline
\hline
240 deg  & 60 deg  & 405.75 m & 96.86 m & 264.46 m & 1340.41 m \\
\hline

\end{tabular}}

\caption{Parameters for the collider ring.}
\label{tab:MEIC3}
\end{table}

\subsection{Interaction Region}
\subsubsection{Detector}

The primary detector of the MEIC will be unique in its ability to provide almost full acceptance for the produced particles of electron-ion collisions. To accomplish this, a high level of integration with the interaction region of the accelerator is required. The central detector will be built around a solenoid barrel, providing tracking, particle identification, and calorimetry for all particles, and two end-caps focusing on the detection of electrons and hadrons respectively.  See section~\ref{sec:detector} for further discussion.


\subsubsection{Crab Crossing}

With a 749 MHz bunch repetition rate for both colliding beams of MEIC, the bunch spacing is about 40 cm. A crab crossing of the colliding beams provides a simple way to separate quickly the two colliding beams near an IP to avoid undesired parasitic collisions. In the present MEIC IR design, the horizontal crabbing angle is 50 mrad. There are two ways currently under consideration to tilt the orientation of the electron or ion bunches in horizontal plane by a half crab crossing angle in order to restore head-on collisions. The first approach is placing crab cavities on each side of an IP. Such approach has been proved recently at KEK-B factory, which led to a record-high luminosity. It has been estimated that we need approximately 1.26 and 16.2 MV integrated transverse kicking voltage for 5 GeV electron and 60 GeV protons respectively. While a KEK-type squashed crab cavity should be readily adopted for the MEIC electron ring, the ion ring needs a set of such KEK crab cavities to achieve the design goal. At Jefferson Lab, a new type of crab cavity, which is more compact and promises much higher field, has been recently conceptually designed using transverse electromagnetic field (TEM), as another candidate of crab cavities for the MEIC. An alternative approach is dispersive crabbing, in which tilting of a bunch is achieved through purposely leaking the horizontal dispersion in the normal accelerating RF cavities. 

\subsubsection{Interaction Region Design}

As mentioned earlier, the two collider rings of MEIC are stacked vertically. In an early version of the IR design, the electron beam was vertically bent into a crab crossing, while the ion ring remains in a plane. Such design layout was abounded due to several problems: (i) bending electrons will generate excessive synchrotron radiation near an IP, which, consequently, interferes with the detector and degrades the background; (ii) the synchrotron radiation and quantum excitations in a beam extension area (betatron values are still very large) could enlarge the electron beam emittance by an order of magnitude; (iii) polarization decrease caused by vertical bend could be also very significant. The present MEIC IR design now requires the ions to undergo a vertical excursion to facilitate a horizontal crab crossing at an IP. For a 50 mrad crab crossing angle, the required bending field of the dipoles is quite modest for our ion energy range.    

A typical magnet lattice layout of the MEIC interaction region is illustrated in figure~\ref{fig:ir_and_central_detector_layout}. At a medium-energy region, the ion beams are modestly asymmetric in two transverse dimensions, as a result of balance of electron cooling and intra-beam scatterings. For instance, the emittance aspect ratio of a 60 GeV proton bunch is about 5. For such modestly flat beams, a final focusing quad doublet is a good choice. 

For the MEIC IR design with a 2 cm or less $\beta^{*}$, chromatic aberration of the final focusing quads is one important issue that special attention must be paid. The chromaticity, defined as a ratio of betatron tune shift and momentum spread, could be as high as 110 per IR. A dedicated chromaticity compensation block, consisting of a set of sextupoles, will be inserted in the beam extension area on both sides of an IR to mitigate the problem. The initial studies indicate that, with proper values of these sextupoles, the chromaticity can be reduced dramatically to single digits. Particle tracking simulations for dynamic aperture are currently underway.

\subsection{Electron Cooling}

Cooling of the ion beam is essential to achieve high luminosity in MEIC. At low energy, DC electron cooling is employed to help stacking of the ions in the pre-booster. At the collider ring, we rely on a concept of staged cooling of bunched ion beams of medium energies. Electron cooling is first called in the injection energy for reduction of the area that the ion beam occupies in the 6D phase space. After ions are accelerated to the collision energy, electron cooling will be utilized again for conditioning the beam to the design values. And most importantly, electron cooling will be continued during the collision mode to suppress the intra-beam scattering induced beam heating and emittance growth. Shortening the bunch length (down to 1 cm or less) that results from electron-cooling of the ion beam captured in a high voltage SRF field, in particular, it is critical for the high luminosity in MEIC, since it facilitates two important advances: an extreme focusing of the colliding beams and implementation of crab crossing at the IPs for achieving the highest bunch collision rate (up to 1.5 GHz) and luminosity.

\begin{figure}[htp]
\begin{center}
      \includegraphics[width=0.75\textwidth]{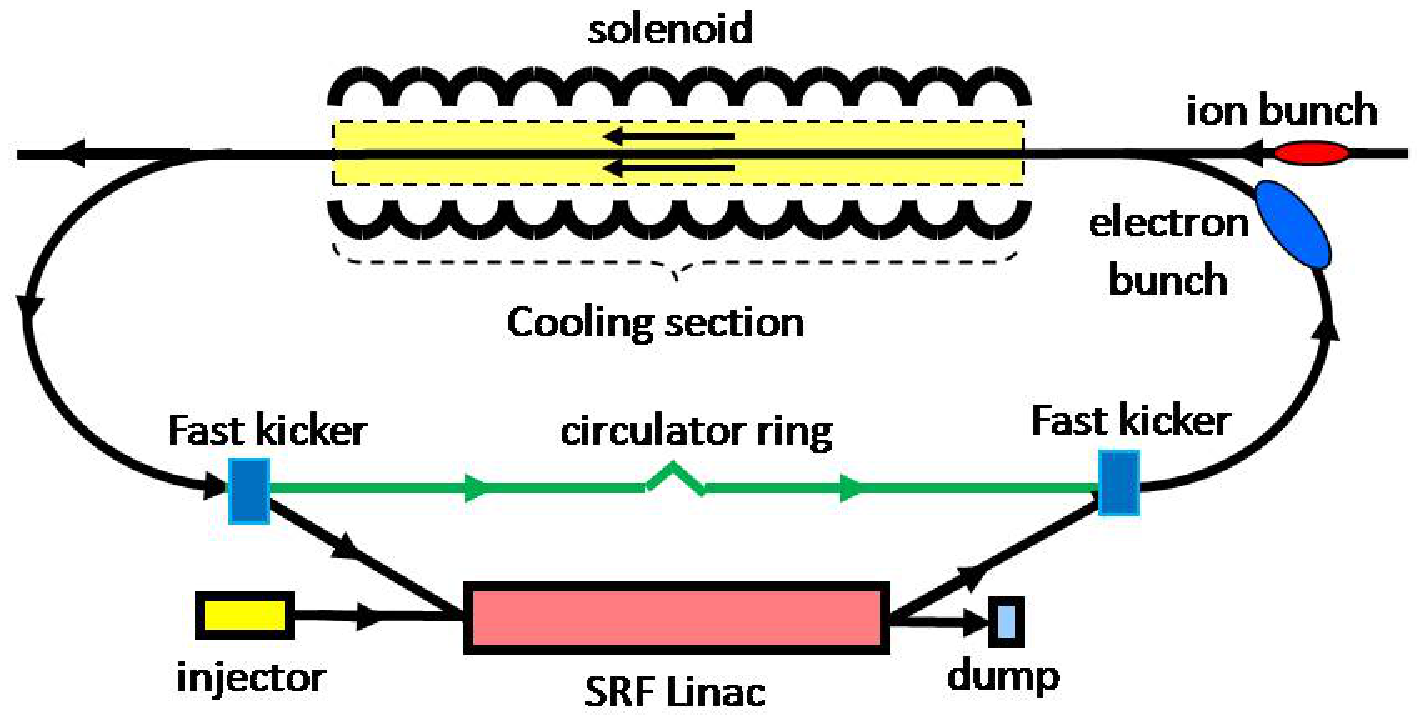}  
\end{center}
\caption{\small{Schematic of electron cooling for the MEIC.}}
\label{MEIC:fig_7}
\end{figure}

A schematic drawing of the MEIC ERL based electron cooler is shown in figure~\ref{MEIC:fig_7}. Two technologies, namely, energy recovery linac (ERL) and circulator ring, play critical roles to the success of this facility. A high-charge electron bunch from a photo-cathode is accelerated in a SRF ERL linac to required energy 10 to 50 MeV and then sent to a specially designed circulator cooler ring, with optics matching the cooling channel for cooling of a proton or ion bunch. The photo-injector and SRF linac ensure a high quality of the injected cooling bunch. An individual bunch circulates a large number of revolutions (up to a few hundred) in the ring before its quality is degraded by intra- and inter-beam scatterings, after which it returns to the same SRF linac for energy recovery. The recovered energy will be used for accelerating a new electron bunch. This circulator ring reduces the average current from a photo-cathode by a factor equal to the number of recirculation. Therefore, it provides a near-perfect solution for two bottlenecks of the facility: the high current and high power of the cooling electron beam. For example, a 3 A, 50 MV (e.g. 150 MW of power) cooling beam can effectively be provided by 30 mA, 2 MV (e.g 60 kW of active beam power) from the electron injector.

\subsection{R\&D}

For an advanced accelerator design like MEIC, there are many R\&D issues needed to be completed to solidify the design. We have identified a list of critical R\&D issues: electron cooling of the bunch ion beams at medium energy; crab crossing and crab cavity; polarization life time and spin tracking; beam-beam effects; non-linear collective beam effects and feedback systems; interaction region design and dynamic aperture, etc. The Jefferson Lab accelerator team and its collaborators are currently working on each of these issues. The details of the ongoing research are reported in a number of publications: overall design \cite{Z11,C11}, lattice design \cite{B11,S11}, ion complex \cite{E11,AEM11}, beam-beam simulations \cite{TZ10,TZ10b}, crab cavity \cite{A11}, electron cloud \cite{AKY11}, beam instabilities \cite{AKY11b}, and others. For brevity, here we only highlight one topic from our R\&D list -- beam-beam simulations.

Beam-beam interactions present a key limitation to collider performance, because they may lead to appreciable emittance growth of colliding beams and rapid reduction of luminosity. Such nonlinear collective beam effects can pose a significant design challenge when the machine parameters are pushed into a new regime. In order to lend credibility to the conceptual design, we use computer simulations to examine beam-beam instabilities, to optimize and explore limits of machine parameters. 

A first phase of the beam-beam simulations of MEIC at Jefferson Lab, featuring a simplified model with linear transfer map, head-on collisions, and perfect chromaticity correction, has been already carried out for the current medium-energy configuration \cite{TZ10,TZ10b,TKJ11} using the state-of-the-art beam collision code BeamBeam3D \cite{QFR02}. These studies established that both designs were safely away from coherent beam-beam instabilities.
	
\begin{figure}[htp]
\begin{center}
      \includegraphics[width=0.7\textwidth]{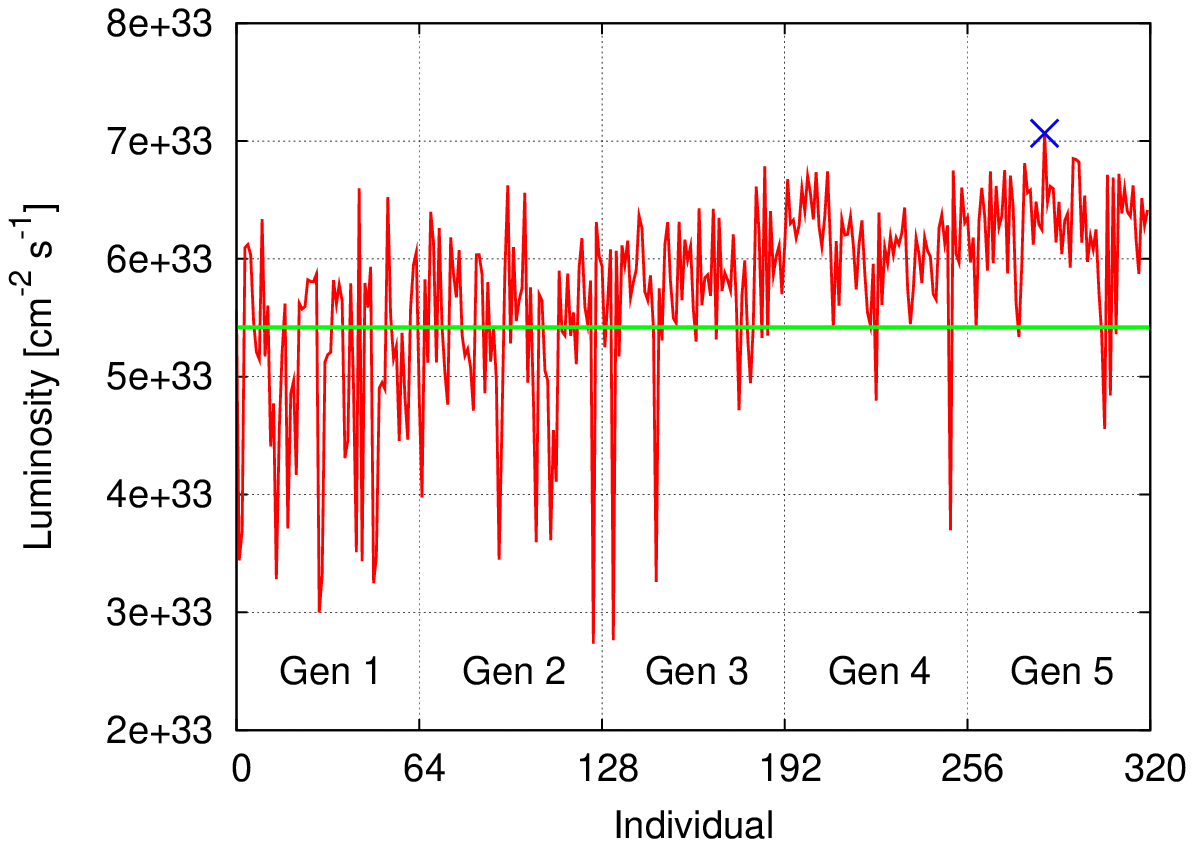}  
\end{center}
\caption{\small{MEIC beam-beam simulation with Evolutionary algorithm. Green line represents the design luminosity. The simulation locates a (near-)optimal working point within only 320 simulations (blue x).}}
\label{MEIC:fig_8}
\end{figure}

Furthermore, we use an evolutionary (genetic) algorithm \cite{ZLT01,BS05} to search for the optimal working point in the tune space, and demonstrated that such an approach is orders of magnitude more efficient than the simple tune scans \cite{KJT10}. Figure~\ref{MEIC:fig_8} illustrates how the evolutionary algorithm successfully navigates the 4D betatron tune space (2 tunes for each beam) to find a (near-)optimal working point for which the luminosity exceeds the design luminosity by about 30\%.

\subsection{Summary}

The MEIC is the future of nuclear physics at Jefferson Lab. It is optimized to collide a wide variety of polarized light ions and unpolarized heavy ions with polarized electrons. It covers an energy range matched to the science program proposed by the Jefferson Lab nuclear physics community ($\sim$ 4200 GeV$^2$), with luminosity exceeding $10^{34} {\rm cm}^{-2}{\rm s}^{-1}$. An upgrade path to higher energies ($250\times 20$ GeV$^2$) has been developed and should provide luminosity of close to $10^{35} {\rm cm}^{-2}{\rm s}^{-1}$. The design is based on a figure-8 ring for optimum polarization, and an ion beam with high repetition rate, small emittances and short bunch lengths. 

We reported on the status of the design for the MEIC at Jefferson Lab. Our design is both mature, having addressed all the required aspects of the design in the various level of detail, and flexible, being able to accommodate revisions in design specifications and advances in accelerator R\&D. We have identified the critical accelerator R\&D topics for the MEIC, and are presently working on them.

\vskip15pt
\noindent {\bf{Acknowledgment}}
\vskip5pt

We would like to thank members of the Jefferson Lab EIC nuclear science study group as well as the CEBAF user community for working with us on developing the MEIC design. In particular, we are grateful to Rolf Ent for coordinating the collaboration between the accelerator and nuclear physics groups, Alberto Accardi for generating Figure~\ref{MEIC:fig_8}, Pawel Nadel-Turonski and Tanja Horn for their contributions on detector and interaction region design.



\renewcommand{\textfraction}{0.2}
\section{Kinematics and detector designs for the different EIC machine designs}
\label{sec:detector}

\hspace{\parindent}\parbox{0.92\textwidth}{\slshape
  E.C. Aschenauer, R. Ent, T. Horn, P. Nadel-Turonski, H. Spiesberger}

\index{Aschenauer, Elke}
\index{Ent, Rolf}
\index{Horn, Tanja}
\index{Nadel-Turonski, Pawel}
\index{Spiesberger, Hubert}


\subsection{\label{sec:det.kin} Kinematics and Requirements for an EIC Detector}

The physics program of an EIC imposes several challenges on the design of a detector, and
more globally, the extended interaction region, as it spans 
a wide range in center-of-mass energy, different combinations of both beam 
energy and particle species, and several different physics processes. The various
physics processes encompass inclusive measurements ($ep/A \rightarrow e'+X$), which require
detection of the scattered lepton and/or the full scattered hadronic debris with high precision;
semi-inclusive processes  ($ep/A \rightarrow e'+h+X$), which require detection in coincidence
with the scattered lepton of at least one (current or target region) hadron; and exclusive processes 
($ep/A \rightarrow e'+N'/A'+\gamma/m$), which require detection of all particles in the reaction.
The following figures in this section demonstrate the differences in 
particle kinematics of some representative examples of these reaction types, as well as differing beam 
energy combinations. For these plots, the directions of the beams are defined as for HERA at DESY: 
the hadron beam is in the positive z direction (0$^o$) and the lepton beam is 
in the negative z-direction (180$^o$). 
\begin{figure}[hbtp]
\begin{center}
\includegraphics[width=0.95\textwidth]{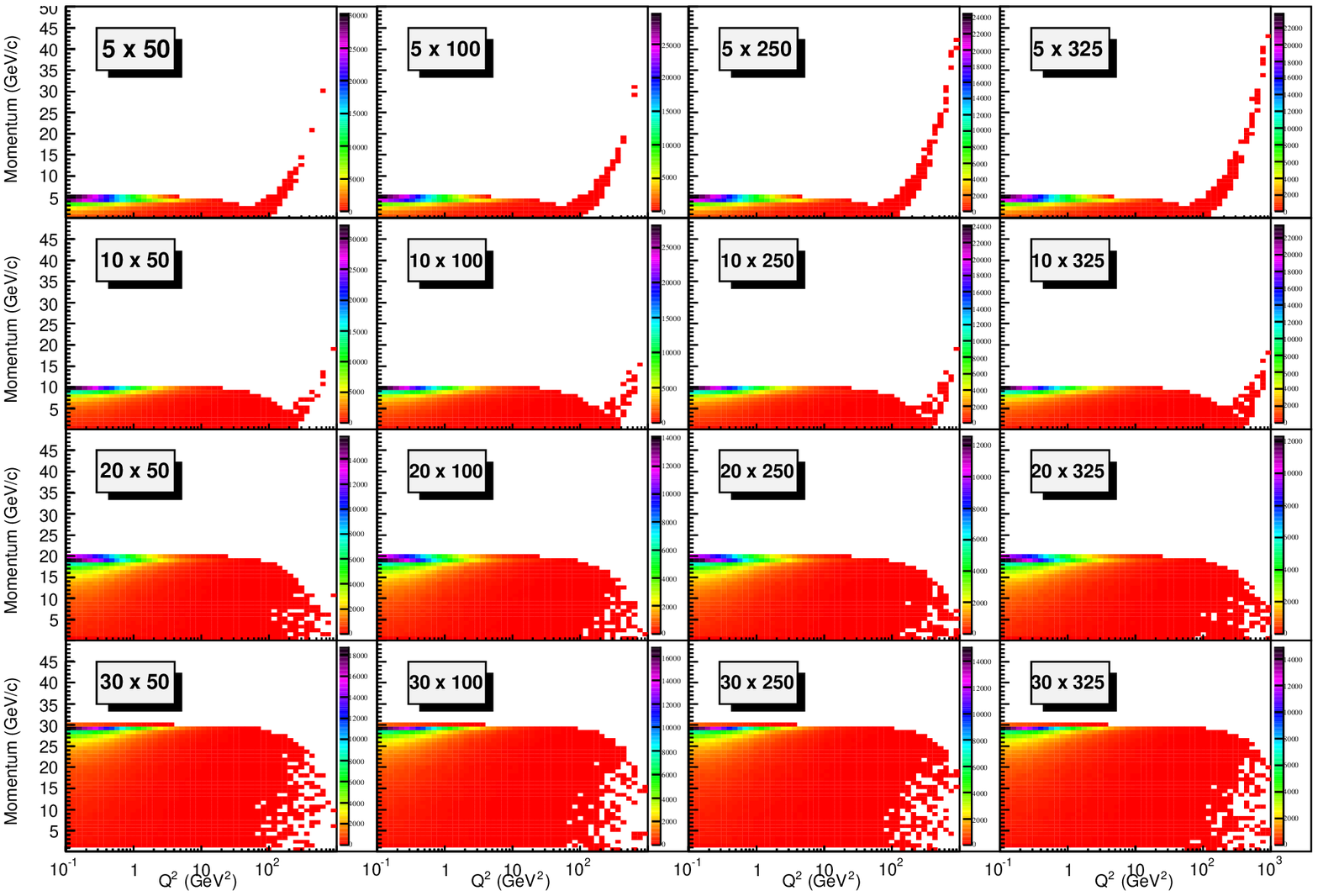}\\
\includegraphics[width=0.95\textwidth]{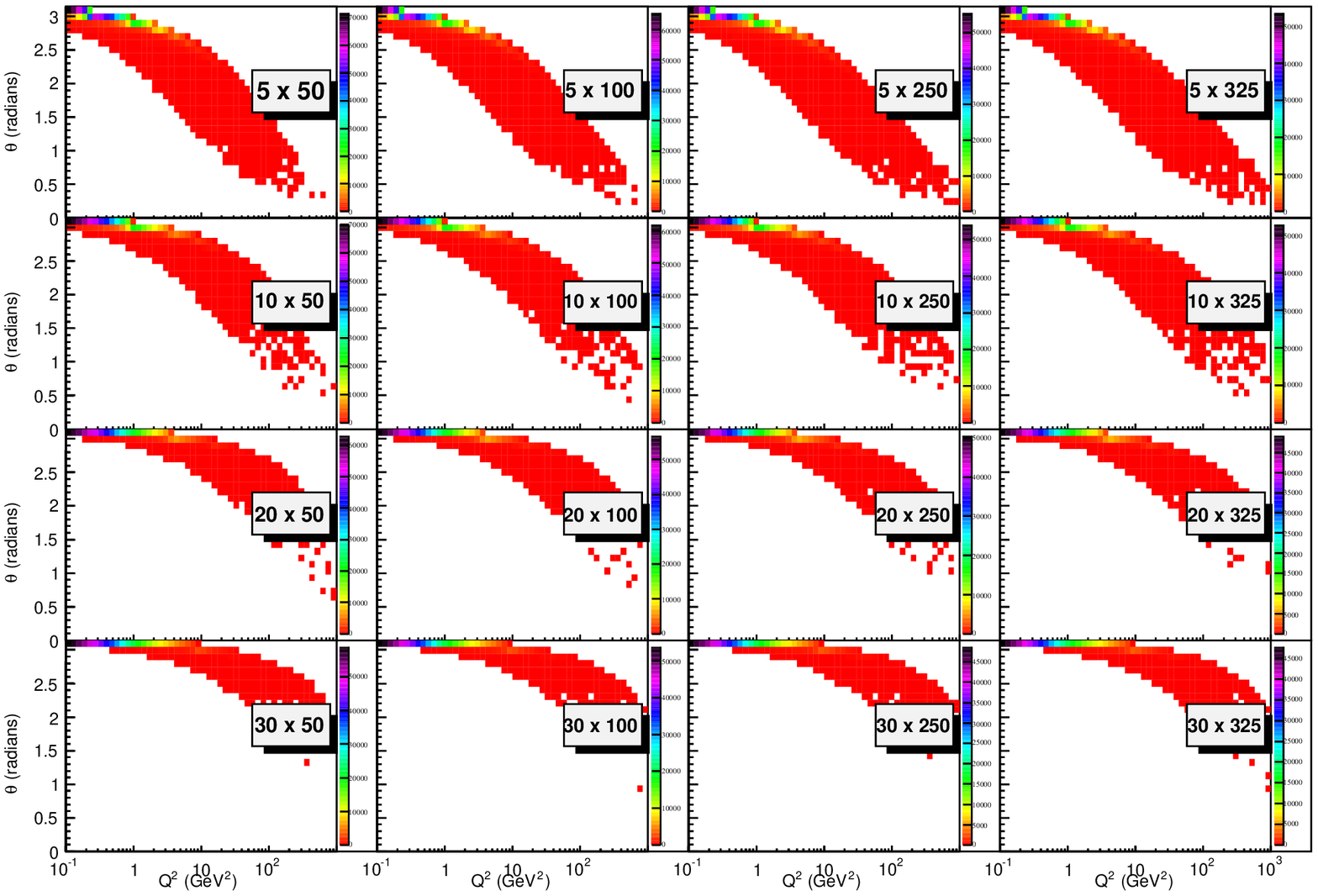}
\end{center}
\caption{\small Q$^2$ vs. momentum (upper panel) 
and Q$^2$ vs. scattering angle (lower panel) of the scattered 
lepton in the laboratory frame. The following cuts have been applied in both figures: 
Q$^2 > 0.1$ GeV$^2$, 0.01 $<$ y $<$ 0.95. The lepton-hadron beam energy combinations are indicated 
by the panel in each individual plots}
\label{fig:scatlep.kinematic} 
\end{figure}
The upper panel of fig. \ref{fig:scatlep.kinematic} illustrates that the lower Q$^2$ is,
the closer the momentum of the scattered lepton is to the original lepton beam energy.
For all lepton-hadron beam energy combinations (indicated by the panel in each of the plots), 
the scattered lepton goes in the direction of the original 
lepton beam for low Q$^2$ and more and more into a central detector acceptance for higher 
Q$^2$. For a fixed hadron beam energy the lepton scattering angle becomes smaller 
at a fixed Q$^2$ with increasing lepton energy.

\begin{figure}[hbtp]
\begin{center}
\begin{tabular}{cc}
\includegraphics[width=0.475\textwidth]{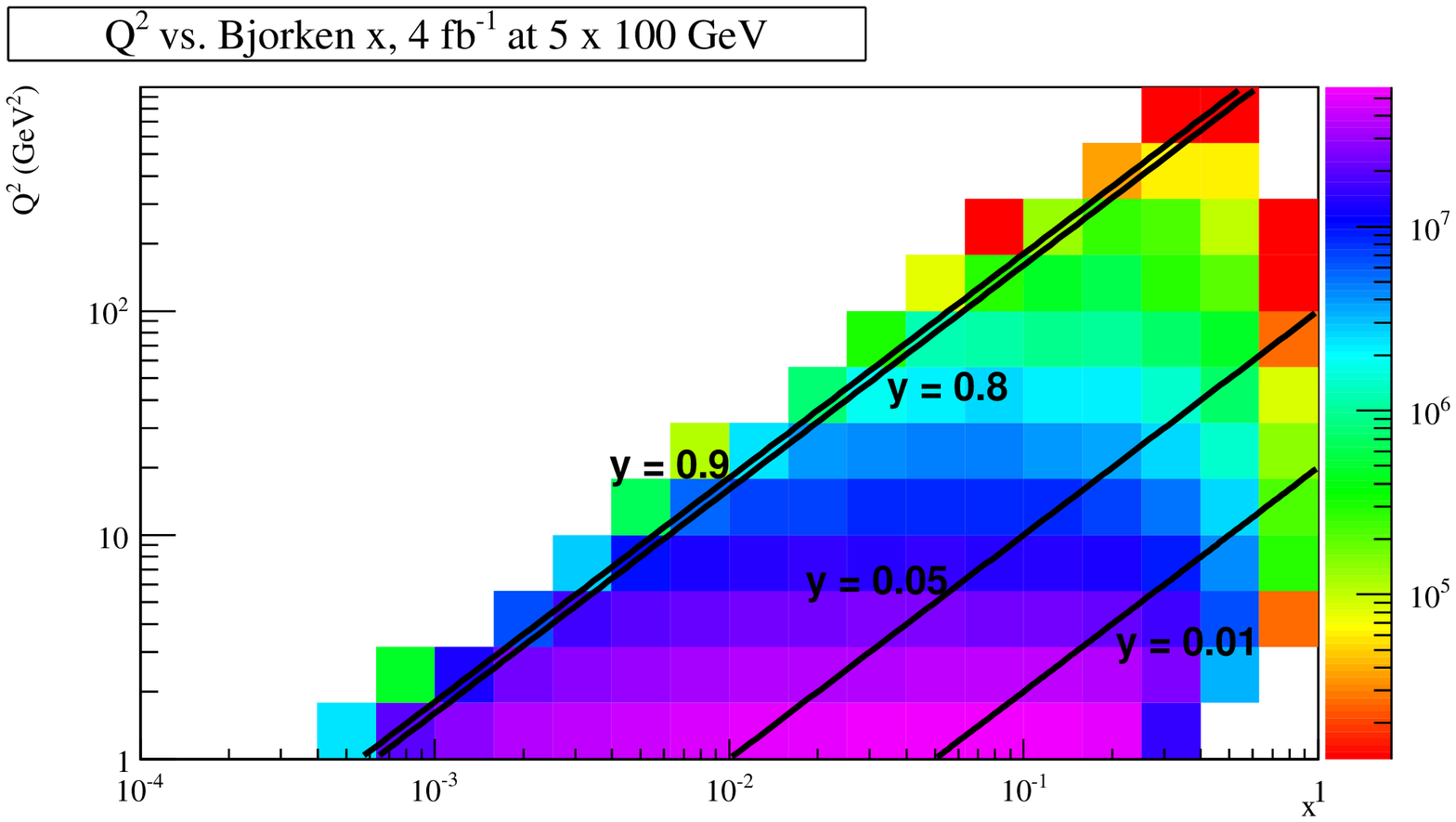} &
\includegraphics[width=0.475\textwidth]{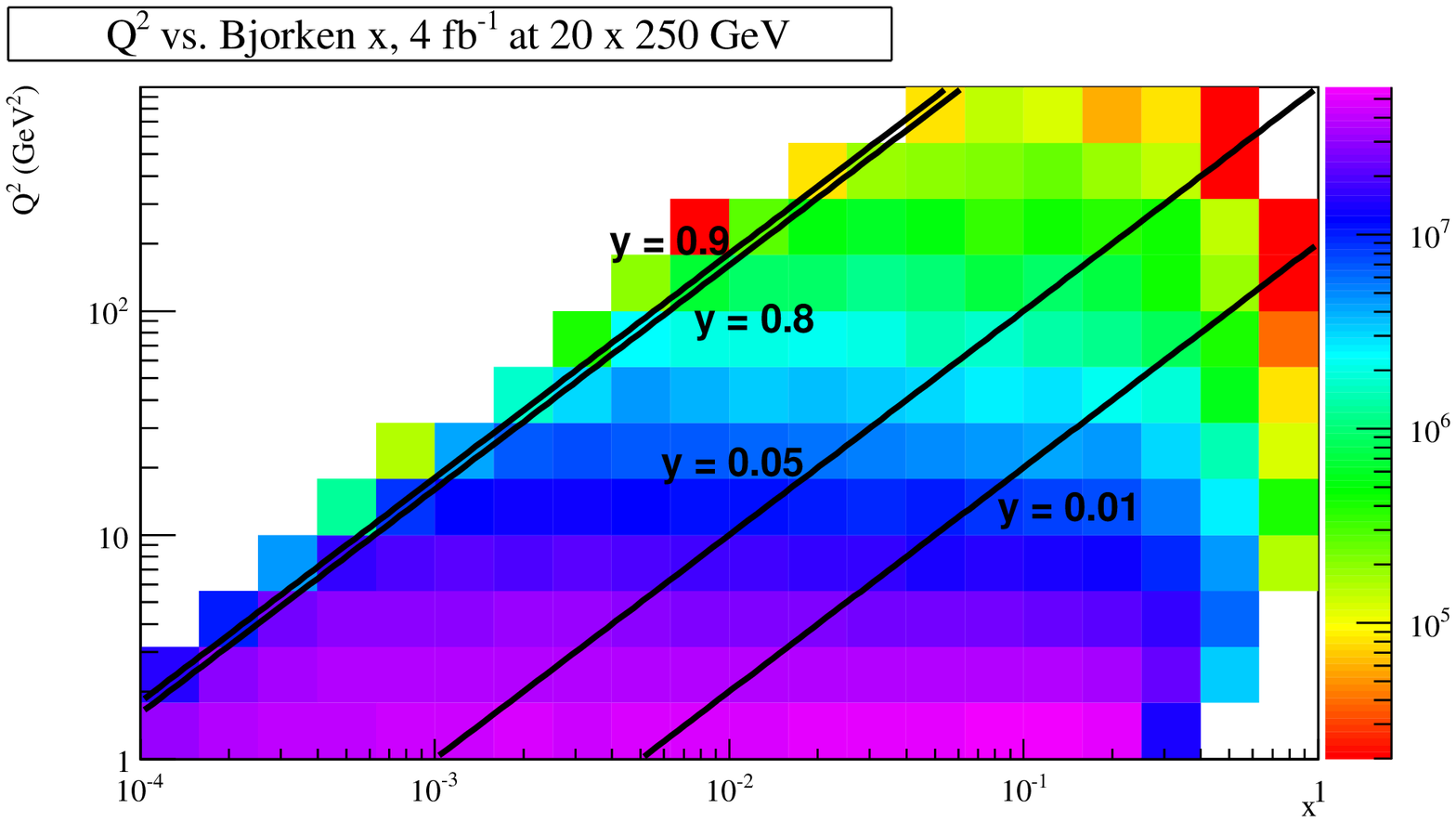}\\
\end{tabular}
\end{center}
\caption{\small The x-Q$^2$ plane for center-of-mass
energy 45 GeV (left) and 140 GeV (right). The black lines indicate different y-cuts 
placed on the scattered lepton kinematics.}
\label{fig:xQ2y}
\end{figure}

Fig. \ref{fig:xQ2y} shows the x-Q$^2$ plane for two different center-of-mass energies.
In general, the correlation between $x$ and $Q^2$ for a collider environment is weaker 
than for fixed target experiments. Nonetheless, it becomes stronger for small
scattering angles or corresponding small inelasticity $y$, and momentum and scattering angle
resolution for the scattered lepton become an issue, at HERA roughly at $y = 0.1$.
To circumvent this problem, HERA reconstructed
the lepton kinematics from the hadronic final state using the Jacquet-Blondel method
\cite{Jacquet:1979jb,Bassler:1994uq}, and has reached successful measurements down to $y$ of 0.005.
The main reason why this hadronic method renders better resolution at low $y$ follows from 
the equation $y_{JB} = E-P_z^{had} / 2 E_e$, 
where $E-P_z^{had}$ is the sum over the energy minus the longitudinal momentum of all 
hadronic final-state particles and $E_e$ is the electron beam energy. This quantity
has no degradation of resolution for $y<0.1$ as compared to the electron method, where
$y_e  = 1 - (1-cos \theta{_e})E'_e / 2 E_e$.
This is directly correlated to the relative resolutions for both quantities:
$\Delta y_{JB}/ y_{JB} \sim$ constant and $\Delta y_e /y_e \sim 1 / y_e$.

Typically, one can obtain for a given center-of-mass energy squared roughly a decade of $Q^2$
reach at fixed $x$ when using only an electron method to determine lepton kinematics, and roughly
two decades when including the hadronic method. If only using the electron method, one can
increase the range in accessible $Q^2$ by lowering the center-of-mass energy, as can be seen
from comparing the two panels of fig. \ref{fig:xQ2y}. This may become relevant
for some semi-inclusive and exclusive processes. The advantages and disadvantages 
of this solution are discussed in the two machine-specific detector sections of this section.

In general, one would like to access as large a range in $Q^2$ at fixed $x$ as possible for a
given beam energy combination, and reach as low $y_{JB}$ as possible. This requirement directly
implies two important considerations for the detector design:
\begin{itemize}
\item good hadronic coverage in the forward direction
\item low noise and/or good noise suppression algorithms in the hadronic calorimeter to allow for
hadron detection down to 0.5 GeV. More detailed detector simulations are needed to confirm these requirements.

\end{itemize}
It is important to point out that the reconstruction of the event kinematics from 
the hadronic final state is also important in suppressing events with radiation 
of a real or virtual photon from the incoming or outgoing lepton (radiative corrections);
for details please see section \ref{sec:radcorr}.

One should keep in mind that there are additional complications at low $y$ for the
measurement of asymmetries and/or polarized cross sections, to for example extract the 
helicity-dependent parton distributions. A depolarization factor, defined in \cite{PhysRevD.62.079902} as:
\begin{align}
D &=
\frac{y[(1+\gamma^2y/2)(2-y)-2y^2m_e^2/Q^2]}{y^2(1-2m_e^2/Q^2)(1+\gamma^2)+2(1+R)(1-y-\gamma^2y^2/4)}
\end{align}
is needed to correct the measured helicity-dependent asymmetries ($A_{||}$).
The depolarization factor corrects for the polarization transfer from the lepton to the virtual photon,
and is small at low $y$. This reduces the effective polarized luminosity and increases the uncertainties
of the measured polarized quantities at low y ($\delta A_1 = \delta A_{||}/D$). Therefore, the
$x-Q^2$-plane of precision polarized cross section measurements will be reduced
as compared to unpolarized ones, for fixed center-of-mass energy.

\begin{figure}[hbtp]
\begin{center}
\includegraphics[width=0.90\textwidth]{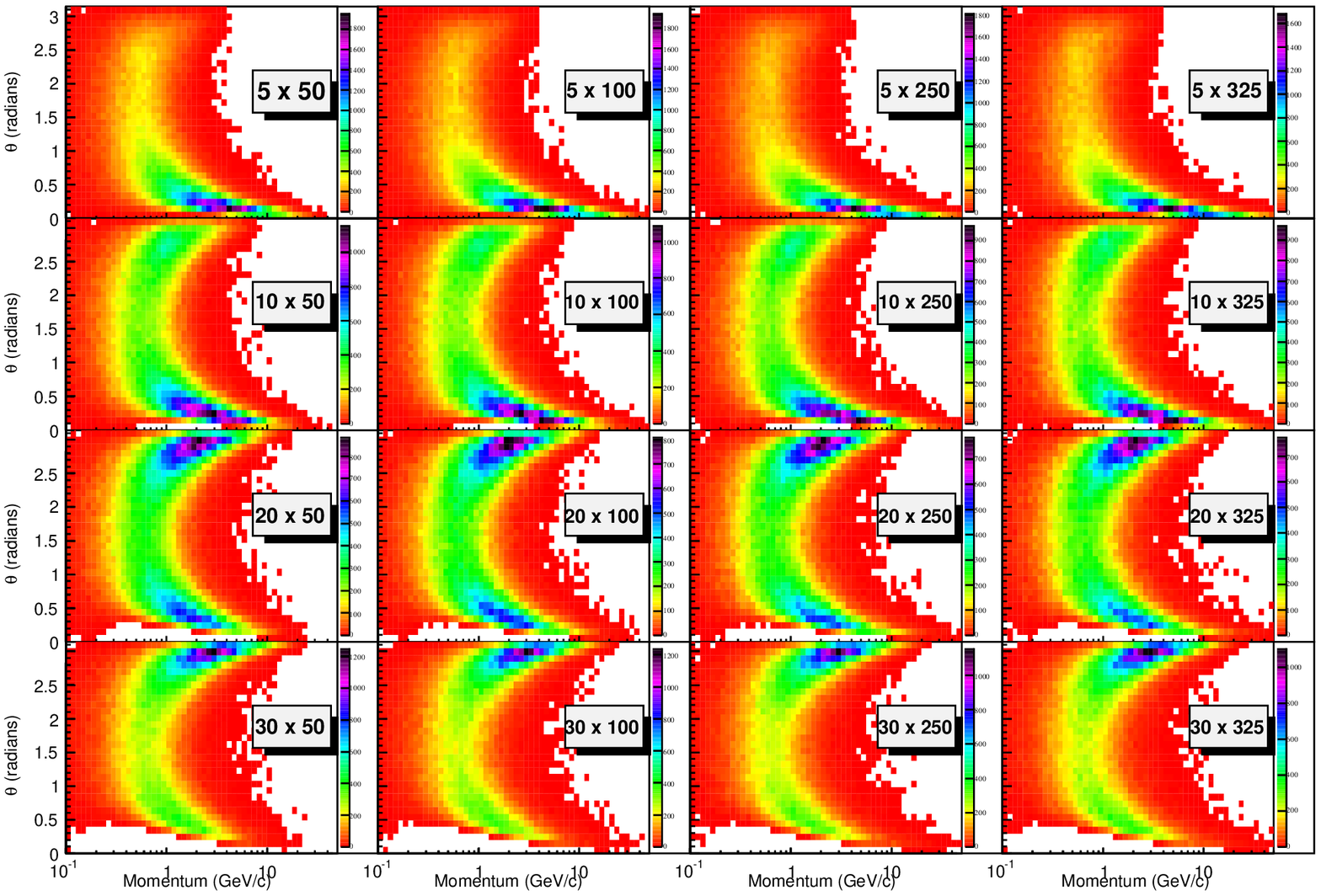}
\end{center}
\caption{\small Momentum vs. scattering angle in the laboratory frames for pions from 
non-exclusive reactions.  The following cuts have been applied: 
Q$^2 > 1$ GeV$^2$, 0.01 $<$ y $<$ 0.95 and 0.1 $<$ z $<$ 0.9 }
\label{fig:pion.kinematic} 
\end{figure}
Fig. \ref{fig:pion.kinematic} shows the momentum versus scattering angle distributions 
in the laboratory frame for pions originating from semi-inclusive reactions, for different 
lepton and proton beam energy combinations. For lower lepton energies, pions are scattered
more in the forward (ion) direction. For fixed low lepton energy of 5 GeV, this pattern
remains more or less constant as a function of proton energy. With increasing lepton beam
energy, the hadrons increasingly populate the central region of the detector, and at the
highest lepton energies, hadrons are even largely produced going backward (i.e. in the lepton
beam direction). The kinematic distributions for kaons and protons, applying the same cuts
as for pions, are essentially identical to those of the pions. The distributions 
for semi-inclusive events in electron nucleus collisions may be slightly altered due to
nuclear modification effects, but the global features will remain. 

Fig. \ref{fig:pion.kinematic} also indicates a shift of the momentum range of pions towards
higher momenta in the central-angle region for higher lepton energy, to typical momenta
of about 10 GeV/c, which has implications for the required particle identification (PID).
To be able to identify the 
different hadron types over a wide momentum and angular range an EIC detector needs 
to have detectors capable of good PID in the forward, central and backward direction. 
For the higher hadron momenta, typically in the forward ion direction and also in the
backward  direction for higher lepton beam energies, the most viable detector technology
is a Ring-Imaging Cherenkov (RICH) detector with dual-radiators. 
In the central detector region a combination of high resolution time-of-flight (ToF)
detectors (preferentially with timing resolutions $\delta$t $\sim$ 10ps), a DIRC, or 
a proximity focusing Aerogel RICH may be adequate detector technologies.

\begin{figure}[htbp]
\begin{center}
\includegraphics[width=0.95\textwidth]{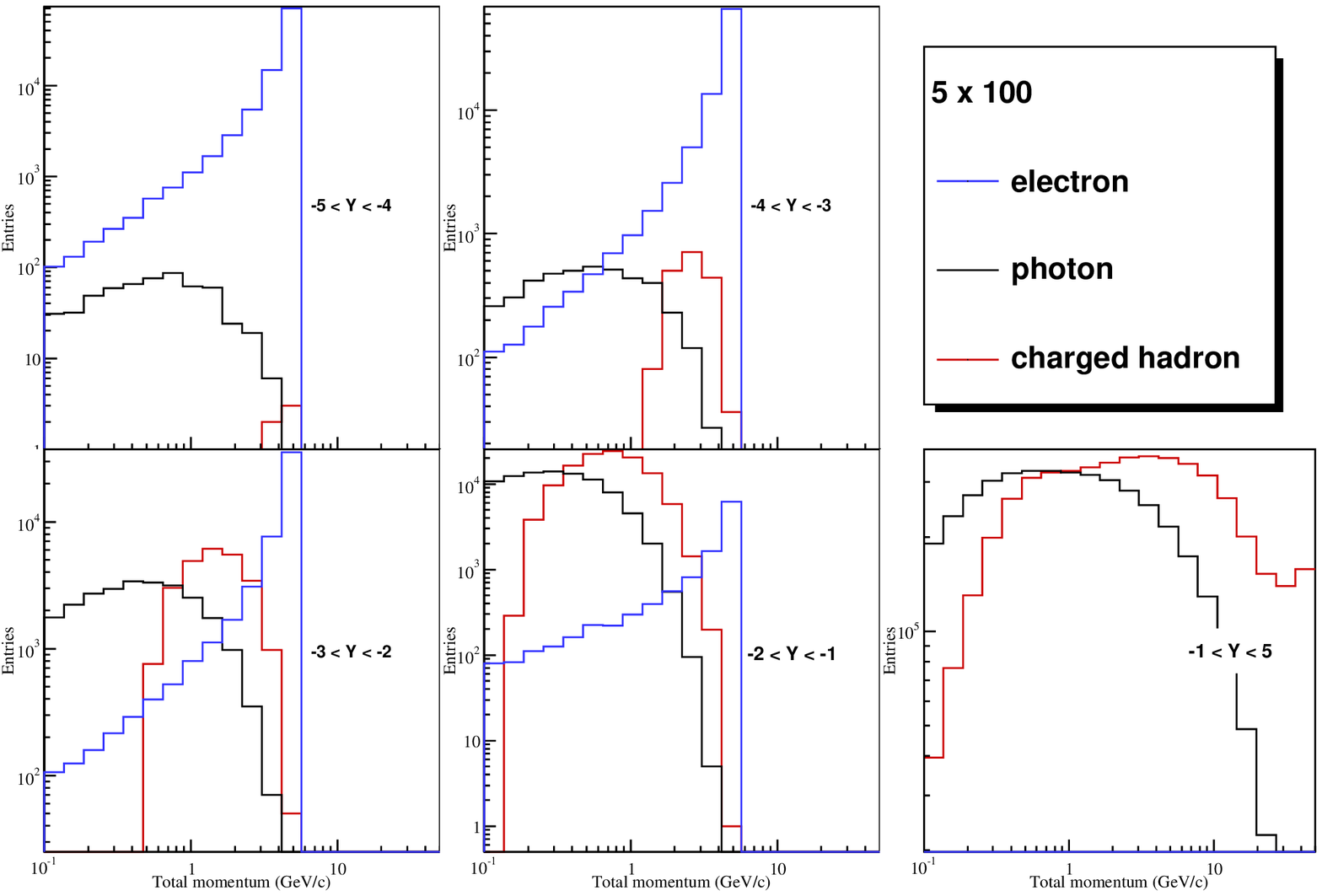}
\includegraphics[width=0.95\textwidth]{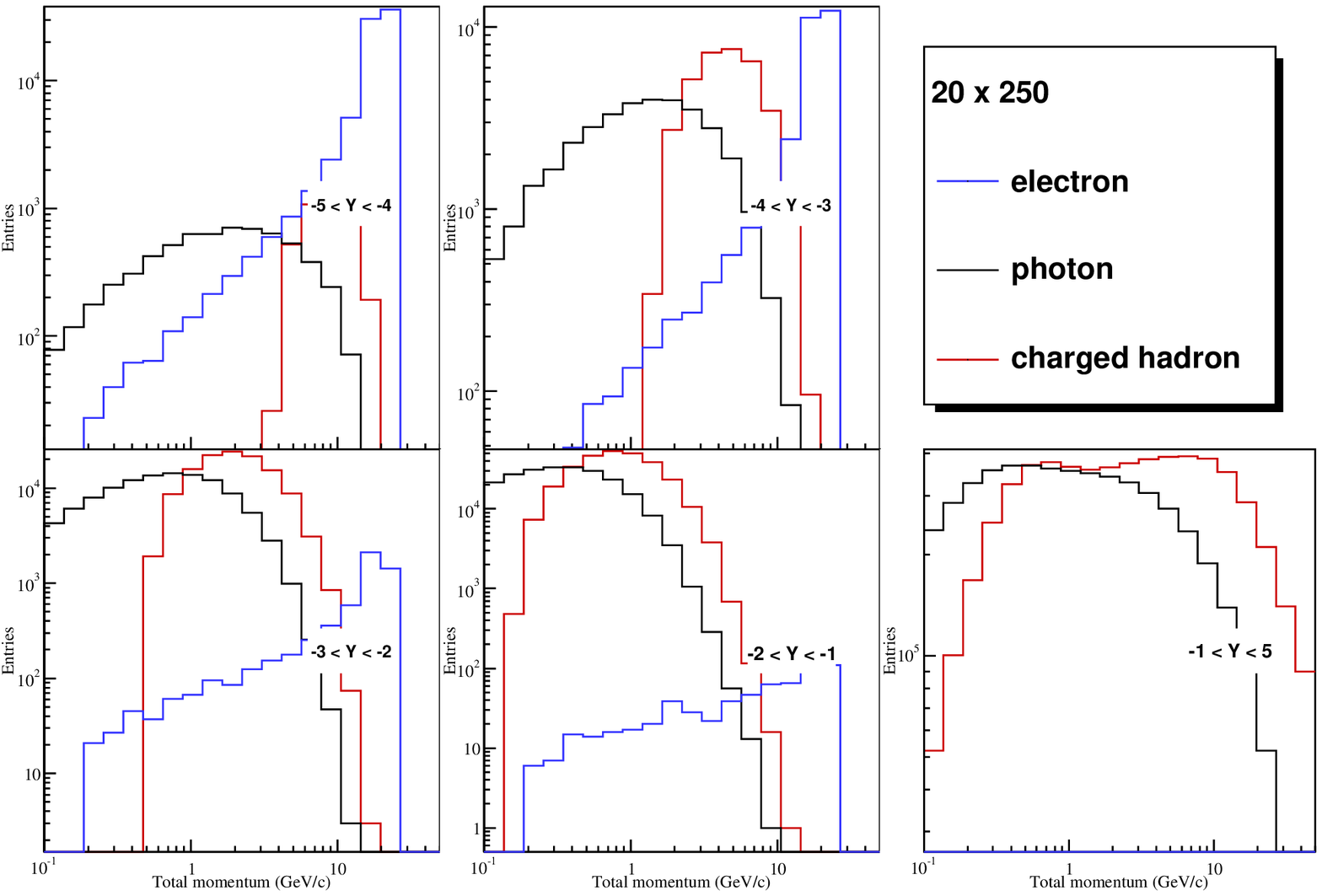}
\end{center}
\caption{\small The number of photons and hadrons as well as the number of 
scattered leptons in a rapidity bin vs momentum having 5 GeV leptons colliding with 
100 GeV protons and 20 GeV leptons colliding with 250 GeV protons. No kinematic cuts 
are applied.}
\label{fig:lephadpho}
\end{figure}

For certain kinematics, the hadrons (both charged and neutral) will be produced in the
backward ion direction (see fig. \ref{fig:lephadpho}) and need to be disentangled from
the scattered leptons. The kinematic region in rapidity $\eta$, over which hadrons and
photons need to be suppressed with respect to electrons, shifts to more negative
rapidity with increasing center-of-mass energy. This can be cross-correlated with the
angular and momentum patterns for scattered leptons of fig. \ref{fig:scatlep.kinematic}.
For the lower center-of-mass combination, electron, photon and charged hadron rates
are roughly comparable at 1 GeV/c total momentum and $\eta$ = -3. For the higher
center-of-mass energy, electron rates are a factor of 10-100 smaller than photon and
charged hadron rates, and comparable again at a 10 GeV/c total momentum.

This adds another requirement to the detector: good electron
identification. The kinematic region in rapidity $\eta$ over which 
hadrons and also photons need to be suppressed, typically by a factor of 10 - 100,
shifts to more negative rapidity with increasing center-of-mass energy.
Measuring the ratio of the lepton energy and momentum, E$_e'$/p$_e'$, typically gives a
reduction factor of $\sim$ 100 for hadrons. This requires the availability of both
tracking detectors (to determine momentum) and electromagnetic calorimetry (to determine
energy) over the same rapidity coverage.
This availability also immediately suppresses the misidentification of photons in the 
lepton sample, by requiring that a track must point to the electromagnetic cluster.
Of course, the availability of good tracking detectors over similar coverage as
electromagnetic calorimetry similarly aids in $y$ resolution at low y from a lepton
method only (see earlier), as the angular as well as the momentum
resolution for trackers are much better than for electromagnetic calorimeters.
The hadron suppression can be further improved by adding a Cherenkov detector to
the electromagnetic calorimetry. Combining the electromagnetic calorimeter response
and the response of Cherenkov detectors may especially help in the region of low-momentum
scattered leptons, about 1 GeV/c. Other detector technologies, such as transition
radiation detectors, may provide another factor 100 hadron suppression for
lepton momenta greater than 4 GeV/c.

\begin{figure}[htbp]
\begin{center}
\includegraphics[width=0.90\textwidth]{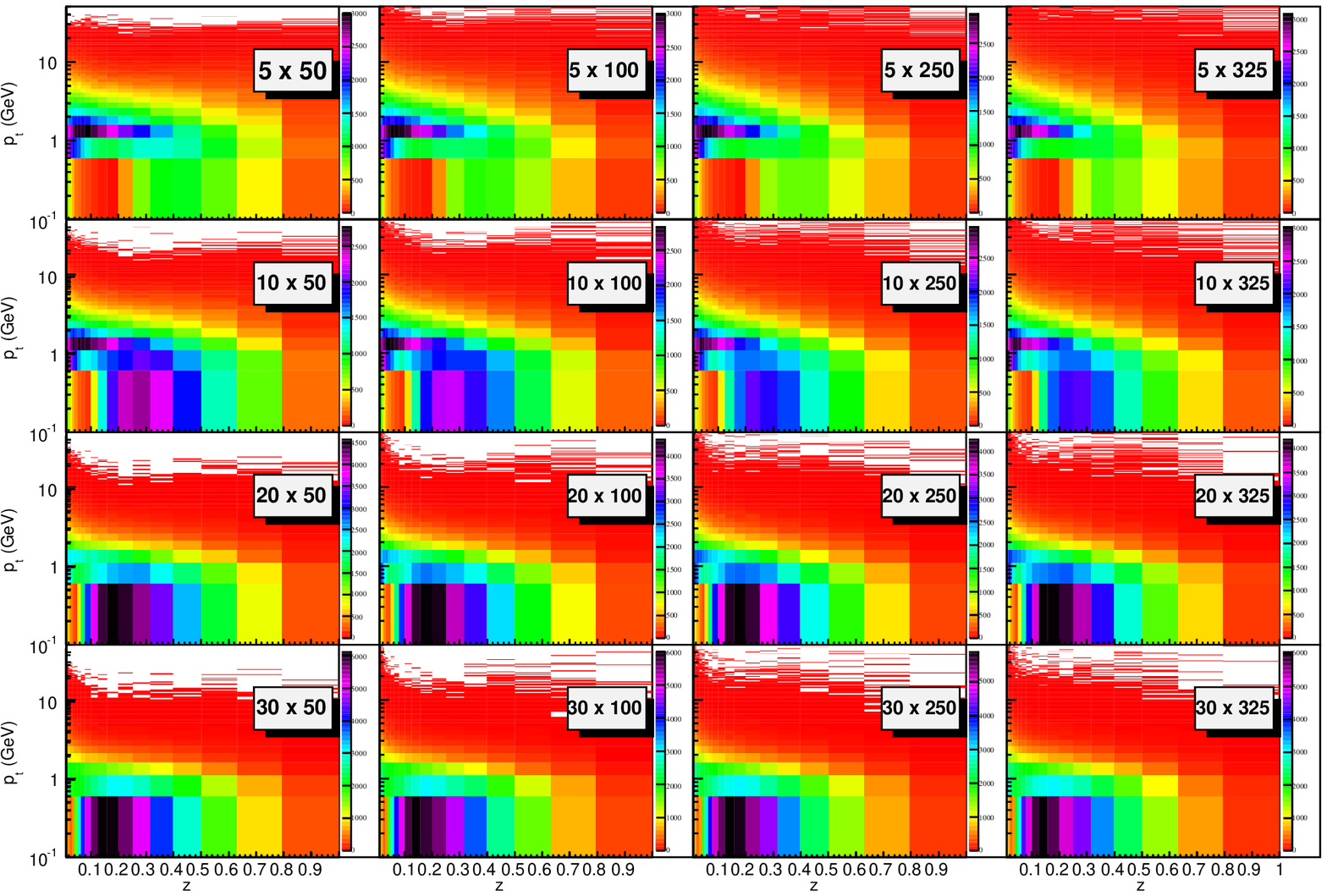}
\end{center}
\caption{\small Transverse momentum vs. z for pions applying the 
following cuts Q$^2 > 1$ GeV$^2$, 0.01 $<$ y $<$ 0.95, 0.5$^o < \theta < 179.5^o$ 
 and p $>$ 1 GeV. A momentum cut is applied to simulate the threshold of 
potential particle-identification-detectors.} 
\label{fig:pion.ptz}
\end{figure}

An additional advantage of a collider detector over a fixed target experiment is the large
coverage in transverse momentum. This is especially important for measurements linking the
perturbative high-transverse momentum $p_T$ region to the region of small transverse momentum,
$p_T \sim \Lambda_{QCD}$, where single-spin asymmetries as functions of $p_T, x, Q^2, z$
and $\phi$ are the prime 
observable to extract TMDs - Transverse Momentum Dependent Parton Distributions
(see chapter~\ref{chap:tmd}), like the Sivers function.
Fig. \ref{fig:pion.ptz} shows the coverage in hadron p$_T$ measured with respect 
to the virtual photon vs. $z=E_h/\nu$ assuming an angular acceptance of a detector 
0.5$^o < \theta < 179.5^o$. One can see that for all beam energy combinations a
large range in transverse momentum is achievable. In general, such physics does not drive
the most forward (or backward) detector requirements, leaving ample phase space in
transverse momentum with respect to the virtual-photon direction - typically more central. 

There is specific interest in detecting events with heavy quarks (charm or bottom).
To measure the inclusive structure functions, F$_2^c$, F$_L^c$, and F$_2^B$ for heavy quarks, 
it is sufficient to tag the charm and the bottom quark content via the detection of additional leptons
(electron, positron, muons) to the scattered lepton. The leptons from charmed mesons
can be identified via a displaced vertex of the second lepton ($<\tau> \sim 150 \mu$m). 
This can be achieved by integrating a high-resolution vertex detector into 
the detector design. For measurements of the charmed (bottom) fragmentation 
functions, or to study medium modifications of heavy quarks in the nuclear environment,
at least one of the charmed (bottom) mesons must be completely reconstructed to have access to the 
kinematics of the parton.
This requires, in addition to measuring the displaced vertex, good particle
identification to reconstruct the meson via its hadronic decay products, e.g. 
$D_0 \rightarrow K^{\pm}+\pi^{\mp}$. 

\begin{figure}[htbp]
\begin{center}
\includegraphics[width=0.90\textwidth]{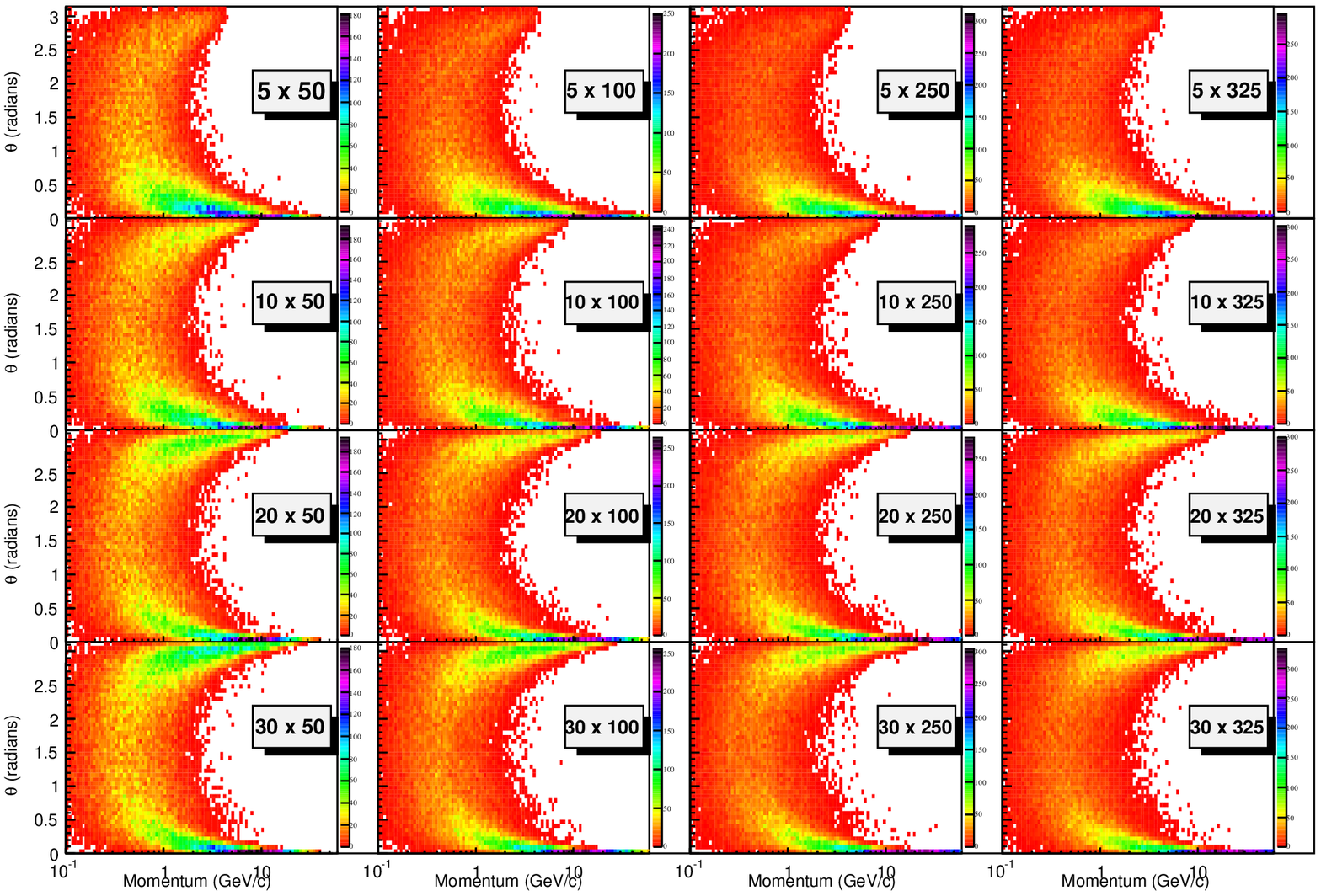} \\
\includegraphics[width=0.90\textwidth]{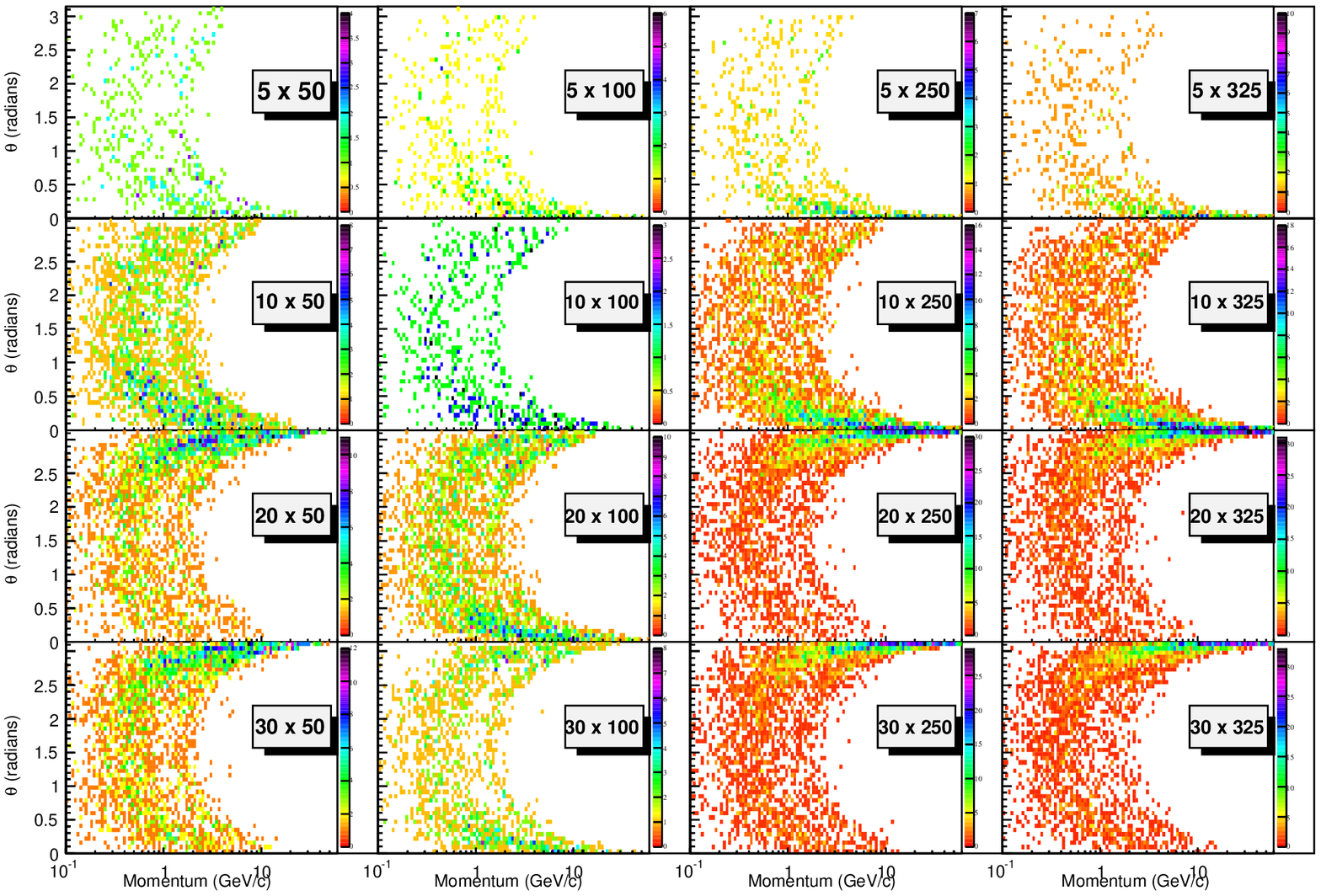}
\end{center}
\caption{\small Upper Panel: Momentum vs. scattering angle in the laboratory frames for pions following
from exclusive $\rho^0$ vector meson production. The following cuts are 
applied: Q$^2 > 1.0$ GeV$^2$, 0.01 $<$ y $<$ 0.95.}
Lower Panel:  Momentum vs. scattering angle in the laboratory frames for muons following
from exclusive J/$\psi$ vector meson production. No cuts on Q$^2$ have been applied as a 
hard scale for the process is given by the J/$\psi$ mass.
\label{fig:exclhadron}
\end{figure}

Fig. \ref{fig:exclhadron} (upper panel) shows the momentum versus scattering angle 
distributions for pions following from an exclusive reaction with a $\rho^0$ vector 
meson production ($Q^2 >$ 1.0 GeV$^2$), in the laboratory frame and for different
beam energy combinations. As in fig. \ref{fig:pion.kinematic}, two familiar patterns
arise. For increasing lepton beam energy, the pion distribution goes from being more
peaked in the forward-angle direction to a distribution with both a peak in the forward
and backward ion direction, and the momentum in the forward-ion direction is slightly
reduced. Most of the forward-ion direction pions in fig. \ref{fig:exclhadron} are correlated
with lower-$Q^2$ processes, though possibly of less interest for these processes. If one
would use a $Q^2 >$ 10 GeV$^2$ cutoff in these exclusive processes, only a peak in the
backward-ion direction would remain and in that sense, lower lepton energies correspond
to lower hadron momenta on average and reduced particle identification requirements.
The distributions for kaons from exclusive $\phi$-mesons production as well as for 
muons/electrons from exclusive $J/\psi$ production look very similar, see lower panel of fig.
\ref{fig:exclhadron}.
The most challenging constraints on the detector design for exclusive reactions compared
to semi-inclusive reactions is, however, not given by the hadrons originating from vector
mesons, but from the detection of the exclusive hadronic state remaining.

\begin{figure}[htbp]
\begin{center}
\includegraphics[width=0.85\textwidth]{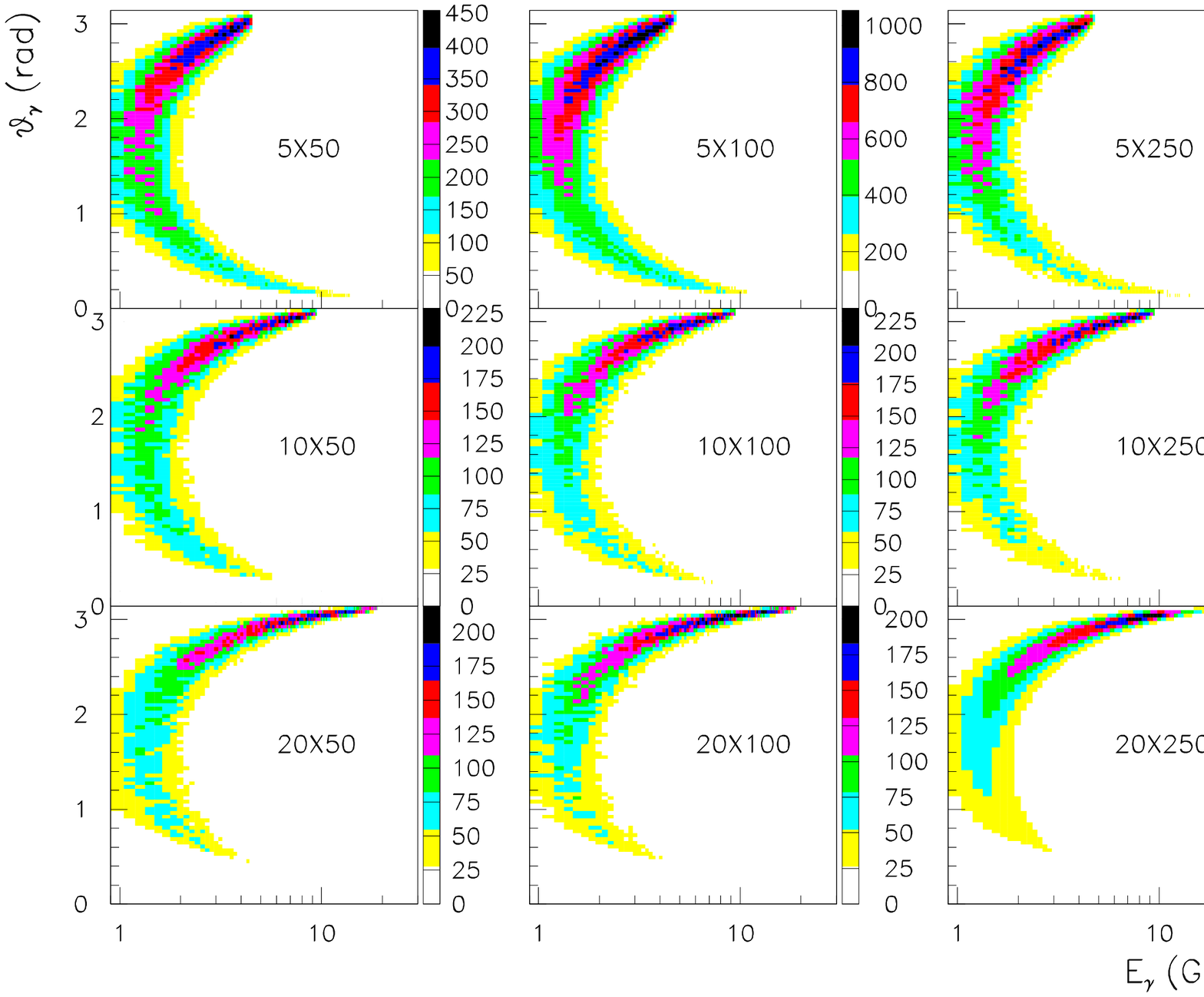} \\
\includegraphics[width=0.85\textwidth]{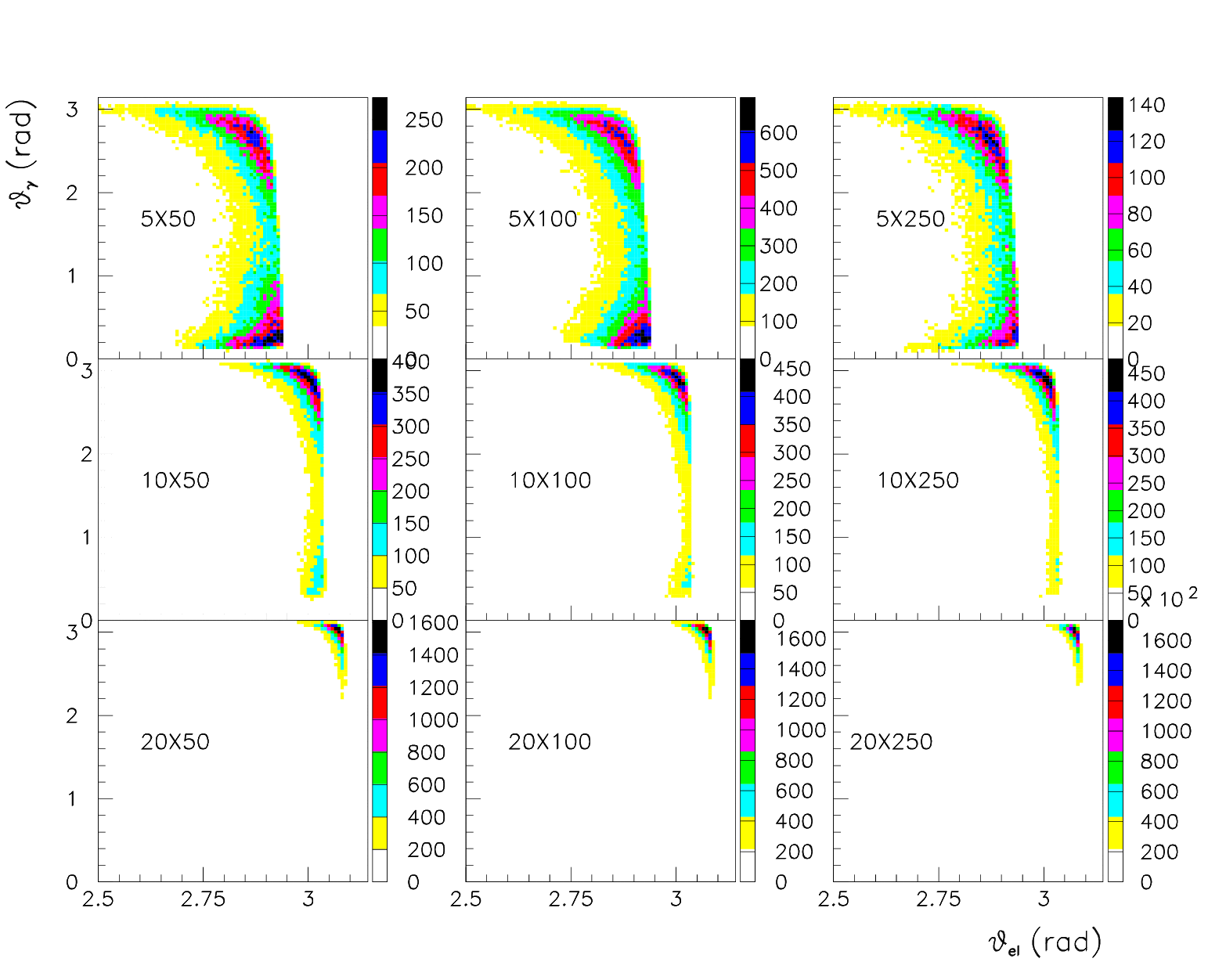}
\end{center}
\caption{\small Upper panel: Energy vs. scattering angle in the laboratory
frame for photons from DVCS. The following cuts have been applied: Q$^2 > 1.0$ GeV$^2$, 
0.01 $<$ y $<$ 0.95 and E$_\gamma >$ 1. GeV.
Lower Panel: The scattering angle in the laboratory frame
of the photon vs. that of the scattered lepton for DVCS events. The following cuts
have been applied: Q$^2 > 1.0$ GeV$^2$, 0.01 $<$ y $<$ 0.95 and E$_\gamma >$ 1. GeV} 
\label{fig:exclgamma}
\end{figure}

As one specific example of an exclusive reaction, deeply virtual compton
scattering (DVCS) was chosen, fig. \ref{fig:exclgamma} (top) shows the energy versus scattering angle
distributions of photons in the laboratory frame, for different beam energy combinations.
A cut of Q$^2 >$ 1 GeV$^2$ is assumed, although larger values of Q$^2$ may be required.
Lower lepton energies show a more symmetric distribution, and higher lepton energies
are more backward-ion angle peaked. The distributions show relatively homogeneous
distributions of the DVCS photons from forward to backward, with a small preference for
the backward direction. The latter is true for all lepton-hadron beam energy combination. 

Fig. \ref{fig:exclgamma} (bottom) correlates the distribution of the photon angle and the
electron scattering angle in the laboratory frame, for different beam-energy combinations.
With increasing lepton beam energy, the photon and scattered lepton tend towards the same
detector hemisphere. Following fig. \ref{fig:exclgamma} (top), electromagnetic calorimetry
is required over the entire rapidity range of the detector. Fig. \ref{fig:exclgamma} (bottom)
illustrates that tracking and electromagnetic calorimetry capabilities covering similar rapidity
range will greatly aid the separation of the photon and lepton, reducing a difficulty encountered
by the ZEUS collaboration in their DVCS event reconstruction.

\begin{figure}[htbp]
\begin{center}
\includegraphics[width=0.90\textwidth]{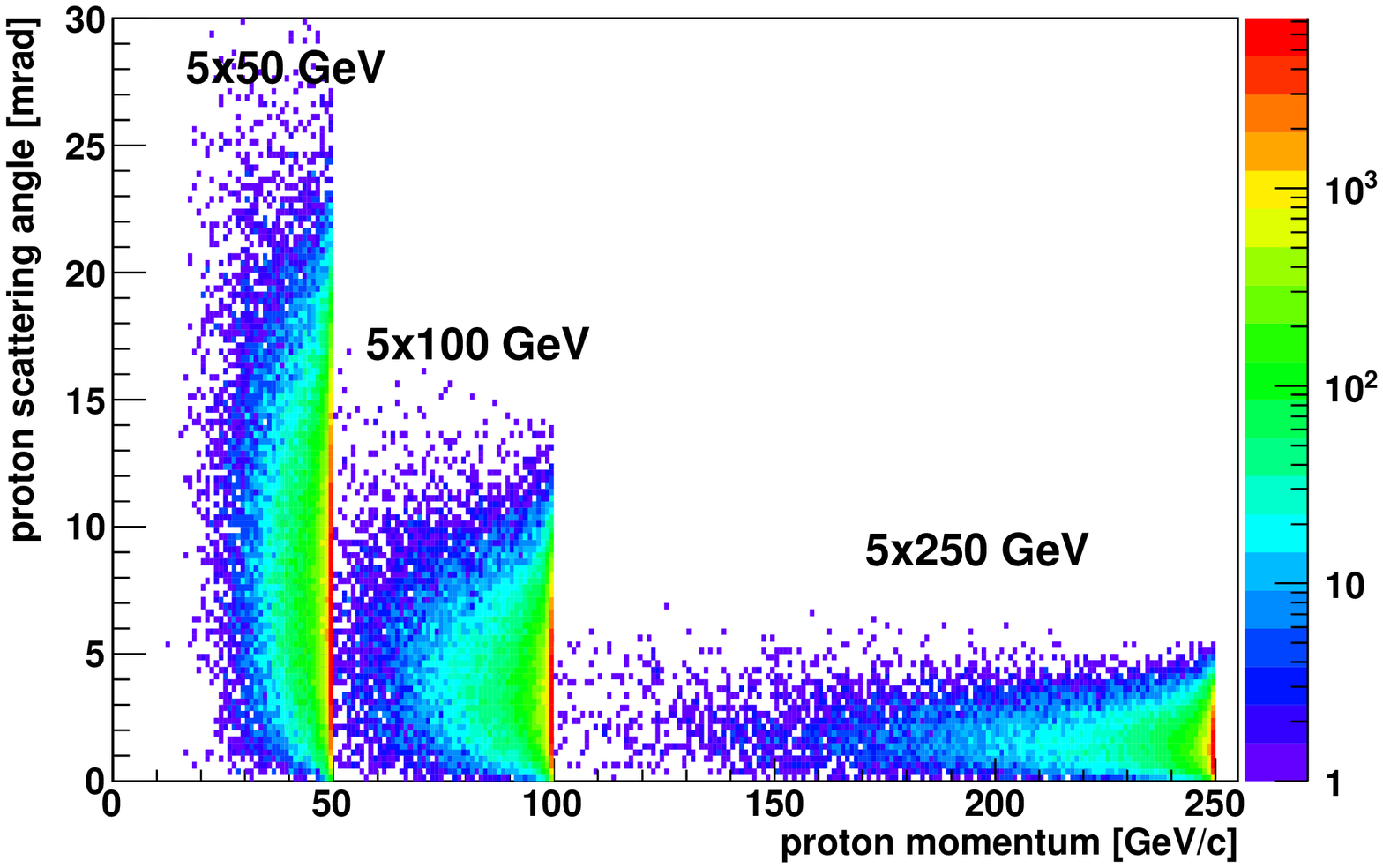}
\end{center}
\caption{\small Scattered proton momentum vs. scattering angle in the laboratory 
frames for DVCS events with different beam energy 
combinations.  The following cuts have been applied: 
1 GeV$^2 < $ Q$^2 < 100$ GeV$^2$, 10$^{-5} < $ x $ < $ 0.7 and 0 $<$ t $<$ 2 GeV$^2$.}
\label{fig:exclproton}
\end{figure}

For exclusive reactions in general, with DVCS as the example above, it is extremely important
to ensure that the remaining nucleon (or the nucleus) remains intact during the scattering process.
Hence, one has to ensure exclusivity by measuring all products. Fig. \ref{fig:exclproton} 
illustrates the kinematic requirements for the DVCS case, showing the 
scattered proton momentum versus its scattering angle for three different beam energy
combinations. In general, for exclusive reactions one wishes to map the four-momentum
transfer (or Mandelstam variable) t to the hadronic system, and then obtain an image
by a Fourier transform, at relatively low t of up to 1-2 GeV. The angle of the recoiling
hadronic system is directly correlated with $t$ and the proton energy $E_p$, as $\sqrt{t}/E_p$.
As can be seen in fig. \ref{fig:exclproton}, the proton scattering angle requirements
indeed linearly (and inversely) scale with proton energy. 

Even at a proton energy of 50 GeV, the proton scattering angles only range to about 1-2$^\circ$.
At proton energies of 250 GeV, this number is reduced to one fifth. In all cases, one obtains
small to extremely small scattering angles, extending to or completely within the
0.5$^\circ$ angular detector cutoff often used above. Because of this, the detection of these 
protons, or more general recoil baryons, is extremely dependent on the exact interaction region
design and will therefore be discussed in more detail in the machine-dependent part of this chapter.

Detection of the intact nucleus following an exclusive reaction in eA collisions is even
more complicated. The binding energy in heavy nuclei is of the order 8 MeV per nucleon.
In general, the smallest measurable outgoing angle of heavy scattered or fragmented nuclei,
$\theta_{min}$, is limited by the beam angular divergence and the requirement to have a
$\sim$ 10 $\sigma$ clearance of any detector element (often `Roman pots') from the beam.
For a beam divergence of say 0.1 mrad and an ion beam of 100 GeV/u, the transverse momentum
required in the nuclear breakup to be beyond the so-called machine `beam-stay-clear' area
of $\sim$ 10 $\sigma$ is 100 MeV, well beyond the 8 MeV (or so) needed for a single nucleon.
This would assume that the transverse momentum is equal to the excitation 
energy of the nucleus. 

The diffractive slope at t = 0 depends on the size of the nucleus. 
Fig. \ref{fig:cmax} shows, for small $t \sim 1/R_A^2$, a very steep t dependence, 
$\sim exp( -t R^2_A / 3)$, and then several diffractive minima 
($R_A = (1.12 fm)A^{1/3}$ - $(0.86 fm)A^{ -1/3 }$, for details see \cite{PhysRevC.81.025203}). 
The incoherent background starts to dominate at $t$ values at which 
the coherent cross section has fallen to $1/e$. These $t$ values can be estimated by 
$exp(-|t| B_0 A^{2/3}) = 1/A$, with $B_0 = (1.12 fm)^{2/3}$. These values of $t$ are much smaller
than the $t$ value corresponding to the first minimum in the coherent cross section and the $t$-values
corresponding to the smallest measurable outgoing angle of scattered heavy nuclei.
Therefore the strategy to ensure exclusive production on a nucleus is to veto nuclear 
breakup, by detecting the neutrons from incoherent events. 

\begin{figure}[htbp]
\begin{center}
\begin{tabular}{cc}
\includegraphics[width=0.45\textwidth]{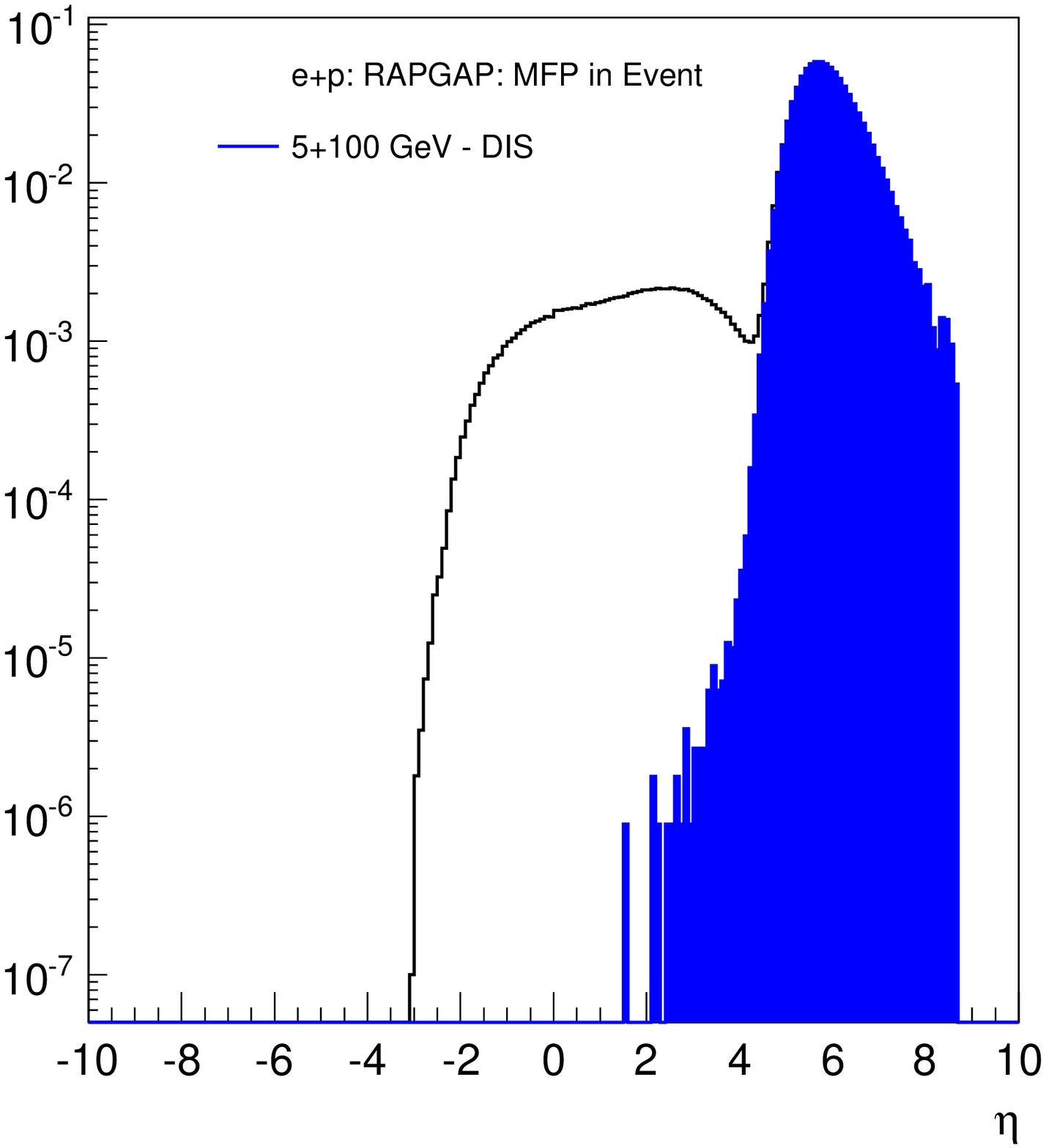} &
\includegraphics[width=0.45\textwidth]{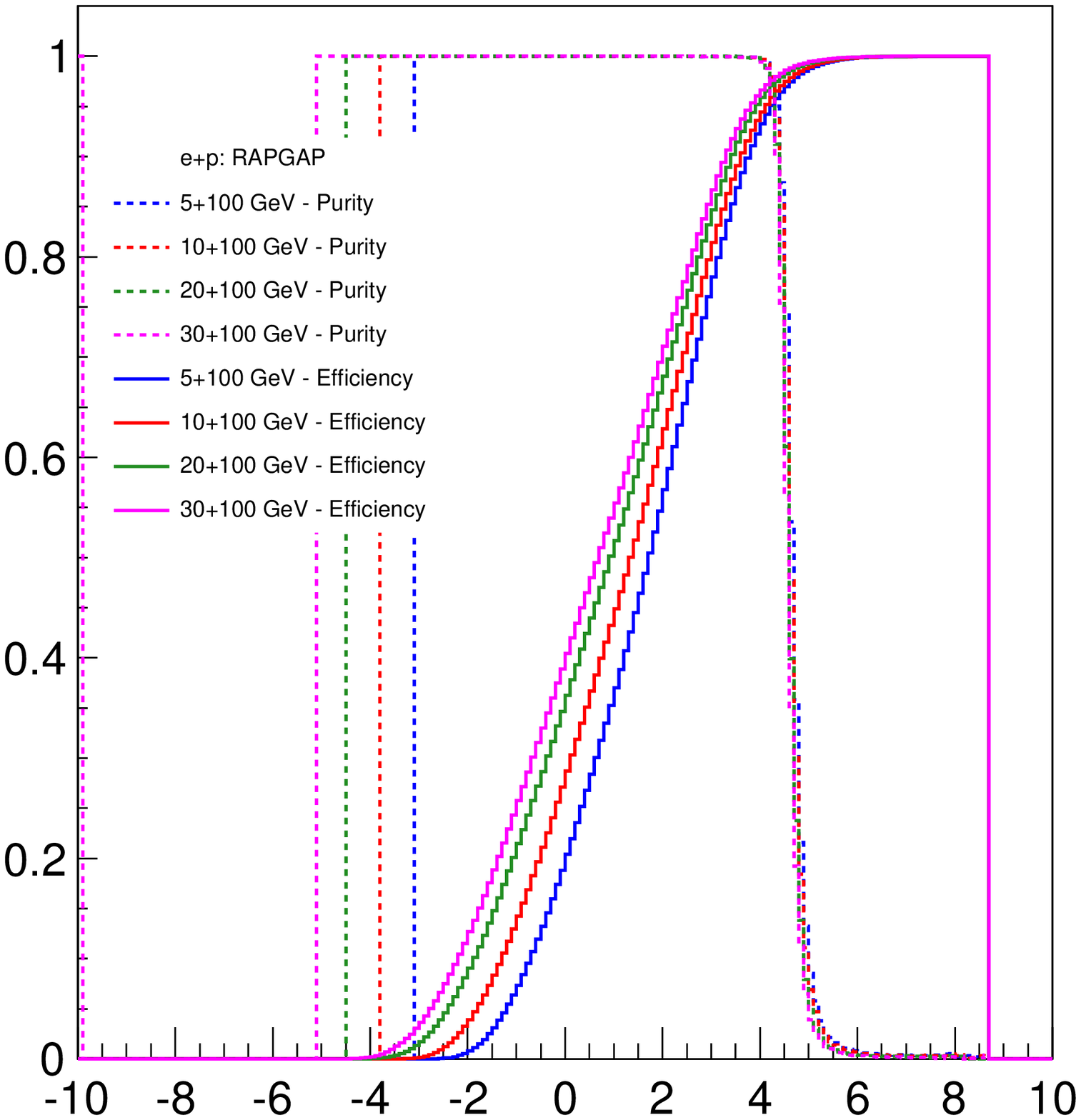} \\
\end{tabular}
\end{center}
\caption{\small Left: Rapidity distribution of DIS and diffractive 
events for the most forward particle (MFP) in the event.  
Right: Efficiency and Purity for diffractive events with respect to 
DIS events (1:1) as a function of the detector rapidity coverage and 
the center-of-mass energy.}
\label{fig:rapgap}
\end{figure}

Another possibility can be to require a rapidity  gap between the hadron beam and the produced jet, 
(vector) meson or real photon (where all events represent the sum of elastic and incoherent events).
The left panel of fig. \ref{fig:rapgap} shows the rapidity distribution of the most forward 
particle in deep-inelastic scattering (blue filled distribution)
and diffractive events (unfilled histogram), respectively, for a 5 GeV electron and a 100 GeV
proton beam energy combination. The 100 GeV is here chosen to mimic the 100 GeV/u ion beam.
The right panel of fig. \ref{fig:rapgap} shows the efficiency and purity 
for diffractive events to DIS events (1:1) as function of rapidity, varying the lepton 
beam energy while keeping the hadron beam energy fixed. If one requires 4 units of rapidity
between the hadron beam and a produced jet, vector meson or real photon, an efficiency of
above 60\% and a purity close to 100\% for diffractive events would be obtained.
A detector with wide rapidity coverage is essential for such events.

\subsection{Radiative Corrections}
\label{sec:radcorr}

\begin{figure}[htbp]
\begin{center}
\includegraphics[width=0.97\textwidth]{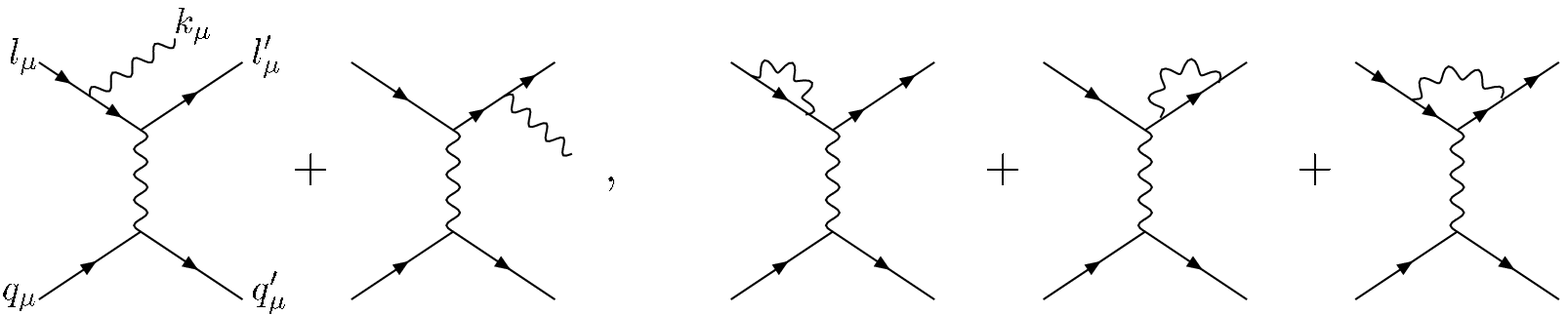}
\end{center}
\caption{\small Feynman diagrams for leptonic radiation in lepton-quark scattering. }
\label{fig:RC:FD}
\end{figure}

The radiation of real and virtual photons leads to large additional contributions to 
the observable cross section of electron scattering at high energies. Precision 
measurements of the nucleon structure require a good understanding of these 
radiative corrections. For neutral-current lepton nucleon scattering, a 
gauge-invariant classification into leptonic, hadronic and interference contributions 
can be obtained from Feynman diagrams. The Feynman diagrams for leptonic corrections 
are shown in fig. \ref{fig:RC:FD}. Leptonic corrections dominate and 
strongly affect the experimental determination of kinematic variables. 

Usually, the cross section is measured as a function of $Q^{2}$ and Bjorken-x, $x_{B}$, defined as
\begin{equation}
Q^2 = - (l - l^{\prime})^2, \quad x_B = \frac{Q^2}{2P\cdot (l - l^{\prime})} \, ,
\label{eq:RC:lepvar}
\end{equation}

\noindent 
\begin{minipage}[t]{0.57\textwidth}
where $l$ and $l^{\prime}$ denote the 4-momenta of the incoming and outgoing lepton, 
respectively, and $P$ is the 4-momentum of the incoming nucleon. The true values of these 
variables seen by the nucleon when a photon with 4-momentum $k$ is radiated are, 
however, given by (see fig.)
\begin{equation}
\tilde{Q}^2 = - (l - l^{\prime}- k)^2, 
\quad
\tilde{x}_B = \frac{\tilde{Q}^2}{2P\cdot (l - l^{\prime} -k)} \, .
\label{eq:RC:truevar}
\end{equation}
If the photon momentum is large and balancing the transverse momentum of the scattered 
lepton, $\tilde{Q}^2$ can be shifted to small values, leading to an enhancement of the 
radiative corrections. This effect is similar to the radiative tail of a resonance. 
\end{minipage}
\hfill
\begin{minipage}[t]{0.4\textwidth}
\vspace*{2mm}
\begin{center}
\includegraphics[width=1.0\textwidth]{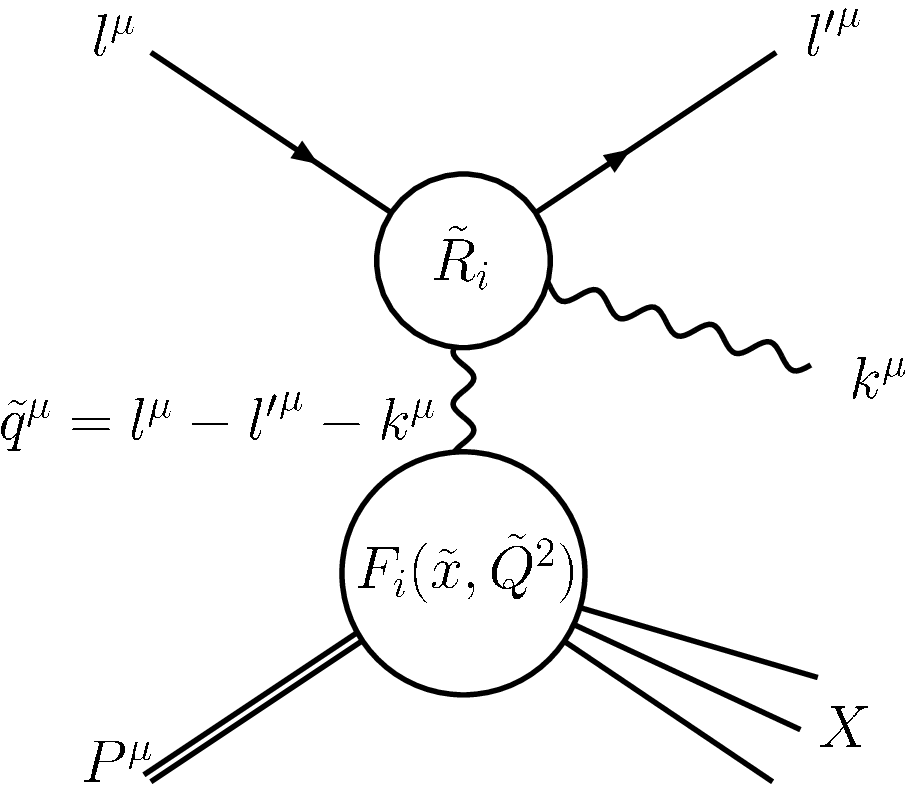} \\[2ex]
{\small Kinematics of leptonic radiation.}
\label{fig:RC:radlep}
\end{center}
\end{minipage}\\
%

The effect of radiation of photons from the lepton can be described with the help 
of radiator functions $\tilde{R}_i(l, l^{\prime}, k)$. There is one $\tilde{R}_i$ for 
every structure function $F_i$, $i=2, L$. The radiator functions comprise both real 
radiation from the initial and the final state as well as the contribution from vertex 
and self-energy diagrams. Using $\tilde{x}_B$ and $\tilde{Q}^2$ from equation 
(\ref{eq:RC:truevar}) to parametrize the integration over the phase space of emitted 
photons, one can express the observed structure functions as convolutions, 
\begin{equation}
F^{\rm obs}_i(x_B, Q^2) = 
\int {\rm d}\tilde{x}_B {\rm d}\tilde{Q}^2 \, 
R_i(x_B, Q^2, \tilde{x}_B, \tilde{Q}^2)
F^{\rm true}_i(\tilde{x}_B, \tilde{Q}^2) \, .
\label{eq:RC:folding}
\end{equation}
The integration limits are determined by the energy allowed for the radiated photon 
which, in the photon-nucleon center-of-mass frame, is given by
\begin{equation}
E_{\gamma}^{\rm max} = \sqrt{\frac{1-x_B}{x_B} Q^2} \, .
\end{equation}
Radiative corrections are, therefore, large at large $Q^2$ and small $x_B$. In contrast, 
at small $Q^2$ and large $x_B$, the phase space for photon emission is restricted and 
negative virtual corrections dominate.

From equation (\ref{eq:RC:folding}) it is obvious that the determination of the true 
structure functions $F^{\rm true}_i(\tilde{x}_B, \tilde{Q}^2)$ requires unfolding, 
a procedure which is in general only possible in an iterative way and with reasonably 
chosen assumptions about the starting values. Moreover, the observed structure 
functions depend on the way in which the kinematic variables are measured. For example, 
if the momentum of the hadronic final state, $p_X$, could be measured, $\tilde{x}_B$ 
and $\tilde{Q}^2$ would be known. In practice this will be difficult to achieve; 
however, any information about the hadronic final state could contribute to a narrowing 
down of the phase space available for photon emission, thereby reducing the size of 
radiative corrections. 

The radiator functions are dominated by peaks in the angular distribution for the 
collinear radiation of photons from the initial state (ISR) or from the final 
state (FSR). At high energies, it is a good approximation to assume that photon 
radiation can be described by a simple rescaling of the lepton momentum, 
$l \rightarrow zl$ for ISR and $l^{\prime} \rightarrow l^{\prime} / z$ for FSR. 
The radiator function in the collinear approximation takes the simple, universal form 
\begin{equation}
R_{\rm coll} = \frac{\alpha}{2\pi} \, \log\frac{Q^2}{m_e^2} 
\, \left(\frac{1 + z^2}{1-z}\right)_+
\end{equation}
so that the cross section is obtained from 
\begin{equation}
{\rm d}\sigma_{\rm ISR} = 
\int \frac{{\rm d}z}{z} R_{\rm coll}(z) \, 
{\rm d}\sigma_{\rm Born}(l^{\mu} \rightarrow zl^{\mu}) \, 
\end{equation}
(and similarly for FSR). The potentially large logarithm 
$\log Q^2 / m_e^2$ may reach the order of 10\,\% at large $Q^2$. 

\begin{figure}[htbp]
\begin{center}
\includegraphics[width=0.44\textwidth]{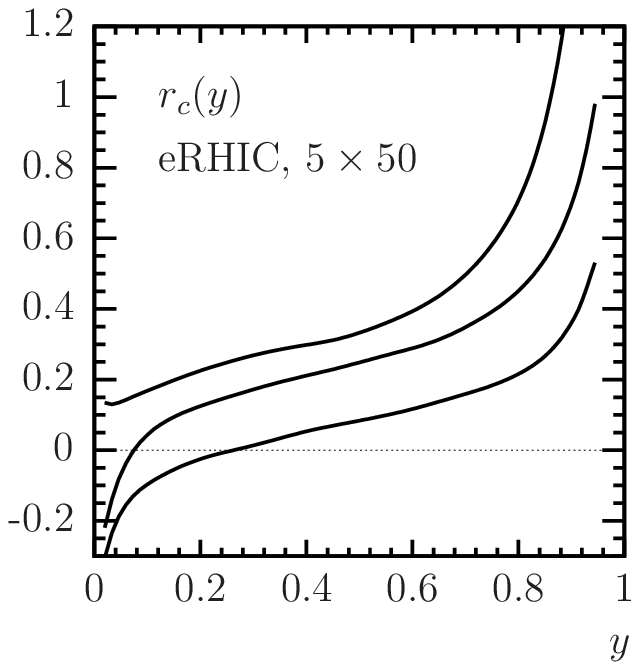}
\includegraphics[width=0.44\textwidth]{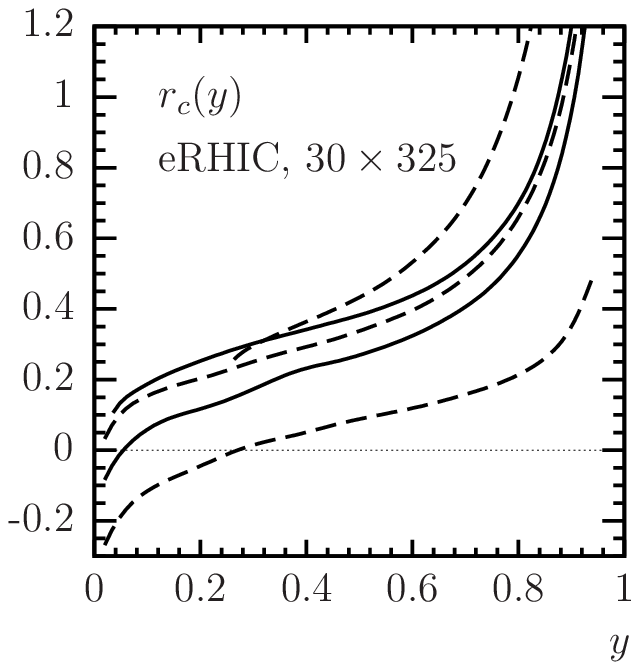}
\end{center}
\caption{\small $y$-dependence of the leptonic radiative correction factor 
for electron proton scattering with different beam energies and in different $x_B$ 
ranges. Left: $E_e=5$ GeV, $E_p=30$ GeV and the curves from the bottom up 
correspond to $0.1 < x_B < 0.4$, $10^{-2} < x_B < 10^{-1}$, 
$10^{-3} < x_B < 10^{-2}$; Right: $E_e=30$ GeV, $E_p=325$ 
GeV and $0.1 < x_B < 0.4$, $10^{-2} < x_B < 10^{-1}$, 
$10^{-3} < x_B < 10^{-2}$, $10^{-4} < x_B < 10^{-3}$, 
$10^{-5} < x_B < 10^{-4}$ (full and dashed lines alternating for better visibility). }
\label{fig:RC:xdep}
\end{figure}

As an example, we show numerical results for electron proton scattering at two 
typical sets of beam energies: $E_e = 5$ GeV with $E_p = 50$ GeV (left panel of 
fig. \ref{fig:RC:xdep}) and $E_e = 30$ GeV with $E_p = 325$ GeV (right panel). The figures show the  
correction factor 
\begin{equation}
r_c(y) = \frac{\left. {\rm d}\sigma / {\rm d}y \right|_{O(\alpha)}}%
{\left. {\rm d}\sigma / {\rm d}y \right|_{Born}} - 1
\end{equation}
where $y = Q^2 / Q^2_{\rm max}$, $Q^2_{\rm max} = x_B S$, $S = 2 l \cdot P$. 
The different curves correspond to different ranges of $x_B$: at the lower 
center-of-mass energy (left panel of fig. \ref{fig:RC:xdep}, from the bottom up): 
$0.1 < x_B < 0.4$, $10^{-2} < x_B < 10^{-1}$ and $10^{-3} < x_B < 10^{-2}$; at the 
higher  center-of-mass energy (right panel, again from the bottom up): 
$0.1 < x_B < 0.4$, $10^{-2} < x_B < 10^{-1}$, $10^{-3} < x_B < 10^{-2}$, 
$10^{-4} < x_B < 10^{-3}$, and $10^{-5} < x_B < 10^{-4}$. 
The general features following from the preceding discussion are clearly visible: 
corrections are large at large $y$ and small $x_B$, while corrections become negative 
at large $x_B$ and small $y$. 

\begin{figure}[htbp]
\begin{center}
\includegraphics[width=0.24\textwidth]{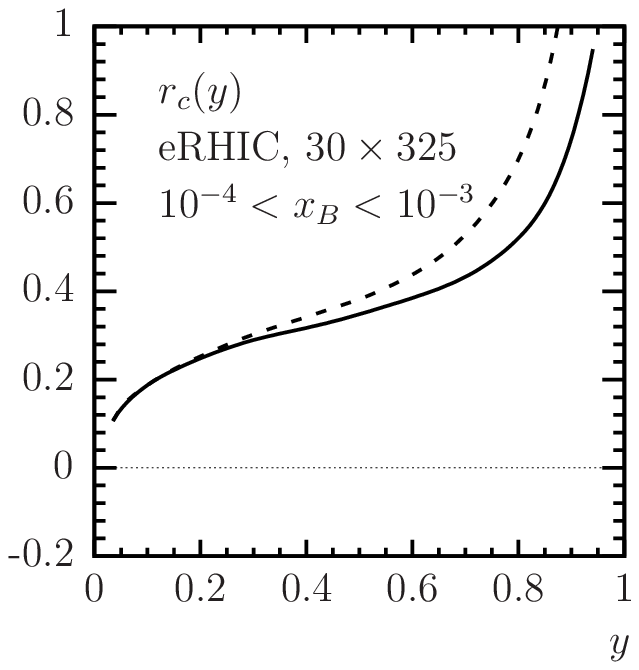}
\includegraphics[width=0.24\textwidth]{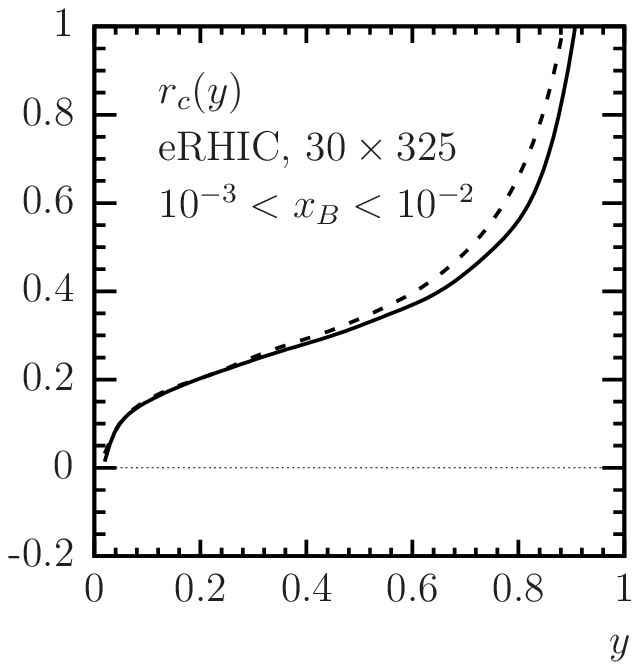}
\includegraphics[width=0.24\textwidth]{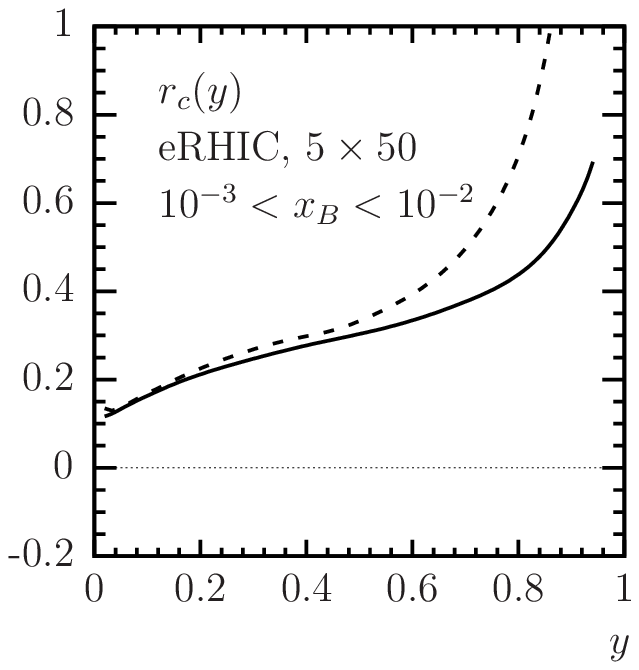}
\includegraphics[width=0.24\textwidth]{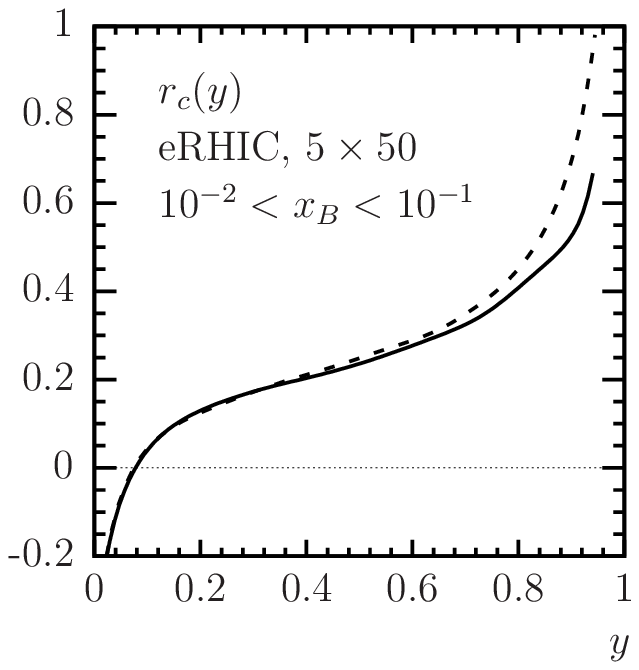}
\end{center}
\caption{\small Influence of a cut on the mass of 
the hadronic final state on the leptonic radiative correction factor for a proton target in 
different $x_B$ ranges and beam energies as indicated in the figures. Dashed curves are 
without a cut, full curves are obtained after a cut of $W_{\rm had} > 1.4$ GeV. }
\label{fig:RC:wcut}
\end{figure}

Lacking a full Monte Carlo event simulation for scattering with 
heavy nuclei at present, we have studied the influence of a simple  
cut on the invariant mass of the hadronic final state. Imposing 
the condition $W_{\rm had} > 1.4$ GeV would remove the elastic tail 
and the contribution from low-lying resonances. A similar effect 
can be achieved cutting on $E-p_z$ from the Jacquet-Blondel method.
The effect of such 
a naive cut is shown in fig. \ref{fig:RC:wcut}. The reduction of 
the radiative corrections is considerable at largest $y$ and at 
small $x_B$, but probably not yet sufficient at larger values of 
$x_B$. From similar studies for electron-nucleus scattering at HERA 
\cite{Spiesberger:1992vu,Akushevich:1995mg,Bassler:1994uq}, 
one can expect to obtain a much stronger reduction of radiative 
corrections, if more refined prescriptions for the measurement of 
kinematic variables are found. 

\begin{figure}[htbp]
\begin{center}
\includegraphics[width=0.44\textwidth]{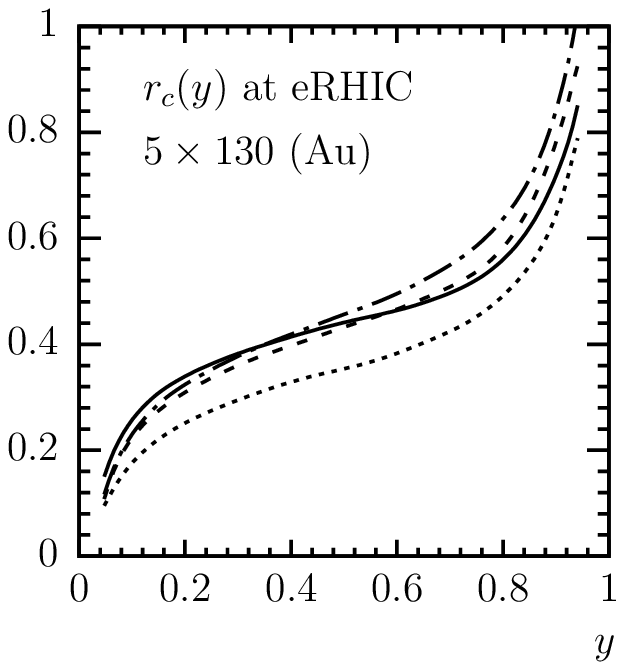}
\includegraphics[width=0.44\textwidth]{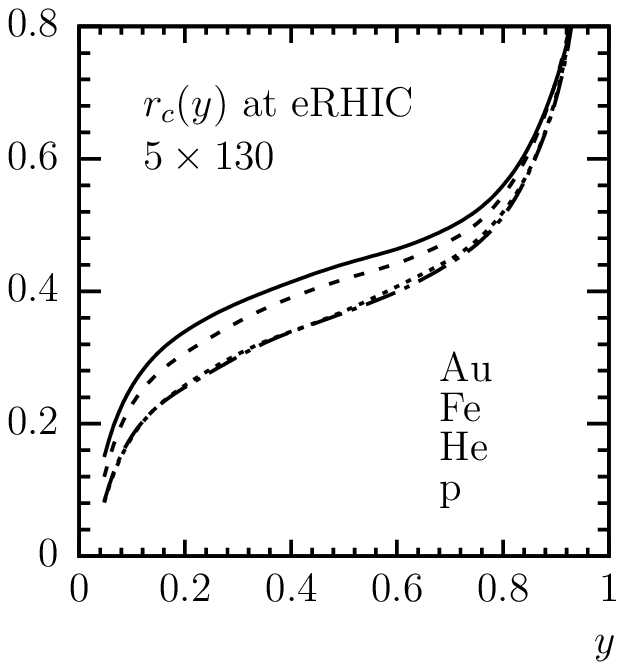}
\end{center}
\caption{\small 
Left: Radiative corrections for electron scattering off a Au nucleus at 
$5\times 130$ GeV$^2$ beam energies, $10^{-3} < x_B < 10^{-2}$, $Q^2>1$ GeV$^2$, 
$W_{\rm had} > 1.4$  GeV with different models for nuclear PDFs: EPS09 (full curve), 
EPS08 (dash-dotted line), EKS98 (dashed line) and HKN (dotted line). 
Right: Radiative corrections for different nuclei with CTEQ61M PDFs modified by the 
EPS09 prescription. Beam energies and kinematic range as in the left figure. From 
the bottom up: proton, ${}^4\mbox{He}$, ${}^{56}\mbox{Fe}$, ${}^{197}\mbox{Au}$. }
\label{fig:RC:nucl}
\end{figure}

Since the determination of the true structure functions requires 
an iterative unfolding procedure, it is important to show that 
the radiative corrections do not depend too strongly on the assumed 
input structure functions. In fig. \ref{fig:RC:nucl}a we show 
the correction factor $r_c(y)$ as defined above for the case of 
electron scattering off an ${}^{197}\mbox{Au}$ nucleus, assuming 
different parameterizations of parton distribution functions corrected 
for nuclear effects, as available in the literature 
\cite{Hirai:2007sx,Eskola:1998df,Eskola:2008ca,Eskola:2009uj}. 
Although differences at the level of 10\,\% are visible, one can 
still observe a similar overall behavior of radiative corrections. 
Finally, in fig. \ref{fig:RC:nucl}b, we show results for scattering 
off different nuclei, again supporting the assumption that 
a common unfolding procedure would allow one to obtain the true 
structure functions. 

Corrections due to the emission of photons from the hadrons, or 
quarks in the deep inelastic regime, require a careful separation 
into contributions which should be considered as a part of the 
hadron structure (leading to an electromagnetic contribution to 
scaling violations \cite{Spiesberger:1994dm}) and contributions 
which can, in principle, be related to the observation of direct 
photons radiated from quarks. The interference of radiation from 
the lepton and the quark is small \cite{Spiesberger:1992vu}. In 
certain phase space regions one may expect higher than one-photon 
corrections to be important. For example, soft-photon exponentiation  
will be necessary at small $y$ and large $x_B$. The procedure is 
well-known and straightforward. Finally, multi-photon radiation 
may become important at large $y$ and small $x_B$. In this case, 
the collinear approximation is sufficient to reach a precision at 
the level of one percent \cite{Kripfganz:1990vm}.

\subsection{Detector Design for eRHIC}
The BNL design of an EIC allows for collisions at three interaction regions: 
one at IP-12 with a new dedicated EIC detector, and at IP-6 and IP-8 
with the current RHIC detectors STAR and PHENIX. 
In the following, first the design considerations for a dedicated EIC detector
are described and then the capabilities of PHENIX and STAR for ep / eA collisions
are discussed.

\subsubsection{A dedicated EIC detector}

\begin{figure}[htbp]
\begin{center}
\includegraphics[width=0.95\textwidth]{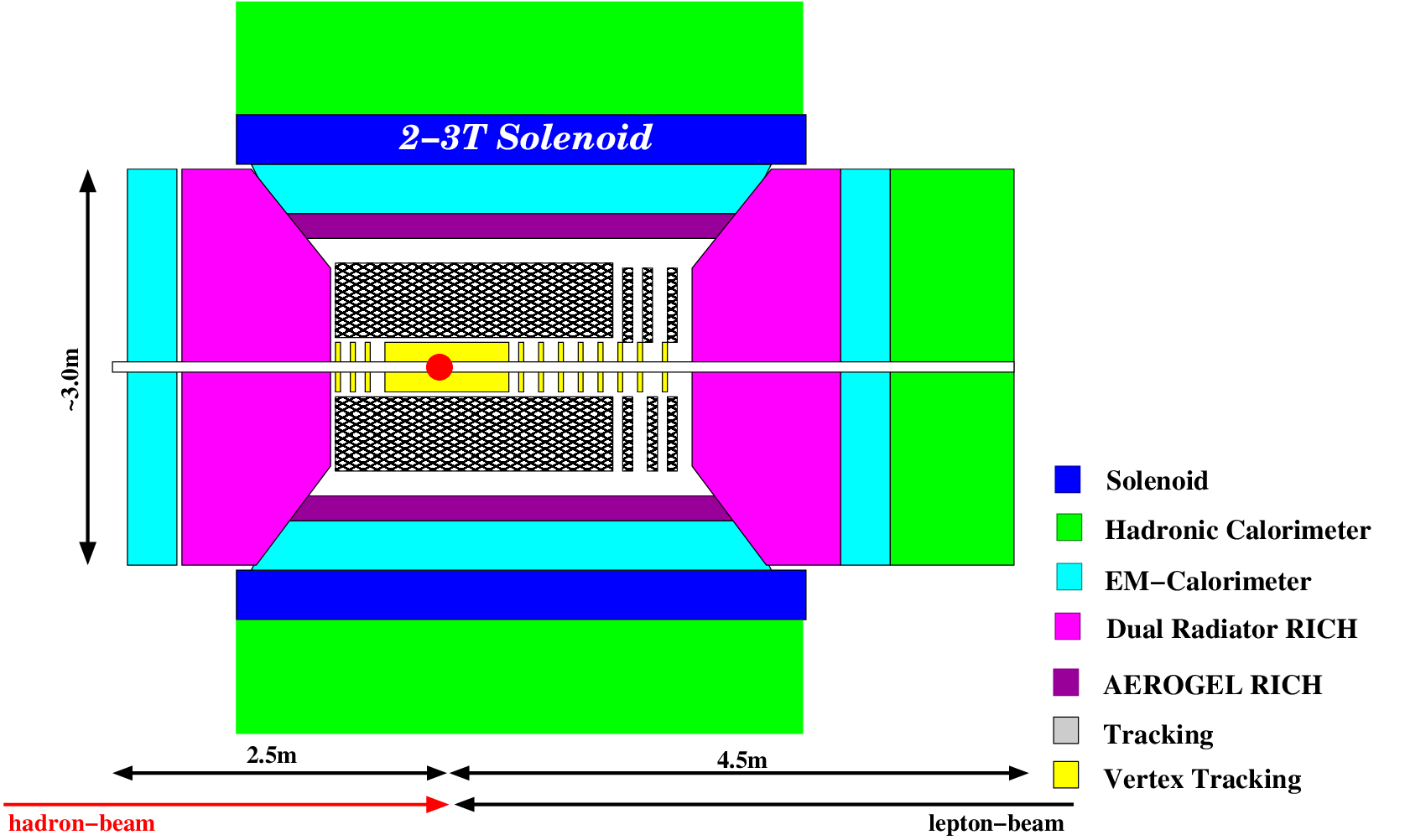}
\end{center}
\caption{\small A schematic view of a dedicated EIC detector. 
Details of the GEANT-3 model can be found at 
\protect{https:$//$wiki.bnl.gov$/$eic$/$index.php$/$Detector\_Design}.}
\label{fig:eRHIC-det}
\end{figure}

Combining all the requirements described in section \ref{sec:det.kin} and in the physics
chapters before, a schematic view of the emerging dedicated eRHIC detector is shown in
fig. \ref{fig:eRHIC-det}.
As already discussed, it is important to have equal rapidity coverage for 
tracking and electromagnetic calorimetry. This will provide good electron identification and
give better momentum and angular resolutions at low inelasticity $y$ than with an 
electro-magnetic calorimeter alone. 

The significant progress in the last decade in the
development of Monolithic Active Pixel Sensors (MAPS), in which the
active detector, analog signal shaping, and digital conversion take
place in a single silicon chip (i.e. on a single substrate; see
\cite{Gaycken:2006se} and references therein), provides a unique opportunity
for a $\mu$-vertex detector for an eRHIC detector. These devices, built using CMOS
technology, use an epitaxial layer as the active sensing element.
Ionization deposited in the epitaxial layer is collected by N+ wells
embedded in the epitaxial layer. The ``pixel'' pitch is determined by
the location of the N wells so there is no need for actual
segmentation of the detector as is done with traditional hybrid pixel
detectors. As a result, CMOS pixel detectors can be built with
high segmentation, limited primarily by the space required for
additional shaping and digital conversion elements.
The key advantage of CMOS MAPS detectors is the reduced material
required for the detector and the (on substrate) on-detector
electronics. Such detectors have been fabricated and extensively
tested (see e.g. \cite{HuGuo:2009zz}) with thicknesses of about 50 $\mu$m,
corresponding to 0.05\% of a radiation length. 

For tracking at larger radii, there are several possibilities which need to 
be investigated first through Monte Carlo studies for position resolution and material 
budget, and later through R\&D and building prototypes. The two most prominent options 
for the barrel tracker are a TPC and a cylindrical GEM-Tracker. For large radii,
forward tracking GEM-Trackers are the most likely option.
The projected rates for a luminosity of 10$^{34}$ cm$^{-2}$ s$^{-1}$ range, depending
on the center-of-mass energy, between 300 and 600 kHz, with an average of 6 to 8 charged 
tracks per event. These numbers do not impose strong constraints on the technology for a 
tracker.

Due to the momentum range to be covered, the only solution for PID in the forward 
direction is a dual radiator RICH, combining either Aerogel with a gas radiator like
C$_4$F$_{10}$ or C$_4$F$_8$O if C$_4$F$_{10}$ is no longer available, or 
combining the gas radiator with a liquid radiator like C$_6$F$_{14}$.

In the barrel part of the detector several solutions are possible, as the momenta of
the majority of the hadrons to be identified are between 0.5 GeV and 5 GeV.
The technologies available in this momentum range are high resolution ToF 
detectors (t $\sim$ 10ps), a DIRC or a proximity focusing Aerogel RICH.

For the electromagnetic calorimetry in the forward and backward direction, a solution 
based on PbWO$_4$ crystals would be optimal. The advantages of such a calorimeter 
would be a small Moli\`ere radius of 2~cm and a factor of two
better energy resolution and higher radiation hardness than, for example, lead-glass. 
To increase the separation of photons and $\pi^0$s to high momenta and 
to improve the matching of charged tracks to the electromagnetic cluster, it would 
be an advantage to add, in front of all calorimetry, a high resolution pre-shower.
We follow for the barrel part of the detector the concept of very compact 
electromagnetic calorimetry (CEMCal). A key feature is to have at least one 
preshower layer with 1--2 radiation lengths of tungsten and silicon strip layers
(possibly with two spatial projections) to allow separation of single photons from
$\pi^0$ to up $p_T \approx 50$\,GeV, as well as enhanced
electron-identification.  A straw-man design could have silicon strips with
$\Delta \eta = 0.0005$ and $\Delta \phi = 0.1$.  The back section for
full electromagnetic energy capture could be, for cost effectiveness and good uniformity,
an accordion Lead-Scintillator Design, which would provide gain uniformity and the 
ability to calibrate the device.  
A tungsten- and silicon-strip-based pre-shower would also be a good solution for the 
forward and backward electromagnetic calorimetry.

\begin{figure}[htbp]
\begin{center}
\includegraphics[width=0.95\textwidth]{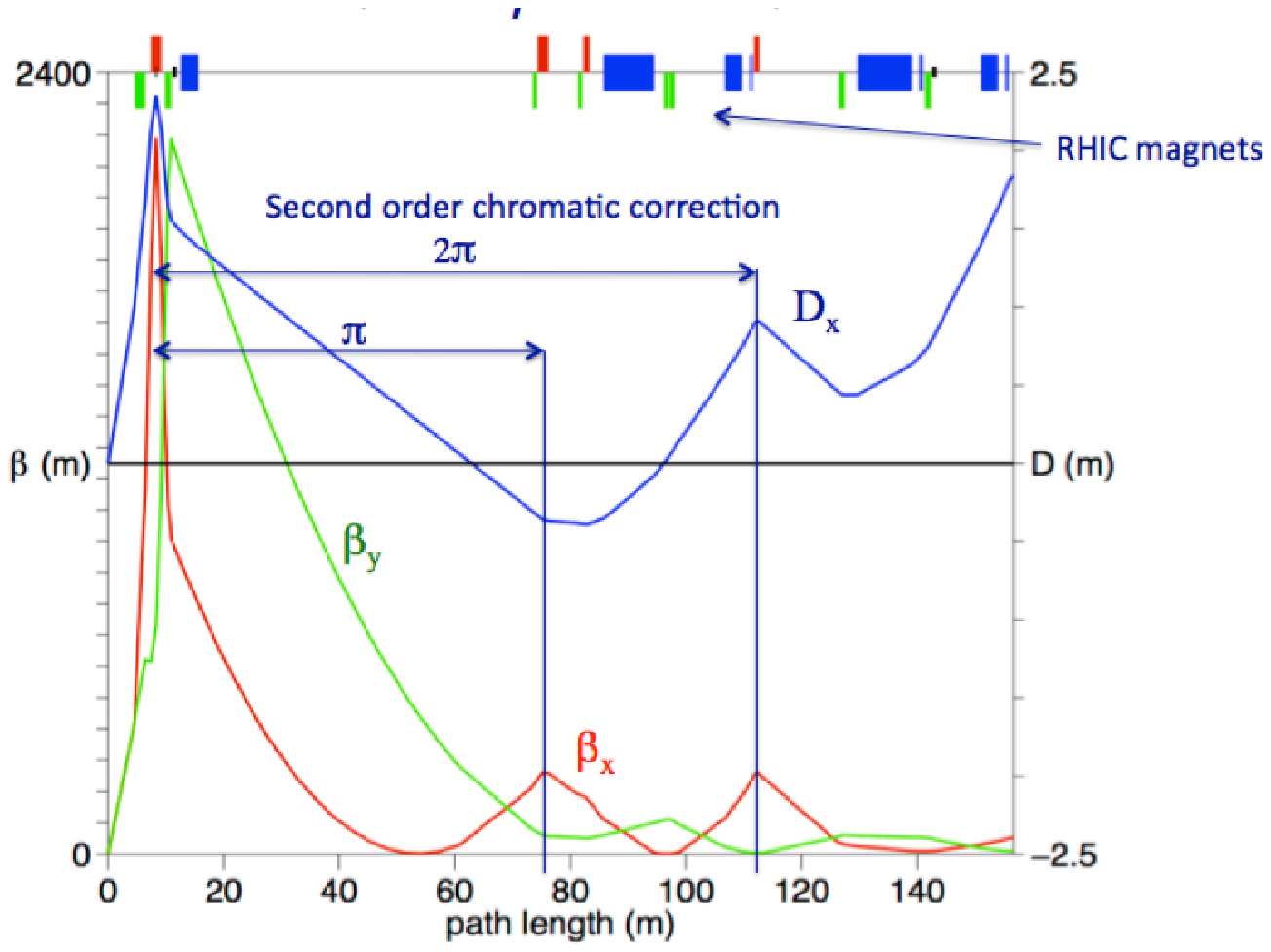}
\end{center}
\caption{\small The Beta-function along the eRHIC hadron ring.} 
\label{fig:ip-1} 
\end{figure}

To achieve the physics program as described in earlier sections, it is extremely important 
to integrate the detector design into the interaction region design of the collider.
As already described, particularly challenging is the detection of forward-going
scattered protons from exclusive reactions, as well as of decay neutrons from
the breakup of heavy ions in non-diffractive reactions.
Previous experience of electron colliders (SLAC, KEK B-factories) and HERA, an 
electron-proton collider, indicated difficulties with synchrotron radiation coming 
from bending the electron beam close to the interaction region (IR). The newest large 
improvements in luminosity at KEK in Japan, by introducing crab cavities, show that 
colliding heavy ions or protons with electrons could be obtained without bending 
the electrons close to the IR, but that it is possible to use the crossing angle 
between the two beams without losing luminosity. This is the path chosen in the 
eRHIC design: a 10 mrad crossing angle between the protons or heavy ions 
during collisions with electrons. This choice removes potential problems for the
detector induced by synchrotron radiation. To obtain luminosities 
higher than 10$^{34}$ cm$^{-2}$ s$^{-1}$, very strong focusing close to the IR
is required to have the smallest beam sizes at the interaction point. A small beam size is only 
possible if the beam emittance is also very small. The focusing triplets 
are 4.5 meters away from the interaction point (IP). The strong focusing quadrupoles 
induce very large chromaticities. The current eRHIC design has its highest values of 
the amplitude betatron functions of the same size as the present operating conditions 
of the RHIC collider. In addition the design allows a correction of the first, 
second and third order chromaticities by using sextupoles at the triplets as well 
as 180 degrees away from the quadrupoles source (as shown in fig. \ref{fig:ip-1}).

\begin{figure}[htbp]
\begin{center}
\begin{tabular}{cc}
\includegraphics[width=0.45\textwidth]{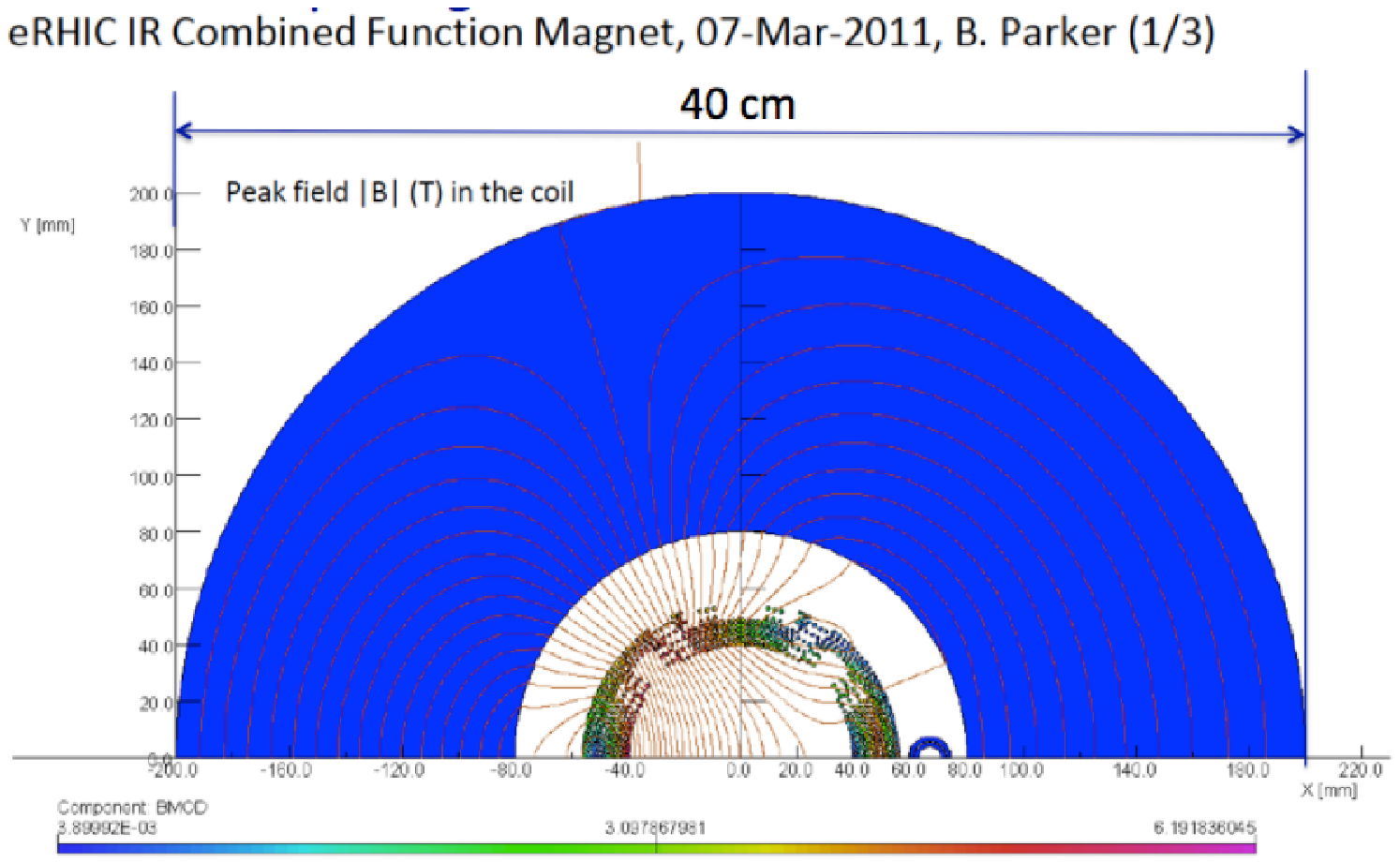} &
\includegraphics[width=0.45\textwidth]{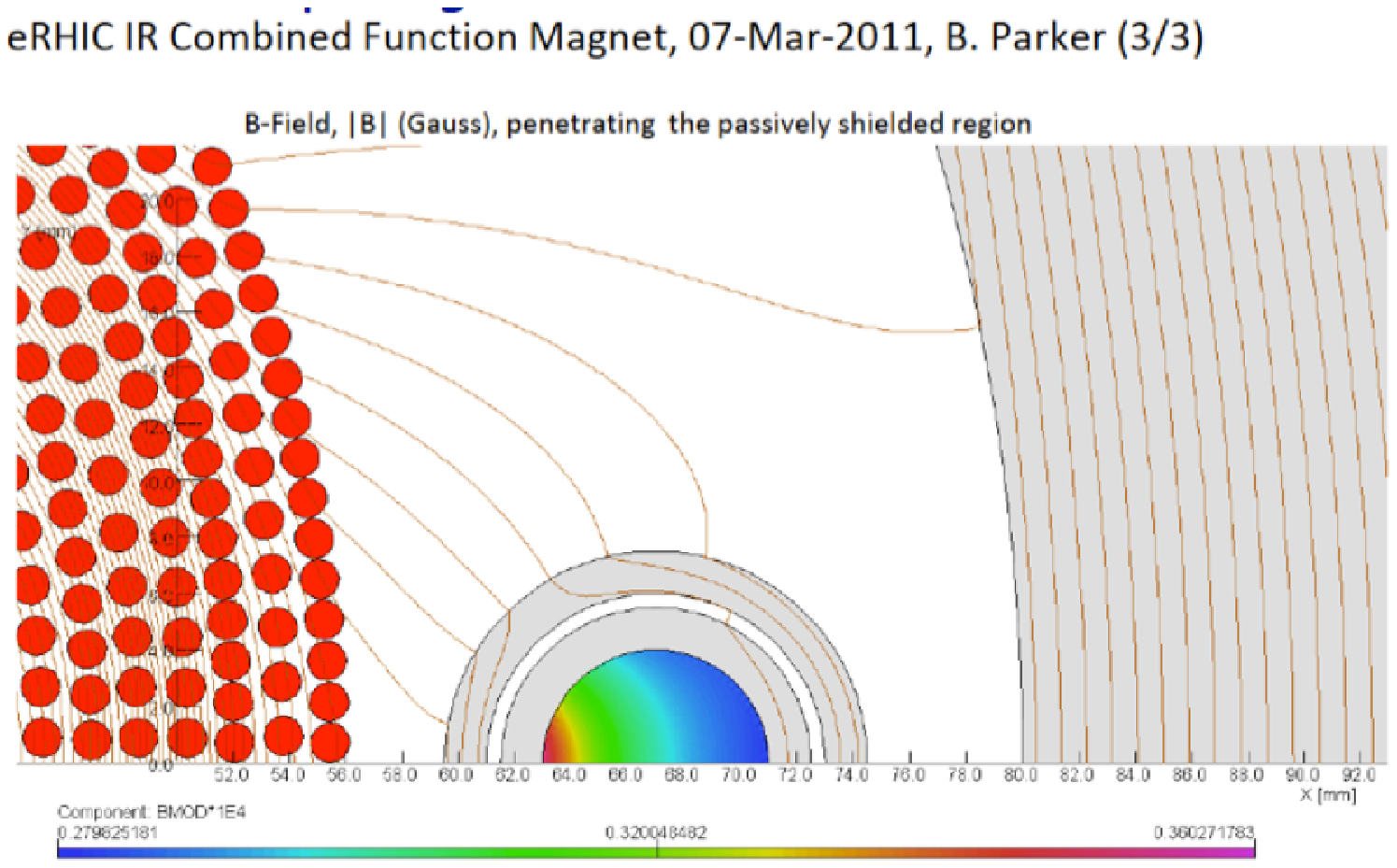} \\
\end{tabular}
\end{center}
\caption{\small Combined-function magnet of the hadron beam high focusing quadrupole 
triplet.} 
\label{fig:ip-2}
\end{figure}

While the above accomplishes a small-emittance electron beam, the ions and protons need 
to be cooled by coherent electron cooling to have small emittance. The eRHIC 
interaction region design relies on the existence of small emittance beams with 
a longitudinal RMS of ~5 cm, resulting in $\beta^*$ = 5 cm. Strong focusing is 
obtained by three high-gradient quadrupole magnets using recent results from the 
LHC quadrupole magnet upgrade program (reaching gradients of 200 T/m at 120 mm aperture).
To ensure the previously described requirements from physics are met, four major 
requirements need to be fulfilled: high luminosity ($>$ 100 times that of HERA), 
the ability to detect neutrons, measurement of the scattered proton from exclusive 
reactions (i.e. DVCS) and the detection of low-momentum protons (p$\sim$p$_0$/2.5) 
from heavy-ion breakup. The eRHIC IR design fulfills all these 
requirements: the first magnet in the high focusing quadrupole triplet is a combined 
function magnet producing a 4 mrad bending angle of the ion/proton beam 
(see fig. \ref{fig:ip-2}). The 120 mm diameter aperture of the last quadrupole 
magnet allows detection of neutrons with a solid angle of $\pm$ 4 mrad, as well as the 
scattered proton from exclusive reactions, i.e. DVCS, up to a solid angle of $\sim$
9 mrad. The electrons are transported to the interaction point 
through the heavy ion/proton triplets, seeing zero magnetic field as shown in 
fig. \ref{fig:ip-2}. 

\begin{figure}[htbp]
\begin{center}
\includegraphics[width=0.95\textwidth]{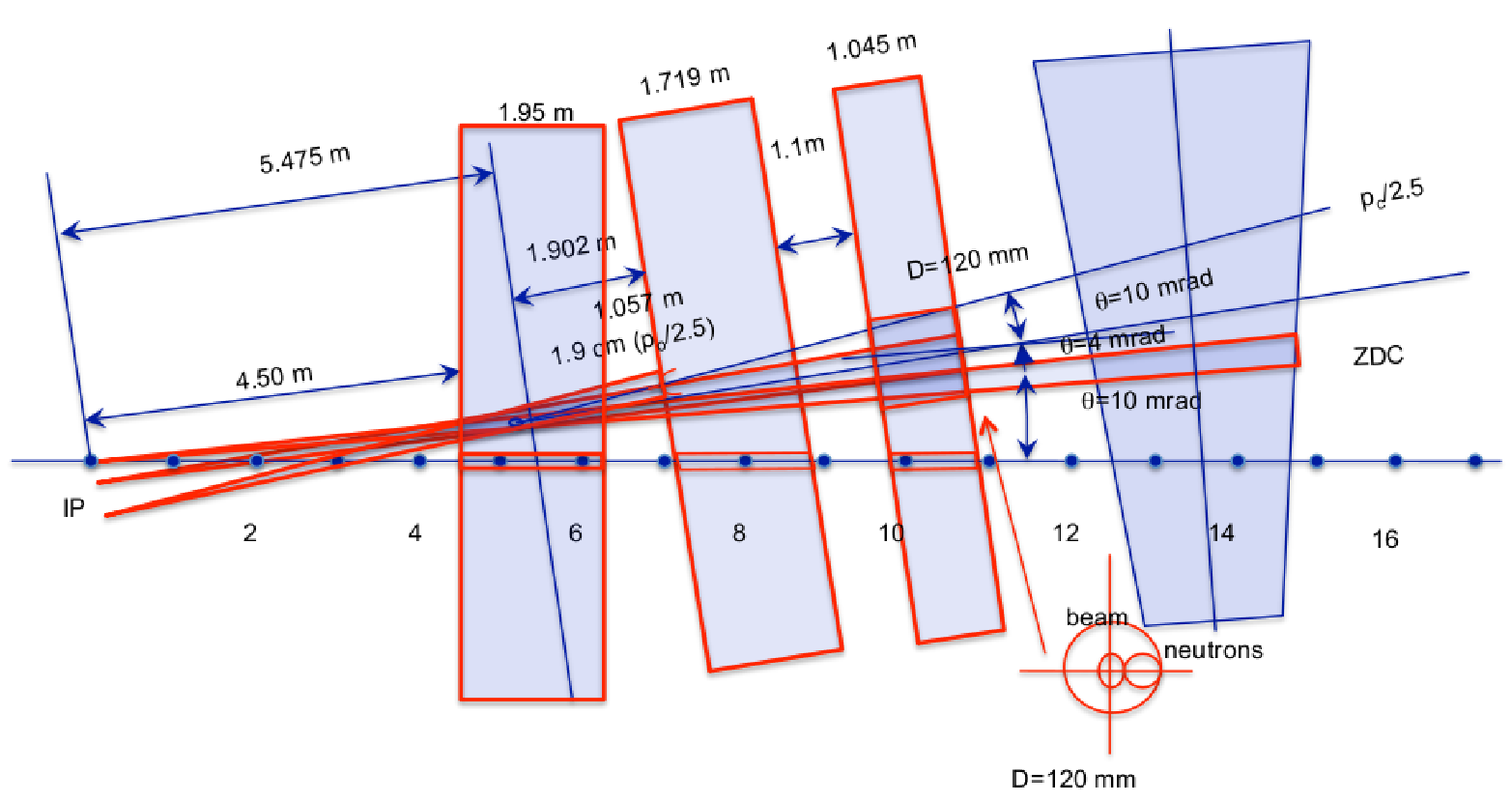} 
\end{center}
\caption{\small Schematic view of the eRHIC interaction region 
design in the direction of the outgoing hadron beam.}  
\label{fig:eRHIC-IR} 
\end{figure}

Fig. \ref{fig:eRHIC-IR} shows the current eRHIC interaction region design in 
the direction of the outgoing hadron beam. The other side of the IR is mirror symmetric
for the incoming hadron beam. For the outgoing lepton beam we are currently 
investigating how to best integrate a low scattering-angle lepton tagger. 
Such a tagger is critical for any low $Q^2$ physics, like elastic $J/\psi$ production in eA collisions 
(see section \ref{sec:ea-monte}).

The scattered proton from DVCS events were tracked through this design and beam optics using HECTOR 
\cite{Jeneret:2007vi}. The DVCS events have been 
generated with MILOU, a MC code dedicated to DVCS \cite{Perez:2004ig}.
\begin{figure}[htbp]
\begin{center}
\begin{tabular}{cc}
5 GeV x 50 GeV & 25 GeV x 250 GeV \\ \hline
\includegraphics[width=0.38\textwidth]{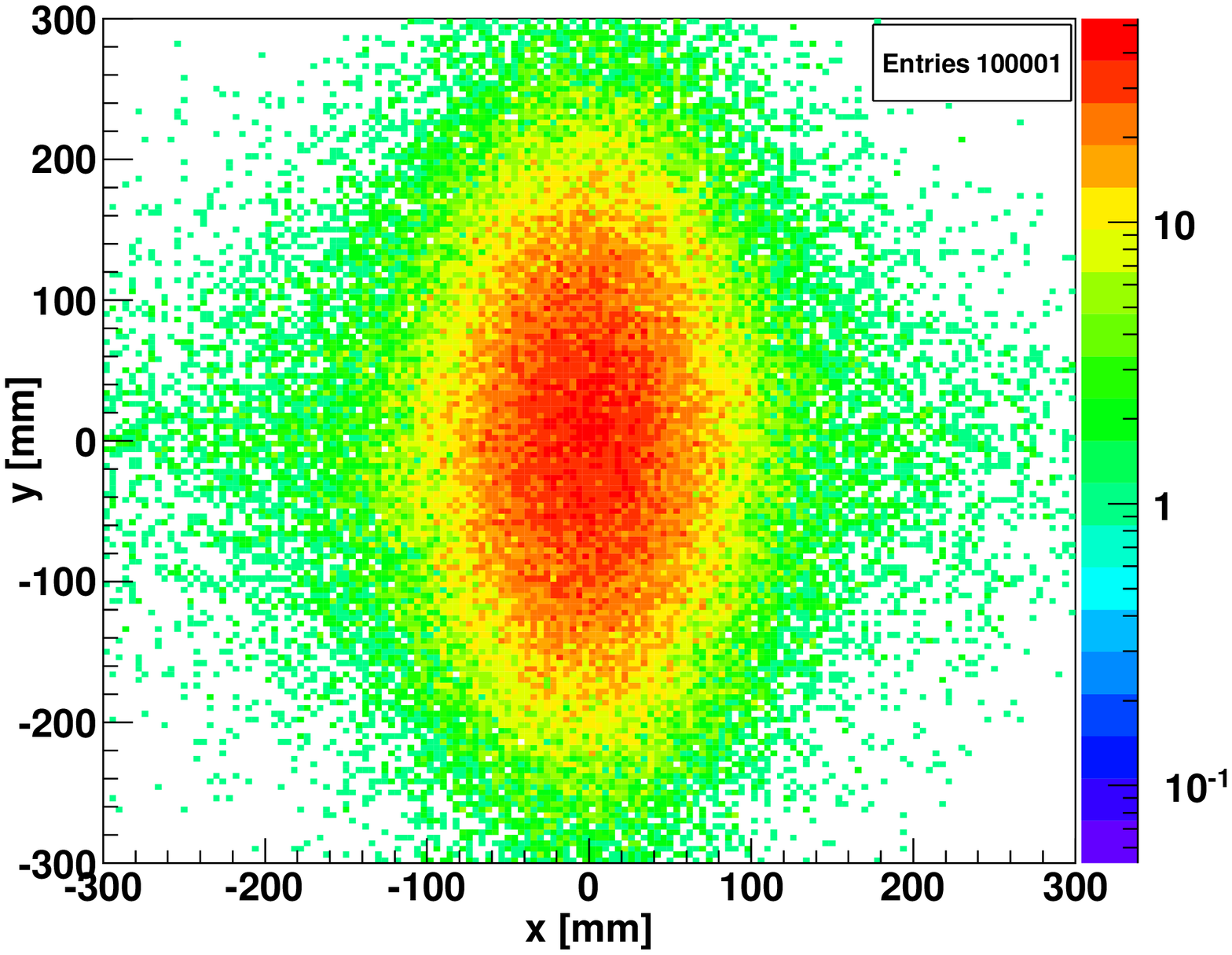} &
\includegraphics[width=0.38\textwidth]{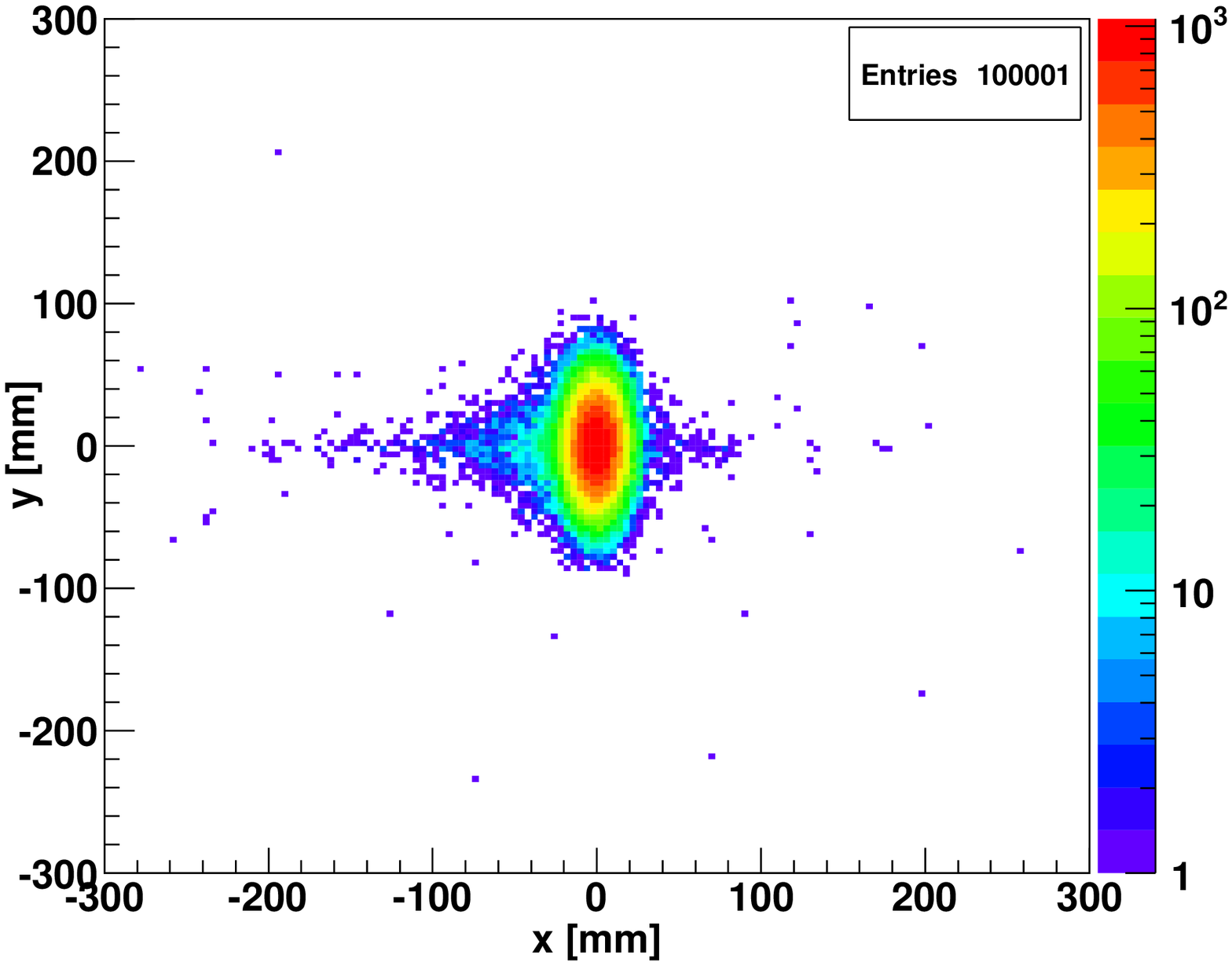} \\[-0.2cm]
\includegraphics[width=0.38\textwidth]{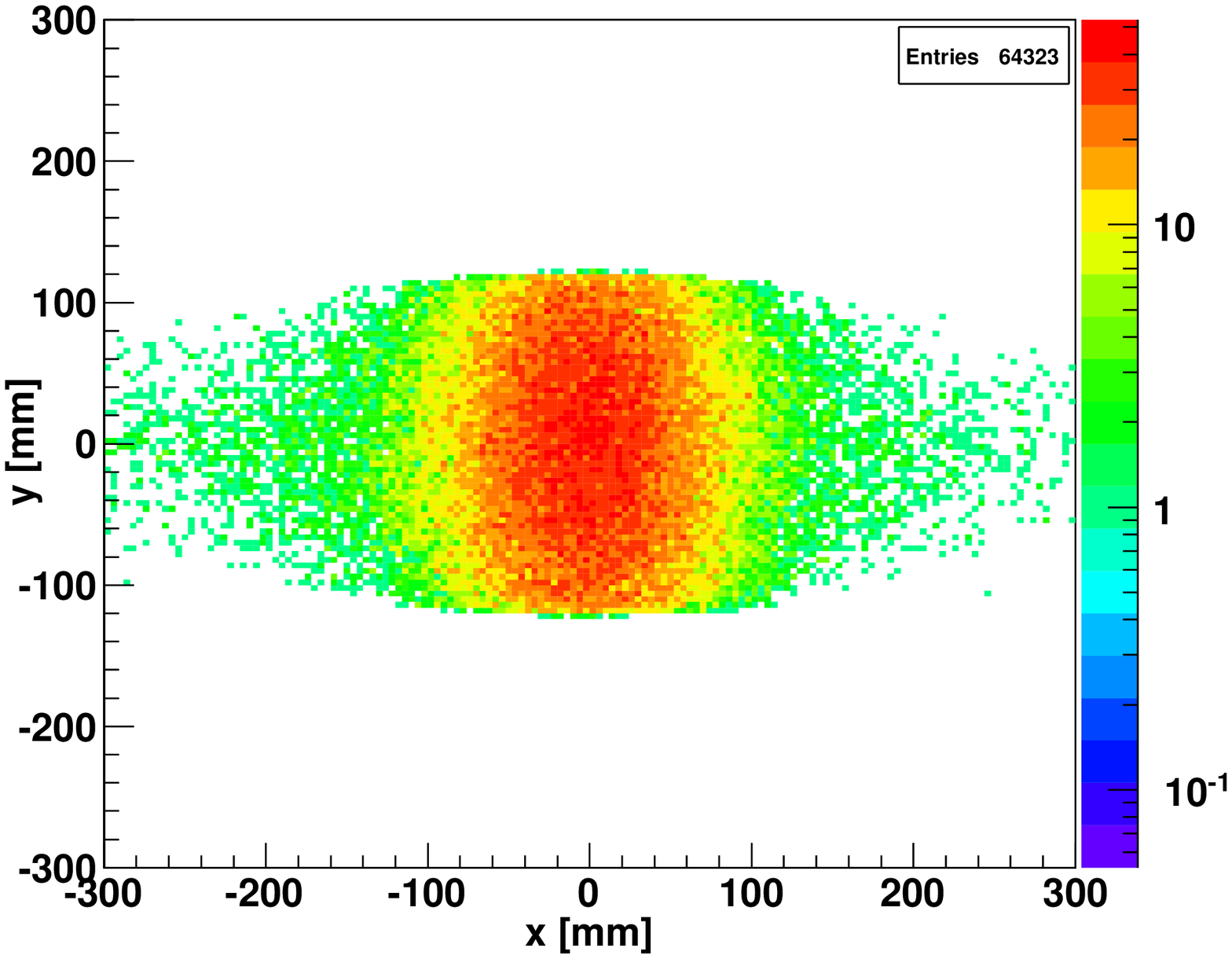} &
\includegraphics[width=0.38\textwidth]{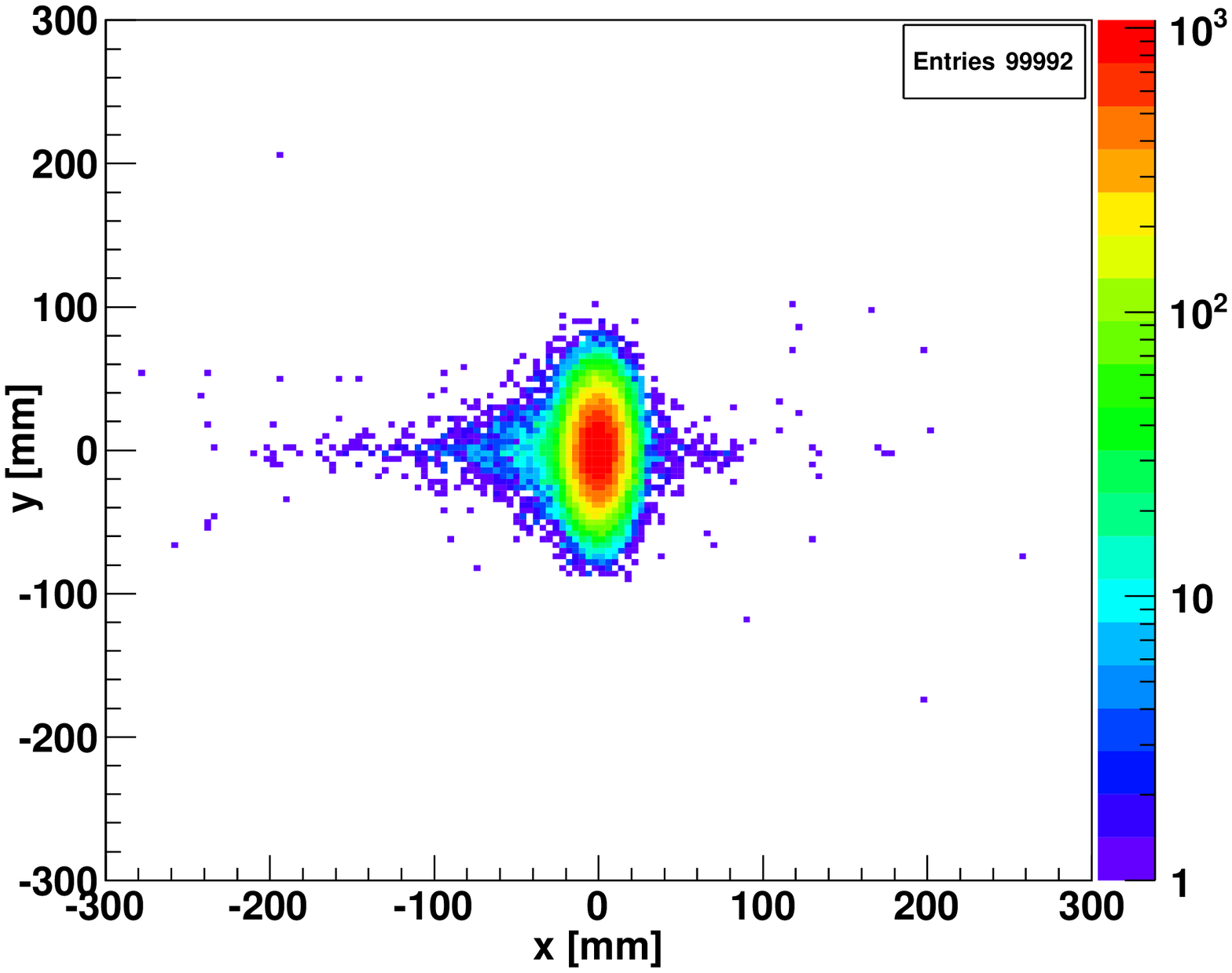} \\[-0.2cm]
\includegraphics[width=0.38\textwidth]{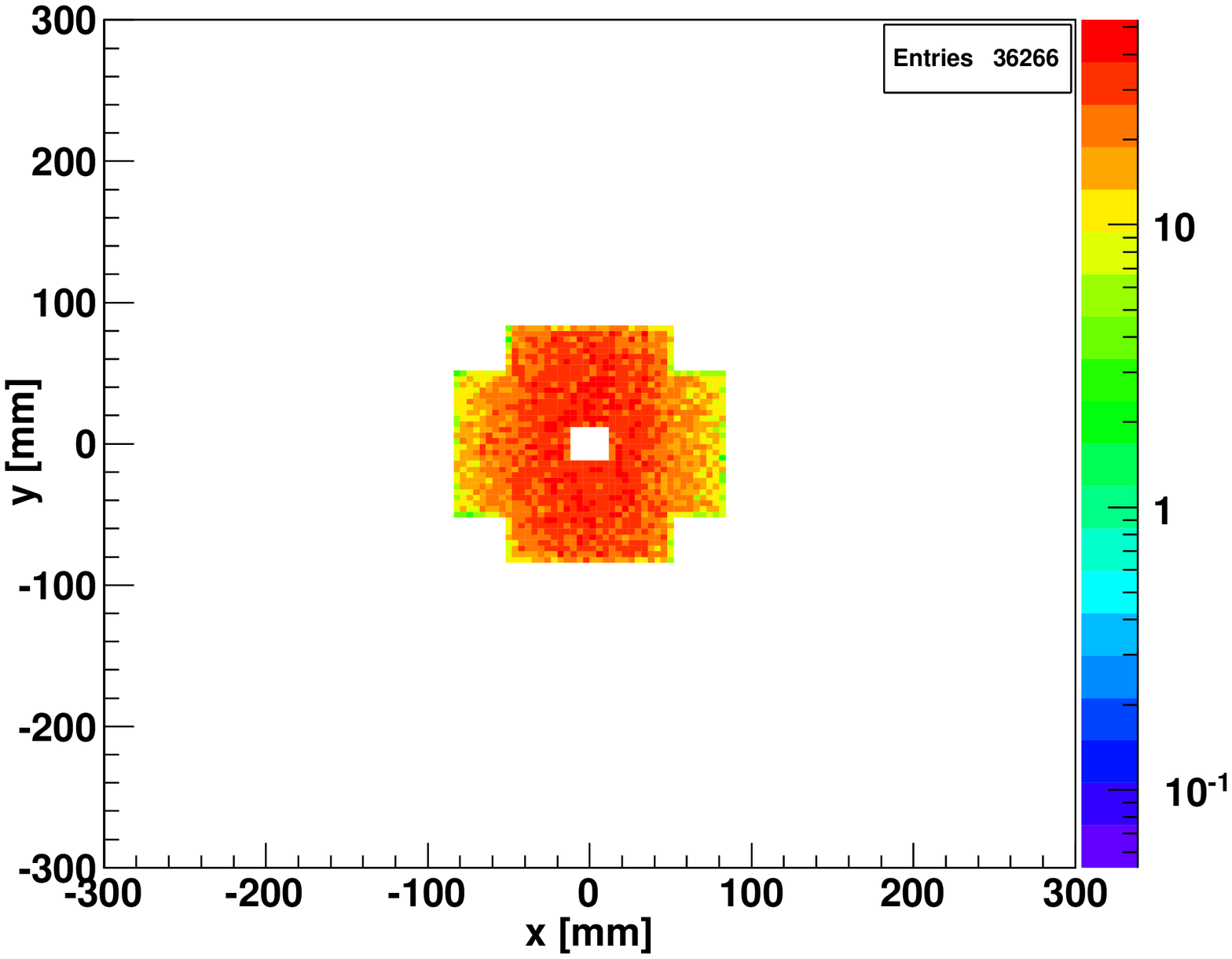} &
\includegraphics[width=0.38\textwidth]{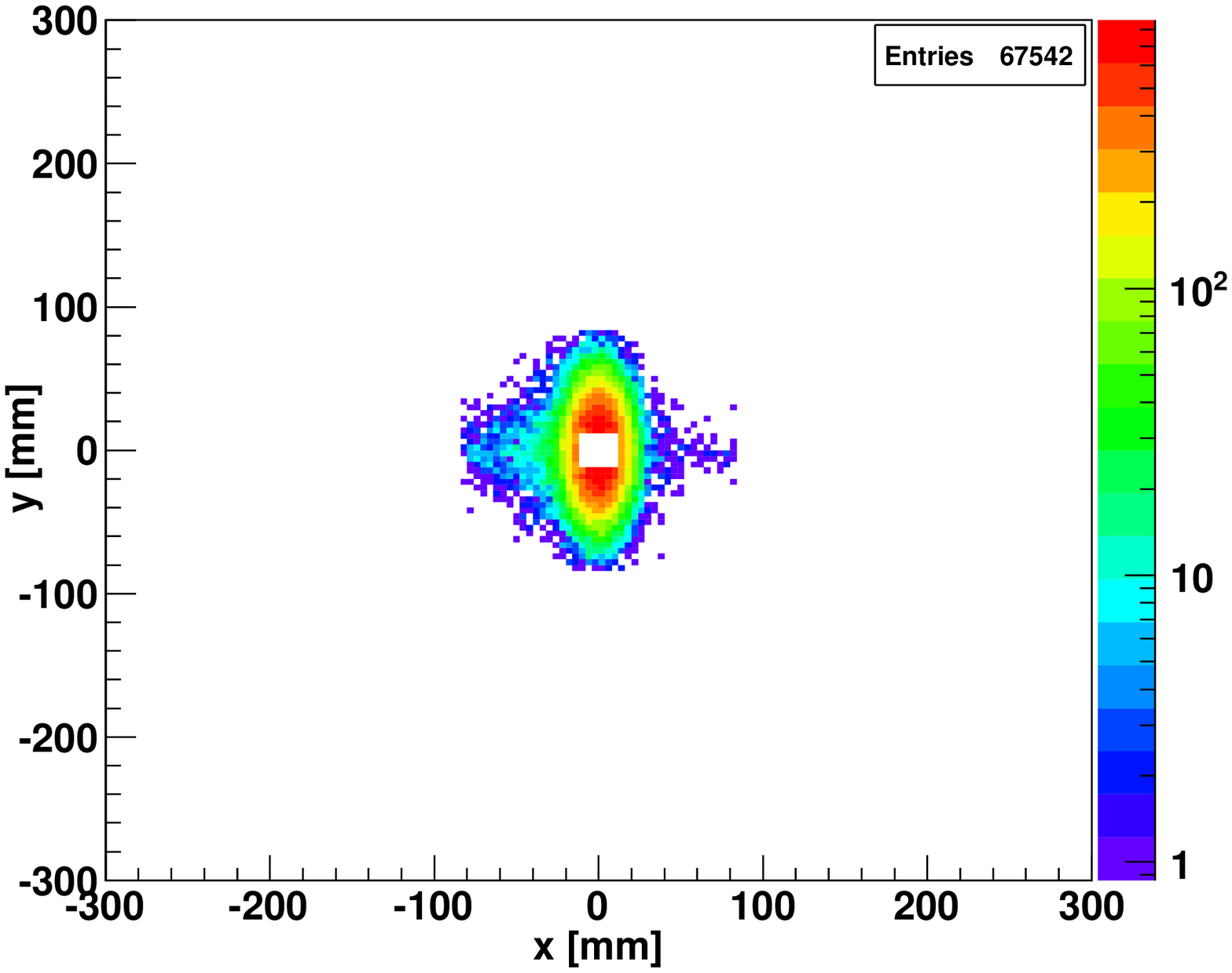} \\[-0.2cm]
\includegraphics[width=0.38\textwidth]{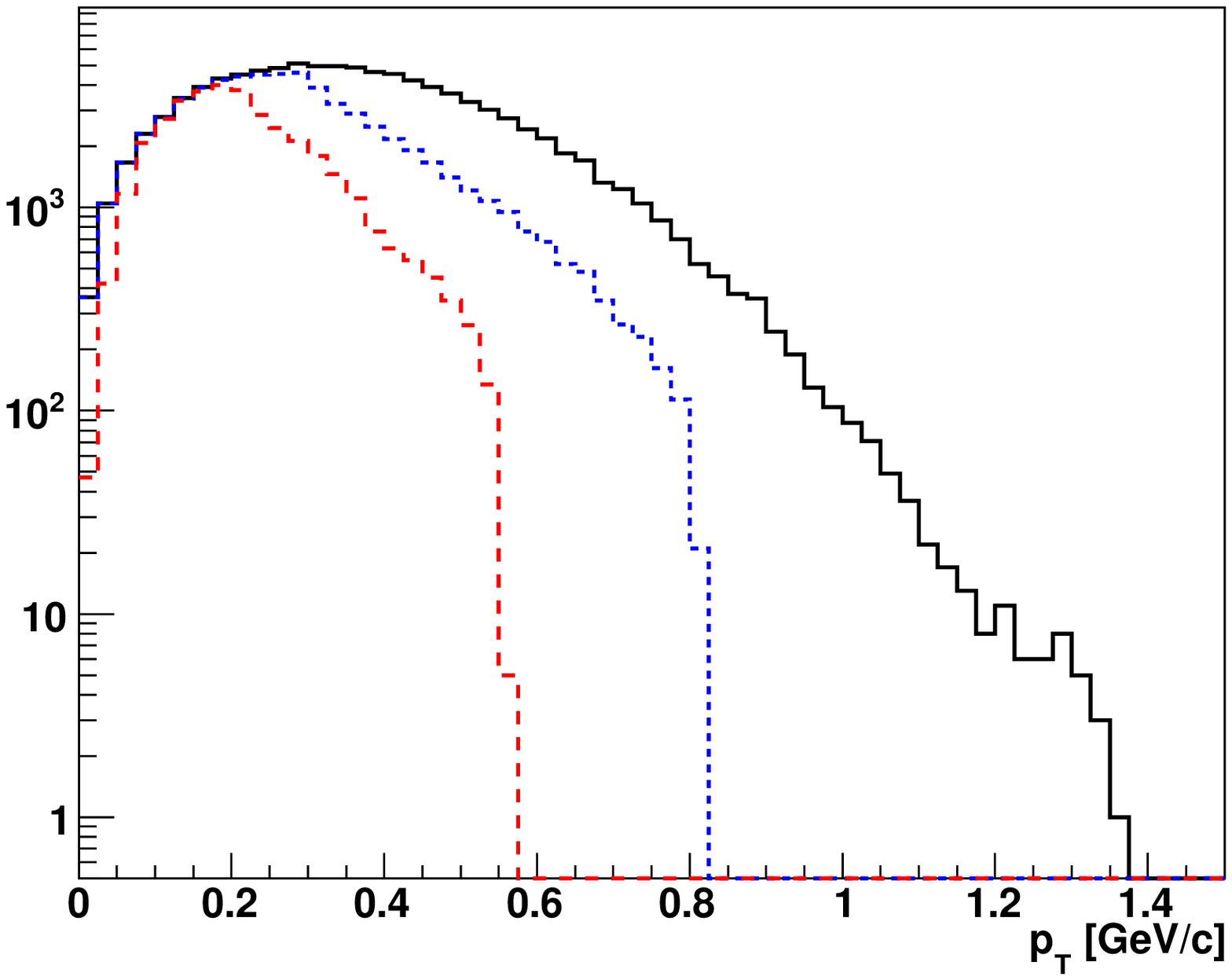} &
\includegraphics[width=0.38\textwidth]{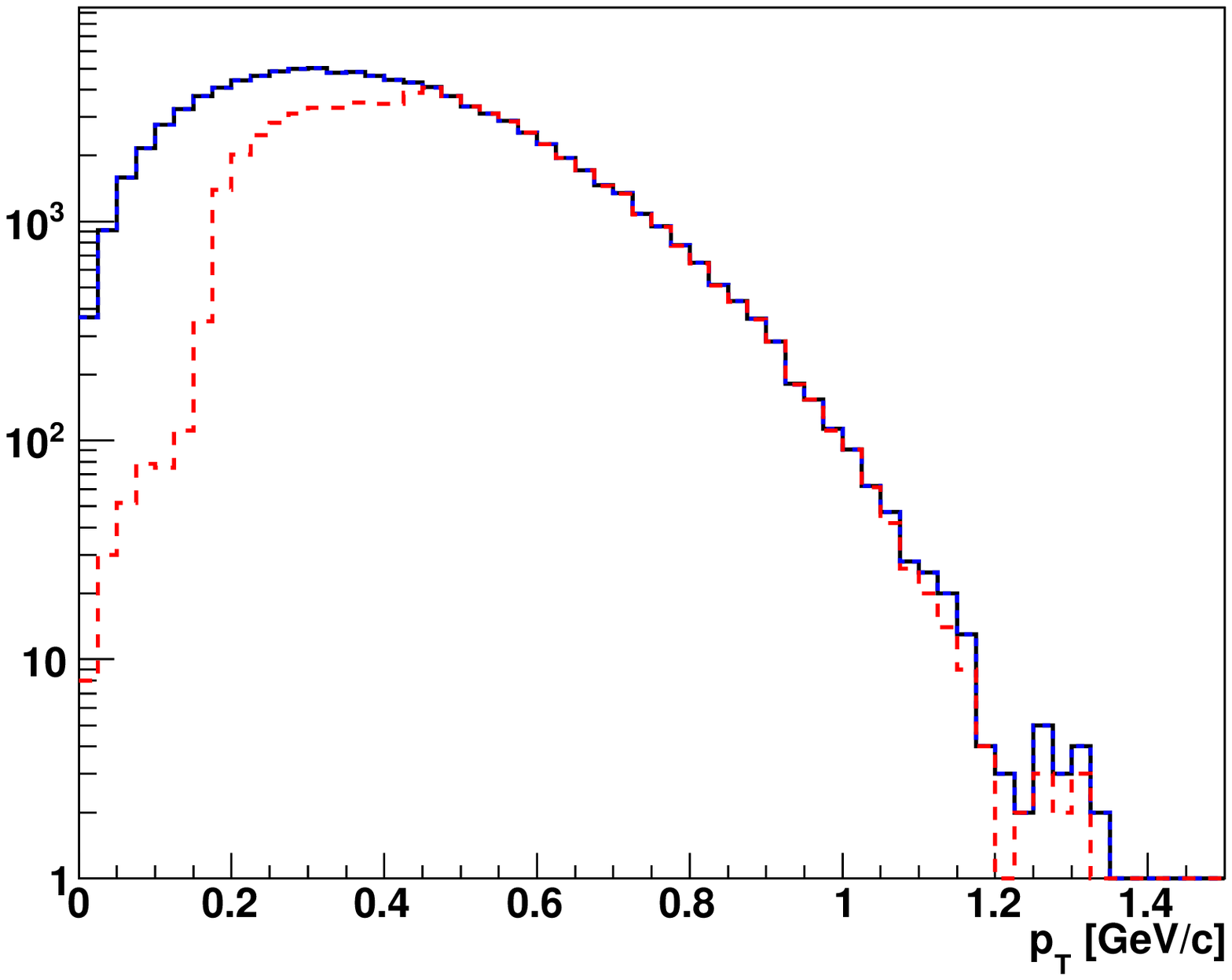} \\[-0.2cm]
\end{tabular}
\end{center}
\caption{\small Row-1: Spatial distribution of the scattered 
protons from DVCS events at 20m from the IP for 2 different beam energy combinations.
Row-2: As Row-1, applying the aperture limitations due to the magnets.
Row-3: As Row-2, applying the limitations due to the 10 $\sigma$ beam
clearance and the acceptance of `Roman Pots' as currently used by pp2pp at STAR.
Row-4: Comparison of the $p_T$ spectrum of generated protons (black), those accepted by 
the quadrupole aperture (blue) and those detected in the `Roman Pots' (red).}
\label{fig:DVCS-prot}
\end{figure}
From fig. \ref{fig:DVCS-prot} it is clear that protons from DVCS events can be measured 
in `Roman Pots' after the high-focusing quadrupole triplet with a high detection efficiency 
for hadron beam energies starting from 100 GeV (as example are shown the results for 50 GeV and 250 GeV). 
More studies are needed to determine the momentum and angular resolution that can be 
achieved depending on the `Roman Pot' design.

\begin{figure}[htbp]
\begin{center}
\includegraphics[width=0.30\textwidth]{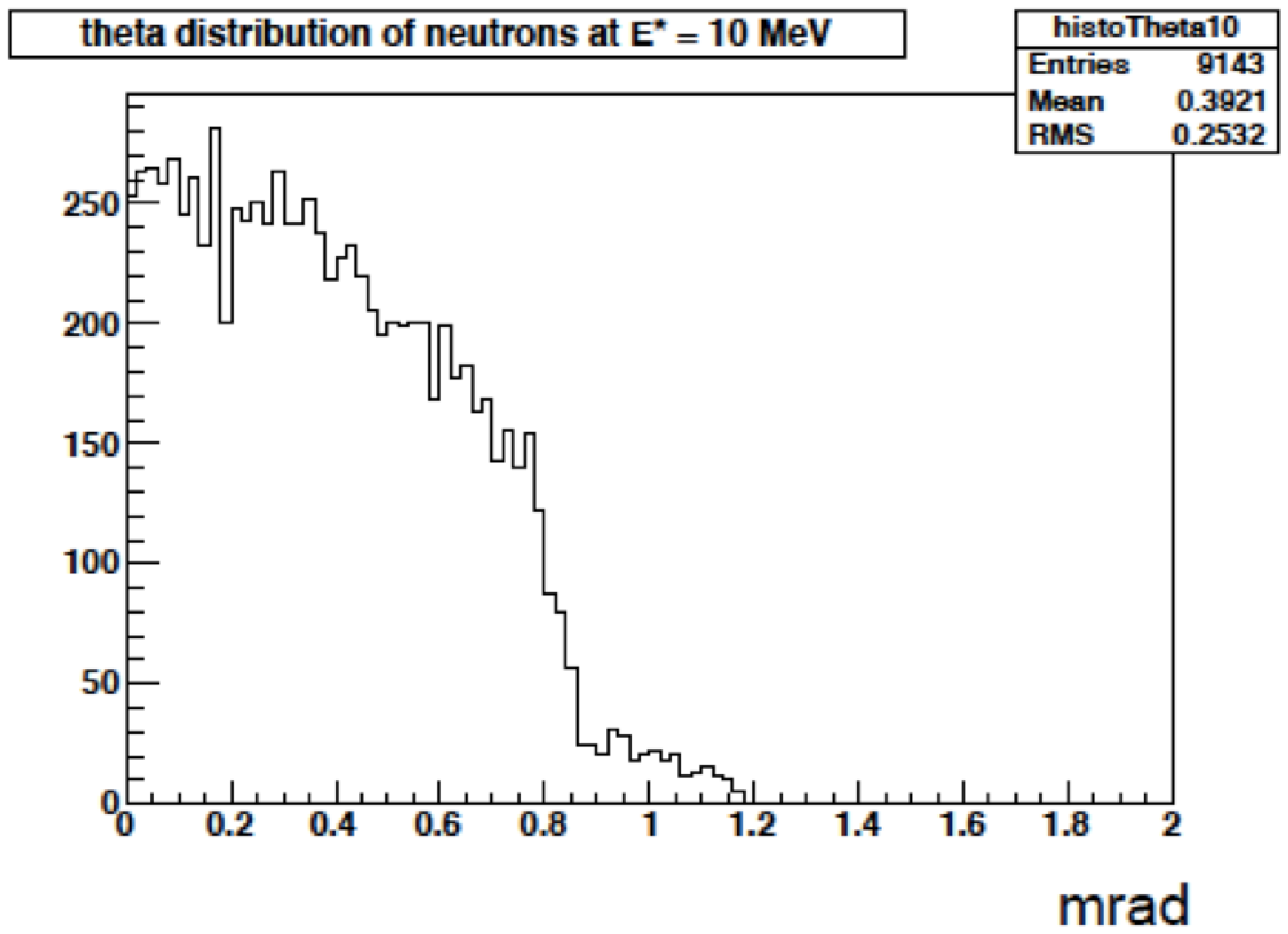}
\includegraphics[width=0.30\textwidth]{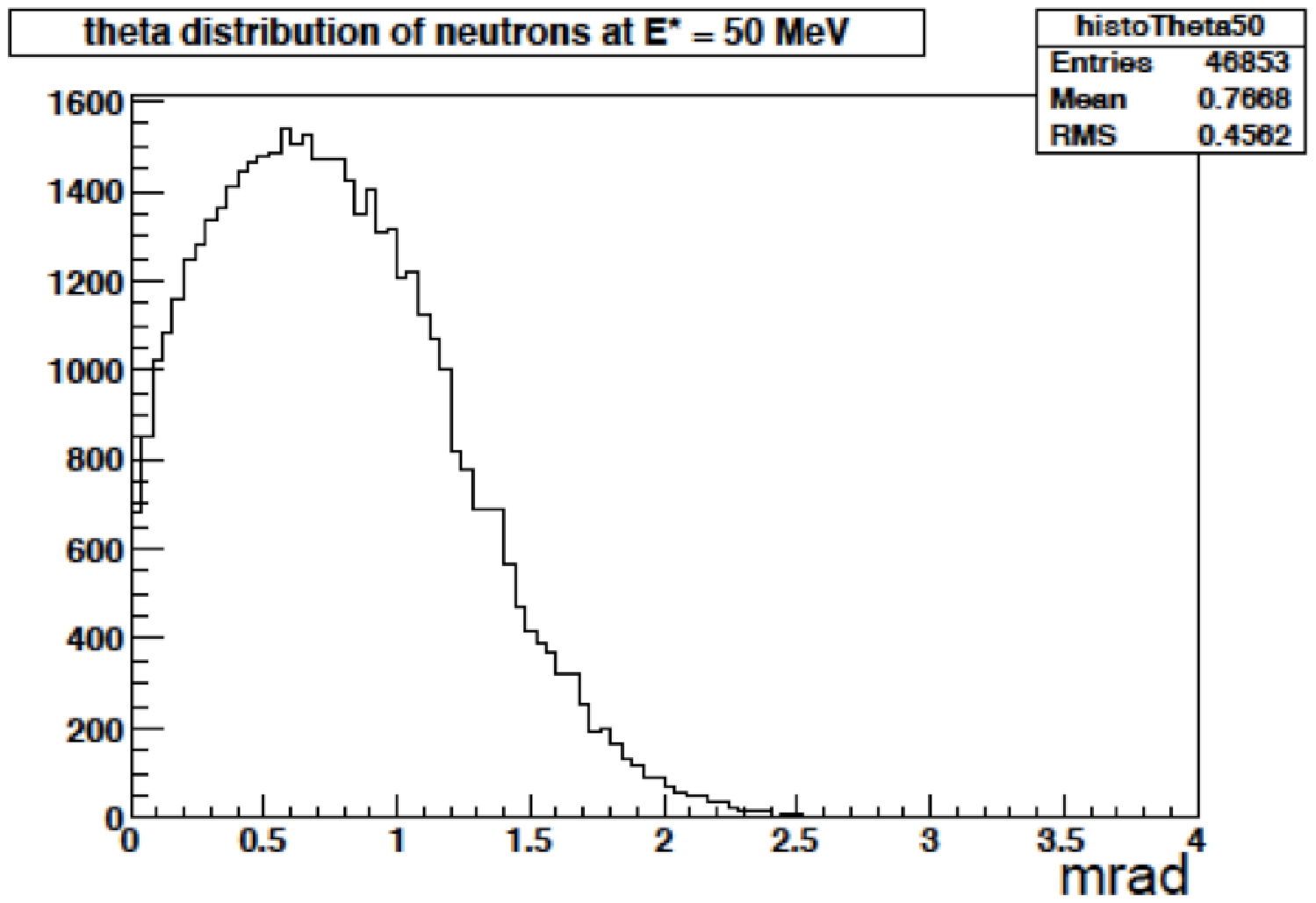}
\includegraphics[width=0.30\textwidth]{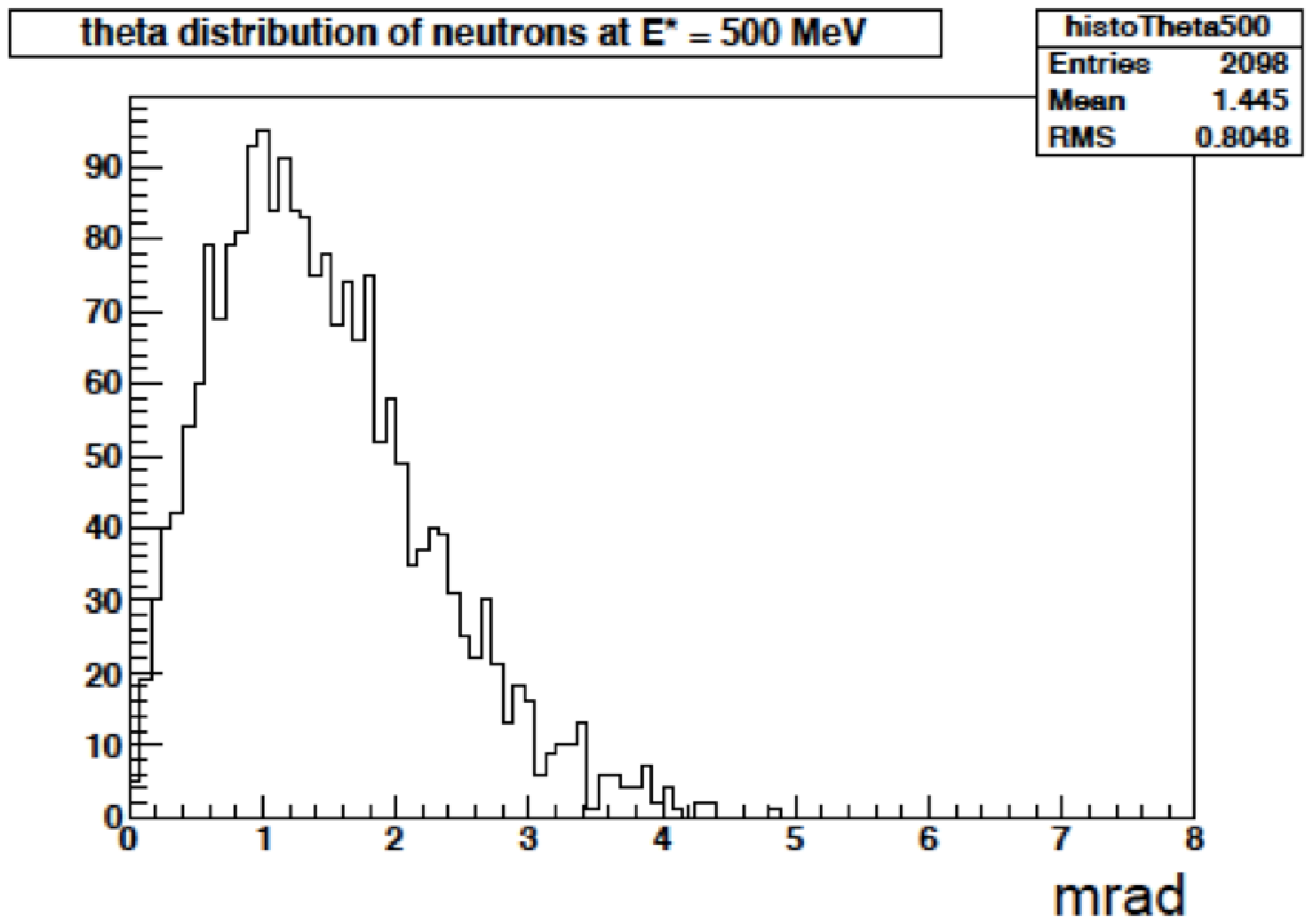}
\end{center}
\caption{\small The angular distribution of neutrons from the 
breakup of a Au nucleus depending on the excitation energy.}
\label{fig:Au:n}
\end{figure}

As pointed out previously, equally challenging is the detection of the breakup neutrons
from heavy ions to veto incoherent events.
The nuclear breakup of Au nuclei depending on the excitation energy E$^*$ was 
simulated using the Monte Carlo generator GEMINI++ \cite{Charity:2010wk} and SMM \cite{Botvina:2008gq}.
The MC simulation showed that whenever the nucleus breaks up there
will be at least one neutron emitted. At very low excitation energies there is the possibility 
that only a photon is emitted, while the nucleus remains intact.
The possibility of detecting these photons still needs 
to be investigated.
Fig. \ref{fig:Au:n} shows the angular distribution of the breakup neutrons
for three different excitation energies. The aperture of 120 mm diameter of the 
last quadrupole magnet allows detection of neutrons with a solid angle of $\pm$4 mrad, 
which is indicated by the simulations to be sufficient.

\begin{figure}[htbp]
\begin{center}
\includegraphics[width=0.30\textwidth]{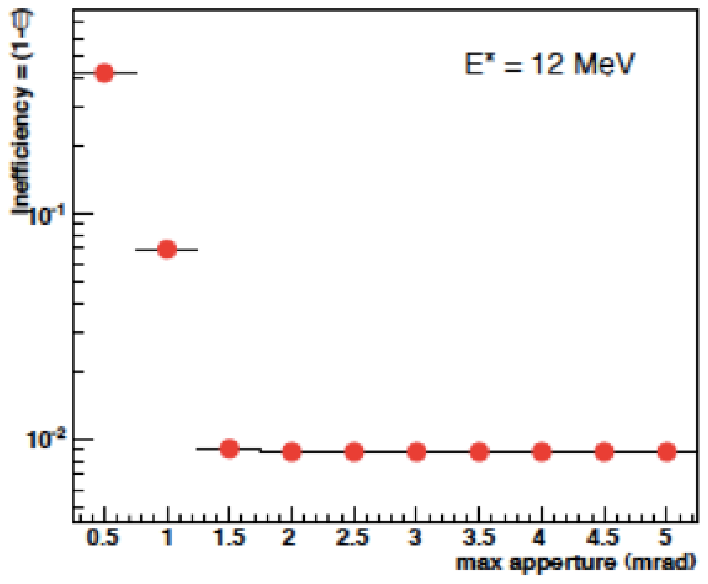} 
\includegraphics[width=0.30\textwidth]{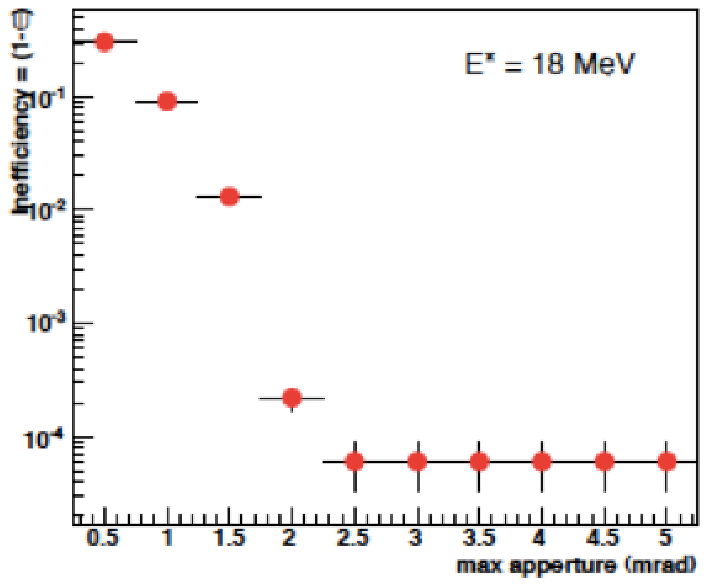}
\includegraphics[width=0.30\textwidth]{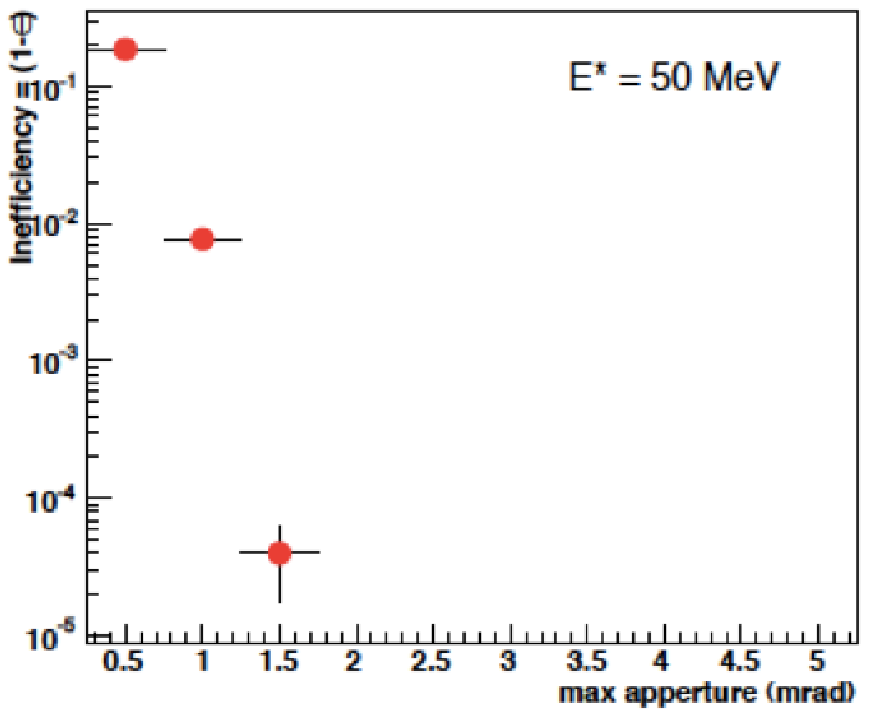} 
\end{center}
\caption{\small The inefficiency to detect the neutrons from the 
breakup of a Au-nucleus as function of the maximal aperture of the last magnet
for different excitation energies.}
\label{fig:Au:n:ineff}
\end{figure}

Fig. \ref{fig:Au:n:ineff} shows the detection inefficiency for these neutrons
for three different excitation energies as function of the maximal aperture of the 
last magnet. For apertures discussed for the IR design the inefficiencies are 
10$^{-2}$ or much lower for all excitation energies. This 
assumes a 100\%-efficient zero degree calorimeter (ZDC).
The critical question is: to suppress incoherent events at high $t$ in eA collisions, 
can the detection inefficiencies be controlled on the 10$^{-3}$ to the 10$^{-4}$ level?

There are many detector, interaction region and machine parameters still to be worked 
out in detail, but one of the hardest questions for an EIC will be to estimate the 
limiting factors for the systematic uncertainties. Due to the high luminosity, 
many inclusive and semi-inclusive physics observables will be systematics-limited 
after a relatively short time of data taking, assuming a 50\% operations efficiency. 
This requires great care to be taken to consider the possible systematic limitations 
from the beginning and to integrate solutions to minimize them into the design.
Only some of the possible limiting systematic effects that will need to be addressed with great care
in the design are listed here. Their impact on key physics observables still needs 
to be studied.

\begin{itemize}
\item Absolute luminosity measurements between different beam energy combinations. This
is extremely important for measurements like the structure function F$_L$.
\item Relative luminosity measurements between bunches with different
bunch helicities, i.e. $++, --, -+$ and $+-$. Here it will be important to investigate whether 
Bremsstrahlung can be used for this measurement, as the Bremsstrahlung
cross section has a term that is dependent on polarization.
\item The measurements of the absolute hadron and electron beam polarization. 
To date the best precision in the measurement of lepton beam polarization at high 
energies in a collider was obtained during HERA-I running with 1.6\% \cite{Beckmann:2000cg}.
At RHIC the best hadron polarization measurement achieved to date is $\sim$5\% \cite{pCarbon,Zelenski2005248,Hjet}
for a polarized proton beam. For high energy polarized $^3$He beams, R\&D is needed to
determine how to measure an absolute polarization.
\end{itemize}

\subsubsection{ePHENIX}
\begin{figure}[htbp]
\begin{center}
\begin{tabular}{cc}
\includegraphics[width=0.45\textwidth]{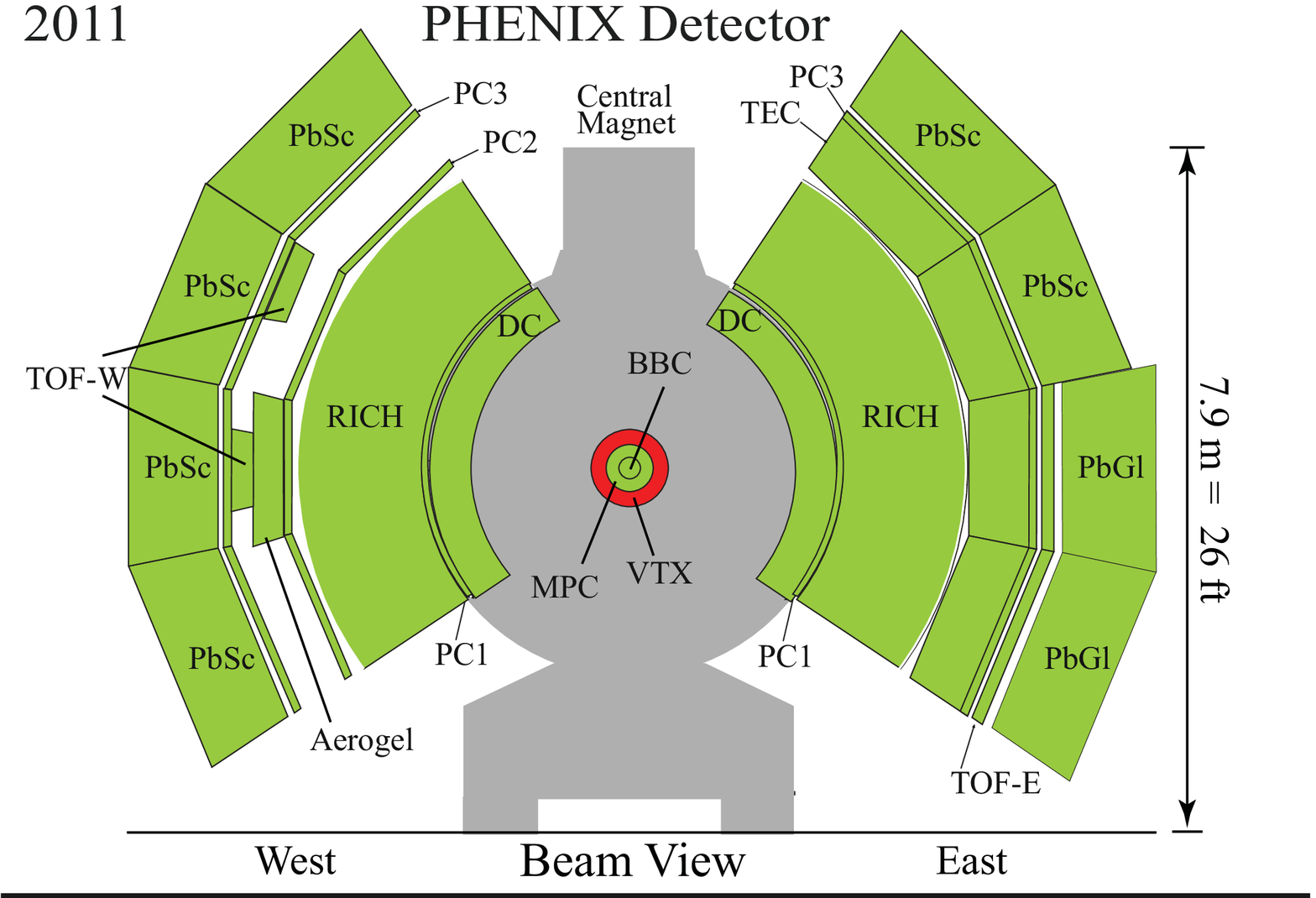} &
\includegraphics[width=0.45\textwidth]{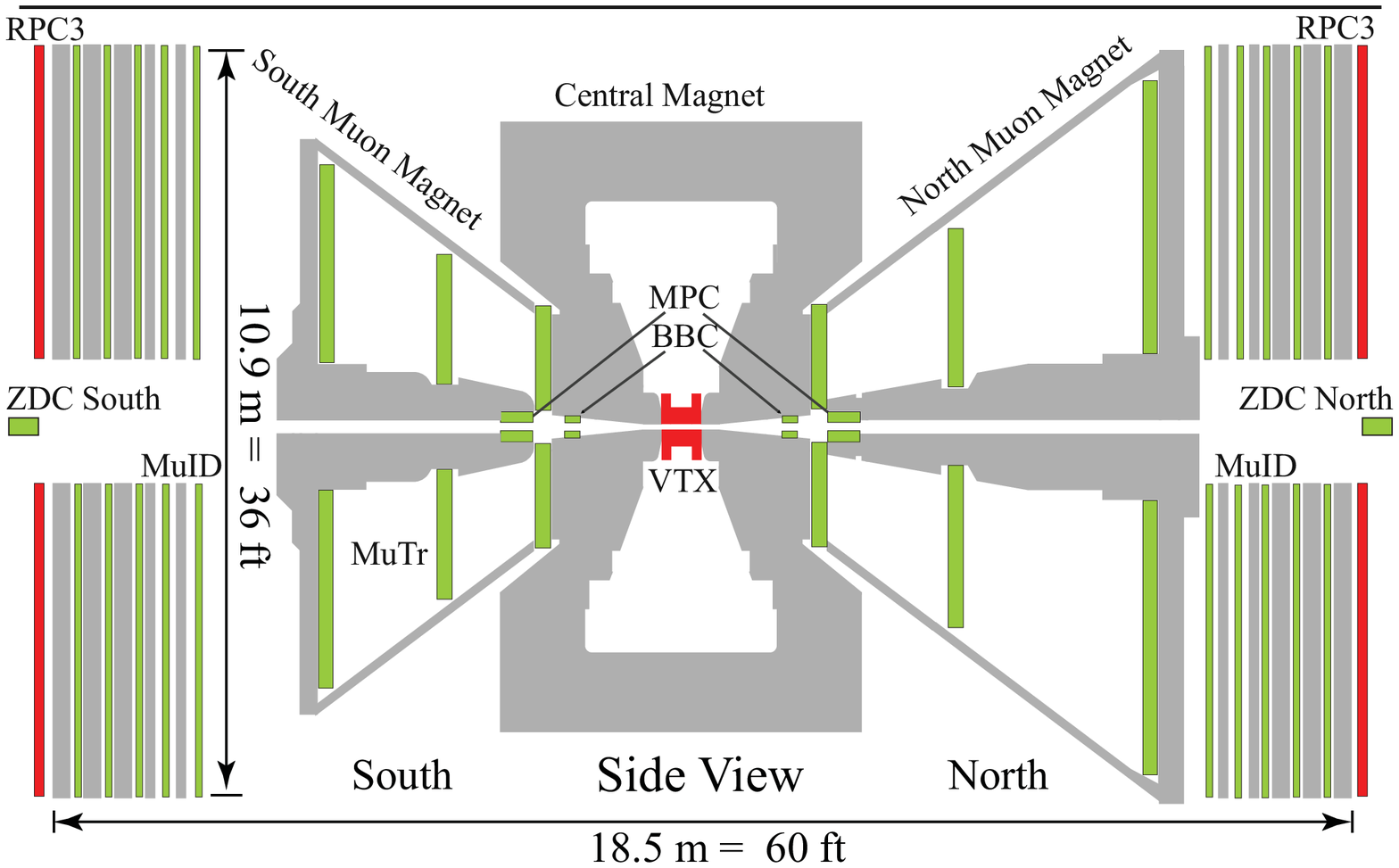}\\
\end{tabular}
\end{center}
\caption{\small A schematic view of the current (2011) configuration of the PHENIX 
detector.}
\label{fig:phenixnow}
\end{figure}

\begin{figure}[htbp]
\begin{center}
\includegraphics[width=0.45\textwidth]{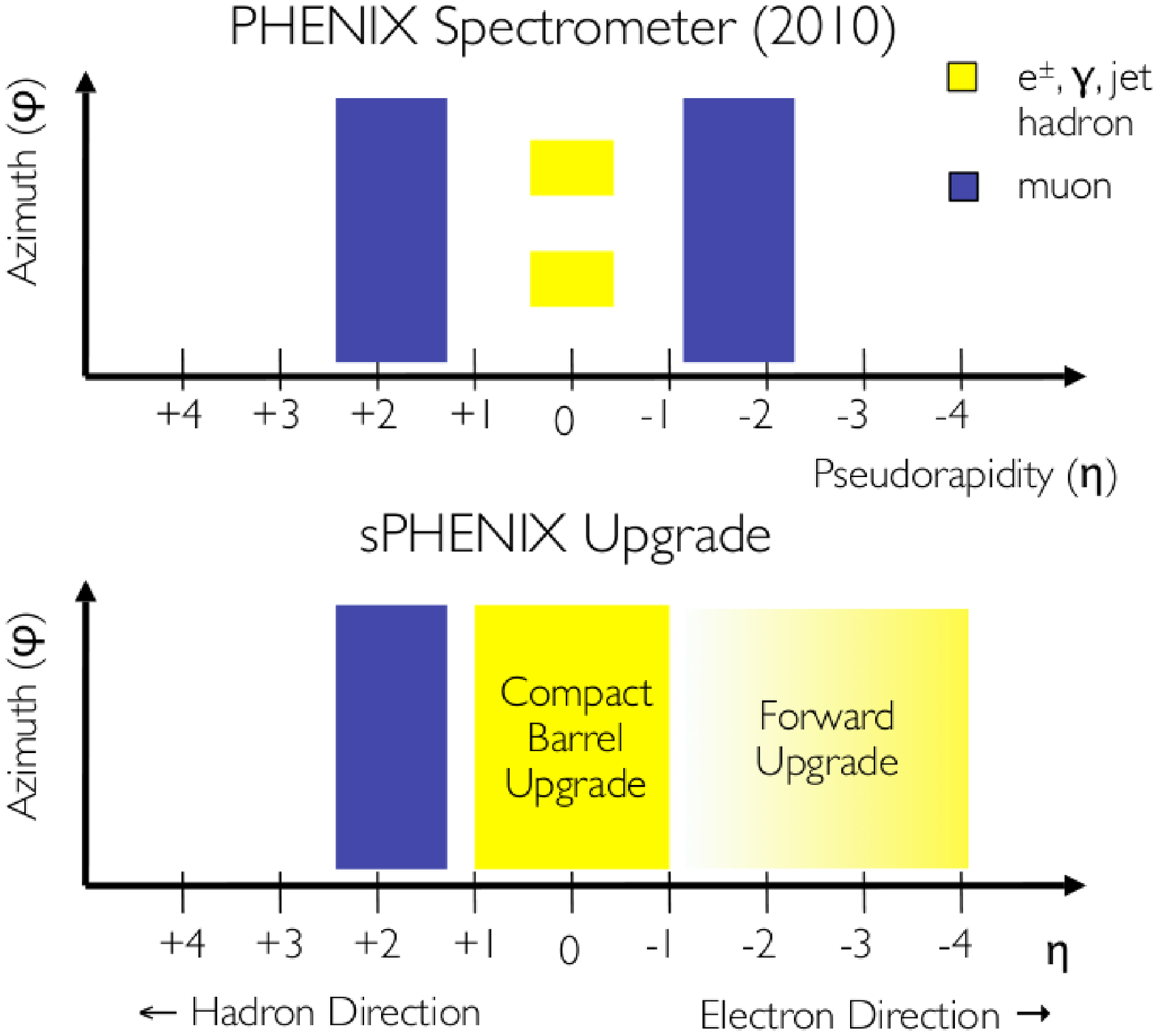} 
\end{center}
\caption{\small Rapidity coverage of the current PHENIX detector compared to the strawman
new PHENIX detector. The central barrel detector covers $|\eta| <$ 1.0; 
the forward detector has tracking coverage for -4 $ < \eta <$ -1, with full 
EMCal and HCAL coverage for -4.0 $< \eta <$ -2.0 - (-1.5)
with the exact range dependent on the final design configuration}
\label{fig:phenixacc}
\end{figure}

PHENIX is one of the two large dedicated RHIC detectors, located at IP-8.
The PHENIX detector consists of two muon spectrometer arms and two central
arms sitting in a 1 tesla solenoid. Over the years the detector has been upgraded
to the configuration shown in fig. \ref{fig:phenixnow}.
Fig. \ref{fig:phenixnow} and the upper plot of fig. \ref{fig:phenixacc} show clearly
that PHENIX in its current configuration has only a very small acceptance 
($|\eta| <$0.35) for the scattered lepton. This makes the current PHENIX detector
basically not usable for DIS physics.

For the RHIC decadal plan covering the period 2010 - 2020, PHENIX has proposed a 
major upgrade of the current detector \cite{phenixdecade}. 
\begin{figure}[htbp] 
\begin{center}
\includegraphics[width=0.95\textwidth]{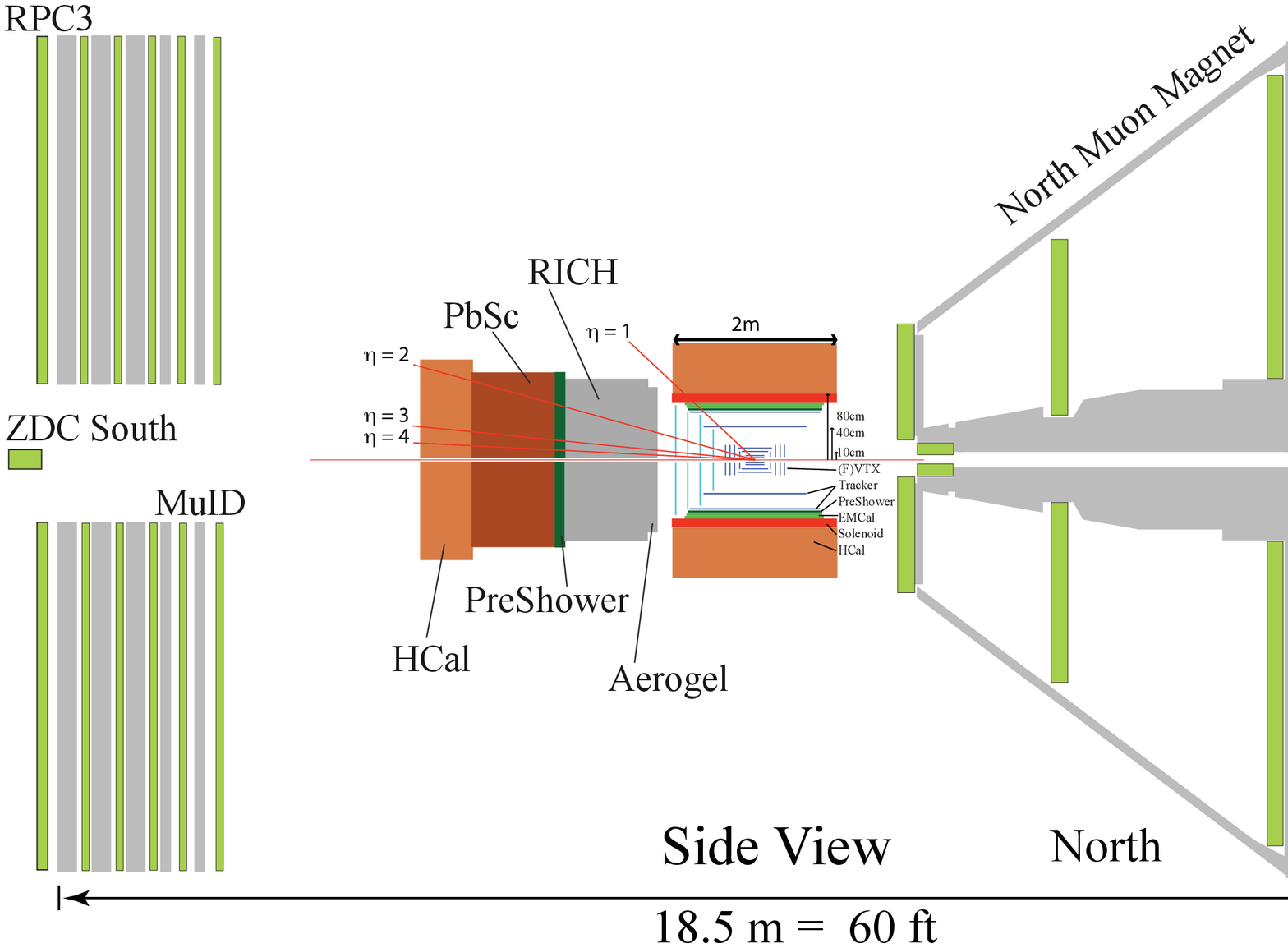}   
\end{center}
\caption{\small Schematic drawing of the new PHENIX detector.}
\label{fig:phenixnew}
\end{figure}
The decadal plan outlines an exciting program in heavy ion and spin physics in polarized 
pp collisions, focused on an investigation of the interplay between perturbative and 
nonperturbative physics in QCD and on the relative importance of strong and 
weak coupling. 
The physics aims have been translated into an extensive set of required physics observables
to answer the key scientific questions, leading to the design of the new PHENIX 
detector.  The upgrade plan involves replacing the PHENIX central magnet with a new 
compact solenoid. The limited aperture provided by the outer central arm detectors 
would be replaced with a compact EMCal and a Hadronic Calorimeter covering two units 
in pseudorapidity and full azimuth, complemented
by the existing VTX and FVTX inner silicon tracking. Two additional tracking
layers would be added. We highlight that the large acceptance and excellent detector
capability is combined with high rate and bandwidth. 
The limited forward coverage of the current PHENIX detector does not allow one to 
adequately address the questions driving the nucleon structure and
cold nuclear matter community, nor does it provide any capabilities for e+p or 
e+A collisions.
Hence, an upgrade is being considered where one muon arm would be replaced by
a new large-acceptance forward spectrometer with excellent PID for hadrons, electrons,
and photons and full jet reconstruction capability. The modified detector layout is 
shown schematically in fig. \ref{fig:phenixnew}. 
The increase in overall acceptance is shown in the lower part of fig. \ref{fig:phenixacc}.
The new compact barrel component at midrapidity is designed for excellent jet 
reconstruction and PID for photons, electrons, and $\pi^0$ in p+p, proton-nucleus, 
through central nucleus-nucleus collisions. The forward upgrade design is driven by 
nucleon structure physics and cold nuclear matter physics.
Such a forward spectrometer added to PHENIX would not only allow measurements of the 
single spin asymmetry at forward rapidity to test the QCD prediction that the 
Sivers function in Drell-Yan and SIDIS is opposite, but would 
also allow the unique possibility to detect the scattered lepton
in e+p/e+A collisions in the era of an eRHIC to virtualities Q$^2 >$ 0.1 GeV$^2$. 
To realize these physics goals it is necessary to upgrade significantly the current 
PHENIX detector to a detector with high acceptance at forward rapidity 1 $< \eta <$ 4.0 .

The strawman design for the central barrel has already been described.
The forward detectors of the strawman design consist of a RICH,
a preshower, an EMCal, an HCal, and additional tracking detectors to provide good
momentum definition of the particles going forward. This combination of detectors is
motivated by both Drell-Yan and e+p/e+A physics to emphasize the detection of
electrons with high efficiency and purity.

It must be stressed again that the PHENIX detector upgrades as discussed above are 
driven by p+p, p+A, A+A physics.  But, comparing the requirements for the physics program 
at an EIC as described in section \ref{sec:det.kin}, it becomes clear that this 
detector upgrade also provides opportunities to carry out an e+p and e+A physics program, 
referred to as ePHENIX. The upgraded PHENIX is well suited for
\begin{itemize}
\item Inclusive e+p physics to measure polarized and unpolarized structure functions.
\item Inclusive e+A physics to measure unpolarized structure functions and derive nuclear
parton distribution functions (nPDFs). 
\item e+p / e+A physics involving charm and bottom 
\item Elastic diffractive physics, i.e. elastic vector meson production and deeply 
virtual Compton scattering. 
These measurements require the addition of `Roman pot' detectors.
\end{itemize}
nearly independent of the center-of-mass energy and lepton and hadron beam combination.
Unfortunately due to the limited PID capabilities of the ePHENIX design, most of the SIDIS 
physics program for an EIC will not be possible.

There are still several open question on the detailed performance of the upgraded
PHENIX detector in ep / eA collisions, which need to be studied in the 
next months. Some examples are given below. Of course, some of these concerns
can easily be solved by addressing them by design changes.
\begin{itemize}
\item How can ePHENIX be integrated in the current IR design of eRHIC?
\item What does the current material budget do to the momentum and angular resolution of
the scattered lepton?
\item Does the current compact solenoid provide enough bending power to achieve sufficient
momentum resolution for the scattered lepton at low Q$^2$?
\item How can a luminosity measurement for ep/eA collisions be integrated in 
the design?
\item Are the currently planned electromagnetic calorimeter designs suited in
energy resolution to separate leptons from hadrons via $E/p$, and to get the required 
resolution for the DVCS photon?
\end{itemize}

\subsubsection{eSTAR}
\begin{figure}[htbp]
\begin{center}
\includegraphics[width=0.95\textwidth]{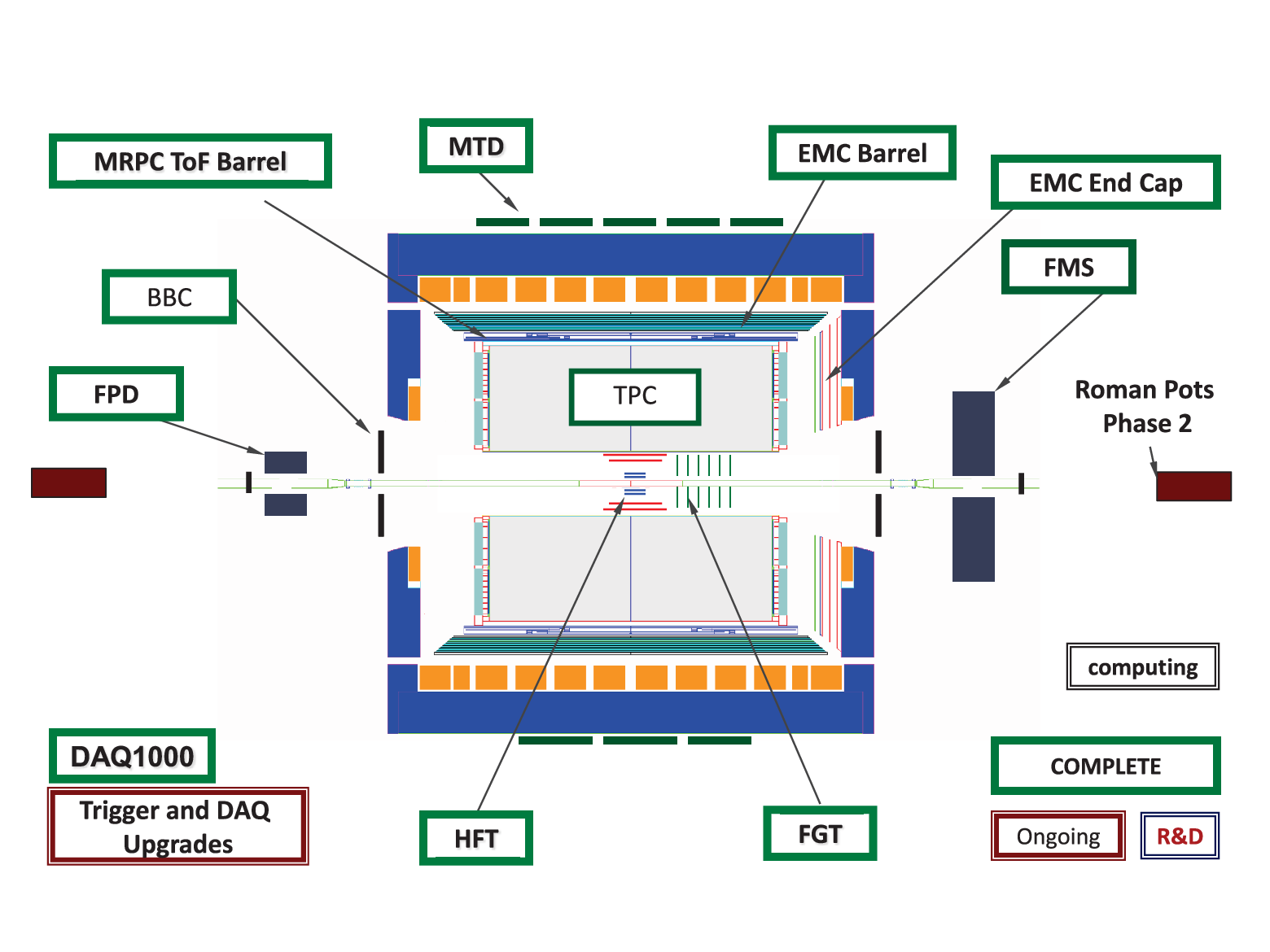}
\end{center}
\caption{\small Schematic drawing of the STAR detector in 2014.}
\label{fig:stardet}
\end{figure}

STAR is the other of the two large dedicated RHIC detectors, located at IP-6.
Fig. \ref{fig:stardet} shows STAR in the configuration anticipated in 2014.

The unique strength of STAR (solenoidal tracker at RHIC) \cite{Ackermann2003624} 
is its large, uniform acceptance capable of measuring and identifying a substantial 
fraction of the particles produced in heavy ion collisions.
The heart of STAR is its main tracking device: a TPC, covering full azimuthal angle
and $\pm$1.5 units of pseudo-rapidity. A dE/dx resolution of $\sim$ 8\% can be
achieved by requiring the tracks of charged particles to have at least 20 out
of a maximum of 45 hits in the TPC. Detailed descriptions of the TPC and
its electronics system have been presented in \cite{Anderson2003659,Anderson2003679}.
The TPC sits in a 0.5 tesla solenoid, surrounded by electromagnetic calorimetry
(EMC Barrel, EMC End Cap, FMS) covering -1 $< \eta <$ 4, muon identification (MTD) covering 
-1 $< \eta <$ 1 and a high-resolution time of flight system (MRPC ToF Barrel) 
covering -1 $< \eta <$ 1.
The tracking in STAR will be further improved by 2014 by adding a forward 
GEM tracker (FGT) covering 1 $< \eta <$ 2 and a high-resolution silicon detector (HFT) covering 
-1 $< \eta <$ 1. The HFT gives the possibility to separate events with charmed mesons from those with 
beauty mesons through the detection of the displaced vertex for charmed mesons.
Identification in the lepton sector will be enhanced with the Muon Telescope Detector (MTD), 
which will tag muons for -1 $< \eta <$ 1.  This will enable dilepton studies in the $\mu-\mu$ 
and $e-\mu$ channels, with a focus on separating the Upsilon states and constraining charm 
backgrounds to the thermal continuum in intermediate mass dileptons.
Another unique feature of STAR is the `Roman Pots' around the main detector; their main 
focus is to detect protons from elastic diffractive events in pp collisions.

\begin{figure}[htbp]
\begin{center}
\includegraphics[width=0.45\textwidth]{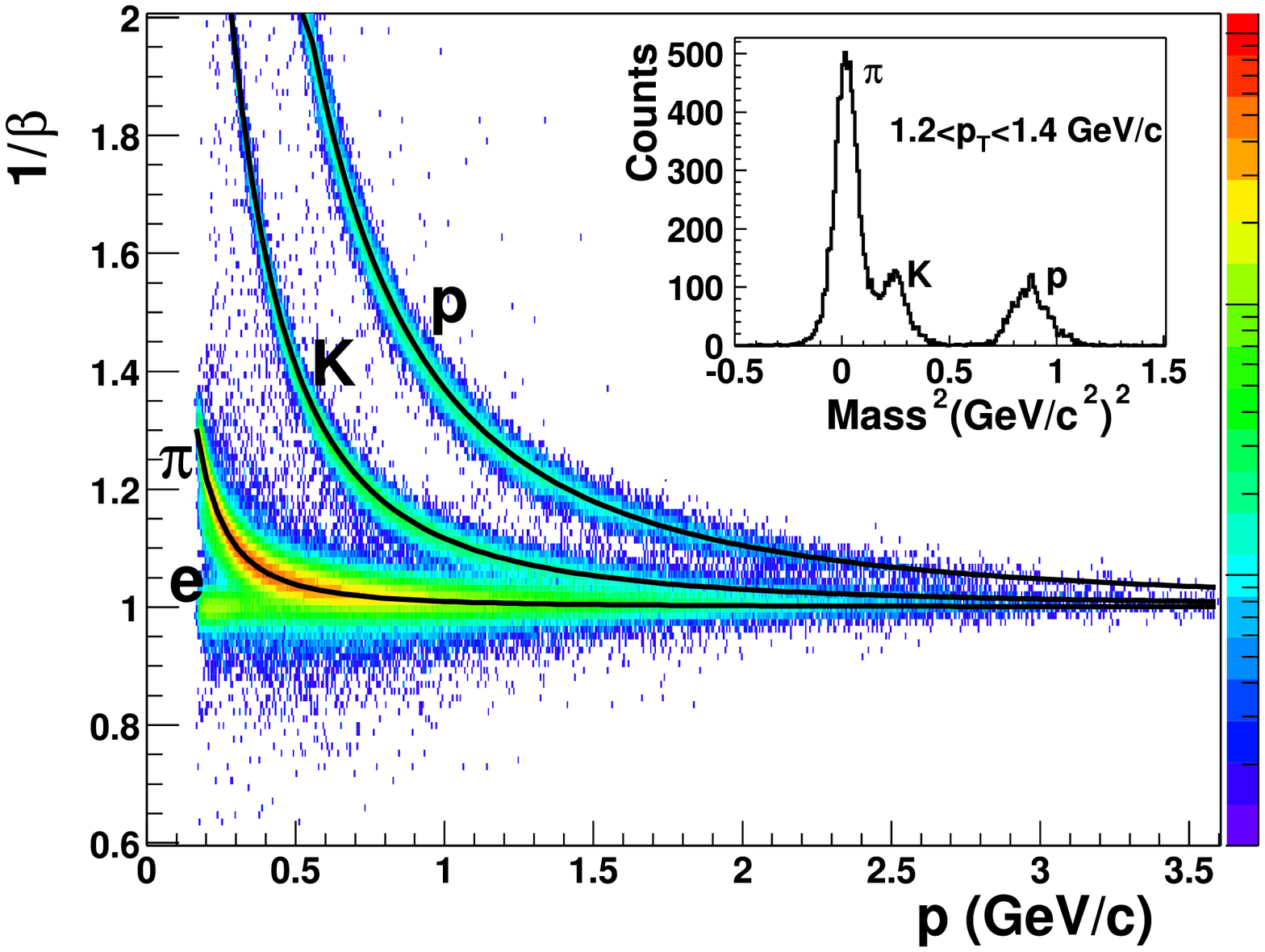}
\includegraphics[width=0.50\textwidth]{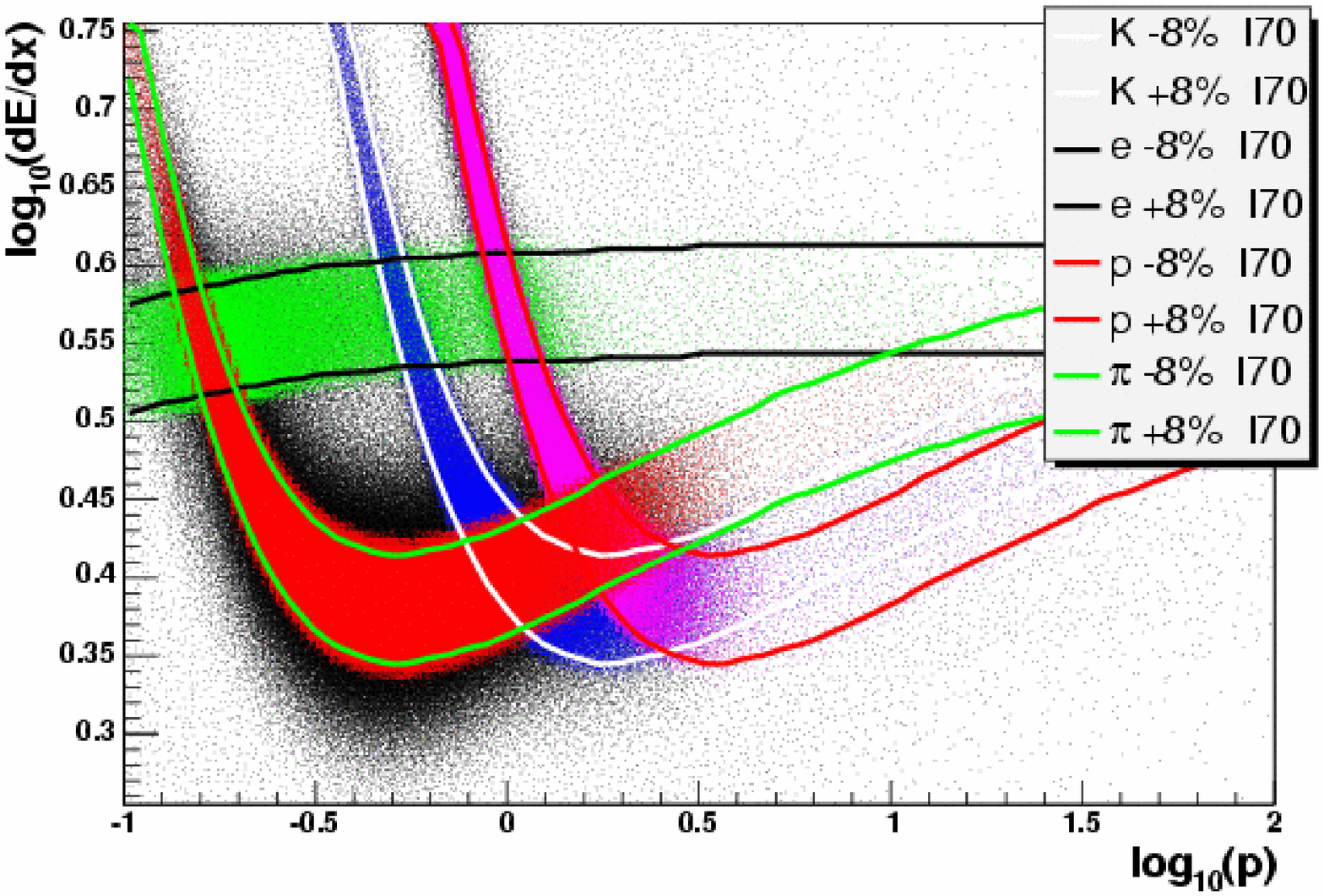}
\end{center}
\caption{\small Left: 1/$\beta$ vs. momentum for $\pi^{\pm}$, $K^{\pm}$, and 
($p\bar{p}$) from 200 GeV d+Au collisions. Separation between pions and kaons 
(kaons and protons) is achieved up to p$_T \sim$ 1.6  (3.0) GeV/c. 
The insert shows $m^2 = p^2(1 / \beta^{2} -1)$ for 1.2 $< p_T <$ 1.4 GeV/c. 
Right: Distribution of $log_{10}(dE/dx)$ as a function of $log_{10}(p)$ for electrons, pions,
kaons and (anti-)protons. The units of $dE/dx$ and momentum ($p$) are keV/cm and
GeV/c, respectively. 
The color bands denote the $\pm 1 \sigma$ $dE/dx$ resolution.}
\label{fig:starpid}
\end{figure}

In addition to large coverage in tracking and electromagnetic calorimetry, STAR has 
good particle identification capabilities. For stable charged hadrons, the TPC provides
$\pi$/K ($\pi$+K/p) identification to p$_T \sim$ 0.7 (1.1) GeV/c by the measurement of ionization energy
loss (dE/dx). The STAR PID capability is further enhanced by the TOF system 
with a time resolution of $<$ 100 ps, which is able to identify $\pi$/K ($\pi$+K/p) to 
p$_T \sim$ 1.6 (3.0) GeV/c, as demonstrated in the left panel of fig. \ref{fig:starpid}. 
In addition, with the relativistic rise of dE/dx from charged hadrons
traversing the TPC at intermediate/high p$_T$  ($>$3 GeV/c) and diminished yields
of electrons and kaons at this p$_T$ range, pions and protons  can be identified 
up to very high p$_T$ ($\sim$ 10 GeV/c) in p+p, p+A and A+A collisions 
(see right panel fig. \ref{fig:starpid}). 

STAR has, like PHENIX, provided a decadal plan outlining the physics program for 
pp, dA and AA collisions in the next 10 years~\cite{stardecade}. Contrary to PHENIX, the STAR upgrade plans are much 
more moderate and focus on forward rapidity (2$<|\eta|<$4). On the side of the STAR detector at 
which the FMS is situated, the
plan is to improve charged particle tracking by adding more tracking planes to the FGT to
cover rapidities 2.5$<\eta<$4. To improve lepton/hadron and $\gamma/\pi^0$ discrimination, as well 
as baryon/meson separation, a RICH detector and a preshower detector will be added in front of the FMS.
The addition of a hadronic calorimeter behind the FMS will further improve the lepton/hadron separation,
as well as give the possibility of measuring the energy due to neutral particles in jet reconstruction.
The motivation for this upgrade is, like in the case of the PHENIX forward upgrade, 
transverse spin physics in pp collisions (Sivers asymmetry in Drell Yan) and 
the study of cold nuclear matter, i.e. parton saturation at small $x$.

The upgrade in rapidity -4$<|\eta|<$-1 is driven solely towards improving the detection capabilities of STAR for the 
scattered lepton in ep/eA collisions during the era of eRHIC.
Currently, proposals include the addition of tracking and electromagnetic calorimetry as well as an additional ToF for
PID. For tracking, it is proposed to combine high-resolution with electron identification by 
for example integrating a Cherenkov detector in the tracking detector.

\begin{figure}[htbp]
\begin{center}
\includegraphics[width=0.48\textwidth]{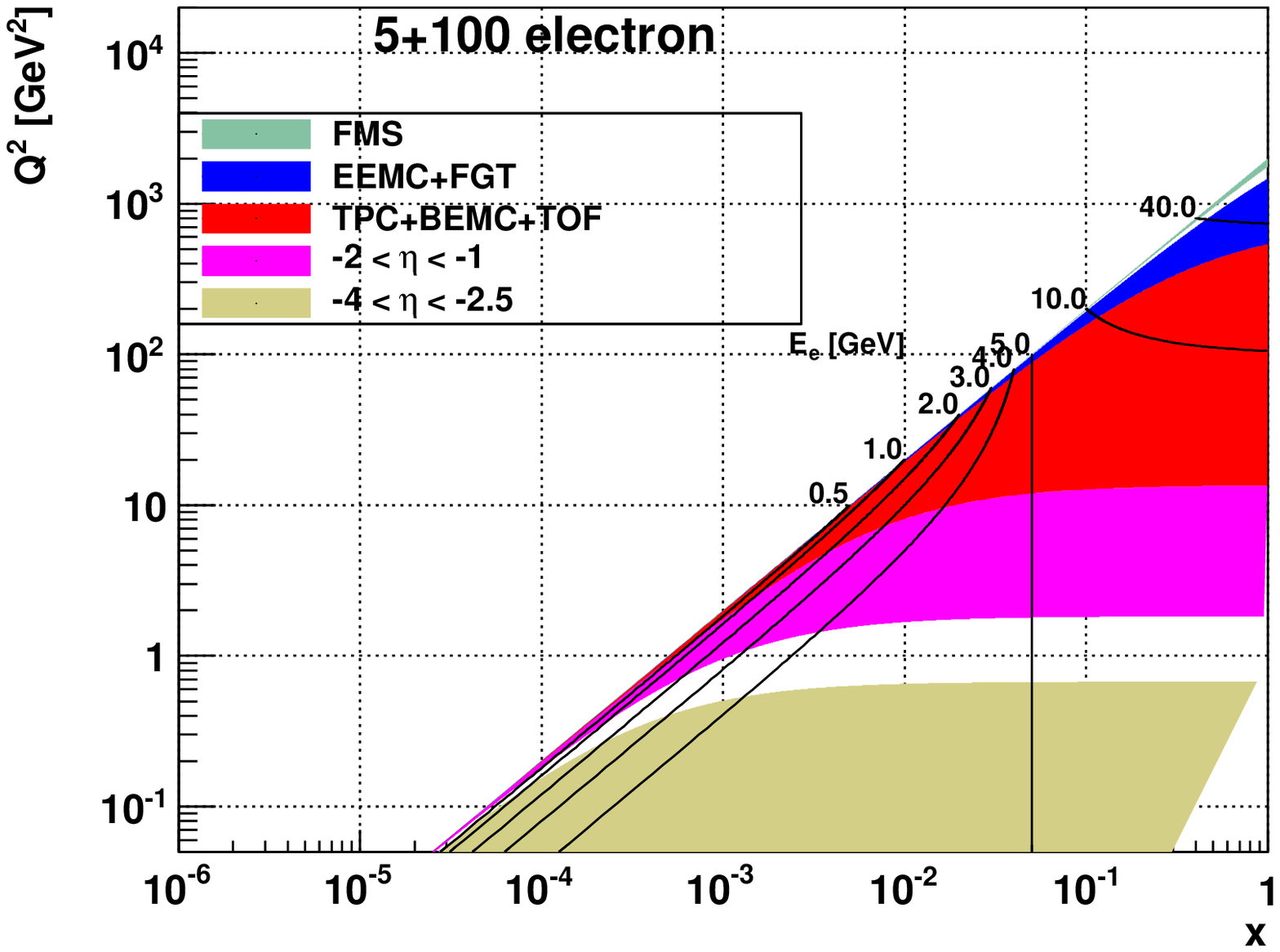}
\includegraphics[width=0.48\textwidth]{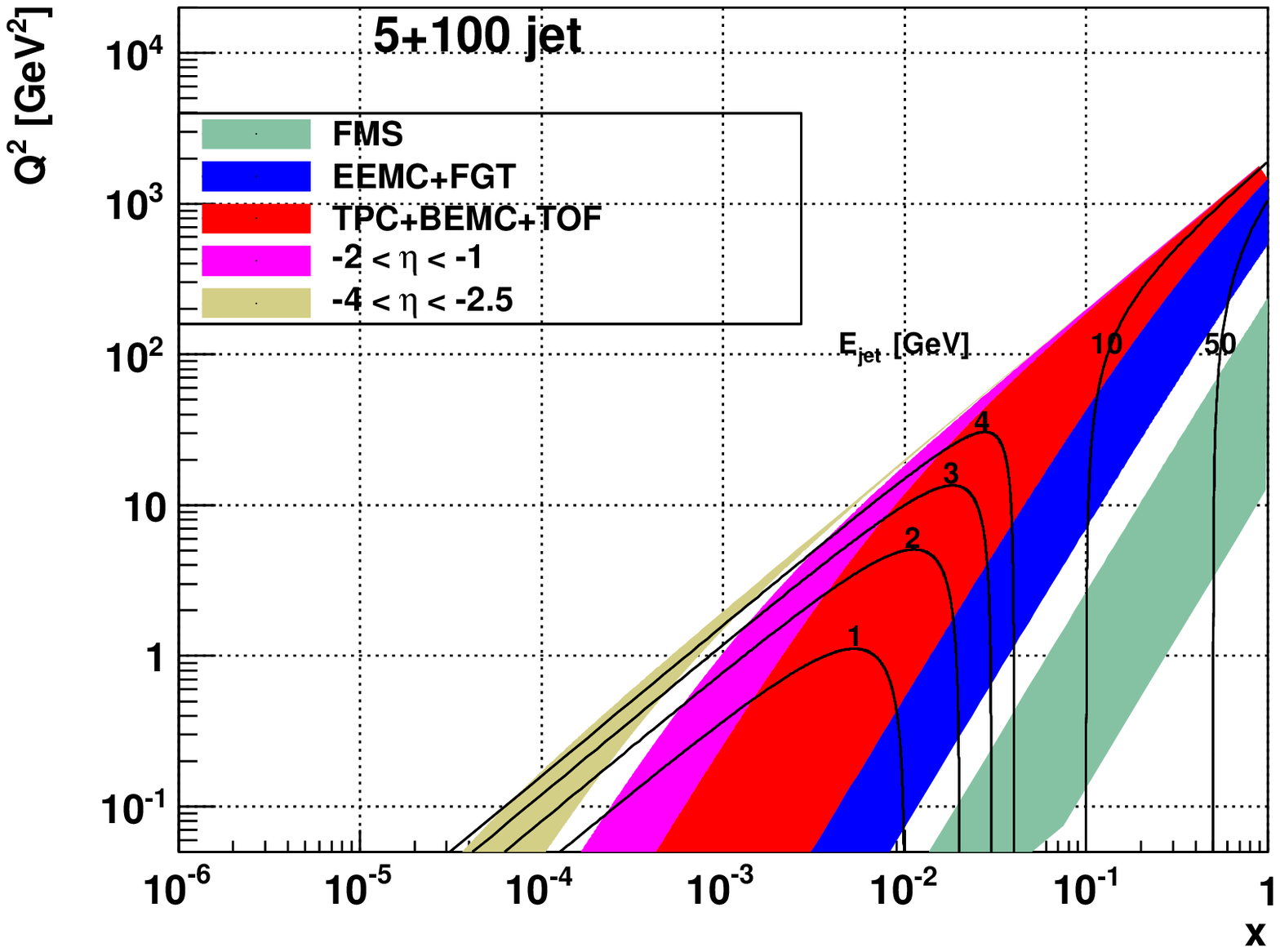}
\end{center}
\caption{\small
Kinematic coverage of the STAR detector in the (x,Q$^2$) plane. Left: electron. Right:
struck quark. The electron beam energy is 5 GeV, and the nucleus beam energy is 100 GeV/u.
Lines of constant laboratory energy of the electron and the struck quark are shown. }
\label{fig:starepeA}
\end{figure}

Combining all these upgrades in fig. \ref{fig:starepeA} shows that STAR will have very good acceptance 
for both the scattered lepton and for the hadrons produced by the current jet at the first stage of eRHIC,
with 5 GeV electron beams colliding with proton beams with energies as high as 325 GeV.
From these figures it is also obvious that the upgrade at negative rapidity is essential to
provide good coverage for the scattered lepton below Q$^2$ of 10 GeV$^2$. 

The list of questions which need to be answered is very similar to that listed in the ePHENIX section,
and many further detailed simulations must be performed to understand in detail the performance of STAR
for ep/eA collisions. However from the first studies it is clear that eSTAR will 
be able to make key measurements such as: 
\begin{itemize}
\item Inclusive e+p physics to measure polarized and unpolarized structure functions.
\item Inclusive e+A physics to measure unpolarized structure functions and derive nuclear
parton distribution functions (nPDFs).
\item Elastic diffractive physics, i.e. elastic vector meson production and deeply
virtual Compton scattering. Here the great advantage is that eSTAR already possesses `Roman pot' detectors.
\item The good particle ID capabilities also open the possibility of studying many of the semi-inclusive
observables in ep/eA collisions, i.e. to do a flavour separation of the quark polarizations to understand both the 
helicity structure and the transverse spin structure (via Sivers 
and Collins functions) of the proton.
\end{itemize}

\subsection{Detector Design for MEIC/ELIC}

The Jefferson Lab design of an EIC is based on a novel figure-8 ring-ring design optimized for polarization 
preservation.  The initial version of this EIC is termed the Medium-Energy EIC, or MEIC, 
which is upgradable to a higher-energy version termed Electron Ion Collider, or ELIC. The MEIC/ELIC will have 
minimal impact on continued operation of the Jefferson Lab (JLab) 12 GeV fixed-target program.

The ring-ring design of the MEIC/ELIC allows simultaneous operation at high luminosity of multiple 
detectors located at different interaction points (IPs). Due to the nature of the figure-8, four IPs are 
foreseen with different functions.
The MEIC detector/interaction region has concentrated on {\sl maximizing acceptance} for deep exclusive processes 
and processes associated with very-forward going particles, which are the most challenging 
from the detector point of view. This section will describe the baseline full-acceptance detector in more detail, 
where it is understood that the various MEIC/ELIC interaction points can house detectors employing different 
technologies and having a slightly different physics focus.

Given that the detailed design of various subsystems does not have to be frozen for another decade or so, and 
dedicated pre-R\&D projects are only now under way, the focus of the JLab effort has been on formulating 
requirements, identifying and addressing critical design issues, and integrating the detector with the interaction 
region of the accelerator. A tentative detector configuration with estimates based on realistic projections 
has been adopted, however, to provide users with input for simulations.

\subsubsection{The Medium-energy EIC (MEIC)}

The current effort is geared towards the MEIC, for which the guiding principle has been based upon science 
motivation and design choices close to present state-of-the-art whenever possible. The exception to the latter 
is the ion beam properties, which have been established for electron-positron colliders but fundamentally depend 
on electron cooling for proton/ion beams. The fundamental choice for the MEIC design has been to assume short 
bunches, each carrying a small charge, and to achieve the requirements for the proton beam quality assume 
extrapolations from conventional electron cooling that have been successfully employed at Fermilab, albeit at 
modest proton energies. Extending this technology may be incremental, rather than transformational in nature.

While ELIC would have a circumference of about 3 km, and support proton energies up to 250 GeV 
(as well as heavy ions up to 100 GeV/A), and electrons post-accelerated up to about 20 GeV, the MEIC 
would be somewhat smaller than the 1.4 km of the CEBAF accelerator, from which it would inject
electron or positron beams between 3 and 11 GeV. The maximum proton energy would be around 100 GeV 
(or 40 GeV/A for heavy ions), but the often quoted design point for which performance parameters are 
being worked out in detail, is 60 GeV.
The choice of a mid-range energy for these studies is primarily based on two considerations. 
On the accelerator side, a proton energy of 60 GeV is a somewhat more conservative value for which one 
could anticipate the performance projections for the electron cooling to become valid at an early stage of 
operations. On the physics side, a range of measurements, for instance related to the 3D structure of the nucleon, 
place strong demands on the resolution in $t$ and the luminosity at modest values of proton energy, corresponding 
to $s \sim 2000$ GeV$^2$. 

\begin{figure}[hbtp]
\begin{center}
\includegraphics[width=0.95\textwidth]{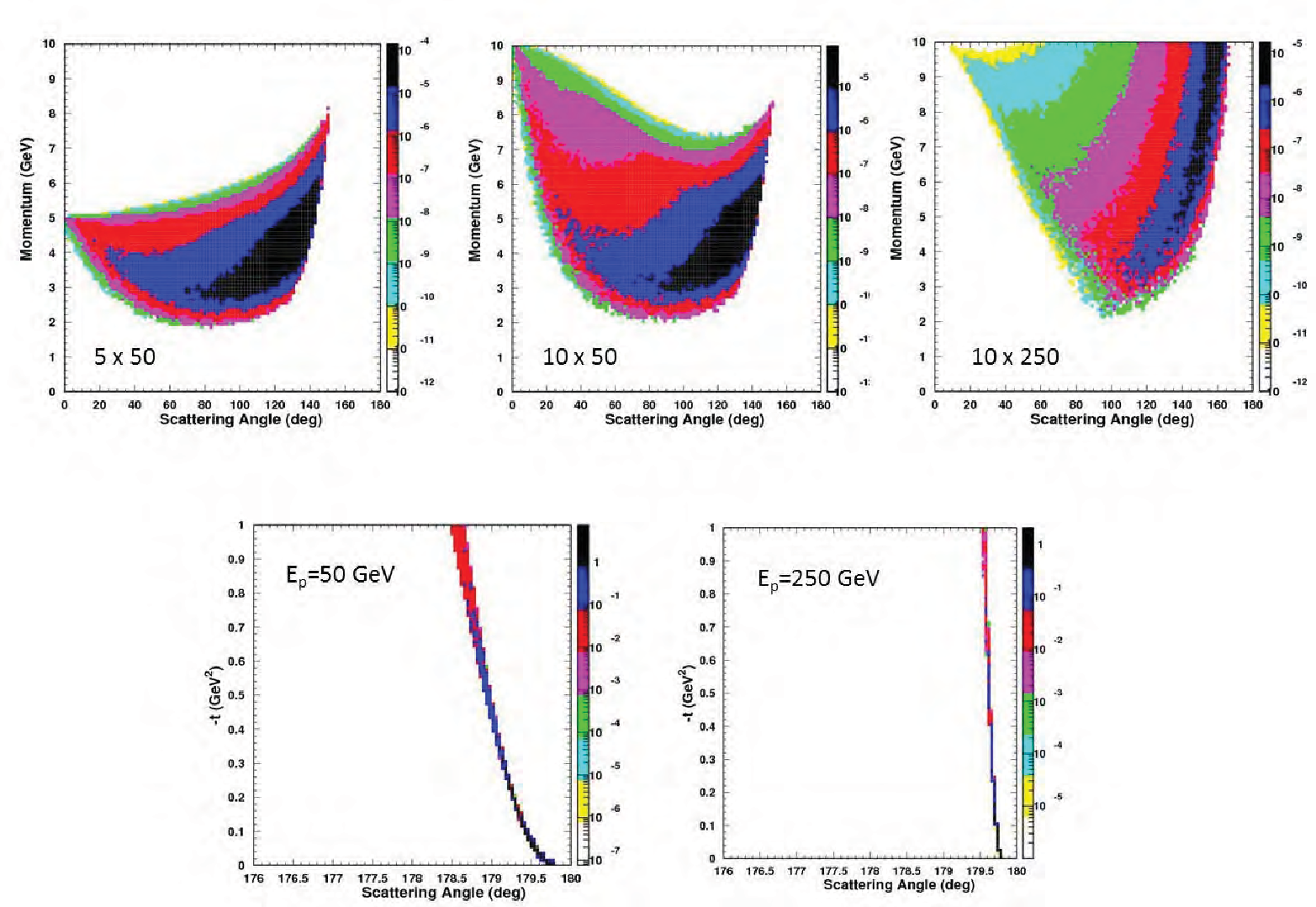}
\end{center}
\caption{The momentum distribution of the exclusive hadronic final state as a function of the scattering angle 
for three different center of mass energies, $\sqrt{s}$=31.6, 44.7, 100 GeV (upper three panels), and the $t$ 
distribution as a function of scattering angle of the recoiling baryon in exclusive reactions for proton beam 
energies $E_p$=50 GeV and 250 GeV (lower two panels). A cut of $Q^2>$ 10 GeV$^2$ is applied to select the 
kinematic range of interest for exclusive processes. For lower center of mass 
energies, the momentum distribution tends towards more central scattering angles and covers lower momenta. 
The angle of the recoiling hadronic system is directly and inversely correlated with the proton energy. It thus decreases with increasing proton energy. For instance, as shown here, the 
baryon scattering angle ranges to about 1-2$^\circ$ at a proton energy of 50 GeV and
is reduced to one fifth of that as the proton energy increases to 250 GeV.}
\label{fig:meson_baryon_kinematics} 
\end{figure}

To further illustrate the importance of a mid-range energy for detailed imaging studies through exclusive reactions,
we come back to the kinematics associated with these processes, but for a cut in $Q^2 >$ 10 GeV$^2$, a likely must
for the valid partonic interpretation of such studies. If one implies a $Q^2 >$ 10 GeV$^2$ cutoff in such exclusive
processes, the kinematic patterns of earlier fig. \ref{fig:exclhadron} drastically change. The upper panels of
fig.~\ref{fig:meson_baryon_kinematics} shows how the momentum distribution of mesons associated with exclusive
pseudoscalar meson production change with lepton and proton energy. Compared to fig. \ref{fig:exclhadron}, the peak in
the forward-ion direction has disappeared completely. Lower lepton energies also push towards lower hadron momenta
in the central-angle region, and thus reduced particle identification requirements. The bottom panels of
fig.~\ref{fig:meson_baryon_kinematics} show one of the most challenging constraints on the detector and interaction
region design for exclusive reactions from the need for detection of the exclusive hadronic state remaining in the
exclusive process. The figures show the direct correlation between $t$ and proton energy, scaling like 1/$E_p$,
and shows the remaining baryonic state goes very much in the forward-ion direction, but far less so (and with lower
momenta) for lower proton energies, which are thus much easier to peel off from any beam-stay-clear area.
Even more, assuming a fixed resolution in $t$, there are obvious benefits of lower proton energies for imaging.
Of course, any high-energy ELIC would in turn greatly benefit from the experience gained from the construction and
operation of the MEIC.

While maintaining a future upgrade path to the high-energy ELIC is important and always folded into the MEIC design,
emphasis has been placed on ensuring that ELIC will not simply supersede the MEIC, but rather provide a complementary
capability. The MEIC is thus designed to excel in the kinematic range that it will cover (i.e., on one hand having an
overlap with JLab 12 GeV, and on the other with HERA data with $y <$ 0.3). Overlap in science goals is in part achieved
by various accelerator features. Perhaps one of the most prominent is the figure-8 shape, which could allow storage of
polarized deuterium beams. By tagging the spectator proton in the small-angle ion spectrometer (discussed below), this
will allow to carry out measurements on quasi-free (polarized) neutrons. A high luminosity over a broad kinematic range
will make it possible to accumulate sufficient statistics for multiple beam energy settings. The capability
to vary the beam energies is essential for some measurements (e.g., $F_L$), but also makes it possible to optimize the
data taking by reducing reliance on data taken at extreme values of $y$, where the systematic uncertainties grow.
This can be achieved by having a lepton beam energy that can be varied continuously, and a series of closely 
spaced discrete ion beam energies. In the MEIC, the latter can be accomplished by changing the number of 
stored ion bunches by one, and the bunch separation distance accordingly - a scheme facilitated by the high bunch 
repetition frequency. Independently varying beam energies also makes it possible to choose the most suitable 
lab kinematics at a certain value of $s$, potentially improving acceptance, resolution, and particle 
identification for the reaction of interest (see also fig.~\ref{fig:meson_baryon_kinematics}).

Having small, short ion bunches with a high bunch repetition frequency also facilitates the use of SRF crab crossing
cavities, which were originally developed for KEKB to allow beams collide at an angle without significant loss of
luminosity.  In the context of an EIC, these were pioneered in the ELIC design, and the possibility of creating a
significant crossing angle (at least 50 mrad) became early on a key feature of the small-angle detection for the MEIC
(see section~\ref{meic-small-angle}).

\subsubsection{Detector Placement and Backgrounds}

The figure-8 ring can support two IPs per straight section, one of which will be a ``high-luminosity'' IP 
with the full crossing angle. In order to minimize backgrounds, the two high-luminosity IPs will be 
located close to where the ion beam exits the arc, and far away from the arc where the lepton beam exits. 
The latter helps to decrease synchrotron radiation (and the secondary neutron flux) at the IP, which is 
anyway already reduced due to the use of crab crossing (with the ion beam, not the electron beam, making 
the horizontal bend correction). The synchrotron background is reduced even further
by lowering the strength of the last arc dipoles. The short distance between the
ion arc and IP suppresses detector backgrounds from interactions of the beam with residual gas in the
beam pipe by providing a smaller ``target'' with line-of-sight to the detector. A shorter section of the 
beamline is also easier to bake and keep at at ultra-high vacuum. A comparison with HERA, also taking into 
account the lower $p-p$ (and $p-A$) cross section and lower hadron multiplicity at the 100 GeV, suggests 
that the hadronic background will be about an order of magnitude lower in the MEIC at comparable vacuum and 
ion beam current, leaving a lot of headroom to increase the latter.
Due to the bends associated with the horizontal crossing, the secondary IPs on each straight section will not have a
line-of-sight along the full straight section, but there this is less of an issue since they are intended
to either have diagnostics equipment (\textit{e.g.}, polarimetry), or special detectors which are less sensitive to
backgrounds or intended to operate at lower beam currents. 

\subsubsection{Detector and Interaction Region Layout}

\begin{figure}[hbtp]
\begin{center}
\includegraphics[width=0.90\textwidth]{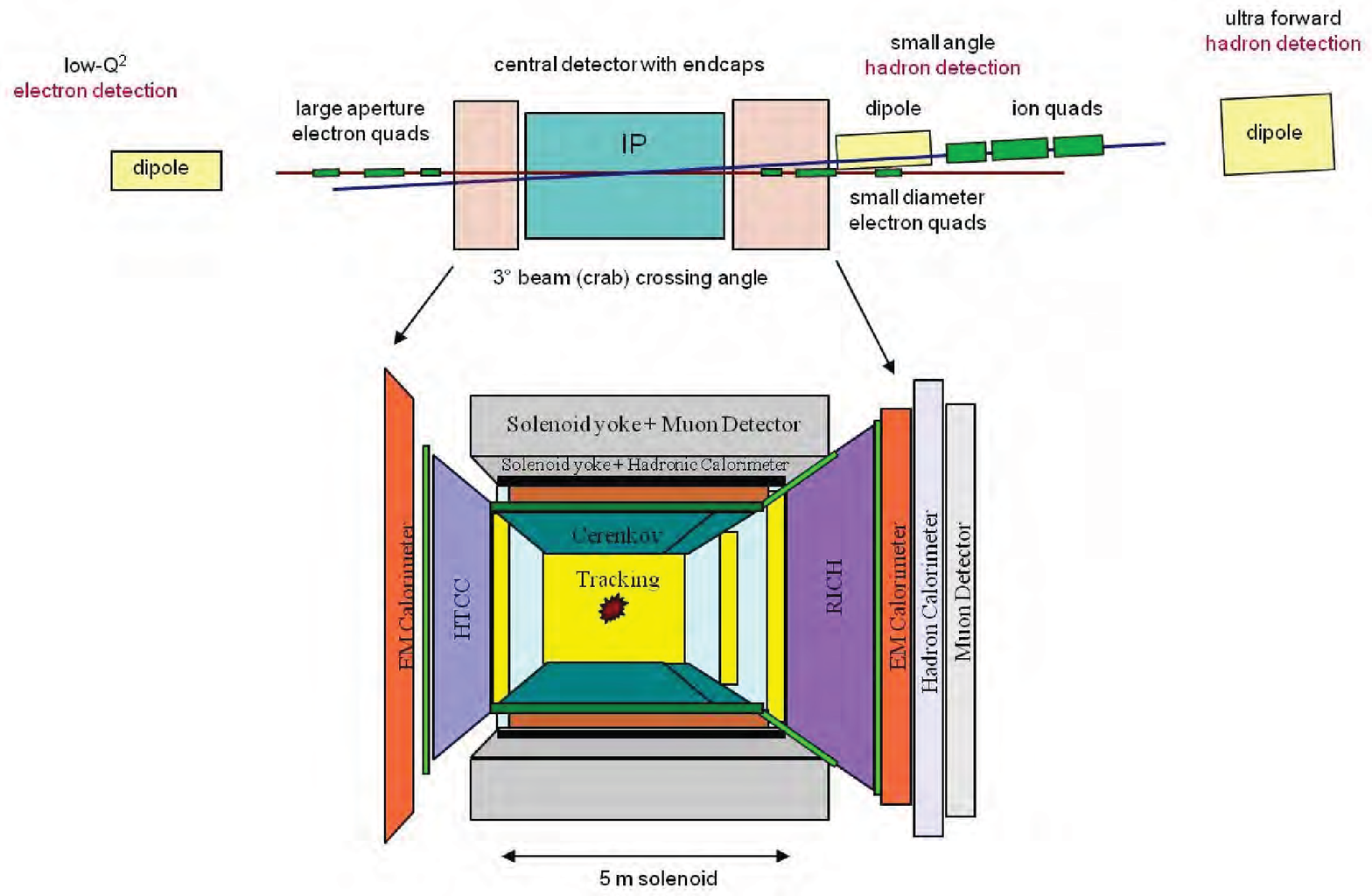}
\end{center}
\caption{Interaction region and central detector layout, and its placement in the general integrated detector and
interaction region. The central detector includes endcaps in both the electron and ion direction.}
\label{fig:ir_and_central_detector_layout} 
\end{figure}

A global outline of the fully integrated MEIC detector and interaction region is given in 
fig.~\ref{fig:ir_and_central_detector_layout}. We will in the subsequent subsections go in more detail over
the central detector region, defined as the region of the detectors operating within the solenoid, the
electron and ion endcaps, and the strategy to accomplish a full-acceptance detector. The latter has two
ingredients, a relatively simple approach to incorporate low-$Q^2$ electron detection and a more challenging
solution to measure forward and ultra-forward (in the ion direction) going hadronic or nuclear fragments.
Here, we make critical use of various ingredients of the MEIC detector/interaction region design:
i) the 50 mrad crossing angle; ii) the range of proton energies;
iii) a small 1-2 Tm dipole field to allow measurement down to 0.5$^\circ$ before the ion final focusing magnets; 
iv) ion final focusing magnets with apertures sufficient for particles with angles up to at least 0.5$^\circ$; and
v) a large 20 Tm dipole field much more downstream to peel off spectator particle and allow for very small-angle 
detection.

The strategy will be that various detector elements, amongst which zero degree calorimeters for neutron detection
and various small-angle detectors, will be placed in the region between the ion final focusing quads and the
20 Tm dipole field, and also beyond this 20 Tm dipole field. This then results in an essentially 100\% full
acceptance detector. The electron beam traverses the center region of the solenoid, while the proton/ion beam
traverses at the crab crossing angle. This choice minimizes any electron steering and synchrotron radiation.
Note that the 50 mr crab crossing angle also facilitates the small-diameter electron final focusing quads to
be moved in to 3.5 meter distance of the interaction point. The lower electron beam energies and hence lower-field
requirements for the electron beam allows the construction of relatively small-sized quadrupoles, much simplifying
the electron optics design.

\subsubsection{Central Detector}

To fulfill the requirement of hermeticity, the central detector will be built around a solenoid magnet (with a 
length of about 5 m). Due to the asymmetric beam energies, the interaction point (IP) will be slightly offset 
towards the electron side (2 m + 3 m). This will allow more distance for the tracking of high-momentum hadrons 
produced at small angles, and a larger bore angle for efficient detection of the scattered beam leptons. 

\begin{figure}[hbtp]
\begin{center}
\includegraphics[width=0.95\textwidth]{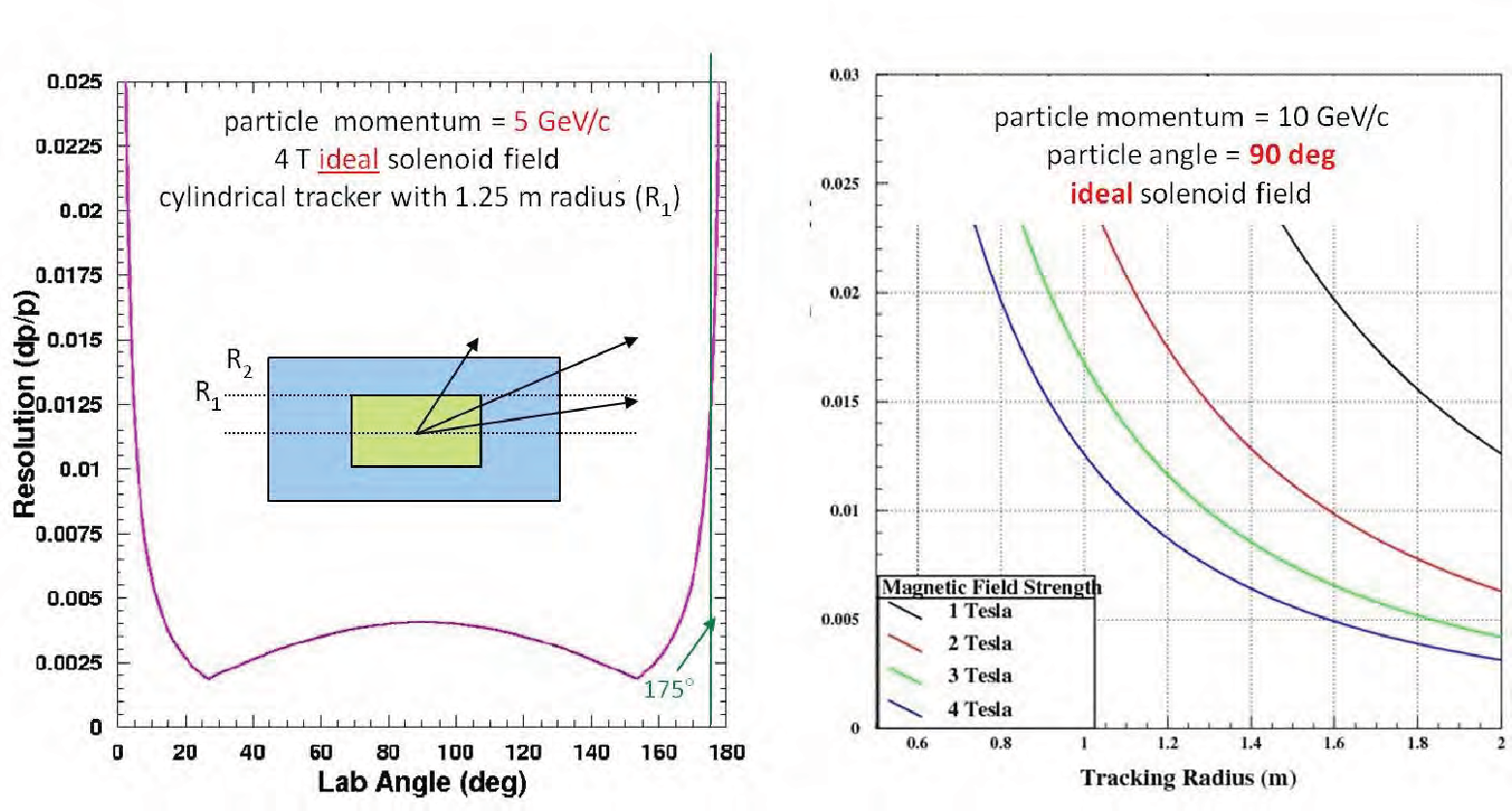}
\end{center}
\caption{(left) The resolution as a function of lab angle for a particle (pion) momentum of 5 GeV/c in a 4 T ideal
solenoidal field and with a cylindrical tracker of radius 1.25 m; (right) The resolution as function of solenoidal
field strength and tracker radius for a particle (pion) momentum of 10 GeV/c and a scattering angle of 90$^\circ$.}
\label{fig:resolution_in_solenoid} 
\end{figure}

The characteristics of the solenoid are guided by the desire to optimize the tracking resolution, 
which at central angles scales like $\Delta p/p \sim \sigma p / BR^2$, where $\sigma$ is the position resolution, 
$p$ the particle momentum, $B$ the magnetic field, and $R$ the radius of the central tracker. At forward angles, 
however, the resolution depends on the scattering angle, but is independent of $R$ as the particle leaves the 
cylindrical central tracking system from the front side (see the left panel of fig.~\ref{fig:resolution_in_solenoid}). 
The resolution will then deteriorate rapidly given the lack of transverse field along the central axis of a solenoid. 
This will later be remedied by adding a small
dipole field, as high ion energies boost the outgoing hadrons to high momenta at forward angles and one wishes to
optimize resolutions also in the forward-ion direction. To obtain a roughly better than 1$\%$ momentum resolution for
central angles and particles in the 5-10 GeV/c momentum range, a field $B$ in the 2-4 T range seems highly desirable. 
This high field requirement suggests a magnet with a reasonably small diameter, preferably not larger than about 4 m,
putting radial space at a premium. Of course, a smaller diameter has the advantage of simplifying the magnet design,
with the additional advantage of reducing detector cost (which scale with the
radius for the barrel calorimeter and roughly as the radius squared for the endcaps). An alternate solution may be
to increase the space for tracking in the central solenoid while reducing the required solenoid field, as illustrated
in the right panel of fig.~\ref{fig:resolution_in_solenoid}. Here, the resolution improvement for pions with 10 GeV/c
momentum and a scattering angle of 90$^\circ$ is shown as a function of the tracking length and solenoidal field.
Thus, there is strong incentive to reduce the space requirements for particle identification detectors within the
central solenoid as much as possible, to use available space for tracking, or reduce the solenoid diameter.

The central detector would contain a tracker, particle identification, and calorimetry.
A three layer configuration of the central tracker was suggested at the JLab EIC detector workshop
(June 4-5, 2010)\footnote{http://conferences.jlab.org/eic2010}. The first layer would consist of a low-mass 
vertex tracker with sufficient resolution to separate primary and secondary vertices in charm production. 
The middle layer would be a Time-Projection Chamber (TPC) with GEM-based readout, and the outer layer would be 
a cylindrical GEM tracker.
The position resolution of the TPC would be about 50 $\mu m$, which is a factor two improvement over the inner drift
chambers of CLAS12. In conjunction with the outer GEM layer, it should provide adequate $(r, \theta, \phi)$ information.
Ongoing R\&D for vertex and micropattern detectors (including GEMs), suggest that such a high-performance tracker could
be built for the EIC detector. Nevertheless, a radius of at least 1 m would be required.

Particle identification in the central detector is the most open design question. At low momenta, dE/dx (in the TPC)
or TOF can be helpful. With precise timing, the momentum range of the latter could be extended somewhat (although this
would require a comparable uncertainty on the track length determination in order to get a good $t_0$). 
The most challenging requirement is, however, for a radially compact detector providing $\pi/K$ identification over 
a sufficiently wide momentum range. Taking up 8 cm of radial space, a BaBar-type DIRC could satisfy this condition, 
providing $3\sigma$ $\pi/K$ separation up to 4 GeV/c, $e/\pi$ separation close to 1 GeV/c, and p/K separation up 
to 7 GeV/c.
An aerogel barrel RICH could provide almost comparable performance. Neither is sufficient for the exclusive (GPD) or
semi-inclusive (TMD) programs. The current baseline design thus includes a Low-Threshold Cherenkov Counter (LTCC) with
$C_4F_{10}$ or $C_4F_8O$ gas in addition to the DIRC. This would provide $e/\pi$ separation between 1 and 3 GeV/c,
and $\pi/K$ separation from 4 to 9 GeV/c, but at a price of 50-70 cm of radial space.
Adding $C_4F_{10}$ to a barrel RICH would increase the radius by at least 80-90 cm, although a RICH could extend the
momentum coverage to 14 GeV/c. Ultimately the allocation of radial space to PID and tracking is a matter of priorities,
and with multiple detectors one could easily imagine that these would offer complementary capabilities.
On the other hand, if one could improve the $\theta_c$ resolution for a DIRC by about a factor of two,
its $3\sigma$ $\pi/K$ separation could be extended to about 6 GeV/c, with the upper limits for the other particle 
species shifting accordingly, eliminating the need for the gas Cherenkov.  Given the size of the EIC detector, 
an all-crystal electromagnetic calorimeter would be financially expensive and only needed in critical regions.
Tungsten powder / scintillating fiber or other technologies may provide a more affordable alternative for the 
barrel without an excessive loss of resolution. If needed, the return yoke of the solenoid magnet can be used 
as part of a hadronic calorimeter, and as an absorber for muon detection (along the lines of CMS).

\subsubsection{Detector Endcaps}
\label{endcaps}

The electron side endcap would face requirements quite similar to those of CLAS12, and it is natural to adopt a similar
design. Due to the offset of the IP, lower particle momenta, and simpler small-angle detection
(see section~\ref{meic-small-angle}), the electron side is not nearly as crowded as the ion one.
For lepton detection at small polar angles ($\theta$), the main priority of the tracking would be to provide good
$\theta$ resolution, as this directly impacts the reconstruction of the event kinematics. The inner part of the endcap
tracker should thus be an extension of the vertex tracker, using semiconductor detectors. At larger angles, the requirements
are not as demanding and the choice of technology is not as crucial. It could include planar GEMs or even cheaper
drift chambers with a small cell size. Given the generous space constraints, a final tracking region could be added
outside of the solenoid itself to improve tracking performance.
Lepton identification will also use an electromagnetic calorimeter and a High-Threshold Cherenkov Counter (HTCC) with
$CF_4$ gas or equivalent. The light can be collected by mirrors, producing a cost-effective readout.
In this endcap region, hadron identification will be partially provided
by a TOF detector, for which the endcap is more suitable than the barrel due to the longer flight path. The $\pi/K$
identification range, again in the electron endcap region, could be extended through the use of a Low-Theshold Cherenkov
Counter (LTCC) with $C_4 F_{8}O$ gas or equivalent, possibly operating slightly above atmospheric pressure to lower the
pion detection threshold. Of course, to push $\pi/K$
identification to larger momenta, $\sim$ 10 GeV/c, a RICH detector may need to be considered, but there does not seem to
be a compelling need in this electron endcap region for the MEIC.
Given the space available on the electron side, there is no strong requirement for a compact electromagnetic calorimeter.
Since the momentum resolution from tracking deteriorates at small angles, where also the rates go up, the ideal
configuration would involve an inner circle of high-resolution, radiation-hard crystals, and a more budget-friendly
outer part. Both could be covered by the same pre-shower calorimeter.

The ion side endcap would have to deal with hadrons with a wide range of momenta, some approaching that of the ion beam.
The forward tracking would thus greatly benefit from good position resolution ({\sl e.g.}, planar GEMs),
at least on par with
the 50 $\mu m$ of the TPC. The smallest angles can be covered by semiconductor detectors as on the electron side.
Of course, a good position resolution will also put significant demands on the detector alignment and field knowledge.
The most important feature of the forward tracker, however, is related to the ion beam crossing angle with respect to the
electron beam. In addition to being a key component of the small-angle detection, this turns the tracking resolution
into a 2D problem. Whereas the momentum resolution in a solenoidal field deteriorates rapidly at small angles
with respect to the axis, the hadron scattering angle is essentially defined with respect to the ion beam line.
Given that the proton/ion beam traverses the solenoid at a 50 mr (crab crossing) angle, so already encounters some
transverse magnetic field component, hadrons scattered away from the electron beam
will end up in a part of the detector with better momentum resolution than those scattered towards the electron beam.
Taking the 2D character of the problem into account, and the significant 50 mr beam crossing angle, the spot of poor
resolution will be moved into the periphery covering and only a small range in the azimuthal angle $\phi$ will be
affected. For most processes, all particle tracks will remain in the zone of good resolution.
In contrast, if the crossing angle is small, all particle tracks at very forward angles will suffer
from poor momentum resolution, as shown in the right panel of fig.~\ref{fig:resolution_in_solenoid}.

To identify particles of various species over the full momentum range, one would ideally want to use several radiators.
A typical combination could include aerogel (perhaps with more than one index of refraction), $C_4F_{10}$ or equivalent gas,
and $CF_4$. This would make some kind of RICH detector an attractive option, in particular if the endcap radius was not
too large. Still, there are several possible approaches which eventually will need to be studied in detail.
One could, for instance, imagine a dual radiator gas RICH combined with a disk DIRC (as in PANDA), with the latter
providing $\pi/K$ identification up to about 4 GeV/c. Having the longest flight path from the IP, the ion endcap is also
where one could achieve the best results with high-resolution TOF (perhaps even integrated with the readout of the RICH).
Regardless of technical solution, the total thickness of the stack of PID detectors is assumed not to exceed 1.5 m.
Calorimetry in the ion endcap will include both electromagnetic and hadronic parts. The main focus of the former
will be to study various reaction products rather than the scattered lepton. However, the same resolution arguments
apply as for the electron endcap, and a solution with an inner high-resolution circle, and a more cost-effective
outer part makes sense here as well. The magnetic enclosure of the endcap can, as in the case of the return yoke
of the central detector, be integrated with a hadronic calorimeter, and serve as an absorber for muon detection.

\subsubsection{Small-angle Detection}
\label{meic-small-angle}

The design for the full-acceptance detector envisions small-angle detection on both sides of the central detector.
The naming convention used here will be that the ``ion side'' or ``ion endcap'' refers to the side of the outgoing ion
and incoming electron beam. The ``electron side" refers to the other one.

\begin{figure}[hbtp]
\begin{center}
\includegraphics[width=0.95\textwidth]{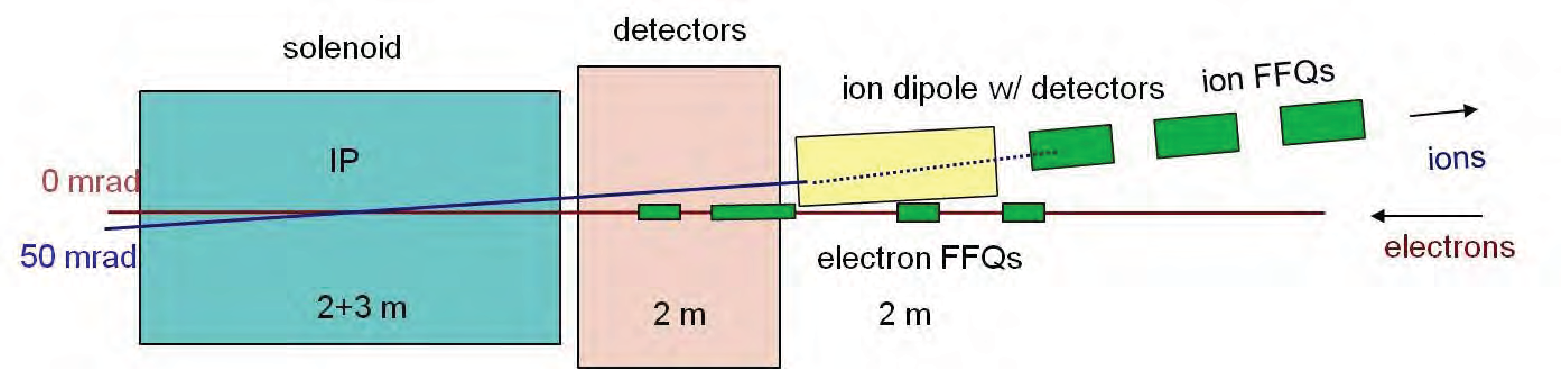}
\end{center}
\caption{Forward ion detection with 50 mrad crossing angle for the full-acceptance detector. Note that the distance
to the final focusing quadrupoles are located 7 m from the IP.}
\label{fig:forward_ion_detection} 
\end{figure}

On the ion side, the detection will be performed in three stages as illustrated in 
fig.~\ref{fig:forward_ion_detection}. The first stage is the endcap (discussed in section~\ref{endcaps}), which 
will cover all angles down to the acceptance of the forward spectrometer. This in turn has two stages, one 
upstream of the ion Final Focus Quadrupoles (FFQs), covering down to 0.5$^\circ$, and one downstream covering 
up to at least 0.5$^\circ$. 
The former will use a 1-2 Tm dipole to augment the solenoid in the range where the resolution is poor.
The magnet will be about 1 m long and cover the distance to the electron beam (corresponding to the horizontal crossing
angle of 50 mrad), and about twice that in the other directions, for a total acceptance of 150 mrad in the horizontal
and 200 mrad in the vertical plane. An important feature of the magnet design is to ensure that the electron beam line
stays field free. The dipole will have trackers at the entrance and exit, and a calorimeter covering the ring-shaped 
area in front of the first ion FFQ. For neutrons, the primary goal of this calorimeter is to have good angular 
resolution.
This intermediate stage is essential for providing good coverage and resolution in $-t$, and to investigate target
fragmentation. The former is of particular importance for the study of exclusive processes, essential for
the 3D imaging of the nucleon, requiring detection of the recoil baryon. Since $t \sim \theta^2_p E^2_p$,
the $t$-resolution depends on the angular resolution that can be achieved. 
With a 50 GeV proton beam, a $-t$ of 1 GeV$^2$ corresponds to about 27 mrad 
(see fig.~\ref{fig:meson_baryon_kinematics}).
With an angular resolution of 1 mrad, the intermediate detection stage would be able to
cover $-t$ up to 2 GeV$^2$ with a resolution of about 40-50 MeV$^2$ a value that would scale with angular 
resolution of the inner silicon forward tracker. Recoil baryons with larger values of $-t$ 
would be detected in the endcap. At higher ion beam energies the $t$-acceptance of the dipole increases, 
but the resolution deteriorates rapidly (due to the $E^2_p$ factor). Going to lower ion energies, the opposite is true.

\begin{figure}[hbtp]
\begin{center}
\includegraphics[width=0.95\textwidth]{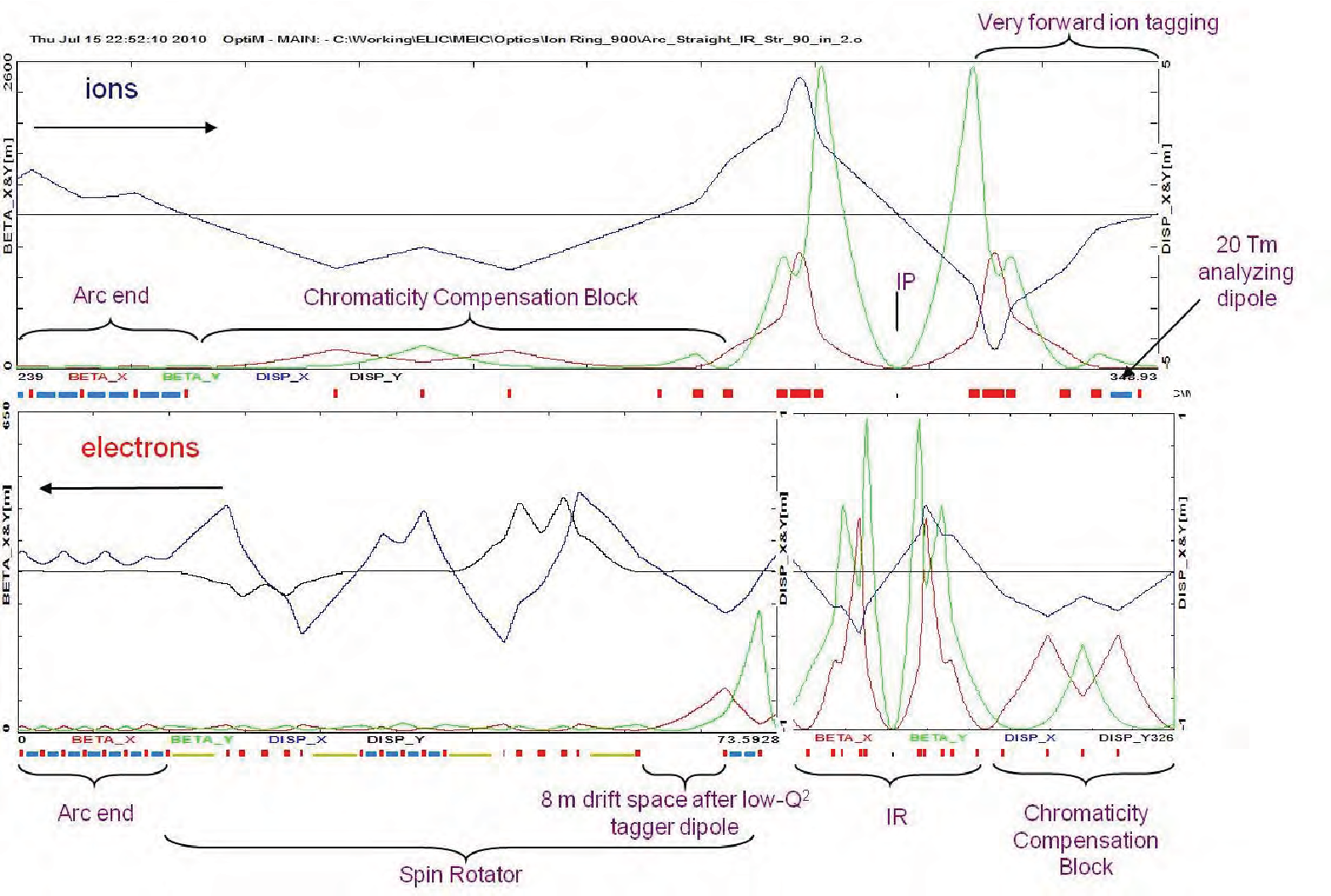}
\end{center}
\caption{The integration of particle detection in the accelerator.}
\label{fig:interaction_region_forward_tagging} 
\end{figure}

The last stage is the ultra-forward detection that is crucial for the tagging of spectator protons in deuterium,
as well as other recoil baryons/nuclei. The design is heavily integrated with the accelerator
(see fig.~\ref{fig:interaction_region_forward_tagging}), using two key features. One is, again, the horizontal 
crossing angle for the ion beam, which needs to be ``corrected'' some distance downstream of the interaction 
point (IP).
For a 50 mrad crossing angle, this corresponds to a bend of close to 100 mrad, and the required 20 Tm dipole(s) can also
serve as a dedicated forward spectrometer, using the long drift space beyond for detection of both charged and neutral
particles. The other feature is a beam optics requiring low quadrupole gradients, allowing large aperture magnets.
In the current design, the maximum quad gradient is less than 65 T/m. With a 10 cm aperture, this creates a 6.5 T 
peak field (simply the product of the aperture radius and gradient), which the magnet design should be able to support
if larger peak fields were acceptable, the apertures would increase accordingly. 
The gradients are further arranged so that they drop off faster than the distance from the IP to that specific 
location, allowing the apertures to become correspondingly larger, and thereby making sure that no bottlenecks 
are created. This defines a geometrical acceptance through the ion final-focusing quads (FFQs) of 10 mrad, 
or well beyond 0.5$^\circ$, on each side of the beam (20 mrad in total). 
To focus the 250 GeV beams in ELIC, the maximum quadrupole gradients would have to be 2.5
times larger than for the 100 GeV of the MEIC, and the apertures reduced accordingly.

The acceptance for charged particles depends on both the polar and azimuthal angles (since quads
focus in one plane and defocus in the other), as well as their momentum. This can be optimized by
placing a dipole spectrometer relatively close to the FFQs. To give a numerical example, a 100 mrad bend
to a deuterium beam would equate to a 200 mrad bend for a spectator proton. Over a drift space of 10 m
(a relatively modest distance), the spectator proton would acquire a transverse separation of 1 m
from the main beam. For heavy nuclei (A/Z = 2.5) with a negligible scattering angle at close to the
beam momentum, this would increase to 1.5 m, while fragments with other A/Z ratios would be lined up
in between (in particular, N=Z would be at 25 cm, while neutron rich fragments would be deflected to
the other side). Ions scattered at zero degrees and having 98\% of the beam momentum would be 2 cm
from the beam after 10 m of drift. Due to the large deflection in a well known field (including the
few preceding elements), the momentum resolution of the spectrometer would be excellent.
Since no position measurements would be possible within the beam-stay-clear area, the angular resolution
would depend on the knowledge of the optics between the IP and detection point. The reconstruction
of the angle would be aided by the scattered particles quickly exiting the beam-stay-clear area
after having passed the spectrometer, with multi-point tracking to be applied in the drift region.
Nevertheless, some some low-momentum particles scattered at large angles will not make it all the way
to the dipole spectrometer. To detect these particles, some \textit{ad hoc} detectors (``Roman pots'')
may be placed along the way, although an interesting idea currently under investigation is to have
a small-diameter compensating solenoid between the FFQs and the 20 Tm dipole. In addition to its
benefits for the accelerator, such a magnet could help in tracking charged particles that do not reach
the final spectrometer dipole.

The low-$Q^2$ tagger on the electron side will complement the electron detection in the central detector and 
electron side endcap. Since the electron quad gradients required for 11 GeV beams are very small compared with what 
is needed for 100 GeV protons, one can make the apertures very large without being constrained by peak fields 
(the different apertures on the incoming and outgoing sides do not affect the optics). The optimal transition 
point from the calorimeter to tagger coverage will ultimately be determined by physics simulations. The quads 
would be followed by a dipole spectrometer with sufficient drift space (8 m in the current layout) 
to detect leptons with a significant fraction of the beam energy.

\subsubsection{Beam Helicity Reversals}
\label{meic-polarization}

The electron and ion beam polarimetry has been given a special ``interaction region'' in the MEIC/ELIC design, in
part due to the often large amount of space needed for Compton polarimetry. With the anticipated work in systematic
understanding of Compton polarimetry in both JLab Halls A and C, and further
plans to cross-calibrate this with atomic beam Moller polarimetry for a future demanding parity-violating
Moller experiment, electron beam polarization determination through Compton polarimetry may well achieve
sub-0.5\% uncertainties. Ion
beam polarimetry remains more complicated, although efforts to reduce uncertainties are underway and
possibilities are studied in elastic and inelastic electron-proton scattering experiments in situ.

The MEIC design will need both fast electron spin helicity reversal or flip for double-spin experiments
and a program of deep-inelastic parity-violating experiments, and fast ion-spin flip for single-nucleon
spin asymmetry experiments. The latter can also be an alternate method for double-spin experiments.
The MEIC design, with its 750 MHz bunch trains, does not assume bunch-to-bunch spin flips, but also
does not need it. A helicity-reversal frequency of 0.1 Hz will be at about the level needed for experiments.

For double-spin experiments, it is to first order equivalent to perform fast helicity reversals of electron
or ions. The choice is a question of detailed precision, as shown later. For single-spin asymmetry experiments,
these techniques are however totally different, and can not replace each other. Single-electron spin asymmetry
(flipping electrons only) is mostly useful for parity violation experiments, while single-nucleon spin asymmetry
(flipping ions) is mostly useful for nucleon transverse-spin and other TMD experiments. Both type of experiments
are routinely performed at JLab, and both will become an important part of the EIC science program.

The rate of the required helicity flips is closely related to the systematic understanding of the precision.
Typically, although already very difficult, one can control the systematic uncertainties between two helicity states
to about 1\%. To further reduce this asymmetry, to a level of 10$^{-8}$ for the case of typical
parity-violating experiments at JLab, or to a level of 10$^{-5}$ for transverse-spin experiments, one has to provide
a suppression faction of 10$^{-6}$ (for electron spin flip) or 10$^{-3}$ (for ion spin flip) by fast spin flip techniques.
The suppression factor by such fast helicity reversal is proportional to $1/\sqrt{N}$,
where $N$ is the number of pairs of spin flip. If we assume a typical single-nucleon spin asymmetry experiment
of 3 months of continuous running (assuming one can keep control of the systematic uncertainties between the two
helicity states at the 1\% level for the full period), one needs to accumulate 10$^{6}$ pairs to reach a suppression
of 1000, or about 8 flips per minute. This is the root of the present 0.1 Hz beam helicity reversal assumption
mentioned above.












\chapter*{Acknowledgments}
\addcontentsline{toc}{chapter}{\numberline{}Acknowledgments}


Major support for this report and the INT program were provided by
Brookhaven Science Associates, LLC under U.S.\ DOE Contract No.\
DE-AC02-98CH10886,
by the Institute for Nuclear Theory at the University of Washington under
U.S.\ DOE Contract No.\ DE-FG02-00ER41132,
and by Jefferson Science Associates, LLC under U.S.\ DOE Contract No.\
DE-AC05-06OR23177. \\

The United States Government retains a non-exclusive, paid-up,
irrevocable, world-wide license to publish or reproduce the published form
of this manuscript, or allow others to do so, for United States Government
purposes. \\

We acknowledge additional support from
ANPCyt, CONICET, and UBACyT (Argentina),
CNPq and FAPERGS (Brazil),
FESR of the Universit\'e de Moncton (Canada),
Fondecyt (Chile), Conicyt-DFG (Chile and Germany),
the Ministry of Science and Education (Croatia), 
the European Union, 
the Academy of Finland,
ANR (France), 
BMBF, DAAD, DFG, GSI, and the Helmholtz Association (Germany),
MIUR and RAS (Italy),
the Ministry of Science and Higher Education (Poland),
STFC (UK),
DOE, the Lightner-Sams Foundation, NSF, and the PSC-CUNY Research Award
Program (USA).






\bibliographystyle{h-physrev5}

\small{
\clearpage
\phantomsection
\addcontentsline{toc}{chapter}{\numberline{}References}

\bibliography{master}{}
}


\clearpage
\phantomsection
\addcontentsline{toc}{chapter}{\numberline{}List of Authors}

\printindex

\end{document}